%% file: main.tex
\documentclass[a4paper,11pt]{article}

\input{General/packages}

\begin{document}

\pagenumbering{roman}
\include{prelude}

\include{wg1}
\include{wg2}
\include{wg3}
\include{wg4}

\include{epilogue}
\end{document}

%% file: General/packages.tex
\usepackage{General/lnfprep}
\usepackage{General/aas_macros}

\usepackage[english]{babel}
\usepackage[T1]{fontenc}
\usepackage[utf8]{inputenc}

\usepackage{amsmath}
\usepackage{amssymb}
\usepackage{amsfonts}
\usepackage{mathtools}
\usepackage{bbm}
\usepackage{bbold}
\usepackage{mathrsfs}
\usepackage{slashed}
\usepackage{braket}
\usepackage{upgreek}

\usepackage{soul}
\usepackage{textcomp}
\usepackage{calligra}
\usepackage{csquotes}
\usepackage[normalem]{ulem}
\usepackage{microtype}

\usepackage{graphicx}
\usepackage[export]{adjustbox}
\usepackage{rotating}
\usepackage{subcaption}

\usepackage[labelfont=bf]{caption}
\usepackage{longtable}
\usepackage{booktabs}
\usepackage{makecell}
\usepackage{xltabular}

\usepackage{enumitem}
\usepackage{subfiles}
\usepackage{titlesec}

\usepackage[square,numbers,sort&compress]{natbib}

\usepackage{siunitx}
\usepackage{xcolor}
\usepackage{authblk}
\usepackage{etoolbox}
\usepackage[switch]{lineno}
\usepackage{comment}
\usepackage{blindtext}

\usepackage{tikz}
\usetikzlibrary{decorations.markings}

\usepackage[colorlinks=true,
  linkcolor=blue,
  citecolor=blue,
  urlcolor=blue]{hyperref}
\usepackage{xurl}
\usepackage{url}

\usepackage{fancyhdr}

\newcommand{\ov}[1]{\overline{#1}}

\makeatletter
\newcommand{\raisemath}[1]{\mathpalette{\raisem@th{#1}}}
\newcommand{\raisem@th}[3]{\raisebox{#1}{$#2#3$}}

\makeatother

\definecolor{MH}{rgb}{0.0,0.9,0}

\titleformat{\part}[display]
  {\normalfont\huge\bfseries}
  {\partname~\thepart}
  {20pt}
  {\Huge}

\titleformat{\section}
  {\normalfont\Large\bfseries}
  {\thesection}
  {1em}
  {}

\makeatletter
\pretocmd{\part}{\thispagestyle{plain}}{}{}
\renewcommand*\l@subsection{\@dottedtocline{2}{1.5em}{3em}}
\makeatother

%% file: prelude.tex
\input{General/titlepages}

{\hypersetup{allcolors=black}
\setcounter{tocdepth}{2}
\thispagestyle{plain}
\tableofcontents
\clearpage
}

\pagestyle{fancy} 
\pagenumbering{arabic}

\input{General/executive_summary}

%% file: General/titlepages.tex
\begin{titlepage}
\begin{table*}[ht]
\vspace{-5em}
\centering

\begin{minipage}[b]{0.3\linewidth}
    \centering
    \includegraphics[height=7em]{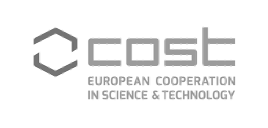}
\end{minipage}
\hfill
\begin{minipage}[b]{0.3\linewidth}
    \centering
    \includegraphics[height=7em]{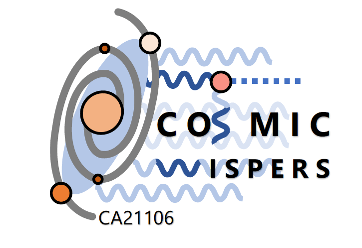}
\end{minipage}
\hfill
\begin{minipage}[b]{0.3\linewidth}
    \centering
    \includegraphics[height=7em]{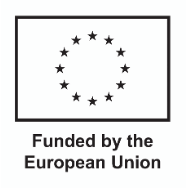}
\end{minipage}
\rule{\linewidth}{0.4pt}

\end{table*}

{\bf\boldmath\LARGE
  \begin{center}
  The COSMIC WISPers   White Paper: \\ 
    The physics case for \\ Weakly Interacting Slim Particles
\end{center}
}

\vspace*{0.5cm}

\begin{abstract}
Axions and other very weakly interacting slim particles (WISPs), with masses below 1 GeV, arise naturally in many extensions of the Standard Model of particle physics.  In particular, they could offer a new framework to explain the nature of dark matter and may help address a range of puzzling observations in astrophysics and particle physics.
This review provides an overview of ongoing WISP searches and outlines the prospects for the next decade, spanning their theoretical motivation, indirect signatures in astrophysical observations, and dedicated laboratory experiments.
It is based on the work carried on by the EU-funded COST Action \emph{“Cosmic WISPers in the Dark Universe: Theory, astrophysics, and experiments”} (CA21106, \url{https://www.cost.eu/actions/CA21106}). This network  plays a key role in coordinating and supporting WISP searches across Europe, while also contributing to the development of a roadmap aimed at securing European leadership in this research area. 
It is  emphasized that Europe is currently pursuing a rich, diverse, and cost-effective experimental program, with the potential to deliver one or more transformative discoveries. 
\end{abstract}

\vspace*{0.5cm}

\vfill
\noindent
Report Numbers: BARI-TH/784-26, CERN-TH-2026-016, hal-05395643, IPPP/26/13, IFT-UAM/CSIC-26-13, KCL-PH-TH/2026-04, KEK-Cosmo-0411, KEK-TH-2804, LAPTH-008/26, MPP-2026-21, RESCEU-5/26, SLAC-PUB-260219, ST/T006994/1, ST/Y004531/1

\clearpage
\thispagestyle{empty}

\newcommand{\KCL}{4}
\newcommand{\TUM}{6}
\newcommand{\IMP}{7}
\newcommand{\CERN}{8}
\newcommand{\UniMainz}{11}
\newcommand{\HelmholtzMainz}{12}
\newcommand{\GSIDarm}{13}
\newcommand{\UCBerkeley}{14}
\newcommand{\GRAPPA}{16}
\newcommand{\LAPTH}{17}
\newcommand{\TUD}{18}
\newcommand{\Zaragoza}{19}
\newcommand{\CAPA}{20}
\newcommand{\Istinye}{23}
\newcommand{\Durham}{24}
\newcommand{\Unibo}{26}
\newcommand{\INFNBO}{27}
\newcommand{\Oxford}{28}
\newcommand{\Madrid}{33}
\newcommand{\UAMCSIC}{34}
\newcommand{\IACTenerife}{35}
\newcommand{\UniTenerife}{36}
\newcommand{\Unipd}{38}
\newcommand{\INFNPD}{39}
\newcommand{\Cartagena}{40} 
\newcommand{\IFPU}{43}
\newcommand{\MPIMun}{44}
\newcommand{\Han}{46}
\newcommand{\INFNRO}{48}
\newcommand{\WPITokio}{49}
\newcommand{\IFIC}{57}
\newcommand{\IFAE}{58}
\newcommand{\LNF}{59}
\newcommand{\DESY}{61}
\newcommand{\MBI}{65}
\newcommand{\Hamburg}{67}
\newcommand{\Uniba}{71}
\newcommand{\INFNBA}{72}
\newcommand{\Weizmann}{90}
\newcommand{\Unito}{91}
\newcommand{\INFNTO}{92}
\newcommand{\Bonn}{101}
\newcommand{\INFNSA}{109}
\vspace*{1.0cm}

\vspace*{0.5cm}
\begin{center}

\textbf{White Paper Authors}

\vskip 2mm

A.~Arza$^{1,2}$,
{\bf D.~Aybas}$^{3,a}$,
S.~Balaji$^{\KCL 
}$,
R.~Balkin$^{5}$,
K.~Bartnick$^{\TUM
}$,
C.~F.~A.~Baynham$^{\IMP 
}$,
I.~M.~Bloch$^{\CERN
,9}$,
C.~Bonati$^{10}$,
D.~Budker$^{\UniMainz
,\HelmholtzMainz
,\GSIDarm
,\UCBerkeley
}$,
C.~Burrage$^{15}$
% ...
\end{center}

\vskip 1cm
    {\footnotesize
      \noindent
$^{*}$ Corresponding authors: Arturo de Giorgi (\url{arturo.de-giorgi@durham.ac.uk}), Maurizio Giannotti (\url{mgiannotti@unizar.es}), Mathieu Kaltschmidt (\url{mkaltschmidt@unizar.es}), Giuseppe Lucente (\url{lucenteg@slac.stanford.edu}), Alessandro Mirizzi (\url{alessandro.mirizzi@ba.infn.it}). \\
$^{a}$ Members of the Editorial Team (names in bold) \\
}

\clearpage

\pagestyle{empty} 
\bigskip

{\it\footnotesize
\begin{description}[style=multiline,
                    leftmargin=!,     
                    labelwidth=1em,  
                    labelsep=0.5em,
                    itemsep=0em,
                    align=left]
  \item[$^{1}$]   Department of Physics and Institute of Theoretical Physics, Nanjing Normal University, Nanjing 210023, China
\item[$^{2}$] Nanjing Key Laboratory of Particle Physics and Astrophysics, Nanjing, 210023, China
  \item[$^{3}$]  Department of Physics, Bilkent University, Ankara 06800, Türkiye 
  \item[$^{4}$] Physics Department, King’s College London, Strand, London, WC2R 2LS, United Kingdom
  \item[$^{5}$] Department of Physics, University of California Santa Cruz and Santa Cruz Institute for Particle Physics, 1156 High St., Santa Cruz, CA 95064, USA
  \item[$^{6}$] Technische Universität München, Physik‑Department, James‑Franck‑Strasse 1, 85748 Garching, Germany
  \item[$^{7}$] Department of Physics, Imperial College London, Blackett Laboratory, Prince Consort Road, London, SW7 2AZ, United Kingdom
  \item[$^{8}$] Theoretical Physics Department, CERN, 1 Esplanade des Particules, CH‑1211 Geneva 23, Switzerland
  \item[$^{9}$] Physics Department, Technion – Israel Institute of Technology, Haifa 3200003, Israel
  \item[$^{10}$] Dipartimento di Fisica dell’Università di Pisa and INFN Sezione di Pisa, Largo Pontecorvo 3, I‑56127 Pisa, Italy
  \item[$^{11}$] Johannes Gutenberg‑Universität Mainz, 55122 Mainz, Germany
  \item[$^{12}$] Helmholtz Institute Mainz, 55099 Mainz, Germany
  \item[$^{13}$] GSI Helmholtzzentrum für Schwerionenforschung GmbH, 64291 Darmstadt, Germany
  \item[$^{14}$] Department of Physics, University of California, Berkeley, CA 94720‑7300, United States of America
  \item[$^{15}$] School of Physics and Astronomy, University of Nottingham, University Park, Nottingham NG7 2RD, UK
  \item[$^{16}$] GRAPPA Institute, Institute for Theoretical Physics Amsterdam, University of Amsterdam, Science Park 904, 1098 XH Amsterdam, The Netherlands
  \item[$^{17}$] Laboratoire d’Annecy de Physique Théorique, CNRS, 9 Chemin de Bellevue, 74940, Annecy, France
  \item[$^{18}$] Fakultät für Physik, Technische Universität Dortmund, Otto‑Hahn‑Straße 4a, 44221 Dortmund, Germany
  \item[$^{19}$] Departamento de Física Teórica, Universidad de Zaragoza, C. Pedro Cerbuna 12, 50009 Zaragoza, Spain
  \item[$^{20}$] Centro de Astropartículas y Física de Altas Energías (CAPA), Universidad de Zaragoza, 50009 Zaragoza, Spain
  \item[$^{21}$] The Oskar Klein Centre, Department of Physics, Stockholm University, Stockholm 106 91, Sweden
  \item[$^{22}$] Institut für theoretische Physik, Universität Heidelberg, Philosophenweg 16, 69120 Heidelberg, Germany
  \item[$^{23}$] Department of Basic Sciences, Istinye University, 34396, Istanbul, Turkiye
  \item[$^{24}$] Institute for Particle Physics Phenomenology, Department of Physics, Durham University, Durham DH1 3LE, U.K.
  \item[$^{25}$] Center for Theoretical Physics of the Universe, Institute for Basic Science, Daejeon 34126, South Korea
  \item[$^{26}$] Dipartimento di Fisica e Astronomia, Università di Bologna, via Irnerio 46, 40126 Bologna, Italy
  \item[$^{27}$] INFN, Sezione di Bologna, viale Berti Pichat 6/2, 40127 Bologna, Italy
  \item[$^{28}$] Rudolf Peierls Centre for Theoretical Physics, Beecroft Building, University of Oxford, Parks Road, Oxford OX1 3PU, U.K.
  \item[$^{29}$] Univ. Lille, CNRS, UMR 8523 – PhLAM – Physique des Lasers Atomes et Molécules, F‑59000, Lille, France
  \item[$^{30}$] Institute of Theoretical Astrophysics, University of Oslo, P. O. Box 1029 Blindern, N‑0315, Oslo, Norway
  \item[$^{31}$] Department of Physics and Helsinki Institute of Physics, PL 64, FI‑00014 University of Helsinki, Finland
  \item[$^{32}$] Laboratoire de physique nucléaire et des hautes énergies (LPNHE), Sorbonne Université, Université Paris Cité, CNRS/IN2P3, Paris, France
  \item[$^{33}$] Departamento de Física Teórica, M‑15, Universidad Autónoma de Madrid, E‑28049 Madrid, Spain
  \item[$^{34}$] Instituto de Física Teórica UAM‑CSIC, Universidad Autónoma de Madrid, C/ Nicolás Cabrera, 13‑15, 28049 Madrid, Spain
  \item[$^{35}$] Instituto de Astrofísica de Canarias, E‑38200 La Laguna, Tenerife, Spain
  \item[$^{36}$] Departamento de Astrofísica, Universidad de La Laguna, E‑38206 La Laguna, Tenerife, Spain
  \item[$^{37}$] The Institute of Physical and Chemical Research (RIKEN), Center for Advanced Photonics, 519‑1399 Aramaki‑Aoba, Aoba‑ku, Sendai, Miyagi 980‑0845, Japan
  \item[$^{38}$] Dipartimento di Fisica e Astronomia, Università degli Studi di Padova, Via Marzolo 8, 35131 Padova, Italy
  \item[$^{39}$] Istituto Nazionale di Fisica Nucleare (INFN), Sezione di Padova, Via Marzolo 8, 35131 Padova, Italy
  \item[$^{40}$] Departamento de Tecnologías de la Información y las Comunicaciones, Universidad Politécnica de Cartagena, Plaza del Hospital, 1, 30202 Cartagena, Spain
  \item[$^{41}$] Max‑Planck‑Institut für Astrophysik, Karl‑Schwarzschild‑Str. 1, D‑85748 Garching, Germany
  \item[$^{42}$] Abdus Salam International Centre for Theoretical Physics (ICTP), Strada Costiera 11, 34151 Trieste, Italy
  \item[$^{43}$] Institute for Fundamental Physics of the Universe (IFPU), Via Beirut 2, 34151 Trieste, Italy
  \item[$^{44}$] Max‑Planck‑Institut für Physik (Werner‑Heisenberg‑Institut), Boltzmannstr. 8, 85748 Garching bei München, Germany
  \item[$^{45}$] Center for Astrophysics and Cosmology, University of Nova Gorica, Vipavska 11c, 5270 Ajdovščina, Slovenia
  \item[$^{46}$] Max‑Planck‑Institut für Gravitationsphysik (Albert‑Einstein‑Institut) and Leibniz Universität Hannover, Hannover, Germany
  \item[$^{47}$] Dipartimento di Fisica, Sapienza Università di Roma, Piazzale Aldo Moro 2, I‑00185 Rome, Italy
  \item[$^{48}$] INFN Sezione di Roma, Piazzale Aldo Moro 2, I‑00185 Rome, Italy
  \item[$^{49}$] Kavli Institute for the Physics and Mathematics of the Universe (WPI), UTIAS, The University of Tokyo, Kashiwa, Chiba 277‑8583, Japan
  \item[$^{50}$] Dipartimento di Fisica, Universit\`a di Milano - Bicocca, Piazza della Scienza 3, I-20126 Milano, Italy
  \item[$^{51}$] INFN Sezione di Milano - Bicocca,
  Piazza della Scienza 3, I-20126 Milano, Italy
  \item[$^{52}$] Deutsches Elektronen‑Synchrotron DESY, Platanenallee 6, 15738 Zeuthen, Germany
  \item[$^{53}$] INFN – Sezione di Napoli, Complesso Universitario Monte Sant’Angelo, 80126 Napoli, Italy
    \item[$^{54}$] Specola Vaticana (Vatican Observatory), V‑00120 Vatican City, Vatican City State
  \item[$^{55}$] INAF/OAS Bologna, via Gobetti 101, I‑40129 Bologna, Italy
  \item[$^{56}$] Laboratório de Instrumentação e Física Experimental de Partículas, LIP Lisbon, Portugal
    \item[$^{57}$] Instituto de Física Corpuscular (IFIC), CSIC – Universitat de València, C/ Catedrático José Beltrán 2, 46980 Paterna (Valencia), Spain
  \item[$^{58}$] Institut de Física d’Altes Energies (IFAE), The Barcelona Institute of Science and Technology (BIST), Campus UAB, 08193 Bellaterra (Barcelona), Spain
  \item[$^{59}$] Istituto Nazionale di Fisica Nucleare, Laboratori Nazionali di Frascati, Frascati, Roma, Italy
  \item[$^{60}$] Physical Sciences, Barry University, 11300 NE 2nd Ave., Miami Shores, FL 33161, USA
  \item[$^{61}$] Deutsches Elektronen‑Synchrotron DESY, Notkestr. 85, 22607 Hamburg, Germany
  \item[$^{62}$] Istituto Nazionale di Fisica Nucleare, Laboratori Nazionali del Gran Sasso, Assergi, 67100, Italy
  \item[$^{63}$] SYRTE, Observatoire de Paris, Université PSL, CNRS, Sorbonne Université, LNE, 75014 Paris, France
  \item[$^{64}$] LTE, Observatoire de Paris, Université PSL, Sorbonne Université, Université de Lille, LNE, CNRS, 61 Avenue de l’Observatoire, 75014 Paris, France
  \item[$^{65}$] Marietta Blau Institute for Particle Physics, Austrian Academy of Sciences, Dominikanerbastei 16, A‑1010 Vienna, Austria
  \item[$^{66}$] Institute for Quantum Optics and Quantum Information, Austrian Academy of Sciences, Vienna, Austria
  \item[$^{67}$] Institut für Experimentalphysik, Universität Hamburg, Luruper Chaussee 149, D‑22761 Hamburg, Germany
  \item[$^{68}$] University of Rijeka, Rijeka, Croatia
  \item[$^{69}$] Faculty of Physics, Sofia University, 5 J. Bourchier Blvd, 1164 Sofia, Bulgaria
  \item[$^{70}$] Technical University of Liberec, Liberec, Czech Republic
  \item[$^{71}$] Dipartimento Interateneo di Fisica “Michelangelo Merlin”, Via Amendola 173, 70126 Bari, Italy
  \item[$^{72}$] Istituto Nazionale di Fisica Nucleare (INFN), Sezione di Bari, Via Orabona 4, 70126 Bari, Italy
  \item[$^{73}$] INFN Sezione di Firenze, Via G. Sansone 1, I‑50019 Sesto Fiorentino, Firenze, Italy
  \item[$^{74}$] SLAC National Accelerator Laboratory, 2575 Sand Hill Rd, Menlo Park, CA 94025, USA
  \item[$^{75}$] INFN Sezione di Genova, 16147 Genova, Italy
  \item[$^{76}$] Department of Physics, University of Florida, Gainesville, FL, USA
  \item[$^{77}$] Department of Physics, Tohoku University, Sendai 980‑8578, Japan
  \item[$^{78}$] Department of Physics, Graduate School of Science, The University of Tokyo, 7‑3‑1 Hongo, Bunkyo, Tokyo 113‑0033, Japan
  \item[$^{79}$] Research Center for the Early Universe, The University of Tokyo, Bunkyo‑ku, Tokyo 113‑0033, Japan
  \item[$^{80}$] ARC Centre of Excellence for Dark Matter Particle Physics, School of Physics, The University of Sydney, NSW 2006, Australia
  \item[$^{81}$] Institute for Mathematics Astrophysics and Particle Physics, Radboud University, Heyendaalseweg 135, 6525 AJ Nijmegen, The Netherlands
  \item[$^{82}$] Institute of Particle and Nuclear Studies, High Energy Accelerator Research Organization (KEK), 1‑1 Oho, Tsukuba, Ibaraki 305‑0801, Japan
  \item[$^{83}$] Physics Department, Eastern Mediterranean University, Famagusta 99628, North Cyprus via Mersin 10, Turkiye
  \item[$^{84}$] Sydney Consortium for Particle Physics and Cosmology, School of Physics, The University of New South Wales, Sydney NSW 2052, Australia
  \item[$^{85}$] University of Vienna, Faculty of Physics, Boltzmanngasse 5, A‑1090 Vienna, Austria
  \item[$^{86}$] CNRS \& Univ. Grenoble Alpes, EMFL, Laboratoire National des Champs Magnétiques Intenses (LNCMI), 25 rue des Martyrs, 38042 Grenoble, France
  \item[$^{87}$] Faculty of Engineering and Natural Sciences, Sabancı University, 34956 Tuzla, İstanbul, Turkiye
  \item[$^{88}$] Astrophysics Research Center (ARCO), the Open University of Israel, Raanana 4353701, Israel
  \item[$^{89}$] Institute for Experimental Particle Physics (ETP), Karlsruhe Institute of Technology (KIT), D‑76131 Karlsruhe, Germany
  \item[$^{90}$] Department of Particle Physics and Astrophysics, Weizmann Institute of Science, Rehovot, Israel, 7610001
  \item[$^{91}$] Università degli Studi di Torino, via P. Giuria 1, 10125 Torino, Italy
  \item[$^{92}$] INFN, Sezione di Torino, via P. Giuria 1, 10125 Torino, Italy
  \item[$^{93}$] School of Physics and Astronomy, University of Glasgow, Glasgow, G12 8QQ, United Kingdom
  \item[$^{94}$] Scuola Normale Superiore and INFN Pisa, Piazza dei Cavalieri 7, 56126 Pisa, Italy
  \item[$^{95}$] Racah Institute of Physics, Hebrew University of Jerusalem, 9190401 Jerusalem, Israel
  \item[$^{96}$] Institute for Theoretical Physics, Kanazawa University, Kakuma‑machi, Kanazawa, Ishikawa 920‑1192, Japan
  \item[$^{97}$] Department of Physics and Astronomy “Ettore Majorana”, University of Catania, Via S. Sofia 64, I‑95125 Catania, Italy
  \item[$^{98}$] INFN‑Sezione di Catania, Via Santa Sofia 64, I‑95123 Catania, Italy
  \item[$^{99}$] Arnold Sommerfeld Center for Theoretical Physics, Ludwig‑Maximilian‑University Munich, Theresienstr. 37, 80333 Munich, Germany
  \item[$^{100}$] Department of Physics, Cornell University, Ithaca, NY 14853, USA
  \item[$^{101}$] Institut für Physik, Rheinische Friedrich‑Wilhelms‑Universität Bonn, 53115 Bonn, Germany
  \item[$^{102}$] Istituto Nazionale di Fisica Nucleare (INFN), Sezione di Pisa, Largo Bruno Pontecorvo 3, 56127 Pisa, Italy
  \item[$^{103}$] Institute of Physics, Theoretical Particle Physics Laboratory, École Polytechnique Fédérale de Lausanne (EPFL), CH‑1015 Lausanne, Switzerland
  \item[$^{104}$] INAF, Osservatorio Astronomico d’Abruzzo, 64100 Teramo, Italy
  \item[$^{105}$] Theoretical Physics Group, Lawrence Berkeley National Laboratory, Berkeley, CA 94720, USA
  \item[$^{106}$] Laboratoire d’Annecy de Physique Théorique (LAPTh), Université Savoie Mont‑Blanc and CNRS, 74941 Annecy, France
  \item[$^{107}$] Jožef Stefan Institute, Jamova 39, 1000 Ljubljana, Slovenia
  \item[$^{108}$] CEICO – FZU, Institute of Physics of the Czech Academy of Sciences, Na Slovance 1999/2, 182 00 Prague, Czech Republic
  \item[$^{109}$] INFN Sezione di Napoli, Gruppo collegato di Salerno, Via Giovanni Paolo II 132, I‑84084 Fisciano (SA), Italy
  \item[$^{110}$] Dipartimento di Fisica “E.R. Caianiello”, Università degli Studi di Salerno, Via Giovanni Paolo II, 132-84084 Fisciano (SA), Italy
  \item[$^{111}$] II. Institute of Theoretical Physics, Universität Hamburg, 22761, Hamburg, Germany
  \item[$^{112}$] Cluster of Excellence PRISMA$^{+}$, Staudingerweg 9, 55128 Mainz, Germany
  \item[$^{113}$] Department of Physics, Tokyo Metropolitan University, Tokyo 192‑0397, Japan
  \item[$^{114}$] University of Patras, Patras, Greece
\end{description}
}

\clearpage
\end{titlepage}

%% file: General/executive_summary.tex
\section*{The COST Action COSMIC WISPers (CA21106)}
\markboth{The COST Action COSMIC WISPers (CA21106)}{}
\addcontentsline{toc}{section}{The COST Action COSMIC WISPers (CA21106)} 
\thispagestyle{plain}

\begin{figure}[!h]
  \begin{center}
    \includegraphics[totalheight=7cm]{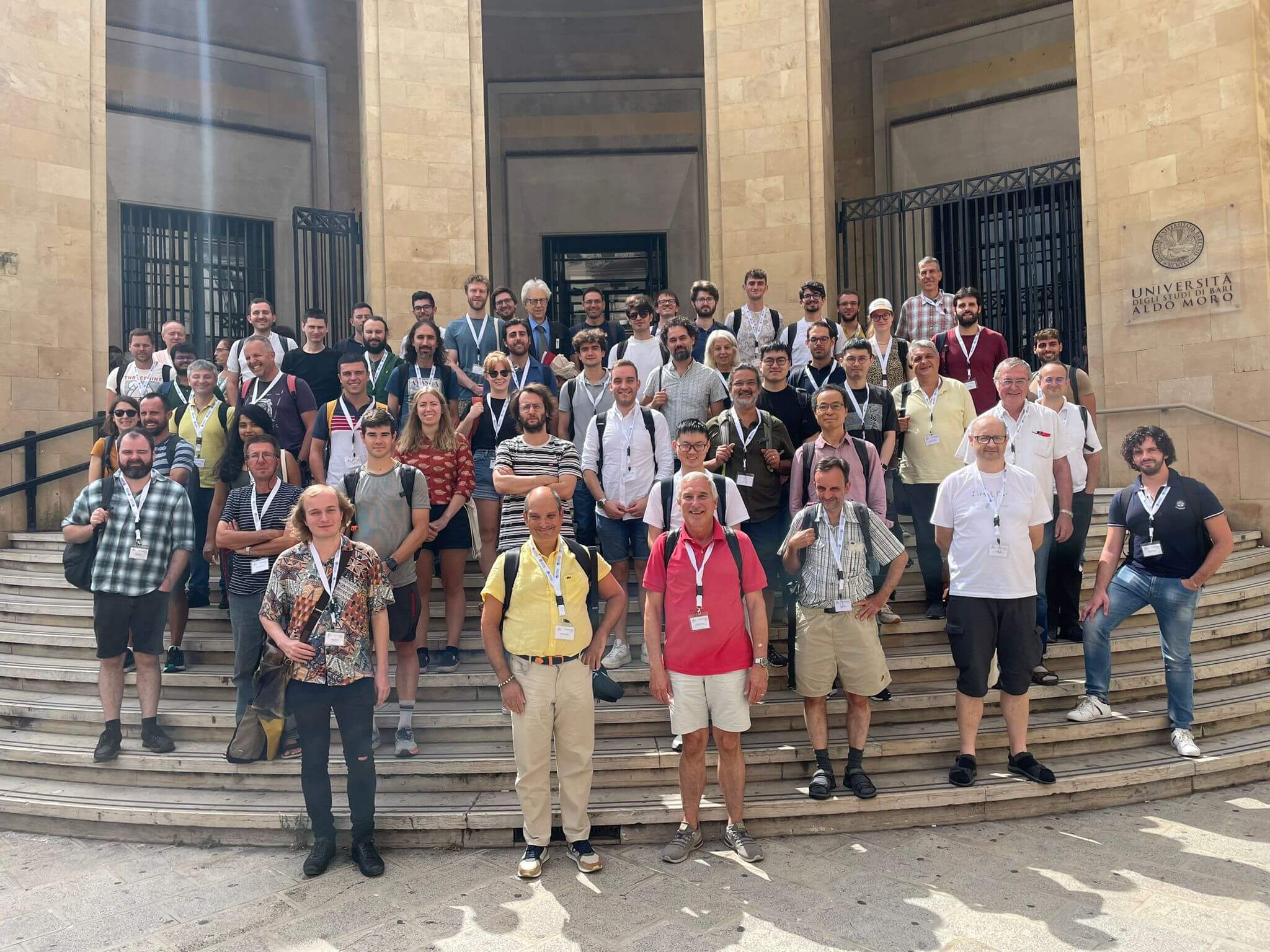}
    \caption{Participants of COSMIC WISPers during the General Meeting in Bari (IT), September 2023.}
    \label{fig:WGpic}
  \end{center}
\end{figure}

\noindent 
COSMIC WISPers is a EU-funded COST Action (CA21106, \url{https://www.cost.eu/actions/CA21106}) focused on very weakly interacting slim ($m< \textrm{GeV}$) particles (WISPs) which emerge in several extensions of the Standard Model of Particle Physics. The aim of this Action is an exhaustive study of these WISPs, notably axions, axion-like particles, and dark photons, ranging from their theoretical underpinning, over their indirect observational consequences in astrophysics, to their search at colliders and beam-dumps, and their direct detection in laboratory experiments.

The main aim of the COSMIC WISPers Action is to promote the formation of a team of scientists with different and complementary expertise to carry out WISP studies on a systematic and coherent basis in the intersecting and interdisciplinary areas of particle physics, astrophysics, and cosmology. The main goals are the following:
\begin{itemize}
\item[-] Provide a discussion forum for the European coordination of WISP Physics activities and express a collective view on the development of WISP Research.
\item[-] Develop a roadmap for WISP Physics in Europe, a description of the status and perspectives of the field within Europe, and linking them to activities in other parts of the world.
\item[-] Develop a common database on WISP theoretical models, experimental, and astrophysical bounds.
Provide cross-community discussions to enable new experiments.
\end{itemize}

\noindent 
The COSMIC WISPers Action started in October 2022 and is expected to end in September 2026. The original proposers were around 70 researchers. Currently (February 2026), more than 500 researchers are registered in the Action and carry out synergistic activities through five Working Groups (WGs):
\begin{itemize}
\item[-] \emph{WG1: WISP THEORY AND MODEL BUILDING.} The aim is to
determine the nature, number, masses and couplings of WISPs that arise in well-motivated theories of fundamental physics, and in particular within string  compactifications that join moduli stabilisation with (semi)-realistic matter sectors.
\item[-] \emph{WG2: WISP DARK MATTER AND COSMOLOGY.} The aim is to obtain precise predictions of the axion and WISP dark matter (DM) relic abundance and to identify distinguishing features of WISP DM in Large Scale Structure data.
\item[-] \emph{WG3: WISPs IN ASTROPHYSICS.} The aim is to deepen the studies of the signatures of WISPs in astroparticle physics. These include WISP oscillations into photons, WISP-induced energy losses in stellar systems, as well as signatures from gravitational waves and from primordial black-hole superradiance.
\item[-] \emph{WG4: WISP DIRECT SEARCHES.} The aim is to
produce a complete, updated and revised summary of the status of WISP searches, highlighting parts of the parameter space, models or couplings that are not under test by current or future searches, to outline a roadmap toward WISP discovery, and to identify strategies to disentangle different WISP models.
\item[-] \emph{WG5: DISSEMINATION AND OUTREACH.} The aim is to enhance the dissemination and communication of the results, and to structure outreach activities to attract public awareness to the challenges and achievements in astroparticle physics.
\end{itemize}

\begin{figure}[!t]
  \begin{center}
    \includegraphics[totalheight=7cm]{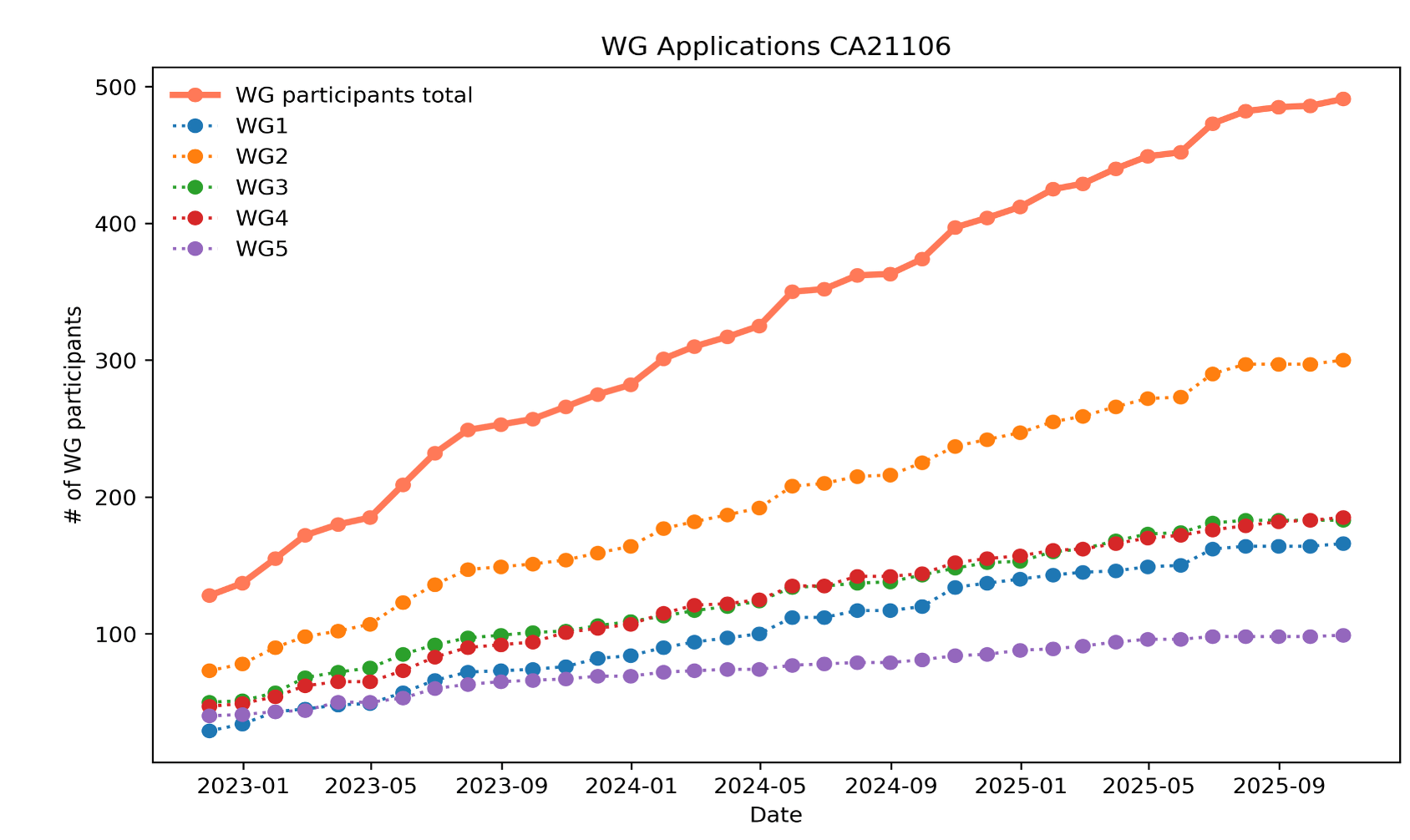}
    \caption{Number of the participants in the different WGs over the lifetime of the Action.}
    \label{fig:WG}
  \end{center}
\end{figure}

\noindent 
Figure~\ref{fig:WG} shows the number of participants in the different WGs over the lifetime of the Action.  Participants in the WGs are from 31 COST countries, of which 16 are Inclusiveness Target Countries (ITCs).
Additionally, there are also five participants from Near Neighbour Countries (NNCs) [AZ], as well as 43 participants from Internal Partner Countries (IPCs) [CL,AE,AU,US,JP,KR,CN,IN,AR,VA,CA,PH,IR,TH].

\begin{figure}[!t]
  \begin{center}
    \includegraphics[totalheight=7cm]{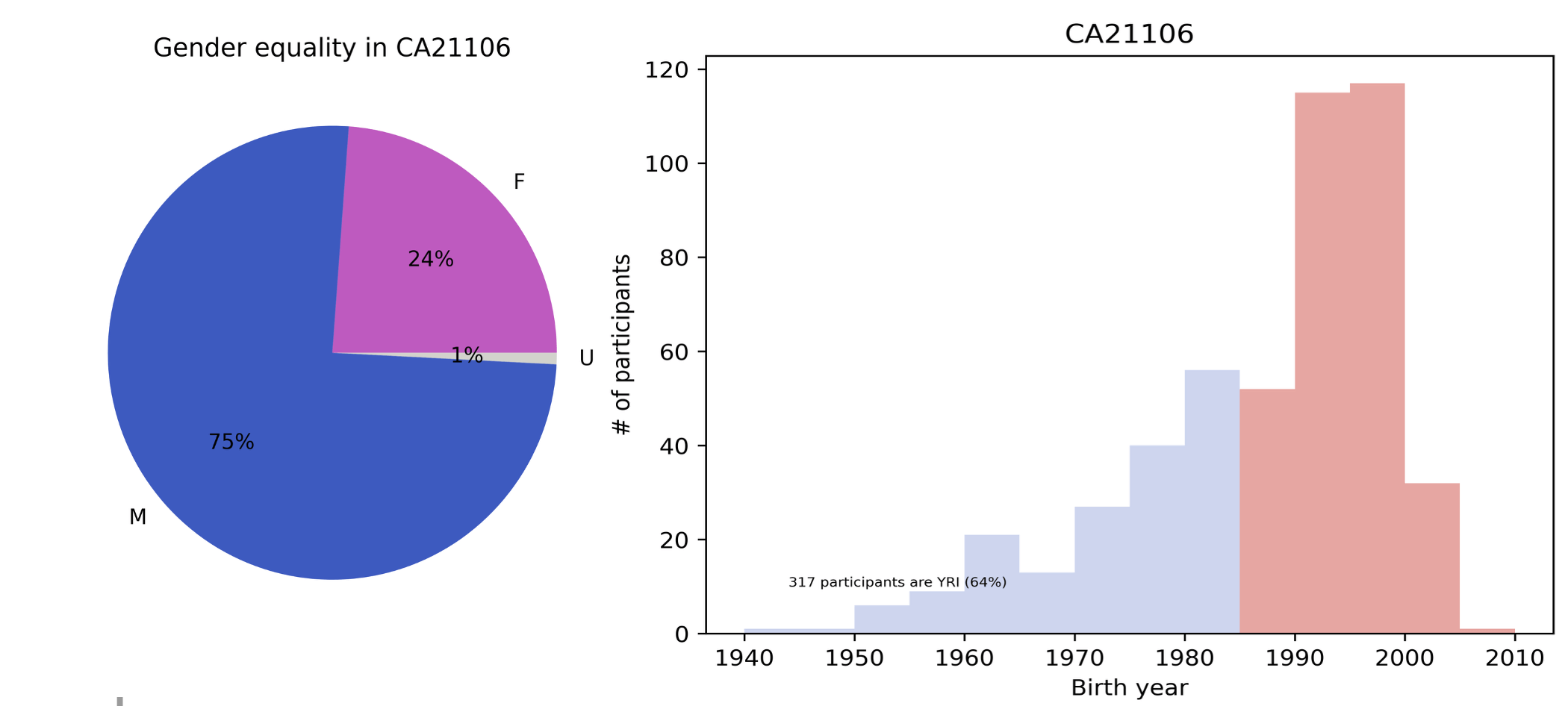}
    \caption{Gender and Young Researchers and  Innovators (YRI) representation in the WGs.}
    \label{fig:gender}
  \end{center}
\end{figure}

\noindent 
Figure~\ref{fig:gender} shows that 24\% of the participants in the Action are women, and 64\% are Young Researchers and Innovators (YRIs), i.e., researchers within eight years of obtaining their PhD.

All the activities of the COSMIC WISPers Action have been possible thanks to the 
crucial role of many people who actively participated to the management and deserve special thanks, starting  with 
Francesca Calore, who acted as Vice-Chair of the Action. Furthermore, the role of the people participating in the Management Committee and in the Core Group of the Action has been crucial. A warm acknowledgment has to be expressed to:
\begin{itemize}
\item[-] 
The Working Group (WG) Leaders:
  Michele Cicoli, Sophie Renner,  Nicole Righi, Arturo de Giorgi (WG1), 
  Edoardo Vitagliano,
Marco Gorghetto, María Benito Castaño, Mathieu Kaltschmidt (WG2), 
Maurizio Giannotti,
Oscar Straniero,  Elisa Todarello,  Giuseppe Lucente (WG3);
 Marin Karuza, 
Claudio Gatti, Deniz Aybas (WG4); 
Olga Mena, Loredana Gastaldo (WG5). 
\item[-] The Grant Evaluation Committee:
Venelin Kozhuharov (Chair),
Mirela Babalic,
Arturo de Giorgi, 
María Benito Castaño,
Serkant Cetin. 
\item[-] The Equity and Inclusiveness Committee:
 Deniz Sunar Cerci (Advisor), María Benito Castaño,
Francesca Calore, Loredana Gastaldo, Olga Mena. 
\item[-] The Colloquium organizers: 
Fuensanta Vilches,
Mario Reig Lopez, Marta Fuentes Zamoro, Giuseppe Lucente, Arturo de Giorgi. 
\item[-] The Journal Club organizers:
Silvana Abi Mershed, 
Xavier Ponce Diaz, Maria Benito Castaño, 
Michele Tammaro. 
\item[-] The Newsletter Editors: 
Damiano Fiorillo,
Alessandro Lella, Giovanni Grilli di Cortona. 
\item[-] The Whispered Tutorials organizers: 
Luca Caloni,
Amedeo Favitta,
Philip Sørensen.
\item[-] The Social media editors:
Silvana Abi Mershed,  
Giuseppe Lucente,
Maria Benito Castaño.
\item[-] The Young Researchers Council (YRC) Representatives: 
Marta Fuentes Zamoro,
Arturo de Giorgi,
Joshua Eby,
Pierluca Carenza.
\item[-] The COST Science Officer Ralph St\"{u}bner and the COST Administrative Officer Rose Cruz Santos. 
\item[-] The administrative staff at Bari University: Federica Camerino, Maria Pia Circella and Vito Tritta (Grant Holder Manager).
\end{itemize}

\noindent 
Special thanks to everyone who contributed their insights and ideas, making this Physics paper possible.
\\
\\
This Physics paper is dedicated to the memory of Prof.~Inmaculada Domínguez (Granada Univ., Spain), who was source of inspiration and encouragement for this Action.
\\
\\

February 2026 $ $ $ $ $ $ $ $ $ $ $ $  $ $  $ $ $ $ $ $ $ $ $ $ $ $  $ $  $ $ $ $ $ $ $ $ $ $ $ $  $ $  $ $ $ $ $ $ $ $ $ $ $ $  $ $  $ $ $ $ $ $ $ $ $ $ $ $  $ $  $ $ $ $ $ $ $ $ $ $ $ $  $ $  $ $ $ $ $ $ $ $    Alessandro Mirizzi (Chair of CA21106)

\clearpage

\section*{Executive Summary \\
\textnormal{\small Author: A. Mirizzi}}
\markboth{Executive Summary}{}
\thispagestyle{plain}
\addcontentsline{toc}{section}{
{Executive Summary \protect\\ \hspace*{-1.0em}\textnormal{Author: A. Mirizzi}}}

The Standard Model (SM) of particle physics is a remarkably successful framework, describing all interactions among known elementary particles observed in laboratory experiments with very high precision. Nevertheless, strong evidence points to the existence of physics beyond the SM (BSM). In particular, the SM does not provide a compelling explanation for the values of its many parameters, offers no consistent way to unify gravity with quantum mechanics, and fails to account for the origin of dark energy and dark matter (DM).
Within this broader context, weakly interacting massive particles (WIMPs) have long represented the leading paradigm for addressing some of these open questions, especially as well-motivated candidates for DM. In this respect, the ability of the Large Hadron Collider (LHC) to probe the electroweak scale has naturally driven DM searches in recent years to concentrate on WIMP-based models.
As a consequence, alternative DM candidates have attracted comparatively less attention from the experimental community. To date, however, neither LHC searches nor (in)direct detection experiments have produced any clear evidence for WIMPs. This lack of unambiguous signals of new physics provides strong motivation for broadening experimental efforts and pursuing approaches beyond the traditional BSM scenarios typically explored at the LHC. Within this framework, bosonic very weakly interacting slim particles (WISPs) are gaining increasing interest as well-motivated dark matter alternatives to WIMPs~\cite{Jaeckel:2010ni}.
Ultraviolet (UV) completions of the SM can give rise to different classes of WISPs, which can be classified according to their spin:

\begin{itemize}
\item[-] \emph{Spin 0:} Pseudo–Nambu–Goldstone bosons can originate from the spontaneous breaking of well-motivated BSM symmetries at energy scales far above the electroweak one. The most prominent example is the pseudo-scalar axion, originally proposed to solve the charge conjugation and parity (CP) problem in Quantum Chromodynamics (QCD) through the Peccei–Quinn mechanism. For suitable masses, the axion can also constitute a viable DM candidate. Beyond the ``QCD axion”, however, there may exist additional pseudo-scalar states with very similar properties. These axion-like particles (ALPs) arise naturally in many theoretically attractive UV completions of the SM, such as the low-energy four-dimensional limit of string theory compactified on Calabi–Yau threefolds \cite{Cicoli:2012sz}. Moreover, UV-complete frameworks of this type typically predict other light scalar fields—such as dilatons and string moduli—which often couple to ordinary matter with gravitational strength. Such particles can play a role in addressing open problems in fundamental physics, including DM, inflation, and dark energy \cite{Cicoli:2023opf}.

\item[-] \emph{Spin 1:} Another common prediction of SM UV completions is the existence of hidden-sector $U(1)$ gauge symmetries under which SM particles carry no charge. If the corresponding gauge boson is massive, the resulting “hidden” or ``dark'' photon can kinetically mix—via loop effects—with the SM hypercharge gauge boson and hence with the ordinary photon \cite{Abel:2008ai}. As a consequence, SM matter acquires a tiny effective charge under the hidden photon. Conversely, hidden-sector fermions, if present, can inherit a small charge under the ordinary electromagnetic group $U(1)_{\rm em}$.

\item[-] \emph{Spin 2:} Massive ``hidden'' or ``dark'' gravitons appear in bigravity, the unique ghost-free bimetric extension of General Relativity. Bigravity can emerge as a low-energy effective description in certain field-theoretic UV completions via double-copy constructions, as well as in string theory. Massive spin-2 states also arise generically in extra-dimensional scenarios, where they correspond to Kaluza–Klein excitations of the metric. More broadly, even focusing on the standard graviton, fundamental physics allows for a wide range of mechanisms capable of generating gravitational waves in early-Universe cosmology.
\end{itemize}

All these WISPs provide a strong physics motivation in connection with the DM puzzle and open up a broad range of experimental and observational discovery avenues. In recent years, interest in WISPs has intensified, accompanied by remarkable progress on both the theoretical and experimental fronts. A rich, diverse, and relatively low-cost experimental program is already underway, with the potential to deliver one or more transformative breakthroughs \cite{Irastorza:2018dyq}. European groups currently play a leading role in this effort, with major particle-physics laboratories—such as CERN, DESY, and LNF—directly involved in WISP searches. This vibrant experimental landscape is matched by equally stimulating theoretical activity, including the development of new ideas and previously unexplored strategies to probe WISPs \cite{Choi:2024ome}. In addition, significant attention has been devoted in recent years to the cosmological implications of WISPs \cite{Arias:2012az,OHare:2024nmr}, as well as to their impact on a range of astrophysical observables \cite{Caputo:2024oqc,Carenza:2024ehj}.

This review presents the current status of WISP searches in 2026 and discusses the prospects for the coming decade. The discussion spans theory and model building (Part~\ref{part:wg1}), cosmology and DM (Part~\ref{part:wg2}), astrophysics (Part~\ref{part:wg3}), and laboratory-based direct searches (Part~\ref{part:wg4}).

This paper builds on work carried out within the EU-funded COST Action \emph{``COSMIC WISPers in the Dark Universe: Theory, astrophysics and experiments''} (CA21106; \url{https://www.cost.eu/actions/CA21106}). The goal of the Action, expected to end in fall 2026, is to establish the scientific basis for the next generation of WISP experiments and searches in Europe. Beyond fostering cooperation and coordination, a key objective has been to strengthen a dedicated European WISP strategy by initiating a roadmap process. The Action currently brings together more than 500 researchers from European and non-European countries, supporting systematic and coherent WISP studies across the intersecting and interdisciplinary domains of particle physics, astrophysics, and cosmology.

\subsection*{Deliverables}

The role of the COSMIC WISPers Action in coordinating and supporting WISP searches in Europe and in shaping a roadmap to ensure European leadership in this field has been outlined in the Input to the European Strategy for Particle Physics~\cite{Aybas:2025pbq}. As a concrete step toward this coordinated strategy, the CA21106 COSMIC WISPers Community has established a centralized digital hub for the field, providing a comprehensive classification of WISP models, astrophysical fluxes from various sources, and an interactive map of existing and planned experiments.
The repository is maintained as a living community resource and is available at \url{https://github.com/CosmicWISPers}.
The concerted community effort to catalog WISP models culminated in a ``WISP-encyclopedia'', the \textit{WISPedia}~\cite{Albertus:2026fbe}. This document compiles a broad range of theoretical frameworks, providing for each a concise summary of its defining features together with the current phenomenological constraints. The contributions to the COSMIC WISPers General Meetings and the Lecture
Notes of the Training Schools have been published in a dedicated series
of Proceedings of Science
\url{https://pos.sissa.it/cgi-bin/reader/family.cgi?code=cosmicwispers}.

\clearpage

%% file: wg1.tex
\part{\texorpdfstring
{WISPs Theory and Model Building \\ \textnormal{\small Editors: A.~de Giorgi, M.~Reig \& N.~Righi}}{WISPs Theory and Model Building (Editors: A.~de Giorgi, M.~Reig \& N.~Righi)}}
\label{part:wg1}

\section{Low-energy WISP models from string theory}
\input{WG1/content/1-1-intro}
\subsection{\texorpdfstring
{Axions from string theory and moduli stabilisation\\
\textnormal{Author: A. Schachner}}{Axions from string theory and moduli stabilisation (Author: A. Schachner)}}
\label{sec:1.1.1}

\input{WG1/content/1-1-1.tex}

\subsection{\texorpdfstring
{Expected number of axions, mass spectrum and decay constants at low-energy from string theory and their statistics  \\ \textnormal{Author: M. Cicoli}}{Expected number of axions, mass spectrum and decay constants at low-energy from string theory and their statistics (Author: M. Cicoli)}}
\label{sec:StringAxionsStats}

\input{WG1/content/1-1-2.tex}

\subsection{\texorpdfstring
{QCD axion from string theory\\ \textnormal{Author: M. Cicoli}}{QCD axion from string theory (Author: M. Cicoli)}}
\label{sec:QCDaxionStringTheory}
\input{WG1/content/1-1-3.tex}

\subsection{\texorpdfstring
{The Axion Quality Problem: EFT vs UV \\ \textnormal{Authors: M. Reig \& N. Righi}}{The Axion Quality Problem: EFT vs UV (Authors: M. Reig and N. Righi)}}
\label{sec:QualityProblem}
\input{WG1/content/quality-problem.tex}

\subsection{\texorpdfstring
{The Dark side of ALPs: phenomenological applications  \\ \textnormal{Authors: F. Pedro, N. Righi \& M. Scalisi}}{The Dark side of ALPs: phenomenological applications (Authors: F. Pedro, N. Righi, and M. Scalisi)}}\label{sec:DarkALPs}
\input{WG1/content/1-1-4.tex}

\subsection{\texorpdfstring
{Dark photons and hidden sector degrees of freedom \\ \textnormal{Author: J.~P. Conlon}}{Dark photons and hidden sector degrees of freedom (Author: J.~P.~Conlon)}}\label{sec:DarkPhotonsST}
\input{WG1/content/1-1-5.tex}

\subsection{\texorpdfstring
{Swampland constraints on WISPs \\\textnormal{Authors: J.~P. Conlon \& M. Scalisi }}{Swampland constraints on WISPs (Authors: J.~P. Conlon and M. Scalisi)}}\label{sec:swampland}
\input{WG1/content/1-1-6.tex}

\section{WISP model building in QFT}
\subsection{\texorpdfstring
{Standard QCD axion models \\ \textnormal{Author: K. Choi}}{Standard QCD axion models (Author: K. Choi)}}\label{subsec:models}
\input{WG1/content/1-2-1.tex}

\subsection{\texorpdfstring
{Quantisation of axion couplings and their implications for IR experiments\\ \textnormal{Authors: M. Reig \& A. Sokolov}}{Quantisation of axion couplings and their implications for IR experiments (Authors: M. Reig and A. Sokolov)}}
\input{WG1/content/1-3-3.tex}

\subsection{\texorpdfstring
{Astrophobic QCD Axion\\
\textnormal{Author: L. Di Luzio}}{Astrophobic QCD Axion (Author: L. Di Luzio)}}
\input{WG1/content/1-2-2.tex}

\subsection{\texorpdfstring
{QCD axions outside the canonical band\\
\textnormal{Authors: K. Choi, L. Di Luzio, M. Ramos \& A. Sokolov}}{QCD axions outside the canonical band (Authors: K. Choi, L. Di Luzio, M. Ramos and A. Sokolov)}}
\input{WG1/content/non-canonical-QCD.tex}

\subsection{\texorpdfstring
{WISPs beyond the QCD axion\\
\textnormal{Authors: A. de Giorgi, L. Merlo, A. \"Ovg\"un, B. Pulice \& A. Ringwald}}{WISPs beyond the QCD axion (Authors: A. de Giorgi, L. Merlo, A. \"Ovg\"un, B. Pulice and A. Ringwald)}}
\input{WG1/content/1-2-4.tex}

\subsection{\texorpdfstring
{Screening mechanisms\\ \textnormal{Author: C. Burrage}}{Screening mechanisms (Author: C. Burrage)}}
\label{sec:1.2.5}
\input{WG1/content/screening.tex}

\subsection{\texorpdfstring
{Time dependence of physical observables\\ \textnormal{Author: M. Reig}}{Time dependence of physical observables (Author: M. Reig)}}
\input{WG1/content/time-dependence.tex}
\section{WISP EFTs}
\label{sec:WISP-EFT}
\input{WG1/content/1-3-intro}
\subsection{\texorpdfstring
{The ALP EFT\\
\textnormal{Authors: S. Renner, A. Ringwald \& L. Merlo}}{The ALP EFT (Authors: S. Renner, A. Ringwald and L. Merlo)}}
\label{eq:ALP-lagrangian}
\input{WG1/content/ALP-Lagrangian.tex}
\input{WG1/content/1-3-5.tex}

\subsection{\texorpdfstring
{Spin-0 EFT\\ 
\textnormal{Author: M. Ramos}}{Spin-0 EFT (Author: M. Ramos)}}
\label{eq:spin0-EFT}
\input{WG1/content/1-3-2.tex}

\subsection{\texorpdfstring
{Spin-1 EFT\\
\textnormal{Author: L. Ubaldi}}{Spin-1 EFT (Author: L. Ubaldi)}}
\label{eq:spin1-EFT}
\input{WG1/content/spin-1.tex}

\subsection{\texorpdfstring
{Spin-2 EFT\\ 
\textnormal{Author: A. de Giorgi}}{Spin-2 EFT (Author: A. de Giorgi)}}
\label{eq:spin2-EFT}
\input{WG1/content/spin-2.tex}

\cleardoublepage

%% file: WG1/content/1-1-intro.tex
The emergence of weakly interacting slim particles (WISPs) is not a fortuitous by-product of string compactifications; it is a structural prediction tied to the geometry and gauge symmetries of higher-dimensional theories. Whenever ten-dimensional superstrings are reduced to four dimensions, the compactification data imprints an organised spectrum of (generically light, or \emph{slim}) fields whose masses, interactions, and multiplicities are dictated by topological invariants, flux choices, and non-perturbative dynamics. In this sense, effective theories of string theory do not resemble minimal extensions of the Standard Model: they contain extended, correlated sectors of axions, moduli, and hidden gauge bosons whose collective properties offer a unique window onto the ultraviolet completion.

Over the last decade, this picture has evolved from a qualitative expectation into a quantitative and increasingly predictive framework. The development of explicit moduli-stabilisation schemes, the construction of large ensembles of Calabi–Yau orientifolds, and the advent of computational techniques capable of exploring high-dimensional configuration spaces have allowed us to treat WISPs in a statistical fashion in the string landscape. Axion mass matrices, decay constants, and the distribution of light modes can now be analysed across thousands of geometries. In parallel, the cosmological role of these light fields, from dark matter to dark radiation, early dark energy or quintessence, has become tightly linked to the underlying moduli dynamics. As a result, a new paradigm has emerged: \emph{WISPs as structured probes of quantum gravity}, whose properties encode specific geometric and dynamical features of the compactification.

This section provides a unified overview of this programme. We begin by reviewing the higher-dimensional origin of axion-like particles and the geometric mechanisms that determine their spectrum, shift symmetries, and couplings (Sec.~\ref{sec:1.1.1}). We then outline how moduli stabilisation through fluxes, perturbative and non-perturbative effects, shapes the landscape of axion masses and the structure of the string axiverse (Sec.~\ref{sec:StringAxionsStats}). In Sec.~\ref{sec:QCDaxionStringTheory} particular attention is devoted to the modelling of viable QCD axions and to the UV aspects of the Peccei–Quinn symmetry, where string theory offers qualitatively different solutions to the axion quality problem than bottom-up EFT approaches, as discussed in Sec.~\ref{sec:QualityProblem}. On the phenomenological side, we discuss the role of string axions and light moduli in cosmology, including their production mechanisms, dark matter abundances, contributions to dark radiation, and signatures in the early universe (Sec.~\ref{sec:DarkALPs}). 

Importantly, the particle spectrum is not limited to pseudo-scalars: string compactifications generically produce dark photons, hidden-sector gauge bosons, and additional ultralight fields whose couplings and spectrum are constrained by the same geometric data, as we show in Sec.~\ref{sec:DarkPhotonsST}. The correlations among these sectors, often overlooked in bottom-up constructions, are a distinctive feature of UV-complete models.

By presenting these ingredients, this section highlights a central message: string theory does not merely produce WISPs; it organises them, constraining their spectrum, interactions, and cosmological footprints in ways that are now increasingly accessible to systematic study. In this sense, the low-energy WISP sector is emerging as one of the most promising arenas to test string theory itself (see Sec.~\ref{sec:swampland}) and through which quantum gravity may leave observable, testable imprints in upcoming astrophysical and laboratory experiments.

%% file: WG1/content/1-1-1.tex
\paragraph{Introduction}

String theory posits that the basic constituents of nature are one-dimensional strings, whose vibrational modes give rise to different particles and forces. A significant prediction of string theory is the existence of extra dimensions, which emerge from the need to cancel anomalies and ensure consistency within a quantum framework. Throughout this section, our focus is on superstring theory in 10 dimensions unless stated otherwise. To find an effective description of a four-dimensional universe at low energies, six dimensions must therefore be curled up at tiny length scales, a process referred to as \emph{Kaluza-Klein (KK) compactification}, rendering them unobservable at macroscopic scales. These compact dimensions critically influence the properties of the resulting four-dimensional effective field theories (EFTs), determining features such as particle masses and interactions. A common choice for the compactification geometry is a \emph{Calabi-Yau (CY) threefold} $X$, which is a complex three-dimensional Kähler manifold admitting a Ricci-flat metric. Indeed, CY threefolds furnish a large class of solutions to the ten-dimensional equations of motion, in particular solving Einstein’s equations in the vacuum.

Upon compactifying superstring theory on such a CY manifold $X$, the resulting four-dimensional EFT is a supergravity theory that, among others, contains a number of chiral multiplets determined by the topology of $X$. In such dimensional reductions, \emph{axion-like particles} (ALPs) appear naturally, particularly originating from the zero-modes of higher-dimensional gauge fields as we briefly explain below. These theories also contain geometric \emph{moduli}, i.e. scalar fields parametrising the shape and size of the compact dimensions, which couple to ALPs in particular through gravitational interactions. Due to the vast number of CY geometries, the resulting string theory landscape of EFTs containing a plethora of ALPs and moduli provides a diverse set of candidates for inflation, dark matter (DM), dark radiation, or other cosmological phenomena \cite{Arvanitaki:2009fg}, making them a significant focus in the search for new physics beyond the Standard Model (see \cite{Cicoli:2023opf} for a recent review).

\paragraph{Axions from string theory}

Conventionally, axions are considered pseudo Goldstone bosons of a spontaneously broken anomalous $U(1)$. In compactified theories, there exists another mechanism\footnote{See \cite{Reece:2025thc} for a more complete discussion of the various mechanisms for generating ALPs.} for generating ALPs, first discussed in \cite{Witten:1984dg}, where the sources are higher-dimensional $p$-forms $C_p$ which enjoy a generalised gauge symmetry $C_{p}\rightarrow C_{p} + \mathrm{d}\Lambda_{p-1}$.\footnote{We focus here mostly on closed string axions. Other types of ALPs arise from open string sectors living on spacetime filling branes hosting gauge sectors, see e.g. Secs.~\ref{sec:StringAxionsStats} and \ref{sec:QCDaxionStringTheory}.} Upon dimensional reduction on a compact manifold $X$, $C_p$ can be expanded as $C_p = \theta^i\omega_i+\ldots$ in terms of a basis of harmonic $p$-forms $\omega_i$, $i=1,\ldots,N_p$, on $X$ where $N_p$ depends on both the choice of $X$ and the value of $p$. The ellipsis contains heavier modes with masses near the KK scale that, at low enough energies, can be safely ignored. The zero-modes $\theta^i$ of these higher-dimensional gauge fields correspond to ALPs in the lower-dimensional theories which we call $C_p$-axions. Their periodicity 
\begin{equation}
    \theta^i \cong \theta^i +2\pi n^i\, ,\quad n^i \in \mathbb{Z} 
\end{equation}
is inherited from the aforementioned gauge invariance under ``large'' gauge transformations of the form $C_p \rightarrow C_p + 2\pi n^i\omega_i$, see e.g. \cite{Reece:2023czb,Choi:2024ome}. Hence, the axions are massless to all orders in perturbation theory and acquire masses only due to non-perturbative effects. This suggests that their phenomenology is highly sensitive to the geometry underlying the compactification.

In what follows, we concentrate on Type IIB compactifications because they arguably lead to the best understood string models when it comes to phenomenological model building in string theory \cite{Cicoli:2023opf}. See \cite{Honecker:2013mya,Petrossian-Byrne:2025mto} for axions in type IIA and \cite{Choi:2011xt,Choi:2014uaa,Buchbinder:2014qca,Agrawal:2024ejr,Loladze:2025uvf,Leedom:2025mlr,Reig:2025dqb} for axions in the heterotic string. 

The ten-dimensional theory contains four different $p$-forms, namely a $4$-form $C_4$, two $2$-forms $B_2,\, C_2$, and one $0$-form $C_0$, which, upon dimensional reduction, give rise to their respective ALPs in the four-dimensional EFT. For simplicity, we focus on $C_4$-axions which have received a lot of attention recently~\cite{Cicoli:2012sz,Cicoli:2013ana,Demirtas:2018akl,Halverson:2019cmy,Mehta:2020kwu,Mehta:2021pwf,Cicoli:2021gss,Demirtas:2021gsq,Cicoli:2022fzy,Gendler:2023kjt,Dimastrogiovanni:2023juq,Sheridan:2024vtt,Benabou:2025kgx,Yin:2025amn,Cheng:2025ggf},\footnote{This can always be achieved by choosing suitable orientifolds, see e.g. \cite{Moritz:2023jdb}.} but note that $B_2$-/$C_2$-axions can be treated similarly and are expected to have rich phenomenologies, see e.g. \cite{Cicoli:2012sz,Cicoli:2021tzt,Cicoli:2021gss} (see also Sect.~\ref{sec:StringAxionsStats}).

We introduce the complex scalars
\begin{equation}
\label{eq:complexKahler}
    T^i = \tau^i + \mathrm{i}\,\theta^i\,,\qquad i=1,\ldots,h^{1,1}\,,
\end{equation}
consisting of the Kähler moduli $\tau^i$ and $C_4$-axions $\theta^i$. Here, $h^{1,1}$ is a Hodge number of $X$ counting the number of such scalars. We also temporarily ignore all other fields in the EFT, see Sect.~\ref{sec:StringAxionsStats} for more details. In Type IIB compactifications to four dimensions, the Lagrangian for the $T^i$ is generically of the form (setting $M_P=1$)
\begin{equation}
\label{eq:LagrangianALPsIIB}
    \mathcal{L} = K_{i\bar{\jmath}}\, (\partial_\mu T^i)(\partial^\mu \overline{T}^{\bar{j}})-V(T^i,\overline{T}^{\bar{i}})+\ldots\, .
\end{equation}
Here, $K_{i\bar{\jmath}}$ is the metric on the field space of the complex scalars $T^i$. The ellipses contain additional interactions involving other sectors, such as couplings between axions and SM gauge bosons or couplings to hidden sectors. At this point, let us note that the QCD axion can in principle be one of the ALPs above, provided that suitable couplings to the gluon field strength arise in Eq.~\eqref{eq:LagrangianALPsIIB}. This is highly model-dependent since the Standard Model (SM) sector has to be carefully engineered, see e.g. \cite{Conlon:2006tq,Cicoli:2012sz,Broeckel:2021dpz,Gendler:2023kjt}. For the moment, we remain agnostic about realisations of the SM in our compactifications and defer the discussion to Sect.~\ref{sec:QCDaxionStringTheory}.

The scalar potential for the axions $\theta^i$ and moduli $\tau^i$ is usually of the form 
\begin{equation}
\label{eq:FtermPotential}
    V(T^i,\overline{T}^{\bar{i}}) = V_{\text{mod}}(\tau^i)+\sum_I \Lambda_I^4\, \cos\bigl (2\pi\mathcal{O}_{Ii}\theta^i+\delta_I\bigl )+\ldots \, .
\end{equation}
The instanton scales $\Lambda_I=\Lambda_I(\tau^i)$ depend on the moduli values $\tau^i$ through (bringing $M_P$ back)
\begin{equation}
\label{eq:instanton_scalesIIB}
    \Lambda_I^4 =  8\pi \, m_{3/2} \frac{\mathcal{O}_{Ii} \tau^i}{\mathcal{V}}\mathrm{e}^{-2\pi\mathcal{O}_{Ii} \tau^i}M_P^3\,,
\end{equation}
where $m_{3/2}$ is the gravitino mass and $\mathcal{V}$ the CY volume in units of the string length. Assuming for simplicity the presence of hierarchies in the instanton scales $\Lambda_{I+1}\ll \Lambda_I$, the masses and decay constants for the canonically normalised fields $a^i = M^{i}\,_{j}\,\theta^j$ can be computed as \cite{Cicoli:2012sz,Gendler:2023kjt}
\begin{equation}
\label{eq:masses_decay_constants_C4}
    f_i = \dfrac{M_P}{2\pi} (\mathcal{O}_{ij}(M^{-1})^{j}\,_{i} )^{-1}\, ,\quad m_i^2 = \frac{\Lambda_i^4}{f_{i}^2}\,,
\end{equation}
where $\mathcal{O}_{ij}$ is the reduced charge matrix for the $h^{1,1}$ leading order contributions to \eqref{eq:FtermPotential}. As argued above, due to higher-dimensional gauge invariance, the axion potential is generated by non-perturbative effects $\sim\mathrm{e}^{-2\pi\tau^i}$, specifically by D-instantons from Euclidean D3-branes wrapping 4-cycles with actions $S_I=2\pi \,\mathcal{O}_{Ii}\, T^i$, where $\tau^i\gtrsim 1$ for computational control. As a result, ultralight axions with masses $m_i\lesssim \mathcal{O}(10^{-20}\text{eV})$ appear quite frequently in such models \cite{Gendler:2023kjt} which could e.g. contribute to cosmic birefringence \cite{Minami:2020odp,Diego-Palazuelos:2022dsq}.

\paragraph{Progress in moduli stabilisation}

The values of the moduli $\tau^i$ setting the instanton scales $\Lambda_I$ in Eq.~\eqref{eq:instanton_scalesIIB} are dynamically fixed through a process referred to as \emph{moduli stabilisation}. This becomes necessary due to experimental bounds on fifth force experiments constraining the presence of light scalar fields in our Universe.\footnote{Note that axions evade these constraints due to their shift symmetry.} Hence, a suitable scalar potential for the moduli, denoted $V_{\text{mod}}(\tau^i)$ in Eq.~\eqref{eq:FtermPotential}, has to be generated, making them heavy. This has profound consequences on the type of cosmologies arising in string theory. For example, provided that this process leads to a de Sitter (dS) minimum with small vacuum energy, it allows us to tackle the cosmological constant problem, thereby providing a fundamental explanation for the accelerated expansion of our Universe. Furthermore, this stabilisation mechanism of moduli affects the masses and couplings of axions \cite{Conlon:2006tq,Cicoli:2012sz}, influencing their potential role as candidates for dark matter and their impact on the dynamics of the early universe, thereby bridging string theory with observable axion physics. Understanding moduli stabilisation is therefore pivotal, as it not only ensures a stable vacuum in the low-energy effective theory, but also determines the features of the axiverse emerging from string theory.

A major obstacle for realising any moduli stabilisation procedure in perturbative string theory is the so-called \emph{Dine-Seiberg problem} \cite{Dine:1985he}. At its core, it states that, because the first non-vanishing contribution in the perturbative expansion always dominates in the potential, there is no non-trivial minimum at weak coupling, but only a runaway. Away out is offered by extra sources of hierarchies by e.g.~balancing terms in the K\"ahler potential $K$ against those in the superpotential $W$, or by tuning fluxes. Various approaches to moduli stabilisation have been developed based on these principles, including mechanisms involving fluxes  \cite{Giddings:2001yu,DeWolfe:2005uu,Douglas:2006es,Denef:2008wq}, non-perturbative effects \cite{Witten:1996bn}, and perturbative corrections \cite{Becker:2002nn,Berg:2004ek,Berg:2005ja,Berg:2007wt,Cicoli:2016chb,Antoniadis:2019rkh}, see \cite{McAllister:2025qwq} for a recent review. In our setting, CY compactifications of Type IIB superstring theory with non-trivial flux backgrounds have been a fertile ground for realising concrete proposals for moduli stabilisation in dS vacua, such as the KKLT scenario \cite{Kachru:2003aw} and the Large Volume Scenario (LVS) \cite{Balasubramanian:2005zx,Cicoli:2008va}. Over the years, such scenarios have profoundly shaped our understanding of the string axiverse \cite{Cicoli:2012sz}. However, fully characterising the landscape of compactifications with $\mathcal{O}(100)$ moduli and axions still remains a daunting task.

In recent years, computer-aided frameworks have been established, pushing the frontiers of string model building in the era of Big Data. Fully automated software for string vacua has been developed in \cite{Dubey:2023dvu,AbdusSalam:2025twp}, combining state-of-the-art automatic differentiation and parallelisation tools with optimisation algorithms specifically tailored to the problem of moduli stabilisation. This implementation has already proven fruitful in identifying biases in sampling methods \cite{Dubey:2023dvu}, characterising the distribution of observables \cite{Ebelt:2023clh,Chauhan:2025rdj}, and finding novel types of solutions in string compactifications \cite{Krippendorf:2023idy,AbdusSalam:2025twp}.

In the same vein, recent advances in computational algebraic geometry, especially through software packages like \texttt{CYTools} \cite{Demirtas:2022hqf} and related extensions \cite{Demirtas:2023als,Moritz:2023jdb}, have been instrumental in unlocking regimes with large numbers ($h^{1,1}\gtrsim 100$) moduli and ALPs. Among others, fully explicit realisations of the KKLT scenario have become feasible \cite{Demirtas:2021nlu,Demirtas:2021ote,McAllister:2024lnt}.\footnote{We refer to \cite{Cicoli:2011qg,Cicoli:2012vw,Cicoli:2013mpa,Cicoli:2013cha,Cicoli:2016xae,Cicoli:2017shd,Cicoli:2017axo,Crino:2020qwk,Cicoli:2024bxw} for similar attempts in the context of the LVS.} The works \cite{Demirtas:2021nlu,Demirtas:2021ote} provided the technologies to stabilise over $100$ scalar fields in supersymmetric anti-dS vacua with vacuum energies $|V_0|\lesssim 10^{-100}\, M_{P}^4$. They built the backbone for the candidate dS vacua with broken supersymmetry recently constructed in \cite{McAllister:2024lnt}, which feature strongly warped throats containing anti-D3-branes at their tips. The solutions of \cite{McAllister:2024lnt} are currently formulated in a leading-order EFT containing known quantum corrections, particularly all $\alpha^\prime$ corrections at string tree level. Thorough control analyses indicate that, to decide whether they represent valid solutions within string theory, effects of additional unknown corrections, such as string loop corrections to the Kähler potential $K$, need to be computed (cf.~\cite{McAllister:2025qwq} for further details). Nonetheless, these advances offer a promising path forward in tackling the cosmological constant problem. Future research will build upon these compactifications to better understand the vacuum structure in cosmological solutions in string theory.

\paragraph{Combinatorial cosmology and the KS axiverse}

The aforementioned numerical advancements pave the way not just for understanding moduli stabilisation beyond traditional analytic methods, but also to find the most explicit string theory constructions to date possessing ultralight axions with a relic abundance large enough to be probed by upcoming cosmological and astrophysical surveys. Specifically, in this so-called \emph{Kreuzer-Skarke (KS) Axiverse}~\cite{Kreuzer:2000xy,Demirtas:2018akl}, axion theories are obtained from Type IIB compactifications on special CY manifolds realised as hypersurfaces in toric varieties determined by triangulations of 4D reflexive polytopes~\cite{Batyrev:1993oya}. These explicit models enable the generation of detailed exclusion plots for physical observables like from black hole superradiance \cite{Mehta:2020kwu,Mehta:2021pwf}, the QCD $\theta$-angle \cite{Demirtas:2021gsq}, or axion-photon couplings \cite{Gendler:2023kjt}. Such efforts are prerequisites for linking string theory to phenomenology and experimental data, see in particular \cite{Jain:2025vfh} for recent progress using Bayesian inference.

A prominent example concerns fuzzy DM, where the masses and decay constants of fuzzy axions must meet observational requirements for DM abundance. Interestingly, a significant relic abundance of fuzzy DM can make such models testable by cosmological measurements. In the context of moduli stabilisation, this has been investigated in \cite{Cicoli:2021gss}, in particular differentiating between LVS and KKLT type models. More recently, \cite{Sheridan:2024vtt} studied a large ensemble of CY compactifications, conducting a systematic analysis of the misalignment production of fuzzy dark matter. Further details on the phenomenological implications of axion DM in type IIB string theory will be discussed in Sect.~\ref{sec:DarkMatterRadiation}.

On the computational side, recent applications of optimisation strategies explored in \cite{MacFadden:2024him} present another promising new direction in studying the KS axiverse. This method excels by optimising a wide class of EFTs derived from Type IIB compactifications on diverse CY manifolds. At its core, \emph{Genetic Algorithms} (GAs) are employed to navigate the intricate landscape of CY compactifications, evolving solutions that satisfy both physical and mathematical constraints. The chosen data encoding reduces computational redundancies, building on \cite{MacFadden:2023cyf}, enabling more efficient and rapid searches, significantly outperforming traditional techniques like Markov Chain Monte Carlo or Simulated Annealing. As shown in \cite{MacFadden:2024him}, GAs can optimise axion decay constants to match phenomenological constraints and enhance axion-photon couplings.

%% file: WG1/content/1-1-2.tex
As we have seen in Sec.~\ref{sec:1.1.1}, the low-energy limit of string compactifications might lead to a plethora of very light closed string axions. However there are a few effects that could remove the axions from the mass spectrum below the TeV-scale. Let us discuss each of them separately:
\begin{itemize}
\item \textbf{Supersymmetric moduli stabilisation:} The axions $\theta_i$ are the imaginary parts of complex scalar fields
$T_i=\tau_i+{\rm i} \theta_i$, where $\tau_i$ are `saxion' fields which control the size of the extra dimensions and many key-features of the EFT like gauge and Yukawa couplings. These saxions typically suffer from the cosmological moduli problem unless their mass respects a severe lower bound: $m_{\tau_i} \gtrsim \mathcal{O}(50)$ TeV \cite{Coughlan:1983ci,Banks:1993en,deCarlos:1993wie}. This implies that, if the potential that stabilises $\tau_i$ above this bound lifts also the axionic directions, the corresponding axions $\theta_i$ would become very heavy. This is the typical situation when the moduli are stabilised supersymmetrically by non-perturbative effects, as in KKLT scenarios. However this situation is highly non-generic in the flux landscape since it requires to make perturbative corrections negligible by tuning the flux superpotential $W_0$ to exponentially small values (see \cite{Chauhan:2025rdj} for a recent exhaustive search of flux vacua which shows the rarity of such solutions). 

On the other hand, the most generic situation is the one where perturbative corrections dominate over non-perturbative ones. Given that the axions $\theta_i$ enjoy a continuous shift symmetry, $\theta_i \to \theta_i + {\rm const}$, which is broken to a discrete one only at non-perturbative level, perturbative corrections generate a potential just for the saxions $\tau_i$. Hence the axions turn out to be hierarchically (exponentially) lighter than the corresponding saxions, $m_{\theta_i}\ll m_{\tau_i}\gtrsim \mathcal{O}(50)$ TeV. In this way the saxions can satisfy the cosmological moduli problem bound while the axions can remain very light.

\item \textbf{Orientifold projection:} Type IIB compactified on a CY threefold yields an $\mathcal{N}=2$ supergravity EFT in four dimensions which is not chiral. To get a chiral $\mathcal{N}=1$ effective supergravity, one has to focus instead on CY orientifolds where some of the original axions are removed from the low-energy spectrum by the orientifold involution. For example, the number of $C_4$-axions is reduced from $h^{1,1}$ to $h^{1,1}_+$ with $1\leq h^{1,1}_+\leq h^{1,1}$. However, this operation does not significantly modify the order of magnitude of the number of closed string axions left, especially since for a typical CY $h^{1,1}\sim \mathcal{O}(100)$.

\item \textbf{St\"uckelberg mechanism:} Even if  axions are not projected out by the orientifold involution and remain light because of moduli stabilisation, they could still disappear from the low-energy EFT by being eaten by an anomalous $U(1)$. This St\"uckelberg mechanism is at play, for example, in the case of a stack of $N$ magnetised D7-branes wrapping the $i$-th divisor $D_i$. The resulting gauge group is $SU(N)\times U(1)$. The $U(1)$ factor is generically anomalous and becomes massive by eating a linear combination of string axions. Let us discuss this process a bit more in detail, following App.~A of \cite{Cicoli:2011yh}. The Chern-Simons part of the brane action contains a term of the form:
\begin{equation}
\int_{\mathbb{R}^{1,3}\times D_i} C_4 \wedge F_2 \wedge F_2\,,
\label{CSaction}
\end{equation}
where $F_2$ is the gauge flux on the brane. The decomposition of the 4-form $C_4$ into KK zero-modes involves two important contributions,
\begin{equation}
C_4 \supset Q_2^j(x) \wedge \hat{D}_j(y) + \theta_j(x) \tilde{D}^j(y)\qquad j=1,\cdots,h^{1,1}\,,   
\label{Expansion}
\end{equation}
where $\hat{D}_j$ and $\tilde{D}^j$ are, respectively, bases of harmonic $(1,1)$- and $(2,2)$-forms. Substituting the expansion (\ref{Expansion}) in the action (\ref{CSaction}), the $j$-th 4D 2-form $Q_2^j$ couples to the $U(1)$ on the $i$-th divisor $D_i$ as
\begin{equation}
q_{ij} \int_{\mathbb{R}^{1,3}} Q_2^j(x) \wedge F_2(x)\,, \quad\text{where}\quad q_{ij} = \int_{D_i} \hat{D}_j \wedge F_2\,.
\label{U(1)charges}
\end{equation}
This coupling induces a St\"uckelberg mechanism: the $i$-th Abelian gauge boson $\gamma'_i$ becomes massive by acquiring an additional degree of freedom given by a linear combination of the axions $\theta_j$ which are 4D dual to all 2-forms $Q_2^j$ with a non-zero coupling $q_{ij}$. The resulting mass $m_{\gamma'_i}^2$ reads \cite{Cicoli:2011yh}
\begin{equation}
m_{\gamma'_i}^2=\frac{q_{ip}\,M_p^2}{2\pi\sqrt{\tau_i}}\left(\frac{\partial^2 K_0}{\partial\tau_p\partial\tau_q}\right)\sum_j\frac{q_{jq}}{\sqrt{\tau_j}}\,, 
\label{U(1)mass}
\end{equation}
where $\tau_k$ is the volume of $k$-th divisor and $K_0=-2\ln\mathcal{V}$ is the tree-level K\"ahler potential. To make a simple estimate, let us consider  $\mathcal{V}=\tau_b^{3/2}-\tau_s^{3/2}$, where $\tau_b$ is a big cycle while $\tau_s$ is a small divisor, $\tau_b\gg \tau_s$. It is straightforward to see that a $U(1)$ on $D_b$ would become massive by eating up $\theta_b$, developing a mass of order the gravitino mass, $m_{\gamma'_b}\sim q_{bb}\,m_{3/2}$, while the Abelian gauge boson on $D_s$ would acquire a mass of order the string scale, $m_{\gamma'_b}\sim q_{ss}\,M_s$, by eating up $\theta_s$.
\end{itemize}
These considerations imply that very light axions can arise in the 4D EFT under three conditions: ($i$) they are not projected out by orientifolding; ($ii$) they are not eaten by an anomalous $U(1)$; ($iii$) the corresponding saxions are fixed perturbatively. As already pointed out, it is rather simple to satisfy each of these conditions in the weak coupling limit where the EFT is under control since in general the orientifold involution leaves several axions, the number of anomalous $U(1)$ is much smaller than the number of axions, and non-perturbative effects are naturally tiny when all volume divisors are large in string units. This analysis leads to the generic expectation of a large number of ultra-light axions whose masses are exponentially suppressed with respect to the gravitino mass that sets the mass scale of the saxions. Note that in order to realise the QCD axion from string theory, stringy instantons or gaugino condensation should not generate a mass for the QCD axion candidate which is larger than the one induced by QCD instantons (see Sec.~\ref{sec:QualityProblem} for a  treatment of the quality problem for the QCD axion).

It is worth pointing out that axions arise also as open strings living on space-time filling branes which wrap some of the extra dimensions and support visible or hidden gauge theories \cite{Choi:2010gm,Cicoli:2013cha,Allahverdi:2014ppa,Cicoli:2017zbx,
Petrossian-Byrne:2025mto, Loladze:2025uvf}. These axions are phases $\zeta$ of matter fields $C = |C|\,e^{{\rm i}\zeta}$ whose radial part
breaks an effective global Peccei-Quinn (PQ) $U(1)$ symmetry by getting a non-zero VEV via D-term stabilisation. At fundamental level, the global PQ $U(1)$ is a local anomalous $U(1)$ that acquires a mass of order the string scale via eating a closed string axion through the St\"uckelberg mechanism described above. At energies below $M_s$, the $U(1)$ appears effectively as global. The number of these open string axions depends on the details of the brane set-up;  it can turn out to be rather large in cases with large numbers of branes (if allowed by tadpole cancellation).

The dynamics of moduli stabilisation determines which axions, either closed or open string modes, can be kept light:
\begin{itemize}
\item \emph{D-term stabilisation}: For the sake of concreteness, let us consider again an anomalous $U(1)$ on a stack of D7-branes wrapping the divisor $D_i$. A non-zero gauge flux $F_2$ generates a moduli-dependent Fayet-Iliopoulos (FI) term in the D-term scalar potential which schematically looks like (see again App.~A of \cite{Cicoli:2011yh} for more details)
\begin{equation}
V_D \simeq g^2 \left(|C|^2-\xi_i\right)^2\,,
\end{equation}
where we assumed just one charged open string mode $C$, and the FI-term $\xi_i$ can be expressed in terms of the K\"ahler form $J = t^j\hat{D}_j$ and the $U(1)$ charges $q_{ij}$ (see (\ref{U(1)charges})) as
\begin{equation}
\xi_i \simeq \frac{1}{\mathcal{V}}\int_{D_i} J\wedge F_2 = -q_{ij}\frac{\partial K_0}{\partial\tau_j}\,.
\end{equation}
Moreover, the gauge coupling $g$ is set by the modulus controlling the volume of the divisor $D_i$, i.e.~$g^{-2}\simeq \tau_i$. Setting the D-terms to zero implies $|C|^2=\xi_i(\tau_j)$. This relation stabilises just one direction in the combined open and closed string moduli space. This saxionic direction corresponds to the supersymmetric partner of the linear combination of axions which is eaten by the anomalous $U(1)$. This, in turn, depends on how the F-terms affect moduli stabilisation. In fact, if F-term stabilisation leads to $\xi_i\neq 0$, a non-zero VEV for $|C|$ breaks the PQ $U(1)$ spontaneously and the phase $\zeta$ of $C$ turns out to be an axion with decay constant
\begin{equation}
\left(f_a^{\rm op}\right)^2=\langle |C|^2 \rangle =\xi_i\,.
\label{Dstab}
\end{equation}
Considering for simplicity just a single K\"ahler modulus $\tau$ charged under the $U(1)$, the mass of this Abelian gauge boson would then get two contributions:
\begin{equation}
m_{\gamma'}^2 \simeq g^2 \left[\left(f_a^{\rm op}\right)^2+\left(f_a^{\rm cl}\right)^2\right]\,,
\end{equation}
where the open and closed string axion decay constants $f_a^{\rm op}$ and $f_a^{\rm cl}$ are given by (see the mass formula (\ref{U(1)mass})):
\begin{equation}
\left(f_a^{\rm op}\right)^2 \simeq \left|\frac{\partial K_0}{\partial \tau}\right|
\qquad\text{and}\qquad \left(f_a^{\rm cl}\right)^2 \simeq \frac{\partial^2 K_0}{\partial \tau^2}\,.
\label{fs}
\end{equation}
If $f_a^{\rm op} \gg f_a^{\rm cl}$, as in the case of D7-branes wrapping cycles in the geometric regime, the combination of moduli fixed by D-terms is mostly $|C|$ and $\zeta$ is eaten by the anomalous $U(1)$. If instead $f_a^{\rm op} \ll f_a^{\rm cl}$, as for the case of D3-branes at singularities, the modulus frozen by D-terms is mostly $\tau$ and the eaten axion is $\theta$ \cite{Allahverdi:2014ppa}.

\item \emph{F-term stabilisation}: Any closed string axion $\theta$ enjoys a continuous shift symmetry, $\theta \to \theta + {\rm const}$, which is broken down to a discrete one only at non-perturbative level. On the contrary, the corresponding saxion $\tau$ is not protected by any symmetry, and so can develop a potential at both perturbative and non-perturbative level, leading to two situations:
\begin{enumerate}
\item If $\tau$ is fixed by perturbative effects, then $\theta$ is exactly massless at this level of approximation and its direction is lifted only by subleading non-perturbative effects. In this case $\tau$
and $\theta$ are stabilised by different effects, and so their masses are in general hierarchically different. In particular, $\tau$ can satisfy the cosmological moduli problem bound $m_\tau\gtrsim \mathcal{O}(50)$ TeV while $\theta$ can remain ultra-light since $m_{\theta}\sim m_\tau\,e^{-a\tau}\ll m_\tau$ for $a\sim\mathcal{O}(1)$ and $\tau\gg 1$.

\item If perturbative effects are made negligible by tuning some parameters, both $\tau$ and $\theta$ are fixed at non-perturbative level. Hence, they get a mass of the same order of magnitude,
rendering the axions rather heavy: $m_\theta \sim m_\tau\gtrsim \mathcal{O}(50)$ TeV. These masses are generically of order the gravitino mass $m_{3/2}$, and so if $m_\theta$ is lowered to smaller values relevant for phenomenology like $m_\theta \ll \mathcal{O}({\rm TeV})$
(assuming a solution to the cosmological moduli problem), one would obtain a tiny scale of supersymmetry breaking that is in tension with observations.
\end{enumerate}
\end{itemize}
A concrete framework where moduli stabilisation has been studied in detail is the type IIB LVS \cite{Balasubramanian:2005zx,Cicoli:2008va}. In this context, the moduli are fixed by the interplay of a number of possible contributions
to the scalar potential: tree-level background fluxes, D-terms, $\alpha'$ and $g_s$ perturbative corrections, and non-perturbative effects. Therefore, focusing on LVS models allows us to illustrate the implications of any moduli fixing effect for the dynamics of axion physics as follows \cite{Cicoli:2012sz}:
\begin{itemize}
\item The dilaton and complex structure moduli are fixed at the semi-classical
level by turning on background fluxes. The VEV of the flux-generated
superpotential is naturally of order unity, $W_0\sim\mathcal{O}(1)$.

\item The $h^{1,1}$ K\"ahler moduli $T_i=\tau_i + {\rm i} \theta_i$, where $\tau_i$ is the volume of the $i$-th internal 4-cycle and $\theta_i$ the corresponding axion, are flat directions at tree-level due to the no-scale cancellation.

\item The scalar potential for the $T$-moduli can be expanded in inverse powers of the CY volume $\mathcal{V}$. For $\mathcal{V}\gg 1$ (as required to trust the EFT), the dominant contribution arises from D-terms.

\item For vanishing open string VEVs, $d$ combinations of $T$-moduli are fixed by the D-term potential, and so $d$ axions get eaten by anomalous $U(1)$s. If $d=h^{1,1}$, the D-term conditions force the CY volume to collapse to zero size. Thus one has to choose a brane set-up and fluxes such that $d< h^{1,1}$. In this case, D-term fixing leaves $h^{1,1}-d \geq 1$ flat directions.

\item When charged open strings develop non-zero VEVs, the number of flat directions after D-term stabilisation is even higher since D-terms would fix open string modes and the eaten axions would be phases of open strings.

\item $n_{\rm np}$ del Pezzo divisors generate
single non-perturbative contributions to the superpotential whose existence is guaranteed by the rigidity of these cycles and the absence of any chiral intersection with the visible sector; $n_{\rm np}$ K\"ahler moduli together with their corresponding axions
develop a mass of order $m_{3/2}$ due to non-perturbative effects.

\item The remaining $n_{\rm ax} = h^{1,1}-n_{\rm np}-d$ moduli tend to be fixed perturbatively by $\alpha'$ or $g_s$ effects. Thus the corresponding axions remain massless and are good QCD axion candidates.
The main example is given by the volume mode $\mathcal{V}$ which develops an exponentially large VEV due to $\alpha'$ corrections: $\mathcal{V} \sim W_0 \,e^{\frac{2\pi}{N g_s}}$ where $N$ is the rank
of an $SU(N)$ theory which undergoes gaugino condensation ($N=1$ for Euclidean D3-instantons). Another example is given by two intersecting local blow-up modes supporting the visible sector, with one combination fixed by D-terms and the other by string loop corrections.

\item The $n_{\rm ax}$ massless axions are lifted by higher-order instanton effects. Given that for an arbitrary CY $h^{1,1}\sim\mathcal{O}(100)$, $n_{\rm ax}$ might turn out to be very large giving rise to an axiverse \cite{Arvanitaki:2009fg}.
\end{itemize}
Let us now comment on the expected values of the axion decay constants. As can be seen from (\ref{fs}), the decay constant of a closed string axion is controlled by the second derivative of the K\"ahler potential. This leads, schematically, to two situations:
\begin{itemize}
\item \textbf{Bulk cycles:} When the saxion $\tau$ is the volume of a bulk cycle, $f_a^{\rm cl}$ turns out to be of order the KK scale $M_{KK}$ since (reintroducing appropriate powers of $M_p$):
\begin{equation}
K_0=-3\ln\tau\qquad\Rightarrow\qquad\left(f_a^{\rm cl}\right)^2 \simeq \frac{\partial^2 K_0}{\partial \tau^2} M_p^2 \simeq \left(\frac{M_p}{\tau}\right)^2\sim M_{KK}^2\,.
\label{fbulk}
\end{equation}

\item \textbf{Local cycles:} When the saxion $\tau_{\rm loc}$ controls the volume of a local blow-up mode resolving a point-like singularity, $f_a^{\rm cl}$ is of order the string scale since:
\begin{equation}
K_0=-2\ln\mathcal{V} + \frac{\tau_{\rm loc}^2}{\mathcal{V}}\qquad\Rightarrow\qquad\left(f_a^{\rm cl}\right)^2 \simeq \frac{\partial^2 K_0}{\partial \tau_{\rm loc}^2} M_p^2 \simeq \left(\frac{M_p}{\sqrt{\mathcal{V}}}\right)^2\sim M_s^2\,.
\label{flocal}
\end{equation}
\end{itemize}
In both cases, $f_a^{\rm cl}$ is of order the cut-off of the effective 4D theory, implying that the PQ symmetry appears to be non-linearly realised. Moreover, if inflation is described within the low-energy theory, inevitably the Hubble scale during inflation is below the axion decay constant, $H_{\rm inf}<f_a^{\rm cl}$, leading to a pre-inflationary axion scenario. 

An interesting aspect is the study of the statistical distribution of the decay constants in the string landscape. One can fix the Calabi-Yau manifold, and so the number of axions, and see how $f_a^{\rm cl}$ of each axion changes when moving in moduli space along a trajectory corresponding to stabilised vacua. This analysis was performed in \cite{Broeckel:2021dpz} finding a logarithmic distribution:
\begin{equation}
N(f_{\theta_i})\sim \ln\left(\frac{f_{\theta_i}}{M_p}\right)\qquad\forall i=1,\cdots,n_{\rm ax}\,,
\end{equation}
where $N(f_{\theta_i})$ gives the number of stabilised flux vacua for a given value of $f_{\theta_i}$. Note that the axion mass spectrum is expected to have a similar logarithmic distribution \cite{Broeckel:2021dpz}. This result is complementary to the ones of \cite{Mehta:2020kwu,Mehta:2021pwf} which studied the distribution of the axion decay constants fixing a point in moduli space (the tip of the stretched K\"ahler cone) and varying the number of axions by considering different CY threefolds. In this case the axion decay constants follow an approximate log-normal distribution with their mean value decreasing when $h^{1,1}$ increases.

The situation is different for open string axions. In this case, as can be seen from (\ref{fs}), $f_a^{\rm op}$ is given by the first derivative of the K\"ahler potential. When $\tau\gg 1$ the first derivative is always larger than the second derivative, and so open string axions tend to be eaten by anomalous $U(1)$s. When instead $\tau$ shrinks to zero size, as for the case of a local blow-up mode $\tau_{\rm loc}$, the situation is opposite and an open string axion can survive in the EFT with a suppressed decay constant:
\begin{equation}
K_0=-2\ln\mathcal{V} + \frac{\tau_{\rm loc}^2}{\mathcal{V}}\quad\Rightarrow\quad\left(f_a^{\rm op}\right)^2 \simeq \frac{\partial K_0}{\partial \tau_{\rm loc}} M_p^2 \simeq \tau_{\rm loc} M_s^2\ll M_s^2\,,\quad \tau_{\rm loc}\ll 1\,.
\label{fopen}
\end{equation}
Explicit realisations of this scenario involve D3-branes at singularities which are sequestered from the fluxes in the bulk that source supersymmetry breaking \cite{Cicoli:2013cha,Allahverdi:2014ppa,Cicoli:2017zbx}. In this case, the axion decay constant can be much smaller than the cut-off scale resulting in a more standard linear realisation of the PQ symmetry that can be spontaneously broken at energies below inflation, realising a post-inflationary axion scenario with $f_a^{\rm op}< H_{\rm inf}$. A precise determination of the statistical distribution of open axion decay constants is still an open question that requires a better understanding.

%% file: WG1/content/1-1-3.tex
In order to find a viable QCD axion from string theory, besides understanding how to keep the axions light, one should also embed QCD in
CY compactifications. More generally, one should build consistent compact models with stabilised moduli and chiral non-Abelian gauge theories.

Type IIB compactifications seem to be a promising framework to achieve this goal because MSSM-like theories can live on D-branes, and moduli stabilisation can be achieved by a combination of fluxes, $\alpha'$ and $g_s$ corrections to the effective action. In this context, globally consistent CY models with QCD-like sectors and moduli stabilisation can be constructed using either intersecting fluxed D7-branes wrapping cycles in the geometric regime \cite{Cicoli:2011qg, Cicoli:2016xae,Cicoli:2017axo,Cicoli:2024bxw} or fractional D3-branes at CY singularities \cite{Cicoli:2012vw,Cicoli:2013cha,Cicoli:2013mpa,Cicoli:2017shd}.

Depending on the D-brane realisation of the QCD sector, the QCD axion can have a specific UV origin and different phenomenological features. To illustrate all the main possibilities, let us focus on type IIB LVS models with a CY volume of the form: 
\begin{equation}\label{eq:LVSvolume}
\mathcal{V}\,= \tau_b^{3/2}-\tau_{\rm np}^{3/2} - \tau_{\rm loc}^{3/2}-\lambda\left(\tau_{\rm int}-\tau_{\rm loc}\right)^{3/2}\,. 
\end{equation}
The modulus $\tau_b$ controls the overall volume, $\tau_{\rm np}$ is a blow-up mode supporting non-perturbative effects required for moduli stabilisation, $\tau_{\rm loc}$ is a modulus associated to a local shrinkable divisor, while $\tau_{\rm int}$ is another blow-up mode which intersects $\tau_{\rm loc}$ if $\lambda\neq 0$. In this scenario, the QCD axion can arise from three possibilities:
\begin{itemize}
\item \textbf{Bulk closed string QCD axion:} The `big' modulus $\tau_b$ is stabilised perturbatively by $\alpha'$ effects, and so the corresponding axion $\theta_b$ is a promising QCD axion candidate. If QCD is built with D7-branes wrapping $\tau_b$, the fact that the value of the QCD gauge coupling at the cut-off scale is set by $\tau_b$ itself, i.e. $\alpha_{\rm QCD}^{-1}\simeq\tau_b\sim \mathcal{O}(25)$, implies a relatively small compactification volume, $\mathcal{V}\sim \mathcal{O}(10^2)$. In turn, using (\ref{fbulk}), the QCD axion decay constant turns out to be very high, $f_a\sim M_p/\tau_b\sim \mathcal{O}(10^{17})\,{\rm GeV}$. Moreover, the contribution to the mass of $\theta_b$ from E3-instantons becomes negligible, $m_{\theta_b} \sim M_p\, e^{-2\pi \tau_b}\sim 0$. Hence $\theta_b$ can acquire mass via QCD instantons and play the role of the QCD axion with mass $m_{\theta_b}\sim \Lambda_{\rm QCD}^2/f_a\sim \mathcal{O}(10^{-10})\,{\rm eV}$. The initial misalignment angle of a QCD axion with such a high decay constant would have to be very small in order to avoid axion DM overproduction. Furthermore, the soft supersymmetry breaking scale in this class of models would be very high, of order $M_{\rm soft}\sim m_{3/2}\sim M_p/\mathcal{V}\sim \mathcal{O}(10^{16})\,{\rm GeV}$, requiring a very high level of tuning to solve the Higgs hierarchy problem. Given that $f_a$ is of order the KK scale, this model would correspond to a pre-inflationary scenario if inflation is described within a 4D EFT. Isocurvature bound would then set a strong upper bound on $H_{\rm inf}$ depening on the QCD axion contribution to DM.

\item \textbf{Local closed string QCD axion:} If the QCD sector is built with magnetised D7-branes wrapping the rigid divisor $\tau_{\rm loc}$, the QCD axion could be realised as the local closed string axion $\theta_{\rm loc}$. There are two ways to obtain a light $\theta_{\rm loc}$ depending on open string VEVs. For vanishing open string VEVs, D-term stabilisation corresponds to setting the FI-term to zero. This condition, in turn, for a suitable choice of gauge flux quanta, would fix $\tau_{\rm int}$ in terms of $\tau_{\rm loc}$ as $\tau_{\rm loc}\propto \lambda^2\left(\tau_{\rm int}-\tau_{\rm loc}\right)$ if $\lambda\neq 0$ \cite{Cicoli:2022fzy}. This relation leaves a flat direction in the ($\tau_{\rm loc}, \tau_{\rm int}$) space which can be lifted at subleading order by string loops \cite{Cicoli:2011qg,Cicoli:2012sz}. The corresponding closed string axion, which we parametrise as $\theta_{\rm loc}$, could then play the role of the QCD axion. The orthogonal closed string axion, $\theta_{\rm int}$, would instead be eaten by the anomalous $U(1)$. Alternatively, if charged open string modes develop non-zero VEVs, one can consider a simpler model with $\lambda=0$. In this case D-term fixing stabilises a charged open string mode in terms of the $\tau_{\rm loc}$-dependent FI-term and the corresponding open string axion is eaten by the anomalous $U(1)$. The local modulus $\tau_{\rm loc}$ is then stabilised, as outlined in \cite{Broeckel:2021dpz}, by a combination of string loops and soft terms for the open string modes. The closed string axion $\theta_{\rm loc}$ remains again a perfect candidate to behave as the QCD axion.

Being an axion associated to a local cycle, as can be seen from (\ref{flocal}), its decay constant is set by the string scale, $f_a\sim M_s\sim M_p/\sqrt{\mathcal{V}}$. Contrary to the case of a bulk closed string axion, the value of $f_a$ is decoupled from the value of the QCD gauge coupling since $f_a$ is controlled by $\mathcal{V}$ while $\alpha^{-1}_{\rm QCD}\sim \tau_{\rm loc}$. Hence in this case $f_a$ can take lower values by considering larger values of the CY volume. Interestingly, $\mathcal{V}\sim \mathcal{O}(10^{14})$ would give $M_{\rm soft}\sim m_{3/2}\sim \mathcal{O}(10)\,{\rm TeV}$ and $f_a\sim\mathcal{O}(10^{11})\,{\rm GeV}$ \cite{Conlon:2006tq}, which would match the observed DM abundance for an $\mathcal{O}(1)$ misalignment angle. However for $\mathcal{V}\sim \mathcal{O}(10^{14})$, the mass of the lightest modulus, which turns out to be $\tau_b$, is well below the cosmological moduli problem bound: $m_{\tau_b}\sim M_p/\mathcal{V}^{3/2}\sim\mathcal{O}(1)\,{\rm MeV}$. Unless a dilution mechanism is at play, $m_{\tau_b}$ should be above the cosmological moduli problem lower bound. This requires $\mathcal{V}\lesssim \mathcal{O}(10^9)$ which would give $m_{\tau_b}\gtrsim \mathcal{O}(50)\,{\rm TeV}$. In turn, the axion decay constant would be pushed to larger values of order $f_a\gtrsim \mathcal{O}(10^{13})\,{\rm GeV}$, requiring a few percent tuning of the initial misalignment angle. Note that this model would lead again to a pre-inflationary scenario with isocurvature constraints since $f_a$ is given by the string scale.

\item \textbf{Open string QCD axion:} When $\lambda=0$ in (\ref{eq:LVSvolume}) and the VEVs of all charged open string modes are vanishing at leading order, D-term stabilisation sets the FI-term $\xi\sim \tau_{\rm loc}\,M_s$ to zero, forcing the local divisor $\tau_{\rm loc}$ to shrink to zero size, $\tau_{\rm loc}\to 0$. This leads to scenarios where the QCD sector can live on fractional D3-branes at the singularity obtained by collapsing $\tau_{\rm loc}$. As explained in Sec.~\ref{sec:StringAxionsStats}, in this case the closed string axion $\theta_{\rm loc}$ gets removed from the low-energy spectrum via the St\"uckelberg mechanism. The remaining candidate to play the role of the QCD axion is the phase $\zeta$ of a charged open string mode $C=|C|\,e^{i\zeta}$. Note that exactly supersymmetric solutions correspond to $\langle|C|\rangle=0$ which would not break the PQ symmetry. However, full vacuum solutions in string compactifications involve supersymmetry breaking effects which can give a shift of the FI-term from zero. In sequestered models with D3-branes at singularities this shift can be a small effect, $\tau_{\rm loc}\sim \mathcal{V}^{-1}\ll 1$, resulting in a spontaneous breaking of the PQ symmetry at energies below the cut-off scale of the 4D low-energy effective theory \cite{Cicoli:2013cha,Cicoli:2017zbx}. In fact, combining (\ref{Dstab}) with (\ref{fopen}) and $\tau_{\rm loc}\sim \mathcal{V}^{-1}\ll 1$, one has
\begin{equation}
f_a^{\rm op}=\langle |C| \rangle =\sqrt{\xi}\simeq \sqrt{\tau_{\rm loc}} \,M_s\simeq m_{3/2}\ll M_s\,.
\end{equation}
Given that $f_a^{\rm op}\ll M_s$, the PQ symmetry associated to an open string QCD axion can be broken after inflation avoiding  problems with isocurvature bounds. In sequestered scenarios, $\mathcal{V}\sim \mathcal{O}(10^7)\,{\rm GeV}$ would lead to low-energy supersymmetry: $M_{\rm soft}\sim M_p/\mathcal{V}^2\sim\mathcal{O}(10)\,{\rm TeV}$ \cite{Blumenhagen:2009gk,Aparicio:2014wxa}. In turn, the mass of the lightest modulus would be well above the cosmological moduli problem bound, $m_{\tau_b}\sim \mathcal{O}(10^7)\,{\rm GeV}$, and the axion decay constant, as well as the gravitino mass, would lie in the intermediate regime, $f_a^{\rm op}\sim m_{3/2}\sim \mathcal{O}(10^{11})\,{\rm GeV}$.
Note that this value is in the right ballpark to reproduce the correct DM abundance in the axion post-inflationary scenario.

Let us finally mention that in this scenario, the decay of the lightest modulus could lead to additional non-thermal contributions to both dark matter (in the form of Wino/Higgsino \cite{Allahverdi:2013noa}) or dark radiation (in the form of ultra-light bulk closed string axions $\theta_b$ \cite{Cicoli:2012aq, Cicoli:2015bpq}). 
\end{itemize}

%% file: WG1/content/quality-problem.tex
The so-called Peccei-Quinn quality problem~\cite{Georgi:1981pu,Lazarides:1985bj,Kamionkowski:1992mf, Ghigna:1992iv, Barr:1992qq, Holman:1992us} refers to explicit CP-violating terms that can shift the QCD axion minimum away from the value $\langle\theta_{\rm QCD}\rangle \lesssim 10^{-10}$ required by the neutron
electric dipole moment measurement~\cite{Abel:2020pzs}. 

In field theory, the PQ symmetry is typically provided in terms of an ordinary (0-form) global symmetry under which a complex scalar, $\Phi$, is charged.  This implies that in the absence of additional structure or symmetries it can be broken by local operators. In the case the axion comes from the phase of a complex scalar, $\Phi$, then one can in principle add a tower of higher-dimensional operators~\footnote{This assumes that only irrelevant operators are able to break the PQ symmetry explicitly. If marginal or relevant operators are allowed to break PQ explicitly, the quality problem would be more severe.}

\begin{equation}\label{eq:PQV_operators}
    \mathcal{L}_{PQV} = \sum_{N>4} \frac{c_N}{\Lambda_{\rm UV}^{N-4}}\Phi^N +\rm{h.c.}
\end{equation}

The UV scale, $\Lambda_{\rm UV}$, is the scale at which new degrees of freedom break PQ explicitly. It is typically associated to the 4-dimensional Planck scale, $M_P$, although in principle might be lower\footnote{For example, one expects that in string theory $\Lambda_{\rm UV}\lesssim M_s$, with $M_s$ being the string scale.}.

The higher-dimensional operators in~\eqref{eq:PQV_operators} may pose a problem for the QCD axion solution to the strong CP because, once $\Phi$ takes a VEV, the minimum is displaced with respect to the CP conserving vacuum.  This can be seen by minimizing the total potential, which includes the new terms in addition to the QCD contribution

\begin{equation}
    V=
    -m_\pi^2 f_\pi^2\cos (\theta_{\rm QCD})-\Lambda_{\rm PQV}^4\cos (\theta_{\rm QCD}+\delta)\,, \text{ with: }\, \Lambda_{\rm PQV}^4= c_N\frac{\langle \Phi \rangle^N}{\Lambda_{\rm UV}^{N-4}}\,.
\end{equation}

For simplicity, we have assumed that the scale $\Lambda_{\rm PQV}$ is dominated by the leading, irrelevant operator and the phase $\delta$ accounts for the fact that the minimum induced by~\eqref{eq:PQV_operators} is not necessarily aligned with the vacuum of the potential induced by QCD. Minimisation of the potential allows us to find the axion field value,

\begin{equation}
    \langle \theta_{\rm QCD} \rangle \approx \frac{\Lambda^4_{\rm PQV}}{m_\pi^2 f_\pi^2}\sin \delta \lesssim 10^{-10}\,,
\end{equation}

where in the last equation we assumed that the new PQ violating contributions are small enough so that the strong CP problem is solved. 
Assuming that $\langle\Phi \rangle \geq 10^8$ GeV, for order 1 Wilson coefficients $c_N\sim O(1)$, the latter requires $N>9$ or 10. The quality problem gets more and more severe as the decay constant approaches $\Lambda_{\rm UV}$.

In recent years, many efforts in field theory model building have been dedicated to finding new mechanisms to solve this problem. These can be classified in two main categories. The first ones involves imposing continuous or discrete gauge symmetries~\cite{Babu:2002ic,Ringwald:2015dsf,Fukuda:2017ylt,DiLuzio:2017tjx,Bonnefoy:2018ibr,Ardu:2020qmo,DiLuzio:2025jhv} that forbid every PQ breaking operator at dimension $N<9$ or $10$. The axion quality is \textit{accidentally emerging} from the fact that only very suppressed, irrelevant operators are both gauge invariant and PQ violating. The second class of models use compositeness~\cite{Randall:1992ut,Dobrescu:1996jp,Redi:2016esr,Lillard:2018fdt,Gavela:2018paw,Contino:2021ayn,Gherghetta:2025fip,Agrawal:2025mke}. In this case, the axion is not an elementary degree of freedom but a composite state, generated by some sort of strong dynamics. In the IR, the effect of compositeness is also to help to forbid PQ breaking operators below a certain dimension, therefore enhancing the axion quality with respect to minimal models where the axion is elementary.
\\

Models with axions arising from higher dimensional gauge fields 
do not admit  a linearly realised
$U(1)_{\rm PQ}$, but yet have a nonlinearly realised $U(1)_{\rm PQ}$
in the low energy limit, which is related to the higher-dimensional gauge symmetry of the model. Such symmetry is locally well-defined, but globally ill-defined, implying that it can be broken {\it only} by non-local effects associated with fields charged under the higher-dimensional gauge symmetry. Those effects are therefore always {\it exponentially suppressed} in the radius of the extra dimension(s). 

More concretely, in string theory, quality-spoiling terms arise from small QCD gauge instantons and stringy instantons. Compliance with bounds on the neutron electric dipole moment implies 
\begin{equation}
\Delta \theta_{\rm QCD} + \Delta\theta_{\rm stringy} \lesssim  10^{-10}\,.
\end{equation}
Thus, for the QCD axion to have high enough quality to solve the strong CP problem, the scale of stringy CP-violating terms must be much lower than the contribution of non-perturbative QCD effects to the potential.   The largest contribution generically comes from QCD gauge instanton effects. However, assuming that CP is not an approximate symmetry of the UV theory, such contribution shifts the minimum of the QCD axion potential. As shown in~\cite{Demirtas:2021gsq}, if one assumes MSSM particle content and no additional gauge sector, the induced shift can be given in terms of the SUSY partner masses as

\begin{equation}
    \Delta\theta_{\rm QCD}\sim 10^{-12}\left(\frac{1\, \text{TeV}}{M_{\rm SUSY}}\right)^3\,,
\end{equation}
where $M_{\rm SUSY}$ is the SUSY-breaking scale. Therefore, for $M_{\rm SUSY}> 1$  TeV, $\Delta \theta_{\rm QCD}$ is within the experimental bound. 

Stringy instanton shifts originate from the non-perturbative contributions to the axion potential due to branes wrapping 
appropriate submanifolds; the magnitude of $\Delta \theta_{\mathrm{stringy}}$ can be computed as
\begin{align}
    \Delta \theta_{\mathrm{stringy}} = \frac{\Lambda^4_{\mathrm{PQV}}}{\chi(0)_{\mathrm{QCD}}}\,,
\end{align}
where $\Lambda^4_{\mathrm{PQV}}$ is the largest CP-violating instanton scale of the model (see Eq.~\eqref{eq:instanton_scalesIIB}), and $\chi_{\rm{QCD}}(T)$ is the (temperature-dependent) topological susceptibility which measures fluctuations of the topological charge in the QCD vacuum, such that \begin{equation}
    m_{a,\rm QCD}(T)= \left(\chi_{\rm QCD}(T)\right)^{1/2}/f_{a,\rm QCD}\,.
\end{equation}

As argued before from non-locality and shown in Eq.~\eqref{eq:instanton_scalesIIB}, $\Lambda^4_{\rm PQV}$ is always exponentially small, making it relatively easy to have $\Delta \theta_{\mathrm{stringy}}\ll  \Delta\theta_{\rm QCD}$. Additionally, models exhibiting  ultralight axions in addition to the QCD one favor a high PQ quality \cite{Sheridan:2024vtt}. 
A typical ordering of scales in well-motivated string models is
\begin{equation}
\Lambda_{\rm PQV}^4 <\Lambda_{\rm lightest \, axion}^4\leq \Lambda_{\rm fuzzy}^4\,,
\end{equation}
where $\Lambda_{\rm fuzzy}^4$ refers to the instanton scale needed to have an axion with fuzzy features (see Sec.~\ref{sec:DarkALPs}).
This hierarchy in turn means
\begin{equation}\label{eq:PQscenario1}
    \frac{\Lambda^4_{\rm PQV}}{\chi(0)_{\mathrm{QCD}}} \lesssim 10^{-10} \,\,\Leftrightarrow\,\,  \frac{m_{\rm lightest\, axion}}{m_\mathrm{QCD}} \lesssim 10^{-5}\,,
\end{equation}
assuming roughly comparable decay constants. We see that 
when \eqref{eq:PQscenario1} holds, 
the quality problem is always solved by requiring a fuzzy axion. On the other hand, if $\Lambda_{\mathrm{PQV}}^4>\Lambda_{\rm lightest \, axion}^4$, it is not guaranteed that the quality problem is solved.  

%% file: WG1/content/1-1-4.tex
\subsubsection{Dark matter and dark radiation}\label{sec:DarkMatterRadiation}

Axions can have numerous observable implications in astrophysics and cosmology. For instance, they might form dark matter (DM) or dark radiation (DR) in our universe, or even dark energy if their mass is close to the Hubble scale today, while an axion with specific features could have played a key role during cosmological inflation. Given that string theory predicts a multitude of axions with masses highly sensitive to the geometry of extra dimensions, it is worthwhile to explore their various cosmological features.

As discussed in Section~\ref{sec:1.1.1}, closed string axions derive from higher-dimensional gauge fields. This implies that there is no symmetry to break as the universe cools down: PQ symmetry is always already broken in 4D, and no PQ phase transition happens. Most notably, there is no point in field space where PQ symmetry can be restored~\cite{Reece:2025thc}.\footnote{Here we discuss only \emph{closed} string axions; other types of axions do arise when compactifying the extra dimension. In particular, QFT-like axions can arise from the open string sector. Since these behave just like the QFT axions, we focus on the implications of closed axions since those are of truly stringy origin.} Therefore, closed string axions effectively belong to the pre-inflationary scenario, irrespective of their decay constant, and are already present in the universe as scalars with a discrete shift symmetry. This should be accounted for when doing model building from string theory. 

First, after inflation happened, the initial angle $\theta_0$ can be drawn from a uniform distribution over the interval $[0,2\pi]$, but the existence of many different Hubble patches before inflation means that $\theta_0$ is effectively a free parameter and values near the extrema of the interval cannot be excluded. Second, there is no domain wall formation after inflation; for string axions, a relevant constraint might come from isocurvature bounds~\cite{Reece:2025thc,Fox:2004kb,Hertzberg:2008wr}. Isocurvature perturbations arise due to quantum fluctuations of the axion field during inflation. The power spectrum of the axion density fluctuations is related to the scale of inflation $H_{\rm inf}$, the decay constant and $\theta_0$.  
If (one of) the string axions behave as DM, typical values of the decay constant in the string landscape~\cite{Demirtas:2018akl,Mehta:2021pwf} predict low-scale inflation (modulo fine-tuned values of $\theta_0$, see~\cite{Visinelli:2017imh} for the inclusion of anharmonicities). However, the isocurvature bound applies under the assumption that axions populate the dark sector and provide a large, possibly total, contribution to the dark matter abundance. Another interesting signal could come from primordial non-Gaussianities, where axions coupled to the inflaton leave a distinct imprint and need not have a non-trivial present-day abundance~\cite{Chakraborty:2023eoq,Chakraborty:2025myb}. Finally, depending on how these axions are coupled to the whole spectrum of particles, their phenomenology can be quite rich but also create problems in reproducing our universe.

A cosmic population of axions can be produced in a model-independent way from the misalignment mechanism, or by model-dependent cosmological features and dynamics. Let us first consider the former. At early times, the axion field has already a potential but it is frozen at its initial value by the Hubble friction and can be misaligned away from its minimum. Then, when $H< m_a$, the field can begin to coherently oscillate around the minimum and its equation of state becomes $w = 0$, thus behaving as matter. An axion heavier than the Hubble expansion rate at matter-radiation equality in $\Lambda$CDM, i.e.~with $m_a \gtrsim 10^{-28}$ eV, begins oscillating during radiation domination and represents a valid dark matter candidate. In particular, one can compute its abundance as 
\begin{equation}
	\label{eq:DMabundance}
	 \Omega_{a}h^2\Bigr|_{\rm r} \approx 0.12\left(\frac{f_a}{10^{16}\,{\rm GeV}}\right)^2\left(\frac{m_a}{4.4\times10^{-19}\,{\rm eV}}\right)^{1/2}\theta_0^2\,.
\end{equation}
Instead, if the axion mass is $m_a \lesssim 10^{-28}$, then it starts to oscillate during matter domination, and its relic density can be computed from
\begin{equation}
    \Omega_a\,h^2\Bigr|_{\rm m} \approx 7.6\times10^{-6}\left(\frac{f_a}{10^{16}\,{\rm GeV}}\right)^2\theta_0^2\,,
\end{equation}
hence, its contribution to DM is negligible. 
Finally, all axions with $m_a \lesssim 3H_0$ have not started to oscillate yet, and can only contribute towards dark energy. 

Assuming $\theta_0\sim\mathcal{O}(1)$, from \eqref{eq:DMabundance} we see that an axion with $m_a \sim 10^{-19}$ eV and $f_a \sim 10^{-16}$ GeV is a natural candidate for constituting the total of DM, and can be embedded naturally in string theory \cite{Hui:2016ltb,Cicoli:2021gss,Sheridan:2024vtt}. This regime makes such axion part of the \emph{fuzzy} DM scenario  \cite{Hu:2000ke, Goodman:2000tg, Peebles:2000yy, Amendola:2005ad}.

String theory statistically predicts the existence of hundreds of axions in the EFT and while the mass is exponentially sensitive to the volume of internal sub-manifolds, the decay constant is only linearly so. Using \eqref{eq:masses_decay_constants_C4}, we can approximate \eqref{eq:DMabundance} as
\begin{equation}
    \frac{\Omega_{a}h^2}{0.12}\propto e^{-S_a/4} f_a^{3/2} \theta_0^2\,.
\end{equation}
Axions with smaller instanton actions $S_a$ are more represented: the abundance is dominated by the heavier axions since big hierarchies among the decay constants are not generic in the string landscape. Therefore, overabundance is a clear obstacle for the string axiverse, as the total abundance due to \emph{all} axions must be
\begin{equation}
     \Omega_{\rm total}h^2 = \Omega_{a,\rm QCD}h^2 + \sum_{a}\Omega_a h^2\Bigr|_{\rm m} + \sum_{a}\Omega_a h^2\Bigr|_{\rm r} + \sum_{a}\Omega_a h^2\Bigr|_{w_R} \leq \Omega_{\rm DM}h^2 \approx 0.12\,,
\end{equation}
where we also included the contributions from the QCD axion and those axions that start oscillating during the reheating phase with equation of state $w_R$.

An eventual detection of a sizable fuzzy DM fraction would imply that to avoid overabundance all heavier axions have been diluted away, their initial angles take values much smaller than the fuzzy axion one,\footnote{See also~\cite{Freivogel:2008qc,Arvanitaki:2009fg,Reig:2021ipa,Kaloper:2024lrr}.} or their decay constant is orders of magnitude smaller than $M_P$.\footnote{This would however generate a scenario where isocurvature bounds on $H_{\rm inf}$ become highly relevant and should be accounted for.} The latter outcome would place the UV completion in a non-generic, special locus of the string landscape. While these requirements represent a challenge for string theory model building, refs.~\cite{Cicoli:2021gss,Sheridan:2024vtt} showed that type IIB compactifications on Calabi-Yau orientifolds can give rise to axion DM with abundances that may be accessible to future
observations. Thus, detecting fuzzy axions from string theory is possible and could inform us about our place in the landscape.

A relic axion population can also be produced via the decay of parent particles and topological defects, through quantum fluctuations during preheating or thermally with the cosmic microwave background. These production mechanisms can deliver relativistic axions; if they stay relativistic until today, they can constitute DR and form a cosmic axion background \cite{Dror:2021nyr}. The cosmic axion background is a generic prediction of string and M-theory compactifications~\cite{Marsh:2015xka,Acharya:2010zx,Cicoli:2012aq,Conlon:2013isa,Higaki:2013lra,Hebecker:2014gka,Cicoli:2022fzy,Cicoli:2018cgu}. The motivation can be traced back to the structure of 4D $\mathcal{N}=1$ supermultiplets. The existence of moduli is generic in any extra-dimensional theory, where the saxions arise from the higher-dimensional modes of the graviton. In superstring theories, they are supersymmetrically partnered with axions (they belong to the same 4D $\mathcal{N}=1$ chiral supermultiplet)~\cite{Grimm:2004uq}, and form the complex field $T$ as in Eq.~\eqref{eq:complexKahler}. Hence, axions and saxions are always coupled via the axion kinetic terms and in the supergravity scalar potential, see Eqs.~\eqref{eq:LagrangianALPsIIB}\eqref{eq:FtermPotential}. Different phenomenologies depend ultimately on the magnitude and UV features of these couplings. 

In general, models of string cosmology are known to suffer from issues of overproduction of DR. An observational signature of (string) axions behaving as DR would arise as a deviation $\Delta N_{\rm eff}$ from the number of relativistic degrees of freedom $N_{\rm eff}$ in the SM, i.e. as a nonzero $\Delta N_{\rm eff}=N_{\rm eff}-3.046$. Current constraints from the CMB place the bound $\Delta N_{\rm eff}<0.226$~\cite{Yeh:2022heq}.

The most studied case for string axions production is through the decay of saxions $\phi^i$~\cite{Higaki:2013lra,Cicoli:2012aq,Conlon:2013isa,Hebecker:2014gka,Angus:2014bia,Acharya:2015zfk, Cicoli:2018cgu,Cicoli:2022uqa,Cicoli:2022fzy}. Because of their coupling, the heavier saxion can easily decay into the (usually) lighter axion, with a rate $\Gamma_{\phi\rightarrow a a}=m_\phi^3/48 \pi M_P^2$.

The production of axions via parametric resonance during preheating~\cite{Traschen:1990sw,Shtanov:1994ce,Kofman:1994rk,Kofman:1997yn} in a string theory context was first explored in \cite{Leedom:2024qgr}. The mass of a string axion depends on the vacuum expectation values of saxions, cf.~Eq.~\eqref{eq:masses_decay_constants_C4}. If one of such saxions is identified with the inflaton, the axion mass varies with time as the inflaton oscillates about the minimum of its potential while undergoing parametric resonance. The inherently stringy exponential dependence on the saxion vev gives rise to a different phenomenology compared to typical EFT studies: the system must be studied with a generalisation of the Mathieu equation, and oscillations in the exponential coupling lead to drastic damping. The axion mass determines the oscillation strength and particle production rate. Parametric resonance produces ultra-light axions primarily through kinetic mixing; this production is unavoidable but is small and well below upcoming experimental probes of $\Delta N_{\rm eff}$. In contrast, when the exponential coupling via instanton contributions dominates, the axion-saxion coupling is highly non-linear. It thus drives strong parametric resonance, resulting in a massive production of superheavy axions. These will then have to decay to avoid issues with BBN.

\subsubsection{Axionic quintessence}

Recent cosmological data points to a universe at the onset of a phase of accelerated expansion, characterized by a Hubble parameter $H_0\sim 10^{-60}M_P^2$ \cite{Planck:2018vyg}. Within Einstein-Hilbert gravity such acceleration is driven by an unknown fluid with negative pressure which is often parametrised as $p=\omega \rho$ with $\omega \approx-1$. The simplest explanation for this observation would be a cosmological constant with an energy density $\rho_\Lambda=10^{-120} M_P^4$ and with $\omega=-1$. Despite its apparent simplicity, much remains to be understood about the cosmological constant, in particular in terms of its description in quantum field theory (for a review see e.g.~\cite{Weinberg:1988cp,Polchinski:2006gy,Padilla:2015aaa}). 

Though the most recent cosmological data present a mild preference for dynamical dark energy (DE) models  \cite{DESI:2025zgx}, a true cosmological constant persists as a possible source for the universe's late time acceleration. The canonical example of these dynamical DE models involves a minimally coupled scalar field endowed with a suitably flat potential and is often called \emph{quintessence}. The dynamics of minimally coupled (both to gravity and to other cosmological fluids) quintessence models is similar to that of inflation in the early universe albeit at a much lower energy scale. 
Dynamical DE models must address the same problems as a cosmological constant, like e.g.~the origin of the DE scale, and must also explain radiative stability of the quintessence potential and provide a mechanism for evading fifth-force constraints that arise from the presence of light scalars in the late universe.

Using axions to drive late-time acceleration presents some important advantages when compared with scalar-driven quintessence. These advantages follow from the axion's shift symmetry, in particular
\begin{itemize}
\item fifth forces: axions couple to matter via the axial vector current and so would require spin-polarised sources to test for fifth forces. Given that current bounds are obtained from macroscopic objects, axionic quintessence models automatically evade these constraints.
\item scale suppression: since the axion is perturbatively flat and its scalar potential is only generated non-perturbatively, it will typically be exponentially suppressed, a feature that can help to accommodate the extremely low value of $H_0$.
\item radiative stability: the perturbative shift symmetry implies a suppression of the radiative corrections that would otherwise lift the axion's mass away from the low values required for quintessence.
\end{itemize}

Let us then consider an axion in the late universe, with a canonical kinetic term and  potential of the form
\begin{equation}
    V=V_0 \left(1-\cos\frac{\phi}{f_a}\right)\,,
\end{equation}
where in order to drive quintessence the scale of the potential must obey $V_0\sim 10^{-120}M_P^4$. Depending on the decay constant $f_a$ the late time dynamics arises when the axion is rolling between the inflection point and the minimum $f_a>M_P$ or between the maximum and the inflection point $f_a<M_P$.

Trans-Planckian decay constants have received much attention in connection to inflation and the weak gravity conjecture\footnote{See Section \ref{sec:swampland} for details on the Swampland program.} and, though attainable, are generically harder to achieve from a UV point of view. This fact coupled with the de Sitter swampland conjecture \cite{Ooguri:2018wrx} has prompted interest in axionic hilltop quintessence models where, by virtue of the smallness of $f_a$, the late time dynamics happens in the vicinity of the maximum of the potential  \cite{Cicoli:2021skd,Cicoli:2024yqh}. 

\begin{figure}[t!]
    \centering
    \includegraphics[width=0.5\linewidth]{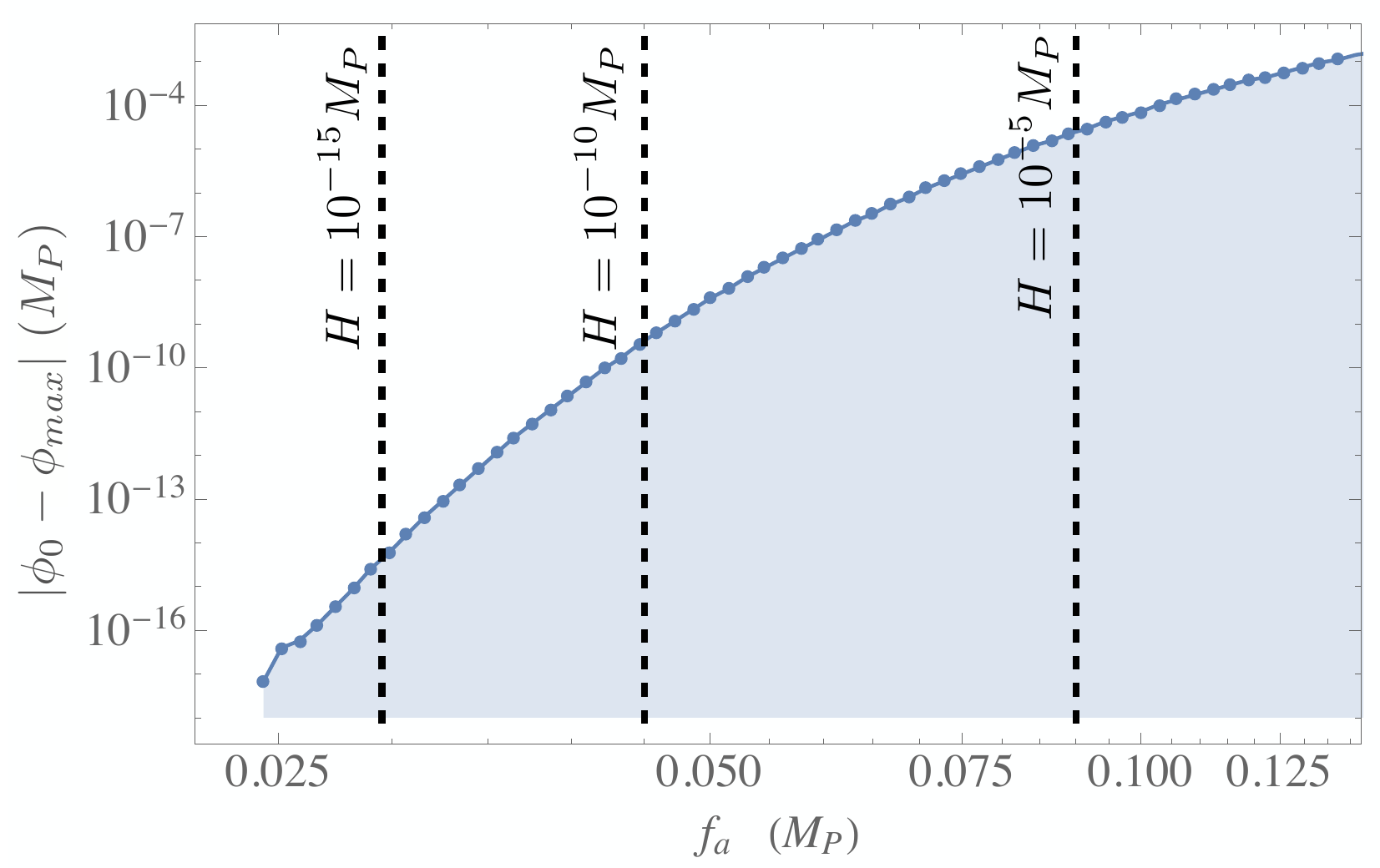}\hfill
        \includegraphics[width=0.5\linewidth]{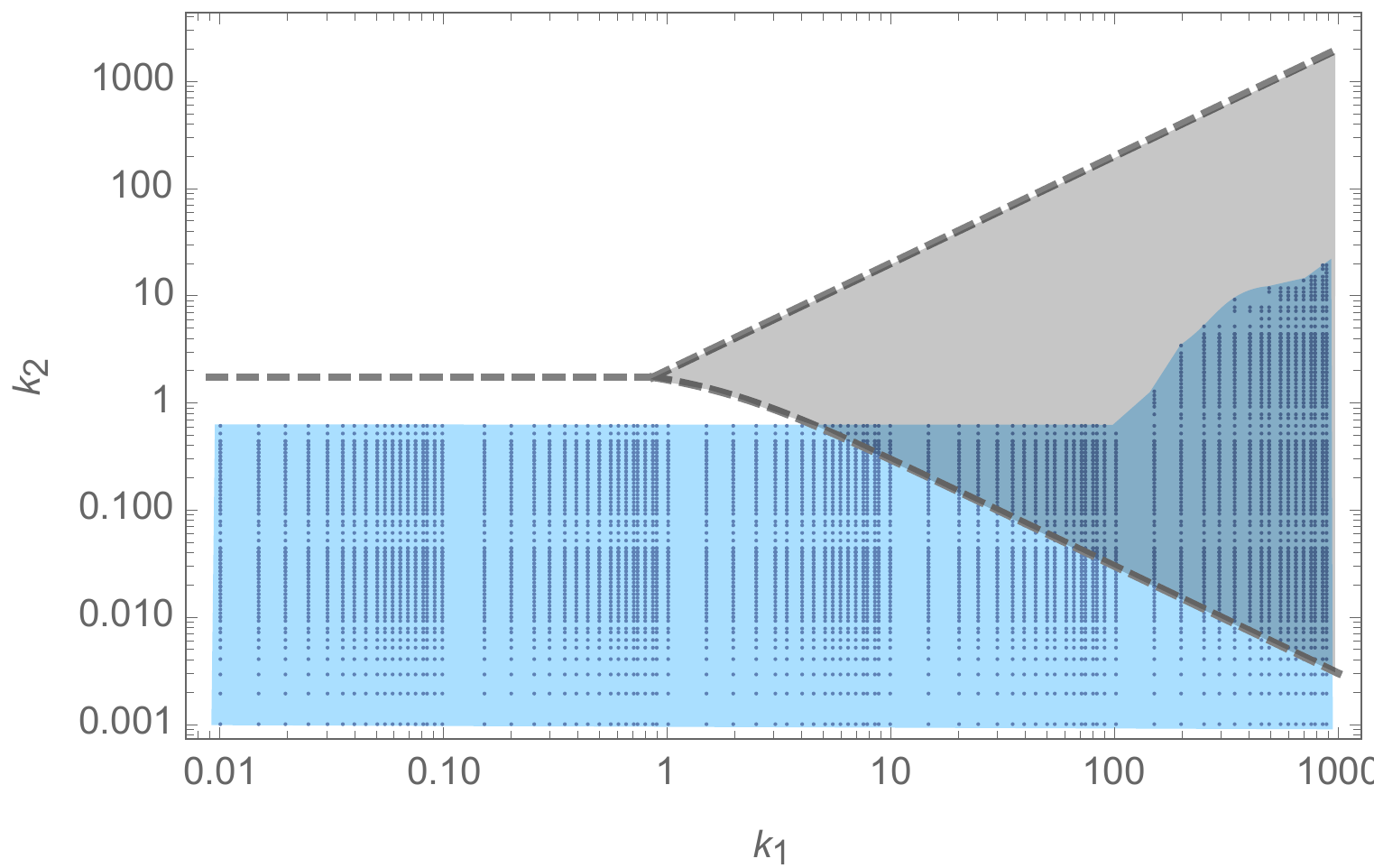}
    \caption{\emph{Left}: Relation between the decay constant $f_a$, the valid initial condition range $|\phi_0-\phi_{max}|$ and the inflationary scale $H_{inf}$ for axionic hill-top quintessence  \cite{Cicoli:2021skd}. \emph{Right}: Observational bounds on the slopes of the scalar potential and kinetic coupling for multifield quintessence models. The solution with non-trivial axionic dynamics is a stable attractor in the grey-shaded region. Adapted from Ref.~\cite{Cicoli:2020noz}.}
\label{fig:axionQuint}
\end{figure}

In axionic hilltop quintessence models there is a trade-off between the smallness of the decay constant and the viable range for the initial misalignment of the axion: the smaller $f_a$, the closer $\phi_0$ needs to be to the hill-top \cite{Cicoli:2021skd, Kaloper:2005aj}. This issue is compounded by stochastic effects in the early universe, in particular during inflation, that tend to blur the choice of initial conditions $\phi_0$ by a factor of $\delta \phi\sim H_{inf}$. For a robust late universe model one needs $|\phi_0-\phi_{max}|>H_{inf}$ which establishes a curious connection between early and late universe physics, c.f.~Fig.~\ref{fig:axionQuint}. For example if the inflationary scale is $10^{-5} M_P$ then $|\phi_0-\phi_{max}|>10^{-5} M_P$ which leads to a viable quintessence model if $f_a>0.08 M_P$ \cite{Cicoli:2021skd}. For a recent analysis constraining hilltop (both scalar and axionic) quintessence models using the latest observational data see \cite{Bhattacharya:2024kxp}.

An alternative way in which axions may play a role in late-time acceleration is by participating in multi-field dynamics together with their saxionic partners/moduli\cite{Boyle:2001du,Sonner:2006yn,Cicoli:2020cfj,Cicoli:2020noz,Brinkmann:2022oxy}. Considering that both scalar and pseudo-scalar originate from the same $\mathcal{N}=1$ supergravity chiral superfield, upon canonical normalisation of the modulus, the action can be written as
\begin{equation}
\frac{\mathcal{L}}{\sqrt{|g|}}=\frac{1}{2}(\partial \xi)^2+\frac{f^2(\xi)}{2}(\partial \phi)^2 -V(\xi)\,,
\end{equation}
where the kinetic coupling function $f(\xi)$ is uniquely determined by the K\"ahler potential and we are assuming that the axions' continuous shift symmetry is unbroken. For constant and  non-vanishing $k_1\equiv-\frac{f_\xi}{f M_P}$  and $k_2\equiv-\frac{V_\xi}{V M_P}$ the system features a solution of the  form \cite{Cicoli:2020noz}
\begin{equation}
        \dot{\xi}=\frac{6 H M_P}{2 k_1+k_2} \qquad \text{and} \qquad f(\xi) \dot{\phi}=\frac{\pm \sqrt{6} H M_P \sqrt{k_2^2+2 k_1 k_2+6}}{2 k_1+k_2}
\end{equation}
if $k_2\ge\sqrt{6+k_1^2}-k_1$. The corresponding equation of state parameter
\begin{equation}
    \omega=\frac{k_2-2 k_1}{k_2+2k_1}
\end{equation}
includes the accelerating regime $\omega<-1/3$ and approaches the observational range $\omega\sim-1$  in the regime of strong kinetic coupling, i.e. for $k_1 \gg k_2$, (a regime that is hard to achieve from UV motivated constructions \cite{Brinkmann:2022oxy}). In Fig.~\ref{fig:axionQuint} we present the observationally viable range for $(k_1,k_2)$. It is interesting to note that the modulus-axion action is of the correct form to implement the recently proposed homeopathy mechanism \cite{Burgess:2021qti,Burgess:2021obw} that can render these models compatible with long-distance tests of gravity. 

\subsubsection{Early dark energy}\label{sec:EarlyDarkEnergy}
Early Dark Energy (EDE) \cite{Poulin:2018dzj,Poulin:2018cxd,Poulin:2023lkg} offers a cosmological mechanism to alleviate the Hubble tension, the discrepancy between local measurements of the Hubble constant and the value inferred from the CMB within $\Lambda$CDM. EDE models posit the existence of a light scalar field that temporarily contributes a few percent of the total energy density of the Universe around matter–radiation equality, raising the expansion rate at recombination and thereby reconciling the datasets. Unlike standard quintessence, the EDE field is dynamically relevant only for a short interval and subsequently dilutes away, avoiding conflicts with late-time cosmology.
 The simplest realizations involve a pseudo-Nambu-Goldstone boson (axion-like field) $\phi$ with a periodic potential of the form
\begin{equation}
V_{\rm EDE}(\phi) = V_0 \left[1 - \cos\left(\frac{\phi}{f}\right)\right]^n\,,
\end{equation}
where $V_0^{1/4}$ sets the characteristic energy scale, $f$ is the decay constant, and $n$ controls the steepness of the potential. At early times, the Hubble friction term in the Klein-Gordon equation
\begin{equation}
\ddot{\phi} + 3H\dot{\phi} + \frac{dV_{\rm EDE}}{d\phi} = 0
\end{equation}
keeps the field frozen, so that it behaves as a cosmological constant with $w_\phi \approx -1$. When the Hubble parameter drops below the effective mass of the field
\begin{equation}
m_\phi^2  \sim \frac{n V_0}{f^2}\,,
\end{equation}
the field begins to oscillate around the minimum of its potential, and its energy density redshifts away. The decay of the EDE field is determined by the exponent $n$ in the potential, and its best-fit value is $n=3$ \cite{Poulin:2018cxd}. In phenomenologically viable models, the fractional contribution to the total energy density reaches a few percent near $z \sim 3000$, sufficient to modify the sound horizon at recombination and partially reconcile the Hubble constant inferred from CMB observations with local measurements.

The mass $m_\phi$ and the energy scale $V_0$ are tightly constrained by both cosmological and particle physics considerations.  Typically, the mass of the EDE field is $m_\phi \sim 10^{-27}\ {\rm eV}$, corresponding to oscillation timescales around matter-radiation equality, while $V_0^{1/4} \sim {\rm eV}$. This implies decay constants of order $f \sim 10^{16} - 10^{17}\,{\rm GeV}$, namely of Planckian size \cite{Poulin:2018cxd,McDonough:2021pdg}. From a particle physics perspective, this identifies the EDE field as a member of the WISP family: extremely light, very weakly coupled, but cosmologically impactful.

Embedding such a setup in a UV-complete theory like string theory is non-trivial, as one must generate the peculiar cosine-power potential while simultaneously stabilizing all moduli. In \cite{McDonough:2022pku}, the authors constructed the first controlled realization within a KKLT scenario. There, the EDE axion originates from a two-form modulus 
G, with the superpotential receiving several non-perturbative contributions,
\begin{equation}
W = W_0 + A e^{-aT} + B e^{-bG} + C e^{-2bG} + D e^{-3bG},
\end{equation}
with $T$ being the volume modulus. The KKLT term $A e^{-aT}$ remains the dominant contribution, ensuring volume stabilization. The other three exponentials generate the first three harmonics of the axion potential, enabling the combination that reproduces the $(1-\cos(\phi/f))^3$ structure.  Although conceptually successful, this construction requires that the Pfaffian prefactors $B$, $C$, $D$ be exponentially small compared to the KKLT scale, reflecting a significant degree of tuning.

A more detailed analysis was carried out in \cite{Cicoli:2023qri}, where the authors developed explicit models of EDE in type IIB string theory by identifying the EDE field with either a  $C_4$  or a $C_2$ axion and achieving full closed-string moduli stabilization within both the KKLT and Large Volume Scenario (LVS) frameworks. The most robust construction arises in LVS with a $C_2$ axion, where three distinct non-perturbative terms in the superpotential are generated by gaugino condensation on D7-branes carrying quantized world-volume flux. These fluxes shift the gauge kinetic function by integer multiples of the odd modulus, producing a superpotential of the form
\begin{equation}
W = W_{\rm LVS} + A_1 e^{-a (T_b + fG)} + A_2 e^{-a (T_b + 2fG)} + A_3 e^{-a (T_b + 3fG)} \,, 
\end{equation}
with $W_{\rm LVS}$ the standard LVS superpotential, $T_b$ the large Kähler modulus and $G$ the orientifold-odd modulus whose imaginary part contains the $C_2$ axion.
Importantly, this structure emerges automatically from standard flux quantization on D7-branes and does not require any exotic ingredients beyond the usual toolkit of type IIB compactifications.

This setup ensures that \emph{all} closed-string moduli, complex structure, dilaton, the large volume modulus $T_b$, and the blow-up moduli, are stabilized at parametrically higher scales, leaving the odd axion as the unique light direction. Moreover, the resulting axion decay constant is determined geometrically, scaling as $
f \simeq 0.2 \sqrt{g_s}\,M_{\rm Pl}\mathcal V^{-1/3}
$, which naturally yields trans-Planckian or mildly sub-Planckian values for moderately large compactification volumes $\mathcal V \sim 10^{4-6}$ (in units of string length).

A crucial achievement of this construction is that the tiny EDE energy scale arises without tuning the Pfaffian prefactors: the fluxed ED3/D7 non-perturbative effects lead to instanton actions $S$ large enough that
\begin{equation}
V_0 \sim M_{\rm Pl}^4\,e^{-S} \simeq M_{\rm Pl}^4\exp\!\big(-\lambda M_{\rm Pl}/f\big)
\end{equation}
automatically reproduces the required suppression for $\lambda \gg 1$. This mechanism sharply improves upon the earlier KKLT-based embedding, where achieving the correct scale required delicate balance among exponentially small prefactors. Hence, the work of \cite{Cicoli:2023qri} provides a globally consistent, moduli-stabilized string vacua in which the axion underlying EDE emerges naturally from the geometry and non-perturbative structure of type IIB compactifications.

%% file: WG1/content/1-1-5.tex
We saw in section~\ref{sec:QCDaxionStringTheory} that axions are one of the best-motivated candidates for WISPs within string theory. Axions arise naturally from the dimensional reduction of form fields, and their masses are protected by the topological properties of the extra-dimensional geometry. However, axions are not the only WISP candidates that arise in string compactifications.

While axions are 0-form fields, a close cousin to them are dark 1-form fields, namely dark photons. Dark photons can arise in (at least) two distinct ways in compactifications.

\subsubsection{Dark Ramond-Ramond Photons}
The first type of dark photon arises in a way very similar to that of axions, with their existence likewise closely tied to the extra-dimensional topology.

Axion-like particles can arise from the dimensional reduction of form fields on cycles in the Calabi-Yau, for example, via an RR $n$-form field dimensionally reduced on an $n$-cycle $\Sigma_i$ to give a 0-form axion,
\begin{equation}
a_i(x^{\mu}) = \int_{\Sigma_{n,i}} C_n(x^{\mu})\,,
\end{equation}
where $x^{\mu}$ refers to space-time coordinates and $dC _n = 0$ as a feature of the extra-dimensional geometry.
In this case, the protection of the axion mass arises from the topological nature of the extra-dimensional form field. For the case of axions, the dimension of the reduced form field matches the dimension of the cycle. One easy way to obtain dark photons in string theory is when the dimension of the form field is one larger than the dimension of the cycle. If we now reduce a RR $(n+1)$-form on an $n$-cycle, we obtain a U(1) gauge boson,
\begin{equation}
A^{\mu}_i = \int_{\Sigma_{n,i}} C_{n+1}\,,
\end{equation}
as the remaining index of the $(n+1)$-form lies in spacetime and so produces a $U(1)$ vector gauge boson in the 4-dimensional effective theory (for example, see \cite{Grimm:2004uq} for details of the dimensional reduction).

In the same way that string compactifications naturally give rise to many axions (as Calabi-Yau manifolds can contain a large number of cycles with $h^{1,1}$ and $h^{2,1}$ easily being $\mathcal{O}(100)$), in a very similar way a large number of RR photons can arise in a compactification.\footnote{Although we use the expression of \emph{RR photon}, corresponding to IIA/IIB compactifications, it is clear that the physics described here can equally well apply to e.g.~M-theory compactifications.} As with axions, the precise number of dark U(1) photons arising from RR forms depends on the number of cycles in the compactification. 

Note that although the number of RR dark photons is related to the number of cycles in the compactification, there is no exact match (neither for axions nor dark photons). The reason is that the number of surviving low-energy particles depends on an orientifold projection. In type IIB and IIA orientifolds, the orientifold action, on dimensional reduction, projects out some number of the form fields depending on their parity under the orientifold projection (leaving e.g. $h^{1,1}_{+}$, $h^{2,1}_{+}$, $h^{1,1}_{-}$ or $h^{2,1}_{-}$ surviving fields depending on the nature of the compactification).

Any U(1) RR gauge bosons that survive into the low-energy theory are automatically light. From a 4-dimensional perspective, this is because they are gauge bosons and so their mass is protected by gauge symmetry. From the higher-dimensional perspective, the lightness arises from the topological nature of the Calabi-Yau and the fact that these are form fields which only appear in the perturbative action via the field strength (e.g.~$F_3 = dC_2$). 

 The darkness of such RR U(1) photons is also set by their action. The only direct couplings of RR form fields are to D-branes, arising via the Chern-Simons part of the DBI $+$ CS coupling,
 \begin{equation}
    S_{DBI+CS} = T_{p} \int_{\Sigma_p} e^{-\phi} \sqrt{g + 2 \pi \alpha' F} +  \int_{\Sigma_p} C_p\,. 
 \end{equation}
 As D-branes are non-perturbative objects in string theory, all states directly charged under RR U(1)s are heavy with string scale masses, and therefore entirely absent from the low-energy theory. 
 
 The other side of this statement is that the light fields present in the effective theory, in particular the Standard Model fields, have no coupling to such dark RR photons. If the Standard Model states arise from fundamental strings, then they are uncharged under RR U(1)s as fundamental strings cannot be charged under RR form fields (and the same is also true of any D-brane hidden sector gauge extensions to the Standard Model). 
 
 Consequently, no Standard Model fields, or light particles in extensions of the Standard Model, will be charged under these RR U(1) photons: such Ramond-Ramond U(1) gauge bosons arising from reduction of the RR fields are truly \emph{dark}.
 
From the perspective of phenomenology, this has positive and negative sides. The positive side is that these provide good and strong reasons for dark U(1) photons to be present in string compactifications; the conditions for the existence of RR photons are quite generic within IIA and IIB compactifications. 

The negative side is that these U(1)s may be \emph{too} dark: much of the most interesting physics of dark photons arises when they have small but non-zero masses and small, but non-zero, mixing with the photon. If no low-energy states \emph{at all} have any couplings to dark RR photons, it is hard either to give such dark RR photons a non-zero mass or generate any mixing between them and the photon. The phenomenology of RR photons is studied in more detail in \cite{Camara:2011jg}.

\subsubsection{Dark Photons from D-Branes}
Another, perhaps more interesting, origin of dark photons in string compactifications can arise from \emph{fundamental} strings starting and ending on D-branes. Open strings in type II string theories starting and ending on D-brane stacks carry U($n$) gauge fields. D-branes at singularities can also split into so-called \emph{fractional branes} which support different gauge groups.

In many phenomenological scenarios in string theory, the Standard Model is realised as a local construction, where all the quarks and leptons arise from degrees of freedom at a particular location in the extra dimensions (which includes the U(1) of electromagnetism).

\begin{figure}[t!]
    \centering
    \includegraphics[width=0.65\linewidth]{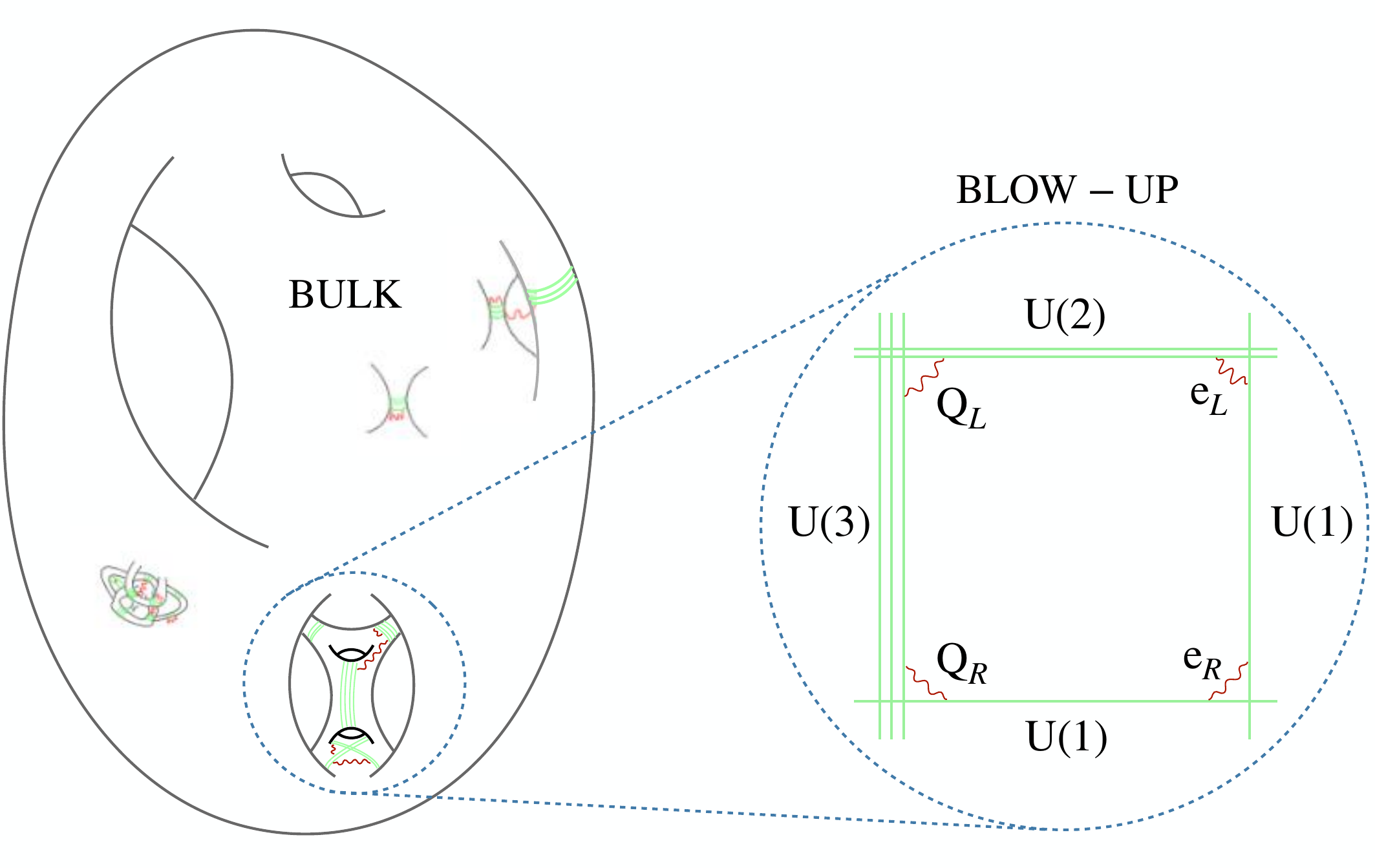} 
    \caption{An illustration of a local string model, with the Standard Model realised at a singularity and other brane stacks present far away in the compactification. Figure adapted from \cite{Conlon:2006wz}.}
    \label{localstringmodel}
\end{figure}

Once we allow the possibility that the Standard Model (including the U(1) of electromagnetism) arises as a local construction in the extra dimensions from a local stack of D-branes, then we can also imagine other local stacks of D-branes at distant points in the compactification, which can also have gauge groups analogous to those of the Standard Model. Such local models for D-brane constructions have received a lot of 
attention -- for examples, see \cite{Aldazabal:2000sa, Beasley:2008kw, Conlon:2008wa,  Cicoli:2013mpa, Cicoli:2021dhg, Quevedo:2014xia} and such a scenario is illustrated in Fig.~\ref{localstringmodel}.

Quite generally, such scenarios result in the presence of hidden U(1) gauge bosons on brane stacks located far from the Standard Model stack.
In weakly coupled string theory, particles charged under the gauge groups arise from fundamental strings ending on D-branes. In this case, there are no light Standard Model states \emph{directly} charged under such distant U(1)s. This is most clearly manifest in weakly coupled D-brane models, where the mass-energies of extended strings grow with length. Any particle arising from a fundamental string which is extended in the extra dimensions has a mass
\begin{equation}
m = \lambda T\,,
\end{equation}
where $T = (2 \pi \alpha')^{-1}$ is the string tension and $\lambda$ is the length of the string in string units. Any extended string with a length greater than the string scale will result in a particle with a string-scale mass and so such states appear irrelevant to low energy physics. However, this point contains a subtlety.

With dark photons, a quantity of great phenomenological import is the magnitude of kinetic mixing between a dark photon and the Standard Model U(1). Generally, if there are two U(1) gauge groups, U(1)$_a$ and U(1)$_b$, and both U(1)s are massless, we can write their kinetic terms using the Lagrangian
\begin{equation}
    \mathcal{L} = - \frac{1}{4 g_a^2} F_{\mu \nu}^{(a)} F^{\mu \nu, (a)} - \frac{1}{4 g_b^2} F_{\mu \nu}^{(b)} F^{\mu \nu, (b)}
    - \frac{\chi_{(ab)}}{2 g_a g_b} F_{\mu \nu}^{(a)} F^{\mu \nu, (b)},
\end{equation}
with $\chi_{(ab)}$ as a measure of the strength of kinetic mixing between the two U(1)s. Kinetic mixing in string theory has been studied in \cite{Abel:2008ai, Goodsell:2009xc, Goodsell:2010ie, Bullimore:2010aj, Goodsell:2011wn, Acharya:2018deu, Hebecker:2023qwl}.    

One of the more interesting features of kinetic mixing is that it is a 1-loop exact quantity. In the context of $\mathcal{N} = 1$ (softly broken) IIB  supersymmetric compactifications, this point is easy to understand. The superfield controlling the dilaton is $S = \frac{1}{g_s} + i c_0$. As the gauge kinetic function has to be holomorphic in the superfields, we can see that both tree-level and 1-loop contributions are consistent with holomorphy (the former leading to dependence linear in $S$ and the latter simply a constant), but that 2-loop (or higher) contributions would be inconsistent with the imaginary part of the superfield being an RR form and with its associated perturbative shift symmetry.

The holomorphy also gives another surprising result in IIB D3/D7 orientifolds, where the K\"ahler moduli, which control the sizes of the internal space, are given by 
\begin{equation}
\label{khtb}
T_i = \frac{{\rm Vol}(\Sigma_i)}{g_s} + i C_{4,i}\,.
\end{equation}
In this context, the real part of the K\"ahler moduli corresponds to the size of the 4-cycles, being precisely given by the volume of the 4-cycle, divided by $g_s$, and their imaginary part is the RR 4-form reduced on that cycle. The significance of this is that the $g_s^{-1}$ factor implies that the K\"ahler moduli cannot enter the 1-loop contribution to the kinetic mixing; by definition, the 1-loop contribution must involve a term of order $g_s^{0}$ (as tree-level in open string computation is $g_s^{-1}$) and so the K\"ahler moduli must therefore be \emph{absent} from the form of the 1-loop correction.

This is somewhat surprising, as it violates an intuition that for a large compactification volume the magnitude of the kinetic mixing could be suppressed by volume-factors; for two separate U(1)s brane stacks separated by ever greater amounts, we might expect the magnitude of the kinetic mixing between the U(1)s to reduce to zero. 
However, calculations supporting this behaviour have been done in \cite{Bullimore:2010aj}. Some ways to avoid these conclusions have been explored in \cite{Hebecker:2023qwl}, for example, involving compactifications without supersymmetry or where the definition of the K\"ahler moduli might be more complex than that of Eq.~\eqref{khtb}. 

What is clear from these explicit calculations is that finite and non-zero kinetic mixing can, under certain conditions, be obtained for separated U(1) gauge groups located on distant brane stacks. From an open string perspective, this is a 1-loop worldsheet calculation. On this perspective, the kinetic mixing is obtained by integrating out heavy states charged under both U(1) factors which are long strings stretching between the two U(1) factors. Although such states are heavy, the fact that they are strings means that they are present in large numbers (due to multiply wound copies and also through the Hagedorn spectrum of strings): integrating them out, they can give a non-zero but finite contribution to kinetic mixing.

Such open string 1-loop diagrams can also be reinterpreted as closed string tree diagrams.
\begin{figure}
    \centering
\includegraphics[width=0.6\linewidth]{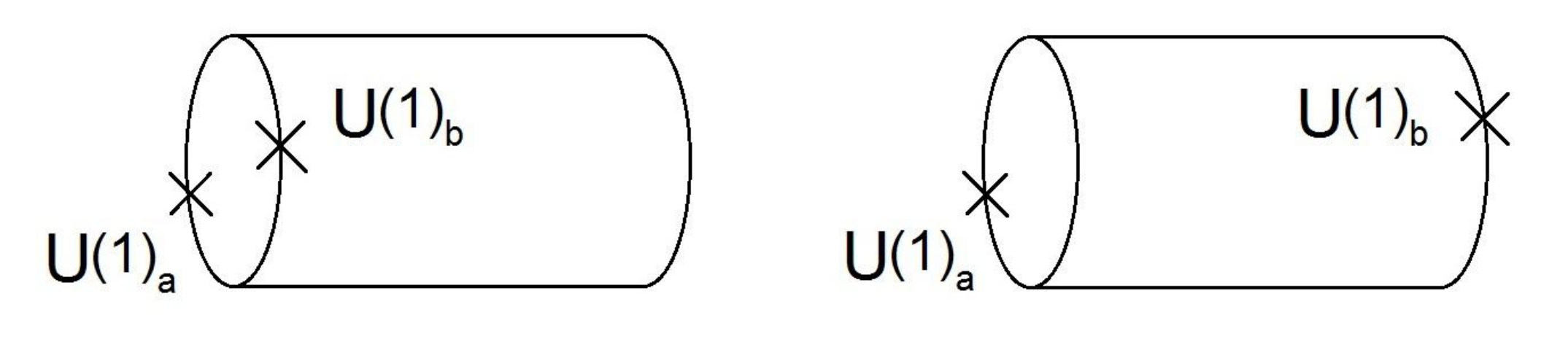} 
    \caption{An open string 1-loop diagram with two insertions of vertex operators for $U(1)$ gauge bosons can be used to compute the kinetic mixing between these $U(1)$ factors. The open string 1-loop diagram can be reinterpreted within string theory as a closed string tree diagram, corresponding to the exchange of fields that couple to both branes.}
    \label{mixing}
\end{figure}From this picture, the diagram corresponds to the propagation of an RR form across the compactification between the two D-branes. The non-zero value of the diagram corresponds to a non-zero amplitude for the exchange of the form. Provided the effective co-dimension for the propagation is two (not four -- the value of two can occur for fractional branes on cycles that are globally but no locally trivial) then there is no volume suppression factor for the exchange of this RR form, and so the kinetic mixing can remain finite even as the volume of the compact space is taken to infinity and the branes carrying the two U(1) factors become arbitrarily separated.

\paragraph{Light Dark Chiral Fermions}
A final source of WISPs in string compactifications are chiral fermions associated with hidden sectors. The Standard Model teaches us that light chiral fermions can be and are realised in nature. The neutrini are the strongest example of this phenomenon, but the remaining fermions of the Standard Model are also very light compared to the Planck scale, with their masses set by their couplings to the Higgs and the size of the Higgs vev.

In local D-brane constructions, as per Fig.~\ref{localstringmodel}, it is easy to imagine the existence of other localised chiral gauge sectors, similar to the Standard Model but realised at other singularities or local cycles in the compactification from distant stacks of D-branes. The chiral fermions from such a sector will be uncharged under the Standard Model gauge sector and -- for the same reasons that the Standard Model fermions are light -- will have small masses that start at zero prior to any Higgs breaking.

Various mechanisms for dynamical symmetry breaking are exponentially sensitive in energy to fundamental parameters in the system (a well-known example is chiral symmetry breaking in QCD, which is set by the scale of $\Lambda_{\rm QCD}$). Such mechanisms can lead to the generation of fermion masses at exponentially small scales, which may include fermion WISPs at mass scales that are cosmologically light. Classic references on how non-perturbatively small Yukawa couplings and fermion masses can be generated in string theory by D-brane instanton effects are \cite{Cremades:2003qj, Blumenhagen:2007zk}.

%% file: WG1/content/1-1-6.tex
A recent idea in fundamental theory has been the idea of the \emph{swampland} \cite{Ooguri:2006in}. This is the notion that theories can appear consistent from a low-energy perspective yet can have no possible embedding into quantum gravity. The \emph{swampland} is the set of such theories that appear valid from all effective field theory arguments but cannot be UV completed into a string theory (or, more broadly, any other quantum gravity theory).

Associated with the swampland are conjectures about the form of low-energy effective theories that are (in)consistent with quantum gravity. Although many such conjectures have been stated, some have much more support than others. Among the swampland conjectures with the strongest support, we can mention the \emph{weak gravity conjecture} and the \emph{swampland distance conjecture}, which we discuss below. For a full review-length description of the swampland conjectures, see the reviews \cite{Palti:2019pca, Harlow:2022ich, vanBeest:2021lhn, Grana:2021zvf, Palti:2022edh}. Here we can only give a very brief overview of the way some of these swampland conjectures may be able to constrain the space of WISP candidates (and we also stress that these are \emph{conjectures}, and that it is often the case that the conjectures with the most far-reaching implications for particle physics are also those with less support).

\paragraph{The Distance Conjecture}

The distance conjecture \cite{Ooguri:2006in, Ooguri:2018wrx} claims that if fields traverse significantly trans-Planckian distances $\Delta \phi \gg M_P$ in field space, then a tower of states descends and becomes exponentially light relative to the cutoff,
\begin{equation}
m_t (\phi) \sim M_0\ e^{-\lambda \Delta \phi /M_P},
\end{equation}
where $\lambda$ is an $\mathcal{O}(1)$ constant and $M_0$ is an initial high scale. 
For arbitrarily large trans-Planckian field excursions, the mass of the tower then becomes exponentially light relative to any other fixed scale.\footnote{Note that in some cases the UV cutoff of the theory may be field-dependent, $\Lambda_{UV} = \Lambda_{UV}(\phi)$, and so the tower may still always remain lower than the cutoff. One example is that of KK modes in the decompactification limit of string compactifications, where the UV string scale also reduces with field displacement.}

How does this affect WISP candidates? This is relevant to the physics of light axions and, in particular, the physics of light axions with large trans-Planckian decay constants, $f_a > M_P$. A genuine WISP candidate is light: the particle mass is much lighter than all the other scales of particle physics. The distance conjecture implies a tension (already previously noticed, see e.g. \cite{Banks:2003sx}) between quantum gravity and axions with large decay constants. On the one hand, a WISPy axion, with a large decay constant, implies a potential that is flat over a trans-Planckian distance in field space. On the other hand, the distance conjecture would require a tower of states to become exponentially lighter as the axion is displaced. In any theory without supersymmetry, this creates a tension as backreaction effects are expected to be induced (see e.g. \cite{Conlon:2011qp, McAllister:2014mpa, Baume:2016psm, Valenzuela:2016yny} for studies of backreaction in models of large-field axions), damaging the flatness of the quantum potential. 

Field excursions are central to the physics of the distance conjecture (see e.g. the universal bounds extracted in \cite{Scalisi:2018eaz,vandeHeisteeg:2023uxj,Cribiori:2025oek}). The existence of a moduli space (at least to a high approximation) allows the notion of a field excursion to be applicable in the present universe. Large-field axions are the most prominent example where a moduli space of vacua can exist today with a trans-Planckian diameter. However, the distance conjecture would also imply constraints on more general scalar WISP candidates.

\paragraph{The Weak Gravity Conjecture}

The weak gravity conjecture originally appeared in \cite{Arkani-Hamed:2006emk} as the notion that `gravity must be the weakest force' -- it is not possible to form consistent theories with arbitrarily weak gauge sectors. Properly formulated, gauge interactions must always be stronger than gravitational interactions.

There has been much work over the last two decades refining this concept: what, precisely, does it mean for gravity to be the weakest force? Certainly, particles can be uncharged under a particular gauge sector. A comprehensive review of the weak gravity conjecture(s) can be found in \cite{Harlow:2022ich} (for a shorter review, see \cite{Palti:2020mwc}).

The weak gravity conjecture comes in two versions, an electric and a magnetic weak gravity conjecture, which can be viewed as the electric weak gravity conjecture applied to the dual of the gauge field present in the electric weak gravity conjecture. The statement of the electric weak gravity conjecture applied to a U(1) gauge theory with a gauge coupling $e$ is that there must exist a particle with mass
\begin{equation}
\label{hyp}
m^2 \leq 2 e^2 q^2 M_P^2\,,
\end{equation}
where $q$ is the charge of the particle. Eq.~\eqref{hyp} is also called the \emph{mild weak gravity conjecture}. The \emph{mildness} is because Eq.~\eqref{hyp} only states that there must exist \emph{a} particle satisfying this bound. As it does not constrain the mass of the particle, the particle could be very massive compared to all the usual particle physics scales.

The magnetic version of the weak gravity conjecture states that the UV cutoff of an effective field theory with a U(1) gauge theory is bounded as
\begin{equation}
\Lambda_{UV} \lesssim e M_P\,.
\end{equation}
Hence, in the presence of very weakly coupled gauge theories, the UV cutoff of the theory must be much lower than the Planck scale.

The centrality of gauge theories in the weak gravity conjecture implies that it may be relevant for ultra-weakly-coupled dark photons \cite{Cheung:2014vva,Reece:2018zvv, Craig:2019fdy, Abu-Ajamieh:2024gaw}. For example, the magnetic weak gravity would imply that a theory with a dark photon with intrinsic gauge coupling $e \lesssim 10^{-18}$ would have a UV cutoff of less than $\sim 1 {\rm GeV}$. This would be especially pertinent for a gauged $\left( B- L \right)$ vector boson, as Standard Model fields would also be directly charged under such a force, and any UV cutoff lower than the scales probed by colliders would be inconsistent.

\paragraph{The Gravitino Conjecture}
The gravitino conjecture \cite{Cribiori:2021gbf,Castellano:2021yye} proposes that the limit of small gravitino mass $m_{3/2}\rightarrow 0$ is obstructed in quantum gravity because it is always accompanied by an infinite tower of states becoming massless. More precisely, the typical mass scale $m_t$ of this tower behaves as
\begin{equation}
    m_t\sim M_P\left(\frac{m_{3/2}}{M_P}\right)^n\,,
\end{equation}
with $n=\mathcal{O}(1)$, as suggested by large classes of string compactifications (e.g.~no-scale models, LVS, KKLT). This relation connects the supersymmetry-breaking scale directly to Kaluza–Klein or winding towers~\cite{Anchordoqui:2023oqm}. Axion-like particles, hidden photons, moduli, and generic light scalars are typically associated with large internal volume $\mathcal{V}$ or approximate shift symmetries, resulting in ultralight masses and tiny couplings; therefore, the gravitino conjecture imposes non-trivial constraints on these WISPs. In generic compactifications, the KK scale behaves as $m_{KK}\sim \mathcal{V}^{-1/3}$, while the gravitino mass scales as $m_{3/2}=e^{K/2}|W|\sim \mathcal{V}^{(\beta-\alpha)/2}$, where $\alpha$ is the modular weight of the Kähler potential $K$ (e.g., $\alpha=2$ in type IIB O3/O7 orientifolds) and $\beta$ encodes the volume-dependence of the flux superpotential $W$. This implies that attempting to make WISP masses parametrically small by increasing the internal volume or decreasing the coupling necessarily correlates with a controlled decrease of the gravitino mass. Pushing WISPs into the deep sub-eV regime thus might drive the theory toward a limit in which entire towers of states become light and the effective field theory cutoff approaches zero. In other words, quantum-gravity consistency imposes a lower bound on the mass of WISPs: arbitrarily light WISPs cannot be obtained without encountering the breakdown of the EFT.

From the low-energy perspective, the gravitino itself can behave as a WISP. Its interactions are Planck-suppressed, making it automatically extremely weakly coupled. In models with low-scale SUSY breaking, the gravitino mass $m_{3/2}=F/(\sqrt{3}M_P)$, with $F$ the SUSY-breaking order parameter, can naturally lie in the eV or sub-eV range. Importantly, LHC bounds constrain soft masses of sparticles but do not impose a lower limit on $m_{3/2}$, allowing the gravitino to be arbitrarily light in low-scale SUSY scenarios. The two scales can in fact be decoupled by raising the messenger scale in gauge mediation. However, the situation changes radically once one incorporates the gravitino conjecture, which forbids the limit of small gravitino mass. As a result, the gravitino mass would become a diagnostic of whether a given WISP scenario can be consistently embedded in quantum gravity, making the gravitino conjecture an important UV-selection principle for WISP model building.

%% file: WG1/content/1-2-1.tex
Historically axion models were first proposed to solve the strong CP problem \cite{Kim:2008hd,DiLuzio:2020wdo}  in the context of 4-dimensional theory 
involving a spontaneously broken anomalous global $U(1)_{\rm PQ}$ symmetry  \cite{Peccei:1977hh,Weinberg:1977ma,Wilczek:1977pj,Kim:1979if,Shifman:1979if,Dine:1981rt,Zhitnitsky:1980tq}. Physical properties of axion crucially depend on the scale where $U(1)_{\rm PQ}$ is spontaneously broken, which is characterized by the axion decay constant $f_a$ defined as the field range of the canonically normalized compact axion field:\footnote{In this subsection, we use this definition of the axion decay constant $f_a$  for the convenience of discussion.}\begin{eqnarray}
a(x)\cong a(x)+2\pi f_a\,. \end{eqnarray}
Considering only the phenomenologically viable models with $f_a$ well above the weak scale,  one can focus on two different types of models,  the Kim-Shifman-Vainshtein-Zakharov (KSVZ)-type models   \cite{Kim:1979if,Shifman:1979if} in which
all Standard Model (SM)  fields are neutral under  $U(1)_{\rm PQ}$ and
the Dine-Fischler-Srednicki-Zhitnitsky (DFSZ)-type models \cite{Dine:1981rt,Zhitnitsky:1980tq} in which some or all of the SM quarks and leptons 
carry nonzero $U(1)_{\rm PQ}$-charge.  There are also two possibilities for the spontaneous breakdown of  $U(1)_{\rm PQ}$, one by a non-zero vacuum value of  $U(1)_{\rm PQ}$-charged elementary scalar field and another by a confining strong dynamics which would result in a composite axion \cite{Kim:1984pt,Choi:1985cb,Redi:2016esr,Gavela:2018paw,Cox:2023dou,Gherghetta:2025fip,Agrawal:2025mke}. In the following, we present a typical example for each of the KSVZ-type, DFSZ-type, and composite axion models. For each example, we also provide a brief discussion of the low-energy axion couplings, which are crucial for experimental searches for axions,  as well as  the number of degenerate vacua   of the axion potential, i.e.~the domain wall number $N_{DW}$ determined by the axion-gluon coupling 
\begin{eqnarray}
\mathcal{L}\supset g_c^2\frac{N_{DW}}{32\pi^2}\frac{a(x)}{f_a} G^a_{\mu\nu}\tilde G^{a\mu\nu}\,,
 \end{eqnarray}
which has important implications for axion cosmology.
 
\subsubsection{KSVZ axion}
The KSVZ-type axion model generically involves an exotic $U(1)_{\rm PQ}$-charged and coloured fermion which obtains a heavy mass as a consequence of the spontaneous breakdown of  $U(1)_{\rm PQ}$.
For the purpose of illustration, let us consider a simple example with  a single flavour of exotic left-handed colour-triplet (but $SU(2)_L$-singlet) fermion $Q$ and anti-triplet  $Q^c$,
which also includes  an elementary scalar field $\sigma$ breaking $U(1)_{\rm PQ}$ spontaneously.
 The relevant part of the Lagrangian density is given by
\begin{eqnarray}
{\cal L}=\partial_\mu\sigma\partial^\mu\sigma^* 
+ i\bar Q \bar\sigma^\mu D_\mu Q +
i\bar Q^c\bar\sigma^\mu D_\mu Q^c
-\,\Big(y \sigma QQ^c+{\rm h.c}\Big) -\lambda \Big(\sigma\sigma^*-\frac{1}{2}f_a^2\Big)^2 \,,\label{eq:ksvz_1.2.1}\end{eqnarray}
with the PQ symmetry acting on the new fields as
\begin{eqnarray} U(1)_{\rm PQ}: \quad \sigma\rightarrow e^{i\alpha}\sigma, \quad (Q,Q^c)\rightarrow e^{-i\alpha/2}(Q,Q^c)\,.\end{eqnarray}
This symmetry is spontaneously broken by the vacuum expectation value 
$\langle \sigma\rangle ={f_a}/{\sqrt{2}}$.
To derive the low-energy axion couplings, one may parameterise $\sigma$ as
\begin{eqnarray} \sigma = \frac{1}{\sqrt{2}}(f_a+\rho) e^{ia(x)/f_a}\,,\label{eq:sigma_1.2.1}
\end{eqnarray}
 and  make the  field redefinition:
 $(Q, Q^c)\rightarrow  e^{-ia(x)/2f_a}(Q,Q^c)$. Integrating out the heavy $\rho$ and $(Q,Q^c)$ while assuming  $m_Q=y f_a/\sqrt{2}\lesssim m_\rho\sim f_a$,
 one finds the axion effective Lagrangian at the scale $\mu=m_Q$,
 \begin{eqnarray}
 {\cal L}_{a}(\mu=m_Q) =\frac{1}{2}\partial_\mu a\partial^\mu a+
\frac{1}{32\pi^2}\frac{a(x)}{f_a} \big(g_c^2G^a_{\mu\nu}\tilde G^{a\mu\nu}+ 3Y_Q^2 g_1^2B_{\mu\nu}\tilde B^{\mu\nu}\big)\,, \end{eqnarray}
where $G^a_{\mu\nu}$ and $B_{\mu\nu}$ denote the $SU(3)_c\times U(1)_Y$ gauge field strength  and $(Y_Q,-Y_Q)$ are the $U(1)_Y$-hypercharges of $(Q,Q^c)$. 

The above KSVZ axion model has $N_{DW}=1$, and therefore is free from the axion domain wall problem even in the post-inflationary scenario in which the spontaneous breakdown of $U(1)_{\rm PQ}$ occurs  in the early Universe, after inflation is over \cite{Marsh:2015xka}. Note that if the model involves multiple ($N_Q>1$) flavors of $(Q,Q^c)$, it predicts $N_{DW}=N_Q$.
A key feature of the KSVZ-type axion is that the couplings to the SM quarks and leptons at  $\mu=m_Q$ are all vanishing.
Quantum corrections to the axion couplings due to the SM gauge interactions and the Yukawa couplings can be taken into account following the analysis of \cite{Choi:2017gpf,Chala:2020wvs,Bauer:2020jbp,Choi:2021kuy,Bonilla:2021ufe}. 
Including only the leading radiative corrections,  the axion couplings at $\mu= 2$ GeV  for the model of  Eq.~\eqref{eq:ksvz_1.2.1}  are given by \cite{Choi:2021kuy}
\begin{eqnarray}
\frac{1}{32\pi^2}\frac{a(x)}{f_a} (g_c^2G^a_{\mu\nu}\tilde G^{a\mu\nu}+ 3Y_Q^2 e^2F_{\mu\nu}\tilde F^{\mu\nu})+\frac{\partial_\mu a}{2f_a}\sum_{\Psi=u,d,e, ...} c_{a\Psi} \bar\Psi\gamma^\mu\gamma_5\Psi\,,\label{eq:ksvzcoupling3_1.2.1}\end{eqnarray}
where \begin{eqnarray}
c_{au}&\simeq& c_{ad}\simeq \frac{6}{25\pi} \left(\alpha_s(\mu=2\, {\rm GeV})-\frac{25}{21}\alpha_s(m_Q)+0.029\right)={\cal O}(10^{-2})\,, \nonumber \\
 c_{ae} &\simeq& \frac{3y_t^2(m_t)}{112\pi^2}\left(1-8\frac{\alpha_s(m_Q)}{\alpha_s(m_t)}+7\left(\frac{\alpha_s(m_Q)}{\alpha_s(m_t)}\right)^{8/7} \right) ={\cal O}(10^{-3})\,,\label{eq:ksvzcoupling4_1.2.1}\end{eqnarray}
for the QCD fine structure constant $\alpha_s=g_c^2/4\pi$ and the top quark Yukawa  coupling $y_t=m_t/v$ ($v=246$ GeV).
As was noticed in  \cite{Choi:2021kuy}, the dominant contribution to the axion-electron coupling in the KSVZ model comes from a three-loop effect involving the exotic heavy quark $(Q,Q^c)$, the top quark, and the SM Higgs doublet as intermediate states, whose leading piece can be obtained from the renormalisation group evolution of the axion-top and axion-Higgs couplings. 
 From these couplings at $\mu=2$~GeV, one can apply the results of  \cite{GrillidiCortona:2015jxo} to derive the axion couplings to the photons and nucleons at scales well below the QCD scale, yielding
\begin{eqnarray}
\frac{1}{4} g_{a\gamma\gamma}a(x)\vec E\cdot \vec B +\frac{\partial_\mu a}{2f_a}(c_{ap}\bar p\gamma^\mu\gamma_5 p +c_{an}\bar n\gamma^\mu\gamma_5 n)\,,\label{eq:ksvzcoupling1_1.2.1}\end{eqnarray}
where
\begin{eqnarray}
g_{a\gamma\gamma}\simeq \frac{\alpha_{em}}{2\pi f_a}(3Y_Q^2-1.92)\,,\quad c_{ap}\simeq -0.47+\Delta c_N\,, \quad 
c_{an}\simeq -0.03+\Delta c_N\,,
\label{eq:ksvzcoupling2_1.2.1}\end{eqnarray}
and
\begin{equation}
    \Delta c_N\simeq 0.047\times \left[\alpha_s(10^{10}\,{\rm GeV})-\alpha_s(m_Q)\right]={\cal O}(10^{-3})\,.
\end{equation}

\subsubsection{DFSZ axion}
In the DFSZ-type model, the  SM quarks and leptons are charged under  $U(1)_{\rm PQ}$.
A simple example  is 
\begin{eqnarray}
{\cal L}&=&\partial_\mu\sigma\partial^\mu\sigma^* 
+ D_\mu H_u^\dagger D^\mu H_u +D_\mu H_d^\dagger D^\mu H_u- \lambda (\sigma\sigma^*-\frac{1}{2}f_a^2)^2\nonumber \\
&-&(\kappa H_uH_d \sigma^{*2} +y_u H_uq u^c + y_d H_d q d^c +y_{\ell}H_d\ell e^c+{\rm h.c})\,,
\end{eqnarray}
where  $q$ and $\ell$ denote the left-handed quark and lepton doublets, with the omitted generation indices,
$u^c, d^c$ and $e^c$ are the left-handed anti-quark and anti-lepton singlets, $H_u$ and $H_d$ are the Higgs doublets generating the masses of the up-type quarks and the down-type quarks, respectively.  
The PQ symmetry of the model can be defined as 
\begin{eqnarray}
U(1)_{\rm PQ}:  \quad 
  \sigma\rightarrow e^{i\alpha}\sigma\,, \quad H_{u,d}\rightarrow e^{i\alpha}H_{u,d}\,, \quad \psi \rightarrow e^{-i\alpha/2}\psi\,,
\end{eqnarray}
where $\psi=(q,u^c,d^c,\ell, e^c)$ stands for the SM quarks and leptons.

For the above DFSZ model, the axion couplings to the SM quarks and leptons are dominated by the tree-level contributions in most of the parameter space, while the axion couplings to the SM gauge fields are determined by the one-loop $U(1)_{\rm PQ}$ anomalies.   To derive the relevant axion couplings,
one can again parameterize $\sigma$ as in Eq.~\eqref{eq:sigma_1.2.1}
 and integrate out $\rho$ while 
making the field redefinition:
$H_{u,d}\rightarrow e^{ia(x)/f_a}H_{u,d}, \,\psi\rightarrow e^{-ia(x)/2f_a}\psi$.
At  $\mu\sim m_{\tilde H}$, where $\tilde H$ denotes the heavier doublet combination of $H_u$ and $H_d$, one can also  integrate out $\tilde H$, which would yield
 \begin{eqnarray}
 H_u = H^* \sin\beta\,,  \quad H_d= H\cos\beta\,, \end{eqnarray}
 where  $H$ is the light Higgs doublet combination and $\tan\beta = \langle H_u\rangle /\langle H_d\rangle$. For convenience, one may eliminate the axion coupling to $H$ through  the additional  field redefinition
 $\psi \rightarrow e^{-2iY_{\psi}\cos 2\beta a(x)/f_a}\psi$, where $Y_\psi$ is the $U(1)_Y$-hypercharge of $\psi=(q,u^c,d^c,\ell,e^c)$.  Ignoring radiative corrections to the axion couplings, the above procedure yields the low energy axion couplings at $\mu=2$ GeV as
\begin{eqnarray}
 \frac{1}{32\pi^2}\frac{a(x)}{f_a} \big(6g_c^2G^a_{\mu\nu}\tilde G^{a\mu\nu}  +16e^2F_{\mu\nu}\tilde F^{\mu\nu} \big) +
 \frac{\partial_\mu a}{2f_a}\sum_{\Psi=u,d,e,...} c_{a\Psi}\bar \Psi\gamma^\mu\gamma_5\Psi\,,
 \end{eqnarray}
 where
\begin{eqnarray}
c_{au} \simeq  2\cos^2\beta\,, \quad c_{ad}   \simeq 
c_{ae} \simeq 2\sin^2\beta\,.\end{eqnarray}
One then finds $N_{DW}=6$, implying that this DFSZ model may suffer from the cosmological domain wall problem in the post-inflationary scenario \cite{Marsh:2015xka}.
It is also straightforward to find 
the resulting axion couplings to photons and nucleons at scales well below the QCD scale. These are given by
\begin{eqnarray}
g_{a\gamma\gamma}\simeq 2.24\frac{\alpha_{em}}{\pi}\frac{1}{f_a}\,,\quad c_{ap}\simeq -1.1-2.56\sin^2\beta\,, \quad
c_{an}\simeq -1.0 + 2.56 \sin^2\beta\,.\end{eqnarray}

\subsubsection{Composite axion}
A key motivation for composite axions is to generate the axion scale $f_a$ by a confining hidden dynamics dubbed as \textit{axicolor}, which would allow the model to avoid the scale hierarchy problem associated with the spontaneous breakdown of $U(1)_{\rm PQ}$ \cite{Kim:1984pt,Choi:1985cb}. Recently it has been noticed that in certain type of composite axion models $U(1)_{\rm PQ}$ appears
as an accidental symmetry which is well protected from quantum gravity by the gauge symmetries of the model, thereby the model also avoids the PQ quality problem 
\cite{Redi:2016esr,Gavela:2018paw,Cox:2023dou,Agrawal:2025mke}. Here, we present two examples of composite axion models, one with a vector-like axicolor sector and another with a chiral axicolor sector.

Our first example  includes the axicolor gauge group
$G_{\rm axicolor}=SU(N_a)$ confining at $\Lambda_a$, together with the vector-like left-handed axicolored fermions, which are assumed to be neutral under  $SU(2)_W\times U(1)_Y$ for simplicity
\cite{Kim:1984pt,Choi:1985cb}:
\begin{eqnarray}
\eta_3=\big(
\psi_{(N_a, 3)},  \psi^c_{(\bar N_a, \bar 3)}\big)\,, \quad \eta_1=\big(\psi_{(N_a,1)},   \psi^c_{(\bar N_a,1)}\big)\,,
\end{eqnarray}
where the subscripts $(N_a,3), ..., (\bar N_a,1)$ denote the representation of the gauge group $SU(N_a)\times SU(3)_c$.  Assuming that these axicolored fermions have  vanishing bare mass, the model has a colour-anomalous, but axicolor anomaly-free $U(1)_{\rm PQ}$ symmetry
under which \begin{eqnarray}  \eta_3\, \rightarrow \,e^{i\alpha/2}\eta_3\,, \quad
\eta_1\,\rightarrow \, e^{-i3\alpha/2}\eta_1\,.
\end{eqnarray}
 This $U(1)_{\rm PQ}$ is spontaneously broken by the fermion condensation
parameterized as
\begin{eqnarray}
\langle \psi_{(N_a,3)}\psi^c_{(\bar N_a,\bar3)}\rangle =\Lambda_a^3 e^{ia(x)/f_a}\,, \quad
 \langle \psi_{(N_a,1)}\psi^c_{(\bar N_a, 1)}\rangle =\Lambda_a^3 e^{-3ia(x)/f_a}\,,
\end{eqnarray}
where $\Lambda_a\sim f_a$.
Integrating out the axicolor sector at $\mu\sim f_a$, the axion effective Lagrangian is simply given by
\begin{eqnarray}
{\cal L}_a (\mu\sim f_a) =\frac{1}{2}\partial_\mu a\partial^\mu a+
\frac{N_a}{32\pi^2}\frac{a(x)}{f_a} G^a_{\mu\nu}\tilde G^{a\mu\nu}\,.
\end{eqnarray}
This shows that $N_{DW}=N_a >1$ and the model may suffer from the cosmological domain wall problem in the post-inflationary scenario \cite{Marsh:2015xka}. Apparently, the axion couplings to the SM quarks and leptons at $\mu\sim f_a$ are all vanishing as in the case of the KSVZ axion. As a consequence, low-energy axion couplings to the photons, nucleons, and electrons
predicted by this composite axion model are similar to those of the KSVZ axion. More explicitly, those couplings  can be obtained from
Eqs.~\eqref{eq:ksvzcoupling3_1.2.1},  \eqref{eq:ksvzcoupling4_1.2.1}, \eqref{eq:ksvzcoupling1_1.2.1}, and \eqref{eq:ksvzcoupling2_1.2.1} by replacing $(f_a, Y_Q)$ with $(f_a/N_a, 0)$. 

Our next example involves the axicolor gauge group $SU(5)_a$
with the axicolored left-handed fermions \cite{Gavela:2018paw}
\begin{eqnarray}
 \eta_{10}=\big(\psi_{(10, 3)} , \psi_{(10,\bar 3)}\big)\,, \quad \eta_{\bar 5}=\big(\psi_{(\bar 5, 3)}, \psi_{(\bar 5,\bar 3)}\big)\,,
 \end{eqnarray}
 where again the subscripts $(10,3),..., (\bar 5,\bar 3)$  denote the representation of the gauge group $SU(5)_a\times SU(3)_c$. Without further assumption, this model  has an accidental colour-anomalous, but axicolor anomaly-free $U(1)_{\rm PQ}$ symmetry under which
 \begin{eqnarray}
 \eta_{10}\,\rightarrow\, e^{i\alpha/10}\eta_{10}\,, \quad \eta_{\bar 5}\,\rightarrow\, e^{-i3\alpha/10}\eta_{\bar 5}\,,\end{eqnarray}
 which is spontaneously broken by the dimension-9 
fermion condensation
\begin{eqnarray}
\langle \eta_{10}\cdot\eta_{\bar 5}\cdot\eta_{\bar 5} \cdot \eta_{10}\cdot \eta_{\bar 5}\cdot \eta_{\bar 5}\rangle\sim \Lambda_a^9 e^{ia/f_a}\,.
\end{eqnarray}
An appealing feature of this chiral axicolor model is that $U(1)_{\rm PQ}$ appears as an accidental symmetry, which is valid up to the dimension-8 operators due to the gauge symmetries of the model \cite{Gavela:2018paw}.  Once the confining axicolor sector is integrated out, the axion effective Lagrangian density at $\mu\sim f_a$ is given by
\begin{eqnarray}
{\cal L}_a (\mu\sim f_a) =\frac{1}{2}\partial_\mu a\partial^\mu a+
\frac{2}{32\pi^2}\frac{a(x)}{f_a} G^a_{\mu\nu}\tilde G^{a\mu\nu}\,,
\end{eqnarray}
showing that $N_{DW}=2$.
Low energy axion couplings to the photons, nucleons, and electrons
in this model also can be obtained from Eqs.~\eqref{eq:ksvzcoupling3_1.2.1},  \eqref{eq:ksvzcoupling4_1.2.1}, \eqref{eq:ksvzcoupling1_1.2.1}, and \eqref{eq:ksvzcoupling2_1.2.1}
by replacing $(f_a, Y_Q)$ with $(f_a/2, 0)$.

%% file: WG1/content/1-3-3.tex
In this section, we will discuss the couplings of axions to gauge bosons. Interestingly, one can often constrain the properties of these couplings without specifying the exact underlying UV-complete theory. Furthermore, in some cases, no knowledge about the underlying degrees of freedom is required at all. In particular, the central property that we will be interested in is the possible quantisation of these couplings.

We will first present the most general, UV-independent arguments and then move towards less general statements. We will also consider different well-motivated UV-complete models and describe the quantisation properties of axion couplings in each of them. Throughout the discussion, we will make as explicit as possible the underlying assumptions on the UV theory.
\\\\
\textbf{UV-independent quantisation}
\vspace{3pt}
\\
Suppose that all that we are given is the IR theory of axion interactions with the gauge bosons such as photons or gluons, i.e. we are fully agnostic about the underlying UV model. Let us determine the necessary conditions that the axion couplings to gauge bosons have to satisfy.

We start with the case of gluons. The axion coupling of interest is
\begin{equation}\label{axgl}
    \mathcal{L}\; \supset \; \kappa \,  a \, \text{tr}\, G_{\mu \nu} \tilde{G}^{\mu \nu} \, ,
\end{equation}
where we normalise the gauge fields such that their kinetic term is proportional to $1/g^2$, $g$ being the gauge coupling, and $\text{tr}\, G_{\mu \nu} \tilde{G}^{\mu \nu}$ = $G^a_{\mu \nu} \tilde{G}^{a\, \mu \nu}/2$. By definition, the axion field is an angular variable, and so $a \rightarrow a+2\pi f_a n$ ($n\in \mathbb{Z}$) should be a symmetry of the theory. On the other hand, there exists a cyclic direction in the Euclidean gauge field configuration space, and the corresponding winding is characterised by the Pontryagin index defined as an integral over the one-point compactified Euclidean space $(\mathbb{R}^{4})^{\raisemath{2pt}{*}}$
\begin{equation}\label{instnum}
    \mathcal{Q} \; = \; \frac{1}{16\pi^2 }\int_{(\mathbb{R}^{4})^{\raisemath{2pt}{*}}} d^4 x \,\, \text{tr}\, G_{\mu \nu} \tilde{G}^{\mu \nu} \; \in \; \mathbb{Z} ,
\end{equation}
which is also known as the instanton number~\cite{Belavin:1975fg}. Note that the one-point compactification arises due to the requirement that the physical fields $G_{\mu \nu}$ vanish at infinity. Due to the existence of instanton processes, that is, tunneling processes that can change the index $\mathcal{Q}$ of a given field configuration, one is not allowed to simply fix $\mathcal{Q}\equiv 0$ for all physical fields, and so the shift $a \rightarrow a+2\pi f_a n$ ($n\in \mathbb{Z}$) in Eq.~\eqref{axgl} is not a symmetry of the theory unless
\begin{equation}\label{quantnonab}
    \kappa \; = \; \frac{N_{\text{DW}}}{16 \pi^2 v_a} \, , \;\; \text{where} \;\; N_{\text{DW}} \in \mathbb{Z} \, ,
\end{equation}
since only in this case the full Euclidean action remains unchanged independently of $\mathcal{Q} \neq 0$.  Eq.~\eqref{quantnonab} thus establishes the general quantisation requirement for the axion-gluon coupling, and normally one introduces the so-called anomaly coefficient as $N = N_{\text{DW}}/2$. Moreover, although we considered the particular case of gluons, the argument presented here can be trivially extended to any kind of non-Abelian gauge bosons.

Let us now proceed to the case of photons, or in general any Abelian gauge boson. The relevant axion coupling $g_{a\gamma \gamma }$ is:
\begin{equation}\label{lagphot}
     \mathcal{L}\; \supset \; -\frac{1}{4}\, g_{a\gamma \gamma }\,  a \, F_{\mu \nu} \tilde{F}^{\mu \nu} \, .
\end{equation}
Contrary to the non-Abelian case, there are no instanton processes for the Abelian field in Minkowski space. Morally, this can be understood from the fact that the only non-zero contribution to $\mathcal{Q}$ in Eq.~\eqref{instnum} comes from the terms cubic in the gauge potentials. For the Abelian field, there is no self-interaction, and so the integral analogous to that in Eq.~\eqref{instnum} will always give zero. 
As such, we see that the shift $a \rightarrow a + c$ ($c \in \mathbb{R}$) is a symmetry of the theory with the Lagrangian Eq.~\eqref{lagphot} for any value of the coupling $g_{a \gamma \gamma}$.

There is therefore no quantisation requirement on the axion-photon coupling in the most general case.

Still, one can obtain the quantisation of the axion coupling in Eq.~\eqref{lagphot} in more specific cases, by restricting the class of IR theories considered. 
One such class is motivated by the observed quantisation of charge in nature, and is represented by the IR theories that feature magnetic monopoles, and/or that describe the spontaneously broken symmetry phase of some non-Abelian theory. In the latter case, quantisation follows directly from Eq.~\eqref{quantnonab}, while in the former case, one can derive quantisation by treating magnetic monopoles as defects in spacetime and considering the boundary contribution to the action associated with such defects~\cite{Sikivie:1984yz, Kapustin:2005py}. Another well-motivated class of IR theories is represented by those descending from String Theory. In the context of String Theory, as reviewed in Sec.~\ref{sec:1.1.1}, one way to obtain axions is from $p$-form gauge fields, $C_p$, integrated over $p$-cycles. In the 4D EFT, their couplings to gauge bosons come from the coupling of $C_p$ to space-filling $D(p+3)$-branes wrapping compact $p$-cycles. The quantisation of axion couplings in this case is naturally explained by the fact that this coupling is topological, contained in the Chern-Simons part of the $D-$brane action. We refer the reader to \cite{Svrcek:2006yi,Reece:2025thc} for more details. For all these specific cases, the quantisation rule for the axion coupling $g_{a\gamma \gamma}$ can be expressed as:
\begin{equation}\label{photqu}
    g_{a \gamma \gamma} \; = \; \frac{E}{4\pi^2 f_a} \, , \;\; \text{where} \;\; E \in \mathbb{Z} \, .
\end{equation}

Besides the above-discussed interactions of axions with the gauge bosons, it is known that at least in the case of axion-photon interactions, the standard coupling Eq.~\eqref{lagphot} is not the only possibility~\cite{Sokolov:2022fvs, Sokolov:2023pos}. In particular, axion can couple to the dual photon field, which leads to the dual axion-photon coupling $g_{a\mbox{\tiny{MM}}}$, and in general also to the CP-violating ``mixed'' coupling $g_{a\mbox{\tiny{EM}}}$. Such couplings can arise in the theories that feature magnetic monopoles: the magnetic four-potential $A_{\mu}$ is no longer continuous everywhere in spacetime and does not represent a good dynamical variable; Thus the structure of the $U(1)$ gauge theory has to be changed, for instance by considering the Schwinger-Zwanziger theory instead, which has the gauge structure of $U(1)_e\times U(1)_m$ and so naturally gives rise to the above new couplings at the EFT level. The dual axion-photon coupling $g_{a\mbox{\tiny{MM}}}$ has to be quantized similarly to the Eq.~\eqref{photqu}, since from the dual perspective, the usual electrically charged particles play effectively the role of magnetic monopoles, and so the above-mentioned argument about quantisation in the presence of monopoles viewed as defects applies. Note however that since the dual potential is normalised by the magnetic charge unit $g$, the physical coupling is now $g_{a\mbox{\tiny{MM}}} \propto g^2$, and so the quantization units for $g_{a\mbox{\tiny{MM}}}$ are different from the ones for $g_{a\gamma \gamma}$ by a factor of $g^2/e^2$~\cite{Sokolov:2021ydn}.

There exists an important caveat to the above quantisation rules: in particular, we assumed that the axion field couples to gauge fields linearly. In general, instead of $a$ in Eqs.~\eqref{axgl} and~\eqref{lagphot}, quantisation arguments allow any function $g(a)$ that transforms similarly to $a$ under the discrete shifts~\cite{Agrawal:2023sbp}. Then, at small values of $a \ll f_a$ one can expand this function to get an additional non-quantised contribution $g'(0)$ to the axion coupling. This is what happens, for instance, if there exist particles that can mix with the axion and couple to $F_{\mu \nu} \tilde{F}^{\mu \nu}$, such as pions~\cite{Fraser:2019ojt}. As expected, this departure from quantised couplings arises only if the axion has a non-zero mass.
\\\\
\textbf{Prospects for measuring anomaly coefficients}
\vspace{3pt}
\\
Measuring $g_{a\gamma\gamma}$ and the mass for the QCD axion will provide information about the value of the ratio of anomaly coefficients $E/N$. 
Obtaining the value of $E/N$ will be the most immediate target following any discovery, but measuring it precisely is very challenging. 
As an example, if the axion decay constant is $f_a \simeq 10^{12}$~GeV, then the axion can be searched at cavity resonance haloscopes~\cite{ADMX:2018gho, ADMX:2019uok, ADMX:2021nhd}. 
In this case, the rate is proportional to,
\begin{align}
    P_{a\to\gamma}
    &\propto
    \rho_{\rm a} g_{a\gamma\gamma}^2
    \,,
\end{align}
where $\rho_a$ is the local axion density~\cite{Sikivie:1983ip, Sikivie:1985yu}. 
Since the mass of the axion could be precisely determined upon discovery at haloscope experiments, these searches are in a position to measure the ratio $g_{a\gamma\gamma}/m_a$ 
Unfortunately, the quantity $\rho_a$ is difficult to estimate even if we assume the axion makes up all of dark matter (DM), as the local DM density is not known precisely~\cite{Frandsen:2011gi, Read:2014qva,Green:2017odb,Guo:2020rcv}. This presents a challenge to using QCD axion couplings in this way as a precision test of UV physics. In principle, however, positive measurements at different experiments involving different couplings (e.g. the axion-gluon coupling) could be employed to get rid of such uncertainty. This poses other challenges, such as a precise theoretical prediction of the axion-induced EDM, which involves complicated QCD calculations~\cite{Pospelov:2000bw}. 
\\\\
\textbf{Quantisation in specific classes of UV-complete theories}
\vspace{3pt}
\\
Due to their topological properties, measuring axion couplings provides information about the far UV limit of the SM, which is otherwise inaccessible. In this section, we review specific cases.\vspace{5pt} \\
\underline{1. The axion-photon coupling and the global structure of the SM}
\vspace{4pt}\\
Measuring precisely the axion-photon coupling can provide us with UV information about the global structure of the SM, that is, about the allowed representations.
The SM matter content is consistent with the SM group being:
\begin{equation}
   G_{SM}= \left( SU(3)\times SU(2)\times U(1) \right) /\Gamma \,,
\end{equation}
with $\Gamma=\mathbb{I},\,\mathbb{Z}_2,\,\mathbb{Z}_3, \text{ or }  \mathbb{Z}_6$. The concrete $\Gamma$ depends on the allowed charges of particles. As an example, simple GUTs with standard embedding of the SM lead to $\Gamma = \mathbb{Z}_6$, which in turn implies that all particles are $\mathbb{Z}_6$ singlets and isolated states carry an integer electric charge.  
The global structure of the SM has an impact on low-energy axion physics, a topic which has attracted interest recently~\cite{Reece:2023iqn,Choi:2023pdp,Agrawal:2023sbp,Cordova:2023her,Dierigl:2024cxm}. The most general coupling of an axion to the SM gauge bosons is given by\footnote{We write the Lagrangian in a non-canonically normalised basis to make more explicit the quantisation of the axion couplings as well as the periodicity of the axion: $\theta \rightarrow \theta +2\pi$.}:
\begin{equation}
    \mathcal{L}=\frac{k_3}{8\pi^2}\theta \text{tr} F_3\wedge F_3+\frac{k_2}{8\pi^2}\theta \text{tr} F_2\wedge F_2+\frac{k_1}{8\pi^2}\theta \text{tr} F_1\wedge F_1\,.
\end{equation}

The allowed values of $k_3$, $k_2$, and $k_1$ depend on the global form of the SM in a non-trivial way. In \cite{Reece:2023iqn}, it has been shown that for $\Gamma = \mathbb{Z}_6$,  the quantised contribution to the axion-photon coupling is correlated to the axion-gluon coupling. The relation reads:
\begin{equation}
    \frac{2}{3}k_3 + k_1 = \mathbb{Z}\,.
\end{equation}

Under the crucial assumption of having a single axion, this result has phenomenological implications. When the axion has no degenerate vacua -- that is, for $k_3=1$ -- the smallest possible value for the axion-photon coupling $g_{a\gamma}$ is saturated by the $E/N=8/3$ line, therefore defining a well-motivated experimental target. Relatedly, in \cite{Choi:2023pdp}, the authors have shown that assuming a single QCD axion scenario, any model lying below the $E/N=8/3$ line faces a DW problem.

As shown in \cite{Cordova:2023her}, the degeneracy in the axion potential leading to the formation of domain walls can be understood with non-invertible symmetries and, in some scenarios, may be broken by exponentially suppressed UV instantons. This kind of mechanism studied in \cite{Cordova:2023her} would explain the smallness of the bias term solution to the DW problem of axion models.\vspace{5pt} \\
\underline{2. Axion-photon coupling in Grand Unified Theories}
\vspace{4pt}\\
In scenarios where the SM gauge group comes from a 4d theory with a simple, unified group $G_{GUT}$, the axion couplings to photons are very restricted~\cite{Agrawal:2022lsp}. The axion couples to the full UV gauge group as:
\begin{equation}
    \mathcal{L} = \mathcal{A} \frac{\alpha_{GUT}}{8\pi } \frac{a}{f_a}  \text{tr} ( \mathcal{G}_{\mu\nu} \tilde{\mathcal{G}}^{\mu\nu})\,,
\end{equation}
where $\mathcal{G}$ is the GUT field strength, $\alpha_{GUT}$ the GUT coupling, and $2\pi f_a$ the periodicity of the axion. In theories with multiple axions, $a$ is in general a linear combination saturating the mixed anomaly between the PQ symmetry of the axion and the GUT gauge group.

$\mathcal{A}$ is the quantised anomaly coefficient and is unaffected by renormalisation group flow. Below the electroweak symmetry breaking scale (irrespective of the hierarchy between $M_{GUT}$ and $F_a$), the axion-photon and axion-gluon couplings are proportional to the anomaly coefficients $E$ and $N$:
\begin{align}\label{eq:4d_axion_couplings}
    \mathcal{L} 
    \supset 
    \frac{a}{8\pi f_a} 
    \left[ \alpha_{\text{EM}} E F_{\mu\nu} \tilde F^{\mu\nu}
    + \alpha_s N G^a_{\mu\nu} \tilde G^{a, \mu\nu})
    \right] \, .
\end{align}

The embedding of the SM into the unified gauge group, GUT,  fixes the ratio $E/N$.\footnote{For most viable 4d unified theories, the ratio is $E/N=8/3$. See the appendix of \cite{Agrawal:2022lsp} for more general cases.} Equation \eqref{eq:4d_axion_couplings} tells us that the linear combination of axions that couples to photons through the anomaly necessarily couples to gluons, obtaining a potential from QCD.

Additional axions could get a coupling to photons at low energies through their mixing with $a$. 

If their mass eigenvalues $m_b$ are smaller than that of the QCD axion, the photon couplings are suppressed relative to those of the QCD axion by the mixing angle, which scales as $\sim m_{b}^2/m_a^2$. For ALP masses $m_b > m_a$, their couplings remain unsuppressed, but the axions are heavier than the QCD axion. In either case, these additional axion-like particles have photon coupling $g_b$ and mass $m_b$ which satisfy the equation

\begin{equation}\label{eq:bound_coupling}
 \frac{g_{a\gamma  \gamma}}{m_a}\leq C \, \frac{\alpha_{\text{EM}}}{2\pi}\frac{1}{m_\pi f_\pi}\,,
\end{equation} 
where the coefficient $C\sim O(1)$ depends on the SM embedding.
Any axion apart from the QCD axion, therefore, falls to the right of the QCD line in the ($g_{a\gamma\gamma},m_a$) parameter space. The same bound holds for orbifold GUT theories, as shown recently in~\cite{Agrawal:2025rbr}. Looking for ALPs that violate the bound \eqref{eq:bound_coupling} offers a way to test GUTs in the laboratory, astrophysics, and cosmology.
\begin{figure}
    \centering
    \includegraphics[width=0.6\linewidth]{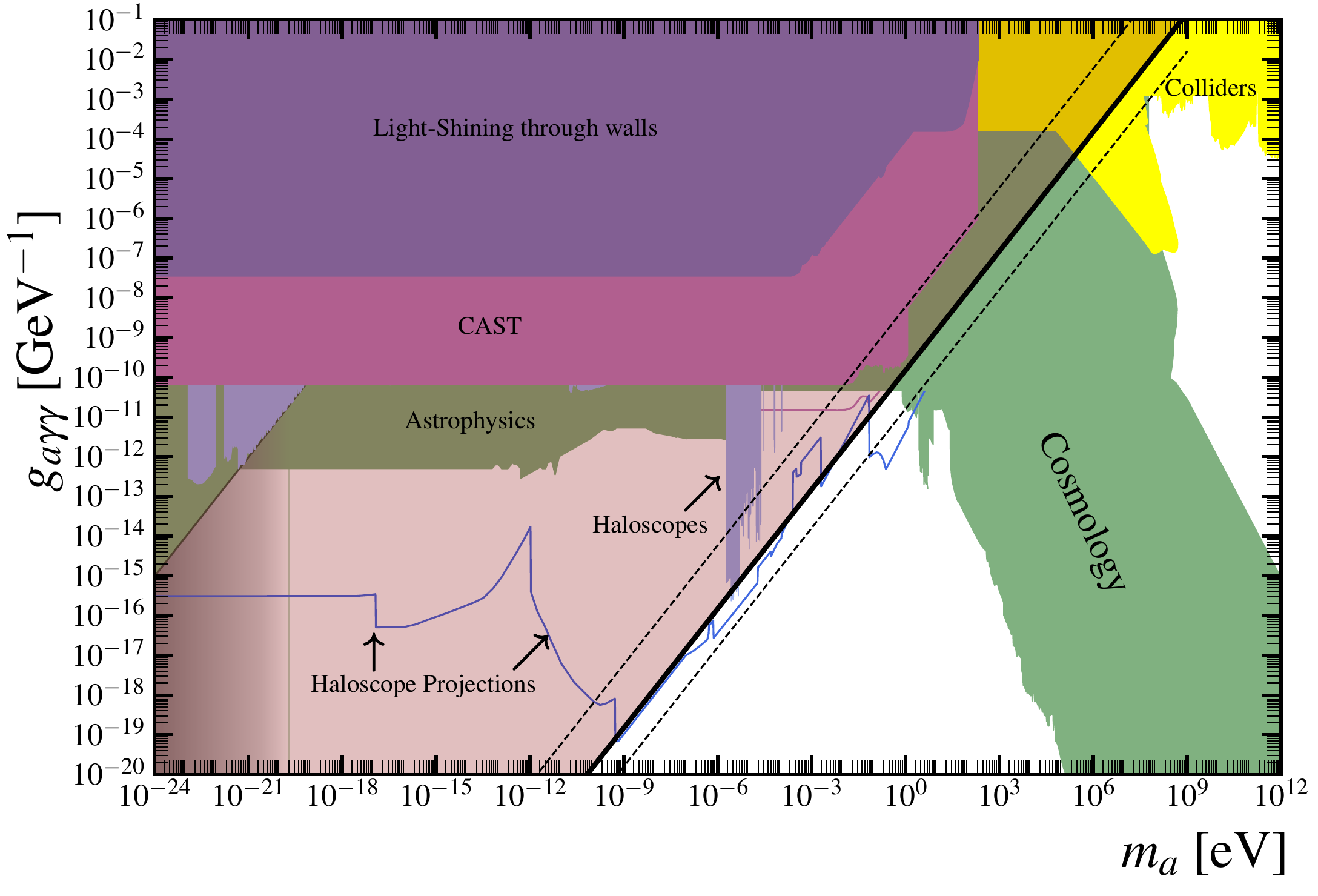} 
    \caption{{Plot summarising the allowed axion parameter space in Grand Unified Theories and heterotic models with standard embedding of the SM, adapted from Refs.~\cite{Agrawal:2022lsp,Agrawal:2024ejr}. The solid black line represents the QCD axion line for the canonical case of $E/N=8/3$, while the dashed lines show the QCD line for $E/N = 2$ and $E/N = 63/2$. The light red shaded area above the QCD axion prediction indicates the region that is not populated in GUT and heterotic string models. The yellow region shows axion constraints from colliders, the green region the constraints from cosmological and astrophysical observations, the purple region shows light-shining through walls experiments, and the pink region is the CAST bound.  Experimental constraints adapted from Ref.~\cite{AxionLimits}.}}
\label{AxionPhoton_heterotic_GUT}
\end{figure} \vspace{5pt}\\

\underline{3. Axion-photon coupling in Heterotic String Theory}
\vspace{4pt}\\
As reviewed in previous sections \ref{sec:1.1.1}, \ref{sec:StringAxionsStats} and \ref{sec:QCDaxionStringTheory}, axions in string theory arise from higher-form gauge fields integrated over extra dimensions. Different string theories include different kinds of higher-form fields. In the case of the heterotic string, axions arise from the $B$ field, a 2-form gauge field under which the superstring is electrically charged. Heterotic axions include \cite{Svrcek:2006yi} the so-called model-independent axion ($a$), model-dependent axions ($b_i$) whose properties, such as the mass and the decay constant, depend on the details of the compact manifold, as well as field-theoretic axions ($c_j$) arising from the phase of complex scalar fields. The latter typically arise in scenarios with pseudo-anomalous $U(1)$ 4d gauge symmetries\footnote{Since they do not change qualitatively the results, we will not consider them explicitly here. See \cite{Agrawal:2024ejr} for details.}. 

The couplings of axion to gauge bosons in the 4d EFT in heterotic strings arise from the Green-Schwarz \cite{Green:1984sg} counter term responsible for 10d anomaly cancellation:
\begin{equation}
   \int_K B\wedge X_8 \,,
\end{equation}
$X_8$ is an eight-form that contains 10d gauge field strength and the curvature 2-form \cite{Polchinski:1998rr}. Consistency of the theory fixes the axion couplings at the 10d level. As shown in \cite{Agrawal:2024ejr,Reig:2025dqb}, axion couplings can be derived by matching mixed anomalies between higher-form global symmetries and the zero-form gauge symmetry in the 10d theory. The 4d axion couplings at the compactification scale are of the form:
\begin{equation}\label{eq:4dEFT_axion_couplings}
   \mathcal{L=}  \frac{\theta_1}{8\pi} \text{tr}_1 F^2 +  \frac{\theta_2}{8\pi} \text{tr}_2 F^2 \,.
\end{equation}
In the equation above the traces $\text{tr}_{1,2} F^2=\sum_jk^{(1,2)}_j\text{tr} \left ( F^{(1,2)}_j\wedge F^{(1,2)}_j\right )$ include the unbroken 4d group factors coming from the $E_8^{1,2}$ group factors. The integers $k_i^{(1,2)}$ are the levels of embedding of the different SM gauge group factors into $E_8\times E_8$. The linear combination of axions $\theta_{1,2}=\theta_{1,2}(a,b_i,c_j)$ in general includes the model-independent, model-dependent, and field-theoretic axions. See also~\cite{Leedom:2025mlr} for a recent study on the expected size of the axion couplings and their masses in the heterotic string.

Quantisation of the axion couplings in 4d implies that these couplings remain unchanged at any energy below the compactification scale. This shows that the coupling to photons at low energies only depends on a few integers $k_i^{(j)}$, which determine the embedding of EM into the 10d gauge group.

Only models where EM is embedded non-trivially into both $E_8$ groups can in principle provide an ALP in the EFT that avoids the coupling to gluons. A light ALP can only be obtained if instantons in the second $E_8$ are sufficiently suppressed. However, a careful study shows that even in this case an upper bound to $g_{a\gamma\gamma}/m_a$ can be obtained~\cite{Reig:2025dqb}, showing that in the weakly coupled heterotic string no ALP above the QCD axion prediction appears. 

Finding an ALP would suppose several model-building challenges for heterotic strings. If $g_{a\gamma\gamma}/m_a\gg 0.01 \text{GeV}^{-2}$, therefore violating the bound~\eqref{eq:bound_coupling} is measured, this would be incompatible with the UV completion of the SM being type-I string theory, heterotic $SO(32)$, and heterotic $E_8\times E_8$ with any kind of SM embedding.\vspace{5pt}\\

\underline{4. KSVZ and DFSZ axion models}
\vspace{4pt}\\
In the standard invisible axion models, such as KSVZ (hadronic)~\cite{Kim:1979if, Shifman:1979if} and DFSZ~\cite{Zhitnitsky:1980tq, Dine:1981rt} models, (partial) quantisation of the axion couplings to photons and gluons follows immediately from integrating out heavy quarks: the result is proportional to the ABJ anomalies of $U(1)_{PQ}$ and therefore the anomaly coefficients are equal to the corresponding group-theoretic factors, and can in fact vary widely depending on the representation in which the integrated out quarks transform.

Specifically, suppose that the new heavy quarks $\mathcal{Q}$ transform in a certain representation \newline $\sum_{\mathcal{Q}} (\mathcal{C}_{\mathcal{Q}}, \, \mathcal{I}_{\mathcal{Q}}, \, \mathcal{Y}_{\mathcal{Q}} )$ of the Standard model gauge group $SU(3)_c\! \times \! SU(2)_L \! \times \! U(1)_Y$, and are also coupled to the PQ field $\Phi$ in the UV via the coupling $(\Phi \bar{\mathcal{Q}}_L \mathcal{Q}_R + \text{h.c.})$. Then, the anomaly coefficients are:
\begin{eqnarray}
    &&N = \sum_Q d(\mathcal{I}_{\mathcal{Q}})\, T (\mathcal{C}_{\mathcal{Q}}) \, , \\ \label{Eanco}
    &&E = \sum_Q d(\mathcal{C}_{\mathcal{Q}})\, \text{tr} \,q_{\mathcal{Q}}^2 \, ,
\end{eqnarray}
where $d(\mathcal{C}_{\mathcal{Q}})$ and $d(\mathcal{I}_{\mathcal{Q}})$ are the dimensions of the colour and weak isospin representations, respectively, $T(\mathcal{C}_{\mathcal{Q}})$ is the Dynkin index of the colour representation, and $q_{\mathcal{Q}}$ is the electromagnetic subgroup generator. One can clearly see that $2N = N_{\text{DW}} \in \mathbb{Z}$ is always an integer, in full consistency with the previously outlined general quantisation argument. On the other hand, the electromagnetic anomaly coefficient $E$ is an integer only if the electric charge of the new quarks is suitably quantised.
\vspace{5pt}\\
\underline{5. Monopole-philic axion model: quantisation of the dual axion-photon coupling}
\vspace{4pt}\\
In the framework of the hadronic axion models, one can also obtain the dual axion-photon coupling $g_{a\mbox{\tiny{MM}}}$, which in particular arises whenever the newly introduced heavy quarks are allowed to carry magnetic charges~\cite{Sokolov:2021eaz, Sokolov:2021ydn, Sokolov:2022fvs, Sokolov:2023pos}. The calculation of the axion-photon coupling is more intricate in this case, and requires worldline path integral methods to correctly account for the Dirac string contributions in the intermediate steps~\cite{Sokolov:2023pos}, however the result is fully in line with the standard expression Eq.~\eqref{photqu}, where the anomaly coefficient is determined by the same group-theoretic factor as in any hadronic axion model. An important difference is that the resulting axion coupling is of a distinct type, and in particular leads to a different form of the axion-Maxwell equations, for instance, qualitatively changing the nature of the signal that one would expect in a given haloscope experiment. Another important distinction, which was mentioned earlier in the general context and which is verified explicitly in the framework of monopole-philic hadronic models, is that the quantisation units for the dual axion-photon coupling differ from the ones for the standard coupling by a factor of $g^2/e^2$.

If the integrated out quark carries both electric and magnetic charges, i.e. a dyon state, an explicit calculation in the framework of monopole-philic axion models shows that there arises a CP-violating $g_{a\mbox{\tiny{EM}}}$ coupling that is also quantised in a way that is standard for the hadronic axion models, however the quantisation units now differ from the standard $g_{a\gamma \gamma}$ case by a factor of $g/e$.

%% file: WG1/content/1-2-2.tex
In this Section, we discuss a specific class of axion models in which it is possible to simultaneously suppress the axion couplings to both nucleons and electrons. 
This realises the so-called astrophobic axion scenario \cite{DiLuzio:2017ogq}, 
wherein the tight bounds on the QCD axion mass from SN1987A \cite{Carenza:2019pxu} and from stellar evolution of 
red giants \cite{Capozzi:2020cbu,Straniero:2020iyi}
and white dwarfs \cite{MillerBertolami:2014rka,Corsico:2019nmr} 
are considerably relaxed compared to benchmark axion models.

It is important to note that achieving the 
nucleophobic condition in a QCD axion context is non-trivial, due to 
irreducible contribution to the axion-nucleon coupling arising from the anomalous axion-gluon interaction.
In fact, it can be seen that in benchmark axion models, such as the KSVZ \cite{Kim:1979if,Shifman:1979if} and the DFSZ \cite{Zhitnitsky:1980tq,Dine:1981rt} model, 
it is not possible to simultaneously suppress the axion coupling to both protons and neutrons. 
On the other hand, the nucleophobic condition necessarily requires a non-universal PQ charge assignment \cite{DiLuzio:2017ogq}, 
where SM fermions from different generations carry distinct PQ charges, also establishing an intriguing link with flavour physics. 

Let us start by considering the nucleophobic condition. 
The axion couplings to nucleons, $N = p,n$, defined via the Lagrangian term 
\begin{equation} 
\label{eq:defcN}
c_{N} \frac{\partial_\mu a}{2 f_a} \bar N \gamma^\mu \gamma_5 N \, , 
\end{equation}
can be computed in the framework of heavy baryon Chiral Perturbation Theory,  
a non-relativistic effective field theory where nucleons are at rest and the axion 
is treated as an external current 
(see \cite{GrillidiCortona:2015jxo,Vonk:2020zfh,Vonk:2021sit} for details). 
In particular, working at leading order with three active flavours, one obtains 
\begin{align}
\label{eq:CpDelta}
c_{p} &=
\left(c_{u} - \frac{1}{1+z+w} \right) \Delta_u 
+ \left( c_{d} - \frac{z}{1+z+w} \right) \Delta_d 
+ \left( c_{s} - \frac{w}{1+z+w} \right)  \Delta_s \, ,
\\
\label{eq:CnDelta}
c_{n} &=
\left(c_{u} - \frac{1}{1+z+w} \right) \Delta_d 
+ \left( c_{d} - \frac{z}{1+z+w} \right) \Delta_u 
+ \left( c_{s} - \frac{w}{1+z+w} \right)  \Delta_s \, ,
\end{align}
where $c_{u,d,s} \equiv c_{u,d,s} (\mu=2 \, \text{GeV})$ 
are low-energy axion couplings to quarks, 
defined analogously to Eq.~\eqref{eq:defcN}, and 
evaluated 
at the scale $\mu = 2 \, \text{GeV}$ 
by numerically solving the renormalisation group (RG) equations  
from the boundary values $c_{u,d,s}(f_a)$
(cf.~discussion below). 
For the quark mass ratios, we have 
$z = m_u(2 \, \text{GeV})/m_d(2 \, \text{GeV}) = 0.473(17)$  
and $w = m_u(2 \, \text{GeV})/m_s(2 \, \text{GeV}) 
=0.0233(9)$ \cite{ParticleDataGroup:2024cfk}. 
$\Delta_q$, with $q=u,d,s$, are hadronic matrix
elements encoding the contribution of a quark $q$ to the spin operator $S^\mu$ of the proton, 
defined as $S^\mu \Delta q = \langle p | \bar{q} \gamma^\mu \gamma^5 q | p \rangle$.
In particular,
$g_A = \Delta_u - \Delta_d = 1.2754(13)$ 
is extracted 
from $\beta$-decays \cite{ParticleDataGroup:2024cfk},  
while $g_0^{ud} = \Delta_u + \Delta_d = 0.44 (4)$ and $\Delta_s = -0.035(9)$ 
are obtained via lattice QCD simulations~\cite{FlavourLatticeAveragingGroupFLAG:2021npn}.

Running effects \cite{Choi:2017gpf,Chala:2020wvs,Bauer:2020jbp}
on the low-energy  axion couplings  
to light quarks 
can be expressed as \cite{Choi:2021kuy}
\begin{align}
\label{eq:CuCdCe}
c_{u,d,s} &\simeq c^0_{u,d,s} + r^t_{u,d,s} \, c^0_t \, , 
\end{align}
where $c^0_{u,d,s,t} \equiv c_{u,d,s,t} (f_a)$ are axion couplings defined at the UV scale $\mu = f_a$, 
the parameters $r^t_{u,d,s}$ encode the RG 
correction approximated by taking the leading one-loop top-Yukawa contribution,
and depend logarithmically on the mass scale of the heavy scalar degrees of freedom in the  UV completion of the axion model, 
which is assumed to be of $\mathcal{O}(f_a)$. 
The values of $r^t_{u,d,s}$, 
obtained
by interpolating the numerical solution to the RG equations, 
are tabulated in Appendix B of \cite{DiLuzio:2023tqe}. 
In the following, we will set as a reference value, 
 $f_a = 10^8$ GeV, 
corresponding to $r^t_{u} = -0.2276$ and $r^t_{d,s} = 0.2290$, 
and neglect the small logarithmic dependence from 
shifts in $f_a$ 
when varying the axion mass. 

In order to discuss nucleophobic axion models, it is convenient to consider the linear combinations 
\begin{align}
\label{eq:cppcn}
c_p + c_n & 
\simeq \left( c^0_u + c^0_d -1 \right) g_0^{ud} + 2 (c^0_s + r^t_{s} \, c^0_t ) \Delta_s \, , \\
\label{eq:cpmcn}
c_p - c_n &
\simeq  \left( c^0_u - c^0_d + (r^t_{u} - r^t_{d}) c^0_t - \frac{1-z}{1+z} \right) g_A \, ,
\end{align}
where 
in the last step we neglected $\mathcal{O}(w)$ corrections and 
employed the approximation $r^t_{u} + r^t_{d} \simeq 0$ (both approximations hold at the per mil level). 
Neglecting also $2 c_s \Delta_s$ in Eq.~\eqref{eq:cppcn}, that amounts to a few per cent (the precise value depending on the specific axion model), nucleophobia requires the following condition between UV axion couplings 
\begin{equation} 
\label{eq:firstnucl}
c^0_u + c^0_d = 1 \qquad \text{(1st nucleophobic condition)} \, .
\end{equation}
As we are going to show, this condition can be enforced exactly in terms of non-universal PQ charges, 
and implies $c_p+c_n\simeq 0$.  
In contrast, to ensure that also $c_p-c_n\simeq 0$, 
requires a tuning of $c^0_u - c^0_d$ against the remainder terms in Eq.~\eqref{eq:cpmcn}: 
\begin{equation} 
\label{eq:secondnucl}
c^0_u - c^0_d = \frac{1-z}{1+z} - (r^t_{u} - r^t_{d}) c^0_t  \qquad \text{(2nd nucleophobic condition)} \, .  
\end{equation}
In such a case, the SN1987A bound on the axion mass 
can be relaxed by 1-2 orders of magnitude 
(e.g.~with respect to the benchmark KSVZ model)
at the cost of a single tuning, 
with the residual limit set by the $2 c_s \Delta_s$ factor in Eq.~\eqref{eq:cppcn}. 

Different realisations of nucleophobic axion models were proposed in Ref.~\cite{DiLuzio:2017ogq} 
(see also \cite{Bjorkeroth:2018ipq,Bjorkeroth:2019jtx,Badziak:2021apn,DiLuzio:2021ysg,Lucente:2022vuo,Badziak:2023fsc,Takahashi:2023vhv,Badziak:2024szg} for model-building variants). 
Here, we consider for definiteness the model denoted as M1 in Ref.~\cite{DiLuzio:2017ogq}, although similar conclusions also apply to other nucleophobic axion models. 

The M1 model features the same scalar sector of the standard DFSZ model \cite{Zhitnitsky:1980tq,Dine:1981rt},  
namely a complex scalar, singlet under $SU(3)_c\times SU(2)_L\times U(1)_Y$, $\Phi \sim (1,1,0)$ and two Higgs doublets $H_{1,2} \sim (1,2,-1/2)$, which are coupled in 
the scalar potential via the non-hermitian operator $H_2^\dag H_1 \Phi$. The vacuum angle is defined by
$\tan\beta = \langle H_2 \rangle / \langle H_1 \rangle$, so that the 
requirement that the PQ current is orthogonal to the hypercharge current fixes the  PQ charges of the two Higgs doublets 
as $X_1=-\sin^2_\beta \equiv -s^2_\beta$ and
$X_2= \cos^2_\beta \equiv c^2_\beta$, 
while we normalize $X_\Phi =1$. 
The M1 model is further characterised by a 2+1 structure of the PQ charge assignments, namely, two generations replicate the same set of PQ charges. Note that, as explained in Ref.~\cite{DiLuzio:2017ogq}, in this case all the entries in the up- and down-type quark Yukawa matrices are allowed, and there are no texture zeros. 
In particular, the Yukawa sector of the M1 model contains the following operators
\begin{align}
&\bar q_\alpha  u_\beta H_1\, ,\ \:\bar q_3  u_3 H_{2}\, , \ \:\bar q_\alpha  u_3 H_{1}\, , \ \:\bar q_3  u_\beta H_{2}\, , \nonumber  \\
  \label{eq:m1}
&\bar q_\alpha d_\beta \tilde H_2\, , \ \:\bar q_3  d_3 \tilde H_{1}\, ,\ \:\bar q_\alpha d_3 \tilde H_{2}\, ,\ \:\bar q_3  d_\beta \tilde H_{1}\, , \nonumber \\
&\bar \ell_\alpha  e_\beta \tilde H_1\, , \ \:\bar \ell_3  e_3 \tilde H_{2}\, , \ \, \:\bar \ell_\alpha  e_3 \tilde H_{1}\, ,\ \, \:\bar \ell_3  e_\beta \tilde H_{2}\, , 
\end{align}
where $\alpha,\beta=1,2$ span over the first and second generations, 
 while $q,\, \ell$ denote left-handed (LH)  
doublets and $u,\, d,\, e$ right-handed (RH) 
singlets. 
Exploiting the invariance under global baryon and lepton number, the PQ charges of SM charged leptons, 
as implied by the Yukawa operators 
in Eq.~\eqref{eq:m1}, can be cast as
\begin{align}
X_{q_i} &= (0, 0, 1)\, , \nonumber \\
X_{u_i} &= (s_\beta^2, s_\beta^2, s_\beta^2)\, , \nonumber \\ 
X_{d_i} &= (c_\beta^2, c_\beta^2,c_\beta^2)\, , \nonumber \\
X_{\ell_i} &=  -X_{q_i}\, ,\: X_{e_i}=-X_{u_i}\, .
\end{align}
The associated anomaly coefficients 
are $E/N = 2/3$ 
and $2N=1$ (cf.~\cite{DiLuzio:2020wdo} for standard notation), while the mixing-independent part of the axion couplings to SM fermions are  
\begin{align}
\label{eq:M1coupl}
&c^0_{u,c} =s_\beta^2\, ,\quad 
\; \ \ c^0_t=-c_\beta^2\, , \nonumber \\ 
&c^0_{d,s}=c_\beta^2\, ,\quad 
\; \ \ c^0_b=-s_\beta^2\, , \nonumber \\
&c^0_{e,\mu}=-s_\beta^2\, , \quad 
c^0_\tau=c_\beta^2\, ,
\end{align}
with 
$\tan\beta \in [0.25, 170]$ set by perturbativity \cite{DiLuzio:2020wdo}.
Since the charges of the RH fields are generation independent, there are no corrections from RH mixings. In the LH sector, mixing effects can play a role because the third generation 
has different charges from the first two. For the quarks, we assume that the LH mixing matrix 
has CKM-like entries so that mixing effects are small, and we will neglect them. 

From Eq.~\eqref{eq:M1coupl} it follows that the first nuclephobic condition in Eq.~\eqref{eq:firstnucl} 
is automatically satisfied, while $c^0_u - c^0_d = s^2_\beta - c^2_\beta$. 
Neglecting RG effects, the second nucleophobic condition in Eq.~\eqref{eq:secondnucl} 
is approximately satisfied for $\tan\beta \simeq \sqrt{2}$. On the other hand, as argued in Ref.~\cite{DiLuzio:2022tyc}, RG effects are relevant for the second nucleophobic condition 
and their role is that of changing the cancellation point,  
which in the M1 model gets shifted to $\tan\beta \simeq 1.19$. 
Hence, although the shift in the parameter space region where the nucleophobic condition is realised is sizeable, running effects do not prevent the possibility of having nucleophobic models. 

With only two Higgs doublets responsible for the breaking of the electroweak symmetry and for providing masses to all the fermions, the lepton sector is unavoidably charged under the PQ symmetry, and electrophobia can only be enforced by an extra tuning between the contribution to the axion-electron coupling proportional to the PQ charge of the electron 
and corrections arising from lepton-flavour mixing \cite{DiLuzio:2017ogq}. 

A more elegant alternative, which was discussed in Ref.~\cite{Bjorkeroth:2019jtx}, 
is to introduce a third Higgs doublet $H_3$ with PQ charge $X_3$ that couples only to leptons in such a way that the condition $X_3 \simeq 0$ can be consistently implemented with a renormalizable scalar potential 
and the axion-electron coupling can be parametrically suppressed. 
Remarkably, this model features a strong correlation between the suppressions of the axion couplings to nucleons and electrons, in such a way that nucleophobia and electrophobia are simultaneously realised in the same region of parameter space, 
at the cost of a single $\mathcal{O}(10 \%)$ tuning, in order to relax typical astrophysical bounds on the axion mass 
by one order of magnitude, say from $0.01$ eV up to $0.1$ eV. 
In such a case, the leading astrophysical constraint stems from stellar evolution in the Horizontal Branch \cite{Ayala:2014pea}
via the axion-photon 
coupling, which in models above remains of standard size 
and in the reach of BabyIAXO and IAXO \cite{IAXO:2020wwp}.

It is also worth commenting on other model variants, 
which have explored the possibility of further relaxing the astrophysical constraints on the QCD axion. 
In Appendix~A of Ref.~\cite{Lucente:2022vuo} it has been shown that the model in \cite{Bjorkeroth:2019jtx} 
can be modified in such a way to obtain $E/N=2$,\footnote{The possibility of suppressing the axion-photon coupling for $E/N=2$ 
was originally observed in Ref.~\cite{Kaplan:1985dv}.} which leads to an accidental (per cent) cancellation in the 
axion-photon coupling. This allows to further relax the Horizontal Branch bound and reach QCD axion masses 
around the eV scale.\footnote{In this region of parameter space, cosmological observations constrain the amount 
of thermally produced axions, which behave as hot dark matter. 
However, since the axion-pion coupling is proportional to $c_p -c_n$, nucleophobia also 
implies pionphobia. Consequently, the cosmological bounds arising from axion thermal production via scattering off pions 
(see e.g.~\cite{Chang:1993gm,Hannestad:2005df,DiLuzio:2021vjd,Notari:2022ffe,DiLuzio:2022gsc}) are relaxed as well.} 
Finally, Ref.~\cite{Badziak:2023fsc} (see also \cite{Badziak:2024szg}) proposed a model with a stronger version of nucleophobia 
in which, by construction, the term proportional to $c_s$ in Eq.~\eqref{eq:cppcn} is absent, and a stronger cancellation of the 
axion-nucleon couplings is possible. This allows for QCD axion masses of a few eV, which 
could be tested by future Cosmic Microwave Background surveys (see \cite{Badziak:2024szg} for further details). 

A recent study \cite{DiLuzio:2024vzg} 
evaluated the 
impact 
on the nucleophobic conditions 
of finite-density corrections to axion-nucleon 
couplings \cite{Balkin:2020dsr,Springmann:2024mjp}, 
which are especially relevant in astrophysical environments near saturation density, such as SNe. 
The findings of Ref.~\cite{DiLuzio:2024vzg} indicate that the nucleophobic solution remains viable also at finite density, and the SN axion bound relaxes significantly in nucleophobic models, even when accounting for the integration over the non-homogeneous environment of the SN core.

%% file: WG1/content/non-canonical-QCD.tex
\label{sec:axions_outside_qcd_band}
In the PQ axion solution to the strong CP problem, the axion receives a potential solely from QCD instanton effects that lead to the well-known mass relation:
\begin{equation}
    g_a \equiv \frac{m_a^2 f_a^2}{\chi_{\rm QCD}} = 1\,,
    \label{eq:gfactor}
\end{equation}
that following the notation in Ref.~\cite{Gavela:2023tzu}  we will refer to as the canonical $g$-factor.

Such predictability associated with the QCD axion gives it a unique role in theory constructions and makes it a narrow target for experimental searches. Moreover, connected to this mass scale, it follows a prediction for axion dark matter,
\begin{equation}
\Omega_a h^2 \sim 0.1 \left(\frac{f_a}{10^{12}\,\rm{GeV}}\right)^{7/6} \theta_i^2\,,
\end{equation} 
in the pre-inflationary scenario, that 
determines particularly well-motivated regions in the QCD axion band upon assumptions on the initial misalignment angle, $\theta_i$.

It is, therefore, a pivotal question whether relation~\eqref{eq:gfactor} can be modified and under which assumptions does it hold. In this section, we will focus on three (recent) attempts that enlarge the QCD axion landscape, providing novel motivated territory for current and future experiments.

\subsubsection{Heavier axions from scalar mass mixing}
The work presented in Ref.~\cite{Gavela:2023tzu} challenged a minimal yet strong assumption of the axion mechanism: that the QCD axion appears alone in Nature. Such an assumption is in contrast with rich theoretical frameworks from which not only the axion but additional ALPs are a ubiquitous prediction, such as string theory compactifications~\cite{Svrcek:2006yi,Arvanitaki:2009fg}. 

\begin{figure}
    \centering
    \includegraphics[width=0.5\linewidth]{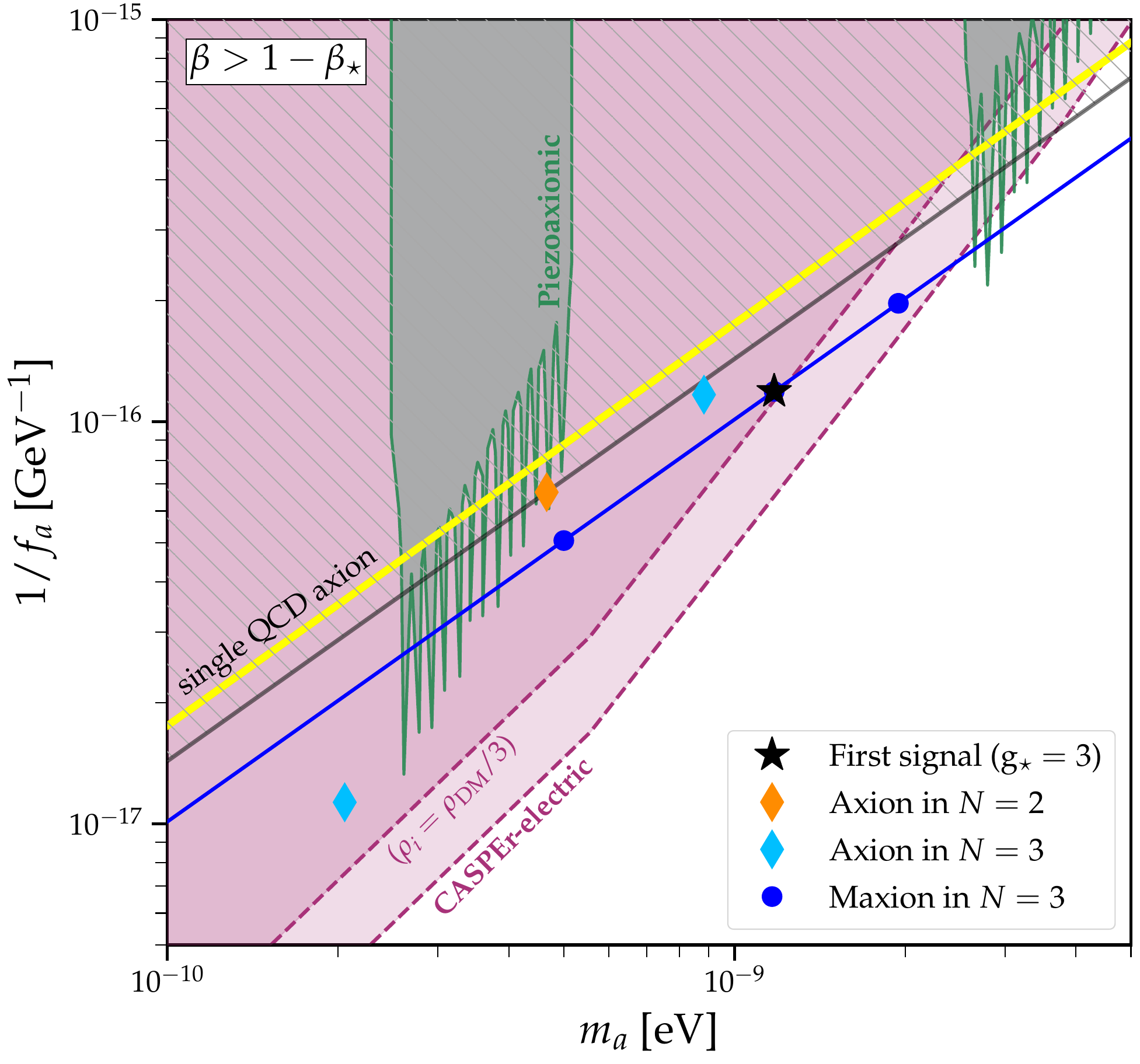}
    \caption{Possible patterns of the multiple QCD axion signals allowed by the sum rule. It is assumed that the first signal is observed at the benchmark point represented by a black star. The gray dashed region
is then excluded if the star describes a true QCD axion (assuming that
the PQ symmetry is only broken by QCD). The projected limits from CASPEr-electric (phase III) and piezoaxionic
experiments are taken from
Refs.~\cite{JacksonKimball:2017elr,Arvanitaki:2021wjk}. Figure adapted from Ref.~\cite{Gavela:2023tzu}.
}
    \label{fig1:sumrule}
\end{figure}
If this is the case, the axion would naturally mix with the other scalars, sourcing extra contributions to the QCD axion potential. These contributions do not spoil the solution to the strong CP problem as long as the PQ symmetry is preserved by the scalar mixing interactions. It has been shown~\cite{Gavela:2023tzu}  that requiring the latter condition on the axion potential is equivalent to an exact sum rule:

\begin{align}
    &\sum\limits_{i=1}^N \frac{1}{g_i} = 1\,, &g_i \equiv \frac{m_i^2 f_i^2}{\chi_{\rm QCD}}\,,
\end{align}

that links the $g$-factors (see Eq.~\eqref{eq:gfactor}) of the $N$ eigenstates in the system that will constitute a \textit{multiple} QCD axion. Notice that the sum rule reduces to~\eqref{eq:gfactor} for $N=1$. Possible patterns of this field are depicted in Fig.~\ref{fig1:sumrule}, which illustrates several of the predictions of the sum rule: {\it (1)} all axion signals appear to the right of the canonical single QCD axion band; 
{\it (2)} the maximum deviation possible for the axion closest to the canonical band is $\sqrt{N}$. In such scenario, all other signals are deviated by the same $\sqrt{N}$ factor and aligned in the same line parallelly to the single axion band. We refer to these signals as \textit{QCD maxions}, which in this limit contribute equally to the axion sum rule.

To discuss the novel features of this mechanism, let us consider a $N=2$ toy model with the following effective Lagrangian:

\begin{equation}
\mathcal{L}_{N=2} = \frac{\alpha_s}{8 \pi}\left(\frac{{\hat a}_1}{\hat f} + \frac{{\hat a}_2}{ \hat f} +\bar \theta\right) G \widetilde{G}-\frac{1}{2}\hat{m}_2^2 \, {\hat a}_2^2\,,
\label{eq:N2model}
\end{equation}
where $\hat a_i$ denote unphysical fields. 
It is easy to see that the PQ symmetry is implemented by shifts of the $\hat a_1$ field, that from here on, we will denote by the PQ field, $\hat a_{\rm PQ}$. Such a shift ensures CP conservation in the QCD vacuum, which can be explicitly checked by studying the induced potential
\begin{equation}
V_{N=2} =  \frac{1}{2}\chi_{\rm QCD} {\left( \frac{{\hat a}_1}{\hat f} + \frac{{\hat a}_2}{\hat f}\right)^2} + \frac{1}{2}\hat{m}_2^2 \, {\hat a}_2^2\,;
\label{eq:toymodel}
\end{equation}
with each term being minimised by the conditions:
\begin{equation}
    { \left< \hat a_1\right>} + { \left< \hat a_2\right>} = 0 \quad \text{\rm and}\quad {\left< \hat a_2\right>} = 0\,.
\end{equation}
While the first condition requires an alignment of the VEVs in field space for a CP-even QCD vacuum, the second one is only selecting a particular direction in this manifold that fixes the VEV of both fields.

\begin{figure}
    \centering
    \includegraphics[width=0.5\linewidth]{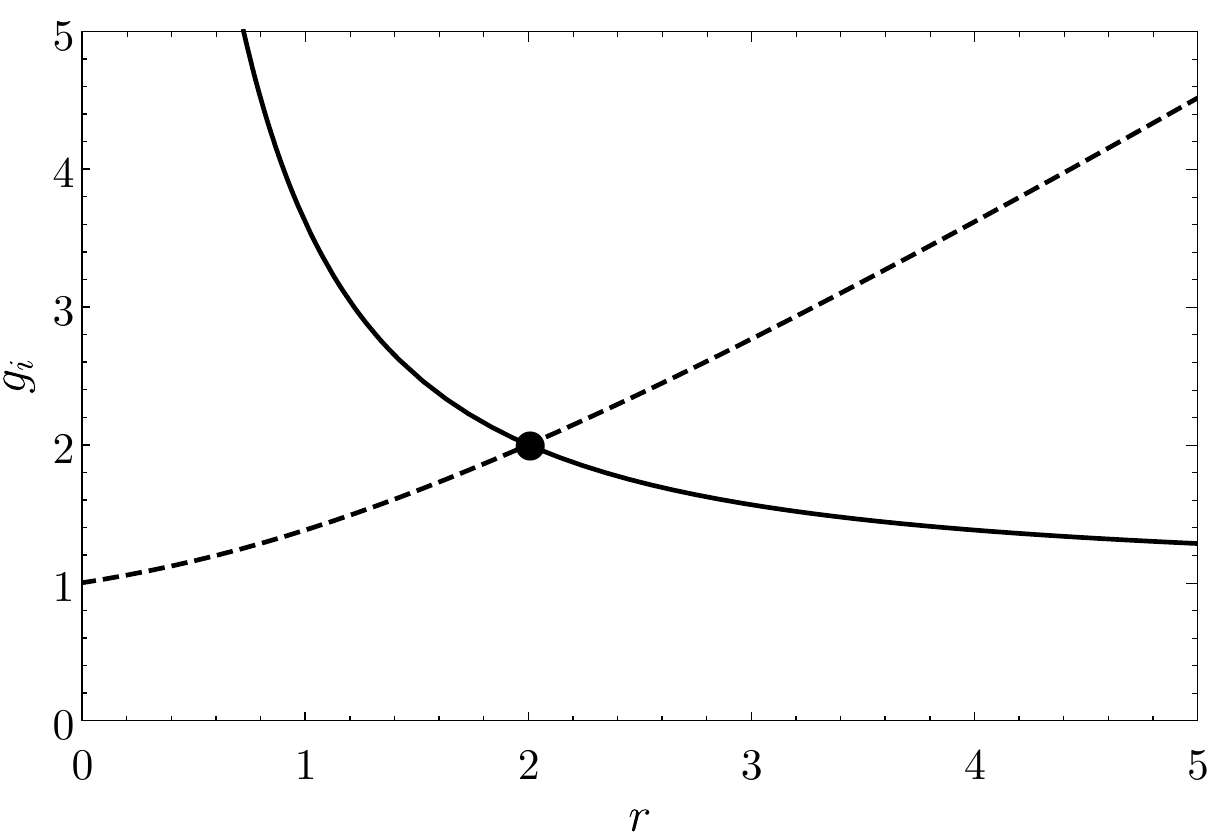}
    \caption{The $g$-factors of the two eigenstates obtained in the model described in Eq.~\eqref{eq:N2model}, with $r \equiv \hat m_2^2 \hat f^2/\chi_{\rm QCD}$; see the text for details. In both $r\to 0$ and $r\to \infty$ limits, one eigenstate decouples such that the model reproduces the single QCD axion case. For $r=2$, we obtain instead the maximal deviation to the standard scenario, with two~\textit{QCD maxions} being produced. Figure adapted from Ref.~\cite{Gavela:2023tzu}.
    }
    \label{fig2:sumrule}
\end{figure}
The original Lagrangian in this example can be rotated to make it explicit that only one combination of fields couples to gluons, ${\hat{a}_{G\widetilde G}}/{F}\equiv  2 ({{\hat a}_1}+ {{\hat a}_2})/{\hat f}$. This field is allowed to mix with an orthogonal combination, i.e. 
\begin{equation}
\label{eq:L2-otro}
\mathcal{L}_{N=2} = \frac{\alpha_s}{8 \pi}\left(\frac{\hat{a}_{G\widetilde G}}{F} +\bar \theta\right) G \widetilde{G}- 
\frac{1}{4}\hat{m}_2^2\, \big(\hat a_{G\tilde G}-\hat a_{\perp}\big)^2\,.
\end{equation}
In the absence of the second term, ${\hat{a}_{G\widetilde G}}$ behaves as the canonical QCD axion (so that $F=f_a$); however, the introduction of an external scale causes modifications to this prediction. This can be checked by diagonalising the system exactly, obtaining the eigenvectors and eigenvalues of the mass matrix, and consequently computing the $g$-factors of the physical fields. The results are shown in Fig.~\ref{fig2:sumrule}. In the two decoupling limits, $\hat m_2 \ll m_{\rm PQ}$ and $\hat m_2 \gg m_{\rm PQ}$ with $m_{\rm PQ}^2 = \chi_{\rm QCD}/\hat f^2$, we recover the canonical QCD axion solution, as one eigenstate has $g\to \infty$ which forces the other to saturate by itself the sum rule. However, for intermediate values of the extra scale, both eigenstates are found with $g_i\neq 1$, being therefore localised outside the canonical QCD axion band. The orange and black points in Fig.~\ref{fig1:sumrule} are one possible representation of these signals.

We can also verify in this toy model that the point of largest deviation from the canonical single axion band is where ${\rm max}\, {\rm min} \left(g_1,\,g_2\right)=2$, which requires that the external scale $\hat m_2$ is comparable to that induced by QCD effects. At this point, both fields contribute actively and equally to the solution to the strong CP problem.

The previous considerations can be better understood by following a more physical interpretation of the sum rule, namely by noting that
\begin{equation}
    \beta_i \equiv \frac{1}{g_i} = \frac{\left< \hat a_{\rm PQ}| a_i\right> \left< a_i | \hat a_{G \widetilde G}\right>}{\left< \hat a_{\rm PQ}|  \hat a_{G \widetilde G}\right>}\,.
\end{equation}
In particular, the decoupling limits follow trivially from this expression: e.g. in the case $\hat m_2 \gg m_{\rm PQ}$, $a_2$ is an eigenstate which does not overlap with $\hat a_1 = \hat a_{\rm PQ}$; this limit must therefore lead to a canonical QCD axion. These limits are, however, just the corner of a landscape of new possibilities for the QCD axion, whenever the projections above take arbitrary values.

Not only the sum rule predicts the possibility of several novel axion signals, but it also provides guidance to test whether these solve effectively the strong CP problem, if ever a signal of an ALP is found in Nature, for example located in the position of the star in Fig.~\ref{fig1:sumrule} (with $\beta_\star = 1/3$). Due to potential scalar mass mixing, we have learnt that such a signal can perfectly be that of a true QCD axion, as long as it is not the only signal observed in Nature. There is therefore a region of the parameter space (represented in dashed black) that can be theoretically excluded, based on the $\beta_\star$ that was observed; if a second signal would be found at the border of this region, with $\beta = 2/3$, one would know that we had found a complete multiple QCD axion system. In other scenarios, distinct patterns are possible, as illustrated by the light and dark blue points assuming $N=3$. Namely, the dark blue points represent 3 QCD maxions, whose pattern can be generalised to any $N$. 

While the QCD axion sum rule relies on a model-independent effective approach, we have obtained the $N$  conditions on the mass matrix (out of which one is always preserving a PQ symmetry) that produce these exotic maxion patterns where no signal is expected close to the canonical axion line. In a $N=2$ scenario, the extra condition is on the trace of the mass matrix (even though there is no bottom limit on the mass of the lightest eigenmode). The toy model in Eq.~\eqref{eq:toymodel} can therefore be simply completed by e.g. considering a double copy of a KSVZ model, with two PQ scalars and two heavy vector-like quarks, with the introduction of mixing in the scalar sector such that the original symmetry $U(1)_{\rm PQ}^A \times U(1)_{\rm PQ}^B$ is explicitly broken to a single $U(1)_{\rm PQ}$~\cite{Gavela:2023tzu}. Another possibility is to introduce an explicit mass for one of the vector-like quarks.
Notwithstanding, the most exotic patterns that deviate substantially from the canonical band require the presence of a large number of scalar fields that mix sizably, possibly stemming from hidden axions in extra dimensions~\cite{Dienes:1999gw,Dienes:2011ja,deGiorgi:2024elx}.

In spite of the gluonic coupling being the most important one to test the solution to the strong CP problem, the coupling of the axions to photons is subject to a substantially larger sample of probes. In scenarios where all axions develop the same model-dependent coupling to photons, namely if they arise from a GUT~\cite{Agrawal:2022lsp}, an analogous sum rule follows and so also the previous conclusions.

Out of the experimental set of probes, haloscope experiments can currently probe important regions of the single QCD axion band and have the projected sensitivity to uncover large regions beyond it. In small $N$ scenarios, and due to model-dependent effects, it could be challenging to distinguish a multiple from a single QCD axion signal. Nevertheless, given that haloscope experiments can distinguish the frequency spectrum very well, the multiplicity of signals could be the smoking gun test to distinguish these two scenarios~\cite{deGiorgi:2025ldc}. The exact localisation of the multiple signals in this parameter space relies, however, on the computation of the dark matter abundance carried by each of the eigenmodes, which might not follow an independent thermal evolution due to multiple level crossings~\cite{Kitajima:2014xla,Ho:2018qur,Cyncynates:2021xzw,Cyncynates:2023esj,Murai:2024nsp,Dunsky:2025sgz}.

Overall, the QCD axion sum rule changes the paradigm of QCD axion physics \textit{within the SM gauge group setting}. It merges the axion and ALP landscape everywhere to the right of the canonical band. Furthermore, it constitutes a counting rule to know how many axions in Nature are required to form a complete solution to the strong CP problem. The reconstruction of the multiple QCD axion system might in turn require the synergy between different experiments in order to be fully probed, as well as the development of a series of novel studies to assess the cosmological and astrophysical behaviour of its multiple signals.

\subsubsection{Photophilic clockwork axion}  The canonical band for QCD axions corresponds to the parameter region with \begin{eqnarray}
{E}/{N_{DW}} ={\cal O}(1),\end{eqnarray}
where $N_{DW}$ and $E$ are the rational numbers parameterising the axion couplings to the gluons and photons before the axion-pion mixing is taken into account, i.e.
\begin{eqnarray}
 N_{DW} \frac{g_c^2}{32\pi^2}\frac{a}{f_a}G^a_{\mu\nu}\tilde G^{a\mu\nu} +
E\frac{e^2}{32\pi^2} \frac{a}{f_a}F_{\mu\nu}\tilde F^{\mu\nu}\end{eqnarray}
for the axion decay constant $f_a$ defined as  $a(x)\cong a(x)+2\pi f_a$.
An interesting example of QCD axion, which is well outside the canonical band, is a photophilic clockwork axion \cite{Farina:2016tgd,Agrawal:2017cmd}
for which $E/N_{DW}\gg 1$ is achieved by the clockwork mechanism involving multiple
axions in the UV limit \cite{Dvali:2007hz,Choi:2014rja,Choi:2015fiu,Kaplan:2015fuy}.
To illustrate the clockwork mechanism, one may start with a simple example of two axions $a_i\cong a_i + 2\pi f_i$ ($i=1,2$) coupled to the gauge fields 
as
\begin{eqnarray}
{\cal L}_{\rm axion}&=& \frac{1}{2}\partial_\mu a_1\partial^\mu a_1 + \frac{1}{2}\partial_\mu a_2\partial^\mu a_2+
\frac{g_c^2}{32\pi^2}\frac{a_1}{f_1}G^a_{\mu\nu}\tilde G^{a\mu\nu}+k\frac{g_1^2}{32\pi^2}\frac{a_2}{f_2}B_{\mu\nu}\tilde B^{\mu\nu}  \nonumber \\
&+&\frac{g_X^2}{32\pi^2}\left(n\frac{a_1}{f_1}+\frac{a_2}{f_2}\right)X^\alpha_{\mu\nu}\tilde X^{\alpha\mu\nu}, \label{eq:twoaxion_1.2.3}
\end{eqnarray}
where  $X^\alpha_{\mu\nu}$,  $G^a_{\mu\nu}$, and $B_{\mu\nu}$ denote the $SU(N)_X\times SU(3)_c\times U(1)_Y$ gauge field strength, and $k$ and $n$ are rational numbers. Assuming $f_1\sim f_2 \gg \Lambda_X\gg \Lambda_{\rm QCD}$, where $\Lambda_X$ is the confining scale of the hidden $SU(N)_X$ gauge interaction, one combination of axions gets a heavy mass $m_{a_H}\sim n\Lambda_X^2/f_1$ and the other combination corresponds to a QCD axion. Then one can integrate out the heavy $a_H$  and consider the low-energy effective theory of the light QCD axion. The axion potential induced by the $SU(N)_X$ gauge interaction enforces $na_1/f_1 +a_2/f_2=0$ which is solved by
\begin{eqnarray}
\frac{a_1}{f_1}=\frac{a}{f_a}, \quad \frac{a_2}{f_2}=-n\frac{a}{f_a},\end{eqnarray}
where $a(x)\cong a(x)+2\pi f_a$ is the QCD axion field. Plugging this solution into  Eq.~\eqref{eq:twoaxion_1.2.3},
one easily finds 
\begin{eqnarray}
{\cal L}_{\rm eff}=  \frac{1}{2}\partial_\mu a\partial^\mu a + \frac{g_c^2}{32\pi^2}\frac{a}{f_a}G^a_{\mu\nu}\tilde G^{a\mu\nu} -nk\frac{e^2}{32\pi^2} \frac{a}{f_a}F_{\mu\nu}\tilde F^{\mu\nu},\end{eqnarray}
with  $f_a^2 =n^2 f_1^2 + f_2^2$, showing that
$E/N_{DW}$ for the QCD axion  is enhanced by the factor $n$. 

To have $E/N_{DW}\gg 1$, one may take the limit $n\gg 1$ which corresponds the Kim-Nilles-Peloso alignment limit \cite{Kim:2004rp}. Alternatively, one can generalise the model by introducing more axions together with the mass mixings between the nearby two axions \cite{Choi:2014rja,Choi:2015fiu,Kaplan:2015fuy}. A simple example \cite{Choi:2015fiu,Agrawal:2017cmd} is 
\begin{eqnarray}
{\cal L}_{\rm CW}&=& \frac{1}{2}\sum_{i=1}^{N}\partial_\mu a_i\partial^\mu a_i + \frac{g_c^2}{32\pi^2}\frac{a_1}{f_1}G^a_{\mu\nu}\tilde G^{a\mu\nu}+k\frac{g_1^2}{32\pi^2}\frac{a_N}{f_N}B_{\mu\nu}\tilde B^{\mu\nu}  \nonumber \\
&+&\frac{1}{32\pi^2}\sum_{j=1}^{N-1}\left(n_j\frac{a_j}{f_j}+\frac{a_{j+1}}{f_{j+1}}\right)X^\alpha_{j\mu\nu}\tilde X_j^{\alpha\mu\nu}, \label{eq:cw_1.2.3}
\end{eqnarray}
where the model involves a tower of confining hidden gauge groups  $\prod_{j=1}^{N-1}SU(N_j)_X$  with the confining scales
$\Lambda_{j}\gg \Lambda_{\rm QCD}$.  This tower of hidden gauge interactions enforces $n_ja_j/f_j + a_{j+1}/f_{j+1}=0$  which is solved by
${a_j}/{f_j}=(-1)^{j-1}\big(\prod_{k=1}^{j-1} n_k\big){a}/{f_a}$,
where $a(x)\cong a(x)+2\pi f_a$ is the lightest axion which can be identified as a QCD axion.  For this QCD axion, it is  straightforward to find \cite{Choi:2014rja,Choi:2015fiu}
\begin{eqnarray}
E/N_{DW}= k\prod_{j=1}^{N-1} n_j \,,\quad \;
f_a/f_i \sim \prod_{j=1}^{N-1} n_j\,, 
  \end{eqnarray}
where all $f_i$ are assumed to be comparable to each other. A particularly interesting feature of this clockwork axion is that \emph{exponentially large} $E/N_{DW}$ and $f_a/f_i$  can be achieved with $n_i={\cal O}(1)$ and a moderately large $N$. It is also straightforward to construct a renormalizable UV completion of the effective lagrangian density Eq.~\eqref{eq:cw_1.2.3}. A simple example involves $N$
PQ-charged complex scalar fields $\phi_i$  developing the vacuum values $\langle \phi_i\rangle=f_ie^{a_i/f_i}/\sqrt{2}$, which are  coupled to the gauge and PQ-charged fermions as  
\begin{eqnarray}
\Delta {\cal L}_{\rm UV} = \kappa \phi_1 QQ^c +\kappa^\prime \phi_N LL^c +\sum_{j=1}^{N-1} \big(y_j \phi_j \lambda_j\lambda_j +y^\prime_j \phi_{j+1}\psi_j\psi_j^c\big) + {\rm h.c.}\,,
\end{eqnarray}
where $(Q,Q^c)=(3,\bar 3)$ of $SU(3)_c$, $(\lambda_j, \psi_j,\psi_j^c)=({\rm Adj}, N_j,\bar N_j)$ of $SU(N_j)_X$, and $(L,L^c)$ 
have the $U(1)_Y$-hypercharge $(Y_L, -Y_L)$. This UV completion  yields  \begin{eqnarray}
n_j = N_j, \quad k=Y_L^2\,.
\end{eqnarray}

\subsubsection{$Z_\mathcal{N}$ axion} 
An alternative approach to obtain a QCD axion that significantly deviates from the canonical band, 
resulting in a universal enhancement of all axion couplings, is offered by the so-called
$Z_\mathcal{N}$ axion model \cite{Hook:2018jle,DiLuzio:2021pxd,DiLuzio:2021gos}. 
This is based on $\mathcal{N}$ mirror copies of the SM,
that are interchanged under 
a $Z_\mathcal{N}$ symmetry, which is non-linearly realised by the axion field: 
	\begin{align}
	Z_\mathcal{N}:\quad &\text{SM}_{k} \longrightarrow \text{SM}_{k+1\,(\text{mod} \,\mathcal{N})} \label{mirror-charges} \, , \\
	     & a \longrightarrow a + \frac{2\pi k}{\mathcal{N}} f_a\,,
	     \label{axion-detuned-charge}
\end{align}
with  $ k=0,\ldots, \mathcal{N}-1$.  
The most general Lagrangian implementing this symmetry reads 
\begin{equation}
\label{Eq: Lagrangian ZN}
\mathcal{L} 
= 
\sum_{k=0}^{\mathcal{N}-1} \left[ \mathcal{L}_{\text{SM}_k} +
\frac{\alpha_s}{8\pi}  \left( \theta_a + \frac{2 \pi k}{\mathcal{N}} \right) G_k \widetilde G_k \right] \, ,
\end{equation}
where we introduced the angular axion field,  
$\theta_a \equiv a / f_a$, defined in the interval $[-\pi,\pi)$. 
A universal (equal for all $k$ sectors) bare theta parameter has been set to zero
via an overall shift of the axion field, 
and the    
SM is identified from now on with the $k=0$ sector. 
 Each QCD$_k$ sector contributes  to the $\theta_a$ potential, 
which in the 2-flavour leading order chiral expansion reads
\begin{align}
V_\mathcal{N}(\theta_a)=-m_{\pi}^{2} f_{\pi}^{2}\sum_{k=0}^{\mathcal{N}-1}  \sqrt{1-\frac{4z}{(1+z)^2} \sin ^{2}\left(\frac{\theta_a}{2}+\frac{\pi k}{\mathcal{N}}\right)}\, ,
\label{Eq:Vsmilga ZN}
\end{align}
with $z = m_u / m_d \sim 1/2$. 
Remarkably, the resulting axion is exponentially lighter than the canonical QCD axion, 
because the non-perturbative contributions to its potential from the $\mathcal{N}$ degenerate QCD sectors conspire 
by symmetry to suppress each other. In the large$-\mathcal{N}$ limit the total axion potential is given by \cite{DiLuzio:2021pxd} 
 \begin{align}
  \label{Eq: fourier potential large N hyper}
V_{\mathcal{N}}\left(\theta_a\right)
  \simeq - \frac{m_a^2 f_a^2}{\mathcal{N}^2} \,\cos (\mathcal{N}\theta_a)\,, 
\end{align}
where now the axion mass obeys
\begin{align}
m_a^2 \simeq \frac{m_{\pi}^2 f_{\pi}^2}{f_a^2} \frac{1}{\sqrt{\pi}} \sqrt{\frac{1-z}{1+z}} \,\,\mathcal{N}^{3 / 2} \, z^{\mathcal{N}} \,,
\label{maZNLargeN}
\end{align}
 which  is  exponentially suppressed ($\propto z^\mathcal{N}$)  in comparison to  the standard QCD axion mass. 
 Equivalently, for fixed axion mass, this results in a universal enhancement of all axion couplings, 
 with strong implications for the prospects of axion-like particle experiments, whose reach is still far from the canonical 
 QCD axion band.

Moreover, the evolution of the $Z_{\mathcal N}$ axion field and its contribution to the dark matter 
relic abundance depart drastically from the standard case. In Ref.~\cite{DiLuzio:2021gos}, it was shown that a variant of the standard misalignment mechanism, named trapped misalignment (see also \cite{Higaki:2016yqk,Kawasaki:2017xwt,Nakagawa:2020zjr,DiLuzio:2024fyt}), 
arises due to the peculiar temperature dependence of the $Z_{\mathcal N}$ axion potential.
The interplay of the different mechanisms together with the implications of the $Z_{\mathcal N}$ reduced-mass axion for axion dark matter searches was studied in detail in \cite{DiLuzio:2021gos}, including the experimental prospects to probe its coupling to photons, nucleons, electrons and the nEDM operator.

%% file: WG1/content/1-2-4.tex
\label{sec:other_ALPs}
In this section, we discuss different extensions of the classical axion models, offering potential solutions to the strong CP problem while providing a link to the fermion masses and/or neutrino mass generation. 

\subsubsection{Flaxion/Axiflavon}
The Flaxion/Axiflavon~\cite{Ema:2016ops,Calibbi:2016hwq} extends the axion framework by incorporating flavour symmetries, which are introduced to explain the observed patterns in the fermion mass spectrum and mixing. In this model, the axion is not only responsible for solving the strong CP problem but also connects to the origin of flavour hierarchies (see \cite{Wilczek:1982rv} for an early discussion). The name ``Flaxion'' or ``Axiflavon'' reflects this dual role in addressing both CP violation and flavour physics.

The model predicts specific couplings between this axion and SM particles, ruled by the Froggatt-Nielsen symmetry~\cite{Froggatt:1978nt}. The axion field is linked to the breaking of a global $U(1)_{FN}$ flavour symmetry, which controls the masses of SM fermions. Focusing on the quark sector and given the freedom when introducing the symmetry, the six flavours of quarks will transform in a different way under this $U(1)_{FN}$. The charges will be labelled as $x_{Q_i},\ x_{u_i},\ x_{d_i},\ x_{L_i},\ x_{e_i}$, associated respectively to the left-handed quark doublet, the right-handed up-type quark, the right-handed down-type quark, the left-handed lepton doublet and the right-handed charged lepton, all of them of the \emph{i-th} generation. While all fermions are charged under $U(1)_{FN}$, the Higgs field $H$ remains neutral for simplicity. Instead, the new symmetry is spontaneously broken at a larger scale by a new complex scalar field $\Phi$, which acquires a vacuum expectation value (VEV) $v_\Phi$. This new field is introduced in the Yukawa couplings as one would usually do in the FN mechanism:
\begin{equation} \label{eq:Axiflavon_Yukawa_Sector}
\mathcal{L}_{Y} = a_{ij}^u \bar{Q}_i u_j H\left (\frac{\Phi}{\Lambda}\right)^{x_{u_j}-x_{Q_i}}+a_{ij}^d \bar{Q}_i d_j H \left(\frac{\Phi}{\Lambda}\right)^{x_{d_j}-x_{Q_i}}+a_{ij}^e \bar{L}_i e_j H\left(\frac{\Phi}{\Lambda}\right)^{x_{e_j}-x_{L_i}}+h.c.\,,
\end{equation}
where $a_{ij}^{u,d,e}$ are complex numbers, $Q_i, \ u_i,\ d_i,\ L_i$ and $e_i$ are the SM Weyl fermions of the \emph{i-th} generation and $\bar{Q}=Q^{\dagger}\gamma_0$ is the usual Dirac adjoint, with $\gamma_i$ being Dirac matrices. For these terms to be invariant under $U(1)_{FN}$ it is needed that the $\Phi$ field has charge $x_\Phi = -1$. When the new scalar field takes a VEV, the usual Yukawa couplings are recovered:
\begin{equation}\label{eq:Axiflavon_Yukawa_Couplings}
Y_{ij}^{u,d,e}=a_{ij}^{u,d,e}\epsilon^{R_j-L_i}\,,
\end{equation}
with $R_i=x_{u_i},\ x_{d_i}$ and ${L_i} = x_{Q_i}$ for quarks and ${L_i}=x_{L_i}$ and $R_i = x_{e_i}$ for leptons. The parameter $\epsilon$ that appears in Eq. \eqref{eq:Axiflavon_Yukawa_Couplings} is defined as:
\begin{equation}\label{eq:Axiflavon_Small_Param}
\epsilon \equiv \frac{v_\Phi}{\sqrt{2}\Lambda}\,.
\end{equation}
The right structure for the quark masses can be obtained when the small parameter is of the order of the Cabibbo angle, i.e., $\epsilon \sim 0.23$, and $U(1)_{FN}$ charges are assigned properly. After doing a fitting to the known Yukawas, the needed values for the charges are:
\begin{equation} \label{eq:Axiflavon_Charges}
x_{Q_i}=\left(3,\ 2,\ 0 \right)\,,\hspace{1cm} x_{u_i} = \left (4,\ 2,\ 0 \right)\,,\hspace{1cm} x_{d_i} = \left( 4,\ 3,\ 3 \right)\,,
\end{equation}
where, for every $x_i$, each entry corresponds to the first, second and third generation, respectively.

The complex scalar field $\Phi$ can be rewritten in the usual polar form as follows:
\begin{equation} \label{eq:Phi_Polar_Form_1}
\Phi = \frac{1}{\sqrt{2}}\left ( v_\Phi + \phi \right) e^{i\frac{a}{v_\Phi}}\,.
\end{equation}
Here, $\phi$ is a CP-even field that will take the role of the flavon from the FN mechanism, whereas $a$ is CP-odd and is the field that will be interpreted as the axion, more in particular, the Flaxion/Axiflavon. By performing now a series expansion in terms of the ratio $\frac{a}{v_\Phi}$ an interaction term between the Flaxion/Axiflavon and the SM fermions is obtained:
\begin{equation}\label{eq:Axiflavon_Fermion_Interaction}
\mathcal{L}_{aff} = \lambda_{ij}^f a \bar{F}_{L_i} F_{R_j} + h.c.\,,
\end{equation}
where $F=u,\ d,\ e$ are the up-type and down-type quarks and charged leptons and $f$ runs through the same indices. In Eq.~\eqref{eq:Axiflavon_Fermion_Interaction} $\lambda_{ij}^f$ stands for the coupling constant between the Flaxion/Axiflavon and the fermions $f$ of the \emph{i-th} and \emph{j-th} generations. The explicit expression of this coupling is:
\begin{equation} \label{eq:Axiflavon_Fermion_Coupling}
\lambda_{ij}^f = i \left (R^f_{j}-L^f_i \right) \frac{v}{v_\Phi}Y_{ij}^f\,,
\end{equation}
with, in general, non-diagonal couplings, meaning that the Flaxion/Axiflavon can mediate flavour-changing processes even at tree level.

Apart from coupling to fermions, the Flaxion/Axiflavon couples to the gauge fields via the chiral anomalies.
In particular, it presents a coupling to gluons that allows it to, via the shift symmetry, solve the Strong CP Problem. These couplings give rise to novel phenomenology, including rare decays and potential signatures at flavour factories like Belle II or the LHCb experiment.

This model also predicts a broader range of possible axion masses compared to the standard QCD axion, as the scale of PQ symmetry breaking is connected to the flavour physics scale. This opens new windows for axion searches in both astrophysical and laboratory-based experiments.

A variation of the Flaxion/Axiflavon model is the Minimal Flavour Violating Axion~\cite{Arias-Aragon:2017eww}, that exploits the PQ symmetry group of the Minimal Flavour Violation setup~\cite{Chivukula:1987py,DAmbrosio:2002vsn,Cirigliano:2005ck,Davidson:2006bd,Alonso:2011jd,Dinh:2017smk}. Here, a symmetry group $U(3)^5$ is identified in the limit of vanishing Yukawas, which can be split into Abelian and non-Abelian parts. The non-Abelian symmetries will take the role of explaining the mass and mixing hierarchy among the three generations, whereas one of the Abelian $U(1)$ will be responsible for the explanation of the ratio between the top quark, bottom quark and tau lepton masses. As another consequence of the symmetry identified in this framework, an axion appears coming from the same Abelian axial $U(1)_{PQ}$ symmetry mentioned before. 
Despite the use of the same tools, a key difference with the Flaxion/Axiflavon model is the absence of any FCNC effects predicted at tree-level. This has a deep impact on the associated phenomenology.

\subsubsection{Majoron}
The Majoron is the (pseudo-)Goldstone boson typically associated with Spontaneous Symmetry Breaking~(SSB) of lepton number~\cite{Chikashige:1980qk,Chikashige:1980ui,Gelmini:1980re,Schechter:1981cv}. Its simplest realisation can be obtained by extending the Type-I seesaw model~\cite{Minkowski:1977sc,Gell-Mann:1979vob,Yanagida:1979as,Mohapatra:1979ia,Schechter:1980gr} by a complex scalar singlet, $\Phi$, charged under Lepton number~(LN). The latter is then assumed to be a robust symmetry of the model, such that the NP Lagrangian reads
\begin{equation}
    \mathcal{L}_\text{NP} =-\ov{L_L}\widetilde{H}Y_DN_R -\dfrac{1}{2}\Phi \ov{N_R^c}Y_N N_R - V(\Phi)\,,
\end{equation}
where $N_R$ are right-handed neutrinos and $V(\Phi)$ is the potential which includes all $\Phi$ dependent terms and is assumed to be LN conserving. If $V(\Phi)$ is properly chosen, SSB occurs. The field can then be conveniently parameterised as
\begin{equation}
    \phi=\dfrac{f_J+\rho_J}{\sqrt{2}}e^{iJ/f_J}\,,
\end{equation}
where $J$ is the Majoron, $f_J$ its decay constant and $\rho_J$ is the radial mode. The mass of the latter is expected to be $m_{\rho_J}\sim f_J \gg v$, so that it is typically neglected in phenomenology. The Majoron is instead massless. For $\mathcal{O}(|Y_D|)\sim 1$, a massless Majoron is phenomenologically viable as long as $f_J\gtrsim3\times 10^{-7}$~GeV~\cite{Capozzi:2020cbu}.
Since global symmetries are not fundamental principles of nature and are believed to be broken by quantum gravity, the assumption of exact LN is typically relaxed, and a small breaking is allowed. An agnostic approach consists in giving the Majoron a ``small'' mass by hand such that $m_J\ll f_J$. Within this approach, no preferred range of values of $m_J$ exists. There exist in the literature attempts to point out preferred ranges. Notable examples are models featuring LN-violating quantum gravity operators~\cite{Akhmedov:1992hi,Rothstein:1992rh}, Axi-Majorons~\cite{Mohapatra:1982tc,Shin:1987xc} and models with explicit LN breaking in the Yukawa sector, which in turn induce radiative Majoron masses via the Coleman-Weinberg potential~\cite{Frigerio:2011in,deGiorgi:2023tvn,deGiorgi:2024str}.

All in all, the tree-level Majoron Lagrangian reads
\begin{equation}
\label{eq:Majoron-Lag}
    \mathcal{L}_\text{NP} \supset\frac{1}{2}\partial^\mu J \partial_\mu J -\frac{1}{2}m_J^2 J^2-\ov{L_L}\widetilde{H}Y_DN_R -\dfrac{1}{2}\ov{N_R^c}M_NN_R e^{iJ/f_J}\,,
\end{equation}
where the Majorana mass is given by $M_J\equiv f_J Y_N/\sqrt{2} $. The above Lagrangian provides a minimal example of a leptophilic ALP model, and thus, they share many (but not all) of their properties (see Sec.~\ref{sec:WISP-EFT}). The main difference arises in the couplings with photons and gluons, as will be discussed later.

The Majoron Lagrangian in Eq.~\eqref{eq:Majoron-Lag} features two important properties, namely, it couples to SM fermions via $Y_D$, and it has no couplings at tree level to charged particles nor gauge bosons. Such couplings are generated at 1- and 2-loops, respectively. 
We report below some of the most relevant couplings for phenomenology; a detailed discussion of traditional Majoron couplings at leading order in $f_J$ can be found, e.g. in Ref.~\cite{Heeck:2019guh}. Coupling of the Majoron to charged leptons in a more general class of seesaw models can be found in Ref.~\cite{Herrero-Brocal:2023czw}. Denoted by $n_i$ the neutral leptons (light and heavy) and by $\ell_i=(e,\mu,\tau)$ the charged ones, the interactions read
\begin{align}
    &\mathcal{L}_{J}\supset\frac{iJ}{2f_J}\left[\sum\limits_{i=1}^N m_{n_i} \ov{n_i}\gamma_5n_i+\frac{1}{16\pi^2}\ov{\ell}\left[M_\ell \text{Tr}(K)\gamma_5+2 M_\ell K P_L-2KM_\ell P_R\right]\ell\right]\,,
\end{align}
where $K\equiv Y_DY_D^\dagger$ and $M_\ell$ is the charged leptons' mass matrix.
Similar results hold for the coupling with quarks and nucleons~\cite{Heeck:2019guh}. All couplings to fermions are proportional to their masses. This implies ultra-weak coupling to light neutrinos, but also $\mathcal{O}(Y_N)$ couplings to the heavy ones. If $||Y_N||\sim 1$, the Majoron-HNLs couplings are the largest in the theory. In recent years, the phenomenological impact of such a coupling has received some attention in the literature in the more general context of ALP physics in beam-dump and collider experiments~\cite{Alves:2019xpc,deGiorgi:2022oks,Abdullahi:2023gdj,Marcos:2024yfm,Wang:2024mrc}, DM and Leptogenesis applications~\cite{Gola:2021abm,Liang:2024vnd,Cataldi:2024bcs,Greljo:2025suh}.

Couplings of the Majoron to gauge bosons are generated at 2-loops. The operators involving chiral gauge bosons take the form $J F\tilde{Z}$, $JZ\tilde{Z}$, and $JW^+\tilde{W}^-$. Concerning vectorial gauge bosons such as $F$ and $G$, the operators $JF\tilde{F}$ and $JG\tilde{G}$ are not generated. Instead, in the limit $m_J\to 0$, the Majoron couples as
\begin{align}
    \mathcal{L}_J\supset -\frac{g_{JVV}}{4}\left(\frac{\Box J}{v^2}\right)V^{\mu\nu}\tilde{V}_{\mu\nu} \approx \frac{g_{JVV}}{4}\left(\frac{m_J}{v}\right)^2 JV^{\mu\nu}\tilde{V}_{\mu\nu}\,, && V=F,G\,.
\end{align}
Intuitively, shift symmetry-breaking contributions to $JV\tilde{V}$ are 1-loop exact due to the topological nature of the operator; as 1-loop contributions are absent, the 2-loop contribution must be shift-symmetric preserving, leading to the $\Box J$ dependence. This constitutes the main difference to a generic ALP\footnote{A detailed discussion about the role of anomalies in ALP models can be found in Ref.~\cite{Quevillon:2019zrd}.}.
Therefore, if the Majoron is much lighter than the EW scale, $m_J\ll v$, the couplings to gauge bosons are suppressed by a two-loop factor and $(m_J/v)^2$, making the Majoron \textit{naturally astrophobic}. This opens up a huge portion of parameter space typically ruled out in generic ALP physics. 

\subsubsection{SM*A*S*H}
``SM*A*S*H'' stands for ``Standard Model*Axion*Seesaw*Higgs-Portal Inflation'' 
-- a minimal extension of the Standard Model (SM) which solves six puzzles 
of particle physics and cosmology in one smash~\cite{Ballesteros:2016euj,Ballesteros:2016xej}: 
vacuum stability, inflation, baryon asymmetry, neutrino masses, strong CP, and dark matter.\footnote{Similar models
have been considered in Refs.~\cite{Boucenna:2014uma,Ballesteros:2019tvf,Sopov:2022bog,Berbig:2022pye,Matlis:2023eli}.} The parameters of SM*A*S*H are constrained by symmetries and 
requirements to solve these puzzles.
This provides various firm predictions for observables, which can be confronted with experiments. 

\begin{table}[h]
\renewcommand{\arraystretch}{1.5}
\begin{center}
$\begin{array}{ccccccccc}
\hline
\hline
  q & u & d & L & N & E    & Q &\tilde Q & \sigma  \\
\hline
 1/2 & -1/2 & -1/2 & 1/2 & -1/2 & -1/2   & -1/2 & -1/2 &1 
\\[.5ex]
\hline
\hline
\end{array}$
\end{center}
\vskip-.3cm
\caption{\small PQ-charge assignments of the fields in SM*A*S*H. 
The remaining SM fields have no PQ charge. 
\label{tab:pq_charges}}
\end{table}

In SM*A*S*H, the field content of the SM is expanded as in the KSVZ model by  
a SM-singlet complex scalar field  $\sigma$ (the Peccei-Quinn (PQ) field) and 
a vector-like quark $Q$ (note the change in notation to follow closely the relevant references). In addition, three SM-singlet neutrinos $N_i$, with $i=1,2,3$, are added. All the new fields, as well as the quarks and leptons of the SM, are assumed to be charged under a global $U(1)_{\rm PQ}$ symmetry, cf. Table~\ref{tab:pq_charges}.
The scalar potential in SM*A*S*H has the general form  
$$V(H,\sigma )= \lambda_H \left( H^\dagger H - \frac{v^2}{2}\right)^2
+\lambda_\sigma \left( |\sigma |^2 - \frac{v_{\sigma}^2}{2}\right)^2
+
2\lambda_{H\sigma} \left( H^\dagger H - \frac{v^2}{2}\right) \left( |\sigma |^2 - \frac{v_{\sigma}^2}{2}\right),$$
where $H$ is the SM Higgs doublet. 
For $\lambda_H, \lambda_\sigma >0$, $\lambda_{H\sigma}^2 <  \lambda_H \lambda_\sigma$, $v_\sigma\gg v \simeq 246$\,GeV, both the electroweak   and the PQ symmetry are broken by the 
vacuum expectation values $\langle H^\dagger H\rangle = v^2/2$, $\langle |\sigma |^2\rangle=v_{\sigma}^2/2$. 
The $U(1)_Y$ hypercharge of $Q$ is required to be 
$-1/3$, such that the most general Yukawa interactions of $Q$ and $N_i$, allowed by SM gauge and PQ symmetries, are  
$${\cal L}\supset 
-[F_{ij}\bar{ N}_j P_L L_i\epsilon H+\frac{1}{2}Y_{ij}\sigma \bar N_i P_L  N_j 
+y\, \sigma \bar Q P_L Q+\,{y_{Q_d}}_{i}\sigma\bar{D}_iP_L Q +h.c.]\,,$$ 
where  $D_i$, $L_i$ denote the Dirac spinors associated with the down quarks and leptons of the $i$th generation.
Electroweak vacuum instability~\cite{Degrassi:2012ry} -- the instability of the Higgs potential at large field values, present for the experimentally preferred value of the top mass -- can be avoided in SM*A*S*H by the stabilising effect of the Higgs portal coupling 
$\lambda_{H\sigma}$~\cite{Lebedev:2012zw,Elias-Miro:2012eoi}. 
This requires $\lambda^2_{H\sigma}/\lambda_\sigma$ to be between $\sim 10^{-2}$ and $\sim10^{-1}$~\cite{Ballesteros:2016xej}. 
Inflation is realised in SM*A*S*H by the dynamics of the PQ and Higgs fields in the presence of their generically present non-minimal gravitational couplings 
to the Ricci scalar $R$~\cite{Spokoiny:1984bd,Futamase:1987ua,Salopek:1988qh,Fakir:1990eg,Bezrukov:2007ep},
$$S\supset - \int d^4x\sqrt{- g}\,\left[
      \xi_H\, H^\dagger H+\xi_\sigma\, \sigma^* \sigma  
  \right] R\,.$$ 
Calculable and observationally viable inflation occurs requires 
$1\gtrsim \xi_\sigma \gg \xi_H$ and $\lambda_{H\sigma}<0$, the inflaton being then a mixture of the real part of the PQ field and the modulus of the Higgs. 
The predicted fluctuations in the cosmic microwave background (CMB) temperature and polarisation agree with the observations if the non-minimal coupling $\xi_\sigma$ and 
the effective quartic coupling of the inflaton, $\tilde\lambda_\sigma=\lambda_\sigma-\lambda_{H\sigma}^2/\lambda_H$, satisfy  
$7\times 10^{-3} \lesssim \xi_\sigma \simeq 4\times 10^{4}\sqrt{\tilde\lambda_\sigma}\lesssim 1$. 
Reheating of the Universe after inflation proceeds then efficiently via the Higgs portal. 
SM*A*S*H predicts a lower bound on the ratio of the power in tensor to scalar fluctuations, $r \gtrsim 0.004$~\cite{Ballesteros:2016euj,Ballesteros:2016xej}, 
 a reheating temperature around $10^{12}$\,GeV~\cite{Ringwald:2022xif}, and a second order PQ phase transition at 
around $T_c \sim 10^8$\,GeV~\cite{Ballesteros:2016euj,Ballesteros:2016xej}.
In SM*A*S*H, as in any KSVZ axion model, the strong CP puzzle is solved  by the PQ mechanism~\cite{Peccei:1977hh}, and the axion~\cite{Weinberg:1977ma,Wilczek:1977pj} is the prime candidate for cold dark matter~\cite{Preskill:1982cy,Abbott:1982af,Dine:1982ah}. 
Since the reheating temperature in SM*A*S*H is predicted to be larger than the 
critical temperature of PQ symmetry breaking, the dark matter axions are not only produced by the misalignment mechanism but also by the decay of topological defects. 
Requiring the produced axions to constitute 100\% of dark matter, its decay constant,  
$f_a = v_\sigma$, is predicted to be in the range
$10^{10}\,{\rm GeV} \lesssim f_a <  2\times 10^{11}\,\mathrm{GeV}$, where the conservative strict upper bound~\cite{Borsanyi:2016ksw} is obtained if one completely neglects the contribution from the 
decay of topological defects, while the approximate lower bound takes into account the 
uncertainties from recent investigations of the contributions from topological defects~\cite{Gorghetto:2020qws,Buschmann:2021sdq,Saikawa:2024bta}. 
The PQ symmetry-breaking scale also gives rise to large Majorana masses for the heavy neutrinos. This can explain the smallness of the masses of the active neutrions through the seesaw mechanism~\cite{Minkowski:1977sc,Gell-Mann:1979vob,Yanagida:1979as,Mohapatra:1979ia} and results in the generation of the baryon asymmetry of the universe via thermal leptogenesis~\cite{Fukugita:1986hr}. 

The prospects and timeline to scrutinise or smash SM*A*S*H by cosmic microwave background polarisation experiments (e.g. CMB-S4~\cite{Abazajian:2019eic} and LiteBIRD~\cite{LiteBIRD:2022cnt}), 
axion haloscopes (ALPHA~\cite{Lawson:2019brd}  and MADMAX~\cite{Beurthey:2020yuq}), and future space-borne gravitational wave detectors (BBO~\cite{Crowder:2005nr,Corbin:2005ny,Harry:2006fi}, DECIGO~\cite{Seto:2001qf,Kawamura:2006up}) have been reviewed recently in Ref.~\cite{Ringwald:2023anj}.

\subsubsection{Geometric $Z^\prime$} 

A new vector boson $Y_\mu$ arising from geometric origins can be regarded as a potential WISP-scale particle. This vector boson emerges naturally from metric-affine gravity (MAG), as discussed in our recent work \cite{Demir2020} which we show that in its minimal formulation, metric-affine gravity (MAG)—where the metric tensor and affine connection are treated as independent fields—dynamically reduces to standard gravity plus a massive vector field.

In metric-affine gravity, the metric tensor and the affine connection are treated as independent fields. This geometric approach leads to a reduction of the theory to general relativity accompanied by a massive vector field $Y_\mu$, which has distinct properties due to its geometric origin. Unlike conventional vector bosons, $Y_\mu$ couples exclusively to fermions and does not interact with scalars or gauge bosons. This unique property of the geometric vector field fundamentally distinguishes it from other vector dark matter candidates found in the literature  \cite{Demir2020}.
(One may check the effects of the geometric vector field on black hole formation and black hole solutions in \cite{Demir2022, Dbh-2,Dbh-3}.) The MAG action is described by \cite{Demir2020}
\begin{eqnarray}
S\left[g, \Gamma, \Psi_m \right] = \int d^4 x\, \sqrt{-g} \left\{\frac{M_{Pl}^2}{2} g^{\mu\nu}{\mathbb{R}}_{\mu\nu}(\Gamma) -  \frac{\xi}{4}  {\overline{\mathbb{R}}}_{\mu\nu}(\Gamma) {\overline{\mathbb{R}}}^{\mu\nu}(\Gamma) + {\mathcal{L}}\left(g, \Gamma, \Psi_m \right)\right\} 
\label{action-mag}
\end{eqnarray}
in which $\mathbb{R}_{\mu\nu}(\Gamma)$ is the Ricci curvature obtained from ${\mathbb{R}}_{\mu\nu}({}^g\Gamma)  =  \partial_{\lambda}{}^g\Gamma^{\lambda}_{\mu\nu} - \partial_{\nu}{}^g\Gamma^{\lambda}_{\lambda\mu} + {}^g\Gamma^{\rho}_{\rho\lambda} {}^g\Gamma^{\lambda}_{\mu\nu}  - {}^g\Gamma^{\rho}_{\nu\lambda} {}^g\Gamma^{\lambda}_{\rho\mu}$
by replacing ${}^g\Gamma$  with $\Gamma$ (affine connection), $\mathcal{L}$ is the Lagrangian of the matter fields $\digamma$ with $\Gamma$ kinetics, and ${{\overline{\mathbb{R}}}}_{\mu\nu}(\Gamma)$ is the second Ricci curvature 
\begin{eqnarray}
\label{ricci2}
{{\overline{\mathbb{R}}}}_{\mu\nu}(\Gamma) =  \partial_{\mu}\Gamma^{\lambda}_{\lambda\nu} - \partial_{\nu}\Gamma^{\lambda}_{\lambda\mu}\, .
\end{eqnarray} 
The Palatini formalism implies that MAG can always be analyzed via the decomposition $\Gamma^{\lambda}_{\mu\nu} =  {}^g\Gamma^{\lambda}_{\mu\nu} + \Delta^{\lambda}_{\mu\nu}$
The MAG action \eqref{action-mag} can be rewritten in a more familiar form by expressing $\Delta^\lambda_{\mu\nu}$ in terms of the non-metricity tensor as follows \cite{Buchdahl1979, Tucker1996,Obukhov1997,Vitagliano2010}, assuming zero torsion throughout $\Delta^{\lambda}_{\mu\nu} = \frac{1}{2} g^{\lambda \rho} ( Q_{\mu \nu \rho } + Q_{\nu \mu \rho } - Q_{\rho \mu \nu} )$
where $\Delta^{\lambda}_{\mu\nu}=\Delta^{\lambda}_{\nu\mu}$ is a symmetric tensor field which is decomposed in terms of the non-metricity tensor $Q_{\lambda \mu \nu} = - {}^\Gamma \nabla_{\lambda} g_{\mu \nu}$.
Using the decomposition of the affine connection in terms of the nonmetricity tensor, one gets general relativity plus a consistent vector field theory
\begin{equation}
S\left[g, Y, \Psi_m \right]\! =\int \!\!d^4 x \sqrt{-g}\! \left\{\! \frac{M_{Pl}^2}{2} R(g) - \frac{1}{4} Y_{\mu\nu} Y^{\mu\nu} - \frac{1}{2} M_Y^2 Y_{\mu} Y^{\mu} + g_Y \overline{f} \gamma^{\mu}\! f Y_{\mu}  + {\overline{\mathcal{L}}}\!\left(g, {}^g\Gamma, \Psi_m\right)\!\right\} 
\label{action-mag-decomp-zprime}
\end{equation}
where the mass of $Y_{\mu}$ is $M_Y^2 = \frac{3 M_{Pl}^2}{2 \xi}$ and its coupling to fermions is $g_Y = \frac{1}{4\sqrt{\xi}}$
and $R(g)\equiv g^{\mu\nu}{\mathbb{R}}_{\mu\nu}({}^g\Gamma)$ is the metrical curvature scalar, $Y_\mu \equiv 2 \sqrt{\xi} Q_{\mu}$ is a vector field generated by the affine connection, and ${\overline{\mathcal{L}}}\left(g, {}^g\Gamma, \digamma\right)$ is part of the matter Lagrangian that does not involve $Y_{\mu}$. 

\begin{figure}[t]
\centering
\includegraphics[width=.8\linewidth]{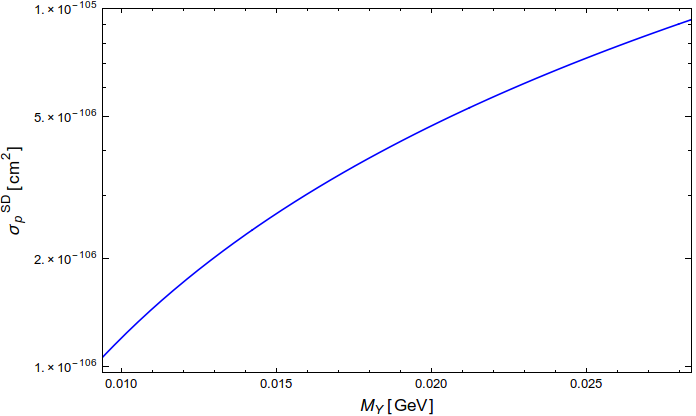} 
\caption{The spin-dependent $Y_{\mu}$--proton cross section as a function of its mass $M_Y$. Figure adapted from Ref.~\cite{Demir:2020brg}.}
\label{fig-xsection}
\end{figure}

The affine connection $\Gamma^\lambda_{\mu\nu}$ or the geometrical tensor $\Delta^\lambda_{\mu\nu}$ contributes to the matter action $\mathcal{L} (g, {}^g \Gamma, \Delta, \Psi)$ through the kinetic terms of fermions. This arises from the spin connection in the covariant derivative of spinor fields in curved spacetime \cite{Hurley94, Fatibene96,Adak2002} ${}^\Gamma \nabla_\mu \psi = (\partial_\mu + \frac{1}{4} \omega_\mu^{a b} \gamma_a \gamma_b) \psi$
in which the flat spacetime Clifford algebra gives $\gamma_a \gamma_b = \eta_{ab} + 2 \sigma_{ab}$ with the Lorentz generator $\sigma_{ab} = \frac{1}{4} [\gamma_a, \gamma_b]$.
$Y_{\mu}$ decays into a fermion $f$ and its anti-particle $\overline{f} $ with a rate
\begin{eqnarray}
\!\!\!\!\!\!\Gamma\left(Y\rightarrow f \overline{f} \right) = \frac{N_c^f}{8\pi}\!\left(\frac{3}{2\xi}\right)^{\frac{3}{2}} 
\!\left(1 +   \frac{4\xi m_f^2}{3 M_{Pl}^2}\right) 
\!\left(1 -  \frac{8\xi m_f^2}{3 M_{Pl}^2} \right)^{\frac{1}{2}}\!  M_{Pl} \, .
\end{eqnarray} 
Here, $m_f$ represents the mass of the fermion, and $N_c^f$ indicates the number of its colours. By summing over the SM quarks (up and down) and leptons (electrons and neutrinos), we obtain its lifetime as $\tau_{\scriptscriptstyle{Y}} = \frac{1}{\Gamma_{tot}}  = \frac{4 \pi}{5} \left( \frac{2}{3} \right)^{3/2} \frac{\xi^{3/2}}{M_{Pl}}$
which is larger than the age of the Universe $t_{U} = 13.8 \times 10^9$ years \cite{Planck:2015fie} if $\xi > 1.1 \times 10^{40}$. On the other side, the decay rate is prevented to go imaginary if $\xi < 1 \times 10^{41}$. These two bounds lead to the allowed mass range 
\begin{align}
9.4\ {\rm MeV} < M_Y < 28.4\ {\rm MeV} \, .
\label{m-range}
\end{align}
Because $Y_\mu$ is a purely geometric vector (not a gauge boson), its mass arises directly from curvature terms rather than symmetry breaking. The large dimensionless parameter $\xi$ simply encodes the mass $M_Y \sim \text{MeV}$ and the coupling $g_Y \ll 1$. The vector field, $Y_\mu$, which interacts only with quarks, leptons, and gravity, remains both neutral and long-lived—surviving well beyond the age of the Universe—when its mass falls within the range given in Eq.~\eqref{m-range}. Furthermore, the vector's scattering cross-section with nucleons is 60 orders of magnitude below current detection thresholds, making the direct detection of this form of dark matter unlikely with today's experiments. This underscores the difficulties encountered by experimental searches, as direct detection efforts like  COUPP \cite{COUPP2012}, SIMPLE \cite{SIMPLE2014}, XENON100 \cite{XENON100-2016}, PICO-2L \cite{PICO2L-2016}, PICO-60 \cite{PICO60-2017}, PandaX-II \cite{PandaX2017}, PICASSO \cite{PICASSO2017} and LUX \cite{LUX2017} have imposed strict upper limits on the spin-dependent cross section for dark matter scattering off the standard model particles. These experiments primarily rule out the existence of WIMPs. The most stringent limit is around $\sigma_p^{SD} \sim \mathcal{O}(10^{-41})\ {\rm cm^2}$. 
This highlights the inherent challenges in current dark matter detection efforts. Additionally, we show that, because of its geometric origin, this vector field interacts exclusively with fermions and does not couple to scalars or gauge bosons. This distinctive characteristic sets it apart from other vector dark matter candidates in the literature. The proposed geometric dark matter model is both minimal and self-consistent from theoretical and astrophysical perspectives, with its weak interactions being essential for its stability and longevity. 
It is evident that the $Y_\mu$--proton spin-dependent cross section given in Fig.~\ref{fig-xsection} is at most ${\mathcal{O}}\left(10^{-106}\right)\ {\rm cm^2}$, which is far too small to be detectable by any of the current experiments. Fig.~\ref{fig-xsection} can be seen as an explanation for why dark matter has not yet been detected in direct searches. While it is possible that future experiments might measure it, it is difficult to envision what advancements in technology could allow access to such an extremely small cross-section.
These experimental constraints, combined with theoretical considerations, provide a compelling explanation for why dark matter has remained undetected in direct searches. This positions geometric dark matter as a viable candidate that is consistent with current astrophysical observations. This $Y_\mu$ can itself be a feebly interacting WISP-scale light particle. 

Moreover, probing the effects of $Y_\mu$ on astrophysical and cosmological phenomena, such as black hole solutions \cite{Demir2022, Dbh-2, Dbh-3}, can provide further insights into its viability as a dark matter candidate. As WISP-scale particles remain a compelling possibility for dark matter, advances in experimental techniques, such as precision measurements in astrophysical observations, indirect detection efforts, or novel collider experiments, may eventually provide indirect evidence for particles like $Y_\mu$.

On the other hand, an extension beyond the Palatini formalism with the geometric Proca field $Q_\mu$ \cite{Demir2020, Vitagliano2010} involves incorporating the metric curvature $R_{\mu\nu}({}^g\Gamma)$ along with the existing affine curvature ${\mathbb{R}}_{\mu\nu}(\Gamma)$. The gravity theory we explore in \cite{Demir2022} is the metric-Palatini gravity extended with the ${\mathbb{R}}_{[\mu\nu]}(\Gamma) {\mathbb{R}}^{[\mu\nu]}(\Gamma)$ invariant, which we call it as "extended metric-Palatini gravity" (EMPG). This framework provides a broader range for the mass of the $Y_\mu$ as well as its coupling to matter so that the $Y_\mu$ can be a weakly interacting WISP-scale light particle. 

The geometric vector field $Y_\mu$ offers a unique dark matter candidate that fits within the broader WISP framework. Its weak interactions and stability make it a theoretically consistent and astrophysically viable solution to the dark matter problem. While direct detection of such a feebly interacting particle remains highly unlikely with current technology, future advancements might enable indirect observations that could confirm the existence of such geometric dark matter particles.

%% file: WG1/content/screening.tex
`Screening' is the name given to a collection of phenomena where the properties of a BSM scalar field (for example, its mass or coupling `constants') can change with the environment. This is often invoked as a way of explaining how a scalar field could have avoided detection to date, particularly in searches for fifth forces,  without fine-tuning the couplings between the scalar and matter to be unnaturally small. 

The phenomenology of screening has mainly been considered in the context of models of dark energy and scalar modifications of gravity \cite{Joyce:2014kja,Koyama:2015vza,Ishak:2018his,Brax:2021wcv}, but also occurs for quadratically coupled ultra-light scalar dark matter models \cite{Hees:2018fpg,Banerjee:2022sqg}. Environmentally dependent phenomenology is possible in any scalar (or pseudoscalar) field theory where the scalar field is sourced by the energy density of its environment, so long as either the scalar potential contains terms beyond $V(\phi)= (1/2) m^2 \phi^2$, or the scalar has quadratic (or higher-order) couplings to matter fields.\footnote{The connection between Higgs-portal models and the conformally coupled scalars often considered as theories of modified gravity is explored in Ref.~\cite{Burrage:2018dvt}.} 

Screening provides both challenges and opportunities for WISP model building and phenomenology. It can suppress the effects of the scalar particle in an experiment, but it is also possible to enhance the probability of detection in a suitably chosen experimental setup. The environmental dependence and the common presence of non-linearities in the scalar field equations can often make calculating the behaviour of the scalar analytically challenging and computationally expensive, but if such calculations can be done, new observable signatures are identified. 

Here, we discuss three commonly studied examples of screening: quadratically coupled scalar dark matter models and the chameleon and symmetron models originally proposed as explanations for dark energy.  

\subsubsection{Quadratically coupled ultra-light scalar dark matter}
Here, we consider a scalar theory of the form
\begin{equation}
    \mathcal{L}\supset -\frac{1}{2}(\partial \phi)^2 -V(\phi)+ \frac{\phi^2}{2\tilde{M}^2}\left( d_e F^2 - d_g(F^A)^2 -\sum d_{m_i}m_i\bar{\psi}_i\psi_i\right)\,,
    \label{eq:quadcouplag}
\end{equation}
where the scalar potential is  $V(\phi)= m^2\phi^2/2$ and the energy scale $\tilde{M}$ controls the coupling of the scalar field to matter, where the dimensionless constants $d_e$, $d_g$ and $d_{m_i}$ allow for different couplings to photons, gluons and fermions respectively (further constants may be introduced when considering the running of these couplings) \cite{Hees:2018fpg}. Quadratic couplings may be the leading interaction between the scalar field and matter if, for example, the theory respects a $\phi \rightarrow -\phi$ symmetry or if the interactions with matter arise from a symmetric Higgs-portal \cite{Binoth:1996au,Patt:2006fw,Ahlers:2008qc,Englert:2020gcp}. Quadratic couplings also arise for axions coupled to gluons \cite{Hook:2017psm,Kim:2023pvt,Bauer:2023czj,Bauer:2024yow}.  Although not the leading order coupling to fundamental particles, for such axions, this may be the leading order coupling to macroscopic objects with zero spin. 

The Lagrangian in equation~\eqref{eq:quadcouplag} leads to an equation of motion for the scalar of the form
\begin{equation}
    \Box \phi = m^2\phi \pm \frac{ \phi}{M^2}\rho\,,
    \label{eq:quad coupeom}
\end{equation}
when matter can be represented as a non-relativistic perfect fluid of density $\rho$ \cite{Burrage:2018dvt,Brax:2023udt,Brax:2021rwk}. Here, for simplicity, we have absorbed all of the different coupling parameters into the energy scale $M$ but we note that the different couplings to different fundamental particles in equation~\eqref{eq:quadcouplag} lead to composition-dependent couplings and violations of the equivalence principle that provide stringent constraints on the theory \cite{Damour:2010rp,Damour:2010rm}. Test particles of unit mass will  experience an acceleration due to the scalar fifth force of the form
\begin{equation}
    \vec{a} = \frac{\vec{\nabla} \phi^2}{M^2}\,.
\end{equation}

To behave as dark matter, the cosmological solution for the field in the late universe oscillates with frequency $m$ (and phase $\delta$).   In the presence of a spherical object of ordinary matter with density profile $\rho(\vec{x}) = \rho \Theta(R-|\vec{x}|)$, with $\rho> M^2 m^2$, the field has the form
\begin{equation}
    \phi(t,\vec{x})= \phi_{\infty}\cos (mt +\delta) \varphi(\vec{x})\,,
\end{equation}
with 
\begin{equation}
    \varphi_-(r) =\left\{\begin{array}{lr}
       \frac{1}{m_- r}\frac{\sin m_- r}{\cos m_- R}  & 0 \leq r \leq R\,,\\
    1 +\frac{1}{m_- r}(\tan m_- R -m_- R) & R\leq r\,,
    \end{array}\right.
    \label{eq:tachsol}
\end{equation}
or 
\begin{equation}
    \varphi_+(r) =\left\{\begin{array}{lr}
       \frac{1}{m_+ r}\frac{\sinh m_+ r}{\cosh m_+ R}  & 0 \leq r \leq R\\
    1 +\frac{1}{m_+ r}(\tanh m_+ R -m_+ R) & R\leq r
    \end{array}\right.
    \label{eq:possol}
\end{equation}
depending on the choice of sign in equation~\eqref{eq:quad coupeom}, with 
\begin{equation}
m_\pm^2=\left|m^2\pm  \frac{\rho}{M^2}\right|\,.
\end{equation}

When $m_{\pm}R\ll1$ outside the source the scalar field grows as $\varphi_{\pm}(r) \approx 1 \mp M_* /4 \pi M^2 r$, where $M_*= 4 \pi \rho R^3/3$  is the total mass of the source.  When $m_{\pm}R\gtrsim 1$, the behaviour of the field is significantly different.  In the case of $\varphi_-$, the factor $\tan m_- R$ may diverge for certain source mass radii, resulting in infinite values for $\varphi_-$ at all distances $r$.  This is a consequence of the field having a tachyonic mass inside the source, and any solution found in this regime is not physically relevant.  When  $m_+ R\gg 1$, we find that outside the source $\varphi \sim 1-R/r$, and the scalar field becomes independent of the density of the source.  The fact that the total mass of the source could be increased without resulting in a greater perturbation of the scalar field is how screening manifests in this theory and is a consequence of the environmental dependence of the scalar's effective mass. 

Experimental constraints on this model come from torsion balance and equivalence principle violation experiments, as well as looking for the effects of the oscillation of the dark matter scalar on the variation of fundamental constants \cite{Hees:2018fpg,Banerjee:2022sqg}.  For experiments considered so far the static solutions described above are sufficient, as the field can adapt to a moving source faster than the timescale of the experiment \cite{Burrage:2024mxn}. Time dependence of the scalar field, $\phi = \phi(t)$, and its experimental consequences will be considered in the next section.

\subsubsection{Chameleon model}
Here, we consider a scalar theory of the form \cite{Khoury:2003rn,Khoury:2003aq,Brax:2004qh}
\begin{equation}
    \mathcal{L}\supset -\frac{1}{2}(\partial \phi)^2 -V(\phi)+ \frac{\phi}{M}\rho\,,
    \label{eq:chamlag}
\end{equation}
where the scalar potential is  $V(\phi)= \Lambda^{4+n}/\phi^n$, where $\Lambda$ is a constant energy scalar and $n$ an integer. The energy scale $M$ controls the coupling of the scalar field to matter and we represent non-relativistic matter by its energy density $\rho$. This coupling to energy density could arise from direct couplings to the fundamental particles, as seen for the quadratically coupled dark matter model in equation~\eqref{eq:quadcouplag}, from a Higgs portal coupling or from a conformal rescaling of the metric that matter particles move on \cite{Burrage:2018dvt}. 

The Lagrangian in equation~\eqref{eq:chamlag} leads to an equation of motion for the scalar of the form
\begin{equation}
    \Box \phi = \frac{\Lambda^{4+n}}{\phi^{n+1}} + \frac{ \rho}{M}\,.
    \label{eq:chameom}
\end{equation}
The field $\phi$ evolves in an effective potential given by
\begin{equation}
    V_{\rm eff}(\phi)= \frac{\Lambda^{4+n}}{\phi^n}+\frac{\phi\rho}{M}\,.
\end{equation}
The minimum of this potential lies at $\phi_{\rm min}(\rho)= (nM\Lambda^{4+n}/\rho)^{1/(n+1)}$
and fluctuations around the minimum 
have a density-dependent mass 
\begin{equation}
    m_{\rm eff}^2 =n(n+1) \Lambda^{-(4+n)/(n+1)} \left(-\frac{\rho}{nM}\right)^{\frac{n+2}{n+1}}\,.
\end{equation}
Test particles of unit mass will  experience an acceleration due to the scalar fifth force of the form
\begin{equation}
    \vec{a} = \frac{\vec{\nabla} \phi}{M}\,.
\end{equation}

The scalar mediated force between two spheres of compact densities $\rho_A$ and $\rho_B$, radii $R_A$ and $R_B$ and total masses $M_A$ and $M_B$, separated by a distance $r$ is \cite{Burrage:2014oza}
\begin{equation}
    F_{\rm chameleon} = \frac{ \lambda_A \lambda_B}{4 \pi M^2 }\frac{M_A M_B}{r^2}\,,
    \label{eq:chamf}
\end{equation}
where
\begin{equation}
    \lambda_i = \left\{ \begin{array}{lc}
      1 \;,  & \rho_iR^2_i < 3M\phi_{\rm bg}\;, \\
      \frac{3M\phi_{\rm bg}}{\rho_iR_i^2}  \;, & \rho_iR^2_i > 3M\phi_{\rm bg}\;,
    \end{array}\right. 
\end{equation}
where $\phi_{\rm bg}=\phi_{\rm min}(\rho_{\rm bg})$ is the value of the field in the background of the experiment, and we have assumed that in this background the Compton wavelength of the scalar $\sim 1/m_{\rm eff}(\rho_{\rm bg})$ is large compared to the scale of the experiment.

In the case of the chameleon we see that the fifth force is suppressed not just by the properties of the object (the same combination of parameters $\rho_i R_i^2$, related to the compactness of the object, that we saw determining the behaviour of the quadratically coupled symmetron) but also on the environment in which an experiment is conducted, the chameleon mediated force is suppressed, $\lambda_i <1$,  around more compact objects (larger $\rho_i R_i^2$), but also more likely to be suppressed when $\phi_{\rm bg}$ is small, corresponding to larger background densities. 

Constraints on the chameleon model come from a variety of laboratory experiments and astrophysical measurements; for a review, see Refs.~\cite{Burrage:2017qrf,Brax:2021wcv,Fischer:2024eic}. For the chameleon model described here, arguably the most well-motivated parameter space, with $\Lambda \sim 10^{-3}\mbox{ eV}$ the dark energy scale, $M\sim M_P$ and $n\sim \mathcal{O}(1)$ is now almost excluded \cite{Yin:2022geb}. 

There has been considerable interest in studying the coupling of the chameleon field to photons, where it behaves as a scalar axion-like particle with a density-dependent mass \cite{Brax:2007ak, Brax:2007hi,Gies:2007su,Ahlers:2007st,Burrage:2008ii,Burrage:2009mj,Pettinari:2010ay,Brax:2010xq,Brax:2010jk,Burrage:2017qrf,CAST:2018bce,Elder:2023oar}.  This relaxes some constraints on axion-like particles, for example the production in high-density regions will be suppressed due to the increased mass of the field. Careful modelling of chameleon production in the sun and the potential for new constraints and the possibility of detection of solar chameleons can be found in Refs.~\cite{Brax:2010xq,CAST:2015npk,ArguedasCuendis:2019fxj,OShea:2024jjw}.

\subsubsection{Symmetron model} 
Here we consider a scalar theory of the form \cite{Hinterbichler:2010es,Hinterbichler:2011ca}
\begin{equation}
    \mathcal{L}\supset -\frac{1}{2}(\partial \phi)^2 -V(\phi)+ \frac{\phi^2}{2M^2}\rho
    \label{eq:symlag}
\end{equation}
where the scalar potential is  $V(\phi)= -\mu^2 \phi^2/2 +\lambda \phi^4 /4$, where $\mu$ is a constant mass scale and $\lambda$ a dimensionless constant. The energy scale $M$ controls the coupling of the scalar field to matter.  As for the chameleon,  we represent non-relativistic matter by its energy density $\rho$. The behavior of the symmetron field is governed by the effective potential:
\begin{equation}
V_{\rm eff}=\left(\frac{\rho}{M^2}-\mu^2 \right)\frac{\phi^2}{2} +\frac{\lambda \phi^4}{4}\,.
\end{equation}
The symmetron model has a quadratic coupling to matter, which means the effective mass of the scalar becomes density-dependent in a similar manner to the quadratically coupled dark matter model described above.  However, in this case, the bare mass of the field is tachyonic, meaning that in regions of low density, $\rho < \mu^2 M^2$,  the field has a non-zero expectation value, but in regions of higher density, the field has zero expectation value. This density-dependent symmetry breaking leads to the formation of domain walls, which lead to distinctive phenomenology for the theory, which could be detected in the laboratory \cite{Llinares:2018mzl,Llinares:2014zxa,Clements:2023bva,Stadnik:2020bfk}, astrophysically \cite{Dai:2021boq,Naik:2022lcn}, or cosmologically \cite{Christiansen:2024uyr,Christiansen:2024vqv}.

The symmetron model is also similar to the chameleon model in that there is an environmental dependence to the suppression of the symmetron-mediated fifth force around compact objects. 
The scalar mediated force between two spheres of compact densities $\rho_A$ and $\rho_B$ and radii $R_A$ and $R_B$, separated by a distance $r$ is \cite{Burrage:2016rkv}
\begin{equation}
    F_{\rm symmetron} = \frac{4 \pi \lambda_A \lambda_B}{r^2 }
    \label{eq:symmf}
\end{equation}
where
\begin{equation}
    \lambda_i = (\phi_{\rm out}-\phi_{\rm in}) R_i \left(\frac{m_{\rm in}R_i-\tanh m_{\rm in}R_i}{m_{\rm in} R_i}\right)
\end{equation}
where $\phi_{\rm out}$ and $\phi_{\rm in}$ are the values of the field that minimize the effective potential respectively outside and inside each sphere, and $m_{\rm in}^2= d^2 V_{\rm eff}/dr^2|_{\phi_{\rm in}}$ is the mass of small fluctuations around $\phi_{\rm in}$. As in the discussion of the chameleon model, we have assumed that, in this background, the Compton wavelength of the scalar $\sim 1/m_{\rm out}$ is large compared to the scale of the experiment. The symmetron-mediated fifth force is suppressed if $m_{\rm in}R \gg 1$. When the density of the sphere is high enough to restore the symmetry $\rho_i> m^2M^2$, the symmetron charge $\lambda_i$ is small when $\rho_i R^2_i \ll M^2$.

In equation~\eqref{eq:symmf}, we see the connection to the quadratically coupled dark matter model, equation~\eqref{eq:possol} in the functional dependence on the effective mass of the field, and the radius of the sphere $R$. We also see a dependence on the environment arising, as in the chameleon model in equation~\eqref{eq:chamf}, through the dependence of how the value of the scalar field changes between the two environments, through the factor $(\phi_{\rm out}-\phi_{\rm in})$.  In both of the previously discussed cases, we see that the fifth force between objects is more likely to be suppressed for objects with larger $\rho_i R_i^2$. 

Constraints on the symmetron model come from a range of laboratory and cosmological measurements.  In contrast to the chameleon model, for the symmetron, there is still much unconstrained parameter space \cite{Brax:2021rwk,Li:2024ynr,Fischer:2024eic,deGiorgi:2025kyp}.

%% file: WG1/content/time-dependence.tex
In the previous section, we reviewed scenarios where the properties of a new, light scalar field change with the environment. We focused on time-independent solutions where, for example, the mass of the scalar -- and therefore the range of the fifth force it mediates -- depends on the ambient energy density.

In this section, we will review the effects of time-dependent background scalar fields. In some scenarios, in particular those coming from string theory, the value of fundamental constants is set by scalar fields (the so-called moduli fields). The value of these fields could change with time, inducing a time-dependent value of some constants, $\alpha = \alpha(\phi (t))$. In the case $\phi$ is the cosmological DM, regardless of its CP properties, the value of this field oscillates as:
\begin{equation}
    \phi(x,t)=\frac{\sqrt{2\rho_{DM}}}{m_\phi}\cos(m_\phi t - k\cdot x)\,.
\end{equation}
If the field $\phi$ couples to SM particles, its VEV sets the value of some parameter, and its oscillations result in a time dependence of such parameter, an effect that can be searched for in experiments.
The phenomenology of these oscillations and the different experiments that can be used to find them depend on the CP properties of the scalar field.

Let us consider first a CP even scalar field coupled to SM fermions and gauge bosons as:
\begin{equation}
    \mathcal{L}_{eff} = \kappa \phi\left (  \sum_f \sum d_{m_i}m_i\bar{\psi}_i\psi_i + \frac{d_e}{4e^2}FF + ... \right )\,,
\end{equation}
where we adopted the conventions of \cite{Damour:2010rm}, and $\kappa=\frac{\sqrt{4\pi}}{M_p}$.
Through these couplings, the relic abundance of $\phi$ induces a temporal dependence of different quantities, including masses of fermions and the value of gauge couplings:
\begin{align}
m_i(t)=m_i\left ( 1+d_{m_i}\kappa\phi(t)  \right )\,,\\
\alpha(t)=\alpha \left ( 1+d_{e}\kappa\phi(t)  \right )\,.
\end{align}
Such oscillations lead to diverse, interesting phenomena that can be searched at different experiments.
For example, one can search for variations of the ratio of different atomic transition frequencies. These frequencies depend on the value of the electron and proton masses as well as the fine structure constant  \cite{Arvanitaki:2014faa,Stadnik:2014tta}:
\begin{equation}
    f_A\propto \left ( \frac{\mu_A}{\mu_b} \right )^{\varsigma_A}\alpha_{EM}^{\xi_A+2}\,,
\end{equation}
with $\mu_A$ the nuclear magnetic moment of A, $\mu_b$ the Bohr magneton, $\varsigma_A=0(1)$ for optical (hyperfine), and $\xi_A$ a calculable parameter. From this equation, one can see that oscillations of $m_e$, $m_p$, and $\alpha_{EM}$ will typically induce oscillatory behaviour in the atomic transition frequencies.
Other experimental strategies to search for the effects of oscillations include accelerometers \cite{Graham:2015ifn}, with atom gradiometers being a promising example  \cite{Badurina:2021lwr}.\\\\
The oscillation of a relic abundance of axions also leads to interesting phenomena. Let us consider an axion field, $a=f_a\theta$, with the following interactions:
\begin{equation}\label{eq:eff_lag_time_dep_axion}
    \mathcal{L}_{eff}= \frac{a}{32\pi^2 f_a}G\tilde{G}+\frac{\partial_\mu a}{f_a} \bar{\psi}\gamma^\mu\gamma_5\psi\,.
\end{equation}
Through the first term, we see that the axion induces an effective $\theta$ angle. The time dependence of the axion will therefore induce a time-dependent EDM for nucleons. For the QCD axion, one gets  \cite{Graham:2013gfa,Budker:2013hfa}:
\begin{equation}
    d_n^{\rm QCD}= 2.4\times 10^{-16}\frac{a}{f_a}\text{ e.cm}\,.
\end{equation}
Interestingly, due to the relation between the mass and the decay constant, if the QCD axion constitutes the dominant part of the DM, the induced EDM is independent of $f_a$, and is given by:
\begin{equation}
    d_n^{\rm QCD}\approx 9\times 10^{-35}  \cos (m_at)\,\text{ e.cm}\,.
\end{equation}
Interestingly, the effect does not decouple as $f_a$ increases, offering an opportunity to test axions with large decay constants. This is the basis, for example, for the CASPER experiment, reviewed in Part IV.

The low-energy limit of the derivative interaction in Eq.~\eqref{eq:eff_lag_time_dep_axion}leads to the following non-relativistic Hamiltonian:
\begin{equation}
    H_a=-g_{aN}\nabla a \cdot \sigma_N \,,
\end{equation}
where $\sigma_N$ is the nuclear spin operator and the nucleon coupling goes roughly as $g_{aN}\sim 1/f_a$. Due to the motion of the Earth through the galaxy, this operator generates an "axion wind" effect \cite{Graham:2013gfa,Budker:2013hfa}, resulting in the time-dependent Hamiltonian,
\begin{equation}
    H_a = g_{aN}a_0 m_a \cos (m_a t)v\cdot \sigma_N\,.
\end{equation}
Such an oscillating signal can also be searched at NMR-type experiments.

Finally, the axion-gluon coupling also leads to a pion mass which depends on the axion field value~\cite{Kim:2022ype},
\begin{equation}
    m_\pi^2(\theta)=B\sqrt{m_u^2+m_d^2+2m_u m_d\cos\theta}\,,
\end{equation}
where $B=-\frac{\langle \bar{q}q\rangle}{f_\pi^2}$ and $m_{u,d}$ are quark masses. This implies that the pion mass  oscillates with time as:
\begin{equation}
    \frac{\delta m_\pi^2}{m_\pi^2} = -\frac{m_u m_d}{2(m_u+m_d)^2}\theta(t)^2 \,.
\end{equation}
Other hadronic quantities have similar behaviour and result into a time dependent atomic energy levels which can be searched with different kinds of ultra-precise clocks (including atomic, molecular, and nuclear clocks)~\cite{Safronova:2017xyt,Peik:2020cwm,Chou:2023hcc}. In the case of the QCD axion, the effect is too small to be searched at existing or near-future experiments, but it can be more sizeable in less minimal models with light QCD axions~\cite{Hook:2018jle,DiLuzio:2021pxd}. 
All in all, the time-dependence of fundamental constants offers exciting opportunities to search for light scalar DM at precision experiments. See part IV for related direct searches of similar phenomena.

%% file: WG1/content/1-3-intro.tex
In the previous sections, we examined several concrete and well-motivated models of new physics. While such models offer detailed and predictive frameworks, the vast landscape of unresolved puzzles in the Standard Model — ranging from the origin of mass hierarchies to the nature of dark matter — suggests caution in committing to any specific UV completion. The risk lies in the potential mismatch between the significant analytical effort required to study a given model and the ultimately low probability that it reflects the true underlying physics.

A more agnostic strategy is to employ Effective Field Theories (EFTs). These frameworks, though less predictive due to their larger parameter spaces, offer greater generality. They allow for systematic capture of the low-energy consequences of broad classes of UV models without assuming a specific microscopic structure. In this sense, EFTs provide a unifying language that can interpolate among multiple theories and encapsulate model-independent constraints.

In this section, we introduce several EFTs relevant to extensions of the Standard Model, organising them by the spin of the light degrees of freedom they describe.

We begin with spin-0 theories in Subsection~\ref{eq:ALP-lagrangian}, focusing first on Axion-Like Particles (ALPs). Following the seminal paper in Ref.~\cite{Georgi:1986df}, the effective description of ALPs~\cite{Choi:1986zw,Salvio:2013iaa,
Brivio:2017ije,Alonso-Alvarez:2018irt,Gavela:2019wzg,Chala:2020wvs,Bauer:2020jbp,DiLuzio:2020oah,Bonilla:2021ufe,Arias-Aragon:2022byr,
Arias-Aragon:2022iwl,Song:2023lxf,DiLuzio:2023cuk} is structurally equivalent to the chiral Lagrangian of a spontaneously broken U(1) symmetry — that is, the low-energy theory of a Goldstone boson. This framework is especially appealing, as it successfully captures the phenomenology of a wide range of models, including those of QCD axions and other pseudo-Nambu–Goldstone bosons arising from global symmetry breaking.
In Sec.~\ref{eq:spin0-EFT}, we generalise the scalar EFT beyond the Goldstone limit by including explicit symmetry-breaking effects and higher-dimensional operators. This allows for the exploration of a broader class of spin-0 particles, including scalar mediators and Higgs-portal-like interactions.

Moving up in spin, Sec.~\ref{eq:spin1-EFT} is dedicated to spin-1 EFTs, where a new vector particle, typically dubbed ``dark'' or ``hidden photon'', appears as a gauge boson of hidden sectors. The dark photon can couple directly to the SM or inherit interactions via kinetic mixing with the SM gauge bosons. It can be a compelling dark matter candidate or mediate interactions with a dark sector. The section reviews some models
building aspects for the Abelian case.

Finally, in Sec.~\ref{eq:spin2-EFT}, we consider the case of a massive spin-2 field — the so-called ``dark graviton''. This can arise in scenarios involving extra dimensions or more general massive gravity theories. The consistent low-energy description of such a particle is subtle due to the well-known difficulties in constructing ghost-free theories with a large cut-off. We comment on the current status and phenomenological applications.

%% file: WG1/content/ALP-Lagrangian.tex
An axion-like particle~(ALP) $a$ can be defined as a pseudo-Nambu-Goldstone boson, with an almost exact shift symmetry invariance. In some cases, they may be considered as an effective description of untraditional Axions. Thus, ALPs may or may not correspond to solutions to the Strong CP problem. The most general ALP interactions with the SM fermions $\psi_F$ and the SM 
$SU(3)_c$, $SU(2)_L$ and $U(1)_Y$ gauge field strengths $G_{\mu\nu}^a$, $W_{\mu\nu}^A$ and $B_{\mu\nu}$, respectively, and their Hodge 
duals (denoted by a tilde), are summarised by the following low-energy effective Lagrangian 
\cite{Georgi:1986df}:
\begin{equation}
\begin{aligned}
   {\cal L}_{\rm eff}
   &= \frac12 \left( \partial_\mu a\right)\!\left( \partial^\mu a\right)  - \frac{m_{a,0}^2}{2}\,a^2
    +  \frac{\partial^\mu a}{f_a}\,
\sum_F \,\bar\psi_F \gamma_\mu C_F \psi_F
\\
   &\quad\mbox{}  -C_{aGG}\,\frac{\alpha_s}{8\pi}\,\frac{a}{f_a}\,G_{\mu\nu}^a\,\tilde G^{\mu\nu,a} 
    -C_{aWW}\,\frac{\alpha_2}{8\pi}\,\frac{a}{f_a}\,W_{\mu\nu}^A\,\tilde W^{\mu\nu,A} \\
    &\quad\mbox{} -C_{aBB}\,\frac{\alpha_1}{8\pi}\,\frac{a}{f_a}\,B_{\mu\nu}\,\tilde B^{\mu\nu}\,,
\end{aligned}
\label{Leff_a}
\end{equation}
where $f_a$ represents the scale of validity of the effective description. Here, $\alpha_s=g_s^2/(4\pi)$, $\alpha_2=g^2/(4\pi)$ and $\alpha_1=g^{\prime\,2}/(4\pi)$ are the respective SM gauge coupling parameters, and 
$F$ denotes the left-handed fermion multiplets in the SM. $C_F$ is a  
Hermitian matrix in generation space with dimensionless entries which, together with the dimensionless coefficients $C_{aGG}$, $C_{aWW}$ and $C_{aBB}$, depend on the specific UV completion featuring the Abelian global symmetry.

The low-energy effective 
Lagrangian \eqref{Leff_a} realises the 
approximate Abelian global symmetry non-linearly through an approximate symmetry under constant shifts, 
$a\to a+\kappa f_a$, broken only by the bare mass term $\propto m_{a,0}^2 a^2$, which parametrises the effect of a possible explicit breaking of the global symmetry, and by the coupling to the topological charge density, $\frac{\alpha_s}{8\pi}\,G_{\mu\nu}^a\,\tilde G^{\mu\nu,a}$, of the $SU(3)_c$ gauge fields.
For the couplings of $a$ to the $SU(2)_L$  
the additional term arising from a constant shift $a\to a+\kappa f_a$ can be removed by field redefinitions~\cite{FileviezPerez:2014xju}. For $U(1)_Y$ it is unobservable in the SM.

The axion, arising from the Peccei-Quinn (PQ) solution of the strong CP problem, is defined via $C_{aGG}\neq 0$ and $m_{a,0}=0$. In fact, the PQ global $U(1)$ symmetry has to be preserved from explicit breaking to a great degree of accuracy, such that $m_{a,0}\approx 0$ to a precision compatible with the 
upper bound on $\bar\theta$ arising from the non-observation of the neutron electric dipole moment. In the literature about axions, it is customary to absorb the coefficient $C_{aGG}$ in the decay constant, $f_A=f_a/C_{aGG}$.

The form of the Lagrangian \eqref{Leff_a} is not unique, since at dimension 5 we can also write down a term which couples the ALP to the Higgs doublet $\phi$:
\begin{equation}
    \mathcal{L}_{\mathrm{eff}}\supset C_\phi O_\phi= C_\phi \frac{\partial_\mu a}{f_a} \left(\phi^\dagger i D^\mu \phi+h.c. \right).
\end{equation}
This operator is redundant with the fermionic operators~\cite{Georgi:1986df}, and can be eliminated through the simultaneous field redefinitions $\phi\to e^{iC_\phi a/f_a}$ and $\psi_F\to e^{-i\beta_F C_\phi a/f_a}$, for all fermion multiplets $\psi_F$, with the $\beta_F$ subject to the conditions:
\begin{equation}
\label{eq:betacombos}
    \beta_u-\beta_Q=-1, ~~\beta_d-\beta_Q=1, ~~\beta_e-\beta_L=1,~~3\beta_Q+\beta_L=0.
\end{equation}
These field redefinitions eliminate $C_\phi O_\phi$ from the Lagrangian, at the expense of shifting the fermionic coupling matrices $C_F$ by $C_F\to C_F+\beta_F C_\phi\mathbbm{1}$. It is consistent (and usual) to thus omit $O_\phi$ from the operator basis. However, its existence plays an important role in the calculation of the renormalisation group equations (RGEs) for the ALP EFT, since it acts as a counterterm for UV divergences in loop diagrams involving the fermionic operators, which hence renormalise all other fermionic operators once reabsorbed into the basis. 

A particular choice of the $\beta_F$ in the above field redefinitions is equivalent to a single linear combination of the five unbroken $U(1)$ global symmetries of the SM. The other four global symmetries can be used in a similar way to eliminate four more coupling parameters, e.g.~the three diagonal components of either $C_L$ or $C_e$, and either $C_{aWW}$ or $C_{aBB}$~\cite{Bauer:2020jbp,Bonilla:2021ufe,Bauer:2021mvw}.

The Wilson coefficients for the gauge boson couplings $C_{aVV}$ (for $V=G,W,B$) are scale independent:
\begin{equation}
    \frac{d}{d\ln \mu}C_{aVV}(\mu)=0\,.
\end{equation}
However, the ALP-fermion couplings are scale-dependent quantities, which obey the RGEs below. At one loop, there are contributions from Yukawa interactions \cite{Bauer:2020jbp,Choi:2017gpf,MartinCamalich:2020dfe} and gauge interactions \cite{Bauer:2020jbp,Izaguirre:2016dfi,Chetyrkin:1998mw}. At two-loop order in the gauge interactions, there are generation-universal contributions from the fermionic couplings, which are included because they are formally of the same order in $\alpha_i$ as the one-loop gauge contributions. Finally, there is a contribution dependent on $X$, defined in Eq.~\eqref{Xdef} below, which comes from diagrams which require the $O_\phi$ counterterm, whose contributions are then rotated back into the basis, following the arguments above \cite{Bauer:2020jbp}. The particular choice of the $\beta_F$ parameters is unphysical; physical quantities always depend only on the combinations \eqref{eq:betacombos}. Overall, the RGEs for the ALP-fermion couplings are:
\begin{align}\label{RGEs}
   \frac{d}{d\ln\mu}\,C_Q(\mu)
   &= \frac{1}{32\pi^2}\,\big\{ Y_uY_u^\dagger 
    + Y_dY_d^\dagger, C_Q \big\}
    - \frac{1}{16\pi^2}\,\big( Y_u\,C_u Y_u^\dagger 
    + Y_d\,C_d Y_d^\dagger \big) \notag\\
   &\quad\mbox{}+ \left[ \frac{\beta_Q}{8\pi^2}\,X
    - \frac{3\alpha_s^2}{8\pi^2}\,C_F^{(3)}\tilde C_{aGG} 
    - \frac{3\alpha_2^2}{8\pi^2}\,C_F^{(2)}\tilde C_{aWW}
    - \frac{3\alpha_1^2}{8\pi^2}\,{\cal Y}_Q^2\,\tilde C_{aBB} \right] \mathbbm{1} \,, \notag\\
   \frac{d}{d\ln\mu}\,C_q(\mu)
   &= \frac{1}{16\pi^2}\,\big\{ Y_q^\dagger Y_q, C_q \big\}
    - \frac{1}{8\pi^2}\,Y_q^\dagger C_Q Y_q 
    + \left[ \frac{\beta_q}{8\pi^2}\,X
    + \frac{3\alpha_s^2}{8\pi^2}\,C_F^{(3)}\tilde C_{aGG} 
    + \frac{3\alpha_1^2}{8\pi^2}\,{\cal Y}_q^2\,\tilde C_{aBB} \right] \mathbbm{1} \,, \notag\\
   \frac{d}{d\ln\mu}\,C_L(\mu)
   &= \frac{1}{32\pi^2} \left\{ Y_eY_e^\dagger, C_L \right\}
    - \frac{1}{16\pi^2}\,Y_e\,C_e Y_e^\dagger 
    + \left[ \frac{\beta_L}{8\pi^2}\,X
    - \frac{3\alpha_2^2}{8\pi^2}\,C_F^{(2)}\tilde C_{aWW} 
    - \frac{3\alpha_1^2}{8\pi^2}\,{\cal Y}_L^2\,\tilde C_{aBB} \right] \mathbbm{1}  , \notag\\
   \frac{d}{d\ln\mu}\,C_e(\mu)
   &= \frac{1}{16\pi^2} \left\{ Y_e^\dagger Y_e, C_e \right\}
    - \frac{1}{8\pi^2}\,Y_e^\dagger C_L Y_e
    + \left[ \frac{\beta_e}{8\pi^2}\,X
    + \frac{3\alpha_1^2}{8\pi^2}\,{\cal Y}_e^2\,\tilde C_{aBB} \right] \mathbbm{1} \,,
\end{align}
where $C_F^{(N)}=\frac{N^2-1}{2N}$ is the eigenvalue of the quadratic Casimir operator in the fundamental representation of $SU(N)$, and we have abbreviated 
\begin{equation}\label{Xdef}
   X = \text{Tr} \left[ 
    3C_Q\big( Y_uY_u^\dagger - Y_dY_d^\dagger \big)
    - 3C_uY_u^\dagger Y_u + 3C_dY_d^\dagger Y_d
    - C_LY_eY_e^\dagger + C_eY_e^\dagger Y_e \right] .
\end{equation}
All quantities on the right-hand side of Eq.~\eqref{RGEs} must be evaluated at the scale $\mu$. The ALP--boson and ALP--fermion couplings entering at ${\cal O}(\alpha_i^2)$ appear in the linear combinations:
\begin{equation}\label{eq:20}
\begin{aligned}
   \tilde C_{aGG} &= C_{aGG} + \text{Tr} \left( C_u + C_d - 2C_Q \right) , \\
   \tilde C_{aWW} &= C_{aWW} - \text{Tr} \left( 3C_Q + C_L \right) , \\[-0.5mm]
   \tilde C_{aBB} &= C_{aBB} + 2\,\text{Tr} \left( \frac43\,C_u + \frac13\,C_d
    - \frac16\,C_Q + C_e - \frac12\,C_L \right) .
\end{aligned}
\end{equation}

The running and mixing of the coefficients of the ALP Lagrangian can have important consequences for phenomenology. For example, an ALP coupling to top quarks at the high scale will mix into a sizeable ($\propto y_t^2$) ALP-electron coupling. This can then open up important production or decay modes for the ALP, even if an ALP-electron coupling is zero at the UV scale.

Other important interactions, for example, the ALP-photon coupling and flavour-changing ALP-quark couplings, can instead be generated via finite loops~\cite{Bauer:2017ris,Bauer:2021mvw}. 

%% file: WG1/content/1-3-5.tex
\subsubsection{The ALP-HNL Portal}
Recent research~\cite{deGiorgi:2022oks,Marcos:2024yfm} explored the intriguing role of ALP interactions with heavy neutral leptons (HNLs) and their potential implications in colliders. The coupling of ALPs to HNLs provides a compelling avenue for detecting BSM physics, especially given the distinct signatures HNLs can produce in high-energy experiments.

At the level of the ALP EFT, the ALP-HNL portal can be constructed in full analogy with the other fermionic operators
\begin{equation}
    \mathcal{L}\supset \frac{\partial^\mu a}{f_a}\,
\ov{N_R} \gamma_\mu C_N N_R\,,
\end{equation}
where $N_R$ is a right-handed neutrino.
In Ref.~\cite{deGiorgi:2022oks}, ALPs are assumed to couple to HNLs via dimension-5 operators that involve both ALP-HNL interactions and gauge-invariant terms with gluons. The coupling strength is generally parameterised by a scale $f_a$, which governs the ALP’s interaction rate with HNLs. Given the nature of the ALP and, in particular, its derivative couplings to fermions, the ALP-HNLs interaction is proportional to the HNL masses, representing a potential enhancement with respect to the other ALP-fermion couplings. Indeed, it introduces a portal between the ALP and the HNL sector, which may be critical in understanding the nature of HNL masses, as well as their possible role as mediators in dark sector physics.

One of the prominent experimental avenues explored in Ref.~\cite{deGiorgi:2022oks} involves collider searches for ALP-induced HNL production. In particular, the ALP-HNL coupling allows for processes where an ALP decays into an on-shell HNL pair that decay according to the standard channels. This interaction provides unique signatures, such as events with missing transverse energy (from undetected HNLs) and secondary particle production through HNL decays. These signatures could be visible in the high-luminosity phase of the LHC and future collider experiments, where enhanced sensitivity to rare decay events could enable the detection of ALP-induced HNL production.

This proof of concept setup has been extended in Ref.~\cite{Marcos:2024yfm}, where other production mechanisms have been taken into consideration, as well as the possible coupling of the ALP with more than one HNL, as in realistic Majoron models.
Future experimental searches may offer further insight into the nature of these couplings, potentially illuminating connections between neutrino physics, ALPs and fundamental symmetries.

\subsubsection{ALP Searches in Meson Decays}
While high-energy colliders remain a powerful probe for ALPs, low-energy processes and particularly those involving meson decays have emerged as a critical window into ALP physics. The energy scales and flavour transitions accessible in meson decays allow for stringent constraints on ALP couplings to SM fermions and gauge bosons, including flavour-violating and CP-violating effects. Processes such as $B \to K^{(\ast)} a$ and $K \to \pi a$ are highly sensitive to small deviations from SM expectations. Depending on the mass and couplings of the ALP, these decays can lead to prompt visible signals, displaced vertices, or invisible signatures, each of which provides complementary information.

Many studies have appeared on this topic~\cite{Izaguirre:2016dfi,Merlo:2019anv,Aloni:2019ruo,Bauer:2019gfk,Bauer:2020jbp,Bauer:2021mvw,Guerrera:2021yss,Gallo:2021ame,Bonilla:2022qgm,Bonilla:2022vtn,deGiorgi:2022vup,Guerrera:2022ykl,Bonilla:2023dtf,Arias-Aragon:2023ehh,DiLuzio:2024jip,deGiorgi:2024str,Alda:2024cxn,Alda:2024xxa,Arias-Aragon:2024qji,Arias-Aragon:2024gdz} and a recent comprehensive analysis~\cite{Alda:2025uwo} systematically explored the parameter space of ALP couplings to SM fields by combining EFT techniques with UV model realizations. The study incorporates renormalisation group evolution, electroweak symmetry breaking, and matching to chiral perturbation theory ($\chi$PT)~\cite{Aloni:2018vki,Bauer:2021wjo,Bai:2024lpq,Ovchynnikov:2025gpx,Bai:2025fvl,Balkin:2025enj}, enabling consistent predictions across all relevant energy regimes. Importantly, the framework accounts for running-induced flavour non-universalities and includes loop-level effects relevant for matching at the electroweak scale.

One notable result is the sensitivity of flavour observables to parameter combinations that remain unconstrained in collider or beam-dump experiments. In particular, loop-induced flavour-changing couplings generated via renormalisation group evolution can lead to observable effects even in models where tree-level flavour violation is absent. Furthermore, the chiral structure of the ALP interactions implies that off-diagonal couplings can be induced through mixing with mesons, requiring a proper treatment via $\chi$PT at low energies.

As a concrete application is the potential explanation of the Belle~II excess in $B \to K \nu \bar{\nu}$ in terms of ALP contributions. There are regions in parameter space where such an anomaly can be accommodated consistently with other existing bounds.
All calculations carried out in Ref.~\cite{Alda:2025uwo} were implemented within a consistent computational framework, \texttt{ALP-aca}~\cite{Alda:2025nsz}, the ALP Automatic Computing Algorithm, enabling the automatic inclusion of quantum corrections and up-to-date experimental constraints. This open-access code allows for flexible inclusion of user-defined constraints. It accurately computes ALP decay widths and branching ratios across the mass range $m_a \in [0.01\,, 10]$~GeV, and distinguishes between prompt, displaced, and invisible decay regimes. It also includes the full implementation of chiral perturbation theory effects, ensuring reliable predictions in the non-perturbative QCD regime. \texttt{ALP-aca} provides a user-friendly tool
for ongoing and future analyses, which may play a central role in the global effort to test the properties of ALPs.

While no signal of ALPs (or any other type of New Physics) emerged experimentally, the results of the studies on ALP and mesons demonstrate that current and near-future experiments, especially Belle II and NA62, will significantly narrow the viable ALP parameter space.

\subsubsection{Signatures of ALPs at Colliders}
At colliders, ALPs can be produced via several mechanisms, primarily depending on their couplings to SM particles. Two of the most prominent production channels are gluon fusion and photon-photon fusion, each exhibiting distinct characteristics based on the ALP mass and couplings.\\

Gluon fusion is one of the most promising channels for ALP production at the LHC. This process is mediated by the ALP's coupling to gluons through a dimension-5 operator. The production cross-section for this process can be expressed as:
\begin{equation}
\sigma(gg \to a) = \frac{\alpha_s^2}{f_a^2} \cdot \mathcal{F}(m_a)\,,
\end{equation}
where $\alpha_s$ is the strong coupling constant, $f_a$ is the ALP decay constant, and $\mathcal{F}(m_a)$ is a model-dependent function that incorporates the ALP mass dependence. Studies, such as those outlined in~\cite{Brivio:2017ije,Bauer:2017ris,Bauer:2018uxu} among others, explore the sensitivity of the gluon fusion process to the parameter space of ALPs, demonstrating its importance in both theoretical and experimental contexts.\\

On the other hand, photon-photon fusion is another notable mechanism for ALP production, particularly for ALPs that couple strongly to photons. The production cross section in this channel depends sensitively on the ALP-photon coupling $g_{a\gamma\gamma}$ and is more pronounced for lighter ALPs:
\begin{equation}
\sigma(\gamma\gamma \to a) \propto g_{a\gamma\gamma}^2 \cdot \mathcal{G}(m_a)\,,
\end{equation}
where $\mathcal{G}(m_a)$ describes the mass dependence of the production rate. This mechanism has been explored in experimental proposals, especially in the context of photon-rich environments, as discussed in~\cite{Bauer:2017ris}.\\

The detection of ALPs at colliders~\cite{Jaeckel:2012yz,Mimasu:2014nea,
Jaeckel:2015jla,Alves:2016koo,Knapen:2016moh,Brivio:2017ije,Bauer:2017nlg,Mariotti:2017vtv,Bauer:2017ris,
Baldenegro:2018hng,Craig:2018kne,Bauer:2018uxu,Gavela:2019cmq,Haghighat:2020nuh,Wang:2021uyb,deGiorgi:2022oks,
Bonilla:2022pxu,Ghebretinsaea:2022djg,Vileta:2022jou,Liu:2021lan,Marcos:2024yfm,Arias-Aragon:2024gpm} is highly reliant on their decay products and the experimental signatures they produce. ALPs can decay into various SM particles depending on their couplings, with the following being the most prominent experimental signatures:
\begin{itemize}
    \item \textbf{Diphoton resonances:} ALPs decaying into two photons produce a sharp peak in the invariant mass distribution of photon pairs, making them an excellent target for diphoton searches in electromagnetic calorimeters. The sharp invariant mass peak from ALP decays is distinguishable from SM background processes such as direct photon production or QCD jets misidentified as photons. Analyses like those in~\cite{Brivio:2017ije,Bauer:2018uxu} emphasise the sensitivity of diphoton resonance searches to ALPs in both high-luminosity and high-energy collider setups.
    \item \textbf{Monojet + missing energy:} For ALPs with invisible decay channels, missing transverse energy accompanied by a high-energy jet offers a distinctive signature.
    \item \textbf{Displaced vertices:} Weakly coupled or light ALPs may have long lifetimes, causing them to decay at distances significantly removed from the interaction point. Such events manifest as displaced vertices, with tracks originating away from the primary vertex. These decays create a unique experimental signature and are pivotal in expanding the accessible parameter space of ALP searches, as they allow probing of regions inaccessible to prompt decay signatures~\cite{Bauer:2017ris}.
\end{itemize}
All in all, one could try to summarise the present constraints from colliders on the ALP couplings by performing a global fit. A recent example of such a study can be found in Ref.~\cite{Biekotter:2023mpd}. 

%% file: WG1/content/1-3-2.tex
The analysis of the previous section assumes an approximate shift symmetry in the ALP interactions broken only by the mass term. This assumption, while motivated by the QCD axion paradigm and solutions to the strong CP problem, does not generally describe a scalar singlet. Indeed, not only {can} shift breaking effects in ALP couplings arise in motivated BSM models,\,\footnote{One example are non-minimal composite Higgs models~\cite{Gripaios:2009pe,Bellazzini:2014yua,Chala:2017sjk} where extra pNGBs arise commonly, and in many setups as singlets under the SM gauge group. Under the partial compositeness scenario~\cite{KAPLAN1991259}, shift-symmetry breaking interactions can produce both a mass and several other terms in the potential of the pNGBs at the same order in the spurions expansion. These terms have important phenomenological consequences, namely for dark matter~\cite{Balkin:2017aep,Wu:2017iji,Ballesteros:2017xeg,Balkin:2018tma,Ramos:2019qqa}, baryogenesis~\cite{Espinosa:2011eu,Chala:2016ykx,Chala:2018opy} and flavor probes~\cite{Castro:2020sba,Blance:2019ixw,CidVidal:2022mrx,Bonilla:2022vtn}. Moreover, the couplings of the pNGBs to fermions are not expected to preserve, in general, the original shift symmetry.} but consistency arguments from quantum gravity lead us to expect shift symmetry-breaking effects in the UV~\cite{Banks:2010zn} that propagate to the IR theory. These are strong reasons to consider the impact of shift-breaking interactions in the ALP EFT and, in particular, their evolution across energies.

This endeavour was carried out first in Ref.~\cite{Chala:2020wvs}, assuming only CP-even interactions. The inclusion of CP-odd terms in the RGEs of the ALP EFT was later computed in Ref.~\cite{DasBakshi:2023lca}. In this review, we highlight the most relevant findings of these works that provided the first systematic study of the entire anomalous dimension matrix of the ALP EFT up to mass-dimension five and one-loop accuracy.

 The most general renormalizable Lagrangian of the SM extended with a real pseudoscalar singlet~\cite{OConnell:2006rsp,Barger:2007im}, that to differentiate from the shift-symmetric ALP particle we will denote by $s$, reads:
\begin{align}
	\mathcal{L}_{\text{SM}+s} &= \frac{1}{2} (\partial_{\mu} s) (\partial^{\mu} s) -\frac{1}{2} m_s^2 s^2 -\frac{\kappa_{s}}{3!}s^3 - \frac{\lambda_{s}}{4!} s^4 - \kappa_{s\phi} s \phi^\dagger \phi - \frac{\lambda_{s \phi}}{2} s^2 \phi^\dagger \phi  \nonumber\\
	& - \frac{1}{4} G^A_{\mu\nu} G^{A \mu\nu} - \frac{1}{4} W^a_{\mu\nu} W^{a \mu\nu} - \frac{1}{4} B_{\mu\nu} B^{\mu\nu} + \theta_{\rm QCD} \frac{g_s^2}{32 \pi^2} G^A_{\mu\nu} \widetilde{G}^{A \mu\nu}\nonumber \\
& + \sum\limits_{\psi=q,l,u,d,e} \overline{\psi}^\alpha i \slashed{D} \psi^\alpha -\left(y_{\alpha\beta}^u \overline{q_L}^\alpha \tilde{\phi} u_R^\beta + y_{\alpha\beta}^d \overline{q_L}^\alpha \phi d_R^\beta + y_{\alpha\beta}^e \overline{l_L}^\alpha \phi e_R^\beta +\text{h.c.} \right)\nonumber \nonumber \\
	&+(D_{\mu} \phi)^\dagger (D^{\mu} \phi) + \mu_{\phi}^2 \phi^\dagger \phi - \lambda (\phi^\dagger \phi)^2  \,.
\label{eq:renorm}
\end{align}
This Lagrangian must be supplemented with additional operators to improve the accuracy of the ALP theory predictions. Up to mass-dimension five (and ignoring lepton number violating operators), a minimal basis is given by~\cite{Chala:2020wvs} 
\begin{align}
\label{eq:basis}
\mathcal{L}_{{\rm SM}+s}^{\rm eff} &\supset s \bigg[ i \,  \overline{q_L} a_{su\phi} \widetilde{\phi} u_R + i \,  \overline{q_L} a_{sd\phi} {\phi} d_R +i \,  \overline{l_L} a_{se\phi} {\phi} e_R + \text{h.c.} \bigg]  + a_{s^5} s^5  \\ 
& + a_{s ^3} s^3 (\phi^\dagger \phi) + a_{s} s (\phi^\dagger \phi)^2  
+  a_{s\widetilde{G}} s G_{\mu\nu}^{A}\widetilde{G}^{A\mu\nu} + a_{s\widetilde{W}} s W_{\mu\nu}^{a}\widetilde{W}^{a\mu\nu}  \nonumber \\
& +  a_{s\widetilde{B}} s B_{\mu\nu}\widetilde{B}^{\mu\nu}+ a_{s{G}} s G_{\mu\nu}^{A}{G}^{A\mu\nu} + a_{s{W}} s W_{\mu\nu}^{a}{W}^{a\mu\nu} +  a_{s{B}} s B_{\mu\nu}{B}^{\mu\nu}\,,
\nonumber
\end{align}
where $a_{s\psi\phi} \equiv a_{s\psi\phi}^0/\Lambda$ are dimension-full and complex matrices in flavour space; $\Lambda$ denotes the cutoff scale of this EFT.
\begin{table}[t]
	\centering{}
	\renewcommand{\arraystretch}{1.5}
	\begin{tabular}{c|c c c c c c }
		\hline
        \hline
		& $s^5$ & $s^3 \phi^\dagger \phi$ & $s (\phi^\dagger \phi)^2$ & $s\overline{\Psi_L} \phi \psi_R$ & $s XX$ & $s X \widetilde{X}$ \\ \hline
	
		$s^5$ & $\textcolor{blue}{\lambda_s}$ & $\lambda_{s\phi}$ & 0 & 0 & 0 & 0 \\
		
		$s^3 \phi^\dagger \phi$ & $\textcolor{blue}{\lambda_{s\phi}}$ & $\lambda_s + \textcolor{blue}{\lambda_{s\phi}} +\textcolor{blue}{\lambda} + y_t^2  $ & $\lambda_{s \phi}$ & $\lambda_{s\phi} y_t$ & $\lambda_{s\phi} g_2^2$ & 0 \\
		
		$s (\phi^\dagger \phi)^2 $ & 0 & $\lambda_{s\phi}$ & $\lambda_{s\phi}$ + $\textcolor{blue}{\lambda} +\textcolor{blue}{ y_t^2}$ & $\textcolor{blue}{ y_t^3 } + \textcolor{blue}{ \lambda y_t}  $ & $\textcolor{blue}{\lambda g_2^2}$ & 0 \\
		
		$s\overline{\Psi_L} \phi \psi_R$ & 0 & 0 & 0 & $\lambda_{s\phi}+\textcolor{blue}{y_t^2}$ & $ \textcolor{blue}{g_3^2 y_t}$ & $ \textcolor{blue}{g_3^2 y_t}$ \\
		
		$s XX$ & 0 & 0 & 0 & 0 & $\textcolor{blue}{g_3^2}$ & 0  \\
		
		$s X \widetilde{X}$ & 0 & 0 & 0 & 0 & 0 & $\textcolor{blue}{g_3^2}$\\
        \hline
        \hline
	\end{tabular}
	\caption{Structure of the dimension-five anomalous dimension matrix. Only BSM and the leading SM contributions are kept. For instance, the top-quark Yukawa coupling $y_{33} \equiv y_t$ and the weak and strong gauge couplings $g_{2,3} $ are the largest contributions in their respective sub-classes. Terms in blue show the contributions that deviate significantly from naive dimensional analysis; see the text for details.}
	\label{tab:uv5}
\end{table}

To order $v/\Lambda$, the dimension-five Wilson coefficients are only renormalised by other dimension-five couplings:
\begin{equation}
\label{eq:AD-5}
16 \pi^2 \mu \frac{\text{d} a_i}{\text{d} \mu} = \gamma_{ij}^{(1)} a_j\,.
\end{equation}
On the contrary, the RGEs of renormalizable couplings can receive contributions from higher-order operators due to the presence of light scales (namely $m=m_s,\,\mu_{\phi}$) in the EFT:
\begin{align}
\label{eq:AD-3}
16 \pi^2 \mu \frac{\text{d} \kappa_i}{\text{d} \mu} & = \gamma_{ij}^{(2)} \kappa_j + \gamma_{ij}^{(3)} m^2 a_j \, , \\
16 \pi^2 \mu \frac{\text{d} \lambda_i}{\text{d} \mu} & = \gamma_{ij}^{(4)} \lambda_j + \gamma_{ijk}^{(5)} \kappa_j a_k \,. 
\end{align}
Considering the CP-even sector of the theory in isolation, the (non-) renormalizable  ALP interactions are (odd) even under a $\mathcal{Z}_2$ {transformation}. The different mass dimensions can therefore only mix in the presence of CP-odd interactions. 

We have renormalised the ALP EFT by computing the divergences generated by one-particle-irreducible diagrams at one-loop with off-shell momenta, and up to one insertion of the dimension-five Wilson coefficients. We have worked in the background field method in dimensional regularization with $\text{d}=4-2 \epsilon$ spacetime dimensions; the counterterms obtained in this approach are therefore gauge invariant. We did most of the computations by hand, but also relied on~\texttt{Feynrules}~\cite{Alloul:2013bka}, \texttt{FeynArts}~\cite{Hahn:2000kx} and \texttt{FormCalc}~\cite{Hahn:1998yk} for the calculations. Additional double checks were performed using~\texttt{Matchmakereft}~\cite{Carmona:2021xtq}. We matched the divergences onto the Green's basis obtained in Ref.~\cite{Chala:2020wvs}, finally projecting the results into the physical basis in Eq.~\eqref{eq:basis}.

The complete expressions for the RGEs were included in a new Mathematica package~\texttt{ALPRunner}\,\footnote{\url{https://github.com/sdbakshi13/ALPRunner}.} together with functions to solve the running numerically. In Tabs.~\ref{tab:uv5} and~\ref{tab:uv4}, we show parts of the full anomalous dimension matrix, in particular the running induced by the dimension-five interactions. The contributions highlighted in blue deviate from naive dimensional analysis by at least one order of magnitude, therefore potentially having a large impact on the ALP phenomenology.

Such RGEs accurately describe the ALP phenomenology above the EW scale. Below such scale, the Higgs develops a    VEV, $v$, and the ALP phenomenology must therefore be described by a different low-energy EFT (LEFT), organised in inverse powers of $v$, which is obtained by integrating out the massive EW bosons of the SM and the top quark. To dimension-five, the most general and minimal Lagrangian for such an EFT reads: 
\begin{align}
	\mathcal{L}_{\text{LEFT}} &= \frac{1}{2} (\partial_{\mu} s) (\partial^{\mu} s) -\frac{1}{2} \Tilde{m}_s^2 s^2 -\frac{\Tilde{\kappa}_{s}}{3!}s^3 - \frac{\Tilde{\lambda}_{s}}{4!} s^4  - \frac{1}{4} G^A_{\mu\nu} G^{A \mu\nu} - \frac{1}{4} A_{\mu\nu} A^{\mu\nu} + \Tilde{\theta}_{\rm QCD} G_{\mu\nu}^A \widetilde{G}^{A\mu\nu} \nonumber\\
	& + \sum\limits_{\psi=u,d,e} \bigg[\overline{\psi} i \,  \slashed{D} \psi -  \overline{\psi_L} \Tilde{m}_\psi \psi_R + i \,  s\overline{\psi_L} \Tilde{c}_\psi \psi_R +s^2 \overline{\psi_L} \Tilde{a}_{\psi} \psi_R + \text{h.c.} \bigg] + {\Tilde{a}_{s^5}}s^5\nonumber\\
		& + {\Tilde{a}_{sA}} s A_{\mu\nu} A^{\mu\nu} + {\Tilde{a}_{sG}}s G_{\mu\nu}^A G^{A\mu\nu} + {\Tilde{a}_{s\Tilde{A}}} s A_{\mu\nu} \widetilde{A}^{\mu\nu} + {\Tilde{a}_{s\Tilde{G}}}s G_{\mu\nu}^A \widetilde{G}^{A\mu\nu} \nonumber\\
		\label{eq:LEFT}
	& + \sum\limits_{\psi=u,d,e} \bigg[  \overline{\psi_L} \Tilde{a}_{\psi A} \sigma^{\mu\nu}\psi_R A_{\mu\nu}+ \overline{\psi_L}\Tilde{a}_{\psi G} \sigma^{\mu\nu} T_A \psi_R G_{\mu\nu}^A +\text{h.c.} \bigg] \,,
\end{align}
where again the non-renormalizable couplings are dimensionfull, $\Tilde{a}\equiv \Tilde{a}^0/v$. The parameters in this LEFT basis can be fixed at the scale $\mu = v$ by requiring that both theories, before and after EWSB, 
describe the same physics. We have performed the complete matching of these theories at tree level under the assumption that the UV physics is well described by the SM+$s$ EFT (Eqs.~\eqref{eq:renorm} and~\eqref{eq:basis}). In particular, we have taken into account the full mixing effects of the singlet with the Higgs boson~\cite{DasBakshi:2023lca}. The results of this matching for the CP-even sector read:
\begin{align}\label{eq:matching}
  &\tilde{e} = g_2 s_w=g_1 c_w\,,&
&\tilde{m}^2 = m^2 + \frac{\lambda_{s\phi}}{2} v^2\,,&  \\
  & \tilde{g}_3 =g_3 \,,&
  & \tilde{\lambda}_s = \lambda_s - 3\frac{v^2}{m_h^2}
  \lambda_{s\phi}^2\,,&
\\
& (\tilde{m}_u)_{\alpha\beta} =\frac{v}{\sqrt{2}} (y^u)_{\alpha\beta}
\,,&
  & (\tilde{c}_u)_{\alpha\beta} = \frac{v}{\sqrt{2}} (a_{{su\phi}})_{\alpha\beta}  \,,&\\ 
& (\tilde{m}_d)_{\alpha\beta} = \frac{v}{\sqrt{2}}
(y^d)_{\alpha\beta}
\,,&
& (\tilde{c}_d)_{\alpha\beta} = \frac{v}{\sqrt{2}} (a_{{sd\phi}})_{\alpha\beta}\,,&\\
& (\tilde{m}_e)_{\alpha\beta} = \frac{v}{\sqrt{2}} (y^e)_{\alpha\beta}
\,,&
& (\tilde{c}_e)_{\alpha\beta} = \frac{v}{\sqrt{2}} (a_{se\phi})_{\alpha\beta}\,,&\\
& \tilde{a}_{s\widetilde{G}} = a_{s\widetilde{G}}\,,&
& \tilde{a}_{s\widetilde{A}} = a_{s\widetilde{W}} s_w^2 + a_{s\widetilde{B}} c_w^2\,,& \label{eq:matching2}
\end{align}
where $c_w$ and $s_w$ are the cosine and sine of the Weinberg angle, respectively.
Note that the coefficients $\tilde{a}_{u,d,e}$ are higher-order in our power counting, since they are triggered by the Higgs boson, which couples to both $s^2$ and to fermionic currents with overall
strength $\sim \lambda_{s\phi} y^\psi/v\sim \lambda_{s\phi} m_\psi/v^2$.
All other Wilson coefficients not explicitly shown vanish at the order we are computing.
At energies below a given fermion mass threshold, the effective
Lagrangian takes exactly the same form as in Eq.~\eqref{eq:LEFT}, as well as the matching equations, but with $\alpha$ and $\beta$ running over a different number of flavors\,\footnote{The only exception arises
if $\tilde{c}_\psi$ is unsuppressed, in which case
integrating out a massive fermion would result in the following
matching condition:
\begin{equation}
  (\tilde{a}_\psi)_{\alpha \beta} = -\frac{(\tilde{c}_\psi)_{\alpha
      \gamma} (\tilde{c}_\psi)_{\gamma \beta}}{(\tilde{m}_\psi)_\gamma}
  \,,
\end{equation}
where $\gamma$ corresponds to the flavour that is being integrated out.}.
\begin{table}[t]
	\centering{}
	\renewcommand{\arraystretch}{1.5}
	\begin{tabular}{c|c c c c c c}
		\hline
        \hline
		& $s^5$ & $s^3(\phi^\dagger \phi)$ & $s (\phi^\dagger \phi)^2$ & $s\overline{\Psi_L} \phi \psi_R$ & $s XX$ & $s X \widetilde{X}$ \\ \hline
		
		$s^3$ & $\textcolor{blue}{m_s^2}$ & $\textcolor{blue}{\mu^2}$ &  $0$  & $0$ & $0$ & $0$\\
		
		$s(\phi^\dagger \phi)$ & $0$ & $m_s^2$ &  $\textcolor{blue}{\mu^2}$ & $y_t \mu^2$ & $ \textcolor{blue}{g_2^2 \mu^2} $ & $0$ \\
		
		$s^4$ & $\textcolor{blue}{\kappa_s}$ & $\textcolor{blue}{\kappa_{s\phi}}$ & $0$ & 0  & $0$ & $0$\\
		
		$s^2(\phi^\dagger \phi)$ & 0 & $\textcolor{blue}{\kappa_s} + \textcolor{blue}{\kappa_{s\phi}}$ & $\textcolor{blue}{\kappa_{s\phi}}$  & $\textcolor{blue}{y_t \kappa_{s\phi}}$ & $\textcolor{blue}{ g_2^2 \kappa_{s\phi} }$ & $0$\\
		
		$(\phi^\dagger \phi)^2$ & 0 & $0$ & ${\kappa_{s\phi}}$  & $0$ & $0$ & $0$\\
        \hline
        \hline
	\end{tabular}
	\caption{Structure of the renormalizable anomalous dimension matrix induced by effective interactions. 
  }
	\label{tab:uv4}
\end{table}
The renormalisation of the ALP LEFT follows the same approach described before. The corresponding results, together with those obtained in the high-energy theory, allow us to describe in detail the ALP phenomenology, from very high to very low energies, up to QCD confining effects. Based on these results, let us emphasise the following points:

\begin{enumerate}

\item The mixing structure in this more general EFT is significantly more complicated than the one explored in the last section, which reflects the breaking of the ALP periodicity.  Particularly, in the shift-symmetric basis, the $2\pi$ periodicity imposes that $c_X \in \mathbb{Z}$. Consequently, the gauge operators in the shift-symmetric ALP EFT are scale-invariant\,\footnote{This result has been checked explicitly up to two-loops in gauge couplings~\cite{Bauer:2020jbp}.}. Since the ALP periodicity is broken in the basis of Eq.~\eqref{eq:basis}, no quantisation rule applies \textit{a priori}. Nevertheless, at one-loop and up to insertions of dimension-five Wilson coefficients, we found that the scale-invariance of these operators is preserved (once we factorize the gauge couplings), simply because there is no diagram that can be drawn with the insertion of the ALP couplings to the Higgs\,\footnote{The analysis presented in Ref.~\cite{Bonnefoy:2022rik}, where the flavour invariants of this EFT have been computed, leads us to expect that this result holds at higher-loop order since the only CP-even invariant obtained at mass dimension-five involves several powers of Yukawa couplings.}.

\item There are several other directions in Tabs.~\ref{tab:uv5} and~\ref{tab:uv4} that are not renormalized; however most of these zeros are trivial. For example, it is not possible to insert one $s^5$ operator in a one-loop diagram without having at least 3 external singlet
legs, which explains the non-renormalisation results in the first column of Tab.~\ref{tab:uv4}. On
the other hand, operators with $n$ singlet fields can renormalize others with $m > n$
singlet fields via the insertion of renormalizable new physics interactions. 

\item The structure of the anomalous dimension matrix is enriched in the LEFT. For example, the RGE of the gauge operators becomes non-trivial:
\begin{align}
\label{eq:leftsG}
\beta_{\Tilde{a}_{s\widetilde{G}}} & =  \left(-\frac{46}{3} \Tilde{g}_3^2 + 2\text{Tr} \left[\Tilde{c_e} \Tilde{c_e}^\dagger+3\Tilde{c_d} \Tilde{c_d}^\dagger  + 3\Tilde{c_u} \Tilde{c_u}^\dagger    \right] \right)    \Tilde{a}_ {s\widetilde{G}}   \nonumber \\
&~~~~ - 2 g_3 \text{Tr} \left[  \Tilde{a}_{d G} \Tilde{c_d}^\dagger+  \Tilde{a}_{u G} \Tilde{c_u}^\dagger +\text{h.c.} \right]\,,
\end{align}
Another generic effect is that the singlet, as well as fermion masses, receive new contributions:
\begin{align}
  \beta_{\tilde{m}_s^2} &= \Tilde{\kappa}_s^2 + \tilde{\lambda}_s \tilde{m}_s^2
  + 12\tilde{m}_s^2\text{Tr}(\tilde{c}_u\tilde{c}_u^{\dagger}) -24\text{Tr}\left(\tilde{c}_u\tilde{c}_u^{\dagger}\tilde{m}_u\tilde{m}_u^\dagger - \tilde{m}_u^\dagger \tilde{a}_u \tilde{m}_u^\dagger \tilde{m}_u \right) +\dots 
\end{align}
All these vanish, however, if the UV physics above EWSB
can be described by the SM$+s$ EFT.

\begin{figure}[t]
    \centering
    \includegraphics[width=0.7\linewidth]{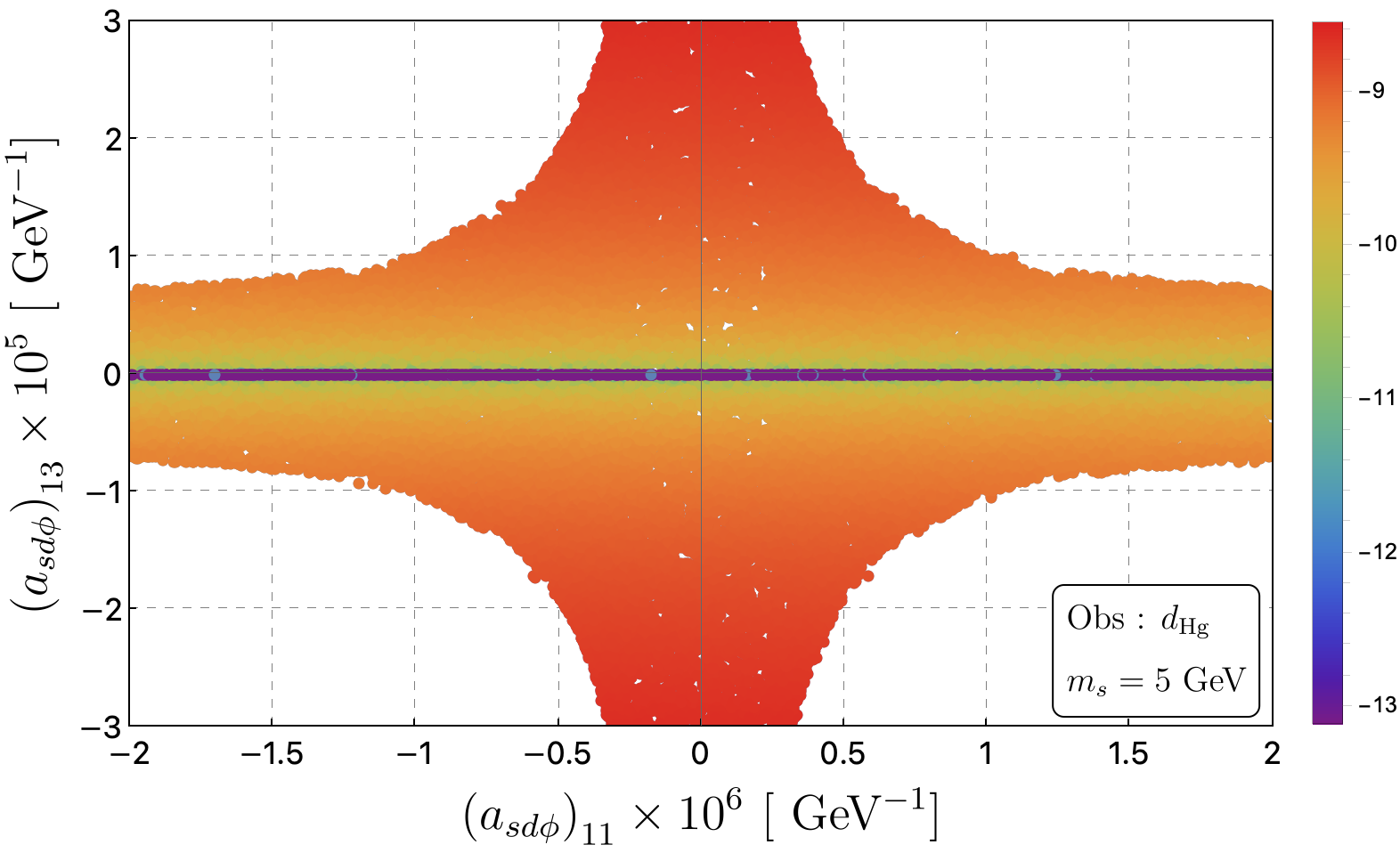}
    \caption{EDM constraints on the UV scenario discussed in point 6 of the text. The bounds result from resumming the RGEs from $\Lambda = 1$ TeV until the experimental scale at which the observable is measured, in this case the neutron EDM of the ${}^{199}{\mathrm{Hg}}$ atom~\cite{Graner:2016ses}.  
    The colorbar shows the size (in log scale) of the shift-breaking invariants identified in Ref.~\cite{Bonnefoy:2022rik}. Figure taken from Ref.~\cite{DasBakshi:2023lca}.}
    \label{fig:edm}
\end{figure}

\item The singlet-fermion interactions produce at low energies several of the interactions between the ALP and the Higgs, which are absent in shift-symmetric
scenarios. 
The induced scalar operators provide new contributions for the ALP-Higgs mixing beyond leading order. Although not explicitly shown in Eqs.~\eqref{eq:matching} to~\eqref{eq:matching2}, we found, e.g. that the mixing angle can be non-zero even for a vanishing singlet VEV. This occurs only in an aligned scenario where $a_s \sim \kappa_{s \phi}/v^2$. Moreover, we found that the SM correlation between the triple and quartic Higgs couplings is only broken by dimension-five interactions in this EFT.

\item The mixing between different classes of operators provides novel (indirect) ways to test secluded scenarios that are difficult to probe directly. For example, the fact that ALP-gauge operators renormalise fermion interactions can be used to provide the best constraints on ALP-gauge couplings. Consider, for instance, a photophobic\,\footnote{Note that by point 1, this condition is stable and
$a_{s\tilde{A}}$ vanishes at all scales.} ALP of mass $\mathcal{O}(\mathrm{keV})$, coupled only to $Z$ bosons at some high-energy scale:
\begin{equation}
  \mathcal{L}
  = \mathcal{L}_{\mathrm{SM}}+ \frac{1}{2}\partial_\mu s\partial^\mu s + \frac{1}{2}\tilde{m}^2 s^2 
 + \frac{a_{s\widetilde{Z}}}{c_\omega^2 - s_\omega^2} s \left( c_\omega^2 W_{\mu\nu} \widetilde{W}^{\mu\nu} - s_\omega^2 B_{\mu\nu} \widetilde{B}^{\mu\nu} \right)
\,.
\end{equation}
Collider bounds from mono-$Z$ searches constrain $a_{s\widetilde{Z}} <(0.04)\,0.2\, \mathrm{TeV}^{-1}$~\cite{Brivio:2017ije} at the (HL--)LHC\,\footnote{This is a very conservative bound, since the corresponding experimental analyses, as well as the recast, rely on a minimum missing energy cut to identify the signal events, that targets $\mathcal{O}$(MeV) ALPs.}. However, this same coupling is generated by running the coupling of the ALP to electrons, which itself is strongly constrained by the modification of Red Giants cooling~\cite{Raffelt:2006cw}. 
Translating this experimental constraint into a bound on the $a_{s\widetilde{Z}}$ coupling results in a limit 4 orders of magnitude stronger than the collider bound~\cite{Chala:2020wvs}. Other phenomenologically relevant examples, in the shift-symmetric scenario, are discussed in Refs.~\cite{Bauer:2021mvw,Bonilla:2021ufe,Bonnefoy:2022rik}.

\item The interesting feature of our shift-breaking ALP basis is that it can be used to eventually pinpoint the nature of a potential ALP signal. In fact, we have worked out the conditions to preserve shift-symmetry in our basis (at the classical level)~\cite{Chala:2020wvs}:
\begin{equation}
\label{eq:aSS}
a_{su\phi} = \frac{i}{f_s} \,  \left(y^{u}c_{u} - c_{q} y^u \right)\,,\quad a_{sd\phi}= \frac{i}{f_s} \,  \left(y^d c_{d} - c_{q} y^d \right)\,,\quad a_{se\phi} = \frac{i}{f_s} \,  \left({y^{e}} c_{e} - c_{l} y^{e} \right)\,.
\end{equation}
These conditions are implicit and do not, in general, allow to identify the $c$-matrices given a set of $a_{s\psi\phi}$ couplings. This is only possible under specific assumptions, like Minimal Flavour Violation or the presence of just one fermion generation coupled to the ALP. Recently, the shift-breaking invariants of this EFT were computed~\cite{Bonnefoy:2022rik} so that we can study the fate of shift-symmetry effects under running without having to refer implicitly to the matrices $c_\psi$. To give an example, let us consider Fig.~\ref{fig:edm} that shows the EDM constraints on a given UV scenario where only two Wilson coefficients $(a_{sd\phi})_{13}$ and $(a_{sd\phi})_{11}$ are turned on in the flavour basis\,\footnote{The CP violation arises in this case when rotating to the physical basis and due to running effects by mixing with the SM Yukawas.}. By computing the size of the shift-breaking invariants for each point in the parameter space, we find easily which regions are more or less compatible with an ALP shift symmetry. This illustrates the importance of our results: if an anomaly in data could be explained by an ALP in some of these specific regions, we would be able to learn about the degree of shift-symmetry of this particle and therefore its UV origin.

\end{enumerate}

%% file: WG1/content/spin-1.tex
A massive spin-1 boson beyond the Standard Model, also referred to in the literature as ``dark" or ``hidden" photon, can be a compelling dark matter candidate or the force
mediator in a dark sector. In this section, we review some aspects of model building for the Abelian case. We refer the reader to Sec.~\ref{subsec:Pradler} for astrophysical constraints on spin-1 particles.

It is quite common in the literature to take as a starting point the Proca Lagrangian~\cite{Proca:1936fbw},
\begin{equation} \label{Proca}
\mathcal{L}_{\rm Proca} = - \frac{1}{4} V^{\mu\nu} V_{\mu\nu} + \frac{1}{2} m^2 V^\mu V_\mu \, .
\end{equation}
Here $V_\mu$ denotes the vector field (i.e. the spin-1 boson), $V_{\mu\nu} = \partial_\mu A_\nu - \partial_\nu A_\mu$.
This Lagrangian correctly describes a vector with mass $m$, however, it is not gauge invariant. 
It is instructive to consider first an Abelian Higgs model, derive from it the Stueckelberg Lagrangian,
 then the Proca Lagrangian. 
Let us start with~\cite{Higgs:1964ia}
\begin{equation} \label{Higgs1}
\mathcal{L}_{\rm Higgs} = -\frac{1}{4} F^{\mu\nu} F_{\mu\nu} + \eta^{\mu\nu} (D_\mu \varphi)^* (D_\nu \varphi) - \frac{\lambda}{4} \left(\varphi^* \varphi - \frac{1}{2} v^2 \right)^2 \, ,
\end{equation}
where
\begin{equation}
F_{\mu\nu} = \partial_\mu A_\nu - \partial_\nu A_\mu \, , \qquad   D_\mu = \partial_\mu - i g A_\mu \, ,
\end{equation}
with $g$ the gauge coupling.
This Lagranagian is invariant under a $U(1)$ gauge symmetry:
\begin{align}
A_\mu(x) & \to A_\mu(x) +  \partial_\mu \alpha(x) \\
\varphi(x) & \to e^{i g \alpha(x)} \varphi(x) \, .
\end{align}
The scalar potential spontaneously breaks this symmetry. We can parametrize the complex 
Higgs field $\varphi$ in terms of two real scalar fields,
\begin{equation} \label{varphisub}
\varphi(x) = \frac{1}{\sqrt{2}} (v + \rho(x)) e^{i \frac{\chi(x)}{v}} \, .
\end{equation}
Under the gauge symmetry, they transform as $\rho(x)  \to \rho(x) \, , \ \chi(x)  \to \chi(x) + g v \alpha(x)$.
We can rewrite the Lagrangian as
\begin{equation}  \label{quadraticAchi}
\begin{split}
\mathcal{L}_{\rm Higgs} = &  -\frac{1}{4} F^{\mu\nu} F_{\mu\nu} + \frac{1}{2} \eta^{\mu\nu} \partial_\mu \rho \partial_\nu \rho \\
& +\frac{1}{2} \eta^{\mu\nu}  g^2(v^2 + 2v\rho +\rho^2) \left( A_\mu - \frac{\partial_\mu \chi}{gv} \right) \left( A_\nu - \frac{\partial_\nu \chi}{gv} \right)  \\
& - \frac{1}{2} \frac{\lambda v^2}{2} \rho^2 - \frac{\lambda}{4} v \rho^3 - \frac{\lambda}{16} \rho^4 \, .
\end{split}
\end{equation}
This is still manifestly gauge invariant.
The gauge boson has a mass
\begin{equation}
m = gv \, ,
\end{equation}
the radial field a mass $m_\rho = v \sqrt{\lambda / 2}$, while $\chi$, the massless goldstone, is the mode
that gets ``eaten'' to become the longitudinal part of the gauge boson.

Now, consider taking $\lambda \to \infty$. In this limit, the radial mode becomes infinitely heavy, it
cannot be excited and does not propagate. 
In \eqref{quadraticAchi} we can then drop the terms with $\rho$. We are left with
\begin{equation} \label{Stueck1}
\mathcal{L}_{\rm Stueck} =  -\frac{1}{4} F^{\mu\nu} F_{\mu\nu}  + \frac{1}{2} m^2 \eta^{\mu\nu} \left( A_\mu - \frac{1}{m} \partial_\mu \chi \right) \left( A_\nu - \frac{1}{m} \partial_\nu \chi \right) \, .
\end{equation}
This is the Stueckelberg Lagrangian~\cite{Stueckelberg:1938hvi, Stueckelberg:1938zz, Ruegg:2003ps}, invariant under the gauge transformations
\begin{align}
A_\mu(x) & \to A_\mu(x) + \partial_\mu \alpha(x)  \label{gauge1}\,, \\
\chi(x) & \to \chi(x) + m \alpha(x) \, . \label{gauge2}
\end{align}
The equations of motion derived from \eqref{Stueck1} are:
\begin{align}
 \partial^2 A^\beta - \partial^\beta \partial^\mu A_\mu + m^2 \left(A^\beta - \frac{1}{m} \partial^\beta \chi \right) & = 0 \, , \\
    \partial^2 \chi - m \partial^\mu A_\mu & = 0 \, .
\end{align} 
If we choose the unitary gauge, $\chi = 0$, from the second equation, we get $\partial^\mu A_\mu = 0$. Then the first equation becomes $(\partial^2 + m^2) A^\beta = 0$, which indeed correctly describes a massive spin-1 particle.

Defining
\begin{equation}
V_\mu \equiv A_\mu - \frac{1}{m} \partial_\mu \chi \, ,
\end{equation}
from the Stueckelberg Lagrangian, we get that of Proca's in \eqref{Proca}.
The extra scalar degree of freedom is not present anymore, and we do not have gauge invariance any longer. The equation of motion from the Proca Lagrangian is
\begin{equation}
\partial_\mu V^{\mu \nu} + m^2 V^\nu = 0 \, .
\end{equation}
Hitting it with $\partial_\nu$, we get 
$m^2 \partial_\nu V^\nu  = 0$.
This looks like the Lorenz gauge condition, but it is not (there is no gauge to fix). It is just a constraint, which we get due to the presence of the mass term, and which reduces the initial four degrees of freedom of
$V_\mu$ down to three. Thanks to it, the equation of motion becomes
\begin{equation}
(\partial^2 + m^2) V^\nu = 0 \, .
\end{equation}

\paragraph*{Higher-dimensional operators}
In principle, nothing prevents adding higher-dimensional operators to the Proca Lagrangian,
\begin{equation} \label{higherdimop}
\lambda_1 (V^\mu V_\mu)^2 + \kappa (V^\mu V_\mu)^3 + \cdots \, . 
\end{equation}
These could be problematic in scenarios where the vector is a light-dark matter candidate and has a high occupation number, which is a large expectation value\footnote{
Here we can think of $V^\mu$ as a quantum operator, and  $\langle V^\mu V_\mu \rangle \equiv \langle 0 | V^\mu V_\mu | 0 \rangle$.}
 $\langle V^\mu V_\mu \rangle$. In such scenarios, it is required to have a hierarchy in
 energy densities
 $\rho_m \gg \rho_{\lambda_1} \gg \rho_\kappa$, where $\rho_m \sim m^2  \langle V^\mu V_\mu \rangle$,
 $\rho_{\lambda_1} \sim \lambda_1  \langle (V^\mu V_\mu)^2 \rangle$,
  $\rho_\kappa \sim \kappa  \langle (V^\mu V_\mu)^3 \rangle$.
  The reason is that it would cost more energy to populate the dark vectors if the higher-dimensional operators
  were dominant. 
Before checking what this requirement leads to, let us elaborate on how one can think of the higher-dimensional operators.
It is useful to get back to the Higgs Lagrangian \eqref{quadraticAchi}. 
In the language of Feynman diagrams,
the term linear in $\rho$ in \eqref{quadraticAchi} gives a vertex with one $\rho$ and two $\left( A_\mu - \frac{1}{gv} \partial_\mu \chi \right)$. We can draw a diagram with two external gauge fields connected to a propagating $\rho$, which ends on another two external gauge fields. If $\rho$ is heavy, we can integrate it out and obtain the dimension 4 operator
\begin{equation}
\frac{g^4 v^2}{-m_\rho^2} \left( A_\mu - \frac{1}{gv} \partial_\mu \chi \right)^4 = - \frac{2g^4}{\lambda}  \left( A_\mu - \frac{1}{gv} \partial_\mu \chi \right)^4 \, .
\end{equation}
Comparing to \eqref{higherdimop}, we have
\begin{equation}
\lambda_1 \sim \frac{g^4 v^2}{m_\rho^2} \approx \frac{g^4}{\lambda} \, .
\end{equation}  
With the same reasoning, one finds
\begin{equation}
\kappa \sim \frac{g^6 v^2}{m_\rho^4} \approx \frac{g^6}{\lambda^2 v^2} \, .
\end{equation}  
Imposing $\rho_m \gg \rho_{\lambda_1} \gg \rho_\kappa$ amounts to requiring
\begin{equation} \label{gconstr}
g^2 \langle V^\mu V_\mu \rangle \ll \lambda v^2 \, .
\end{equation}
In the Abelian case, which is the focus here, there is no problem in taking a very large $\lambda$,
as we did above to get the Stueckelberg Lagrangian from the Higgs Lagrangian. In such a case 
the higher dimensional operators are very suppressed and \eqref{gconstr} is easily
satisfied.
If instead the vector boson resulted from the spontaneous symmetry breaking of a non-Abelian
gauge group, then the Higgs mass would be subject to the unitarity bound and could not be arbitrarily 
large. In that case, the requirement \eqref{gconstr} would put severe bounds~\cite{Agrawal:2018vin} on the gauge coupling $g$.

\paragraph*{Kinetic mixing}
The vector $V_\mu$ can mix with the visible photon. The relevant terms in the Lagrangian are
\begin{equation} \label{kinmix}
\mathcal{L} = - \frac{1}{4} F^{\mu\nu} F_{\mu\nu} - \frac{1}{4} V^{\mu\nu} V_{\mu\nu}  - \frac{1}{2} \epsilon F^{\mu\nu}V_{\mu\nu}  + \frac{1}{2} m^2 V^\mu V_\mu \, ,
\end{equation}
where $F_{\mu\nu} = \partial_\mu A_\nu - \partial_\nu A_\mu$ is the field strength of the visible photon.

In order to understand the effects of the kinetic mixing, one could do two different field redefinitions~\cite{Jaeckel:2012mjv}.
The first is $A_\mu \to A_\mu - \epsilon V_\mu$. This removes the kinetic mixing term and couples
the dark vector boson to the Standard Model fermions, which on top of their charge $Q$
under electromagnetism now acquire a millicharge $\epsilon Q$ under the dark $U(1)$ force.
There are several ways to probe this new force experimentally, see~\cite{Cline:2024qzv} for
a recent review.
Alternatively, one could shift $V_\mu \to V_\mu - \epsilon A_\mu$, which also gets rid of the
kinetic mixing term and introduces mass mixing between visible and dark photon. This leads
to oscillations between the two vector bosons. 
The physics does not depend on the field redefinition choice clearly, and one can simply choose the most convenient depending on the phenomenon under study.

%% file: WG1/content/spin-2.tex
From the point of view of QFT, General Relativity~(GR) is the leading order effective theory of a propagating massless spin-2 particle, the graviton. The theory is defined at lowest order by the Einstein-Hilbert action (we neglect the cosmological constant in the following)
\begin{equation}
    S_\text{EH}=\frac{\bar{M}_{P}^2}{2}\int d^4x\sqrt{-g}\,\mathcal{R}\,,
\end{equation}
where $g_{\mu\nu}$ is the metric, $g$ is its determinant, $\mathcal{R}$ is the Ricci scalar, $\bar{M}_{P}\equiv\frac{1}{\sqrt{8\pi G}}\approx 2.4\times 10^{18}$~GeV is the \textit{reduced} Planck-mass and $G$ is the Newton constant. The particle content of the theory can be obtained by expanding the metric around the Minkowski vacuum solution
\begin{align}
    \label{eq:kappa}&g_{\mu\nu}=\eta_{\mu\nu}+\kappa h_{\mu\nu}\,, &\kappa\equiv \frac{2}{\bar{M}_{P}}=\sqrt{32\pi G}\,,
\end{align}
where $h_{\mu\nu}$ is the graviton field. The Lagrangian at $\mathcal{O}(\kappa^0)$ reads
\begin{align}
\label{eq:graviton-kinetic}
    \mathcal{L}_0=-\frac{1}{2}\left(h_{\mu\nu,\alpha}h^{\mu\nu,\alpha}-h_\alpha h^\alpha+2h_{,\mu}h^{\mu\nu}_{,\nu}-2h^{\mu\nu}_{,\nu}h_{\mu\alpha}^{,\alpha}\right)\,,
\end{align}
where $h\equiv h^{\mu\nu}\eta_{\mu\nu}$ and $h_{,\alpha}\equiv \partial_\alpha h$.
Notice that the action does not show a mass term for the graviton, i.e. the GR-graviton is massless.
The unique combination of operators that can provide the graviton a mass without introducing instabilities in the theory was found long ago by Fierz and Pauli~(FP)~\cite{Fierz:1939ix}. The FP mass term reads
\begin{equation}
\label{eq:graviton-FP}
    \mathcal{L}_\text{FP}=-\frac{m_\text{FP}^2}{2}\left(h_{\mu\nu} h^{\mu\nu}-h^2\right)\,.
\end{equation}
Experimentally, stringent and complementary upper bounds on the graviton mass from gravitational-wave dispersion and solar system planetary trajectory measurements have been placed
\begin{align}
    m_\text{FP}\leq \begin{cases}
        1.76\times10^{-23}~\text{eV} &\text{at 90\% C.L.}~\cite{LIGOScientific:2020tif}\,,\\
        3.16\times10^{-23}~\text{eV} &\text{at 90\% C.L.}~\cite{Bernus:2020szc}\,.
    \end{cases}
\end{align}
The Planck mass further suppresses the graviton's self-interactions as they appear at a higher order in $\kappa$, and thus, they are irrelevant to (most) particle physics phenomenology.
The GR's diffeomorphism invariance of the action dictates the interactions of the graviton with matter. At leading order in $\kappa$, they read
\begin{equation}
\label{eq:graviton-interaction}
    \mathcal{L}_\text{int.}= -\frac{\kappa}{2}\,h_{\mu\nu}T^{\mu\nu}= -\frac{1}{\bar{M}_{Pl}}\,h_{\mu\nu}T^{\mu\nu}\,,
\end{equation}
where $T^{\mu\nu}$ is the energy-momentum tensor of the theory. It can be derived from the matter action $S$ via
\begin{equation}
    T_{\mu\nu}\equiv \left.-\frac{2}{\sqrt{-g}}\frac{\delta S}{\delta g^{\mu\nu}}\right|_{g_{\mu\nu}=\eta_{\mu\nu}}=\left.2\frac{\partial\mathcal{L}}{\partial g^{\mu\nu}}-g_{\mu\nu}\mathcal{L}\right|_{g_{\mu\nu}=\eta_{\mu\nu}}\,.
\end{equation}
For example, in QED with a flat background, it reads
\begin{align}
    T_{\mu\nu}^\text{QED}=\frac{\eta_{\mu\nu}}{4}F^{\alpha\beta}F_{\alpha\beta}-F_{\mu\alpha}F_{\nu}{}^\alpha+\frac{i}{4}\ov{\psi}\left[\gamma_{\mu}\overset{\leftrightarrow}{D}_\nu+\gamma_{\nu}\overset{\leftrightarrow}{D}_\mu\right]\psi\,.
\end{align}

From a bottom-up perspective, it comes naturally to wonder what would happen if we introduced a new weakly interacting spin-2 particle, $\varphi_{\mu\nu}$. This amounts to employing the quadratic action defined by Eq.~\eqref{eq:graviton-kinetic} and \eqref{eq:graviton-FP} with free mass parameter, $m_\varphi$.
From a top-down perspective, massive gravitons naturally arise in multimetric gravity~\cite{deRham:2010ik,deRham:2010kj,Hassan:2011zd,Hassan:2011vm,deRham:2013tfa,Lust:2021jps} (for a pedagogical introduction see e.g. Ref.~\cite{Schmidt-May:2015vnx}), in double copy of massive Yang-Mills theory in four dimensions~\cite{Momeni:2020vvr}, and in extra-dimensional theories as Kaluza-Klein (KK) modes of the extra-dimensional metric~\cite{Kaluza:1921tu,Klein:1926tv,Antoniadis:1990ew,Arkani-Hamed:1998jmv,Appelquist:2000nn,Randall:1999ee,Randall:1999vf}.
\bigskip

Massive gravitons are, phenomenologically, some of the most intriguing BSM particles.~\footnote{For a comprehensive review of theoretical aspects of massive gravity see e.g. Ref.~\cite{Hinterbichler:2011tt}.}
The interactions of $\varphi$ are inherited by the GR graviton of \eqref{eq:graviton-interaction} with new coupling strength $\kappa'$
\begin{equation}
\label{eq:gravNP-interaction}
    \mathcal{L}_\text{int.}=-\frac{\kappa'}{2}\,\varphi_{\mu\nu}T^{\mu\nu}\equiv-\frac{1}{\Lambda}\,\varphi_{\mu\nu}T^{\mu\nu}\,.
\end{equation}
where $\Lambda$ is the cut-off of the theory. Alternatively, both a dark Newton's constant $G'$ and dark reduced Planck mass $M'$ are used in the literature, in analogy with those defined for GR in Eq.~\eqref{eq:kappa}.

All couplings of the new graviton to the SM are simultaneously present and controlled by a single parameter, $\kappa'$. A deviation from the universality of $\kappa'$ is possible only at the price of more complex UV modelling. A notable example lies in extra-dimensional solutions where the SM particles can propagate in the bulk, yielding different wave-functions' profiles, and thus different interactions with the massive KK-gravitons (see e.g. Refs.~\cite{Fitzpatrick:2007qr,Gherghetta:2000qt}). 
The massive graviton propagator can be written as
    \begin{align}
        &i\Delta^{\mu\nu\alpha\beta}(k)=\frac{iP^{\mu\nu\alpha\beta}(k)}{k^2-m_\varphi^2}\,, &&P^{\mu\nu}=-\eta^{\mu\nu}+\frac{k^\mu k^\nu}{m_\varphi^2}\,,
    \end{align}
    where
    \begin{align}
        &P^{\mu\nu\alpha\beta}=\sum\limits_{s=0,\pm 1\pm 2} \epsilon_s^{\mu\nu}(k)\epsilon_s^{\alpha\beta}(k)=\frac{1}{2}\left(P^{\mu\alpha}P^{\nu\beta}+P^{\mu\beta}P^{\nu\alpha}-\frac{2}{3}P^{\mu\nu}P^{\alpha\beta}\right)\,.
    \end{align}
The limit $m_\varphi \to 0$ is not well defined and leads to a discontinuity in theory space. We will comment more on such an effect later on when talking about the energy validity of the theory.

Massive dark gravitons come with a very rich phenomenology due to the abundance of interactions and great predictivity. For example, massive spin-2 fields have been exploited in the literature in the context of DM, both as a DM candidate~\cite{Aoki:2014cla,Babichev:2016bxi,Aoki:2017cnz,Marzola:2017lbt,Chu:2017msm,GonzalezAlbornoz:2017gbh,Manita:2022tkl,Gonzalo:2022jac,Kolb:2023dzp} as well as a DM portal~\cite{Lee:2013bua,Lee:2014caa,Rueter:2017nbk,Bernal:2018qlk,Folgado:2019gie,Folgado:2019sgz,Bernal:2020fvw,deGiorgi:2021xvm,deGiorgi:2022yha,Chen:2023oip,Koutroulis:2024wjl,Chivukula:2024nzt}.
Experimentally, searches for massive gravitons can be performed at very different energies. For a review of bounds, see e.g. Ref.~\cite{Cembranos:2017vgi}. For masses below $m_\varphi \lesssim 10^{-5}$ ~eV, the main bounds come from the study of long-ranged effects such as modifications of the Earth's gravitational field and anomalous procession of the Moon and planetary orbits~\cite{DeRujula:1986ug,Talmadge:1988qz}. These bounds strongly bound $G'\ll G$ and peak at $m_\varphi\sim 10^{-15}$~eV where $G'\lesssim 10^{-10} G$. Moving up to $m_\varphi\lesssim$~eV, the constraints relax and allow for $G'>G$. The firsts to appear are fifth-force searches~\cite{Hoskins:1985tn,Bordag:2001qi,Mostepanenko:2001fx,Chiaverini:2002cb,Long:2003dx,Chen:2014oda,Tan:2016vwu,Lee:2020zjt}, which rely on short-distance modifications of the Newtonian potential in the form
\begin{equation}
    V_N=-G\frac{m_1 m_2}{r^2}\left(1+\alpha e^{-m_\varphi r}\right)\,,
\end{equation}
where $\alpha\sim G'/G$.
Astrophysical constraints~\cite{Hannestad:2003yd, Cembranos:2017vgi,Garcia-Cely:2025ula,Hardy:2025ajb} place the strongest limits, $G'\lesssim 10^{15}G$, for masses up to $m_\varphi\sim 100$~MeV. A more detailed discussion on astrophysical constraints on bimetric massive gravitons can be found in Sec.~\ref{subsec:Garcia_Cely_Ringwald}. For even larger masses, beam-dump experiments and colliders become most relevant~\cite{Giudice:1998ck,Agashe:2007zd,Kraml:2017atm,deGiorgi:2021xvm,Jodlowski:2023yne,dEnterria:2023npy,Im:2024kuw,Voronchikhin:2024ygo,CMS:2024nht} and allow for even greater $G'$.
\bigskip

Finally, a warning must be placed on the limits of applicability of the theory. Naively, the massive graviton EFT breaks down at energies comparable to the cutoff $\Lambda$. However, explicit computations of spin-2 scattering amplitudes reveal a rapid growth that signals the loss of perturbative control well below the fundamental scale of the theory.  This is due to the large energy growth of the polarization tensor of massive gravitons' longitudinal modes
\begin{align}
\epsilon_0^{\mu\nu}(m,E)\propto \frac{E^2}{m_\varphi^2}\,.
\end{align}
By studying the $2\to 2$ scattering of massive gravitons’ longitudinal modes, one may estimate in the high-energy regime the amplitude's leading order to naively grow as
\begin{align}
\label{eq:unitarity-amplitude}
\mathcal{M}\sim \frac{s^5}{ m_\varphi^8\Lambda^2}\,,
\end{align}
where $\sqrt{s}$ is the center-of-mass energy. This would imply a much smaller effective cutoff $\Lambda_\text{eff.}\sim (m_\varphi^4 \Lambda)^{1/5}\ll \Lambda$. Therefore, the limit $m_\varphi\to 0$ effectively pushes the theory in the strongly coupled regime, obstructing any predictive statement.
The phenomenon is in complete analogy with the SM unitarity problem associated with the scattering of longitudinal modes of EW gauge bosons that led to the quest for the Higgs. This problem has been extensively studied in the context of massive gravity \cite{Arkani-Hamed:2002bjr,Schwartz:2003vj,deRham:2010ik,deRham:2010kj}, yet finding consistent formulations that circumvent these issues remains an active area of research \cite{Bonifacio:2019mgk,Gabadadze:2019lld,Falkowski:2020mjq}.
While an effective theory of a massive graviton suffers from such an issue, UV theories may raise the value of $\Lambda_\text{eff.}$ closer to $\Lambda$ compared to the estimation of Eq.~\eqref{eq:unitarity-amplitude}.
For example, extra-dimensional gravity is expected to remain well-behaved up to the fundamental cutoff in analogy with GR, meaning that such unitarity issues should not emerge in the corresponding 4D effective theories. However, an individual KK-graviton cannot escape the general conclusions drawn for massive spin-2 fields and sketched in Eq.~\eqref{eq:unitarity-amplitude}. Restoring unitarity at high energies involves the inclusion in the computations of the full spectrum of particles generated upon integrating out the extra dimensions, especially the infinite tower of KK massive spin-2 states. The exact mechanism depends on the geometry of the extra dimensions~\cite{Schwartz:2003vj,Bonifacio:2019ioc,Bonifacio:2019mgk}. An explicit model-dependent verification of such a statement has been presented in the context of flat and warped extra-dimensional models~\cite{Chivukula:2020hvi,Chivukula:2023qrt,Chivukula:2024vgg}. It was found that, upon including the full particle content of the theory, a delicate series of cancellations considerably raise the effective scale to $\Lambda_\text{eff.}\sim (m_\varphi \Lambda^2)^{1/3}$.
Similar studies have been performed in the presence of matter fields, yielding analogous results~\cite{deGiorgi:2020qlg,Chivukula:2023sua,deGiorgi:2023mdy}.

Complementary and somewhat more stringent conclusions can be reached by studying positivity bounds of the EFT~\cite{Bellazzini:2023nqj,Dong:2025dpy}. Massive gravity violates positivity unless the effective cut-off and the graviton's mass lie relatively close to each other
\begin{equation}
    \Lambda_\text{eff.}\lesssim \mathcal{O}(10)\,m_\varphi\,.
\end{equation}
The above constraint applies to a theory which contains solely a massive graviton and can be lifted if the theory features multiple gravitons close in mass scale. Theories with KK-towers of spin-2 represent the optimal scenario discussed previously.

%% file: wg2.tex
\part{\texorpdfstring
{WISP Dark Matter \& Cosmology
\\ \textnormal{\small Editors: A.~Drew, S.~Gasparotto, M.~Gorghetto, M.~Kaltschmidt \& E.~Vitagliano}}{WISP Dark Matter and Cosmology (Editors: A.~Drew, S.~Gasparotto, M.~Gorghetto, M.~Kaltschmidt and E.~Vitagliano)}}
\label{part:wg2}
 
Axions, ALPs, dark photons and dark gravitons are  compelling candidates for dark matter and may also contribute to the dark energy, which together account for approximately 95\% of the Universe’s total energy. Indeed, these particles  can be produced abundantly as cosmological relics and  may naturally behave as an effective vacuum energy component. However, a detailed and quantitative understanding of their production mechanisms, their evolution in the early Universe and their effects during structure formation remains incomplete to this day. Progress on these questions is essential to put such WISP models on a firmer footing, confront their signatures with current and upcoming cosmological data, and potentially uncover new pathways for WISP discovery. This section aims to present recent advances in the cosmology of WISPs, outlining key theoretical and numerical developments, as well as observational constraints.
Understanding the properties of WISPs requires a combination of quantum field theory calculations, numerical simulations and theoretical modeling. Specifically, lattice QCD calculations are needed to refine predictions for the QCD axion, which arises as a possible solution to the strong $CP$ problem. On the other hand, large-scale simulations of string and domain wall networks aim to reduce uncertainties and narrow the viable range of axion masses, as well as determine the dark matter structure at subgalactic scales of these particles. 
Beyond axions, dark photons have emerged as an intriguing dark matter candidate, with potential imprints ranging from altered structure formation to CMB birefringence and gravitational-wave signatures. Meanwhile, spin-2 theories (i.e. dark gravitons) predict qualitatively new gravitational effects that can impact cosmic expansion and anisotropies. They also feature distinctive production mechanisms and detection prospects, including searches with pulsar timing arrays (PTAs) and (atom) interferometers. 
This section will highlight the key open problems in these areas and thereby lay the groundwork for future research directions. It will also clarify the interdisciplinary efforts needed to explore the cosmology of WISPs.
\clearpage
\section{QCD Axion Cosmology}
The first contributions in this part review key topics in QCD axion cosmology.
\input{WG2/content/misalignment_lattice} 
\input{WG2/content/strings_and_dws}
\input{WG2/content/miniclusters}
\input{WG2/content/hot_axions}

\section{WISP Cosmology (ALP, ULDM, Dark Photons and Dark Gravitons)}
The second part of this overview focuses on WISP cosmology beyond QCD axions, especially on axion-like particles (ALPs), ultra-light dark matter, dark photons and dark gravitons. 
\input{WG2/content/ul}
\input{WG2/content/dark_energy_new}
\input{WG2/content/dark_photons}
\input{WG2/content/spin2}

\cleardoublepage

%% file: WG2/content/misalignment_lattice.tex
\subsection[Axion Production via the Misalignment Mechanism: Role of Lattice QCD\\ \textnormal{Authors: C. Bonati \& M.P. Lombardo}]{QCD Axion Production via the Misalignment Mechanism: Role of Lattice QCD\\ \textnormal{Authors: C. Bonati \& M.P. Lombardo (with an introduction from M. Kaltschmidt)}}
\label{subsec:misalignment}

\subsubsection{Introduction}
\label{intro:misalignment}
\begin{figure}[t]
    \centering
    \includegraphics[width=0.85\linewidth]{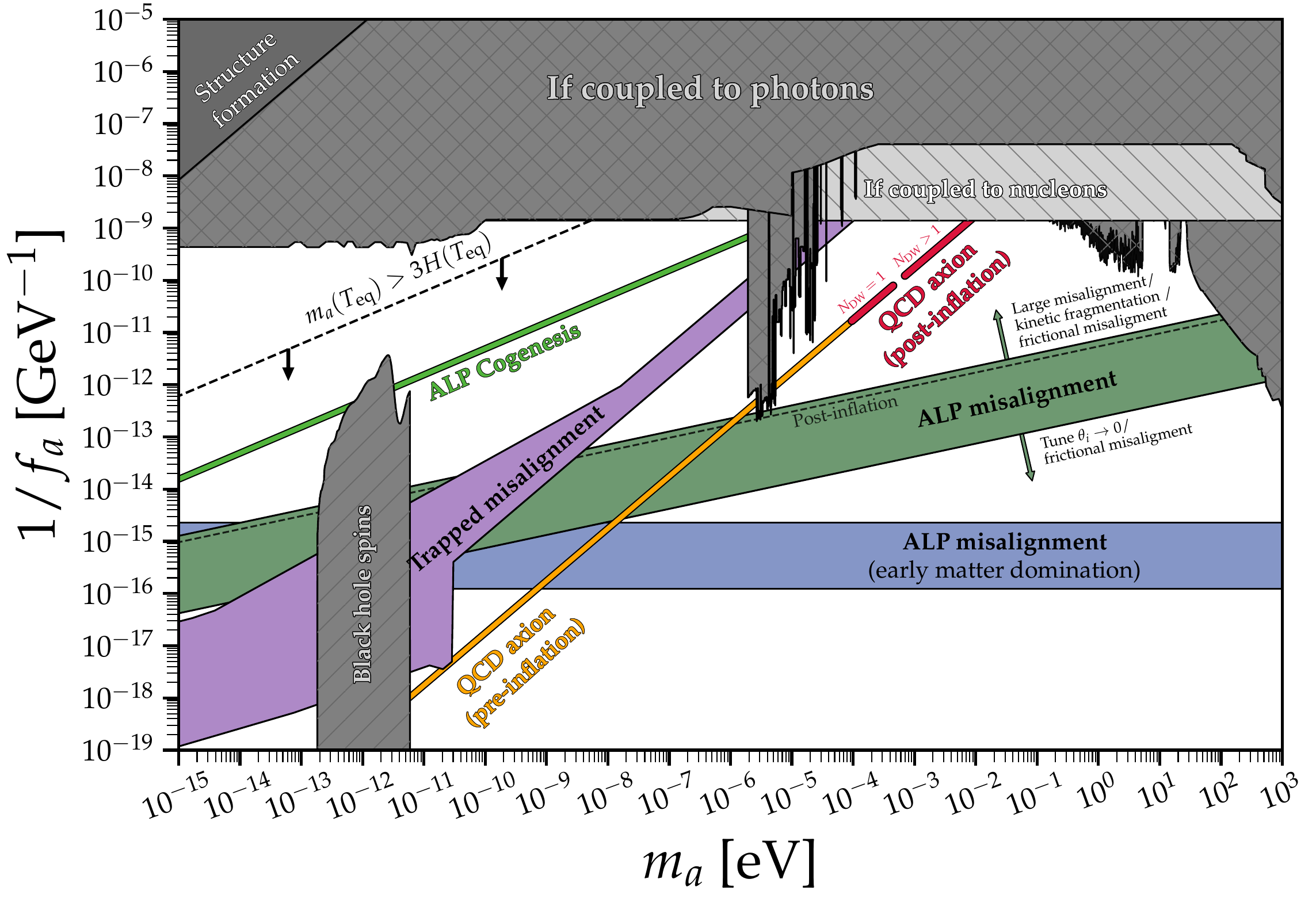}
    \caption{Relevant parameter space for the non-thermal production mechanisms for QCD axion dark matter (and extensions to ALPs) as discussed in Secs.~\ref{intro:misalignment} and~\ref{subsec:strings_and_dws}. The current bounds for axion models with specific couplings are highlighted in different shades of gray.  Figure reproduced from the AxionLimits repository~\cite{AxionLimits}.}
    \label{fig:misalignment_predictions}
\end{figure}

If the PQ symmetry is broken \emph{before} inflation, the dominant non-thermal cosmological production mechanism for the QCD axion is the \emph{misalignment mechanism}~\cite{Preskill:1982cy,Abbott:1982af,Dine:1982ah}. After the PQ symmetry breaking, the axion field can take an initial value $\theta_{\rm i}$ that is generically misaligned with respect to the minimum of its potential at $\theta=0$. While the Hubble expansion rate $H$ is large compared to the axion mass, the field is overdamped and remains frozen at its initial value. Once the expansion rate drops below the axion mass,
\begin{equation}
H(T_{\rm osc}) \simeq m_a(T_{\rm osc}),
\end{equation}
the field begins coherent oscillations around the minimum of its potential. These oscillations behave as a condensate of non-relativistic particles with an adiabatically conserved number density, 
\begin{equation}
n_a^{\rm mis}\left(T_0\right)=n_a^{\rm mis}\left(T_{\mathrm{osc}}\right)\left(\frac{R_{\mathrm{today}}}{R_{\mathrm{osc}}}\right)^{-3},
\end{equation}
with
\begin{equation}
    n_a^{\rm mis}\left(T_{\mathrm{osc}}\right)=\frac{\rho_a\left(T_{\mathrm{osc}}\right)}{m_a\left(T_{\mathrm{osc}}\right)} \overset{(\dot{\theta}_{\rm i}\approx 0)}{\approx} \frac{1}{2} m_a\left(T_{\mathrm{osc}}\right) f_a^2 \theta_{\rm i}^2,
\end{equation}
that redshifts as cold dark matter.\footnote{For a discussion of the case of ultra-light axions (ULAs), we refer to Sec.~\ref{subsec:ul}.} The axion mass $m_a$ is related to the decay constant $f_a$ via the QCD topological susceptibility $\chi(T)$, 
\begin{equation}
m_a^2(T) f_a^2 = \chi(T),
\end{equation}
which encodes the non-perturbative QCD dynamics governing the axion potential~\cite{Borsanyi:2016ksw}. At high temperatures, $\chi(T)$ decreases rapidly with temperature, delaying the onset of oscillations and modifying the relic abundance. For a complete discussion of the properties of $\chi(T)$, cf. the following Sec. \ref{sec:lattice_challenges}. Using the result from Ref.~\cite{GrillidiCortona:2015jxo} for the axion mass,
\begin{equation}
    m_a=(5.70 \pm 0.07)\ \mu \mathrm{eV}\left(\frac{10^{12} \mathrm{GeV}}{f_a}\right),
\end{equation}
the relic axion abundance from the misalignment mechanism is given by\footnote{For a complete pedagogical derivation we refer to Ref.~\cite{OHare:2024nmr}.}
\begin{equation}
   \Omega_a h^2= \frac{m_a n_a^{\mathrm{mis}}\left(T_0\right)}{\rho_{\text {crit }} / h^2} = 0.12\left(\frac{\theta_{\rm i}}{2.155}\right)^2\left(\frac{28\  \mu \mathrm{eV}}{m_a}\right)^{1.17}.
\end{equation}
For an initial misalignment angle $\theta_{\rm i} \sim \mathcal{O}(1)$, this points towards axion masses in the $\mathcal{O}(10)\ \mu\mathrm{eV}$ range as those yielding the observed dark matter abundance. This mass window is currently probed most sensitively by haloscope experiments, as illustrated in Fig.~\ref{fig:misalignment_predictions}.

Beyond this very general picture, non-standard misalignment scenarios can significantly modify axion cosmology and broaden the viable parameter space. Some of them have already been introduced in Secs.~\ref{sec:DarkALPs} and~\ref{subsec:models}, but we aim to give a compact overview of these models here, to conclude this introduction.

A quite general feature of different extensions of the misalignment mechanism is the \emph{delayed onset of oscillations}, which can arise for example from large initial misalignment angles, $\theta_{\rm i}\approx\pi$~\cite{Arvanitaki:2019rax, Co:2018mho, Huang:2020etx}. In these scenarios, the axion remains overdamped for longer, reducing dilution from cosmic expansion and therefore enhancing the relic abundance relative to the standard misalignment prediction~\cite{Co:2019jts}. 

Another important extension is the \emph{kinetic misalignment mechanism}\footnote{These models are sometimes referred to as \emph{spinning} or \emph{rotating} axion models.}~\cite{Co:2019jts, Chang:2019tvx, Eroncel:2024rpe}, where the axion field starts with a large initial velocity, $\dot{\theta}_{\rm i}\gg 0$. The field can traverse multiple periods of the axion potential before becoming \emph{trapped} once the kinetic energy becomes of the order of the potential energy. This leads again to a delayed onset of oscillations and an enhanced dark matter abundance.

Such dynamics can result in the formation of structures such as \emph{miniclusters}~\cite{Barman:2021rdr, Eroncel:2022efc, Chathirathas:2026}, cf. Sec~\ref{subsec:miniclusters} for an overview, explain \emph{baryogenesis}~\cite{Co:2019wyp}, induce a period of \emph{kination} in the early Universe~\cite{Gouttenoire:2021jhk}, or source \emph{gravitational-wave backgrounds}~\cite{Gouttenoire:2021wzu}. 

In this context, a rotating complex scalar condensate 
carries a conserved $U(1)~$ charge whose sound-wave–like fluctuations can be  excited by cosmic perturbations~\cite{Bodas:2025acm,Eroncel:2025qlk}. Once the underlying condensate is sufficiently diluted by expansion, these fluctuations behave as axion dark matter, potentially exceeding the abundance from both conventional and kinetic misalignment and even yielding warm dark matter signatures that 
could impact structure formation and experimental prospects~\cite{Bodas:2025acm}.
This new source of axion production has been denoted \enquote{acoustic misalignment} or \enquote{curvature-induced misalignment}. The key point is that the rotating scalar field sources scalar field fluctuations through the coupling to primordial curvature perturbations. These axion fluctuations  can contribute dominantly to the dark matter abundance in a wide region of parameter space, especially for the QCD axion, and produce complementary constraints and potential  gravitational-wave signatures tied to the expansion history and kination phases of the early Universe~\cite{Eroncel:2025qlk}.

As seen in Sec.~\ref{sec:DarkMatterRadiation}, QCD axion models with discrete $\mathbb{Z}_N$ symmetries and mirror-sector constructions, in which additional gauge sectors reduce the effective topological susceptibility and shift the QCD axion band to lower masses have been proposed in Refs. ~\cite{Hook:2018jle,DiLuzio:2021pxd, DiLuzio:2021gos}. These scenarios interpolate between standard QCD axions and generic axion-like particles, motivating experimental searches across a broad range of masses and couplings.

In summary, all these variants of the misalignment mechanism provide interesting frameworks for the non-thermal production of QCD axion dark matter, with important phenomenological implications for the early-Universe dynamics of the axion field. A somewhat complete overview of the relevant  parameter space for the axion models discussed in this overview can be found in Fig.~\ref{fig:misalignment_predictions}.\footnote{The post-inflationary production of QCD axions from the decay of topological defects is summarised  in a dedicated contribution, cf. Sec.~\ref{subsec:strings_and_dws}.} In the following sections, we focus on the ``standard'' misalignment mechanism and on the role of lattice simulations in the precise determination of the topological susceptibility, $\chi(T)$.

\subsubsection{Status and Challenges of Lattice
Simulations on Topological Susceptibility and Axion Potential}
\label{sec:lattice_challenges}
Already in the seminal papers ~\cite{Preskill:1982cy,Abbott:1982af, Dine:1982ah} 
it was realized that to compute the present day abundance of axions produced by the misalignment mechanism the main property of the strongly interacting matter that we need to know is the topological susceptibility, and more precisely its temperature dependence in the deconfined high-temperature regime of QCD.

The topological susceptibility $\chi(T)$ is the second derivative of the QCD free energy density $f(\theta, T)$  computed at
vanishing $\theta$-angle, at
finite temperature and $\theta$, but in general not much is known about the
specific form of $f(\theta, T)$, see e.~g.  Ref.~\cite{Vicari:2008jw}.  In the high temperature limit, the Dilute Instanton Gas Approximation (DIGA) is expected to become reliable, and it can be shown that~\cite{Gross:1980br,Schafer:1996wv}
\begin{equation}\label{eq:DIGA}
f_{\mathrm{DIGA}}(\theta,T)\simeq f(0,T)+\chi(T)(1-\cos\theta)\,.
\end{equation} 
Using perturbation theory in the one instanton background, on top of DIGA, it is
also possible to estimate $\chi(T)$ in the high temperature limit $T\gg
\Lambda_{QCD}$, obtaining an expression of the form
\begin{equation}\label{eq:chi_pert}
\chi(T)\simeq \frac{A}{T^{\beta_1-4}}\prod_{i=1}^{N_\ell} \frac{m_{i}}{T},
\end{equation}
where $\beta_1=11N_c/3-2N_{\ell}/3$, and the product extends over the $N_{\ell}$
light (with respect to the temperature) quarks. See Refs.~\cite{Gross:1980br,
Boccaletti:2020mxu} for more details on the numerical value of prefactor. 

The study of $\theta$-dependence in finite-$T$ gauge theories dates back to mid
90s (see, e.g., Ref.~\cite{Alles:1996nm}), however only in~Ref.~\cite{Berkowitz:2015aua} it was suggested to use lattice simulations to obtain first principles estimates of the present day axion abundance.  By that time, the $\theta$-dependence of finite-$T$ SU(3) Yang-Mills theory (sometimes also
called \textit{quenched} or \textit{pure glue} approximation of QCD) was quite well understood. For $T\gtrsim 1.1T_c$ the free energy density is well approximated by the functional form in Eq.~\eqref{eq:DIGA}, see
Refs.~\cite{Bonati:2013tt, Borsanyi:2015cka, Vig:2021oyt}, and the topological susceptibility $\chi(T)$ can be reliably computed using standard techniques up to temperatures of a few $T_c$, see Refs.~\cite{Berkowitz:2015aua, Borsanyi:2015cka, Kitano:2015fla}.  

This is not the case for real-world QCD: numerical simulations of QCD with physical, or close-to-physical, quark masses are much more computationally demanding than those of Yang-Mills theory, and even the numerical problems already encountered in Yang-Mills theory are more severe in this case. From the early results obtained by different groups, cf. Refs.~\cite{Trunin:2015yda,
Bonati:2015vqz, Petreczky:2016vrs, Borsanyi:2016ksw, Burger:2017xkz}, it thus became clear that a brute force approach is of limited use for QCD, and more refined methods are needed, see e.\,g. Ref.~\cite{Lombardo:2020bvn}, and section
VI of Ref.~\cite{Aarts:2023vsf} for recent reviews.

Let us now summarize the main problems encountered in studying the topological
susceptibility (or, more generally, the $\theta$ dependence) in the high
temperature regime of QCD with dynamical quarks, to understand the motivations
of some of the simulation choices that will be discussed in the following.

The first numerical problem we discuss is known since long time, and often goes under the name of ``topological freezing'' problem~\cite{Alles:1996vn, DelDebbio:2004xh, Schaefer:2010hu}. This is a purely algorithmic problem, unrelated to almost all the physical details of the system under investigation, nevertheless its origin is quite easy to understand. In numerical simulations,
configurations are sampled by Markov chain Monte Carlo algorithms, and new configurations are generated by applying small changes to old ones. In practice, the lattice field evolves in an almost continuous way, and this, at small lattice spacing, prevents the topological properties of the field to
change, thus breaking the ergodicity of the sampling algorithm, and making it unusable or very inefficient.

To make the freezing problem even worse, topological observables have typically very large lattice artifacts in QCD simulations with light quarks, and very fine values of the lattice spacing are (or would be) required to perform a reliable continuum extrapolation. This is not unexpected, due to the fact that topology and chirality are tightly connected by the Atiyah-Singer index theorem \cite{Atiyah:1968mp} and all
discretizations of the fermion fields commonly used explicitly break at the lattice scale the chiral symmetry of massless QCD. In particular at high
temperature the low-laying modes of the Dirac operator expected in the
background of an instanton can be significantly affected by the explicit lattice breaking of the chiral symmetry, and these low-lying modes are the ones which provide the largest contribution to $\chi(T)$~\cite{Borsanyi:2016ksw}. 

In trying to mitigate this problem, different groups have used different
definitions of the topological observables, which all agree in the continuum
limit but have different lattice artifacts, with the aim of identifying the one
which is less sensitive to finite lattice spacing effects. Definitions that
have been used include: the standard gluonic discretization of the topological
charge, which requires the use of smoothing algorithms to remove UV divergences, cf. Refs.~\cite{Alles:2000sc, Bonati:2014tqa, Alexandrou:2015yba, Alexandrou:2017hqw},
those based on the lattice index theorem, cf. Refs.~\cite{Neuberger:1997fp,
Hasenfratz:1998ri, Luscher:1998pqa}, the so called spectral definitions, cf.
Refs.~\cite{Giusti:2008vb, Luscher:2010ik, Bonanno:2019xhg}, and those based on the
relation between $\chi(T)$ and the disconnected chiral susceptibility, cf.
Refs.~\cite{Petreczky:2016vrs, Burger:2017xkz}, a relation which is
true only in the chiral symmetric limit, hence this definition becomes reliable
at high temperature, as it has been explicitly checked in
Ref.~\cite{Petreczky:2016vrs}.

Note also that, although simulations at quark masses larger then the physical
ones can provide useful insights into the physics of this problem, it is not
simple to use them to extrapolate results at the physical point. On the
contrary of what happens at zero temperature, in which chiral perturbation
theory is a solid guide to the extrapolation, the determination of the function
to be used to extrapolate to physical quark masses at finite temperature is
part of the problem to be solved. In particular, the temperature at which the
linear dependence on each light quark mass in Eq.~\eqref{eq:chi_pert}
becomes reliable is not known a priori.

The last problem encountered could be called the ``small volume'' problem.
Finite temperature simulations are usually performed at fixed aspect ratio,
i.e. at fixed $L/T$, where $L$ is the spatial extent of the lattice and $T$ the
temperature; because screening lengths typically scale with the temperature as $1/T$.  However, the typical amount of topological charges to be
found in a lattice of four-dimensional volume $\mathcal{V}=L^3/T$ is
$\sqrt{\chi(T) \mathcal{V}}$, and $\chi(T)$ rapidly approaches zero as $T$ increases (see Eq.~\eqref{eq:chi_pert}). This means that high-temperature
simulations almost always sample the zero topological charge sector. The net
effect is analogous to that of the freezing problem, but in this case the
source of the problem is not algorithmic but physical, and more precisely, the
smallness of the topological susceptibility. Configurations with nonzero
topological charge are thus rare events, and once again we have to worry about
the ergodicity and the efficiency of numerical algorithms.

\subsubsection{The State of the Art and possible Developments}

As already anticipated, the results of the first studies of QCD with physical,
or almost-physical, quark masses suggested that a direct approach, analogous to
the one used in Yang-Mills theory, can be hardly carried out in that case: all
the technical difficulties encountered indeed make it very difficult to perform a
reliable continuum extrapolation of the results.

In Ref.~\cite{Borsanyi:2016ksw} several technical improvements have been
introduced to avoid the numerical difficulties discussed in the previous
section. If the probability of observing nonvanishing values of the topological
charge is small, then it is simple to show that
\begin{equation}\label{eq:chi_small}
\chi(T)\simeq \frac{2Z_1(T)}{\mathcal{V}Z_0(T)}\,,
\end{equation}
where $\mathcal{V}=L^3/T$, and $Z_Q(T)$ is the partition function at
temperature $T$ restricted to the sector of topological charge $Q$. If
the lattice spacing is small enough, freezing prevents the topological charge to
change in numerical simulations, and it is thus possible to estimate the ratio
$Z_1(T)/Z_0(T)$ for $T>T_0$, given its value at $T=T_0$, by using an analogue of the thermodynamic integration method
commonly adopted in the computation of the equation of state, cf. Refs.~\cite{Borsanyi:2016ksw, Frison:2016vuc}. Some integrations performed by changing the quark masses were also used in
Ref.~\cite{Borsanyi:2016ksw}. This method however has two possible drawbacks:
on one hand the value of the topological susceptibility estimated by using
Eq.~\eqref{eq:chi_small} is the correct one, i.\,e. the thermodynamic value corresponding to
$L\to\infty$, only if instantons behave as a perfect
gas, which is just the hypothesis underlying the DIGA approximation. On the
other hand the sampling performed in such a way is clearly non ergodic, and
ergodicity is a fundamental ingredient to prove the stochastic exactness of the
computed average values. For this reason some
consistency checks have been performed in the simpler Yang-Mills theory in Ref.~\cite{Borsanyi:2016ksw}.

To reduce the lattice artifacts, in Ref.~\cite{Borsanyi:2016ksw} a Dirac
operator eigenvalue reweighting has also been implemented. This means that, for
all the gauge configurations generated, the low-lying eigenmodes of the Dirac
operator have been measured, and average values have been computed by rescaling
a posteriori the weight of a configuration with topological charge $Q$ by
\begin{equation}\label{eq:rew}
\prod_f \prod_{n=1}^{|Q|}\frac{m_f^2}{m_f^2+|\lambda_n|^2}\ ,
\end{equation}
where the first product extends over the quark flavours, $\lambda_n$ is the
$n$-th eigenvalue of the Dirac operator on the given configuration
($|\lambda_n|\le |\lambda_{n+1}|$ is assumed), and the anti-Hermiticity of the
Dirac operator has been taken into account.\footnote{Some subtleties related to
the staggered discretization has been neglected, see Ref.~\cite{Borsanyi:2016ksw}
for the precise definition.} The reason for such a reweighting is that the
Dirac operator should have, in the continuum and for a gauge configuration of
topological charge $Q$, at least $Q$ zero modes, but the explicit lattice
breaking of chiral symmetry prevents these would-be zero  modes from being
zero on the lattice. Using Eq.~\eqref{eq:rew} we are effectively forcing some
would-be zero modes to be zero modes in the weight of the configuration. This
procedure drastically reduces the size of the lattice artifacts, however also
changes the estimate of $\chi(T)$ by about two orders of magnitudes, so it is
important to investigate its possible systematics. In particular, the form
Eq.~\eqref{eq:rew} assumes that exactly $|Q|$ near-zero modes of the massless
Dirac operator exist, and that they have been correctly identified in the
spectrum, something which is nontrivial, especially in the presence of
instanton-antiinstanton pairs.

Using these techniques, a surprisingly similar behavior for $\chi(T)$ to the one expected from Eq.~\eqref{eq:chi_pert} was found in Ref.~\cite{Borsanyi:2016ksw},
which gives $\chi(T)\propto T^{-\alpha}$ with $\alpha=8\div 8.33$, already
starting practically from $T_c\approx 155\mathrm{MeV}$. The numerical prefactor, however, does not match the perturbative/semiclassical one in this case. The
corresponding bound for the axion mass obtained is $m_a\ge 28(2)\ \mu$eV, obviously assuming a post-inflationary breaking of the Peccei-Quinn symmetry.

Later studies tried to estimate $\chi(T)$ using computational strategies
different from the ones put forward in Ref.~\cite{Borsanyi:2016ksw}. Since
different techniques are likely to exhibit different systematic errors, a
concordance between the results would provide definitive proof of their
robustness. 

A strategy that has been proposed to cope with the small volume problem is a
variant of the so called multicanonical algorithm, cf.  Refs.~\cite{Bonati:2017woi,
Jahn:2018dke, Bonati:2018blm}, which is an algorithm introduced in statistical
mechanics studies of strong first order transitions. The idea is to explicitly
modify the distribution sampled by the Monte Carlo to enhance the occurrence of
rare events (transitions between the coexisting phases in its original
formulation, configurations with nonvanishing topological charge in the present
context).  Since the modification of the distribution to be sampled is known
analytically, it is possible, at the end of the simulation, to reweigth the
sample in order to estimate the expectation value with respect to the original
distribution.  The modification of the weights is thus just a way to make the
Monte Carlo sampling more efficient, which does not affect its stochastic
exactness.

The multicanonical algorithm has been used in Ref.~\cite{Athenodorou:2022aay}
to estimate $\chi(T)$ by using the spectral definition of the topological
charge. This definition uses the IR part of the Dirac spectrum (see
Refs.~\cite{Bonanno:2019xhg, Athenodorou:2022aay} for more details) and has a smaller
dependence on the lattice spacing then the gluonic definition. A possible
explanation of this fact is that the same explicit chiral symmetry breaking is
present both in the ensemble generation and in the observable, so there is a
partial cancellation. Since the dependence on the lattice spacing is quite
smooth, it is possible to perform a reliable continuum extrapolation, although
very fine lattice spacings can not be simulated due to the freezing problem.
Data obtained in this way display a DIGA-like behavior starting from a
temperature with lies between $T_c$ and $2T_c$, slightly larger than
the one observed in Ref.~\cite{Borsanyi:2016ksw}. The values of $\chi(T)$ in the
region between $2T_c$ and $3T_c$ are also systematically larger, by roughly an
order of magnitude, than the corresponding data reported in
Ref.~\cite{Borsanyi:2016ksw}.

The data discussed so far has been obtained using the so called staggered fermion discretization, which is the one most commonly used in QCD computations at finite temperature, since it is less computationally demanding then the other discretizations. Simulations adopting Wilson-like fermions, more precisely twisted mass fermions, have been presented in Refs.~\cite{Burger:2018fvb,
Kotov:2021ujj}. Although the continuum limit has been properly investigated only for quark masses larger than the physical one, also in this case a DIGA-like behavior emerges only for $T\gtrsim 2T_c$. When tentatively rescaled with the mass dependence predicted by Eq.~\eqref{eq:chi_pert}, the data falls
in the same ball-park of the results obtained using staggered fermions, see Ref.~\cite{Aarts:2023vsf}, suggesting that a consistent picture is starting to
emerge. 

This consistency is however challenged by the recent results presented in
Ref.~\cite{Chen:2022fid}, obtained by using domain-wall fermions, in which the
topological susceptibility is almost constant up to $\simeq 1.5T_c$, and slowly
decreases for larger temperatures. This behavior of $\chi(T)$ is somehow
analogous to the one previously found in Ref.~\cite{Bonati:2015vqz}, which was
however later ascribed to the presence of systematical errors in the continuum
extrapolation. As the Authors of Ref.~\cite{Chen:2022fid} correctly remark, the
control of the systematic errors is the main problem to be addressed in all
future determinations of $\chi(T)$.

From the lattice QCD side it is clear that new studies will appear, trying to
estimate $\chi(T)$ with complete or at least better control of the systematics,
and in this respect the development of algorithms to overcome the topological
freezing, like the one introduced in Ref.~\cite{Hasenbusch:2017unr} and applied to
QCD in Ref.~\cite{Bonanno:2024zyn}, will be most beneficial. Another important topic
to be investigated is the actual functional form of the free-energy density
(related to the axion potential). In Yang-Mills theory the functional
form in Eq.~\eqref{eq:DIGA} is known to be valid for $T\gtrsim 1.1T_c$, but in
QCD with physical quark masses we still do not know the precise range of
temperatures in which this functional form is reliable.

\subsubsection{Contemporary Challenges}

We  briefly summarize, and mention here some developments that may be significant in the context of axion physics. 

\begin{itemize}

\item 
We aim at a good control on the limits of the QCD axion mass in the overclosure limit: to this end, it is important to understand how errors on the topological susceptibility affects the limits on the  axion mass. A
preliminary study may be found in Ref.~\cite{Burger:2018fvb}, see in particular
Fig. 13, but more work is needed.

\item 
From the phenomenological axion side, it would  be very interesting to compare
the effect of the systematics associated to the different contributions to the
present day axion abundance, to understand the level of precision that lattice
simulations have to reach to really be useful for this purpose.

\item 
The axion potential: so far studies of the $\theta$-dependence in QCD mainly
relied on the computation of just the first nontrivial cumulant of the
topological charge, i.~e. the topological susceptibility. It is however also
possible to estimate higher cumulants of the topological charge, or use the
distribution of topological charge to reconstruct the Grand Canonical
Partition Function as a function of $\theta$ and temperature, hence of the
axion field.  In these ways we can have direct access to the axion potential, and
to higher order axion couplings, in the entire range of temperatures. It would
be very interesting to contrast and compare model predictions with lattice
results. Some results obtained with staggered fermions have been discussed in
Refs.~\cite{Bonati:2015vqz, Bonati:2018blm}, while results with twisted mass fermions
are presented in Refs.~\cite{Lombardo:2020bvn,Kotov:2025ilm}.

\item Thermal Axions: it has been pointed out that a thermal
production of relativistic axions in the early universe is unavoidable,
for recent reviews see e.~g. Ref.~\cite{DEramo:2023asj} and the corresponding Sec. \ref{subsec:hot_axions}. Thermal production rates of axions are then needed in a broad range of
temperatures and possibly momenta~\cite{Notari:2022ffe}. This brings in the
challenge of real time physics on the lattice, see e.~g. Ref.~\cite{Bonanno:2023thi}
for the determination of the sphaleron rate in QCD, and discussions with
phenomenologists are particularly welcome in order to shape these new studies.

\end{itemize}

%% file: WG2/content/strings_and_dws.tex
\subsection{Comparison of String Network and Domain Wall Simulations: Range of the Axion Mass\\ \textnormal{Authors: M. Buschmann, J.R. Correia, A. Drew, M. Kaltschmidt \& K. Saikawa}}
\label{subsec:strings_and_dws}
\subsubsection{Motivation}
A central unresolved issue of axion dark matter models that relies on theory input is the precise prediction of the QCD axion mass. Despite the known theoretical constraints and bounds from astrophysical and cosmological observations, the allowed mass range spans many orders of magnitude. 

A precise theoretical prediction of the number density $n_a(m_a)$, and therefore the relic abundance $\Omega_a(m_a)$ of axion dark matter, under the assumption that axions make up all dark matter, i.\,e. $\Omega_{\mathrm{DM}} \equiv \Omega_a(m_a)$, would allow us to solve for the unique mass that matches the dark matter abundance observed by \textsc{Planck},  $\Omega_{\mathrm{DM}}h^2 = 0.1198 \pm 0.0012$ \cite{Planck:2018vyg}, where $h= \frac{H_0}{100 \mathrm{km}/s/\mathrm{Mpc}}$ is the dimensionless Hubble constant. This would in turn enable direct detection experimental efforts to target the predicted mass, which could significantly accelerate a potential discovery.

Such a prediction for the axion dark matter mass $m_a$ is only possible in certain scenarios. In the case where the Peccei-Quinn (PQ) symmetry is broken \textit{after} inflation, 
the initial axion field values $a_i$
lack coherence beyond the causal horizon $H^{-1}$, with significant consequences for the phenomenology of the cosmological evolution of the axion field in the early universe.\footnote{We refer the reader to Refs.\,\cite{Marsh:2015xka,OHare:2024nmr} for further details.} This section outlines the physics of this post-inflation symmetry-breaking scenario, and details the current state of the art and future directions for analytic and numerical predictions of $m_a$.

\subsubsection{Axion Strings and Domain Walls}\label{subsubsec:axionstrings}
\begin{figure}[t]
    \begin{subfigure}[c]{0.49\textwidth}
    \centering
    \includegraphics[width=\textwidth]{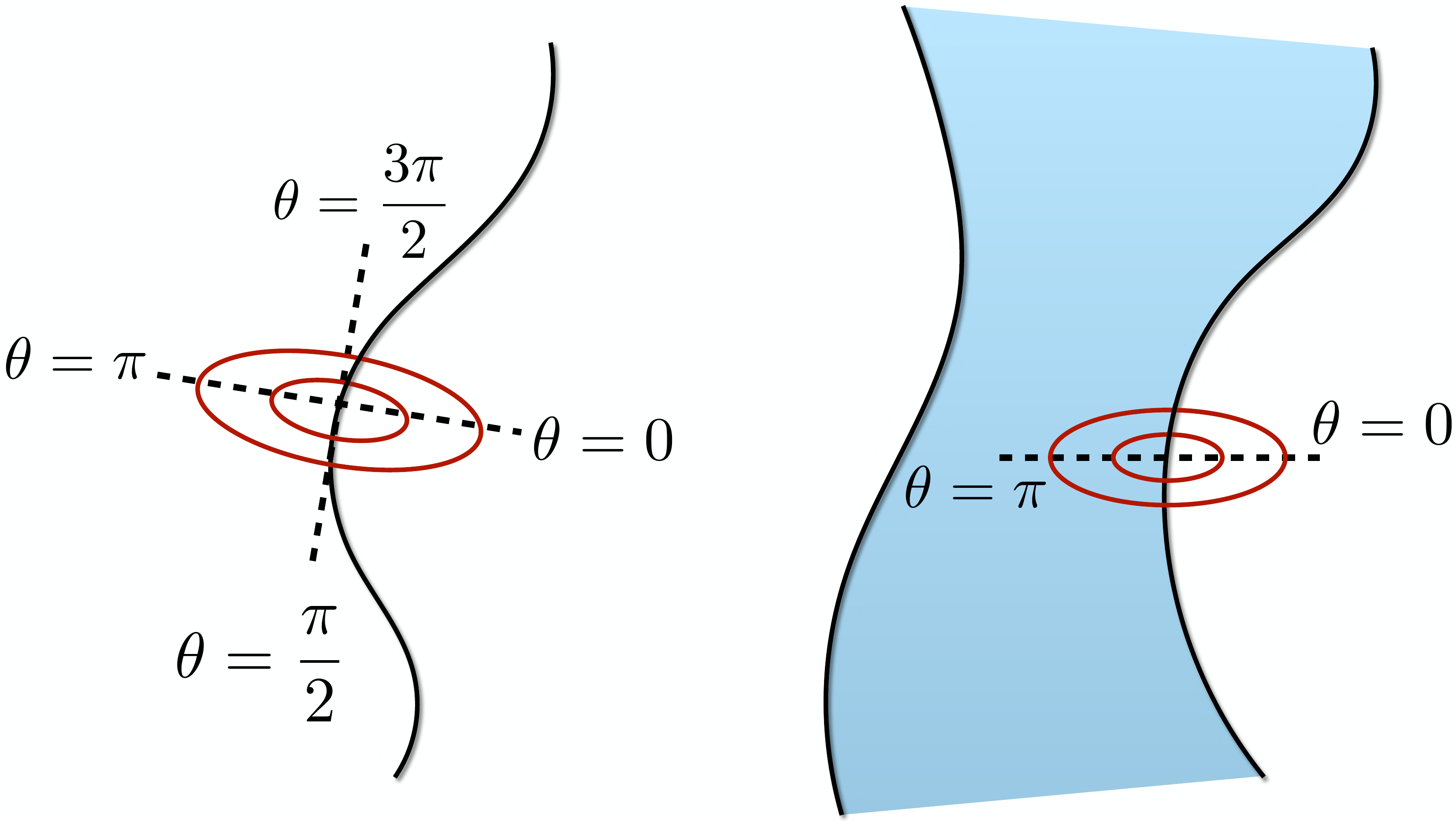}
    \end{subfigure}
    \hfill
     \begin{subfigure}[c]{0.49\textwidth}
        \centering
        \includegraphics[width=\textwidth]{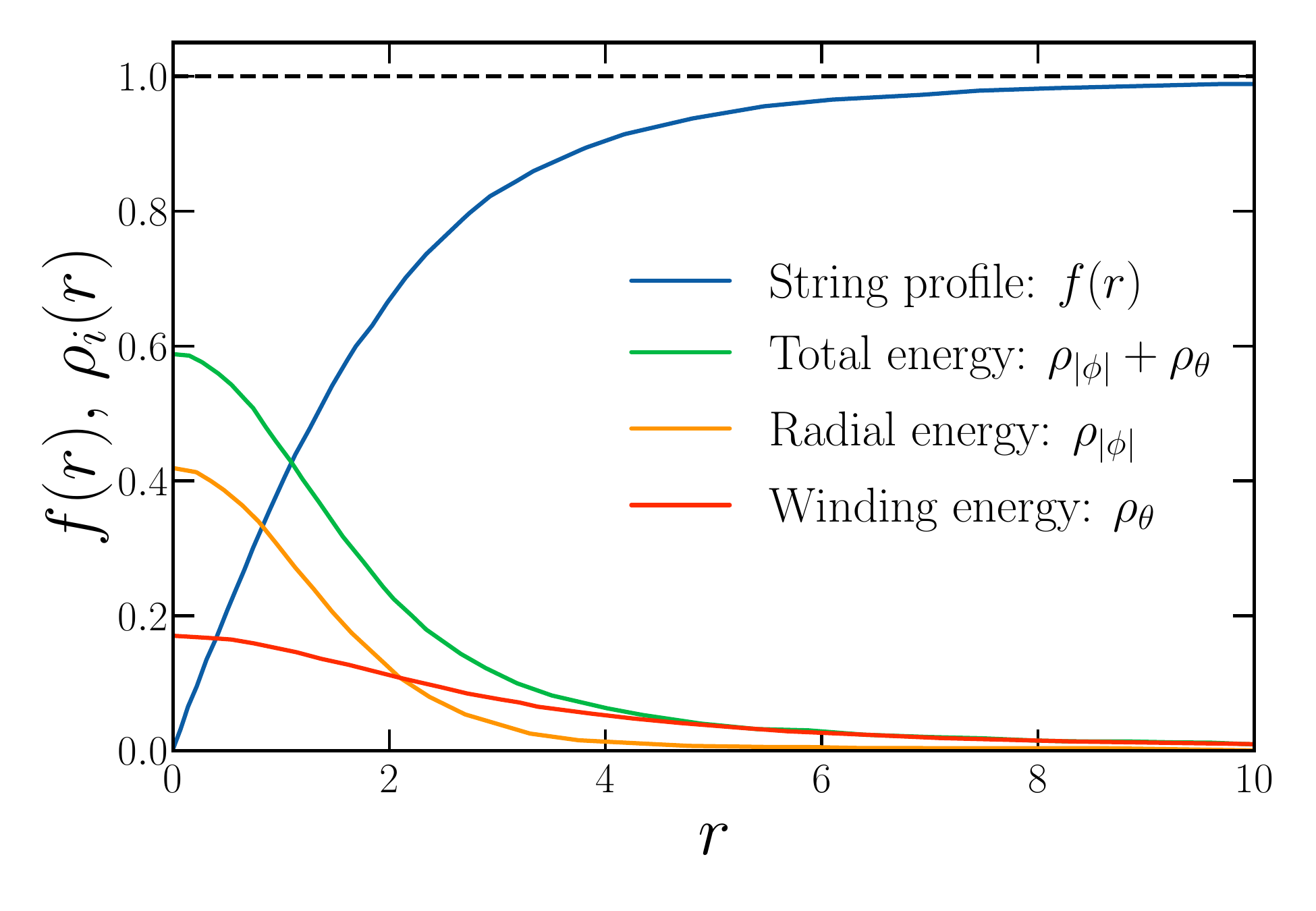}
    \end{subfigure}
    \caption{\textit{Left:} Visualisation of strings (black lines) and domain walls (blue region) as topological defects, relevant in the post-inflation scenario. \textit{Right:} String field profile and  components of the energy density as a function of radial distance from the string core.}
    \label{fig:visualisation}
\end{figure}

The spontaneous breaking of a global continuous symmetry inevitably leads to the formation of a network of global cosmic strings through the Kibble-Zurek mechanism \cite{Kibble:1976sj,Zurek:1985qw}.\footnote{
The general condition for the existence of topologically stable cosmic strings is that the underlying vacuum manifold $\mathcal{M}$ has a non-trivial fundamental group, $\pi_1(\mathcal{M}) \neq \mathbb{1}$. For the axion case, where the relevant PQ symmetry is $U(1)$, the vacuum manifold is a circle, $\mathcal{M}=S^1$. The different homotopy classes of loops are distinguished by their winding number $n$, i.e. how many times the field wraps around the circle of minima, so the fundamental group is simply the additive group of integers, $\pi_1\left(S^1\right)=\mathbb{Z}$. For a review of the more formal aspects of the classification of topological defects, cf. Ref. \cite{Kibble:1999yk}.} As the Universe cools down and the temperature falls below $ T \sim f_a \gtrsim 10^{10} \, \text{GeV}$, the axion field $a(x)$ takes on random values\footnote{There are different scenarios for the initial hierarchy, $H_I > f_a$ or $H_I < f_a$, imposing a model-dependent relation between the scale of inflation $H_I$ and the reheating temperature $T_{\rm reheating}$. In the second case, the correlation length $l_\mathrm{corr} \sim 1/T \ll H^{-1}$, for the case of thermal symmetry restoration ($T_{\rm reheating} > f_a$), but in the case of inflationary fluctuations ($H_I > f_a$) it would become $l_{\rm corr} \lesssim H^{-1}$. In both cases we expect an inhomogeneous field distribution, but the initial condition would be somewhat different.} 
within the fundamental domain \( (0, 2\pi f_a] \) in each Hubble patch.
In the post-inflation scenario, the axion field configuration remains highly inhomogeneous, with the field values `wrapping' around the $U(1)$ vacuum manifold, forming non-contractible loops in physical space, cf. the left panel of Fig. \ref{fig:visualisation} for a visualisation.

In general, global strings can be described by certain configurations of a complex scalar field $\phi = |\phi|e^{i\vartheta(x)}$. The strings correspond to vortex-like regions where the modulus $|\phi|$ approaches zero, due to the phase $\vartheta(x)$ winding through a full circle. Since the field values vary continuously, there must be a point within each $\vartheta$-loop where $\vartheta$ is undefined, necessitating $|\phi| = 0$. In field space, these regions correspond to local maxima of the potential $V_{\mathrm{PQ}}(\phi) = \frac{\lambda}{4}\left(|\phi|^2-v^2\right)^2$, indicating that the strings retain some residual potential energy from the unbroken phase $|\phi|=0$. The vacuum expectation value of the field is $v\equiv \langle\phi\rangle\simeq f_a$.

The Lagrangian of a complex scalar field with potential $V(\phi)$ reads
\begin{equation}
	\mathcal{L} = \frac{1}{2}\partial_{\mu}\phi^*\partial^{\mu}\phi - V(\phi), \label{eqn:Lagrangian}
\end{equation}
where we set $V(\phi)\approx V_{\mathrm{PQ}}(\phi)$. It is invariant under a $U(1)$ global symmetry of the form $\phi(x)\rightarrow e^{i \alpha}\phi(x)$. This symmetry is broken in the vacuum state such that $\vartheta(x) \rightarrow \vartheta(x) + \alpha$ under a $U(1)$ transformation. Anticipating that couplings to the axion field $a(x)$ are suppressed by the symmetry breaking scale $v$, we identify $\vartheta(x) = a(x)/v$ as the phase of the complex scalar field. Now, $a(x) = v \arg \phi(x)$ has a residual shift symmetry, $a(x) = a(x) + 2\pi v n$, where we have introduced the winding number $n$.

One can construct a simple ansatz for a static string along the $z$-direction in terms of polar coordinates $r$ and $\theta$,
\begin{equation}
	\phi(r,\theta) = vf(r)e^{in\theta}\,,\label{eq:phi_ansatz}
\end{equation} 
where $f(0)=0$ and $f(r\rightarrow\infty) = 1$ respectively. The string profile for the simplest case ($n=1$) is depicted in the right panel of Fig. \ref{fig:visualisation}.

Resolving the complex dynamics of individual, and especially a network of cosmic strings is highly intricate, so dedicated numerical simulations are necessary. These are computationally challenging due to the large hierarchy of scales in the system. In order to make this statement more apparent, one can compute the energy density $\rho$ stored in strings. For a general complex scalar field model with potential $V_\mathrm{PQ}(\phi)$, this is given by
\begin{equation}
    \rho(r)=\frac{1}{2}\left((\nabla|\phi|)^2+|\phi|^2(\nabla\vartheta)^2\right)+\frac{\lambda}{4}(|\phi|^2-v^2)^2. \label{eqn:rho_string}
\end{equation}
By substituting the string ansatz Eqn. (\ref{eq:phi_ansatz}) and integrating over the polar coordinates one obtains the \textit{energy per unit length} or \textit{tension} of the string, 
\begin{equation}
\begin{aligned}
   \mu &\equiv 2 \pi \int_0^{\infty} r \mathrm{~d} r\left(\frac{1}{2}\left(\partial_r f\right)^2+\frac{1}{2}\frac{f^2}{r^2}+ \frac{\lambda v^2}{4}(f^2-1)^2\right)\\
   &= \mu_0+\pi v^2\int_{c_{\mathrm{UV}}}^{c_\mathrm{IR}}\frac{\mathrm{d}r}{r}\approx\mu_0+\pi v^2\ln\left(\frac{c_\mathrm{IR}}{c_{\mathrm{UV}}}\right). \label{eq:string_tension}
   \end{aligned}
\end{equation}
Note that $\mu_0\approx 4.9v^2$ \cite{Drew:2019mzc} comes from the integration of the first and third term in Eqn. (\ref{eq:string_tension}) and the divergent contribution from the winding energy requires proper regularisation. In the cosmological scenario of the QCD axion, the UV cutoff $c_{\mathrm{UV}}$ is given by the width of the string core $c_{\mathrm{UV}}\approx m_r^{-1}$, where $m_r\sim \sqrt{\lambda}v$ is the radial (or \textit{saxion}) mass, and the IR cutoff by the inter-string distance $c_\mathrm{IR} \sim H^{-1}$, as strings are expected to enter a \textit{scaling} regime with $\mathcal{O}(\text{few})$ strings per Hubble patch \cite{Kibble:1976sj,Yamaguchi:1998iv}.

Furthermore, at around the QCD scale, when $T\sim \Lambda_{\mathrm{QCD}} \approx 1$ GeV, the potential $V(\phi)$ becomes dominated by the QCD potential $V(\phi)\approx V_{\mathrm{QCD}}(\phi)$. Although the computation of the exact low-energy form of the QCD potential is very complicated, cf. Sec. \ref{subsec:misalignment}, at high $T$ it can be derived using the dilute instanton gas approximation (DIGA) \cite{Borsanyi:2015cka}. This yields
\begin{equation}
V_{\mathrm{QCD}}(a)=\chi\left(1-\cos \left(N_{\mathrm{DW}} a / v\right)\right), 
\end{equation}
where $\chi$ is the \textit{topological susceptibility} and $N_{\mathrm{DW}}$ is the domain wall number, which is linked to the colour anomaly of QCD and depends on the concrete UV completion of the model under consideration, cf. Sec.~\ref{subsec:models}. The axion acquires a temperature-dependent mass,\footnote{We use $f_a = v/N_{\rm DW}$ for the remainder of this section. 
}
\begin{equation}
    m_a^2(T) = \frac{\chi(T=0)}{f_a^2}\left(\frac{T_{*}}{T}\right)^{n_{\mathrm{QCD}}} = m_{a,0}^2\left(\frac{T_{*}}{T}\right)^{n_{\mathrm{QCD}}},
\end{equation}
with the present-day value $m_{a,0}^2$, $\chi(T=0) \approx (76$ MeV$)^4$, the critical temperature of the QCD crossover $T_{*}\approx 150$ MeV and $n_{\mathrm{QCD}}\sim 7-8$ \cite{Borsanyi:2016ksw}. 
The PQ residual shift symmetry is explicitly broken, $a\rightarrow a + 2\pi f_a$, resulting in the formation of another class of topological defect, \textit{domain walls}, attached to the strings, cf. Fig. \ref{fig:visualisation}. In this string-wall system, one has to take account of three different scales $H \ll m_a \ll m_r$, where the domain wall thickness $\mathord\sim m_a^{-1}$.
This  makes simulations even more challenging than string-only simulations with 
two scales $H\ll m_r$.\footnote{Another caveat is that the post-inflation scenario in models with $N_{\mathrm{DW}}>1$ suffers from the so-called domain wall catastrophe. Here, the domain wall contribution to the energy density of the Universe becomes dominant, leading to overclosure \cite{Zeldovich:1974uw}. This can be seen by looking at the analogue of the string tension, which for the domain walls is a \textit{surface tension} $\sigma_{\rm DW}$,
\begin{equation}
    \rho_{\mathrm{DW}}(t)\sim\frac{\sigma_{\mathrm{DW}}(t)}{t}\propto\frac1t \quad \text{with}\quad \sigma_{\rm DW}(t)= 8m_a(t)f_a^2.
\end{equation}
The scaling $\sim 1/t$ shows that the energy density of domain walls $\rho_{\mathrm{DW}}$ must dominate that of the Universe, which is in conflict with observations. This assumes that the domain wall evolution also enters a scaling regime. Without that, we would have $\rho_{\rm DW} \propto \mathrm{area}/\mathrm{volume} \propto R^2/R^3 \propto 1/R$, where $R(t)$ is the scale factor of the Universe.
The problem can be avoided by introducing a bias term in the potential to make the overwise stable domain walls collapse at late times~\cite{Sikivie:1982qv}. This introduces yet another branch of axion cosmology to be explored.} 

To quantify the contribution to $\Omega_a$ from strings, we compare the axion number density from strings $n_a^{\mathrm{str}}$ with the number generated via the misalignment mechanism $n_a^{\mathrm{mis}}$ (cf. Sec. ~\ref{subsec:misalignment}). In units of the entropy density $s$, $n_a^{\mathrm{str}}$ is given by \cite{Saikawa:2024bta}
\begin{align}
Y_a \equiv \frac{n_a^{\mathrm{str}}}{s} = \frac{Kc_{\rm mis}H(T_{\rm osc})f_a^2}{\displaystyle \left[\frac{2\pi}{45}g_{*s}(T_{\rm osc})T_{\rm osc}^3\right]} = Kc_{\rm mis}\frac{\sqrt{45\pi g_{*\rho}(T_{\rm osc})}}{g_{*s}(T_{\rm osc})}\frac{f_a^2}{m_{\rm Pl}T_{\rm osc}}\,,
\label{eq:DMyield}
\end{align}
where we have used $H(T_{\rm osc}) = \sqrt{\frac{8\pi^3g_{*\rho}(T_{\rm osc})}{90}}\frac{T_{\rm osc}^2}{m_{\rm Pl}}$, $g_{*s}(T)$ and $g_{*\rho}(T)$ are the effective degrees of freedom for the entropy density and energy density, respectively, and $m_{\rm Pl}$ is the Planck mass. It is expected that $n_a$ is ``frozen" after the axion field starts to oscillate around the potential, $m_a(T_{\rm osc}) \approx H(T_{\rm osc})$.
The factor $c_{\rm mis} \approx 2.31$ is the numerical result of the standard calculation of the misalignment production~\cite{Borsanyi:2016ksw,GrillidiCortona:2015jxo}, and the prefactor $K \equiv n_{\mathrm{a,\,str}}/n_{\mathrm{a,\,mis}}\sim 30-200$ is the \textit{axion production efficiency}, a numerical factor that parameterises the enhancement (or suppression) of the axion abundance according to the details of the dynamics of strings and domain walls~\cite{Saikawa:2024bta}. For QCD axions, the decay constant can be related to the axion mass as $f_a \propto 1/m_a$, and the relic density scales as $\Omega_a \propto m_aY_a \propto T_{\rm osc}^{-1}m_a^{-1} \propto m_a^{-(n_{\mathrm{QCD}}+6)/(n_{\mathrm{QCD}}+4)}$, implying that the dark matter abundance increases for smaller $m_a$.\footnote{\label{fn:abundance_ALPcase}For axion-like particles having no definite relation between $f_a$ and $m_a$, we instead have $\Omega_a \propto T_{\rm osc}^{-1}m_af_a^2$, and the scaling of $\Omega_a$ with $m_a$ could be the other way around.} Since the contribution from the decay of strings becomes most relevant at around $T \sim T_{\rm osc}$, we expect that the parameter dependence given by Eq.~\eqref{eq:DMyield} holds in
this case. 
Most of the simulations discussed in this contribution are performed in the \enquote{simplest} setting with $N_{\mathrm{DW}}=1$. In this case, since the saddle point $a=\pi f_a$ is only quasi-stable, the string-wall network is expected to collapse shortly after the QCD phase transition with values of $H_{\mathrm{QCD}} \sim 10^{-19}$ GeV and $m_r \sim f_a \sim 10^{11} $ GeV, leading to a separation of scales $\log(m_r/H_{\mathrm{QCD}})\approx 70$ that is inaccessible to numerical simulations. In practice, one must therefore simulate networks at significantly lower tensions and carefully extrapolate the results to the \enquote{realistic} value. This is challenging, as axion emission from strings is expected to depend on $\log(m_r/H)$~\cite{Gorghetto:2018myk,Kawasaki:2018bzv,Buschmann:2019icd,Gorghetto:2020qws,Hindmarsh:2021vih,Buschmann:2021sdq,Saikawa:2024bta,Kim:2024wku,Correia:2024cpk,Kim:2024dtq,Benabou:2024msj,Correia:2025nns}. 

Improvement in simulations of axion domain walls, both for $N_{\rm DW}>1$ and $N_{\rm DW}=1$, will be important, given their relevance to the details of axion models (e.g. \cite{DiLuzio:2016sbl,Lu:2023ayc}), potential observational consequences (e.g. \cite{Ferrer:2018uiu,Gelmini:2022nim,Gelmini:2023ngs,Kitajima:2023cek, Gouttenoire:2023gbn, Ferreira:2024eru}) and their potential contribution to the axion mass prediction~\cite{Benabou:2024msj}. Simulations with $N_{\mathrm{DW}}>1$ have been performed, for example in Refs. \cite{Hiramatsu:2012sc,Kawasaki:2014sqa,Gorghetto:2022ikz}, and possible effects of the extrapolation to the large scale separation have been discussed recently in \cite{Gorghetto:2022ikz}. As additional axions are produced from collapsing domain walls, scenarios with $N_{\mathrm{DW}} > 1$ are likely to generate higher axion masses than the $N_{\mathrm{DW}}=1$ case. However, Ref.~\cite{Benabou:2024msj} provides preliminary evidence that domain wall contributions may be significant even for $N_{\mathrm{DW}}=1$, potentially increasing the axion abundance by a factor of $\sim5$ compared with a string-only network. Nevertheless, we will focus on the production of axions from the $N_{\mathrm{DW}}=1$, string-only network dynamics in the following sections.

We will discuss which parameters can be used to characterise the axion emission from topological defects, how one can model the complex dynamics (semi-)analytically, which systematic effects can affect the interpretation of the results, and how the dynamical range of the simulations can be extended.

\subsubsection{Axion String Network Modelling}
\label{subsubsec:AxionMassPrediction}
An important quantity for calculating the axion mass is the number of strings per Hubble patch $\xi$. Given that current simulations do not possess sufficient dynamic range to simulate all the way to the QCD phase transition, one must extrapolate this quantity from simulations of smaller scale separations. We review below the methods employed in the literature, which fall broadly into two categories: semi-analytic modelling using the Nambu-Goto action, and extrapolation of measurements of $\xi$ directly from simulations.

\paragraph{Semi-analytic Modelling.}\label{par:semianalytic}

The cosmological evolution of axion string networks has been modelled semi-analytically using extensions \cite{Yamaguchi:2005gp,Martins:2018dqg,Klaer:2019fxc} of the Velocity-dependant One-Scale (VOS) model for \textit{local} cosmic strings \cite{Martins:1996jp} (itself an extension of Kibble's one-scale model \cite{Kibble:1984hp}). Local strings are similar to the global strings presented in Sec. \ref{subsubsec:axionstrings}, but with the addition of a vector gauge field which removes the long-range, logarithmically divergent contribution to the energy density in \eqref{eq:string_tension}. This makes local strings significantly easier to model using the Nambu-Goto string action, under certain assumptions. The VOS model evolves two macroscopic properties of the network: the average string velocity $v_s$ and the average rest-frame string separation $L$.\footnote{$L^3$ is defined to be the volume that contains one string on average, hence $L$ is often referred to as a \enquote{correlation length}.} This approach treats the evolution of so-called \textit{long strings}, considering any loops smaller than the Hubble size $d_H$ to be separate from the network. In the radiation era, $L$ can be related to the number of long strings per Hubble patch $\xi$ via $(L/t)^2 = 1/\xi$ for cosmic time $t$.\footnote{By construction, a volume $L^3$ contains one long string of length $L$. Therefore, in a Hubble volume $d_H^3$, the total string length is $(d_H^3/L^3)L$. If we define the length of a long string crossing a Hubble patch to be $d_H$, the number of long strings per Hubble volume $\xi = (d_H^3/L^2) / d_H = d_H^2/L^2$, where $d_H \propto t$ in the radiation era.}

In an expanding FLRW background with scale factor $R(\tau)$ at conformal time $\tau$, 
the equations of motion of a Nambu-Goto string are given by \cite{Vilenkin:1991zk}
\begin{align}
    \mathbf{\ddot{x}} + 2\frac{\dot{R}}{R}(1-\mathbf{\dot{x}}^2)\mathbf{\dot{x}} &= \frac{1}{\mathbf{\epsilon}}\left(\frac{\mathbf{x'}}{\mathbf{\epsilon}}\right)'\,, \\ \label{NGepsilon}
    \mathbf{\dot{\epsilon}} + 2\frac{\dot{R}}{R}\mathbf{\dot{x}}^2\mathbf{\epsilon} &= 0\,.
\end{align} Here, $\mathbf{\epsilon} = \sqrt\frac{\mathbf{x}'^2}{1-\mathbf{\dot{x}}^2}$ is the \textit{coordinate energy per unit length}, $\mathbf{x}$ is the position in space of the string core and the dash and dot denote differentiation with respect to the spacelike and timelike string worldsheet coordinates, $\sigma$ and $\tau$ respectively.\footnote{See e.g. \cite{Vilenkin:2000jqa} for further details.} 
Considering the time evolution of the energy density of the network $\rho = \mu/L^2$ and following the approximations made in \cite{Kibble:1984hp,Martins:1996jp}, we obtain the evolution of the string separation 
\begin{align}
&2\frac{dL}{dt} = 2HL(1+v_s^2) + cv_s,
\end{align}
where the constant $c$ comes from assuming a scale-invariant distribution of loop sizes and $v_s^2 = \frac{\int\dot{\mathbf{x}}^2\mathbf{\epsilon}\mathrm{d}\sigma}{\int\mathbf{\epsilon}\mathrm{d}\sigma}$. The evolution of $v$ can be determined by considering the motion of a relativistic string segment. Differentiating our expression for $v_s^2$, we obtain
\begin{equation}
    \frac{dv_s}{dt} = \bigg( 1-v_s^2 \bigg) \bigg( \frac{k(v_s)}{L} -2Hv_s \bigg)\,
\end{equation}
under the assumption that the average curvature scale is approximately equal to $L$ \cite{Martins:1996jp}. This system evolves towards a fixed point attractor in phase space
\begin{align}\label{phasespace}
    \left(\frac{L}{t}\right)^2 \equiv \frac{1}{\xi} = k(k+c)\,, && v_s^2 = \frac{k}{k+c}\,
\end{align}
which is usually referred to as \textit{linear scaling}, because $L \propto t$. Numerical simulations are necessary to calibrate the model by measuring the parameters $k$ and $c$ (see e.g. \cite{Moore:2001px,Allen:1990tv,Bennett:1989yp}). It is expected that calibration for smaller simulations can be safely used to extrapolate to the late-time behaviour of the network. 

Despite its success for local strings, the VOS model has limitations. Based solely on the Hubble expansion of the network, the probability of (infinitely thin) strings intersecting and the acceleration of the strings, it does not include effects of radiation backreaction on the network or, for global strings, long-range inter-string forces. It also requires phenomenological parameters to be calibrated by simulations, rather than deriving them from first principles. Extensions to the model have been proposed for global strings which attempt to take some of these effects into account. These include the re-derivation of the model for time-dependent string tension \cite{Yamaguchi:2005gp,Martins:2018dqg,Revello:2024gwa}, adding a term for energy loss via axion emission backreaction $\propto v_s^6$ \cite{Martins:2000cs} and proposing analytic forms for $k$ \cite{Klaer:2019fxc}. These extensions modify the value of the fixed point \eqref{phasespace} by adding a small time-dependent correction. This class of models is still under development and it is unclear whether a complete analytic treatment will predict local-string-like linear scaling for global strings. Recent large-scale simulations of global string networks \cite{Correia:2024cpk,Hindmarsh:2021vih} have shown evidence of convergence of $\xi_r$ (measured in the string rest frame)\footnote{This is computed via the energy density of the network and should correspond to what the VOS predicts as $\xi$. Most authors measure the quantity $\xi_\mathrm{w}$ measured in the Universe frame, derived from summing the total string length in the simulation box (see Sec. \ref{par:simulations}). This differs from $\xi_r$ by a boost factor, ie. $\xi_r = \xi_w \gamma^{-1}$.} to a fixed point for dense initial conditions.

\paragraph{Simulation-based Modelling.}\label{par:simulations}

$\xi$ at the QCD phase transition can also be estimated by tracking its evolution through axion string network simulations and extrapolating the observed behaviour to later times. Most simulations measure $\xi=\ell t^2/\mathcal{V}$ at cosmic time $t$, where $\ell$ is the physical string length within the physical simulation volume $\mathcal{V}$. The majority of simulations (aside from $\xi_r$ measured in \cite{Hindmarsh:2021vih,Correia:2024cpk}) observe $\xi$ to grow approximately linearly with the natural logarithm of the core-to-horizon ratio, $\log(m_r / H)$, reaching $\xi \sim \mathcal{O}(1)$ at $\log(m_r / H) \sim 8 \;\mathrm{or}\; 9$ \cite{Gorghetto:2018myk,Kawasaki:2018bzv,Buschmann:2019icd,Gorghetto:2020qws,Buschmann:2021sdq,Saikawa:2024bta,Kim:2024wku,Kim:2024dtq,Benabou:2024msj}. Several different extrapolations based on this observed dependence on $\log(m_r / H)$ have been employed. 

In \cite{Gorghetto:2018myk}, the time-evolution of the network is postulated to be consistent with a constant velocity and linear growth of $\xi$ with $\log(m_r / H)$ at late times. In \cite{Gorghetto:2020qws}, the same authors fit polynomials of the form 
\begin{equation}
f_\xi(m_r / H) = c_0 + c_1\log(m_r / H) + c_{-1}\log^{-1}(m_r / H) + c_{-2}\log^{-2}(m_r /H)
\end{equation}
to multiple curves of $\xi$ (for different initial conditions), finding that $c_1 \approx 0.24$ provides a good fit and interpreting this to be the long-term attractor of the network. We will refer to this as \textit{log scaling} in order to distinguish it from the VOS-based modeling described in Sec. \ref{par:semianalytic}. Other groups \cite{Kawasaki:2018bzv,Buschmann:2019icd,Kim:2024wku} have used similar extrapolation methods. In the case of \cite{Buschmann:2021sdq}, physical motivation for the $\log(m_r / H)$ dependence is derived from simulations of singular loops decaying into axions (see \cite{Davis:1986xc,Hagmann:1990mj,Davis:1985pt,Vilenkin:1986ku} and references therein), although \cite{Saurabh:2020pqe,Baeza-Ballesteros:2023say} suggest contrasting decay rates. 

Another approach has been to extrapolate $\xi$ using a range of possible fits that take into account our incomplete understanding of the late-time network attractor. In \cite{Klaer:2019fxc}, a method of extrapolation is devised by assuming that the real value of $\xi$ trails and eventually plateaus to the evolution of so-called `conformal' string networks.\footnote{Here, the scalar coupling is modified to scale with conformal time $\tau$, i.e. $\lambda=m_r^2 \tau^2$. Given the implicit dependence of the comoving string width on $\lambda$, this implies that the string radius grows with time \cite{Bevis:2006mj}. The Lagrangian then has a conformal symmetry, which leads to an argument for the existence of scaling. For conformal strings, the string tension becomes fixed (via the ratio $m_r/H$ being fixed) and one can change the mass to inspect its effect on the fixed point. This is related to the so-called `PRS' (for Press-Ryden-Spergel \cite{Press:1989yh}) or `fat string' trick, where a time-dependent scalar coupling is introduced to force a constant comoving radius. This is often used in field theory simulations to prevent the string width from becoming smaller than the comoving lattice resolution. This can be done either by (pre-)evolving the network using core-growth or conformal evolution \cite{Bevis:2006mj,Klaer:2019fxc}, evolving only PRS strings, or using adaptive mesh refinement, which will be discussed in Sec. \ref{subsubsec:AMR}.} If the number of strings per Hubble length at a given ratio $m_r / H$ is given by $\xi_c(\log(m_r/H))$, the following tracking formula is proposed:
\begin{equation}
\frac{d\xi}{dt} = \frac{C}{t} \bigg[ \xi_c(\log(m_r/H)) - \xi(t) \bigg],\label{eq:model_conformal_scaling}
\end{equation}
with  $C$ being some $\mathcal{O}(1)$ constant. In \cite{Saikawa:2024bta}, this is used to describe the long-term behaviour of the network by upgrading the constant $C$ to a function $C(x)=x/(1+\sqrt{x}/c_0)$ with $x=\xi/\xi_c$ and $c_0\approx 1.5^{+0.8}_{-0.4}$, phenomenologically chosen to interpolate between two opposing limits: one where $C(\xi/\xi_c \rightarrow 0) \rightarrow \xi/\xi_c$ corresponding to a frozen network and one where $C(\xi/\xi_c \gg 1) \approx c_0 \sqrt{\xi/\xi_c}$. Two functional forms are tested and fit to $\xi_c$:
\begin{align}
\xi_{c,\mathrm{log}} = -0.19 + 0.205\log(m_r / H), && \text{and} && \xi_{c,\mathrm{sat}} = \frac{-0.25 + 0.23\log(m_r / H)}{1 + 0.02\log(m_r / H)}\,,
\end{align}
approximately capturing the two primary late-time behaviours proposed above, and used to extrapolate to the time of the QCD phase transition at $\log(m_r / H) \approx 70$.\footnote{For consistency, we use slightly different notation here than \cite{Saikawa:2024bta}; our $\xi_{c,\mathrm{log}}$ is $\xi_{c,\mathrm{lin}}$ in \cite{Saikawa:2024bta}.} The saturation approach is chosen to more closely mimic scaling than direct logarithmic extrapolation, but plateaus at a larger value of $\xi \sim 7$ than predicted by VOS modeling approaches. We will refer to these extrapolations as \textit{logarithmic conformal scaling} and \textit{saturated conformal scaling}. Further work is underway to test another extrapolation method detailed in \cite{Klaer:2017qhr,Klaer:2019fxc} which uses additional fields to imitate strings with a large tension.

\begin{table}[h]
\renewcommand{\arraystretch}{1.5}
\centering

\begin{minipage}{\textwidth}
\begin{center}
\scalebox{0.81}{
\begin{tabular}{l c c c}
\hline\hline
Reference & Type of scaling & Extrapolation of $\xi$ & $\xi_\mathrm{QCD}$ (extrapolated) \\
\hline 
Correia {\it et al}.~\cite{Correia:2024cpk} & Linear scaling      & $\xi \approx \xi_r \gamma^{-1} \approx \mathrm{const.}$ &  $\approx 1.220\pm0.057$  \\
Gorghetto {\it et al}.~\cite{Gorghetto:2020qws} & Log scaling         & $0.24\log(m_r/H)+0.02$                               &  $\approx 15 \pm \mathcal{O}(1)$        \\
Buschmann {\it et al}.~\cite{Buschmann:2021sdq} & Log scaling         & $0.25\log(m_r/H)-1.82$                               &  $\approx 15\pm3$        \\
Benabou {\it et al}.~\cite{Benabou:2024msj} & Log scaling         & $0.21\log(m_r/H)-0.8$                               &  $\approx 13\pm2$        \\
Saikawa {\it et al}.~\cite{Saikawa:2024bta}     & Conformal scaling Log & 
Eq.~\eqref{eq:model_conformal_scaling} with $\xi_c = \xi_{c,\mathrm{log}}$
&  $\approx 13.8\pm0.5$ \\
Saikawa {\it et al}.~\cite{Saikawa:2024bta}     & Conformal scaling Sat &     
Eq.~\eqref{eq:model_conformal_scaling} with $\xi_c = \xi_{c,\mathrm{sat}}$
&  $\approx 7 \pm 3$    \\
Kim {\it et al}.~\cite{Kim:2024wku}             & Log scaling         & $0.26\log(m_r/H)-1.15$                                &  $\approx 16.6$ (no error est.)       \\ \hline\hline
\end{tabular}
}
\caption{Summary of extrapolations used by different simulation groups to model the string density at the QCD phase transition, $\xi_\mathrm{QCD}$.} 
\label{tab:xi_extrap_comparisons}
\end{center}
\end{minipage}
\end{table}

A summary of recent extrapolations of the late-time strings-per-Hubble length parameter is given in Table~\ref{tab:xi_extrap_comparisons}. As has been demonstrated, string network evolution, and the degree to which a time-varying tension impacts it, is a delicate topic. Even small differences in the interpretation of simulation results, when extrapolating over many orders of magnitude, can lead to significantly different results for $\xi$ at late times. As computational resources, new techniques and different extrapolation approaches are tested, the efforts and views of all groups tackling this endeavor play a key role. 

\subsubsection{Measurement of the Axion Emission Spectrum}

The estimation of the relic dark matter abundance and determination of the axion mass depends in large part on the interpretation of numerical results on the spectrum of axions radiated by strings, in addition to the modeling of the string density $\xi$. The issue of understanding the axion emission spectrum has been a long-term controversy, lasting for three decades or more.
The recent generation of simulations continues to tackle this issue with more sophisticated methods and fresh perspectives.

In modern terminology, the axion radiation efficiency is described in terms of the instantaneous emission spectrum~\cite{Gorghetto:2018myk} defined by
\begin{align}
F\left(\frac{k}{RH},\frac{m_r}{H}\right) = \frac{H}{\Gamma_a}\frac{1}{R^3}\frac{\partial}{\partial t}\left(R^4\frac{\partial\rho_a}{\partial k}\right),
\label{eq:instantaneous_spectrum}
\end{align}
where $k$ is the comoving wavenumber of radiated axions, $R$ is the scale factor, $\rho_a$ is the energy density of axions, and $\Gamma_a$ is their net energy density emission rate. Its dependence on the comoving wavenumber $k$ is often parameterised in terms of a simple power law function $F \propto k^{-q}$. The index $q$ is a crucial indicator of whether the axion radiation is UV- or IR-dominated and its evolution is thus the most important factor for predicting the QCD axion mass. An IR-dominated spectrum would imply a large axion number density, where each axion carries only a small amount of energy. A UV-dominated spectrum would be the opposite - a small axion number density where each axion carries a large amount of energy. This ultimately ties the symmetry breaking scale $f_a$ and thus the axion mass $m_a$ to the observed axion relic density.
To lower the relic density, a lower breaking scale (larger axion mass) is required (see Eq.~\eqref{eq:DMyield}).
Several different groups~\cite{Gorghetto:2020qws,Buschmann:2021sdq,Saikawa:2024bta,Kim:2024wku,Benabou:2024msj} have tried to measure the value of $q$ by fitting the numerical results in a range of $\log(m_r/H)$ available in the simulations 
and extrapolating the outcomes to a realistic value of $\log(m_r/H)\sim 70$. 
The setups and results of recent simulations are summarised in Table~\ref{tab:sim_comparisons}.
The conclusion on the value of $q$ differs among different groups, which becomes the main source of the discrepancy in the axion dark matter mass prediction. Note also that the different interpretation of $q$ could lead to different parametric dependence of the correction factor $K$ in Eq.~\eqref{eq:DMyield}~\cite{Gorghetto:2020qws,Buschmann:2021sdq,Saikawa:2024bta},
\begin{align}
K \propto \left\{
\begin{array}{ll}
[\xi\log(m_r/H)]^{\frac{1}{2}+\frac{1}{n_{\mathrm{QCD}}+4}} & \text{for}\quad q \gg 1, \\[1.5mm]
\langle RH/k\rangle\xi\log(m_r/H) & \text{for} \quad q \lesssim 1,
\end{array}
\right.
\label{eq:K_parametric_dependence}
\end{align}
where $\langle RH/k\rangle$ is the mean inverse momentum (in units of $H$) of radiated axions.

\begin{table}
\renewcommand{\arraystretch}{1.5}
\centering

\begin{minipage}{\textwidth}
\scalebox{0.81}{
\begin{tabular}{l c c c c c}
\hline\hline
Reference & Method & Grid size\footnote{The size of the AMR simulation in Ref.~\cite{Buschmann:2021sdq} (\cite{Benabou:2024msj}) is given by $2048^3$ ($8193^3$) uniform grids with up to $l=5$ levels of refinement, which effectively amounts to $65536^3$ ($262144^3$) lattice sites.} & $\log(m_r/H)$ & $(m_ra)_{7.5}$ & $q$ (extrapolated) \\
\hline 
Gorghetto {\it et al}.~\cite{Gorghetto:2020qws} & Static lattice & $4500^3$ & $\lesssim 8.01$ & 0.776 & $ q \gg 1$ \\
Buschmann {\it et al}.~\cite{Buschmann:2021sdq} & AMR & $2048^3 \times (2^l)^3$ $(l\le 5)$ & $\lesssim 9.00$ & 0.185 & $q \approx 1$ \\
Benabou {\it et al}.~\cite{Benabou:2024msj} & AMR & $8192^3 \times (2^l)^3$ $(l\le 5)$ & $\lesssim 9.75$ & 0.154 & $q \approx 1$ \\
Saikawa {\it et al}.~\cite{Saikawa:2024bta} & Static lattice & $11264^3$ & $\lesssim 9.08$ & 0.567 & $q \approx 1$ or $q \gg 1$\\
Kim {\it et al}.~\cite{Kim:2024wku} & Static lattice & $4096^3$ & $\lesssim 7.86$ & 0.837 & $q \gg 1$ \\ 
Correia {\it et al}.~\cite{Correia:2025nns} & Static lattice & $12288^3$ & $\lesssim 8.75$ & 0.536 & $q \approx 1$ \\ \hline\hline
\end{tabular}
}
\end{minipage}
\caption{Setups of recent simulations of global string networks and 
their conclusions on the value of $q$ at a physically relevant value of $\log(m_r/H)$.
The largest value of $\log(m_r/H)$ reached at the final time of the simulation and the value of $m_r\frac{L_{\rm sim}}{N}\equiv m_ra$ at $\log(m_r/H)=7.5$ (denoted as $(m_ra)_{7.5}$) are also shown in the fourth and fifth column, respectively.}
\label{tab:sim_comparisons}
\end{table}

The fact that the conclusions on the value of $q$ are discrepant between different groups implies that the measurement of this quantity is a quite delicate issue. Indeed it has recently been found that there are several systematic effects that could potentially bias the value of $q$~\cite{Saikawa:2024bta}:
\begin{enumerate}
\item \emph{String density}: 
It has been reported based on the simulations with different initial string densities that there is some correlation between the values of the string density $\xi$ and the spectral index $q$~\cite{Saikawa:2024bta,Kim:2024wku}.
For simulations with over-dense string networks, there are more contributions from small scales which make the spectrum harder and give rise to smaller values of $q$,
while those with under-dense networks tend to give larger values of $q$.
Finding the genuine value of the string density on the attractor is crucial to avoid this kind of bias, which may be feasible by improving the modeling of the string network evolution.

\item \emph{Oscillations in the spectrum}:
The evolution of the system inevitably leads to the production of a \emph{coherently} oscillating axion field 
when the corresponding mode enters the horizon or the parametric resonance effect induced by the radial field becomes relevant.
The existence of these oscillations can lead to some contamination in the instantaneous emission spectrum and induce a systematic error if they are not handled properly in the analysis.

\item \emph{Discretisation effects}:
A poor resolution of the string core due to the finiteness of the lattice spacing can lead to a distortion of the axion emission spectrum.
The effect is conveniently parameterised by the quantity $m_ra$~\cite{Fleury:2015aca}, the product of the mass $m_r$ of the radial component of the PQ field (inverse of the string core radius), and the lattice spacing $\frac{L_{\rm sim}}{N}\equiv a$, where $L_{\rm sim}$ is the size of the simulation box.
The limitation of the computing resources prevents us from performing simulations in the vicinity of the continuum limit $m_ra \to 0$, and
typically a larger value of $m_ra$ leads to an overestimate of the value of $q$.
\end{enumerate}

Among the possible systematic effects described above, the discretisation error seems to be the most serious issue.
Most of the simulations executed on the static lattice end up with $m_ra \approx 1$ at larger values of $\log(m_r/H)$ (see Table~\ref{tab:sim_comparisons}),
which implies that the bias due to the discretisation effects is not completely suppressed.
As a consequence, the behavior of $q$ at larger values of $\log(m_r/H)$ remains uncertain.
Intriguingly, recent simulations are getting closer to a scale-invariant value $q \sim 1$ within the simulated range of $\log(m_r/H)$, 
which motivates us to have two different possibilities:
one is to expect that $q$ continues to increase with $\log(m_r/H)$
such that $q \gg 1$ at a physical relevant value of $\log(m_r/H)$~\cite{Gorghetto:2020qws,Kim:2024wku}, 
and the other is the scenario where $q$ remains constant around $q\approx 1$ ~\cite{Buschmann:2021sdq,Benabou:2024msj,Correia:2025nns}.\footnote{Reference~\cite{Saikawa:2024bta} concluded that both interpretations are possible within the range of uncertainty.}
Axion spectra from some individual string configurations have been investigated in detail \cite{Davis:1986xc, Davis:1985pt, Vilenkin:1986ku, Hagmann:1990mj, Battye:1993jv, Saurabh:2020pqe, Baeza-Ballesteros:2023say, Benabou:2023ghl, Dine:2020pds, Drew:2025txu, Kaltschmidt:2026}, but the implications for a cosmological network are unclear. More accurate measurements from networks at larger values of $\log(m_r/H)$ to distinguish these scenarios are therefore highly warranted.

\subsubsection{Future Directions and Computational Advances}\label{subsubsec:AMR}
As we have seen, in order to measure $q$ and $\xi$ accurately, it is crucial to simulate axion strings for as long as possible to improve the extrapolation to very large scale separations. This, however, is computationally intensive. Consider, for example, two simulations with a final scale separation of $\log(m_r/H)$ and $\log(m_r/H)+1$. Naïvely, one could think this is simply a matter of running one of the simulations for longer, yielding a computational resource increase factor of $\sqrt{e}\sim 1.65$. However, in reality, this factor is much closer to $e^{\frac{1}{2}(1+3+3+1)}\sim 55$ for a standard static lattice simulation. The factors in the exponent have the following origin: one power comes from the increase in total simulation time, three powers come from the fact that we need a larger simulation volume to have the same number of Hubble patches in the final state, another three powers are due to the smaller string width in the final state, requiring us to use a higher spatial resolution $\Delta x$, and the last power is indirectly a result of the higher spatial resolution, since smaller $\Delta x$ usually requires a smaller time step size as well. Since simulating from  $\log(m_r/H)$ to $\log(m_r/H)+1$ is increasing the computational cost by about a factor of $\sim55$, static lattice simulations become quickly unfeasible.

To alleviate this problem, a technique called {\sl adaptive mesh refinement} (AMR) is sometimes employed. This takes advantage of the fact that high spatial resolution is primarily needed only within the vicinity of axion strings. In AMR simulations, it is possible to focus computational resources on only certain parts of the simulation volume by imposing high spatial and temporal resolution, whereas the resolution in the space in between strings can be dropped. The method is {\sl adaptive}, as these refined regions can be adjusted during the simulation to track axion strings and their radiation. This increases the complexity of the simulation massively but reduces the computational cost, allowing us to access larger scale separations than on a static lattice. This has been employed in some recent studies of axion string networks \cite{Buschmann:2019icd,Buschmann:2021sdq,Benabou:2023ghl,Benabou:2024msj} which make use of the GPU-enabled code \texttt{sledgehamr} \cite{Buschmann:2024bfj}. The most recent work from the same authors reaches an effective number of gridpoints of $262,144^3$ on the finest grid, allowing for simulations of up to $\log{(m_r/H)} \approx 10$. AMR simulations of individual axion strings \cite{Drew:2023ptp,Drew:2022iqz,Drew:2019mzc} have also been performed using an adapted version of \texttt{GRChombo} \cite{Clough:2015sqa,Andrade:2021rbd}.

In addition to the memory-saving benefits of AMR, another key computational development has come in recent years from significant growth in the efficiency of GPUs, driven in particular by artificial intelligence and machine learning. GPUs provide the ability to parallelise codes over several thousand \enquote{threads} in comparison to CPUs, which make more use of OpenMP or MPI parallelism. This has the potential to significantly speed up network simulations, particularly for fixed grids. Recent works have utilised this capability of GPUs to run simulations on a fixed grid with $11,264^3$ gridpoints using the code \texttt{jaxions} \cite{Vaquero:2018tib,jaxions, Saikawa:2024bta} and even  $16,384^3$ gridpoints using the code \texttt{AxHILA} \cite{HILA,Correia:2024cpk,Correia:2025nns}, facilitating runs up to $\log{(m_r/H)} \approx 9$.

\subsubsection{Current Status of Axion Dark Matter Mass Predictions}

\begin{figure}[t]
    \centering
    \includegraphics[width=\textwidth, trim=90 0 0 800, clip]{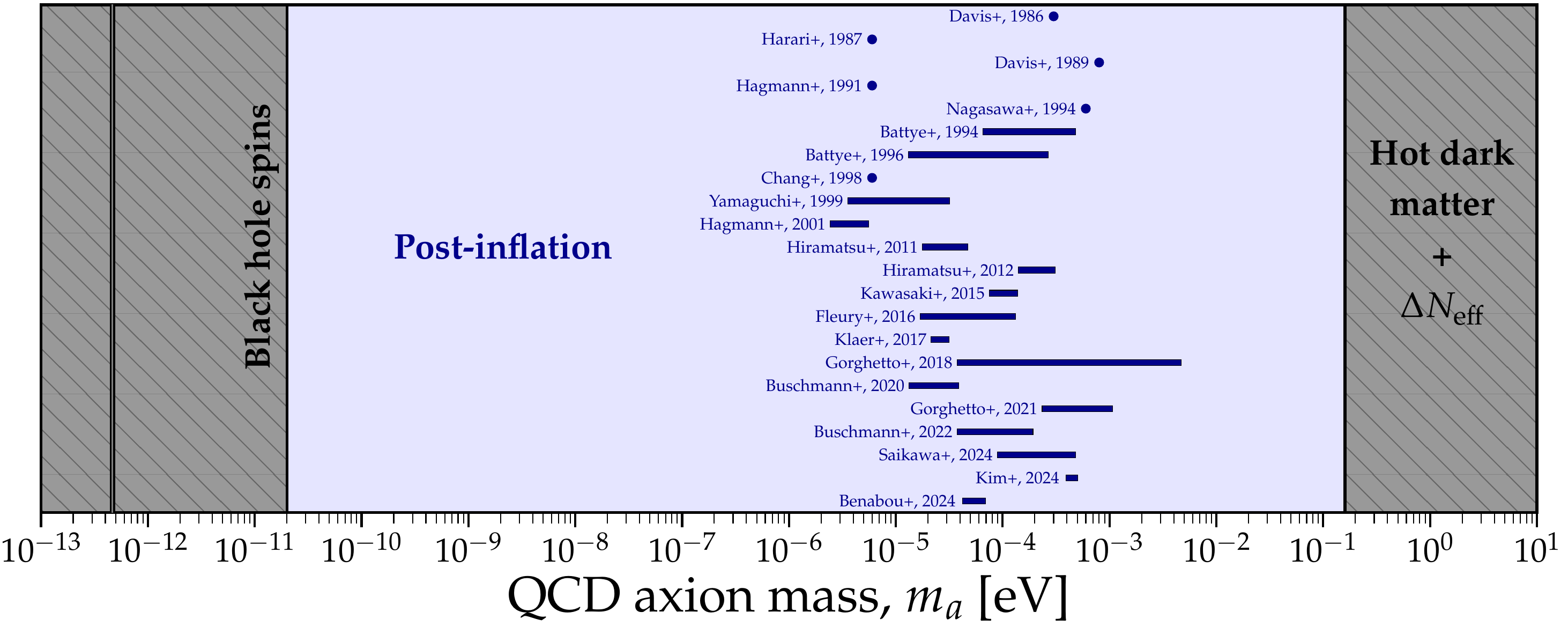}
    \caption{Overview of predictions for the axion dark matter mass from string simulations in the post-inflationary scenario. The figure was adapted from Ref. \cite{Kaltschmidt:2025nkz}. The original template is courtesy of Ciaran O'Hare~\cite{AxionLimits}. 
    }
    \label{fig:mass_predictions}
\end{figure}

To conclude this discussion, we will summarise the current state-of-the-art of axion dark matter mass predictions from string simulations in the post-inflationary scenario. A graphical overview is presented in Fig. \ref{fig:mass_predictions}. 

For the most recent simulations, depending on the concrete interpretation of the value of $q$, the mass predictions range from $m_a \simeq 65 \pm 25\,\mu\mathrm{eV}$ with $q\simeq 1$ \cite{Benabou:2024msj} to $m_a \gtrsim (250-1000)\,\mu\mathrm{eV}$ with $q\gg 1$~\cite{Gorghetto:2020qws,Kim:2024wku}.\footnote{The actual upper bound is related to the assumption that there is \textit{no} contribution from the decay of domain walls. As Ref.~\cite{Benabou:2024msj} points out, however, domain wall contributions may be significant, potentially increasing the mass range prediction to as much as $300\,\mu\mathrm{eV}$ for the $q\simeq 1$ case.} Additionally, the extensive study of systematic effects on the determination of the spectrum in Ref. \cite{Saikawa:2024bta} points towards the region in between the previous works, explicitly $95\,\mu$ev $\leq m_a \leq 450\,\mu$eV. 
Another independent mass prediction comes from simulations of the effective high-tension model of Klaer and Moore with additional degrees of freedom \cite{Klaer:2017qhr}, which lead to $m_a \simeq 26\,\mu\mathrm{eV}$ and contribute valuable insight into the dynamics of string networks at high tension.

The ultimate goal of predicting the axion dark matter mass is of central importance for the rich experimental search program throughout the whole mass range and can ultimately help to  drastically speed up a potential discovery, for a complete overview of the experimental program, we refer to Part~\ref{part:wg4}.\\
At the lower end of the mass range predicted by string simulations in the post-inflationary scenario, there are (and will be) many different haloscope experiments in operation,  for example FLASH \cite{Alesini:2023qed}, (Baby)IAXO-RADES \cite{Armengaud:2014gea,Melcon:2018dba}, ADMX \cite{ADMX:2020ote} and QUAX \cite{QUAX:2020adt}. The higher masses are experimentally more challenging to probe, but there are also some experiments planned that will focus on this mass range, e.\,g.  ALPHA \cite{ALPHA:2022rxj}, MADMAX \cite{Caldwell:2016dcw}, ORGAN \cite{McAllister:2017lkb} or CADEx \cite{Aja:2022csb}.

To summarise, there has been a lot of progress in refining the axion dark matter mass prediction from string simulations in the post-inflationary scenario in recent years. Computational advances and new approaches in the (semi-)analytic modeling of the network evolution will allow us to tackle the relevant problems and open questions discussed in this contribution. This will ultimately help us to guide experimental searches towards the favored mass range of post-inflationary axions and also  improve our understanding of the subsequent cosmological history, including the formation of axion dark matter substructure, such as axion minclusters and axion stars. These are discussed in the following contribution, cf. Sec. \ref{subsec:miniclusters}.

%% file: WG2/content/miniclusters.tex
\subsection{Axion Dark Matter Substructure: Miniclusters and Axion Stars\\ \textnormal{Authors: C.A.J. O'Hare, G. Pierobon \& L. Visinelli}}
\label{subsec:miniclusters}

\subsubsection{Introduction}
For both indirect and direct searches for axion dark matter, it is impossible to make predictions of observational signatures without some assumption about the distribution of dark matter inside halos. The standard assumption adopted in the field is to describe galactic halos as smooth and fully-virialised, modelled by a standard set of density profiles and velocity distributions. However, it has been known since some very early works discussing the production of axion dark matter in the early Universe that this common assumption may not be appropriate for all axion cosmologies. Particularly in the so-called post-inflationary scenario for QCD axions, $\mathcal{O}(1)$ inhomogeneities in the density inherited from dynamics occurring at the QCD scale lead to the formation of miniclusters on extremely small sub-galactic scales~\cite{Hogan:1988mp, Kolb:1994fi, Kolb:1995bu}. Additionally, the field's gradient energy, as well as any self-interactions, can both interplay with the effects of gravity to seed long-lived field configurations on even smaller scales called axion stars~\cite{Tkachev:1991ka}. These structures are anticipated to survive, at least to some extent, in galaxies today, leading to a slew of additional indirect probes of the existence of axion dark matter, as well as potentially major modifications to the expected signals for haloscopes searching for axions in the post-inflationary mass window. 

Although the possibility for miniclusters and axion stars was identified many years ago~\cite{Hogan:1988mp,Tkachev:1991ka, Kolb:1994fi, Kolb:1995bu}, there has been a flurry of developments recently, both in terms of the predicted properties of these structures, as well as methods of detecting them. There is a strong overlap between these studies and the work involved in predicting the axion mass in the post-inflationary scenario because the final state of the field from the early-Universe lattice simulations required to make those predictions forms the \textit{initial} conditions for the Schr\"odinger-Poisson or N-body simulations needed to follow the gravitational collapse of the field into cosmic structure. So, as well as generating many more opportunities for the eventual discovery of axions, if these substructures are observed, they provide one of the only observational windows on the axion's early-universe dynamics. This topic is, therefore, worthy of much more research.

\subsubsection{Axion miniclusters}
Axion miniclusters are a consequence of the so-called post-inflationary scenario, in which the PQ symmetry breaking that leads to the axion as a Goldstone boson occurs after inflation.\footnote{See Sec. \ref{subsec:strings_and_dws} for further discussion of the distinction between pre-inflationary and post-inflationary axions.} In this scenario, the topological defects that result from this breaking---namely a cosmic string-domain wall network---are not inflated away, and this impacts the resulting abundance and distribution of dark matter. In particular, the inherent sub-horizon inhomogeneities present in the distribution of initial axion field values in this scenario, as well as the decay of the string-wall network, leads to large enhancements in dark matter perturbations on a specific range of scales. 

\begin{figure}
    \centering
    \includegraphics[width=0.85\linewidth]{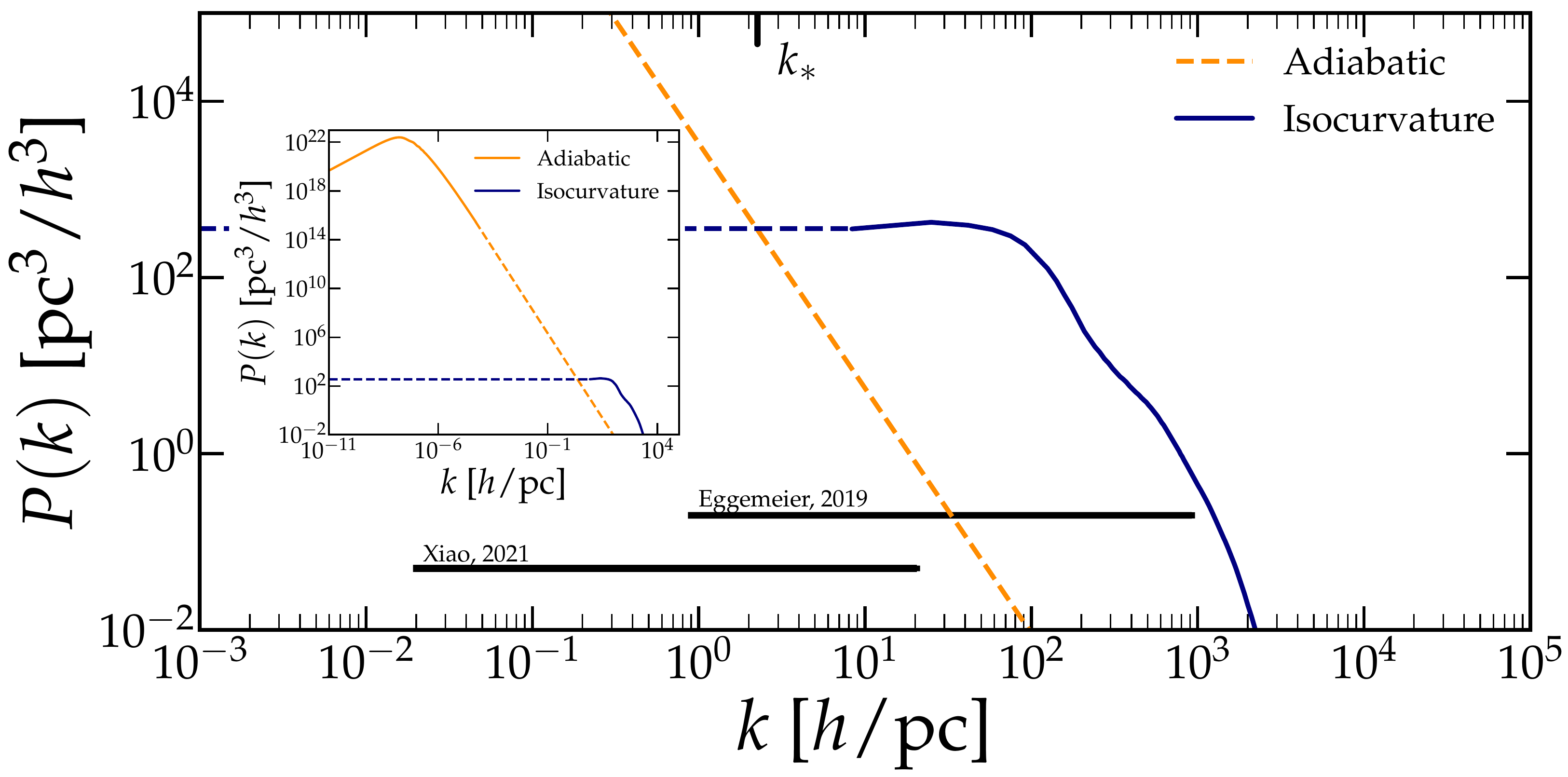}
    \caption{Comparison between matter power spectrum originating from standard inflationary adiabatic fluctuations and from post-inflationary axion dark matter isocurvature fluctuations. The isocurvature spectrum (solid blue line) is taken from Ref.~\cite{Vaquero:2018tib} and extrapolated to large spatial scales (small $k$) as uncorrelated white noise $P(k) \sim const$. We highlight the scale below which the isocurvature fluctuations associated with miniclusters dominates: $k>k_*\sim $ few pc$^{-1}$. The black bars represent range of this power spectrum which has been fed into N-body simulations: Eggemeier et al.~\cite{Eggemeier:2019khm} and Xiao et al.~\cite{Xiao:2021nkb}.}
    \label{fig:spectra}
\end{figure}

The QCD axion becomes dark matter when it acquires its zero-temperature mass, which occurs around the QCD scale. The quantity of dark matter that happens to be within the horizon at that time sets the typical mass scale of eventual substructures that form when the inhomogeneities collapse under gravity. This mass scale is many orders of magnitude below the halo masses forming from the gravitational collapse and mergers from adiabatic inflationary fluctuations (see Fig.~\ref{fig:spectra}), and so post-inflationary axion dark matter is said to form `minihalos', more commonly referred to in the axion literature: \textit{miniclusters}. 

Before we discuss this in more detail, it is important to highlight here that axion dark matter substructure is necessarily a highly model-dependent phenomenon. Most importantly, it is a generic consequence of only one of the two prevailing narratives for axion dark matter production. The conventional wisdom is that inflation will act to homogenise the axion field before it becomes dark matter, and so neither the inhomogeneity in initial values nor the topological defects that lead to substructure are present inside the horizon in the pre-inflationary scenario. That said, it is always possible to engineer models where substructure can emerge in a pre-inflationary scenario, just as it is possible to engineer non-standard cosmological histories and non-QCD ALP models where the properties of the miniclusters are not tied to QCD-scale dynamics---we refer to Refs.~\cite{Arvanitaki:2019rax, Hardy:2016mns,Visinelli:2018wza,Blinov:2019jqc,Hertzberg:2020hsz,Eroncel:2022efc,Gorghetto:2023vqu} for a selection of alternative models. Therefore, while we discuss miniclusters fairly generally here, any tests of their existence must be performed using predictions that are aligned with their expected properties given a particular cosmological scenario. So far, there have only been detailed simulation studies of the formation, properties, and survival of QCD axion miniclusters specifically, meaning this is an obvious open area of research where more work is needed. 

\paragraph{Formation and properties.}

The emergence of the axion's mass around the QCD scale provides the mechanism through which the string-wall network decays. Once this has happened, the axion field now describes a massive particle and is characterised by a highly inhomogeneous distribution, with large and mostly uncorrelated \emph{isocurvature}\footnote{Isocurvature perturbations represent spatial variations in the axion number to photon number $\delta(n_a/n_{\gamma})\neq 0$.} density fluctuations. These fluctuations will collapse gravitationally around the matter-radiation equality epoch and form miniclusters~\cite{Hogan:1988mp}. Their typical size and mass are given by the overdensity parameter $\delta_a=\rho_a/\bar{\rho}_a-1$, defined with respect to the cosmological average density $\bar{\rho}_a$. The largest overdensities, $\delta_a>\mathcal{O}(1)$ decouple from the Hubble flow in advance of matter-radiation equality at scale factor: $R_{\rm coll}=(1+\delta_a)^{-1}R_{\rm eq}$. The naive estimate of the minicluster mass is established by counting the axion density inside the Hubble horizon at the time axions acquire a mass, $\ell_{\rm osc}=H^{-1}(R_{\rm osc})$, is written as~\cite{OHare:2024nmr}\begin{equation}
    M=\frac{4\pi}{3}(1+\delta_a)\bar{\rho}_a\ell^3_{\rm osc}\simeq 2.4\times 10^{-12}~M_{\odot}\left(\frac{10^{-4}~\text{eV}}{m_a}\right)^{1/2}.
\end{equation} 
where $R_{\rm osc}$ is the standard labelling of the scale factor when the axion field begins its damped oscillations, which one can think of as the time when the  axion becomes dark matter. This occurs roughly around the QCD scale; however, the precise value depends on the axion mass, hence the appearance of $m_a$ on the right-hand side of the equation above.

The naive assumption is that miniclusters appear in the matter power spectrum at wavenumbers $k\sim \ell^{-1}_{\rm osc}$, but this conclusion is sensitive to the initial momentum distribution of the axions around the time when the miniclusters are forming. This is where we find the first connection with early-Universe axion dark matter lattice simulations. A recent study found that miniclusters form on smaller scales, $k\sim \mathcal{O}(10)\ell^{-1}_{\rm osc}$ for models where the string network emission index is scale-invariant $q\sim 1$~\cite{Vaquero:2018tib}, and even smaller scales for IR-dominated spectra with $q>1$~\cite{Pierobon:2023ozb,Gorghetto:2024vnp}---we refer to the discussion on the index $q$ in Sec. \ref{subsec:strings_and_dws}. This uncertainty has a knock-on effect, leading to smaller minicluster masses at formation with respect to the naive estimate. Generally speaking, the uncertainty on the string emission spectrum $q$ is also expected to propagate to other minicluster properties, such as the halo-mass function (distribution of masses) and their internal density profiles~\cite{Pierobon:2023ozb}. 

The first estimate of the physical sizes and densities of miniclusters was performed by Kolb and Tkachev~\cite{Kolb:1994fi} using a spherical collapse model. By including fluctuations that can collapse during the radiation domination epoch, the main result from Ref.~\cite{Kolb:1994fi} (see also Ref.~\cite{Ellis:2020gtq} for additional details) is an estimate of the typical core density of the minicluster as a function of the amplitude of fluctuation that seeded it: $\rho_{\rm MC}\simeq 136 \rho_a(R_{\rm eq})\delta^3_a(1+\delta_a)$. Smaller perturbations with $\delta_a\lesssim 1$ collapse during the matter-dominated epoch and can be modelled using linear perturbation theory. Following the growth of the density fluctuations in Fourier space, the linear growth of modes $\delta_{a,k}\equiv\delta_k$ is then described by the equation of motion \begin{equation}
     \ddot{\delta}_k+2H\dot{\delta}_k+\left(\frac{c^2_sk^2}{R^2}-4\pi G\bar{\rho}_a\right)\delta_k=0, \label{eq:deltapt}
\end{equation} 
where $c_s$ is the speed of sound of the cosmological fluid, which for fluctuations larger in size than the axion Compton wavelength can be approximated as $c^2_s\approx k^2/4m_aR^2$. The last term in Eq.~\eqref{eq:deltapt} defines the so-called Jeans scale $k_J(R)=(16\pi G\bar{\rho}_a(R)R^4)^{1/4}$, which demarcates modes with $k<k_J$ as being those which are able to grow, while modes with $k>k_J$ have their growth prevented due to the outward pressure associated with the gradient energy in the field.   

Following recent advances in lattice simulations of axion dark-matter production~\cite{Vaquero:2018tib, Buschmann:2019icd, OHare:2021zrq, Pierobon:2023ozb}, the range of minicluster masses and sizes can be obtained from the correlations in energy density fluctuations, such as the power spectrum. As seen in Fig.~\ref{fig:spectra}, the matter power spectrum shows a white-noise behaviour, $P(k)=$ const., for uncorrelated modes $k\ll k_{\rm osc}$ and a sharp cut-off at the Jeans scale $k>k_J$ beyond which structures cannot grow. Either this power spectrum or the full final state of the field after a string/wall decay simulation can equivalently serve as the starting point for modelling the non-linear growth of this distribution as they merge into minicluster halos. This can be achieved with semi-analytical techniques~\cite{Fairbairn:2017sil, Enander:2017ogx, Blinov:2019jqc}, the `Peak-Patch' method~\cite{Ellis:2020gtq, Ellis:2022grh}, or a given realisation can be simulated directly using a gravitational N-body simulation~\cite{Zurek:2006sy, Eggemeier:2019khm, Xiao:2021nkb, Shen:2022ltx, Eggemeier:2022hqa,Eggemeier:2024fzs}. In Fig.~\ref{fig:spectra}, we highlight the dynamical range of existing high-resolution N-body simulations of minicluster halos. The simulation in Eggemeier et al.~\cite{Eggemeier:2019jsu} evolves the small-scale correlated part of the axion dark-matter spectrum, which encodes the structure left behind by the early-Universe dynamics, while Xiao et al.~\cite{Xiao:2021nkb} instead focused on the larger-scale statistics of halos coming from the uncorrelated white noise part of the spectrum.

\begin{figure*}[t]
    \centering
    \includegraphics[width=0.99\textwidth]{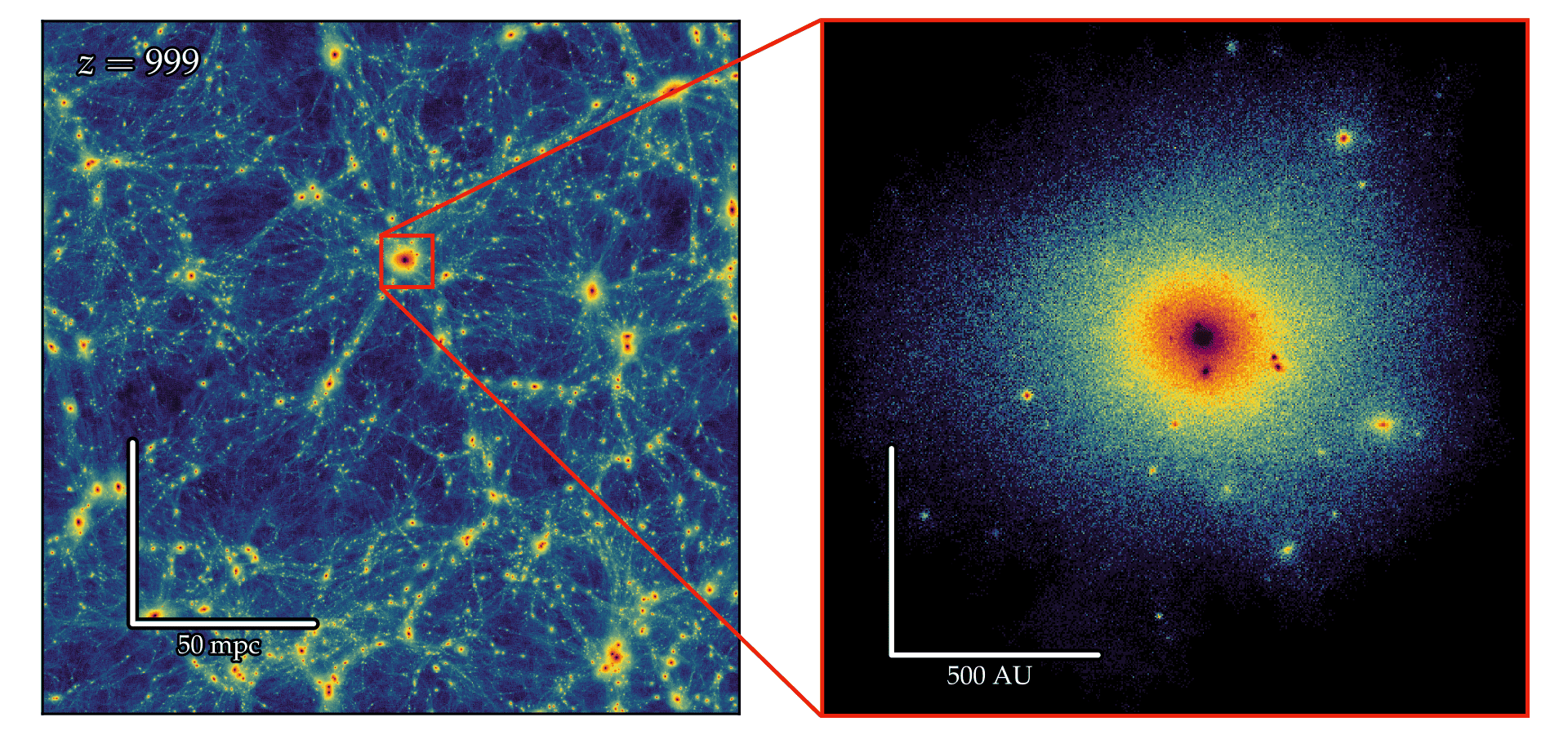}
    \caption{Snapshot of some merging axion miniclusters in the N-body simulations performed for Ref.~\cite{Eggemeier:2022hqa}. The right-hand panel zooms in on the largest minicluster halo in the (0.2~pc)$^3$ box. The colour is scaled according to logarithmic mass density. Figure reproduced from Ref.~\cite{OHare:2024nmr}.}
    \label{fig:miniclusters}
\end{figure*}

A visualisation of axion dark matter in this scenario is shown in Fig.~\ref{fig:miniclusters}. Although the structure of the network of halos shares a resemblance with large-scale cosmic structure, the physical scale shown on the figure is key. In this scenario, gravitationally bound structures are forming on significantly smaller scales than galactic halos, and so although miniclusters will gradually merge to form larger stcutures, this entire distribution is still expected to be embedded inside usual galactic halos giving all galactic dark matter an intrinsic granularity. The scale of this structure is the reason why these objects are so important when it comes to the present-day signals of axion dark matter. A significant portion of the total mass of dark matter is expected to be concentrated inside miniclusters which means it turns out that the typical value of the density at any given location is greatly suppressed compared to the expected large-scale average density in any given region of a galaxy. Miniclusters only make up a percent fraction of the $\mathcal{O}({\rm pc})$ volume in simulations, while underdense regions---so-called ``minivoids''~\cite{Eggemeier:2022hqa}---cover close to 80\% of the space between bound objects. The statistics and density of dark matter inside these minivoids was performed in Ref.~\cite{Eggemeier:2022hqa}, who found a typical suppression of around 10\% in the density of axions, compared to the average value in the simulation. Unlike standard cosmic voids found between galaxies, minivoids are important here because they are sub-galactic structures---we are most likely inside one right now. 

Finally, we note that miniclusters are unlikely to form for an ALP, since the characteristic momentum scale of the order-one density fluctuations, $k \sim \mathcal{O}(10)\,\ell_{\rm osc}^{-1}$, lies below the Jeans scale defined above~\cite{Gorghetto:2022ikz,Gorghetto:2025uls}.

\paragraph{Survival and detection.}
\label{par:AMCsurvival}

The existence of small-scale substructure in the late universe is a critical issue when it comes to the potential discovery of axion dark matter. At a minimum, the presence of miniclusters and minivoids greatly modifies the expected distribution of dark matter within galactic halos on sub-parsec scales, and so predictions for direct and indirect signatures of axion dark matter in the relevant mass window for the post-inflationary scenario ($m_a\sim\mathcal{O}(100)~\mu$eV) must be modified. More optimistically however, several additional observational tests are enabled \textit{only if} these substructures persist to the present day. The ones that have been explored so far in the literature include gravitational microlensing~\cite{Kolb:1995bu, Fairbairn:2017dmf, Fairbairn:2017sil,Ellis:2022grh} as well as transient radio signals due to interactions with neutron stars~\cite{Edwards:2020afl}---the latter is a relevant signature of axion dark matter in general, but the presence of miniclusters would turn a continuous radio signal into a series of transient ones. We will discuss some of these opportunities for detection here, but it should be kept in mind that they are all likely to be quite challenging, even if our galaxy is rife with miniclusters. Although miniclusters represent large localised overdensities of dark matter compared to the usually assumed values of the dark matter density inside galaxies, they are still very diffuse structures, with an asteroid's mass of dark matter spread across the volume of the solar system.\footnote{Axion stars, however, are much denser than miniclusters and have interesting additional properties that allow them to generate many more signals, including electromagnetic ones, as discussed at the end of this chapter.}

The critical uncertainty underpinning the potential success of all of these probes is the properties of miniclusters at $z=0$. As discussed in the previous section, there are still uncertainties about their mass function and density profiles even before galaxy formation ($z>20$), and so these unresolved issues will undoubtedly impose a knock-on uncertainty when it comes to their properties inside modern-day halos (including the question of whether they survive at all). Fortunately, once an initial mass function is known from N-body simulations, the propagation of miniclusters to the present day can be much more straightforwardly understood using a semi-analytic treatment of their tidal disruption during infall into halos and after repeated encounters with stellar objects; so these investigations have already begun.

There have been several works now detailing the tidal stripping of miniclusters in the Milky Way, beginning with the first identification of this problem and some quantitative estimates in Refs.~\cite{Tinyakov:2015cgg, Dokuchaev}. These works were followed by a more detailed Monte-Carlo approach in Ref.\cite{Kavanagh:2020gcy}. Other works include the recent Refs.~\cite{Dandoy:2022prp, Shen:2022ltx, OHare:2023rtm}, and in particular Ref.~\cite{DSouza:2024flu} which refined earlier treatments of the disruption process to properly account for the relaxation of the minicluster profile in between encounters---this has revealed that degree of mass loss for each minicluster is currently underestimated. With the uncertainties related to minicluster properties aside, the general conclusion is that miniclusters can lose a substantial fraction of their mass, but there will be many surviving remnant cores. 

The amount of mass lost and the survival probability of a given minicluster are both dependent on its initial properties as well as the orbit it takes through the galaxy. Generally speaking, the miniclusters losing the majority of their mass and with the lowest chance of survival are those that have the largest masses, NFW-like density profiles, and orbits that spend the most time passing through the galactic disk or bulge. Minicluster initial mass functions are generally expected to follow a falling power law over a large range of masses. Low-mass miniclusters persisting to the present day with masses less than around $10^{-13}\,M_\odot$ are those that have been generally left alone during the merger process, and so they will possess the sharper power-law density profiles they were born with, whereas higher mass miniclusters than this will have NFW profiles that generically emerge following successive mergers. Power-law profile substructures are more concentrated and more resilient, and so by \textit{number}, there is expected to still be a large number of small surviving miniclusters locally, even if by \textit{mass} they do not contribute a substantial amount of the total dark matter.

If compact enough, axion miniclusters could induce a measurable change in the brightness of lensed objects, making them potential targets for microlensing experiments. These experiments aim to detect subtle magnifications in the brightnesses of some set of background stars caused by the gravitational influence of compact objects passing in front of the line of sight. Both current and historical surveys, such as MACHO~\cite{Macho:2000nvd}, Kepler~\cite{Griest:2013aaa}, EROS~\cite{EROS-2:2006ryy}, OGLE~\cite{Niikura:2019kqi}, and HSC~\cite{Smyth:2019whb}, have contributed significantly to constraining the abundance and characteristics of such objects. 
Microlensing as a probe of the minicluster abundance in the galaxy was proposed in Refs.~\cite{Kolb:1995bu}. The heaviest miniclusters would undergo substantial numbers of mergers, potentially pulling their masses up to the sensitivity range of surveys like HSC~\cite{Fairbairn:2017dmf, Fairbairn:2017sil}. The interpretation of these microlensing results in terms of QCD axions requires knowing both the fraction of dark matter contained in miniclusters and the mass distribution of these structures. 
Unfortunately, early estimates based on the Press-Schechter modelling of miniclusters proved too optimistic. Given an updated understanding of the typical minicluster masses and density profiles gained from more sophisticated treatments such as the Peak-Patch method in conjunction with results from N-body simulations, it is more likely that most miniclusters will not be centrally concentrated enough to lens effectively~\cite{Ellis:2022grh}, apart from a narrow window in parameter space. However, this conclusion needs to be balanced against more recent work by Ref.~\cite{Eggemeier:2024fzs} suggesting that a fraction of the most massive miniclusters can have even sharper central density profiles than previously found, and so this work should be revisited for future high-cadence microlensing surveys.

In Refs.~\cite{Pshirkov:2007st, Huang:2018lxq, Hook:2018iia, Safdi:2018oeu}, neutron star magnetospheres were proposed as prime sites for enabling axion dark matter conversion into photons which could then be searched for using radio telescopes. This led to several first search attempts as well as a large theoretical literature on the treatment of axion electrodynamics in the vicinity of neutron stars, e.g.~\cite{Battye:2019aco, Leroy:2019ghm, Witte:2021arp, Millar:2021gzs, Battye:2021xvt, McDonald:2023ohd,Tjemsland:2023vvc,Bhura:2024jjt}. One important signature of an axion dark matter distribution infalling onto a population of neutron stars is the expected forest of radio lines towards regions with high dark matter as well as large numbers of neutron starw i.e.~the galactic centre. The lines would be close to the axion mass in frequency, but each would be Doppler shifted and broadened by an amount particular to a given neutron star. Calculations of this observable signature have been presented in e.g.~Refs.~\cite{Foster:2020pgt,Battye:2021yue,Foster:2022fxn,McDonald:2023ohd}, but importantly in the case where axions are bound in substructures, these estimates require modifications to account for the large localised enhancements in the density as well as an understanding of the inherent transient nature of the radio lines which would last for approximately a few days to a month at a time. Calculations relating to this scenario were presented in Ref.~\cite{Edwards:2020afl,Witte:2022cjj}, and a preliminary search in radio data taken on the Andromeda galaxy was presented recently in Ref.~\cite{Walters:2024vaw}.

Finally, we remark upon the implications of miniclusters for \textit{direct} detection. It is expected that there is less chance that miniclusters will improve discovery prospects because the frequency with which our Earth-bound laboratories would physically encounter an axion miniclusters is, at best, once in a few thousand years. That is the case even if all forming miniclusters survive to the present day in their entirety, which is unlikely to be the case. But while the possibility of traversing the huge density enhancement that would come with passing through a minicluster is tantalising~\cite{Dandoy:2023zbi}, a much more likely scenario is that direct detection efforts must rely on whichever axions are \textit{not} bound inside miniclusters. 

The Earth traverses the galaxy at a rate of 0.2 milli-parsecs per year---a negligibly small distance from the perspective of galactic astronomy. For example, inferences of the local dark matter density $\sim0.4$~GeV/cm$^3$~\cite{Read:2014qva} are based on tracer stellar populations on scales of 100 parsecs at least. These are the smallest-scale probes of dark matter we have currently and they have essentially no sensitivity at all to the question of whether galactic dark matter is bound inside asteroid-sized clumps or not. The actual density sampled by a haloscope at the position of the Earth can only equal $0.4$~GeV/cm$^3$ if the distribution on sub-100-pc scales is homogeneous. While for most dark matter models, this is a safe assumption, it is not in the minicluster scenario. Predicting the expectation value of the density here requires knowing the initial abundance of miniclusters and how the axions in those miniclusters are re-distributed after tidal disruption~\cite{Tinyakov:2015cgg,OHare:2023rtm}.

At minimum the density of dark matter available to haloscopes will be around $\sim10\%$ coming from the left-over minivoid density of axions that never get bound inside miniclusters~\cite{Eggemeier:2022hqa}. However, this must be balanced against the fact that tidal disruption inevitably re-distributes axions that were at one point bound inside miniclusters. A crude estimate of how much re-filling of the distribution occurs can be performed by taking the typical disruption timescale to be that of the age of the galaxy $t \sim 10$~Gyr and multiplying this by the minicluster's internal velocity dispersion, $\sigma \sim \sqrt{GM/R}$ the combination of these two numbers $\ell \sim \sigma t$ gives a typical length scale that the tidal tail of some disrupted minicluster is expected to grow~\cite{Tinyakov:2015cgg}. For canonical QCD axion miniclusters this is around the scale of a few parsecs, which means that the volume-filling fraction of miniclusters by the present day can grow by around a factor of $10^4$ or more. This significantly enhances the prospects for discovery because the expectation value of the density at any given point is brought much closer to the large-scale averaged value for the density due to there being now many (hundreds to thousands) overlapping tidal tails in the vicinity of the Earth. An exact statement depends upon the assumptions made about the mass function and density profiles of miniclusters, as discussed in Ref.~\cite{OHare:2023rtm}. However, given that the majority of the dark matter mass is contributed by the most massive miniclusters---and these are the ones which get the most of their mass tidally stripped---it seems fairly clear that the majority of the \textit{mass} of dark matter will not be in the form of bound miniclusters by the present day.

While the formation of miniclusters does not seem to immediately doom prospects for direct detection, this scenario still comes with a very distinct prediction for the nature of the signal in those experiments. Namely, it now implies that the majority of detectable axions should possess a kinematic signature of the fact they originated from initially small clumps of dark matter, as opposed to the smooth virialised distribution of the main halo of the Milky Way. The lineshape in this scenario is then expected to be composed, in part at least, of discrete narrow features rather than the usual assumption of a smooth Maxwellian lineshape~\cite{Turner:1990qx,Evans:2018bqy} (see also Ref.~\cite{Sikivie:1995dp,OHare:2018trr,OHare:2019qxc} for a different idea but with similar conclusions). There are already some studies in the literature by haloscope collaborations like ADMX where dedicated high-resolution re-analyses of data are performed, which is what is required to pick up extremely fine-grained signatures like this~\cite{ADMX:2024pxg}. Due to their narrowband nature, minicluster streams are also expected to undergo a non-negligible Doppler oscillation in frequency due to Earth's rotation~\cite{Knirck:2018knd,ADMX:2023ctd}, and the entire lineshape is expected to also have some variability on days to months due to the finite physical size of the individual streams~\cite{OHare:2023rtm,OHare:2017yze}. These signatures, while still somewhat speculative and needing more refined predictions, are intriguing new avenues for axion direct detection. See Ref.~\cite{OHare:2025jpr} for more discussion. As with the indirect signals mentioned above, they provide one of the only ways to probe the early-universe cosmology of axion dark matter in an experiment.

\subsubsection{Axion stars}

\paragraph{Properties.}

Axion stars are gravitationally bound configurations composed of QCD axions. These objects form when a balance is achieved between the gravitational attraction that drives the star's collapse and the quantum pressure arising from the wave-like nature of the boson field. This equilibrium is predicted as a solution to the Klein-Gordon equation within the framework of general relativity, known as the Einstein-Klein-Gordon equation~\cite{Kaup:1968zz, Ruffini:1969qy}. The self-gravitational energy of the axion field is sourced by both the spatial gradients and the time derivative of the field. This balance is closely linked to the Heisenberg uncertainty principle, which implies that a particle of mass $m_a$ confined within an object of size $2R$, where $R$ is the radius of the star, must have a velocity $v \sim (2R m_a)^{-1}$. The total kinetic energy is proportional to the gradient of the wavefunction, with the wave-like behaviour of the field playing a crucial role in maintaining stability against gravitational collapse.

The axion field, being a real pseudoscalar field, lacks the U(1) symmetry required to sustain a conserved charge, which typically ensures stability for solitonic solutions. However, soliton-like solutions involving real scalar bosons can still exist, even without a conserved Noether current, leading to ``oscillaton'' solutions over by time-dependent, oscillatory spacetime metrics~\cite{Seidel:1991zh, Copeland:1995fq, Urena-Lopez:2001zjo}.

Axion stars are typically considered in the so-called ``dilute'' regime, where quantum pressure counteracts gravity, preventing further collapse and leading to a relatively large radius~\cite{Chavanis:2011zi, Chavanis:2011zm}. Although more compact configurations could theoretically exist when self-interactions are included, these ``dense'' axion stars are generally unstable due to the attractive nature of the QCD axion's quartic coupling~\cite{Visinelli:2017ooc, Schiappacasse:2017ham}. The formation of axion stars is of significant theoretical interest, particularly because of their potential connection to dark matter and the distinctive observational signatures they may produce, such as gravitational lensing or emissions through axion-photon coupling.

\paragraph{Formation and detection.}

Axion star formation can occur both in astrophysical and cosmological contexts. In dense regions, such as axion miniclusters or compact halos surrounding primordial black holes, axion stars are thought to nucleate via the thermalisation mediated by gravitational interactions among axions in the condensate~\cite{Levkov:2016rkk, Eggemeier:2019jsu, Hertzberg:2020hsz, Dmitriev:2023ipv, Yin:2024xov}. This nucleation process has been numerically studied, with the gravitational timescale well-understood from a theoretical perspective. For long-range gravitational interactions, thermalisation follows a self-similar behaviour, so that the structure of the axion star retains universal characteristics during its evolution~\cite{Dmitriev:2023ipv}. 

Alternatively, axion stars may form through parametric resonance when the initial value of the misalignment angle is $\mathcal{O}(1)$~\cite{Arvanitaki:2019rax}. In addition to axion stars, cosmological evolution can also lead to the formation of related configurations known as axitons~\cite{Kolb:1993hw}. These are oscillating solutions of the Klein-Gordon equation that appear briefly at times on the order of the inverse axion mass, $\sim m_a^{-1}$. Axitons have been observed in simulations of the early Universe, suggesting they are transient but common features of the axion field dynamics soon after the QCD phase transition~\cite{Kolb:1994fi, Vaquero:2018tib, OHare:2021zrq}. Together with axion miniclusters, axion stars and axitons offer environments where the density of axion dark matter is enhanced with respect to its cosmological abundance, potentially providing observable signatures in dense astrophysical environments.

In fact, axion stars could be part of galactic halos and lead to several detectable signatures such as microlensing, as described in Sec.~\ref{par:AMCsurvival}. While stars composed of QCD axions are generally too light to produce detectable microlensing effects, more massive and compact boson stars are within the sensitivity of these surveys. For this, the microlensing method remains a valuable tool in the search for boson stars, which may provide insights into the properties of axion-like particles and their role as DM~\cite{Croon:2020wpr, Croon:2020ouk,Prabhu:2020pzm}. The collapse of axion stars above a critical mass can lead to bursts of relativistic axions~\cite{Levkov:2016rkk,Eby:2021ece, Fox:2023xgx} or photons~\cite{Amin:2020vja,Amin:2021tnq, Hertzberg:2020dbk,Chung-Jukko:2023cow,Escudero:2023vgv,Di:2024snm}, and, under certain conditions, may even result in the formation of black holes~\cite{Chavanis:2016dab, Helfer:2016ljl}.

%% file: WG2/content/hot_axions.tex
\subsection{Axions as Hot/Warm Dark Matter: Status of the Calculation of the Axion Thermalization rate, Bounds for Planck\\ \textnormal{Author: F. D'Eramo}}
\label{subsec:hot_axions}
In contrast to the non-thermal production of QCD axions via the misalignment mechanism (Sec.~\ref{intro:misalignment}) or from the decay of topological defects (Sec.~\ref{subsec:strings_and_dws}), this contribution summarises the thermal production of QCD axions and the related phenomenology. 

\subsubsection{Introduction}

Dimension 5 contact interactions between the axion and standard model fields can be divided into two main classes: anomalous interactions with gauge bosons and derivative couplings with fermions. For the specific case of the QCD axion solving the strong CP problem, the coupling to gluons is not optional and is model-independent. Interactions with other gauge bosons and fermions are model-dependent and present more often than not. These interactions are suppressed by the axion decay constant $f_a$ which is typically much larger than the weak scale. Furthermore, the axion mass is protected by the shift symmetry inherited by the axion field itself. Axions are light and feebly-interacting particles, and these two properties are the premises for rich cosmological consequences.  

We cannot observe the early universe directly before the last scattering surface due to its high opacity. However, the astonishing success of light element predictions by Big Bang Nucleosynthesis (BBN) makes the radiation-dominated epoch very plausible back to temperatures at the MeV scale. This phase cannot extend arbitrarily back to the past since inflationary dynamics must kick in at some point. The highest temperature achieved by the thermal bath during this radiation-dominated epoch is conventionally known as the reheating temperature. Extrapolating the radiation-dominated epoch up to temperatures just a few orders of magnitude above the MeV is quite reasonable, and it provides us with one of the most intriguing particle physics laboratories: a primordial bath populated by all standard model degrees of freedom in thermal equilibrium. Scattering and/or decays of bath particles dump relativistic axions in the early universe. These interactions can happen often enough to bring axions in thermal equilibrium or may not produce enough axions to have the inverse processes at any appreciable rate. Once one specifies the sizes of axion couplings, it is only a matter of tracking how often these processes happen and the resulting asymptotic axion population that survives until today.

This cosmic population of \textit{thermally produced} axions could manifest itself in different ways. Given their thermal origin, they are typically produced with a momentum dispersion of the size of the bath temperature even if they never achieve equilibrium. Whether they thermalize or not, at some point they stop interacting with other bath particles, propagate along FLRW geodesics, and lose energy proportionally to the inverse FLRW scale factor. 

It is reasonable that they are still relativistic at the BBN epoch. Such an amount of additional \textit{dark radiation} is historically quantified in terms of an effective number of additional neutrino species $\Delta N_{\rm eff}$. The predictions of light nuclei abundances through BBN from standard Big Bang cosmology~\cite{Steigman:2007xt,Pisanti:2007hk,Pospelov:2010hj,Consiglio:2017pot,Pitrou:2018cgg} are consistent with observational data and this puts quite severe constraints on $\Delta N_{\rm eff}$~\cite{Yeh:2020mgl,Pisanti:2020efz,Yeh:2022heq}. We probe the radiation content of the early Universe also at the time of the Cosmic Microwave Background (CMB) formation when the temperature of the primordial bath was slightly below the eV scale. If the axion mass is such that the thermally produced population is still relativistic, we get a complementary bound on $\Delta N_{\rm eff}$~\cite{Planck:2018vyg,ACT:2020gnv,SPT-3G:2021wgf} which is at the moment competitive with the one from BBN. Thus the presence of dark radiation in the early universe is at the moment strongly constrained with BBN and CMB both giving the upper bound $\Delta N_{\rm eff} \lesssim 0.2$. Excitingly, significant advancement is expected for the CMB sensitivity on $\Delta N_{\rm eff}$ in the near future. The first improvement will come from the Simons Observatory~\cite{SimonsObservatory:2018koc} with a sensitivity $\sigma(\Delta N_{\rm eff}) \approx 0.05$, and will be potentially followed by CMB-S4~\cite{CMB-S4:2016ple,Abazajian:2019eic,CMB-S4:2022ght} ($\sigma(\Delta N_{\rm eff}) \approx 0.03$),  LiteBIRD~\cite{LiteBIRD:2022cnt}, and other futuristic proposals~\cite{Sehgal:2019ewc, Sehgal:2020yja} ($\sigma(\Delta N_{\rm eff}) \approx 0.01$). Testing the amount of dark radiation is and will be a robust probe for  physics beyond the standard model. In the coming decades, we will limit the presence of additional relativistic degrees of freedom in the early universe and we could potentially open up new windows into the dark sector with conclusive evidence for $\Delta N_{\rm eff} \neq 0$. For these reasons, having accurate and precise predictions is paramount. 

The case in which the axion mass is not negligible at the time of recombination deserves a separate discussion. If this is the case, axions provide a \textit{hot dark matter} component with cosmological signatures similar to massive neutrinos: they suppress cosmological perturbations on length scales smaller than their free-streaming length and they also modify the CMB temperature anisotropy spectrum via the early integrated Sachs-Wolfe effect~\cite{Beutler:2011hx,Font-Ribera:2013rwa,Ross:2014qpa,BOSS:2016wmc,Baumann:2017gkg,Planck:2018lbu,Planck:2019nip,Xu:2021rwg,DESI:2024mwx}. Given the undeniable evidence for non-vanishing neutrino masses, in this case we have a mixed hot dark matter scenario with anti-correlated axion and total neutrino masses. 

We review the state-of-the-art predictions for axion thermalization rates in Sec.~\ref{subsubsec:rates}. The bounds on axion couplings to gauge bosons and fermions are presented in Sec.~\ref{subsubsec:bounds}.

\subsubsection{Axion production rates}
\label{subsubsec:rates}

The cosmological evolution of thermally produced axions is described by an integro-differential Boltzmann equation that tracks the evolution of every momentum bin as a function of the FLRW cosmic time $t$
\begin{equation}
\omega \frac{df_a(k, t)}{dt}  = C[f_a(k, t)] \ .
\label{eq:BEforF}
\end{equation}
Here, the axion energy $\omega$ and spatial momentum $k$ are related by the usual dispersion relation $\omega = \sqrt{k^2 + m_a^2}$, and they are both physical (i.\,e. not comoving) quantities. The left-hand side accounts for the geometry of the expanding Universe, and the right-hand side takes care of axion number-changing processes. The latter includes scatterings and decays, and the collision operator $C[f_a(k, t)]$ acting on the axion phase space distribution $f_a(k, t)$ is the sum of several contributions that account for all possible processes impacting the axion number density.

Solving the integro-differential Boltzmann equation in Eq.~\eqref{eq:BEforF} is computationally expensive thus approximate methods are desirable. If one focuses on \textit{thermal production} only, it is quite reasonable to make a few assumptions that simplify that expression into an ordinary differential equation. Bath particles participating in the production processes are in both chemical and kinetic equilibrium thanks to standard model interactions that have rates significantly larger than the Hubble rate. We need to make some assumptions on the axion distribution. Given the thermal origin of the produced axion particles, it is reasonable to assume kinetic equilibrium. Furthermore, at the high temperatures of interest, we neglect quantum degeneracy effects and we describe axion particles with Maxwell-Boltzmann statistics. Besides being reasonable, these assumptions are harmless if one focuses only on the present data, but they could lead to significant errors in the predictions if one aims at exploiting the sensitivity of future missions~\cite{DEramo:2023nzt}.

We focus on the most generic production process with only one axion in the final state. This is good enough for our purposes due to a severe price to pay in terms of $1/f_a$ in the transition amplitude for each final state axion. Besides that, we do not make any further assumption about the total number of particles participating in the production process and we focus on the broad class described schematically by the following expression
\begin{equation}
\underbrace{\mathcal{B}_1 + \ldots + \mathcal{B}_n}_{n} \longleftrightarrow \underbrace{\mathcal{B}_{n+1} + \ldots + \mathcal{B}_{n+m}}_{m} + a \, .
\label{eq:genprocess}
\end{equation}
We integrate the Boltzmann equation in Eq.~\eqref{eq:BEforF} over the phase space to identify the axion number density. Under the assumptions made in the previous paragraph for the axion phase space distribution (kinetic equilibrium and Maxwell-Boltzmann statistics, for details of when these assumptions enter into the derivation, see App.~C of Ref.~\cite{DEramo:2023nzt}), the integro-differential Boltzmann equation can be integrated over the phase space to get an ordinary differential equation tracking the number density
\begin{equation}
\frac{dn_a}{dt} + 3 H n_a = \gamma_a \left(1 - \frac{n_a}{n_a^{\rm eq}} \right)   \, . 
\label{eq:BEforn}
\end{equation}
Here, $n_a^{\rm eq}$ is the equilibrium axion number density, and the production rate $\gamma_a$ is defined as the number of interactions per unit time and volume. We have not made any assumptions about the bath particle statistics yet, and the expression for $\gamma_a$ still contains the Bose-enhancement and Pauli-blocking factors $1\pm f_{\mathcal{B}_{j}}$ that correctly treat bath particles quantum mechanically. However, the calculations get significantly simplified once one drops these statistical factors and makes the approximation $1\pm f_{\mathcal{B}_{j}} \simeq 1$. This is an approximation widely employed in the literature, and it is also a reasonable one for the same reasons described above. This further approximation leads to the expression 
\begin{equation}
\gamma_a \equiv \int \prod_{i = 1}^{n} d\Pi_i \, f_{\mathcal{B}_{i}}  \prod_{j = n+1}^{n+m}  d\Pi_j  (2\pi)^4\delta^4 
\left(P_{\rm fin} - P_{\rm init}  \right) |\mathcal{M}_{n \rightarrow m + a}|^2   \, .
\label{eq:gamman}
\end{equation}
The Lorentz invariant phase space factors for bath particles are defined as $d\Pi_i = g_i d^3 p_i / [(2\pi)^3 2 E_i]$, where $g_i$ is the number of internal degrees of freedom of the bath particle $\mathcal{B}_i$, and the squared matrix element $|\mathcal{M}_{n \rightarrow m + a}|^2$ is averaged over both initial and final states. This expression can be integrated over all bath particles momenta to find the production rate as a function of the bath temperature. We report here the explicit expressions for $n=1$ (decays) and $n=2$ (scatterings) that explicitly read (see, e.g., App.~A of Ref.~\cite{DEramo:2017ecx})
\begin{equation}
\begin{split}
\text{DECAYS:} & \, \qquad \, \gamma_a^{\rm d} = n_a^{\rm eq} \, \Gamma_{\mathcal{B}_1 \rightarrow  \mathcal{B}_2 \ldots  \mathcal{B}_{1+m} a} \, \frac{K_1[m_{\mathcal{B}_1}/T]}{K_2[m_{\mathcal{B}_1}/T]} \, , \\
\text{SCATTERINGS:} & \, \qquad  \gamma_a^{\rm s} =  \frac{g_1 g_2}{32 \pi^4}  \, T \, \int_{s^{\rm min}_{12}}^\infty ds \, \frac{\lambda(s, m_{\mathcal{B}_1}, m_{\mathcal{B}_2})}{s^{1/2}} \, \sigma_{\mathcal{B}_1 \mathcal{B}_2 \rightarrow  \mathcal{B}_3 \ldots  \mathcal{B}_{2+m} a}(s) \, K_1[\sqrt{s} / T]  \, .
\end{split}
\label{eq:gammaexplicit}
\end{equation}
The production rate for decays is proportional to the width in the mother particle rest frame with the Bessel functions taking care of the lifetime Lorentz dilatation. The scattering rate arises from an integral of the cross section over the Mandelstam variable $s$ with lower integration extreme $s_{12}^{\rm min} = (m_{\mathcal{B}_1} + m_{\mathcal{B}_2})^2$ and $\lambda(x, y, z) \equiv [x - (y + z)^2] [x - (y - z)^2]$.

Axion production rates due to interactions with standard model fermions can be evaluated via Eq.~\eqref{eq:gammaexplicit}. For practical applications, we are interested in processes (decays and scatterings) with only two particles in the final state that have the least amount of phase space suppression. Explicit results for the production rates valid in the phase where electroweak symmetry is broken, both due to flavor flavor-diagonal and flavor-violating couplings, can be found in Refs.~\cite{Brust:2013ova,Baumann:2016wac,Ferreira:2018vjj,DEramo:2018vss,Arias-Aragon:2020qtn,Green:2021hjh,DEramo:2021usm}. Restricting to the broken phase is usually enough since these interactions lead to an IR-dominated production (in spite of being dimension 5 operators); this is easily recognized upon integrating the derivative axion coupling by parts and factoring out fermion masses as resulting from the equation of motion. The top quark is somewhat unique since its mass is at the weak scale and a more accurate analysis is recommendable. This was done by Ref.~\cite{Arias-Aragon:2020shv} where the production rates were evaluated both above and below the weak scale, and it was shown how the two results were smoothly connected across the electroweak phase transition. The expressions in Eq.~\eqref{eq:gammaexplicit} are also applicable to other cases such as axion thermalization via scatterings with pions~\cite{Berezhiani:1992rk,Chang:1993gm}, heavy-colored and PQ-charged fermions within the KSVZ framework~\cite{Turner:1986tb,DEramo:2021lgb}, and heavy Higgs bosons within the DFSZ framework~\cite{DEramo:2021lgb,Sakurai:2024cbi}.

\begin{figure*}[!t]
	\centering
		\includegraphics[width=0.45\textwidth]{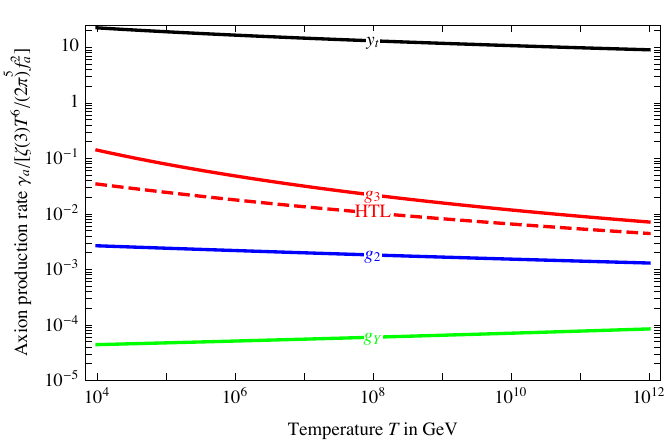} $\qquad$
	\includegraphics[width=0.45\textwidth]{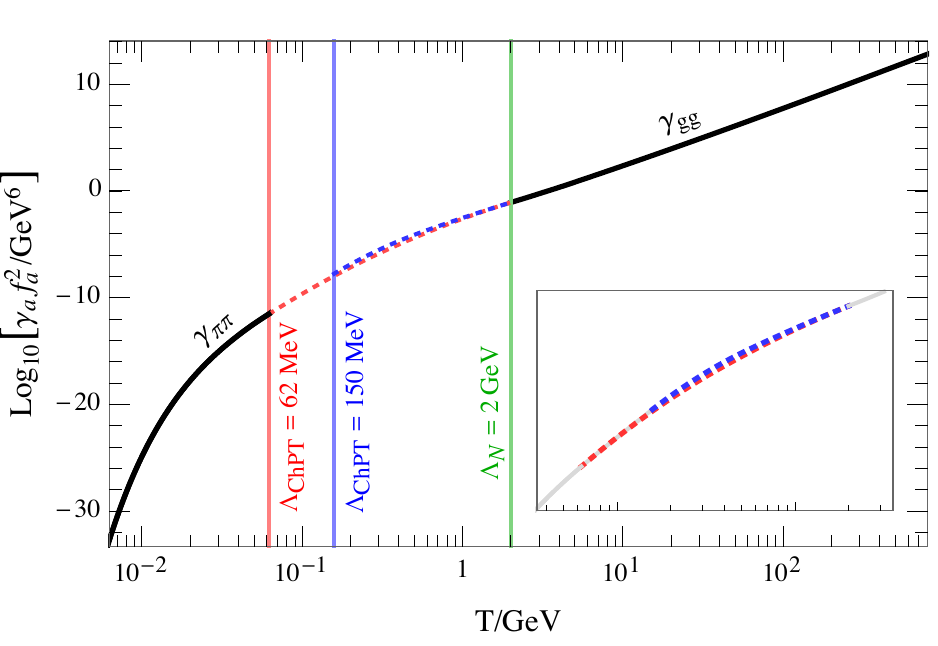}
	\caption{Axion production rate $\gamma_a$ due to the coupling to gluons above the weak scale (\textit{left}, reproduced from Ref.~\cite{Salvio:2013iaa}) and below the weak scale (\textit{right}, reproduced from Ref.~\cite{DEramo:2021psx}).}
	\label{fig:RateGluon}
\end{figure*}

Axion thermalization via its coupling to gluons is the most interesting case given its solid motivation from the strong CP problem, but it is also the case for which theoretical predictions are most challenging. This is due to the unpleasant IR behavior of long-range interactions mediated by massless gluons. The first calculation was performed by Ref.~\cite{Masso:2002np} and the resulting divergences were regulated by a gluon Debye mass put by hand. If one employs a proper thermal field theory treatment, the axion production rate results in~\cite{Weldon:1990iw,Gale:1990pn}
\begin{equation}
\gamma_a = - 2\int \frac{d^3 k}{\left(2\pi\right)^3 2 E} f_a^{\rm eq}(k)\,{\rm Im} \,\Pi_a(k)  \, ,
\label{eq:AxionProductionRate}
\end{equation}
where the axion self-energy $\Pi_a$ contains thermal corrections. The first analysis within the hard thermal loop (HTL) regime was provided by Ref.~\cite{Graf:2010tv}, but its regime of validity was restricted by the domain of applicability of the HTL approximation itself. An important step forward was performed by Ref.~\cite{Salvio:2013iaa} that extended the calculation beyond the HTL regime but still in the electroweak unbroken phase; the resulting rate is shown in the left panel of Fig.~\ref{fig:RateGluon}. The rate below the weak scale was computed by Ref.~\cite{DEramo:2021psx}, extrapolated down to scales of a few GeV, and smoothly interpolated across the QCD crossover to match the known results due to pion scatterings as shown in the right panel of Fig.~\ref{fig:RateGluon}. Intriguingly, the QCD scale is also the frontier where the current Planck bound lies as we will show in the next section. This makes a throughout investigation of the axion production rate across the QCD crossover an absolute priority. There are engaging open questions on both sides of this threshold. On the pion side, leading order chiral perturbation theory breaks down at scales around $100 \, {\rm MeV}$~\cite{DiLuzio:2021vjd}, and this ignited a recent revival of axion-pion scattering rate calculations~\cite{Notari:2022ffe,DiLuzio:2022gsc,Wang:2023xny}. However, this issue is confined to the tiny temperature range of $100 \, {\rm MeV} \lesssim T \lesssim 150 \, {\rm MeV}$ because we cannot treat the primordial plasma within the hadron resonance gas (HRG) approximation~\cite{Hagedorn:1984hz,Huovinen:2009yb,Megias:2012hk,Venumadhav:2015pla} at larger temperatures. On the other side, it is important to ensure the convergence of perturbative QCD at finite temperature down to scales of few GeV~\cite{Notari:2022ffe,Bouzoud:2024bom}. Finally, there are no explicit rate calculations in the intermediate region $150 \, {\rm MeV} \lesssim T \lesssim 2 \, {\rm GeV}$ with the smooth interpolation of Ref.~\cite{DEramo:2021psx} being the only attempt. 

All the rate computations presented so far were for an axion coupling switched on at the time. We conclude this section with a list of studies where the axion production rate was provided for explicit UV complete models. Ref.~\cite{DEramo:2021lgb} considered both axion couplings to bosons and fermions and evaluated the rate for the KSVZ and DFSZ frameworks. Thermal production for UV complete models of the so-called astrophobic axion framework was analyzed in Ref.~\cite{Badziak:2024szg}. Furthermore, the assumption of being in a radiation-dominated universe is not necessary, and thermal production is viable also when the thermal bath is a sub-dominant contribution to the energy budget at early times. Examples of this situation include thermal axion production in low reheating scenarios~\cite{Arias:2023wyg,Carenza:2021ebx}.

\subsubsection{Bounds on axion couplings}
\label{subsubsec:bounds}

\begin{figure*}[!t]
	\centering
		\includegraphics[height=13em]{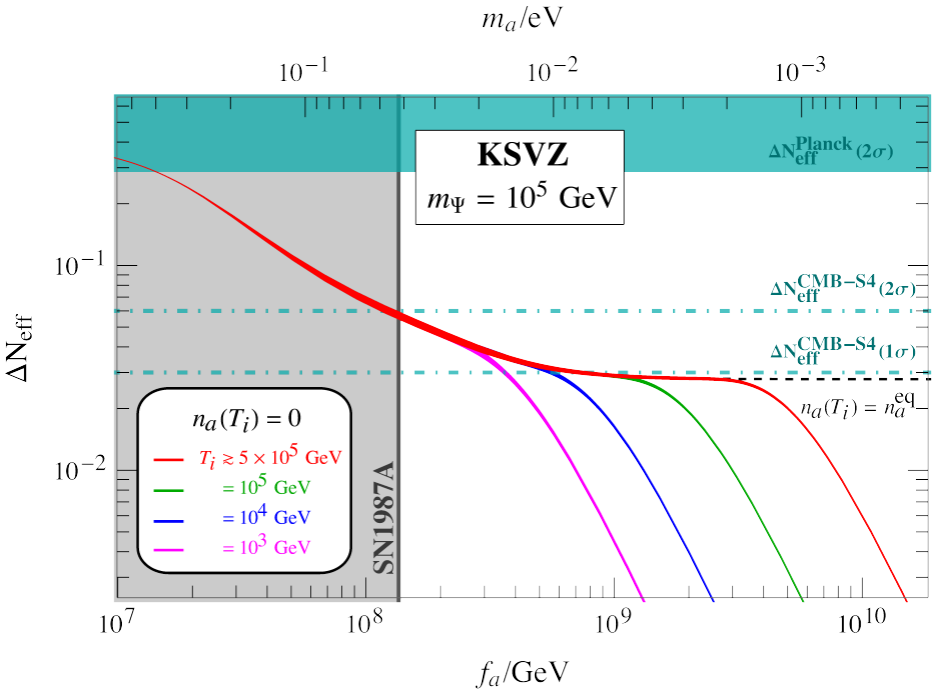} 
	\hfill\includegraphics[height=13em]{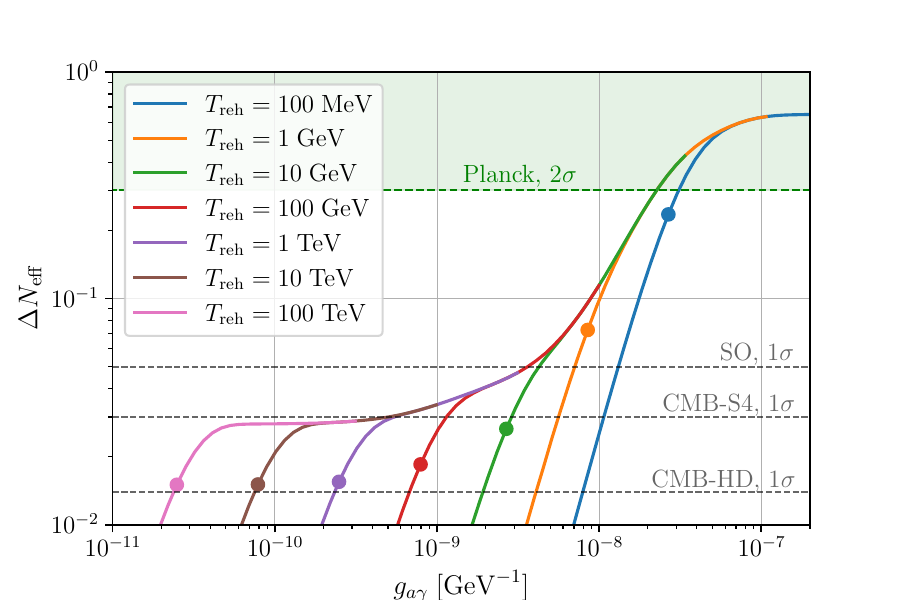}
	\caption{Predicted $\Delta N_{\rm eff}$ as a function of the coupling to gluons (\textit{left}, reproduced from Ref.~\cite{DEramo:2021lgb}) and photons (\textit{right}, reproduced from Ref.~\cite{Caloni:2024olo}).}
	\label{fig:BoundsG}
\end{figure*}

We now turn to the observable effects of thermally produced axions via anomalous interactions with standard model gauge bosons, and we start by considering couplings to gluons. The commonly employed strategy for this case is to solve the Boltzmann equation in Eq.~\eqref{eq:BEforn} tracking the axion number density with the rate $\gamma$ as shown in Fig.~\ref{fig:RateGluon}. The asymptotic result for the axion number density $Y_a = n_a /s$ (with $s$ the entropy density) is then converted into an effective number of additional neutrino species via the relation $\Delta N_{\rm eff} \simeq 75.6 \, Y_a^{4/3}$. This last step introduces a further approximation because it assumes a thermal profile of the distribution function to convert from axion number density to the corresponding energy density. The resulting predictions for $\Delta N_{\rm eff}$ are shown in the left panel of Fig.~\ref{fig:BoundsG}. We notice how the region not excluded by astrophysics is potentially within reach of future missions. Furthermore, the region probed by Planck corresponds to an axion mass (identified by the upper horizontal axis) that is of the order of the CMB temperature. This is the minimum axion mass that we expect barring fine tuning between the QCD contributions and other unknown dynamics. Thus finite axion mass effects have to be taken into account if one wants to exploit the Planck data. Ref.~\cite{Ferreira:2020bpb} considered the DFSZ axion in the QCD confined phase. Ref.~\cite{Caloni:2022uya} considered finite mass effects for axion couplings to gluons and photons. Ref.~\cite{DEramo:2022nvb} performed a throughout study for the KSVZ and DFSZ frameworks with all finite axion mass effects by using the interpolated rates of Ref.~\cite{DEramo:2021lgb}. A follow-up analysis for the KSVZ axion can be found in Ref.~\cite{DiValentino:2022edq}. Refs.~\cite{Notari:2022ffe,Bianchini:2023ubu} analyzed finite axion mass effects by solving the Boltzmann equation in momentum space but only for pion scatterings in the QCD confined phase. The ballpark for the axion mass bound identified by these studies is $m_a \lesssim 0.2 \, {\rm eV}$. 

Interactions with electroweak gauge bosons are possible but not mandatory. There is no connection between the axion mass and couplings in these cases. Thus the resulting observable effect could be either dark radiation or hot dark matter. Ref.~\cite{Caloni:2022uya} studied axion produced via the Primakoff effect and the resulting finite axion mass effects. On the contrary, the work in Ref.~\cite{Caloni:2024olo} neglected the axion mass and computed the expected $\Delta N_{\rm eff}$ due to axion interactions with photons as shown in the right panel of Fig.~\ref{fig:BoundsG}.
 
\begin{figure*}[!t]
	\centering
		\includegraphics[width=0.49\textwidth]{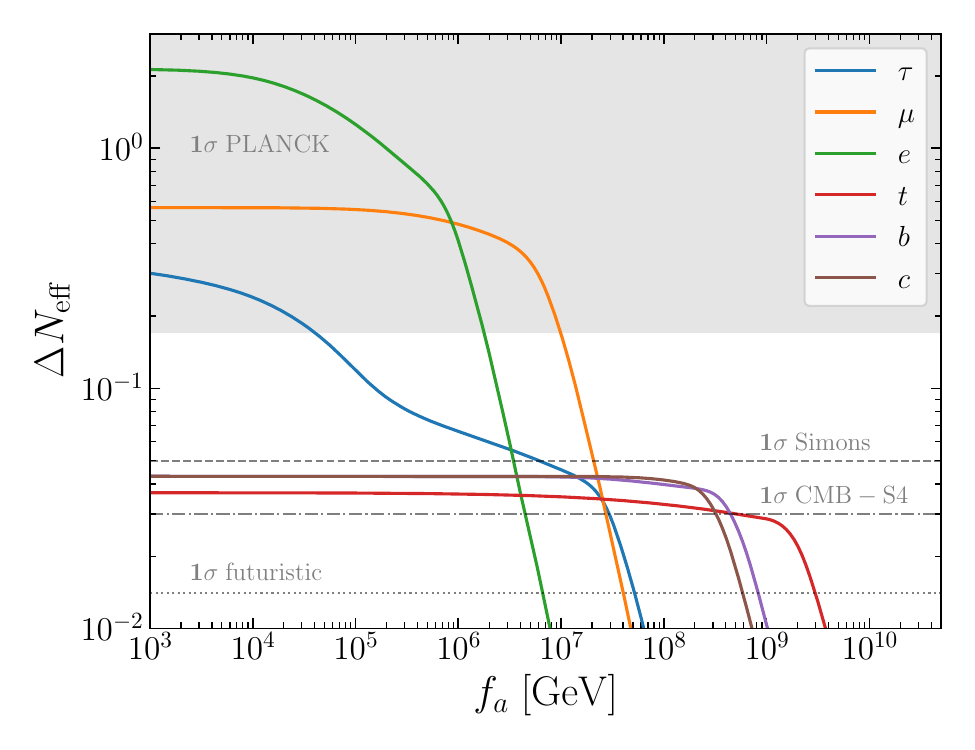} \hfill\includegraphics[width=0.49\textwidth]{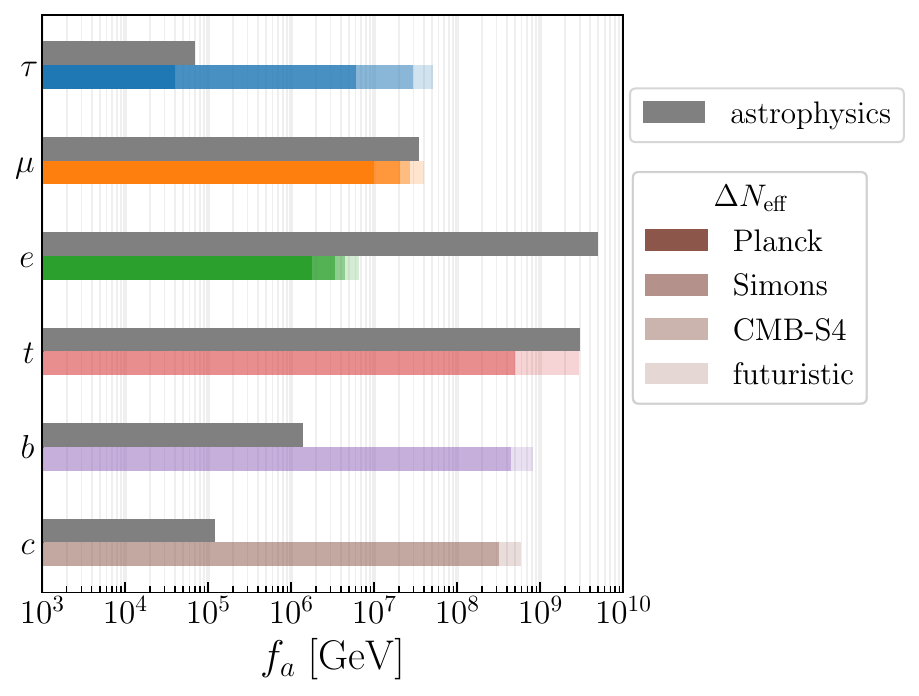}
	\caption{Predicted $\Delta N_{\rm eff}$ as a function of the coupling to fermions (\textit{left}) and comparison between astrophysical and cosmological bounds (\textit{right}). Figures reproduced from Ref.~\cite{DEramo:2024jhn}. }
	\label{fig:BoundsF}
\end{figure*}

Axion couplings to fermions are not responsible for generating any contribution to the axion mass. We neglect $m_a$ for this case and limit ourselves to the prediction of $\Delta N_{\rm eff}$. Furthermore, this case is immune from theoretical issues due to IR divergences and this is the reason why the problem has been completely solved in phase space. In other words, the integro-differential equation in Eq.~\eqref{eq:BEforF} has been employed to track the full phase space axion distribution and the resulting $\Delta N_{\rm eff}$ was evaluated from the appropriate phase space integral over the asymptotic comoving result. Remarkably, Refs.~\cite{Badziak:2024qjg,Badziak:2025mkt,DEramo:2024jhn, Barbieri:2026ewj} identified spectral distortions due to non-instantaneous axion decoupling that affected the resulting predictions by an amount larger than the sensitivity of future CMB surveys. Thus these phase space analyses will be more and more relevant as the resolution improves over the coming years. Refs.~\cite{Badziak:2024qjg, Badziak:2025mkt} analyzed both flavor-conserving and flavor-violating couplings to second- and third- generation charged leptons. Ref.~\cite{DEramo:2024jhn} investigated the cosmological consequences of flavor-conserving couplings to charged leptons and the three heaviest quarks. The quark choice was due to the dangerous QCD crossover. The top quark case is quite safe, the bottom quark is somewhat intermediate, and the charm case is rather concerning. Uncertainties due to the QCD crossover were discussed in the reference. Fig.~\ref{fig:BoundsF} reports the predicted $\Delta N_{\rm eff}$ as a function of the axion couplings (left panel) and a comparison between astrophysical and cosmological bounds (right panel). As we can see from the figure, astrophysics is hard to beat for couplings to electrons. However, for all the other fermions the future bounds set by CMB observations will be either competitive with astrophysical ones or they will even supersede them.

%% file: WG2/content/ul.tex
\subsection{Ultra-light Dark Matter: Theory and Cosmological Constraints\\ \textnormal{Author: K.~K. Rogers}}
\label{subsec:ul}

\subsubsection{Motivation and production mechanism}

The dark matter (DM) can (entirely or partly) compose of ultra-light bosonic fields, where we define ultra-light as having masses \(m_a \leq 10^{-17}\,\mathrm{eV}\).\footnote{Fermions cannot compose the dark matter for masses \(\lesssim 1\,\mathrm{MeV}\) owing to the Pauli exclusion principle \citep{PhysRevLett.42.407} unless there is a vast number of species \citep{Davoudiasl:2020uig}.} Indeed, light (pseudo-)scalar fields are abundantly produced in high-energy theories and so are an excellent DM particle candidate. A concrete example is the axion, a pseudoscalar with a broken angular degree of freedom. Originally postulated as a solution to the strong charge-parity (CP) problem, it was soon identified as a DM candidate \citep{Peccei:1977hh,Weinberg:1977ma,Wilczek:1977pj}. Relaxing the relation between axion mass and axion decay constant \(f_\mathrm{a}\) (the high-energy scale at which the axion forms) that specifically solves the strong CP problem, axion-like particles (ALPs) can have any mass from the Hubble to the Planck scale \citep[e.g.,][]{Antypas:2022asj}. ALPs can also form dark matter. Indeed, e.g., in string theory compactifications, tens or hundreds of ALPs are formed and one or more of these may contribute significantly to the dark sector \citep[a scenario dubbed the ``axiverse'',][]{Svrcek:2006yi,Arvanitaki:2009fg}. Recent work has calculated the masses and decay constants of axions in a corner of the string landscape in order to compare to astrophysical limits \citep{Sheridan:2024vtt,Jain:2025vfh}. Here, we typically refer to ultra-light axions as a particularly well-motivated example, though, unless specified, the limits discussed apply to any ultra-light scalar.

\begin{figure}
\includegraphics[width=0.9\columnwidth]{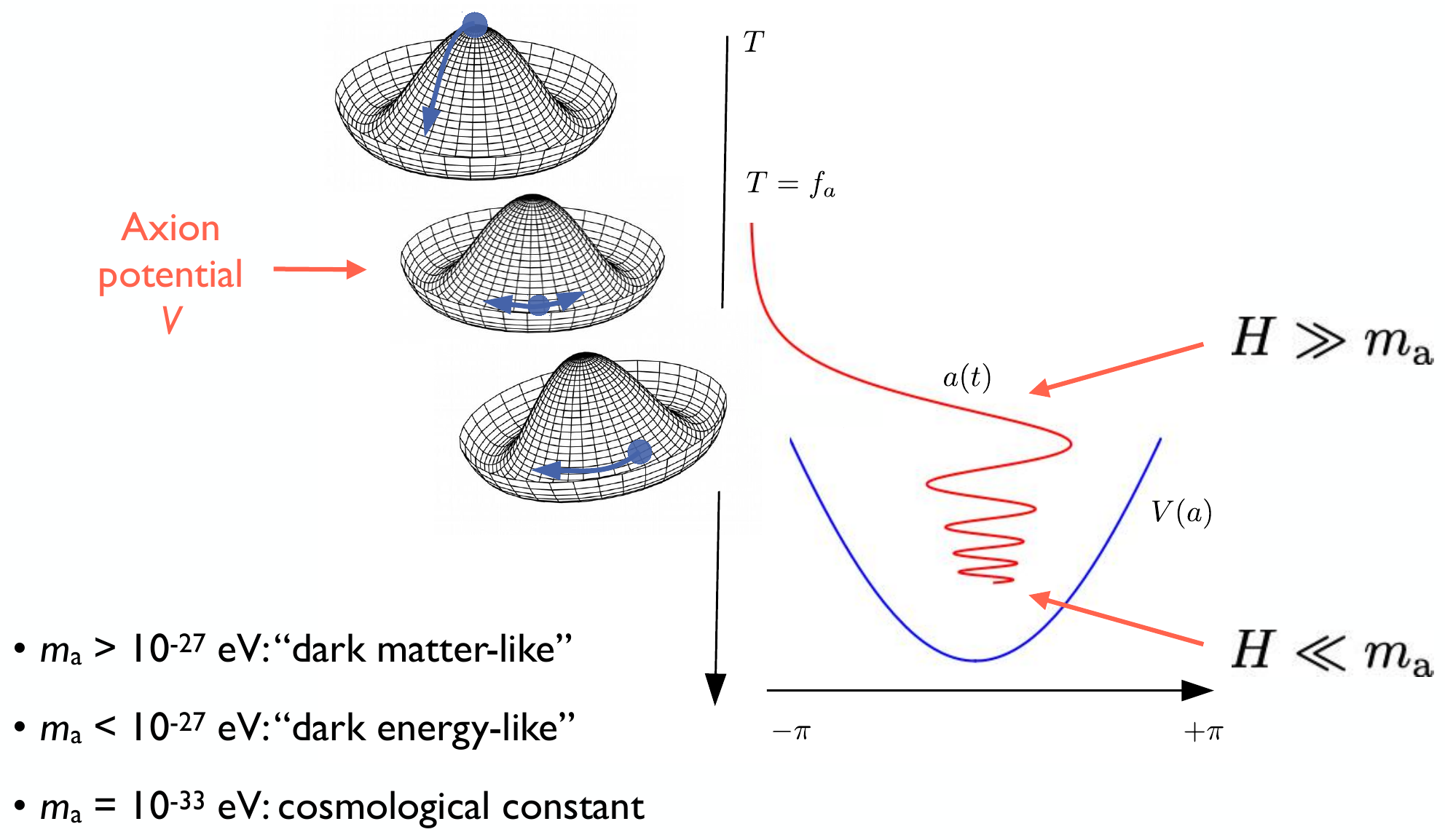}
\caption{A schematic view of the misalignment production of ultra-light axions with a definition of the transition from dark energy-like to dark matter-like ULAs.\label{fig:misalignment}}
\end{figure}

Axions form through the misalignment mechanism (see Fig.~\ref{fig:misalignment} and the discussion in Sec. ~\ref{intro:misalignment}). 
For the ultra-light regime discussed here, \(m_a \gtrsim 10^{-27}\,\mathrm{eV}\), the relevant time scale for the misalignment mechanism always occurs before the redshift of matter-radiation equality \(z_\mathrm{eq}\), meaning that these axions are DM-like in their effect on the cosmic microwave background and structure formation. For \(m_a \lesssim 10^{-27}\,\mathrm{eV}\), this transition occurs after \(z_\mathrm{eq}\), meaning that these axions are DE-like up until late times and then start to contribute to the matter content upon the onset of oscillations. We dub these axions ``DE-like.'' For \(m_a \lesssim 10^{-33}\,\mathrm{eV}\), the axion has not begun oscillating by today and it behaves like a cosmological constant for all of our cosmic history. The consequence is that, depending on the axion mass, different cosmological data across redshifts and scales will be constraining.\footnote{As introduced in Sec.~\ref{intro:misalignment}, the relic energy density of axions \(\Omega_\mathrm{a}\) depends on both axion mass and the initial field displacement \(\theta_\mathrm{i}\) \citep[see, e.g.,][]{Ferreira:2020fam}.}  In cosmological data analyses, both \(m_a\) and the relic density \(\Omega_\mathrm{a}\) are treated as free parameters and inferred from data.

Thus, axions are excellent candidates for cold dark matter (CDM). 
However, ultra-light axions (ULAs) are distinguishable from ``standard'' CDM because of a key phenomenon: the so-called ``quantum pressure'' \citep{Hu:2000ke}. We will discuss this effect in more detail below. The main point is that these particles are sufficiently light that, by the uncertainty principle, they have astrophysically-sized de Broglie wavelengths (kpc to Gpc). The ULAs cannot cluster at or below a Jeans scale where the quantum pressure balances gravitational collapse. In other words, if ULAs contribute to the DM, there will be a suppression in DM perturbations at high wavenumber, where the Jeans wavenumber increases as the axion mass increases.

This property means that axions have been invoked \citep{Hu:2000ke,Hui:2016ltb} to address the so-called CDM ``small-scale crisis'' \citep{Bullock:2017xww}. The small-scale crisis is the collection of discrepancies between the predictions of CDM-only simulations and observations at sub-Galactic scales. E.g., the cusp-core problem is that DM-only simulations of dwarf galaxies produce cuspy, divergent density profiles at the centre, whereas shallower cored profiles (\(\sim\) kpc) are inferred from observation. ULAs with \(m_a \sim 10^{-22}\,\mathrm{eV}\) address this issue as they form bosonic structures called solitons with cored profiles \(\sim\) kpc. The missing satellites problem refers to the historical underabundance of observed Milky Way satellite galaxies relative to expectations from DM-only simulations. ULAs with \(m_a \sim 10^{-22}\,\mathrm{eV}\) address this issue as low-mass sub-halos (\(\lesssim 10^9\,M_\odot\)) which host low-luminosity satellites do not form owing to Jeans suppression. It has been shown that cores can be produced by astrophysical processes like supernovae feedback \citep{2012MNRAS.421.3464P}, although it is not entirely clear how effective this process will be in the lowest-mass dwarfs. Further, more complete, deeper observations have identified the satellite population and inferred the existence of sub-halos down to \(\sim \mathrm{few} \times 10^7\,M_\odot\) \citep{DES:2019vzn,DES:2020fxi}, although we have no direct evidence yet of DM structure forming on smaller scales. The improved astrophysical modelling and observation on sub-Galactic scales means that we can now set more robust limits on the abundance of ULAs.

ULAs are also invoked to address other discrepancies in the standard cosmological model. The \(S_8\) tension refers to a \(2 - 3 \sigma\) discrepancy in inference of the cosmological parameter \(S_8 \equiv \sqrt{\frac{\Omega_\mathrm{m}}{0.3}} \sigma_8\) between CMB and galaxy weak lensing data \citep{Abdalla:2022yfr}, where \(\Omega_\mathrm{m}\) is the relic matter energy density and \(\sigma_8\) measures the root mean square matter perturbation at a scale of 8 Mpc. The low value of \(S_8\) inferred from weak lensing can indicate suppressed clustering below 8 Mpc, which can be explained by ULAs with \(m_a \sim 10^{-25}\,\mathrm{eV}\) \citep{Rogers:2023ezo}. Ref.~\cite{Rogers:2023upm} measured a \(\sim 5 \sigma\) discrepancy in inference of the small-scale matter power spectrum (wavenumber \(k \sim 1\,h\,\mathrm{Mpc}^{-1}\)) at redshift \(z = 3\) between CMB and Lyman-alpha forest data \citep{eBOSS:2018qyj}, which can also be explained by ULAs with \(m_a \sim 10^{-25}\,\mathrm{eV}\). There is a \(3 - 4 \sigma\) hint of parity-violating birefringence (rotation of the angle of polarisation) in the CMB \citep{Minami:2020odp,Eskilt:2022cff}. Axion-like particles with \(m_a \lesssim 10^{-27}\,\mathrm{eV}\) are invoked to explain this effect by their Chern-Simons coupling to CMB photons. As upcoming observatories are poised to increase the statistical precision of these measurements, it is critical to couple more accurate modelling of axion structure formation with astrophysical processes like baryonic feedback \citep[which can also suppress small-scale power,][]{Chisari:2019tus,Amon:2022azi} and Galactic foregrounds \citep[which can also cause polarisation rotation,][]{Diego-Palazuelos:2022cnh}.

\subsubsection{Suppression of cosmological perturbations}

As mentioned above, if ULAs contribute to dark matter, they will suppress cosmological perturbations in a scale-dependent way. In a comoving gauge, the equation of motion of the ULA density perturbation \(\delta = \frac{\delta \rho}{\rho}\) (\(\rho\) is background density) \citep{Hwang:2009js} is \citep[here I follow the notation in the review of][see references therein]{Ferreira:2020fam}
\begin{equation}
    \ddot\delta + 2H\dot\delta + \omega_k^2\delta = 0,
\end{equation}
where the dispersion relation
\begin{equation}
    \omega_k^2 = \frac{k^4}{4m_a^2R^4} - 4\pi G\rho,
\end{equation}
\(R\) is the scale factor and \(G\) is the gravitational constant. It follows that the dispersion of density perturbations is scale-dependent, i.e., the sound speed is scale-dependent. A Jeans wavelength \(\lambda_\mathrm{J} = \frac{2\pi R}{k_\mathrm{J}}\) forms at the transition where \(\omega_k(k_\mathrm{J})\) = 0:
\begin{equation}
    \lambda_\mathrm{J} = 9.4\,(1 + z)^\frac{1}{4} \left(\frac{\Omega_\mathrm{a}h^2}{0.12}\right)^{-\frac{1}{4}} \left(\frac{m_a}{10^{-26}\,\mathrm{eV}}\right)^{-\frac{1}{2}} \mathrm{Mpc},
\end{equation}
where \(h\) is the dimensionless Hubble constant. For wavelengths \(> \lambda_\mathrm{J}\), the perturbations grow; for wavelengths \(< \lambda_\mathrm{J}\), the perturbations oscillate and are suppressed \citep{Marsh:2010wq}. The full effect of ULAs on the cosmic microwave background (CMB) and the linear matter power spectrum, including the case of a mixture of ULAs and a standard CDM component, is calculated in modified Boltzmann codes \citep[\texttt{AxionCAMB},][]{Hlozek:2014lca} \citep[\texttt{AxiECAMB},][]{Liu:2024yne} \citep[\texttt{AxiCLASS},][]{Poulin:2018dzj}.

The lightest DE-like axions (\(m_a \lesssim 10^{-27}\,\mathrm{eV}\)) are most degenerate in CMB data with a standard DE component. These axions have two primary effects on the CMB. First, by transitioning from DE to DM behaviour after recombination, they change the age of the Universe and thus the angular size of the sound horizon and the CMB acoustic peaks. Second, they change the integrated effect of dark energy and thus the amplitude of the large-scale integrated Sachs-Wolfe effect. DM-like axions (\(m_a \gtrsim 10^{-27}\,\mathrm{eV}\)) transition from DE behaviour during the radiation epoch and so change the height of acoustic peaks, with the effect diminishing as mass increases since the transition occurs earlier. In the linear matter power spectrum, there are two non-degenerate effects. First, the Jeans effect suppresses high-wavenumber modes \(> k_\mathrm{J}\), with \(k_\mathrm{J} \propto \sqrt{m_a}\). Second, at a fixed \(m_a\), as the fraction of DM in ULAs increases, the strength of the suppression at a given \(k\) increases, i.e., the step-like drop in the power spectrum gets deeper. These matter power spectrum effects also manifest in CMB lensing anisotropies.

\subsubsection{Axion structure formation}
The full non-linear evolution of the ULA field follows a coupled set of Schr\"{o}dinger-Poisson equations \citep{Ferreira:2020fam}:
\begin{equation}
\begin{aligned}
    i\dot{\psi} &= -\frac{3}{2}iH\psi - \frac{\nabla^2 \psi}{2m_aR^2} + m_a\Phi\psi,\\
    \nabla^2\Phi &= 4\pi G \delta\rho,
\end{aligned}
\end{equation}
where \(\Phi\) is the gravitational potential. The wavefunction \(\psi\) comes from a factorisation of the field \(\phi\) in the non-relativistic limit:
\begin{equation}
    \phi = \frac{\psi e^{-im_at} + \psi^* e^{im_at}}{\sqrt{2m_aR^3}}.
\end{equation}
Simulations that solve these equations in a cosmological volume have revealed a rich and distinctive phenomenology. The de Broglie wavelength
\begin{equation}
    \lambda_\mathrm{dB} \approx 0.2\,\left(\frac{m_a}{10^{-22}\,\mathrm{eV}}\right)^{-1} \left(\frac{v}{V_{200}}\right)^{-1}\,\mathrm{kpc},
\end{equation}
where the virial velocity
\begin{equation}
    V_{200} \sim 85\,\left(\frac{M}{10^{12}M_\odot}\right)^\frac{1}{3}\sqrt{1 + z_\mathrm{vir}}\,\frac{\mathrm{km}}{\mathrm{s}},
\end{equation}
\(M\) is the virialised halo mass and \(z_\mathrm{vir}\) is the redshift of virialisation. At scales around \(\lambda_\mathrm{dB}\), interference patterns in the ULA density arise from overlapping wavefunctions \citep{Schive:2014dra}. Quantum pressure (arising from higher-order gradients of the density) prevents the gravitational collapse of ULA DM on scales below \(\lambda_\mathrm{dB}\). At the centre of a halo, this effect leads to the formation of a dense core or soliton \citep{LEE1992251}. In the surrounding regions, transient granules of DM form with density oscillating with a coherence time-scale
\begin{equation}
    t_\mathrm{c} = \frac{\lambda_\mathrm{dB}}{2v} \approx 1\,\mathrm{Myr}\,\left(\frac{m_a}{10^{-22}\,\mathrm{eV}}\right)^{-1} \left(\frac{v}{250\,\frac{\mathrm{km}}{\mathrm{s}}}\right)^{-2}.
\end{equation}
These granules are a source of dynamical heating to the stellar halo.

These explicit wave-like phenomena are still being explored through numerical simulations \citep{Schive:2014dra,PhysRevLett.123.141301,May:2021wwp,May:2022gus,Lague:2023wes}. Progress is limited by the computational cost of these simulations. There is a vast dynamic range between the need to resolve the de Broglie wavelength within halos and the cosmological volumes needed to do systematic studies. Further, there is a computationally-depending time-step requirement to resolve wavefunction fluctuations. Adaptive mesh refinement has been successfully implemented to focus resources on resolving ULA granules inside halos \citep{Schwabe:2020eac}. The Gaussian beams approach to solving the Schr\"{o}dinger-Poisson equations, where the wavefunction is decomposed into coherent Gaussian wave packets, can also speed up calculations \citep{Schwabe:2021jne}. An intriguing possibility is to ``paint-in'' interference fringes into steady-state background systems, e.g., filaments \citep{Zimmermann:2024vng}. Nonetheless, much of the suppression effect of ULAs into the non-linear regime is already captured by either making a hydrodynamical \citep[Madelung,][]{1926NW.....14.1004M} transformation to fluid variables, which captures quantum pressure but without interference \citep{Nori:2018hud}, or modifying a CDM-like simulation to have ULA-like initial conditions, which captures most of the suppression effect on scales relevant to current observations. As with CDM, comparing simulations to data in statistical inference requires machine learning (ML) emulation \citep{Heitmann:2013bra} or simulation-based inference \citep[SBI,][]{2016arXiv160506376P} techniques, which have been developed for ULA cosmologies \citep[e.g.,][]{Rogers:2020cup,Ma:2025srn}. A complementary approach is to develop semi-analytical structure formation models, e.g., the effective field theory of large-scale structure (EFT of LSS) for mildly non-linear, redshift-space-distorted scales \citep{Lague:2021frh,Rogers:2023ezo} or a mixed axion and cold halo model with baryonic feedback for fully non-linear modes \citep{2016arXiv160505973M,Vogt:2022bwy,Winch:2024mrt,Dome:2024hzq}.

\subsubsection{Cosmological limits}

\begin{figure}
\includegraphics[width=\columnwidth]{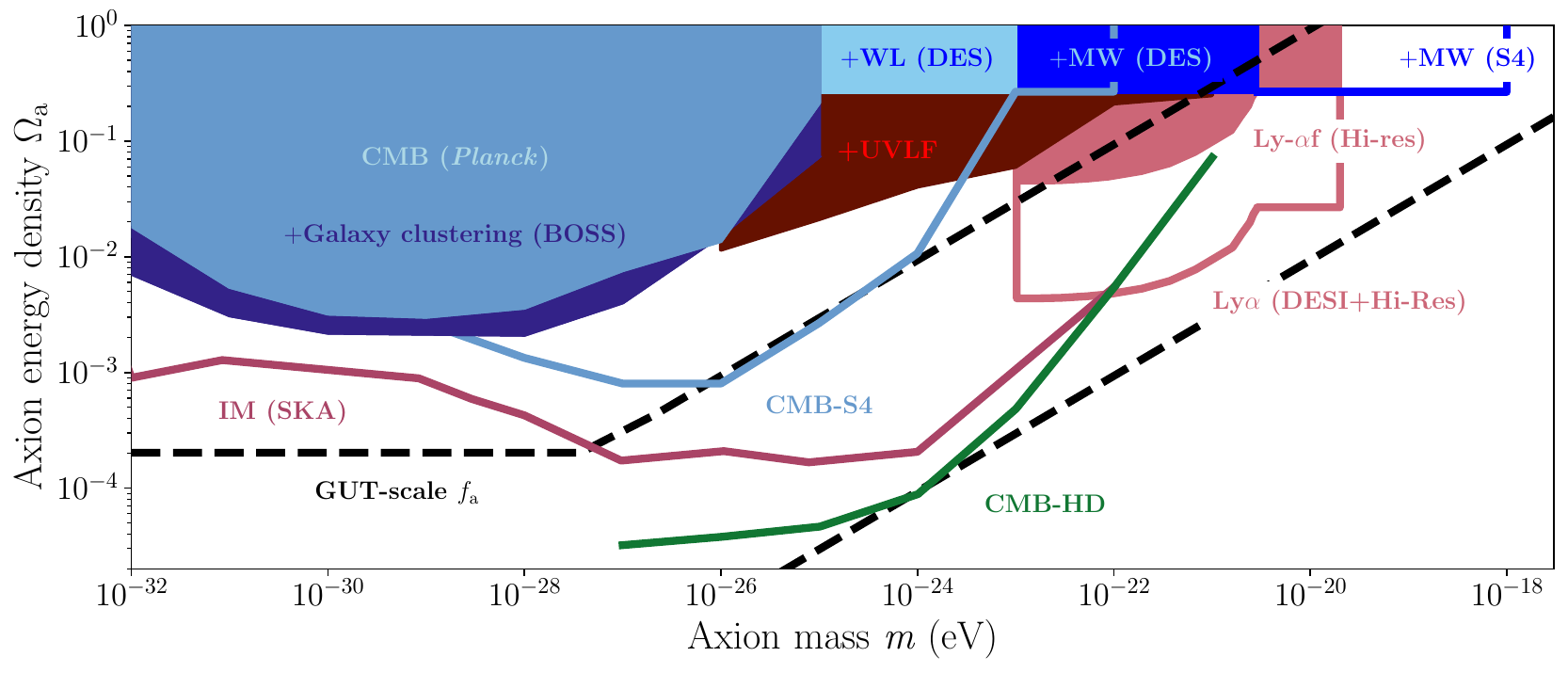}
\caption{A summary of current cosmological bounds (shaded regions) on ULA mass \(m_a\) and relic density \(\Omega_\mathrm{a}\) with projected sensitivities (solid lines). See text for more details.\label{fig:limits}}
\end{figure}

Figure \ref{fig:limits} summarises current cosmological bounds on the mass and relic energy density of axions across the ultra-light regime: shaded regions are already excluded at 95\% c.l., while solid lines indicate the 95\% c.l. sensitivity of upcoming observations (each of these forecasts has different levels of robustness; we show each to indicate the direction of progress). The region between the black dashed lines has \(f_\mathrm{a}\) at the Grand Unified Theory (GUT) scale. For \(m_a < 10^{-26}\,\mathrm{eV}\), the strongest limits come from a combination of \textit{Planck} 2018 CMB temperature, polarisation and lensing anisotropies and the Baryon Oscillation Spectroscopic Survey (BOSS) galaxy power spectrum and bispectrum \citep{Rogers:2023ezo}. The galaxy power spectrum, as a biased tracer of the matter power spectrum, is modelled on mildly non-linear scales (\(k < 0.4\,h\,\mathrm{Mpc}^{-1}\)) using the EFT of LSS \citep{Carrasco:2012cv}. For these lightest masses, axions cannot be more than (1 - 10)\% of the DM today. Competitive limits in this regime come from the number counts of galaxy clusters as inferred from X-ray emission in the eROSITA All Sky Survey (eRASS1) and gravitational weak lensing from the Dark Energy Survey (DES), Kilo-Degree Survey and Hyper Suprime-Cam \citep{Zelmer:2025gln}.

ULAs also source isocurvature perturbations (if the symmetry breaking that sets the relic density happens during inflation) with amplitude set by the inflationary Hubble parameter \(H_\mathrm{I}\) and axion mass \(m\) \citep{AXENIDES1983178}. \(H_\mathrm{I}\) sets the amplitude of the inflationary gravitational wave background, detectable as tensor modes in CMB B-mode polarisation \citep{Seljak:1996gy}. Electromagnetic ULA couplings can also source CMB spectral distortion signatures \citep{Mukherjee:2018oeb}. The kinetic Sunyaev-Zel'dovich (kSZ) effect in the CMB is another probe of the scale-dependent growth arising in an ULA cosmology \citep{Farren:2021jcd}. In Fig.~\ref{fig:limits}, we show the estimated sensitivity of the proposed CMB-S4 observatory (a CMB experiment with 1 \(\mu\)K-arcmin map noise) through the high-resolution lensing anisotropies it would measure \citep{Hlozek:2016lzm,Dvorkin:2022bsc}. The Simons Observatory \citep[SO,][]{SimonsObservatory:2018koc} will already achieve stronger sensitivity (6 \(\mu\)K-arcmin map noise), while the planned CMB-HD \citep{Sehgal:2019ewc} observatory will probe the CMB lensing power spectrum up to multipoles \(L \sim 40,000\). In Fig.~\ref{fig:limits}, we show the estimated sensitivity from the kSZ Ostriker-Vishniac effect in CMB-HD \citep{Farren:2021jcd}. Sensitivity to isocurvature and tensor modes will come from CMB polarisation measurements from SO, CMB-S4 and the LiteBIRD \citep{2018JLTP..193.1048S} satellite.

For \(10^{-26}\,\mathrm{eV} < m_a < 10^{-22}\,\mathrm{eV}\), CMB and spectroscopic galaxy data lose sensitivity as the Jeans suppression manifests on smaller scales than currently probed. Galaxy weak lensing (WL) and its combination with galaxy clustering (3 \(\times\) 2-pt) from DES have been used to set a lower limit on a single axion being all the DM for \(m_a > 10^{-23}\,\mathrm{eV}\) (Fig.~\ref{fig:limits}) \citep{Dentler:2021zij}. Extensions to the more general case of a mixture of axion and cold DM require fully non-linear modelling up to \(k \sim 1\,h\,\mathrm{Mpc^{-1}}\) including the gravitational interaction between ULAs and CDM and galactic feedback. This can be achieved either through hydrodynamical simulations and/or a semi-analytical halo model, with prospects for next-generation photometric observatories \textit{Euclid}, \textit{Rubin} and \textit{Roman} probing up to \(m_a \sim 10^{-20}\,\mathrm{eV}\) \citep[for a \textit{Rubin} galaxy weak lensing forecast, see][]{Preston:2025tyl}. Jeans suppression of DM halos also changes the UV luminosity function (UVLF) of high-redshift galaxies \citep{Bozek:2014uqa}. The UVLF from the Hubble Space Telescope limits ULAs to be no more than 22\% of the DM for \(10^{-26}\,\mathrm{eV} < m_a < 10^{-23}\,\mathrm{eV}\) (Fig.~\ref{fig:limits}) \citep{Winch:2024mrt}, with further expected sensitivity as the James Webb Space Telescope observes fainter and higher-redshift galaxies \citep{Sipple:2024svt}. Intensity mapping (IM) of emission lines like the 21 cm neutral hydrogen transition by surveys like HIRAX \citep{Newburgh:2016mwi} and the Square Kilometre Array \citep{Weltman:2018zrl} will trace small-scale DM fluctuations with high precision (see Fig.~\ref{fig:limits}) \citep{Bauer:2020zsj}. Measurement of the velocity acoustic oscillation feature in the 21 cm power spectrum with HERA \citep{HERA:2021noe} could probe \(m_a \sim 10^{-18}\,\mathrm{eV}\) \citep{Hotinli:2021vxg}.

For \(m_a > 10^{-23}\,\mathrm{eV}\), the Lyman-alpha forest (Ly-\(\alpha\)f), spectral absorption tracing the intergalactic medium (IGM) in filaments and voids, sets the strongest limits on the relic axion density (Fig.~\ref{fig:limits}) \citep{Kobayashi:2017jcf,Rogers:2020ltq} by probing the matter power spectrum away from the highest-density regions (halos) affected by strong non-linearities \citep{Croft:1997jf,Viel:2005qj,Irsic:2017yje}. Ref.~\cite{Rogers:2020ltq} sets a lower limit using high-resolution (hi-res) Lyman-alpha forest from the \textit{Keck} and VLT telescopes on a single axion being all the DM for \(m_a > 2 \times 10^{-20}\,\mathrm{eV}\). An analysis of \textit{Planck} CMB, baryon acoustic oscillations, supernovae magnitudes and eBOSS Lyman-alpha forest data finds a preference for a (1 - 5)\% DM contribution of axions with \(m_a \sim 10^{-25}\,\mathrm{eV}\) \citep{Rogers:2023upm}. Larger sets of spectra (including from DESI, PFS, WEAVE, 4MOST and future Extremely Large Telescopes), coupled with control over systematic effects \citep[e.g., from spatial reionisation fluctuations in simulations,][]{Fernandez:2023grg}, are expected to increase sensitivity to lower axion densities. However, it will be challenging to probe heavier axions as their wave effects will manifest below the pressure smoothing scale of the IGM.

A promising avenue is to trace sub-structure within halos, which will have a suppressed sub-halo population (from Jeans suppression) with the wave-like features (interference, solitons, granules) discussed above. The luminosity function of Milky Way (MW) satellite galaxies from DES and Pan-STARRS sets a lower limit \(m_a > 2.9 \times 10^{-21}\,\mathrm{eV}\) (Fig.~\ref{fig:limits}) \citep{DES:2020fxi}. The combination of sub-structure in strong gravitational lenses \citep{Hsueh:2019ynk,Laroche:2022pjm,2023MNRAS.524L..84P} and sub-halos perturbing MW stellar streams as observed by Gaia and Pan-STARRS \citep{Banik:2019smi} sets competitive limits \citep{Schutz:2020jox}. However, ignoring axion wave effects in the analysis of stellar streams is incorrect \citep{Dalal:2020mjw}. Future (stage-IV, S4) wide- and deep-field photometric surveys \citep[\textit{Rubin},][]{LSSTDarkMatterGroup:2019mwo} \citep[\textit{Euclid},][]{Euclid:2024yrr} \citep[\textit{Roman},][]{10.1093/mnras/stab1762}, coupled with spectroscopic follow-up \citep[e.g., the VIA survey,][]{Lu:2025qbp}, will discover hundreds of new satellites, thousands of lenses and thousands of stream member stars with projected axion sensitivity up to \(m_a \sim 10^{-18}\,\mathrm{eV}\). Galaxy rotation curves from the SPARC database are used to exclude ULA DM for \(m_a \sim 10^{-22}\,\mathrm{eV}\) by the absence of soliton cores \citep{Bar:2018acw,Bar:2021kti}. Refs.~\cite{Marsh:2018zyw} and \cite{Dalal:2022rmp} exclude ULA DM for \(m_a \sim 10^{-19}\,\mathrm{eV}\) by the absence of dynamical heating in low-mass dwarf galaxies from ULA wave effects. These latter analyses are sensitive to the detailed modelling of the interaction of ULA dynamics and galaxy formation.

\subsubsection{Open questions}

Figure \ref{fig:limits} illustrates that, with current cosmological data, we are already probing the relic density of axions for twelve orders of magnitude in mass across the ultra-light regime. Further, it is forecast that we can significantly extend our sensitivity in two aspects: smaller densities and higher masses. In particular, we have the prospect of probing the theoretically-appealing parameter space with GUT-scale \(f_\mathrm{a}\). To achieve this sensitivity, we must combine information from CMB, large-scale structure (Lyman-alpha forest, weak lensing, IM) and sub-structure (MW satellites, streams, lenses) probes. These analyses require the complex modelling of ULA Jeans suppression and wave effects in different astrophysical systems. This effort will require dedicated programmes of ULA simulations combining full Schr\"odinger-Poisson solvers, approximate methods using modified initial conditions and detailed hydrodynamics of, e.g., galactic feedback and cosmic reionisation. In parallel, we must develop ML methods to compare simulations to data (emulators, SBI) and more efficient semi-analytical models (e.g., EFT of LSS, halo model). In order to distinguish ULA effects convincingly from other DM models that suppress small-scale power, e.g., warm or interacting DM, we must accurately model explicit wave effects and find observables that probe these effects (e.g., dynamical heating in stellar streams or halos).

Another future direction of interest is to consider extensions beyond the simplest ultra-light scalar with a quadratic potential that is considered here. Extreme axions drop the assumption that the initial field displacement is small, finding that power enhancement can occur at certain scales, potentially relaxing some of the above bounds \citep{Winch:2023qzl}. Above, we ignored possible non-gravitational self-interactions in the axion field which will introduce an additional pressure term in the equations of motion and change the behaviour of the soliton \citep{Chavanis:2011zi,Glennon:2023gfm}. Further, it is well motivated from fundamental theory to have more than one axion field, two or more of which may lie in the ultra-light mass range. It has already been shown in early simulations that the wave effects can differ significantly in this setting \citep{Luu:2023dmi}. We also restricted ourselves to a quadratic potential whereas, in general, the full potential is cosine and it is already known that different ALP potentials can drive very different cosmological evolution, e.g., early dark energy \citep{Poulin:2018cxd}. Moreover, in fundamental theory, ALPs are accompanied by `dilaton’ scalar fields whose expectation values determine the gauge coupling. A cosmic phase transition of such a dilaton field can lead to an exponential increase in the ALP mass, rendering the true vacuum kinematically inaccessible to any initial ALP population. This way, ALPs become trapped and compressed into ``axion relic pockets'' \cite{Carenza:2024tmi}: regions of false vacuum stabilised from collapse by the pressure of the kinematically trapped, hot ALP gas. Axion relic pockets are naturally long-lived and could comprise all dark matter. Their sizes range from point-like to brick-sized, and their phenomenology is distinct from standard WISP paradigms. Finally, we consider here only ultra-light scalars but ultra-light vectors \citep{Amin:2022pzv} and tensors \citep{Unal:2022ooa} can also be produced. The extra degrees of freedom will again change cosmological phenomena at the quantitative level.

%% file: WG2/content/dark_energy_new.tex
\subsection{ALPs as Dark Energy: Inflaton, Early Dark Energy and Quintessence\\ \textnormal{Authors: E. Ferreira, S. Gasparotto \& I. Obata}}
\label{subsec:darkenergy}

\subsubsection{Introduction and motivation}

The origin of Dark Energy (DE), the component of the universe that drives its accelerated expansion, is one of the greatest challenges of modern physics. From a model-building point of view, a scalar field whose energy density ($\rho_a$) is dominated by the potential term is the simplest way to model the accelerated expansion. 
Among various candidates, the axion, or ALP, stands out as a promising dynamical DE component, potentially explaining both the inflaton field in the early universe and the quintessence field in the late universe. The primary motivation is that the axion, being a pseudo-scalar field, enjoys an approximate shift symmetry which restricts the allowed couplings making it naturally weakly coupled with other fields and protecting the axion mass and potential energy density, which are otherwise UV-sensitive quantities (see Refs~\cite{Frieman_1995, Kolda_1999,Svreck2006} for foundational discussions, and Refs.~\cite{Kim1999,Kim_2003,Choi2000,Kim2009} for initial model proposals). 
Given the different mechanisms where ALPs arise,  their mass range spans several orders of magnitude, from quintessence to the inflationary regime.
This is explained because the axion potential, coming from highly suppressed non-perturbative effects, is exponentially sensitive to the parameters of the theory. The typical axion potentials can be expressed as
\begin{equation}
V(a) = \Lambda^4\left[ 1 - \cos\left(\dfrac{a}{f_a}\right)\right]^n \qquad \text{with}\qquad \Lambda^4=m^2_af^2_a\simeq M^4_{\rm Pl}e^{-S}\ ,
\label{eq:ALP_pot}
\end{equation}
where $S$ is the non-perturbative action.
The potential with an index $n=1$ is a usual cosine form and has been used for inflation~\cite{Freese:1990rb} and late-time dark energy~\cite{Frieman_1995}, and $n>1$ have been used for Early Dark Energy (EDE) models.  
With the potential at hand, the axion evolution in a Friedmann-Lemaître-Robertson-Walker (FLRW) flat Universe is given by the Klein-Gordon equation together with the Friedman equation determining the evolution of the scale factor $R$
\begin{equation}\label{eq:axevol}
\Ddot{a}+3H\Dot{a}=-V'(a) \quad \text{with}\qquad H^2= \left(\frac{\Dot{R}}{R}\right)^2= \frac{\rho_{\rm tot}}{3} .
\end{equation}
When the Hubble term $3H\Dot{a}$, which acts as a drag force, dominates over the potential term $V'(a)={\rm d}V/{\rm d}a$, the field remains almost frozen at its initial value, generically different to the potential minimum. In this over-damped regime, the kinetic energy is suppressed compared to the potential one, and the equation of state (EOS), defined in terms of the pressure $p_a$ and energy density $\rho_a$ of the axion field, is
\begin{equation}\label{eq:EOS}
    w=\frac{p_a}{\rho_a}=\frac{\Dot{a}^2/2-V}{\Dot{a}^2/2+V}\approx -1 + \frac{\Dot{a}^2}{V},
\end{equation}
and this fulfills the condition for an accelerated expansion $w<-1/3$. Note that the EOS $w\geq -1$ and its deviation from the cosmological constant value, i.e. $w_\Lambda=-1$, is proportional to the axion kinetic energy. Eventually, when the Hubble friction term becomes smaller than the potential term, the field exits from this DE regime and begins to oscillate around the potential's minimum, with an EOS and transition time given by the potential chosen and mass of the field. In the quadratic approximation of the potential~\eqref{eq:ALP_pot}, the onset of oscillation started when $H_{\rm osc}\sim m_a$~\cite{Marsh:2015xka}, behaving like dark matter with an effective EOS $\langle w\rangle\simeq 0$.
However, in the simplest scenario of explaining DE with a single axion field without fine-tuning the initial conditions, one typically needs a super-Planckian decay constant $f_a\gtrsim M_{\rm Pl}$ both for inflation~\cite{Freese:2004un} and late DE~\cite{Kaloper:2005aj}. 
Such a large decay constant may spoil a global symmetry by a quantum gravity effect~\cite{Kallosh:1995hi}, or be challenging to be realized from controlled string models~\cite{Banks:2003sx}. 
Although the minimal scenario presents some challenges, the axions remain one of the best-motivated candidates and many different models have addressed these difficulties and found successful ways out either in inflation or late DE regime, as we discuss in the next sections. Moreover, the axion EDE models may alleviate some of the modern puzzles in cosmology as the Hubble tension.  

This chapter is structured to examine the axion role across different cosmological epochs, from more massive fields potentially active during inflation to scenarios involving early or late DE models. In each section, we discuss the theoretical models, their observable consequences, their current bounds, and prospects. 

\vspace{0.5cm}
\subsubsection{ALPs in the early universe}
ALP is a (pseudo) Nambu-Goldstone boson that naturally arises in theories with spontaneously broken symmetries, making it an elegant and theoretically motivated candidate for the inflaton; called natural inflation \cite{Freese:1990rb}.
However, for successful natural inflation scenarios, a super-Planckian decay constant is required \cite{Freese:2004un}; hence, several extensions of natural inflation have been developed to resolve this issue. One is a scenario of multiple fields realizing an effective flat direction due to an alignment of cosine potentials with sub-Planckian ALP fields \cite{Kim:2004rp,Choi:2014rja,Higaki:2014pja,Kappl:2014lra,Peloso:2015dsa}, N-flation mechanism \cite{Liddle:1998jc,Dimopoulos:2005ac,Easther:2005zr}, hierarchies between mass/couplings \cite{Ben-Dayan:2014zsa,Kehagias:2016kzt}, considering kinetic and/or Stueckelberg mixing \cite{Shiu:2015uva,Shiu:2015xda,Fonseca:2019aux,Choi:2019ahy}, or friction forces due to the non-minimal coupling to gravity \cite{Germani:2010hd}, the backreaction of the gauge field \cite{Anber:2009ua,Notari:2016npn,Ferreira:2017lnd}, and so on.
    There have been many studies to construct such inflationary scenario from string compactifications and supergravity models \cite{Long:2014dta,Ben-Dayan:2014lca,Gao:2014uha,Grimm:2007hs,Cicoli:2014sva,Kenton:2014gma,Abe:2014xja,Palti:2015xra,Kallosh:2014vja}.

    Another attempt to this caveat is to break a discrete shift symmetry of ALP background: $a \rightarrow a + 2\pi f_a$.
    One such scenario is a monodromy mechanism \cite{Silverstein:2008sg,McAllister:2008hb,Kaloper:2008fb,Flauger:2009ab,Berg:2009tg,Kaloper:2011jz,Kaloper:2014zba,Marchesano:2014mla,Blumenhagen:2014gta,Hebecker:2014eua,Blumenhagen:2014nba,Kaloper:2016fbr,DAmico:2021vka}, where the ALP discrete shift symmetry is broken by a wrapping brane or flux compactifications in string theory and ALP gets a power-law potential.
    The idea of breaking ALP shift symmetry has been also introduced in the context of a relaxion mechanism \cite{Graham:2015cka,Espinosa:2015eda,Hardy:2015laa,Patil:2015oxa,Jaeckel:2015txa,Gupta:2015uea,Batell:2015fma,Flacke:2016szy,Fonseca:2019lmc, Chatrchyan:2022pcb}, where the dynamics of ALP (or we could say it QCD-like axion) field scan the Higgs mass and solve a hierarchy problem of electroweak theory. 
    For an interplay between ALP and Higgs field in this mechanism, a large field excursion of the relaxion field over Planck scale is required. Therefore, to relax this issue, several theoretical frameworks such as Clockwork axion models \cite{Choi:2015fiu,Kaplan:2015fuy,Giudice:2016yja,Coy:2017yex,Long:2018nsl,Agrawal:2018mkd} or introducing backreaction of gauge quanta \cite{Hook:2016mqo,Choi:2016kke,Tangarife:2017rgl, Fonseca:2019lmc} have been suggested.
    
    Recent CMB observations disfavor natural inflation or simple power-law potentials from axion monodromy inflation \cite{Planck:2018jri}.
    From a point of view of UV completion, however, quantum gravity effects such as gravitational instanton \cite{Giddings:1987cg,Montero:2015ofa,Hebecker:2016dsw,Alonso:2017avz} or application of Weak Gravity Conjecture \cite{Arkani-Hamed:2006emk} may not be overlooked in many of axion inflationary models \cite{Rudelius:2014wla,delaFuente:2014aca,Rudelius:2015xta,Brown:2015iha,Bachlechner:2015qja,Hebecker:2015rya,Brown:2015lia,Junghans:2015hba, Heidenreich:2015wga} or axion monodromy \cite{Hebecker:2015zss,Ibanez:2015fcv}.
    After all, it may force to modify the original ALP potentials and such extended inflationary potentials may fit with Planck constraint \cite{Kobayashi:2015aaa,CaboBizet:2016uzv,Nomura:2017ehb,Nomura:2017zqj}.
    Intriguingly, it potentially predicts a modulation in the potential and generates an oscillatory scalar power spectrum \cite{Kappl:2015esy,Choi:2015aem}. This feature may give us interesting observables such as primordial black holes and dark matter on small scales \cite{Ozsoy:2020kat}.
    For a review of ALP inflation in string theory, see \cite{Cicoli:2023opf}.

    The parity-odd nature of ALP may drive rich cosmological phenomena related to parity-violation, depending on the dynamics of its coupling to the matter sector, such as the gauge field.
    The gauge field is conformal-invariant and dilutes away due to the expansion in the standard cosmological scenario. However, in the presence of topological coupling to ALP during inflation
    \begin{equation}
    \mathcal{L} \supset -\dfrac{\alpha}{4}\dfrac{a}{f_a}F\tilde{F} \ ,
    \end{equation}
    one helicity mode of gauge quanta experiences a tachyonic instability around horizon crossing and is non-adiabatically amplified.
    This particle production provides a variety of cosmological observables, such as the generation of primordial gravitational waves.
    
    The magnitude of particle production depends on the background motion (velocity field) of ALP and a mechanism of enhancing other cosmological perturbations is different between models of ALP coupled to Abelian and non-Abelian fields. 
    We will describe those two classes of inflationary models with ALP-gauge dynamics below.
    
    \paragraph{Abelian model.} Abelian ($U(1)$) gauge field $F_{\mu\nu} = \partial_\mu A_\nu - \partial_\nu A_\mu$ sources coupled cosmological perturbations at second order level.
    The equation of motion for the gauge field in Fourier space is written as follows:
    \begin{equation}
    \left[\dfrac{d^2}{d\eta^2} + k^2 \pm k\dfrac{\alpha}{f_a}\dfrac{da}{d\eta}\right]A_\pm = 0 \ .
    \end{equation}
    Recalling $\eta = -(HR)^{-1} \rightarrow 0$ in de Sitter spacetime, the effective square mass term for one of two helicity states becomes negative and the corresponding mode function grows exponentially: $A_+ \propto e^{\pi\xi}$ where $\xi \equiv \alpha\dot{a}/(2fH)$ is a dimensionless speed of background ALP field.
    Then, one helicity mode sources tensor modes $A_+ \times A_+ \rightarrow h_{ij}$ via one-loop interaction and predicts parity-violating gravitational waves \cite{Sorbo:2011rz}.
    Subsequent studies investigated the detectability of scale-dependent GWs sourced by ALP-gauge dynamics \cite{Barnaby:2011qe,Cook:2011hg,Mukohyama:2014gba,Namba:2015gja,Domcke:2016bkh,Garcia-Bellido_2016,Obata:2016oym,Ozsoy:2020ccy}.
    This helicity mode also sources a curvature perturbation at second order of the gauge field $A_+ \times A_+ \rightarrow \zeta$ and predict a large non-Gaussianity of curvature perturbation \cite{Barnaby:2010vf,Barnaby:2011vw,Barnaby:2012xt,Cook:2013xea}.
    Therefore, to meet a constraint on the non-detection of non-Gaussianity in CMB observations, a model with a large amount of GWs and small scalar non-gaussianity has been developed \cite{Barnaby:2012xt}.
    One such scenario is a model of axion as a spectator field during inflation, where the curvature perturbation is sourced only via a gravitational interaction and its generation is therefore much suppressed \cite{Ferreira:2014zia}.
    Not only on large scales, depending on the background motion of ALP on its potential, particle production occurs at an intermediate stage during inflation, on scales much smaller than CMB. 
    Then, it may lead to a generation of primordial black holes and the secondary-induced gravitational waves after inflation \cite{Linde:2012bt,Garcia-Bellido:2017aan}, giving an implication for a recent signal of gravitational wave background signals in pulsar-timing measurements \cite{Niu:2023bsr,Unal:2023srk}.
Regarding theoretical consistency of such models, to evaluate the magnitude of GW spectrum for a large value of axion velocity field, an accurate treatment for perturbativity \cite{Ferreira:2015omg,Peloso:2016gqs} and backreaction of the gauge field \cite{Cheng:2015oqa,Sobol:2019xls,Domcke:2020zez,Gorbar:2021rlt,Gorbar:2021zlr,Durrer:2023rhc,vonEckardstein:2023gwk} are important.
    After things become a fully non-linear system, lattice simulation will be crucial as a non-perturbative approach to solving such ALP-gauge dynamics during inflation \cite{Caravano:2022epk}.

    The grown vector field by ALP inflation might also explain a primordial magnetic field in our Universe \cite{Garretson:1992vt,Anber:2006xt,Caprini:2014mja,Fujita:2015iga,Anber:2015yca,Adshead:2016iae,Caprini:2017vnn,Fujita:2019pmi,Patel:2019isj}. The generated magnetic field is helical. Hence, even if the helical magnetic field was generated on microphysical scales, the energy is transferred to large scales and its coherent length grows to cosmological scales by an inverse cascade process due to a conservation of magnetic helicity in the primordial plasma \cite{Son:1998my,Field:1998hi,Vachaspati:2001nb,Sigl:2002kt,Campanelli:2004wm}.
    The large scale helical magnetic field also potentially provides a baryon asymmetry in our Universe \cite{Fujita:2016igl,Jimenez:2017cdr}. 
    Recently, however, it was pointed out that any magnetogenesis scenario beyond the electroweak scale overproduces a baryon isocurvature and therefore should be ruled out \cite{Kamada:2020bmb}.
     
   \paragraph{Non-Abelian model.} In this scenario, originally  known as gauge-flation \cite{Maleknejad:2011jw} or chromo-natural inflation \cite{Adshead:2012kp} (\cite{Maleknejad:2012fw} for the review), the non-Abelian (SU(2)) gauge field $F^a_{\mu\nu} \equiv \partial_\mu A^a_\nu - \partial_\nu A^a_\mu + g\epsilon^{abc}A^b_\mu A^c_\nu$ has an isotropic and homogeneous vacuum expectation value (vev)
    \begin{equation}
        \langle A^a_i \rangle = R(t)Q(t)\delta^a_i, \,\,
    \end{equation}
    and acts as a new friction force even with a steep ALP cosine potential, namely with a sub-Planckian decay constant $f_a < M_{\rm Pl}$, and supports inflationary attractor solution.
    This kind of isotropic configuration is possible from $SU(2)$ algebra or even larger $SU(N)$ gauge group \cite{Fujita:2021eue,Fujita:2022fff,Murata:2022qzz}, and is a stable attractor solution against anisotropic background \cite{Maleknejad:2011jr,Maleknejad:2013npa,Wolfson:2020fqz,Wolfson:2021fya}.
    
    One of the interests in this scenario is that, due to the presence of SU(2) vev, fluctuations of the gauge field have a tensor mode $\delta A^a_i \supset T_{ai}$. Namely, it sources gravitational waves at tree-level, and therefore the generation of scalar mode does not correlate with that of tensor modes \cite{Maleknejad:2011sq,Dimastrogiovanni:2012ew}. The amplification of gravitational waves is characterized by a model parameter $m_Q \equiv g Q/H$ which is associated with the effective mass of the background gauge field. By perturbation analysis, however, chromo-natural inflation was found to be inconsistent with CMB observation because it provides a too large tensor-to-scalar ratio for an allowed value of scalar spectral index \cite{Adshead:2013qp,Adshead:2013nka,Namba:2013kia}. Subsequent studies expanded this original scenario to be consistent with observation, such as ALP as spectator field \cite{Dimastrogiovanni:2016fuu,Iarygina:2021bxq}, happening chromo-natural inflation on small scales \cite{Obata:2014loa,Obata:2016tmo,Obata:2016xcr,Domcke:2018rvv}, replacing cosine potential with an arbitrary potential realizing a small $SU(2)$ vev \cite{Maleknejad:2016qjz}, introducing Higgs mechanism \cite{Adshead:2016omu,Adshead:2017hnc}, a non-minimal coupling to gravity \cite{Dimastrogiovanni:2023oid}, and so on.
    The resultant scale-dependent tensor power spectrum \cite{Fujita:2018ndp} is potentially testable with CMB, PTA, or interferometer scales \cite{Thorne:2017jft,Campeti:2020xwn,Campeti:2022acx} (see \cite{Komatsu:2022nvu} for the review).
    Moreover, due to the uncorrelation between the generation of tensor modes and that of scalar modes, this model allows to generate a large tensor non-Gaussianity such as tensor-tensor-tensor \cite{Agrawal:2017awz,Agrawal:2018mrg}, scalar-tensor-tensor \cite{Dimastrogiovanni:2018xnn,Fujita:2018vmv,Ozsoy:2021onx}, or 4-point tensor trispectrum \cite{Fujita:2021flu}. 
    Still, for a large value of $m_Q$ leading to a large tensor non-Gaussianity, the curvature perturbation is also sourced non-linearly by the gauge field and violates the CMB constraint \cite{Papageorgiou:2018rfx,Papageorgiou:2019ecb}. 
    Similar to the case of the Abelian model, the backreaction effect of gauge quanta to the inflationary background dynamics becomes relevant and a careful treatment becomes necessary \cite{Fujita:2017jwq,Maleknejad:2018nxz,Ishiwata:2021yne,Iarygina:2023mtj}.
    
\vspace{0.5cm}

    The resultant primordial gravitational waves are chiral and hence potentially predict the parity-odd CMB angular correlation function \cite{Lue:1998mq,Saito:2007kt,Gluscevic:2010vv}.
    Since current CMB measurements are yet to detect the tensor-to-scalar ratio and accordingly do not have much sensitivity to constrain the chirality at the level of two-point power spectrum, testing higher-order statistics with clean information for the parity-odd correlations will become important in future surveys \cite{Shiraishi:2013kxa,Shiraishi:2014ila,Gerbino:2016mqb,Shiraishi:2016yun,Shiraishi:2019yux}.
    In addition to gravitational waves, these models predict the parity-odd primordial density fluctuation.
    Contrary to the tensor modes, however, it is a directionless scalar quantity and hence does not have parity-odd information in 2- and 3-point correlation functions. We can access its parity-odd nature from the 4-point correlation function, and recently the parity-odd trispectum from the ALP inflation has been developed \cite{Niu:2022fki,Fujita:2023inz}. This might be measurable with the parity-odd 4-point function in CMB \cite{Shiraishi:2016mok,Philcox:2023ffy,Philcox:2023ypl} or large-scale structure surveys \cite{Hou:2022wfj,Philcox:2022hkh,Cabass:2022oap}.

    After inflation ends, the ALP field starts to oscillate at the bottom of the potential.
    In the presence of ALP-gauge coupling, the ALP dynamics could trigger a resonant production of gauge fields and lead to preheating \cite{Adshead:2015pva}.
    Such a gauge preheating predicts a gravitational wave spectrum \cite{Adshead:2018doq,Adshead:2019lbr,Adshead:2019igv}, with a frequency peaked at MHz-GHz.
    Although testing such a high-frequency gravitational wave background is challenging, ALP preheating may give us another gravitational wave source. 
    For example, if the ALP field behaves as dark matter with mass $10^{-14}$ eV, the gauge field resonant process could generate nano-frequency gravitational wave background \cite{Kitajima:2020rpm}.
    Even in the absence of gauge sector, owing to a specific curvature of ALP potential, an oscillatory motion of ALP field could predict scalar field fragmentation, generating localized objects called oscillons \cite{Bogolyubsky:1976nx,Gleiser:1993pt,Copeland:1995fq,Kasuya:2002zs,Amin:2011hj}.
    Then, gravitational waves are emitted through their formation process \cite{Zhou:2013tsa,Hong:2017ooe,Soda:2017dsu,Kitajima:2018zco,Kawasaki:2019czd,Kawasaki:2020jnw,Lozanov:2022yoy,Lozanov:2023knf}.
    Interestingly, depending on the mass of the ALP field, the resultant gravitational wave spectrum has a peak on broad frequency regimes.
   
\subsubsection{ALPs as early dark energy}

One of the contexts in which ALPs are also used in cosmology is to generate models with an early dark energy period. EDE is a class of models where there is a period in the evolution of the universe where DE has a sizeable contribution to the energy density budget of the universe, much earlier than in the present. This is given by either having an additional component in the universe that behaves like DE or from phenomenological models where DE energy density is important at different epochs.

This idea first appeared in the context of quintessence-like models where ALPs are the quintessence field~\cite{Albrecht:1999rm,Doran:2000jt,Wetterich:2004pv,Doran:2006kp}, leading to phenomenological parameterizations of DE with an early period where DE has a large density. 

The EDE period can be pre- or post-recombination, with pre-recombination models being more prevalent since the current observations impose strong constraints on having a sizable post-recombination EDE~\cite{Knox:2019rjx}.

Recently EDE models have received renewed interest in light of the Hubble tension. The Hubble ($H_0$) tension ~\cite{Knox:2019rjx,DiValentino:2020zio,Schoneberg:2021qvd,Kamionkowski:2022pkx,Verde:2023lmm} is the name given to the discrepancy between direct and indirect measurements of the Hubble constant, the current rate of expansion of our universe. This tension reaches $5 \sigma$ when comparing the $H_0$ value inferred from the Cosmic Microwave Background (CMB) for the $\Lambda$CDM model using data from the \textit{Planck} satellite~\cite{Planck:2018vyg} and the direct measurement from the Cepheid-calibrated Type Ia supernovae (SNIa) of the SH0ES collaboration~\cite{Riess:2021jrx}. The origin of this tension is still unknown with the possibility of being systematics in one of the measurements or, more excitingly, new physics beyond the $\Lambda$CDM model. If one assumes that we need new physics to solve this tension, we can try to change the physics at early or late times in order to get inferred values of $H_0$ that match the direct measurements. EDE was proposed as an early-time solution~\cite{Karwal:2016vyq,Poulin:2018cxd,Poulin:2018dzj} (see also the reviews \cite{Kamionkowski:2022pkx,Poulin:2023lkg,McDonough:2023qcu}) and considered as one of the most promising models for this task~\cite{Knox:2019rjx,Schoneberg:2021qvd}. To solve the tension, we need an additional DE component that contributes to around $10\%$ of the total energy density after matter-radiation equality, rapidly decaying before recombination. This EDE phase decreases the size of the sound horizon at recombination leading to an increase in $H_0$, in principle enough to make the inferred value of $H_0$ from \textit{Planck} compatible with the SH0ES value.

\paragraph{EDE models with ALPs.} The simplest way to implement a DE component that acts at early times is to use very light scalar fields, such as ALPs. The most studied EDE model is given by an ALP field with potential~(\ref{eq:ALP_pot}), a generalization of the axion potential allowing different index $n$. This is also known as the `canonical' EDE model.
The mass of the ALP dictates when this field will behave like DE and for a pre-recombination EDE, $m_a \lesssim H_{\mathrm{eq}}$, where $H_{\mathrm{eq}}$ is the Hubble parameter at matter-radiation equality. The shape of the potential dictates how fast DE decays and $n=3$ presents the best fit to the data, allowing larger values of EDE and given its faster decay~\cite{Smith:2019ihp, Poulin:2023lkg}. To better describe its cosmological evolution, this model can be rewritten with phenomenological quantities: the maximum fraction of EDE $f_{\mathrm{EDE}} = \Omega_{\mathrm{EDE}} (z_c)$ at a certain redshift $z_c$, the initial value of the ALP field $\theta_{\rm i} = a_{\rm i}/f$ and $n$. 

There are many variants of the EDE model.
In the class of quintessence-like EDE models, we have ALPs with different potentials~\cite{Yin:2020dwl,Ye:2020btb,Albrecht:1999rm, Adil:2022hkj}, like the Rock ’n’ Roll EDE with a power-law potential~\cite{Agrawal:2019lmo}; Acoustic DE~\cite{Lin:2019qug}; or scaling EDE~\cite{Gomez-Valent:2021cbe, Doran:2000jt, Doran:2006kp, Pettorino:2013ia}, and tracker EDE using exponential potentials or K-essence ALP field~\cite{Copeland:2023zqz} (for a review of dynamical system analysis of DE see~\cite{Wetterich:1987fm,Copeland:1997et,Barreiro:1999zs}).
The predictions of these EDE models can be obtained by Boltzmann codes like \texttt{AxionCAMB}~\cite{Hlozek:2014lca}\footnote{\href{https://github.com/dgrin1/axionCAMB}{https://github.com/dgrin1/axionCAMB}}, \texttt{AxiCLASS}~\cite{Smith:2019ihp,Poulin:2018dzj}\footnote{\href{https://github.com/PoulinV/AxiCLASS}{https://github.com/PoulinV/AxiCLASS}} and \texttt{CLASS EDE}~\cite{Hill:2020osr}\footnote{\href{https://github.com/mwt5345/class_ede}{https://github.com/mwt5345/class\_ede}}.

Another class of EDE models is called new EDE (NEDE)~\cite{Niedermann:2023ssr}, where a scalar field ($M \sim \mathrm{eV}$), undergoes a phase transition between BBN and recombination which can be triggered by an ultra-light scalar field ($m_a \sim 10^{-27} \, \mathrm{eV}$), in the Cold NEDE model~\cite{Niedermann:2019olb,Niedermann:2020dwg}, or by the temperature of the dark sector, the Hot NEDE~\cite{Niedermann:2021vgd,Niedermann:2021ijp}. There is also a Boltzmann code to calculate the predictions of this model called \texttt{TriggerCLASS}~\cite{Niedermann:2020dwg}\footnote{\href{https://github.com/NEDE-Cosmo/TriggerCLASS}{https://github.com/NEDE-Cosmo/TriggerCLASS}}.

There are also efforts to construct UV complete ALP-like EDE models from string theory~\citep{Alexander:2019rsc,Cicoli:2023qri,McDonough:2022pku}, string-inspired Chern-Simons gravity~\cite{Guendelman:2022cop,Gomez-Valent:2023hov,Gomez-Valent:2024tdb,Braglia:2020bym} or higher order instanton corrections~\citep{Kappl:2015esy}.

One can also couple the EDE field to other particles. EDE coupled with neutrino was investigated in~\cite{Sakstein:2019fmf,CarrilloGonzalez:2020oac,deSouza:2023sqp,CarrilloGonzalez:2023lma}.  Alternatively, EDE can couple to DM, studied in different contexts~\cite{Karwal:2021vpk, McDonough:2021pdg, Lin:2022phm,Gomez-Valent:2022bku,Liu:2023kce,Liu:2023wew,Garcia-Arroyo:2024tqq}; or in the Waterfall DE model~\cite{Talebian:2023lkk}. In~\cite{Berghaus:2019cls} an ALP field couples to a dark non-Abelian gauge group, mimicking EDE. There are also models where the same ALP field is responsible for EDE and for late DE~\cite{Chowdhury:2023opo,Ramadan:2023ivw}.

Some of the variants cited here are proposed to address some shortcomings of the phenomenological EDE model, like the fine-tuned shape of the potential of the canonical model or the specific values of $n$, $\theta_i$ and $z_c$ necessary to solve the tension. Some of these extensions aim to address other tensions in cosmology, like the $S_8$ tension~\cite{DiValentino:2020vvd} for EDE coupled to DM or with massive neutrinos~\cite{Reeves:2022aoi}. 

The list of EDE models presented here is not extensive and given the scope of this review, we focused on models that use ALPs to describe the EDE component.

\paragraph{Observables and constraints.} 

Having a non-negligible DE component at early times, either pre- or post-recombination, changes the evolution of the universe and this change can be measured with cosmological observations. 

Previous to the Hubble tension, in~\cite{Doran:2000jt,Wetterich:2004pv} it was pointed out that CMB can be used to put strong constraints in post-recombination EDE models. CMB anisotropies are sensitive to the fraction of EDE at last scattering. The presence of EDE can influence the CMB peaks and its separation is related to the amount of DE before recombination, and lead to a suppression of structure growth that can be constrained with CMB and LSS.  
In~\cite{Doran:2006kp} using CMB data from WMAP~\cite{WMAP:2003elm}, the Very Small Array~\cite{Dickinson:2004yr}, Cosmic Background Imager~\cite{Readhead:2004gy} and Boomerang~\cite{Montroy:2005yx};  LSS data from the Sloan Digital Sky Survey (SDSS)~\cite{SDSS:2003eyi}; and SNIa data from~\cite{SupernovaSearchTeam:2004lze}, find that for EDE has to be less than $4\%$ if present around $z \sim 10$. Later, in~\cite{Pettorino:2013ia} using CMB data from WMAP and the South Pole Telescope (SPT)~\cite{Simard:2017xtw} find $f_\mathrm{EDE}< 5\%$ for $z \lesssim 100$
More recently, in~\cite{Gomez-Valent:2021cbe} using data from Planck 2018~\cite{Planck:2018vyg}, the Pantheon SNIa~\cite{Pan-STARRS1:2017jku}, a prior on the absolute magnitude of SNIa by SH0ES~\cite{Riess:2020fzl}, and weak lensing data from KiDS+VIKING-450 and DES-Y1~\cite{Joudaki:2019pmv}, found that $f_\mathrm{EDE} \lesssim 1.5\%$ for $z\in (100,1000)$ at $2\sigma$ CL, alleviating the $H_0$ tension.

The most well-studied and constrained EDE models are the ones proposed to solve the Hubble tension. 
In this section, we are going to focus mostly on the canonical EDE model. The first constraints in EDE in the context of the Hubble tension~\cite{Poulin:2018cxd,Smith:2019ihp} used data from \textit{Planck}, BAO \cite{Beutler:2011hx,Ross:2014qpa,Mueller:2016kpu}, SNIa \cite{Pan-STARRS1:2017jku}, and $H_0$ from the SH0ES collaboration \cite{Riess:2018byc, Riess:2019cxk}, finding $f_{\text{EDE}} \sim [5\%, 10\%] $ at $z_c \sim [5000, 3700]$, respectively, showing that EDE can solve the $H_0$ tension. 

However, as pointed out in~\cite{Hill:2020osr}, a large $f_\mathrm{EDE}$ increases the early Integrated Sachs-Wolfe effect, and to maintain a good fit to CMB we need a higher dark matter density $\Omega_\mathrm{cdm}$, and the clustering amplitude $\sigma_8$~\cite{Smith:2019ihp, Vagnozzi:2021gjh, Ye:2021nej, Pedreira:2023qqt}. LSS observations constrain the $\Omega_\mathrm{cdm}$ and $\sigma_8$, and including these observations strongly bounds $f_\mathrm{EDE}$ to small values that do not solve the Hubble tension. This was shown in many studies: in ~\cite{Hill:2020osr} for \textit{Planck} data, weak lensing data from DES-YI 3x2pt~\cite{DES:2017myr}, and $S_8$ priors from KiDS and HSC \cite{Hildebrandt:2016iqg,Hildebrandt:2018yau,HSC:2018mrq}, they find $f_\mathrm{EDE} > 0.060$ and $H_0 = 68.92^{+0.57}_{-0.59}\, \mathrm{km/s/Mpc}$; 
for a full-shape analysis of BOSS data \cite{BOSS:2015npt,BOSS:2015fqm,Beutler:2011hx,Ross:2014qpa,eBOSS:2019ytm,eBOSS:2019qwo} different authors~\cite{Ivanov:2020ril, DAmico:2020ods} find $f_\mathrm{EDE} < [0.053, 0.08]$ and $H_0 = [68.73^{+0.42}_{-0.69}, 68.57^{+0.48}_{-1.0}]\, \mathrm{km/s/Mpc}$, respectively. \cite{Gsponer:2023wpm} does a similar analysis using Gaussian and Jeffreys priors, obtaining  $f_\mathrm{EDE} = [0.1179^{+0.025}_{-0.022}, 0.1399^{+0.023}_{-0.022}]$ and $H_0 = [71.73^{+0.82}_{-0.86}, 72.03^{+0.82}_{-0.87}]\, \mathrm{km/s/Mpc}$. 
Although it is still true that the LSS observations strongly constrain the EDE model, it was pointed out in~\cite{Murgia:2020ryi, Smith:2020rxx} that the statistical analysis above had strong prior volume effects which were affecting the constraints on $f_\mathrm{EDE}$ and $H_0$, leading to lower values of $H_0$ from this statistical artifact. This was shown explicitly in~\citep{Herold:2021ksg, Herold:2022iib} (later in \cite{Gomez-Valent:2022hkb}) using the frequentist approach based on the profile likelihood for \textit{Planck} CMB data, and BOSS full-shape power spectrum finding $f_\mathrm{EDE} = 0.087 \pm 0.037$ and $H_0 = 70.57 \pm 1.36 \,\mathrm{km/s/Mpc}$ which is statistically compatible with SH0ES within $1.4\sigma$. The role of priors in the previous analysis was also studied in~\citep{Gsponer:2023wpm}.
For NEDE this was shown in~\cite{Niedermann:2020dwg}, finding $H_0 = 69.56^{+1.16}_{-1.29} \, \mathrm{km/s/Mpc}$ and $f_\mathrm{NEDE} = 0.076^{+0.040}_{-0.035}$ at $ 68 \% $  c.l. for \textit{Planck}, BAO and Pantheon data; or also with additional LSS~\cite{Niedermann:2020qbw, Cruz:2023cxy} and ground-based CMB data~\cite{Poulin:2021bjr, Haridasu:2022dyp, Cruz:2022oqk}, showing equivalent results.

This shows that not only does the EDE model have a complex parameter structure that plays a role in this discrepancy, but that the data sets available for these analyses were not constraining enough to properly bound all model parameters. The frequentist and the Bayesian analysis should agree when enough data is available.

A result that points in this direction is in \cite{Efstathiou:2023fbn} analyzing EDE using the alternative \textit{Planck} likelihood
\texttt{CAMSPEC}~\cite{Efstathiou:2019mdh} with the \textit{Planck} PR4 maps \cite{Planck:2020olo}, plus BAO data from BOSS, SDSS, and 6dFGS \cite{Beutler:2011hx} and SNIa from Pantheon+, finding that the frequentist and Bayesian statistical analysis yield compatible constraints, namely $H_0=68.37 \pm 0.75\,\mathrm{km/s/Mpc}$ and $f_\mathrm{EDE}<0.094$ and
$H_0=68.11^{+0.47}_{-0.82}\,\mathrm{km/s/Mpc}$ and $f_\mathrm{EDE}<0.061$, repctively. This analysis is robust to prior volume effects and shows that for this data set the EDE model is in tension with SH0ES at $3.7\sigma$. A similar result was found in \cite{McDonough:2023qcu} using the \texttt{Hillipop}~\cite{Tristram:2023haj} likelihood.

There are other constraints in the 'canonical' EDE model.  The Atacama Cosmology Telescope (ACT) DR4 data \cite{ACT:2020frw} together with \textit{Planck} large scale, and BAO yields $f_\mathrm{EDE}=0.091^{+0.020}_{-0.036}$ and $H_0=70.9^{+1.0}_{-2.0}\,\mathrm{km/s/Mpc}$  \cite{Schoneberg:2021qvd,Hill:2021yec,Poulin:2021bjr}; or $f_\mathrm{EDE}=0.158^{+0.015}_{-0.094}$ and $H_0=73.43^{+0.200}_{-0.093}\,\mathrm{km/s/Mpc}$ for ACT, WMAP, BAO, and Pantheon. 
The same preference for high $f_\mathrm{EDE}$ appears for SPT-3G TEEE 2018 \cite{SPT-3G:2021eoc, LaPosta:2021pgm}, and \textit{Planck}~\cite{Planck:2018vyg}. However, it was pointed out in \cite{Benevento:2020fev, Camarena:2021jlr,Efstathiou:2021ocp,Smith:2023oop} that this preference comes from ACT data, and more analysis of this is required. There are also bounds on EDE coming from Lyman-$\alpha$ forest data from SDSS eBOSS \cite{eBOSS:2018qyj} and MIKE/HIRAS \cite{Viel:2013fqw} in ~\citep{Goldstein:2023gnw} finding $H_0=67.9\pm 0.4  \,\mathrm{km/s/Mpc}$ ($68\%$ CL), although some questions remain on this results given the contested validity of using this data set given it discrepancy with \textit{Planck} data \cite{Rogers:2023upm} even for $\Lambda$CDM.

Other EDE models, with different potentials, were also constrained with observations showing that they can only alleviate the tension~\cite{Smith:2019ihp, Poulin:2023lkg,Agrawal:2019lmo,Gomez-Valent:2021cbe}. 
There are also other observables proposed to constrain EDE or associated effects of axion-like EDE. One example is cosmic birefringence (see Section \ref{sec:Gasparotto} for more details on that), an effect that would be present if EDE is described by an ALP. In~\citep{Choi:2021aze, Murai:2022zur, Eskilt:2023nxm, Yin:2023srb} this was considered showing that EDE ALP is not a good fit to the \textit{Planck} data (however see \citep{Kochappan:2024jyf}). Other probes that were suggested to be used to constraint EDE are: spectral distortions~\citep{Hart:2022agu}, or high-z galaxies from JWST~\citep{Boylan-Kolchin:2021fvy}.

The above analyses show that there is still no consensus if the EDE model can solve the Hubble tension and more data is needed. However, besides the Hubble tension, the presence of an ALP EDE  field with an abundance of $\mathcal{O}(1\%)$ is allowed by CMB, LSS, and local data, showing that this is a possible scenario for the evolution of our universe.

\subsubsection{ALPs as late-time dark energy}

We now move to the regime where the ALP drives the current accelerated universe expansion. The idea of a pseudo-Nambu-Goldstone boson as a quintessence field was first proposed in Refs.~\cite{Frieman_1995, Carroll:1998zi}. These works explained as an ALP field with a mass smaller than the current Hubble constant $m_a\lesssim H_0$ could naturally serve as a DE candidate with the following attractive properties: \textit{(i)} a naturally small potential which is stable from UV radiative correction and \textit{(ii)} evade the most stringent spin-independent fifth force bounds~\cite{Adelberger:2003zx}. Such a field is not the QCD axion, since the $G\tilde{G}$ term would introduce corrections to the potential of the order $\mathcal{O}(\Lambda_{\rm QCD})$, thereby spoiling the smallness of the potential~\cite{Carroll:1998zi}. To circumvent this, several UV-complete models were proposed to accommodate a DE axion~\cite{Kim1999, Kim_2003, Choi2000,Nomura:2000yk}, and the possibility of two axion-like fields to explain a common origin of DE and dark matter~\cite{Kim1999}. 

Looking for the existence of an axion quintessence field is particularly motivated within string theory that generically predicts the existence of many ultralight ALPs with masses down to the DE scale~\cite{Arvanitaki:2009fg}. Axions could potentially address many challenges associated with constructing a successful quintessence model in string theory~\cite{Cicoli:2021skd}, however, some difficulties are present for the minimal single-field model with the usual cosine potential  (Eq.~\eqref{eq:ALP_pot} with $n=1$). 

\paragraph{DE difficulties in minimal model.} 
For instance, to match the energy scale of DE, i.e. $\Lambda_a^4=\rho_{\rm DE} \sim (0.003\, \text{eV})^4$, the non-perturbative action should be $S \sim \mathcal{O}(100)$, therefore, from the scaling $f_a \sim M_{\rm Pl}/S$~\cite{Arvanitaki:2009fg} one finds $f_a \sim 10^{16}$ GeV.  Moreover, to ensure the field is currently in an over-damped regime, the local curvature of the potential should be smaller than the Hubble parameter today, i.e. $\sqrt{V''(a)} \lesssim H_0$ which implies $\Lambda_a^4 \sim H_0^2 f_a^2$. Therefore, one finds that to account for the entirety of DE, $\rho_a \sim H_0^2 f_a^2 \sim H_0^2 M_{\rm Pl}^2$, a Planckian decay constant $f_a \sim M_{\rm Pl}$ is required~\cite{Kaloper:2005aj}. This is not only in tension with the earlier estimate, but it also challenges the consistency of the theory, where a sub-Planckian decay constant is expected $f_a \lesssim M_{\rm Pl}/S$~\cite{Arkani-Hamed:2006emk}. 

\paragraph{Different models.}
One potential resolution is the \textit{Hilltop} scenario, where the axion initially resides very close to the top of the potential~\cite{Dutta:2008qn}. In a single-field scenario, the initial displacement from the maximum, $\pi - a_i/f_a$, is exponentially sensitive to the ratio of the decay constant to the Planck scale, $a_i/f_a \sim e^{-M_{\rm Pl}/f_a}$~\cite{Kaloper:2005aj}. For $f_a \sim 10^{16}$ GeV, this would require extreme fine-tuning of the initial condition, $a_i/f_a \sim e^{-100}$. This fine-tuning is further exacerbated by quantum diffusion during inflation, which induces an inevitable displacement from the potential maximum of the order $a_i/f_a \sim H_{\rm inf}/2\pi f_a$~\cite{Kaloper:2005aj}. Some attempts to dynamically explain this maximal misalignment have been proposed~\cite{Reig:2021ipa}. In the context of \textit{multi-axion fields}, one can estimate the probability of randomly realizing a universe resembling ours~\cite{Kamionkowski:2014zda} given the underlying distributions of the axion parameters. In particular, if all of them have decay constants around $f_a \sim 10^{16}$ GeV, then $\mathcal{O}(10^4)$ axions would be necessary to match the current DE density~\cite{Svrcek:2006yi}. Another possible solution is the \textit{alignment} scenario between two or more axion fields that through mixing in the non-perturbative potential could produce an effective Planckian decay constant making one direction of the potential sufficiently flat to account for DE~\cite{Kim:2004rp} and the other field combination making possibly explaining DM.
A crucial insight for resolving these issues is to maintain $\Lambda_a^4 \sim \rho_{\rm DE}$ while achieving $f_a \sim M_{\rm Pl}$. This is feasible even for a single axion in minimal supersymmetric models, where the axion potential originates from instantons of the electroweak gauge group, which takes the name of \textit{electroweak quintessence axion}~\cite{Nomura:2000yk,Ibe:2018ffn}. 
Another approach involves considering non-periodic potentials, such as \textit{monodromies}, where the field value is not limited by the decay constant. In these models, initially proposed for inflation~\cite{Silverstein:2008sg}, the potential can take a monomial form $V \sim a^n$~\cite{McAllister:2014mpa}, with variations including linear cases~\cite{Panda:2010uq} or more complex scenarios with multiple fields and higher powers~\cite{DAmico:2016jbm, DAmico:2018mnx}. For additional successful quintessence models in string theory, typically with multifields and possible interaction with dark matter look at~\cite{Cicoli:2012tz,Brinkmann:2022oxy,Cicoli:2024yqh,Poulot:2024sex,Aboubrahim:2024spa}. 
Recently, in~\cite{Muursepp:2024mbb} a new mechanism to explain DE and DM was proposed by considering a QCD axion as DM coupled to a dark pNGB associated with a hidden gauge sector that is still under confinement today. In this model, the temperature dependence of the dark pNGB explains the DE behaviour, while the interaction with QCD, which allows energy transfer between these two axions, is made to match the current DE density. 

\paragraph{Observables and constraints.}
Considering the cosmological evolution of the field, the axion is one of the best examples of the ``thawing'' class of quintessence models~\cite{Caldwell:2005tm,Scherrer:2007pu,Dutta:2008qn,Linder:2015zxa}, which characterize models that are typically frozen at an early epoch and slowly start rolling down the potential, thereby deviating from $w \sim -1$ at later times, but always remain above the line $w = -1$. Recent measurements of baryon acoustic oscillations (BAO) by the Dark Energy Spectroscopic Instrument (DESI), when combined with observations of the cosmic microwave background (CMB) and Type Ia supernovae, reveal intriguing hints of an evolving DE component~\cite{DESI:2024mwx}. While DESI data alone remain consistent with a cosmological constant, their combination with supernova data shows a preference for dynamical DE. In Ref.~\cite{DESI:2024mwx}, this behaviour was illustrated using the linear parametrization of the DE equation of state (EOS),
$w(R) = w_0 + w_a (1-R)$,
which introduces two additional parameters $(w_0,w_a)$ relative to the $\Lambda$CDM model. Notably, the preference for evolving DE persists across different EOS parameterizations~\cite{Notari:2024rti,Giare:2024gpk,Berghaus:2024kra}. For the class of so-called ``thawing'' models, however, these parameters are correlated, motivating dedicated analyses such as those presented in Refs.~\cite{DESI:2024kob,Gialamas:2024lyw,Berghaus:2024kra,Shlivko:2024llw}.

The second DESI data release (DR2) strengthens the evidence for dynamical DE and, in particular, favours an EOS $w<-1$ both at high and low redshift, indicating a ``phantom-crossing'' behaviour~\cite{DESI:2025zgx,DESI:2025fii}. Constructing a theoretically consistent DE model with $w<-1$, corresponding to an increasing DE density with time, is notoriously challenging~\cite{Carroll:2004hc} and typically requires the presence of ghosts~\cite{Carroll:2003st,Cline:2003gs} or superluminal propagation~\cite{Dubovsky:2005xd}. An effective phantom behaviour can nevertheless emerge from interactions between dark energy and dark matter (DM), whereby the DM energy density dilutes more slowly and mimics a phantom EOS when the evolution is parametrized solely in terms of an effective DE EOS~\cite{Khoury:2025txd,Guedezounme:2025wav,Chen:2025ywv}. 

Minimal axion dark energy models have recently been confronted with DESI DR2 data in Refs.~\cite{Urena-Lopez:2025rad,Lin:2025gne}, yielding slightly different constraints on the axion decay constant $f_a$ and mass $m_a$, primarily driven by different treatments of the initial conditions. Extensions of this framework have also been explored: Ref.~\cite{Berbig:2024aee} studies axion DE with a nonzero initial field velocity, while Ref.~\cite{Nakagawa:2025ejs} considers a scenario in which the total DE component is a mixture of a cosmological constant and a dynamical axion field, and Ref.~\cite{Goldstein:2025epp} considers the monodromic case.

More generally, the existence of multiple axion-like fields is well motivated, particularly in string-theoretic constructions, where they may naturally account for both DM and DE and interact with one another. For reviews of the cosmological implications of DM--DE interactions in this context, see Refs.~\cite{Wang:2016lxa,Wang:2024vmw}. Such interacting DE models offer a promising avenue to simultaneously address the $H_0$ and $S_8$ tensions (see Ref.~\cite{Abdalla:2022yfr} for a review) while providing a good fit to DESI observations~\cite{Giare:2024smz}.

\paragraph{Hints from CMB and Cosmic Birefringence.}
Cosmological observables such as CMB and galaxy clustering put stringent bounds on the axion abundance that have a significant evolution from recombination and today, for instance, they constraint ALPs with masses $10^{-32}\lesssim m_a\lesssim 10^{-28}$ to constitute less than $1\%$ of DM~\cite{Hlozek:2014lca,Rogers:2023ezo}. Late evolution of DE has two main effects on the CMB power spectrum, it affects the peak location and it enhances the power at low-multiples through the late-time Integrated Sacks-Wolde effect~\cite{Coble:1996te}. Since these effects are very small when the axion mass is $m_a\lesssim 10^{-33}$ eV CMB cannot distinguish between a dynamical or a constant DE energy. However, if the ALP is directly coupled to the photon through the axion-photon interaction $g_{a\gamma\gamma} a F_{\mu\nu}\Tilde{F}_{\mu\nu}$ this would modify the propagation properties for the left- and right-handed photon. 
One relevant effect is the rotation of the linear polarization of the CMB photon as they pass through a dynamic axion field which is known as \textit{Cosmic Birefringence}~\cite{Carroll:1989vb, PhysRevD.43.3789, Harari:1992ea, Lue:1998mq} (see Ref.~\cite{Komatsu:2022nvu} for a review). Such a rotation would induce a non-zero cross-correlation in the parity-odd $EB$ power spectrum which, in the absence of parity-violating phenomena, should vanish.  The angle associated with a clockwise is called \textit{birefringent angle} and it is given by $\beta = g_{a\gamma\gamma}(a_0-a_{\rm LS})/2$, then just depending on the field difference between the last scattering surface and today and the axion-photon coupling.

This topic received renewed interest recently with a new analysis of \textit{Planck} data finding a $2.4 \sigma$ evidence for a non-zero birefringence angle~\cite{Minami:2020odp}, subsequently confirmed by other groups~\cite{Diego-Palazuelos:2022dsq, Eskilt:2022wav, Eskilt:2022cff, Cosmoglobe:2023pgf, Ballardini:2025apf,Remazeilles:2025wzd,Yin:2025fmj,Sullivan:2020wnv}, with a measured value of $\beta \sim 0.3^\circ$, currently excluding the zero value by more than $3.6\sigma$.
This finds an intriguing explanation in the axion quintessence scenario, naturally uniform overall sky, where the field displacement is directly connected with the EOS parameter $\Delta a\propto \sqrt{w+1}$ or in an axion which started oscillating between CMB and today, i.e. in the mass regime $10^{-32}~{\rm eV}\lesssim m_a \lesssim 10^{-27}~{\rm eV}$~\cite{Fujita:2020aqt, Sigl:2018fba}. 

Theoretical implications for the axion quintessence regime are discussed in Refs.~\cite{Fujita:2020aqt, Fujita:2020ecn} for the quadratic and cosine potentials, in~\cite{Gasparotto:2022uqo} for the linear monodromy potential and in~\cite{Choi:2021aze, Lin:2022niw, Lin:2025gne} for the electroweak quintessence models. Moreover, Ref.~\cite{Obata:2021nql} considered the two ALP fields, one being DE and the other one explaining dark matter potentially leaving a birefringence signal in terrestrial experiments.  
In Ref.~\cite{Sherwin:2021vgb} they suggested a way to separately infer the birefringent angle from the recombination and reionization epochs which could distinguish between a dynamic DE or dark matter origin of the birefringent angle~\cite{Nakatsuka:2022epj}. Moreover, different models of axion quintessence could be differentiated by examining the anisotropic counterpart of the cosmic birefringence~\cite{Greco:2024oie, Greco:2022xwj, Greco:2022ufo} and the cross-correlation of cosmic birefringence with the CMB temperature~\cite{PhysRevD.84.043504} or with the spatial distribution of galaxies~\cite{Arcari:2024nhw}. 
The possible connection between the DESI indications of dynamical DE and observations of cosmic birefringence has been investigated in Refs.~\cite{Tada:2024znt,Barman:2025ryg,Lin:2025gne,Nakagawa:2025ejs,Berbig:2024aee,Lee:2025yvn}. These works explore a variety of dynamical axion dark energy scenarios that can consistently accommodate both the DESI data and current constraints on cosmic birefringence.

\vspace{0.5cm}

Since their proposal, it was noticed that ALP can behave like dark energy during its cosmological evolution. This has been shown to be very powerful and widely used in describing many effects in cosmology from the early universe, to the late time expansion, and even having a role in solving cosmological tensions. Discovering what role ALPs have in the dark sector of cosmology is one of the main goals of cosmology in the next few years.

%% file: WG2/content/dark_photons.tex
\subsection{Dark Photon Dark Matter\\ \textnormal{Authors: M. Gorghetto, W. Ratzinger \& L. Ubaldi}}
\label{subsec:dark_photons}

\subsubsection{Production mechanisms} \label{subsubsec:dp_production}

A (light) dark photon can be a dark matter candidate. Appealing features are the simplicity and the minimality of the model, and the fact that we can devise experiments
to probe it. In this section we review some mechanisms for producing dark photon dark matter in the early universe.
We take the Friedmann-Robertson-Walker (FRW) metric $g_{\mu\nu} ={\rm diag}(-1,  R^2(t), R^2(t), R^2(t))$ and write the action for a free massive dark vector
boson $A_\mu$ as
\begin{equation} \label{DPfreeac}
S = \int {\rm d}^4 x \sqrt{-g} \left[ -\frac{1}{4} g^{\mu\nu}g^{\rho\sigma} F_{\mu\nu}F_{\rho\sigma} - \frac{1}{2} m_A^2 \ g^{\mu\nu} A_\mu A_\nu \right] \, ,
\end{equation}
where $F_{\mu\nu} = \partial_\mu A_\nu - \partial_\nu A_\mu$.

It was first proposed that a massive dark photon could be produced via coherent oscillations~\cite{Nelson:2011sf}, in analogy with the axion
misalignment mechanism. However, this does not work out of the box. The reason is that the vector cannot remain misaligned
at some value $\langle A_\mu \rangle \neq 0$, because its energy density redshifts as $\rho \sim R^{-2}$. This should be contrasted to
the scalar field case, for which the energy density remains constant as the universe expands, once the field is displaced from its minimum. The difference traces back
to the $g^{\mu\nu}$ factor in the mass term of the vector, which dominates the energy density and reads $\frac12m^2g^{\mu\nu}A_{\mu}A_{\nu}\propto R^{-2}$, given that $g_{ij}\propto  R^2 \delta_{ij}$. 
One could think of adding ${\cal L} = \frac{1}{2} \xi \mathcal{R} g^{\mu\nu}A_\mu A_\nu$ to the lagrangian in \eqref{DPfreeac}, as advocated in 
Ref.~\cite{Arias:2012az}, with $\mathcal{R}$ the Ricci scalar. Choosing $\xi = 1/6$ one obtains the same equation of motion as for the scalar, thus allowing the vector to remain 
misaligned and eventually oscillate. However, this scenario is problematic as well, due to a ghost instability~\cite{Dvali:2007ks, Himmetoglu:2008zp, Himmetoglu:2009qi}: with the introduction of the term proportional to $\mathcal{R} $, the kinetic term of the longitudinal mode has 
the wrong sign for some comoving momentum modes $k$, which ``calls into question the health of the theory'' \cite{Karciauskas:2010as}.
Another option proposed~\cite{Nakayama:2019rhg} to rescue the misalignment scenario is to 
multiply the kinetic term in \eqref{DPfreeac} by a function 
$f^2(\phi)$, with $\phi$ a scalar field (which could be the inflaton).
In this case $f^2(\phi)$ must be proportional to $R^\alpha(t)$, with the power $\alpha$ equal to either $-4$ or $+2$. This requires $f(\phi)$ to
be exponential in $\phi$, which in turn calls for some heavy model building~\cite{Martin:2007ue}, departing 
significantly from simplicity. Moreover, while the introduction of $f(\phi)$ saves the misalignment, it introduces severe issues with
isocurvature and statistically anisotropic curvature perturbations~\cite{Kitajima:2023fun}.
Some of these issues are related to the Abelian nature of the dark photon. In Ref.~\cite{Fujita:2023axo} the authors showed that considering
instead a non Abelian $SU(2)$ gauge group, one can write a simple model involving gauge fields and inflaton, which succeeds in generating
vector dark matter via misalignment, while avoiding constraints from curvature and isocurvature perturbations.

With the simple action of \eqref{DPfreeac} the dark photon has another interesting  production mechanism, occurring via inflationary fluctuations and distinct from misalignment, as shown in Ref.~\cite{Graham:2015rva}.
In the small mass limit, $m_A \ll H_I$, with the $H_I$ the Hubble scale during inflation, 
the longitudinal mode of the vector $A_L$ behaves like a massless scalar in the relativistic limit, which breaks 
conformal invariance. As a consequence, $A_L$ is produced during inflation via
quantum fluctuations. 
As the comoving momentum modes $k$ of $A_L$ exit the horizon they quickly become `classical' 
(see Appendix A in Ref.~\cite{Arvanitaki:2021qlj} for a clear heuristic explanation of the quantum-to-classical transtition)
and behave like a dark electromagnetic field. When a mode $k$ becomes comparable to $R(t) m_A$, it
becomes non relativistic. Eventually one ends up with non relativistic light dark photons with a very large occupation number.
While scalars and tensors are produced during inflation with a nearly scale invariant spectrum, the vector has the virtue 
of developing a power spectrum peaked at intermediate wavelength. This suppresses dangerous isocurvature 
perturbations at large, cosmological scales. Moreover, at long wavelengths the vector inherits the adiabatic perturbations of the inflaton, a 
necessary condition for a good dark matter candidate. Given current bounds on $H_I$, this mechanism can produce the observed dark matter relic abundance for $m_A\gtrsim 10^{-5}$~eV. 
In the work of Ref.~\cite{Graham:2015rva} the calculation is done under the assumption of instantaneous 
reheating. This is a good approximation for ultra light dark photons, but is not applicable for heavier ones
(roughly for $m_A >$ GeV if the reheating temperature is $10^9$~GeV). 
In Refs.~\cite{Kolb:2020fwh, Ema:2019yrd,Ahmed:2020fhc} the analysis was extended by allowing a finite duration of reheating. 
More recently, it was observed~\cite{Capanelli:2024pzd, Capanelli:2024rlk} that if one includes nonminimal couplings to gravity, 
$\xi_1 \mathcal{R}  g^{\mu\nu} A_\mu A_\nu$ and $\xi_2 \mathcal{R} ^{\mu\nu} A_\mu A_\nu$, with $\mathcal{R} $ the Ricci scalar,
$\mathcal{R} ^{\mu\nu}$ the Ricci tensor, $\xi_1, \xi_2$ dimensionless constants, the inflationary production of
the longitudinal dark photon mode suffers from an instability which leads to runaway production
of high-momentum modes. The authors point out~\cite{Capanelli:2024rlk} that in the 
instability-free regime there is still an enhancement of dark photon production. This can allow the dark photon to account for the observed dark-matter abundance even for $m_A\lesssim 10^{-5}$\,eV, opening up parameter space. However, Ref.~\cite{Lebedev:2025snd} disputes these conclusions, pointing out that UV completions of such effective models do not show any enhanced growth of
high-momentum modes.

The mechanism just described falls within the broader class of gravitational particle–production. It has the remarkably simple Proca action, Eq.~\eqref{DPfreeac}, as the starting point and no interactions between the dark photon and the 
visible sector, aside from gravitational. 
The mass term in Eq.~\eqref{DPfreeac} is often referred to as Stueckelberg mass, because it is straightforward to make it
gauge invariant by writing $-\frac{1}{2} g^{\mu\nu} m_A^2 \left(A_\mu -  m_A^{-1} \partial_\mu \theta  \right)
\left(A_\nu - m_A^{-1} \partial_\nu \theta  \right)$, with $\theta$ the Stueckelberg field. It is important
for the vector to be massive throughout the cosmic evolution for the mechanism of inflationary fluctuations of Ref.~\cite{Graham:2015rva} 
to work. This is the case with the Proca and Stueckelberg lagrangians, but is not necessarily true if the mass is generated via a Higgs mechanism, because one can transition between the broken and the unbroken phase of the theory. 
In the Higgs case it is still possible to produce dark photon dark matter, if the corresponding $U(1)$ symmetry was broken throughout the entire cosmological history. In the opposite case, as studied in Ref.~\cite{Redi:2022zkt}, the dark sector
contains extra degrees of freedom and interactions that lead to non-trivial dynamics, including thermalization, phase
transitions and cosmic strings production. The predictions depend strongly on the coupling of the Higgs to the curvature, and
on the Hubble scale of inflation relative to the scale of spontaneous symmetry breaking, and can differ significantly
from the predictions of the pure Stueckelberg case of  Ref.~\cite{Graham:2015rva}.

The Higgs model can lead to different production mechanisms as well~\cite{Dror:2018pdh, Long:2019lwl, Cline:2024wja}. 
It is possible that the radial mode $h$ of the Higgs field at the end of inflation is displaced from the zero-temperature
expectation value $v$ by an amount $\phi_0$ such that $\phi_0 \gg v$, thereby storing a significant amount
of energy density. When $h$ starts oscillating, in the radiation dominated era, it can efficiently produce dark photons
via parametric resonance. As shown in~\cite{Dror:2018pdh}, this can lead to the correct abundance of dark photon dark matter for a wide range of masses. 

In Ref.~\cite{Long:2019lwl} the authors studied the production from near-global Abelian-Higgs cosmic strings.
In the limit $m_A \ll m_h$, with $m_h$ the Higgs mass, realized when the the gauge coupling is much smaller than the quartic coupling
in the Higgs potential, the cosmic strings produce dominantly the longitudinal polarization of the vector. 
Following the evolution of the string network and of the radiated dark photons it was found that
one can obtain the correct relic abundance for $m_A $ as low as $10^{-22}$ eV.

One could slightly deviate from the minimal action in Eq.~\eqref{DPfreeac} and add the term
\begin{equation} \label{FFtilde}
 \int {\rm d}^4 x \sqrt{-g} \left[ -\frac{\alpha}{4f} \phi \frac{1}{2} \epsilon^{\mu\nu\rho\sigma} F_{\mu\nu} F_{\rho\sigma} \right]  \, ,
\end{equation}
where $\phi$ is a real scalar field, $\alpha$ a dimensionless coupling, $f$ a constant
of mass dimension one, $\epsilon^{\mu\nu\rho\sigma}$ the antisymmetric tensor. This operator arises typically in axion models. It leads to the following equations for the vector
\footnote{
See Appendix A in Ref.~\cite{Bastero-Gil:2021wsf} for a detailed derivation.
}
\begin{align} \label{ALeq}
\ddot A_L + \frac{3k^2 + R^2(t)m_A^2}{k^2 + R^2(t)m_A^2} H \dot A_L + \left(\frac{k^2}{R^2(t)} + m_A^2 \right) A_L & = 0 \, , \\
\ddot A_\pm + H \dot A_\pm + \left( \frac{k^2}{R^2(t)} \mp \frac{\alpha}{f} \dot \phi \frac{k}{R(t)} + m_A^2  \right) A_\pm & = 0 \, .
\label{ATeq}
\end{align}
Here a dot denotes a derivative with respect to cosmic time $t$, and $H \equiv \dot R / R$ is the Hubble parameter. 
$A_L$ is again the longitudinal mode, while $A_\pm$ are the transverse modes. The former is unaffected by the operator in Eq.~\eqref{FFtilde}, whereas the latter are. Considering a negligibly small mass $m_A$, we see that for 
$\frac{k}{R(t)} < \frac{\alpha}{f} \dot \phi$ the solution for one of the two transverse helicities is exponential. This implies a tachyonic production of those small momentum modes, 
which quickly leads to  large occupation numbers.
In this way a classical dark electromagnetic field, coherent over long wavelengths, is generated, at the expense
of the kinetic energy density of $\phi$ [note it is the time derivative of $\phi$ that enters in Eq.~\eqref{ATeq}].

This tachyonic production has been exploited in Refs.~\cite{Agrawal:2018vin, Co:2018lka, Bastero-Gil:2018uel,Adshead:2023qiw}.
In Refs.~\cite{Agrawal:2018vin, Co:2018lka} the scalar field $\phi$ is an axion (not necessarily the QCD one)
with a sinusoidal potential, and the relevant dynamics take place during the radiation dominated epoch.
The axion is initially misaligned from the minimum of its potential. When $H \sim m_a$, with $m_a$ the axion
mass, it starts oscillating. Due to the coupling in Eq.~\eqref{FFtilde} it dissipates some of its energy density into dark photons.
For large couplings, $\alpha$ between 35 and 50, a decay constant $f_a = 10^{14}$~GeV, an axion mass $m_a = 10^{-8}$~eV,
a dark photon mass $m_A \sim 0.1 m_a$, within a few axion oscillations most of the energy density initially stored in $\phi$
is efficiently transferred to dark photons~\cite{Agrawal:2018vin}. The system quickly enters a non-linear regime
which requires a numerical study on the lattice~\cite{Agrawal:2018vin}. At the end of the day one has a small residual  energy density in axions and a larger dominant one in dark photons. They both contribute to the dark matter
abundance. An important point to reach these conclusions is that the masses $m_a$ and $m_A$ are comparable. 
From the model building point of view this could be flagged as tuning, since $m_a$ and $m_A$ typically have 
different origins. When these masses are very different, the axion only transfers a small fraction of its energy density
during oscillations to dark photons; this could still easily be enough for the dark photon to match the relic abundance today~\cite{Co:2018lka}.
However, in such a scenario one needs to introduce more operators in order to make the axion decay into the visible 
sector and deplete its energy density, after it has produced enough dark photons~\cite{Co:2018lka}. Tachyonic amplification in axion models with shallow potentials can also lead to efficient dark photon production, with large initial misalignment driving parametric resonance~\cite{Zhang:2025pgb}.

Another possibility is that $\phi$ in Eq.~\eqref{FFtilde} is the inflaton, in which case dark photon production happens efficiently
towards the end of inflation~\cite{Bastero-Gil:2018uel, Bastero-Gil:2021wsf}.
In this scenario, inflation must end with $\phi$ reheating only the visible sector. After reheating the long-wavelength dark electromagnetic
field constitutes only a small fraction of the energy density budget, but eventually, when the mass
$m_A$ matches the inverse of the physical redshifting wavelength, the dark photon energy density redshifts like 
non relativistic matter, becomes dominant, and can match the observed dark matter density today.
One can study the power spectrum in detail~\cite{Bastero-Gil:2021wsf} and show that it does not conflict with isocurvature constraints on cosmological scales.

Instead of the term in Eq.~\eqref{FFtilde}, which is of the form $\phi F \tilde F$,
one could consider a coupling of the form $f^2(\phi) F F$. The former has the virtue of generating 
inevitably the tachyonic growth of one transverse polarization, as we saw. The latter can also induce a tachyonic growth, with both transverse polarizations treated on equal footing. However, it can do so efficiently
only under certain conditions:
$f^2(\phi)$ has to be proportional to $R^\alpha(t)$ with a negative power $\alpha$. 
In Ref.~\cite{Nakai:2020cfw} it was pointed out that, taking $\phi$ to be the inflaton field, 
the $f^2(\phi) F F$ interaction leads to the production of large-wavelength dark photons
which can account for the entire dark matter for masses as low as $10^{-21}$ eV. However, this is the same model
of Ref.~\cite{Nakayama:2019rhg} we mentioned above, and it suffers from the same issues and severe 
constraints.
As an alternative, a model which makes use of both couplings of the inflaton $\phi$ to $FF$ and to $F\tilde F$ was proposed 
in Ref.~\cite{Salehian:2020asa}. The authors also introduce a complex scalar which serves both as the Abelian Higgs and the waterfall field. The gain is that during inflation the dark photon is massless, so it
is efficiently produced. At the end of the cosmological evolution one can achieve the correct relic abundance for a very wide
range of dark photon masses, from $10^{-20}$ eV to $10^{15}$ GeV.

Most of the mechanisms described above aim at the production of very light dark photons. These scenarios often require unnaturally small couplings, which call for an explanation and typically demand additional model building; see, for example, the discussion in Appendix A of Ref.~\cite{Agrawal:2018vin}.
Some regions of parameter space can also violate some versions of weak gravity conjectures. For example, it was argued in Ref.~\cite{Reece:2018zvv} that such conjectures would require the dark photon produced 
via the mechanism of Ref.~\cite{Graham:2015rva} to be heavier than the eV scale. Some may simply dismiss these objections as conjectural rather than theorem-based. The unnaturally small couplings might be viewed as an aesthetic concern rather than a physical one.

A further subtlety in these mechanisms was highlighted in Ref.~\cite{East:2022rsi}. The authors showed that the large densities stored in the dark electromagnetic fields can lead to the restoration of the $U(1)$ symmetry associated to the dark photon and
the formation of topological defects (vortices). This would transform the dark photon energy density into vortex strings,
thus eradicating the coherent dark photon dark matter field. Strings dissipate part of their energy via relativistic dark photons and gravitational waves emission, providing a different and new target for experimental
 searches. 
On the other hand, this would get rid of the coherent non-relativistic dark photons, and hence of the dark
matter candidate.

What are the conditions for vortices to form?
Consider the Higgs model and the potential $\frac{\lambda}{4} \left( |\Phi|^2 - v^2 \right)^2$. 
The radial mode of the Higgs has mass $m_h = \sqrt{2\lambda} v$, while the dark
photon has mass $m_A = g v$, with $g$ the gauge coupling of the dark $U(1)$. When the energy 
density in dark photons, given roughly by $m_A^2 \langle A_\mu A^\mu \rangle$, exceeds the 
threshold $\lambda v^4$, then vortices form and the dark matter production mechanism is spoiled. 
One way to avoid this is to take a tiny gauge coupling, $g \ll \lambda$. However, this would generically
imply a small kinetic mixing $\epsilon \sim e g / (16\pi^2)$ between visible (with gauge coupling $e$) 
and dark photon.
In Refs.~\cite{Cyncynates:2023zwj, Cyncynates:2024yxm},
the authors showed that the constraints on the kinetic mixing, derived from avoiding vortex formation,
are so strong to exclude most of the parameter space being probed by the experiments dedicated to dark photon searches. They also proposed novel models, e.g. based on the clock-work mechanism, to get around this issue and
motivate experimental searches.

It is not clear that vortex formation poses a fundamental obstacle for production mechanisms based on the Stueckelberg action. After all, the Stueckelberg theory can be obtained as the large-$\lambda$ limit of the Higgs theory, and this limit still defines a well-behaved quantum field theory. In that regime, it is plausible that the usual vortex constraint can be evaded.

 Another possibility to avoid the vortex catastrophe is to add either scalars of fermions charged
 under the dark $U(1)$ Abelian gauge group.
This opens up at least two new avenues. In the first scenario, the coherent dark-photon field produced by the mechanism of Ref.~\cite{Graham:2015rva} dissipates upon horizon re-entry during the radiation era through Schwinger pair production, strong-field electromagnetic cascades, and plasma dynamics~\cite{Arvanitaki:2021qlj}. 
In the second, the dark electromagnetic field produced during inflation~\cite{Bastero-Gil:2018uel, Bastero-Gil:2021wsf}, can in turn generate scalars or fermions via the Schwinger mechanism close to the end of inflation~\cite{Bastero-Gil:2023htv, Bastero-Gil:2023mxm}. 
In both cases, the produced charged particles, not the dark photon, can account for the dark matter 
abundance.

\subsubsection{Cosmological signatures of dark photon dark matter}

This section presents the main effects related to the cosmological production of dark-photon dark matter and identifies some interesting open questions.

\paragraph{Gravitational Waves.}
The dark-photon production mechanism in which the dark photon is generated through scalar couplings of the form $\phi FF$ or $\phi F\tilde F$, is typically accompanied by the generation of a large anisotropic stress. While the energy is initially in the homogeneous scalar field $\phi$ and its velocity, the tachyonic instability leads to the production of relativistic, finite momentum dark photon modes. The resulting highly inhomogeneous photon field strength and energy density in this configuration imply a sizable transverse–traceless contribution to the energy–momentum tensor. This leads to the production of gravitational waves (GWs). This phenomenon was studied in Refs.~\cite{Anber:2012du,Domcke:2016bkh,Adshead:2018doq,Adshead:2019lbr} for scenarios where the scalar drives inflation and in Refs.~\cite{Machado:2018nqk,Machado:2019xuc,Salehian:2020dsf} where the scalar is identified with an ALP. The resulting stochastic GW background may be observed in PTAs \cite{Ratzinger:2020koh,Kitajima:2020rpm,Madge:2023dxc,Figueroa:2023zhu} or the next generation of interferometers.

Since tachyonic production becomes efficient once the tachyonic growth rate is larger than the Hubble rate, the resulting GW spectra are typically peaked at a wavelength $\mathcal{O}(10-100)$ smaller than the horizon at the time of dark photon and GW production \cite{Machado:2018nqk}. In the inflationary case, the GW emission can be prolonged, yielding spectra that extend over many orders of magnitude in frequency~\cite{Domcke:2016bkh}. 
 The spectrum’s amplitude is determined by the source strength, which scales with the energy initially in the scalar field and subsequently transferred to the dark photon. If the scalar is massive, the amount of GWs that can be produced is limited, because excessive scalar dark matter relics could overclose the universe. This poses a challenge for generating an observable GW background after inflation in minimal models
\cite{Ratzinger:2020oct}. Various alternatives to enhance the scalar field’s energy have been proposed \cite{Banerjee:2021oeu,Madge:2021abk}.

As discussed in Sec.~\ref{subsubsec:dp_production},  with the coupling $\phi F \tilde F$ only one dark-photon polarization becomes tachyonic, depending on the sign of the velocity of the scalar. Even in scenarios where the scalar oscillates, this results in a highly polarized dark-photon spectrum, as most photons are produced when the field is moving the fastest \cite{Adshead:2018doq,Machado:2018nqk,Bastero-Gil:2022fme}. Consequently, the resulting GW background is also polarized. This distinctive polarization, along with the unique spectral shape, allows the GW signal to be distinguished from other scenarios. The effect is most pronounced in regions of parameter space where the dark photon’s back-reaction on the scalar is weak or negligible, since prolonged field interactions tend to wash out the polarization \cite{Adshead:2018doq,Ratzinger:2020oct}.

\paragraph{Dark photon dark matter substructure and compact objects.}

Unlike the scalar misalignment, dark photon dark matter is typically not produced homogeneously. Instead, it has order-one density fluctuations on small scales, e.g. set by inflationary fluctuations that are not damped the subsequent cosmological evolution~\cite{Graham:2015rva}, or by parametric resonance/tachyonic instabilities, often related to $m_A$.

For dark photon dark matter produced by inflationary fluctuations, the power spectrum peaks at inverse comoving momenta $\sim 3 \times 10^{-3}\,\mathrm{pc}\,(10^{-5}\,\mathrm{eV}/m_A)^{1/2}$~\cite{Graham:2015rva}. Similar to the QCD axion in the post-inflationary scenario discussed in Chapter~\ref{subsec:miniclusters}, these isocurvature fluctuations undergo gravitational collapse around matter–radiation equality. Due to a coincidence between their size and the Jeans scale at matter-radiation equality, they give rise to a nontrivial dark matter substructure in which an order-one fraction of the dark matter resides in dark photon stars (or `vector solitons')~\cite{Brito:2015pxa}.  These are localized, gravitationally bound, nonrelativistic, stationary objects, analogous to axion stars, with size comparable to the de Broglie wavelength of their constituent particles~\cite{Gorghetto:2022sue}. They are extremely light, with masses $\sim 3 \cdot 10^{-16} M_{\odot} \,(10^{-5}\,{\rm eV}/m_A)^{3/2}$, as determined by numerical simulations, and can be estimated analytically from the mass contained in an order-one fluctuation at matter–radiation equality~\cite{Gorghetto:2022sue}. Such objects provide a potential avenue to detect dark photon dark matter purely via gravitational effects, as they likely remain undisrupted in our Galaxy and frequently pass through the Solar System.

Although not yet explored in detail, other scenarios—such as that of Ref.~\cite{Bastero-Gil:2018uel}—also predict dark matter production with small-scale fluctuations that can collapse into compact objects before standard structure formation. This may have significant implications for detection strategies and deserves further investigation. Dark photon (or Proca) stars have been studied in broader contexts~\cite{Brito:2015pxa,Jain:2023qty}. If extremely dense, dark photon stars formed during cosmological evolution can decay into gravitons via gravity alone~\cite{Nakayama:2023jhg} or collapse into black holes~\cite{Wang:2023tly}. Their properties are further modified in the presence of multiple fields and rotation~\cite{Zhang:2023rwc}, self-interactions~\cite{Zhang:2024bjo}, asphericities~\cite{Herdeiro:2023wqf}, or non-minimal couplings to gravity~\cite{Zhang:2023fhs}.

Regardless of whether dark photon dark matter is produced with more (or less) power on small scales, fuzzy vector dark matter—i.e., with $m_A \simeq 10^{-21}$\,eV or slightly higher—is expected to form macroscopic solitonic cores at the centers of galaxies, analogous to scalar fuzzy dark matter~\cite{Schive:2014dra}. These cores correspond to the same solitonic solutions as the dark photon stars discussed earlier but are larger and reside at galactic centers. The formation mechanism of dark photon solitonic cores is still largely unknown, but they may develop via gravitational condensation. In this case, formation occurs on timescales similar to the scalar case, although uncorrelated or partially correlated configurations condense more slowly~\cite{Chen:2023bqy}.  

Unlike scalar solitons, vector solitonic cores can be linearly polarized (with vanishing total intrinsic spin) or circularly polarized (with a macroscopic intrinsic spin)~\cite{Zhang:2021xxa}. Additionally, properties such as the core spin, degree of wave interference, core-to-halo mass ratio, and central halo shape can help distinguish fuzzy dark photon solitons from scalar ones~\cite{Amin:2022pzv}. Finally, Ref.~\cite{Chen:2024vgh} demonstrates that dark photon dark matter can form halos and central dark photon stars with polarization density fluctuations tracking those of the dark matter density in the halo.

\paragraph{Kinetic mixing.}

Dark photon dark matter may couple to the SM, e.g. via kinetic mixing. While such mixing is not required for its production -- and can, in fact, complicate or inhibit inflationary production 
-- alternative production mechanisms may remain viable. Therefore, here we only briefly discuss the resulting cosmological effects, which place upper limits on the mixing parameter $\chi$ defined by the Lagrangian term $\chi F_{\rm SM}^{\mu\nu}F_{\mu\nu}$; see also Refs.~\cite{McDermott:2019lch,Caputo:2021eaa} for a more complete summary.

Independent of whether dark photons constitute the dark matter, kinetic mixing allows CMB photons to convert into dark photons. As in the MSW effect for neutrinos, this mixing is resonantly enhanced when the dark photon mass matches the photon plasma mass, which follows the free-electron density. The resulting effect is an increase in the effective number of relativistic degrees of freedom~\cite{Jaeckel:2008fi}, along with a depletion of the CMB spectrum at observed frequencies. Constraints on CMB spectral distortions from COBE/FIRAS give the most stringent bounds on the kinetic mixing for $m_A \sim 10^{-14}\!-\!10^{-6}\,\mathrm{eV}$~\cite{Mirizzi:2009iz,Chluba:2024wui,Arsenadze:2024ywr} at the level of $\chi\lesssim 10^{-6}$, with recent studies tightening these limits by accounting for plasma-mass inhomogeneities during resonant oscillations~\cite{Caputo:2020bdy,Caputo:2020rnx,Witte:2020rvb,Aramburo-Garcia:2024cbz,Garcia:2020qrp}, and anisotropic (patchy) screening~\cite{Pirvu:2023lch,McCarthy:2024ozh}.

Instead, if (light) dark photons are produced as a nonrelativistic relic and make up a substantial dark matter fraction, conversion into the SM photon bath dominates due to their high number density. Resonant processes can be efficient enough to convert nearly all of the dark photon dark matter into radiation~\cite{Arias:2012az}. In order for the dark matter to survive, one needs to require a large dark matter density in the early universe. To preserve successful nucleosynthesis, however, the matter density at $T \sim \mathrm{MeV}$ must not exceed the energy density of additional light degrees of freedom.
BBN and CMB constraints on these impose the strongest limits on kinetic mixing for dark photon masses $m_{A} \sim 10^{-4}\,\mathrm{eV}$~\cite{Arias:2012az}, at the level of $\chi\lesssim 10^{-8}$.

In addition, after recombination, dark photon dark matter can be depleted through mixing and inverse bremsstrahlung processes. Consistency with observations requires that the density of the decaying dark matter particles does not change substantially after matter-radiation equality, leading to a bound on the mixing parameter for $m_A\gtrsim10^{-12}$ eV~\cite{Arias:2012az}; see also Ref.~\cite{McDermott:2019lch} for a more stringent bound, obtained requiring at most a $\lesssim 2\div3\%$ change in the dark matter density after matter-radiation equality. Moreover, conversion of dark photon dark matter -- both resonant and nonresonant -- into low-energy photons can inject energy during the dark ages, induce helium reionization, or heat the intergalactic medium in the post-reionization epoch. These effects have been tightly constrained~\cite{McDermott:2019lch,Witte:2020rvb}, including by Lyman--$\alpha$ observations studied via hydrodynamical  simulations~\cite{Trost:2024ciu,Bolton:2022hpt}. The resulting limits are  $\chi\lesssim10^{-13}$ for $m_A\simeq 10^{-14}{\rm eV}-10^{-10}{\rm eV}$. Finally, ultra-light dark photon dark matter can efficiently heat the ionized interstellar medium~\cite{Dubovsky:2015cca}, or heat/cool the gas in the gas-rich Leo T dwarf galaxy at a rate exceeding the ultra-low radiative cooling rate of the gas~\cite{Wadekar:2019mpc}.  These effects lead to strong bounds on kinetic mixing down to $m_A \simeq 10^{-20}\,\mathrm{eV}$.

A recent claim challenges all the above-mentioned dark photon dark matter constraints from resonant photon conversion, arguing that the resonance saturates due to plasma nonlinearities~\cite{Hook:2025pbn}.

Finally, dark photon dark matter may also couple to photons via higher-dimensional mixing operators. The results of Ref.~\cite{Amin:2023imi} could be used to derive constraints on dimension-6 operators, as these induce dark photon–photon conversion inside vector solitons (dark photon stars).

%% file: WG2/content/spin2.tex
\subsection{Dark Graviton Dark Matter\\ \textnormal{Author: F. Urban}}
\label{subsec:spin2}

\newcommand{\td}{\mathrm{d}}
\newcommand{\Tr}{\mathrm{Tr}}
\newcommand{\Det}{\mathrm{det}}

\newcommand{\mpl}{m_\mathrm{Pl}}
\newcommand{\mfp}{m_\mathrm{FP}}
\newcommand{\Lag}{\mathcal{L}}

\subsubsection{Theory}

\paragraph{Massive spin-2 fields.}

Theories of massive spin-2 fields have a long history starting with the development of the first consistent, Lorentz-invariant, linear theory which propagates the five physical helicity states of a massive spin-2 field~\cite{Fierz:1939ix}. Non-linear extensions of this theory propagate an additional ghost degree of freedom~\cite{Boulware:1972yco}, except in the unique construction of ghost-free massive gravity~\cite{deRham:2010kj}, which can be further generalised to include both a massless and a healthy massive spin-2 field~\cite{Hassan:2011zd}. In what follows we will focus primarily on the standard form of bigravity, but a host of alternative theories of massive gravity and bigravity have been developed: generalised massive gravity~\cite{DeRham:2014wnv}, projected massive gravity~\cite{Gumrukcuoglu:2020utx}, minimal massive gravity~\cite{DeFelice:2015hla}, minimal bigravity~\cite{DeFelice:2020ecp}, chameleon bigravity~\cite{DeFelice:2017oym} and Lorentz-violating massive gravity theories~\cite{Dubovsky:2004sg,Rubakov:2008nh}. While there are significant differences between these theories, the phenomenology and cosmology of a massive spin-2 dark matter particle, and in particular in the case of a spin-2 ultra-light dark matter, has been developed mostly in the context of bigravity.

The theory contains two tensor fields: \(g_{\mu\nu}\) and \(f_{\mu\nu}\), and is defined by the action~\cite{Hassan:2011zd}
\begin{align}\label{bgaction}
S=m_g^2\int\td^4x\biggl[&\sqrt{|g|}\mathcal{R} (g)+\alpha^2\sqrt{|f|}\mathcal{R} (f)-2\alpha^2m_g^2\sqrt{|g|}\,V\left(X;\beta_n\right)\biggr]
,\end{align}
where \(m_g\) is a mass scale related to the physical reduced Planck mass as 
\(\mpl^2=m_g^2(1+\alpha^2)\) where \(\mpl \approx 2.4\times10^{18}\,\mathrm{GeV}\).
The dimensionless \(\alpha\) (the ``ratio of Planck masses'') measures the relative interaction strength of the two tensor fields as well as the mixing between the mass eigenstates relative to the interacting states \(g_{\mu\nu}\) and \(f_{\mu\nu}\). The interaction potential \(V(X;\beta_n)\) contains five dimensionless parameters \(\beta_n\): \(\beta_0\) and \(\beta_4\) act as bare cosmological constants for \(g_{\mu\nu}\) and \(f_{\mu\nu}\), respectively (and, because they are irrelevant for spin-2 dark matter, we will ignore them in what follows), while the remaining parameters \(\beta_1,\,\beta_2,\,\beta_3\) encode the nonlinear interactions between the two tensor fields:
\begin{equation}
\label{Vdef}
V\left(X;\beta_n\right) = \sum_{n=0}^4\beta_n e_n(X),
\end{equation}
where the \(e_n(X)\) are the elementary symmetric polynomials defined in terms of the eigenvalues of the matrix \(X\) defined through \(X^\rho_{~\sigma}X^\sigma_{~\nu}=g^{\rho\mu}f_{\mu\nu}\). Explicitly they can be obtained via tracing the unit-weight totally anti-symmetric products \(e_n(X)=X^{\mu_1}_{~[\mu_1}\cdots X^{\mu_n}_{~\mu_n]}\) such that \(e_4(X)=\det(X)\) and \(e_n(X)=0\) for all \(n>4\).

For a covariantly conserved source, the standard Bianchi identities in all give five independent constraint equations which can be used to remove five degrees of freedom; diffeomorphism invariance removes \(2\times4=8\) more. Therefore, bigravity propagates seven degrees of freedom which, on backgrounds when such a split makes physical sense, correspond to a massless spin-2 field (two helicity states) and a massive spin-2 field (five helicity states).

In order to see where the massive spin-2 field comes from, it is useful to linearise the metric around proportional metrics as
\begin{align}
g_{\mu\nu} &\rightarrow g_{\mu\nu} + h_{\mu\nu}\,, \\
f_{\mu\nu} &\rightarrow f_{\mu\nu} + l_{\mu\nu}\,.
\end{align}
The canonically normalised mass eigenstates are then defined through
\begin{align}
G_{\mu\nu} &= \frac{\mpl}{1+\alpha^2} \left(h_{\mu\nu}+\alpha^2l_{\mu\nu}\right), \label{masseigG} \\
M_{\mu\nu} &= \frac{\alpha\,\mpl}{1+\alpha^2} \left(l_{\mu\nu}-h_{\mu\nu}\right), \label{masseigM}
\end{align}
which imply the inverse relations
\begin{align}
h_{\mu\nu} &= \frac{1}{\mpl} \left(G_{\mu\nu}-\alpha M_{\mu\nu}\right), \label{invgrel} \\
l_{\mu\nu} &= \frac{1}{\mpl} \left(G_{\mu\nu}+\alpha^{-1}M_{\mu\nu}\right)\,.
\end{align}
From these expressions we also see that the parameter \(\alpha\) quantifies the mixing between the fluctuations. In terms of the mass eigenstates \(G_{\mu\nu}\) and \(M_{\mu\nu}\) the quadratic part of the action diagonalises into
\begin{equation}\label{bgaction2}
S^{(2)} = \frac12 \int\td^4x\sqrt{|g|}\,\biggl[ G_{\mu\nu}\mathcal{E}^{\mu\nu\rho\sigma}G_{\rho\sigma} + M_{\mu\nu}\mathcal{E}^{\mu\nu\rho\sigma}M_{\rho\sigma} - \frac12 \mfp^2\left(M_{\mu\nu}M^{\mu\nu}-M^2\right)\biggr]\,,
\end{equation}
where \(M=g^{\mu\nu}M_{\mu\nu}\) and the Lichnerowicz operator \(\mathcal{E}^{\mu\nu\rho\sigma}\) is defined by
\begin{equation} \mathcal{E}^{\mu\nu}_{~~\rho\sigma} = \frac12 \left(\, \delta^\mu_\rho \delta^\nu_\sigma \Box - g^{\mu\nu} g_{\rho\sigma} \Box + g^{\mu\nu} \nabla_\rho \nabla_\sigma  + g_{\rho\sigma} \nabla^\mu \nabla^\nu - \delta^\mu_\sigma \nabla^\nu \nabla_\rho - \delta^\mu_\rho \nabla^\nu \nabla_\sigma\right)\,. \label{eq:lich}
\end{equation}
The Fierz-Pauli mass of the massive spin-2 field is defined as
\begin{equation}\label{mfpdef}
\mfp = \sqrt{\beta_1 + 2\beta_2 + \beta_3}\,\mpl\,.
\end{equation}

Finally, the theory couples to matter via the action
\begin{equation}
S_\mathrm{m} = \int\td^4x\sqrt{|g|}\,\Lag_\mathrm{m}(g,\Phi)\,,
\end{equation}
where the matter Lagrangian \(\Lag_\mathrm{m}\) contains the Standard Model matter fields \(\Phi\), which we have taken to be minimally coupled only to \(g_{\mu\nu}\). This choice is without loss of generality since the theory treats the metrics symmetrically and, in the bimetric theory, standard matter fields can only couple minimally to one of the tensor fields without introducing ghost instabilities. When expanded in terms of the mass eigenstates \(G_{\mu\nu}\) and \(M_{\mu\nu}\) this action reveals the universal coupling of the massive spin-2 field to matter fields (with coupling \(\alpha\)):
\begin{equation}
S_\mathrm{m}^{(2)} = \frac{1}{\mpl}\int\td^4x\sqrt{|g|}\,\left(G_{\mu\nu}-\alpha M_{\mu\nu}\right) T^{\mu\nu}\,.
\end{equation}
One of the peculiarities of the (Lorentz-invariant) spin-2 dark matter model compared to lower spins is that the universal, direct coupling to matter is non-negotiable. Indeed, if we wish to tune away the coupling \(\alpha\) at the same time the \(M_{\mu\nu}\) field becomes infinitely strongly coupled because at higher orders in the linear expansion the action contains terms such as \(\alpha^{-1}M^3\) or \(\alpha^{-2} M^4\) and so on. In other words, if the dark matter does not interact with matter, it also completely decouples from the theory. This can be relaxed in more exotic scenarios, for instance Lorentz-violating theories.

\paragraph{Ultra-light spin-2 dark matter.}

The massive spin-2 field of bigravity has originally been proposed as a candidate for dark matter (also sometimes called ``dark tensor'' or ``dark graviton'') in the works of~\cite{Babichev:2016hir,Babichev:2016bxi,Aoki:2016zgp}, with production mechanisms ranging from freeze-in~\cite{Babichev:2016hir,Babichev:2016bxi} to bubble collision in the preheating era~\cite{Aoki:2016zgp}. These works have since been extended to include gravitational production~\cite{GonzalezAlbornoz:2017gbh,Kolb:2023dzp}, dark freeze-out of 3-to-2 processes~\cite{Chu:2017msm} and production during inflation~\cite{Gorji:2023cmz}. In all those models, in order for the massive spin-2 field to account for all of the dark matter, the field must be heavy, with masses ranging from order of hundreds of~MeV all the way up to the scale of inflation.

The cosmology of ultra-light massive spin-2 particles has been considered in~\cite{Aoki:2017cnz,Marzola:2017lbt,Aoki:2017ixz,Manita:2022tkl,Lopez-Sanchez:2025osk}. The spin-2 dark matter in~\cite{Aoki:2017cnz} is generated via its coupling to an anisotropic source that must be coherent over large scales in the early Universe. One possible candidate is cosmological magnetic fields, which blazar observations suggests that may be omnipresent at very large coherence lenghts, permeating the whole Universe including large voids. If these fields are primordial, i.e.~generated in the very early Universe, possibly during inflation, they can sustain the production of a massive spin-2 condensate. The production in this case takes place mostly around \(H \simeq \mfp\), where \(H\) is the Hubble parameter. In order for this field to be the whole of the dark matter its mass must satisfy
\begin{equation}
\mfp \simeq 10^{-4} \left(\frac{0.1\,\mathrm{nG}}{B_0}\right)\,\mathrm{eV}\,,
\end{equation}
from which we see that, given that current limits on the strength of a cosmological magnetic field today that stretches across the whole Universe are around \(1\,\mathrm{nG}\), the spin-2 dark matter could be as light as \(10^{-4}\,\mathrm{eV}\).

In~\cite{Marzola:2017lbt} it was shown that, because of the structure of the higher-order vertices in the regime where \(\mfp \gg H\), that is, at late times, the theory can be seen as a massive spin-2 field \(M_{\mu\nu}\) in a standard Friedman-Lema\^itre-Robertson-Walker (FLRW) cosmology described by a resummed, non-linear metric \( \mathcal{G}_{\mu\nu} = g_{\mu\nu} + G_{\mu\nu}/\mpl \). In this case one can show that the Bianchi identities for covariantly conserved sources enforce the transversality and traceleness conditions for \(M_{\mu\nu}\) as \( \nabla^\mu M_{\mu\nu} = 0 = g^{\mu\nu} M_{\mu\nu}\). When the massive spin-2 field is non-relativistic, the transversality and tracelessness conditions in a homogeneous and isotropic FLRW background allow the choice of the six \( M_{ij} \) components, subject to the additional tracelessness constraint \( M^i_{~i}=0 \), as the five tensor physical degrees of freedom. Indeed, the mixed time-space components must satisfy \(M_{0i} \propto k^j M_{ji} \ll M_{ji}\) and the time-time component is \(M_{00} \propto k^i M_{0i} \ll M_{0i} \ll M_{ji}\).

The equations of motion for the spin-2 dark matter field read
\begin{align}
  \ddot M_{ij} + 3H\dot M_{ij} -\bigtriangleup M_{ij}+ \mfp^2 M_{ij} = 0\,. \label{eq:eom}
\end{align}
This equation is reminiscent of that for scalar ULDM and it has the same solutions. The homogeneous background solution is given by
\begin{equation} 
  M_{ij} = \frac{\hat{M}_{ij}}{R^{3/2}} \cos{(\mfp t+\Upsilon)} \varepsilon_{ij} = \frac{\sqrt{2\rho_\mathrm{DM}}}{\mfp R^{3/2}}\cos{(\mfp t+\Upsilon)}\varepsilon_{ij} \,,\label{eq:hij}
\end{equation}
where \(R\) is the scale factor of the Universe, the overall amplitude has been fixed so that the dark matter energy density matches the observed dark matter density \(\rho_\mathrm{DM}\), \(\Upsilon\) is a random phase and \(\varepsilon_{ij}\) is an angular quadrupole matrix with unit norm, zero trace and is symmetric.  With this solution the dark matter energy density scaled as dust and the pressure averages to zero on the large time-scales relevant for the cosmological background evolution. The parallels with the scalar and vector case suggest that even in this case the dark matter could be produced via the misalignment mechanism, as suggested in~\cite{Marzola:2017lbt}.

The possible existence of solutions called massive graviton geons was proposed in~\cite{Aoki:2017ixz}. These massive graviton geons are nonspherically symmetric solutions to the Schrödinger-Poisson equation in the non-relativistic limit with (total, orbital) angular momenta \((j,l)=(2,0)\) as well as spherically symmetric solution with \((j,l)=(0,2)\). While spin-2 geons in principle decay, their life-time can be long enough for them to comprise all of the dark matter. When these geons are comprised of ultra-light particles, they are the analogues of the more common scalar (and vector) ultra-light dark matter solitons. The existence of spin-2 soliton solutions has been also derived in~\cite{Jain:2021pnk}, where it was shown that such solitons are extremally polarised, with macroscopically large spin, but no orbital angular momentum. Some of the possible unique effects of cosmological spin-2 solitons were studied in~\cite{Lopez-Sanchez:2025osk}, where simulations of varying number soliton mergers revealed that the resulting spin-2 halo has a distinct profile from mergers of spin-0 and spin-1 solitons with equivalent properties.

Lastly, the production of spin-2 dark matter from a double phase transition in the early Universe was developed in~\cite{Manita:2022tkl}. In the first step, the Universe goes from an unstable isotropic state to an anisotropic axisymmetric Bianchi type-I fixed point of bigravity, thereby providing a vacuum expectation value for the anisotropic stress which feeds the massive graviton. In a second stage, thanks to a dynamical change in the mass \(\mfp(\chi)\) driven by, for example, a chameleon-like field \(\chi\), the anisotropic fixed point becomes unstable and flips back to an isotropic fixed point, thereby releasing the energy density stored in the anisotropic stress as a spin-2 dark matter field. This mechanism can also be applied to the case of the (Lorentz-violating) minimal theory of modified gravity, in which only the helicity-two components of the massive spin-2 field are present.

The possibility of an ultra-light dark matter candidate of spin-2 in a Lorentz-violating theory was considered in~\cite{Dubovsky:2004ud}. This is achieved via introducing additional Goldstone bosons which generate a mass for the graviton in the form
\begin{equation}
\Lag_\mathrm{LV} = \frac12 \left(m_0^2 M_{00}^2 + m_1^2 M_{0i}^2 + m_2^2 M_{ij}^2 + m_3^2 M_{ii}^2 + m_4^2 M_{00} M_{ii} \right)\,,
\end{equation}
where in general all the mass parameters \(m_k\) can be all different. In the specific case of~\cite{Dubovsky:2004ud} \(m_1=0\) whereas the remaining terms are non-zero. While the specific production mechanism depends on the details of the UV completion of the low-energy theory, we can expect the massive spin-2 field to be generated during inflation and acquire a vacuum expectation value proportional to \(H\) similarly to the misalignment mechanism for the production of scalar ALPs.

Spin-2 ULDM can be detected in pulsar timing data (see below) as well as with gravitational wave interferometers and atom interferometers, see the corresponding sections in Part~\ref{part:wg4}.

\subsubsection{Spin-2 ULDM signatures in pulsar timing data}

Spin-2 ultra-light dark matter has an effect on pulsar timing data in a similar but distinct way compared to the scalar case. The main difference with the scalar case is that, at least for values of \(\alpha\) that are not too small, the indirect gravitational effects caused by the oscillating dark matter pressure and energy density, which in turn generates oscillating gravitational potentials, are generally outshined by the effects of the direct interaction between bodies through the dark matter.

\begin{figure}[tbhp!]
\centering
    \includegraphics[width=0.8\linewidth]{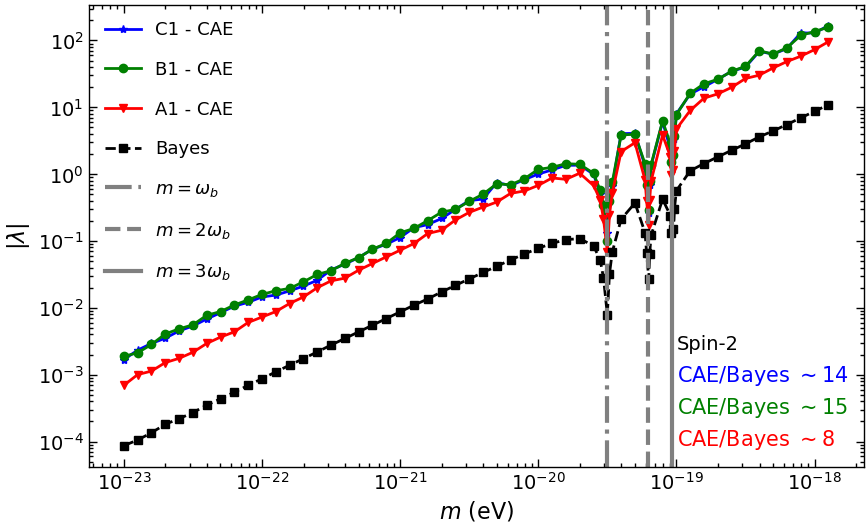}
    \caption{Limits for three 99\% accurate autoencoders: A1 (red triangles), B1 (green squares) and C1 (blue circles) for spin-2 ULDM, compared to the Bayesian limit (black squares). The vertical grey lines mark the resonant frequencies. Figure reproduced from Ref.~\cite{Kus:2025bqz}.}
    \label{fig:sensitivity_plots_autoencoder}
\end{figure}

One of the most prominent effects of the ultra-light dark matter of any spin is the secular drift of the orbital parameters for a Kepler orbit whose period is in resonance with the dark matter oscillations; this can be measured taking advantage of the incredibly precise pulsations of pulsars in a binary system~\cite{Blas:2016ddr,LopezNacir:2018epg,Blas:2019hxz}. The spin-2 case is unique in this landscape because, whereas in the scalar case only binary systems with a large eccentricity can feel the secular drift, if the dark matter has spin-2, because of the quadrupolar nature of the oscillations, any type of orbit will be perturbed. This fact vastly enlarges the pool of possible binaries that can be used to test this model, thereby allowing to probe a wider set of masses. The constraints from these resonances can be the most competitive ones (even compared to fifth-force tests) in the range \(10^{-23}\,\mathrm{eV} \lesssim \mfp \lesssim 10^{-18}\,\mathrm{eV}\), reaching \(\alpha \lesssim 10^{-5}\) for current precision~\cite{Armaleo:2019gil,Kus:2025bqz}. New techniques have also been developed in order to take advantage of the full ULDM effects on the orbital motion, even beyond the resonances~\cite{Kus:2024vpa,Foster:2025nzf,Foster:2025csl,Kus:2025bqz}. As an example specific to the spin-2 case, we show in Fig.~\ref{fig:sensitivity_plots_autoencoder} the projected sensitivities from the one-step Bayesian approach developed in~\cite{Kus:2024vpa} and the machine-learning-based approach of~\cite{Kus:2025bqz}---the three autoencoders are trained including three noise types: (A) pure white Gaussian noise; (B) white Gaussian noise plus nuisance effects; (C) equal-weight combination of white and red Gaussian noise, along with nuisance effects.

\begin{figure}[tbhp]
\centering
	\includegraphics[width=0.8\textwidth]{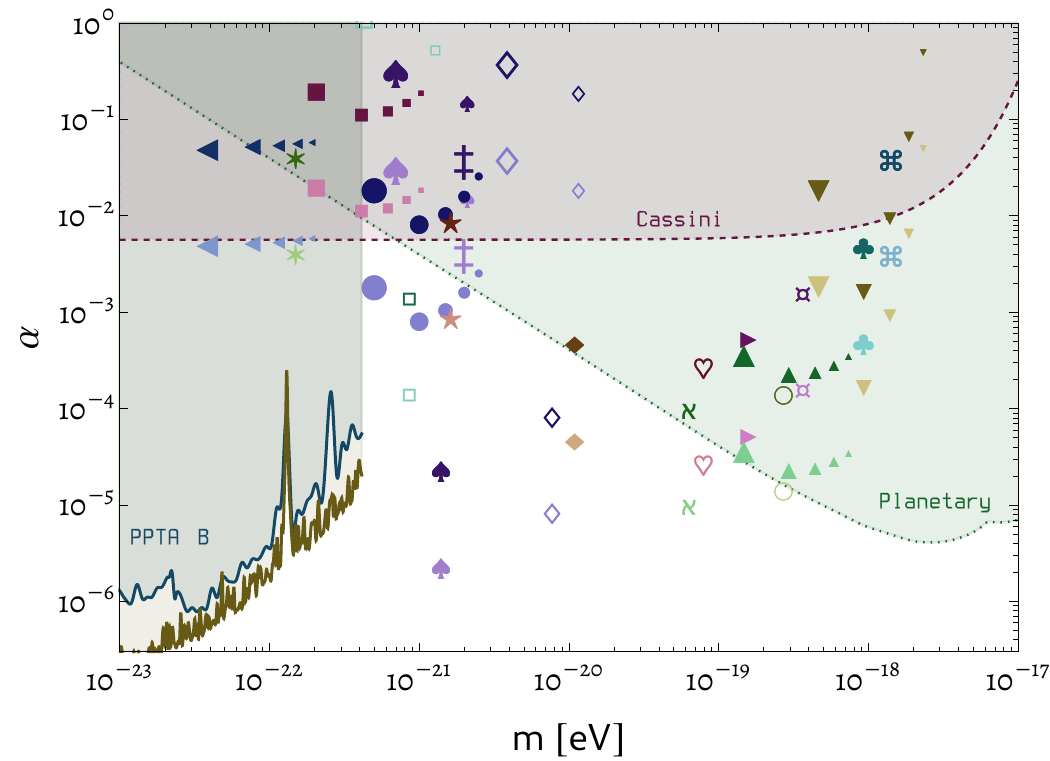}
	\caption{PPTA bounds on the strength of the spin-2 dark matter coupling \(\alpha\) versus the dark matter mass \( \mfp \). The markers show the constraints derived from the non-observation of secular variations in the orbital parameters of a binary pulsar system, see~\cite{Armaleo:2019gil}. Figure reproduced with permission from Ref.~\cite{Armaleo:2020yml}}\label{fig:armaleo}
\end{figure}

Another effect, once again in analogy but with distinct features to the scalar case, is the direct perturbation of the times-of-arrival of the pulses of pulsars (isolated or in binary systems)~\cite{Armaleo:2020yml}. This can be most easily seen by performing a redefinition of the metric that can do away with the interaction term altogether
\begin{equation}\label{eq:coord}
    \tilde{g}_{\mu\nu} = g_{\mu\nu} + \frac{\alpha}{\mpl} M_{\mu\nu}\,.
\end{equation}
In this frame the Earth, the pulsars, and photons no longer interact directly with the dark matter. However, photons travelling from the pulsar to the Earth will follow the geodesics of the new metric \(\tilde{g}\), which explicitly depends on the ultra-light dark matter field \(M_{\mu\nu}\). In other words, it is possible to ``reabsorb'' the massive spin-2 field into the metric. Considering only the Earth term, the time residual in the radio pulses caused by the dark-matter-induced frequency shifts is given by
\begin{equation}\label{eq:shift_earth}
	t_r(t) = -\frac{\alpha\sqrt{\rho_\mathrm{DM}}}{\sqrt2\mfp^2\mpl} \varepsilon_{ij}n^in^j \sin\left(\mfp t+\Upsilon\right)\,,
\end{equation}
where the \(n^i\) unit vector points at the \(i\)-th pulsar. This time variation can be constrained with PTA observations, and the results are shown in Fig.~\ref{fig:armaleo}.

\begin{figure}[tbhp]
\centering
	\includegraphics[width=0.8\textwidth]{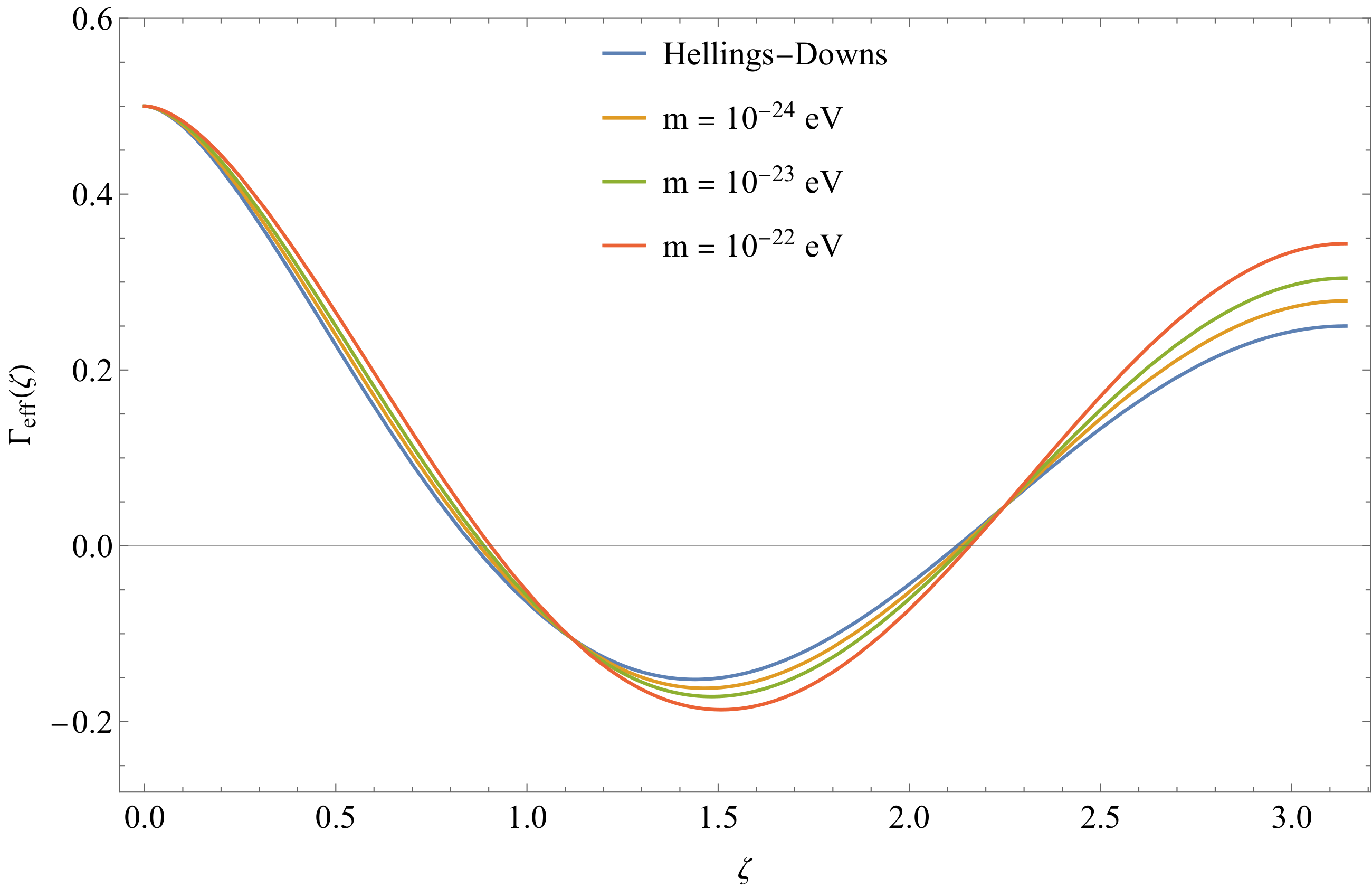}
	\caption{Cross-correlations curves between pulsars separated by an angle \(\zeta\) in the sky with \(\alpha = 10^{-6}\) for different values of \(\mfp\). Figure reproduced with permission from Ref.~\cite{Cai:2024thd}}\label{fig:cai}
\end{figure}

Moreover, the peculiar geometry of the spin-2 dark matter case imprints a specific shape to the angular correlation of the time residual signals of different pulsars in different locations in the sky, offering yet one more opportunity to distinguish the signal from, e.g., a stochastic gravitational wave background. The effects of the individual polarisations follow distinct angular patters~\cite{Armaleo:2020yml}, and, if all polarisations are assumed to be equally populated, as expected in a virialised halo, the overall angular correlation between pulsars separated by an angle \(\zeta\) depends on the strength of the coupling \(\alpha\) as well as the mass of the spin-2 dark matter \(\mfp\)~\cite{Cai:2024thd}, see Fig.~\ref{fig:cai}. This curve deforms the well-known Hellings-Downs curve that is expected for a stochastic background of gravitational waves in which only the two standard (massless) polarisations of the graviton appear; what is more, in the case of dark matter the oscillations are confined to one (very narrow band around a) peak frequency that corresponds to the mass \(\mfp\).

Besides the direct coupling to the pulsars in the form of the \(M_{\mu\nu} T^{\mu\nu}\) term, where \(T_{\mu\nu}\) is the energy-momentum tensor or matter, the mere presence of the spin-2 ultra-light dark matter causes oscillations in the gravitational potentials sourced by the oscillating energy and pressure density of the dark matter itself, in the same way as spin-0 and spin-1 dark matter would do. In this case the perturbations felt by the pulsar-Earth system oscillate with frequency \(2\mfp\) because they are sourced by the energy and pressure density of the dark matter field, which are quadratic in the field. This also means that the effect drops as \(\mfp^{-2}\) (or \(\mfp^{-3}\) for the time residuals) instead of inverse-linearly (\(\mfp^{-2}\) for the time residuals)~\cite{Wu:2023dnp}.

The case in which the spin-2 dark matter is comprised of a wideband spectrum of masses, as for instance would be the case for string compactification scenarios or ``clockwork'' multi-gravity constructions~\cite{Marzola:2017lbt}, has been studied in~\cite{Sun:2021yra}. This is analogous to the case in which the mass distribution of a large number of axions is described by random matrix theories, for instance by the Marcenko-Pastur distribution or Rayleigh distribution. By performing a Bayesian analysis on the first five frequency bins of the NANOGrav 12.5\,yr dataset it is possible to set a limit on the coupling constant of order \(\alpha\lesssim10^{-6}\), similarly to the monochromatic case.

\begin{figure}[tbhp]
\centering
	\includegraphics[width=0.8\textwidth]{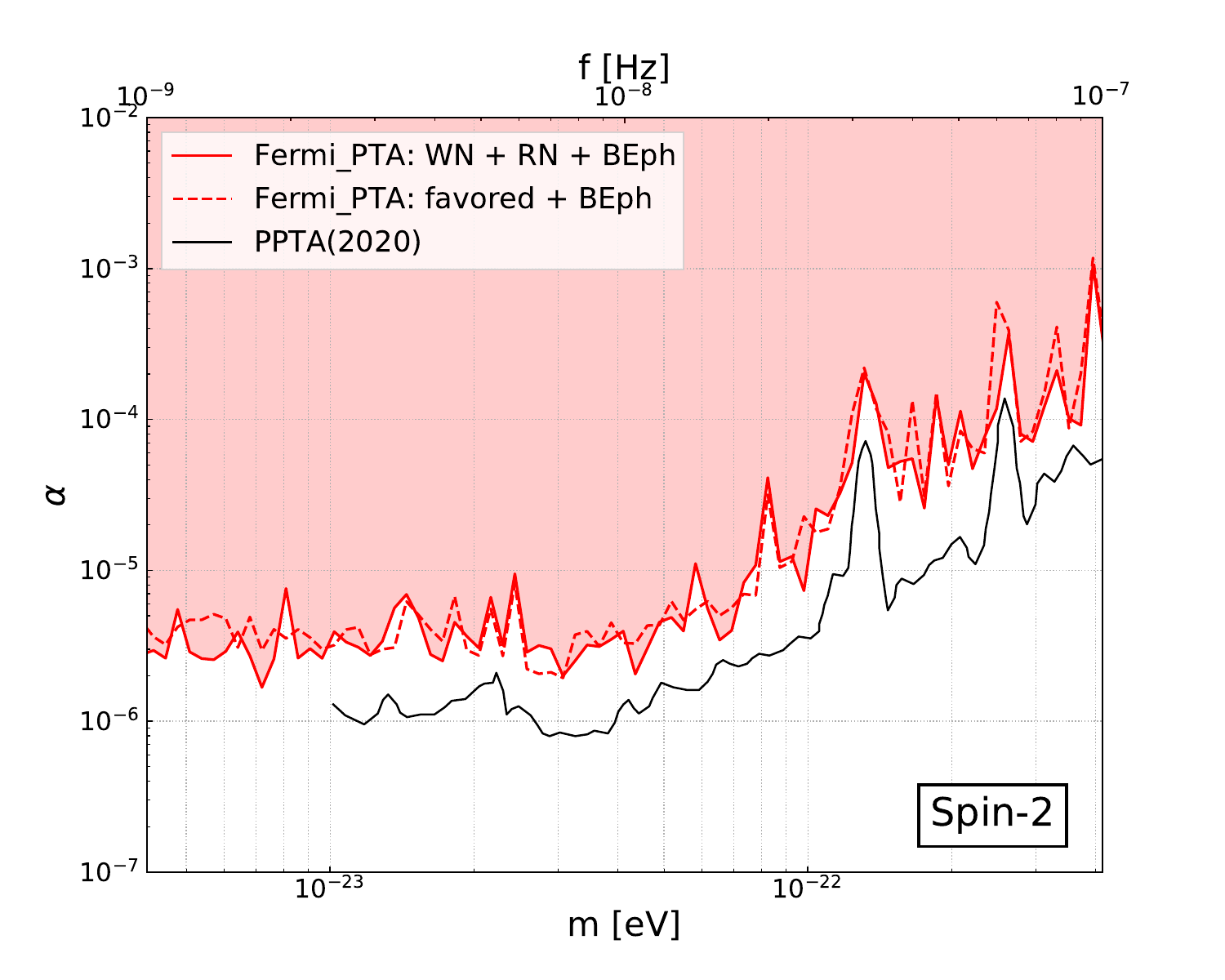}
	\caption{95\% confidence level upper limits of the coupling strength \(\alpha\) for the spin-2 ultra-light dark matter model. The black line indicates the upper limit set by the PPTA data in~\cite{Armaleo:2020yml}. Figure reproduced with permission from Ref.~\cite{Xia:2023hov}} \label{fig:xia}
\end{figure}

A search for the monochromatic time-residuals induced by a with the first gamma-ray pulsar timing array build from the Fermi Large Area Telescope (Fermi-LAT)~\cite{Xia:2023hov}. This approach offers an independent verification of time residual measurements obtained from radio pulsar timing observations, one that is mostly immune to dispersion and scattering effects caused by the interstellar medium. The results of~\cite{Xia:2023hov}, obtained by examining the pulse arrival times from 29 millisecond pulsars, are shown in Fig.~\ref{fig:xia} and are seen to be slightly less constraining than existing limits obtained in~\cite{Armaleo:2020yml} from PPTA data.

%% file: wg3.tex
\thispagestyle{plain}
\part{\texorpdfstring
{WISPs in Astrophysics: from Production to Detection
\\ \textnormal{\small Editors: M.~Giannotti, A.~Lella, G.~Lucente, O.~Straniero \& E.~Todarello}}{WISPs in Astrophysics: from Production to Detection (Editors: M.~Giannotti, A.~Lella, G.~Lucente, O.~Straniero and  E.~Todarello)}}
\label{part:wg3}

\label{sec:Introduction}
\input{WG3/content/Introduction}
\clearpage

\section{Solar Production}
\label{sec:SolarProduction}

\subsection{\texorpdfstring
{Solar production of pseudoscalar particles \\ \textnormal{\small Author: P. Carenza}}{Solar production of pseudoscalar particles (Author: P. Carenza)}}
\label{subsec:Carenza}
\input{WG3/content/Pierluca_Carenza}

\subsection{\texorpdfstring
{Solar production of scalar particles\\ \textnormal{\small Author: T. O'Shea}}{Solar production of scalar particles (Author: T. O'Shea)}}
\label{subsec:OShea}
\input{WG3/content/Tomas-OShea}

\subsection{\texorpdfstring
{Solar production of spin-2 particles \\ \textnormal{\small Authors: C. García-Cely \& A. Ringwald}}{Solar production of spin-2 particles (Authors: C.~García-Cely and A. Ringwald)}}
\label{subsec:Garcia_Cely_Ringwald}
\input{WG3/content/Garcia_Cely_Ringwald}

\section{Stellar Production}
\label{sec:StellarProduction}

\subsection{\texorpdfstring
{Pseudoscalars from stars \\ \textnormal{\small Authors: P. Carenza \& D.F.G.~Fiorillo}}{Pseudoscalars from stars (Authors: P. Carenza and D.F.G.~Fiorillo)}}
\label{subsec:Carenza2}
\input{WG3/content/Pierluca_Carenza_2}

\subsection{\texorpdfstring
{Scalars from stars  \\ \textnormal{\small Authors: S. Balaji \& A. Lella}}{Scalars from stars (Authors: S. Balaji and A. Lella)}}
\label{subsec:Balaji}
\input{WG3/content/Shyam_Balaji}

\subsection{\texorpdfstring
{Vectors from stars \\ \textnormal{\small Author: J. Pradler}}{Vectors from stars (Author: J. Pradler)}}
\label{subsec:Pradler}
\input{WG3/content/Josef_Pradler}

\section{Compact Objects}
\label{sec:CompactObjects}

\subsection{\texorpdfstring
{White Dwarf cooling \\ \textnormal{\small Author: G. Lucente}}{White Dwarf cooling (Author: G. Lucente)}}
\label{subsec:Lucente}
\input{WG3/content/Giuseppe_Lucente}

\subsection{\texorpdfstring
{Supernova Neutrinos: SN~1987A Signal Duration \\ \textnormal{\small Author: G. Raffelt}}{Supernova Neutrinos: SN~1987A Signal Duration (Author: G. Raffelt)}}
\label{subsec:Raffelt}
\input{WG3/content/Georg_Raffelt}

\subsection{\texorpdfstring
{Astrophysical transient constraints beyond the SN~1987A cooling bound \\ \textnormal{\small Author: E. Vitagliano}}{Astrophysical transient constraints beyond the SN~1987A cooling bound (Author: E. Vitagliano)}}
\label{subsec:Vitagliano}
\input{WG3/content/Edoardo_Vitagliano}

\subsection{\texorpdfstring
{WISP production within neutron stars \\ \textnormal{\small Author: M. Buschmann}}{WISP production within neutron stars (Author: M. Buschmann)}}
\label{subsec:Buschmann}
\input{WG3/content/Malte_Buschmann}

\subsection{\texorpdfstring
{Superradiance \\ \textnormal{\small Authors: F. Chadha-Day \& T.~K. Poddar}}{Superradiance (Authors: F. Chadha-Day and T.~K. Poddar)}}
\label{subsec:Chadha-Day}
\input{WG3/content/Francesca_Chadha-Day}

\subsection{\texorpdfstring
{Scalar field modified stars \\ \textnormal{\small Authors: K.~Springmann, R.~Balkin, K.~Bartnick, J.~Serra, S.~Stelzl \& A.~Weiler}}{Scalar field modified stars (Authors: K.~Springmann, R.~Balkin, K.~Bartnick, J.~Serra, S.~Stelzl and A.~Weiler)}}
\label{sec:Springmann}
\input{WG3/content/Konstantin_Springmann}

\section{Conversion of DM WISPs in magnetic fields}
\label{sec:Conversion_DM_WISPs}

\subsection{\texorpdfstring
{Dark Matter conversion in the solar atmosphere \\ \textnormal{\small Authors: M. Taoso \& E. Todarello}}{Dark Matter conversion in the solar atmosphere (Authors: M. Taoso and E. Todarello)}}
\label{subsec:TaosoTodarello}
\input{WG3/content/Taoso_Todarello}

\subsection{\texorpdfstring
{Low-Energy Signatures of Axions near Neutron Stars \\ \textnormal{\small Author: S. Witte}}{Low-Energy Signatures of Axions near Neutron Stars (Author: S. Witte)}}
\label{subsec:Witte}
\input{WG3/content/Samuel_Witte}

\section{\texorpdfstring
{Extragalactic sources \\ \textnormal{\small Authors: F. Calore \& C. Eckner}}{Extragalactic sources (Authors: F. Calore and C. Eckner)}}
\markboth{\thesection.\ Extragalactic sources}{}
\label{sec:Extragalactic_sources}
\input{WG3/content/Calore_Eckner}

\section[Birefringence \\ \textnormal{\small Authors: P. Diego-Palazuelos, E. Ferreira, S. Gasparotto, K. Murai, T. Namikawa, F. Naokawa, \& I. Obata}]{Birefringence \\ \textnormal{\small Authors: P. Diego-Palazuelos, E. Ferreira, S. Gasparotto, K. Murai, T. Namikawa, \\[-.5em] F. Naokawa, \& I. Obata}}
\markboth{\thesection.\ Birefringence}{}
\label{sec:Gasparotto}
\input{WG3/content/Silvia_Gasparotto}

\section{\texorpdfstring
  {Astrophysical searches from radio to millimeter \\ \textnormal{\small Author: M. Regis}}{Astrophysical searches from radio to millimeter (Author: M. Regis)}}
\markboth{\thesection.\ Astrophysical searches from radio to millimeter}{}
\label{sec:Radio}
\input{WG3/content/Marco_Regis}

\section{\texorpdfstring
  {X-ray telescopes \\ \textnormal{\small Authors: F. R. Cand\'{o}n, D. D\'{i}ez-Ib\'{a}\~{n}ez, C. Margalejo, J. Ruz \& J. K. Vogel}}{X-ray telescopes (Authors: F. R. Cand\'{o}n, D. D\'{i}ez-Ib\'{a}\~{n}ez, C. Margalejo, J. Ruz and J. K. Vogel) }}
\markboth{\thesection.\ X-ray telescopes}{}
\label{subsec:RuzVogel}
\input{WG3/content/Julia_Vogel}

\section{\texorpdfstring
  {Fermi-LAT and Cherenkov Telescopes \\ \textnormal{\small Authors: F. Calore \& C. Eckner}}{Fermi-LAT and Cherenkov Telescopes (Authors: F. Calore and C. Eckner)}}
\markboth{\thesection.\ Fermi-LAT and Cherenkov Telescopes}{}
\label{subsec:CaloreEckner2}
\input{WG3/content/Calore_Eckner_2}

\section{\texorpdfstring
  {The future of the MeV-range astrophysics \\ \textnormal{\small Author: P. De la Torre Luque}}{The future of the MeV-range astrophysics (Author: P. De la Torre Luque)}}
\markboth{\thesection.\ The future of the MeV-range astrophysics}{}
\label{subsec:deLaTorre}
\input{WG3/content/Pedro_De_la_Torre_Luque}

\section{Other Messengers}
\label{sec:OtherMessangers}

\subsection{\texorpdfstring
  {Neutrinos  \\ \textnormal{\small Author: M. Wurm}}{Neutrinos (Author: M. Wurm)}}
\label{subsec:Wurm}
\input{WG3/content/Michael_Wurm}

\subsection{\texorpdfstring
  {Gravitational Waves \\ \textnormal{\small Author: R. Vicente}}{Gravitational Waves (Author: R. Vicente)}}
\label{subsec:Vicente}
\input{WG3/content/Rodrigo_Vicente}

\cleardoublepage

%% file: WG3/content/Introduction.tex
As discussed in the previous chapters, WISPy fields, such as axions, axion-like particles (ALPs), and hidden photons, represent a frontier in the exploration of fundamental physics, bridging particle physics, cosmology, and astrophysics. In the following sections, we will explore their role in astrophysics and summarize the effort to search for their signatures in astrophysical observations.

In the previous section, dedicated to cosmology, we discussed how WISPs are essential for explaining large-scale phenomena such as dark matter, the evolution of the universe and, consequently, the origin of structures. In this section, we shift our focus to smaller scales, exploring the impact of WISPs in extragalactic sources and compact objects. Specifically, we investigate their role in shaping the behavior of extreme astrophysical environments, from high-energy emissions in distant galaxies to the energy loss mechanisms in stars and compact remnants. By bridging the link between cosmological-scale phenomena and localized astrophysical processes, this section highlights how WISPs can provide unique insights across vastly different scales of the universe.

As we shall see, astrophysics, with its natural laboratories of extreme temperatures, densities, and magnetic fields, plays a pivotal role in constraining WISPs' properties and uncovering their potential signatures. Astrophysical environments provide unique opportunities to test WISP theories due to their sensitivity to small interaction strengths, extremely rare events, and their ability to probe energy scales unattainable in terrestrial experiments. Stars, as well as extragalactic sources such as active galactic nuclei offer diverse conditions where WISPs might be produced, leaving observable imprints through exotic energy losses, modified stellar evolution, or direct detection of secondary signals.

This section explores these aspects comprehensively, examining stars and the broader universe as laboratories and factories for WISPs---from the production mechanisms in astrophysical environments such as the Sun, red giants, neutron stars, and supernovae, to the resulting constraints derived from energy-loss arguments and other observational data. We review recent updates on stellar and supernova bounds, alongside insights into the role of compact objects and cosmological sources in advancing our understanding of WISP physics. Alongside this, we discuss detection strategies for WISPs, highlighting the role of current and future astronomical tools and techniques, including helioscopes, interferometers, and high-energy observatories capable of probing WISP interactions. Emphasis is placed on how cutting-edge instrumentation and observational campaigns can refine our knowledge of WISP astrophysics and open new avenues for discovery.
Our ultimate aim is the presentation of \emph{state-of-the-art results} and the future \emph{strategic directions}. Among the state-of-the-art review, we highlight several recent astrophysical constraints on WISP properties. In particular, a rich and successful investigation of the role of compact objects in the study of WISPs has produced a plethora of new results in the last few years. These include several improvements in the axion production processes in supernova environments, with significant revision of the production channels and even the identification of novel production channels, updated constraints from SN1987A, revision of the axion production in neutron stars, and especially a revision of the WISP production mechanisms near the very strong magnetic fields produced by very compact objects. On the experimental side, extensive studies have analyzed various indirect detection methods, such as gamma-ray and neutrino observations from supernovae, which complement traditional laboratory experiments (discussed in Part~\ref{part:wg4}).

The emergence of new instruments in the last few years, and the next generation plans for instruments in the entire electromagnetic spectrum, as well as neutrino and gravitational waves telescopes, provide encouraging perspectives for the future of WISP astrophysics. Aligning efforts with major astrophysics missions like JWST, Athena, and CTA also promises novel avenues to complement laboratory searches for WISPs.

Concerning the strategic directions which emerge from this study, one can identify some pressing needs. These include the necessity to improve the experimental coverage of the entire electromagnetic spectrum, especially in the MeV region where several WISP signatures could be hiding. This \emph{MeV Gap} remains a critical frontier for understanding photon-WISP interactions. Developing instruments sensitive to MeV-scale photons, such as space-based telescopes, could transform our understanding of WISP physics. In addition, several theoretical efforts are required to refine models of WISP production and propagation in astrophysical environments, especially in strong and highly inhomogeneous magnetic fields, with the related conversion mechanisms. The substantial advancements in this respect in the last few years have also highlighted difficulties and needs for the next generation of studies.

It is evident that astrophysics provides unparalleled opportunities to uncover the properties of WISPs, especially in light of the technological advancements discussed in this review. These advancements, combined with the unique observational capabilities of astrophysics, enable us to explore regions of parameter space that are otherwise inaccessible through laboratory experiments.
To fully harness this potential, it is crucial to assess the current status of research, identify the next essential steps, and foster collaborations between the astrophysical and particle physics communities. Such integrated efforts can maximize the complementarity between astrophysical constraints and laboratory experiments, enhancing our ability to explore unexplored parameter spaces.
This section aims to synthesize theoretical advances, observational data, and computational models to highlight the interplay between astrophysical phenomena and the search for WISPs. By defining the current constraints on WISP properties, identifying promising observational targets, and outlining future directions for experimental and theoretical investigations, we seek to advance our understanding of these elusive particles. Viewing the universe as a dynamic and versatile laboratory, astrophysics not only refines our theoretical frameworks but also broadens the discovery potential of WISPs.

%% file: WG3/content/Pierluca_Carenza.tex
\subsubsection{Introduction} 
\begin{figure*}[t!]
    \centering  \includegraphics[width=0.495\columnwidth]{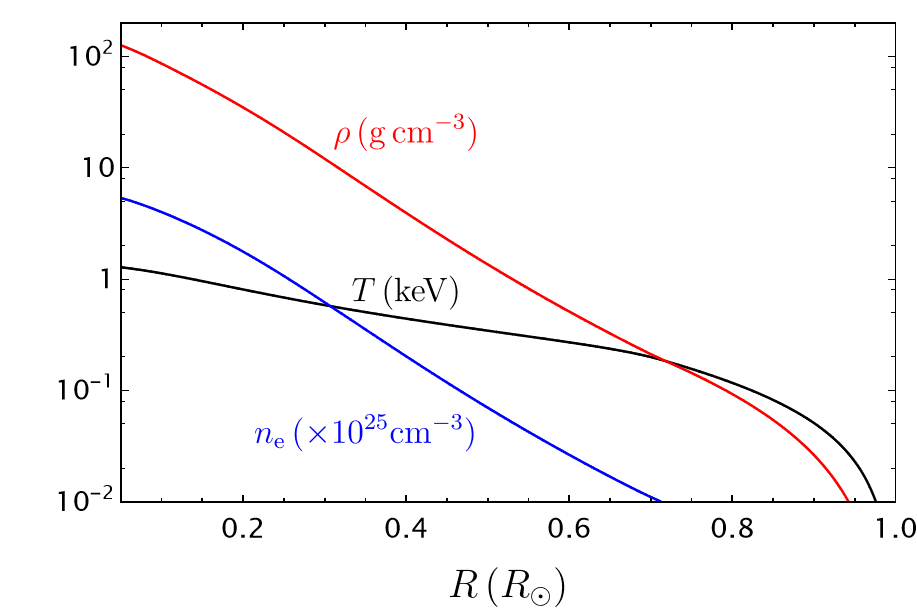}   \includegraphics[width=0.495\columnwidth]{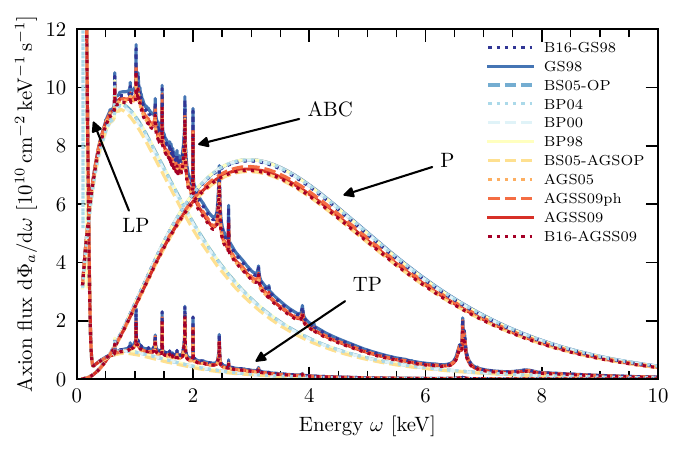}  
    \caption{{\it Left:} Profile of the solar properties. Temperature (black line), density (red line) and electron number density (blue line) are shown as a function of the solar radius for the photospheric model in Ref.~\cite{Magg:2022rxb}.\\{\it Right:} Total solar axion fluxes for various solar models, for $g_{ae} = 10^{-12}$, $g_{a\gamma\gamma} = 10^{-10}~{\rm GeV}^{-1}$ and $m_{a}\ll {\rm keV}$. Figure taken from Ref.~\cite{Hoof:2021mld} with permission.}
        \label{fig:Sunmodel}
\end{figure*}

The Sun is the most well-known star, therefore it is a promising laboratory to probe new physics. The Sun is a main sequence star with a mass $M_{\odot}=1.989\times 10^{33}$~g, a radius $R_{\odot}= 6.9598 \times 10^{10}$~cm, a core temperature of $\sim1.3$~keV and a core density of $\sim 150~{\rm g}~{\rm cm}^{-3}$.
The chemical composition of the Sun varies from an innermost region composed of $\sim 36\%$ of hydrogen ($^{1}$H), $\sim 62\%$ of helium ($^{4}$He); to a more external layer composed of $\sim 76\%$ of $^{1}$H and $\sim 23\%$ of $^{4}$He. The profiles of some important quantities for axion physics are shown in Fig.~\ref{fig:Sunmodel} (left panel).
The Sun emits photons with a luminosity $L_{\odot}= 3.8418 \times 10^{33}$ erg\,${\rm s}^{-1}$~\cite{Serenelli:2009yc,Magg:2022rxb}, not considering a percent contribution of energy lost into neutrinos~\cite{Bahcall:1989ks}.
Neutrinos are the messengers from nuclear reactions fueling the Sun. As a matter of fact, in the Sun protons undergo fusion into helium to produce  radiation pressure~\cite{Vinyoles:2016djt}. The so-called $pp$ chain is a series of processes initiated by weak interactions, whose first step is the proton fusion to produce deuterium $p+p\to \,^{2}{\rm H}+e^{+}+\nu_{e}$. The other steps of the chain involve strong interactions converting predominantly deuterium into $^{4}$He and releasing $26.73$~MeV of energy per chain, with $0.59$~MeV of this energy lost into neutrinos. Neutrinos stemming from $pp$ chains dominate the neutrino flux at Earth in the MeV range~\cite{Vitagliano:2019yzm} and
have been detected in several experiments~\cite{GALLEX:1992gcp,SAGE:1994ctc,BOREXINO:2014pcl,BOREXINO:2018ohr,PandaX:2024jjs}. The second most important energy source for the Sun is the CNO cycle, in which the proton fusion is catalyzed by carbon, nitrogen and oxygen, which are consumed and completely regenerated within the cycle. Recently, CNO neutrinos were detected, in agreement with the standard solar model~\cite{BOREXINO:2020aww}. Finally, neutrinos can be produced by thermal processes, resulting in a keV-range flux~\cite{Vitagliano:2017odj,Vitagliano:2019yzm}.

Given this comprehensive and deep knowledge of the solar physics, it is reasonable to expect that any deviation from the expected behavior might be a signal of new physics. In the following, we will focus on using the Sun as a probe of sub-keV pseudoscalar particles, discussing their production and phenomenological implications. In particular, we will consider an axion coupled to photons, electrons and nucleons as 
\begin{equation}
   \mathcal{L}=-\frac{1}{4}g_{a\gamma\gamma}a\,F_{\mu\nu}\tilde{F}^{\mu\nu}+\sum_{i=e,p,n}\frac{g_{ai}}{2m_{i}}\bar{\Psi}_{i}\gamma^{\mu}\gamma^{5}\Psi_{i}\partial_{\mu}a\,,
\end{equation}
where $g_{a\gamma\gamma}$ is the axion-photon coupling, $g_{ai}$ the axion-fermion coupling (where the fermion can be an electron and/or a nucleon, with mass $m_{i}$), $F_{\mu\nu}$ is the electromagnetic field strength tensor and $\tilde{F}^{\mu\nu}$ its dual.

\subsubsection{Production processes}
\emph{Primakoff process.---} The axion-photon coupling makes possible to convert thermal photons into axions in the microscopic electromagnetic fields of electrons and ions. Initially studied for the photo-production of neutral pions~\cite{Primakoff:1951iae}, this process was then applied to axion production in stars~\cite{Dicus:1979ch}. The most recent evaluations report an almost thermal axion flux peaked at $\sim3$~keV and with a flux at Earth $\Phi_{a}=g_{10}^{2}\,3.58\times10^{11}~{\rm cm}^{-2}{\rm s}^{-1}$, where $g_{10}=g_{a\gamma\gamma}/10^{-10}~{\rm GeV}^{-1}$~\cite{Wu:2024fsf}

\emph{Plasmon conversion in the solar magnetic field.---} In a similar spirit, macroscopic magnetic fields allow for the plasmon conversion into axions, as originally proposed in the context of magnetars~\cite{Mikheev:2009zz} and then refined and applied to the Sun~\cite{Caputo:2020quz,OHare:2020wum,Guarini:2020hps}.
Longitudinal plasmons (LP) resonantly convert into axions when their energy matches the solar plasma frequency; while transverse (TP) ones, when the axion mass is equal to the plasma frequency.
LP conversions give rise to a low-energy axion flux peaked at $\sim0.1$~keV~\cite{Caputo:2020quz,OHare:2020wum}. TP axion fluxes peak at $\sim1$~keV, with a luminosity that strongly depends on the axion mass, which has to be in the range $10~{\rm eV}\lesssim m_{a}\lesssim300~{\rm eV}$ to achieve the resonance~\cite{Guarini:2020hps,Hoof:2021mld}. 

\emph{ABC processes.---} Axions coupled to electrons are produced in the Sun via several processes: Atomic recombination and de-excitation~\cite{Dimopoulos:1986mi,Dimopoulos:1986kc}, electron-electron~\cite{Raffelt:1985nk} or electron-ion~\cite{Zhitnitsky:1979cn,Krauss:1984gm} Bremsstrahlung and Compton scattering  (ABC)~\cite{Mikaelian:1978jg,Fukugita:1982ep,Fukugita:1982gn}.
The most updated calculation of the ABC flux is given in Ref.~\cite{Hoof:2021mld}, improving on the comprehensive work of Ref.~\cite{Redondo:2013wwa}. 
In the solar core hydrogen and helium are fully ionized, therefore recombination and de-excitation processes will involve metals, i.e. any other element heavier than helium, which are only partially ionized. For instance, $\sim20\%$ of iron is not fully ionized in the core~\cite{Hoof:2021mld}. 
Electron-ion bremmstrahlung was recently revisited to include degeneracy effects~\cite{Carenza:2021osu}. As a rough estimate, these three processes contribute to the solar axion number flux with the following proportions: $65\%$ of bremsstrahlung, $28\%$ of atomic processes and $7\%$ of Compton scattering~\cite{Redondo:2013wwa}.

\emph{Nuclear reactions.---} As we discussed, nuclear reactions belonging to the $pp$ chain fuel the Sun. The second step of the chain is the process $p+\,{2}{\rm H}\to \,^{3}{\rm He}+\gamma$, producing a 5.5~MeV photon. It was noted that axions coupled to nucleons will allow for an analogous process, where the photon is replaced by an axion~\cite{Raffelt:1982dr}. Experimental searches for this line signal were performed in neutrino detectors~\cite{Borexino:2012guz} and future experiments will improve the current sensitivities~\cite{Lucente:2022esm}. Axions might also show up in steps of the CNO cycle, leading to de-excitations of some metals~\cite{Massarczyk:2021dje}. The largest contribution comes from the $^{57}{\rm Fe}$ de-excitation, emitting a $14.4$~keV axion with a flux $\Phi_{a}=5.06 \times 10^{23}\ (g_{aN}^{{\rm eff}})^2 \ \rm{cm}^{-2}\rm{s}^{-1} \,\ ,$
where  $g_{aN}^{{\rm eff}}=0.16\, g_{ap} +1.16\, g_{an}$~\cite{Avignone:2017ylv}. Experimental searches to detect axion fluxes focused on decays of $^{7}{\rm Li}$~\cite{Krcmar:2001si,CAST:2009klq,Borexino:2008wiu,Belli:2012zz}, $^{57}{\rm Fe}$~\cite{Avignone:2017ylv}, $^{83}{\rm Kr}$~\cite{DiLuzio:2021qct}  and $^{169}{\rm Tm}$~\cite{Derbin:2023yrn}.

\begin{figure}[t!]
    \centering  \includegraphics[width=\columnwidth]{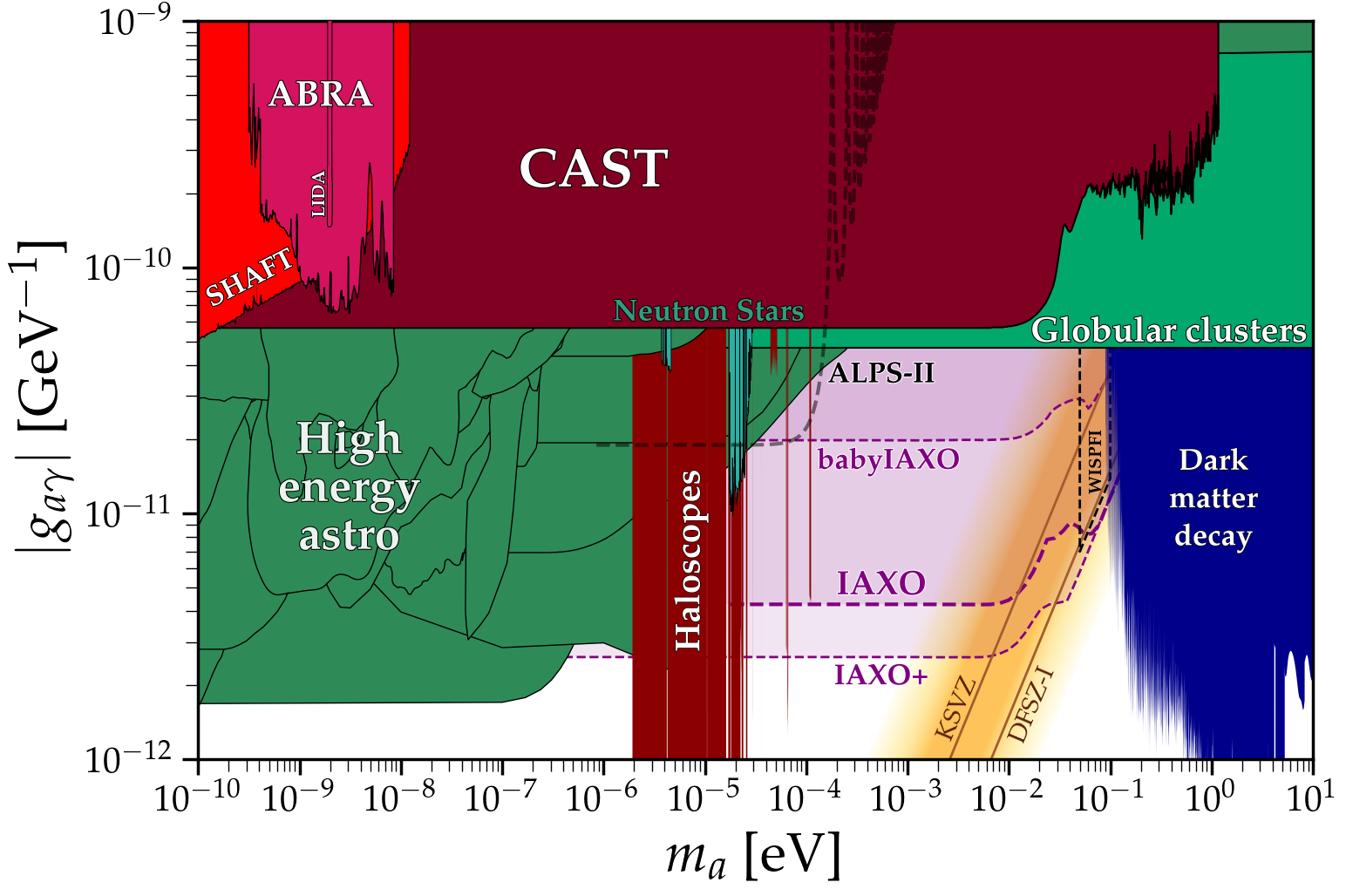}
        \caption{Expected IAXO sensitivities (dashed purple), for various stages: babyIAXO, IAXO  and IAXO+. For comparison, we show astrophysical bounds in green, experimental bounds in red, including the CAST and haloscopes constraints, and dark matter decay bounds in blue. Together helioscope and haloscope searches will probe a large portion of the parameter space, including the QCD axion band (yellow region). Figure taken from Ref.~\cite{AxionLimits}.
        }
        \label{fig:IAXO}
\end{figure}

\subsubsection{Phenomenological implications} 
The most updated calculation of the solar axion fluxes was performed in Ref.~\cite{Hoof:2021mld}, resulting into Fig.~\ref{fig:Sunmodel} (right panel). Here, the fluxes are calculated for $g_{ae}=10^{-12}$, $g_{a\gamma\gamma}=10^{-10}~{\rm GeV}^{-1}$ and $m_{a}\ll {\rm keV}$. It is important to notice that these fluxes are quite well understood, making solar axion fluxes a primary target for several direct and indirect searches. The most important arguments for indirect searches are summarized here. Helioseismological sound-speed profiles are able to exclude energy losses beyond the standard ones if the axion luminosity is $L_{a}\gtrsim 0.2L_{\odot}$~\cite{Schlattl:1998fz}. A more sensitive probe of the solar interior is given by neutrino fluxes, which exclude any loss with $L_{a}\gtrsim 0.1L_{\odot}$~\cite{Gondolo:2008dd}.
Putting together helioseismological and neutrino signals allowed losses with $L_{a}\gtrsim 0.03L_{\odot}$ to be excluded, resulting in a constraint on the axion-photon coupling $g_{a\gamma\gamma}\le4.1\times10^{-10}~{\rm GeV}^{-1}$~\cite{Vinyoles:2015aba} (neglecting the TP/LP plasmo conversion); an upper limit on the axion-electron coupling $g_{ae}\le6.6\times10^{-12}$~\footnote{From a private communication with Maurizio Giannotti.} and on the axion-nucleon coupling combination $g_{aN}^{\rm eff}\le 1.9\times10^{-6}$~\cite{DiLuzio:2021qct}.

Several direct searches aim to convert the solar axion flux into photons in controlled laboratory magnetic fields. Then, X-ray technologies are used to reveal the produced photons. This is the idea behind helioscopes, such as the famous CERN Axion Solar Telescope (CAST) which sets a robust bound on the axion-photon coupling $g_{a\gamma\gamma}<0.66\times10^{-10}~{\rm GeV}^{-1}$ at $95\%$ CL for $m_{a}\lesssim20~{\rm meV}$~\cite{CAST:2017uph} (see the red region in Fig.~\ref{fig:IAXO}. Moreover, CAST sets a bound on the combination of electron and photon couplings, $g_{a\gamma\gamma}g_{ae}<8.1\times10^{-23}~{\rm GeV}^{-1}$ at $95\%$ CL for $m_{a}\lesssim10~{\rm meV}$~\cite{Barth:2013sma}. We mention Refs.~\cite{CAST:2009jdc,CAST:2009klq} with similar bounds on the combination of nucleon and photon couplings.

An improved version of CAST, with a stronger magnetic field, is the International Axion Observatory (IAXO), designed to reach $g_{a\gamma\gamma}\sim10^{-12}~{\rm GeV}^{-1}$~\cite{Armengaud:2014gea,IAXO:2019mpb}. The projected sensitivity of IAXO is shown in Fig.~\ref{fig:IAXO} (thick dashed purple line), including the prototype version babyIAXO and a futuristic improved IAXO+ (thin dashed purple lines). Remarkably, this experiment will explore a good portion of the QCD axion band (yellow band) for axion masses from tens to hundreds of meV. In particular, around a few tens of meV, IAXO will explore a portion of the parameter space that cannot be probed by future haloscopes~\cite{BREAD:2021tpx,Baryakhtar:2018doz}.
The role of IAXO will be of primary importance in case of a discovery, since this helioscope can distinguish between axion models~\cite{Jaeckel:2018mbn} and determine the axion mass, for masses above $20$~meV~\cite{Dafni:2018tvj,Hoof:2021mld}. In addition, axions will constitute another messenger of stellar properties, allowing us to probe the innermost solar structure~\cite{Jaeckel:2019xpa,OHare:2020wum}.

\subsubsection{Conclusions} 
The Sun is the closest star and, therefore, it is an excellent source of axions. Over the years there have been several strategies to probe solar axions, spanning indirect to direct searches. Even though no positive signal has been found so far, solar bounds are among the most robust axion constraints, providing precious information on the nature of axions. In the coming years, new technologies will push the boundaries of solar axion searches into a theoretically motivated region of the parameter space. While hoping for a revolutionary discovery, the axion community continues to investigate the Sun and its properties with increasing precision. At the same time, the design of new experiments is stimulating the technological advancement. Now, more than ever, the future of solar axion searches looks bright.

%% file: WG3/content/Tomas-OShea.tex
Similar to the pseudoscalar case of axions, it is theoretically possible for scalar particles to be produced in the solar interior. This can occur through a number of processes depending on the couplings of the particular scalar model to standard model (SM) particles.

For scalar models with a photon coupling, production by the Primakoff process is possible~\cite{OShea:2024jjw}. A well known example of such a model is the chameleon screened scalar (for a detailed description of the chameleon mechanism see Refs.~\cite{Khoury:2003rn,Vagnozzi:2021quy} and Sec.~\ref{sec:1.2.5}). Light scalar fields mediate unobserved long-range forces, so for a hypothetical scalar field model to be viable it must include a screening mechanism to suppress long range 5th forces in our local vicinity. The chameleon mechanism achieves this by including a coupling to matter fields that gives the scalar a density-dependent effective mass, suppressing the field in regions of high density such as the solar system whilst allowing it to have an effect on intergalactic scales and to behave as a dynamical dark energy driving the accelerated expansion of the universe. It has been shown that scalar fields that interact with matter in such a way must include a 2-photon coupling~\cite{Brax:2010uq} and as such can be produced by the Primakoff process in a similar way to axions. However there is a difference in the allowed processes due to the difference in the form of the photon coupling between scalars and pseudoscalars. The interactions have the form
\begin{equation}
\label{eq:photon_interactions}
\begin{split}
    \mathcal{L}_a \supset& \, \frac{g_{a\gamma\gamma}}{4} a F_{\mu\nu} \tilde{F}^{\mu\nu} 
    \propto a\, \mathbf{E}\cdot\mathbf{B}\\
    \mathcal{L}_\phi \supset& \, \frac{g_{\phi\gamma\gamma}}{4} \phi F_{\mu\nu} F^{\mu\nu}
    \propto \phi\, \left(B^2 - E^2\right)\,,
\end{split}
\end{equation}
where $a$ is the axion field, $\phi$ is the chameleon-screened scalar field, $g_{a\gamma\gamma}$ is the axion-photon coupling, $g_{\phi\gamma\gamma}$ is the chameleon-photon coupling, $F^{\mu\nu}$ is the electromagnetic field strength and $\tilde{F}^{\mu\nu} = \epsilon^{\mu\nu\alpha\beta}F_{\alpha\beta}$ is its Hodge dual.

Clearly the Primakoff production of both scalars and pseudoscalars from 2 photons is allowed, but the production resulting from plasma processes differs slightly. In a plasma, the properties of electromagnetic excitations are affected by the presence of charged particles~\cite{Raffelt:1996wa,Kapusta:2023eix}. The result is a spin-1 quasiparticle with 1 longitudinal and 2 transverse modes. The transverse modes can be interpreted as standard photons with modified dispersion relations due to the plasma. The longitudinal mode, however, does not exist in a vacuum and can be interpreted as a direct result of the collective motion of the charged particles in the plasma~\cite{Braaten:1993jw}. These are commonly referred to as `plasmons' with the transverse modes simply labelled photons. As in a vacuum, photons consist of oscillating electric and magnetic fields perpendicular both to each other and to the direction of propagation. The plasmon, however, consists only of an electric field oscillating along the direction of propagation with no magnetic component.

\begin{figure}[t!]
    \centering \includegraphics[width=\linewidth,keepaspectratio]{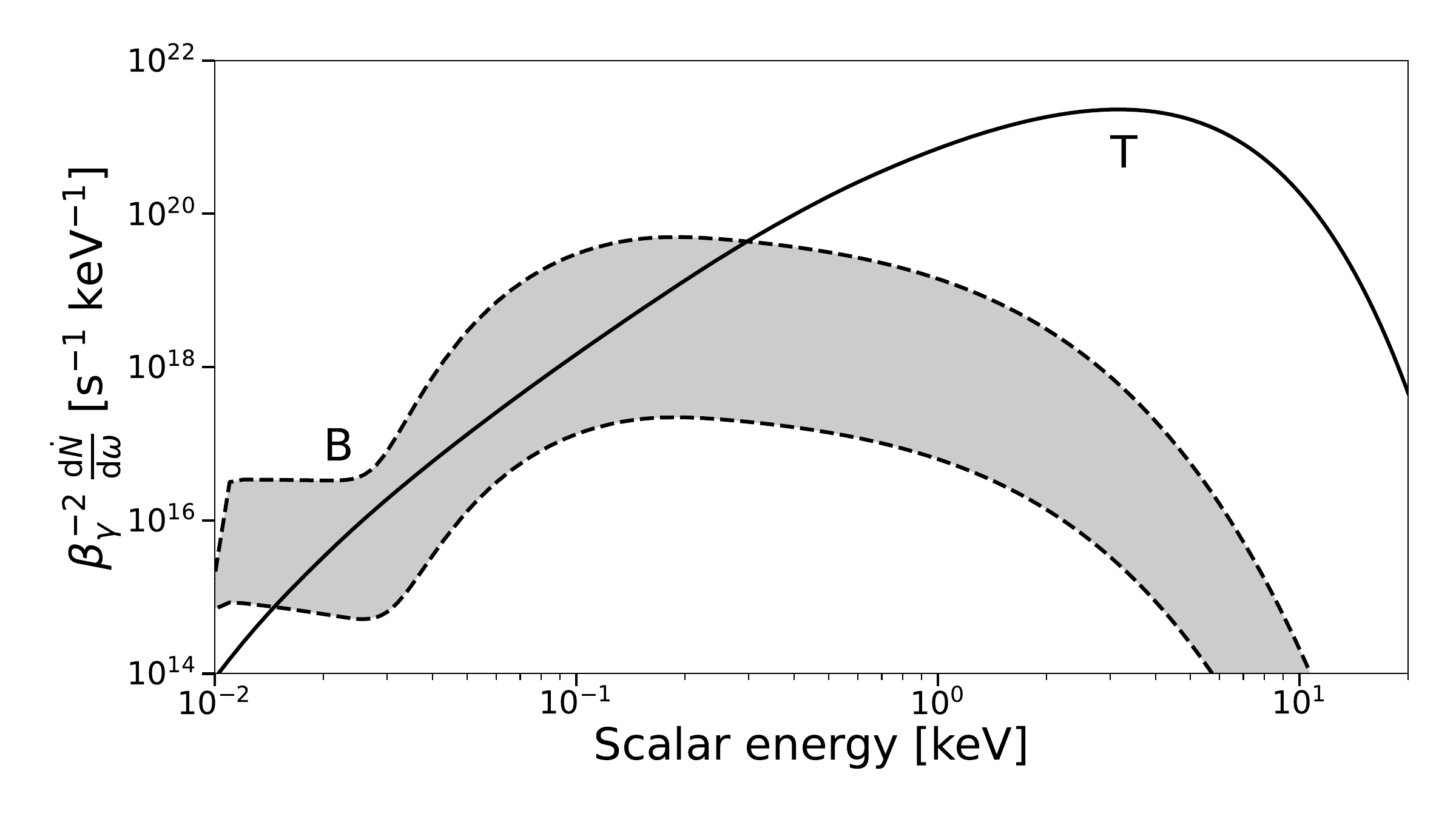}
    \caption{Comparison of the emission spectra (differential particle production rate per unit energy) of solar chameleons arising via Primakoff production from (transverse) photons in the presence of charged particles (``T'', solid curve), as well as from production in the bulk magnetic field (``B'', dot-dashed curve). The width of the ``B'' band reflects the uncertainty on the value of the magnetic field strength~\cite{OHare:2020wum,Hoof:2021mld}. Note that the production rate is normalised by $\beta_{\gamma}^2$, in order to factor out the dependency on the chameleon-photon coupling. We have adopted the AGSS09 stellar model~\cite{Asplund:2009fu}, whereas the chameleon parameters are fixed to $\beta_m = 10^2$ and $n=1$, with the height of the potential set to the dark energy scale $\Lambda_{\text{DE}} = 2.4$\,meV (with this choice of parameters the spectra are indistinguishable from that of a fixed low mass scalar $m_\phi \ll \omega$, as $m_{\mathrm{eff}}\ll\omega$ everywhere in the Sun - see main text or Ref.~\cite{OShea:2024jjw} for details). }
    \label{fig:chameleon-spectrum}
\end{figure}

This has an effect when it comes to the production of (pseudo)scalars in the bulk magnetic fields of the Sun. A photon interacting with a magnetic field can provide both $\mathbf{E}$ and $\mathbf{B}$, so according to Eq.~\eqref{eq:photon_interactions} both scalars and pseudoscalars can be produced~\cite{OShea:2024jjw,Raffelt:1987im,Guarini:2020hps}. This gives the well known spectrum for axions produced in solar magnetic fields (see Fig.~5 and Fig.~7 of Ref.~\cite{Guarini:2020hps}), and a very similar spectrum for scalars (the ``B'' curve in Fig.~\ref{fig:chameleon-spectrum}). However, since a plasmon can only supply an electric field, only psuedoscalars can be produced by the interaction of plasmons with an external magnetic field, making this an example of a production process exclusive to axions that is not shared by scalars.

Similar arguments can be used for the Primakoff process in the electric fields of charged particles in the solar plasma. Since photons can supply both $\mathbf{E}$ and $\mathbf{B}$, both scalars and pseudoscalars can be produced through the interaction of photons with charged particles. This gives the classic axion Primakoff spectrum (see eg. Fig.~9 of Ref.~\cite{Irastorza:2018dyq}) and a very similar spectrum for scalars (the ``T'' line in Fig.~\ref{fig:chameleon-spectrum}).
However, only scalars can be produced by the interaction of a plasmon with an electric field. This process - exclusive to scalars - is expected to dominate the flux at low energies.
The calculation of this flux is complicated by the limits of the interpretation of the plasmon as a quasiparticle: for high momenta, the plasmon cannot propagate~\cite{Braaten:1993jw,Raffelt:1996wa}, but even without the collective motion of electrons this production process still exists in the form of $e^- e^- \to e^- e^- \phi$~\cite{OShea:2024jjw}. The full calculation of the flux from this process is the subject of a work currently in progress.

The production of scalars in the Sun is not limited to the Primakoff process. Depending on the particular model and its interactions with standard model particles, other processes may be possible, such as bremsstrahlung or Compton production. The production of chameleons, which by definition must include electron and nucleon couplings, from these processes has not yet been calculated. However, studies on the solar production rate of Higgs-coupled scalars - massive scalars coupling to the SM only by a mixing angle with the SM Higgs~\cite{OConnell:2006rsp} - have been performed~\cite{Dev:2020jkh,Balaji:2022noj}.
In a comparison of the bremsstrahlung processes $e^- N \to e^- N S$ and $N N \to N N S$, with $N$ being a nucleon and $S$ being the Higgs-coupled scalar; the Compton process $e^- \gamma \to e^- S$; the loop-level Primakoff process $\gamma X \to S X$, with $X$ being a charged particle; and resonant mixing with the plasmon; it was found that the dominant production channel for all scalar masses is $e^-N$ bremsstrahlung, with the scalar $S$ predominantly coupling to the nucleon. The solar emission spectra for these processes can be found in Fig.~8 of Ref.~\cite{Balaji:2022noj}.

Using solar energy-loss arguments, limits on the relevant parameter space can be calculated from the production rate. In the case of the Higgs-coupled scalar, this results in constraints on the values of the scalar mass $m_S$ and the mixing angle, usually expressed as $\sin\theta$. Requiring that the exotic energy loss is no more than $3\%$ of the solar luminosity, stringent constraints can be placed on the mixing angle for $m_s \lesssim 10$ keV. This is shown in Fig.~10 of Ref.~\cite{Balaji:2022noj}.
The chameleon parameter space is more complicated. Typical chameleon models with a $\phi^{-n}$ potential depend on 4 parameters: the dimensionless matter coupling $\beta_\mathrm{m}$, the dimensionless photon coupling $\beta_\gamma \equiv g_{\phi\gamma\gamma} M_\mathrm{Pl}$, the height of the potential $\Lambda$, and the potential parameter $n$~\cite{Khoury:2003rn,CAST:2018bce,Vagnozzi:2021quy}.
Analysing the region of parameter space accessible to solar chameleons shows that solar chameleon studies are complimentary to 5th force experiments and can probe a previously unexplored region of the parameter space~\cite{OShea:2024jjw}.
Setting the `standard' values of $n=1$ and $\Lambda=2.4$ meV (the dark energy scale), the solar energy-loss bounds provide the most stringent published constraints on the value of $\beta_\gamma$, i.e. $\beta_\gamma \lesssim 10^{10}$ (see Fig.~5 of Ref.~\cite{OShea:2024jjw}). However, it should be noted that the previous best limit comes from the CAST collaboration~\cite{CAST:2018bce} using only the chameleon flux from the magnetic field in the tachocline~\cite{Brax:2010xq}, and an updated CAST analysis using the more complete flux of Ref.~\cite{OShea:2024jjw} is expected to improve on the solar energy-loss bound.
These standard parameters appear to be excluded by recent experiments~\cite{Burrage:2014oza,Upadhye:2012qu,Yin:2022geb}. However, moving away this choice of parameters these bounds drop off very quickly, whereas the solar bounds are largely insensitive to such changes. This is because apart from $\beta_\gamma$, which scales the production rate, the chameleon parameters only enter into the production rate through the effective mass term. The constraints on $\beta_\gamma$ calculated in Ref.~\cite{OShea:2024jjw} hold as long as $m_\mathrm{eff} \ll \omega$ is satisfied.
The result is that the study of solar chameleons is able to search for chameleons in a parameter space inaccessible to 5th force experiments (see Fig.~6 of Ref.~\cite{OShea:2024jjw}). This is of great interest for future solar axion experiments, for example IAXO~\cite{IAXO:2012eqj} and its predecessor BabyIAXO~\cite{IAXO:2020wwp}, which as a result of their design will be sensitive to solar chameleons and have the possibility of detecting them.

%% file: WG3/content/Garcia_Cely_Ringwald.tex
\subsubsection{Introduction}

Current gravitational wave (GW) observations~\cite{LIGOScientific:2017vwq,LIGOScientific:2016aoc,EPTA:2023fyk,Reardon:2023gzh,NANOGrav:2023gor,Xu:2023wog} primarily target signals with frequencies below a few kilohertz. However, there has been growing theoretical and experimental interest in the high-frequency regime~\cite{Aggarwal:2025noe}, driven by the ambitious goal of probing signatures from the Early Universe. Within this context, a natural question arises regarding the gravitational wave backgrounds generated by known Standard Model sources, particularly the Sun, which is the most prominent astrophysical object in the vicinity of the Earth.

In a foundational study, Weinberg~\cite{Weinberg:1965nx} estimated the total GW power output of the Sun to be approximately $6 \times 10^7$ W.  While this estimate remains robust, his analysis was restricted to  soft gravitons produced via proton and electron bremsstrahlung, and therefore focused only on one part of the spectrum. We start this brief contribution, by summarizing our findings presented in Ref.~\cite{Garcia-Cely:2024ujr}, which provided a comprehensive derivation of the solar GW spectrum. 

Drawing on techniques originally developed for modeling axion emission, Ref.~\cite{Garcia-Cely:2024ujr} constructed a detailed framework for understanding GW production within the Sun. This study reveals a striking similarity between the behavior of gravitons and axions building upon this insight. The framework may be naturally extended to include the emission of massive spin-2 particles from the Sun, which—depending on their coupling strength and mass—could constitute viable dark matter candidates in the Universe~\cite{Babichev:2016hir,Babichev:2016bxi,Chu:2017msm}. This is the second part of this contribution. For a complementary theoretical discussion on massive spin-2 fields see Sec.~\ref{eq:spin2-EFT}.

We conclude by summarizing the results and offering an outlook on future directions.

\subsubsection{Gravitational wave emission from the Sun}

The Sun is modeled as a non-relativistic plasma composed of electrons and nuclei, primarily hydrogen and helium, with contributions from heavier elements considered negligible. Temperature and density profiles are obtained using the B16-GS98 solar model~\cite{Vinyoles:2016djt}. In this environment, GW emission is understood to arise from both microscopic (particle collisions) and macroscopic (hydrodynamical) sources. The distinction between these regimes is established through the evaluation of the collision frequencies $\omega_c^{(i)}$ for each species~\cite{Weinberg:1972kfs, Weinberg:2019mai}, which define the frequency thresholds delineating the applicable domains of the respective regimes.

\emph{Hydrodynamical fluctuations.---}
At frequencies below all plasma collision rates, GW emission is dominated by collective hydrodynamic modes. Following Ref.~\cite{Ghiglieri:2015nfa}, the corresponding spectrum is computed using the shear viscosity $\eta$ of the solar plasma:
\begin{equation}
\frac{\mathrm{d}P}{\mathrm{d}\omega}\Bigg|_\mathrm{Hydrodynamics}= \frac{16 G \omega^2}{\pi} \int \mathrm{d}^3 \mathbf{r} \, \eta\, T\,.
\end{equation}
Due to the proton-electron mass hierarchy, the dominant contribution to $\eta$ is attributed to proton momentum transfer, with values estimated through fits to state-of-the-art simulations~\cite{PhysRevE.90.033105}.

\emph{Microscopic contributions.---}
Microscopic contributions arising from particle collisions are computed by thermally averaging the graviton emission rate:
\begin{equation}
\frac{\mathrm{d}P}{\mathrm{d}\omega}\Bigg|_\mathrm{Collisions}= \int \mathrm{d}^3 \mathbf{r} \sum_i \omega \left\langle \frac{\mathrm{d}\Gamma^{(i)}}{\mathrm{d}\omega \mathrm{d}V} \right\rangle\,.
\label{eq:Collisions}
\end{equation}
The emission rate per unit volume of one spin-2 particle in the collision of two particles is given by
\begin{equation}
\frac{d \Gamma}{ \mathrm{d} \omega \mathrm{d} V}\left(1+2 \rightarrow h_\lambda \cdots\right)=\int d n_1 d n_2|{\cal M }(\lambda)|^2{\rm d(  P S)} (2 \pi)^4 \delta^{(4)}\left(p_1+p_2-p-\sum_k p_k\right)
\label{eq:master}
\end{equation}
where $p=(\omega, \bf{p})$ and $\lambda$ are respectively the momentum and helicity of the graviton. 
These rates are detailed in Ref.~\cite{Garcia-Cely:2024ujr} for the following processes, which are  the dominant contributions:

\begin{itemize}
\item \emph{Bremsstrahlung.} This is the emission of gravitons from  electron-nucleon and electron-electron collisions. 
The effects of plasma screening are incorporated by following the treatment applied to axion bremsstrahlung~\cite{Raffelt:1985nk}. The full rates obtained in this way  respect soft theorems~\cite{Weinberg:1965nx} and are shown to match classical predictions in the relevant limit~\cite{Gould:1985, Steane:2023gme}. Consistency with earlier treatments~\cite{Barker:1969jk, Papini:1977fm} is confirmed in overlapping regimes of validity. 

\item \emph{Photoproduction.} These are processes of the type $\gamma\, e\to  e\,h$ or $\gamma\, Z\to  Z\,h$. 
Plasma-induced pair correlations are taken into account, yielding a form factor that regularizes collinear divergences, analogous to axion Primakoff and Compton processes~\cite{Voronov:1973kga, Raffelt:1996wa}. Corrections associated with the plasma frequency of the initial-state photon are shown to be negligible under solar conditions.
\end{itemize}

\emph{Transition regime and interpolation.---}
In the regime between hydrodynamical and collision-dominated limits, the GW power spectrum is observed to interpolate from a $\omega^2$ to a flat $\omega^0$ behavior. A phenomenological interpolation factor of the form $\omega^2/(\omega^2+\omega_0^2)$ is implemented, with $\omega_0 \approx (\omega_c^{(p)} \omega_c^{(e)})^{1/2}$. This approach is found to be consistent with analytical studies of energy-momentum tensor fluctuations in similar contexts.

\subsubsection{Beyond gravitational waves: massive spin-2 particles}
There is a clear similarity between the emission of gravitons and axions. Building upon this observation, one may extend the analysis to include the emission of massive spin-2 particles from the Sun, which—depending on their coupling and mass—could potentially serve as viable dark matter candidates in the Universe~\cite{Babichev:2016hir,Babichev:2016bxi,Chu:2017msm}. These hypothetical particles are typically described by traceless symmetric tensor fields, $h_{\mu\nu}$, satisfying the Fierz-Pauli formalism~\cite{Fierz:1939ix}. At the linear level, their interaction with matter is governed by the coupling
\begin{equation}
{\cal L} =  (8\pi G')^{\frac{1}{2}}h_{\mu\nu} T^{\mu\nu}\,,
\label{eq:Lspin2}
\end{equation}
where $T^{\mu\nu}$ denotes the energy-momentum tensor of the SM. This universal coupling structure is realized in several theoretical scenarios, including Kaluza-Klein gravitons and bimetric gravity~\cite{Schmidt-May:2015vnx}. Such interactions lead to modifications of Newton's inverse square law, which are stringently constrained by short-distance gravity experiments~\cite{Geraci:2008hb,Geraci:2010ft,Chen:2014oda}. Moreover, if these particles are stable over cosmological timescales, they may decay into SM photons, resulting in astrophysical signatures observable with terrestrial telescopes.

As for ordinary gravitons, the solar emission of massive spin-2 particles can be calculated using Eqs.~\eqref{eq:Collisions} and \eqref{eq:master}. Interestingly, taking the limit of vanishing mass, we find the decoupling of the $h_{\pm1}$ polarizations, consistent with energy-momentum conservation, while the $h_{\pm2}$ and $h_0$ modes remain coupled. This behavior confirms the presence of the van Dam–Veltman–Zakharov (vDVZ) discontinuity~\cite{vanDam:1970vg,Zakharov:1970cc}, which implies that the massless limit of a massive spin-2 field does not reproduce the dynamics of a pure massless graviton as in linearized General Relativity, but instead retains an additional scalar (spin-0) degree of freedom. 

\begin{figure}[t]
\includegraphics[height=0.36\textheight]{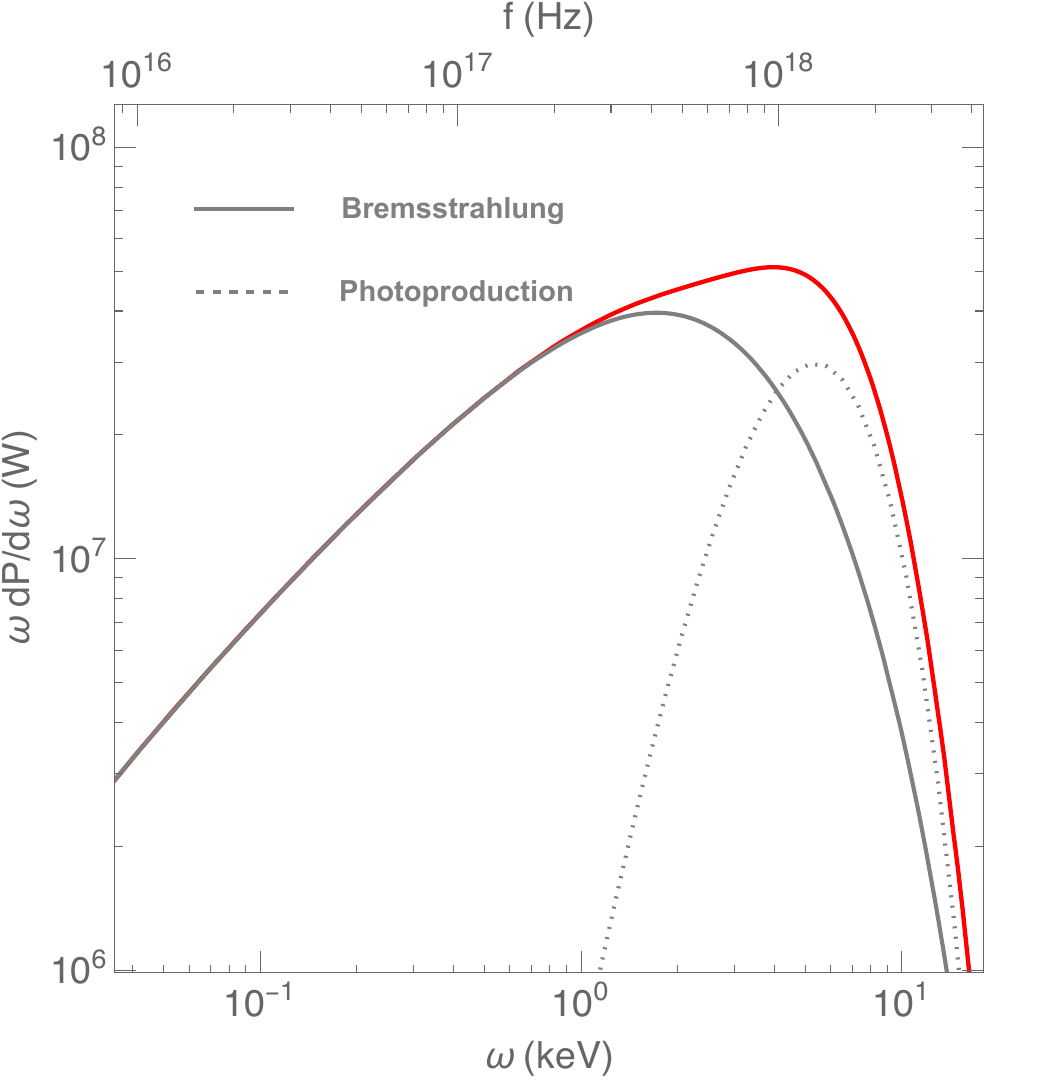}%
\includegraphics[height=0.36\textheight]{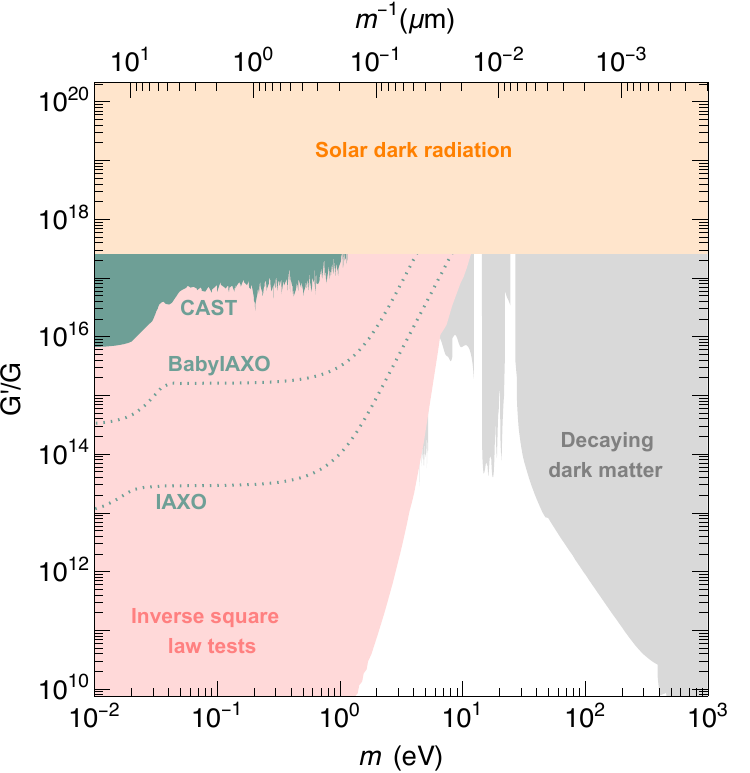}\\
\caption{\emph{Left:} The spectrum of spin-2 particles from the Sun is shown in red, taking $G'=G$. The individual contributions of Bremsstrahlung and photoproduction are shown for comparision. In general, the spectrum of light spin-2 particles produced in the solar plasma closely follows that of ordinary gravitons, differing only by an overall normalization factor $G'/G$~\cite{Garcia-Cely:2025ula}.
\emph{Right:} Constraints on the coupling ratio $G'/G$ as a function of the mass, $m$, of the spin-2 particle. 
The bound from the non-observation of exotic radiation produced by the Sun is shown in orange, together with limits on dark matter decay into photons~\cite{AxionLimits}. The figure also includes bounds on deviations from Newtonian gravity~\cite{Chen:2014oda}. For convenience, the upper axis shows the corresponding interaction range for each mass.}
\label{fig:fig1}
\end{figure}

Following the approach discussed above for gravitational waves, we calculate the spectrum of very light spin-2 patricles produced by the Sun taking $G'=G$ as benchmark. The result is shown in the left panel of Fig.~\ref{fig:fig1}.  There is a remarkable spectral similarity between solar gravitational waves and the emission of sub-keV spin-2 particles, differing essentially by a coupling normalization.  Helioseismological data impose stringent constraints on anomalous energy loss from the Sun. In particular, no more than approximately 10\% of the solar luminosity may be emitted in non-standard forms of radiation. This constraint—referred to here as the solar dark radiation bound—places a model-independent upper limit on the total power emitted in sub-keV spin-2 particles and, consequently, bounds the effective coupling strength $G'$, as illustrated in the right panel of Fig.~\ref{fig:fig1}. Our findings are in partial disagreement with Ref.~\cite{Cembranos:2017vgi}, likely due to differences in the treatment of screening effects in the solar plasma. \\

Furthermore, in analogy with the Gertsenshtein effect, spin-2 particles can convert into photons in the presence of magnetic fields~\cite{Biggio:2006im}. This conversion proceeds with a modified prefactor compared to the axion-photon case. In Fig.~\ref{fig:fig1}, we use current CAST data to place bounds on $G'$ and present future sensitivity projections for IAXO and its prototype, BabyIAXO. For most of the parameter space, especially at higher masses, the CAST limits are subdominant to those from inverse square law tests.  Heavier spin-2 particles are expected to decay on cosmological timescales and therefore cannot account for the observed dark matter abundance in galactic halos. In Fig.~\eqref{fig:fig1} we show the correspoding excluded region according to compilation in Ref.~\cite{AxionLimits}. A more detailed discussion of this synergy and other stellar bounds is presented in Ref.~\cite{Garcia-Cely:2025ula}.

\subsubsection{Conclusions}

Our calculation of solar GW emission reveals a striking similarity between the behavior of gravitons and axions in stellar environments. Both particles are sourced through analogous processes, such as bremsstrahlung and Compton-like scattering, and are sensitive to the same underlying plasma physics. This parallel motivates the consideration of generalized scenarios involving other weakly interacting particles with gravitational couplings.
In particular, one can extend this framework to include the emission of massive spin-2 particles, which arise in various extensions of general relativity and may serve as viable dark matter candidates~\cite{Babichev:2016hir,Babichev:2016bxi,Chu:2017msm}.

By adapting the methods discussed for gravitational waves, we find that their emission spectrum from the Sun closely mirrors that of massless gravitons, differing primarily by an overall normalization governed by the strength of the coupling to SM fields. The spectral shape and frequency dependence remain effectively unchanged, as the dominant contributions stem from the same thermal and dynamical processes.

%% file: WG3/content/Pierluca_Carenza_2.tex
\subsubsection{Introduction}
Axions, pseudoscalar particles with feeble interactions with matter, are among the most studied exotic particles. They were originally introduced to solve the strong-CP problem, i.e. the unnatural smallness of the observed CP-violating interactions in Quantum Chromodynamics (QCD)~\cite{Weinberg:1977ma,Wilczek:1977pj}, and soon recognized as good dark matter candidates~\cite{Abbott:1982af,Preskill:1982cy,Dine:1982ah} if their mass is at the $\mu$eV scale~\cite{Borsanyi:2015cka,Ringwald:2016yge}. 
Pseudoscalars with similar properties emerge as low-energy manifestations of String Theory~\cite{Svrcek:2006yi,Cicoli:2012sz,Halverson:2019kna}. For a complementary theoretical discussion on axion models we refer the reader to Part~\ref{part:wg1}.   Despite the variety of scenarios in which pseudoscalars, or axions, emerge, we will discuss them in a model-independent fashion based on their interactions with Standard Model particles.\\
Generically, the phenomenology of axions depends primarily on their coupling structure. The coupling to photons has the structure
\begin{equation}
    \mathcal{L}=-\frac{g_{a\gamma\gamma}}{4}a\,F_{\mu\nu}\tilde{F}^{\mu\nu}\,,
    \label{eq:agg}
\end{equation}
where $a$ is the axion field, $F_{\mu\nu}$ is the field strength tensor and $\tilde{F}^{\mu\nu}$ its dual. Its primary phenomenology consists in the axion-photon conversion in an external electromagnetic field, e.g. in the electric field of an ion with charge $Z\,e$ (the so-called Primakoff conversion $\gamma+Z\,e\to a+Z\,e$) or in a large-scale magnetic field. Heavy axions can also be produced in hot environments through photon-photon coalescence $\gamma\gamma\to a$, which does not require any external field.\\ The coupling to fermions is
\begin{equation}
    \mathcal{L}=\frac{g_{af}}{2m_{f}}\bar{\Psi}\gamma^{\mu}\gamma^{5}\Psi\,\partial_{\mu}a\,,
    \label{eq:fermion}
\end{equation}
where $\Psi$ is any fermion field with mass $m_{f}$. In dense stellar cores, this interaction is responsible for copious axion production, either via bremsstrahlung $ff\to ffa$ at low temperatures, or via semi-Compton $f\gamma\to f a$ at high temperatures. In partially ionized environments, such as the core of the Sun, free-bound and bound-bound transitions for electronic coupling may also contribute. We now turn to discuss axion phenomenology in concrete stellar environments.

\subsubsection{Main sequence stars}

Stars are generated by the gravitational collapse of a self-gravitating portion of a molecular cloud, mostly composed by molecular hydrogen. The collapse heats up the so-called protostar and makes it dense enough that photons cannot escape. This process continues until radiation pressure halts the gravitational collapse. If, at some point, the radiation pressure is generated via nuclear fusion of hydrogen, a Main Sequence (MS) star is born. 
In general, for protostars with a mass $M\lesssim 0.08~M_{\odot}$, the temperature will not be high enough to ignite the nuclear hydrogen fusion, resulting in a brown dwarf instead of a MS star~\cite{Nakajima:1995sv}. On the other extreme, there is no clear consensus on the upper mass limit for MS stars, which should be $\sim 150~M_{\odot}$~\cite{Crowther:2010cg}. Therefore, MS stars able to ignite nuclear fusion in their core, and visible today, have masses in the $0.08~M_{\odot}\lesssim M\lesssim 150~M_{\odot}$ range. 

However, more massive stars formed in the past and they are not visible anymore today due to their short lifetime. For example, the first stars to form, the pop-III stars, formed in a low-metallicity environment by the first elements produced during the Big-Bang nucleosynthesis. Their masses could reach up to hundreds of solar masses~\cite{2009ApJ...706.1184O} (see Ref.~\cite{2024ApJ...967L..42W} for a possible pop-III star detection in a high-redshift galaxy) and they produced elements heavier than helium, which were expelled in their death via a pair-instability supernova explosion. Axions coupled to photons are produced in pop-III stars and their energy loss may affect stars more massive than $80~M_{\odot}$~\cite{Choplin:2017auq}.
Another potential impact is on the lower-mass limit of stellar black holes, although it appears that, within the known constraints, photophilic or electrophilic axions are unable to explain the heaviest black holes observed within the mass gap~\cite{Croon:2020oga,Sakstein:2020axg}.
Another potential probe is offered by the radiative axion decay, producing a diffuse keV flux, which lead to constraints on photophilic axions stronger than CAST~\cite{Nguyen:2023czp}, but generically excluded by astrophysical arguments~\cite{Langhoff:2022bij,DeRocco:2022jyq,Beaufort:2023zuj,Candon:2024eah}. Of course, a powerful source of constraints belonging to MS stars is our Sun, which is reviewed in Sec.~\ref{subsec:Carenza}.

\subsubsection{Red giants} 
When a MS star depletes the hydrogen in the core, no more energy can be produced to halt the gravitational collapse. Therefore, stars with a mass $M\lesssim2M_{\odot}$ will develop a helium core at a temperature too low to burn helium, surrounded by a burning hydrogen shell. Such a star is called Red Giant (RG), characterized by expanding outer layers that increase the stellar luminosity because of the larger radius, even though the surface temperature lowers. While the hydrogen shell burns, the produced helium is continuously accumulated in the core, increasing its density and the electron degeneracy. Being a degenerate system, the core shrinks to increase the Fermi pressure and contrast the collapse. But in this process also the gravitational attraction on the core surface increases, heating up this region. Together with the external layer, also the inner core is heated up until it can ignite the helium burning. This very rapid process induces a characteristic flash, due to the steep temperature dependence of the helium burning rate. This helium flash singles out a characteristic location, the so-called RGB tip, in the color-magnitude diagram of globular clusters (GCs).

The location of the tip is sensitive to possible exotic energy losses, competing with the standard plasmon decay into neutrinos $\gamma^{*}\to \nu\bar{\nu}$,  because the helium flash would be delayed in their presence making the star at the tip brighter, as first proposed by Ref.~\cite{Dearborn:1985gp}. The resulting constraints are among the most stringent ones for axions coupled to electrons, mainly produced via bremsstrahlung~\cite{Raffelt:1994ry,Catelan:1995ba,Straniero:2018fbv,Straniero:2020iyi,Capozzi:2020cbu,Carenza:2024ehj,Troitsky:2024keu}. The dominant source of uncertainty in the derived bound arises from the distance determination of GCs (see, e.g., the discussion in \cite{Carenza:2024ehj}). The most recent analyses constrain $g_{ae}<0.95\times 10^{-13}$~\cite{Carenza:2024ehj} and $g_{ae}<0.52\times 10^{-13}$~\cite{Troitsky:2024keu} at $95\%$~CL, based on distance measurements to 21 and 7 GCs, respectively, derived from the third Gaia data release.

\subsubsection{Horizontal Branch stars} The stellar configuration reached when the star burns helium in the core, surrounded by a burning hydrogen shell, is called Horizontal Branch (HB) star.
Due to the high core temperature, $\sim10$~keV, electrons are non-degenerate, making axion production more efficient especially through the axion-photon coupling, due to the more copious photons which can convert via Primakoff processes into axions. In turn, HB star evolution is a much more sensitive probe for axion-photon coupling compared to RGs~\cite{Raffelt:1987yu}. The main observational feature of a non-standard cooling of HB stars is a change in the R-parameter, defined as the ratio among the number of stars in the HB and in the RG phase, $R=N_{\rm HB}/N_{\rm RGB}$. The $R$ parameter is accessible by counting stars in GCs, gravitationally bound systems of stars with same age and composition, differing only in their initial mass. The most recent evaluation led to $R=1.39\pm0.03$ from the analysis of 39 GCs~\cite{Ayala:2014pea}. Copious emission of photophilic axions from HB stars would shorten the HB phase and leave the RG phase unaffected, therefore reducing $R$. This leads to strong constraints on photophilic axions, excluding $g_{a\gamma}>0.65\times10^{-10}~{\rm GeV}^{-1}$ at $95\%$ CL~\cite{Ayala:2014pea} for masses $m_a \lesssim 10$~keV. The bound can be extended to masses $m_a \simeq 400$~keV when accounting for coalescence processes $\gamma\gamma\to a$~\cite{Carenza:2020zil}; such heavy axions not only cool the star, but also transfer energy across it via their radiative decay~\cite{Lucente:2022wai}.

Finally, after a HB star has burned helium, it enters a particularly bright phase called the Asymptotic Giant Branch (AGB), with a carbon-oxygen core surrounded by a shell of burning helium. Photophilic axions shorten the duration of this phase even more efficiently than the HB, as first proposed in Ref.~\cite{Dominguez:1999gg}; indeed, the so-called $R_2$ parameter, defined as the ratio of AGB to HB stars in a GC, was recently shown to lead to the best probe of $g_{a\gamma}$~\cite{Dolan:2022kul}.

\subsubsection{White Dwarfs} 

White dwarfs (WDs) are the final phase of stars lighter than $8\,M_\odot$, forming degenerate, nearly isothermal cores supported by electron pressure. They cool over timescales of Gyr timescales, initially by volumetric neutrino emission in plasmon decays, and later by surface photon emission. This cooling could be accelerated by axions, especially if coupled to electrons, mainly produced via electron bremsstrahlung, similarly to red-giant cores~\cite{Isern:1992gia}. A robust probe is the WD Luminosity Function (WDLF), namely the relation between WD mass and luminosity, extracted from the stellar population. By requiring consistency with the WDLF, values of the axion-electron coupling $g_{ae}\gtrsim 1.4\times10^{-13}$~\cite{MillerBertolami:2014rka} are excluded. Additional probes come from pulsating WDs (see, e.g., \cite{Corsico:2012sh,Isern:2019nrg,Corsico:2019nmr}, whose period evolution constrains $g_{ae}\lesssim7\times10^{-13}$~\cite{Corsico:2016okh}. Recently, it was also proposed to use the  WD Initial-Final Mass Relation (IFMR) is sensitive to exotic physics~\cite{Dolan:2021rya}. In this approach, the relation between the initial MS mass and final WD mass is used to look for axion signatures. This is mostly important for heavy axions, $m_{a}\sim300$~keV, giving constraints competitive with the HB ones. We refer to Sec.~\ref{subsec:Lucente} for a detailed discussion.

A particular sub-class of WDs, Magnetic WDs (MWDs), can host extremely strong magnetic fields reaching up to $10^9\,\mathrm{G}$. Compared to non-compact stars, they exhibit the largest product of magnetic field times the radius, which makes them one of the most efficient laboratories for axion-photon conversion~\cite{Dessert:2022yqq}; the main observable consequence would be the generation of linear polarization for surface photons passing through the field, leading to some of the most stringent constraints on $g_{a\gamma}$~\cite{Benabou:2025jcv}.

\subsubsection{Massive pre-supernova stars}
Massive stars above $8\,M_\odot$ ultimately end their lives as supernovae, as we discuss later. Before reaching this stage, they undergo a phase lasting several Myr when they expand significantly, developing outer radii of tens to hundreds of solar radii. Their core temperatures rise dramatically during this period, making them promising probes of axions, particularly those coupled to photons. Although massive stars have not traditionally played a central role in axion searches, recent studies have exploited their late evolutionary stages to constrain axions. For example, Ref.~\cite{Friedland:2012hj} examined stars in the $8$–$12\,M_\odot$ range as axion probes. During helium burning, these stars undergo a characteristic excursion in the color–magnitude diagram, where the surface temperature first increases and then decreases at nearly constant luminosity. This brief evolutionary phase, known as the blue loop, is supported observationally by the existence of Cepheid variables. Excessive axion energy losses can suppress the blue loop, excluding $g_{a\gamma} \gtrsim \mathcal{O}(10^{-10})$~GeV$^{-1}$~\cite{Friedland:2012hj}. Although this overlaps with already excluded regions (e.g., by Ref.~\cite{Ayala:2014pea}), improvements in the modeling and observation of massive Cepheids may allow probing lower couplings~\cite{Anderson:2024sfq}.

In addition to their impact on stellar structure, axions produced in massive stars could give rise to an observable X-ray signal on Earth. If the axion is light, it might then convert back in the Galactic magnetic field leading to observable signals; this has been proposed for the neighboring case of Betelgeuse~\cite{Xiao:2020pra}. This strategy is even more efficient when applied to an entire stellar population within Starburst Galaxies (SBGs) where axions can convert in the strong field of the host galaxy, leading in the case of the SBG M82 to strong constraints from the non-observation of hard X-rays~\cite{Ning:2024eky}. Finally, these hot stars are a particularly sensitive probe of heavy axions, due to their large temperatures; these particles are much easier to probe observationally, because they decay radiatively, with strong constraints on $10-100\,\mathrm{keV}$ axions~\cite{Candon:2024eah}. See Sec.~\ref{subsec:RuzVogel} for more details.

\subsubsection{Supernovae} 

When the stellar core of a massive star becomes dominated by iron, nuclear fusion cannot proceed any further, leading to the core collapse, interrupted only when nuclear densities $\rho\sim 10^{14}\,\mathrm{g/cm}^3$ are reached. The halted collapse triggers a shock wave, which stalls after some tens of milliseconds due to photodissociation losses. The innermost core is a Proto-Neutron Star (PNS), a nuclear-density core with a radius of $\sim 10\,\mathrm{km}$. Its huge temperatures, reaching up to tens of MeV, are powered by the accreting matter on the surface, and make it a WISP factory. In the Standard Model, the resulting neutrino outflow rejuvenates the shock wave and finally causes the SN explosion. Detecting these neutrinos requires a sufficiently close SN: the paradigmatic case was SN~1987A, from which a handful of neutrinos were observed at Kamiokande II~\cite{Kamiokande-II:1987idp}, Baksan~\cite{ALEXEYEV1988209}, and the Irvine-Michigan-Brookhaven~\cite{Bionta:1987qt} detectors. The time-integrated properties of the signal are in good agreement with the expectations from current SN simulations~\cite{Fiorillo:2023frv}.\\
Axion emission from the PNS would of course be very efficient due to the large temperatures, especially if the axion couples to nucleons. Compared to other stars, here even determining the amount of produced axions entails considerable challenges, since dense nuclear matter is not easily modeled. The dominant emission channel is nucleon-nucleon bremsstrahlung $NN\to NNa$, depending sensitively on how the nuclear interaction potential is modeled, with multiple treatments over the years trying to model it in terms of pion and heavier-meson exchange~\cite{Raffelt:1987yt,Turner:1987by,ericson1989axion,Carenza:2019pxu,Lella:2022uwi}, as well as on the effect of multiple-nucleon scattering~\cite{Raffelt:1991pw}. All these uncertainties have been reviewed and rediscussed in Ref.~\cite{Fiorillo:2025gnd}, concluding that a detailed model of the nuclear potential---which should include usually neglected short-range interactions and potentially a non-perturbative treatment---can probably be foregone in favor of a parametric treatment where all nuclear uncertainties are encoded in the spin flip rate of nucleons. The latter can be extracted directly from the measured nucleon-nucleon cross section. Significant uncertainties persist, connected with medium renormalization of nucleon-nucleon interactions (e.g. Refs.~\cite{Schwenk:2003pj,Schwenk:2003bc,Shternin:2018dcn}) and nucleon-axion couplings~\cite{Springmann:2024mjp}, but this is all the more motivation to favor a parametric method, which turns out to be quite consistent with previous approaches. Recently, the role of pionic Compton-like processes $\pi^- p\to na$~\cite{Turner:1991ax,Raffelt:1993ix,Keil:1996ju} has also been reconsidered~\cite{Fore:2019wib,Carenza:2020cis,Fischer:2021jfm}, although it appears that the strong suppression of thermal $\pi^-$ due to their energy shift in the nuclear medium likely reduces this contribution significantly~\cite{Fiorillo:2025gnd}.\\
SNe are also an efficient factory of axions coupled to photons and electrons, although in this case they become competitive with other stellar probes only for heavy particles which cannot be produced in colder stars. For photons, Primakoff conversion and photon-photon coalescence are the dominant reactions~\cite{Lucente:2020whw}, while for electrons bremsstrahlung and $e^+ e^-$ coalescence are usually included~\cite{Carenza:2021pcm,Ferreira:2022xlw}, although the dominant reaction turns out to be Compton emission $\gamma e^-\to a e^-$~\cite{Fiorillo:2025sln}.\\
Once produced, the question of how to identify axion emission turns out to be surprisingly intricate. As further discussed in Sec.~\ref{subsec:Raffelt}, the best indirect probe is the duration of the neutrino burst from SN~1987A, which would be shortened by a non-standard PNS cooling~\cite{Raffelt:2006cw}; this was observed in numerical simulations of PNS cooling~\cite{Burrows:1988ah, Burrows:1990pk}, which however did not include the role of PNS convection, now understood to be crucial in shaping the duration of the neutrino burst~\cite{Fiorillo:2023frv}. Other indirect probes might include the impact on nucleosynthesis in the pre-SN stage~\cite{Aoyama:2015asa,Dominguez:2017mia} and the relation between the progenitor mass and the explosion energy~\cite{Straniero:2019dtm,Dominguez:2025bgg}, although even in the standard case there remain several open issues on what exactly this relation would be (see, e.g., the reviews in Ref.~\cite{Janka:2025tvf,Raffelt:2025wty}).\\
In view of this situation, there has been considerable interest in identifying direct probes, at least for those cases where axions can decay, such as axions coupling to electrons or photons. The situation is reviewed in Sec.~\ref{subsec:Vitagliano}: broadly speaking, rapidly decaying axions could deposit their energy in the progenitor powering the visible explosion~\cite{Caputo:2022mah,Fiorillo:2025yzf} or triggering the production of 511-keV photons~\cite{Chauhan:2025xqr}; longer-lived axions would decay outside, producing visible gamma-rays~\cite{Jaeckel:2017tud,Hoof:2022xbe,Muller:2023vjm} (potentially visible even from extragalactic SNe~\cite{Muller:2023pip}, and giving leading bounds especially for Type Ic SNe with very compact progenitors~\cite{Candon:2025ypl}) or reprocessed X-rays within a fireball~\cite{Diamond:2023scc}. For lighter axion, direct detection might still be possible through the conversion in the Galactic magnetic field~\cite{Payez:2014xsa,Hoof:2022xbe,Calore:2023srn,Lella:2024hfk,Carenza:2025uib} or at larger masses $m_a\lesssim 10^{-3}\,\mathrm{eV}$ even in the magnetic field of the progenitor star~\cite{Manzari:2024jns}. This possibility has recently attracted considerable attention since it might be so sensitive as to even probe the QCD axion for a next galactic SN, although the detection prospects seem to require magnetic fields on the very large side of those expected for most progenitors~\cite{Fiorillo:2025gnd}. The compact progenitors of Type Ibc SNe might host fields large enough for this task~\cite{Candon:2025sdm}. Finally, for axions coupled to nucleons even direct detection might have been possible for very large axion-nucleon couplings~\cite{Lella:2023bfb,Carenza:2023wsm,Alonso-Gonzalez:2024ems,Alonso-Gonzalez:2024spi}.

\subsubsection{Neutron Stars} 

The remnant of the SN explosion is a Neutron Star (NS), with a core composed of degenerate neutrons, slowly cooling by volumetric neutrino emission first, and surface photon emission later, similar to WDs. With their nuclear densities, NSs are probes of axion-nucleon coupling comparable to SNe, since non-standard cooling would affect the relation between surface temperature and age, especially at the transition between neutrino and photon cooling. An analysis of the NS in the SN remnant HESS J1731-347 gives a constraint on the axion-neutron coupling $g_{an}<2.8\times10^{-10}$ at $90\%$ CL for $m_{a}\lesssim 10$~keV~\cite{Beznogov:2018fda}. Even though it is based on a single observation, the advantage of this study is that the young NS analyzed is hotter than NSs in a more advanced phase, maximing the axion production. The leading bounds come however from the Magnificent Seven NSs, roughly at the same level of SN~1987A~\cite{Buschmann:2021juv}; this is not a coincidence, since axion and neutrino emission both come from axial current emission and therefore scale similarly with the temperature, so the axion-to-neutrino luminosity ratio is quite similar for a PNS and a NS~\cite{Fiorillo:2025zzx}. See Sec.~\ref{subsec:Buschmann} for more details. Finally, the detection of gravitational waves from the Neutron Star Merger (NSM) GW170817 has triggered a new line of works on axion probes, since the Hypermassive Neutron Star (HMNS) formed in the merger has properties similar to the PNS of a SN. Light axions produced therein might be directly detected through their conversions in the Galactic~\cite{Fiorillo:2022piv} or in the remnant field~\cite{Manzari:2024jns}, although none of these probes appear to be competitive when realistic properties of the HMNS are included~\cite{Lecce:2025dbz,Fiorillo:2025gnd}. Heavy axions might instead be detected from their radiative decay: while the direct gamma rays would not have been observable by Fermi-LAT~\cite{Dev:2023hax}, their reprocessing into X-rays in the fireball naturally formed from the dense plasma would have made them visible, leading to stringent constraints~\cite{Diamond:2023cto}.

%% file: WG3/content/Shyam_Balaji.tex
\subsubsection{Introduction}

A light CP-even scalar $S$ is a well-motivated and frequently studied extension of the SM, appearing in a broad range of theoretical frameworks. Such particles can stabilize the SM vacuum~\cite{Gonderinger:2009jp, Gonderinger:2012rd, Lebedev:2012zw, Elias-Miro:2012eoi, Khan:2014kba, Falkowski:2015iwa, Ghorbani:2021rgs}, address the hierarchy problem in relaxion models~\cite{Graham:2015cka, Flacke:2016szy, Frugiuele:2018coc, Banerjee:2020kww, Brax:2021rwk}, participate in baryogenesis~\cite{Espinosa:1993bs, Choi:1993cv, Ham:2004cf, Profumo:2007wc, Espinosa:2011ax, Barger:2011vm, Profumo:2014opa, Curtin:2014jma, Kotwal:2016tex, Chen:2017qcz, Balaji:2020yrx}, arise from radiative breaking of classical scale invariance~\cite{Foot:2011et, Heikinheimo:2013fta, Wang:2015cda}, or act as mediators between the dark and visible sectors~\cite{Pospelov:2007mp, Baek:2011aa, Baek:2012uj, Baek:2012se, Schmidt-Hoberg:2013hba, Krnjaic:2015mbs, Beniwal:2015sdl, Balaji:2018qyo}. At the keV scale, $S$ can even constitute a viable dark matter candidate~\cite{Babu:2014pxa, Balaji:2019fxd}. 

\paragraph{Generic couplings.}
At low energies, the interactions of a light scalar with Standard Model fields can be expressed through the effective Lagrangian
\begin{equation}
\label{eq:Lscalar}
\mathcal{L}_{\rm int}
= \sum_{f} g_{S f}\, S\, \bar{\psi}_f \psi_f
+ \frac{g_{S\gamma}}{4}\, S\, F_{\mu\nu}F^{\mu\nu}
+ \frac{g_{Sg}}{4}\, S\, G_{\mu\nu}^a G^{a\,\mu\nu}\,,
\end{equation}
where $g_{S f}$, $g_{S\gamma}$, and $g_{Sg}$ denote the couplings to fermions, photons, and gluons, respectively.
These operators capture a wide class of theoretical scenarios and provide a convenient, model-independent way to express experimental and astrophysical limits.

\paragraph{Higgs-portal realisation.}
A particularly well-studied case is the Higgs-portal model, in which $S$ mixes with the SM Higgs boson $h$ through a small angle $\sin\theta$. 
In this framework, all couplings to SM particles are induced from the Higgs interaction,
\begin{equation}
g_{S f} = \frac{m_f}{v}\sin\theta\,, 
\qquad 
g_{S\gamma,g} \propto \frac{\alpha_{(\!s\!)}}{2\pi v}\sin\theta\,,
\end{equation}
where $m_f$ is the fermion mass, $v\simeq246$ GeV is the electroweak vacuum expectation value and $\alpha_{(\!s\!)}$ the electromagnetic or strong coupling constant.
This relation implies a universal pattern of suppressed interactions, connecting stellar, supernova, and collider probes through the common parameter $\sin\theta$.

\paragraph{Astrophysical motivation.}
Light scalars coupled to ordinary matter can be produced copiously in stellar interiors, potentially altering well-measured energy-loss rates and evolutionary timescales. 
Astrophysical systems therefore provide some of the most stringent constraints on such particles. 
In the following sections, we summarise the main bounds arising from core-collapse supernovae, red giants (RGs), horizontal branch (HB) stars, white dwarfs (WDs), the Sun, and neutron stars (NSs). 
These environments together probe scalar masses from the eV to hundreds of MeV range and couplings down to $g_{S e}\sim10^{-15}$ and $g_{S N}\sim10^{-14}$, well beyond the reach of current laboratory searches in this mass range.

\subsubsection{Red Giants, Horizontal Branch Stars, and White Dwarfs}

Among the most sensitive stellar probes of new light scalars are RGs, HB stars, and WDs. The basic idea, established in Raffelt’s classic monograph~\cite{Raffelt:1996wa}, is that any exotic energy-loss channel must remain below the level allowed by observed stellar lifetimes and ignition conditions. For RG cores just prior to helium ignition, the standard cooling rate is set by neutrino emission at $\epsilon_\nu \sim 4$~erg~g$^{-1}$~s$^{-1}$, so additional losses must satisfy $\epsilon_{\rm new} \lesssim 10$~erg~g$^{-1}$~s$^{-1}$ to avoid delaying helium ignition~\cite{Viaux:2013lha}. In HB stars during helium burning, the fusion rate $\epsilon_{3\alpha} \sim 80$~erg~g$^{-1}$~s$^{-1}$ constrains exotic losses at a similar level, $\epsilon_{\rm new} \lesssim 10$~erg~g$^{-1}$~s$^{-1}$, otherwise the helium-burning lifetime would be shortened beyond what observations allow~\cite{Ayala:2014pea}.

Ref.~\cite{Hardy:2016kme} demonstrated that previous kinetic-theory estimates (e.g. the methodology of Refs.~\cite{Carlson:1986cu,Grifols:1988fv,Dent:2012mx,Rrapaj:2015wgs}) would yield underestimated constraints by neglecting plasma mixing effects. In dense stellar plasmas, longitudinal photon modes (plasmons) cross the light cone, enabling resonant production scalars for $m_S \lesssim \omega_p$, where $\omega_p$ is the photon plasma frequency. Unlike vectors, scalar production is not suppressed by small $m_S$, so resonant processes can dominate. This strengthens the limits on scalar–electron and scalar–nucleon couplings by up to three orders of magnitude compared to earlier literature. For Higgs-portal scalars, the resulting constraint is $\sin\theta \lesssim 10^{-10}$ for $m_S \lesssim $ a few keV.

WDs provide an additional testbed: although degeneracy suppresses some emission channels, their long-term cooling history and low photon luminosities make them sensitive to exotic energy loss. Some more recent works have used WD luminosity functions in the Galactic disk to extract constraints~\cite{MillerBertolami:2014rka,Bottaro:2023gep}, finding limits around $\sin\theta \lesssim 10^{-10}$ for $m_S \lesssim 1$ keV. Independently, Ref.~\cite{Yamamoto:2023zlu} derived comparable bounds in the range $\sin\theta \sim 10^{-10}$–$10^{-9}$ for $m_S \lesssim 10$ keV. While WD constraints are broadly similar to those from RGs and HBs, they are generally somewhat weaker once plasma mixing effects are included, and RG/HB stars continue to set the tightest stellar limits \cite{Hardy:2016kme}.

Overall, the combined picture from RG, HB, and WD cooling implies
\begin{equation}
    g_{Se} \lesssim 10^{-15}, \qquad g_{SN} \lesssim 10^{-12},
\end{equation}
for scalar masses below the relevant plasma frequency of the stellar cores ($\omega_p \sim \text{keV}$). These constraints are model-independent and can be mapped to Higgs-portal scenarios via $g_{Se} = (m_e/v)\sin\theta$.

\subsubsection{The Sun}

The Sun provides an excellent laboratory for testing light scalar particles, since its internal structure is known with exceptional precision through helioseismology and solar neutrino measurements. 
Any exotic energy loss exceeding a few percent of the solar luminosity would disrupt the successful agreement between solar models and observations~\cite{Raffelt:1996wa,Vinyoles:2015aba}. 
This makes solar physics a sensitive probe of scalar couplings to electrons and photons at levels relevant for Higgs-portal and other feeble-interaction scenarios.

In the solar core ($T_c \simeq 1.3$ keV, $\rho_c \simeq 150~{\rm g\,cm^{-3}}$), light scalars can be produced via electron bremsstrahlung $e\,Z \to e\,Z\,S$, Compton-like scattering $e\,\gamma \to e\,S$, and Primakoff conversion $\gamma + Z \to Z + S$. 
Detailed calculations of these processes, originally developed for axions~\cite{Raffelt:1985nk}, apply directly to scalars with appropriate coupling replacements. 
The solar luminosity constraint requires that the exotic scalar luminosity remain below about $L_S \lesssim 0.1 L_\odot$~\cite{Raffelt:1996wa}, corresponding to $g_{Se} \lesssim 10^{-13}$ for light scalars. 
More refined analyses combining helioseismology and solar neutrino data~\cite{Vinyoles:2015aba} confirm that deviations at this level would be detectable.

Overall, solar observations provide robust and complementary constraints: they are less sensitive than RG/HB stars to ultra-weak couplings, but their high precision, continuous monitoring, and direct detection channels make them a uniquely clean probe of light scalar emission and energy transport in stellar interiors. Together with the RG, HB, and WD limits discussed above, solar observations complete the picture of scalar production in ordinary stars.

\subsubsection{Supernovae}
\label{sec:SN}

Core–collapse supernovae represent the most extreme astrophysical environments accessible to observation, with core densities $\rho \sim 10^{14}\,{\rm g\,cm^{-3}}$ and temperatures $T \sim 30$~MeV. The neutrino burst detected from SN~1987A by Kamiokande, IMB, and Baksan~\cite{Bionta:1987qt,Kamiokande-II:1987idp,Alekseev:1987ej} provides a unique window into the physics of such dense plasmas. The observed $\mathcal{O}(10\,{\rm s})$ duration of the neutrino signal is in good agreement with the delayed neutrino-heating mechanism~\cite{Bethe:1985sox}, implying that any additional energy loss channel must not significantly shorten the burst. This requirement can be formalised through the Raffelt criterion~\cite{Raffelt:1987yt,Turner:1987by,Raffelt:1990yz}, which constrains the luminosity in new particles at $1\,$s after the core bounce to be smaller than the neutrino one.

Early studies on scalar emission in supernovae examined nucleon–nucleon bremsstrahlung $NN \to NN S$ within the one–pion–exchange (OPE) approximation~\cite{Ishizuka:1989ts,Diener:2013xpa,Krnjaic:2015mbs}. This simple treatment, while historically valuable, tends to overestimate nuclear scattering rates and neglects multi–pion exchange and nuclear correlation effects~\cite{Carenza:2019pxu}. The situation mirrors that encountered in the axion literature, where improved calculations using soft–theorem methods and empirical nucleon–nucleon cross sections revealed that OPE-based emissivities could be overestimated by roughly an order of magnitude~\cite{Raffelt:1996wa,Chang:2018rso}. These findings emphasised the need for more accurate nuclear modelling in setting scalar limits.

Subsequent analyses extended this picture by including scalar decay and reabsorption within the protoneutron star, as well as variations in progenitor profiles~\cite{Balaji:2022noj}. These refinements proved important for identifying the “trapping regime,” in which scalars are efficiently produced but cannot escape, thereby defining the shape of the excluded region in the $(m_S,\sin\theta)$ parameter space.  

A major conceptual advance came from recognising that light scalars can mix with the in–medium longitudinal photon, or plasmon. This resonant mixing, first studied in the context of dark photons~\cite{Redondo:2013lna,An:2013yfc,Chang:2016ntp}, also applies to CP–even scalars coupled to charged fermions. Whenever the scalar mass satisfies $m_S < \omega_p(r)$, where $\omega_p$ is the local plasma frequency, resonant conversion enhances production dramatically. The process was initially analysed for HB and RG stars~\cite{Hardy:2016kme}, and has recently been generalised to supernova conditions~\cite{Hardy:2024gwy}. Since it depends only on well-understood plasma properties rather than uncertain nuclear cross sections, this mechanism provides the most reliable bounds for light scalars with $m_S \lesssim 10$~MeV.

At higher scalar masses, $m_S \gtrsim \omega_p \sim 10$–20~MeV, resonant production is suppressed and nucleon bremsstrahlung becomes dominant once again. Modern treatments, building upon soft–theorem approaches ~\cite{Hanhart:2000ae,Rrapaj:2015wgs,Chang:2018rso,Carenza:2019pxu}, have used empirical nucleon–nucleon scattering data to compute conservative scalar emission rates~\cite{Hardy:2024gwy}. These improved calculations yield slightly weaker but significantly more robust limits than the earlier OPE-based results~\cite{Dev:2020eam,Balaji:2022noj}.

The emerging picture is that for $m_S \lesssim 10$~MeV, resonant plasmon mixing dominates; for $m_S \gtrsim 30$~MeV, nucleon bremsstrahlung controls the emission; and in the intermediate regime, both processes contribute with details depending on trapping, decay, and progenitor modelling. The resulting excluded band in coupling–mass space remains one of the most stringent constraints on light scalar particles, complementary to those derived from RG, HB and WD cooling at lower masses~\cite{Hardy:2016kme,Bottaro:2023gep,Yamamoto:2023zlu,Lella:2023bfb,Fiorillo:2025zzx} and collider searches at higher energies~\cite{Egana-Ugrinovic:2019wzj,Dev:2019hho,Dev:2021qjj}.

Open issues include progenitor dependence, potential collapse–induced thermonuclear explosion scenarios~\cite{Bar:2019ifz}, and the behaviour of exotic energy transport in the trapping regime. Ultimately, full three-dimensional supernova simulations incorporating new scalar degrees of freedom will be required to refine these limits. For now, the combined application of resonant mixing and soft–theorem bremsstrahlung provides the most complete and reliable framework for constraining light scalar emission from core–collapse supernovae.

\begin{figure}[t]
    \centering
    \includegraphics[width=0.7\linewidth]{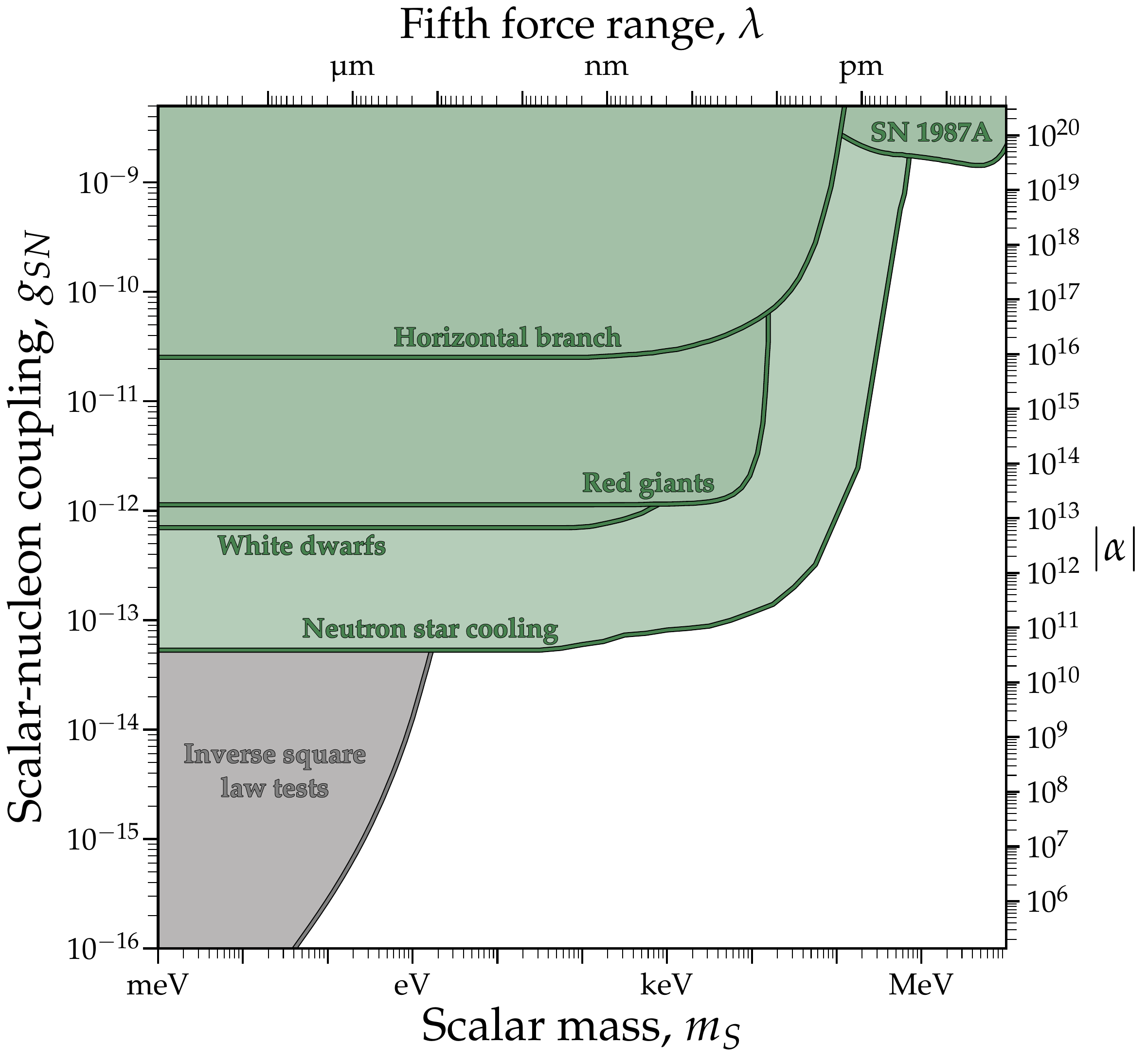}
    \caption{Constraints on the scalar–nucleon coupling $g_{SN}$, assuming equal couplings to protons and neutrons. Astrophysical limits~\cite{Hardy:2016kme,Bottaro:2023gep,Hardy:2024gwy,Fiorillo:2025zzx} and laboratory tests of the inverse-square law~\cite{Chen:2014oda} are shown in green and gray, respectively. For comparison with fifth-force parameterizations, the equivalent Yukawa strength relative to gravity, $|\alpha|$, is given on the right-hand axis, and the force range $\lambda = m_S^{-1}$ is shown on the top axis. Figure courtesy of Ciaran A.~J.~O'Hare, adapted from Ref.~\cite{Fiorillo:2025zzx}.}
    \label{fig:ScalarsNS}
\end{figure}

\subsubsection{Neutron Stars}

Compact remnants of supernova explosions, NSs, are powerful laboratories for constraining light scalar particles. Their cooling is well understood, and even modest exotic contributions to the thermal luminosity would spoil the observed agreement with standard neutrino-driven models~\cite{Yakovlev:2000jp,Page:2006ud}.  

The key point is that scalar emission differs qualitatively from pseudoscalar or neutrino emission in nucleon bremsstrahlung processes. For axions, emission requires a spin flip of the nucleon, so both axion and neutrino luminosities are suppressed by the emitted particle momentum, leading to similar scaling with temperature~\cite{Iwamoto:1984ir}. For scalars, however, no spin flip is required. Instead, the emission rate is proportional to the quadrupole moment of the nucleon system, scaling as $(p_N/m_N)^4$, where $p_N$ is the nucleon momentum.  

In supernova cores, where nucleons are only mildly degenerate, the nucleon momentum is expected to be in the order $p_N \sim \sqrt{m_N T}$ with $T \sim 30$ MeV. By contrast, in the interior of cold neutron stars at $T_{\rm NS} \sim 10$ keV, nucleons are highly degenerate, with momenta set by the Fermi momentum $p_F \sim 200$ MeV. The ratio of scalar to neutrino luminosities is then enhanced by several orders of magnitude:
\begin{equation}
    \left(\frac{L_\phi}{L_\nu}\right)_{\rm NS}
    \sim \left(\frac{L_\phi}{L_\nu}\right)_{\rm SN}
    \frac{p_{F}^{4}}{m_N^2 T_{\rm NS}^2}
    \sim 10^7 \left(\frac{L_\phi}{L_\nu}\right)_{\rm SN}\,,
\end{equation}
implying that cold NSs can yield bounds on scalar couplings up to 3–4 orders of magnitude stronger than SN1987A.  

Following the approach pioneered for axions in~\cite{Buschmann:2021juv}, a recent work~\cite{Fiorillo:2025zzx} has analyzed the cooling of five isolated NSs with kinematic ages $\sim 10^5$ yr, for which thermal luminosity data are available~\cite{Potekhin:2020ttj,Suzuki:2021ium}. The results exclude scalar–nucleon couplings
\begin{equation}
    g_{SN} \gtrsim 5 \times 10^{-14}
\end{equation}
for scalar masses below $m_S \lesssim 1$ MeV. These bounds are currently the strongest astrophysical constraints on light scalars in the range ${\rm eV} \lesssim m_S \lesssim {\rm MeV}$, surpassing even the RG/HB and WD limits discussed above. They also imply stringent bounds on axion CP-violating couplings~\cite{Raffelt:2012sp,OHare:2020wah,Fiorillo:2025zzx}.

For Higgs-portal models, this translates to
\begin{equation}
    \sin\theta \lesssim 6 \times 10^{-11}\,,
\end{equation}
which improves significantly over stellar bounds based on scalar–electron couplings. Figure~\ref{fig:ScalarsNS} illustrates these constraints in comparison with other astrophysical and laboratory probes. The NS limits probe effective Yukawa strengths as small as $|\alpha| \sim 10^{-15}$ relative to gravity at ranges $\lambda = m_S^{-1}$ in the micron to millimeter regime.

\subsubsection{Conclusions}

Astrophysical environments provide a broad and complementary set of probes of light scalar particles. 
At the lowest masses, RGs, HB stars, and WDs set the strongest limits on scalar–electron couplings through resonant plasma production, reaching $g_{Se} \sim 10^{-15}$ for $m_S \lesssim \mathrm{keV}$. 
The Sun provides an independent laboratory, with helioseismology and solar neutrino fluxes constraining exotic energy losses at the percent level of the solar luminosity. 
At higher temperatures and densities, SN~1987A remains the key probe in the MeV–100 MeV mass range, where resonant plasmon mixing and soft bremsstrahlung now provide the most reliable bounds. 
Finally, cold neutron stars surpass all other stellar systems for light scalar–nucleon couplings, improving sensitivity by several orders of magnitude relative to supernovae, excluding $g_{SN}\gtrsim 5\times 10^{-14}$ for $m_S\lesssim 1$~MeV. Taken together, these astrophysical bounds cover the scalar mass window from the eV to hundreds of MeV, reaching effective couplings far beyond current laboratory reach. They stand as robust, largely model–independent constraints, and demonstrate the unique power of stellar astrophysics in charting the viable parameter space of Higgs-portal and related scalar models.

%% file: WG3/content/Josef_Pradler.tex
\subsubsection{Dark vector particles probed in stars}

A ``dark vector'' $V^\mu$ generically arises  by introduction of an additional Abelian gauge symmetry, $\mathrm{U}(1)_V$, to the SM gauge group $\mathrm{SU}(3)_c \times \mathrm{SU}(2)_L \times \mathrm{U}(1)_Y$. For pedagogical reasons, we review the basics of dark vector models useful for phenomenology. We refer the reader to Sec.~\ref{eq:spin1-EFT} for a complementary theoretical discussion. In the absence of any direct gauge interactions of SM particles under $\mathrm{U}(1)_V$,
the interaction between the SM and the new physics sector occurs through kinetic mixing with parameter $\epsilon$ between the field strength tensor of the hypercharge, $F_{\mu \nu}^Y$, and the field strength tensor $V_{\mu \nu}$ associated with the $\mathrm{U}(1)_V$ symmetry. The stellar production of $V^\mu$ particles involves energy scales much lower than the electroweak scale,
where the effective Lagrangian describing the relevant low-energy physics takes the form~\cite{Holdom:1985ag}
\begin{equation}
\label{eq:L}
\mathcal{L} =\mathcal{L}_{\rm SM} - \frac{1}{4} V_{\mu \nu}^2 - \frac{\varepsilon}{2} F_{\mu \nu} V^{\mu \nu} + \frac{m_V^2}{2} V_\mu V^\mu  .
\end{equation}
Here, $\mathcal{L}_{\rm SM} $ is the SM Lagrangian, $F_{\mu \nu} = \partial_\mu A_\nu - \partial_\nu A_\mu$ is the photon field strength and 
$\varepsilon = \cos \theta_W \epsilon$ is the kinetic mixing parameter where the cosine of the weak mixing angle $\cos \theta_W $ has been absorbed.

The stellar energy-loss rates are sensitive to the origin of dark vector mass~$m_V$ in~\eqref{eq:L}.
The simplest option is that~$m_V$ is of the Stückelberg type, which naturally preserves its smallness due to the conservation of the Abelian vector current. This protection against sensitivity to ultraviolet (UV) scales renders this option attractive as it allows to study astrophysical sensitivity to dark vectors with the minimal set of parameters~$\varepsilon$ and~$m_V$.
Another possibility is that $V^\mu$ is ``Higgsed'', involving of a new scalar field $\phi$ that is charged under $\mathrm{U}(1)_V$ with gauge coupling~$g'$, amending~\eqref{eq:L} by
\begin{align*}
\mathcal{L}_{\text {higgs }}=\left|\left(\partial_\mu-i g^{\prime} V_\mu\right) \phi\right|^2-V(\phi). 
\end{align*}
The potential $V(\phi)$ leads to spontaneous breaking of $\mathrm{U}(1)_V$ by the vacuum expectation value $v'$ with  $\phi = \sqrt{1/2}(h'+v')$ in unitary gauge from which $m_V = g' v'$ arises. This scenario introduces a new interaction term in the Lagrangian, $g^{\prime} m_V h^{\prime} V_\mu^2$, where the physical hidden Higgs field, $h^{\prime}$, of mass $m_{h'}$ couples to the vector field $V_\mu$, among additional self-interactions of $h^{\prime}$ and an interaction $h'^2 V^2$. In the regime where $m_V$ and $m_{h^{\prime}}$ are much smaller than the stellar temperature~$T$, the production of dark sector states is then predominantly driven by dark Higgsstrahlung and by pair production of $\mathrm{U}(1)_V$-charged Higgs scalar fields~\cite{Pospelov:2008jk}.

The kinetic mixing term in~\eqref{eq:L} is not the only option to connect $V^\mu$ to the SM. If SM fermions $f$ are gauged under $\mathrm{U}(1)_V$, one has additional interactions $\Delta \mathcal{L} = \sum_f \bar{f} q_f g' \gamma^\mu  f V_\mu $ that are part of the gauge covariant derivative terms with~$V^\mu$. Prominent anomaly-free options are the gauged combinations
$B-L$, $L_{\mu} - L_\tau$, or $L_{e} - L_{\mu(\tau)}$ of baryon number $B$ and lepton numbers $L_{e,\mu,\tau}$ with $L$ being the latters' sum. For example, under $B-L$, quarks carry $q_f = 1/3$ while SM leptons (including right hand neutrinos) carry $q_f =-1/3$. In a non-relativistic stellar environment made of protons and electrons, this choice translates into photon-like effective couplings $g_V^{(p)} = - g_V^{(e)}$, such that stellar energy-loss rates become similar to the dark photon case. The limits on the dark photon kinetic mixing parameter $\varepsilon$ can then be translated into the $B-L$ gauge coupling $g'$ through the replacement~$\varepsilon e \to g'$~\cite{An:2020bxd}. For other, more general choices and its implications for stellar energy loss, see~\cite{Li:2023vpv}.

Finally, {\it complex} dark vector particles $V^\mu \neq (V^\mu)^\dag$ may arise in non-Abelian extensions of the SM gauge group.  For example, replacing $\mathrm{U}(1)_V \to \mathrm{SU}(2)_V $ implies three vector fields $V_a^\mu$ ($a=1,\dots,3$) in the adjoint representation of the dark gauge group $\mathrm{SU}(2)_V$ which become massive after spontaneous symmetry breaking. Depending on the concrete UV setup, the vector particles can then couple to the SM photon with (radiatively induced) kinetic mixing through $V_0^\mu$ or with electromagentic multipole interactions through $V^\mu \equiv\left(V_1+i V_2\right)^\mu / \sqrt{2}$~\cite{Gabrielli:2015hua,Hisano:2020qkq,Chu:2023zbo}.
 Exemplary low energy effective Lagrangian terms read
\begin{equation}
\label{eq:LeffVcomplex}
\mathcal{L}/e  \supset   i \kappa V_\mu^{\dagger} V_\nu F^{\mu \nu}+\frac{i \lambda}{\Lambda^2} V_{\lambda \mu}^{\dagger} V^\mu{}_\nu F^{\nu \lambda} .
\end{equation}
Here, the first term is discussed in the context of millicharged vectors~\cite{Gabrielli:2015hua} and both terms play into the context of magnetic dipole and electric quadrupole interactions of vector particles with respective strengths~\cite{Hagiwara:1986vm}
\begin{equation}
\label{eq:multipoles}
\mu_V=\frac{e}{2 m_V} \kappa+\frac{e m_V}{2 \Lambda^2} \lambda, \quad Q_V=-\frac{e}{m_V^2} \kappa+\frac{e}{\Lambda^2} \lambda ,
\end{equation}
where $\Lambda $ relates to the UV scale. Such couplings are then astrophysically constrained, by the pair production through plasmon decay $\gamma^* \to V^\dag V$ among other production channels~\cite{Chu:2023zbo}.

\subsubsection{Stellar production rates of vector particles}

The dominant stellar production channels of dark photons with St\"uckelberg mass are originally computed in~\cite{Redondo:2008aa} for transversely polarized particles using an equations of motion approach  and in~\cite{An:2013yfc} for longitudinally polarized particles using a Feynman-diagrammatic approach at finite temperature. Qualitatively new additions to these calculations considering additional production channels, model variations, or using alternative theoretical formulations are made in~\cite{Redondo:2013lna,Redondo:2015iea,Hardy:2016kme,Lasenby:2020goo,Li:2023vpv}. Calculations of supernova cooling constraints (see Sec.~\ref{subsec:Raffelt} for more details) have their own somewhat separate history~\cite{Bjorken:2009mm,Kazanas:2014mca,Rrapaj:2015wgs,Chang:2016ntp}, involving nucleon bremsstrahlung and optical depth considerations; see also~\cite{Croon:2020lrf}. Below, we review the stellar production of dark photons in ordinary stars, focusing on its essential features.

The differential production rate of transverse $(T)$ and longitudinal ($L$) dark photons~$V$ per energy interval and volume, $d \Gamma_{T, L}^{V}/d \omega d V|_{\rm prod} $, is closely related to the corresponding production rate of in-medium photons, $d \Gamma_{T, L}/d \omega d V|_{\rm prod} $~\cite{An:2013yfc},
\begin{equation}
\label{eq:Vproduction}
\left. \frac{d \Gamma_{T, L}^{V}}{d \omega d V}\right|_{\rm prod}  = \varepsilon_{T, L}^2 \left. \frac{d \Gamma_{T, L}}{d \omega d V}\right|_{\rm prod} ,
\end{equation}
 where $\varepsilon_{T, L}^2$ is the in-medium induced mixing angle, 
\begin{equation}
\label{eq:mixingangle}
\varepsilon_{T, L}^2=\varepsilon^2  \frac{ m_V^4}{\left(m_V^2-\operatorname{Re} \Pi_{T, L}\right)^2+\left(\operatorname{Im} \Pi_{T, L}\right)^2} . 
\end{equation}
In this expression, $\Pi_{T, L}$ are the photon self-energy functions of transverse and longitudinal polarization as they appear in the in-medium photon polarization tensor, $ \Pi^{\mu \nu}=\Pi_T \sum_{i=1,2} \epsilon^{\mu}_{T,i} \epsilon^{\nu}_{T,i }+\Pi_L \epsilon_L^{\mu} \epsilon_L^{\nu}$ where $\epsilon^{\nu}_{T,i} $ and  $\epsilon_L^{\mu}$ are the two transverse and longitudinal polarization vectors, respectively.

Equation~\eqref{eq:mixingangle} reveals the possibility of resonance emission for $m_V^2 = \operatorname{Re} \Pi_{T, L}$.
Using the leading  functional dependence of  $\Pi_{T}(\omega, \vec k) \simeq  \omega_p^2$ and $\Pi_{L}(\omega, \vec k) \simeq  \omega_p^2 m_V^2/\omega^2$
on photon energy $\omega$ and momentum $\vec k$  for an isotropic non-relativistic  plasma~\cite{Braaten:1993jw,Raffelt:1996wa}, yields as resonance conditions $\omega^2 = \omega^2_p(r_{\rm res})$ for $L$-resonant emission and $m_V^2 = \omega^2_p(r_{\rm res})$ for $T$-resonant emission. Here, $\omega^2_p(r_{\rm res})$ is the squared plasma mass of photons at the resonance radius $r_{\rm res}$. This shows that for $m_V$ within the dynamical range of plasma frequencies inside the star, the transverse resonance condition is fulfilled at a single radius $r_{\rm res}$, while for $L$-modes, $r_{\rm res}$ becomes a function of dark photon energy.
Using then that the optical theorem at finite temperature relates $\operatorname{Im} \Pi_{T, L}$ with the rate for attaining thermal equilibrium in media, isolating from it the production rate by detailed balancing, and taking the volume integral turns the above rate formula~\eqref{eq:Vproduction} into the star's total energy-differential production rate,
\begin{equation}
\left. \frac{d \Gamma_{T, L}^{V}}{d \omega }\right|_{\rm prod} \simeq\left(\frac{2 \varepsilon^2  r^2}{e^{\omega / T(r)}-1} \frac{\sqrt{\omega^2-m_V^2}}{\left|\partial \omega_p^2(r) / \partial r\right|}\right)_{r=r_{\mathrm{res}}} \times \begin{cases}m_V^2 \omega^2 & L\text{-resonance}\\  m_V^4 & T\text{-resonance}\end{cases}.
\end{equation}
As can be seen, the production becomes independent of the details of the photon producing channel, and the whole process can be viewed as a gradual transition of the photon bath into the dark photon sector. Most importantly, the decoupling for $T$- and $L$-modes as a function of diminishing $m_V^2$ is different, such that for $m_V^2 \ll \omega_p^2$ the $L$-emission starts dominating by an increasingly large factor; note that the onshell condition for photon-dark photon conversion $\omega^2 - \vec k^2 = m_V^2$ holds. Off-resonance contributions, such as the emission of longitudinally polarized bremsstrahlung photons have been calculated as well and expressions are found in the original papers cited above. 

In the above cases of a ``hard'' St\"uckelberg photon mass, we observed the decoupling with $m_V^4$ for $T$-modes and with $m_V^2$ for $L$-modes. Turning to the Higgsed case, we have another process at our disposal, the decay of transverse plasmons in to a $V + h^{\prime}$ pair. This Higgsstrahlung process scales as $m_V^0$, which in the limit of $m_V\ll \omega$ is equivalent to the emission of two Goldstone bosons. The production rate is given by~\cite{An:2013yua},
\begin{align*}
\left.\frac{d \Gamma^{Vh'}}{d V d \omega}\right|_{\rm prod}=\frac{(g' \varepsilon)^2 \omega_p^2}{4 \pi^3} \int_{\omega+\frac{\omega_p^2}{4 \omega}}^{\infty} \frac{d q^0\left(\omega q^0-\omega^2-\omega_p^2 / 4\right)}{\left(e^{q^0} / T-1\right)\left[\left(q^0\right)^2-\omega_p^2\right]},
\end{align*}
where $q^0$ is the plasmon energy. A similar non-decoupling with $m_V^0$ is also observed for other light gauge bosons, such as $L_e-L_{\mu(\tau)}$~\cite{Li:2023vpv}.

\begin{figure}[t]
\includegraphics[width=0.5\textwidth]{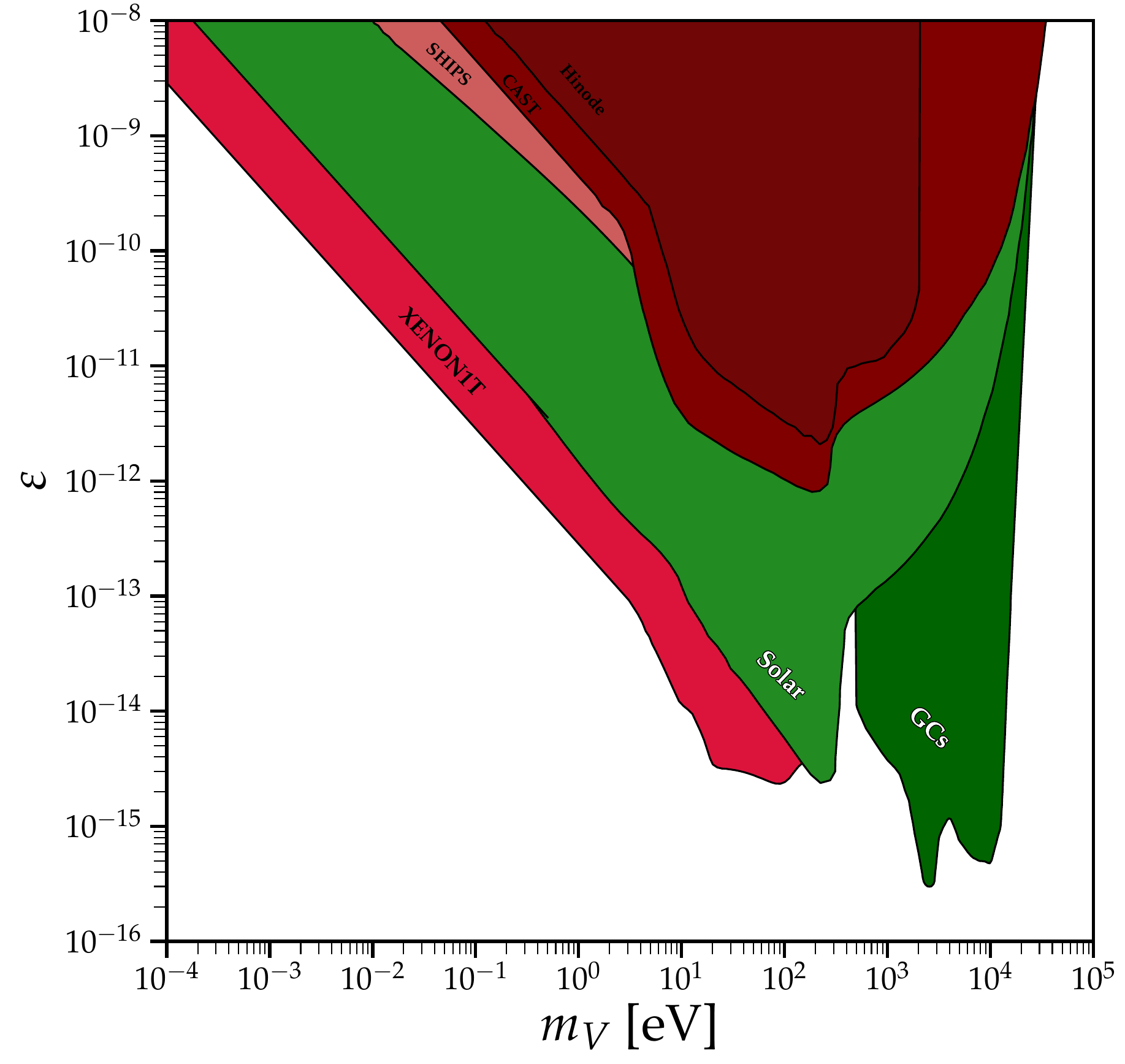}\includegraphics[width=0.5\textwidth]{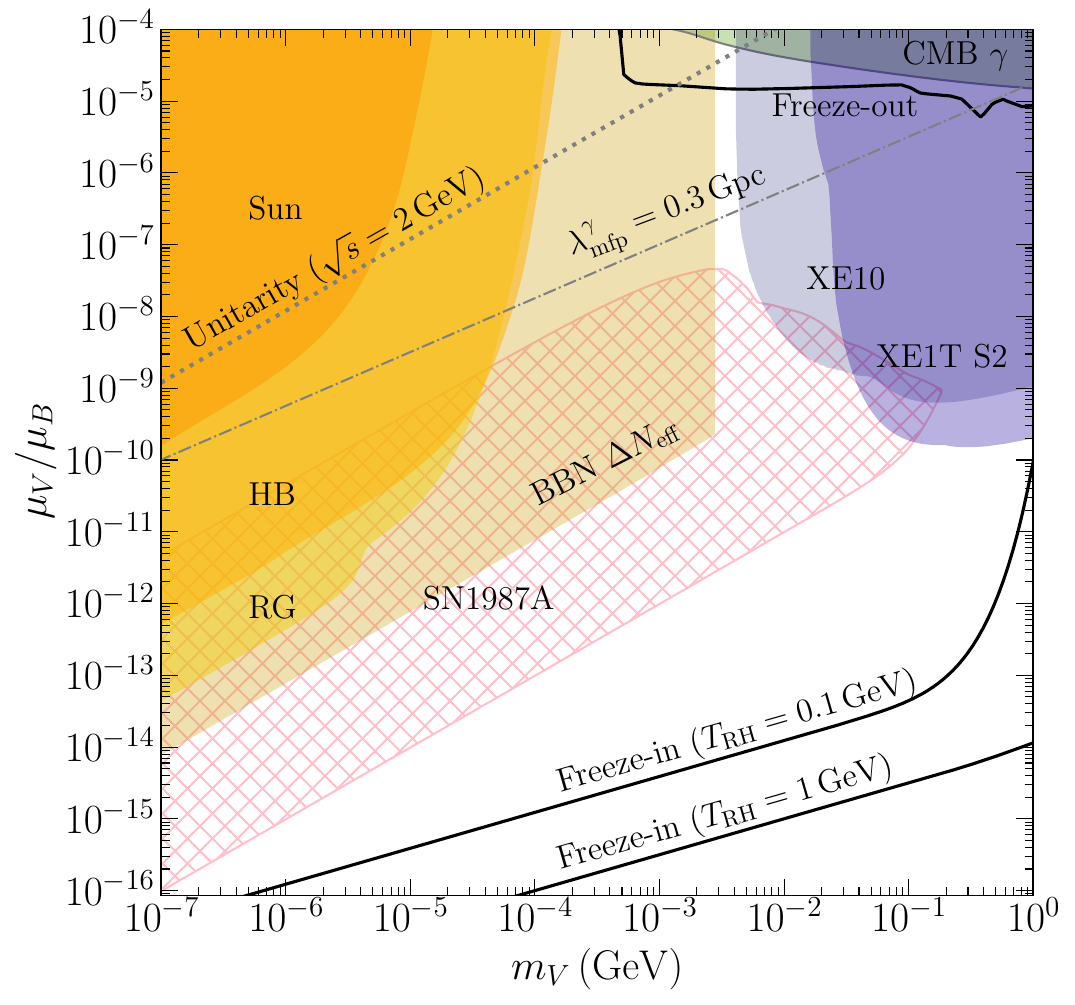}
\caption{
{\it Left:} Limits on the dark photon kinetic mixing parameter $\varepsilon$ that arise from stellar emission. The green shaded regions are energy-loss constraints from the Sun~\cite{Redondo:2008aa,An:2013yfc,Redondo:2013lna}) and from globular cluster stars~\cite{Dolan:2023cjs} as labeled. The solar dark photon emission is also probed through the ionization signal from dark photon absorption by atomic electrons in dark matter direct detection experiments shown in red~\cite{XENON:2021qze} (updated from Refs.~\cite{An:2013yua,An:2020bxd,Bloch:2016sjj}). Additional laboratory sensitivity from solar emission in this mass range is shown for SHIPS~\cite{Schwarz:2015lqa}, CAST~\cite{Redondo:2008aa} and Hinode~\cite{Frerick:2022mjg}. The plot is produced using the python scripts assembled for~\cite{Caputo:2021eaa} and does not rely on any dark matter abundance in form or dark photons.
{\it Right:} Limits on the magnetic moment interaction of complex vector particles in units of the Bohr magneton. For low-mass particles, the strongest stellar energy-loss constraints are from SN1987A and from RG stars; taken from Ref.~\cite{Chu:2023zbo}.
}\label{fig:DarkPhotonReview}
\end{figure}

Finally, when considering instead the pair production $V V^\dag$ of complex vector particles (effectively) coupled to the photon through interactions such as~\eqref{eq:LeffVcomplex}, one may again cast the production rate in terms of photon self-energies~$\Pi_{T,L}$. In the kinematically unsuppressed regime $m_V\ll T$~\cite{Chu:2023zbo},
\begin{align}
\left.\frac{d \Gamma^{VV^\dag}}{dV d k^2} \right|_{\rm prod} = & -\sum_{i=\mathrm{T}, \mathrm{L}} \int \frac{d^3 \vec{k}}{(2 \pi)^3} \frac{g_i  \operatorname{Im} \Pi_i(\omega, \vec{k})}{\omega\left(e^{\omega / T}-1\right)} \frac{f\left(k^2\right)}{16 \pi^2\left|k^2-\Pi_i\right|^2} .
\end{align}
Here, $f\left(k^2\right)$ is the squared amplitude for $\gamma^* \rightarrow V^\dag{V}$, integrated over the 2-body phase space of the vector pair. For example, $f(k^2) \simeq \mu_V^2 k^6/4m_V^2$ for magnetic dipole and $Q_V^2 k^6/16$ for the electric quadrupole interactions given in Eq.~\eqref{eq:multipoles};  $k^2 = \omega^2 - \vec k^2$ is the invariant mass of $VV^\dag$ and $g_i = 1\ (2)$ for the production from $L$ $(T)$-modes. Once again, the last denominator signals the possibility of resonance emission, and for $\operatorname{Re} \Pi_{T, L} =k^2$ the formula turns into the rate for $T$- or $L$-polarized plasmon decay into a vector-pair, $\gamma_{T,L} \to VV^\dag$, and otherwise contains any emission process to leading order in the dark coupling constant, such as  Compton- and Bremsstrahlung-type vector pair-production. When probing such higher-dimensional interactions, the scaling of emission rates with powers of $m_V$ is entirely different, and constraints typically strengthen with diminishing~$m_V$ for $\sqrt{k^2} / m_V \gg 1$, as observed in the right panel of Fig.~\ref{fig:DarkPhotonReview}. The discussion of the $m_V\to 0$ limit (and its assumptions) is found~\cite{Chu:2023zbo}.

\subsubsection{Constraints on vectors from stellar emission}

Having access to the dominant stellar energy loss rates enables one to derive limits on the SM-dark sector coupling as a function of $m_V$, based on observationally inferred permissible energy-loss rates; see~\cite{Raffelt:1996wa} for a textbook discussion. Figure~\ref{fig:DarkPhotonReview}
illustrates the resulting constraints from stellar energy loss for dark photons (left panel) and for electric quadrupole interactions of a complex vector particle (right panel); details are found in the figure caption. Remarkably, the sensitivity of laboratory direct detection experiments has now reached a level where the solar emission of dark photons is best probed through their subsequent absorption by atomic electrons, with the total rate given by~\cite{An:2013yua}
\begin{equation}
\Gamma_{T, L}=-\frac{\varepsilon_{T, L}^2 \operatorname{Im} \Pi_{T,L}}{\omega} .
\end{equation}
Here, $\varepsilon_{T, L}$ and $\operatorname{Im} \Pi_{T, \mathrm{L}}$ now represent the effective mixing angle and the photon self-energies  within the {\it detector's} medium, respectively. In the context of ton-scale experiments, such as XENON1T~\cite{XENON:2021qze},  $\varepsilon_{T, L}$ and $\operatorname{Im} \Pi_{T, \mathrm{L}}$ are then obtained from the measured optical properties and total photoabsorption cross sections of liquid xenon.

%% file: WG3/content/Giuseppe_Lucente.tex
\subsubsection{Introduction}
White dwarfs are the final stage of the evolution of low and intermediate-mass stars, $M \lesssim 8-9$~M$_\odot$. Their structure is composed by a degenerate core containing the bulk of mass and a partially degenerate mantle made of hydrogen and helium, absent in $\sim 20\%$ of cases. The core mass determines the chemical composition of WDs. In particular, WDs with $M \gtrsim 1.05$~M$_\odot$ are made of a mixture of oxygen and neon (O/Ne-WDs), those with a mass $0.4\,{\rm M}_\odot \lesssim M \lesssim 1.05$~M$_\odot$ are composed by carbon and oxygen (C/O-WDs) and the ones with $M \lesssim 0.4$~M$_\odot$ are made of helium (He-WDs).

Because of the high electron degeneracy in the core, WDs cannot obtain energy from nuclear reactions and their evolution is just a gravo-thermal process~\cite{2010A&ARv..18..471A}. As sketched in Fig.~\ref{fig:wdsketch}, the WD cooling process can be divided into four phases: neutrino cooling, fluid cooling, crystallization and Debye cooling. At very early times ($t \sim 10^4$~yrs), the main contribution to the nascent WD luminosity comes from hydrogen burning via CNO cycle. After this stage, nuclear reactions stop and neutrino emission via plasmon decay becomes dominant~\cite{Winget:2003xf}. After $10^7-10^8$~yrs, neutrino emission rapidly decreases due to its steep dependence on the core temperature and photon losses become dominant. At the same time, the WD core behaves like a fluid that can be described as a Coulomb plasma not very strongly coupled~\cite{1994ApJ...434..641S,2010CoPP...50...82P,2021ApJ...913...72J}. When the temperature is low enough, the plasma crystallizes into a classical body centered crystal~\cite{Isern:1997na}. Finally, when almost all the star is solid, the specific heat follows the Debye's law and the WD cools down very quickly. However, since the outermost layers are far from the Debye regime, their energy content prevents the sudden disappearance of the WD~\cite{1989ApJ...347..934D}.  

WDs showing hydrogen in their spectra are called DAs, while those without hydrogen are non-DAs. The most common interpretation is that the DAs have an envelope made of hydrogen and helium, while the non-DAs have just a helium layer or an extremely thin hydrogen layer on the top. The DAs are $\sim 80\%$ of the total but this proportion changes along the cooling sequence~\cite{1997ASSL..214..165S}. The most important difference between these two families is that the DAs cool down more slowly than non-DAs since hydrogen is more opaque than helium. Therefore, the nature of the envelope is very important during the cooling stage dominated by photons. 

The WD cooling is an efficient tool to probe WISPs. Indeed, the WISP production can modify the WD cooling, affecting the secular drift of the period of pulsation of variable WDs as well as the shape of the luminosity function (see Refs.~\cite{Isern:2019nrg,Isern:2022vdx,Carenza:2024ehj} for recent reviews).

\subsubsection{WISPs bounds from the secular drift of the pulsation period}
During their evolution, WDs experience episodes of variability~\cite{2010A&ARv..18..471A,2019A&ARv..27....7C,2020FrASS...7...47C,2021RvMP...93a5001A}, characterized by a frequency spectrum which depends on the structure of the star. The pulsation modes can be radial, when they maintain the spherical symmetry, and non-radial, which can be classified as toroidal and spheroidal. Depending on the restoring force, spheroidal modes can be divided into g- and f-modes, if the restoring force is gravity, and p-modes, if it is pressure gradient. There are at least six classes of pulsating WDs~\cite{2019A&ARv..27....7C}, but WISPs have been probed by considering only DAVs, DBVs and DOVs. Due to their multifrequency character and their period of pulsation ($10^2-10^3$~s), these WDs are classified as g-mode pulsators, with buoyancy as driven mechanism. In such a context, as the cooling proceeds, degeneracy increases, the buoyancy decreases and the star contracts inducing a secular change in the period of pulsation $P$. The temporal evolution of the period can be computed as~\cite{1983Natur.303..781W}
\begin{equation}
\frac{\dot P}{P}= -a\frac{\dot T}{T}+ b\frac{\dot R}{R}\,,
\label{pdot}
\end{equation}
where $a$ and $b$ are positive constants of the order of unity. The first term of the r.h.s. represents the decrease of the Brunt-V\"ais\"al\"a frequency with the temperature, while the second one is the increase of the frequency induced by the residual gravitational contraction.

The secular drifts can be used to test the predicted evolution of WDs and to test any physical effect able to change the pulsation period of these stars. In particular, the change in the luminosity and in the pulsation period due to the inclusion of an extra cooling can be estimated as
\begin{equation}
\frac{{{L_{0}} + {L_x}}}{{{L_{0}}}} \approx \frac{{{{\dot P}_{obs}}}}{{{{\dot P}_{0}}}}
\label{eqise92}
\end{equation}
where $\dot P_{obs}$ is the observed period drift,  $L_0$ and $\dot P_0$ are obtained from standard models and $L_x$ is the extra luminosity necessary to fit the observed period \cite{Isern:1992gia}.

\begin{figure}[t!]
\center
  \includegraphics[width=0.9\linewidth,clip=true,trim=4.5cm 9.3cm 2cm 10cm]{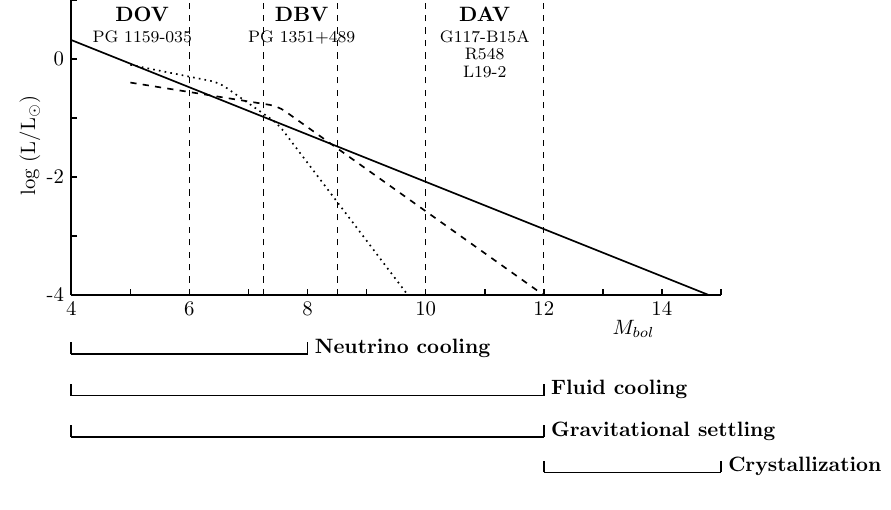}
\caption{Sketch showing the different phases of the WD cooling, including photon (solid), neutrino (dotted) and axion (dashed) luminosities, as well as the regions occupied by the different sources of energy. The position of the variable WDs is also shown. Figure taken from Ref.~\cite{Carenza:2024ehj}.}\label{fig:wdsketch}
\end{figure}

As shown in Fig.~\ref{fig:wdsketch}, the cooling rate of DOVs and DBVs is determined by the photon and neutrino emission, while DAVs cool only by photon emission since the neutrino bremsstrahlung emission is rapidly suppressed with the temperature.

The first class of pulsating WDs to be discovered were DAVs~\cite{1968ApJ...153..151L}, which constitute the most numerous group, including more than 400 samples~\cite{2022MNRAS.511.1574R}. Being of the order of $10^{-15}$~s$^{-1}$, the secular drift has been measured only in three DAV stars (G117-B15A, R548, and L19-2). The monitoring of G117-B15A started in 1974 and is still continuing. The first obtained value~\cite{1991ApJ...378L..45K} was larger than the one predicted by the existing models at the epoch~\cite{1991ASIC..336..153F} and no conventional ways to solve this discrepancy were accepted~\cite{1991ASIC..336..143K,1991ASIC..336..153F}. Axions were proposed as a possible solution to this puzzle~\cite{Isern:1992gia}. At temperatures and densities typical of hot DAVs and assuming they can interact with electrons, axions are emitted via electron bremsstrahlung $e + Ze \to e + Ze + a$. The axion production rate scales as $\epsilon_a \propto T^4$~\cite{Nakagawa:1987pga,Nakagawa:1988rhp,Raffelt:1996wa,Carenza:2021osu} while for neutrinos it scales as $\epsilon_\nu \propto T^7$ in this phase~\cite{1983ApJ...275..858I}. Therefore, they can still be an important sink of energy at  the luminosity domain of DAVs. Using Eq.~(\ref{eqise92}) and a simple WD model, Ref.~\cite{Isern:1992gia} showed that DFSZ axions with a mass $m_a\cos^2 \beta \approx 8.5 $~meV ($ g_{ae} \approx 2.36\times 10^{-13}$) could cause the drift. Both the measured values~\cite{1991ApJ...378L..45K,2000ApJ...534L.185K,2005ApJ...634.1311K,2012ASPC..462..322K,2021ApJ...906....7K} and the ones predicted by theoretical models~\cite{2001NewA....6..197C,2008ApJ...675.1512B,2012MNRAS.424.2792C,2012MNRAS.420.1462R} have evolved with time. The most recent values are ${(5.12 \pm 0.82)\times 10^{-15}}$~s$^{-1}$~\cite{2021ApJ...906....7K} for the observed secular drift and ${(1.25 \pm 0.09)\times 10^{-15}}$~s$^{-1}$ \cite{2021ApJ...906....7K} for the theoretical one, obtained using fully evolutionary models. This implies that the best value of the axion-electron coupling to account for discrepancies is $g_{ae}=(5.66 \pm 0.57)\times 10^{-13}$ ($m_a \cos^2 \beta =20 \pm 2 $~meV if DFSZ type)~\cite{2021ApJ...906....7K}. Similar results~\cite{2008ApJ...675.1512B,2018phos.confE..28B,2012JCAP...12..010C,2016JCAP...07..036C} have been obtained by observing the secular drifts of R548~\cite{2013ApJ...771...17M} and L19-2~\cite{2015ASPC..493..199S}. An analogous analysis was performed with PG 1351 + 489, a DBV star with an observed drift of $\dot P= (2.0 \pm 0.69) \times 10^{-13}$~s$^{-1}$ for a period of 489~s~\cite{2011MNRAS.415.1220R} and a predicted one of $\dot P= (0.81 \pm 0.5) \times 10^{-15}$~s$^{-1}$ \cite{2016JCAP...07..036C}. This discrepancy would be solved with $g_{ae} < 5.5 \times 10^{-13}$ \cite{2016JCAP...08..062B}, concordant with the values obtained with the DAVs. However, in this case DBVs are characterized by a large neutrino emission and the result could be affected by an extra emission if neutrinos have a magnetic moment (NMM). In such a context, the secular drift of PG 1351 + 489 constrains values of the NMM $\mu_\nu \lesssim 7\times 10^{-12}~\mu_{\rm B}$~\cite{Corsico:2014mpa}, with $\mu_{\rm B}$ the Bohr magneton.

Thus, all observations suggest the introduction of an extra cooling term in the energy balance of WDs. However, due to both observational and theoretical uncertainties, it is not possible to assess if the anomaly has a systematic origin or it is caused by WISPs or anything else~\cite{Isern:2022vdx}. In this context, it is worth to mention that the values of the axion-electron coupling suggested by the secular drift of the WD pulsation period are in tension with constraints from the brightness of red giant tip, excluding $g_{ae}\gtrsim 1.48 \times 10^{-13}$~\cite{Capozzi:2020cbu,Straniero:2020iyi} (see Sec.~\ref{subsec:Carenza2} for more details).

\subsubsection{WISPs bounds from the luminosity function}
The luminosity function of a WD ensemble (WDLF) is the number of WDs of a given magnitude per unit of magnitude interval, i.e. their distribution in magnitudes or, equivalently, in luminosities. Assuming that WDs are not destroyed and the ensemble is closed, the number $N$ that have an absolute magnitude within the range  $M_{abs}\pm 0.5\Delta M_{abs}$ per unit of magnitude interval is given by~\cite{Isern:2022vdx,Carenza:2024ehj}
\begin{equation}
N\left( M_{abs}\right) = \int\limits_{{M_l}}^{{M_u}} {\Phi \left( M \right)\Psi \left( {{T_G} - {t_{cool}} - {t_{ps}}} \right){\tau _{cool}}\,dM} 
\label{eq:lf}
\end{equation}
where $t_{cool}$ is the cooling time down to the magnitude $M_{abs}$, $\tau_{cool} = dt/dM_{abs}$ is the characteristic cooling time of the WD at this magnitude, $M$ and $t_{ps}$ are the mass and the lifetime of the WD progenitor, $T_G$ is the age of the Galaxy or the population under study, and $M_u$ and $M_l$ are the maximum and the minimum mass of the main sequence stars able to produce a white dwarf. Thus, $M_l$ satisfies the condition $T_G=t_{cool}(M_{abs},M_l)+t_{ps}(M_l)$. Additionally, $\Phi(M)$ is the initial mass function (IMF) and $\Psi(t)$ is the star formation rate (SFR) of the population under consideration. Moreover, not explicitly written, there is the initial-final-mass-relation (IFMR) function, connecting the progenitor and WD properties. In order to compare theory with observations, and since the total density of WDs is not yet  well known, the computed luminosity function is usually normalized to a bin with a small error bar, usually $\log L/L_\odot \simeq 3$ or the corresponding magnitude. 
We stress that Eq.~\eqref{eq:lf} contains observational [$N(M_{abs})$] and stellar terms [$t_{cool},\tau_{cool}, t_{PS}, M_{u}, M_{l}$], as well as the IMF and the galactic terms $\Phi$ and $\Psi$.

The first luminosity function was derived by Weidemann~\cite{Weidemann:1968uw} and by the end of the nineteens, before the advent of the large cosmological surveys, the samples contained few hundreds of stars~\cite{1988ApJ...332..891L,1992MNRAS.255..521E,1996Natur.382..692O,1998ApJ...497..294L,1999MNRAS.306..736K}. The WDLFs obtained from the large cosmological surveys, such as the Sloan Digital Sky Survey (SDSS)~\cite{2006AJ....131..571H} and the Super COSMOS Sky Survey (SCSS)~\cite{2011MNRAS.417...93R}, increased the samples to several thousands WDs, improving the precision of the WDLF so that it is possible to compare the observational and theoretical shapes. Recently, the WDLF for a statistically complete 100~pc WD sample has been obtained with data provided by the Gaia mission~\cite{Isern:2022vdx}.

The WDLF can be used to test the existence of an extra source or sink of energy since it depends on the characteristic cooling time. Since the bright branch of the luminosity function is dominated by WDs coming from low-mass main sequence stars, Eq.~\eqref{eq:lf} can be written as
\begin{equation}
{N \propto \left\langle {{\tau _{cool}}} \right\rangle \int\limits_{{M_l}}^{{M_u}} {\Phi \left( m \right)\Psi \left( T_G-t_{ps}-t_{cool}\right)} } \,dm\, .
\label{eq:wdlf2}
\end{equation}

Bright WDs have had no time to cool down, $t_{cool}$ is small, and, as a consequence of the strong dependence of the main sequence lifetimes with mass, $M_l$ is almost constant and independent of the luminosities under consideration. Thus, the slope essentially depends on the averaged characteristic cooling time, dominated by low-mass WDs, and only weakly on the star formation history. 

Therefore, one can use the slope of the bright branch of the WDLF to detect the presence of unexpected additional sinks or sources of energy in WDs. This technique was used to constrain axions for the first time in \cite{2006AJ....131..571H} and then in \cite{2014JCAP...10..069M}, where the WDLFs obtained with the SCSS catalogue \cite{2011MNRAS.417...93R}, the SDSS catalogue \cite{2006AJ....131..571H} and from UV-excesses \cite{2009A&A...508..339K} were used, as well as a self consistent treatment of the neutrino cooling. This analysis suggested that the agreement between theoretical and observed WDLFs could be improved if DFSZ axions with $g_{ae}\in (0.7-2.1)\times 10^{-13}$ exist. A recent analysis of the cooling of WDs in 47 Tucanae, based on observations with the Hubble Space Telescope, excludes $g_{ae}\gtrsim 0.81 \times 10^{-13}$ at $95\%$~CL~\cite{Fleury:2025ahw}, providing the strongest constraint on the axion-electron coupling.

Due to the degeneracy between the stellar and galactic terms of Eq.~(\ref{eq:wdlf2}), the shape of the WDLF may be also a consequence of changes in the SFR. To break this degeneracy, the WDLFs of populations with different formation histories can be considered. Indeed, if axions modify the cooling rate of WDs, their impact will be observable in all the WDLFs at roughly the same luminosity. Effectively, axions with $g_{ae}\in(2.24-4.48)\times 10^{-13}$ \cite{Isern:2018uce} would reduce the discrepancies between the theoretical and observed luminosity functions of the thin and thick disks and halo~\cite{2011MNRAS.417...93R}, and a value of $g_{ae}\approx 2.24\times 10^{-13}$ is favored after improving the evaluation of the disk WDLF assuming a constant WD scale height~\cite{2017AJ....153...10M} or different variable scale heights~\cite{2017ApJ...837..162K} above the Galactic plane. 

Another possibility to break the degeneracy between the stellar and galactic terms in Eq.~\eqref{eq:wdlf2} is provided by massive WDs. In this case, the lifetime of their progenitors is much shorter than the cooling time and their luminosity function closely follows the temporal variation of the SFR. The luminosity function of WDs with a mass in the range of $0.9-1.1$~M$_\odot$ and within a distance of 100~pc has been obtained using Gaia data~\cite{2019Natur.565..202T}. This would exclude values $g_{ae}\gtrsim 4\times 10^{-13}$. However, if a time-dependent WD height scale is included, values $g_{ae} \approx 2.24\times 10^{-13}$ are acceptable and $g_{ae} \approx 4.5\times 10^{-13}$ are rejected~\cite{Isern:2019nrg}. Here we mention again that hints from the WDLFs, similarly to the ones from the secular drift of the pulsation period, are in tension with constraints from red giants (see Refs.~\cite{Capozzi:2020cbu,Straniero:2020iyi,Carenza:2024ehj,Troitsky:2024keu} and the discussion in Sec.~\ref{subsec:Carenza2}) and with the recent bound from the cooling of WDs in 47 Tucanae~\cite{Fleury:2025ahw}.

Luminosity functions have been exploited to constrain also the NMM. In particular, using the WDLF of Refs.~\cite{2009A&A...508..339K,2011MNRAS.417...93R} and WD models computed in~\cite{Renedo:2010vb}, values $\mu_n \gtrsim 5 \times 10^{-12}\,\mu_{\rm B}$ have been excluded~\cite{MillerBertolami:2014oki}. Additionally, the hot WDLF in 47~Tucanae led to the constraint $\mu_\nu \lesssim 3.4\times 10^{-12}\,\mu_{\rm B}$~\cite{Hansen:2015lqa}.

Recently, the same WDLFs used in Ref.~\cite{2009A&A...508..339K} have been employed to probe interactions between scalars $\phi$ and fermions, described by the Lagrangian $\mathcal{L} \supset g_{\psi} \phi \psi \bar{\psi}$, with $\psi=p,e$. In this case, the emission rate via bremsstrahlung process $e + p \to e + p + \phi$ scales as $\epsilon_\phi \propto T^2$. This implies that scalars are relevant for the cooling of colder, i.e. older, WDs, compared to axions. At this stage, the WD cooling is dominated by photon emission from the surface. Therefore, by taking into account the WDLF only in the range of magnitudes corresponding to old WDs cooling mainly via photon emission, values $g_{\phi p}\gtrsim 2.3\times 10^{-12}$ and $g_{\phi e}\gtrsim 1.4\times 10^{-15}$ were excluded~\cite{Bottaro:2023gep}. Similar results were obtained by \cite{Yamamoto:2023zlu}, assuming a one-zone model for WDs and requiring that the scalar luminosity must be lower than the stellar one.

The studies discussed above neglect the potential impact of magnetic fields in white dwarfs, which can reach strengths up to $10^9$~G in the so-called Magnetic White Dwarfs. Such intense fields may affect particle emission from the stellar plasma~\cite{Brahma:2025vdr}, as suggested in the context of neutrino emission~\cite{Drewes:2021fjx}.

%% file: WG3/content/Georg_Raffelt.tex
\subsubsection{Introduction}

Stars with masses exceeding around $8\,M_\odot$ go through all nuclear burning phases until they develop a degenerate iron core that eventually collapses until it reaches nuclear density. The subsequent rebound and concomitant shock wave is thought to explode the star, although in up to a quarter of all cases, the explosion fails and a stellar-mass black hole forms instead of a NS remnant. The binding energy of the newly formed NS, some 20\% of its rest mass or 2--$4\times10^{53}~{\rm erg}$, emerges almost exclusively in the form of neutrinos of all flavors over a time scale of a few seconds. Neutrino energy deposition behind the stalling shock wave is what leads to a successful explosion. The hydrodynamical evolution is characterized by large-scale convective motions and sometimes the dipolar SASI (Standing Accretion Shock Instability) and LESA (Lepton-number Emission Self-sustained Asymmetry) instabilities. Fully three-dimensional numerical simulations provide strong support for the now-standard neutrino-driven supernova (SN) explosion mechanism (also called delayed explosion mechanism or Bethe-Wilson mechanism), for contemporary reviews see Refs.~\cite{Janka:2012wk,Mirizzi:2015eza,Janka:2016fox,Janka:2017vcp,Muller:2020ard,Mezzacappa:2020oyq,Burrows:2019zce,Burrows:2020qrp,Raffelt:2025wty}. Besides NS mergers and the early universe, SN cores provide the only astrophysical environments where neutrinos cannot stream freely and actually thermalize. However, their mean-free path is much larger than that of photons, explaining their dominant role for radiative energy transfer. Within standard physics, a third form of radiation is gravitational, but gravitons interact so feebly that they are irrelevant for radiative energy transfer and carry away only a small fraction of a SN's thermal energy. 

New feebly interacting particles, depending on their exact properties, can play a role intermediate between standard neutrinos and gravitons and significantly affect SN physics by the transfer or loss of energy and/or lepton number or by producing new signatures through decays into visible states \cite{Raffelt:1990yz,Raffelt:2006cw,Caputo:2024oqc}.
Based on the SN~1987A neutrino signal, simultaneous $\gamma$-ray observations, all past SNe as a source of diffuse cosmic radiations, or the energetics of SN explosions, many constraints on new particles have been derived and will be reviewed in Sec.~\ref{subsec:Vitagliano} below. Here we focus on the traditional SN~1987A cooling argument based on the signal duration that was advanced shortly after the observations with a focus on axions \cite{Raffelt:1987yt,Turner:1987by,Mayle:1987as,Burrows:1988ah,Choi:1988xt,Turner:1989wa,Ericson:1988wr,Carena:1988kr,Mayle:1989yx,Ishizuka:1989ts,Burrows:1990pk,Engel:1990zd,Altherr:1990tf,Raffelt:1991pw,Chang:1993gm,Raffelt:1993ix,Janka:1995ir,Keil:1996ju,Giannotti:2005tn,Stoica:2009zh,Stoica:2012zz,Fischer:2016cyd,Lee:2018lcj,Bar:2019ifz,Carenza:2019pxu,Vonk:2020zfh,Ge:2020zww,Lucente:2020whw,Carenza:2020cis,Hook:2021ous,Calore:2021klc,Carenza:2021pcm,Fischer:2021jfm,Choi:2021ign,Asai:2022pio,Lucente:2022vuo,Betranhandy:2022bvr,Ferreira:2022xlw,Foguel:2022fef,Lella:2022uwi,Ho:2022oaw,Diamond:2023scc,Mori:2023mjw,Carenza:2023lci,Carenza:2023wsm,Li:2023thv,Anzuini:2023whm,Fiorillo:2023frv,Zhang:2023vva,Lella:2023bfb,Cavan-Piton:2024ayu,Springmann:2024mjp,Springmann:2024ret,Chakraborty:2024tyx,Lella:2024dmx}, later applied to active-sterile neutrino mixing \cite{Kainulainen:1990bn,Mukhopadhyaya:1992dg,Raffelt:1992bs,Shi:1993ee,Pastor:1994nx,Kolb:1996pa,Goldman:1999hg,Dolgov:2000pj,Dolgov:2000jw,Hidaka:2006sg,Hidaka:2007se,Raffelt:2011nc,Arguelles:2016uwb,Syvolap:2019dat,Mastrototaro:2019vug,Suliga:2020vpz,Carenza:2023old}, the mass of Dirac neutrinos \cite{Gaemers:1988fp,Perez:1989xm,Grifols:1990jn,Gandhi:1990bq,Maalampi:1991bv,Altherr:1991rp,Natale:1990yx,Kolb:1991id,Cline:1991hk,Pantaleone:1991zr,Turner:1991ax,Cline:1991bv,Babu:1991rc,Lam:1992nq,Burrows:1992ec,Dodelson:1992tv,Mayle:1992sq}, large extra dimensions \cite{Cullen:1999hc,Barger:1999jf,Hanhart:2000er,Hanhart:2001fx,Hannestad:2002ff,Satheeshkumar:2008fb}, unparticles \cite{Hannestad:2007ys,Das:2007nu,Dutta:2007tz}, light supersymmetric particles \cite{Ellis:1988aa,Grifols:1988fw,Grifols:1997jd,Dicus:1997sw,Dreiner:2003wh}, majorons and other bosons that couple to neutrinos \cite{Aharonov:1988ee,Aharonov:1988ju,Choi:1987sd,Aharonov:1989ik,Grifols:1988fg,Choi:1989hi,Chang:1993yp,Kachelriess:2000qc,Farzan:2002wx,Heurtier:2016otg,Brune:2018sab,Fiorillo:2022cdq,Akita:2023iwq,Cerdeno:2023kqo}, dark photons \cite{Dent:2012mx,Rrapaj:2015wgs,Chang:2016ntp,Shin:2021bvz,Caputo:2025avc}, muonic couplings on new bosons \cite{Bollig:2020xdr,Croon:2020lrf,Caputo:2021rux,Manzari:2023gkt}, and assorted other hypothesis \cite{Grifols:1989qm,Grifols:1989tk,Davidson:1993sj,Haghighat:2009pv,Das:2013zba,Keung:2013mfa,Kadota:2014mea,Zhang:2014wra,Mahoney:2017jqk,Cembranos:2017vgi,Tu:2017dhl,Sung:2019xie,Dev:2020eam,Camalich:2020wac,Fischer:2024ivh}, and has been vigorously revived for the past few years.

\subsubsection{SN~1987A Cooling Argument}

\begin{figure}[t!]
    \centering
    \includegraphics[width=1.0\textwidth]{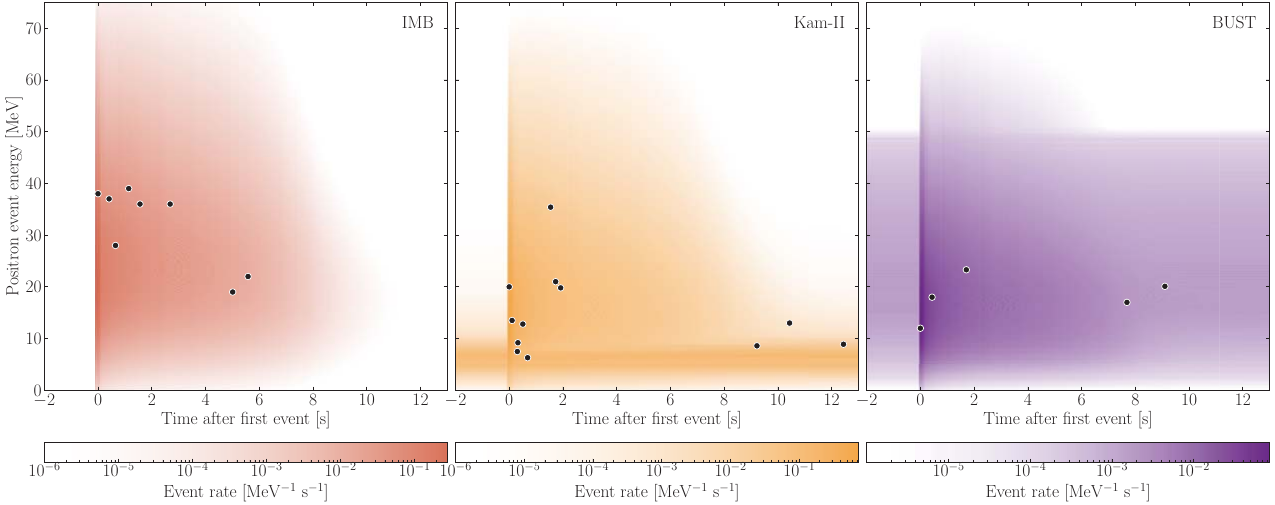}
    \caption{SN~1987A events as measured by IMB~\cite{Bionta:1987qt,IMB:1988suc}, Kam-II~\cite{Kamiokande-II:1987idp,Hirata:1988ad}, and BUST~\cite{Alekseev:1987ej,Alekseev:1988gp}, overlaid with a contour plot of the expected event rate from a typical proto-neutron star (PNS) cooling calculation of the Garching group. Figure taken from Ref.~\cite{Fiorillo:2023frv} with permission. For Kam-II and BUST, the horizontal bands of expected event rates represent the background, whereas IMB is essentially background free. The three late events in Kam-II and the two late ones in BUST are difficult to explain with PNS cooling in the presence of convection that accelerates the cooling speed.}
    \label{fig:SN1987A-Signal}
\end{figure}

The neutrino signal from a core-collapse SN was observed only once, namely in  connection with SN~1987A on 23rd February 1987 in the IMB~\cite{Bionta:1987qt,IMB:1988suc} and Kamiokande~II (Kam-II)~\cite{Kamiokande-II:1987idp,Hirata:1988ad} water Cherenkov detectors and the still-running Baksan Underground Scintillator Telescope (BUST)~\cite{Alekseev:1987ej,Alekseev:1988gp}. Many details concerning the data and their ``anomalies'' were recently reviewed~\cite{Fiorillo:2023frv} as well as their interpretation in the context of modern numerical simulations~\cite{Fiorillo:2023frv,Li:2023ulf,Olsen:2021uvt}. The data are shown in Fig.~\ref{fig:SN1987A-Signal}, i.e., the detected positron energy from inverse beta decay $\bar\nu_e+p\to n+e^+$, as a function of time, in each case after the first event. The relative timing between the different detectors was uncertain to within about a minute due to clock problems both at Kam-II and BUST. The general $\bar\nu_e$ signal characteristics with 10-MeV-range energies, the spread over several seconds, and the implied overall energy emitted in neutrinos correspond well to expectations and so these measurements were taken as a confirmation of our overall understanding of core-collapse~SNe.

The binding energy of the collapsed SN core is carried away primarily by neutrinos of all flavors that emerge from the weak decoupling region around the proto-neutron star (PNS) surface, often called neutrino sphere, but really a geometrically thick layer of energy-dependent neutrino optical depth around one. If axions or other feebly interacting particles exist that can escape directly from the PNS interior, they carry away some of the energy that would otherwise power neutrino emission after transport to the PNS surface. Therefore, the late-time part of the SN neutrino signal is particularly sensitive to this effect. After collapse and during the initial phase, the neutrino signal is largely powered by accretion, not by energy from deeply inside the PNS. Also, the inner PNS is initially relatively cold and only heats up in the course of deleptonization, so axion volume emission takes time to fully develop. Therefore, the effect of energy loss by axions (or other FIPs) will not strongly affect the signal, say, during the first second so that the total emitted energy in neutrinos is much less sensitive to novel energy losses than the signal duration. Of course, if the new particles interact so strongly that they are themselves trapped, they also will be emitted from a thick region that can be pictured as an axion-sphere, and they will contribute to radiative energy transport within the PNS \cite{Burrows:1990pk,Caputo:2022rca}. Overall, axion cooling will have a strong effect in an interval of coupling strength such that axion emission is not too small and thus irrelevant (like graviton emission) and on the other hand the mean-free path is not so short that the additional contribution to energy transfer is small, more similar to what photons do. 

Early numerical simulations of PNS evolution that included axion losses or axion energy transfer confirmed this picture \cite{Burrows:1988ah,Burrows:1990pk}. On this basis, a rough criterion was developed to estimate when axion (or similar) energy losses would become important, i.e., the new energy loss rate should be smaller than about $10^{19}~{\rm erg}~{\rm g}^{-1}~{\rm s}^{-1}$ for a medium at nuclear density of $3\times10^{14}~{\rm g}~{\rm cm}^{-3}$ and temperature $T=30~{\rm MeV}$ \cite{Raffelt:1990yz,Raffelt:2006cw}. In this form, the SN~1987A cooling argument was applied to many cases as cited earlier. Sometimes the novel energy losses were computed from a numerical SN model, but the argument has always remained somewhat schematic in that the criterion was that the new energy loss should not exceed a certain threshold, without direct comparison with SN~1987A data.

\subsubsection{Emission Rates}

The SN~1987A cooling argument is especially useful for axions, new particles that derive from QCD and thus generically couple to nucleons and nuclear matter. On the other hand, here the nuclear-physics uncertainties concerning the emission rates are large \cite{Choi:1988xt,Turner:1989wa,Ericson:1988wr,Carena:1988kr,Raffelt:1991pw,Raffelt:1993ix,Janka:1995ir,Giannotti:2005tn,Stoica:2009zh,Stoica:2012zz,Vonk:2020zfh,Carenza:2019pxu,Carenza:2020cis,Choi:2021ign,Ho:2022oaw,Li:2023thv,Anzuini:2023whm,Springmann:2024mjp,Springmann:2024ret,Cavan-Piton:2024ayu} and intertwined with the question of the nuclear equation of state (EoS). Of course, neutrino interaction rates are also affected by some of the same uncertainties. Traditional nuclear bremsstrahlung emission $NN\to NN a$ may be outdone by pionic emission of the form $\pi^-+p\to n+a$, once more with large EoS uncertainties. Core-collapse SN physics with fully consistent treatment of the EoS, neutrino interaction rates, and possible new physics ingredients remains elusive for now.

\subsubsection{Some Doubts}

The often-used SN~1987A cooling argument is also subject to doubts, one obvious issue being the sparse neutrino statistics. More fundamentally, one often-cited critique \cite{Bar:2019ifz} descends from the non-detection of a compact remnant even nearly 40~years after the SN explosion. Have we really seen PNS cooling in the neutrino signal or was a black hole formed after all? The underlying motivation for these doubts derive from an alternative picture of the core-collapse SN explosion mechanism as a collapse-induced thermonuclear explosion (CITE) \cite{Kushnir:2014oca,Blum:2016afe} rather than the traditional neutrino-driven scenario. As the CITE mechanism has not been further pursued by the original authors or the community at large, such ideas remain largely unexplored. The success of numerical core-collapse SN simulations provides little motivation for this community to seek alternative scenarios, which however does not prove that other scenarios might not be viable. Arguments against the CITE interpretation of SN~1987A were summarized in a recent review~\cite{Caputo:2024oqc}.

A separate issue concerns the actual comparison of the SN~1987A signal with theoretical predictions, i.e., the question of how well the observed signal agrees with model predictions. A recent systematic study based on 1D numerical models of the Garching group with different cases of nuclear EoS and different final NS masses revealed that the expected signal duration was always much shorter than indicated by the data \cite{Fiorillo:2023frv}. One example is shown in Fig.~\ref{fig:SN1987A-Signal}, where the signal is overlaid with the expected time-energy distribution, including the background expectation of the Kam-II and BUST detectors. The late events in these detectors cannot be easily explained by either signal or background except as an extremely rare upward fluctuation. The key physical ingredient causing this discrepancy is PNS convection that speeds up cooling relative to neutrino diffusive transport. PNS convection is a standard effect and physically unavoidable, but not always implemented even in modern simulations. While the late events may be caused by fallback or a nuclear phase transition, it is at present not clear what exactly to make of this situation with respect to the additional effect of energy loss by new particles. Dedicated models that include both PNS convection and particle losses have yet to be constructed.

\subsubsection{Summary}

The SN~1987A cooling bound, in its different manifestations, remains a powerful tool to understand the impact of new particles on SN physics, but of course has systematic limitations, like any other astrophysical or cosmological argument. Observing a high-statistics neutrino signal from the next galactic SN would go a long way to providing a sounder basis and perhaps would be the most important particle-physics harvest from such an event. Meanwhile, it remains important to develop complementary arguments. Most intriguingly, if one were to detect axions or other FIPs in violation of the cooling bound would point to a crucial role of these particles in SN physics---a potential paradigm~shift.



%% file: WG3/content/Edoardo_Vitagliano.tex
\subsubsection{Introduction}
As discussed in Sec.~\ref{subsec:Raffelt}, the existence of new WISPs implies  an additional cooling channel, and affects the duration of the neutrino signal of SN~1987A. The SN cooling argument excludes the existence of novel WISPs, unless their couplings are so small they are not produced efficiently (free-streaming regime) or they are so strongly interacting that they cannot escape, so no additional cooling channel exists (trapping regime). In the free-streaming regime, a simple criterion is that the new energy loss should not exceed the neutrino energy loss, $10^{19}\,{\rm erg} \,{\rm g}^{-1} \,{\rm s}^{-1}$, or an overall luminosity of around few times $10^{52}\,{\rm erg}\,{\rm s}^{-1}$, to be calculated at nuclear density $\rho=3\times10^{14}\,{\rm g}\,{\rm cm}^{-3}$ and $T=30~{\rm MeV}$ \cite{Raffelt:1996wa}.

Owing to their high temperature and density, the cores of PNSs and the metastable hypermassive neutron star left behind respectively by core-collapse supernovae and neutron star mergers can be factories of particles with mass of up to several hundreds MeV. While in the massless limit WISPs with extremely large lifetimes can be copiously produced in astrophysical environments, in the keV-GeV mass range the very same coupling that triggers the novel particle production can reduce its lifetime. For example, ALPs with a two-photon coupling $g_{a\gamma}a F\Tilde{F}$, produced through Primakoff scattering and photon coalescence, can decay back to two photons. Although short lifetimes prevent such WISPs from being viable dark matter candidates, they can act as a portal between a stable DM candidate and the SM (see e.g.~\cite{Pospelov:2007mp,Knapen:2017xzo,Fitzpatrick:2023xks,Allen:2024ndv}). Moreover, heavy WISPs can arise in top-down inspired scenarios~\cite{Hook:2019qoh,Lillard:2018fdt,Cox:2019rro,Dimopoulos:1979pp,Holdom:1982ex,Dine:1986bg,Flynn:1987rs,Agrawal:2017ksf,DiLuzio:2020wdo} and be releated to many particle physics conundra, e.g., new bosons~\cite{Foot:1990mn,He:1990pn,He:1991qd,Bauer:2018onh} as an explanation of recent $g_\mu-2$ measurements~\cite{Pospelov:2008zw,Gninenko:2001hx,Chen:2017awl}, or as secret neutrino interaction mediators~\cite{Blinov:2019gcj,DeGouvea:2019wpf,Hansen:2017rxr,Kelly:2020pcy}. For concreteness, we will focus here on ALPs with a two-photon coupling (Fig.~\ref{fig:heavymassALPbounds}), though many of these results apply similarly to any particle that can decay to charged leptons and photons.

\subsubsection{Astrophysical constraints on sub-GeV particles: recent developments}

\begin{figure}[t!]
    \centering \includegraphics[width=1\linewidth]{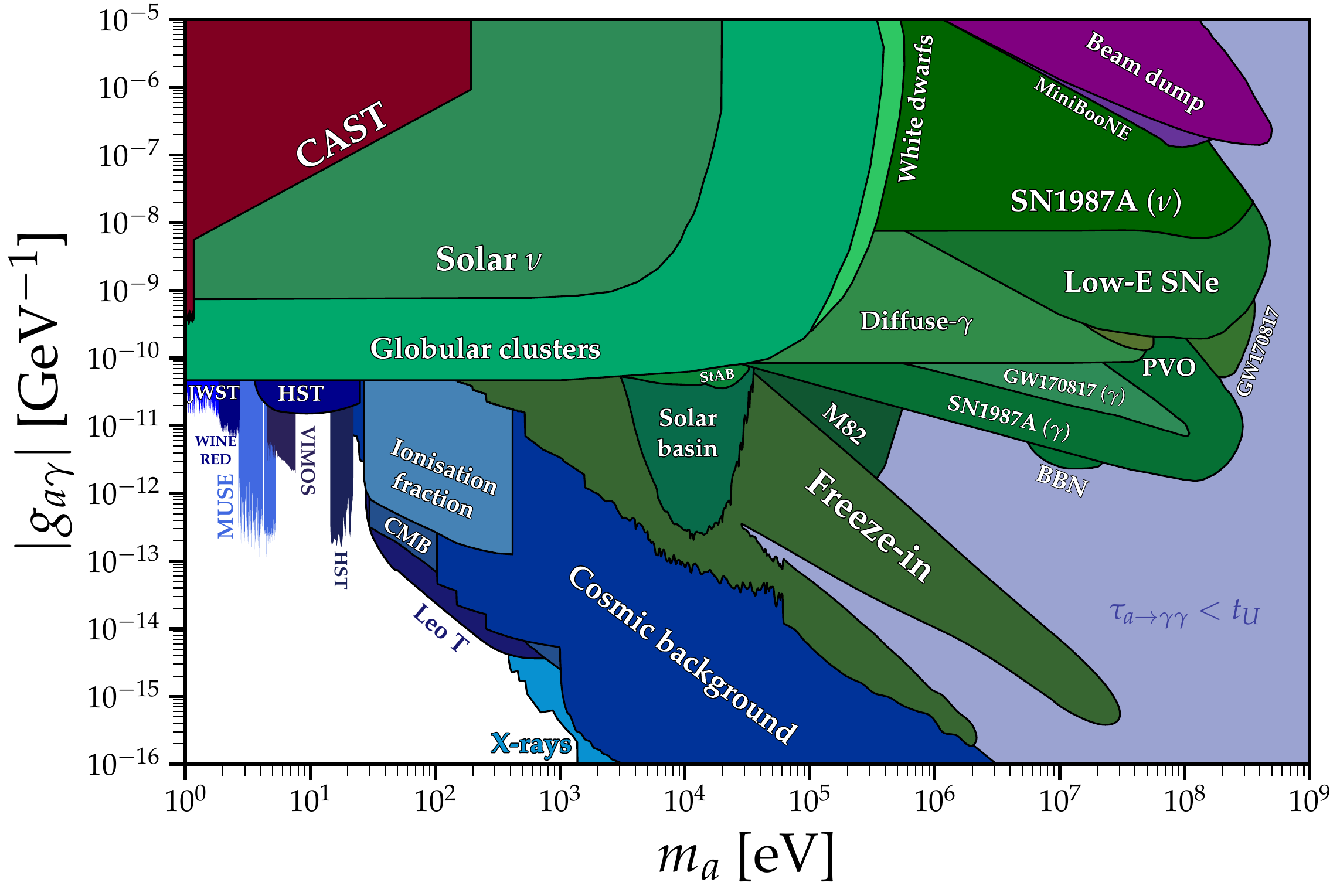}
    \caption{Close-up of the heavy ($m_a>1\,\rm eV$) axion-like particle parameter space for the coupling $g_{a\gamma}F\Tilde{F}$. The constraints showed here include  helioscope~\cite{CAST:2007jps,CAST:2017uph,CAST:2024eil}, stellar~\cite{Vinyoles:2015aba,Dolan:2021rya,Dolan:2022kul,Beaufort:2023zuj,Nguyen:2023czp,Candon:2024eah}, and cosmological~\cite{Grin:2006aw,Cadamuro:2011fd,Wadekar:2021qae,Carenza:2023qxh,Todarello:2023hdk,Janish:2023kvi,Yin:2024lla,Porras-Bedmar:2024uql,Todarello:2024qci} bounds. However, unless the reheating temperature of the Universe is large enough, the cosmological constraints reduce to the irreducible axion background (``Freeze-in'', $T_{\rm RH}=5\,\rm MeV$)~\cite{Langhoff:2022bij}. Therefore, the region $\tau_{a\rightarrow \gamma\gamma}<t_U$ is ruled in, except from laboratory~\cite{CHARM:1985anb,Riordan:1987aw,Dolan:2017osp,Blumlein:1990ay,NA64:2020qwq,Capozzi:2023ffu} and astrophysical transient constraints~\cite{Caputo:2021rux,Caputo:2022mah,Hoof:2022xbe,Diamond:2023cto,Diamond:2023scc,Dev:2023hax}, described in the main text. Figure credit: Ciaran O'Hare~\cite{AxionLimits}.}
    \label{fig:heavymassALPbounds}
\end{figure}
The best probe to discover these particles depends on the decay length, that could be as small as a hundred meters, or be large on cosmological scales. At large couplings, ALPs with mass $m_a\lesssim 1\,\rm GeV$ are constrained by laboratory searches~\cite{CHARM:1985anb,Riordan:1987aw,Dolan:2017osp,Blumlein:1990ay,NA64:2020qwq,Capozzi:2023ffu}. The smallest couplings are constrained the most by cosmology: Ref.~\cite{Cadamuro:2011fd} put some bounds assuming a population produced with freeze out, unavoidable when the reheating temperature of the Universe is large enough, that eventually decay affecting cosmological observables such as $N_{\rm eff}$. Such bounds have been revisited more recently in Refs.~\cite{Depta:2020wmr,Depta:2020zbh}. If the reheating temperature is small enough, the parameter space of ALPs with mass $m_a\gtrsim 100\,\rm keV$ is largely unconstrained, barring a region still excluded by an irreducible axion background produced through freeze-in~\cite{Langhoff:2022bij}. For a thorough discussion of the topic, cf. Sec. \ref{subsec:hot_axions}. Analogous considerations can be carried over to, e.g., dark photons~\cite{Redondo:2008ec,Fradette:2014sza,Linden:2024fby} and scalars mixed with the Higgs boson~\cite{Fradette:2018hhl,DEramo:2024lsk}.
If the particle has decay length in-between the extremely short beam-dump scale and extremely long cosmological scale, one can probe its existence through astrophysical transient observations.

Once produced, ALPs leaving the core can decay back in the mantle, lighting up the SN. One can think of the SN as onion-shaped, with a PNS at its center and a mantle surrounding the PNS. Most of the energy released during the collapse, around $3\times 10^{53}\,\rm erg$, is in neutrinos, which are emitted from the neutrinosphere---the region where the density becomes small enough for them to not be trapped anymore. Neutrinos deposit a small amount of their energy in the explosion ($10^{51} \,\rm erg$).
The cooling bounds correspond to asking that the luminosity in ALPs is comparable to the one of neutrinos. If the decay length is roughly smaller than the SN progenitor radius, ALPs would deposit 100 times the energy normally deposited by neutrinos leaving from the proto-neutron star. This means that the luminosity in new particles must be 1\% the one of neutrinos, much more stringent than the one set by cooling bounds. The possibility for particle decays to light up SNe was already noticed in 1978~\cite{Falk:1978kf} in the context of radiative neutrino decays. While one might be tempted to consider SN explosions triggered by ALP decays, this possibility is excluded  by the observation of low-energy Type II-P SNe such as SN~2008bk, which have an explosion energy of less than $10^{50} \,\rm erg$~\cite{Valenti:2009ps,2014MNRAS.439.2873S}. Therefore, the total energy emitted in ALPs must be smaller than $10^{50} \,\rm erg$, rather than $3\times 10^{53} \,\rm erg$~\cite{Caputo:2022mah,Fiorillo:2025yzf}.
While some doubts have been casted on SN 1987A bounds (see Sec.~\ref{subsec:Raffelt}), energy-deposition bounds, being calorimetric, seem very robust, and could be improved by using  low-energy SNe light-curve shapes and spectral line
velocities. These constraints might apply also in the trapping regime~\cite{Caputo:2021rux}, though one should take the latter case with a grain of salt, since they were obtained by use of an unperturbed SN model, i.e., without accounting for the effect that the novel particle would have on the energy transport inside the PNS (see also Ref.~\cite{Fiorillo:2025yzf} for recent developments). Low-energy SN bounds can be applied to any novel particle decaying to lighter SM particles as charged leptons or photons~\cite{Sung:2019xie,Shin:2022ulh,Chauhan:2023sci,Carenza:2023old,Chauhan:2024nfa,Fiorillo:2025yzf}. Intriguingly, one can also constrain self-interacting dark particles~\cite{Fiorillo:2024upk}, which could emit photons through, e.g., bremsstrahlung.

A second class of bounds come from multimessenger observations of NS mergers~\cite{Diamond:2021ekg,Diamond:2023cto,Dev:2023hax}. Before collapsing into black holes, the hypermassive unstable remnants produced by NS mergers are indeed also efficient factories of heavy ALPs. After being produced, the latter propagate with energies $\mathcal{O}(100\,\rm MeV)$. The photons produced by the ALP decay can annihilate into $e^+ e^-$ pairs, which in turn can also emit photons through bremsstrahlung. The subsequent electromagnetic cascade would produce an axion-sourced fireball in correspondence of GW signals, e.g., from GW170817~\cite{LIGOScientific:2017ync,LIGOScientific:2017zic}. Such fireball would eventually feature photons with energies below $\mathcal{O}(1\,\rm MeV)$, low enough to be picked up by gamma-ray burst monitors such as CALET CGBM~\cite{CALET}, Konus-Wind~\cite{Konus}, and Insight-HXMT/HE~\cite{Insight}. In the case of GW170817, the energy emitted in ALPs before the collapse of the remnant into a black hole, i.e., over a time interval of 1~s, needed to be smaller than $3\times 10^{46} \,\rm erg$.

Considering yet longer lifetimes, ALPs et similia can get outside of the SN mantle and decay to photons outside of the progenitor radius. Therefore, one would have observed a gamma-ray flux of photons with energy $\mathcal{O}(100\,\rm MeV)$ at the Solar Maximum Mission telescope~\cite{forrest1980gamma}, which measured the fluence in the interval $4.1-100$ MeV during a 223.2 s interval coincident with the SN 1987A neutrino signal~\cite{Jaeckel:2017tud,Hoof:2022xbe}, an idea first applied to putative MeV-mass $\tau$ neutrino radiative decays~\cite{Oberauer:1993yr}. However, as recently pointed out~\cite{Diamond:2023scc}, in a region of the parameter space the density of the photons produced by ALP decay is so large that they would have cascaded into a fireball~\cite{Diamond:2023scc}. The rather low-energy photons, $\mathcal{O}(1\,\rm MeV)$,
would have shown up at the Pioneer Venus Observatory~\cite{PVO,colin1980pioneer,colin197711}, closing again
the parameter space. Similar more stringent bounds in the large coupling regime are obtained with stripped-envelope SNe, whose progenitors have lost their hydrogen or even helium envelopes and feature small radii~\cite{Candon:2025ypl}.
One could also consider the diffuse x-ray and gamma-ray background produced by the so called ``diffuse SN axion background'', created by all the SNe which happened during the history of the Universe. 
The idea dates back to an early paper~\cite{Cowsik:1977vz}, and it
has been revived recently in the context of ALPs~\cite{Calore:2020tjw,Caputo:2021rux,Lella:2022uwi}.

At small masses, the large Lorentz boost factor suppresses the decay rate of ALPs produced in astrophysical transients.
The take-away lesson is that the constraints are the strongest for ALPs light enough to be produced, and heavy enough that
the Lorentz boost does not impede their decay---as can be seen in Fig.~\ref{fig:heavymassALPbounds} for SN~1987A, GW170817, and low-energy SN bounds. One way to avoid this drawback is by relying on colder sources. In Ref.~\cite{Candon:2024eah} it has been proposed that starburst galaxies such as M82 can copiously produce ALPs as stars can reach temperatures in the tens of keV during part of their life. Such ALPs would have shown up in M82 observations of x-ray telescopes such as NuSTAR, as further discussed in Sec.~\ref{subsec:RuzVogel}.

In absence of a direct coupling to photons, couplings of scalars and pseudo-scalars to charged leptons can also be similarly probed since the novel particle can decay to photons at one-loop. This idea was used to constrain muonic bosons~\cite{Caputo:2021rux}, and was later applied to the electron case~\cite{Ferreira:2022xlw,Fiorillo:2025sln}. Vectors cannot decay to two photons as it is forbidden by the Landau-Yang Theorem~\cite{Landau:1948kw,Yang:1950rg}, but can decay to $e^+ e^-\gamma$, see, e.g., \cite{Kazanas:2014mca,DeRocco:2019njg}. If the novel particle interacts with neutrinos, e.g., in the case of heavy majorons, very stringent bounds arise from the non observation of a 100-MeV neutrino flux from SN 1987A~\cite{Fiorillo:2022cdq}. Majorons would be produced in the core with $100\,\rm MeV$ energies, and decay outside of the SN into $100\,\rm MeV$ neutrinos. Since the cross section of neutrinos in a detector scales as the energy squared and no $100\,\rm MeV$ neutrino was observed by Kam-II~\cite{Kamiokande-II:1987idp,Hirata:1988ad} and IMB~\cite{Bionta:1987qt,IMB:1988suc}, one requires the majoron luminosity to be 1\% of the neutrino luminosity. Other models can also be constrained by the same token~\cite{Akita:2022etk}. On the other hand in the trapping regime, while it has been suggested otherwise~\cite{Manohar:1987ec,Shalgar:2019rqe,Chang:2022aas}, SNe leave large part of the parameter space unconstrained in the large coupling regime~\cite{Fiorillo:2023cas,Fiorillo:2023ytr} (see also Refs.~\cite{Dicus:1988jh} for an early discussion).

\subsubsection{Summary}
In recent years we have witnessed an ever increasing interest in astrophysical transients such as core-collapse supernovae and neutron star mergers. The development of multimessenger astronomy, of new detectors, and the advancements in numerical simulations (e.g.,~\cite{Kastaun:2016yaf, Hanauske:2016gia,Janka:2016fox,Foucart:2016rxm,Bollig:2017lki,Wu:2017drk,Ardevol-Pulpillo:2018btx,Hanauske:2019qgs,Shibata:2019wef,Perego:2019adq,George:2020veu,Radice:2023zlw,Pajkos:2024iry,Foucart:2024cjr,Bernuzzi:2024mfx,Chatziioannou:2024jsr}), makes these phenomena extraordinary astroparticle physics laboratories.
Future events will allow us to explore uncharted part in the parameter space of WISPs~\cite{DeRocco:2019jti,Muller:2023vjm,Akita:2022etk,Telalovic:2024cot}, and could even tell us something about the most extreme astrophysical events~\cite{Caputo:2021kcv}. Strong constraints could also come from previous events. For example, if extreme low-energy SNe such as the hydrogen-deficient SN 2008ha~\cite{Valenti:2009ps} are confirmed as core-collapse SNe from stripped-envelope progenitors new constraints could be obtained. While we have focused on heavy WISPs, light axions are also powerfully probed, since they can be copiously produced in core-collapse supernovae and neutron star mergers, and later convert in the local, galactic, and extragalactic magnetic fields~\cite{Brockway:1996yr,Grifols:1996id,Payez:2014xsa,Meyer:2016wrm,Ge:2020zww,Hoof:2022xbe,Lella:2024hfk,Manzari:2024jns,Carenza:2025uib,Fiorillo:2025gnd,Candon:2025sdm} (see Sec.~\ref{sec:Extragalactic_sources} and Sec.~\ref{sec:constraints-space} for more details).

%% file: WG3/content/Malte_Buschmann.tex
Neutron stars are the densest and hottest class of stellar objects. Their extreme environment makes them excellent laboratories for searches for WISPs since such particles can be abundantly produced within the hot neutron star core~\cite{Iwamoto:1984ir,Brinkmann:1988vi,Iwamoto:1992jp}. Crucially, WISPs can escape the neutron star unhindered due to their weakly interacting nature. This generally has two implications: (1) The energy carried away by escaping WISPs cools the neutron star anomalously~\cite{Leinson:2014ioa,Sedrakian:2015krq,Sedrakian:2018kdm,Hamaguchi:2018oqw,Leinson:2021ety,Buschmann:2021juv}. And (2), certain types of WISPs may convert to Standard Model particles within the strong magnetosphere surrounding the neutron star, leading to a potentially observable X-ray signature~\cite{Morris:1984iz,Fortin:2018ehg,Buschmann:2019pfp}. Note that the latter will also happen if the neutron star encounters WISPs that are part of a relic abundance or have been produced in magnetosphere gaps~\cite{Prabhu:2021zve} (see Sec.~\ref{subsec:Witte}). However, WISPs produced in the neutron star core will generally produce X-ray photons upon conversion, whereas relic WISPs leave a lower-energy radio signature behind~\cite{Pshirkov:2007st,Safdi:2018oeu,Hook:2018iia,Leroy:2019ghm,Battye:2019aco,Foster:2020pgt,Witte:2021arp}.

The following sections summarize some of the WISPs production mechanisms within the neutron star and the implications on neutron star cooling and potential X-ray signatures. While neutron stars may be relevant for other types of WISPs (see e.g.~\cite{Giannotti:2015kwo}) we will focus primarily on the most well-studied case of ALPs.

\subsubsection{Production mechanisms}
Despite their weakly interacting nature, WISPs can be efficiently produced by a variety of different mechanisms within a neutron star. Their small coupling strength to neutrons, protons, and leptons is compensated by the extremely hot nature of the core with temperatures of around $10^8$ K. Here we will touch on the two dominant production channels, nucleon bremsstrahlung and production through Cooper-pair formation and breaking.

For an ALP and a non-superfluid neutron star core, the leading emission mechanism is nucleon bremsstrahlung: $nn\rightarrow nna$, $np\rightarrow npa$, and $pp\rightarrow ppa$~\cite{Iwamoto:1984ir,Brinkmann:1988vi,Iwamoto:1992jp}. The production rate for pure neutron scattering can be estimated as~\cite{Buschmann:2021juv} 
\begin{equation}
\begin{split}
\epsilon(nn\rightarrow nna)\simeq
(7.373\times 10^{11} \text{ erg/cm}^3\text{/sec})
\left(\frac{g_{an}}{10^{-10}}\right)^2
\left(\frac{p_{F,n}}{1.68 \text{ fm}^{-1}}\right)
\\\times
\left(\frac{F(x_n)}{0.601566}\right)\left(\frac{T}{10^8 \text{ K}}\right)^6\left(\frac{\beta_{nn}}{0.56}\right)
\left(\frac{\gamma_{nn}}{0.838}\right)
\left(\frac{\gamma}{1}\right)^6
\left(\frac{\mathcal{R}_{nn}}{1}\right)\,.
\end{split}
\end{equation}
Here, $g_{an}$ is the ALP coupling to neutrons, $T$ is the core temperature, $p_{F,n}$ is the neutron Fermi momentum, and $F(x_n)=1-3x_n/2 \arctan(1/x_n)+x_n^2/(2+2x_n)$ with $x_n=m_{\pi^\pm}/(2p_{F,n})$. We furthermore have various suppression factors: $\beta_{nn}$ accounts for short-range correlations induced by the hard core of the $nn$ interactions, $\gamma_{nn}$ corrects for multi-pion exchange, and $\gamma$ describes the medium-dependence of nucleon couplings in the high-density core. Lastly, should the neutron star core be in a superfluid state, the bremsstrahlung rate will be suppressed by $R_{nn}$ below the superfluid critical temperature $T<T_c$. A detailed discussion of these rates can be found in~\cite{Buschmann:2021juv}.

If the neutron star core is in a superfluid state the bremsstrahlung rate is suppressed as most nucleons are bound in Cooper pairs and unavailable for scattering. However, Cooper pair formation and breaking allows for another production mechanism which produces ALPs that carry away the liberated binding energy. This process takes place if the local core temperature is not far below the critical temperature, in which case this mechanism dominates over other ALP production channels. The emission rate depends on the spin and flavor of the paired nucleons and their pairing gap, which are typically spin-0 $S$-wave neutron-neutron pairings, spin-0 $S$-wave proton-proton pairings, and two different types of spin-1 $P$-wave neutron-neutron pairings. WISPs may also be produced through other forms of matter such as leptons, more exotic hyperon superfluids or pionic and kaonic Bose-Einstein condensates.

The bremsstrahlung emission spectrum can be described by a modified thermal distribution~\cite{Iwamoto:1984ir}
\begin{equation}
    \frac{dF}{dE}\propto z^3\frac{z^2+4\pi^2}{e^z-1} \,,
\label{eq:spectrum}
\end{equation}
where $z=E/T$, $E$ is the local axion energy, $T$ the local core temperature, and $F$ is flux. For a typical old neutron star, this spectrum peaks at $\mathcal{O}(\text{keV})$. The emission spectrum for ALPs produced through Cooper pair breaking and formation is slightly more complex due to its dependence on the critical temperature. Typically, this spectrum is slightly harder than the spectrum of the bremsstrahlung process with a peak emission at $\mathcal{O}(10\text{s of keV})$.

\subsubsection{Neutron star cooling}
Since WISPs produced in the neutron star core can leave the interior unhindered they carry away energy and contribute to anomalous cooling of the star. This extra cooling mechanism is subleading for most of the neutron stars' early existence compared to cooling through neutrinos~\cite{Yakovlev:2000jp}, but ALP cooling becomes dominant after a few thousand years~\cite{Buschmann:2021juv}. After about $10^5$ years cooling through photon emission is most effective. Thus, the ideal age for a neutron star to observe anomalous cooling is around $10^5$ years and hence the Magnificent Seven neutron stars (M7) are often considered. The M7 are a set of seven X-ray dim and isolated neutron stars with ages around $\sim 10^5-10^6$ years. Their ages are known from kinematic considerations and their surface luminosity is well-measured.

In~\cite{Buschmann:2021juv} the M7 were used to set constraints on the axion parameter space, limiting the QCD axion mass to be below $\sim 10-30$ meV, depending on the details of the UV completion. Note that this method probes the coupling relevant for production, i.e. the coupling to nucleons $g_{aN}$, rather than the axion-photon coupling $g_{a\gamma\gamma}$. Other objects like the proto-neutron star from SN 1987A~\cite{Raffelt:2006cw,Fischer:2016cyd,Chang:2018rso,Carenza:2019pxu,Carenza:2020cis} and the young neutron star located at the center of the supernova remnant Cassiopeia A~\cite{Leinson:2014ioa,Sedrakian:2015krq,Sedrakian:2018kdm,Hamaguchi:2018oqw,Leinson:2021ety} have been used as well to constrain the ALP parameter space. Note, however, that these results may not be as robust due to possible systematic bias in the experimental data~\cite{Posselt:2018xaf} and difficulties with the self-consistent modeling of these young objects~\cite{Fischer:2016cyd,Bar:2019ifz}.

\subsubsection{X-ray signatures}
The energy spectrum of emitted ALPs is set by the core temperature, see Eq.~\eqref{eq:spectrum} and thus the produced X-ray photons are harder than the surface emission. The surface emission can be described by a (double) blackbody spectrum that typically peaks at or below $\mathcal{O}(100\text{ eV})$ but falls off steeply at higher energies. This may make an observation of X-ray photons from ALP production possible if one targets X-ray dim neutron stars~\cite{Morris:1984iz,Fortin:2018ehg,Buschmann:2019pfp} such as the M7.
Tantalizingly, an X-ray excess in the 2-8 keV range has been observed in some of the M7 neutron stars that currently defies explanation~\cite{Dessert:2019dos}. This excess may be interpreted as ALP in origin with a mass below $2\times 10^{-5}$ eV and best fit-coupling of $g_{aN}g_{a\gamma\gamma}\in(2\times 10^{-21},10^{-18})$ GeV$^{-1}$~\cite{Buschmann:2019pfp}, but more data is required to properly test this hypothesis.

%% file: WG3/content/Francesca_Chadha-Day.tex
Superradiance is the amplification of incident radiation by a dissipative system. This amplification has been observed experimentally in fluids \cite{Torres:2016iee}. If the radiation is also trapped within the system, a superradiant instability may be triggered in which the radiation is exponentially amplified. This effect can occur when a bosonic field is gravitationally bound to a rotating Black Hole (BH) or a star. As discussed below, BH superradiance is a powerful probe of Beyond Standard Model (BSM) physics. A comprehensive review of superradiance can be found in \cite{Brito:2015oca}.

\subsubsection{Black Hole Superradiance}

It is well known that radiation passing through the ergoregion of a Kerr BH is amplified, i.e. it may escape the vicinity of the BH with increased energy. This is an example of superradiant amplification. Furthermore, massive radiation may be gravitationally confined in the vicinity of the BH and hence be superradiantly amplified by repeated interaction with the ergoregion. The energy source for this amplification is the rotational kinetic energy of the BH.

To calculate the rate of the superradiant growth of a bosonic field $\phi$ of mass $\mu$ around a Kerr BH, we must consider the bound states of $\phi$ in the Kerr background. These have a similar form to the hydrogen wavefunctions:

\begin{equation}
\psi_{nlm}(r) \simeq e^{-i \omega_{nlm} t+i m \varphi} Y_{lm}(\theta) R_{nlm}(r)+\text { h.c. },
\end{equation}

where $R_{nlm}(r)$ and $Y_{lm}(\theta)$ represent the radial and angular components of the wavefunction, respectively, with $n,l$ and $m$ labeling the corresponding eigenmodes and \text { h.c.} denotes the Hermitian conjugate. The eigenfrequencies of the bosonic states $\omega_{nlm}$ are complex valued, where the real part of the frequency has the same form as the hydrogenic spectrum, $\mathcal Re(\omega_{nlm})\approx\mu(1-(\alpha^2/2n^2))$ and the sign of $\mathcal Im(\omega_{nlm})$ determines the superradiant condition. The parameter $\alpha=M \mu$ is called the gravitational fine-structure constant, which determines the system's dynamics. The gravitational potential generated by the BH is approximately $V(r) \sim 1/r$, except for very close to the horizon. We find that the corresponding eigen-energies have an imaginary component, corresponding to the scalar field being dissipated into the BH or to the superradiant amplification of $\phi$. These can be calculated from the action

\begin{equation}
S = \int d^4 x \sqrt{-g} (- \frac{1}{2} \triangledown_{\mu} \phi \triangledown^{\mu} \phi - \frac{1}{2} \mu^2 \phi^2),
\end{equation}

whose equations of motion can be solved numerically and in some cases analytically. For example, in the non-relativistic regime we find \cite{Hoof:2024quk}

\begin{equation}
{\mathcal Im} \left( \omega_{n l m} \right)  \approx \mu\left(m a_*-2 r_{+} \mu\right) C_{n l} \Pi_{l m} \alpha^{4 l+4},
\end{equation}
with
\begin{equation}
\quad C_{n l}  =\frac{2^{4 l+1}(n+l)!}{n^{2 l+4}(n-l-1)!}\left[\frac{l!}{(2 l)!(2 l+1)!}\right]^2,
\end{equation}
and 
\begin{equation}
\quad \Pi_{l m}=\prod_{k=1}^l\left[k^2\left(1-a_*^2\right)+\left(m a_*-2 r_{+} \mu\right)^2\right],
\end{equation}

where $m$ is the magnetic quantum number, $k$ is the bosonic momentum, $a_{\mathrm{BH}}$ is the BH's spin, $r_{+} = M(1+\sqrt{1-a_*^2})$ is the radius of the outer horizon and $a_* = a_{\mathrm{BH}}/M$ is the dimensionless spin parameter, with $M$ denotes the mass of the BH.

The timescale of the superradiant instability is then given by $\tau = \frac{1}{{\mathcal Im} \left( \omega_{n l m} \right)}$. In general, a superradiant instability is possible when ${\mathcal Re} \left( \omega_{n l m} \right) \simeq \mu < m \Omega_H$, where $\Omega_H$ is the angular velocity of the BH, given by $\Omega_H=a_*/(2M(1+\sqrt{1-a^2_*}))$. The instability is most efficient when the BH's gravitational radius is similar to the boson's Compton wavelength and is less efficient for higher $l$ and $m$ modes, where the wavefunction's maxima are further from the BH. Similar results also apply to vector superradiance \cite{Baryakhtar:2017ngi}.

When a superradiant instability is active, the occupation number of the corresponding field $\phi$ around the BH will grow exponentially as long as $\tau_{\mathrm{SR}}<\tau_{\mathrm{ch}}$, where $\tau_{\mathrm{ch}}$ is called the characteristic timescale over which the angular momentum of the BH changes due to other physical processes. These characteristic timescales are $\tau_{\mathrm{Sal}}\approx 4.5\times 10^{7}~\mathrm{yr}$ for astrophysical BH \cite{Salpeter:1964kb,Shakura:1972te} and $\tau_\mathrm{BH}\approx 10^9~\mathrm{yr}$ for supermassive BH \cite{Davoudiasl:2019nlo}. The observation of highly spinning BH excludes boson mass at $M^{-1}$. For example, astrophysical [supermassive] BHs of mass $(1-100)~M_\odot~[(10^6-10^9)~M_\odot]$ exclude boson mass for $(10^{-11}-10^{-13})~\mathrm{eV}~[(10^{-17}-10^{-20})~\mathrm{eV}]$. The initial seed for this growth could be DM or some other astrophysically produced particle, or a quantum fluctuation. If $\phi$ has significant self-interactions, these can lead to level mixing, and must be included in the calculation. Annihilation of $\phi$ into other particles can also decrease the superradiance rate. Furthermore, for large initial seeds, if both superradiant and non-superradiant modes are populated, the instability may not occur \cite{Ficarra:2018rfu}.

\subsubsection{Axion bounds}

Black-hole superradiance may be efficient with a BSM boson. The SM bosons are either massless or too heavy to lead to an observable effect. (Although there has been some discussion on the subtle issue of whether a photon with a plasma mass may lead to superradiance \cite{Conlon:2017hhi,Wang:2022hra,Cannizzaro:2023ltu}.) Notably, BH superradiance is a very promising effect to search for axions, as first noted in \cite{Arvanitaki:2009fg}. 

If axion superradiance occurs, axions may build up around Kerr BH from an initial quantum fluctuation. How might we observe this axion cloud? Firstly, we may observe a `bosenova' as the axion cloud collapses \cite{Arvanitaki:2010sy}. The energy of a cloud of size $R$ with $N$ axions is approximated by:

\begin{equation}
V(R) \sim N \frac{l (l+1) +1}{2 \mu R^2} - N \frac{G M \mu}{R} + \frac{N^2}{32 \pi f_a^2 R^3},
\end{equation}

where $\mu$ and $1/f_a$ are the axion's mass and self-coupling. At large $N$, the gradient energy of the axion field makes the cloud unstable. The collapse may be observed as a gravitational wave and potentially a $\gamma$-ray burst.

Secondly, we may observe depletion of the BH's spin as its rotational energy is taken by the axion cloud. We cannot measure BH spin down directly, but we can measure current BH spins using the X-ray spectra of BH binaries and using gravitational wave emission from BH mergers. The superradiant instability of a BSM boson would lead to a gap in the the BH mass vs spin plot corresponding to the boson's mass. Therefore, the observation of highly spinning BHs may be used to bound BSM bosons such as axions. These bounds only apply to axions with sufficiently weak self-interactions. If $f_a$ is too {\it low}, bosenova collapse prevents significant spin depletion. Stellar mass BH spin measurements exclude $6 \times 10^{-13} \, {\rm eV} \lesssim \mu \lesssim 2 \times 10^{-11} \, {\rm eV}$ for $f_a \gtrsim 10^{13}$~GeV \cite{Arvanitaki:2014wva}. Advanced LIGO will be sensitive to $\mu \lesssim 10^{-10} \, {\rm eV}$ \cite{Arvanitaki:2016qwi}. Further discussion of the statistical subtleties of these bounds can be found in \cite{Hoof:2024quk}.

Finally, various indirect methods of detecting the axion cloud have been proposed. These include  the observation of GWs from atomic transitions between eigenstates \cite{Arvanitaki:2016qwi}, birefringence \cite{Plascencia:2017kca}, lasing \cite{Ikeda:2018nhb}, the cloud's effect on orbits in binary systems \cite{Kavic:2019cgk,Tomaselli:2024dbw} and the effect of superradiance on the spectrum of the accretion disk \cite{Sarmah:2024nst}.

\subsubsection{Background effects on black hole superradiance}

Ultralight scalar bosons can interact with the ubiquitous Cosmic Neutrino Background (C$\nu$B), thereby altering BH superradiance and enabling constraints on scalar-neutrino couplings from observations of highly spinning BHs. Part of this parameter space is already restricted by neutrino oscillation experiments (e.g., SNO, Super-Kamiokande) \cite{Berlin:2016woy} and by cosmological probes such as the Cosmic Microwave Background (CMB). Additional bounds on ultralight bosons arise from Lyman-$\alpha$ data \cite{Kobayashi:2017jcf} and from stellar heating in ultra-Faint Dwarf (UFD) galaxies \cite{Dalal:2022rmp}, though these apply under the assumption that the bosons constitute DM, unlike the superradiance limits which do not require this.

Cosmic neutrinos decoupled at $T\sim 1~\mathrm{MeV}$, earlier than photons, and today have a temperature $T_\nu\simeq 1.95~\mathrm{K}$ and number density $n_{\nu,\mathrm{tot}}\simeq 336~\mathrm{cm^{-3}}$ \cite{Dolgov:1997mb,Mangano:2005cc,Bauer:2022lri}. At least two neutrino mass eigenstates are non-relativistic at present. Direct detection is extremely challenging due to their low energies ($10^{-4}-10^{-6}$~eV); the PTOLEMY experiment aims to observe them via inverse $\beta$ decay of tritium \cite{PTOLEMY:2019hkd}, while indirect signatures appear in the CMB \cite{Vagnozzi:2017ovm,Planck:2018vyg,Jiang:2024viw}.

The interaction between relic active Dirac neutrinos $\nu_\alpha$ and a scalar field $\phi$ is described by the Lagrangian

\begin{equation}
\mathcal{L}\supset \tfrac{1}{2}\partial_\mu \phi \partial^\mu \phi - \tfrac{1}{2}\mu^2\phi^2 - m_{\alpha\beta}\bar{\nu}_\alpha \nu_\beta - y_{\alpha\beta}\phi \bar{\nu}_\alpha \nu_\beta,
\label{super1}
\end{equation}

where $m_{\alpha\beta}$ is the neutrino mass matrix and $y_{\alpha\beta}$ the Yukawa coupling, often assumed universal ($y_{\phi\nu}$). In non-SUperSYmmetry (SUSY) theories, neutrino loops radiatively generate a quartic self-interaction for $\phi$ via the Coleman-Weinberg mechanism \cite{Coleman:1973jx},

\begin{equation}
\lambda^{(0)} \sim \frac{y_{\phi\nu}^4}{16\pi^2}\ln\Big(\frac{\mu^2}{m_\nu^2}\Big),
\label{super2}
\end{equation}

where $m_\nu$ is the neutrino mass. In SUSY scenarios, such contributions cancel at zero temperature, but thermal effects reintroduce both an effective scalar mass and self-interaction \cite{Nadkarni:1988fh,Baier:1991dy,Thoma:1994yw}.

The thermal correction to the scalar mass from the C$\nu$B in the limit $\mu\ll m_\nu,~T_\nu$ is \cite{Babu:2019iml}

\begin{equation}
\Delta \mu^2=\frac{y^2_{\phi\nu}}{\pi^2}\int ^{\infty}_{m_\nu} d\epsilon \,\sqrt{\epsilon^2-m^2_\nu}\, f_\nu(\epsilon),
\label{super3}
\end{equation}

yielding $\Delta \mu^2 \simeq 1.2\times 10^{-10}~y^2_{\phi\nu}\,\mathrm{eV^2}$ for inverted hierarchy ($m_1=m_2=50~\mathrm{eV},\, m_3=10~\mathrm{meV}$), and $f_\nu(\epsilon)$ denotes the Fermi-Dirac distribution of neutrinos at energy $\epsilon$. Thus, the effective scalar mass is $\mu^2_{\mathrm{eff}}=\mu^2+\Delta \mu^2$. 

Loop corrections also induce a quartic interaction, $\mathcal{L}_{\mathrm{int}}=\tfrac{1}{4!}\lambda\phi^4$, with $\lambda=\lambda^{(0)}+\Delta\lambda^{(T)}$. In the limit $m_\nu\gg T_\nu$, the thermal contribution is about \cite{Lambiase:2025twn}

\begin{equation}
\Delta \lambda^{(T)} \sim \frac{y^4_{\phi\nu}\, n_{\nu,\mathrm{tot}}}{m^3_\nu}.
\label{super6}
\end{equation}

The induced self-interaction term $\mathcal{L}_{\mathrm{int}}$ leads to non-linear dissipation effects at values of $y_{\phi\nu}$ where mass corrections can be safely neglected, making the efficiency of energy extraction governed by the non-linear dynamics. Consequently, BH spin measurements exclude scalar-neutrino couplings as large as $y_{\phi\nu}\sim 10^{-16}\,(10^{-20})$ for astrophysical (supermassive) BHs, with the precise bounds depending on $\mu$ \cite{Lambiase:2025twn}. These results do not change appreciably for the normal mass hierarchy.

Another possible background is the Diffuse Supernova Neutrino Background (DSNB), with a number density of only $n_\nu \sim 10^{-11}~\mathrm{cm^{-3}}$ \cite{DeGouvea:2020ang,Ando:2023fcc,MacDonald:2024vtw}, too low to significantly affect the above results. The previous estimates assume a uniform cosmic neutrino density; however, massive neutrinos can cluster gravitationally near BHs and form spikes, which would weaken the superradiance bounds by roughly an order of magnitude \cite{Lambiase:2025twn}.

In addition to neutrinos, ultralight scalars may also couple to electrons in the BH accretion disk, generating an effective mass correction \cite{Lambiase:2025twn},

\begin{equation}
\Delta \mu^2\simeq 3\times 10^{-10}~\mathrm{eV^2}\, y^2_{\phi e}\Big(\frac{n_e}{10^{10}~\mathrm{cm^{-3}}}\Big),
\label{super7}
\end{equation}

where $y_{\phi e}$ is the scalar-electron Yukawa coupling. In thin or thick disks, electron densities can reach $n_e \sim 10^{19}~\mathrm{cm^{-3}}(M_\odot/M_{\mathrm{BH}})$ \cite{Narayan:1994is}, several orders of magnitude above the interstellar background ($\sim 10^{-2}~\mathrm{cm^{-3}}$ \cite{Dima:2020rzg}), allowing meaningful constraints even for small $y_{\phi e}$. Further progress requires detailed spectral and timing studies of X-ray emission from accretion disks.

\subsubsection{Stellar superradiance}
We emphasize here that horizons are not a necessary condition for superradiance. Superradiance requires a source of dissipation, which changes sign in the superradiant regime. For BHs, this dissipation is provided by the horizon. Stellar superradiance can occur if there is a dissipative non-gravitational interaction between a bosonic field and the star. Examples include dark photon superradiance around Neutron Stars (NSs) \cite{Cardoso:2017kgn}, in which dissipation is provided by a `dark conductivity', and axion superradiance around NSs \cite{Day:2019bbh}, in which dissipation is provided by the photon's dissipative interaction with the magnetosphere followed by mixing with the axion. 

More generally, many different BSM interactions could lead to stellar superradiance. Stellar environments are complex, with many more degrees of freedom than BHs. However, NS superradiance offers the potential advantage that the spin down can be observed directly. A more general method for computing stellar superradiance rates from a BSM Lagrangian by combining thermal field theory and the worldline effective field theory formulation of superradiance \cite{Endlich:2016jgc} is presented in \cite{Chadha-Day:2022inf}.

%% file: WG3/content/Konstantin_Springmann.tex
We review the properties of a light (Lorentz-)scalar field $\phi$ with non-derivative coupling to matter fields $\psi_i$, which can dramatically change the structure of compact objects such as white dwarfs (WD) and neutron stars (NS) as recently discussed in \cite{Balkin:2022qer,Balkin:2023xtr}.
At finite matter densities 
$n_i = \langle \psi^\dagger \psi \rangle \simeq \braket{\bar{\psi}_i\psi_i }\neq 0$, the potential of $\phi$ (and the derivative coupling to matter \cite{Balkin:2020dsr,Springmann:2024mjp,Springmann:2024ret}) gets modified such that it acquires a non-zero expectation value inside these objects \cite{Balkin:2020dsr,Hook:2017psm,Balkin:2021wea,Balkin:2021zfd}. 
This comes with a strong backreaction on the system, which is the focus of this review. 
In certain cases, the backreaction drastically alters the composition of stellar objects, providing a new probe for previously unexplored parameter space of these theories \cite{Balkin:2022qer,Gomez-Banon:2024oux,Kumamoto:2024wjd,Bartnick:2025lbg,Bartnick:2025lbv}.

For illustration, we follow \cite{Balkin:2023xtr} and consider the scenario of an axion-like particle (ALP) with a potential
\begin{equation}
    V(\theta)=-\Lambda^4\left(\cos\theta-1\right),\quad \theta\equiv\phi/f,
\end{equation}
where $f$ is the typical scale of the scalar field $\phi$ and $\Lambda$ sets the overall scale of the potential. 
We consider a non-derivative coupling to nucleons of the form
\begin{equation}
    \mathcal{L}_\text{int}=-\frac{g\, m_N}{2}\left(\cos\theta-1\right)\bar{N}N,
\end{equation}  
where $g>0$ is a coupling constant, $m_N$ the nucleon mass and $N=(p,n)^{\operatorname{T}}$ the nucleon field.
In the presence of a nucleon density $n_s=\braket{\bar{N}N}$, the axion potential is minimized at $\theta=\pi$ for densities larger than the critical density $n_s^c\equiv2\Lambda^4/(g\,m_N)$.
For a compact object with finite size, the scalar field tracks its minimum if gradient effects are negligible \cite{Balkin:2022qer,Balkin:2023xtr}.
This is the case if the typical scale of $\phi$ is much smaller than the radius of the object $R\gg m_\phi^{-1}(n_s)$, given by the density-dependent inverse mass.
In this limit, a modified equation of state can be used to solve the hydrostatic equilibrium equations.

\begin{figure}[t!]
    \centering
    \includegraphics[width=0.7\linewidth]{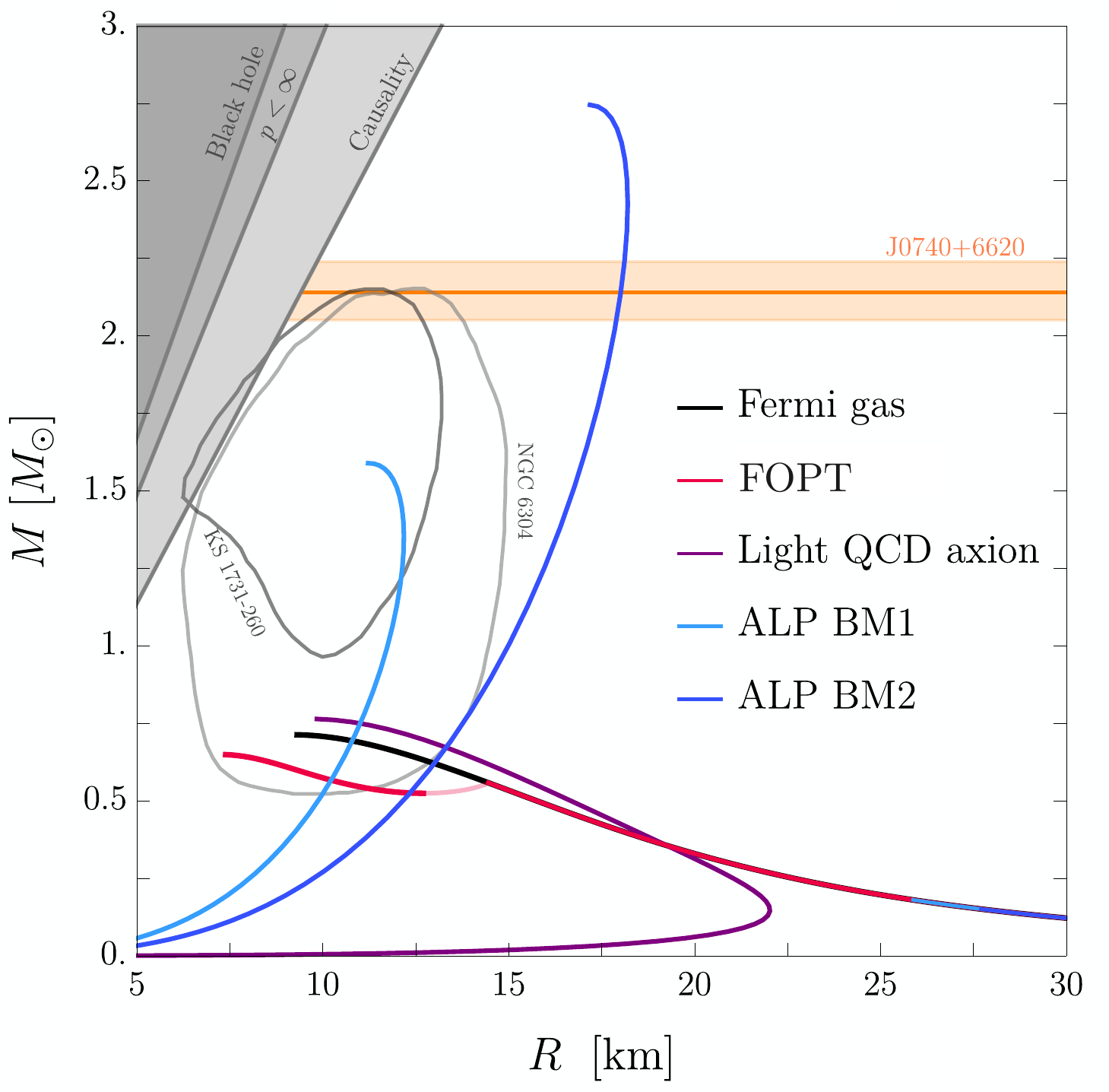}
    \caption{Mass radius curves for several equations of state. For details see the text and Ref.~\cite{Balkin:2023xtr}, from which this figure is reproduced.}
    \label{fig:MRCurves}
\end{figure}
Following \cite{Balkin:2023xtr}, we employ for simplicity a free Fermi gas of neutrons.
The equation of state is found by minimizing the total energy density $\varepsilon(\theta,k_f)=\varepsilon_N(\theta,k_f)+V(\theta)$ with respect to $\theta$ for each value of $k_f$. Here $\varepsilon_N$ is the energy density of the nucleon fluid and $k_f$ is the nucleon Fermi momentum.
Once the total energy is minimized at $\theta=\pi$, the nucleon mass is 
reduced by the factor $\delta m_N=-gm_N$.
This has striking consequences as it can lead to a minimum in the energy per particle, $E_N(n)=\varepsilon/n$ with the number density $n=k_f^3/3\pi^2$, that lies lower than the nucleon mass itself at some density $n_s^\ast$.
We have hence found a new ground state of nuclear matter, which is energetically favorable compared to ordinary nuclear matter, reminiscent of strange quark matter \cite{Witten:1984rs}. 
This effectively separates the equation of state into two branches, a meta-stable branch at densities $n_s<n_s^c$ and an absolutely stable branch in the new ground state at $n_s>n_s^\ast$ \cite{Balkin:2022qer,Balkin:2023xtr}.

On the other hand, if the minimum lies above the nucleon mass a first-order phase transition (FOPT) is found. 
Performing a Maxwell construction, one finds the thermodynamically stable equation of state. 
A third scenario is possible in which no minimum in the energy per particle is induced by the ALP, in which little effect occurs.

Let us discuss the stars resulting from these equations of state, focusing on the case of a new ground state.
In this case, the solutions will follow the prediction of a theory without ALP
until central densities of $n_s^c$ are reached. 
At densities $n_s^c<n_s<n_s^\ast$ the energy per particle reduces from values larger than the nucleon mass to lower values.
The slope, which is related to the total pressure of the system is hence negative in this region.
This implies that no stable configuration can be found in this regime and translates to a gap in the mass-radius curve.
At densities $n_s>n_s^\ast$ the absolutely stable branch is populated with zero or positive total pressure. We show example mass-radius curves for different scenarios in Fig. \ref{fig:MRCurves}. 
Interestingly, these objects are completely self-bound and absolutely stable, even in the absence of gravity.
We therefore predict macroscopic objects with densities $n_s^\ast$ and minimal size $\lambda_\text{min}\simeq f/\Lambda^2$ \cite{Balkin:2023xtr}.

For objects that resemble NSs we also find stunning results. 
The scalar field in the new ground state can stiffen the equation of state.
Typical NS radii and masses both scale like $\propto  1/m_N^2$. 
Our nucleon mass reduction hence implies that NSs can be increasingly larger and more massive, as shown in Fig. \ref{fig:MRCurves} for two benchmark points (BM1) with $g=0.5$ and (BM2) $g=0.75$.

\begin{figure}[t!]
    \centering
    \includegraphics[width=1\linewidth]{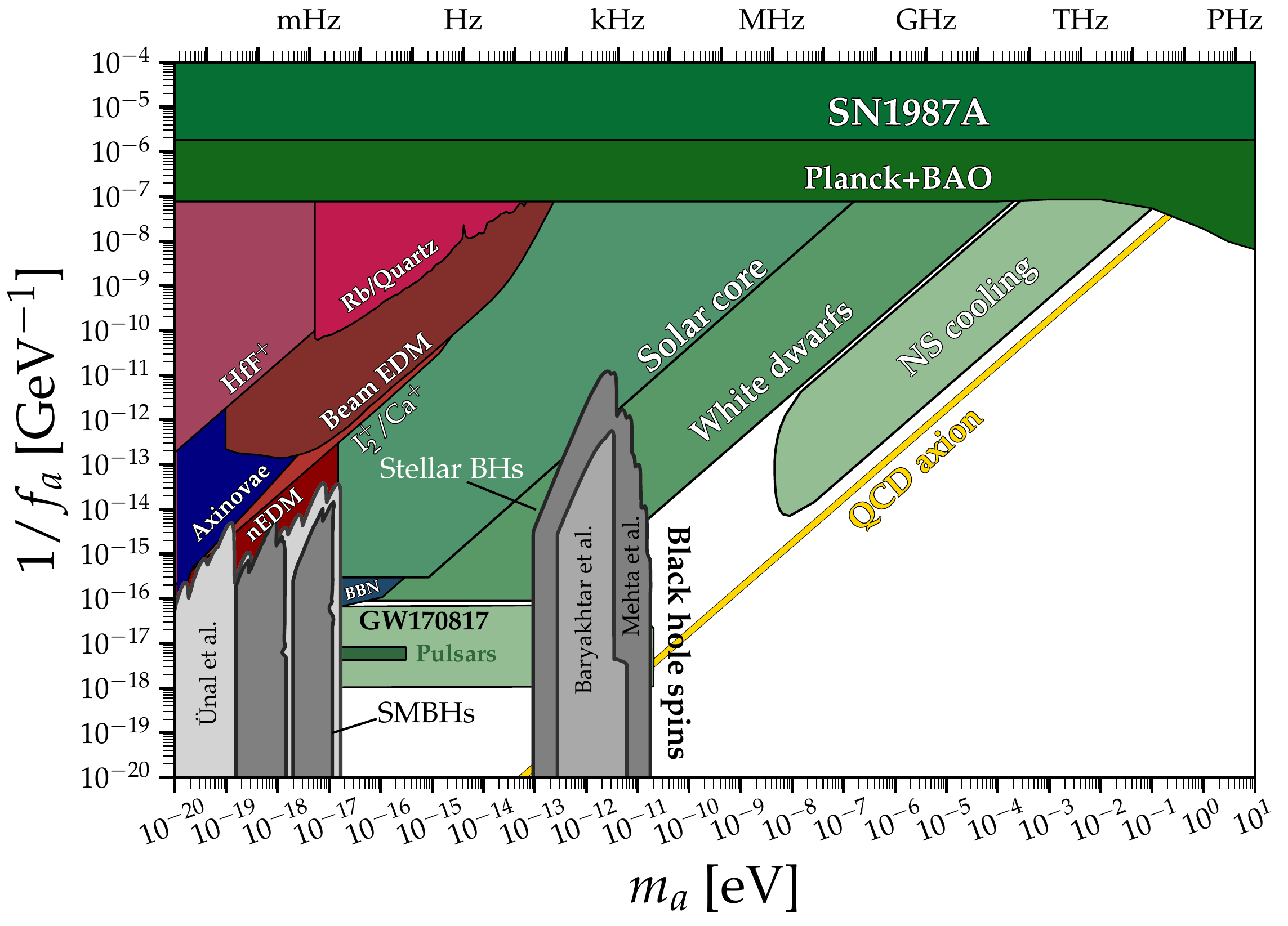}
    \caption{Axion parameter space. Adapted from Ref.\cite{AxionLimits}, other bounds from Refs.~\cite{Springmann:2024ret,Hook:2017psm,Schulthess:2022pbp,Abel:2017rtm,Roussy:2020ily,Madge:2024aot,JEDI:2022hxa,Zhang:2022ewz,Fox:2023xgx,Blum:2014vsa,Mehta:2020kwu,Baryakhtar:2020gao,Unal:2020jiy,DiLuzio:2021pxd,DiLuzio:2021gos,Zhang:2021mks,Caloni:2022uya,Witte:2024drg,Kumamoto:2024wjd,Banerjee:2025dlo,Gue:2025nxq}}
\label{fig:AxionParamSpace}
\end{figure}

In the case of a FOPT, the stiffening of the equation of state due to the nucleon mass reduction is exceeded by the softening due to the large increase in energy density at the critical pressure of the phase transition.
Hence, this results in hybrid stars which are less massive than without the ALP.
This effect exacerbates the existing difficulty of explaining the heaviest NSs observed. 
As an example, we show the case of the QCD axion in Fig. \ref{fig:MRCurves} for which such a behavior naively occurs.

Finally, we would like to discuss how aspects of these changes have been used to probe lighter than expected QCD axion models.
Lighter than expected QCD axions are approximately described by the ALP model with $\Lambda^4=\epsilon m_\pi^2f_\pi^2$, where $\epsilon\leq1$ is a small parameter, and $g m_N=2\sigma_{\pi N}$ where $\sigma_{\pi N}\simeq 50\,\text{MeV}$ is the nucleon sigma term.
In \cite{Balkin:2022qer} it was found that the modified equation of state leads to a gap in the mass radius relationship which is incompatible with observed WD mass and radius data.
Recently, bounds on light QCD axions were also put forward in the case where the potential leads to a FOPT within white dwarfs \cite{Bartnick:2025lbv}.
Similarly, in \cite{Gomez-Banon:2024oux} the sourced axion leads to a compression in the heat-blanketing envelope that reduces the NS isolation and results in faster cooling, incompatible with NS cooling data.
In \cite{Witte:2025ilt} it was found that due to the coupling of the axion to electromagnetism, an axion gradient outside the star leads to a large modification of Maxwell's equations, which in turn leads to a significant shift of the pulsar death line, in disagreement with observations.
These effects probe large parts of previously unexplored axion parameter space. The parameter space is shown in Fig. \ref{fig:AxionParamSpace} including both the WD bound \cite{Balkin:2022qer} and the NS cooling bound \cite{Gomez-Banon:2024oux}.

%% file: WG3/content/Taoso_Todarello.tex
\subsubsection{Introduction}

Axions, ALPs and dark photons are among the leading light DM candidates. From a phenomenological perspective, an important feature of these particles is their mixing with the photon.
ALPs, a generalization of the QCD axion, couple to photons through the Lagrangian term 
$\mathcal{L}=-g_{a\gamma}\,a\, \mathbf{E}\cdot\mathbf{B},$ where $g_{a\gamma}$ is
the strength of the axion-photon coupling, $a$ is the ALP field, and $\mathbf{E}$ and $\mathbf{B}$ denote the electric and magnetic fields, respectively. \\
From this interaction, one can appreciate that, in the presence of an external field, an ALP and a photon can be converted into each other. This phenomenon is at the basis of most experimental ALP searches, which typically employ an external magnetic field.
The dark photon is a massive vector boson coupled to the photon via the kinetic mixing term $\mathcal{L}=-\epsilon/2\,F_{\mu\nu}F^{\prime}_{\mu\nu}$, where $\epsilon$ is the kinetic mixing coupling and $F_{\mu\nu}$ ($F^{\prime}_{\mu\nu}$) is the field strength of the photon (dark photon). This interaction can lead to dark photon-photon conversion, similar to the case of ALPs, but with the important difference that an external electromagnetic field is not required in the process. For a theoretical discussion about a broad class of axion and dark-photon models, see Part~\ref{part:wg1}.

In astrophysical environments, the conversion of ALP or dark photon DM leads to an almost monochromatic photon signal since the energy of the photon is equal to the energy of the DM, and DM is non-relativistic. Crucially, this conversion process can be greatly enhanced at resonance, namely if the plasma frequency in the medium matches the ALP/dark photon mass. In fact, in a plasma, the photon dispersion relation is modified, and the photon acquires an effective mass equal to the plasma frequency.
In the context of dark photon DM, Ref.~\cite{An:2020jmf} demonstrated that the solar atmosphere is a promising target to search for this photon signal from resonant DM conversion. Indeed, the plasma frequency and the transparency of this medium allow dark photon masses $\lesssim\mathcal{O}(\mu{\rm eV})$ to be tested. 
Furthermore, the solar atmosphere is permeated by magnetic fields, therefore enabling ALP-photon conversion,  as investigated in~\cite{Todarello:2023ptf,An:2023wij,An:2023mvf}.
Interestingly, on the solar surface there exist regions, the sunspots, with enhanced magnetic fields, with an intensity that can be as large as of $\mathcal{O}(10^3)$ G \cite{2003A&ARv..11..153S,2019ApJ...880L..29A}.

In the following we will review the detection prospects of ALPs and dark photon signals from the solar atmosphere with current and near future observations.

\subsubsection{Signal}

The ALP-photon resonant conversion probability in a weakly magnetized isotropic plasma is given by
\begin{equation}
P(a\rightarrow\gamma) \simeq \frac{\pi}{2}\frac{g_{a\gamma}^2\,B_{\bot }^2}{v_c \, |\omega_p^{\prime}|}\,,
\end{equation}
where $B_{\bot }$ is the component of the magnetic field transverse to the direction of ALP propagation, $v_c$ is the ALP velocity, and $\omega_p^{\prime}$ is the gradient $d\omega_p/dr$ along the ALP/photon trajectory of the plasma frequency $\omega_p.$ All these quantities are evaluated at the point of the solar atmosphere $r_c$ where the resonant condition $\omega_p(r_c)=m_a$ is met, with $m_a$ being the ALP mass.
In the case of the dark photon, the conversion probability is obtained through the substitution~\cite{An:2020jmf}: $g_{a\gamma}\,B_{\bot } \rightarrow \sqrt{2/3}\,\epsilon\, m_{A^{\prime}},$ with $m_{A^{\prime}}$ the dark photon mass and $m_{A^{\prime}}=\omega_p(r_c).$ The plasma frequency is given by $\omega_p(r)= 1.17\, \mu{\rm eV}\sqrt{n_e(r)/(10^9~{\rm 
 cm}^{-3})},$ where $n_e$ is the density of free electrons in the solar atmosphere.
The photon signal from the conversion is a narrow line at an energy equal to the energy of the DM particle, therefore close to the DM mass, and with a relative width given by the DM velocity dispersion squared $\delta v^{2}\simeq10^{-6}.$ Given the range of plasma frequencies 
in the solar atmosphere (see Fig. 1 of~\cite{An:2020jmf})
the conversion signal falls in the radio band.

The photon flux per unit frequency on Earth can be
estimated as
\begin{equation}
S=\int \frac{d\Omega}{4\pi\Delta\nu} \rho_c\, v_c\, P(a/A^{\prime}\rightarrow\gamma)\, e^{-\tau}\,,
\end{equation}
where $\Delta\nu$ is the bandwidth and $\rho_c$ is the DM density at the conversion surface. The latter can be estimated from the local
DM density $\rho_{\infty} \simeq 0.3\, {\rm GeV\, cm}^{-3}$ far from the Sun and accounting for the gravitational focusing of the Sun, giving $\rho_{c} \simeq 1\, {\rm GeV\, cm}^{-3}$. The integration is performed over the angular region of the solar atmosphere explored by a specific observation.  
The optical depth $\tau$ takes into account absorption processes. 
The optical depth associated with thermal bremsstrahlung is large in the chromosphere (see Fig.1 of~\cite{Todarello:2023ptf}), implying that the signal is strongly absorbed for masses $\gtrsim 2\times10^{-6}~{\rm eV}.$
Gyro-resonance absorption occurs when $\omega=n\,\Omega_B,$ where $\omega$ is the photon energy, $\Omega_B=eB/m_e \simeq 0.012 \,\mu{\rm eV} B/G$ is the cyclotron frequency and the integer $n$ refers to the different energy levels.
As described in~\cite{Todarello:2023ptf}, the optical depth associated with $n\leq 4$ is strong. In practice, one can conservatively derive an upper limit on the magnetic field at the conversion point in order to avoid strong absorption, see Fig. 1 of~\cite{Todarello:2023ptf}. In fact, for a sufficiently small and monotonically decreasing magnetic field, gyro-resonance absorption (for $n\leq 4$) does not occur after the DM-photon conversion layer.

Finally, radio photons are affected by scattering processes during their propagation in the solar atmosphere, leading to an angular broadening of the emission. For the entire Sun, the effect is modest, while for a point-like source at the solar surface, it has been estimated in~\cite{2019ApJ...884..122K}. For example, at 100 MHz, the Full width at half maximum (FWHM) is 0.1 deg.

\subsubsection{Current limits and future prospects}

\begin{figure}[t!]
    \centering
    \includegraphics[width=\textwidth]{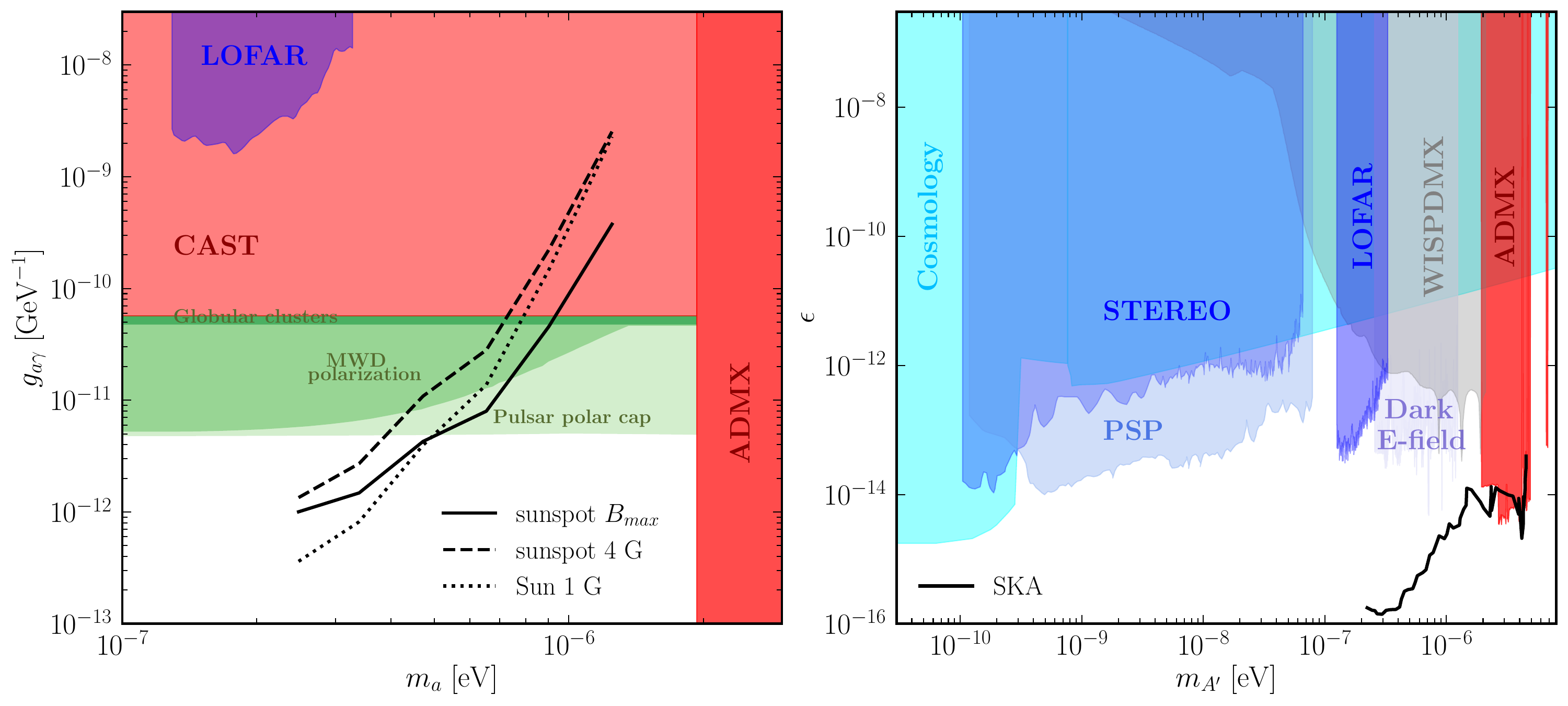}
    \caption{
    Existing limits (colored areas) and projected sensitivities (black lines) on the ALP-photon coupling (left panel) and dark photon kinetic mixing (right panel).
    \emph{Left:} Figure adapted from Ref.~\cite{Todarello:2023ptf}. Forecasts assume 100 hours of observations with SKA-1 Low. The LOFAR bounds are from Ref.~\cite{An:2023wij}.
    \emph{Right:} Limits from STEREO and the Parker Solar Probe (PSP)~\cite{An:2024wmc}, LOFAR~\cite{An:2023wij},
    WISPDMX~\cite{Nguyen:2019xuh}, Dark E-field~\cite{Levine:2024noa} and cosmology~\cite{Arias:2012az,McDermott:2019lch}. SKA sensitivities from Ref.~\cite{An:2023mvf} also assuming 100 hours of observations.
    }
    \label{fig:sensitivity}
\end{figure}

A search for the aforementioned conversion signal with current observations has been performed in~\cite{An:2023wij}, analyzing a 17-minute dataset from the LOFAR radio telescope in the frequency range $30-80$ MHz.
Upper limits have been derived, which in the case of the dark photon lead to the strong constraints shown in Fig.~\ref{fig:sensitivity}, significantly improving existing ones from other probes. For ALPs, the bounds from this search are instead weaker than existing limits from laboratory experiments, see Fig.~\ref{fig:sensitivity}.
Ground-based radio observations at low frequencies are prevented by the reflection of radio waves by the Earth's ionosphere. To circumvent this limitation, Ref.~\cite{An:2024wmc} exploited data from the Parker Solar Probe (PSP), a spacecraft that entered into the low solar corona, and from the STEREO (Solar TErrestrial RElations Observatory) satellite, covering in total the frequency range $\simeq$ 3~kHz$\,-\,$20~MHz.
Competitive upper limits on the dark photon kinetic mixing have been obtained, extending those from LOFAR (Low Frequency Array) observations to lower DM masses, see Fig.~\ref{fig:sensitivity}.

%
%

A huge improvement in the sensitivity of radio observations will be available in the near future with the Square Kilometre Array (SKA), which is currently under construction and will be the largest and most sensitive radio observatory. 
In Fig.~\ref{fig:sensitivity} we present the projected reach to the ALP conversion signal for two targets: the observation of the entire Sun (assuming a magnetic field $B=1$ G) or a sunspot with a size of $4\times10^4$ km. For the latter case, we explore two scenarios: we assume that the sunspot hosts either the maximum magnetic field which avoids strong gyro-resonance absorption, or (more conservatively) a magnetic field of $B=4$ G. 
Details on the derivation of the sensitivities are in~\cite{Todarello:2023ptf}. As shown in Fig.~\ref{fig:sensitivity},
for ALP masses in the range $\sim 2\times10^{-7}-10^{-6}\,{\rm eV}$ future SKA observations can improve over existing laboratory bounds and be competitive with other astrophysical probes, testing currently unexplored regions of the parameter space.
The SKA sensitivity to the dark-photon conversion signal
is shown in Fig.~\ref{fig:sensitivity}, taken from~\cite{An:2023mvf}. The reach on the dark photon kinetic mixing can be improved by several orders of magnitudes with respect to current bounds.

In conclusion, the Sun is a promising target to search for ALP or dark photon radio conversion signals. 
In the case of the dark photon, current searches already provide leading constraints.
In recent years, neutron stars have been extensively studied in the context of ALPs, since their large magnetic fields can strongly enhance photon conversion signals, leading to excellent prospects for detection.
The Sun can be considered as a complementary and likely more robust target, since its properties relevant for the DM signal are much better known than in the case of neutron stars.

%% file: WG3/content/Samuel_Witte.tex
The goal of this Section is to review the recent progress in the field of indirect low-energy searches for axions in the magnetospheres of neutron stars, which has become an increasingly active area of research over the last six years, emerging as a probe which is highly complementary to conventional laboratory searches for axions. 

At face value, the motivation for searching for axions in neutron star magnetospheres appears to be relatively straightforward: these environments host the largest magnetic fields in the Universe (with magnetic field strengths more than ten orders of magnitude in excess of what is typically used in the laboratory), suggesting the local environments may serve to dramatically enhance interactions between axions and electromagnetism. In reality, however, the story is slightly more subtle.  One of the most interesting aspects of this idea stems from the fact that these objects host an ambient plasma with a characteristic density that is optimal for resonantly enhancing the mixing of low-energy axions with radio photons (with the resonance being triggered when there is a degeneracy between the four-momentum of the axion and the four-momentum of the photon, i.e. $k_\mu^\gamma \simeq k_\mu^a$) -- if the plasma density were significantly lower, resonant mixing would produce photons which are absorbed in the interstellar medium (and/or the atmosphere of Earth) before they can be detected, and if the density were significantly higher, the large backgrounds in the sub-mm and infrared frequencies would likely prohibit this from being a useful technique (with axions at yet higher frequencies being severely constrained via other probes). In this sense, these environments appear to have conspired to create an idyllic axion laboratory. The following sections are dedicated to outlining the most promising signatures that are expected to emerge from these systems, and observational efforts that have been made toward demonstrating the viability of these proposals. A summary of the current constraints derived from these techniques, and projected sensitivity to future observational probes is provided in Fig.~\ref{fig:lim_axionNS}.

\subsubsection{Radio lines from axion dark matter}

One of the first proposals for using neutron star magnetospheres to detect axion dark matter dates back to 2007, when Ref.~\cite{Pshirkov:2007st} demonstrated that the presence of axion dark matter in the galaxy should lead to the emergence of a narrow spectral line at radio frequencies that could be detectable with modern telescopes. The paper, however, was before its time, and went largely unnoticed (receiving only around two citations per year) until 2018, when Refs.~\cite{Hook:2018iia,Huang:2018lxq} revived the idea, demonstrating that the next generation radio array SKA may be able to probe regions of parameter space of the QCD axion. These papers emerged at a time when the interest of the astroparticle community was shifting toward low-energy searches (and specifically toward the QCD axion), and thus played an important role in generating interest and facilitating discussions around how these objects could be used to detect axions.

\begin{figure}[t!]
    \centering
    \includegraphics[width=0.85\linewidth]{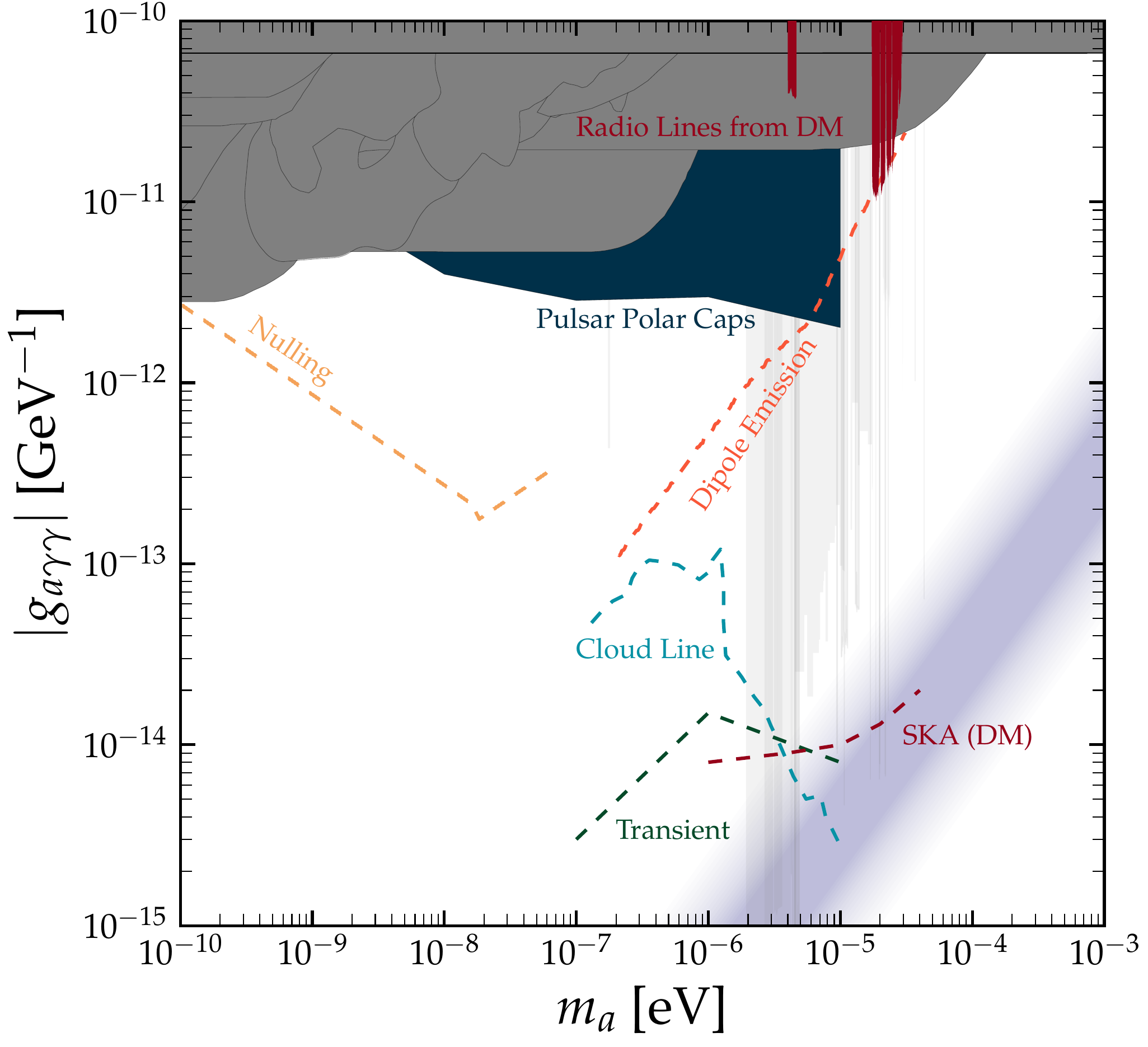}
    \caption{Overview of current constraints derived from searches for radio lines from axion dark matter `DM' (red)~\cite{Foster:2022fxn,Battye:2023oac} and excess radio emission sourced from axions produced in the polar caps of pulsars (dark blue)~\cite{Noordhuis:2022ljw}. Projected sensitivity (using currently available radio telescopes) to pulsar nulling (yellow)~\cite{Caputo:2023cpv}, incoherent dipole emission (orange)~\cite{Caputo:2023cpv}, the spectral end-point produced by the axion cloud (blue)~\cite{Noordhuis:2023wid}, and the transient decay of the axion cloud (green)~\cite{Noordhuis:2023wid}, is shown with dashed lines. Projected SKA sensitivity to spectral lines from axion dark matter is also shown in red (dashed)~\cite{Foster:2022fxn}. Existing constraints are shown in grey, and the purple band roughly demarcates the QCD axion parameter space. }
    \label{fig:lim_axionNS}
\end{figure}

The original proposals of Refs.~\cite{Hook:2018iia,Huang:2018lxq} framed the problem in the following manner. One starts by assuming there exists axion dark matter in the galaxy, 
and that there exists a neutron star which is at rest with respect to the dark matter halo. Ambient axion dark matter will then naturally fall into and out of the gravitational potential of the neutron star -- during this traversal of the magnetosphere, the axion will encounter a spatially varying plasma density, with the plasma frequency near the stellar surface potentially reaching values as large as $\omega_p^{\rm max} \simeq \mathcal{O}({\rm few} \times 10^{-5})$ eV (at least for rapidly rotating stars hosting large magnetic fields). The axion and photon dispersion relations are nearly degenerate when the axion mass and plasma frequency are approximately equal, i.e. $m_a \simeq \omega_p$\footnote{For relativistic axions this resonance condition is shifted by a factor which depends on the relative orientation of the momentum and the magnetic field. For relativistic axions with energies $\omega \simeq \mathcal{O}({\rm meV})$ in extremely strong field environments, one can also induce a secondary resonance that arises from the fact that the photon dispersion relation receives an additional correction from the Euler-Heisenberg four-photon interaction; unfortunately, the observational prospects for detecting such a resonance do not, at present, appear promising~\cite{Long:2024qvd}.} -- at this point, axion-photon mixing is resonantly enhanced, with conversion probabilities even capable of saturating to $\mathcal{O}(1)$ values in some cases~\cite{Tjemsland:2023vvc}. Notice that the variation of $\omega_p$ over many orders of magnitude implies that this resonance will be encountered for any axion with $\omega_p^\infty \lesssim m_a \lesssim \omega_p^{\rm max}$ (with the `asymptotic' plasma frequency $\omega_p^\infty \simeq \mathcal{O}(10^{-12})$ eV for typical pulsars in the Galaxy) -- in this sense, the magnetosphere acts like a broadband resonator.
Photons are then assumed to escape the gravitational potential radially with no absorption or energy broadening (thereby retaining the initially narrow asymptotic energy distribution of the in-falling axions). The resultant signal was argued to be an extremely narrow $\delta f / f \sim \mathcal{O}(10^{-6})$ radio line centered on the axion mass\footnote{Ref.~\cite{Todarello:2023xuf} has also proposed looking for narrow radio lines (with width $\delta f / f \sim \mathcal{O}(10^{-3})$)  near pulsars produced from the stimulated decay of local axion dark matter. In general, however, decay searches favor large dark matter column densities which are most naturally realized in large-scale environments (see e.g.~\cite{Caputo:2018ljp,Caputo:2018vmy,Ghosh:2020hgd,Buen-Abad:2021qvj,Sun:2021oqp,Dev:2023ijb,Sun:2023gic,Roach:2022lgo,Foster:2021ngm,Roy:2023omw,Janish:2023kvi,Calore:2022pks}). Pulsar-targeted decay searches are unlikely to probe new parameter space in the foreseeable future.}. The original studies focused on targeting isolated neutron stars, but one can also point telescopes in regions with dense neutron star populations, such as globular clusters or the galactic center, and search for the collective axion signal arising from the full population -- this signal instead manifests as an `axion forest', comprised of many spectral lines centered about the axion mass and with a cumulative width fixed by the Doppler broadening of the stellar population~\cite{Safdi:2018oeu,Foster:2022fxn,Bhura:2024jjt} (note that such a signal also offers a wealth of information in the time domain, see e.g. ~\cite{Foster:2022fxn}). This technique offers a way to marginalize over unknown systematics associated to individual objects, and enhances the net signal relative to what could be achieved with any signal object.

The prospect of detecting the QCD axion (and of having a distinctive signature that directly provides the axion mass, thereby allowing terrestrial searches to perform targeted follow-ups) brought this idea to the forefront of the astroparticle community; however, the large number of assumptions present in these analyses (e.g. perfectly radial in-falling/out-going axion/photon phase space distributions, a simplified one-dimensional treatment of axion-photon mixing, perfect preservation of the asymptotic axion energy distribution, magnetospheres described by force-free dipolar field configurations, no photon absorption, etc.) raised important questions about the extent to which the projected sensitivity was valid. Early follow-up work demonstrated the importance of appropriately tracking the axion and photon phase space (rather than assuming fully radial trajectories), illustrating how this could be naturally accomplished using techniques from geometric ray tracing~\cite{Leroy:2019ghm,Witte:2021arp,Battye:2021xvt,McDonald:2023shx,Tjemsland:2023vvc,Satherley:2024wsz}. Ray tracing also provides a natural way to accurately trace energy exchange between the photon and plasma, which occurs shortly after photon production and can induce a broadening of the spectral line at the order of magnitude level~\cite{Witte:2021arp,Battye:2021xvt}, local absorption processes~\cite{Witte:2021arp}, and the time evolution of the spectral line (which is intrinsically periodic on the timescale of the pulsar period, but with a time variance which strongly depends on e.g. the pulsar, axion mass, and viewing angle)~\cite{Witte:2021arp,Battye:2021xvt,Battye:2023oac}. The question of axion-photon mixing also proved to be an interesting topic of debate, with a number of groups calling into question the validity of the one-dimensional approximation~\cite{Battye:2019aco,Witte:2021arp,Millar:2021gzs}. In the last year, two independent analytic methods ~\cite{McDonald:2024uuh,McDonald:2023ohd} have yielded convergent results, and these expressions appear to be in excellent agreement with direct numerical simulations of the classical equations~\cite{Gines:2024ekm}, suggesting this may be now a resolved issue. 

Enormous progress has been made in the last five years; nevertheless, there still exist open questions which need to be addressed, the most notable of which is the question of how plasma is distributed in the closed zones of active pulsars \footnote{Open field lines are negligibly small in the region of interest, and plasma distributions in the near-field regime of dead pulsars are reasonably well understood (see e.g.~\cite{krause1985electrosphere,Spitkovsky:2003um}). It is worth mentioning that one concern which is often raised is the question of whether non-dipolar magnetic fields appearing near the neutron star could significantly alter the signal -- the answer is that neglecting higher order multipoles is always guaranteed to reflect a conservative treatment, so long as the signal is not dominated by one (or a very small number) single object.}. While one does not expect strong deviations from the baseline assumption of a Goldreich-Julian (GJ) plasma in standard pulsars, obtaining a direct probe of charge densities in this regime is observationally challenging. An important exception applies to the case of magetnars, which are expected to have localized twists in the magnetic field lines which can sustain enhanced charge/current densities (such an effect would allow for probes of high frequency signals, but would likely suppress the emission at standard radio frequencies, see ~\cite{McDonald:2023shx}). Ref.~\cite{Roy:2025mqw} has recently performed an in-depth analysis of the two leading magnetar models, demonstrating that the expectations are highly sensitive to the location of the twisted field lines and the state of the local plasma near the surface~\cite{Roy:2025mqw}.

Numerous efforts have been made to detect such radio lines observationally\footnote{Note that similar ideas have also been applied to search for dark photons~\cite{Hardy:2022ufh}.
}. Early attempts relied upon simplified analytic approximations of the expected radiated power and line properties, which tend to result in an overestimation of the signal strength~\cite{Darling:2020plz,Darling:2020uyo,Foster:2020pgt,Zhou:2022yxp}\footnote{Refs.~\cite{Darling:2020plz,Darling:2020uyo} also suffered from an incorrect translation between frequency and energy. }; moreover, many of these further relied upon observations of the Galactic Center magnetar~\cite{Darling:2020plz,Darling:2020uyo,Battye:2021yue}, where the assumption the GJ charge density is expected to be significantly less reliable. Only two attempts to date have applied state-of-the-art numerical techniques to more conventional systems; this includes Ref.~\cite{Foster:2022fxn}\footnote{It is worth noting that the fiducial constraint derived in ~\cite{Foster:2022fxn} is likely significantly underestimated. This limit was derived before a firm understanding of the axion-photon conversion probability was obtained (see discussion above), and thus the authors chose to artificially suppress photon production so as to avoid any potential over-estimation of the derived limit.}, which searched for the collective population of radio lines coming from the neutron star population in the Galactic Center, and Ref.~\cite{Battye:2023oac}, which searched for a time-oscillating line in a nearby pulsar. While these constraints are currently competitive with terrestrial and astrophysical searches (see red regions of  Fig.~\ref{fig:lim_axionNS}), the true power of this technique will come with the arrival of SKA.

It is worth mentioning that the above discussion has been framed under the assumption that axion dark matter is smoothly distributed in the Universe; this may or may not be a reasonable assumption, depending on the production channel of axion dark matter, the evolutionary dynamics in the early Universe, and the position of the pulsar (or pulsar population) in the  Galaxy. For example, in the post-inflationary scenario, axions are expected to form small-scale gravitationally bound objects such as axion miniclusters and axion stars. These objects will undergo significant tidal disruption via stellar encounters~\cite{Kavanagh:2020gcy,Dandoy:2022prp,Shen:2022ltx,OHare:2023rtm}, implying the axion distribution in the center of galaxies is likely smooth, while at larger radii (e.g. $r \geq \mathcal{O}(5)$ kpc) axions are more likely to be found in confined self-bound objects (again, assuming post-inflationary axion production). 

Various groups have suggested that rare transient encounters between such self-bound objects and neutron stars might give rise to more explosive transient signatures at radio frequencies~\cite{Iwazaki:2014wka,Bai:2017feq,Dietrich:2018jov,Buckley:2020fmh,Edwards:2019tzf,Edwards:2020afl,Prabhu:2020yif,Nurmi:2021xds,Bai:2021nrs,Witte:2022cjj,Kouvaris:2022guf,Maseizik:2024qly} (with some early works even suggesting similarities between these events and fast radio bursts~\cite{Iwazaki:2014wka,Buckley:2020fmh}  -- although recent observational advances in the fast radio burst community~\cite{Zhang:2020qgp,Zhang:2022uzl}, and improved estimations of the transient encounter rate~\cite{Maseizik:2024qly}, are strongly suggestive that this is not viable). Such an event, should it occur, would surely give rise to striking signatures (see e.g.~\cite{Witte:2022cjj}). Early estimations of the minicluster-neutron star encounter rate suggested the encounter rate in the Milky Way could be sufficiently large to probe the QCD axion~\cite{Edwards:2020afl}; the transient rate estimation, however, was severely over estimated by orders of magnitude\footnote{This over estimation arose from the fact that Ref.~\cite{Edwards:2020afl} adopted neutron star properties drawn from fitted distributions at birth (which are inferred using population synthesis). Neutron star spin down and magnetic field decay dramatically alter this distribution for neutron stars older than $\tau \gtrsim 1$ Myr (i.e. $\sim 99.998 \%$ of the population). More specifically, the plasma density in older neutron stars is lower, implying these objects often do not allow for resonant mixing -- this problem is further exacerbated by the fact that miniclusters and axion stars preferentially form in the post-inflationary scenario where the axion mass is expected to be $m_a \gtrsim 26$~ $\mu$eV, implying an above average plasma density is needed for resonant mixing.}, and thus in the absence of a severe amount of extremely exotic new physics, transient encounters are likely sufficiently rare to make them effectively un-observable~\cite{Maseizik:2024qly} .

\subsubsection{Local production of axions from pair cascades}
The ideas discussed in the preceding section intrinsically assume that axions comprise a sizable fraction of dark matter. A complementary idea emerged in 2021 which focused instead on identifying a novel mechanism which leads to the unavoidable production of light axions~\cite{Prabhu:2021zve}\footnote{This idea is somewhat related to an earlier idea outlined in~\cite{Garbrecht:2018akc}, which focused on axion production from neutron star electromagnetic fields in vacuum. Neutron stars, however, are known to be surrounded by a dense plasma capable of efficiently screening large-scale electric fields, and thus `axion configurations' discussed in Ref.~\cite{Garbrecht:2018akc} are more representative of an intriguing thought experiment.}, effectively removing any assumption about the origin of these particles (and their contribution to the dark matter abundance). The idea here is that pulsar magnetospheres inherently support small regions with dynamically evolving $(\vec{E} \cdot \vec{B})$ fields\footnote{The need for $(\vec{E} \cdot \vec{B}) \neq 0$ can be seen from multiple perspectives. Observationally, we see radio emission from the inner magnetosphere -- emission requires acceleration, acceleration requires electric fields, and in strong $\vec{B}$ environments particles are forced to move along field lines themselves. From a theory perspective, one can see that the large-scale twist of the open field lines demands a current, and the charges in this current cannot locally screen everywhere the induced electric field. The requirement that $(\vec{E} \cdot \vec{B})$ is dynamical instead stems from the fact that the induced electric field is unstable to pair production, which in turn generates a quasi-periodic screening.} -- since $(\vec{E} \cdot \vec{B})$ arises as a source term in the axion equation of motion, the dynamical evolution of these fields leads to local axion production. This can effectively be thought of as an indirect coupling of the axion to the rotational energy of the pulsar, where a small amount of the energy budget is stolen by the axion field (rather than being dumped into electromagnetic radiation). For large, but unconstrained, axion-photon couplings, this production mechanism can source a local axion population that is far more numerous than what one would expect from axion dark matter. As these locally-sourced axions traverse the  magnetosphere, they can resonantly mix with low-energy photons, producing a broadband radio flux (which appears as an excess on the intrinsic radio emission of the pulsar itself).

The initial work of Ref.~\cite{Prabhu:2021zve} demonstrated that this approach could prove to be quite promising across a broad range of axion masses, but it relied on a number of simplified approximations, including a one-dimensional analytic model of the gap dynamics, and an approximate low-dimensional treatment of the axion/photon propagation and mixing. This work was soon followed-up by Ref.~\cite{Noordhuis:2022ljw}, which significantly improved the modeling of the axion spectrum by: $(1)$ developing a 2+1 dimensional semi-analytic model of the gap dynamics, and $(2)$ re-adapting a numerical particle-in-cell code which directly simulates the non-linear evolution of the dynamical screening of $(\vec{E} \cdot \vec{B})$ in the polar cap. Ref.~\cite{Noordhuis:2022ljw} coupled this to a relativistic ray tracing algorithm which could then carefully track the three dimensional evolution of the axion/photon trajectories and the mixing, thereby generating quantitative estimates of the radio emission. By ensuring the period-averaged radio emission produced from locally sourced axions does not exceed the on-pulse radio emission observed in nearby pulsars, Ref.~\cite{Noordhuis:2022ljw} set extremely stringent limits on the axion-photon coupling across a wide range of parameter space (see blue region of Fig.~\ref{fig:lim_axionNS}). Despite appearing to be a complex problem, these limits are, perhaps surprisingly, stable against many modeling uncertainties -- this is partially because the limit scales with $g_{a\gamma}^4$, and partially because the fundamental scales at which axion production is most efficient are dictated by linear, rather than non-linear, electrodynamics~\cite{Noordhuis:2022ljw}.

It was recently pointed out that if the axion mass happens to lie in, or slightly below, the radio band, a non-negligible fraction of the energy density produced from the polar cap cascades will go into gravitationally bound axions. Since axions interact very feebly, they can accumulate on long (astrophysical) timescales, forming the so-called axion clouds~\cite{Noordhuis:2023wid,Witte:2024akb}. The formation of these objects is quite generic, with typical densities exceeding the local dark matter density by (in some cases) more than 20 orders of magnitude. Various observational signatures of these clouds have been identified, including $\mathcal{O{\%}}$-level spectral lines~\cite{Noordhuis:2023wid,Caputo:2023cpv}, a short-lived quasi-periodic nulling of the intrinsic radio emission generated by the pulsar itself (occurring on $\sim \mu$s-ns timescales)~\cite{Caputo:2023cpv}, and highly-time variable radio emission~\cite{Noordhuis:2023wid}. No search for these signals has yet been performed, but the prospects are extremely promising (see  Fig.~\ref{fig:lim_axionNS}, and also~\cite{Berghaus:2025kvn}), potentially even allowing for the detection of the QCD axion in some regions of parameter space.

A related, but alternative, idea to the one discussed above is to exploit the variational timescale of the pulsar itself (rather than the small-scale time evolution generated by the dynamical screening). This idea was proposed in Ref.~\cite{Khelashvili:2024sup}, however the detection prospects are not as promising as those discussed above.

%% file: WG3/content/Calore_Eckner.tex
\subsection{Introduction}
The local universe is populated by a large variety of astrophysical objects, mostly hosted by galaxies and clusters of galaxies. 
These objects include Active Galactic Nuclei (AGN), SNe, star-forming galaxies, and galaxy clusters. 
The environments of these objects are rich in magnetic fields and energetic particles, making them ideal candidates for studying axion(-like) 
particles and other exotica. In addition to discrete sources, the universe is permeated by several cosmic photon backgrounds
 that provide further avenues to explore axion-related physics.

\subsection{A condensate of high-energy astrophysical sources}
The observation of individual, multi-wavelength emitting, astrophysical objects helps us understand the acceleration mechanisms in a variety of extreme environments, as well as in building population models for these emitters.
Here, we review the main classes of high-energy astrophysical sources, highlighting the characteristics that make them suitable for studying axions and the like.

{\bf Active galactic nuclei.} AGN are powered by supermassive black holes that accrete matter from their surroundings, for a review we refer the reader to~\cite{2008NewAR..52..227T}. 
A significant fraction of the total luminosity of an AGN is non-thermal, resulting from two key components:
(i) Emission from the accretion disc surrounding the black hole, and 
(ii) emission from highly collimated relativistic jets extending from sub-parsec to kiloparsec scales.
The magnetic fields within AGN jets have a complex structure, consisting of both poloidal (aligned with the jet axis) and toroidal (perpendicular to the axis) components~\cite{2021Galax...9...58G}. 
The strength of the magnetic field in these jets can reach $B \sim 10^3$ G. 
The coherence length of these fields ranges from sub-parsec near the central black hole to 
several parsecs along the jet's extension. These magnetic fields are essential for synchrotron radiation and could play a role in axion-photon interactions through the (inverse) Primakoff effect.

{\bf Supernovae and their remnants.}
SNe are violent explosions marking the end of a star's life, see e.g.~\cite{Orlando:2023alz}. They come in two main types:
(i) Type I SNe result from the accretion of matter onto a white dwarf in a binary system, leading to a thermonuclear runaway, while (ii) Type II SNe occur from the gravitational collapse of the core of a
massive star ($M \gtrsim 8 \, M_{\odot}$).
The explosion expels the outer layers of the star, forming a supernova remnant (SNR) that expands into and shocks the surrounding interstellar medium (ISM). 
The magnetic fields in these remnants reach strengths ranging from $25 \, \mu$G to $1000 \, \mu$G~\cite{2012SSRv..166..231R}. Supernovae are powerful factories of axions and alike, cf.~Sec.~\ref{subsec:Raffelt}. Besides, shocks in SNRs are ideal sites
 for cosmic-ray acceleration via diffusive shock acceleration, and the magnetic fields also provide a medium for potential axion-photon conversion.

{\bf Star-forming galaxies.}
Star-forming galaxies are characterised by high rates of star formation, particularly in regions known as star-forming or starburst regions. 
These regions are sites of intense stellar activity and can host large-scale magnetic fields ranging from a few $\mu$G to hundreds of $\mu$G, see e.g.~\cite{2006ApJ...645..186T}. The strong 
magnetic fields are thought to be generated by dynamo processes, and they influence the transport of cosmic rays as well as photon-axion mixing in these environments.

{\bf Galaxy clusters.}
Galaxy clusters are the largest gravitationally bound structures in the universe, containing hundreds to thousands of galaxies. 
The intracluster medium (ICM) is composed of hot, X-ray-emitting, highly ionized gas with temperatures reaching $T \sim 10^7$ to $10^8$ K. 
The magnetic fields in galaxy clusters are relatively weak, typically ranging from $0.1 \, \mu$G to a few $\mu$G, for a review see~\cite{Govoni:2004as}.
These magnetic fields are often chaotic and tangled due to the turbulent nature of the ICM, although large-scale ordered fields can also exist. 
The coherence lengths of magnetic fields in galaxy clusters range from 10 to 100 kpc. 

Besides individually resolved objects, the universe is also filled with various cosmic photon backgrounds, originating from different astrophysical processes. 
These backgrounds are mostly of extragalactic origin, formed from a superposition of faint photon emitters or truly diffuse processes -- mostly from cosmic-ray interactions. The cosmic photon background extends from radio wavelengths up to $\mathcal{O}(100)$ GeV. The main components at low energies are the Cosmic Microwave Background (CMB), a relic of the Big Bang, with a temperature of 2.7255 K after the subtraction of Galactic foregrounds, the 
Cosmic Infrared Background (CIB), which represents the emission from dust heated by stars in unresolved galaxies,
the Cosmic Optical Background (COB), mainly due by starlight from galaxies and difficult to clean from Galactic contamination, and the 
Cosmic Ultraviolet Background (CUB), which originates from ionizing sources such as star-forming galaxies and quasars.
At the highest energies, the Cosmic X-ray Background (CXB) is believed to be dominated by bremsstrahlung radiation from hot accretion discs around AGN, and the Cosmic Gamma-ray Background (CGB) originates from the superposition of large populations of AGNs, star-forming galaxies, and galactic sources. For a dedicated review of these cosmic photon backgrounds, we refer the reader to~\cite{Hill:2018trh}.
Each of these backgrounds offers a unique probe of the large-scale structure of the universe and the potential interactions between photons and axions in extragalactic environments.

At higher energies, diffuse gamma-ray emission has been detected by ground-based telescopes such as the High Energy Spectroscopic System (H.E.S.S.)~\cite{HESS:2014ree, HESS:2017tce}, the High-Altitude Water Cherenkov Gamma-Ray Observatory (HAWC)~\cite{HAWC:2021bvb, HAWC:2023wdq}, the Tibet AS$\gamma$ experiment~\cite{2021PhRvL.126n1101A}, and the Large High Altitude Air Shower Observatory (LHAASO)~\cite{LHAASO:2023gne, LHAASO:2024lnz}.
This emission is primarily of Galactic origin, constrained by the horizon of very high-energy (VHE) photons, which are absorbed through pair production interactions with the extragalactic background light (EBL), made up
by the CIB, COB, and CUB~\cite{Franceschini:2021wkr}. 

\subsection{Fundamentals of ALP-photon mixing}
The physics of photon-ALP mixing, governed by the Primakoff process, can be derived from the Lagrangian~\cite{Raffelt:1987im}
\begin{equation}
\label{eq:ALP_lagrangian}
\mathcal{L}_{a\gamma}=-\frac{1}{4} g_{a\gamma} F_{\mu\nu}\tilde{F}^{\mu\nu}a=g_{a\gamma} {\bf E}\cdot{\bf B}\,a \, ,
\end{equation}
leading to the well-known conversion of ALPs into photons in the presence of an external magnetic field ${\bf B}$.
We summarise here the key equations and components necessary to compute the probability of photon-ALP conversions. For comprehensive discussions on the theoretical and mathematical aspects, we refer to~\cite{Raffelt:1987im, DeAngelis:2007dqd, Mirizzi:2009aj, DeAngelis:2011id, Meyer:2014epa, Kartavtsev:2016doq}.

Photon-ALP conversion requires an external magnetic field $\bf{B}$ with a non-zero transverse component $\bf{B}_{\perp}$ relative to the propagation direction. Assuming the initial state propagates along $\bf{\hat{z}}$ and $\bf{B}_{\perp}$ is expressed as $B(\cos\theta, \sin\theta, 0)^T$, where $\theta$ is the angle between $\bf{B}_{\perp}$ and the $\bf{\hat{x}}$-axis, the photon polarisation states $A_x$ and $A_y$ combine to form $A_{\perp}$ (perpendicular to $\bf{B}_{\perp}$) and $A_{\parallel}$ (parallel to $\bf{B}_{\perp}$). Only $A_{\parallel}$ can convert into ALP states.

The evolution of a photon-ALP state propagating along $\bf{\hat{z}}$ is described by~\cite{Raffelt:1987im, Kartavtsev:2016doq}:
\begin{equation}
\label{eq:ALP-photon-prop-pure}
i\frac{\mathrm{d}\mathcal{A}}{\mathrm{d} z} = \left(\mathcal{H}_{\mathrm{dis}} - \frac{i}{2} \mathcal{H}_{\mathrm{abs}}\right) \mathcal{A},
\end{equation}
where $\mathcal{A} = \left(A_{\perp}, A_{\parallel}, a\right)^T$ represents the photon polarisation states and the ALP state $a$. 
The dispersion Hamiltonian $\mathcal{H}_{\mathrm{dis}}$ accounts for photon-ALP mixing and is expressed as~\cite{Raffelt:1987im, Mirizzi:2009aj}:
\begin{equation}
\label{eq:mixing_matrix}
\mathcal{H}_{\mathrm{dis}} = \begin{pmatrix}
\Delta_{\perp} & 0 & 0 \\
0 & \Delta_{\parallel} & \Delta_{a\gamma} \\
0 & \Delta_{a\gamma} & \Delta_{a}
\end{pmatrix},
\end{equation}
with components defined as~\cite{Mirizzi:2009aj, Kartavtsev:2016doq}:
\begin{align}
\Delta_{\perp} &= \Delta_{\mathrm{pl}} + 2\Delta_{\mathrm{B}} + \Delta_{\gamma\gamma}, \\
\Delta_{\parallel} &= \Delta_{\mathrm{pl}} + \frac{7}{2}\Delta_{\mathrm{B}} + \Delta_{\gamma\gamma}, \\
\Delta_{a} &= -\frac{m_a^2}{2\omega}, \\
\Delta_{a\gamma} &= \frac{g_{a\gamma}}{2}B.
\end{align}
Here, $\Delta_{\mathrm{pl}}$, $\Delta_{\mathrm{B}}$, and $\Delta_{\gamma\gamma}$ describe dispersion due to plasma density, magnetic field, and photon-photon interactions, respectively~\cite{Dobrynina:2014qba}.
The absorption Hamiltonian $\mathcal{H}_{\mathrm{abs}}$ models photon absorption effects and is given by~\cite{Kartavtsev:2016doq}:
\begin{equation}
\label{eq:absorption_matrix}
\mathcal{H}_{\mathrm{abs}} = \begin{pmatrix}
\Gamma & 0 & 0 \\
0 & \Gamma & 0 \\
0 & 0 & 0
\end{pmatrix},
\end{equation}
where $\Gamma$ quantifies photon-photon absorption strength~\cite{Mirizzi:2009aj}. The importance of the absorption term depends on the particular environment's conditions, and it can describe e.g.~Galactic absorption and/or absorption on the CMB and EBL.

\begin{figure}[t!]
    \centering
    \includegraphics[width=0.8\linewidth]{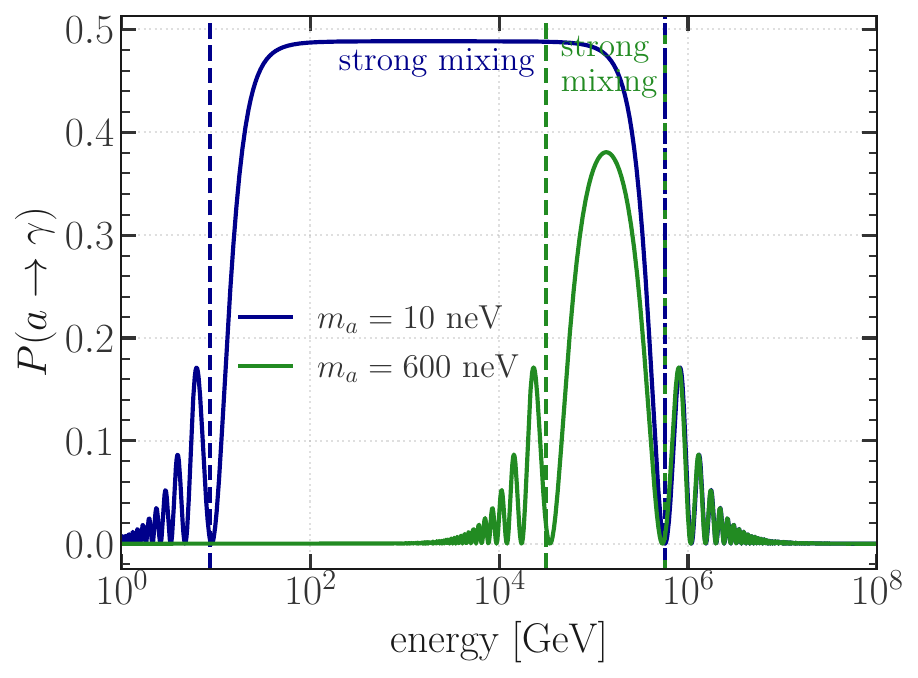}
    \caption{Dependence of the ALP-photon conversion probability $P(a \rightarrow \gamma)$ on the ALP mass. We display the cases $m_a = 10$ neV (blue) and $m_a = 600$ neV (green) for a coupling strength of $g_{a\gamma} = 3\times10^{-11}$ GeV$^{-1}$. The dashed vertical lines mark the onset and end of the strong mixing regime with energy. Note that the upper boundary of this regime is the same for both ALP masses. The conversion probability was calculated for a single domain featuring a coherent magnetic field with parameters as defined in the tutorial of the ALP propagation software \texttt{gammaALPs} \cite{Meyer:2021pbp}. }
    \label{fig:pag_oscillation_with_ALPmass}
\end{figure}

In a homogeneous or slowly varying $B$-field, a photon beam
develops a coherent axion component.
In the limit of a purely polarised photon beam in a single magnetic domain $L$ with a coherent $B$-field, the propagation equations reduce to a 2-dimensional problem, where one can define~\cite{Mirizzi:2009aj}:
\begin{equation}
    \frac{1}{2} \tan( 2\theta )= \frac{\Delta_{a \gamma}}{\Delta_{\parallel} - \Delta_a }  \, ,
\end{equation}
as the angle that diagonalises the mixing matrix. 
In analogy with neutrino oscillations then, the probability for a purely polarised photon beam to oscillate into an ALP after a distance $L$ 
can be written as:
\begin{equation}
    P(a \rightarrow \gamma) =\sin^2(2 \theta) \sin^2 (\Delta_{\rm osc} L /2 )    \, , 
\end{equation}
where $\Delta_{\rm osc} \equiv [(\Delta_\parallel - \Delta_a)^2 + 4 \Delta_{a\gamma}^2)]^{1/2}$.
Therefore, the mixing is maximum when $\Delta_{a\gamma} \gg \Delta_\parallel - \Delta_a$ (strong mixing regime). On the other hand, the probability is suppressed by plasma or CMB effects when $\Delta_{a\gamma} \ll \Delta_\parallel - \Delta_a$.
In the limit of massless ALPs, sufficiently low couplings and low energies, the probability reduces to: 
\begin{equation}
    P(a \rightarrow \gamma) = \Delta_{a \gamma}^2 \frac{\sin^2 (\Delta_{\rm osc} L /2 )}{(\Delta_{\rm osc} L /2 )^2}    \, .
\end{equation}
The oscillation length is $L_{\rm osc} = 2\pi/\Delta_{\rm osc}$, with coherent oscillations realised in the case $L \ll L_{\rm osc}$.
In this case, the plasma term dominates and one can express: 
\begin{equation}
    \Delta_{\rm osc} = \left[\frac{(m_a^2 - \omega_{\rm pl}^2)^2}{4 \omega^2} + (g_{a\gamma} B)^2 \right]^{1/2} \, ,
\end{equation}
with $\omega_{\rm pl}^2 \equiv (4 \pi \alpha_{\rm em} n_e)/m_e$ depending on the 
free electron density $n_e$ in the environment under study.
One can then write:
\begin{equation}
    \Delta_{\rm osc} = 2 \Delta_{a\gamma} \sqrt{1 + \frac{\omega_c^2}{\omega^2}} \, ,
\end{equation}
where we have defined the critical energy, $\omega_c$:
\begin{equation}
    \omega_c = 2.5\,   \mathrm{GeV} \Bigl( \frac{m_{\it a}^2 - \omega_{\mathrm{pl}}^2}{1 \, \rm neV^2} \Bigr) \Bigl(\frac{1 \, \rm \mu G}{B}\Bigr) \Bigl(\frac{10^{-11} \, \rm GeV^{-1}}{g_{{\it a}\gamma}}\Bigr) \, .
\end{equation}
Depending on the ALP mass and free electron density, the critical energy varies, and it is equivalent to $\sim$ GeV for neV masses and $\sim$ keV for $10^{-12}$ eV masses.
The critical energy defines two main regimes: (a) the strong mixing regime ($\omega \gg \omega_c$), where the conversion probability is energy independent and it is approximated as $P(a \rightarrow \gamma)\propto g_{a\gamma}^2 B^2 L^2$; and the (b) oscillation regime ($\omega \simeq \omega_c$) where the conversion probability displays an oscillatory, energy-dependent, behavior, which can imprint specific patterns in the energy distribution of gamma-ray sources. 
In Fig.~\ref{fig:pag_oscillation_with_ALPmass}, we display the conversion probability over a wide range in energy to exemplify the dependence of its oscillatory pattern on the ALP mass (here: $m_a = 10$ neV and 600 neV) at a fixed value of $g_{a\gamma} = 3\times10^{-11}$ GeV$^{-1}$.
For a study of the conversion probability as a function of ALP mass, we refer the reader to~\cite{Calore:2023srn}.

For realistic scenarios, the density matrix formalism provides a better description. The density matrix $\bf{\rho}$, defined as $\bf{\rho} = \mathcal{A}\otimes\mathcal{A}^{\dagger}$, evolves as~\cite{Mirizzi:2009aj, Kartavtsev:2016doq}:
\begin{equation}
\label{eq:ALP-photon-prop-density}
i\frac{\mathrm{d}\bf{\rho}}{\mathrm{d} z} = \left[ \mathcal{H}_{\mathrm{dis}}, \bf{\rho} \right] - \frac{i}{2} \{\mathcal{H}_{\mathrm{abs}}, \bf{\rho}\}.
\end{equation}

\subsection{Main observables for ALPs}
The search for ALPs in astrophysical environments relies on key observables that can provide indirect evidence for their existence. 
These observables stem from both individual sources and large-scale diffuse emissions, each offering unique insights into the interaction of axions with photons in the presence of magnetic fields.

One of the most important observables in astrophysics is the spectral energy distribution (SED) of individual sources. 
The SED represents the energy output of a source across different wavelengths or photon energies, and it is typically measured in $\rm erg/cm^2/s$. 
It provides detailed information about the physical processes occurring within the source, as for example
synchrotron radiation, inverse Compton scattering, and bremsstrahlung emission. 
In the context of axion and ALP searches, deviations from the expected, astrophysical, SED can indicate the presence of axion-photon conversion. 

Complementary to the SED is the time-dependence of the photon flux, where variations in the photon output over time can also signal potential axion-photon conversions. 
Such time-dependent observations are crucial for identifying transient or variable phenomena that may be linked to axions. 

Another key observable is the diffuse large-scale flux, which provides spatial information about photon emission across different regions of the sky -- the energy flux is in this case averaged over a given region of interest, carrying an additional $1/\rm sr$. 
This flux arises from a variety of unresolved sources or truly diffuse processes and forms the cosmic radiation background described above.
By studying the diffuse emission at different positions in the sky and its energy dependence, it is possible to extract information about the distribution of photon sources and their interaction with cosmic magnetic fields. 

The combination of spectral, temporal, and spatial observables allows for a comprehensive approach to axion searches. SEDs provide detailed insight into individual sources, while large-scale diffuse flux 
measurements offer a broader view of photon behavior across the universe. Together, they form a powerful set of tools for probing the existence of axions and ALPs in a variety of astrophysical environments.

\subsection{Production of ALPs and rates}

Axions and alike can be efficiently produced in various astrophysical systems, either through direct production mechanisms in stellar cores or through the interaction of photons with magnetic fields, which induces photon-ALP conversion. Both processes are relevant in extragalactic environments where strong magnetic fields and energetic particle emissions are present. We review here the main production mechanisms in high-energy astrophysics. 

\subsubsection{Direct production in stars}
ALPs can be efficiently produced in stellar cores, particularly in core-collapse SNe. The dominant production process in such environments is the Primakoff effect, assuming that the only relevant coupling is the photon-ALP coupling and the ALP mass is lower than the stellar temperature (see Sec.~\ref{subsec:Raffelt} and Sec.~\ref{subsec:Vitagliano} for a more detailed discussion on the ALP production in the SN core). In this process, thermal photons scatter off charged particles in the plasma, converting them into ALPs in the presence of strong electromagnetic fields. \\
The production of ALPs is not limited to SNe. Other stellar environments, from main sequence stars~\cite{Nguyen:2023czp} to neutron stars (NS), can be laboratories for ALPs and other light particles, cf. Sec~\ref{subsec:Carenza2}. The cumulative emission from all stars in external galaxies, and especially starburst ones like M82 and M87, can fuel significant ALP rates~\cite{Ning:2024eky}. The dense stellar populations and active star-forming regions in these galaxies provide conditions for the production of ALPs via similar mechanisms. For a recent review about the production of ALPs in stellar systems also considering couplings with electrons and nucleons, we refer the reader to \cite{Carenza:2024ehj}.

\subsubsection{ALP production in neutron star mergers}
Another key extragalactic event that can lead to ALP production is the merger of NSs. In these highly energetic events, strong magnetic fields and extreme conditions can give rise to ALP production. Neutron star mergers produce a burst of electromagnetic radiation, and in the presence of the intense magnetic fields generated during the merger, photons may convert into ALPs and other light particles~\cite{Diamond:2021ekg, Fiorillo:2022piv}. These mergers also create an environment rich in energetic particles, further enhancing ALP production rates.
The observational signatures of ALP production in neutron star mergers are an active area of research. They could manifest as anomalies in the electromagnetic spectrum, or even in gravitational wave signals when correlated with gamma-ray bursts~\cite{Dev:2023hax,Diamond:2023cto}.

\subsubsection{In-situ photon-ALP conversion}
\begin{figure}[t!]
    \centering
    \includegraphics[width=0.8\linewidth]{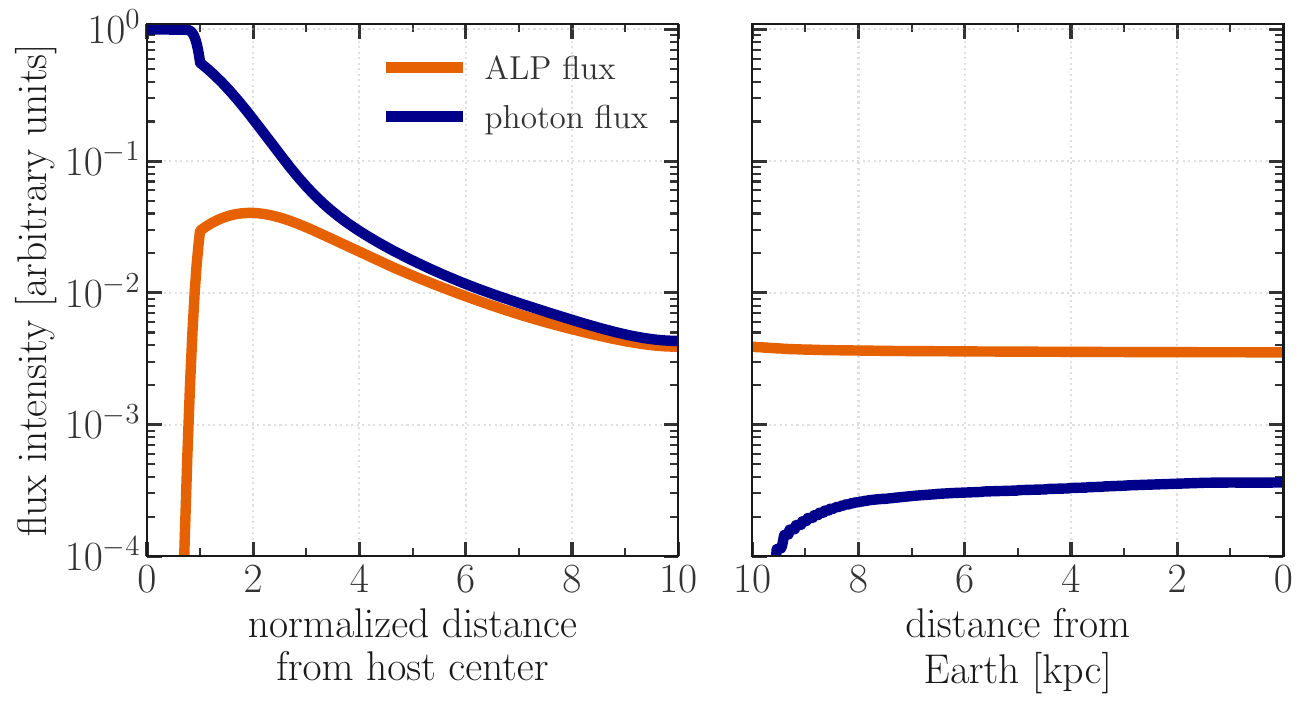}
    \caption{Schematic illustration of the evolution of the photon (blue) and ALP (orange) fluxes as a function of the distance traveled within either a distant extragalactic host source (left panel) or the Milky Way (right panel). The displayed photon flux has been normalized to one, and the ALP flux scales accordingly. The distance in the left panel is stated in units of the coherence length of the magnetic field of the host. We also exemplify the impact of photon attenuation on EBL and CMB in the left panel leading to a reduction of the overall gamma-ray flux beyond the part being converted into ALPs. Both the photon and ALP fluxes were computed via the ALP propagation software \texttt{gammaALPs}. }
    \label{fig:conversion_prob}
\end{figure}

Extragalactic gamma-ray emitters, such as AGN, SNR, and star-forming galaxies, are ideal candidates for in-situ photon-ALP conversion due to their strong magnetic fields and the high-energy photons they emit. In these systems, particle acceleration processes, such as hadronic interactions (e.g., $pp$ and $p\gamma$ collisions) or leptonic mechanisms (e.g., inverse Compton scattering), produce high-energy photon spectra. These photons can then convert into ALPs in the presence of strong magnetic fields through the Primakoff effect, and leave an imprint in the individual source SED and in diffuse backgrounds~\cite{Vogel:2017fmc}. In the left panel of Fig.~\ref{fig:conversion_prob}, we schematically illustrate how the initial gamma-ray flux of a distant extragalactic host source is reduced through either attenuation on the CMB or EBL and conversion into axions or alike via the Primakoff effect. The distance from the host's center is expressed in units of the coherence length of the surrounding magnetic field. As exemplified here, the photon survival probability can drop significantly due to photon-photon absorption on EBL and CMB.

However, the photon-ALP conversion efficiency in such environments is subject to significant uncertainties, stemming from: the properties of the ISM and intergalactic radiation fields; and the strength, geometry, and coherence length of the magnetic fields.
These uncertainties introduce challenges in predicting the ALP production rates and in identifying clear observational signatures.

Among other systems, galaxy clusters, with their turbulent magnetic fields,  provide a favourable environment for photon-ALP conversion. The magnetic fields in galaxy clusters are typically chaotic and exhibit random configurations across different magnetic domains~\cite{Govoni:2004as}. The conversion of photons to ALPs in these fields requires an ensemble average over all possible realizations of the magnetic domains, typically denoted as $N$ domains~\cite{Mirizzi:2009aj}.
In the absence of photon absorption, the maximum value of the photon-ALP transfer function is $\sim 2/3$. However, the stochastic nature of the turbulent magnetic fields leads to a large variance in the transfer function's values, implying that there is a substantial spread in the potential observational signatures of ALPs. This stochasticity also affects the transparency of VHE photons, which may be influenced by photon-ALP conversions in these environments. As a result, galaxy clusters are important targets for studying ALP effects on the propagation of VHE photons and testing models of ALP production in large-scale extragalactic magnetic fields.

\subsection{Production of Photons from Extragalactic ALPs}

ALPs can interact with the cosmic environment during their propagation through extragalactic space, leading to the production of photons across various wavelengths. The primary processes responsible for this photon production are the photon-ALP conversion in magnetic fields and the spontaneous decay of ALPs.

\subsubsection{ALP-photon conversion in magnetic fields}

ALPs can convert into photons (and vice versa) in the presence of magnetic fields, with the probability of conversion depending on the properties of the ALPs as well as the strength and coherence length of the magnetic field. This process can occur in both extragalactic and Galactic magnetic fields, influencing the observable photon spectra from distant astrophysical sources. 
The intergalactic magnetic field strength, eventually connected to primordial magnetism, is quite low, with upper limits of the order of a few nG or even pG ~\cite{AlvesBatista:2021sln}, implying that the rate of conversion is negligible in these environments (but see the discussion in~\cite{AlvesBatista:2023jky}). 
Conversions in the Galactic magnetic field, on the other hand, are always present and sizable, depending on the incident line of sight.  
The right panel of Fig.~\ref{fig:conversion_prob} schematically demonstrates how an extragalactic ALP flux is partially converted into photons within the Galactic magnetic field. Note that the exact yield of photons depends on the direction from which the ALP flux originates because of the particular structure of the Milky Way's magnetic field.
In particular, besides the regular magnetic field components, also the turbulent one has been shown to affect the computation of the probability conversion, even in cases when its correlation length is
 smaller than the ALP oscillation length~\cite{Carenza:2021alz}.

\paragraph{Signatures of ALP-photon conversion.}

Photon-ALP conversion can lead to different observable signatures from extragalactic sources. Here, we distinguish signals from 
individual astrophysical emitters, which typically require 
point-source analysis pipelines and contribution to cosmic photon backgrounds, where the ALPs imprint can be detected by studying large-scale diffuse emission. 

\begin{itemize}
\item \textit{Point-source prompt gamma-ray emission.} 
In core-collapse SNe the presence of ALPs can alter the typical expectations for SN explosion. 
The ejecta of the explosion directly after the core bounce forms a very dense environment that typically prevents the escape of promptly generated gamma rays. Only feebly interacting particles like neutrinos may pass this explosion stage to yield information about the first seconds after the core collapse. ALPs may escape as well when produced in the protoneutron star matter due to couplings to photons, leptons or nuclear matter. The prompt ALP burst translates into a prompt gamma-ray burst (GRB) via the conversion of ALPs into photons inside the Milky Way, see for e.g.~\cite{Payez:2014xsa} for SN1987A, and, eventually, also in the progenitor stellar magnetic field~\cite{Manzari:2024jns,Fiorillo:2025gnd,Candon:2025sdm}. Observing a prompt GRB coincident with the neutrino burst is a strong indicator of the presence of ALPs, the study of the total emission yields further information about the particle properties of the ALP (and the protoneutron star)~\cite{Raffelt:1996wa}. 
ALPs production in core-collapse SNe should produce $\mathcal{O}(10)$ second bursts of photons, with a characteristic energy spectrum peaking around 60 -- 80 MeV, see e.g.~\cite{Calore:2021hhn,Caputo:2022mah}.

\item \textit{Point-source spectral distortions.} As discussed above, the conversion probability of ALPs into photons transitions from an energy-independent strong mixing regime to an energy-dependent regime at an energy scale depending on the magnetic field strength of the host and the mass of the ALP. The energy-dependent regime is characterized by rapid oscillations of the conversion probability that imprint on the SED of any gamma-ray emitting source within this energy range. Conventional astrophysical sources shine in gamma rays due to cosmic-ray acceleration processes, which do not produce oscillating spectra but rather smooth and regular profiles. Knowing the standard expectation for the intrinsic spectrum of a target source, it is possible to search for ALPs by looking for irregularities in the spectra that resemble the oscillations of the photon-ALP conversion probability. More theoretical details are provided in \cite{Kachelriess:2023fta} while the authors provide specific code in \cite{Kachelriess:2022you}.
 Noteworthy examples include spectral distortions looked for in the gamma-ray spectra of objects like NGC1275~\cite{Fermi-LAT:2016nkz} and Mrk421~\cite{Gao:2023dvn}, which may result from photon-ALP conversion along the line of sight. The same formalism also applies to X-rays (e.g., \cite{Schallmoser:2021sba}).
    
\item \textit{Point-source spectral hardening.} The ALP-induced transparency of the universe at VHE can result in the hardening of spectra for distant gamma-ray sources~\cite{Biteau:2015xpa}. Indeed, if some of these VHE photons convert into ALPs during propagation through the magnetic fields of galaxy clusters or intergalactic space, they can travel greater distances before converting back into detectable photons. This process enhances the transparency of the universe to VHE photons, providing a potential indirect signature of ALPs~\cite{Dominguez:2011xy,Horns:2012kw,Montanino:2017ara}. The degree of this effect is, again, dependent on the configuration of the magnetic fields and the ALP properties. A more theoretical overview is given in \cite{AlvesBatista:2023jky}. Examples include quasars \cite{Davies:2022wvj}, requiring improved jet models and photon-photon dispersion in blazar jets \cite{Davies:2021wqw}. Besides spectral hardening, another correlated observable is the maximal photon energy from a source, which can be exploited to set constraints on the photon-ALP coupling~\cite{Buehler:2020qsn}.

\item \textit{Diffuse emission signals.} 
ALPs can create diffuse large-scale gamma-ray emissions in multiple ways. Any sufficiently gamma-ray bright and magnetized source in the universe may serve as an ALP source as well; for instance, AGNs, SNe or star-forming galaxies. Depending on the number density of the respective source class as a function of redshift, i.e.~the evolution of the universe, the population's cumulative emission may provide a considerable ALP flux. 
In a seminal paper, the possibility of a diffuse axion background from core-collapse SNe was proposed~\cite{Raffelt:2011ft}. Later, the concept 
was expanded to ALPs from SNe~\cite{Calore:2020tjw} and other astrophysical objects, such as star-burst galaxies~\cite{Eckner:2022rwf, Mastrototaro:2022kpt}. 
Besides the brightening of diffuse photon background, thanks to conversion in the Galactic magnetic field, this ALP emission is partially re-converted into gamma rays via the Primakoff effect. Hence, the ALP background receives a spatial morphology that traces the Milky Way's magnetic field structure. 
For a full spectral and spatial analysis of the diffuse ALPs SN background we refer the reader to~\cite{Calore:2021hhn}.
\end{itemize}

\paragraph{Best targets for photon-ALP conversion searches.}

The best targets for detecting photon-ALP conversion are bright gamma-ray sources, which allow for precise spectral measurements. The sources should be located behind or within regions of strong transverse magnetic fields, such as those associated with galaxy clusters. Accurate knowledge of the magnetic field properties is crucial for constraining ALP models since ALP searches are sensitive to the product of the coupling constant $g_{a\gamma}$ and the magnetic field strength $B$. As a result, uncertainties in the magnetic field configuration limit the precision of ALP constraints.

In comparison to Galactic targets, extragalactic sources offer several advantages, such as the fact that the observational signatures depend primarily on the latitude and longitude of the source, and that spectral determination is highly accurate, as there is minimal contamination from Galactic diffuse emission. However,
the modelling of multiple magnetic fields (intra-cluster, intergalactic, Galactic) is required as well as the one of EBL absorption. 

\subsubsection{Spontaneous ALP decays}
For heavier ALPs with masses greater than $\sim$keV, spontaneous decay into photons can become a significant channel for photon production. In this regime, photon-ALP conversion is typically suppressed, but the decay rate is non-negligible.
The decay rate of ALPs writes as:
\begin{equation}
    \Gamma_{a\gamma}=\frac{g_{a\gamma}^2 m_a^3}{64 \pi} \,.
\end{equation}

The final product is a pair of photons that are emitted back-to-back in the rest frame of the 
decaying ALPs, each photon carrying energy equivalent to half of the ALP energy. 
If the mother particle is at rest (e.g. dark matter ALPs in the halo of the Milky Way) the signature of this decay is a narrow photon line at an energy which is strongly correlated with the ALP mass. 
On the other hand, if the ALP is boosted and/or has a specific spectral energy distribution, the signal 
is broader and the exact spectral shape depends on the ALP production process.

\paragraph{ALP signatures from decay.}

The spontaneous decay of ALPs can produce several distinctive observational signatures, which, again can be classified into point-like source and large-scale diffuse contributions.
The decay of ALPs can produce photons from a variety of astrophysical sources:
\begin{itemize}
    \item \textit{Decay from past SNe and other stars} ALPs produced in SN events can decay into photons over time, leading to observable signals from SN explosions. As for conversion from core-collapse SNe, 
    we can look for time-dependent gamma-ray bursts, or the cumulative emission of these decays over time (and reshift). The strongest constraints on MeV ALPs comes indeed
    by the non-observation of gamma-ray decay signals from SN1987A~\cite{Muller:2023pip}, low-energy SNe and their contribution to the gamma-ray extragalactic diffuse emission~\cite{Calore:2020tjw,Caputo:2022mah}. This topic is further discussed in Sec.~\ref{subsec:Vitagliano}. 
    
    \item \textit{Decay from ALPs dark matter.} If ALPs make up a fraction of dark matter, their decay could generate photon emission from distant galaxies or galaxy clusters, and, more in general, from targets characterized by high dark matter densities. The spectral signature is a narrow gamma-ray line, opportunely redshifted due to the expansion of the universe. Spatially, the signal should be correlated with the dark matter spatial distribution of the target. 
    Searches for such signals, including in satellite galaxies but also large-scale structures at different wavelengths, could provide constraints on ALP dark matter models. In particular, since the ALP mass dictates the electromagnetic waveband that is sensitive to its decay into photons, different telescopes and instruments are necessary to probe the full ALP dark matter parameter space. 
\end{itemize}
These searches for decaying ALPs can be applied to other light particle candidates, like sterile neutrinos and dark photons, and can be similarly extended to final states other than photons (e.g. electrons).
We cite for example~\cite{Calore:2021lih, DelaTorreLuque:2024zsr, Carenza:2023old} for an application to feebly interacting particles produced in SNe and decaying into electron-positron pairs, and~\cite{Boyarsky:2006fg} for searches of decaying sterile neutrino dark matter.

\paragraph{Best targets for ALPs dark matter decay searches.}

\begin{itemize}
\item \textit{Satellite galaxies of the Milky Way.} The Milky Way is surrounded by many satellite galaxies of which those with substantial total masses, i.e.~baryonic plus dark matter mass, are viable targets for the search for decaying ALPs, as the signal scales with the mass enclosed in a given search volume. Refs.~\cite{Caputo:2018vmy, Caputo:2018ljp} developed the theory of stimulated axion decay in multiple targets including Milky Way satellites but also the Galactic center or galaxy clusters while focusing on the radio spectrum. There has been large interest in satellite galaxies in recent years. For instance, Ref.~\cite{Blout:2000uc} has provided the first search for nearly monochromatic microwave photons from Local Group dwarf galaxies with the Haystack Observatory using its 37-meter radio telescope. MUSE spectroscopic observations in the optical range were exploited in an early work \cite{Regis:2020fhw} and later in a more refined analysis \cite{Todarello:2023hdk} by the same set of authors to constrain ALPs with masses around the eV scale with data from five galaxies, composed of both classical and ultra-faint dwarf spheroidal galaxies of the Milky Way. Ref.~\cite{Yin:2023uwf} explores how the SUBARU telescope can improve such constraints based on observations of similar targets, while Ref.~\cite{Yin:2024lla} sets bounds using WINERED observations of two dwarf galaxies. Additionally, Ref.~\cite{Guo:2024oqo} used 2-hour radio observations of Coma Berenices taken by the Five-hundred-meter Aperture Spherical radio Telescope (FAST) to set limits on the ALP parameter space, which are weaker than those of the CAST helioscope (see Sec.~\ref{subsec:RuzVogel} and Sec.~\ref{sec:CAST-MM} for more details).

\item \textit{Dark matter halo of the Milky Way.} Although the Milky Way's dark matter halo is not a genuine extragalactic object, it is a viable search target for ALP decay due to the large mass of our Galaxy. Using blank sky X-ray observations of XMM Newton, the authors of \cite{Foster:2021ngm} search for ALP decays in the Galactic halo and provide an update in \cite{Dessert:2023vyl} using Hitomi data and future XRISM (X-ray Imaging and Spectroscopy Mission) data. Similar searches have been performed based on eROSITA data \cite{Dekker:2021bos} and INTEGRAL (INTErnational Gamma-Ray Astrophysics Laboratory) data (reaching the hard X-ray spectrum), specifically targeting the center of the Milky Way \cite{Calore:2022pks}. A forecast of the James Webb Space Telescope's sensitivity (infrared waveband) to ALP decays in the Galactic halo was presented in \cite{Roy:2023omw}. In contrast, \cite{Janish:2023kvi} used blank sky observations intended for sky subtraction of James Webb to set constraints on ALPs with masses from 0.8-3 eV. Refs.~\cite{Saha:2025any, Pinetti:2025owq} analyzed a larger set of JWST observations, constraining ALPs with masses $\lesssim 1$~eV. Dedicated radio observations of the Vela supernova remnant, acquired with the FAST telescope, were used in Ref.~\cite{Yang:2025yvh} to search for the signal of stimulated ALP decay.

\item \textit{Galaxy clusters.} Galaxy clusters are the most massive virialised objects in the universe. As such, they are excellent targets to search for signatures from decaying dark matter. The authors of \cite{Battye:2019aco} analyzed radio data of the Virgo cluster to search for the spontaneous decay of ALPs, while Ref.~\cite{Grin:2006aw} searched for this process in optical observations of Abell 2667 and 2390. Ref.~\cite{Chan:2021gjl} turned to radio data of the Bullet cluster (1E 0657-55.8) with the same objective and found constraints on the ALP parameter space tighter than the limit obtained by CAST. Refs.~\cite{Todarello:2024qci, Todarello:2025rff} analyzed ultra-violet data from the Hubble Space Telescope and the International Ultraviolet Explorer to place bounds on ALPs with masses of order 10~eV.

\item \textit{Large-scale structure of the universe.} The filaments of dark and baryonic matter creating the universe's large-scale structure may also be exploited in the search for decaying ALPs. The EBL as a cumulative observable of the population of galaxies in the universe was used in various ways to look for ALPs \cite{Bernal:2022xyi}. Ref.~\cite{Caputo:2020msf} focused on the near-infrared background angular power spectrum (including emissions from galaxies and contributions from the intra-halo light) to constrain an additional component generated from ALP decay by comparing it to measurements from the Hubble Space Telescope and Spitzer. The authors of \cite{Libanore:2024hmq} demonstrate that future broadband ultraviolet surveys, like GALEX (Galaxy Evolution Explorer) or the upcoming ULTRASAT (Ultraviolet Transient Astronomy Satellite) satellite, can improve current ALP limits by exploiting the ultraviolet component of the EBL. The optical component of the EBL was the target of \cite{Nakayama:2022jza} and \cite{Carenza:2023qxh} to explore whether ALP decay can be related to an anisotropy in the COB flux, in relation to the LORRI excess. This excess from LORRI images is an observational inconsistency with our expectations about the flux of the COB, exceeding the estimates based on deep galaxy counts by about a factor of two \cite{Bernal:2022wsu}.  The authors of \cite{Shirasaki:2021yrp} constrained eV-mass ALPs by cross-correlating line intensity and weak lensing maps exploiting optical and infrared observations.
By analyzing the cosmic photon background at multiple wavelengths, 
\cite{Porras-Bedmar:2024uql} could set constraints over a wide ALP mass range, also challenging the LORRI excess ALPs interpretation.

\item \textit{Distant galaxies.} Observations of distant galaxies like quasars can also be directly used to search for ALPs. Ref.~\cite{Wang:2023imi} analyzed ALP dark matter decay by stacking dark-sky spectra from the Dark Energy Spectroscopic Instrument (DESI) at the redshift of nearby galaxies from DESI’s Bright Galaxy and Luminous Red Galaxy samples. Finally, Ref.~\cite{Sun:2023wqq} showed how inconsistencies with the currently available datasets of high-redshift quasars can be remedied by invoking ALPs.
\end{itemize}

%% file: WG3/content/Silvia_Gasparotto.tex
\subsection{Introduction}
\label{Gasparotto_sec_1}
In this Section, we explore the fascinating phenomenon of axion birefringence, a unique effect on photon propagation arising from the interaction between axions and photons described by the coupling term $g_{a\gamma\gamma} a F_{\mu\nu}\tilde{F}^{\mu\nu}$~\cite{Lue:1998mq}. This interaction implies that, in the presence of a dynamical axion field, the vacuum behaves like a birefringent medium where the circular polarisation states of photons propagate at different velocities, resulting in a net rotation of the polarisation vector. This effect was first discussed in Refs.~\cite{Carroll:1989vb, Carroll:1991zs, Harari:1992ea} (see Ref.~\cite{Komatsu:2022nvu} for a review) and represents a key signature of axion physics that can be probed observationally. 
This topic received renewed interest recently with a new analysis of \textit{Planck} data finding a $2.4 \sigma$ evidence for a non-zero birefringence angle~\cite{Minami:2020odp}. This has been followed up by a huge effort from the community to explore the presence of birefringence in our universe.

To understand this phenomenon, let us start by considering the axion-photon interaction, which exhibits a distinct behavior under parity transformations $\Vec{x} \xrightarrow{} -\Vec{x}$, where $F \xrightarrow{} F$ and $\tilde{F} \xrightarrow{} -\tilde{F}$. This makes the interaction parity-odd, leaving a parity-violating imprint on the electromagnetic field~\cite{Lue:1998mq}. Specifically, when a plane wave with vector potential $\Vec{A}$ propagates in a homogeneous axion field $a = a(\eta)$, where $\eta$ is the conformal time in the Friedmann–Lema\^{i}tre–Robertson–Walker metric ${\rm d}s^2 = R^2(\eta)(-{\rm d}\eta^2 + {\rm d}\Vec{x}^2)$, the equation of motion for the two helicity states of $\Vec{A}$, denoted as $A_\pm = (A_x \mp iA_y)/\sqrt{2}$, can be written as:
\begin{equation}
    \left[\frac{{\rm d}^2}{{\rm d}\eta^2} + \left(
    k^2 \mp k g_{a\gamma\gamma} \frac{{\rm d}a}{{\rm d}\eta}
    \right)\right]A_k^\pm = 0,
\end{equation}
where we have applied the gauge conditions $A_0 = 0$ and $\Vec{\nabla} \cdot \Vec{A} = 0$.

In the regime where ${g_{a\gamma\gamma}}{\rm d}a/{\rm d}\eta$ varies slowly compared to the wavelength of the radiation, the two helicity states propagate with different phase velocities,
\begin{equation}
    \omega_{\pm} = \sqrt{k^2 \mp k g_{a\gamma\gamma} \frac{{\rm d}a}{{\rm d}\eta}} \approx k \mp \frac{g_{a\gamma\gamma}}{2} \frac{{\rm d}a}{{\rm d}\eta},
\end{equation}
where higher-order corrections in $\mathcal{O}(g_{a\gamma\gamma})$ are negligible~\cite{Eskilt:2022wav}. This effect is frequency-independent, distinguishing it from other phenomena like Faraday rotation or potential quantum gravity effects~\cite{Gleiser:2001rm,Myers:2003fd,Gubitosi:2009eu,Gubitosi:2010dj}. If the axion field varies spatially, $a = a(\eta, \Vec{x})$, the derivative must be modified as ${\rm d}a/{\rm d}\eta \rightarrow {\rm d}a/{\rm d}\eta + \Vec{k} \cdot \Vec{\nabla}a$~\cite{Harari:1992ea}.

The differential propagation speeds of the photon helicities cause an initially linearly polarised photon to undergo a rotation of its polarisation plane, described by the birefringence angle,
\begin{equation}
    \beta = -\frac{1}{2} \int_{\eta_{\rm source}}^{\eta_{\rm obs}} \mathrm{d}\eta (\omega_+ - \omega_-) = \frac{g_{a\gamma\gamma}}{2}\left[a(\eta_{\rm obs}, \hat{n}) - a(\eta_{\rm source}, \hat{n})\right],
    \label{eq:alpha_formula}
\end{equation}
where $\beta$ represents the rotation angle of the polarisation vector. This effect can manifest as a net phase shift of the electromagnetic radiation (a direct current DC effect) if the axion field changes significantly over the propagation distance. Alternatively, if the local oscillation of the axion field dominates, a periodic modulation of the polarisation vector occurs with frequency $\omega=m_a$ (an alternating current AC effect)~\cite{Fedderke:2019ajk}.

In the 1990s, pioneering studies~\cite{Carroll:1989vb, Nodland:1997cc, Carroll:1997tc, Harari:1992ea} sought to constrain the DC birefringence effect by examining correlations between the polarisation vectors and elongation axes of high-redshift polarised sources, such as radio galaxies and quasars. Later, it was recognized that this phenomenon induces a parity-violating signal in the cosmic microwave background (CMB) polarisation~\cite{Lue:1998mq}, commonly referred to as \textit{cosmic birefringence}. The CMB, with its linearly polarised photons from the last scattering surface, provides an ideal probe for detecting the slow background evolution of the axion field with masses $m_a\lesssim \mathcal{O}(10^{-28})~{\rm eV}$. 
As the CMB polarisation can be decomposed into $E$ and $B$ modes, which are respectively even and odd under parity transformation~\cite{Zaldarriaga:1996xe, Kamionkowski:1996ks}, any rotation of the polarisation plane due to axion birefringence would mix these modes and generate non-zero parity-odd cross-correlations, such as $EB$ and $TB$ (correlation between temperature $T$ and $B$ modes) which vanish in the standard parity-invariant case~\cite{Lue:1998mq, Liu:2006uh, Finelli:2008jv,Feng:2004mq}. Here, we will focus on the $EB$ signal as it is the one expected to provide higher sensitivity for ongoing and future CMB experiments~\cite{Planck:2016soo}. When generated by the evolution of a homogeneous axion field, the observed $EB$ power spectrum is related to the $EE$ and $BB$ power spectra of a $\Lambda$CDM universe with $g_{a\gamma\gamma}=0$ as follows~\cite{Lue:1998mq, Liu:2006uh, Finelli:2008jv,Feng:2004mq}
%
\begin{equation}\label{eq:observed_EB}
    C_\ell^{EB,\mathrm{obs}} = \frac{\sin(4\beta)}{2} \left(C_\ell^{EE,\Lambda\mathrm{CDM}} - C_\ell^{BB,\Lambda\mathrm{CDM}}\right).
\end{equation}

Recent analyses have reported accumulating evidence for non-zero cosmic birefringence, providing tantalising hints for the existence of an ultralight axion field. In response to these developments, this review will first examine the current observational evidence from CMB data and explore future prospects in Sec.~\ref{Gasparotto_sec_2}. We then consider the implications of these observations for different axion models in Sec.~\ref{Gasparotto_sec_3}. Finally, we discuss searches for axion birefringence from astrophysical sources and in laboratory experiments in Sec.~\ref{Gasparotto_sec_4}, which are generally sensitive to higher axion masses compared to the CMB. This structure allows us to review the observables of axion birefringence across the broadest range of scales, from cosmological distances sensitive to $m_a \sim \mathcal{O}(10^{-33})~\mathrm{eV}$ axions, potentially contributing to dark energy, to terrestrial scales probing axion dark matter with masses up to $m_a \sim \mathcal{O}(10^{-9})~\mathrm{eV}$.

\subsection{Observational evidence from CMB polarization}
\label{Gasparotto_sec_2}
We start by reviewing the state-of-the-art constraints and the experimental challenges of measuring the cosmic birefringence produced by $m_a\lesssim \mathcal{O}(10^{-28})$~eV axions with CMB polarisation data. First measured by Ref.~\cite{Feng:2006dp} in $2006$, it has become customary to employ CMB polarisation data to measure the static and isotropic rotation produced by the background evolution of the axion field~\cite{QUaD:2008ado, WMAP:2012nax, Planck:2016soo, POLARBEAR:2019kzz, ACT:2020frw, SPT:2020cxx,Yin:2025fmj,Namikawa:2025sft}.

\begin{figure}
    \centering
    \includegraphics[width=1\textwidth]{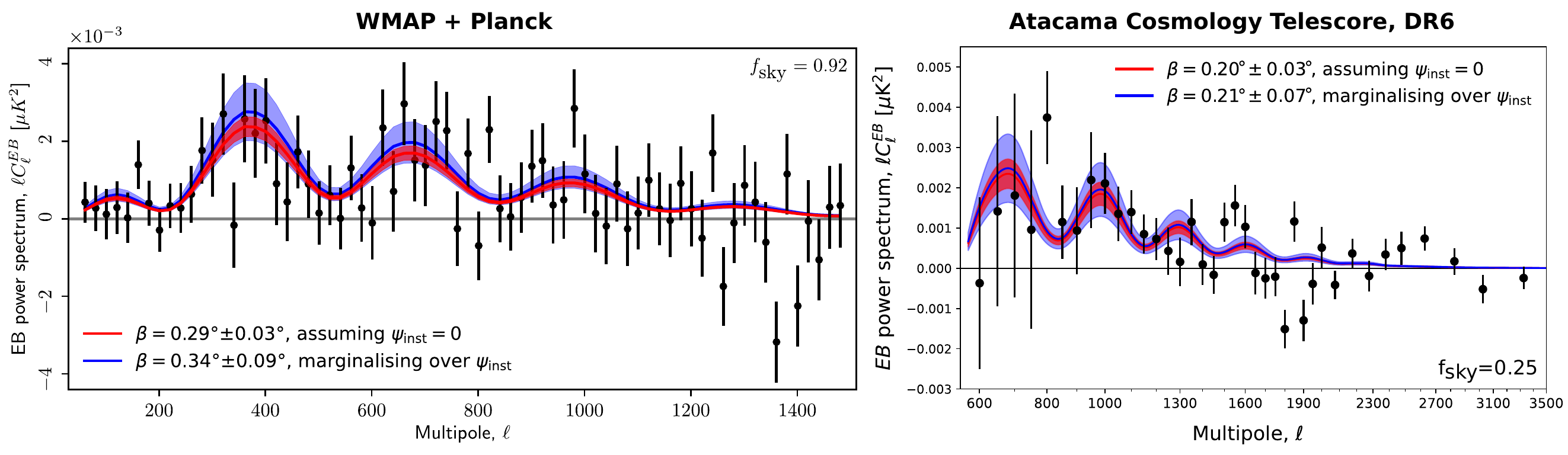}
    \caption{$EB$ power spectrum stacked from the $12$ polarised frequencies of $9$-year WMAP~\cite{WMAP:2012fli} and \textit{Planck} \texttt{NPIPE} data~\cite{Planck:2020olo} (\textit{left}), and the PA4\,f220, PA5\,f090, PA5\,f150, PA6\,f090, and PA6\,f150 array-bands from ACT DR6~\cite{AtacamaCosmologyTelescope:2025blo} (\textit{right}). The spectrum is calculated over a $92\%$  and $25\%$ fraction of the sky for WMAP+\textit{Planck} and ACT, respectively. Red contours show $68\%$ C.L. from fitting $\beta$ with Eq.~\eqref{eq:observed_EB}, assuming no miscalibration of polarisation angles. Blue contours show the $68\%$ C.L. when marginalising over polarisation angles with Eq.~\eqref{eq: MK observed EB}, using a model of Galactic emission in the case of WMAP+\textit{Planck}, and a model of the instrument optics in the case of ACT. Figures adapted from Refs.~\cite{Eskilt:2022cff,Diego-Palazuelos:2025dmh}.
    }
    \label{fig: planck + wmap EB spectrum}
\end{figure}

Around 2020, a series of works~\cite{Minami:2020odp, Diego-Palazuelos:2022dsq, Eskilt:2022wav, Eskilt:2022cff, Cosmoglobe:2023pgf} 
reanalysed the data from the \textit{Planck}~\cite{Planck:2018nkj, Planck:2020olo} and WMAP~\cite{WMAP:2012fli} missions and found exciting evidence of a non-null birefringence angle of roughly $\beta\approx 0.3^\circ$. 
The findings of these works are summarised by the stacked $EB$ power spectrum shown in the left panel of Fig.~\ref{fig: planck + wmap EB spectrum}, where fluctuations resembling the acoustic oscillations of the CMB $EE$ spectrum are visible around the $200 \lesssim \ell \lesssim 900$ angular scales. 
Fitting the observed spectra with Eq.~\eqref{eq:observed_EB} yields $\beta=0.29^\circ\pm 0.03^\circ$ ($68\%$ C.L.)~\cite{Eskilt:2022cff}, a remarkable $\sim 9\sigma$ measurement.
The other astrophysical sources of microwave-range polarised radiation within our Galaxy, namely synchrotron and thermal dust emission, cannot produce such an oscillating signal as their power spectra behave mostly like power laws~\cite{Martire:2021gbc, Planck:2018gnk}. 
Such fluctuations remain present (albeit with a lesser statistical significance due to the reduced sky coverage) after masking most of the Galactic plane~\cite{Eskilt:2022cff}, attesting to the isotropy of the signal and dismissing Galactic emission as its source.

However, instrumental systematics cannot be entirely ruled out as the origin of the signal due to the degeneracy between a DC isotropic birefringence rotation and the miscalibration of instrumental polarisation angles, $\psi_\mathrm{inst}$~\cite{Shimon:2007au, Miller:2009pt}. 
The insufficient precision of polarisation angle calibration was the main uncertainty limiting previous studies~\cite{QUaD:2008ado, WMAP:2012nax, Planck:2016soo, POLARBEAR:2019kzz, ACT:2020frw, SPT:2020cxx}.
Refs.~\cite{Minami:2020odp, Diego-Palazuelos:2022dsq, Eskilt:2022wav, Eskilt:2022cff, Cosmoglobe:2023pgf} were able to break the $\beta$-$\psi_\mathrm{inst}$ degeneracy by calibrating polarisation angles against Galactic emission. 
This method is grounded on the assumption that, for slowly-rolling axion fields, locally emitted Galactic photons will not suffer a significant birefringence compared to CMB photons~\cite{Minami:2019ruj}. 
After including Galactic emission, Eq.~\eqref{eq:observed_EB} transforms into~\cite{Minami:2019ruj, Minami:2020xfg, Minami:2020fin}
\begin{equation}\label{eq: MK observed EB}
\begin{aligned}
    C_\ell^{EB,\mathrm{obs}} = & \frac{\tan (4\psi_\mathrm{inst})}{2} \left( C_\ell^{EE,\mathrm{obs}} - C_\ell^{BB,\mathrm{obs}}\right) + \frac{C_\ell^{EB,\mathrm{Gal}}}{\cos (4\psi_\mathrm{inst})} \\
    & + \frac{\sin (4\beta)}{2\cos (4\psi_\mathrm{inst})}\left( C_\ell^{EE,\Lambda\mathrm{CDM}} - C_\ell^{BB,\Lambda\mathrm{CDM}}\right),
\end{aligned}
\end{equation}
where ``$\Lambda$CDM'' denotes the CMB power spectra for a $\Lambda$CDM universe with $g_{a\gamma\gamma}=0$, ``Gal'' the spectra from Galactic emission, and ``obs'' the total signal observed by CMB instruments at a given frequency band. 
Galactic synchrotron emission does not present a significant global $EB$ correlation~\cite{Martire:2021gbc, Rubino-Martin:2023fya}, but the misalignment between dust filaments and the plane-of-sky orientation of the Galactic magnetic field produces $TB$ and $EB$ correlations~\cite{Huffenberger:2019mjx, Clark:2021kze, Cukierman:2022kei} that can bias birefringence measurements, if not accounted for~\cite{Diego-Palazuelos:2022dsq, Diego-Palazuelos:2022cnh}.
Two approaches to model Galactic emission have been explored so far: reproducing Galactic emission through Bayesian analyses that fit parametric spectral energy distribution models to CMB data~\cite{Diego-Palazuelos:2022dsq, Diego-Palazuelos:2022cnh, Jost:2022oab, LiteBIRD:2025yfb} or directly predicting the dust $EB$ spectrum from different estimators of the misalignment between dust filaments and the Galactic magnetic field~\cite{Diego-Palazuelos:2022dsq, Eskilt:2022wav, Eskilt:2022cff, Cosmoglobe:2023pgf, Hervias-Caimapo:2024ili}.
Adopting the latter, Ref.~\cite{Eskilt:2022cff} obtained $\beta=0.34^\circ\pm0.09^\circ$ ($68\%$ C.L.) in a nearly full-sky analysis of \textit{Planck} and WMAP data, the tightest constraint to date with a $3.6\sigma$ ($99.987\%$ C.L.) statistical significance. Acknowledging the current limitations of dust models, independent high-precision measurements from ongoing and future CMB experiments like the Simons Observatory (SO)~\cite{SimonsObservatory:2025wwn,Jost:2022oab}, AliCPT~\cite{Dou:2025luz}, and LiteBIRD~\cite{LiteBIRD:2022cnt, LiteBIRD:2025yfb}, and a better understanding of polarised dust emission~\cite{Hervias-Caimapo:2024ili, Cukierman:2022kei, Vacher:2021heq, Vacher:2022xdw} are needed to obtain a definitive birefringence measurement with this methodology. Further tests, without explicitly modeling $\psi_\mathrm{inst}$, have confirmed the presence of a $\approx0.3^\circ$ rotation in \textit{Planck} data~\cite{Ballardini:2025apf,Remazeilles:2025wzd,Yin:2025fmj,Sullivan:2020wnv}.

Recently, the Atacama Cosmology Telescope (ACT) provided independent evidence for a non-null global rotation of $\beta+\langle\psi_\mathrm{inst}\rangle=0.20^\circ \pm 0.08^\circ$ in their data release 6 (DR6)~\cite{AtacamaCosmologyTelescope:2025blo}. Using the knowledge about instrumental systematics provided by the ACT optics model~\cite{Murphy:2024fna}, Ref.~\cite{Diego-Palazuelos:2025dmh} isolated the birefringence contribution to be $\beta= 0.21^\circ\pm 0.07^\circ$ (68\% C.L.) after marginalizing over $\psi_\mathrm{inst}$. However, there are still unresolved systematics in the ACT DR6, such as the moderate discrepancy between the rotations derived from the 90 and 150\,GHz data in PA5~\cite{AtacamaCosmologyTelescope:2025blo}. The corresponding stacked $EB$ spectrum is shown in the right panel of Fig.~\ref{fig: planck + wmap EB spectrum}. Assuming that the ACT and WMAP+\textit{Planck} results are statistically independent due to the limited overlap in the angular scales used in each analysis, the combined measurement gives $\beta = 0.264^\circ\pm 0.058^\circ$, which excludes $\beta=0$ with a $4.6\sigma$ statistical significance.

Although the consistency seen between independent experiments is compelling, we can not draw strong cosmological conclusions yet, as neither WMAP, \textit{Planck}, nor ACT has an absolute calibration of polarisation angles without relying on Galactic emission or instrument models. Definitive confirmation might come soon from experiments using artificial polarization sources as calibrators, such as BICEP3~\cite{Cornelison:2022zrc,BICEPKeck:2024cmk}, CLASS~\cite{Coppi:2025fmt}, and the SO~\cite{Murata:2023heo}. Calibration techniques continue to develop to reach accuracies of the order of $0.01^\circ$~\cite{Casas:2021sjb,Ritacco:2024kug, Johnson:2015tga, Nati:2017lnn, Navaroli:2018zds, Dunner:2020}. The full shape of the $EB$ spectrum can also be used to break the degeneracy between $\beta$-$\psi_\mathrm{inst}$ angles and constrain axion birefringence without relying on precise instrument calibration or dust modelling, as will be further discussed in Sec.~\ref{Gasparotto_sec_3}.

In addition to the isotropic DC rotation, a spatially-dependent rotation of CMB polarisation, $\beta(\vec{n})$, will induce correlations between $E$ and $B$ modes across different angular scales. 
These correlations can be used to reconstruct anisotropic cosmic birefringence \cite{Kamionkowski:2008fp, Gluscevic:2009mm,Namikawa:2020ffr}. 
Anisotropic birefringence also produces a distinct $BB$ power spectrum signal~\cite{Namikawa:2024dgj,Lonappan:2025hwz,Balkenhol:2025xas} but, unlike isotropic birefringence, it generates no $EB$ cross-power spectrum~\cite{Cai:2021zbb}. We will discuss the anisotropic birefringence signature originating from different axion models in Sec.~\ref{Gasparotto_sec_3}. Multiple works have constrained anisotropic birefringence by reconstructing $\beta(\vec{n})$ \cite{Gluscevic:2012me, POLARBEAR:2015ktq, BICEP2:2017lpa, Contreras:2017sgi, Namikawa:2020ffr, Gruppuso:2020kfy, SPT:2020cxx, Bortolami:2022whx, BICEPKeck:2022kci, Zagatti:2024jxm}, or through CMB polarisation power spectra \cite{Li:2014oia,diSeregoAlighieri:2014tvt,Liu:2016dcg,Namikawa:2024dgj,Lonappan:2025hwz,Balkenhol:2025xas}. 
So far, CMB experiments have only been able to put upper bounds on anisotropic birefringence, with the amplitude of the scale-invariant power spectrum currently constrained to be 
$A_{\rm CB}\equiv [\ell(\ell+1)/2\pi] C_\ell^{\beta\beta}\lesssim 0.04\times 10^{-4}$ at $2\sigma$ from $\ell=2$ to $2000$~\cite{BICEPKeck:2022kci}.
These constraints will be greatly improved by ongoing and future CMB experiments such as the  SO and LiteBIRD~\cite{CMB-HD:2022bsz}.

\subsection{Implications and forecasts for axion models}
\label{Gasparotto_sec_3}
In this section, we discuss the possible explanations for the reported $\beta \approx 0.3^\circ$ isotropic birefringence angle by axion birefringence.
In Eq.~\eqref{eq:alpha_formula}, the birefringence angle experienced by CMB photons is independent of the photon frequency and determined only by the axion-field values at the times of observation and last scattering.
While axion fluctuations at the last scattering surface result in anisotropic cosmic birefringence, isotropic cosmic birefringence is sourced by the evolution of
the background field
and the fluctuation at the observer's location. 

First, we focus on explaining $\beta \approx 0.3^\circ$ with the time evolution of the axion background field.
This possibility has been studied in Refs.~\cite{Fujita:2020ecn, Fung:2021wbz, Mehta:2021pwf, Nakagawa:2021nme, Reig:2021ipa, Fung:2021fcj, Choi:2021aze, Obata:2021nql, Nakatsuka:2022epj, Gasparotto:2022uqo, Lin:2022niw, Murai:2022zur, Greco:2022xwj, Galaverni:2023zhv, Girmohanta:2023ghm, Eskilt:2023nxm, Yin:2023srb, Naokawa:2023upt, Gasparotto:2023psh, Gendler:2023kjt, Greco:2024oie, Tada:2024znt, Naokawa:2024xhn, Mahmoudi:2024wjy, Murai:2024yul, Kochappan:2024jyf,Namikawa:2025doa,Berbig:2024aee,Lee:2025yvn,Nakagawa:2025ejs,Lin:2025gne,Barman:2025ryg}.
To explain the reported signal, the axion must evolve between the times of last scattering and observation.
For simplicity, we assume the mass potential for the axion, i.e., $V(a) = m_a^2 a^2/2$, and obtain the time evolution of the field by solving the equation of motion $\Ddot{a}+3H\Dot{a}+{\rm d}V(a)/{\rm d}a=0$.

If $m_a \lesssim \mathcal{O}(10^{-33})$\,eV, the axion slowly rolls down the potential even in the present time.
In this regime, the axion-photon coupling to explain isotropic cosmic birefringence is given by~\cite{Fujita:2020ecn}
\begin{align}
    g_{a \gamma \gamma} 
    =
    2.2 \times 10^{-18} \, \mathrm{GeV}^{-1}
    \times 
    \left( \frac{\beta}{0.3^\circ} \right)
    \left( \frac{\Omega_a}{\Omega_\Lambda} \right)^{-1/2}
    \left(  
    \frac{m_a}{10^{-35}\,\mathrm{eV}} \right)^{-1}
    \ ,
\end{align}
where $\Omega_a$ and $\Omega_\Lambda$ are the current axion and dark energy densities.
In particular, axions can account for dark energy with $m_a \lesssim 8.5 \times 10^{-34}$\,eV, which satisfies \textit{Planck}'s constraints on the equation of state of dark energy, $w < -0.95$~\cite{Planck:2018vyg}.
Recent results from the Dark Energy Spectroscopic Instrument (DESI), however, provide strong evidence for dynamical dark energy and favour $w>-1$ at late times, thereby allowing axion models with larger masses than previously permitted~\cite{DESI:2025zgx,DESI:2025fii}. 
This dark energy regime is also discussed in Refs.~\cite{Fujita:2020ecn} for the cosine potentials, \cite{Gasparotto:2022uqo} for the linear monodromy potential, and \cite{Choi:2021aze, Lin:2022niw} for the electroweak quintessence models. The possible connection between the DESI indications of dynamical DE and observations of cosmic birefringence has been investigated in Refs.~\cite{Tada:2024znt,Barman:2025ryg,Lin:2025gne,Nakagawa:2025ejs,Berbig:2024aee,Lee:2025yvn}.

If $\mathcal{O}(10^{-33})$\,eV $ \lesssim m_a \lesssim \mathcal{O}(10^{-28})$\,eV, the axion field starts to oscillate after the recombination epoch and before the current time.
The amplitude of the field today becomes then much smaller than the initial value, $a_\mathrm{ini}$, and we obtain $\Delta a \approx a_\mathrm{ini}$, which leads to the relation of~\cite{Fujita:2020ecn}
\begin{align}
    g_{a \gamma \gamma} 
    =
    3.2 \times 10^{-18} \, \mathrm{GeV}^{-1}
    \times 
    \left( \frac{\beta}{0.3^\circ} \right)
    \left( \frac{\Omega_a h^2}{0.001} \right)^{-1/2}
    \ ,
\end{align}
where $\Omega_a h^2 \sim 0.001$ corresponds to the typical upper bound on axion density for $\mathcal{O}(10^{-31})\,\mathrm{eV} \lesssim m_a \lesssim \mathcal{O}(10^{-28})\,\mathrm{eV}$ obtained from a joint analysis of the CMB and galaxy clustering data~\cite{Rogers:2023ezo}.
For this particular mass range,
the axion field significantly evolves between the epochs of recombination and reionisation and imprint distinct rotations into the small ($\ell\gtrsim20$) and large ($\ell\lesssim20$) angular scales of the $EB$ spectrum~\cite{Sherwin:2021vgb, Galaverni:2023zhv, Greco:2022xwj, Nakatsuka:2022epj}. 
Such an angular scale dependence breaks degeneracy between the $\beta$ and $\psi_\mathrm{inst}$ angles for these axion masses~\cite{Sherwin:2021vgb, Nakatsuka:2022epj,Diego-Palazuelos:2024lym}.
In this case, the above relation for $g_{a \gamma \gamma}$ and $\beta$ is only valid for the CMB photons scattered in the recombination epoch.
Furthermore, the shape of the $EB$ spectrum at high-$\ell$ is sensitive to $m_a\approx \mathcal{O}(10^{-28})\mathrm{eV}$ axions as it significantly depends on the axion-field evolution during recombination~\cite{Nakatsuka:2022epj, Murai:2022zur}. Thus, with an accurate theoretical prediction for the $EB$ power spectrum, including the gravitational lensing effect~\cite{Naokawa:2023upt}, future CMB experiments can constrain axions at this mass range without relying on precise instrument calibration or dust modelling. As a complementary probe, one could also probe the axion-field evolution after reionisation with the polarised Sunyaev-Zel'dovich effect \cite{Namikawa:2023zux}. 

If the axion mass is larger than $m_a\gtrsim\mathcal{O}(10^{-28})$\,eV, axions start oscillating before recombination, and the birefringence angle depends on the photons' time of last scattering.
Thus, we will observe photons with both positive and negative $\beta$.
As a result, the $EB$ spectrum is suppressed, and its shape depends on the multipole in a different way than the $EE$ spectrum so that Eq.~\eqref{eq:alpha_formula} no longer is a good description of the effect.
In addition, the oscillation of $\beta$ reduces the $EE$ spectrum~\cite{Finelli:2008jv}, which is called the \textit{washout} effect~\cite{Fedderke:2019ajk}.
Ref.~\cite{Fedderke:2019ajk} gives an analytical formula for the washout effect, whose validity is numerically confirmed in Ref.~\cite{Murai:2024yul}.
From the washout effect, the axion-photon coupling for axion dark matter is constrained as~\cite{Fedderke:2019ajk}
\begin{align}
    g_{a \gamma \gamma} 
    \lesssim
    9.6 \times 10^{-17} \, \mathrm{GeV}^{-1}
    \times 
    \left( \frac{m_a}{10^{-25}\,\mathrm{eV}} \right)
    \left(  
    \frac{\kappa \Omega_c h^2}{0.11933} \right)^{-1/2}
    \ ,
\end{align}
where $\kappa$ is the fraction of dark matter in the form of axions at the average redshift of decoupling. 

We can also consider axion potentials beyond the mass potential.
For example, we can consider $V \propto a^n$ with $n \geq 4$ around the potential minimum~\cite{Murai:2022zur,Galaverni:2023zhv,Eskilt:2023nxm,Yin:2023srb,Kochappan:2024jyf}, which is often adopted in the early dark energy (EDE) models~\cite{Karwal:2016vyq,Poulin:2018dzj} motivated by the Hubble tension (see Refs.~\cite{Kamionkowski:2022pkx,Poulin:2023lkg} for reviews).
To relax the Hubble tension, the EDE field must start to oscillate before the recombination epoch.
Since the axion with $n \geq 4$ oscillates slower than that with the mass potential, such an EDE field oscillates once or less during the recombination epoch. 
As a result, the resultant $EB$ spectrum can be as large as the reported signal. We can distinguish EDE models from constant $\beta$ via the shape of the $EB$ spectrum, which depends on the EDE parameters~\cite{Eskilt:2023nxm,Yin:2023srb, Kochappan:2024jyf}. 
Moreover, we can consider an asymmetric potential~\cite{Murai:2024yul}.
If the axion experiences an asymmetric oscillation during the recombination epoch, the cancellation of the birefringence angle becomes incomplete, and a substantial $EB$ spectrum can be produced. Multiple fields could also contribute to the reported value of $\beta$. Ref.~\cite{Obata:2021nql} considered two axion fields, one being dark matter, potentially giving rise to a birefringence signal in laboratory searches (see Sec.~\ref{Gasparotto_sec_4}) and the other dark energy. Ref.~\cite{Gasparotto:2023psh} discussed the possibility of using $\beta$ to constrain the probability distribution of the axion parameters in light of the Axiverse.

In these scenarios where the axion background dynamics explain the reported isotropic birefringence, anisotropic birefringence is necessarily predicted from the axion fluctuations. 
In the pre-inflationary scenario, i.e. $f_a>H_I$ where $H_I$ is the Hubble parameter during inflation, axion fluctuations can originate from both entropy and adiabatic modes.
The former arises from quantum fluctuations during inflation, $\delta a= H_I/2\pi$, which are uncorrelated with primordial density fluctuations and present even for massless axions~\cite{Pospelov:2008gg}. 
The corresponding power spectrum is almost scale invariant  $P^a_k=H_I/2k^3$. Hence, from their bound on $C_\ell^{\beta\beta}$, Ref.~\cite{BICEPKeck:2022kci} constrained $g_{a\gamma\gamma}\leq 2.6\times10^{-2}/H_I$.
The latter is sourced by adiabatic fluctuations of the gravitational potential and is fully correlated with primordial density fluctuations.
The auto- and cross-spectra between CMB anisotropies and anisotropic birefringence have been studied for massless fields~\cite{Pospelov:2008gg,Li:2008tma,Yadav:2009eb,Caldwell:2011pu,Zhao:2014yna,Namikawa:2024dgj}, axions with mass or cosine-type potentials~\cite{Caldwell:2011pu,Liu:2016dcg,Zhai:2020vob,Greco:2024oie}, and EDE models~\cite{Capparelli:2019rtn,Greco:2022xwj}. 
Furthermore, the cross-correlation of birefringence and large-scale structure~\cite{Arcari:2024nhw}, the effect of the time evolution of anisotropic birefringence~\cite{Greco:2022xwj}, and its higher-order correlations with CMB anisotropies~\cite{Greco:2022ufo}, have also been studied for axion models.
If axions couple not to photons but to dark photons, dark photons may experience cosmic birefringence, which eventually propagates to the CMB photons via the kinetic mixing between photons and dark photons~\cite{Lee:2023xzt}. In this scenario, the cosmic birefringence seen in CMB photons also has both isotropic and anisotropic components.

The spatial dependence of the axion field can also induce isotropic cosmic birefringence, e.g., through topological defects~\cite{Takahashi:2020tqv, Kitajima:2022jzz, Gonzalez:2022mcx, Kitajima:2023kzu, Ferreira:2023jbu}.
In the presence of domain walls without cosmic strings, $\beta$ is determined by the domains at the observer and on the last scattering surface and both isotropic and anisotropic birefringence are induced with correlated magnitudes, $C_\ell^{\beta\beta}\propto \beta^2$.
On the other hand, in the presence of cosmic strings, $\beta$ depends on the photon trajectory, and the anisotropic component becomes dominant~\cite{Agrawal:2019lkr,Jain:2021shf,Yin:2021kmx,Jain:2022jrp,Hagimoto:2023tqm,Hagimoto:2024sgw}, unless a string loop is close enough to cover a large fraction of the sky~\cite{Ferreira:2023jbu}. Moreover, the motion of these topological defects would inevitably generate a stochastic gravitational wave background that can be looked for in the CMB $B$ modes~\cite{Ferreira:2023jbu}. 

\begin{figure}[t!]
    \centering
    \includegraphics[scale=0.47]{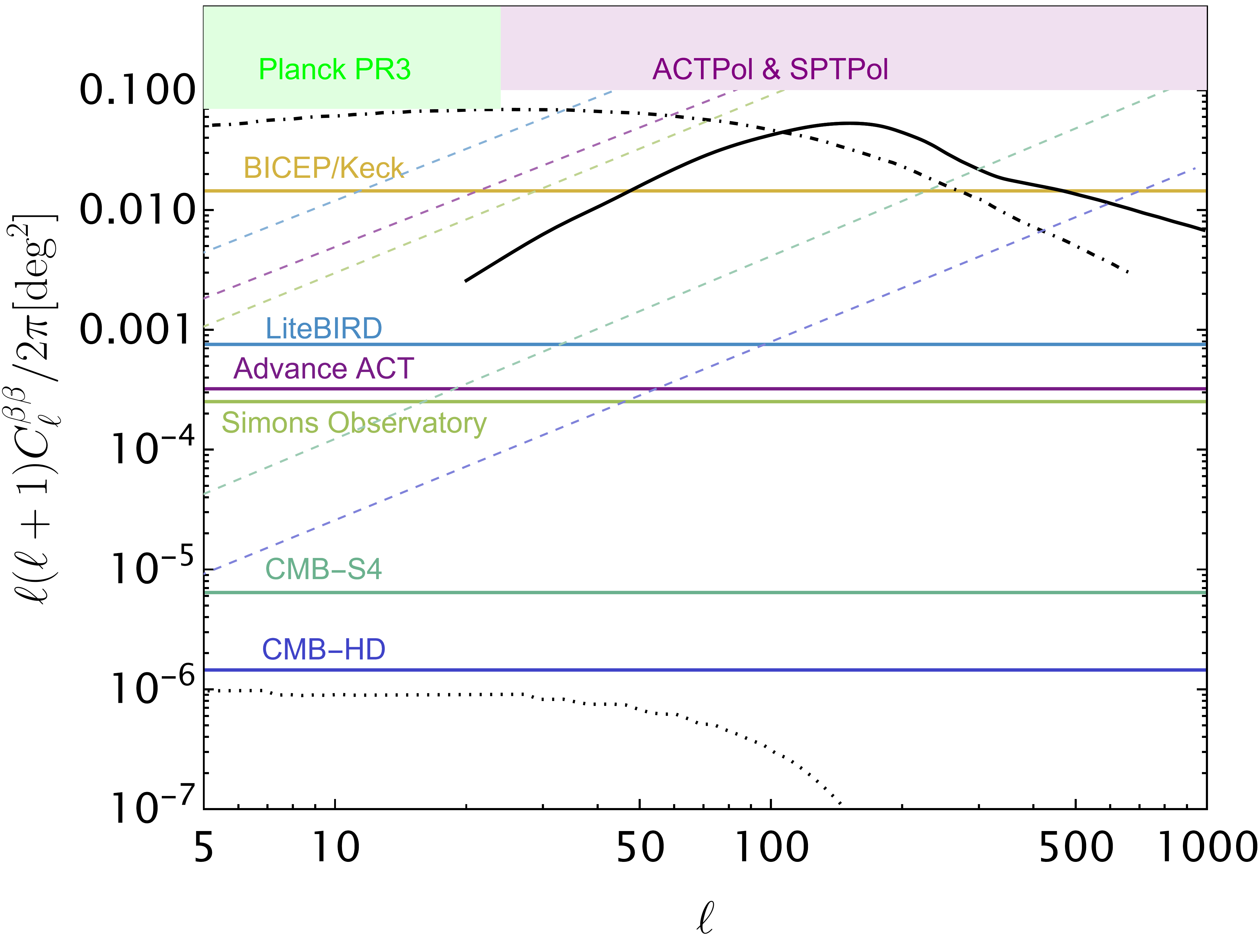}
    \caption{Current and future constraints on anisotropic cosmic birefringence, compared with the typical spectra from representative models. For a scale-invariant power spectrum, \textit{Planck} PR3~\cite{Bortolami:2022whx} excludes the green-shaded region, ACTPol~\cite{Namikawa:2020ffr} and SPTPol~\cite{SPT:2020cxx} the purple-shaded region, and BICEP/\textit{Keck}~\cite{BICEPKeck:2022kci} the amplitudes above the yellow line. The remaining horizontal solid lines show projected constraints on the scale-invariant power spectrum from future experiments according to Ref.~\cite{CMB-HD:2022bsz}, with same-coloured dashed lines indicating their projected sensitivities. The solid black line shows the $C_\ell^{\beta\beta}$ of a domain wall network taken from Ref.~\cite{Ferreira:2023jbu}, corresponding to $\beta=0.3^\circ$. The black dash-dotted line shows the $C_\ell^{\beta\beta}$of a long-lived axion string network taken from the string loop model of Ref.~\cite{Jain:2022jrp}. 
    The dotted black line shows the $C_\ell^{\beta\beta}$ from the adiabatic perturbation for the most promising case of Ref.~\cite{Greco:2022xwj}, corresponding to $m_a=10^{-27}$ eV.}
    \label{fig:anisotropic-biref-forecast}
\end{figure}

In addition, the axion fluctuation at the observer also sources isotropic cosmic birefringence, enhancing $\beta$ if $m_a \gtrsim 10^{-25}$\,eV and axion dark matter forms the non-linear structures in the late universe. However, the oscillations of such axions on the last scattering surface would also result in a washout effect incompatible with observational constraints. Consequently, the local amplitude of the axion dark matter cannot account for the reported isotropic cosmic birefringence signal~\cite{Zhang:2024dmi}. 
In Fig.~\ref{fig:anisotropic-biref-forecast}, we show the $C_\ell^{\beta\beta}$ spectra of three representative models where anisotropic birefringence is generated from the adiabatic fluctuations (dotted) from Ref.~\cite{Greco:2022xwj}, a network of domain walls (solid) from Ref.~\cite{Ferreira:2023jbu}, and a persistent network of string loops (dash-dotted) from Ref.~\cite{Jain:2022jrp}. The different characteristic features will help distinguish between the various production mechanisms. 

Up to here, we assumed that the birefringence angle is $\beta \approx 0.3^\circ$.
However, $\beta$ is estimated from the linear polarisation of the CMB photons, which is described via spin-2 quantities.
Thus, $\beta$ has an ambiguity of $n\pi$ ($n\in\mathbb{Z}$)~\cite{Naokawa:2024xhn}.
This ambiguity can be partly broken by the shape of the CMB and $C_\ell^{\beta\beta}$ spectra.

Finally, we comment on other possible non-axion explanations for the reported birefringence signal.
When linearly polarised photons propagate in a magnetised medium (either at Galactic~\cite{2022A&A...657A..43H} or cosmological scales~\cite{Campanelli:2004pm, Subramanian:2015lua}), the plane of the linear polarisation rotates via Faraday rotation with a frequency dependence of $\beta \propto \nu^{-2}$. 
Instead, quantum gravity models predict a $\beta\propto\nu^2$ rotation~\cite{Gleiser:2001rm,Myers:2003fd,Gubitosi:2009eu,Gubitosi:2010dj}. 
None of these frequency-dependent $\beta$ are consistent with the reported signal~\cite{Eskilt:2022wav,Eskilt:2022cff}.
Even when extending the particle model within the framework of the effective theory of the Standard Model of particle physics, the resulting birefringence cannot account for the reported isotropic cosmic birefringence~\cite{Nakai:2023zdr}.
Alternatively, we could also consider the possibility of the CMB $EB$ spectrum originating form primordial chiral gravitational waves~\cite{Lue:1998mq, Saito:2007kt, Sorbo:2011rz} rather than birefringence.
However, this scenario does not work due to the overproduction of CMB $B$ modes~\cite{Fujita:2022qlk}.

\subsection{Astrophysical and laboratory searches for axion birefringence}
\label{Gasparotto_sec_4}
Thus far, we reviewed recent observational constraints on cosmic birefringence from CMB data and their theoretical explanation. However, axion birefringence has been studied with many different observations and experiments over the years. This section reviews some examples of astrophysical observations and laboratory experiments aiming to detect a birefringence signal.\,\footnote{Note that in the present section we follow the CMB convention on linear polarization angle: polarization angle increasing clockwise when looking at the source. This should be kept in mind when comparing CMB constraints with other astrophysical sources~\cite{Galaverni:2017blq,Galaverni:2014gca,2017ExA....43...19D}.} 

As mentioned in Sec.~\ref{Gasparotto_sec_1}, the pioneering measurements of birefringence were done over 30 years ago by S. Carroll, G. Field and R. Jackiw with observations of radio galaxies~\cite{Carroll:1989vb}. When using astrophysical sources to test the rotation of linear polarization, the main difficulty is to know the intrinsic direction of the sources' polarization angle. Ref.~\cite{Carroll:1989vb} used radio galaxies with extended jets to overcome this problem as their synchrotron radiation is linearly polarized and the difference between the polarization and position angles of the jets is known to be around $90^\circ$~\cite{Gardner_1963, Gardner_1966, 10.1093/mnras/190.2.205}. Ref.~\cite{Carroll:1989vb} reported no birefringence detection after statistically studying the polarization rotation over many radio galaxies. In the same ages, optical photons from radio galaxies were also used to test birefringence~\cite{Cimatti_1994}. These methods using astrophysical indicators - standard crosses~\cite{Naokawa:2025shr} - have not produced a reliable detection so far~\cite{Nodland:1997cc,Carroll:1997tc, Wardle:1997gu, diSeregoAlighieri:2010wuw}, but results will improve with the recent increase of extended radio galaxies samples by new generation radio surveys \cite{Dabhade:2020,Oei:2022fsk,Andernach:2021,Simonte:2022rye,Mostert:2024ued,Vanderwoude_2024}. New methods exploiting polarized star-forming galaxies have also been proposed~\cite{Yin:2024fez, Zhou:2025bvz}. Under such circumstances, Ref.~\cite{Naokawa:2025shr} found universal redshift profiles of cosmic birefringence in low-redshift and proposed a model-independent test for the dynamical DE scenario using such standard crosses.

High-$z$ sources such as the CMB or distant galaxies are needed to detect cosmological time scale variations in the axion field for $m_a\lesssim \mathcal{O}(10^{-25})~\mathrm{eV}$ masses. At higher masses, the DC effect from high-$z$ sources would be washed out by the many oscillations along the line of sight~\cite{Fedderke:2019ajk}.
For these heavier axions, which may make up the whole of dark matter, the field value oscillates rapidly as $a(t)=a_0 \cos ( m_a t + \gamma )$, where $\gamma$ is just a constant here, leaving a periodic modulation of the polarisation angle $\beta(t)$ that can be seen from low-$z$ polarised astrophysical sources~\cite{Fedderke:2019ajk}.  
As an example, Ref.~\cite{Fujita:2018zaj} considers the case of a protoplanetary disk where the light from the central star is linearly polarised in a well-defined concentric distribution after scattering from the dust particles within the protoplanetary disk so that deviations from such distribution can be used to constrain the axion-photon coupling.
Another example is proposed in Ref.~\cite{Basu:2020gsy}, which uses the time delay between two strongly lensed images of a polarised quasar to measure the differential birefringence rotation, requiring no prior knowledge of the source's intrinsic polarisation angle. 
Polarisation measurements of M87$^*$ and SgrA$^*$ with the Event Horizon Telescope have also been proposed as birefringence probes since the dense axion cloud expected to form around supermassive black holes through the superradiance mechanism would enhance the sensitivity to photon birefringence~\cite{Chen:2019fsq,Chen:2022oad,Gan:2023swl}. The polarised emission from the Crab Nebula~\cite{POLARBEAR:2024vel} and individual pulsars~\cite{Caputo:2019tms, Castillo:2022zfl} has also been used to search for birefringence in the mass range $\mathcal{O}(10^{-23})~{\rm eV}\lesssim m_a\lesssim\mathcal{O}(10^{-19})~{\rm eV}$~\cite{Liu:2021zlt}, and polarised measurements from Pulsar Timing Array could be used to place stringent constraints for masses $\mathcal{O}(10^{-27})~{\rm eV}\lesssim m_a\lesssim\mathcal{O}(10^{-19})~{\rm eV}$~\cite{Liu:2021zlt}.
CMB polarisation can also be used to detect this oscillation by measuring the magnitude of the washout effect~\cite{Fedderke:2019ajk,Zhang:2024dmi} or studying time-domain $\beta(t)$ data~\cite{BICEPKeck:2021sbt,SPT-3G:2022ods,POLARBEAR:2023ric}.

\begin{figure}[t!]
    \centering
    \includegraphics[width=0.80\textwidth]{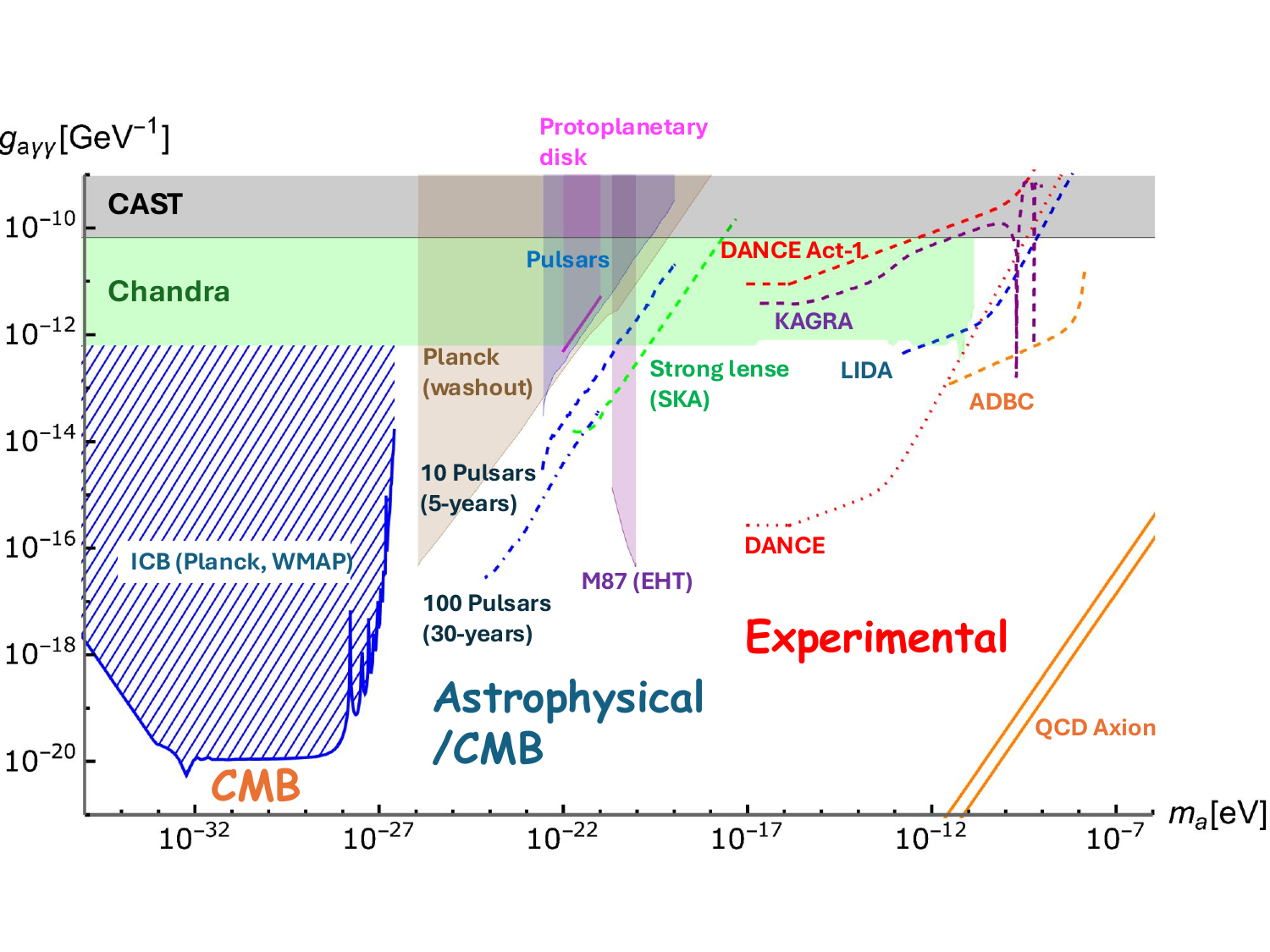}
    \caption{
    Overview of constraints from axion birefringence. 
    Measurements of axion-photon conversion from CAST~\cite{CAST:2004gzq,CAST:2017uph,CAST:2024eil} exclude the gray-shaded region. Blue lines show the over 3$\sigma$ evidence of isotropic cosmic birefringence from \textit{Planck}/WMAP~\cite{Fujita:2020aqt,Fujita:2020ecn}. The blue hatched region represents the parameter space that can explain the isotropic cosmic birefringence signal either through a small abundance of axion or through the phase ambiguity of the rotation angle \cite{Naokawa:2024xhn}. The remaining shaded regions are excluded by studying the protoplanetary disk of AB Aur~\cite{Fujita:2018zaj} (pink), pulsars~\cite{Castillo:2022zfl} (blue), and \textit{Planck}'s washout effect~\cite{Fedderke:2019ajk,Winch:2024mrt,Zhang:2024dmi} (brown), assuming that axions are the dominant component of dark matter. Observations of M87's polarisation with the EHT~\cite{Chen:2019fsq,Chen:2022oad} constrain the axion cloud formed via a black-hole superradiance (purple). Dotted lines show the parameter space within reach of future cosmological/astrophysical searches with, e.g., SKA's strong gravitational lenses~\cite{Basu:2020gsy} (light green), or long-term pulsars observations~\cite{Castillo:2022zfl,Liu:2021zlt} (dark blue), as well as laboratory experiments like DANCE~\cite{Obata:2018vvr,Michimura:2019qxr} (red), KAGRA~\cite{Nagano:2019rbw,Nagano:2021kwx} (purple), LIDA~\cite{Heinze:2023nfb} (blue) and ADBC~\cite{Pandey:2024dcd} (orange). Orange lines show the parameter region of QCD axion models~\cite{Kim:1979if,Shifman:1979if,Dine:1981rt,Zhitnitsky:1980tq}.}
    \label{fig: birefringence plots}
\end{figure}

The rapid polarisation oscillation produced by axions within the $\mathcal{O}(10^{-17})~\text{eV} \lesssim m_a \lesssim \mathcal{O}(10^{-9})~\text{eV}$ mass range is challenging to observe in astrophysical data but can be detected in the laboratory. Optical interferometry is able to provide the high-accuracy measurements of tiny deviations in the phase velocity of laser beams needed for gravitational wave detectors, quantum optomechanics, and precision tests of general relativity. We can also apply this technology to search for the photon's birefringence produced by axion dark matter ~\cite{Melissinos:2008vn, DeRocco:2018jwe,Obata:2018vvr,Liu:2018icu,Nagano:2019rbw,Martynov:2019azm,Nagano:2021kwx}. Among them, resonant ring cavity experiments consisting of four mirrors are more efficient and allow a broader axion-mass range coverage than linear cavity designs, such as Michelson interferometers, since they avoid the parity-flipping of polarisation orientation produced by the reflection onto mirrors~\cite{Obata:2018vvr,Michimura:2019qxr,Liu:2018icu,Martynov:2019azm}. Their prototypes implementing such instrumental configuration recently obtained the first-ever constraints on axion dark matter from the birefringence laboratory search ~\cite{Oshima:2023csb,Heinze:2023nfb,Pandey:2024dcd}. The long arm length of the linear cavities used in gravitational wave interferometers can compensate for the hindrance of the parity-flipping effect, allowing a significant sensitivity to photon birefringence if special optics are installed into their detection ports~\cite{Nagano:2021kwx}. Such an optical system has been installed recently in the KAGRA site~\cite{Michimura:2021hwr}.

Finally, Fig.~\ref{fig: birefringence plots} summarizes current and future constraints on the axion-photon coupling from the measurement of photon birefringence with cosmological, astrophysical, and laboratory searches. 
They cover a wide range of masses, from ultra-light $m_a \sim\mathcal{O}(10^{-33})$~eV axions acting as dark energy to $m_a \gtrsim \mathcal{O}(10^{-22})$~eV dark matter axions, and are competitive with conventional axion-photon conversion experiments/observations.

%% file: WG3/content/Marco_Regis.tex
As discussed in other Sections, there are several mechanisms that can be exploited to look for a radio signal of WISPs. If WISPs make a significant fraction of DM, i.e., if they are present in DM halos with non-relativistic velocities and small dispersion (depending on the specific structure), the associated signals of decay and conversion to photons are given by nearly monochromatic spectral features.
For this kind of searches, the spectral line capability of a telescope is clearly crucial.
On the other hand, WISPs can be produced inside sources, e.g., pulsars, with a continuum spectrum which in turn gives raise to a broad band radio spectrum of emission from the conversion signal. Continuum radio searches are in order for this case. We refer the reader to Sec.~\ref{sec:Conversion_DM_WISPs} for a more detailed discussion on this topic.
Another interesting signal for ultra-light ALPs is given by the birefringence, namely, an achromatic rotation of the plane of polarization of linearly polarized light, due to the left and the right polarization traveling at different velocities in the ALP field (see Sec.~\ref{sec:Gasparotto} for more details).
Birefringence searches thus require precise polarization measurements.
The above signals can show time variability, with cases on very short time-scales requiring excellent time resolution, as for fast radio burst lasting in milliseconds, as well as signals varying on long time scales which require long duration observations. The latter is the case of the birefringence rotation of the pulsar polarization, which oscillate over periods of years.
Moreover, above signals might come both from very compact isolated sources, e.g., neutron stars, requiring excellent angular resolution, and diffuse signals, e.g., from the Galactic halo or the cosmological background, requiring large fields of view (FoV).

It should be clear from this short overview that the relevance of the specific capability of a radio telescope depends on the type of signal we are searching for. 
In the following, we provide a broad but shallow overview of the radio facilities that are already available to conduct WISP searches or will be operating soon.

Table~\ref{tab:tele} summarizes telescope capabilities, computed at a reference frequency, with a fractional bandwidth $\Delta \nu/\nu=0.3$, and for 1h of observation time. One can infer the sensitivity for a different setup through the scaling of the signal-to-noise (SNR) ratio
\begin{equation}
    \mathrm{SNR}=\frac{g\,S\,\sqrt{\Delta \nu\,\Delta t}}{T_{sys}}\;,
\end{equation}
where $g$ is the telescope gain, $S$ is the flux density of the source, $\Delta \nu$ is the bandwidth, $\Delta t$ is the observing time, and $T_{sys}=T_{sky}+T_{inst}$, with $T_{sky}$ being the contribution of the background sky brightness in the direction of the observation and $T_{inst}$ being the noise temperature of the instrument (given by different contributions).

\begin{figure}[t!]
    \centering
    \includegraphics[width=0.9\linewidth]{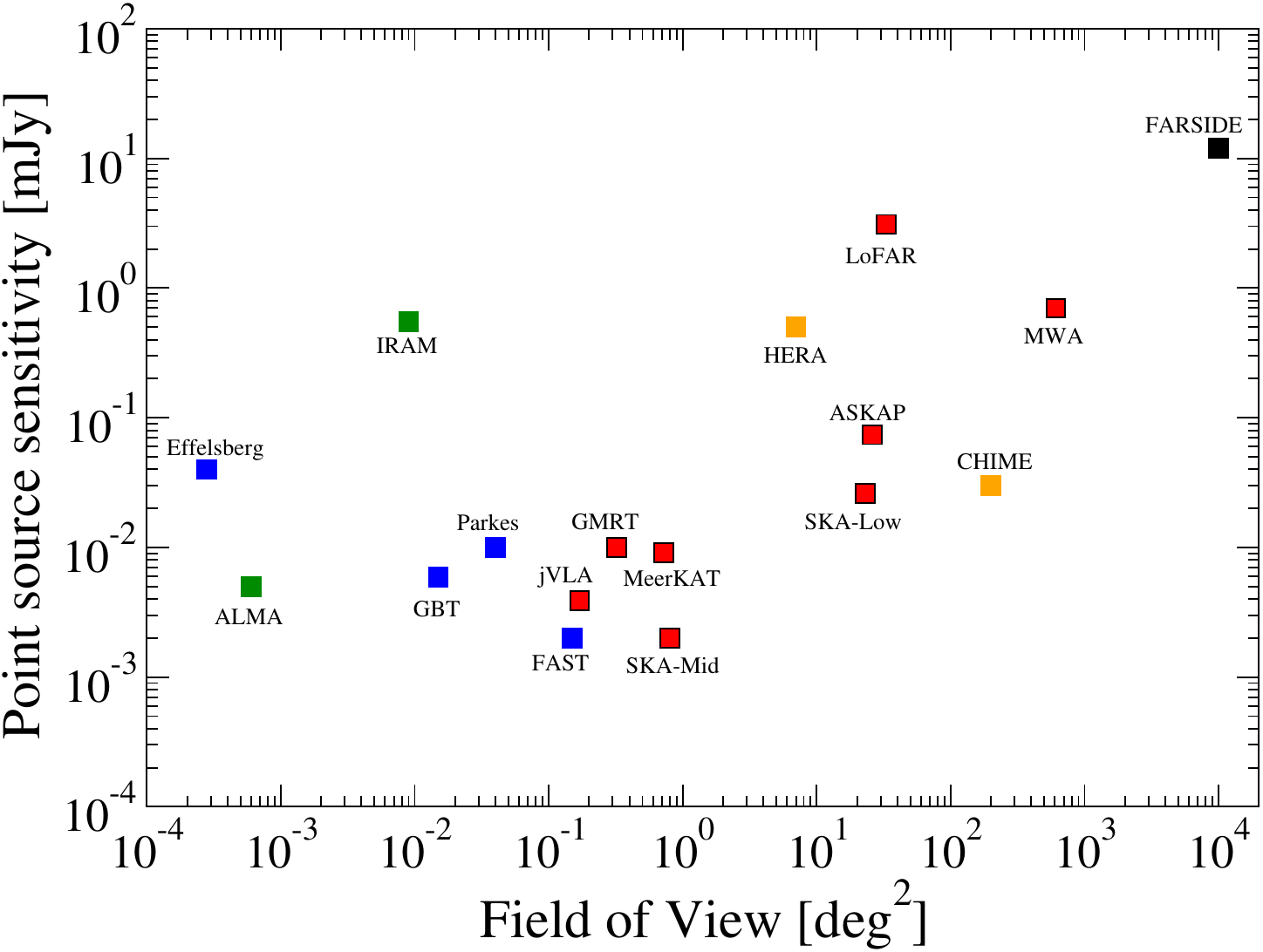}
    \caption{Summary of sensitivity and FoV for a selection of main radio telescopes, computed at the frequency reported in Tab.~\ref{tab:tele}.}
    \label{fig:teleradio}
\end{figure}

For an outlook of the sensitivity of radio telescopes, see Fig.~\ref{fig:teleradio}.

\subsection{Below 10 MHz}
At the lowest radio frequencies, the terrestrial atmosphere is not transparent
because of free electrons in the ionosphere. Indeed, the transmission is not possible when the frequency of the radiation is below the plasma frequency $\nu_p$ of the atmosphere 
\begin{equation}
    \frac{\nu_p}{\mathrm{kHz}}\simeq 9\,\sqrt{\frac{n_e}{\mathrm{cm^{-3}}}}\,,
\end{equation}
where $n_e$ is the electron density of the plasma. The precise value of $n_e$ depends on solar activity, hour of the day, space weather, etc., but the order of magnitude is $n_e\sim10^6\,\mathrm{cm^{-3}}$, which means that below $\sim 10$ MHz it is not possible to perform radio astronomy from Earth.

Satellites would have small collecting area. On the other hand, the possibility to deploy a radio telescope on the Moon is becoming more and more real~\cite{Burns:2021pkx}. In 2024, and despite some issues, the first radio astronomy experiment was performed by the prototype ROLSES~\cite{Burns:2021psj}. In 2026, an upgraded ROLSES-2 experiment and the LuSEE-Night~\cite{Bale:2023} should operate on the Moon surface. These experiments will pave the way for telescopes like FARSIDE, which is currently under development~\cite{Burns:2021pkx}. The technical requirements of the proposed FARSIDE design are reported in Tab.~\ref{tab:tele}.

\begin{sidewaystable}
\renewcommand{\arraystretch}{1.5}
\centering
\begin{tabular}{ccccccccc}
 \hline
 \hline
Telescope & Frequency & Reference &RMS & Angular  & Field of & Frequency & Largest & Ref.   \\

& range & frequency &sens. & resol. & view & channels & scale &\\
 \hline
FARSIDE & $0.1-40$ MHz & 200 kHz & $12$ mJy & $10$ deg & $10^4$ deg$^2$ & 64 &  N/A & \cite{Burns:2021pkx}\\
\hline
FAST & $1.05-1.45$ GHz & 1.25 GHz & $2\,\mu$Jy & $2.9'$ & $0.15$ deg$^2$ & 1024 &  N/A &\cite{2020RAA....20...64J}\\
Green Bank & $0.3-100$ GHz & 1.4 GHz & $5.9\,\mu$Jy & $9'$ & $0.015$ deg$^2$ & $5\times 10^5$ &  N/A & \cite{gbt}\\
Parkes & $0.7-26$ GHz & 1.4 GHz & $0.01$ mJy & $14'$ & $0.04$ deg$^2$ & $2.6\times 10^5$ &  N/A& \cite{parkes}\\
Effelsberg & $0.3-89$ GHz & 10 GHz & $0.04$ mJy & $1.15'$ & 1 arcmin$^2$ & 65536 &  N/A&\cite{effelsberg} \\
 \hline
MeerKAT & $0.58-3.5$ GHz & 1.5 GHz & $9.1\,\mu$Jy & $4"$ & $0.72$ deg$^2$ & $32000$ &  $23'$&\cite{2016mks..confE...1J} \\
ASKAP & $0.7-1.8$ GHz & 0.94 GHz & $74$ $\mu$Jy & $10"$  & $26$ deg$^2$ & 16384 &  0.5 deg& \cite{2021PASA...38....9H}\\
JVLA & $0.74-50$ GHz & 1.5 GHz & $3.9$ $\mu$Jy & $1.3"$ & $0.17$ deg$^2$ & $4\times 10^6$ &  0.32 deg& \cite{vla}\\
GMRT & $0.12-1.46$ GHz & 700 MHz & $10\,\mu$Jy & $4"$ & $0.32$ deg$^2$ & 16384 &  15'& \cite{gmrt} \\
MWA & $80-300$ MHz & 150 MHz & $0.7$ mJy & $1'$ & $610$ deg$^2$ & 3072 &  14 deg& \cite{Tingay:2012ps}\\
LOFAR & $10-240$ GHz & 45 MHz & $3.1$ mJy & $1.1"$ & $33$ deg$^2$ & 512 &9 deg & \cite{LOFAR:2013jil}\\
SKA-Low & $50-350$ MHz & 114 MHz & $0.026$ mJy & $12.3"$ & $23$ deg$^2$ & $10^5$ &  $10.2'$ & \cite{Braun:2019gdo} \\
SKA-Mid & $0.35-15.3$ GHz & 1.4 GHz & $2.0$ $\mu$Jy & $0.6"$ & $0.8$ deg$^2$ & $10^5$ &  $78"$& \cite{Braun:2019gdo}\\
\hline
CHIME & $400-800$ MHz & 600 MHz & $0.03$ mJy & $0.3$ deg & $200$ deg$^2$ & 1024 &  N/A& \cite{CHIME:2022dwe} \\
HERA & $50-250$ MHz & 150 MHz & $0.5$ mJy & $11'$ & $7$ deg$^2$ & 1024 &  7.8 deg&\cite{DeBoer:2016tnn} \\
\hline
ALMA & $31-1000$ GHz & 100 GHz & $5\,\mu$Jy & $0.062'$  & $2$ arcmin$^2$ & 7680 &  66.7"& \cite{alma}\\
IRAM & $80-370$ GHz & 290 GHz & $0.55$ mJy & $11"$ & $0.009$ deg$^2$ & 12000 &  N/A& \cite{iram}\\
\hline
\hline
\end{tabular}
\caption{Selection of radio telescopes with their capabilities. The reference frequency in the third column is the one at which all the subsequent columns are evaluated. The sensitivity is computed for 1h of integration time and a fractional bandwidth $\Delta \nu/\nu=0.3$. The sensitivity to spectral lines can be computed, in first approximation, considering a scaling as $\sqrt{\nu/\Delta \nu}$.
DISCLAIMER: This Table provides approximate values for some specific configuration; for more precise estimates, see the references (last column) and the related sensitivity calculators. }
\label{tab:tele}
\end{sidewaystable}

\subsection{From meters to centimeters}
The regime of wavelengths from above the cm to few tens of meters is the ``traditional" range of radio astronomy where atmospheric absorption does not play a relevant role. There are several facilities currently running or under construction and we separate them in three different classes, that can access to quite different angular scales.  
\subsubsection{Single-dishes}
The simplest radio telescope is a parabolic reflector, namely, a ``single dish" with a feed.  
Considering a telescope with aperture of width $D$ at wavelength $\lambda$, the angular beamwidth is $\theta\approx \lambda/D$. For long wavelengths, the finite size of the dish thus represent a significantly limit to obtain superior angular resolutions. Examples of giant parabolas on a valley are Arecibo ($D=300$ m) and FAST ($D=500$ m), while steerable ones are Effelesberg ($D=100$ m) and Green Bank ($D=100$ m).
There is currently a dozen of single dishes used for radio astronomy around the world.
An overview of an arbitrary selection is reported in Tab.~\ref{tab:tele}.

An important difference between a single-dish and an interferometer is that the former collects emissions from any angular scale, i.e., it is sensitive down to zero spacing in the Fourier plane. Therefore its observations are, on one hand, contaminated by diffuse foreground, but, on the other hand, able to measure diffuse emissions on large scales.

\subsubsection{High-sensitivity Interferometers}
An interferometer is an array of radio telescopes. 
The sensitivity is determined by the sum of the collecting areas of each antenna, with the advantage that the instrumental noise typically does not severely affects observations, since the noise in each antenna is uncorrelated and averages out when combining the different elements of the array.
The angular resolution is set by the longest baseline between the two antennas, while the largest spatial scale is determined by the shortest baseline.
Therefore, they represent the best radio technique for high angular resolution studies, but interferometers need to be complemented by single-dishes if searching for extended emissions. 
Thanks to the SKA and its precursors, listed in Tab.~\ref{tab:tele}, radio interferometry is currently undergoing a very exciting era.

Connecting different telescopes and orbiting antennas across the world provides the longest possible baselines, and thus the highest achievable angular resolution. The technique is known as very long baseline interferometry (VLBI), with a few projects currently running, and capable to observe scales well below the mas resolution, see, e.g., Ref.~\cite{Venturi:2020vfm}.

\subsubsection{Large-FoV Interferometers}
In recent years, as a sort of compromise between single-dishes and high-sensitivity interferometers, radio interferometers with large instantaneous field of view have been developed~\cite{CHIME:2022dwe,DeBoer:2016tnn}. They can provide surveys of large fraction of the sky in a short observation period, with possibly excellent frequency and time resolutions. They currently cover sub-GHz frequencies since originally devised for intensity mapping of the 21-cm line, but future similar experiments at mm wavelength looking for CO or CII lines are foreseen.

\subsection{Millimeter telescopes}
The absorption of mm waves depends on the amount of water vapour in the air, and requires telescopes to be located on high mountains in desert regions, or at the South Pole, or on-board of satellites. In this contribution we do not discuss CMB-oriented experiments, which can be however relevant for WISP searches at mm wavelengths (see, e.g., \cite{Chang:2022tzj} for an overview).
In Tab.~\ref{tab:tele}, the capabilities of the ALMA interferometer and IRAM 30m single-dish telescope are reported.

%% file: WG3/content/Julia_Vogel.tex
\subsection{Introduction}

X-ray optics play a crucial role in the search for weakly interacting slim particles (WISPs), particularly axions and axion-like particles (ALPs). The importance of this technology in the field arises from two key aspects: their ability to focus photons produced through the inverse Primakoff effect and their capability to significantly enhance the signal-to-noise ratio (SNR) of axion/ALP search experiments. This Primakoff mechanism, which describes the conversion of  axions into photons and vice versa in the presence of strong electromagnetic fields, is central to the majority of axion search strategies, since it is a generic feature of axion models.

The interplay between astrophysical magnetic fields and potential axion-photon conversion presents a compelling case for X-ray telescopes. In astrophysical environments, we encounter a trade-off between magnetic field strength and distance scales: while the magnetic fields may be relatively weak (typically of the order of micro-gauss), the enormous distances, over which these fields extend, create favorable conditions for axion-photon conversion. This makes X-ray telescopes particularly sensitive to axions and ALPs with very small masses ($<1$ eV), as the conversion probability depends on both the magnetic field strength and the path length through the magnetized region.

Nevertheless, also in ground-based searches for axions produced through the Primakoff effect in the solar core, the potential signal is expected predominantly in the X-ray band, with energies ranging from approximately $1$ to $10$ keV. This energy range aligns well with the capabilities of modern X-ray astronomy observatories, whose advanced optics can efficiently focus and detect such photons in what is often referred to as the soft X-ray band. The ability to concentrate these potential axion-induced X-rays onto small detector areas significantly improves the axion sensitivity by reducing background levels.

The development of increasingly sophisticated X-ray optics has thus become instrumental in advancing axion searches, enabling new constraints on axion-photon coupling strengths and mass ranges that complement traditional ground-based experiments. This technological progress, combined with the unique characteristics of astrophysical environments, has established X-ray astronomy as a vital tool in the ongoing search for these elusive particles.

\subsection{X-ray optics technologies}

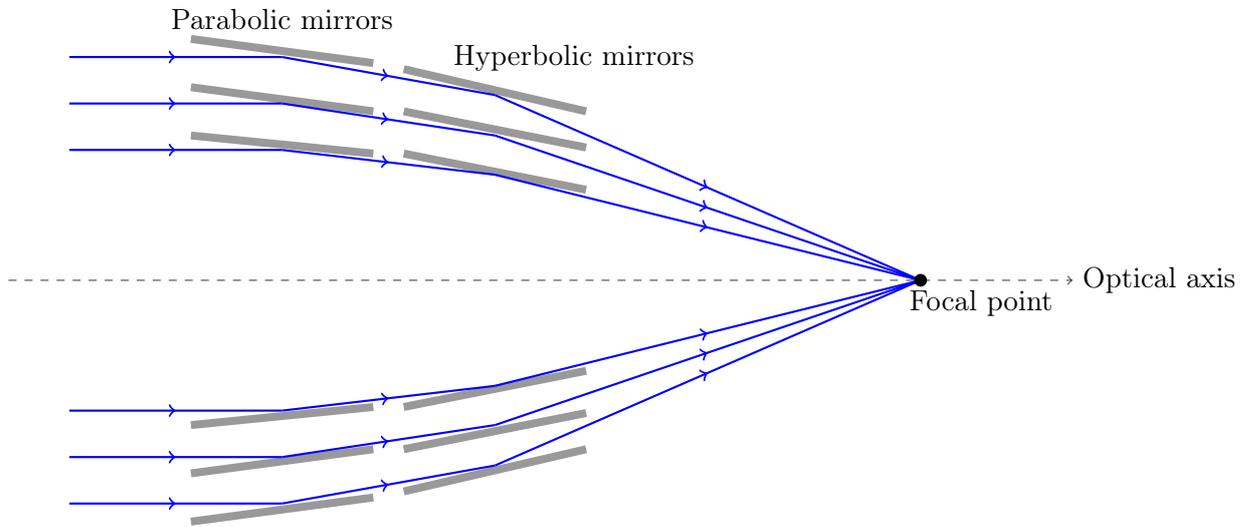
\begin{figure}[t!]
    \centering
    \begin{tikzpicture}[scale=0.8]
        \tikzset{
            ray/.style={thick, blue, 
                decoration={
                    markings,
                    mark=at position 0.5 with {\arrow{>}}
                },
                postaction={decorate}
            },
            mirror/.style={line width=3pt, gray!80}  
        }
        \draw[->, dashed, darkgray] (-1,0) -- (16.5,0) node[right, black] {Optical axis};
        \draw[mirror] (2,4) -- (5,3.6);
        \draw[mirror] (2,3.2) -- (5,2.8);
        \draw[mirror] (2,2.4) -- (5,2.1);
        \draw[mirror] (2,-4) -- (5,-3.6);
        \draw[mirror] (2,-3.2) -- (5,-2.8);
        \draw[mirror] (2,-2.4) -- (5,-2.1);
        \draw[mirror] (5.5,3.5) -- (8.5,2.8);
        \draw[mirror] (5.5,2.8) -- (8.5,2.2);
        \draw[mirror] (5.5,2.1) -- (8.5,1.5);
        \draw[mirror] (5.5,-3.5) -- (8.5,-2.8);
        \draw[mirror] (5.5,-2.8) -- (8.5,-2.2);
        \draw[mirror] (5.5,-2.1) -- (8.5,-1.5);
        \draw[ray] (0,3.7) -- (3.5,3.7);
        \draw[ray] (0,2.93) -- (3.5,2.93);
        \draw[ray] (0,2.16) -- (3.5,2.16);
        \draw[ray] (0,-3.7) -- (3.5,-3.7);
        \draw[ray] (0,-2.93) -- (3.5,-2.93);
        \draw[ray] (0,-2.16) -- (3.5,-2.16);
        \draw[ray] (3.5,3.7) -- (7,3.07);
        \draw[ray] (3.5,2.93) -- (7,2.4);
        \draw[ray] (3.5,2.16) -- (7,1.75);
        \draw[ray] (3.5,-3.7) -- (7,-3.07);
        \draw[ray] (3.5,-2.93) -- (7,-2.4);
        \draw[ray] (3.5,-2.16) -- (7,-1.75);
        \draw[ray] (7,3.07) -- (14,0);
        \draw[ray] (7,2.4) -- (14,0);
        \draw[ray] (7,1.75) -- (14,0);
        \draw[ray] (7,-3.07) -- (14,0);
        \draw[ray] (7,-2.4) -- (14,0);
        \draw[ray] (7,-1.75) -- (14,0);
        \fill (14,0) circle (3pt) node[below, xshift=0.8cm] {Focal point};
        \node[above] at (3.5,4) {Parabolic mirrors};
        \node[above] at (8.3,3.3) {Hyperbolic mirrors};
        \end{tikzpicture}
        \caption{Schematic of a typical Wolter I geometry. X-rays enter the telescope from the left side and can be focused onto a detector placed in the focal plane. 
        }
        \label{fig:wolterI}
\end{figure}
In the early days of X-ray astronomy, instruments like \textit{UHURU}~\cite{1972ApJ...178..281G} relied on non-focusing, collimated detectors, providing only moderate sensitivity and spatial resolution. The development of true focusing X-ray telescopes marked a transformative step forward. Unlike visible wavelengths, X-ray photons are readily absorbed at typical (i.e. large) incidence angles, thus requiring reflections at very shallow grazing angles. This condition enables total external reflection from high-density materials (e.g. Ir or Au), whose critical angle depends on the electron density and the photon wavelength. By applying such coatings via vacuum deposition, modern X-ray mirrors achieve high reflectivity and deliver sharp imaging capabilities. These advances, based on carefully crafted mirror geometries and multilayer coatings, have paved the way for large fields of view, greater sensitivity, and higher angular resolution in contemporary X-ray observatories than achieved in the early missions.

Today’s instruments mostly employ grazing-incidence mirror geometries, notably the Wolter-I configuration~\cite{wolter1952spiegelsysteme}, which utilizes nested paraboloid and hyperboloid mirror shells to focus X-rays onto compact focal plane detectors~(see Fig.~\ref{fig:wolterI}). This approach reduces background by confining the source photons to a smaller detector area, thereby improving the signal-to-noise ratio~\cite{Christensen2024}. Efficient focusing requires mirrors composed of extremely smooth surfaces — in the order of a few angstroms — to minimize scattering of high-energy photons. The mirrors are typically coated with metals of high atomic number, such as iridium or gold, to maximize reflectivity up to energies of a 10-15 keV~\cite{giacconi1960telescope}. To extend the energy range beyond the limit of single-layer coatings, advanced missions like the NASA's Nuclear Spectroscopic Telescope Array (\textit{NuSTAR}) incorporate multilayer coatings composed of alternating materials (e.g., Pt/C, W/Si), thereby increasing the critical angle for reflection at higher energies and extending effective sensitivity up to nearly 80 keV~\cite{NuSTAR:2013yza}. Multilayers can also be employed to tailor and maximize the optical performance to match expected spectra, as in the case of solar axions.

Improved angular resolutions, as achieved by \textit{Chandra} (0.5 arcseconds), stem from meticulous polishing and coating processes. In contrast, segmented mirror technologies (such as those in the \textit{NuSTAR} telescopes), which are lighter and easier to fabricate in larger effective areas, accept a modest compromise in angular resolution (e.g., $\sim$18 arcseconds FWHM) to gain unprecedented sensitivity in the hard X-ray regime. Instruments like \textit{Swift}-XRT \cite{SWIFT:2005ngz} use a simpler single-layer mirror approach, which is sufficient for softer (0.2--10~keV) energies and excel in rapid localization of transient X-ray sources.

From a design and operational standpoint, requirements for axion searches and traditional X-ray astrophysics differ. While astronomical missions optimize collecting area, energy band coverage, and angular resolution for diverse cosmic sources, experiments aiming to detect hypothetical particles (e.g., axions) may focus on minimizing background and employing narrow energy bands or tailored multilayers to optimize reflectivity at very specific energies. These specialized configurations often require adjustments to the mirror geometry, coatings, and detection strategies, building upon the same fundamental technologies yet driven by distinct scientific objectives.

\subsubsection{Technology transfer of multilayers}

Coating techniques developed to produce multilayer X-ray optics may have further applications in WISP searches. Haloscope experiments have been profiting from recent advances in coating developments to increase the performance of resonant cavities in axion and dark photon searches~\cite{Posen:2022tbs}. Single thin layers of superconducting materials have the potential to increase the quality factor $Q$ of these cavities, which in turn allows for faster scans of the parameter space. The challenge for haloscope-type experiments is mostly to maintain the properties of the microwave cavities when placed inside a strong magnetic field as needed to convert axions into detectable photons. 

Regarding multilayer coatings, several proposals of  multilayer dielectric haloscopes have been recently designed and operated successfully~\cite{Baryakhtar:2018doz,Manenti:2021whp}. They use a pattern of layers with different thicknesses to optimize the conversion probability inside the almost periodic dielectric medium, thus promising significant increases in sensitivity for next-generation searches.

\subsection{Ground based applications in axion searches}

Experimental axion searches mostly rely on the coupling of axions to photons. Other model-dependent couplings, such as axion-electron axion-nucleon couplings, could also be present and several experiments use these in their search efforts.

To detect axions and ALPs arriving to Earth from astrophysical sources, experiments must provide a strong, transverse magnetic field, in which the inverse Primakoff effect can occur, enabling the back-conversion of axions. Solar axions, produced in the core of the Sun from blackbody photons of a few keV in the strong fields of the solar plasma, are expected to convert back to X-ray photons in laboratory experiments dubbed axion helioscopes. Typical energies of photons from reconversion of solar axions are in the 1-10 keV range, peaking around 3 keV, following the thermal distribution of photons in the solar core. However, the expected signal is extremely low, requiring experiments with low background but high signal-to-noise ratio. This makes X-ray focusing techniques particularly valuable for solar axion searches, as they can concentrate the converted X-ray photons from a large detection area onto a much smaller detector surface, significantly improving the signal-to-noise ratio by reducing the effective detector area exposed to background events while preserving the signal. 

Observatories such as CAST~\cite{Zioutas:1998cc} and the future BabyIAXO~\cite{IAXO:2020wwp} and IAXO~\cite{Armengaud:2014gea,IAXO:2019mpb} exemplify these experiments. The state-of-the-art is CAST, which concluded its physics runs in 2021 and set the most stringent limit to the solar axion-photon coupling, $g_{a\gamma}<5.8\times10^{-11}$\,GeV${^{-1}}$ for axion masses below $0.02\,$eV~\cite{CAST:2024eil}, thanks to the use of X-ray optics optimized for solar axion searches along with latest generations of ultra-low background detectors. CAST has also investigated other axion production channels such as the axion-electron coupling~\cite{Barth:2013sma}, as well as high-energy axions using calorimeters~\cite{CAST:2009klq}, dark photons~\cite{Redondo:2015iea} or chameleons~\cite{CAST:2015npk}. It has also contributed to scanning other regions of the parameters space as a haloscope searching for dark matter axions~\cite{Adair:2022rtw, CAST:2020rlf}. 

\begin{figure}[t!]
    \centering
    \includegraphics[width=0.5\linewidth]{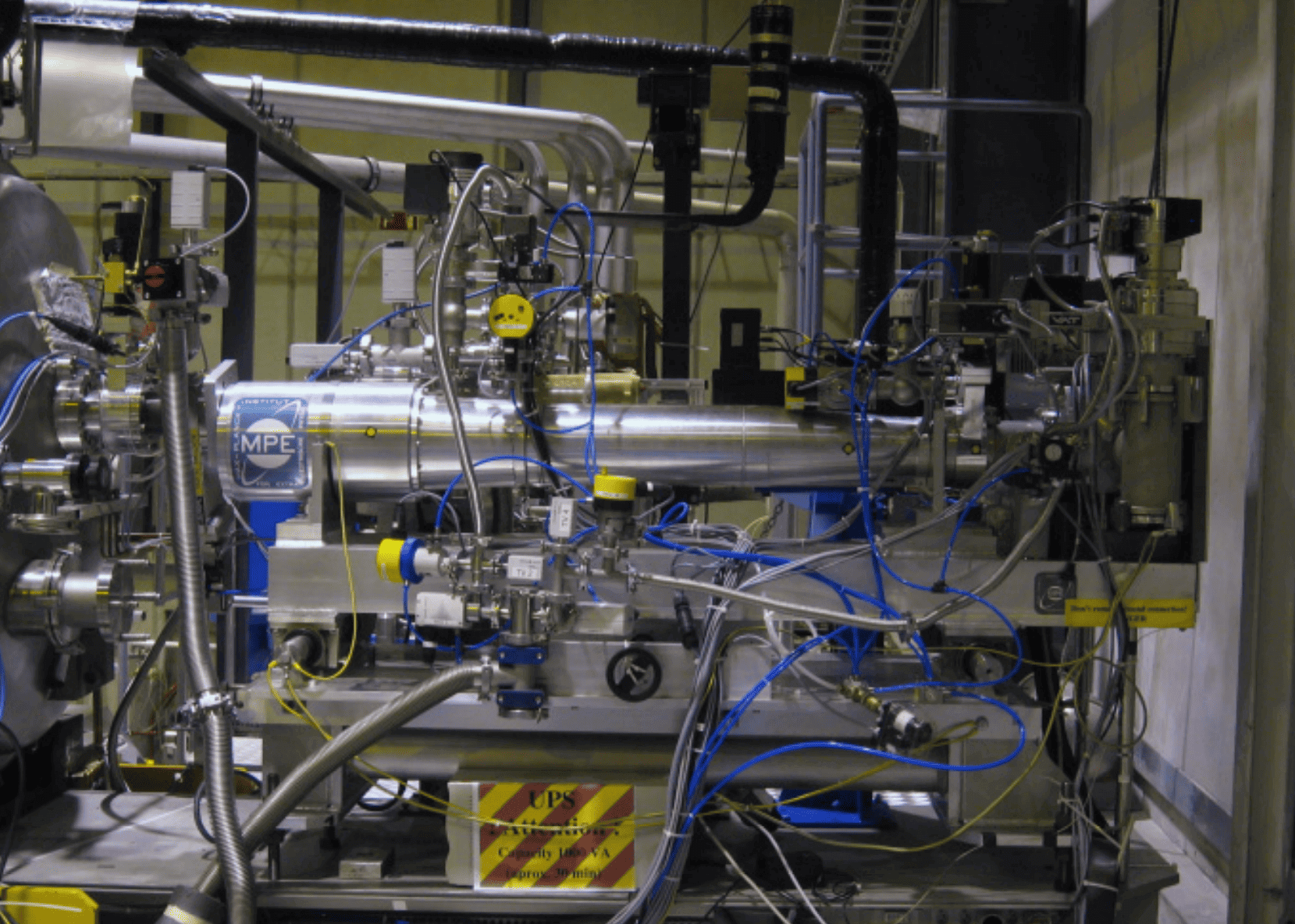}
    \includegraphics[width=0.38\linewidth]{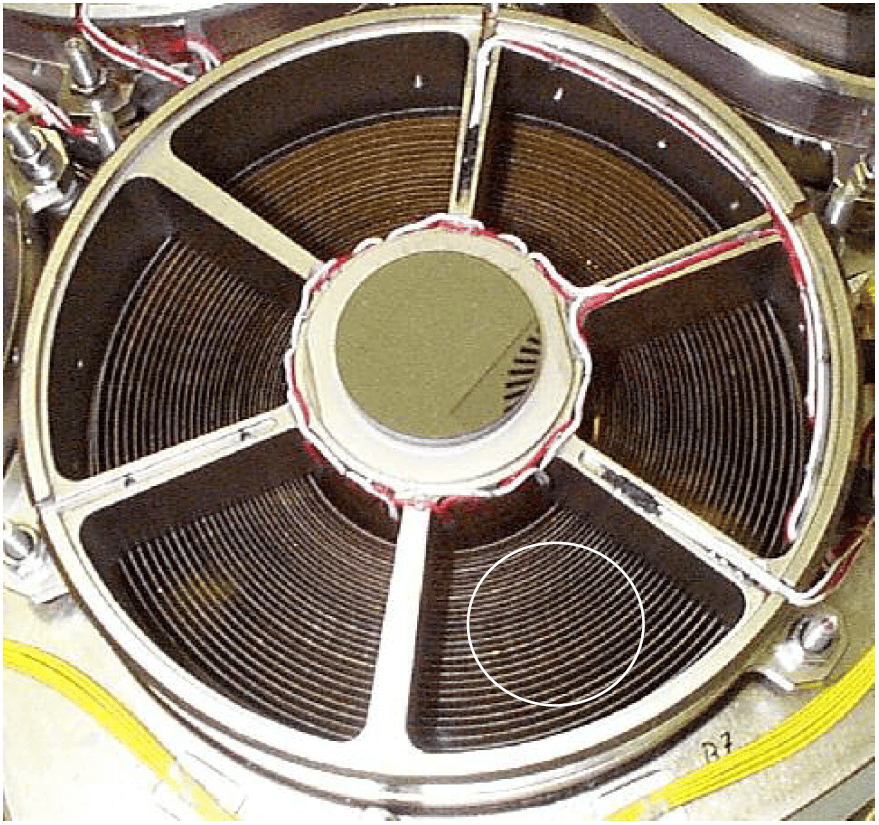}
    \caption{The ABRIXAS X-ray telescope in CAST. {\it Left.} telescope installed at CAST. The metallic tube is the mirror housing and it is coupled to the magnet bore on the left and to the detector on the right. {\it Right.} Image of the telescope mirror system. The area exposed to the magnet bore is indicated by a white circle. Images adapted from Ref.~\cite{Kuster:2007ue}.}
    \label{fig:abrixas}
\end{figure}

\begin{figure}[t!]
    \centering
    \includegraphics[width=0.8\linewidth]{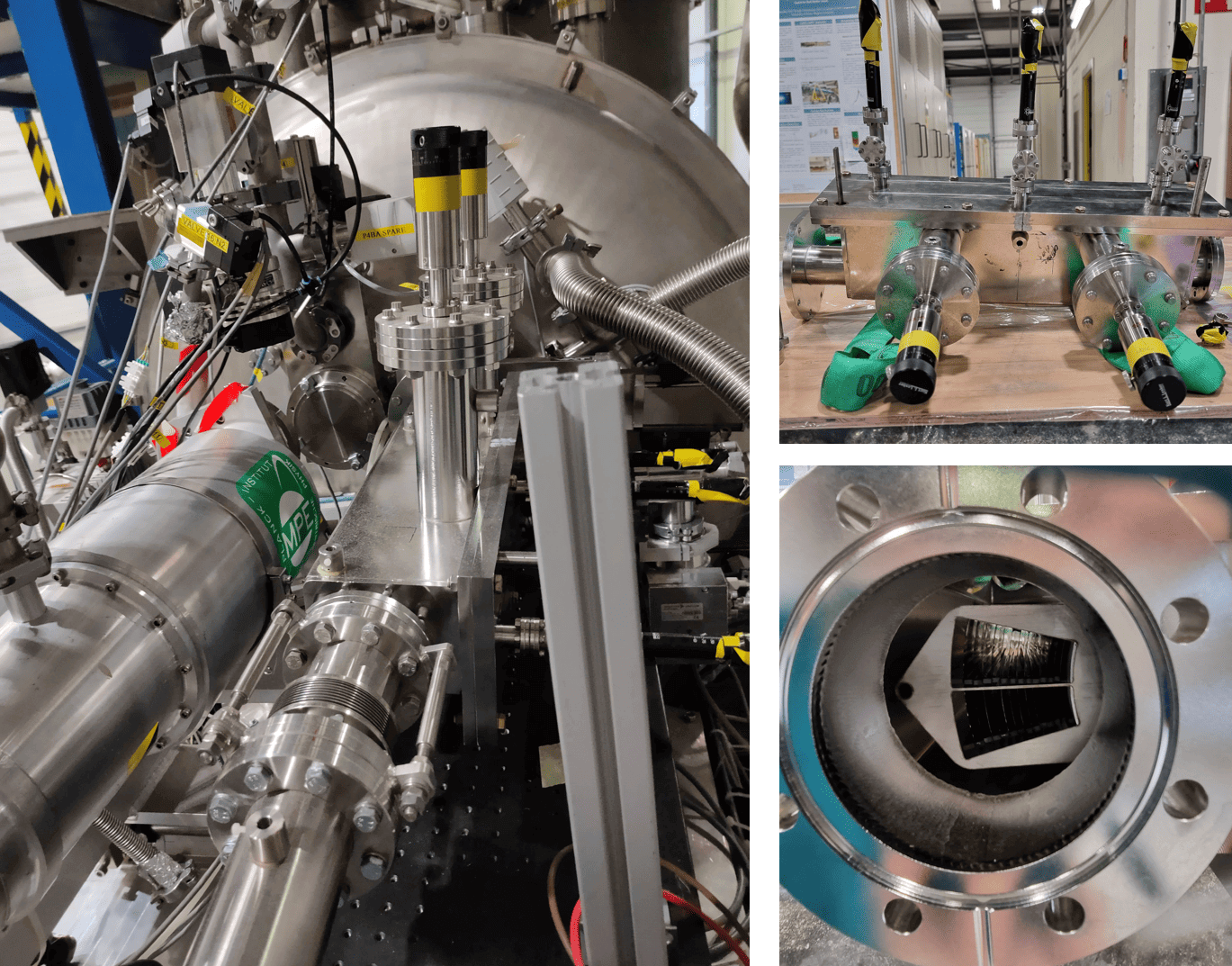}
    \caption{X-ray focusing system of the CAST experiment. {\it Left.} Integration of the X-ray telescope in CAST, showing the optics housing mounted between the magnet bore (rear) and the detector beamline (front). {\it Upper right.} View of the telescope housing with vacuum flanges. {\it Lower right.} Direct view of the optics segment through the vacuum pipe, which matches the 43-mm diameter of the magnet bore. Original pictures taken by the author on site. Credits: CAST/IAXO collaboration.}
    \label{fig:LLNL}
\end{figure}

CAST employed two X-ray telescopes during its operational lifetime. The first was a spare Wolter-I module from the ABRIXAS mission~\cite{ABRIXAS-I:1998,ABRIXAS-II:1998}, consisting of 27 nested gold-coated mirror shells with a 1.6~m focal length. Due to the 43~mm magnet bore aperture, only a single sector of the telescope was exposed (see Fig.~\ref{fig:abrixas}). 
In 2014, a second telescope specifically optimized for solar axion searches was installed~\cite{Aznar:2015iia}. This optic, designed to maximize throughput at the expected solar axion flux peak ($\sim 3$~keV), utilized NuSTAR fabrication techniques but comprised a 13-layer Wolter-I segment rather than full-revolution mirror modules (see Fig.~\ref{fig:LLNL}). This pathfinder demonstrated the viability of segmented glass optics for helioscopes, paving the way for the BabyIAXO and IAXO telescope designs. For further details on the CAST experiment, see Sec.~\ref{sec:CAST-MM} in WG4.

CAST first telescope was a spare module built for the X-ray mission ABRIXAS~\cite{ABRIXAS-I:1998,ABRIXAS-II:1998}. It is a Wolter I type telescope consisting on 27 nested, gold-coated mirror shells. It has a focal length of 1.6 m and its diameter ranges from 76 mm to 163 mm, with the aperture  divided into six azimuthal sectors. The aperture of the CAST bore is 43 mm, so the telescope was positioned off-axis such that only a circular section of one of the sectors is exposed~(see Fig. \ref{fig:abrixas}). The chosen sector was the one with the best measured energy-dependent effective area~\cite{Kuster:2007ue}.
In 2014 a new telescope specifically designed for axion searches is installed in CAST. This telescope is optimized to maximize throughput at energies where the peak of the solar axion flux is expected ($\sim 3$\,keV)~\cite{Aznar:2015iia}. 
Its design and fabrication utilized the same techniques and tools developed for the NASA's NuSTAR mission, but consisted of a segment of a 13-layer Wolter I optic with a focal length $f$ of 1500 m rather than the full-revolution NuSTAR mirror modules (133 layers, $f=10.14$ m). An important feature for rare-event searches is the possibility of nesting mirror shells, since it increases the reflective surface available and therefore the overall telescope efficiency, which is essential for experiments where low signal is expected. The NuSTAR-like telescope placed in CAST during 2014-2021 is shown in Fig.~\ref{fig:LLNL}.

\begin{figure}[t!]
    \centering
    \includegraphics[width=1\linewidth]{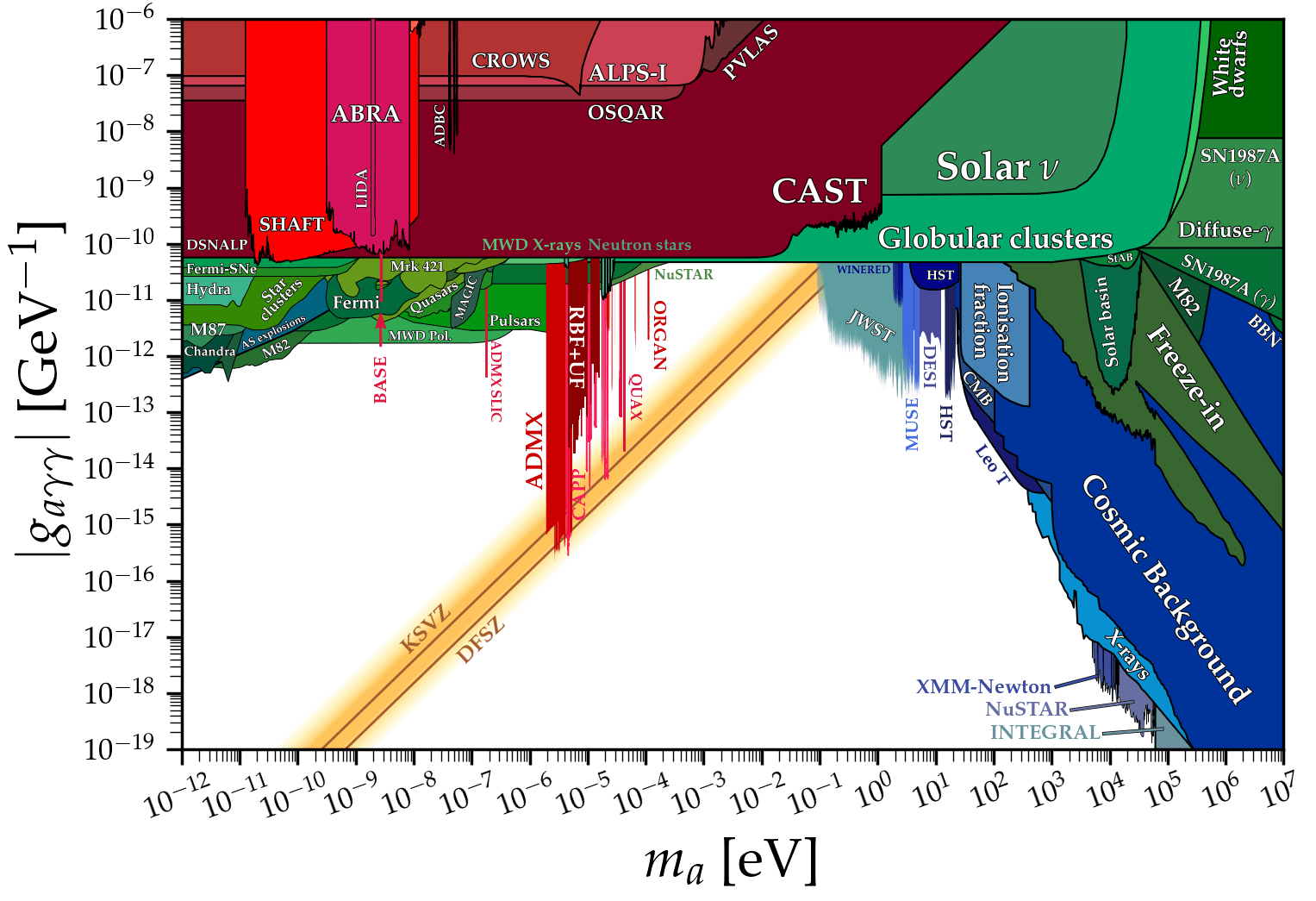}
    \caption{The axion-photon coupling $(g_{a\gamma})$ versus axion mass $(m_a)$ parameter space. Excluded regions from various experiments and astrophysical bounds are shown. The yellow band represents the QCD axion models, with KSVZ and DFSZ being specific theoretical predictions. The projected sensitivities of the future experiments BabyIAXO, and IAXO are shown with dashed lines, demonstrating their potential to probe unexplored axion parameter space with significantly better sensitivities than CAST. This plot was made with the help of the AxionLimits repository \cite{AxionLimits}. 
    }
    \label{fig:panorama_2024}
\end{figure}

The future IAXO experiment~\cite{Armengaud:2014gea,IAXO:2019mpb} will feature eight 60-cm diameter bores, each equipped with X-ray telescopes optimized for solar axion searches. IAXO will probe over an order of magnitude beyond CAST in $g_{a\gamma}$ sensitivity (see Fig.~\ref{fig:panorama_2024}). For detailed experiment specifications, see Sec.~\ref{sec:IAXO} in WG4.

The future experiment IAXO (International Axion Observatory) 
\cite{Armengaud:2014gea,IAXO:2019mpb} is a fourth generation helioscope that will build upon all the knowledge and experience acquired with CAST. It will feature a purpose-built 20 m long superconducting magnet with a magnetic field of up to 5.4 T, vastly improving the figure of merit of the CAST magnet. The magnet will have 8 conversion bores of 60 cm diameter, each equipped with X-ray telescopes and low background detectors. As shown in Fig.~\ref{fig:panorama_2024}, IAXO will probe large unexplored ALP space, QCD axion models in the meV to eV mass band and astrophysically hinted regions. Compared to CAST, it will be more than 1 order of magnitude more sensitive to the axion-photon coupling $g_{a\gamma}$ for a wide range of axion masses.
As a preliminary step towards IAXO, the BabyIAXO experiment \cite{IAXO:2020wwp} has been proposed as a technological prototype. It builds on mature technology developed, implemented and tested in CAST, which used a ~9 m long magnet with a single 43-cm aperture bore and a magnetic field of up to 9 T. The experiment will feature two 70-cm aperture bores, each operating at a magnetic field strength of 2.5 T, and each will use an X-ray focusing system. BabyIAXO will be able to produce relevant physical outcome as it will probe part of the QCD axion band as well as some of the astrophysical hints (see Fig.~\ref{fig:panorama_2024}), and it is expected to improve the signal-to-noise ratio of CAST by a factor $>10^2$.

For the BabyIAXO project, specific requirements drive the design. The system needs maximized throughput efficiency between 40-60\%, which can be enhanced using multilayer coatings optimized for the region of interest and low energy response. The focal spot area must be minimized to 0.2 cm$^2$ (radius less than 2.5 mm) to optimally match and enhance the detector performance, though only modest spatial resolution at the arcminute level is needed to cover the solar disc. 

\begin{figure}[t!]
    \centering
    \includegraphics[width=0.35\linewidth]{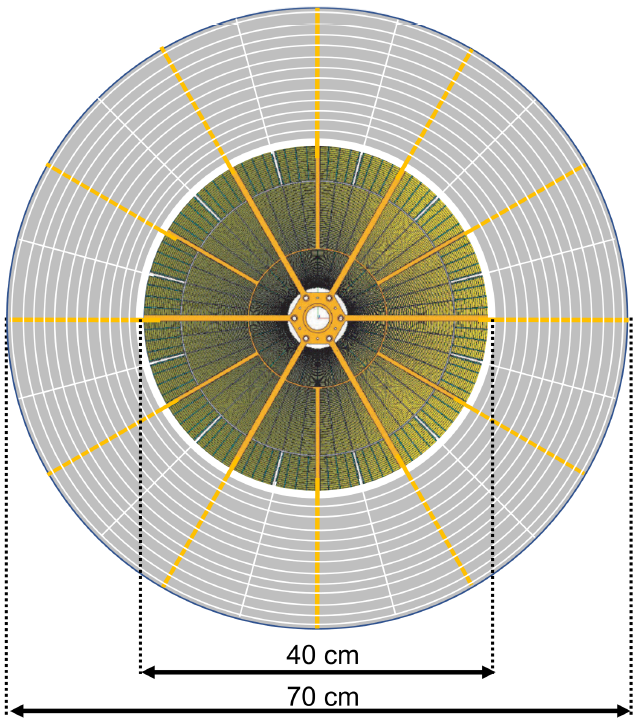}
    \hspace{1cm} 
    \includegraphics[width=0.35\linewidth]{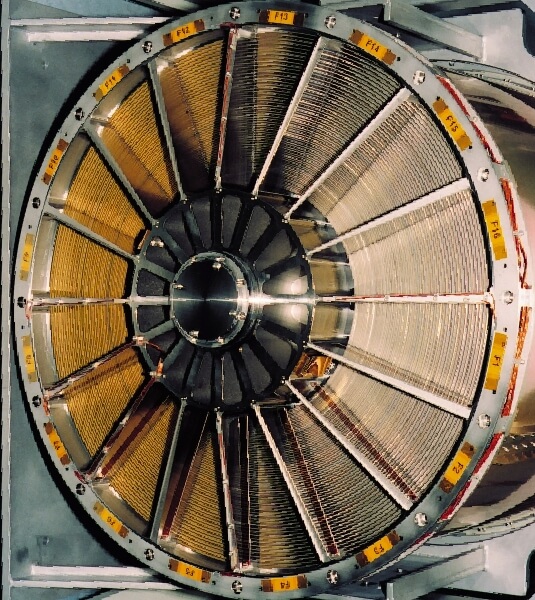}
    \caption{Proposed X-ray optics for the BabyIAXO experiment. \emph{Left:} Schematics of the custom optics showing the differentiated inner core and the corona. Figure taken from Ref.~\cite{IAXO:2020wwp} with permission. \emph{Right:} Image of the XMM optics. Credit: ESA/XMM~\cite{ESA_XMM_Mirrors}.}
    \label{fig:bIAXO_optics}
\end{figure}

Segmented glass optics (or aluminum foil alternatives) have emerged as an ideal solution for BabyIAXO, building on their successful implementation in major space missions like NuSTAR, XRISM, and Athena.
The baseline configuration comprises a custom IAXO optic~(Fig. \ref{fig:bIAXO_optics} left) covering one bore, and a flight spare module of the XMM telescope~(Fig. \ref{fig:bIAXO_optics} right) covering the second bore. 
The XMM~\cite{XMM:2001haf} is a true Wolter-I telescope made using replication technology, where each mirror shell is formed by being placed on top of a highly precise and polished mandrel to replicate its shape. Its specifications closely align with IAXO's design requirements~\cite{Irastorza:2011}. 
The development of custom IAXO optics features an innovative design that combines two different technologies, one used for the core optic and another one for the outer telescope part. The core optic, which will cover the inner $\sim 40$ cm diameter, uses hot-slumped, multi-layer-coated segmented glass technology, similar to that used in NuSTAR. A NuSTAR-like pathfinder has already undergone successful testing at CAST, demonstrating the feasibility of this approach. The outer corona, which completes the full 70-cm diameter optic, employs cold-slumped glass optics (CSGO)~\cite{Civitani:2016}, a technology initially developed in the frame of the Athena mission~\cite{ATHENA:2022hxb}.  This technique involves shaping thin glass sheets at room temperature against a precisely figured mandrel, followed by a thermal cycle to release internal stresses, enabling cost-effective production of large-area mirror segments while maintaining the required optical performance. For further details on the BabyIAXO experiment, including magnet specifications and sensitivity projections, see Sec.~\ref{sec:BabyIAXO}.

\subsection{Space-based X-ray telescopes for WISP searches}
High-energy astrophysical environments provide ideal conditions for producing ALPs through their coupling with nucleons, photons and electrons. ALPs can be detected by X-ray telescopes in two primary ways: either through their decay into photons before reaching the detector or via their conversion into photons by an external magnetic field, typically the galactic or solar magnetic field. On the other hand, one can use precision X-ray polarization studies to place bounds on the ALP parameter space. Each X-ray telescope is unique, and depending on the specific case under study, one may select one instrument over another. A selection of the most important characteristics of some of the X-ray telescopes mentioned in this review is provided in Table~\ref{tab:telescopes}.

Hereafter, we will review the most recent and strongest bounds on the parameter space for ALPs, for the cases in which X-ray telescopes were used. Then, we will summarize the most important achievements in searches for other WISPs candidates.

\begin{sidewaystable}
\renewcommand{\arraystretch}{1.5}
\centering
\begin{tabular}{cccccc}
\hline
\hline
Telescope &\,\,\,Energy range\,\,\,& Energy resolution & Angular resolution & FOV & Ref.\\
\hline 

\begin{tabular}[c]{@{}c@{}}Chandra \\ (ACIS)\end{tabular}
 & 0.1--10\,keV
 & \begin{tabular}[c]{@{}l@{}}95\,eV @ 1.49\,keV\\150\,eV @ 5.9\,keV\end{tabular}
 & 0.4'' / 1.0''
 & \begin{tabular}[c]{@{}l@{}}$16.9' \times 16.9'$ 
 \end{tabular}
 & \cite{Weisskopf:2000tx}
\\ \hline

\begin{tabular}[c]{@{}c@{}}XMM\\ (EPIC)\end{tabular}
 & 0.15--15\,keV
 & 70\,eV @ 1\,keV
 & 6.0'' / 15''
 & $30' \times 30'$
 &\cite{XMM:2001haf}
\\ \hline

\begin{tabular}[c]{@{}c@{}}Swift \\ (XRT)\end{tabular}
 & 0.2--10\,keV
 & 135\,eV @ 6\,keV
 & 7.0' / 18'
 & $23.6' \times 23.6'$
 & \cite{SWIFT:2005ngz}
\\ \hline

eROSITA
 & 0.2--8\,keV
 & 80\,eV @ 1\,keV
 & 15'' / ~--
 & $1.03^\circ$ (diameter) 
 & \cite{eROSITA:2020emt}
\\ \hline

IXPE$^{*}$
 & 2--8\,keV
 & 598\,eV @ 6\,keV
 & --~ / 30''
 & $12.9' \times 12.9'$
 & \cite{weisskopf2016imaging}
\\ \hline

NuSTAR
 & 3--79\,keV
 & \begin{tabular}[c]{@{}l@{}}400\,eV @ 10\,keV\\0.9\,keV @ 60\,keV\end{tabular}
 & 18'' / 56''
 & $13' \times 13'$
 & \cite{NuSTAR:2013yza}
\\ \hline

\begin{tabular}[c] {@{}c@{}}INTEGRAL\\SPI \end{tabular} 
 & \begin{tabular}[c]{@{}c@{}}20\,keV--\\8\,MeV \end{tabular}
 & \begin{tabular}[c]{@{}c@{}}2.2\,keV FWHM\\@ 1.33\,MeV\end{tabular} 
 & $2.5^\circ$ / $3.0^\circ$
 & \begin{tabular}[c]{@{}c@{}}14\textdegree\ flat-to-flat\\16\textdegree\ corner-to-corner\end{tabular}
 & \cite{Winkler:2003nn}\\
\hline
\hline
\end{tabular}

\begin{flushleft}
\small 
\(^\dagger\)~First value is the FWHM and the second one is the HPD. \\
\small 
\(^*\)~IXPE is the only telescope in the list that is able to measure X-ray polarization.
\end{flushleft}
\caption{Summary of the most important characteristics of the X-ray Telescopes mentioned in the text: 
Chandra, XMM-Newton, Swift, eROSITA, IXPE, NuSTAR, INTEGRAL.
}
\label{tab:telescopes}
\end{sidewaystable}

\subsubsection{Constraints from X-ray Telescopes}

\begin{figure}[t!]
  \centering
  \begin{tikzpicture}
    \node[anchor=center] (main) 
      {\includegraphics[width=0.55\textwidth]{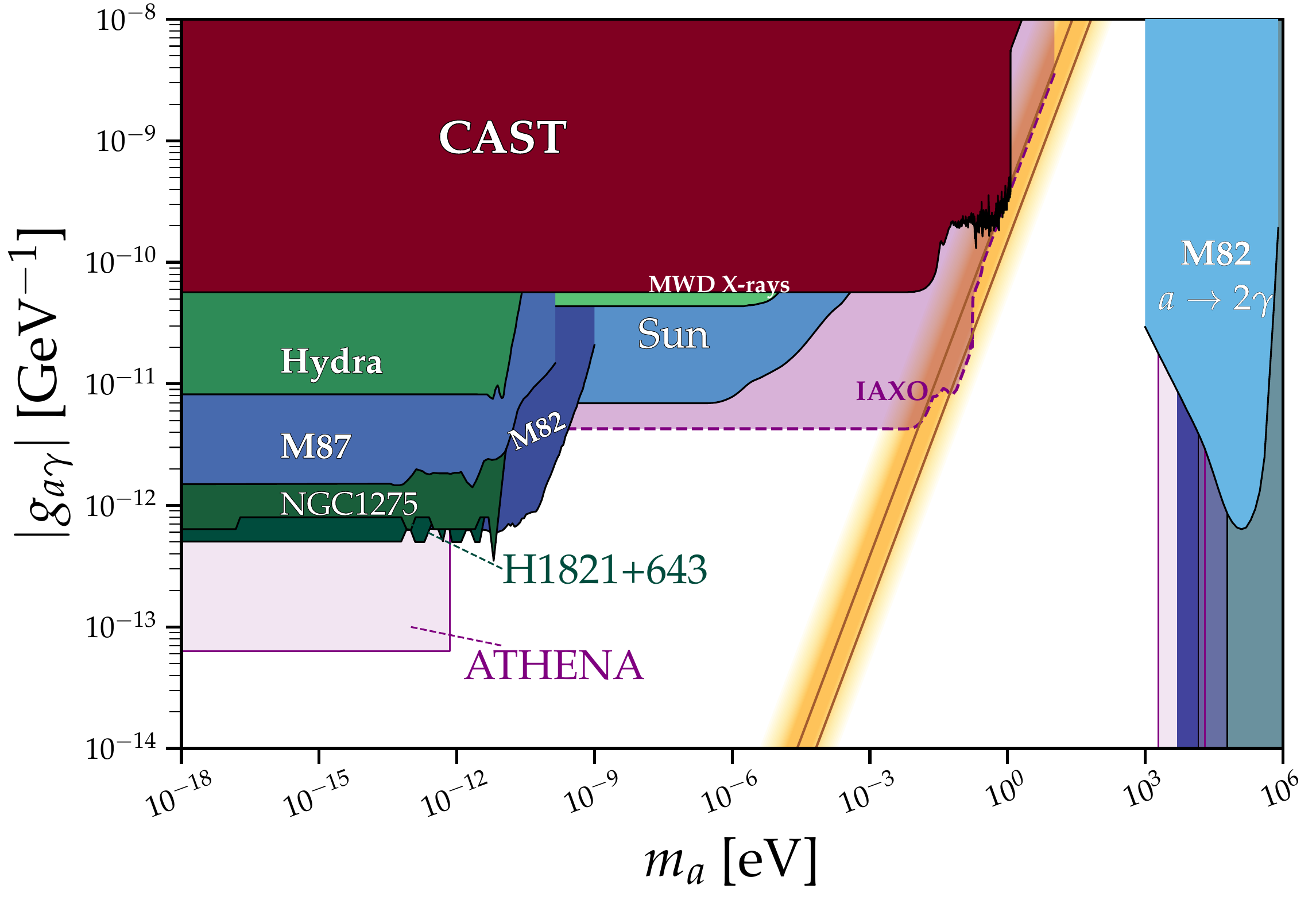}};

    \node[anchor=north, inner sep=0] (second) 
      at ([yshift=-0.15cm]main.south) 
      {\includegraphics[width=0.55\textwidth]{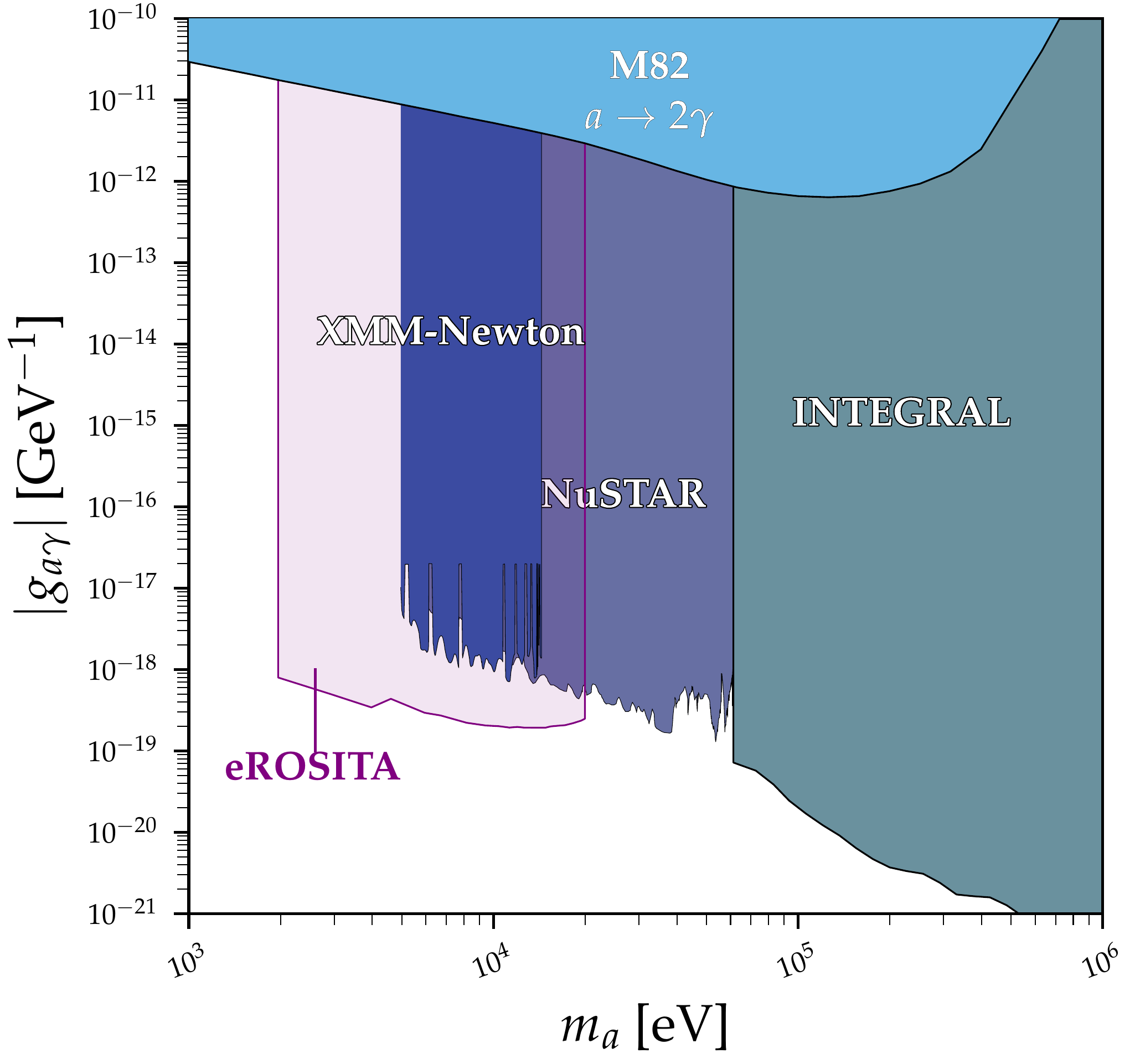}};
  \end{tikzpicture}
\caption{
Existing constraints on the axion-photon coupling, $g_{a\gamma}$, versus the axion mass, $m_a$, from X-ray telescopes are shown, alongside the bounds established by CAST (red) \cite{CAST:2024eil} and the projected sensitivity of IAXO (dashed purple) \cite{IAXO:2024wss}.
Chandra constraints in the low-mass regime from H1821+643, NGC1275, Hydra, and MWD X-rays are shown in green tones \cite{Reynes:2021bpe,Reynolds:2019uqt,Wouters:2013hua,Dessert:2021bkv}.
In blue tones, we have the NuSTAR constraints of M82, M87, Solar, and M82 axion decay \cite{Ning:2024eky,Ruz:2024gkl,Candon:2024eah}.
In the lower panel, the $g_{a\gamma}$ values are extended to better illustrate the parameter space explored by Dark Matter decay studies \cite{Foster:2021ngm,Calore:2022pks,Roach:2022lgo,Dekker:2021bos}.
Finally, in purple, the projections of ATHENA in the low-mass regime \cite{Sisk-Reynes:2022sqd} and eROSITA in the heavy mass regime \cite{Dekker:2021bos} are displayed. This plot was made with the help of the Axion Limit repository \cite{AxionLimits}.}
  \label{fig:axion-photon}
\end{figure}

First, we review the existing literature that constrains the axion-photon coupling $g_{a\gamma}$ (see Sec.~\ref{sec:Extragalactic_sources} for more details). Figure~\ref{fig:axion-photon} presents a summary of the most restrictive and recent constraints derived from X-ray telescopes.
In the light-ALP regime, the most restrictive limits come from the \textit{Chandra} \cite{Weisskopf:2000tx} observation of the quasar H1821+643 \cite{Reynes:2021bpe}, excluding $g_{a\gamma} > 6.3 \times 10^{-13} ~\mathrm{GeV}^{-1}$ for masses $m_a < 10^{-12}$~eV. This analysis examines potential spectral distortions that could be explained by photon-ALP oscillations. Rich clusters of galaxies hosting bright active galactic nuclei (AGNs) are powerful photon-ALP converters. These clusters are permeated by a magnetized plasma known as the intracluster medium (ICM), which typically has magnetic field strengths ranging from approximately 0.1 to 1.0 $\upmu$G. The capability of these clusters to constrain ALP properties is characterized by the product $( B \times L )$, where $B$ is the magnetic field strength $\sim 0.1 - 1 ~\upmu$G and  $L$ is the distance traveled by AGN photons through the cluster along the line of sight ($\sim 0.1 - 1 $ Mpc). In galaxy clusters, this product can significantly exceed the values achievable in laboratory experiments, making them excellent environments for studying and constraining ALPs.

A hard X-ray (30--70~keV) observation of the M82 starburst galaxy conducted by \textit{NuSTAR}~\cite{NuSTAR:2013yza} achieves a comparable sensitivity, setting an upper limit on the axion-photon coupling constant of \( g_{a\gamma} <6.4 \times 10^{-13}~\mathrm{GeV}^{-1} \) for axion masses \( m_a < 10^{-10}~\mathrm{eV} \) \cite{Ning:2024eky}. In their analysis, the authors aggregated the entire stellar population within these galaxies to calculate the axion luminosity. Additionally, they account for the conversion of axions to photons by utilizing magnetic field profiles derived from simulated IllustrisTNG (an ongoing series of large, cosmological magnetohydrodynamical simulations of galaxy formation \cite{Nelson:2018uso}) analogue galaxies.

Solar observations conducted by NuSTAR were able to exclude ${g_{a\gamma} > 6.9 \times 10^{-12}\,\mathrm{GeV}^{-1}}$  for $m_a  \lesssim 5 \times 10^{-4}~ \mathrm{eV}$~\cite{Ruz:2024gkl}, by using state-of-the-art magnetic field models of the solar atmosphere as the converter of axions produced in the solar core into detectable X-rays. 

The high-mass regime is highly constrained by high-energy observations, where we differentiate between two approaches: the decay of massive axions produced in stellar environments and the decay of halo dark matter into X-rays. In the mass range from 30 keV to 500 keV, NuSTAR observations of M82 were used to constrain the axion-photon coupling by modeling the axion emission from the starburst galaxy. In this case, the process considered was the decay of heavy axions into photons, $a \rightarrow \gamma + \gamma$, excluding $g_{a\gamma} > 10^{-10} - 10^{-12}$~\cite{Candon:2024eah}. On the other hand, observation from  \textit{XMM-Newton}~\cite{Foster:2021ngm}, INTEGRAL~\cite{Calore:2022pks}, NuSTAR~\cite{Roach:2022lgo}, and eROSITA~\cite{Dekker:2021bos} of narrow spectral features in X-ray data coming from the decay of dark matter into photons have been used to place bounds on $g_{a\gamma}$ (see the lower panel of Fig.~\ref{fig:axion-photon}). Dark matter profiles in galaxies can be modeled using profiles such as the commonly used Navarro-Frenk-White (NFW)~\cite{Navarro:1996gj}. Given the decay rate  for a given process $\Gamma_a$, one can compute the expected decay flux of the DM and compare it with the observational data. 

X-ray telescopes can constrain also other couplings, different from the axion-photon one. The product of the electron-coupling $g_{ae}$ with the axion-coupling $g_{a\gamma}$ is constrained by NuSTAR observations of the red supergiant star, Betelgeuse. ALPs can be produced by thermal processes in the stellar interior, escape from the star, and, if sufficiently light, be converted into photons in the external galactic magnetic field. The non observation of observable signatures related to this scenario exclude for masses $m_a \leq (3.5 - 5.5) \times 10^{-11}~$eV, the product of the coupling by $g_{a\gamma} g_{ae} \geq (0.4 - 2.8 ) \times 10^{-24}~\mathrm{GeV}^{-1}$~\cite{Xiao:2022rxk}. 

Finally, the product of axion-photon coupling and axion-nucleon coupling $g_{a\gamma} g_{aN}$ was constrained by using IXPE's X-ray polarization observation of the magnetars 4U 0142+61 and 1RXS J170849.0-400910 by \textit{IXPE}, excluding $g_{a \gamma} g_{a N} \geq 2.1 \times 10^{-20}~\mathrm{GeV}^{-2}$ for masses $m_a < 10^{-5}$~eV~\cite{Gau:2023rct}. Complementary probes of axion–nucleon interactions can be obtained from X-ray searches for emission lines from nuclear de-excitations, which directly constrain \enquote{effective} axion–nucleon couplings. The most extensively studied case is $^{57}\mathrm{Fe}$,  with $g_{aN}^{\mathrm{eff}} = 0.16g_{ap} + 1.16g_{an}$, due to its low-lying excited state at $E^{*}=14.4\ \mathrm{keV}$. For $^{57}\mathrm{Fe}$, the same NuSTAR observations of Betelgeuse discussed earlier in this section were used to set a bound of $|g_{a\gamma}g_{aN}^{\mathrm{eff}}|<(1.2-2.7)\times 10^{-20}\ \mathrm{GeV}^{-1}$ for $m_a\lesssim 10^{-10}\ \mathrm{eV}$~\cite{Candon:2025vpv}. Recently, this strategy was extended to additional isotopes, e.~g.  $^{61}\mathrm{Ni}\ (E^{*}=67.4\ \mathrm{keV})$ and  $^{73}\mathrm{Ge}\ (E^{*}=68.8\ \mathrm{keV})$,  and to X-ray observations of M82 and M87, further strengthening the constraints, tightening bounds to $|g_{a\gamma}g_{an}|\lesssim 1.1\times 10^{-22}\ \mathrm{GeV}^{-1}$ for $m_a\lesssim 10^{-10}\ \mathrm{eV}$~\cite{Ning:2025kyu}. 

\subsubsection{Other WISP searches}

The dark photon is a proposed massive spin-1 particle that appears in numerous beyond-the-Standard-Model theories featuring an extra $U(1)$ gauge symmetry. In these models, its gauge boson interacts with Standard Model fermions through kinetic mixing with the ordinary photon~\cite{fabbrichesi2021physics}. For more details see Sec.~\ref{eq:spin1-EFT}. The primary difficulty in constraining dark photon decays stems from the extremely diffuse morphology of the resulting X-ray signal. For a standard NFW galactic dark matter profile, the most sensitive decay searches are those which observe as much of the sky as possible. This makes investigations with high-resolution telescopes
such as Chandra or XMM-Newton difficult. However, wide-field telescope missions, such as INTEGRAL, are highly effective for dark matter decay searches. In~\cite{Linden:2024fby}, values of the kinetic mixing $\varepsilon > 10^{-14} - 10^{-18}$ were excluded for masses $m_{A'}$ in the range $m_{A'}\in [0.1,\,10]$~MeV.

We already mentioned the decay of galactic dark matter into X-rays as a possible search line via axion decays. However, analogous constraints can be placed by searching for sterile neutrino decay lines in the X-ray band. 
Sterile neutrinos (commonly denoted $\nu_s$) are another well-motivated dark matter candidate that may undergo radiative decay, producing a monochromatic photon line via $\Gamma_{\nu_s}$~\cite{Foster:2021ngm,Calore:2022pks,Roach:2022lgo,Dekker:2021bos}. The latest and stronger constrain was obtained in~\cite{Calore:2022pks}, where for sterile neutrino masses $m_{\nu_s}$ between $60~\mathrm{keV} - 2~\mathrm{MeV}$ the mixing angle values $\sin^2\theta > 10^{-14} - 10^{-29}$ were excluded. 

The chameleon gravity model introduces a scalar field  that mediates a fifth force (see Sec.~\ref{sec:1.2.5} for more details), influencing the hot X-ray--emitting gas in galaxy clusters but leaving their weak-lensing signals unaffected. By comparing X-ray and weak-lensing profiles of 58 clusters (drawn from the XMM Cluster Survey and The Canada-France-Hawaii Telescope Lensing Survey CFHTLenS), constraints for the strength of a fifth force were obtained~\cite{Wilcox:2015kna}. The results were, however, consistent with general relativity, showing no evidence for a chameleon-mediated fifth force. In the special case of $f(R)$ gravity, this analysis sets a bound 
$|f_{R0}| < 6 \times 10^{-5}$ (95\% CL), representing one of the strongest constraints to date on cosmological scales.

\subsection{Future Missions}
Building on current-generation observatories, two next-generation X-ray missions, \textit{Athena} and \textit{AXIS}, are expected to significantly advance our ability to probe ultralight axions. Both missions will leverage bright active galactic nuclei (AGN) in galaxy clusters, such as NGC 1275 in the Perseus cluster, to place stringent constraints on the photon--ALP coupling~\cite{Sisk-Reynes:2022sqd}.

\subsubsection{Athena (Advanced Telescope for High-ENergy Astrophysics)}

Selected by the European Space Agency (ESA) as its second large-class mission in the Cosmic Vision Program, Athena aims to address the ``Hot and Energetic Universe'' theme~\cite{Nandra:2013jka}. Its single X-ray telescope, made from iridium-coated silicon pore optics with a 12\,m focal length, will achieve an angular resolution of 5'\ (half-energy width) and an effective area of 1.4\,m$^2$ at 1\,keV. Two focal-plane instruments, (i) the Wide Field Imager (WFI) based on DePFET active pixel sensors, and (ii) the cryogenically cooled X-ray Integral Field Unit (X-IFU), will enable high-sensitivity and high-resolution X-ray spectroscopy~\cite{Barret:2019qaw}.

In particular, X-IFU's unprecedented spectral resolution ($\sim$2.5\,eV) will allow precise measurements of the intrinsic AGN emission and nearby intracluster medium, making Athena exceptionally well suited to improved constraints on very-light ALPs. Simulations indicate that Athena/X-IFU can potentially strengthen current bounds on the photon--ALP coupling by up to an order of magnitude for masses $m_a \lesssim 10^{-12}\,\mathrm{eV}$, provided its detector calibration is well characterized~\cite{Sisk-Reynes:2022sqd}. These projected constraints are illustrated in Fig.~\ref{fig:axion-photon}. Longer exposures (e.g., 1\,Ms on NGC~1275) would tighten these bounds further.

\subsubsection{AXIS (Advanced X-ray Imaging Satellite)}

Proposed in response to NASA's Astrophysics Probe Explorer Program and highlighted by the Astro2020 Decadal Survey, AXIS is designed for launch in the early 2030s and will offer powerful complementarity with Athena~\cite{Mushotzky:2018wio}. AXIS focuses on high spatial resolution, achieving 1'\ HPD imaging across a wide $24^\prime \times 24^\prime$ field of view. Its mirror system, employing precision-cut monocrystalline silicon foils, provides an effective area of about 0.7\,m$^2$ (0.57\,m$^2$ at 1\,keV after filtering and quantum efficiency factors).

Although its spectral resolution is more modest than Athena/X-IFU, AXIS can exploit its superb angular resolution to obtain AGN spectra largely uncontaminated by cluster emission. Simulated observations of bright AGNs in clusters indicate that AXIS can improve existing ALP limits by up to a factor of $\sim3$ in short exposures (200\,ks), and potentially exceed current bounds by more than an order of magnitude in deeper exposures (1\,Ms)~\cite{Sisk-Reynes:2022sqd}.

\subsubsection{HEX-P (High-Energy X-ray Probe)}

The high energy X-ray probe (HEX-P) is a proposed NASA probe-class mission that aims to combine high angular resolution with a broad X-ray bandpass, delivering a transformative capability for addressing key astrophysical questions in the next decade~\cite{Madsen:2023tlr}. The payload consists of a suite of co-aligned X-ray telescopes designed to cover the 0.2–80\,keV bandpass. In particular, the Low Energy Telescope (LET) operates over 0.2–20\,keV, while the High Energy Telescope (HET) covers 2–80\,keV. The HET offers a $13.7'\times13.7'$ field of view, primarily set by the detector size, and provides an on-axis half-power diameter of roughly $10''$ at 6\,keV and $23''$ at 60\,keV. The expected spectral resolution of the CZT detectors is on the order of a few hundred eV in the 10--60\,keV range, closely resembling NuSTAR’s performance. HEX-P is envisioned as a natural follow-up to NuSTAR, promising significant advances in both conventional high-energy astrophysics and the search for new physics phenomena such as ALPs.

%% file: WG3/content/Calore_Eckner_2.tex
\subsection{Review of gamma-ray instruments}
\label{sec:gammarayInstruments}
Gamma-ray telescopes can be deployed on Earth and in space. Their location strongly impacts the instrument's performance in terms of angular resolution, energy dispersion, and sensitivity because they require different gamma-ray detection strategies. Space-borne instruments can directly detect an incident photon. On the other hand, on the ground the gamma ray needs to penetrate Earth's atmosphere resulting in extensive air showers whose secondary emission is exploited to indirectly detect and reconstruct the properties of the primary gamma ray. We give a brief overview of current- and next-generation gamma-ray facilities and their specific event detection methods.

\subsubsection{Current telescopes}
\label{sec:currentInstruments}

The Large Area Telescope (LAT) mounted on the \textit{Fermi} Gamma-Ray Space Telescope is the only dedicated gamma-ray instrument currently operating in space. It was launched on the 11$^{\rm th}$ of June 2008 and can detect gamma rays of energies from 20 MeV up to energies of a few TeV. It exhibits a ``Field of View'' (FoV) -- the visible fraction of the sky per instant of time -- of more than 2 sr, an energy resolution better than $10\%$ as well as an angular resolution of around $3.5^{\circ}$ at 100 MeV and less than $0.15^{\circ}$ above 10 GeV. The gamma-ray detection strategy of \textit{Fermi} LAT is comparable to a modern detector at a particle accelerator. An anti-coincidence detector -- made from plastic scintillator tiles -- surrounds the innermost part of the instrument which itself hosts a silicon strip tracker. A gamma-ray event is detected if no electromagnetic signal was registered in the anti-coincidence detector while in the tracker two particle trajectories were recorded that originate in an $e^+e^-$-pair conversion of the incident photon inside the tracker material. A calorimeter on one end of the tracker measures the energy of the lepton pair which is directly related to the energy of the incoming primary gamma-ray~\cite{Fermi-LAT:2009ihh}. 

There are multiple currently operating ground-based gamma-ray telescopes. They are typically classified into Imaging Atmospheric Cherenkov Telescopes (IACTs) and Water Cherenkov Telescopes reflecting the two main strategies of gamma-ray detection. On one side, IACTs are mirror-based, optical telescopes that employ the imaging atmospheric Cherenkov telescope technique exploiting the Cherenkov effect caused by the particles of the extensive air shower initiated by the primary photon while they are still propagating through the atmosphere. The direction and yield of Cherenkov photons contain the information to reconstruct the primary gamma ray's origin, energy and momentum. Current-generation instruments of this kind are the Major Atmospheric Gamma Imaging Cherenkov array (MAGIC)~\cite{MAGIC:2014zas}, the Very Energetic Radiation Imaging Telescope Array System (VERITAS)~\cite{2015ICRC...34..771P} and the High Energy Spectroscopic System (H.E.S.S.)~\cite{Holler:2015uca}. Conversely, Water Cherenkov Telescopes use water tanks to detect Cherenkov light generated by secondary particles inside the tank itself, i.e.~from secondary air shower particles that survive until they reach the Earth's surface, e.g. muons. The High-Altitude Water Cherenkov Gamma-Ray Observatory (HAWC)~\cite{Abeysekara:2017mjj} and the Large High Altitude Air Shower Observatory (LHAASO)~\cite{Aharonian:2020iou, LHAASO:2021zta} currently use this technique. 

To improve the reconstruction quality of the primary events, both IACTs and Water Cherenkov Telescopes are typically operated in stereo mode, i.e. having an array of at least two telescopes or a large area filled with individual water tanks. For this reason, IACTs are sensitive to gamma rays from a few 100 GeV to $\mathcal{O}(100)$ TeV with a sweet spot of enhanced sensitivity at $\mathcal{O}(1)$ TeV. Extending their sensitivity to even higher energies would require a larger surface area covered by the individual telescopes which is monetarily prohibitive. The individual water tanks of a water-Cherenkov telescope are less costly allowing for a wider area to be covered and, thus, they achieve a better sensitivity at the highest energies up to the PeV scale, like in the case of LHAASO. Typically, their prime sensitivity is achieved for energies larger than 30 TeV.

In direct comparison, IACTs outperform water-Cherenkov instruments in angular and energy resolution\footnote{In the case of VERITAS, the angular resolution is better than $0.1^{\circ}$ above 1 TeV while the energy resolution can be around $\Delta E/E = 15\%$, both at 68\% containment level.} but they are limited to operation at night and under favorable conditions (reduced Moonlight contamination, clear skies, etc.). Water-Cherenkov telescopes in return are less prone to atmospheric conditions and have a much larger duty cycle during day and night while they observe a large fraction of the sky in an instant of time. We refer to the provided studies for further details about each telescope's performance and specifications. Common to both telescope types is the presence of a cosmic-ray background originating in misclassified extensive air showers triggered by charged primary cosmic rays. This background is mitigated differently depending on the telescope type. The background from charged cosmic rays is much easier to eliminate in the case of \textit{Fermi} LAT because of the anti-coincidence detector that only fires for charged particles. 

\subsubsection{Future instruments}
\label{sec:futureInstruments}

The next decade of gamma-ray astronomy will be dominated by ground-based instruments exploring the TeV regime. Currently under construction, the Cherenkov Telescope Array Observatory (CTAO) \cite{ScienceWithCTA2019} will be the next-generation IACT surpassing the performance of all currently operating instruments of that kind. CTAO will be located on two experiment sites in the northern (CTAO-N) and southern (CTAO-S) hemispheres. The array is designed as a combination of differently sized individual telescopes featuring large-sized (LST, 23-m diameter dish), medium-sized (MST, 12-m diameter dish) and small-sized telescopes (SST, 4-m diameter dish). CTAO-N is comprised of 4 LSTs and 9 MSTs on the \textit{Roque de los Muchachos} on La Palma, Canary Islands; where also the MAGIC array is located. CTAO-S in contrast is designed with 14 MSTs and 37 SSTs in Paranal, Chile, as the initial array layout. Taken together, both array sites will cover an energy range from 20 GeV to about 300 TeV with unprecedented sensitivity while bridging the gap to the LAT's GeV gamma-ray regime. CTAO will also feature an angular resolution about twice as good as comparable current-generation IACTs in the TeV range and an energy resolution that surpasses the operating telescopes by about a factor of three.\footnote{See, for instance, \url{https://www.ctao.org/for-scientists/performance/} for a comparison of the sensitivity of different current- and next-generation gamma-ray telescopes.}

The driving idea behind the development of the 
Southern Wide-field Gamma-ray Observatory (SWGO) \cite{Abreu:2019ahw} is to establish a high-duty-cycle water-Cherenkov instrument in the southern hemisphere to complement similar facilities in the northern hemisphere (HAWC and LHAASO). Such an endeavor would finally render the Galactic centre of the Milky Way accessible to this type of ground-based gamma-ray observatory. In August 2024 it was decided to construct this telescope at the Atacama Astronomical Park in Chile at an altitude of a bit less than 4800 m. The design philosophy around SWGO is directly adopted from the experience gained with HAWC~\cite{Albert:2019afb}. SWGO will feature a high-fill core region and a less densely filled outer array. Besides studies of the Milky Way's centre, it will allow for efficient transient studies as well as a strong contribution to multi-wavelength and multi-messenger astronomy.

A future space-borne instrument will be hosted aboard the Tiangong Space Station and is called the High Energy cosmic-Radiation Detection (HERD; the launch is potentially scheduled for 2027)~\cite{Kyratzis:2022rxa}. In fact, HERD is not a specialised gamma-ray telescope but rather a combination of different detectors enabling the detection of charged cosmic rays and gamma rays. In terms of gamma-ray astronomy, HERD will achieve an improved angular and energy resolution compared to the LAT. Yet, it will exhibit an acceptance or effective area about five times smaller than \textit{Fermi} LAT while covering in principle the same energy range. In terms of scientific objectives, this telescope can bridge the gap between the GeV and TeV energy regime allowing for broadband studies of gamma-ray sources in synergy with ground-based facilities beyond the operation time of the LAT. 
Recently, propositions for constellations of telescopes satellites have been put forward~\cite{Manzari:2024jns}.
These instruments will be particularly suited for detections of SN events, given their large instantaneous filed of view.

\subsection{Generalities of constraints on WISPs}
\label{sec:general-gamma-search-strategies}
The existing and future gamma-ray instruments come with certain peculiarities that render them suitable tools to search for WISPs, particularly ALPs. ALPs leave traces in different observables accessible to gamma-ray astronomy due to their coupling with photons that induce ALP decays into photons as well as their mutual conversion in the presence of external magnetic fields known as the Gertsenshtein effect. Broadly speaking, the presence of ALPs in the universe may generate a large-scale diffuse emission or it may imprint itself in the spectral energy distribution~(SED) of a gamma-ray emitting source of Galactic or extragalactic origin. These ALP signatures may either be time-dependent or -independent. We provide a brief description of these signatures as well as the instrument specifications that render certain gamma-ray telescopes suitable to search for such effects. More details about generic ALPs signatures from extragalactic sources can be found in Sec.~\ref{sec:Extragalactic_sources}.

\subsubsection{Imprint in gamma-ray diffuse backgrounds}
The cumulative emission of ALPs from astrophysical objects throughout the universe at all redshifts is expected to 
provide sizable imprints in the diffuse large-scale gamma-ray emission, see discussion in Sec.~\ref{sec:Extragalactic_sources}.
This signal can be looked for as a brightening of the gamma-ray diffuse background (overall spectral intensity), or for 
specific patterns in the spatial distribution of photons, which, if of ALPs origin, would trace primarily 
the Galactic magnetic field structure.
This diffuse ALP background can be searched for by contrasting it with the expected other types of diffuse large-scale gamma-ray emission like the Milky Way foreground or the isotropic gamma-ray background~\cite{Fermi-LAT:2014ryh}. If ALPs are the dark matter of the universe, their decay into photons can induce a diffuse large-scale emission following the DM density profile of a suitable target like the Milky Way dark matter halo. However, such an emission would only generate gamma rays for very heavy ALPs. 

Detecting and efficiently exploiting large-scale gamma-ray emissions is possible only with instruments that exhibit a large instantaneous FoV like the LAT or water-Cherenkov telescopes. In addition, it is favorable to use datasets of instruments with a good angular resolution in order to profit from the spatial morphology of the diffuse ALP signature generated by the Galactic magnetic field to break potential degeneracies. At sub-PeV energies, this requirement can be slightly relaxed because of the opacity of the universe to such very high-energy gamma rays. Any diffuse emission component besides the expected Galactic contribution would require a certain amount of exotic physics. In this sense, it is possible to leverage ground-based water-Cherenkov facilities in the search for diffuse emissions from ALPs.

\subsubsection{Individual high-energy sources: Time-integrated SED}
Besides the cumulative large-scale emission from a population of sources, individual bright sources provide another avenue for ALP searches themselves. Suitable targets are, e.g., AGNs, SNe, and galaxy clusters. As discussed and introduced in Sec.~\ref{sec:Extragalactic_sources} for extragalactic sources, the environmental conditions realized in Galactic and extragalactic gamma-ray sources can lead to the production of ALPs \textit{in situ}. Among other properties, the target source's magnetic field strength and structure, electron density, radiation fields and spatial extension play a crucial role in generating a sufficiently high ALP yield that can be exploited in indirect searches. To enhance the sensitivity to ALPs, it is suitable to examine the time-integrated SED of a target source, which may reveal different types of ALP imprints.
\begin{itemize}
    \item \textit{Prompt gamma-ray emission.} The presence of ALPs can alter the standard astrophysical expectations regarding the gamma-ray emission of core-collapse SNe, leading to a prompt GRB. 
    Instruments suitable for this type of emission should exhibit a large instantaneous FoV to maximize the chances of capturing such a transient event. Otherwise, they need to be pointed to the SN, which first requires an alert triggered by the neutrino burst. \textit{Fermi} LAT and HERD in the future are particularly useful in this regard. Yet, the typical gamma-ray energies of a SN GRB are of the order of 100 MeV rendering specialized space-borne MeV telescopes the ideal search facilities.
    
    \item \textit{Spectral distortions.} As discussed in Sec.~\ref{sec:Extragalactic_sources}, ALP-photon mixing can cause spectral irregularities in high-energy sources. These can be of extragalactic or galactic origin.
    To search for spectral irregularities in practice, a suitable gamma-ray instrument must, first and foremost, provide an adequate energy resolution to resolve such intricate modulations. \textit{Fermi} LAT and IACTs are thus ideal search tools.
    
    \item \textit{Spectral hardening.} The SED of extragalactic gamma-ray sources is softened by attenuation of very-high-energy gamma rays on the extragalactic background light (EBL) (see Sec.~\ref{sec:Extragalactic_sources}). In fact, above a few 100 TeV, only Galactic gamma rays of such energies can reach Earth. However, if gamma rays convert into ALPs within the respective target source, a fraction of the original intrinsic gamma-ray emission is restored when the ALP re-converts into photons close to Earth. This leads to a spectral hardening of the target source compared to naive expectations from EBL models and gamma-ray propagation. Any individual extragalactic objects with a hard intrinsic spectrum are suitable for this type of search. In addition, one may use the combined emission from many extragalactic objects and look for a statistically robust hardening in a stacked analysis indicating an altered overall transparency of the universe.  Therefore, facilities with sensitivities beyond 10 TeV, i.e.~water-Cherenkov instruments, are particularly useful for this type of analysis.
\end{itemize}

\subsubsection{Individual high-energy sources: Time-dependent SED}

Violent GRBs, for instance, triggered by SNe can outshine a whole galaxy for a short amount of time. Due to their high luminosity, GRBs yield enough photon statistics to render feasible searches for time-dependent ALP signatures that contain even more information about the ALP itself, its production processes and the conditions within the host. Key to such searches is a good temporal resolution and short dead time because of the plethora of photon events.

\subsection{Constraints from space-based instruments}
\label{sec:constraints-space}

\begin{figure}
    \centering
    \includegraphics[width=0.5\linewidth]{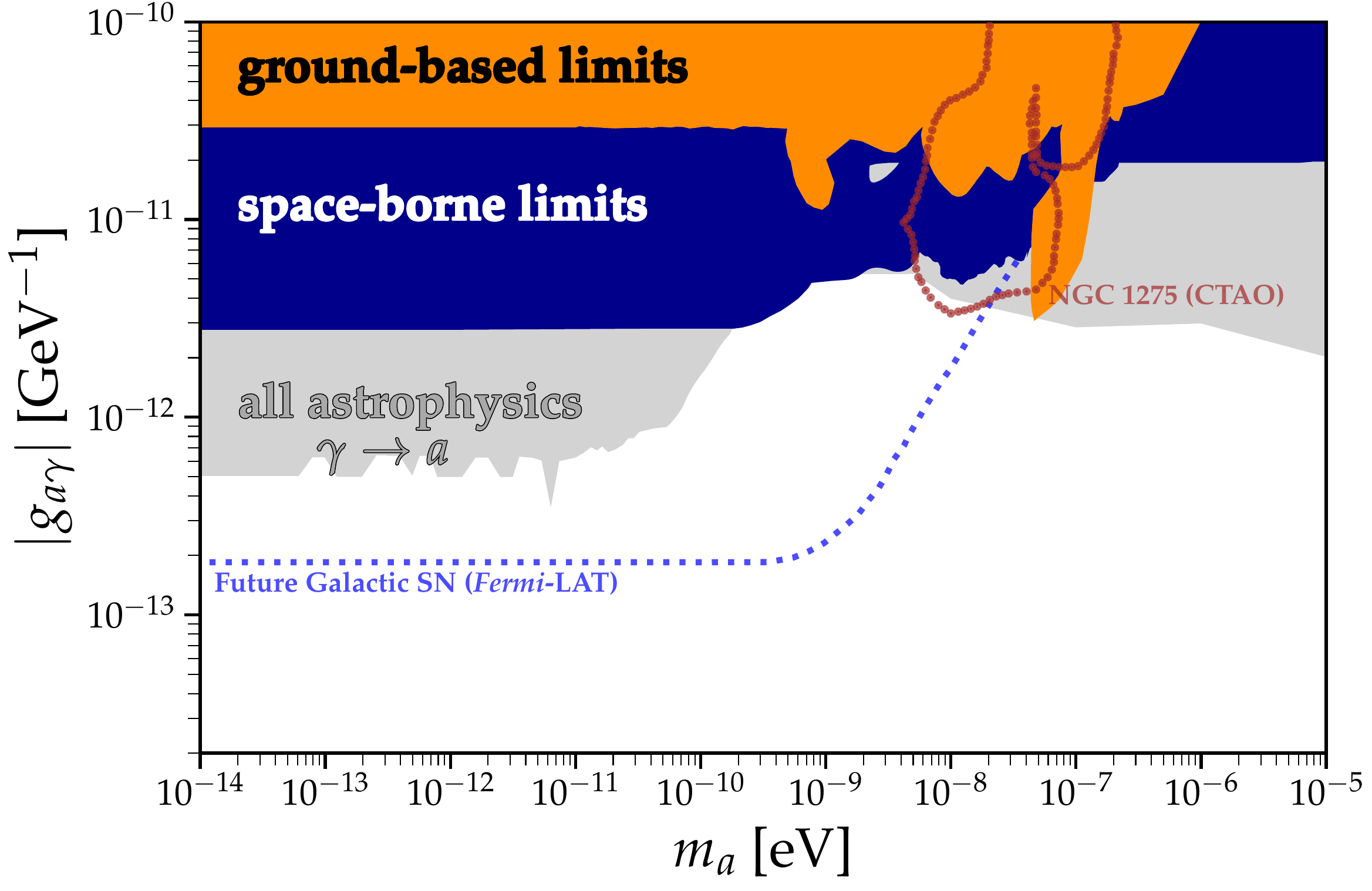}\hfill
    \includegraphics[width=0.5\linewidth]{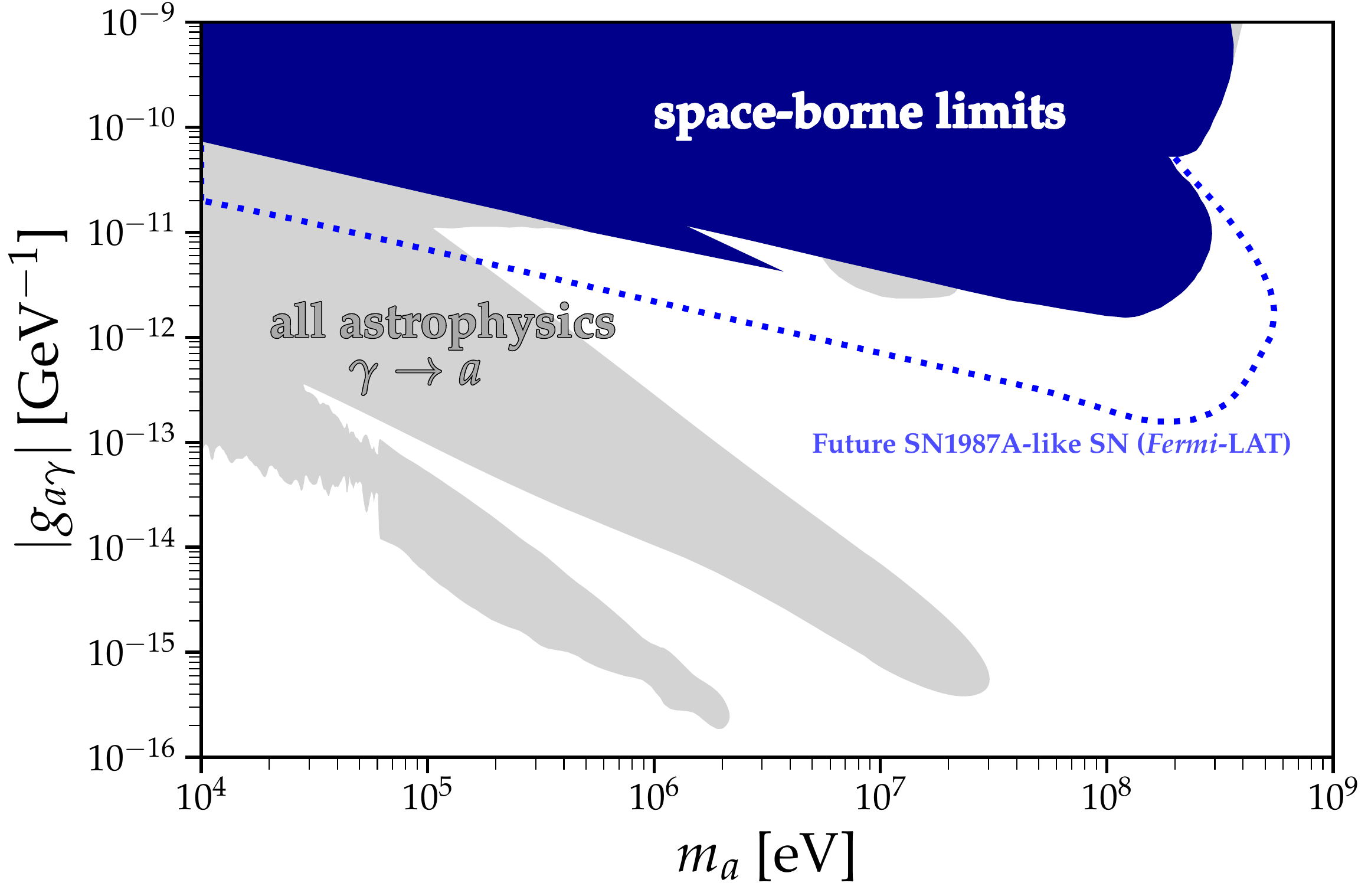}
    \caption{Summary of astrophysical constraints from gamma-ray datasets obtained for the ALP-photon coupling constant $g_{a\gamma}$ as a function of the ALP mass $m_a$ for light ALPs up to $m_a = 10^{-5}$ eV (\textit{Left}), and heavy ALPs from $m_a = 10^{4}$ eV onwards (\textit{Right}). We distinguish bounds derived from datasets taken with space-borne instruments (blue) and ground-based gamma-ray telescopes (orange). Sensitivity prospects are shown with dotted lines. The full landscape of exclusion limits from any type of astrophysical observation is displayed as gray shaded region. The shown constraints do not assume that the ALP is the dark matter in the universe. These plots were adapted from \cite{Eckner:2022rwf} and realized with the help of \texttt{AxionLimits} \cite{AxionLimits}.}
    \label{fig:constraint-figure}
\end{figure}

We provide a brief review of the existing body of literature deriving constraints on the ALP parameter space from datasets of space-borne instruments. A plot of these upper limits is given in Fig.~\ref{fig:constraint-figure}, which displays the space-borne constraints shaded in blue. Projections are shown with dotted lines. Note that these constraints do not assume that ALPs are the dark matter in the universe.

\subsubsection{Present}

\textit{\textbf{Supernovae.}} Several studies have explored gamma-ray signatures of ALPs from core-collapse SNe using \textit{Fermi}-LAT observations. The authors of~\cite{Calore:2021hhn} performed a comprehensive analysis of all-sky \textit{Fermi}-LAT data to identify morphological features linked to ALP-photon conversion within the diffuse gamma-ray background from supernova-generated ALPs. This study builds upon earlier work presented in~\cite{Calore:2020tjw}.

Ref.~\cite{Jaeckel:2017tud} provided further insights employing gamma-ray data from SN 1987A to investigate heavy ALP decays, with an improved forecast presented by~\cite{Hoof:2022xbe} that leverages the temporal data from multiple active gamma-ray instruments (see Sec.~\ref{subsec:Vitagliano} for more details). 
Expanding on the analysis of potential gamma-ray signatures from core-collapse SNe,~\cite{Meyer:2016wrm} explored the sensitivity of \textit{Fermi}-LAT to future SN explosions, examining ALP-photon conversion and its effects on gamma-ray light curves. This line of research has been extended in subsequent studies, including~\cite{Calore:2023srn} on light ALPs through conversion,~\cite{Muller:2023vjm,Benabou:2024jlj}, on heavy ALPs via decay, and~\cite{Manzari:2024jns}, which specifically investigated the role of the progenitor's magnetic field in the conversion process. In this body of literature,~\cite{Calore:2023srn, Lella:2024hfk} included the time-dependent spectrum of the SN's prompt GRB to exemplify the potential such data can have to reconstruct the ALP parameters. In addition,~\cite{Lella:2024hfk} added an ALP-nucleon coupling that sizably boosts the expected GRB signal and yields information about the properties of the proto-neutron star formed after the core bounce. This additional coupling was also considered in Ref.~\cite{Carenza:2025uib} to assess the potential of next-generation helioscopes to detect SN ALPs using a dedicated MeV detector.
As for extragalactic core-collapse SNe, Ref.~\cite{Meyer:2020vzy} analysed their time-dependent SED to assess ALP effects on their gamma-ray signatures. The authors of~\cite{Crnogorcevic:2021wyj} performed the reverse analysis and looked into the spectra of recorded long LAT GRBs -- which may or may not be originating in SNe -- to search for potential ALP signatures. Ref.~\cite{Muller:2023pip} complements these studies by examining the heavy ALP decay signatures in the recent observation of SN 2023ixf. Together, these studies leverage \textit{Fermi} LAT's capabilities to constrain ALP properties and advance our understanding of ALP interactions in supernova environments.

\textit{\textbf{Other Galactic targets.}}
Ref.~\cite{Majumdar:2018sbv} analyzed a sample of six, bright Galactic pulsars, finding a
4.6$\sigma$ significance for common ALP-photon mixing corresponding to $20\%-40\%$ spectral variation despite 3\% experimental systematic uncertainty. This analysis was followed by a re-assessment in uncertainties on up-to-date magnetic field models and distance measurements~\cite{Pallathadka:2020vwu}. In this same work, it was shown as well how other astrophysical and laboratory constraints can be evaded to rescue the pulsars' hint. See also~\cite{Choi:2018mvk} for another possible mechanism. 
Spectral oscillations from a sample of SN remnants were first looked for by~\cite{Xia:2019yud}, combining \textit{Fermi}-LAT with HESS/MAGIC/VERITAS data and finding a mild hint for IC443. 
This was not confirmed in a previous analysis on an extended sample~\cite{Liang:2018mqm}.

\textit{\textbf{Active galaxies and other extragalactic targets.}} Numerous studies have utilized \textit{Fermi}-LAT data to investigate ALP constraints from extragalactic sources beyond SNe, focusing on the distinctive signatures in the respective sources' SEDs. The authors of~\cite{Fermi-LAT:2016nkz} analyzed the gamma-ray spectrum of NGC 1275, the central galaxy of the Perseus cluster, to search for spectral irregularities from ALP-photon conversion, setting stringent limits on ALP-photon coupling for ALP masses between 0.5 and 5 neV. This work was revisited by~\cite{Cheng:2020bhr} and~\cite{Pallathadka:2020vwu}, who refined the methodology by assessing generic spectral irregularities and considering uncertainties in the magnetic field structure, further tightening ALP parameter constraints.
A systematic study of a large sample of AGNs was performed in~\cite{Yu:2022psh}, by looking for common spectral features and finding mild evidence for ALP-induced spectral modulations. See also a previous analysis on blazar spectra~\cite{Zhou:2021usu}.
In addition, Ref.~\cite{Davies:2022wvj} studied three bright, flaring flat-spectrum radio quasars, employing an improved jet model to probe ALP effects on gamma-ray opacity. This study hinges on prior theoretical considerations about the importance of photon-photon dispersion in blazar jets~\cite{Davies:2021wqw}. Likewise,~\cite{Buehler:2020qsn} searched for ALP signatures in the highest-energy photons from hard gamma-ray blazars, with a focus on how ALP-photon conversion could impact transparency at very high energies. An analysis targeting the cumulative effect in the extragalactic gamma-ray background was presented in~\cite{Liang:2020roo}, finding compatibility between gamma-ray data and EBL model expectations.

\textit{\textbf{Neutron star mergers.}} As a more recent source of ALPs, neutron star -- neutron star (NS-NS) mergers were considered. While neutron stars may themselves copiously produce ALPs, mergers of binary neutron stars offer a multi-messenger window to study exotic physics as evidenced by the GRB coincident with the gravitational wave event GW170817~\cite{LIGOScientific:2017ync} detected by the \textit{Fermi} satellite and LIGO/VIRGO, respectively. Ref.~\cite{Fiorillo:2022piv} discusses prospects for the search for ALPs regarding NS-NS merger events. ALP constraints were derived in \cite{Dev:2023hax} using the LAT gamma-ray data collected from the GRB triggered by GW170817. Additionally, as further discussed in Sec.~\ref{subsec:Vitagliano}, Ref.~\cite{Diamond:2023cto} studies how the formation of ALP-sourced fireballs around the merged proto-neutron star can alter the standard expectations invoking also signatures in the X-ray spectrum to improve constraints.

\subsubsection{Future}
As pointed out in~\cite{Meyer:2016wrm, Calore:2023srn, Muller:2023vjm, Manzari:2024jns}, \textit{Fermi} LAT offers exquisite potential to probe a large portion of the ALP parameter space for either light or heavy ALPs in case of a future Galactic core-collapse SN. An example of its sensitivity is shown in both panels of Fig.~\ref{fig:constraint-figure}. However, the future of the LAT is uncertain since its operation time already passed the initially intended period. HERD will continue observing gamma rays with a large FoV after 2027. Yet, its reduced effective area compared to the LAT limits the prospects of this instrument. In addition, both the LAT and HERD operate best at energies around 1 GeV while losing sensitivity towards the MeV range where prompt SN GRBs will shine bright. Therefore, an optimal exploitation of Galactic SN data may only be achieved with a space-borne MeV mission. Such a mission would massively improve the energy resolution of the prompt ALP-induced gamma-ray GRB's spectrum, hence, extending the projections derived regarding \textit{Fermi} LAT. Besides, such a mission would close the so-called ``MeV gap'' in gamma-ray astronomy. This term succinctly paraphrases the fact that current-generation instruments leave a pronounced gap of sensitivity between X-ray and GeV energies~\cite{2021ExA....51.1225D}. Closing this gap does not only yield crucial information for standard astrophysics but can help to constrain primordial black holes or other WISPs like sterile neutrinos and dark photons~\cite{DelaTorreLuque:2023huu, DelaTorreLuque:2024zsr}. We refer the reader to Sec.~\ref{subsec:deLaTorre} for more information on this energy range.

\subsection{Constraints from ground-based instruments}
\label{sec:constraints-ground}

We continue with a brief review of the existing body of literature deriving constraints on the ALP parameter space from datasets of ground-based instruments. Their targets are mostly of extragalactic origin as these instruments operate best within the TeV range and beyond. A plot of these upper limits is given in Fig.~\ref{fig:constraint-figure}, which displays the ground-based constraints shaded in orange or as dotted lines in the case of projections.

\subsubsection{Present}
\textit{\textbf{Galactic and extragalactic targets.}} Several studies have leveraged currently operating ground-based gamma-ray observatories to constrain ALP properties. A general review of ALP searches using IACTs is provided by~\cite{Batkovic:2021fzr}, which covers methods and challenges across current facilities. Specific observational efforts include~\cite{Liang:2018mqm}, who analysed spectral modulations in ten Galactic TeV-bright sources detected by H.E.S.S., discussing photon-ALP oscillations and prospective sensitivity improvements with CTAO. Likewise exploiting H.E.S.S.~data,~\cite{HESS:2013udx} investigated spectral irregularities in the gamma-ray emission from PKS 2155-304, imposing constraints on ALP-photon conversion within this source. 
The authors of~\cite{Li:2022pqa} examined the VERITAS source VER J0521+211 in combination with LAT data for irregularities and spectral hardening in the source's SED while accounting for the uncertain redshift determination. VERITAS observations of the flat spectrum radio quasar 4C+21.35 were exploited in~\cite{Li:2022jgi} to set further constraints on ALP properties.
MAGIC has also contributed substantially to ALP research, as seen in~\cite{MAGIC:2024arq}, which examined the Perseus Galaxy Cluster for spectral distortions due to ALP-photon conversions. Further,~\cite{Pant:2023omy} explored photon-ALP oscillations in FSRQ QSO B1420+326 using MAGIC data, establishing valuable ALP constraints. Meanwhile,~\cite{Li:2023qyr} studied 10 years of gamma-ray emission from 1ES 1215+303, yielding robust limits on ALP-photon couplings by examining high-energy spectral features. Ref.~\cite{Gao:2023dvn} analysed a combined MAGIC and \textit{Fermi}-LAT dataset to search for spectral irregularities in the emission of Mrk 421 with a bespoke statistical method adapted to the flaring behaviour of this blazar.

Water-Cherenkov facilities have also proven essential. Ref.~\cite{Eckner:2022rwf} estimated the cumulative sub-PeV emission from star-forming galaxies due to photo-hadronic interactions and ALP conversion, thus, contributing an exotic large-scale diffuse component at these energies, which they constrained via diffuse gamma-ray measurements of the Galactic plane from HAWC and Tibet AS$\gamma$. The same idea was applied in \cite{Mastrototaro:2022kpt} to early diffuse data obtained with LHAASO. Ref.~\cite{Jacobsen:2022swa} employed HAWC data to detect possible ALP-induced spectral irregularities in blazars. LHAASO observations provided constraints on Galactic ALP sources, as discussed in~\cite{Li:2024ivs}, while~\cite{Li:2024zst} combined \textit{Fermi}-LAT and HAWC data for Mrk 421 and Mrk 502, investigating spectral irregularities and hardening effects.
Studies like~\cite{Li:2020pcn} and~\cite{Gao:2023dvn} analysed Mrk 421 regarding spectral distortions using ARGO-YBJ, and \textit{Fermi}-LAT yielding stringent limits on photon-ALP conversions from its SED. 

\textit{\textbf{GRB 221009A.}} On the 9th of October 2022, GRB 221009A was detected and observed for about ten minutes~\cite{Frederiks:2023bxg}. It was soon realised that this GRB event is the ``Brightest of All Times'' (BOAT) regarding its gamma-ray luminosity with an estimated occurrence rate of once every 10,000 years~\cite{Burns:2023oxn}. Observations with the James Webb Space Telescope showed that the GRB was caused by a SN in a star-forming galaxy at a redshift of $z = 0.151$~\cite{Blanchard:2023zls}. While this event broke several records, it is also intriguing from an ALP point of view. GRB 221009A is the only burst so far with confirmed gamma-ray emission above 10 TeV despite its distance from Earth. The LHAASO collaboration issued a telegramme~\cite{2022ATel15669....1D} to have detected a gamma-ray event of 18 TeV associated with the GRB, which was later corrected to an energy of 13 TeV~\cite{LHAASO:2023lkv}. The Carpet-2 array at Baksan (Russia) claims to have observed an event of around 251 TeV associated with GRB 221009A. Observing gamma-ray events above 10 TeV raises questions about the transparency of the universe and the role ALPs can play in these measurements.

Ref.~\cite{Gao:2023und} examined LHAASO observations of GRB 221009A, investigating whether the transparency of very high-energy gamma rays could be attributed to photon-ALP oscillations, thereby placing constraints on ALP models. Building on this,~\cite{Galanti:2022chk} considered a broader range of very high-energy gamma-ray observations, including Carpet-2 data, suggesting that ALPs could account for the signal but may require modified mixing parameters. Meanwhile,~\cite{Carenza:2022kjt} evaluated ALP explanations for the LHAASO and Carpet-2 detections, incorporating host-internal gamma-ray absorption models and finding that the ALP hypothesis could not fully account for both events simultaneously. Finally,~\cite{Baktash:2022gnf} assessed multiple extragalactic EBL models to examine the 18 TeV gamma-ray detected by LHAASO (before the correction), concluding that ALP explanations remain plausible if adjustments are made to standard mixing conditions.

\subsubsection{Future}
In the future, CTAO is expected to play a pivotal role in the search for WISPs and ALPs. For instance,~\cite{DeFranco:2017wdr} examined the expected number counts of extragalactic sources detectable in CTAO's extragalactic survey, noting that ALP-photon oscillations could alter these predictions by mitigating absorption from the EBL and potentially reducing the impact of intergalactic magnetic fields. Additionally,~\cite{Galanti:2018upl} focused on CTAO’s sensitivity to ALP-induced spectral modulations in Mrk 501, predicting that CTAO, alongside other gamma-ray observatories, could identify spectral irregularities indicative of ALPs.
Further studies by the CTAO Consortium, specifically~\cite{CTA:2020hii}, highlight the array’s full Omega configuration and its anticipated sensitivity to ALP-induced spectral distortions in AGN, particularly within the Perseus cluster. This configuration aims to detect both opacity changes and spectral distortions due to photon-ALP conversions, setting the stage for stringent ALP parameter constraints. Additionally,~\cite{Viana2024:aaa} explored CTAO’s potential in detecting ALP signals from decaying dark matter within the Milky Way. By focusing on ALP couplings to photons and electrons, this study suggests that CTAO’s observations above 100 GeV could uncover high-energy gamma rays resulting from ALP interactions with cosmic rays, broadening the scope of ALP research into dark matter contexts.

Prospects for water-Cherenkov facilities are similarly promising. For example,~\cite{Long:2021udi} evaluated LHAASO’s capacity to detect ALP-induced opacity in the gamma-ray spectra of AGN, emphasizing that LHAASO’s sensitivity to high-energy photons could provide key insights into ALP properties across extragalactic distances. Together, CTAO and LHAASO will contribute complementary capabilities, advancing the detection prospects for ALPs through spectral modulation and opacity measurements in high-energy astrophysics.

%% file: WG3/content/Pedro_de_La_Torre_Luque.tex
\subsection{Minding the MeV gap for WISPs}
\label{sec:WhyMeVGap}

A variety of WISPs in the MeV mass range, such as dark photons, sterile neutrinos or ALPs can fit as DM candidates and can be naturally produced in astrophysical events, for example in supernovae (SNe) with typical core temperatures of $\mathcal{O}(10)$~MeV. This has motivated the search for signals of sub-GeV FIPs in the MeV gamma-ray domain, which, in turn, is the most unexplored window, among recent multiwavelength observations in astrophysics, to look for signatures of new physics. At the MeV, the cross section of interactions of photons with matter reach a minimum,\footnote{As comparison, at X-ray energies photon-matter cross sections are $\sim10^4$ times larger than at the MeV.} being dominated by Compton scattering (which makes localizing the direction of the incoming photon troublesome).
 Hence, gamma-ray measurements at the MeV suffer from limited resolution and sensitivity, along with higher systematic uncertainties 
 than its neighbors, i.e. X-rays and GeV gamma rays (see Sec.~\ref{subsec:RuzVogel} and Sec.~\ref{subsec:CaloreEckner2}, respectively). This is the so-called ``MeV gap'', where very limited measurements exist and the few ones present large uncertainties.  
Besides their unique capabilities to study a variety of astrophysical processes, recent investigations have shown the high potential to probe signatures of new physics from current MeV observations, that allow us to probe a parameter-space region not well constrained yet. This has triggered a renewed interest to exploit the MeV sky and to propose new projects of MeV gamma-ray telescopes for the near future.

The following sections briefly review the main strategies used to probe different kinds of WISPs in the MeV range, either from their direct photon production or from the photon production generated by their products, consisting of continuum or ``line'' emissions.

\subsubsection{Constraining WISPs from direct photon production in the MeV}

Generically, different kinds of MeV DM particles can undergo annihilation or decay into two photons, leading to a distinctive line-like photon emission that can be searched for in a similar way as done at GeV energies.

Using SPI (INTEGRAL) data,~\cite{Siegert:2024hmr} presented constraints on $\langle \sigma v \rangle$ for both the $\gamma\gamma$ and e$^+$e$^-$ (through final state radiation -- FSR) channels, covering a mass range from a few hundred keV up to $\sim300$~MeV. These result in leading constraints for the $\gamma\gamma$ channel, but FSR lead to slightly weaker constraints on $\langle \sigma v \rangle_{ee}$ than cosmological constraints or others from the secondary interactions of the e$^+$e$^-$. Ref.~\cite{Gonzalez-Morales:2017jkx} also discussed the potential of future MeV missions to constrain annihilation of sub-GeV DM.
The extra-galactic (isotropic) MeV emission from DM annihilation/decay, which consists of a continuum emission as a result of the red-shift of the line, was also studied in a model independent way by~\cite{Yuksel:2007dr}.

Decay of specific FIP models have been also recently investigated in~\cite{Calore:2022pks,DelaTorreLuque:2025zjt}, where the authors set constraints on the decay rate of ALPs and sterile neutrinos, which are then used to constrain the ALP coupling with electrons and photons as well as the mixing angle for sterile neutrinos.
While ALPs decay into two photons, sterile neutrinos preferentially decay into a visible neutrino and a photon (also resulting in a line emission), with typical rates~\citep{Barger:1995ty, Picciotto:2004rp}:
\begin{equation}
     \frac{1}{\tau_{\gamma\gamma}} = \frac{g_{a\gamma}^2 }{64\pi}m_a^3 \,\,\,\,\,\,  \text{(ALPs)}\,, \,\,\,\,\,\,\, \,\,\,\,\,\,\,\,\,\,\,  \frac{1}{\tau_{\nu\gamma}} \approx \frac{ 9 \alpha G_F^2 m_{\nu_s}^5 \sin^2(2\theta)}{1024 \pi^4} \,\,\,\,\,\,  \text{(Sterile $\nu$)}\,.  
\end{equation}

Given that dark photons can not decay into one or two photons, they cannot produce a similar line-like feature. 
While for masses above twice the electron mass the dominant channel of decay will be to e$^+$e$^-$ (whose secondary emissions can be used to probe dark photons as described in the next section), at lower energies they decay preferentially into three photons, producing a continuum spectrum. Ref.~\cite{Linden:2024fby} recently used SPI data and obtained improved constraints on dark photons for masses in the keV-MeV range.

Then, beyond DM, FIPs are expected to be copiously produced in core-collapse SNe, creating a sort of cosmic diffuse SN FIP flux emitted by all past SNe in the Universe, and with peak energy at around $\sim30$~MeV~\citep{Calore:2021hhn}. In the case of ALPs, one can obtain constraints for masses below the meV from the isotropic gamma-ray flux at energies above a few tens of MeV from their conversion in the Milky Way magnetic field, as it was done in~\cite{Calore:2020tjw, Calore:2021hhn}, using Fermi-LAT data.
A similar strategy was used by~\cite{Caputo:2021rux} to obtain strong constraints on $g_{a\gamma}$ for MeV ALPs. The authors used isotropic extragalactic data, in the $2$-$200$~MeV domain, to constrain the gamma-ray flux produced by ALPs (and muonic bosons) decaying after leaving the SN volume.
The same reasoning can be used to get constraints for dark photons or sterile neutrinos producing gamma rays in their decay~\cite{ Linden:2024fby,Nguyen:2024kwy,Nguyen:2025tkl}.

In this regard, detection of gamma rays from a SN can set the strongest constraints (from direct photon observation) for FIPs in the MeV mass range, as discussed also in previous Sections (see, e.g., Sec.~\ref{subsec:Vitagliano} and Sec.~\ref{sec:Extragalactic_sources}). The Solar Maximum Mission~(SMM) gamma-ray observations from SN 1987A have been extensively studied and have led to stringent limits on $g_{a\gamma}$ for MeV ALPs~\citep{Chupp:1989kx, Jaeckel:2017tud,Hoof:2022xbe, Muller:2023vjm}.~\cite{Muller:2023vjm} also estimated the sensitivity of the Fermi-LAT experiment to the gamma-ray signal produced from a future nearby SN, showing that if observation is achieved, it can improve the current constraints  the observation of SN 1987A by more than an order of magnitude for masses of up to $600$~MeV.
Another interesting possibility related to a SN event was explored in \cite{Carenza:2022som}, where the authors showed that neutrino-boosted axion DM signals would generate a peculiar MeV gamma-ray signal that can be probed by next generation gamma-ray missions covering the MeV gap.

\subsubsection{Secondary photon production from WISPs}
\begin{figure}[t!]
\includegraphics[width=0.58\linewidth]{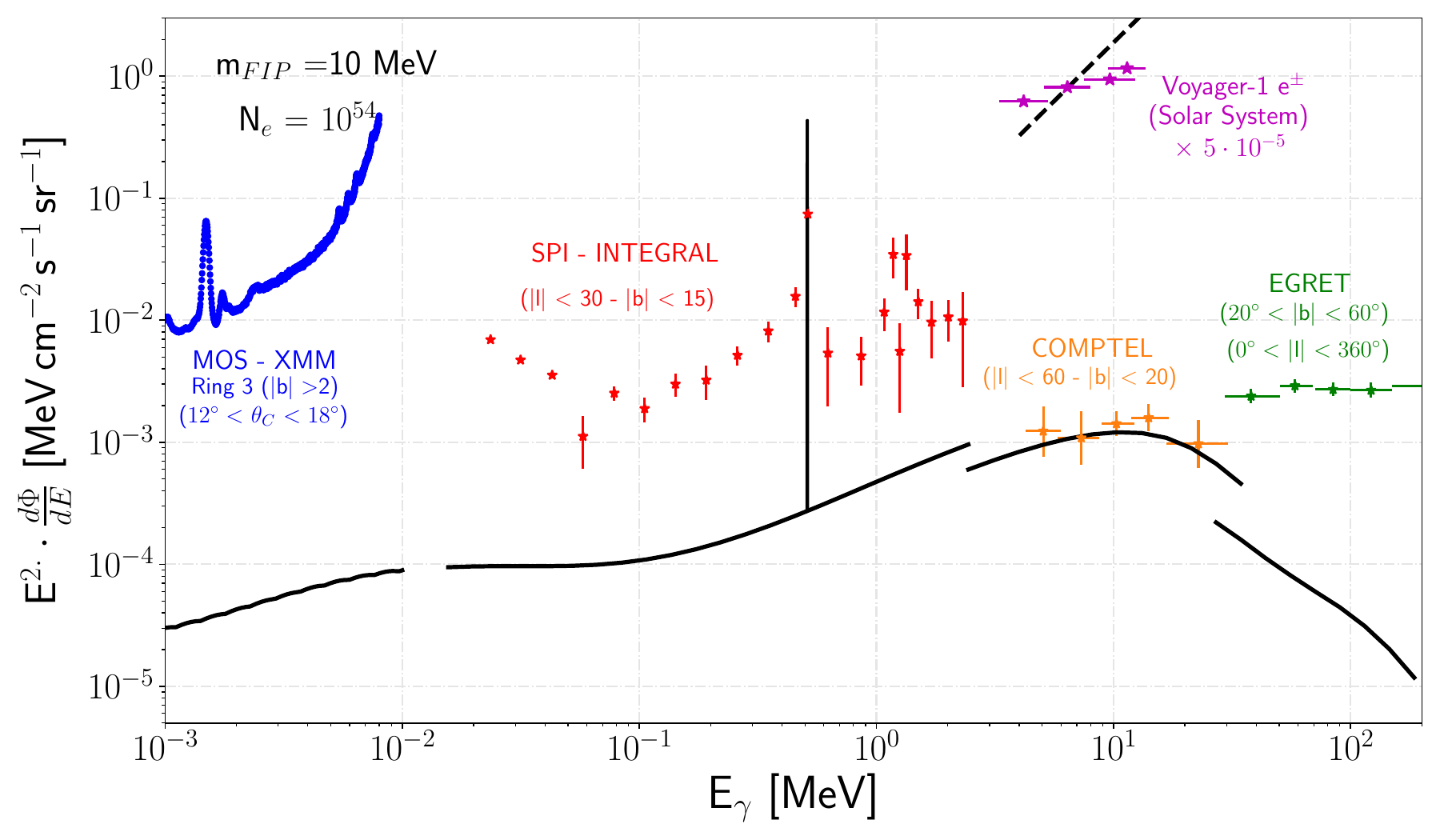} 
\includegraphics[width=0.42\linewidth]{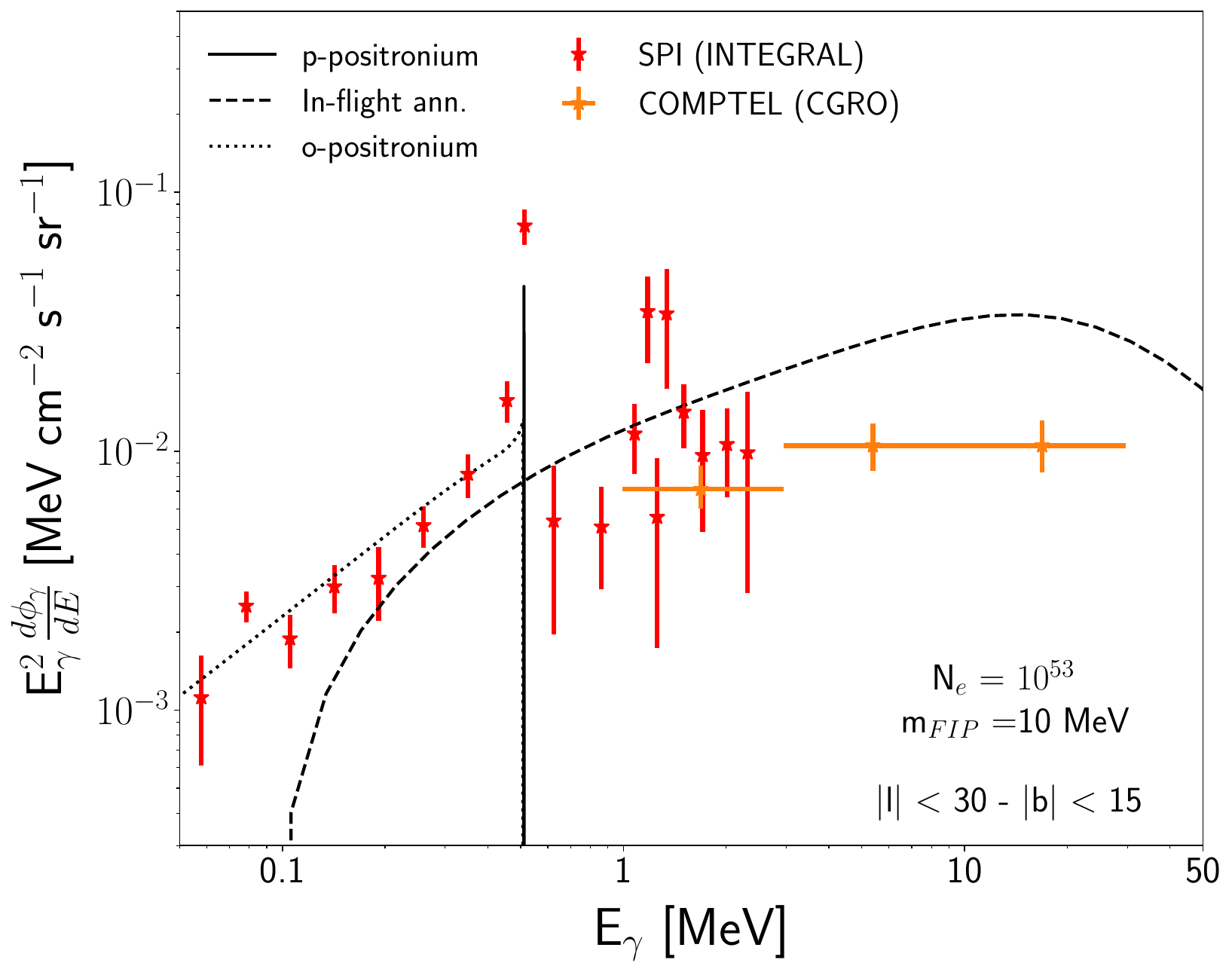} 
\caption{Predicted secondary  X-to-$\gamma$-ray signals for $10$~MeV FIPs produced in SNe. \emph{Left:} Inverse Compton, bremsstrahlung and $511$~keV line emission for N$_e = 10^{54}$. Figure adapted from \cite{DelaTorreLuque:2023huu}. \emph{Right:} Secondary emissions associated to electron-positron interaction for N$_e = 10^{53}$. Figure adapted from \cite{DelaTorreLuque:2024zsr}.}
\label{fig:Obs}
\end{figure}

The MeV gap have resulted tremendously interesting to constrain WISPs coupling to electrons. 
The decay or annihilation of any kind of sub-GeV DM particle into e$^+$e$^-$ pairs will produce a continuous diffuse sea of electrons and positrons that will travel in the Galaxy for even Gyrs, following a sort of diffusive propagation. During this time, these leptons interact with the Galactic environment producing gamma rays that can be used to probe the properties of the parent particles. A similar sea of secondary radiations would be produced by FIPs that are generated in SNe and can escape the SN envelope and decay in the interstellar medium~\citep{DelaTorreLuque:2023huu}, acting as continuous sources of e$^{\pm}$. 

The e$^{\pm}$ propagation can be approximated by
\begin{equation}
\label{eq:CRtransport}
- \nabla\cdot\left(D\vec{\nabla} f_e - \vec{v_{\omega}}\right) + \frac{\partial}{\partial p_e} \left[ p_e^2 D_{pp} \frac{\partial}{\partial p_e}\left(\frac{f_e}{p_e^2}\right)\right] = Q_e + \frac{\partial}{\partial p_e}\left( \dot{p}_e f_e + \frac{p_e}{3}\left( \vec{\nabla} \cdot \vec{v_{\omega}}\right) f_e \right) \;. 
\end{equation} 
The term $Q_e$ contains the spectrum of injected e$^{\pm}$ and the distribution of their sources (in this case, featuring the spatial distribution of Galactic SNe or the DM density distribution). The diffusion term, $D$, characterises the transport of charged particles in the Galaxy and has to be determined from local observations of cosmic rays at the GeV, which is an important limitation~\citep{DelaTorreLuque:2023olp}.
Once the diffuse distribution of e$^{\pm}$ is computed, their interactions with the gas (producing bremsstrahlung emission), interstellar fields (leading to inverse Compton emission) and magnetic fields (producing synchrotron radio emission) can be (often numerically) calculated and used to obtain constraints.
Using SPI data in the keV range, \cite{Cirelli_2021} derived competitive bounds on sub-GeV DM from inverse Compton scattering on galactic radiation fields and the CMB, for masses above $\sim 10$ MeV. This strategy has recently been extended to the X-ray band (slightly above $1$~keV) in \cite{Cirelli:2023tnx, DelaTorreLuque:2023olp, Balaji:2025afr}, resulting in limits that can be even stronger than cosmological ones. 
\cite{Bartels:2017dpb} showed that bremsshtralung and in-flight positron annihilation are dominant for low mass DM indirect searches in the MeV, and projected limits for future MeV missions.

In turn, injection of $e^\pm$ by FIPs produced in SNe can be modeled through a modified black-body spectrum. 
In this scenario, constraints from inverse-Compton and bremsstrahlung emissions from keV energies to hundreds of MeV were investigated in~\cite{DelaTorreLuque:2023huu, DelaTorreLuque:2023nhh}, finding that they can place competitive constraints on the FIP parameter space. Nevertheless,  limits from the $511$~keV line, produced when positrons thermalize and annihilate with interstellar electrons, lead to the strongest constraints for typical propagation parameters. This was previously investigated in~\cite{Calore:2021klc} and~\cite{Calore:2021lih}, considering not only the line originated in the Galaxy, but also the extra-galactic line emission.
These works used a general prescription that enables to study the emission of $e^\pm$ from FIPs produced in SNe in a model independent way, setting constraints that can be translated into an approximate bound on the coupling constant of any specific FIP models in the weak-mixing regime (see~\cite{DelaTorreLuque:2023huu} and~\cite{Calore:2021lih} for details).
The left panel of Fig.~\ref{fig:Obs} illustrates the secondary emissions associated to inverse Compton, bremsstrahlung and the 511 keV line in the keV to MeV range, compared to Galactic diffuse MeV data.
More recently,~\cite{DelaTorreLuque:2024zsr} and~\cite{Balaji:2025alr} showed that the in-flight positron annihilation emission leads to even stronger constraints on FIPs than the $511$~keV line, due to the high energies of the positrons injected. In particular, the authors discussed that by applying this scenario to the case of sterile neutrinos, ALPs and dark photons produced in SNe and considering COMPTEL observations, it is possible to improve previous limits by $1$-$2$ orders of magnitude. The right panel of Fig.~\ref{fig:Obs} shows the secondary emissions associated to the injection of positrons in the Galaxy considering in-flight positron annihilation, the $511$~keV line that results from para-positronium production and the continuum associated to orto-positronium production~\cite{Guessoum:2005cb}.

\subsection{Future improvements and opportunities in the MeV gap}

The renewed interest in the MeV domain has motivated a variety of proposed missions to bridge the gap. The upcoming COSI~\cite{COSI} experiment is expected to significantly improve the line and continuum sensitivity of the SPI instrument, covering the $0.2$-$5$ MeV region. Other proposed missions such as the All-sky Medium Energy Gamma-ray Observatory eXplorer (AMEGO-X)~\cite{AMEGO}, e-Astrogam~\cite{eASTROGAM}, the Advanced Energetic Pair Telescope (AdEPT)~\cite{AdePT} or the Advanced Particle-astrophysics Telescope (APT)~\cite{APT} are planned to cover the MeV-to-GeV range with much higher effective area and sensitivity than previous missions. The CRYSTAL EYE~\cite{CRYSTALEYE} ($0.01$-$30$~MeV), the Gamma-Ray and AntiMatter Survey (GRAMS) project~\cite{GRAMS} ($0.5$-$20$~MeV) or the Galactic Explorer with a Coded Aperture Mask Compton Telescope (GECCO)~\cite{GECCO} ($0.1$-$10$~MeV) are other interesting missions proposed in this domain. Besides the fact that these experiments are expected to improve current constraints by more than one order of magnitude, one of the most promising prospects of future MeV missions relies on the measurement of photon polarization.
From the point of view of the modeling, improving our knowledge of backgrounds for both the diffuse Galactic and extra-galactic MeV gamma-ray emission, as well as for the $511$~keV emission, would significantly enhance constraints from different experiments. Furthermore, improving the description of e$^{\pm}$ propagation at MeV energies will lead to a significant reduction of uncertainties affecting present constraints.

%% file: WG3/content/Michael_wurm.tex
\subsubsection{Introduction}

While not directly sensitive to WISPs, large-scale neutrino observatories looking for astrophysical neutrinos are able to provide constraints on WISP parameters, e.g.~ALP coupling and mass ranges (see, e.g., Ref.~\cite{Caputo:2024oqc} for a recent review). This is mostly due to the fact that -- if WISPs exist -- the hot and ultra-dense environments realized in the cores of stellar objects will not only produce neutrinos but also axions, hidden photons and other hypothetical particles. Their production will reduce neutrino luminosities as compared to the expectation as well as change properties of the spectra or time behavior (see Sec.~\ref{subsec:Raffelt} for more details). Therefore, the sensitivity of these observations to WISPs does not only depend on the neutrino data itself but also on the degree to which we understand their stellar sources and thus the expectation for the Standard-Model neutrino signal.

This contribution is structured as follows: Section \ref{sec:solar} reviews the possibilities to set constraints on WISP parameters using the existing observations of neutrinos from solar fusion. Section \ref{sec:sn} brushes the existing limits from SN 1987A and the possibility for deriving even stronger constraints from the next galactic Supernova before turning the attention to the question how -- in the absence of a galactic Supernova -- a future (non-)detection of the Diffuse Supernova Neutrino Background (DSNB) might be used to derive limits on WISP properties. Both sections describe as well the potential of currently running and future neutrino detectors, primarily present Super-Kamiokande, JUNO, and soon-to-be Hyper-Kamiokande experiments, to advance measurements of astrophysical neutrinos and thus further tighten the existing constraints on WISP parameters.

\subsubsection{Solar Neutrinos}
\label{sec:solar}

Since the early days of solar neutrino detection, the Standard Solar Model (SSM) has been very successfully used to predict the amount of neutrinos produced in the thermonuclear fusion reactions at the center of our Sun~\cite{Serenelli:2016dgz}. After some initial controversy about the mismatch of the solar neutrino measurements by Homestake, KamiokaNDE and gallium experiments with respect to SSM expectations, with the arrival of the SNO it became clear that this discrepancy was originated by neutrino flavor oscillations~\cite{Bellerive:2016byv,Wurm:2017cmm}. If those are taken into account, solar neutrino measurements and the SSM expectation are in excellent agreement (see, e.g.,~\cite{BOREXINO:2022abl}).

The SSM is crucially dependent on the measured photon output of the solar surface that very directly constraints the amount of energy to be released by thermonuclear fusion processes. If additional WISPs (e.g.~axions) were to be thermally produced in the core region of the Sun and thus contribute to its cooling, their existence would impact the energy balance of the star \cite{Caputo:2024oqc,Vinyoles:2015aba}. Fusion (and hence neutrino production) rates would have to be higher to balance the energy loss caused by WISP production. Higher fusion rates can be easily achieved by a slight increase in the solar core temperature $T_c$, facilitating fusion reactions. However, this would impact some fusion reactions more than others; for instance, the reaction $^7$Be+p$\to$$^8$B features a steep dependence ($\sim T_c^{18}$) and hence the resulting $^8$B neutrino flux would be more significantly enhanced compared to e.g.~the basic $pp$ fusion rate \cite{Vinyoles:2015aba}.

Limits on axion or hidden photon emission can hence be derived by evaluating the level of agreement between the SSM predictions for solar neutrino fluxes and the measured rates of individual solar neutrino species (i.e.~neutrinos resulting from different fusion reactions)~\cite{Caputo:2024oqc,Vinyoles:2015aba,Gondolo:2008dd}. This comparison has to take into account uncertainties in the SSM prediction, uncertainties in our knowledge of the oscillation parameters as well as the accuracy of the neutrino rate measurement. These observational bounds can be combined with results from helioseismology which provides additional intricate information on the conditions in the solar core. 
The limit based on present solar neutrino observations, excluding $g_{a\gamma}\geq 5\times 10^{-10}$\,GeV$^{-1}$~\cite{Vinyoles:2015aba}, is somewhat weaker than the ones derived from the life time of Horizontal Branch stars~\cite{Ayala:2014pea}. Note that for other WISP species, as for hidden photons, the limits are fully competitive~\cite{Vinyoles:2015aba}. 

Can these limits be improved by future observations? Yes and no. It is worth noting that the precise measurements of the $^8$B neutrino flux by the SNO and Super-Kamiokande experiments are providing uncertainties on the (2-3)\% level, as is true for the $^7$Be neutrino result of Borexino~\cite{Bellerive:2016byv,Super-Kamiokande:2016yck,BOREXINO:2018ohr}. However, in all of these cases the uncertainties in the SSM prediction are on the 10\% level and thus limit the sensitivity. Therefore, improvements in the $^8$B flux measurement expected from the recently operational JUNO and future Hyper-Kamiokande experiments will not directly translate to stricter limits on WISP parameters~\cite{JUNO:2023zty,JUNO:2022jkf,Yano:2020aap}. On the other hand, the SSM does very good in predicting the $pp$ and $pep$ neutrino fluxes, e.g.~0.6\% in the case of the basic $pp$ fusion reaction --  but here experiments are struggling to match this level of accuracy. Borexino's final result on the $pp$ neutrino flux featured an uncertainty of 10\%~\cite{BOREXINO:2018ohr}. Here, there are several ways conceivable of how to improve the measurement accuracy, ranging from small-scale scintillator experiments optimized for a $pp$ neutrino detection~\cite{Bieger:2021sas} to the next-generation of liquid-xenon WIMP detectors (DARWIN, XLZD) that will provide great sensitivity on the (sub-)per cent level~\cite{DARWIN:2020bnc}. As pointed out in Ref.~\cite{Vinyoles:2015aba}, a more precise $pp$ measurement will primarily improve the robustness of the existing limits on WISPs derived from solar neutrino observations, rendering them more model-independent, but not significantly impact the allowed parameter space.

\subsubsection{Supernova Neutrinos}
\label{sec:sn}

Core-collapse SNe not only optically outshine their host galaxies but are as well an incredibly bright neutrino source. The ten-second neutrino burst is mainly caused by the cooling of the emerging PNS. The ultra-dense matter of a PNS is mostly impenetrable to all particles other than neutrinos that are thus most effective at cooling via thermal production of $\nu\bar\nu$ pairs. Quite similarly to neutrinos, WISPs may be thermally produced, e.g.~via coupling to hadrons, and will be able to leave the PNS despite the high matter density. In this case, they will contribute to the energy loss of the PNS~\cite{Caputo:2024oqc}. Therefore, standard WISPs might affect in the neutrino signal in two different ways: either via an overall reduction in the observed neutrino flux compared to the gravitational energy released, or by a visible shortening (or deformation) of the exponential cooling time profile of the neutrino signal. Via both methods, stringent limits can be derived using the information encoded in the neutrino signal observed from Supernova SN 1987A (e.g.~\cite{Caputo:2024oqc,Carenza:2021pcm}). They are reported in detail in Sec.~\ref{subsec:Raffelt}. 

How much will existing limits on WISPs be improved by the observation of a future galactic Supernova? If the SN will happen in the central region of our galaxy, the number of neutrino events collected in larger and smaller scale neutrino detectors around the world will reach into the tens of thousands. This will provide the possibility for a much more detailed study of neutrino energy output and cooling curves than what has been possible on the $\sim$20 events detector from SN 1987A \cite{Caputo:2024oqc}. For this purpose, large-scale observatories like Super- and Hyper-Kamiokande, JUNO and DUNE will provide the most detailed information on the energies, rates and flavors of neutrinos arriving at Earth~\cite{GalloRosso:2020qqa,Hyper-Kamiokande:2021frf,DUNE:2020zfm}. Their data will be complemented by the information available from high-energy neutrino telescopes like IceCube that will provide the time-dependent envelope of the neutrino luminosity based on the increase in ``noise rate'' of the Digital Optical Module~(DOM) photo sensors~\cite{IceCube:2011cwc}. Further information will be given by very specialized experiments like HALO (Helium and Lead Observatory) that will provide data on $\mu$ and $\tau$ neutrino flavors~\cite{GalloRosso:2020qqa}. Individually and combined, information of all those experiments will be used to study the SN collapse and explosion, greatly refining uncertainties on astrophysical parameters like the overall explosion energy and the neutrino light curve. Those will be useful to further scrutinize the signal for signatures of exotic processes like axion emission competing with the standard neutrino cooling scenario. So while it can be expected that these results may be used to expand the current exclusion regions, the extent of this strengthening will largely depend on the SN progenitor, neutrino signal observed and the modeling of uncertainties connected to both.

On the other hand, there is an alternative SN neutrino signal available that -- while not yet observed -- could as well be used to infer information on WISP properties. The Diffuse Supernova Neutrino Background (DSNB) is emitted by all SNe occurring throughout the universe; they all contribute to a faint and nearly isotropic neutrino background not yet observed but by now very much in the range of the largest of current-day detectors. It seems very likely that a combination of gadolinium-loaded Super-Kamiokande~\cite{Harada:2023apz} already running and the JUNO experiment~\cite{JUNO:2022lpc} coming online will be able to reach at the discovery of the DSNB before the end of the decade~\cite{Li:2022myd}, maybe even inferring first limits on the underlying astrophysics. Statistics will be collected even faster once Hyper-Kamiokande comes online, even if without gadolinium loading.

In which way will this inform our knowledge on WISPs? For starters, the same SNe that contribute to the DSNB potentially create as an analogous isotropic background flux of WISPs. In case of axion-like particles, this has been coined the Diffuse Supernova ALP Background (DSAB)~\cite{Raffelt:2011ft}. Up to now, constraints have been derived from the non-observation of the decay of those ALPs into gamma-rays, that can be translated into limits on the diffuse gamma and hence DSAB flux. The upper limit corresponds to a coupling strength of $g_{a\gamma}\leq6\times10^{-13}\,{\rm GeV}^{-1}$ in the mass range below a few $10^{-10}$\,eV~\cite{Calore:2020tjw} (see Sec.~\ref{subsec:Vitagliano} and Sec.~\ref{sec:Extragalactic_sources} for more details). In the context of neutrino detection, it can be expected that an actual measurement of the DSNB neutrino flux will tighten the parameter range for the calculation of the original ALP flux. This will set the calculations of ALP production rates on a somewhat more solid basis and thus lead to an improvement in the existing limits on the coupling strengths by diffuse gamma observations.

\begin{figure}[t]
\centering
\includegraphics[width=0.6\linewidth]{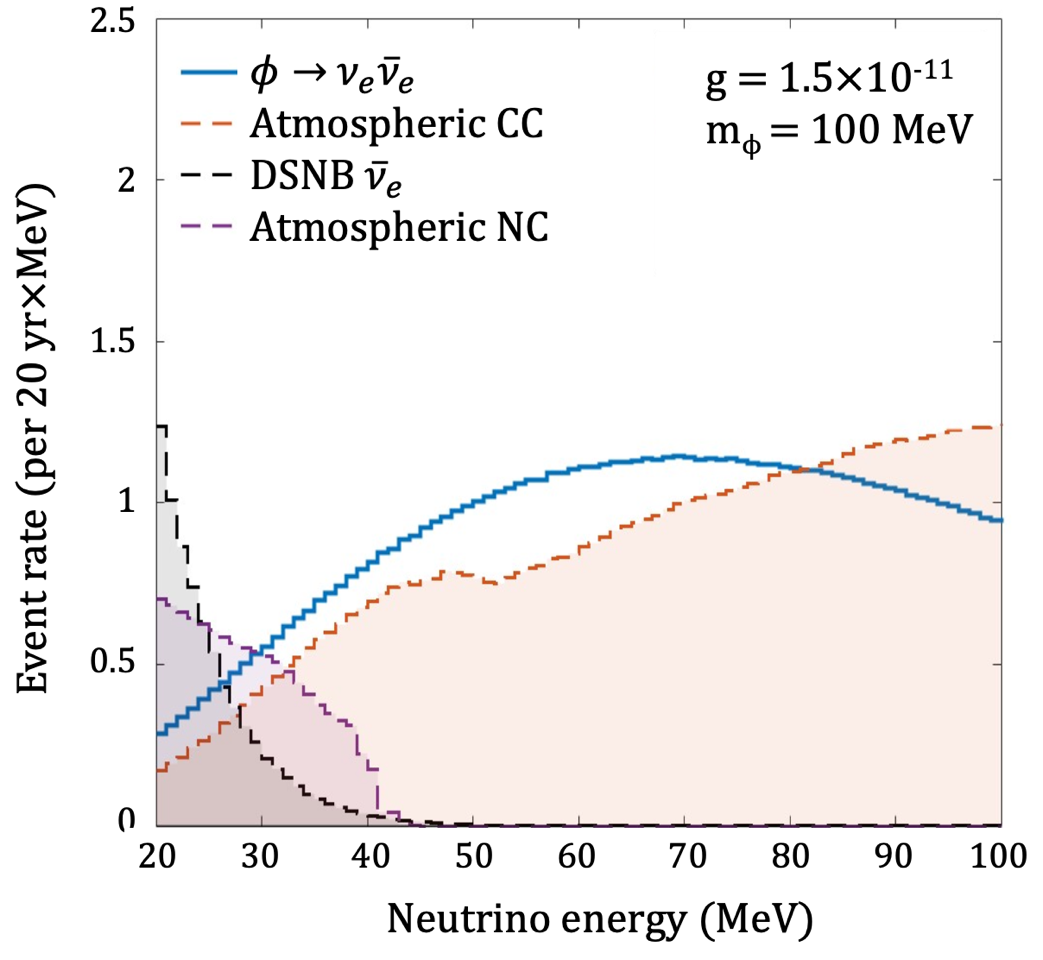}
\caption{\label{fig:dsnb} JUNO has the potential to observe the DSNB in the energy range of about $10-30$\,MeV. At higher energies, the signal is dominated by atmospheric neutrino backgrounds. A light boson produced in the cooling proto neutron star and coupling to neutrinos could create an excess signal enhancing both flux and spectrum of the DSNB. Figure taken from Ref.~\cite{Akita:2022etk}.}
\end{figure}

From the point of view of neutrino observations, the existence of WISPs could modify the expectation for flux and energy of the DSNB signal. As laid out in Ref.~\cite{Akita:2022etk}, if neutrinos feature a non-vanishing coupling with a hypothetical light boson in the mass range of 100\,MeV, these bosons will be produced in core-collapse SNe in neutrino pair annihilation. Upon leaving the source, these bosons may later on re-decay into $\nu\bar\nu$ pairs, and by this add a high-energy component to the standard SN neutrino spectrum caused by PNS cooling. 
The absence of such a high-energy component in the SN~1987A neutrino data strengthens the corresponding SN cooling limits by roughly an order of magnitude~\cite{Fiorillo:2022cdq}. A future Galactic SN neutrino burst, with far higher statistics, would be sensitive to this additional high-energy component, improving current SN 1987A constraints~\cite{Telalovic:2024cot}. Crucially, it would modify as well the energy spectrum expected for the DSNB. As shown in Fig.~\ref{fig:dsnb} for the JUNO detector, depending on coupling strength and boson mass the contribution to the $\bar\nu_e$ component can exceed the basic DSNB signal and also extend the spectrum to considerably higher energies above the standard $10-30$ MeV range. A similar effect would be visible in both Super- and Hyper-Kamiokande. Apart from a general enhancement in DSNB detection rates, the only signature would be a surplus of events at energies in the $30-100$ MeV range where the main background will be interactions of atmospheric neutrinos, sensitivity likely being limited by the uncertainty of their spectra. 
The upcoming DSNB measurements may strengthen the traditional SN~1987A cooling bound by up to a factor of two~\cite{Akita:2022etk} but they would not surpass the tighter constraints set by the absence of high-energy events in the SN~1987A neutrino data~\cite{Fiorillo:2022cdq}. Nonetheless, they remain valuable because they offer an independent channel sensitive to the same high-energy neutrino component.

\subsubsection{Conclusions}

Given the possibility to produce both neutrinos and WISPs in the dense and hot environments of stellar objects, neutrino observations have a considerable potential to constrain the allowed parameter space for all kinds of species of WISPs based on arguments about energy loss and cooling times. To the present day, competitive limits have been derived both using observations of solar neutrinos and the neutrino signal detected from SN 1987A. It is expected that these existing constraints will be further tightened by future experiments that provide higher-statistics measurements, e.g.~JUNO, Hyper-Kamiokande but also DARWIN and DUNE. In this context, of special interest will be a precision measurement of the solar $pp$ neutrino flux and a first detection of the Diffuse Supernova Neutrino Background since they will be able to provide qualitatively new ways to put with WISP constraints.

%% file: WG3/content/Rodrigo_Vicente.tex
\subsubsection{Introduction}

The most well-established property of dark matter (DM) is that it couples to gravity, but only very feebly (if at all) with the Standard Model. So, it is natural to try probing DM in the most extreme gravitational fields in the Universe: the vicinity of \emph{black holes} (BHs) and \emph{neutron stars} (NSs). 
Gravitational waves (GWs) provide us with a clean channel into such strong gravity regions, and we wonder: what can we learn about DM with GW observations? DM structures can either source GWs, or affect the sourcing and propagation of GWs from other conventional astrophysical sources.
Due to their collective coherent behavior on astrophysical scales, ultra-light dark matter (ULDM) models are specially suited to be probed via GW observations (for a review, see~\cite{Ferreira:2020fam,Hui:2021tkt}).
Here, we provide a brief overview of recent work on the topic. We refer the reader to Sec.~\ref{subsec:ul} for more details on the theory and cosmological constraints on ULDM.

While ULDM can be made of possibly several species of ultralight bosons\footnote{Or fermions, for a sufficiently large number of species~\cite{Davoudiasl:2020uig}.} of arbitrary spin, for concreteness, we will have in mind the minimal model of a single real (pseudo)scalar~$\Phi$ described by the Lagrangian density
\begin{equation}
	\mathcal{L}=-\frac{1}{2}\partial_\mu\Phi \partial^\mu \Phi-V(\Phi)\,,
\end{equation}
with~$V(\Phi)$ its self-interaction potential and~$\mu_\Phi\equiv\sqrt{ d^2V/d\Phi^2}|_{\Phi=0}$ its mass. We take~$\Phi$ to be minimally coupled to the spacetime metric~$g_{\mu \nu}$, sourcing its gravitational field via~$T_{\mu \nu}\supseteq \nabla_{\mu} \Phi \nabla_{\nu} \Phi-g_{\mu \nu}[\tfrac{1}{2}\nabla^\alpha \Phi\nabla_\alpha \Phi+V(\Phi)]$ in the Einstein equations. The ULDM field dynamics is described by\footnote{We use natural units~$G=c=\hbar=1$ and a mostly positive metric signature~$(-+++)$.}
\begin{equation}
	\Box_{\boldsymbol{g}}\Phi=dV/d\Phi\,, \qquad \qquad G_{\mu \nu}[\boldsymbol{g}]=8\pi T_{\mu \nu}[\boldsymbol{g},\Phi,\, \cdot\,]\,,
\end{equation}
with~$G_{\mu \nu}$ the Einstein tensor. In most astrophysical systems, the DM density is dilute enough to justify the use of~$V\approx \mu_\Phi^2 \Phi^2/2$; in some cases, one may need to consider a correction~$\lambda \Phi^4/24$.\footnote{For an axion-like ULDM particle~$\lambda \approx -\mu_\Phi^2/f_\Phi^2$, with a grand-unification or string symmetry breaking scale~$f_\Phi\sim 10^{16}\text{--}10^{18}\,\mathrm{GeV}$~\cite{Arvanitaki:2009fg, Hui:2016ltb}.} The most striking feature of ULDM models (and that distinguishes them from heavier \emph{particle} DM models) is that their de Broglie length scale~$\lambda_{\rm dB}\equiv 2\pi/(\mu_\Phi \bar{v}_\Phi)$ is macroscopic; as a reference, in the Galactic halo,~$\lambda_{\rm dB}\sim 1.5 {\rm \, km}$ for~$\mu_\Phi\sim 10^{-6}\,{\rm eV}$, or~$\lambda_{\rm dB}\sim 0.5 {\rm \, kpc}$ for~$\mu_\Phi\sim 10^{-22}\,{\rm eV}$. 

\subsubsection{GWs from ULDM structures}

Long-lived structures of ULDM are allowed by the balance between the field's self-gravity and wave pressure. These structures are known as \emph{solitons}\footnote{They are also referred to as \emph{oscillatons} (for a real scalar) or \emph{boson stars} (for a complex scalar); for reviews, see Refs.~\cite{Jetzer:1991jr, Liebling:2012fv, Visinelli:2021uve}.}~\cite{Kaup:1968zz, Ruffini:1969qy, Colpi:1986ye, Friedberg:1986tq, Seidel:1991zh} and have a characteristic radius~$R\lesssim\lambda_{\rm dB}/2\sim 1/(M \mu_\Phi^2)$, with~$M$ the mass of the ULDM structure. The minimum-energy (ground-state) field configurations are spherical and form dynamically via gravitational cooling (i.e., by expelling scalar field to highly excited/unbound states)~\cite{Seidel:1993zk}. Due to their morphology, these soliton structures cannot decay into GWs.\footnote{The sphericity of minimum-energy structures originates from the spin-0 assumption for the ULDM particle. It has recently be shown that the ground-state of a spin-1 field configuration is prolate (dipolar)~\cite{Herdeiro:2023wqf}, which for a real field (dark photon) leads to GW emission.} However, during their formation process (e.g., via Jeans instability of a diffuse scalar cloud), asymmetries and transient structures with more complex multipolar content may develop (like excited states, or perturbed ground-state solitons) which can relax through GW emission~\cite{Ferrell:1989kz,Kojima:1991np,Yoshida:1994xi}. However, for these GWs to be loud enough, the solitons must be very compact and, so, need to form in the early universe~\cite{Liddle:1992fmk, Arvanitaki:2019rax, Eroncel:2022efc}. Additionally, if (ultra)compact solitons exist they can form binaries and merge, emitting similar signals to the ones of standard astrophysical compact binaries~\cite{CalderonBustillo:2020fyi}. Still, in principle, these can be distinguished through, e.g., their tidal Love numbers or quasi-normal modes~\cite{Cardoso:2019rvt}.

A more robust constraint to GW phenomenology is derived for another type of ULDM structures, those sustained by the gravity of astrophysical compact objects. Such field configurations are dubbed ``gravitational atoms'' as their spatial distribution are akin to hydrogenoid wave functions. In the simplest instance, the existence of these overdense structures relies only on BH superradiance~\cite{1972JETP...35.1085Z,Starobinsky:1973aij,Brito:2015oca}, which provides a mechanism to rapidly grow dense \emph{boson clouds} around BHs~\cite{Detweiler:1980uk, Dolan:2007mj, Arvanitaki:2010sy, Brito:2014wla}. It is important to remark that this mechanism does not rely on the putative ultralight boson being all (or any) of the DM, as even vacuum fluctuations could serve as seed for an instability. In recent years, there has been much work exploring the GW signals sourced by superradiant boson clouds and assessing their detectability by current and future observations (see e.g.~\cite{Siemonsen:2022yyf}). In the following we summarize some of the important ideas and highlight the most recent observation/constraining prospects, to give the reader a flavor of the field. We note that, as discussed in Sec.~\ref{subsec:Chadha-Day}, BH superradiance can also be used to constrain the existence of ultralight bosons through other means.

\begin{figure}[t!]
	\centering
		\includegraphics[width=0.82 \textwidth]{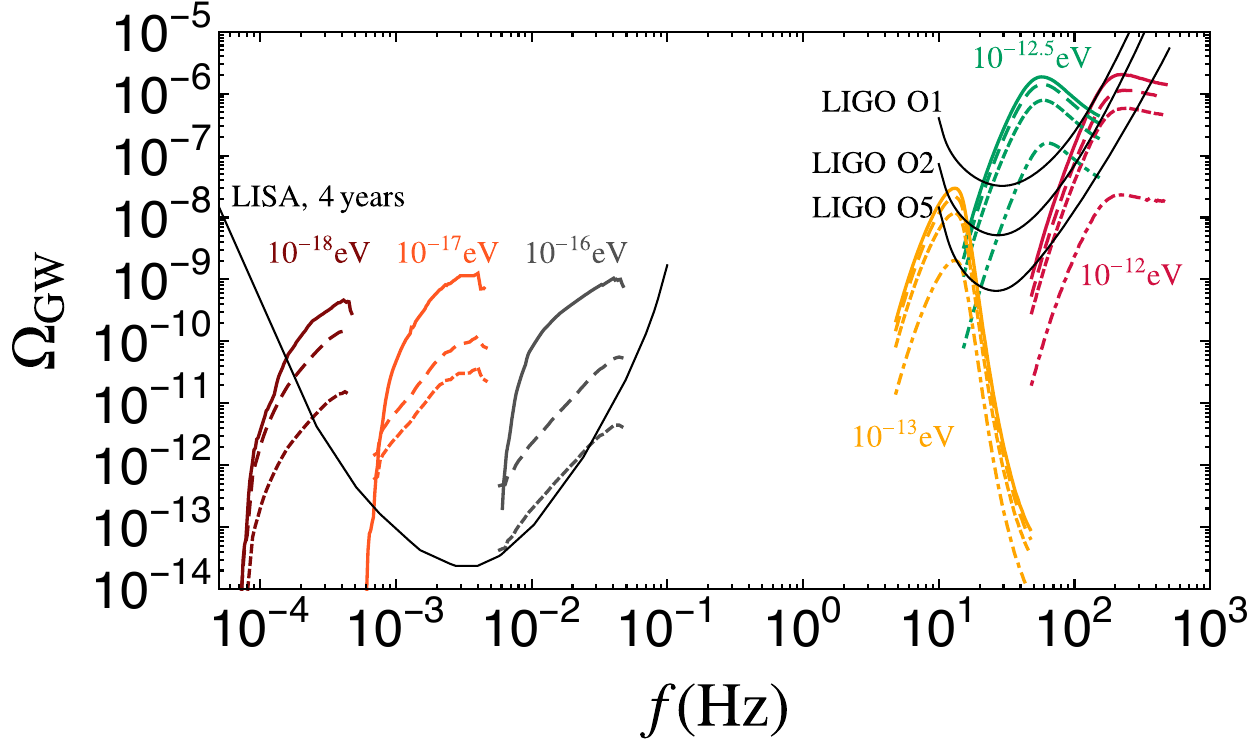}\\
\includegraphics[width=0.82\textwidth]{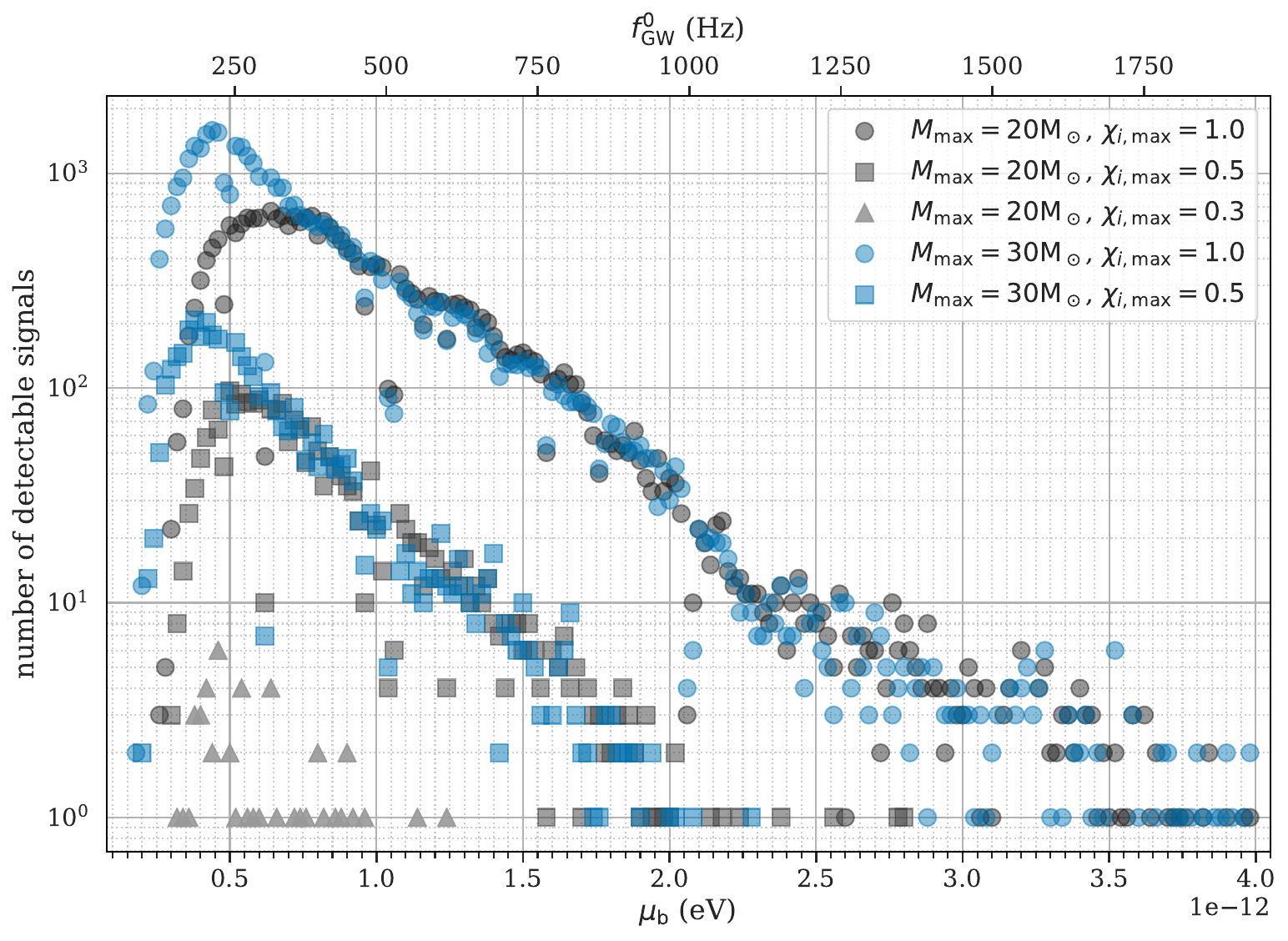}
	\caption{\small Constraining/observation prospects of blind searches for boson annihilation signals. These rely on assumptions about the astrophysical population of BHs, whose effect is illustrated by the different line patterns in the left panel, and different symbols/colours in the right panel. \emph{Upper panel:} Stochastic GW background from boson annihilation in LIGO and LISA bands. Figure taken from Ref.~\cite{Brito:2017wnc} (see also Ref.~\cite{Brito:2017zvb}). \emph{Lower panel:} Expected signals from boson annihilation above the sensitivity of current all-sky searches for continuous GWs. Figure taken from Ref.~\cite{Zhu:2020tht} (see also Ref.~\cite{Brito:2017zvb}). }
	\label{fig:annihilation}
\end{figure}

\begin{figure}[h]
\centering
	\includegraphics[width=0.8 \textwidth]{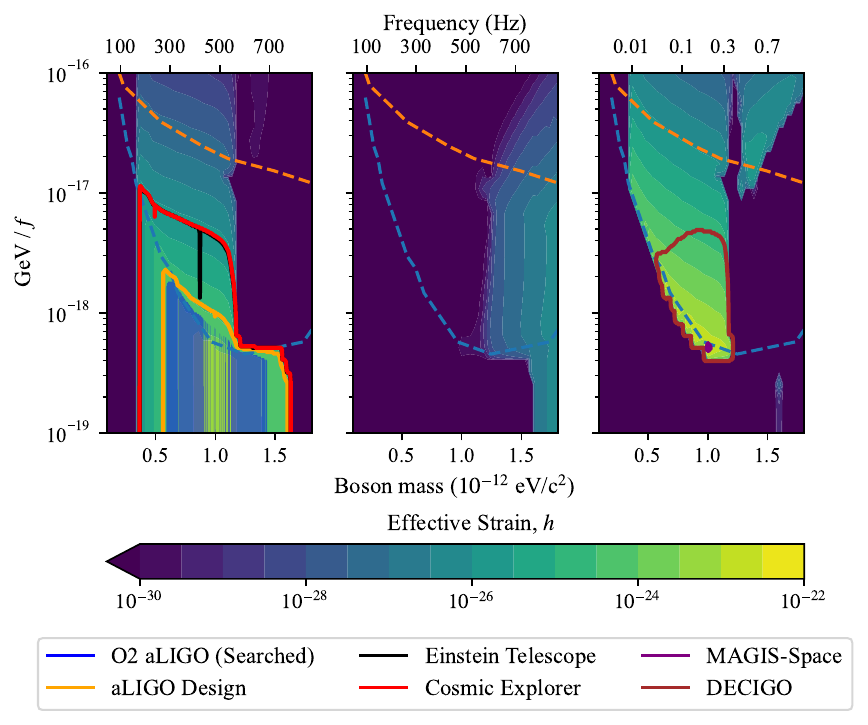}
	\caption{Parameter space probed by current/future direct searches targetting Cygnus X-1. Panels show the effective GW strain from $(211)$ annihilation (left), $(322)$ annihilation (middle), and~$(322)\to(211)$ transition (right). Figure taken from Ref.~\cite{Collaviti:2024mvh} (see also Refs.~\cite{Isi:2018pzk, Sun:2019mqb}).}
	\label{fig:multiband}
\end{figure}

Due to the specifics of BH superradiance, boson clouds are grown with a characteristic multipolar structure. In particular, for a gravitational coupling~$\alpha\equiv \mu_\Phi M_{\rm BH}\lesssim 0.22$ and for BHs born with spin higher than~$\chi_c \approx 4\alpha/(1+4\alpha^2)$, only the $(n\ell m)=(211)$ and $(322)$ states of a scalar cloud are grown fast enough~\cite{Collaviti:2024mvh}, where $n-1$ is the number of radial nodes of the field and $(\ell m)$ the spherical-harmonic numbers. Now, a scalar cloud state~$(n\ell m)$ has the form~$\Phi_{n \ell m}\propto \mathrm{Re}[e^{-i\omega_{n \ell m} t}Y_{\ell m}(\theta, \varphi)]$, which for~$\ell\geq 1$ leads to a time-periodic, non-vanishing, quadrupolar moment of the energy density~$\rho=T_{tt}$, sourcing GWs of frequency~$\omega_{\rm gw}=2 \omega_{n \ell m}\sim 2 \mu_\Phi$. Heuristically, this can be seen as the annihilation of scalars into GWs. 
When self-interactions are weak, the cloud is in the gravitational regime (i.e., the cloud states grow only via superradiance). In this regime, GWs originate only from boson annihilation and the signals are (quasi-)monochromatic with frequency\footnote{There is a tiny positive frequency drift originating from corrections associated to the non-linear self-binding energy of the cloud, which is proportional to the square of the boson cloud mass~\cite{Zhu:2020tht}.}
\begin{equation*}
	f_{\rm gw}^{\rm annih}\sim 50 \,{\rm Hz}\,\Big(\frac{\mu_\Phi}{10^{-13}\, {\rm eV}}\Big)\approx 0.5 \,{\rm mHz}\,\Big(\frac{\mu_\Phi}{10^{-17}\, {\rm eV}}\Big)\,.
\end{equation*}
Such annihilation signals can be used to probe masses~$\mu_\Phi \sim 10^{-13}\text{--}10^{-12}\,{\rm eV}$ with LIGO and~$\mu_\Phi \sim 10^{-19}\text{--}10^{-15}\,{\rm eV}$ with LISA (see Fig.~\ref{fig:annihilation}). 
For moderate self-interactions,\footnote{For the precise conditions on~$f_\Phi$, see Ref.~\cite{Baryakhtar:2020gao, Collaviti:2024mvh}.} the quartic self-interactions can appreciably populate the $(322)$ state (via a $211\,\times\,211\,\to\,322\,\times\,{\rm BH}$ process), with the~$(211)$ remaining highly populated~\cite{Collaviti:2024mvh}. Such system tends to a transient quasi-equilibrium regime~\cite{Omiya:2022gwu}, with a constant ratio of the occupation number of~$(211)$ to~$(322)$, due to the simultaneous induced transition~$322\,\times\,322\,\to\,211\,\times\,{\infty}$. The large occupation of the two levels leads to mode mixing in~$\rho=T_{tt}$, which results in the emission of GWs with frequency~$\omega_{\rm gw}=\omega_{322}-\omega_{211}\sim \alpha^2 \mu_\Phi $~\cite{Arvanitaki:2014wva, Siemonsen:2019ebd}; equivalently, 
\begin{equation*}
	f_{\rm gw}^{\rm trans}\sim 2.5 \times 10^{-3}\, {\rm Hz}\,\Big(\frac{\alpha}{0.1}\Big)^2\Big(\frac{\mu_\Phi}{10^{-13}\, {\rm eV}}\Big).
\end{equation*}
This process can be interpreted as a level transition~$322\,\to\,211\,\times\,{\rm GW}$, and happens simultaneously with the boson annihilation.\footnote{Such level transition process also happens for complex fields.} The much lower frequency of GW signals from level transitions compared to annihilation allows one to probe heavier boson masses, or a multi-band probe for a given boson mass (see Fig.~\ref{fig:multiband}). For large self-interactions, the superradiance instability is suppressed and no substantial power into GW is expected~\cite{Collaviti:2024mvh}.

GWs from superradiant clouds can be resolvable or stochastic (due to many superposed signals), and the searches may either be directed (to a system with known parameters) or blind. While blind searches have the largest constraining power in terms of ultralight boson masses, they rely on assumptions about the astrophysical population of BHs, e.g., distribution of distances, velocities, masses, spins, ages, which are not well known. On the other hand, directed searches have less constraining power, but, depending on our knowledge about the parameters and history of the system, they can be much more robust. Some of the most recent constraining/observation prospects are shown in Figs.~\ref{fig:annihilation} and~\ref{fig:multiband}, respectively, for blind and directed searches. Most of the GW phenomenology of superradiant clouds of ultralight vectors is qualitatively similar to scalars; for the specifics, we refer the reader to Refs.~\cite{Siemonsen:2019ebd, Siemonsen:2022yyf}. On the other hand, ultralight tensors have a very distinct dynamics around BHs~\cite{East:2023nsk}.

\subsubsection{ULDM effects on the sourcing of GWs}

Astrophysical environments where compact binaries evolve may leave imprints on their GW waveforms depending, among others, on the environment density and number of cycles in band~\cite{Kocsis:2011dr, Barausse:2014tra, Cardoso:2019rou, Cole:2022yzw, CanevaSantoro:2023aol, Roy:2025qaa, Vicente:2025gsg}. Such effects originate from the binary interaction with their environments through accretion, dynamical friction, resonant torques, etc. Thus, it is expected that the same happens for compact binaries (like BHs and NSs) inspiralling within the overdense ULDM structures described in the previous section. In particular, the tidal field of a companion object can induce level transitions in a (superradiant) boson cloud leading to energy/angular momentum exchange, resulting in a rich phenomenology (e.g., floating/sinking orbits) which has been explored in many works (e.g.,~\cite{Baumann:2018vus, Berti:2019wnn, Baumann:2019ztm, Takahashi:2021yhy, Baumann:2021fkf, Baumann:2022pkl, Takahashi:2023flk}). However, the tidal induction of level transitions is most effective early in the inspiral phase much before coalescence, and their effect might be more easily observed in the distribution of binary parameters~\cite{Tomaselli:2024bdd, Boskovic:2024fga, Tomaselli:2024dbw}. When a binary is in band and close enough to coalescence, if the ULDM structure is not significantly depleted by then, we expect dynamical friction~\cite{Lancaster:2019mde, Annulli:2020ilw, Annulli:2020lyc, Traykova:2021dua, Vicente:2022ivh, Buehler:2022tmr, Traykova:2023qyv, Tomaselli:2023ysb, Brito:2023pyl, Duque:2023seg} (also called gravitational \emph{ionization} in this context) to be the dominant environmental effect from ULDM. 

Dynamical friction from ULDM environments is expected to cause a de-phasing in the waveform with respect to the ``chirping'' in vacuum~\cite{Aurrekoetxea:2023jwk}, which can be modelled (in the adiabatic approximation) through angular momentum conservation
\begin{equation}
	\frac{d}{dt}L^z_{\rm orb}=\tau^z_{\rm gw}+\tau_{\rm df}^z,
\end{equation}
where~$L^z_{\rm orb}$ is the orbital angular momentum (the contribution of accretion may be included in its derivative),~$\tau^z_{\rm gw}$ is the torque from GW radiation reaction, and~$\tau^z_{\rm df}$ the one from dynamical friction, in which relativistic corrections can also be included. From the last equality, one can find the evolution of the orbital angular frequency and use then the stationary phase approximation~\cite{Droz:1999qx} to obtain the waveform in frequency-domain. Ref.~\cite{Roy:2025qaa} showed that such procedure captures well the numerical relativity simulations of Refs.~\cite{Bamber:2022pbs, Aurrekoetxea:2023jwk}; the resulting waveforms were then used to derive upper bound on the density of scalar fields around binary BHs observed by LIGO-Virgo-Kagra.

\begin{figure}[t!]
\centering
	\includegraphics[width=0.8 \textwidth]{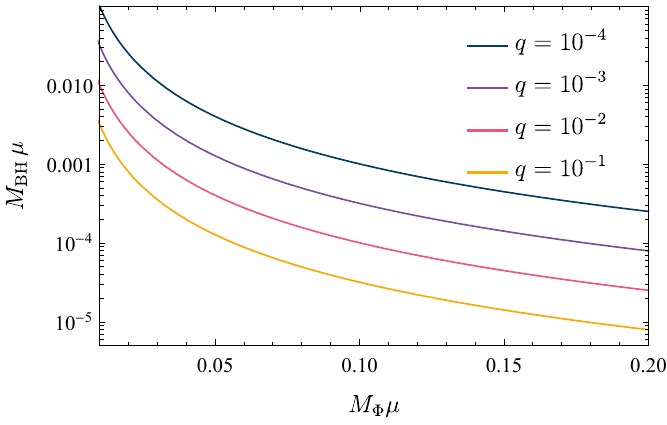}
	\caption{Curves show the parameter space for a one-cycle de-phasing of an EMRI in a ULDM soliton (above the de-phasing is larger), with~$M_{\rm BH}$ the mass of the primary BH and~$M_\Phi$ the mass of the soliton. Figure from Ref.~\cite{Duque:2023seg}.}
	\label{fig:EMRI}
\end{figure}

Various studies forecast that near-future GW detectors (like LISA or B-DECIGO) can be sensitive to the densities of ULDM structures, and might even be able to distinguish them from other astrophysical or particle DM environments (see, e.g.,~\cite{Cole:2022yzw}). As an example, using a fully-relativistic analysis, Ref.~\cite{Duque:2023seg} has shown that boson clouds with mass~$M\gtrsim 0.01 M_{\rm BH}$ and gravitational coupling~$\alpha \gtrsim 0.08$ lead to (at least) one cycle of de-phasing in extreme-mass-ratio inspirals with mass-ratio~${q\gtrsim 5 \times 10^{-5}}$, which can be used as a simple rule-of-thumb for the detectability with LISA~\cite{Kocsis:2011dr}. For a ULDM soliton the parameter space leading to (at least) one cycle of de-phasing is shown in Fig.~\ref{fig:EMRI}. 
The first cosmological simulations~\cite{Schive:2014dra, Schive:2014hza} indicate that~$M_\Phi \mu_\Phi \sim 10^{-3}$ for halos of mass~$M_h\sim 10^{12} M_\odot$, but there has been considerable dispersion in the literature~\cite{Chan:2021bja} allowing up to~$M_\Phi \mu_\Phi \lesssim 0.02(\mu_\Phi/10^{-22}\,{\rm eV})^{1/2}$. This indicates that the de-phasing is potentially observable for sufficiently compact solitons with~$\mu_\Phi\gtrsim 10^{-20}{\rm eV}$.
While we have focused on the imprints in the inspiralling, which can accumulate over many cycles, we remark that an ULDM environment can also alter the merger-ringdown waveform~\cite{Bamber:2021knr, Chung:2021roh, Choudhary:2020pxy, Zhang:2022rex}. 

\subsubsection{ULDM effects on the propagation of GWs}

It has been shown that ULDM structures induce a change in the speed of GWs, which can be described through an effective refractive index depending on the mass and self-interactions of the ULDM particle, and the GW frequency~\cite{Dev:2016hxv}. Such deviation in the speed of GWs could be potentially observed with multimessenger astronomy. The analysis of Ref.~\cite{Banerjee:2022zii} indicates that such deviations cannot be probed with LIGO-Virgo-Kagra, but suggests that a meaningful constraint could be obtained with much lower~$\lesssim\mathrm{nHz}$ frequencies using future (and perhaps current) PTA experiments.
 
Interestingly, while the energy density~$\rho=T_{tt}$ of a non-relativistic ULDM soliton is static at leading order, its pressure~$p=T_{ii}$ is time-periodic. This sources an oscillating post-Newtonian contribution to the ULDM gravitational potential with frequency~$\omega=2 \mu_\Phi$. When high-frequency electromagnetic or gravitational waves propagate in a oscillating background they are subjected to a frequency modulation due to a Sachs-Wolfe effect~\cite{Laguna:2009re}. The effect on pulsar timing was studied in Ref.~\cite{Khmelnitsky:2013lxt} (see also~\cite{DeMartino:2017qsa}), and has already been used to constrain the local ULDM density for masses~$\mu_\Phi\lesssim 10^{-23}\,{\rm eV}$~\cite{EuropeanPulsarTimingArray:2023egv}, and conformal ULDM couplings to baryonic matter~\cite{Smarra:2024kvv}.
A similar phenomenology can be explored with GWs~\cite{Wang:2023phr, Brax:2024yqh, Blas:2024duy}, which provide a cleaner channel to galactic centres, where the DM density can be much larger. Ref.~\cite{Blas:2024duy} considered several near-futures detectors and GW sources using state-of-art astrophysical population models and concluded that the most optimistic case corresponds to high-frequency~$(\sim 500\,{\rm Hz})$ GWs from young spinning NSs at the Galactic centre. Then observations of the Einstein Telescope or Cosmic Explorer could be used to exclude a soliton at the Galactic centre for all~$\mu_\Phi\lesssim 10^{-22}\,{\rm eV}$.

%% file: wg4.tex
\part[WISP Direct Searches \\ \textnormal{\small Editors: D.~Aybas, N.~Ferreiro~Iachellini, C.~Gatti, M.~Karuza, A.~Rettaroli, O.~M.~Ruimi \& M.~Staelens}]{WISP Direct Searches
\\ \textnormal{\small Editors: D.~Aybas, N.~Ferreiro~Iachellini, C.~Gatti, M.~Karuza, A.~Rettaroli,\\[-.5em] O.~M.~Ruimi \& M.~Staelens}}\label{part:wg4}

\input{WG4/content/Introduction}
\label{sec:WG4Introduction}

\section{European WISP Experiments}
\label{sec:EUexperiments}

\subsection{\texorpdfstring
  {Introduction to Haloscopes \\ \textnormal{\small Authors: D. Aybas \& C. Gatti}}{Introduction to Haloscopes (Authors: D. Aybas and C. Gatti)}}
\label{sec:IntroductionHaloscopes}
\input{WG4/content/Intro-Haloscopes}

\subsection{\texorpdfstring
  {Precision Metrology for Ultralight ALP Dark Matter \\ \textnormal{\small Author: S. Chakraborti}}{Precision Metrology for Ultralight ALP Dark Matter (Author: S. Chakraborti)}}
\input{WG4/content/AxionULDMQuantumSensors}

\subsection{\texorpdfstring
  {Optical Cavity Techniques for Axion Dark Matter Detection \\ \textnormal{\small Authors: A. Ejlli \& S. Kunc}}{Optical Cavity Techniques for Axion Dark Matter Detection (Authors: A. Ejlli and S. Kunc) }}
\input{WG4/content/Intro-DarkMatterOpticalHaloscopes}

\subsection{\texorpdfstring
  {European haloscopes in progress or completed \\ \textnormal{\small Authors: D.~Aybas, I.~M.~Bloch, D.~Budker, A.~D\'iaz-Morcillo, C.~Gatti, D.~Gavilan-Martin, D.~Horns, A.~Lindner, O.~Maliaka, M.~Maroudas, L.~H.~Nguyen \& O.~Ruimi}}{European haloscopes in progress or completed (Authors: D.~Aybas, I.~M.~Bloch, D.~Budker, A.~D\'iaz-Morcillo, C.~Gatti, D.~Gavilan-Martin, D.~Horns, A.~Lindner, O.~Maliaka, M.~Maroudas, L.~H.~Nguyen and O.~Ruimi)}}
  
\subsubsection{The BRASS experiment}
\input{WG4/content/BRASSExperiment}

\subsubsection{The CASPEr Experiment}
\label{sec:caper}
\input{WG4/content/CASPErExperiment}

\subsubsection{The CAST-CAPP experiment}
\label{sec:cast-capp}
\input{WG4/content/CAST-CAPPExperiment}

\subsubsection{The MADMAX experiment}
\input{WG4/content/MADMAXExperiment}

\subsubsection{The QUAX experiment}
\label{Sec:QUAX}
\input{WG4/content/QUAXExperiment}

\subsubsection{The RADES-CAST experiment}
\label{sec:cast-rades}
\input{WG4/content/RADES-CASTExperiment}

\subsubsection{The WISPDMX experiment}
\input{WG4/content/WISPDMXExperiment}

\subsubsection{The GNOME Experiment}   
\input{WG4/content/GNOMEExperiment}
        
\subsubsection{The NASDUCK Experiment}
\input{WG4/content/NASDUCKExperiment}

\subsection{\texorpdfstring
  {European haloscopes planned or under construction \\ \textnormal{\small Authors: C.~Baynham, J.~De Miguel, A.~D\'iaz-Morcillo, A.~Ejlli, G.~Higgins, M.~Maroudas, G.~Mueller, P.~Pugnat, A.~Rettaroli, Q.~Rokn, K.~Schmieden, M.~Schott \& C.~Toni}}{European haloscopes planned or under construction (Authors: C.~Baynham, J.~De Miguel, A.~D\'iaz-Morcillo, A.~Ejlli, G.~Higgins, M.~Maroudas, G.~Mueller, P.~Pugnat, A.~Rettaroli, Q.~Rokn, K.~Schmieden, M.~Schott and C.~Toni)}}
  
\subsubsection{The AION experiment}
\label{sec:AionExperiment}
\input{WG4/content/AIONExperiment}

\subsubsection{APE Experiment}
\input{WG4/content/APE}

\subsubsection{The CADEx experiment}
\input{WG4/content/CADEx_Experiment}

\subsubsection{The DALI experiment}
\input{WG4/content/DALIExperiment}

\subsubsection{The FLASH experiment}
\input{WG4/content/FLASHExperiment}

\subsubsection{The GrAHal experiment}
\input{WG4/content/GrahalExperiment}

\subsubsection{The MaglevHunt experiment}
\input{WG4/content/MaglevHuntExperiment}

\subsubsection{The RADES-BabyIAXO experiment}
\input{WG4/content/RADES-BabyIAXOExperiment}

\subsubsection{The RADES-LSC experiment}
\input{WG4/content/RADES-LSCExperiment}

\subsubsection{The RadioAxion experiment}
\input{WG4/content/RadioAxionExperiment}

\subsubsection{The SUPAX experiment}
\input{WG4/content/SupaxExperiment}

\subsubsection{The WISPLC experiment}
\input{WG4/content/WISPLCExperiment}

\subsection{\texorpdfstring
  {Introduction to Helioscopes \\ \textnormal{\small Author: K. Zioutas}}{Introduction to Helioscopes (Author: K. Zioutas)}}
\label{sec:IntroHelioscopes}
\input{WG4/content/Helioscopes_intro}

\subsection{\texorpdfstring
  {European helioscopes in progress or completed \\ \textnormal{\small Authors: C.~Margalejo \& J.~K.~Vogel}}{European helioscopes in progress or completed (Authors: C.~Margalejo and J.~K.~Vogel)}}

\subsubsection{The CAST experiment}
\label{sec:CAST-MM}
\input{WG4/content/CASTHelioscopeExperiment}
        
\subsection{\texorpdfstring
  {European helioscopes planned or under construction \\ \textnormal{\small Authors: F.~R.~Candón \& J.~K.~Vogel}}{European helioscopes planned or under construction (Authors: F.~R.~Candón and J.~K.~Vogel)}}

\subsubsection{The IAXO experiment}
\label{sec:IAXO}
\input{WG4/content/IAXOexperiment}

\subsubsection{The BabyIAXO experiment}
\label{sec:BabyIAXO}
\input{WG4/content/BabyIAXOExperiment}

\subsection{\texorpdfstring
  {Introduction to Pure Laboratory Experiments \\ \textnormal{\small Authors: A. Ejlli \& S. Kunc}}{Introduction to Pure Laboratory Experiments (Authors: A. Ejlli and S. Kunc)}}
\label{sec:IntroPureLab}
\input{WG4/content/Pure_Lab_intro}

\subsection{\texorpdfstring
  {Fifth Force Experiments \\ \textnormal{\small Author: L. Cong}}{Fifth Force Experiments (Author: L. Cong)}}
\input{WG4/content/Intro-PureLab5thForce}

\subsection{\texorpdfstring
  {Searching for Spin-Dependent Exotic Interactions with Exotic Atomic Spectroscopy  \\ \textnormal{\small Author: L. Cong}}{Searching for Spin-Dependent Exotic Interactions with Exotic Atomic Spectroscopy (Author: L. Cong)}}
\label{Sec:EAS}
\input{WG4/content/ExoticAtomExperiments}

\subsection{\texorpdfstring
  {European pure lab WISP experiments in progress or completed \\ \textnormal{\small Authors: A. Ejlli, S. Kunc, A.~Lindner \& M.~Maroudas}}{European pure lab WISP experiments in progress or completed (Authors: A. Ejlli, S. Kunc, A.~Lindner and M.~Maroudas)}}

\subsubsection{The ALPS II experiment}
\input{WG4/content/ALPSIIExperiment}

\subsubsection{OSQAR Experiment}
\input{WG4/content/OSQARExperiment}

\subsubsection{PVLAS Experiment and VMB Theory Summary}
\input{WG4/content/PVLAS}

\subsubsection{The WISPFI experiment}
\input{WG4/content/WISPFIExperiment}

\subsection{\texorpdfstring
  {European pure lab WISP experiments planned or under construction \\ \textnormal{\small Authors: L. Cong, A. Ejlli, S.~Kunc, G.~Mueller, L.~Roberts \& P.~Spagnolo}}{European pure lab WISP experiments planned or under construction (Authors: L. Cong, A. Ejlli, S.~Kunc, G.~Mueller, L.~Roberts and P.~Spagnolo)}}

\subsubsection{ALP searches with nEDM experiments}
\label{Sec:EDM}
\input{WG4/content/EDMexperiments}
    
\subsubsection{The STAX experiment}
\input{WG4/content/STAXExperiment}

\subsubsection{The VMB@CERN experiment}
\label{sec:vmbcern}
\input{WG4/content/VMB_CERN}

\subsubsection{High-Frequency Rotating-Field VMB Polarimeter}
\input{WG4/content/VAMBI}

\subsection{\texorpdfstring
  {Introduction to Accelerated Particle-Beam-Driven WISP Experiments: Fixed-Target and Beam-Dump  \\ \textnormal{\small Author: V. Kozhuharov}}{Introduction to Accelerated Particle-Beam-Driven WISP Experiments: Fixed-Target and Beam-Dump (Author: V. Kozhuharov)}}
\label{sec:IntroBeamDump}
\input{WG4/content/Intro-BeamDump}

\subsection{\texorpdfstring
  {European Accelerated Particle-Beam-Driven WISP Experiments \\ \textnormal{\small Authors: B.~D\"obrich, V.~Kozhuharov, L.~Marsicano \& R.~Quishpe}}{European Accelerated Particle-Beam-Driven WISP Experiments (Authors: B.~D\"obrich, V.~Kozhuharov, L.~Marsicano \& R.~Quishpe)}}

\subsubsection{The PADME experiment}
\input{WG4/content/PADMEexperiment}

\subsubsection{The LUXE-NPOD experiment}
\input{WG4/content/LUXE-NPOD}

\subsubsection{The NA62 experiment in beam-dump mode}
\input{WG4/content/NA62Experiment}

\subsubsection{The NA64 experiment}
\input{WG4/content/NA64Experiment}

\subsection{\texorpdfstring
  {Introduction to Non-WISP-Focused Experiments with Ability to Detect WISPs  \\ \textnormal{\small Author: S. Cetin}}{Introduction to Non-WISP-Focused Experiments with Ability to Detect WISPs (Author: S. Cetin)}}
\input{WG4/content/Intro-NonWISP}

\subsection{\texorpdfstring
  {Gravitational-wave and atom interferometers  \\ \textnormal{\small Author: F. Urban}}{Gravitational-wave and atom interferometers (Author: F. Urban)}}
\input{WG4/content/GWInterferometers}

\subsection{\texorpdfstring
  {Gravitational-Wave Detectors for Axion Searches  \\ \textnormal{\small Authors: A. Ejlli \& S. Kunc}}{Gravitational-Wave Detectors for Axion Searches (Authors: A. Ejlli and S. Kunc)}}
\input{WG4/content/Intro-GWOptical}

\subsection{\texorpdfstring
  {European Non-WISP-Focused Experiments with Ability to Detect WISPs \\ \textnormal{\small Authors: C. De Dominicis, M. Gallinaro, G. Grilli di Cortona, V.~A.~Mitsou \& M.~Staelens}}{European Non-WISP-Focused Experiments with Ability to Detect WISPs (Authors: C. De Dominicis, M. Gallinaro, G. Grilli di Cortona, V.~A.~Mitsou and M.~Staelens)}}

\subsubsection{The ATLAS and CMS experiments at the LHC}
\label{sec:LHCexp}
\input{WG4/content/LHCexperiments}

\subsubsection{The MoEDAL-MAPP experiment}
\input{WG4/content/MAPPexperiment}

\subsubsection{The DarkSide program}
\input{WG4/content/DarkSideExperiment}

\subsubsection{The DAMIC-M Experiment}
\input{WG4/content/DAMICM}

\section{Summary of the European contribution to WISP searches}
\input{WG4/content/SummaryPlotSection}

\subsection{\texorpdfstring
{High magnetic field facilities in Europe - EMFL \\ \textnormal{\small Author: P. Pugnat}}{High magnetic field facilities in Europe - EMFL (Author: P. Pugnat)}}
\label{sec:EMFL}
\input{WG4/content/EMFLFacility}

\subsection{\texorpdfstring
{The cryoplatform at DESY in Hamburg \\ \textnormal{\small Authors: A.~Lindner \& J.~Schaffran}}{The cryoplatform at DESY in Hamburg (Authors: A.~Lindner and J.~Schaffran)}}
\label{sec:DESY}
\input{WG4/content/DESYFacility}

\subsection{\texorpdfstring
{The Frascati National Laboratory of INFN \\ \textnormal{\small Author: C. Gatti}}{The Frascati National Laboratory of INFN (Author: C. Gatti)}}
\label{sec:LNF}
\input{WG4/content/LNFFacility}

\section{New experimental schemes for WISP searches}
\input{WG4/content/IntroNewSchemes}

\subsection{\texorpdfstring
{Searching for ultralight bosonic dark matter with molecular spectroscopy \\ \textnormal{\small Author: F. Constantin}}{Searching for ultralight bosonic dark matter with molecular spectroscopy (Author: F. Constantin) }}
\input{WG4/content/MolSpecExp}

\subsection{\texorpdfstring
{Search for Dark Photon in microwave cavities with Rydberg atoms \\ \textnormal{\small Author: J. Gué}}{Search for Dark Photon in microwave cavities with Rydberg atoms (Author: J. Gué)}}
\input{WG4/content/RydbergDarkPhotons}

\subsection{\texorpdfstring
{Directly Deflecting Particle Dark Matter \\ \textnormal{\small Author: S. Ellis}}{Directly Deflecting Particle Dark Matter (Author: S. Ellis)}}
\input{WG4/content/DirectDeflection}
        
\subsection{\texorpdfstring
{Searching for axion forces with spin precession in atoms and molecules \\ \textnormal{\small Author: L. Cong}}{Searching for axion forces with spin precession in atoms and molecules (Author: L. Cong)}}
\input{WG4/content/AtomsMoleculesExperimentalScheme}

\subsection{\texorpdfstring
{Searching for axion forces from dark matter by the daily modulating magnetic field \\ \textnormal{\small Author: W. Yin}}{Searching for axion forces from dark matter by the daily modulating magnetic field (Author: W. Yin)}}
\input{WG4/content/cosmicaxionforce}

\subsection{\texorpdfstring
{Searching for undulator WISPs \\ \textnormal{\small Author: W. Yin}}{Searching for undulator WISPs (Author: W. Yin)}}
\input{WG4/content/UndulatorAXION}

\subsection{\texorpdfstring
{Searching for eV Dark Matter via Infrared Spectrographs \\ \textnormal{\small Author: W. Yin}}{Searching for eV Dark Matter via Infrared Spectrographs (Author: W. Yin)}}
\input{WG4/content/eVaxion}
        
\subsection{\texorpdfstring
{Searching for Cosmic Walls directly with Paleo Detectors \\ \textnormal{\small Author: W. Yin}}{Searching for Cosmic Walls directly with Paleo Detectors (Author: W. Yin)}}
\input{WG4/content/PaleoWall}

\subsection{\texorpdfstring
{Dark Photon Dark Matter Radio Signal from the Milky Way Electron Density \\ \textnormal{\small Author: A. Arza}}{Dark Photon Dark Matter Radio Signal from the Milky Way Electron Density (Author: A. Arza)}}
\input{WG4/content/darkphotonelectron}

\subsection{\texorpdfstring
{Antiferromagnets for light dark matter detection \\ \textnormal{\small Author: A. Esposito}}{Antiferromagnets for light dark matter detection (Author: A. Esposito)}}
\input{WG4/content/Antiferromagnets}

\section{Summary Table of European Experiments}
\input{WG4/content/IntroExperimentsTables}

\cleardoublepage

%% file: WG4/content/Introduction.tex
{Introduction Author: C. Gatti}\\

\begin{figure}[htbp]
  \begin{center}
    \includegraphics[totalheight=8cm]{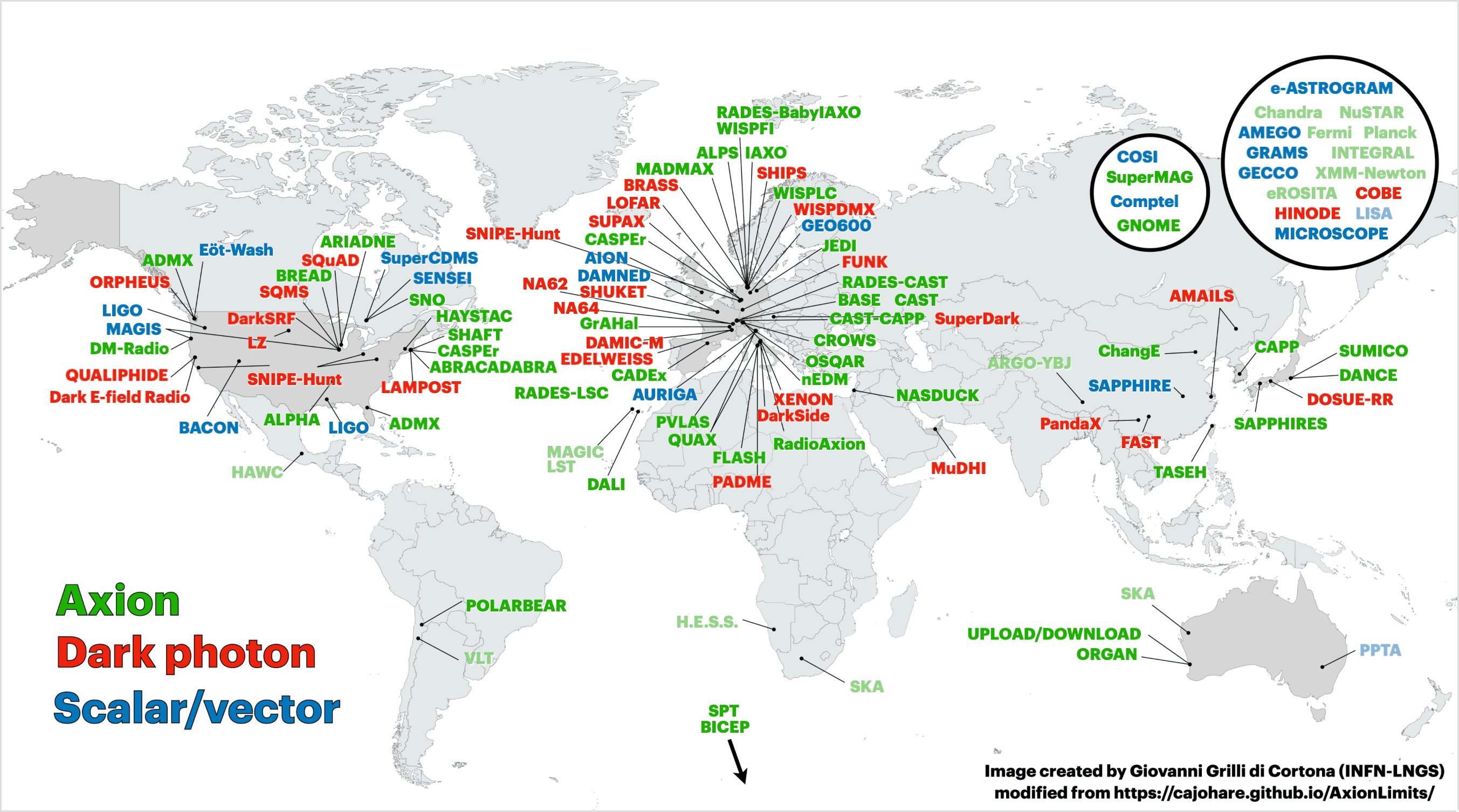}
    \caption{Map of existing and planned WISPs experiments~\cite{AxionLimits}. An interacting version is available at this link \url{https://cosmicwispers.github.io/wisps-experiments-map/}.
    }
    \label{fig:AxionMapp}
  \end{center}
\end{figure}

The last 15 years have seen an increasing interest in the direct search for Weakly Interacting Slim Particles (WISPs), particularly light bosons.
These are well-motivated candidates for physics beyond the Standard Model (SM) and potential constituents of Dark Matter. Spin-0 particles—such as scalars (dilatons) or pseudoscalars (axions)—can arise from spontaneous symmetry breaking. Spin-1 particles, like dark photons, originate from a hidden-sector $U(1)$ gauge symmetry under which SM particles remain neutral. Spin-2 particles, such as hidden gravitons, emerge in various high-energy theories, including extra-dimensional models and extensions of General Relativity. Importantly, particles predicted by these models can all be naturally slim, meaning they possess a small mass. These well-founded theoretical frameworks have inspired a broad range of new experiments designed to detect the non-zero coupling to SM particles and can be classified in:
\begin{description}
    \item[Haloscopes] Experiments designed to detect the WISP dark matter halo in our galaxy.
    \item[Helioscopes] Experiments designed to detect WISPs produced in the sun.
    \item[Pure lab experiments] Experiments designed to detect effects induced by WISP generated in the lab.
    \item[Fixed target and beam dump experiment]  Experiments designed to detect WISPs generated in the collision of an accelerated beam on a target.
\end{description}
Beside these, there are non-WISP-focused experiments with ability to detect WISPs. These are either collider experiments, experiments designed to detect WIMPs, astrophysical neutrino detectors, gravitational wave interferometers, or high precision experiments able to detect small deviation from theoretically well predictable observables. 
The liveliness of this sector is particularly evident in Europe where there is a large and diverse WISPs research program with great impact and discovery potential. Dozens of experiments are, in fact, underway or being set up in universities and national laboratories, as highlighted in Fig.~\ref{fig:AxionMapp}. 
The list of existing European WISP experiments is summarized in Tab.~\ref {tab:experiments1}.
This more than decade-long program of complementary research will probe much of the parameter space for various light dark matter models. As an example, the future impact of European research on the search for the QCD axion, one of the most theoretically motivated models, is exemplified in Fig.~\ref{fig:AxionLimits-agg-present} (bottom panel). 

%% file: WG4/content/Intro-Haloscopes.tex
The idea of a Haloscope detecting the ``invisible'' axion was introduced in 1983 by Pierre Sikivie (Fig.~\ref{fig:Pierre}) in a seminal paper~\cite{Sikivie:1983ip}. 
Sikivie's haloscope is sensitive to axions within a halo originated at the galactic center. 
It exploits the resonant conversion of axions into a monochromatic electromagnetic signal inside a microwave cavity immersed in a large static magnetic field, called the ``inverse Primakoff effect''.
The first search for galactic axions were performed by two pilot experiments at Brookhaven National Laboratory (BNL)~\cite{PhysRevLett.59.839,PhysRevD.40.3153} and at the University of Florida (UF)~\cite{PhysRevD.42.1297}, setting a sensitivity baseline useful for detector designs in the future experiments. 
A few years later, the CARRACK experiment~\cite{MATSUKI1996213} in Kyoto, Japan, operated inside a dilution refrigerator and used a detection method based on Rydberg atoms, offering a novel approach.
Since then, many new experiments with different detection methods have been proposed and realized by different research groups and collaborations. 
Some experiments are based on Sikivie's original haloscope concept to search for a photon; while others use designs based on ferro- and para-magnets, dielectrics, dish-antenna, lumped-element LC oscillators and more, to search for a fictitious magnetic field.
Haloscopes of different designs are naturally sensitive to different and mostly narrow axion mass ranges limited by experimental capabilities and sensor bandwidths. Therefore, an international community effort is required to cover all of the possible axion masses, ranging from \( 10^{-22}\text{ eV}\) to \( 10^{5}\text{ eV}\).
The non-European haloscopes include ADMX~\cite{ADMX:2021nhd}, ALPHA~\cite{ALPHA:2022rxj}, HAYSTAC~\cite{HAYSTAC:2020kwv}, ORGAN~\cite{PhysRevLett.132.031601}, IBS-CAPP~\cite{PhysRevX.14.031023}, TASEH~\cite{TASEH:2022vvu},  DMRadio/ABRACADABRA \cite{DMRadio:2022pkf}, BREAD~\cite{BREAD:2021tpx} and MuDHI~\cite{Manenti:2021whp}.

%% file: WG4/content/AxionULDMQuantumSensors.tex
A different approach to searches for ultralight axion-like particle (ALP) dark matter in the mass range $m_a \lesssim \text{eV}$~\cite{Bauer:2024hfv} is to look for small but temporally varying modifications to Standard Model (SM) parameters and interactions induced by the dark field coherent oscillations.
While linear pseudoscalar couplings to SM fields are typical in many axion models, quadratic couplings emerge naturally in parity-conserving frameworks and are well motivated by ultraviolet (UV) completions.

\subsubsection*{Quadratic ALP interaction at low energy}

At quadratic order in the ALP decay constant $f$, ALPs have scalar-like interactions described by the dimension-six operators:

\begin{eqnarray}
\mathcal{L}_{\text{eff}}^{D=6}(\mu \lesssim \Lambda_{\text{QCD}}) &=& \bar{N} \left(C_N(\mu)\,1 + C_5(\mu)\,\tau \right) N \frac{a^2}{f^2} + C_E(\mu) \frac{a^2}{f^2} \bar{e}e +
\\\nonumber
&& C_\gamma(\mu) \frac{a^2}{4f^2} F_{\mu\nu}F^{\mu\nu},
\end{eqnarray}

where $a$ denotes the ALP field and $C_i$ are the ALP coupling strengths with SM where $i$ denotes nucleon, electron and photon. These can be obtained by renormalization group equation (RGE) running and threshold matching from the UV Lagrangian that has interactions with gluons, SM gauge bosons, and flavor-diagonal couplings with SM fermions. 

The ALP field undergoes coherent oscillations characterised by a high coherence time, $T_{\rm coh} \sim 10^6\, T_{\rm osc}$, evolving as $a(t) \approx a_0 \cos(m_a t)$, where the amplitude $a_0$ is determined by the local DM density, $\rho_{\text{DM}} \sim 1/2\, m_a^2 a_0^2$. The oscillation frequency is given by the ALP’s Compton frequency. These oscillations induce ALP field–dependent variations in fundamental constants (FCs), including the fine-structure constant ($\alpha$), and the electron and nucleon masses ($m_e$ and $M_N$ respectively) as shown below:

\begin{align}
\alpha^{\text{eff}}(a) &= \left(1 + \delta_\alpha(a) \right) \alpha \hspace{2em} \text{with} \quad \delta_\alpha(a) = \frac{1}{12\pi} \left(1 - 32 c_1 \frac{m_\pi^2}{M_N} \right) \delta_\pi(a)  \tag{2} \\
\quad m_e(a) &= m_e \left(1 + \delta_e(a) \right) \hspace{2.8em} \text{with} \quad \delta_e(a) = \frac{3\alpha}{4\pi} C_\gamma \frac{a^2}{f^2} \ln \left(\frac{m_\pi^2}{m_e^2} \right)  \tag{3} \\
M_N(a) &= M_N \left(1 + \delta_N(a) \right) \hspace{3.6em} \text{with} \quad \delta_N(a) = -4c_1 \frac{m_\pi^2}{M_N} \delta_\pi(a) \tag{4}
\end{align}

where 
\begin{align*}
\delta_\pi(a)\approx\frac{a^2}{f^2}\bigg(1-\frac{\Delta_m^2}{\hat m^2}\bigg),
\label{Eq:shift_mpi2}
\end{align*}
with $\hat m =(m_u+m_d)/2$ and $\Delta_m=(m_u-m_d)/2$ and $m_u$ and $m_d$ denoting the light quark masses. Notably, for pseudoscalar DM, such shifts arise only from quadratic couplings due to their spin-independent structure. As a result, these variations in fundamental constants also oscillate coherently in time, with an oscillation frequency of $2m_a$, characteristic of quadratic ALP DM backgrounds.

\subsubsection*{Quantum sensor current sensitivities}

These interactions motivate a class of high-precision quantum sensor experiments designed to detect coherent periodic signals. Three primary strategies are highlighted:

\begin{itemize}
  \item[(1)] \textbf{Atomic clocks}: In atomic clocks, lasers or microwaves are typically used to lock onto optical or hyperfine transitions in different atomic references. Any minute deviation from the nominal transition frequency can be detected with extraordinary precision, often at the fractional level of $10^{-15}$–$10^{-17}$. Owing to this exceptional sensitivity, FC shifts in ultralight dark matter (ULDM) background can leave potentially detectable imprints in atomic clocks when monitored over extended periods. In clock-comparison experiments, the ratio of two transition frequencies, each with different sensitivities to variations in FCs such as $m_e$, $\alpha$, and nuclear magnetic moments, is measured. Networked comparisons between multiple atomic species (e.g., Rb, Cs, Sr and Yb), with distinct dependencies on Standard Model parameters, allow environmental noise to be suppressed and coherent, common-mode oscillations to be extracted. Fountain clocks, such as Rb/Cs~\cite{Hees:2016gop}, and time-evolved atomic spectroscopy setups like Dy/Dy~\cite{VanTilburg:2015oza}, are particularly effective at constraining lower ALP masses around $10^{-23}~\rm{eV}$, as shown in Fig.~\ref{fig:ULDM-ALP-QS}. In contrast, optical clocks such as $\rm{Yb}^+$ ion clocks and Yb/Sr comparisons~\cite{BACON:2020ubh,Filzinger:2023zrs} offer better stability and precision, making them more sensitive to higher mass ranges up to $10^{-18}~\rm{eV}$.

  \begin{figure}[htb!]
    \centering
    \includegraphics[width=1.0\linewidth]{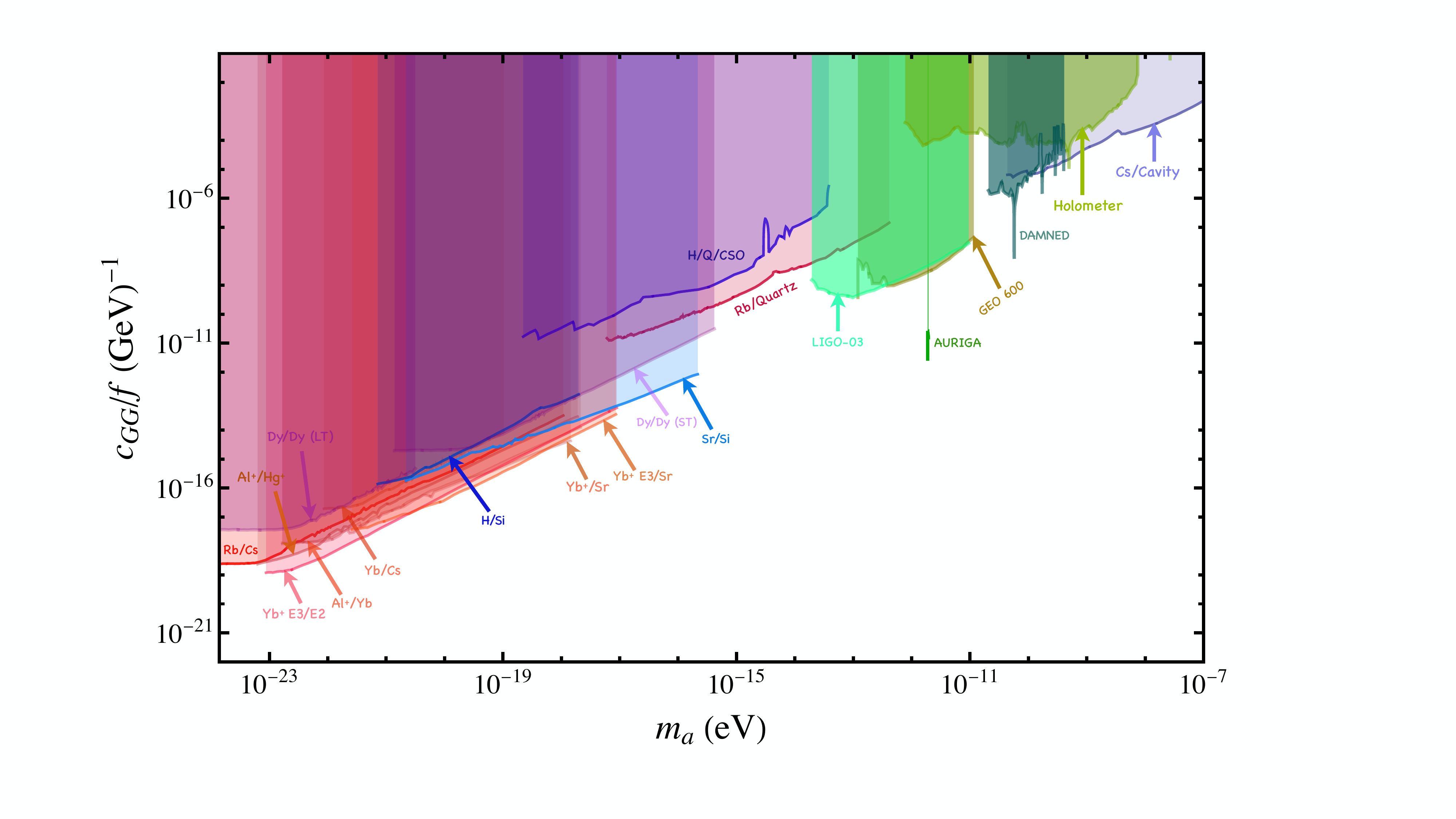}
    \caption{Limits on the ALP-gluon coupling at the UV scale from existing quantum sensors. In shades of
red, we show atomic clocks, atomic spectroscopy and quartz oscillator. In shades of blue are shown clock/cavity comparisons and in
shades of green appear the optical/laser interferometers and mechanical resonators. Figure reproduced from Ref.~\cite{Bauer:2024hfv}.}
    \label{fig:ULDM-ALP-QS}
\end{figure}

  \item[(2)] \textbf{Optical cavities and clock/cavity comparisons}: Optical cavities consist of mirrors forming a resonant structure in which a laser is locked to standing wave modes that define the cavity’s frequency. ULDM oscillations can induce variations in the Bohr radius, leading to fractional changes in the cavity length. Since the cavity eigenfrequencies scale inversely with physical length, this results in shifts in the reference frequency. Comparing a laser stabilised to the cavity resonance with either another cavity or atomic transitions in a clock allows one to search for coherent drifts arising from ALP-induced variations in fundamental constants.
  
One of the best cavity setups compares a silicon optical cavity with a $^{87}$Sr optical lattice clock (Sr/Si)~\cite{Kennedy:2020bac}, primarily probing variations in the fine-structure constant. Its high optical stability allows strong constraints in the ALP mass range $m_a \sim 10^{-17}$–$2 \times 10^{-16}$ eV. A similar setup comparing the Si cavity to a hydrogen maser (H/Si)~\cite{Kennedy:2020bac}, sensitive to both $\alpha$ and $m_e$ performs slightly better at low masses ($m_a \lesssim 10^{-18}$ eV) but is less effective at higher frequencies. Further bounds arise from comparing a quartz bulk acoustic wave oscillator with a hydrogen maser and a cryogenic sapphire oscillator (H/Q/CSO)~\cite{Campbell:2021mnu}, probing $m_a \sim 10^{-16}$–$10^{-14}$ eV.
  
  \item[(3)] \textbf{Mechanical resonators}: Precision force sensors and optomechanical systems can experience tiny displacements due to time-varying mass or strain modulations induced by ULDM. Experiments such as the cryogenic bar detectors and optomechanical oscillators are capable of probing strain-like responses generated by ULDM couplings to matter. These systems are also being developed as potential detectors of high-frequency GW, offering complementary sensitivity to existing GW observatories.

Mechanical resonators, similar to optical cavities, are sensitive to time-dependent strain in solid materials resulting from oscillations in fundamental constants such as $\alpha$ and $m_e$ in the presence of an ALP DM background. Quadratic couplings lead to a coherent mechanical strain of the form $h(t) \sim \delta \alpha(t) + \delta m_e(t)$, which can be resonantly amplified when an acoustic mode of the device is tuned to $2m_a$.

Cryogenic resonant-mass detectors like AURIGA~\cite{Branca:2016rez} operate in narrow frequency bands, corresponding to ALP masses in the peV range, as shown in Fig.~\ref{fig:ULDM-ALP-QS}. In contrast, future compact acoustic resonators~\cite{Manley:2019vxy} made from superfluid helium, sapphire, or quartz can access a broader frequency range (Hz to MHz), enabling sensitivity to ALP masses from below a peV to several hundred neV.
\end{itemize}

With the advent of emerging technologies such as nuclear clocks and atom interferometers, future experiments are expected to achieve sensitivities several orders of magnitude beyond current limits, significantly extending the reach of ALP ULDM searches.

%% file: WG4/content/Intro-DarkMatterOpticalHaloscopes.tex
Optical‑cavity searches target axion‑induced birefringence (polarization rotation/ellipticity) or differential resonance shifts between polarization eigenmodes. Sensitivity is set by the optical cavity finesse, the ellipticity noise floor of the polarimeter, and the modulation scheme that places the axion line within the readout bandwidth while avoiding technical $1/f$ noise.

One realization is the Axion Detection with Birefringent Cavities (ADBC) concept~\cite{Pandey:2024dcd,Liu:2018icu}. Here, a bow-tie optical cavity is built to have a slight inherent birefringence between two orthogonal polarization eigenmodes; an axion field would modulate the resonance condition of one mode relative to the other. The first results from an ADBC prototype were recently reported by a team at MIT, demonstrating the technique's viability. 

Another cutting-edge example is the DANCE project (Dark matter Axion search with riNg Cavity Experiment) in Japan~\cite{DANCE}. DANCE uses an m-scale bow-tie ring cavity to actively amplify a rotational oscillation of linear polarization that an axion field would induce. A recent experiment, the Laser-Interferometric Detector for Axions (LIDA), has demonstrated the feasibility of using high-finesse optical cavities to search for axion-like particles by detecting axion-induced polarization rotation in laser light~\cite{Heinze:2023nfb}. Because no external magnets are used, the dominant noises are photon shot noise and intra‑cavity birefringence fluctuations; performance improves with longer integration and by operating the modulation above the polarimeter’s $1/f$ knee. 

Importantly, all these experiments require no external magnetic field, they are purely optical experiments, making them the first demonstrations of polarization-based axion DM searches without any magnets.

%% file: WG4/content/BRASSExperiment.tex
{\small Author: L.~H.~Nguyen}\\

The dish antenna, as outlined by \cite{Horns13}, provides a sensitive haloscope geometry suitable for effectively searching for WISPs across a high mass range, from microelectronvolts to millielectronvolts, or from gigahertz up to the infrared spectrum. The Broadband Radiometric Axion/ALP Search Setup (BRASS) leverages cutting-edge technology from radio astronomy, including wide-band horn antennas, amplifiers, and ultra-broadband real-time data acquisition systems. These are coupled with cost-effective magnetic panels that facilitate the conversion between axions/ALPs and photons. The prototype of BRASS, BRASS-p, is installed in a radio-quiet laboratory at the University of Hamburg, located in Hamburg, Germany.
\begin{figure}
    \centering
    \includegraphics[width=\linewidth]{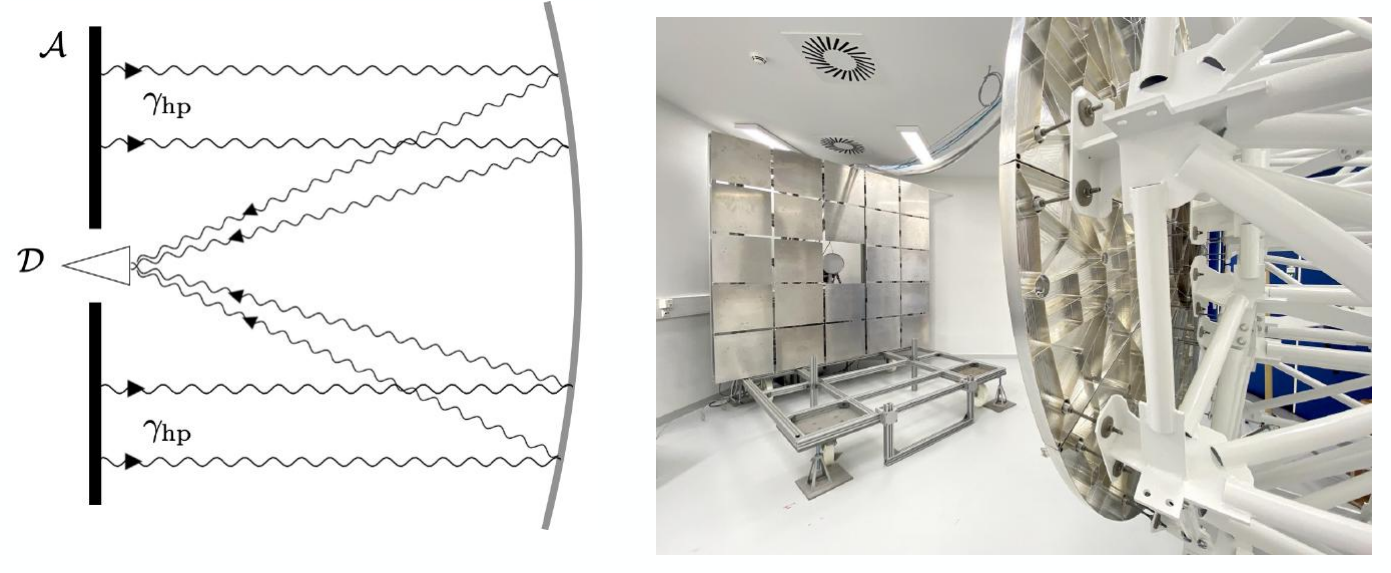}
    \caption{\textbf{Left}: Schematic design of BRASS. Figures reproduced from Ref.~\cite{Nguyen:2023uis}.}
    \label{fig:BRASSDesign}
\end{figure}
The conceptual optical design of BRASS, illustrated in Fig.~\ref{fig:BRASSDesign}-left, includes a shielded chamber and a flat, conductive, permanently magnetized conversion surface ($\mathcal{A}$, for details see \cite{2025JInst..20P2026N}). This surface generates an electromagnetic signal in response to the electric field produced by the dark matter as it propagates through the system. This electric field is either inherent to the dark matter particles themselves, in the case of dark/hidden photons, or induced by the magnetic field near the conversion surface for axions and ALPs. The resultant electromagnetic signal is focused by a parabolic reflector and detected using a heterodyne receiver, ($\mathcal{D}$), equipped with a broadband digital backend. Fig.~\ref{fig:BRASSDesign}-right shows the prototype of the BRASS experiment, detail in \cite{Nguyen:2023uis}, which is sensitive to dark/hidden photons in the mass range of 49.63 to 74.44 µeV. With just 50 hours of data collection, BRASS-p has established the world-leading limit within its sensitive mass range. The position and magnetic-field direction of BRASS are listed in Tab.~\ref{tab:BRASS_orientation}.

\begin{table}[t]
\renewcommand{\arraystretch}{1.5}
  \begin{center}
    \begin{tabular}{c|c}
    \hline
    \hline
    Latitude &  $53^{\circ}$ $34^{\prime}$ $41.7^{\prime\prime}$ \\
      Longitude & $9^{\circ}$ $53^{\prime }$ $9.1^{\prime\prime}$ \\
      Elevation & 20~m\\ 
      $B$-field direction & Vertical  \\
      \hline\hline
    \end{tabular}
    \caption{Position and magnetic-field direction of BRASS experiment.}
    \label{tab:BRASS_orientation}
  \end{center}
\end{table}

%% file: WG4/content/CASPErExperiment.tex
{Authors: O.~Maliaka, O.~Ruimi, D.~Aybas, D.~Budker}\\

The Cosmic Axion Spin Precession Experiment (CASPEr) is an international research program using Nuclear Magnetic Resonance (NMR) techniques to search for axions and axion-like particles (ALPs) based on dark-matter-driven spin precession \cite{Budker:2013hfa}. It is comprised of two primary experiments: CASPEr-gradient and CASPEr-electric, probing the coupling of ALPs to axial nuclear current and the gluon field, respectively. In both experiments the axion field is treated as a pseudo-magnetic field that drives nuclear magnetic spin precession.

\begin{figure}[t!]
  \begin{center}
    \includegraphics[width=0.75\linewidth]{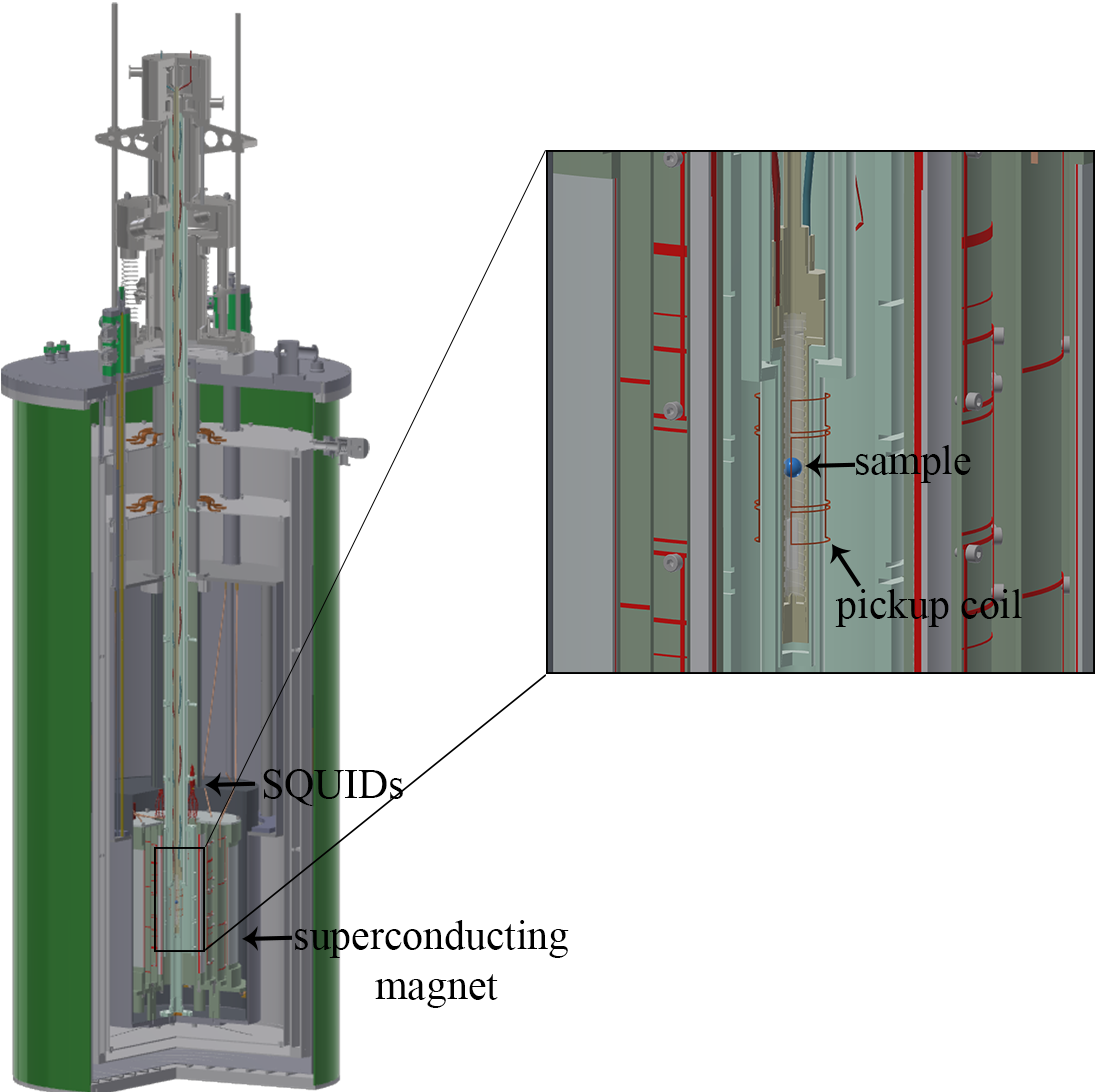}
    \caption{The setup used in CASPEr-gradient low-field for axion and ALP searches. Image credits to the CASPEr Collaboration.}
    \label{fig:casperlowfieldsetup}
  \end{center}
\end{figure}

\paragraph{CASPEr-gradient.}
CASPEr-gradient low-field, located at the Helmholtz Institute Mainz (HIM), Germany, aims to detect axions through their coupling to nuclear spin for masses corresponding to Compton frequencies in the range of $1~\si{kHz}$ up to $4.2~\si{MHz}$ using a tunable 0.1\,T superconducting magnet and a detection system based on Superconducting Quantum Interference Devices (SQUIDs), as shown in Fig.~\ref{fig:casperlowfieldsetup}. 
A leading magnetic field $B_\mathrm{0}$ is applied to a sample with bulk magnetisation or density of nuclear moment $\mathbf{M}$. If a transverse oscillating magnetic field (such as an axion field) of frequency equal to the Larmor frequency $\omega_L=\gamma B_\mathrm{0}$ is present, with $\gamma$ the gyromagnetic ratio, then $\mathbf{M}$ is tilted to the transverse plane and starts to precess. The precessing magnetisation generates a signal that can be measured by a magnetometer such as a SQUID.

To explore higher masses that correspond to frequencies in the range $70-600~\si{MHz}$, CASPEr-gradient high-field is currently under construction at HIM. It relies on a superconducting magnet which was recently installed at HIM and is capable of generating a field of up to 14.1\,T. For these higher frequencies, an NMR probe is currently under development based on inductive pickup with a resonant circuit and a cooled semi-conductor amplifier with support from industry partners.

Previous work includes the implementation of a comagnetometry scheme for measuring ultralow-field NMR signals involving two different nuclei in a liquid-state sample \cite{PhysRevLett.122.191302} and a dark matter search using ultra-low NMR in the mass range from $1.8\times10^{-6}$ to $7.8\times10^{-14}\,~\si{eV}$ \cite{doi:10.1126/sciadv.aax4539}. 
A first proof-of-principle science run using the CASPEr-gradient setup with thermally polarised methanol has recently provided a coupling limit near $1.3~\si{MHz}$ \cite{JulianWalter25}.
Ongoing work involves improvements on the setup \cite{YuzheZhang23,JulianWalter23} and the connection between the operational $\mathrm{{^{129}}Xe}$ Spin Exchange Optical Pumping (SEOP) setup to the low-field setup. Other sample preparation methods under consideration for future experiments using both the low-field and the high-field setup include 1\,T prepolarized proton samples, Parahydrogen Induced Polarization (PHIP), Dynamic Nuclear Polarization (DNP) and $\mathrm{{^{3}}He}$ cells.

\paragraph{CASPEr-electric.}
CASPEr-electric, located at Boston University in the United States, is sensitive to the defining QCD gluon coupling of axion dark matter to the nuclear spins of solids. This interaction induces an oscillating nucleon electric dipole moment (EDM). Via parity (P) and time reversal (T)-violating nuclear forces, it also induces oscillating nuclear Schiff moments~\cite{Sushkov2023b}. CASPEr-electric uses the precision NMR approach with solid samples, carefully chosen to maximize the observable transverse nuclear ensemble magnetization. 

Recent work includes the axion-like dark matter search in the mass range $162-166~\si{neV}$ using $\mathrm{{^{207}}Pb}$ in polarized ferroelectric PMN-PT~\cite{PhysRevLett.126.141802}. 
Current experimental efforts focus on reaching the QCD axion band in a narrow-band search. 
Ongoing work also includes the development of DNP protocols and nuclear spin optical pumping for optimizing nuclear ensemble polarization. 
Other sample materials, such as $\mathrm{EuCl_3\cdot6H_2O}$, are being characterized for future experiments.

%% file: WG4/content/CAST-CAPPExperiment.tex
{Author: M. Maroudas}\\

CAST-CAPP is an axion haloscope detector that operated at CERN in Geneva, Switzerland, inside the CAST dipole magnet from 09/2019 to 06/2021 \cite{adair_capp_2022}. CAST-CAPP follows the Sikivie haloscope principle where dark matter axions in the presence of a strong magnetic field are converted into photons inside a high-quality microwave cavity resonating to the corresponding frequency defined by the unknown axion rest mass. CAST-CAPP has searched for axions in the 19.74 $\mu$eV to 22.47 $\mu$eV mass range.

\begin{figure}[t!]
  \begin{center}
    \includegraphics[width=0.65\linewidth]{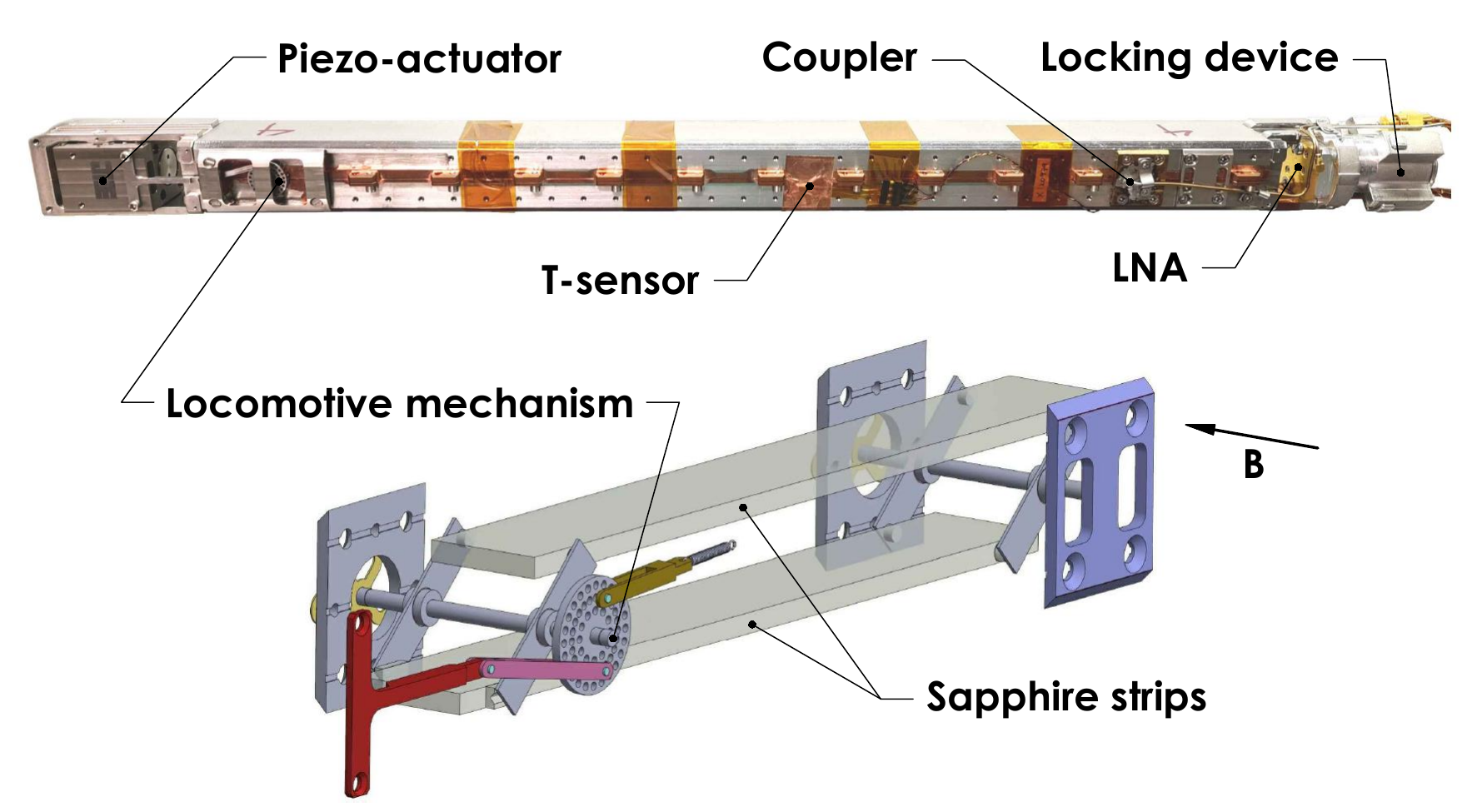}
    \caption{The CAST-CAPP cavity assembly (\emph{top}) with its tuning mechanism (\emph{bottom}). Figure reproduced from Ref.~\cite{adair_capp_2022}.}
    \label{fig:cast_capp_assembly}
  \end{center}
\end{figure}

\begin{figure}[t!]
  \begin{center}
    \includegraphics[width=0.55\linewidth]{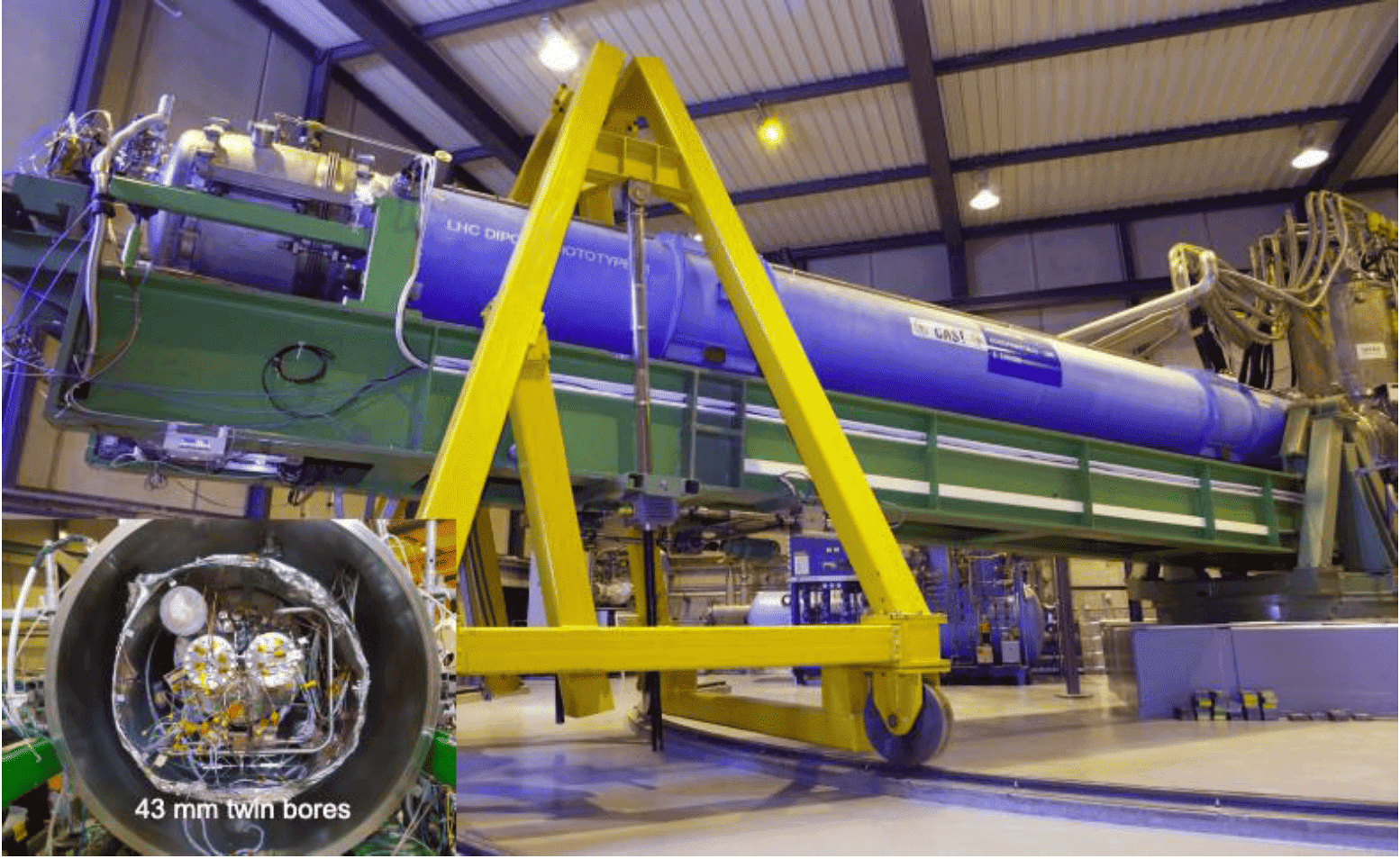}
    \caption{The CAST experiment with a close-up photo of the twin bores where CAST-CAPP cavities were installed.}
    \label{fig:cast_capp_magnet}
  \end{center}
\end{figure}

CAST-CAPP consists of four tuneable $23 \times 25 \times 390$\si{\mm} rectangular cavities, each with a volume of $V = \SI{224}{\centi\meter\cubed}$, as shown in Fig.~\ref{fig:cast_capp_assembly}. Each cavity consists of two pieces of stainless steel, coated with \SI{30}{\micro\meter} of copper. They are installed in series inside one of the two bores, \SI{43}{\milli\meter} diameter, of CAST's superconducting dipole magnet at CERN seen at Fig.~\ref{fig:cast_capp_magnet} \cite{CAST:2017uph}. 

The cavity tuning mechanism consists of two sapphire strips which are displaced by a piezoelectric motor through a locomotive mechanism providing a wide tuning range of about \SI{660}{\MHz} and a tuning resolution of less than \SI{100}{\Hz} in stable conditions. Additionally, the maximum scanning speed reached with CAST-CAPP is \SI[per-mode = symbol]{10}{\MHz\per\minute} with the coverage of the full frequency range taking about \SI{1}{\hour}.

In order to reach higher axion masses, smaller cavities are required which in turn reduces the detection sensitivity. To mitigate this issue, CAST-CAPP uses four identical cavities together with the phase-matching technique to increase the effective volume and detection sensitivity. This technique, introduced for the first time in axion research, improves linearly the signal-to-noise ratio with the number of cavities \cite{jeong_matching_2018}. To achieve this, a coherent combination of the simultaneous power outputs from the four frequency-matched cavities has to be performed in data-taking conditions.

Together with its fast-tuning mechanism, CAST-CAPP is also sensitive to transient events such as axion streams \cite{Vogelsberger:2010gd} and cosmologically motivated axion mini-clusters \cite{Tkachev:1991ka, Kolb:1993zz,Zioutas:2017klh}. These can give rise to temporally enhanced flux densities ($\rho_a$) by several orders of magnitude, in particular when combined with the gravitational lensing effects by the solar system \cite{hoffmann_gravitational_2003}, \cite{patla_flux_2013}. The faster the scanning the shorter the axion bursts that can be utilized, making the fast tuning mechanism of CAST-CAPP an indispensable component.

CAST-CAPP acquired in total about 172 days of data with both single and phase-matched cavities in data-taking conditions with $B=\SI{8.8}{\tesla}$. The scanned frequency range extended from \SIrange{4.77}{5.43}{\GHz} covering a parameter phase space of $\sim\SI{660}{\MHz}$. This allowed CAST-CAPP to exclude axion-photon couplings for virialized galactic axions down to $ g_{a\gamma\gamma} = 8 \times 10^{-14}\,\si{\GeV\tothe{-1}} $ at $90\%$ confidence level for axion masses between \SIrange{19.74}{22.47}{\micro\eV}. Additionally, the recent CAST-CAPP results \cite{CASTCAPP_2025_modulations} validate a proof-of-principle approach to exploiting daily modulation signatures from Axion Quark Nuggets \cite{AQN_2020_modulations}, offering a promising new pathway for relativistic axion searches even though no definitive modulation signal has yet been observed. Finally, an independent analysis procedure is ongoing, which will allow CAST-CAPP to search for transient events from axion streams \cite{Vogelsberger:2010gd} and axion mini-clusters \cite{Tkachev:1991ka}.

In summary, CAST-CAPP was able to set world-class limits on the galactic DM axion-photon conversion and laid the foundations for a search for transient events by making use of the two newly developed techniques of fast scanning and phase-matching, which will help define the future axion searches through the next-generation haloscopes \cite{adair_capp_2022}.

Its position and orientation are shown in Tab.~\ref{tab:cast_capp_orientation}.

\begin{table}[t!]
\renewcommand{\arraystretch}{1.5}
  \begin{center}
    \begin{tabular}{c|c}
    \hline
    \hline
    Latitude &  $46^{\circ}$ $14^{\prime}$ $31.0596^{\prime\prime}$ \\
      Longitude & $6^{\circ}$ $05^{\prime }$ $49.3548^{\prime\prime}$ \\
      Elevation & 479.511m\\ 
      $B$-field direction & Vertical  \\
      \hline\hline
    \end{tabular}
    \caption{Position and magnetic-field direction of CAST-CAPP experiment.}
    \label{tab:cast_capp_orientation}
  \end{center}
\end{table}

%% file: WG4/content/MADMAXExperiment.tex
{Author: A. Lindner}\\

\begin{figure}[tbh]
  \begin{center}
    \includegraphics[width=0.9\linewidth]{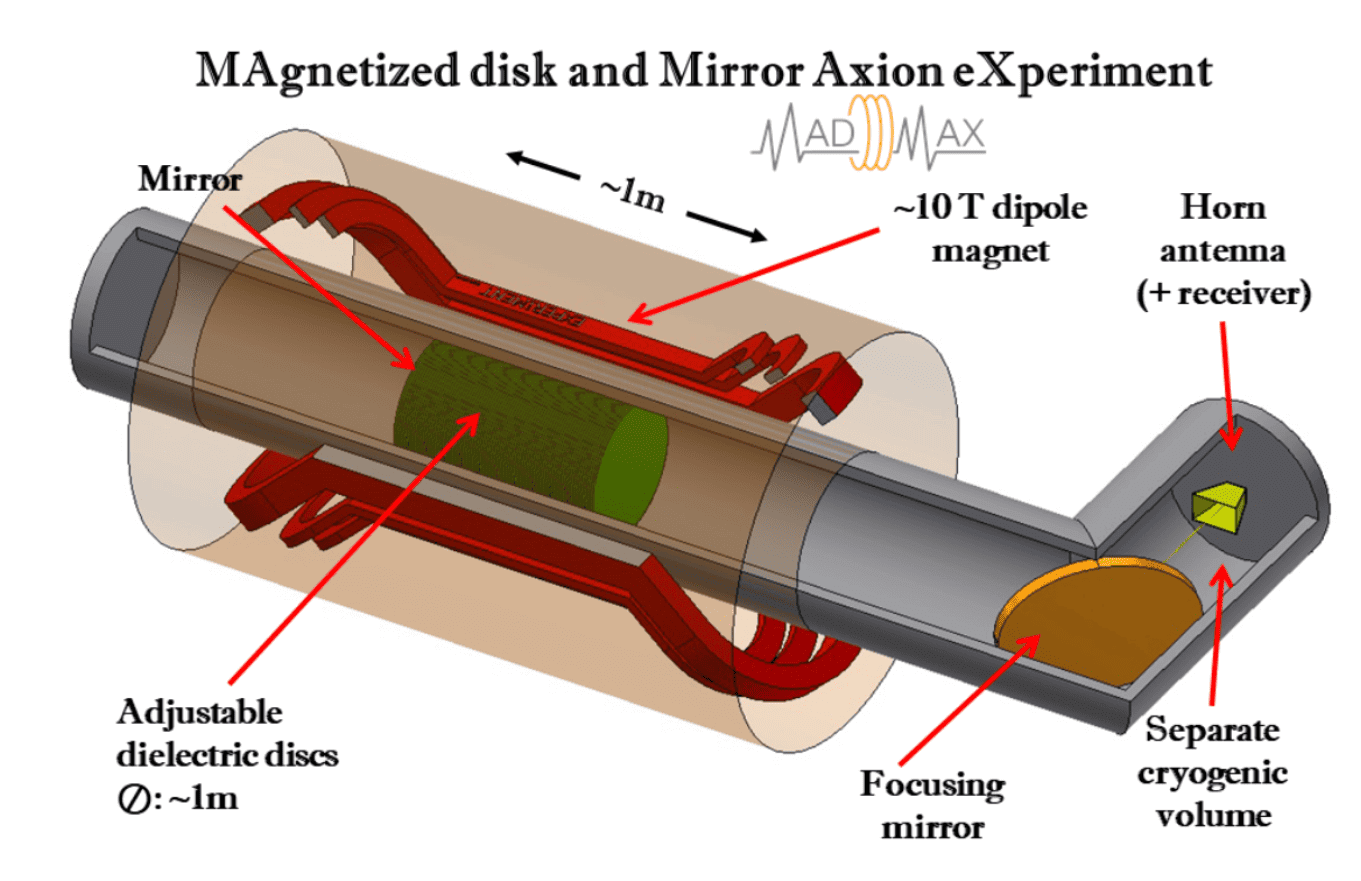}
    \caption{Sketch of the MADMAX baseline design. Adapted from Ref.~\cite{Brun2019}.}
    \label{fig:madmax}
  \end{center}
\end{figure}

MADMAX will search for local dark matter axions. An assembly of movable well-aligned dielectric disks is able to boost a weak electromagnetic wave, expected by axion conversion into photons in the presence of a high magnetic field, and make it detectable \cite{Majorovits:2023kpz}.
The experimental concept is displayed in
Fig.\,\ref{fig:madmax}. 

This technology allows to search for dark matter axions in the mass range around 100\,$\mathrm{\mu eV}$, which is not accessible with haloscopes based on microwave cavities, but strongly motivated by preferred cosmological scenarios. The MADMAX collaboration is led by the Max Planck Institute für Physik and consists of the following institutes:
    Aix Marseille Université, CNRS/IN2P3, CPPM, Marseille, France, DESY in Hamburg, Germany,
    Fermi National Accelerator Laboratory (USA), Max Planck Institute für Physik Garching (Germany),
     MPI für Radioastronomie Bonn (Germany), RWTH Aachen (Germany), Université Paris-Saclay, CNRS/IN2P3, IJCLab, Orsay (France), University of Hamburg (Germany), University of Tübingen (Germany), University of Zaragoza (Spain).

The confirmation of the detection concept, mainly the booster calibration and its verification, has been achieved recently. It provides the basis for first physics results on dark photon searches \cite{MADMAX:2024jnp}, using a so-called open booster, and on axion dark matter searches \cite{Garcia:2024xzc} with a closed booster in CERN's Morpurgo magnet.
In addition, the mechanical functionality of a tunable booster was demonstrated \cite{MADMAX:2024pil}.
Developments towards a huge dipole magnet with about 9\,T field strength and a bore diameter of 1.35\,m have resulted in a first conceptual design (see \cite{Torre:2023ssj} and references therein). \\
In the next years, the collaboration will scale up the booster by increasing its size and the number of disks, incorporate tuning mechanisms, operate prototypes at cryogenic temperatures in a recently delivered dedicated large cryostat, and work on improved sensing schemes.
The final experiment will be operated at DESY's cryoplatform in the early 2030-ties.

%% file: WG4/content/QUAXExperiment.tex
{Author: C. Gatti}\\

The QUAX$_{a\gamma}$ experiment looks for axion dark-matter through the axion-photon coupling with two Sikivie's haloscopes in a mass range around 40~$\mu$eV corresponding to frequencies of operation of about 10~GHz. The two haloscopes, located in the two national laboratories of INFN of Legnaro (LNL) and Frascati (LNF), are both composed by a resonant cavity surrounded by a superconducting solenoid-magnet and inserted in a dilution refrigerator. The experiment, that sees the collaboration of INFN groups (Padua, LNL, Trento, LNF, Salerno) and recently the Institut Néel in Grenoble, is taking data with sensitivity to KSVZ axions with the goal of probing a 1~GHz wide region. The results of the research carried out to date with the LNL and LNF haloscopes are published on~\cite{PhysRevD.108.062005,Alesini:2022lnp,Alesini:2020vny,QUAX:2024fut,infirri2025searchpostinflationaryqcdaxions}. The team has carried out intensive R\&D on superconducting and dielectric cavities~\cite{Alesini:2019ajt,ALESINI2021164641,PhysRevApplied.17.054013}, on quantum amplifiers~\cite{10.1063/5.0098039} and on microwave photon counters that will allow to increase the sensitivity to DFSZ models

\begin{figure}[t!]
  \begin{center}
    \includegraphics[totalheight=5.cm]{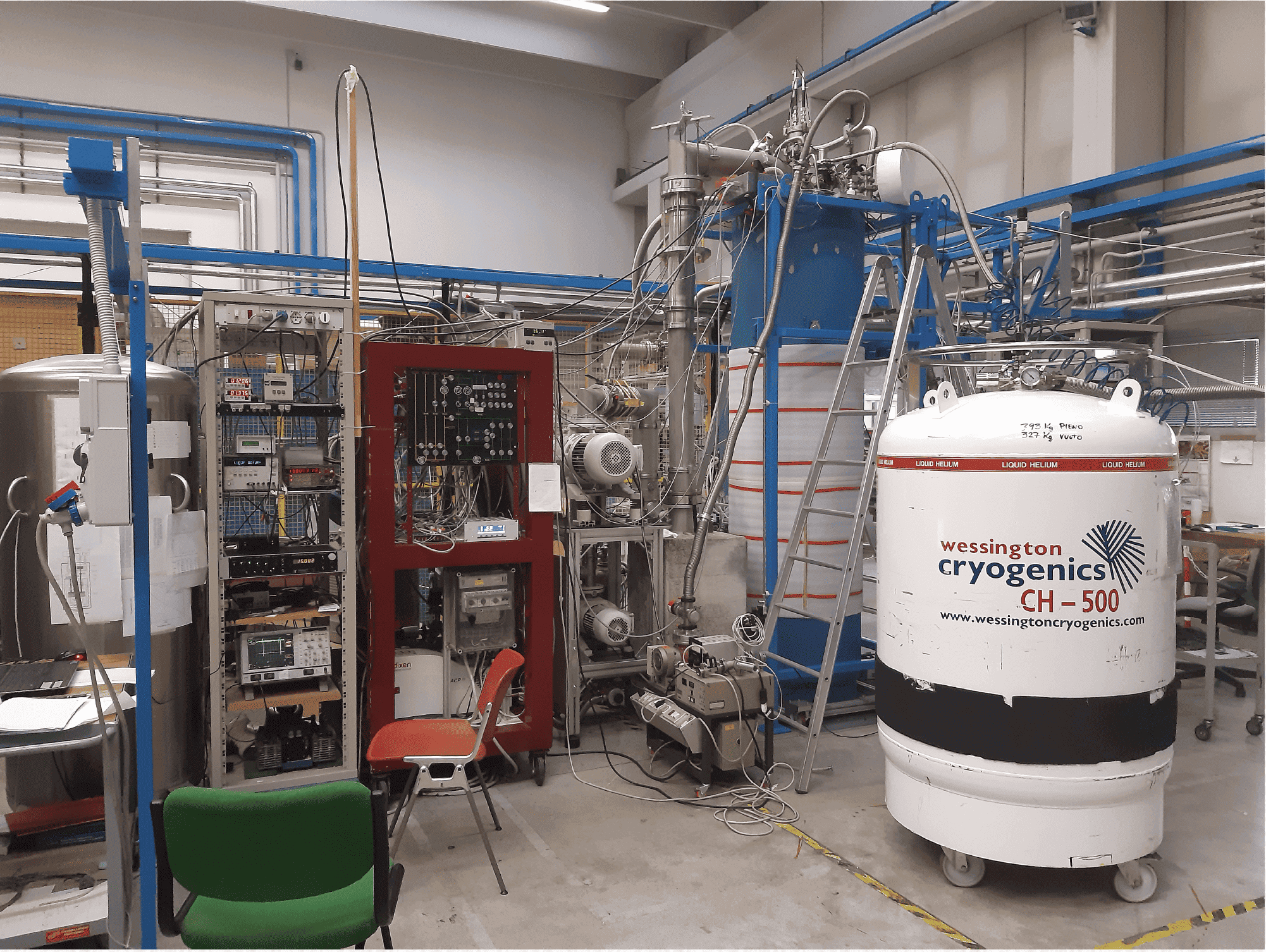}
    \includegraphics[totalheight=5.cm]{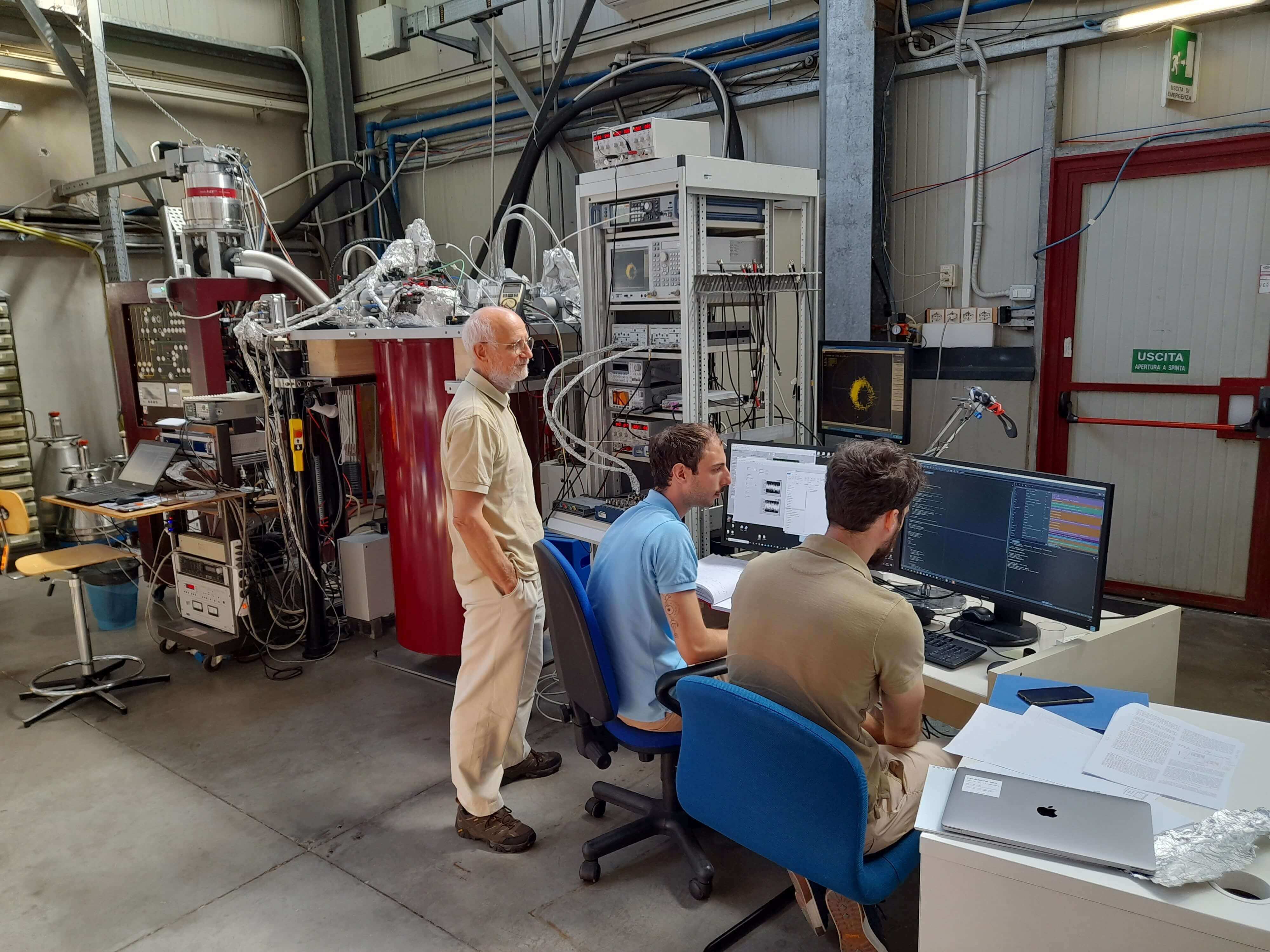}
    \caption{The LNL (\emph{left}) and LNF (\emph{right}) haloscopes of the QUAX experiment.}
    \label{fig:QUAX}
  \end{center}
\end{figure}

Beside QUAX$_{a\gamma}$, two experiments are in an R\&D phase at LNL: the QUAX$_{ae}$ experiment exploits the axion-electron coupling to search axion dark-matter~\cite{QUAX:2020adt}; the QUAX$_{g_pg_s}$ experiment~\cite{CRESCINI2017109,CRESCINI2017677} looks for forces between electron spin and nuclei mediated by axions through the scalar and pseudo-scalar couplings.

Quax haloscopes position and orientation are listed in Tab.~\ref{tab:quaxlnfhorientation} and Tab.~\ref{tab:quaxlnlhorientation}.
\begin{table}[tbh]
\renewcommand{\arraystretch}{1.5}
  \begin{center}
    \begin{tabular}{c|c}
    \hline
    \hline
    Latitude &  $41^{\circ}$ $49^{\prime}$ $26^{\prime\prime}$ \\
      Longitude & $12^{\circ}$ $40^{\prime }$ $13^{\prime\prime}$ \\
      Elevation & 120~m\\ 
      $B$ field direction & vertical  \\
      \hline\hline
    \end{tabular}
    \caption{Position and magnetic-field direction  of QUAX-LNF experiment}
    \label{tab:quaxlnfhorientation}
  \end{center}
\end{table}

\begin{table}[tbh]
\renewcommand{\arraystretch}{1.5}
  \begin{center}
    \begin{tabular}{c|c}
    \hline
    \hline
    Latitude &  $45^{\circ}$ $21^{\prime}$ $9^{\prime\prime}$ \\
      Longitude & $11^{\circ}$ $57^{\prime }$ $0^{\prime\prime}$ \\
      Elevation & 0~m\\ 
      $B$ field direction & vertical  \\
      \hline\hline
    \end{tabular}
    \caption{Position and magnetic-field direction  of QUAX-LNL experiment}
    \label{tab:quaxlnlhorientation}
  \end{center}
\end{table}

%% file: WG4/content/RADES-CASTExperiment.tex
{Author: A. D\'iaz-Morcillo}\\

\begin{figure}[t!]
    \centering
    \includegraphics[width=0.6\linewidth]{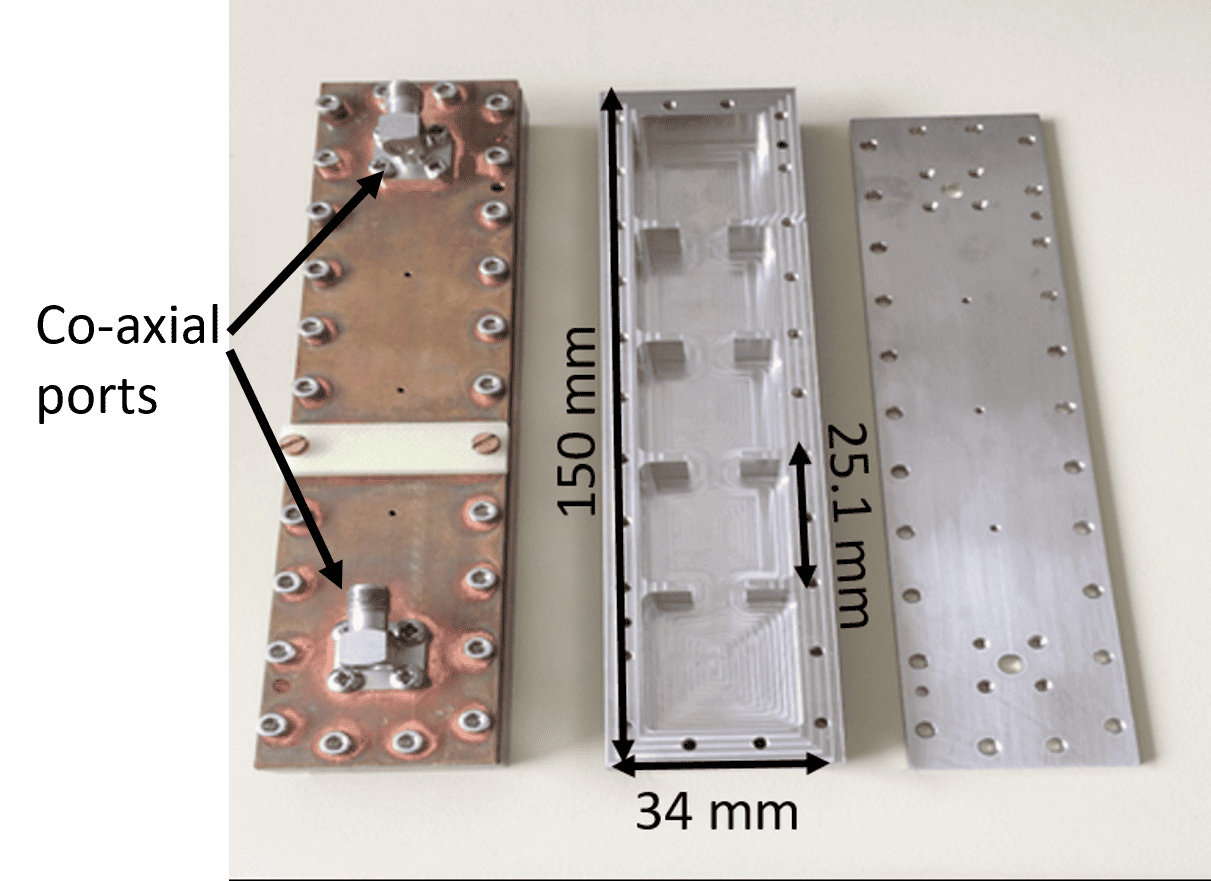 }
    \caption{RADES-CAST cavity after copper coating (closed) and before the copper coating (opened).  Figure reproduced from Fig.~1 in Ref.~\cite{CAST:2020rlf}.}
    \label{fig:RADES-CAST_cavity}
    
\end{figure}

RADES-CAST is part of the CERN Axion Solar Telescope (CAST), searching for axion dark matter in the 34.67 {$\mu$eV} mass range. A radio frequency cavity consisting of 5 sub-cavities coupled by inductive irises (see Fig. \ref{fig:RADES-CAST_cavity}) took physics data inside the CAST dipole magnet for the first time using this filter-like haloscope geometry.
An exclusion limit with a 95\% credibility level on the axion-photon coupling constant of g$_{a\gamma}\gtrsim4\times10^{-13}\,\textup{GeV}^{-1}$ over a mass range of $34.6738\,\mu\textup{eV} < m_a < 34.6771\,\mu\textup{eV}$ is set~\cite{CAST:2020rlf}.

RADES collaboration is composed at that time of Universidad Polit\'ecnica de Cartagena (UPCT), European Organization for Nuclear Research (CERN), Center for Astroparticle and High Energy Physics – University of Zaragoza (CAPA), Institut de Ci\`encies del Cosmos – Universitat de Barcelona (UB-IEEC), Yebes Observatory – National Centre for Radioastronomy Technology and Geospace Applications, Instituto de F\'isica Corpuscular (IFIC) – CSIC-University of Valencia, Laboratorio Subterr\'aneo de Canfranc. The experiment position and orientation is listed in Tab.~\ref{tab:RADES-CASTorientation}.

\begin{table}[t!]
\renewcommand{\arraystretch}{1.5}
  \begin{center}
    \begin{tabular}{c|c}
    \hline
    \hline
    Latitude &  $46^{\circ}$ $14^{\prime}$ $31^{\prime\prime}$ \\
      Longitude & $06^{\circ}$ $05^{\prime }$ $49^{\prime\prime}$ \\
      Elevation & 433~m\\ 
      $B$ field direction & horizontal  \\
      \hline\hline
    \end{tabular}
    \caption{Position and magnetic-field direction  of RADES-CAST experiment.}
    \label{tab:RADES-CASTorientation}
  \end{center}
\end{table}

%% file: WG4/content/WISPDMXExperiment.tex
{Author: L.~H.~Nguyen}\\

The WISPDMX Experiment~\cite{Nguyen:2019xuh}, located at the Institute of Experimental Physics of the University of Hamburg, Germany, is focused on the search for dark photon dark matter (DPDM). Building on the principles of Sikivie's haloscope, WISPDMX diverges from the traditional approach of scanning for WISP's signal solely in the $\mathrm{TM_{010}}$ mode. Instead, it extends its search to multiple resonant modes and off-resonant frequencies where the quality factor is about unity ($Q \approx 1$). To facilitate this approach, WISPDMX is equipped with a broadband receiver chain, enhancing its sensitivity to a wide range of DPDM parameters during each instance of data collection. For an RF resonant cavity that searches for DPDM, the sensitivity is described by
\begin{equation}
    \chi = 6.41 \times 10^{-5} \sigma_\mathrm{noise}^{1/2} \left( \kappa \mathcal{G} \mathrm{ Q V} m_{\gamma'} \rho_\mathrm{DM} \right)^{-1/2}
\end{equation}
 where Q is the quality factor, V the volume, $\kappa$ the antenna coupling, $\sigma_\mathrm{noise}$ the noise-power standard-deviation, while the form factor of the resonant mode with the DPDM is given by 
\begin{equation}
    \mathcal{G} = \frac{|\int dV \mathbf{E}^\mathrm{*cav}(\mathbf{x})\cdot \hat{\mathbf{n}}|^2}{V \int dV |\mathbf{E}^\mathrm{*cav}(\mathbf{x})|^2}
\end{equation}
with $\mathbf{E}^\mathrm{*cav}$ is the electric field of the equivalent resonant modes and $\hat{\mathbf{n}}$ is the direction of the DPDM field. 
\begin{figure}[t!]
  \begin{center}
    \includegraphics[width=0.48\linewidth]{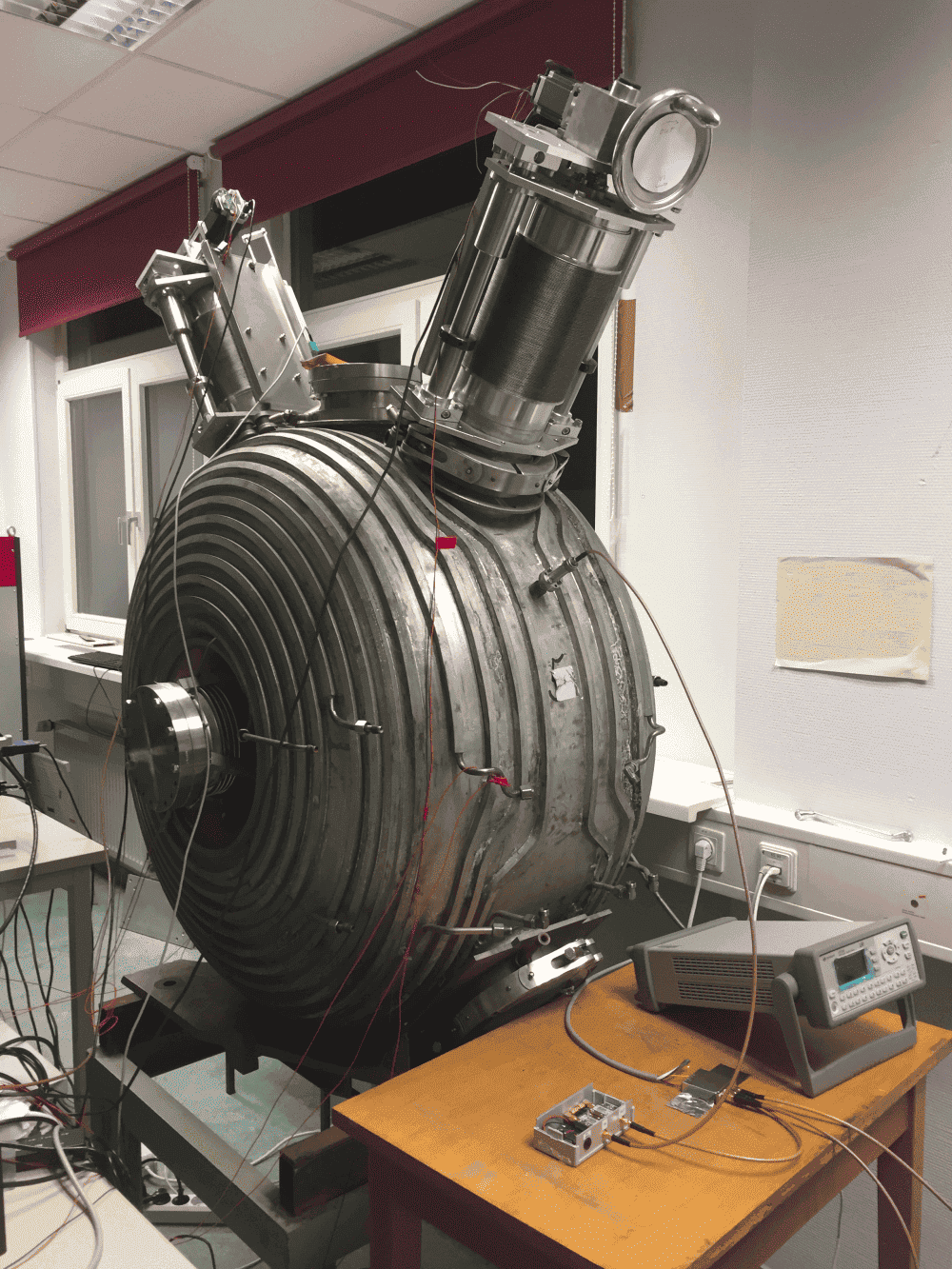}
  \end{center}
      \caption{WISPDMX's 208 MHz HERA resonant cavity and the two tuning plungers. Figure reproduced from Ref.~\cite{nguyen2018search}.}
    \label{fig:wispdmx_setup}
\end{figure}

WISPDMX features a HERA resonant cavity and a tuning mechanism comprising two plungers, as illustrated in Fig. \ref{fig:wispdmx_setup}. The resonant cavity has a diameter of 96 cm and a volume of 447 liters. Across the frequency range of 100 MHz to 500 MHz, encompassing ten resonant modes, four modes are particularly sensitive to DPDM. The frequencies, form factors, and quality factors of these four modes are detailed in Table \ref{tab:wispdmx_resonant}. 

\begin{table}[htbp]
\renewcommand{\arraystretch}{1.5}
\centering
\begin{tabular}{cccc}
\hline
\hline
\textbf{Resonant Mode} &
  \textbf{\begin{tabular}[c]{@{}c@{}}Frequency \\ (MHz)\end{tabular}} &
  \textbf{\begin{tabular}[c]{@{}c@{}}Form Factor \\ $\mathcal{G}$\end{tabular}} &
  \textbf{\begin{tabular}[c]{@{}c@{}}Quality Factor \\ $Q_0$\end{tabular}} \\ \hline
$\mathrm{TM}_{010}$   & 207.99 & 0.433 & 50000 \\ 
$\mathrm{TE}_{111-1}$ & 321.69 & 0.679 & 60000 \\ 
$\mathrm{TE}_{111-2}$ & 322.69 & 0.679 & 60000 \\ 
$\mathrm{TM}_{020}$   & 455.07 & 0.321 & 47000 \\
\hline
\hline
\end{tabular}
\caption{Resonant modes, frequencies and geometrical form factors $\mathcal{G}$ of the four resonant modes that are sensitive to DPDM.}\label{tab:wispdmx_resonant}
\end{table}

The signal receiver chain of the WISPDMX consists of a loop antenna with weak broadband coupling $\bar{\kappa} = 0.007$, and the amplifier chain that provides a balanced gain of approximately 80 dB over the frequency band of 150-600 MHz. The digital acquisition system consists of a high-speed ADC, which provides a sampling speed of 1 GHz; the samples are streamed to the GPU for the Fast Fourier Transform, which provides a high-resolution spectrum at 50 Hz resolution while still maintaining real-time performance. The experiment position and orientation are listed
in Table \ref{tab:wispdmx_orientation}.

\begin{table}[!ht]
\renewcommand{\arraystretch}{1.5}
  \begin{center}
    \begin{tabular}{c|c}
    \hline
    \hline
    Latitude &  $53^{\circ}$ $34^{\prime}$ $41.7^{\prime\prime}$ \\
      Longitude & $9^{\circ}$ $53^{\prime }$ $9.1^{\prime\prime}$ \\
      Elevation & 20~m\\ 
      $Z$-axis pointing & 0.92$\hat{\mathcal{N}}$ + $0.38\hat{\mathcal{W}}$   \\
      \hline\hline
    \end{tabular}
    \caption{Position and magnetic-field direction of WISPDMX experiment.}
    \label{tab:wispdmx_orientation}
  \end{center}
\end{table}

The science run was made during the time period from 23rd October 2017 to 2nd November 2017, comprising a total of 22000 spectra, each produced from a single 10~s data-taking and tuning step. Followed by the signal scan within the resonances and off-resonant. A signal candidate showed a strong correlation with the expected Maxwellian profile. However, this signal was later determined as an RFI. The first science run of WISPDMX set the exclusion limit on the coupling constant of the dark photon at the levels of $10^{-13}$ for the resonant frequency ranges and $10^{-12}$ for broadband mass range 0.2~$\mu$eV--2.07~$\mu$eV, and steadily increasing at masses below 0.2~$\mu$eV~\cite{Nguyen:2019xuh}.

%% file: WG4/content/GNOMEExperiment.tex
{Authors: D. Gavilan Martin, D. Aybas, O. Ruimi, D. Budker}\\

The {\textbf{G}}lobal {\textbf{N}}etwork of {\textbf{O}}ptical {\textbf{M}}agnetometers for {\textbf{E}}xotic physics searches (GNOME) is a distributed network of geographically separated and time-synchronized optically pumped atomic magnetometers \cite{afach2018characterization}. The GNOME aims to detect energy shifts in Zeeman sublevels caused by exotic fields such as axions and WISPs. For the axion the interaction is mediated via an effective coupling of the gradient of the field to the nuclear spins. The experiment targets the hypothesis that the energy density of an Ultra-light Bosonic Dark Matter (UBDM) field may be concentrated in large composite structures. Consequently, Earth would only sporadically pass through regions with significant dark matter gradients, leading to rare and brief signals in the magnetometers.

The sensors utilized in the GNOME make use of the interaction of atomic spins with external fields~\cite{budker2007optical}. Typically, alkali metal vapors, contained in glass cells, are used for the measurements.  The spins of the polarized atoms within the cell precess due to a nonzero magnetic field or an exotic field coupled to atomic spins. The continuous monitoring of the spin ensemble provides quantitative information about the field. 

\begin{figure}[t!]
  \begin{center}
    \includegraphics[totalheight=6.5cm]{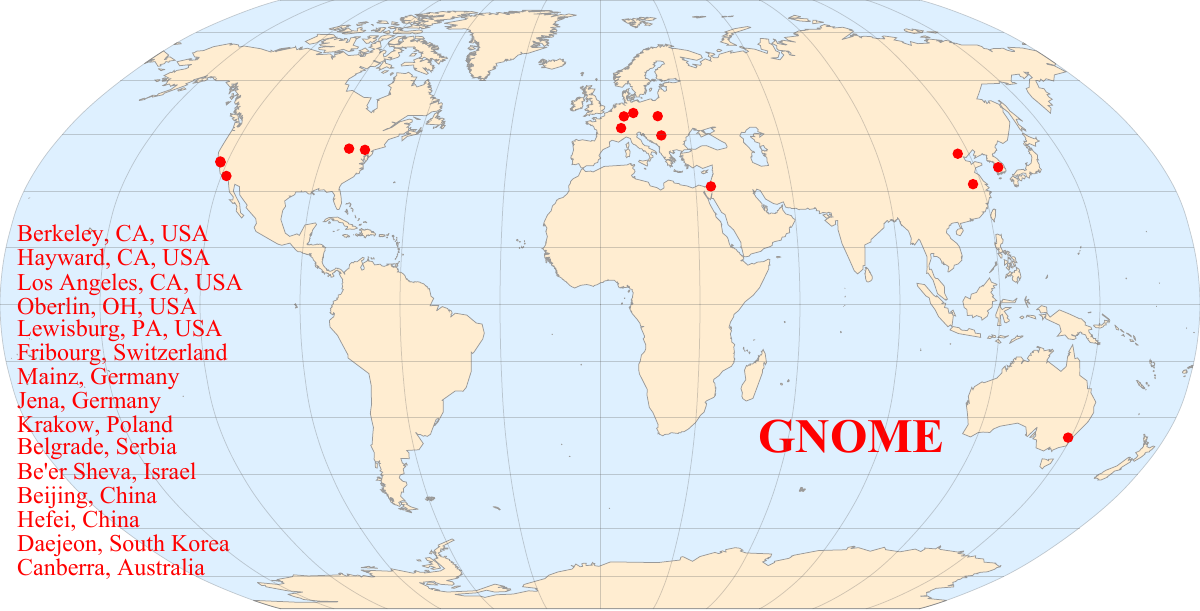}
    \caption{GNOME stations distribution around the world. Figure reproduced from Ref.~\cite{afach2023AP}.}
    \label{fig:GNOME}
  \end{center}
\end{figure}

Currently, the GNOME experiment employs a time-synchronized array of 14 magnetometers distributed globally (Fig.~\ref{fig:GNOME}), enabling the identification of transient signals attributed to dark matter structures. The experiment characteristics are
listed in Table~\ref{tab:GNOME}. The geographical distribution of the network facilitates the identification of spatiotemporal correlated patterns. In this configuration false positive detections can be vetoed and uncorrelated noise suppressed. Therefore, encounters with composite UBDM structures are confidently identified.

The GNOME experiment targets various exotic physics scenarios~\cite{afach2023AP}, with initial searches focusing on axion or axion-like particle (ALP) domain walls~\cite{afach2021search}. These domain walls are topological defects that form between regions of space in which the ALP field possesses different, but energy-degenerate, vacuum states. They are associated with considerable potential energy, and within them the ALP field has a nonzero gradient, which allows for direct detection with the GNOME. In the same line, the GNOME experiment explores the possibility of other composite structures, such as Q-balls, axion stars~\cite{kimball2018searching} and the solar axion halo~\cite{banerjee2020searching}. Besides composite structures, the collaboration also looks for signal patterns corresponding to dark matter field fluctuations~\cite{PhysRevD.108.015003} and bursts of ultrarelativistic scalar fields that can be produced in the course of some high-energy astrophysical event~\cite{dailey2021quantum}.

Current efforts in the collaboration are directed towards implementing self-compensating noble-gas-alkali-metal comagnetometers in the ``Advanced GNOME'' experiment. These sensors improve current sensitivity to exotic couplings because they cancel magnetic noise and have the capability to probe proton, neutron, and electron spin couplings~\cite{padniuk2023universal}. Initial tests of Advanced GNOME sensors demonstrate a sensitivity at the level of 10$^{-21}$\,eV/$\sqrt{\mathrm{Hz}}$ (at 1\,Hz) for exotic fields coupling to neutron spins, and about 10$^{-19}$\,eV/$\sqrt{\mathrm{Hz}}$ (at 1\,Hz) for the proton spin coupling, surpassing that of GNOME magnetometers by a factor of 100~\cite{afach2023AP}. Recently, an ALP halo search was performed with Advanced GNOME sensors in an interferometric configuration that set stringer constraints in the ultra-light mass regime~\cite{Gavilan-Martin:2024nlo}.
\begin{table}[t!]
\renewcommand{\arraystretch}{1.5}
    \centering
\begin{tabular}{lccccll}
\hline 
\hline
Station & Longitude & Latitude & Az & Alt & Type & Probed Transition \\
         \hline
         Beijing & 116.1868$^{\circ}$ E &40.2457$^{\circ}$ N & $251^{\circ}$ & 0$^{\circ}$ & NMOR & $^{133}$Cs D$_2$, $F=4$ \\
         Berkeley & 122.2570$^{\circ}$ W &37.8723$^{\circ}$ N & 0$^{\circ}$ & $+90^{\circ}$ & SERF & $^{87}$Rb D$_1$, $F=2$ \\
         Belgrade & 20.3928$^{\circ}$ W &44.8546$^{\circ}$ N & $300^{\circ}$ & 0$^{\circ}$ & rf-driven & $^{133}$Cs D$_1$, $F=4$ \\
         Beersheba & 34.8043$^{\circ}$ E &31.2612$^{\circ}$ N & 0$^{\circ}$ & $+90^{\circ}$ & SERF & K/Rb D$_1$ \\
         
         Daejeon & 127.3987$^{\circ}$ E &36.3909$^{\circ}$ N & 0$^{\circ}$ & $+90^{\circ}$ & NMOR & $^{133}$Cs D$_2$, $F=4$ \\
         Hayward & 122.0539$^{\circ}$ W &37.6564$^{\circ}$ N & 0$^{\circ}$ & $+90^{\circ}$ & SERF & $^{87}$Rb D$_1$  \\
         Krakow & 19.9048$^{\circ}$ E &50.0289$^{\circ}$ N & 0$^{\circ}$ & $+90^{\circ}$ & SERF & $^{87}$Rb D$_1$, $F=2$ \\
         Lewisburg & 76.8825$^{\circ}$ W &40.9557$^{\circ}$ N & 0$^{\circ}$ & $+90^{\circ}$ & SERF & $^{87}$Rb D$_2$ \\
         Los Angeles & 118.4407$^{\circ}$ W &34.0705$^{\circ}$ N & $270^{\circ}$ & 0$^{\circ}$ & rf-driven & $^{85}$Rb D$_2$, $F=2$ \\
         Mainz & 8.2354$^{\circ}$ E &49.9915$^{\circ}$ N & 0$^{\circ}$ & $-90^{\circ}$ & SERF & $^{87}$Rb D$_1$, $F=2$ \\
         Moxa & 11.6147$^{\circ}$ E &50.6450$^{\circ}$ N & $270^{\circ}$ & 0$^{\circ}$ & rf-driven & $^{133}$Cs D$_1$, $F=4$ \\
         Oberlin & 82.2204$^{\circ}$ W &41.2950$^{\circ}$ N & $300^{\circ}$ & 0$^{\circ}$ & SERF & K/Rb D$_1$\\
         \hline \hline
    \end{tabular}
    \caption{Position, orientation, and characteristics of GNOME stations.}
    \label{tab:GNOME}
    \end{table}

%% file: WG4/content/NASDUCKExperiment.tex
{Author: I. M. Bloch}\\

\begin{figure}[t!]
  \begin{center}
    \includegraphics[width=0.75\linewidth]{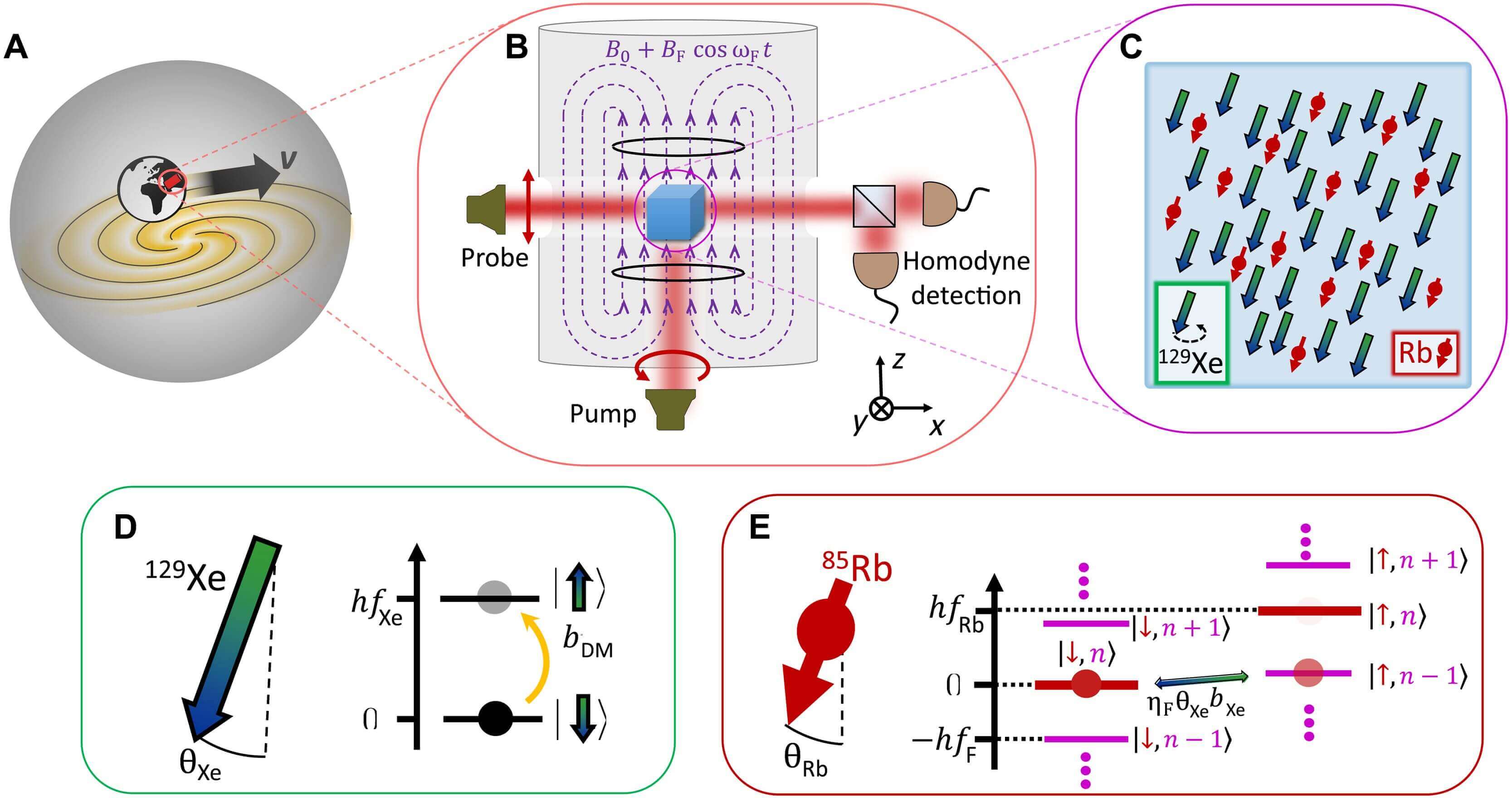}
    \caption{The setup used in NASDUCK Floquet. Figure reproduced from Ref.~\cite{NASDUCK1}.}
    \label{fig:nasduckfloquet}
  \end{center}
\end{figure}

The Noble And Alkali Spin Detectors for Ultralight Coherent darK matter (NASDUCK) experiment is an Israeli experiment magnetometers, and comagnetometers to search for Axion-Like Particles and other spin-coupled ultralight dark matter candidates. As of date, the collaboration has released two publications~\cite{NASDUCK1,NASDUCK2}.
NASDUCK focuses on demonstrating known or novel magnetometry techniques using small demonstrators.

%% file: WG4/content/AIONExperiment.tex
{Author: C. Baynham}\\

The Atom Interferometric Observatory and Network (AION) project~\cite{Badurina:2019hst} is constructing and operating a next-generation atom interferometer, starting with a 10~m device and progressing via a 100~m experiment to a 1~km instrument. AION will enable the exploration of ultra-light DM and Gravitational Waves from the very early universe and astrophysical sources in the mid-frequency band (mHz to a few Hz).
The experiment uses two clouds of cold atoms separated by a long baseline whose wavefunctions are split and recombined in a Mach-Zender interferometry sequence.

Ultimate sensitivity is reached by interoperating and networking with other such instruments around the world,
such as MAGIS~\cite{MAGISLOI,MAGIS100,Graham:2017pmn} in the US, which also targets an eventual km-scale atom interferometer, and contributes strongly to the design study of a mission proposal for an Atomic Experiment for Dark Matter and Gravity Exploration in Space (AEDGE)~\cite{AEDGE}.
AION and MAGIS complement several other terrestrial cold atom experiments, currently being prepared (MIGA~\cite{Canuel:2017rrp} and ZAIGA~\cite{Zhan:2019quq}), or being proposed (ELGAR~\cite{Canuel:2019abg}).

Atom interferometers can measure a distinctive prediction of scalar DM~\cite{Geraci:2016fva,Arvanitaki:2016fyj}, namely oscillations of the electron mass and fine-structure constant in time, with a frequency set by the mass of the scalar DM and an amplitude determined by the local DM density. This in turn leads to temporal variations of atomic transition frequencies,
and a non-trivial signal phase occurs in a differential atom interferometer when the period of the DM wave matches the total duration of the interferometric sequence~\cite{Arvanitaki:2016fyj}.
\begin{figure}[t!]
\centering
\includegraphics[height=11em]{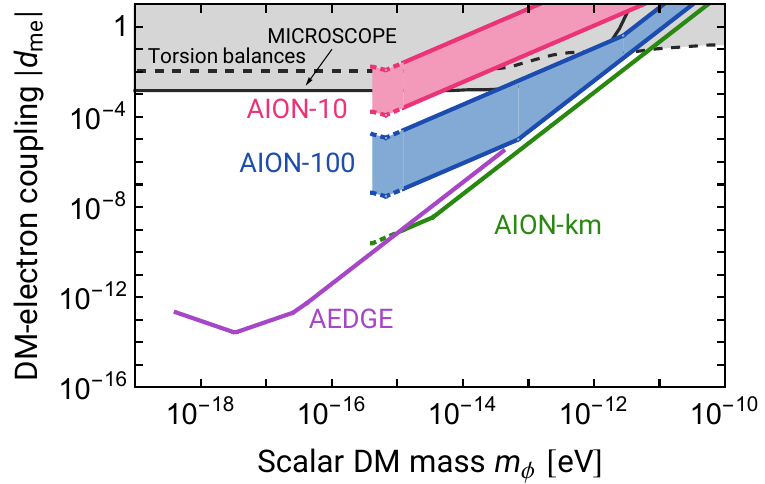} \hfill
\includegraphics[height=11em]{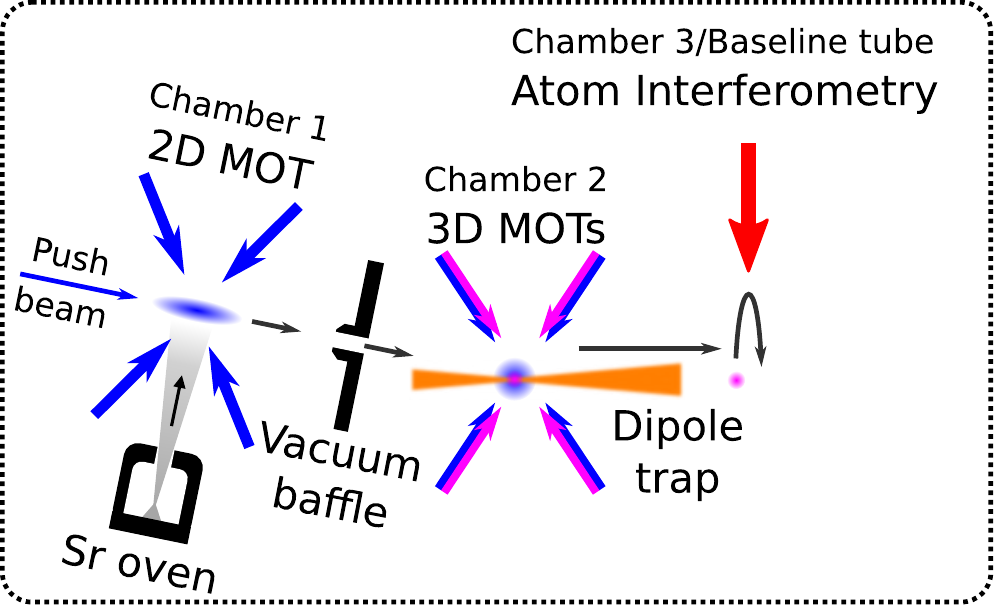}
\caption{
    \textit{Left:} Sensitivity of atom interferometers to scalar DM interactions with electrons. The shadings indicate the sensitivity of the `initial' and `goal' scenarios defined in Ref.~\cite{Badurina:2019hst}.
    The assumptions entering the various projections are outlined in Ref.~\cite{Badurina:2019hst,AEDGE}.
    \textit{Right:}
    Conceptual schematic of an AION science chamber, depicting the strontium laser cooling regions. Strontium atoms are trapped and laser-cooled inside ultra-high vacuum chambers then interrogated with an ultra-stable clock laser.  
    Figure reproduced from Ref.~\cite{AION-centralized}.
}
\label{DMplot}
\end{figure}
The left panel of Fig.~\ref{DMplot} shows the projected sensitivity of atom interferometer experiments to light scalar DM couplings to electrons with a signal-to-noise (SNR) equal to one after an integration time of $10^8$~s.
The grey regions show parameter space that has already been excluded by searches
for violations of the equivalence principle~\cite{Wagner:2012ui} or by MICROSCOPE~\cite{Berge:2017ovy}. A 100m experiment is projected to probe electron couplings for scalar DM masses $\sim 10^{-15}$ eV to $\sim 10^{-12}$~eV. A km-scale experiment extends the sensitivity to smaller couplings while AEDGE further extends the sensitivity to even lower values of the scalar DM mass and electron coupling.
Similar results are obtained for photon and Higgs portal coupling scenarios~\cite{Badurina:2019hst,AEDGE}. These projections 
do not include the possible gravitational gradient noise, which may be measured and subtracted, and also assume that the interferometer runs in broadband mode.
However, it is also possible to operate the interferometer in resonant mode~\cite{Graham:2016plp}, in which case the AEDGE sensitivity could be extended between $10^{-16}$~eV and $10^{-14}$~eV.~\cite{Arvanitaki:2016fyj}

Experiment position and orientation are listed in Tab.~\ref{tab:aion_position}.
\begin{table}[t!]
\renewcommand{\arraystretch}{1.5}
  \begin{center}
    
    \begin{tabular}{c|rrr}
    \hline
    \hline
      & AION-10 & AION-100 & AION-km \\
    \hline
      Latitude & $51^{\circ}$ $45^{\prime}$ $35^{\prime\prime}$ & \textit{t.b.c.} & \textit{t.b.c.} \\
      Longitude & $1^{\circ}$ $15^{\prime }$ $26^{\prime\prime}$ & \textit{t.b.c.} & \textit{t.b.c.} \\
      Elevation & 70~m & \textit{t.b.c.} & \textit{t.b.c.}\\ 
      $B$ field direction & Controllable & Controllable & Controllable \\
      \hline\hline
    \end{tabular}
    \caption{Position and magnetic-field direction of AION.}
    \label{tab:aion_position}
  \end{center}
\end{table}

%% file: WG4/content/APE.tex
{Authors: Q. Rokn, A. Ejlli, G. Mueller}\\

The Axion Polarimetric Experiment (APE) is a search for Galactic Halo dark matter in the form of axions and ALPs~\cite{Rokn:2025kc}. 
The setup employs a Fabry--Pérot cavity with two quarter-wave plates (QWPs) inside to increase sensitivity to measure axion-induced polarization rotation in linearly polarized light. The proposed setup offers opening previously unexplored parameter spaces (mass \(< 10^{-12}\)~eV) in axion research~\cite{Ejlli_2023}.

\begin{figure}[t!]
    \centering
    \includegraphics[width=0.9\linewidth]{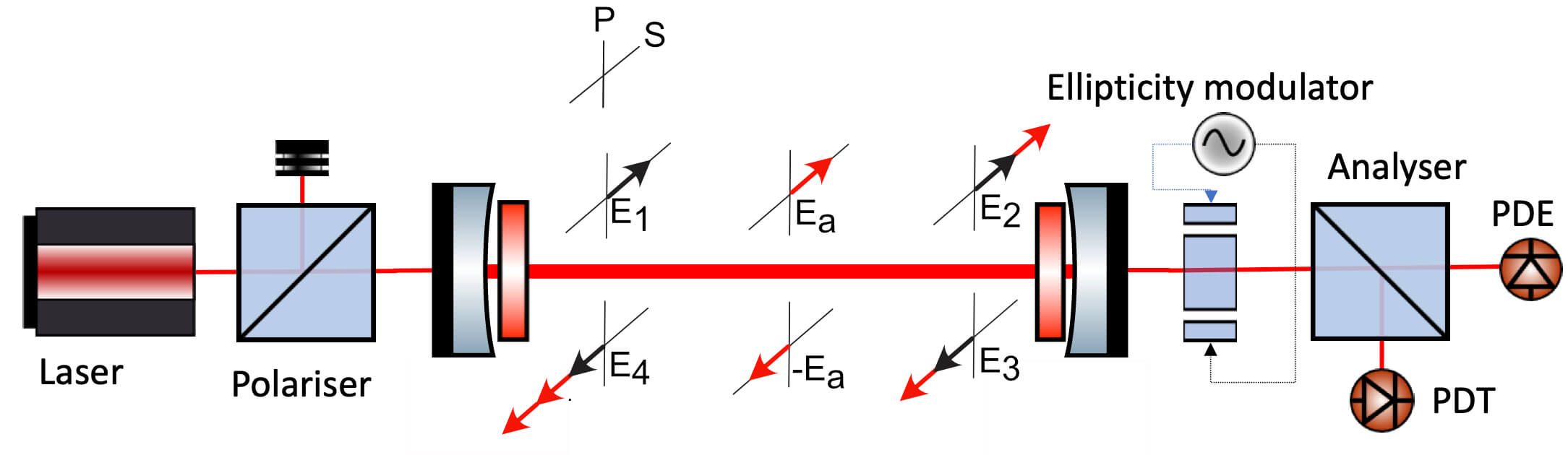}
    \caption{\small Diagram of the polarimetric setup with Fabry--P\'erot cavity and quarter-wave plates. Figure reproduced from Ref.~\cite{Rokn:2025kc}.}
    \label{fig:setup_APE}
\end{figure}

Axions and ALPs are assumed to be oscillating fields that rotate the polarization plane of linearly polarized light. Such effects are quantified by the axion-photon coupling constant \(g_{a\gamma\gamma}\)~\cite{Peccei:1977hh}. The APE setup offers a unique approach by applying high-sensitivity polarimetry to detect this polarization rotation.
The experiment uses a Fabry--Pérot cavity, combined with QWPs, to accumulate axion-induced s-polarization shifts that would otherwise cancel out. A p-polarized laser source enters the cavity, and the QWPs introduce a \(\pi/2\) phase shift between the p- and s-polarized components on each pass. This setup maximizes the accumulation of s-polarization, which is indicative of an axion interaction, and is detected as an axion signal (Fig.~\ref{fig:setup_APE})~\cite{Ejlli_2023}.

The detected intensity \(I^{\text{ext}}\) at the extinguished port is given by:
\begin{equation}
    \frac{I^{\text{ext}}}{I_0}(\beta) \approx \sigma^2 + \eta^2 - \left(\frac{1 + R}{1 - R}\right) \eta \frac{\beta}{2},
\end{equation}
where \(\sigma\) is the extinction ratio of the crossed polarizers, \(\eta = \eta_0 \cos(2\pi \nu_{\text{PEM}} t)\) with \(\eta_0\) the modulation depth of the photoelastic modulator (PEM) and \(\nu_{\text{PEM}}\) its modulation frequency, \(R\) is the mirror reflectivity, and \(\beta\) is the polarization rotation induced by the axion field.

The experimental sensitivity to the axion-photon coupling \(g_{a\gamma\gamma}\) is governed by:
\begin{equation}
    g_{a\gamma\gamma} = \frac{s^{\text{tot}}_{\beta}}{2\tau} \sqrt{\frac{(l_\text{tot} + T_2 + T_1)^2 + 4 \sin^2(\pi \nu_\text{a} \tau)}{2 \rho_\text{local}}},
\end{equation}
where \(\tau\) is the photon storage time in the cavity, \(l_\text{tot}\) the total optical loss, \(T_1\) and \(T_2\) the mirror transmission coefficients, \(\nu_\text{a}\) the axion oscillation frequency, and \(\rho_\text{local}\) the local dark matter density. The measurement's integration time is determined by the anticipated coherence length of the axion field ($10^6/\nu_a$).\\
Noise sources include shot noise, seismic noise, dark current noise, and relative intensity noise (RIN). Shot noise is expected to dominate the sensitivity.
The APE aims to detect axions in the low-mass band \(< 10^{-12}\,\mathrm{eV}\) by exceeding current sensitivity limitations. The experiment is being developed at the Max Planck Institute for Gravitational Physics (Albert-Einstein-Institut), Hannover.
The Axion Polarimetric Experiment presents a feasible and innovative strategy to advance the detection of axions and ALPs, pushing the boundaries of dark-matter research. By capturing minute polarization shifts with high precision, the APE has the potential to unlock new insights into particle physics and astrophysics, with broader implications for understanding dark matter.

%% file: WG4/content/CADEx_Experiment.tex
{Author: A. D\'iaz-Morcillo}\\

\begin{figure}[t!]
    \centering
    \includegraphics[width= 0.9\linewidth]{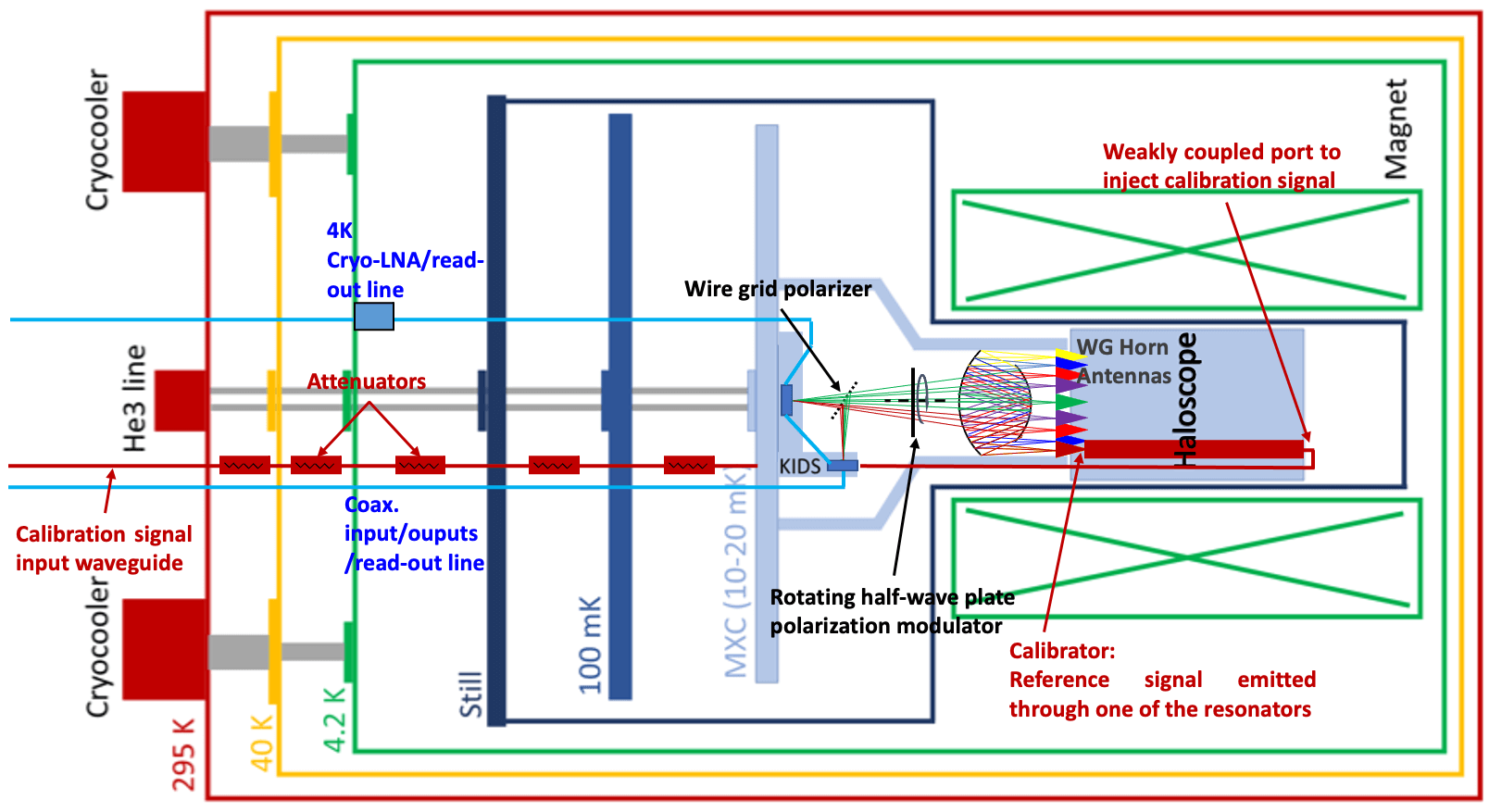}
    \caption{Schematic block diagram proposed for the CADEx's accommodation in the dilution refrigerator of the Canfranc Underground Laboratory. Adapted from Fig.~1 in Ref.~\cite{Aja:2022csb}.}
    \label{fig:CADEx_scheme}
    
\end{figure}

CADEx aims to search for axions in the mass range \SIrange[range-phrase=--, range-units=single]{330}{460}{\micro \eV} within the W-band (\SIrange[range-phrase=--, range-units=single]{75}{110}{\giga \hertz}) by combining the resonant-cavity haloscope approach with an incoherent detection system based on Kinetic Inductor Device (KID) technology \cite{Aja:2022csb}. The incoherent detectors in CADEx will measure the linearly polarized axion signal generated in the haloscope against the unpolarized background emission as a function of the resonant frequency of the haloscope. This consists of an array of 16 rectangular cavities, working with the $TM_{110}$ mode, whose signal is extracted by an iris on a waveguide and transmitted by horn antennas through a quasi-optic system towards the detector. CADEx will be installed in the dilution refrigerator of the Canfranc Underground Lab (LSC) to decrease the impact of cosmic rays on the final sensitivity using broadband incoherent detectors. Fig. \ref{fig:CADEx_scheme} shows a block diagram of the experiment accommodated inside the LSC dilution refrigerator, which includes a solenoid magnet, indicating the location of the main subsystems and their temperature.

CADEx collaboration is composed of LSC, Centro de Astrobiolog\'ia (CAB, CSIC-INTA), Universidad de Cantabria (UC), Universidad Polit\'ecnica de Cartagena (UPCT), Universidad P\'ublica de Navarra (UPNa), Instituto de Física Corpuscular (IFIC, CSIC-UV), Instituto de F\'isica de Cantabria (IFCA), Institut de Ci\`encies del Cosmos (ICCUB), Centro Astronómico de Yebes (CDT-IGN) and IMDEA Nanociencia. Currently, the first cavity prototype and a 64 LEDKID array have been designed and manufactured, and the quasy-optical system is being designed. CADEx operations are foreseen for 2027 - 2035. The experiment position and orientation are listed in Tab.~\ref{tab:CADExlnfhorientation}.

\begin{table}[!th]
\renewcommand{\arraystretch}{1.5}
  \begin{center}
    \begin{tabular}{c|c}
    \hline
    \hline
    Latitude &  $42^{\circ}$ $48^{\prime}$ $21^{\prime\prime}$ \\
      Longitude & $-0^{\circ}$ $33^{\prime }$ $28^{\prime\prime}$ \\
      Elevation & 1,195~m\\ 
      $B$ field direction & vertical  \\
      \hline\hline
    \end{tabular}
        \caption{Position and magnetic-field direction  of CADEx experiment.}
    \label{tab:CADExlnfhorientation}
  \end{center}
\end{table}

%% file: WG4/content/DALIExperiment.tex
{Author: J. De Miguel}\\

\begin{figure}[t!]
  \begin{center}
    \includegraphics[width=1\linewidth]{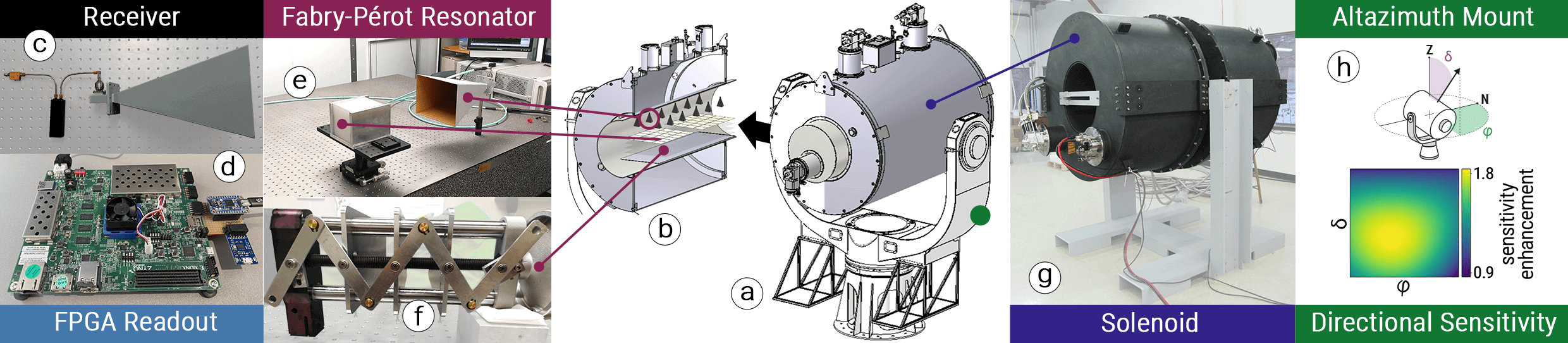}
    \caption{The general layout of the DALI MPA+FP haloscope is shown in insets (a). The cryostat of the experiment, sectioned in (b), is cooled to sub-kelvin temperatures using $^3$He coolers. Inset (c) displays a high-electron-mobility transistor based receiver that operates around 1~K, allowing it to function only a few times above the quantum noise limit. The readout system utilizes a field-programmable gate array (FPGA) architecture (d). The opto-mechanical system has been tested separately: the optics in (e), and the mechanics in (f). This decoupling of uncertainties ensured a more reliable study. The system is now being reintegrated to develop the FP tuner prototype. A standard MRI superconducting magnet with a warm bore of $\sim$1/2 m houses the cryostat (g), and it is mounted on an alt-azimuth mount (h) to enhance sensitivity to the daily modulation of the signal caused by DM streams.}
    \label{DALI_1}
  \end{center}
\vspace{-20pt}
\end{figure}

The Dark Photons \& Axion-Like Particles Interferometer (DALI)~\cite{DeMiguel2021,DeMiguel2023nmz,Cabrera2023qkt,2024JInst19P1022H,Cabrera2024oek,Cabrera2024SPIE} employs a new experimental setup designed to probe wavelike dark matter with masses exceeding 25 $\mu$eV: the magnetized phased array (MPA). Leveraging the know-how transferred from cosmic-microwave background observations, an MPA haloscope features a large flat mirror housed inside a solenoid-type magnet. To enhance the faint signal originating from axion-to-photon conversion via the inverse Primakoff effect, or through the kinetic mixing of dark photons, DALI will use a multi-layer tunable Fabry-P\'erot (FP) resonator, which allows for quality factors of up to $Q\sim 50,000$ in the 6-60 GHz range~\cite{Cabrera2024oek}. The scanning frequency in a FP interferometer is decoupled from the experiment’s cross-section enabling access to a parameter space that remains largely unexplored. An overview of this MPA+FP haloscope is shown in Fig.~\ref{DALI_1}. 

DALI will  benefit of an altazimuth mount, to enhance sensitivity to daily background modulation caused by transient events such as DM streams~\cite{DeMiguel2023nmz}; the capability of scanning two or three resonant frequencies simultaneously within its band, effectively doubling or tripling the scanning speed~\cite{DeMiguel2023nmz,Cabrera2024oek,Cabrera2024SPIE}; readily available equipment, making the experiment cost-effective and enabling faster deployment in the highly competitive axion search landscape~\cite{DeMiguel2021,DeMiguel2023nmz}. DALI's approach to study transient events, such as direct detection of axion streams, could achieve unprecedented sensitivities, far exceeding the coupling strengths predicted by QCD axion models~\cite{DeMiguel2024cwb}. DALI will also impose new constraints on a gravitational wave background at high frequencies~\cite{DeMiguel2023nmz}.

The DALI project began in 2018 as a collaboration between the RIKEN Center in Japan and the Instituto de Astrof\'isica de Canarias (IAC). The full-scale DALI experiment will be installed at the IAC’s Teide Observatory (28$^{\circ}$18'00''N 16$^{\circ}$30'35''W) over 2000 meters above sea level. The site, known as CMBLab, provides a controlled environment shielded from terrestrial microwave sources. 
\begin{figure}[t!]
  \begin{center}
    \includegraphics[width=.6\linewidth]{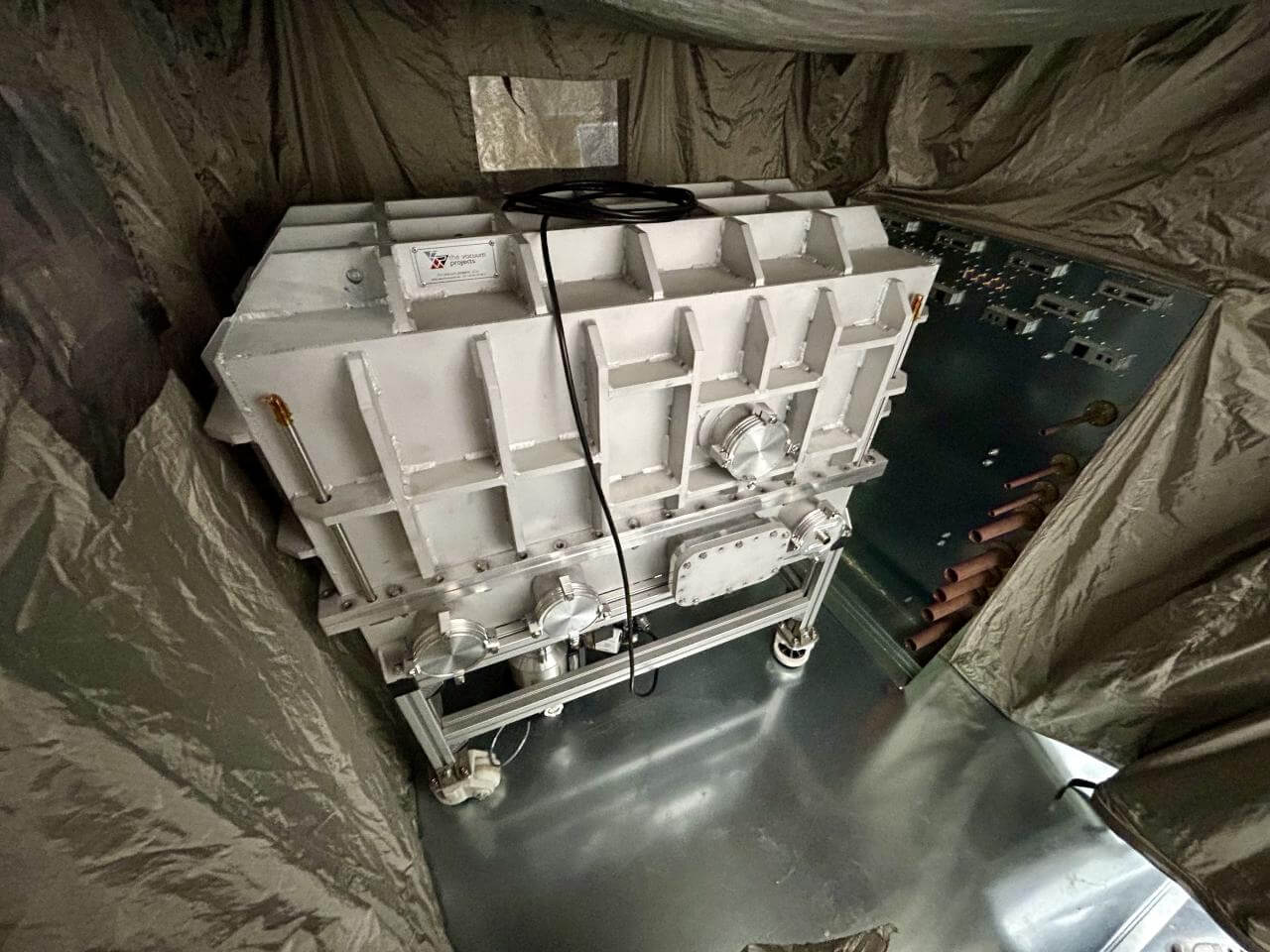}
    \caption{The scaled-down DALI prototype is now in its final  commissioning stage.}
    \label{DALI_2}
  \end{center}
\vspace{-20pt}
\end{figure}

A scaled-down DALI prototype, hosted at the IAC headquarters is depicted in Fig. \ref{DALI_2}. 
This prototype was successfully commissioned in September 2025 and is currently acquiring data. Although smaller in size, this prototype is designed to retain part of DALI’s potential, while alleviating hardware complexity. It will probe two broad bands centered at approximately $30 \, \mu \text{eV}$ and $140 \, \mu \text{eV}$ in the search for axion DM. 

%% file: WG4/content/FLASHExperiment.tex
{Author: A. Rettaroli}\\

\begin{figure}[t!]
  \begin{center}
    \includegraphics[totalheight=5.5cm]{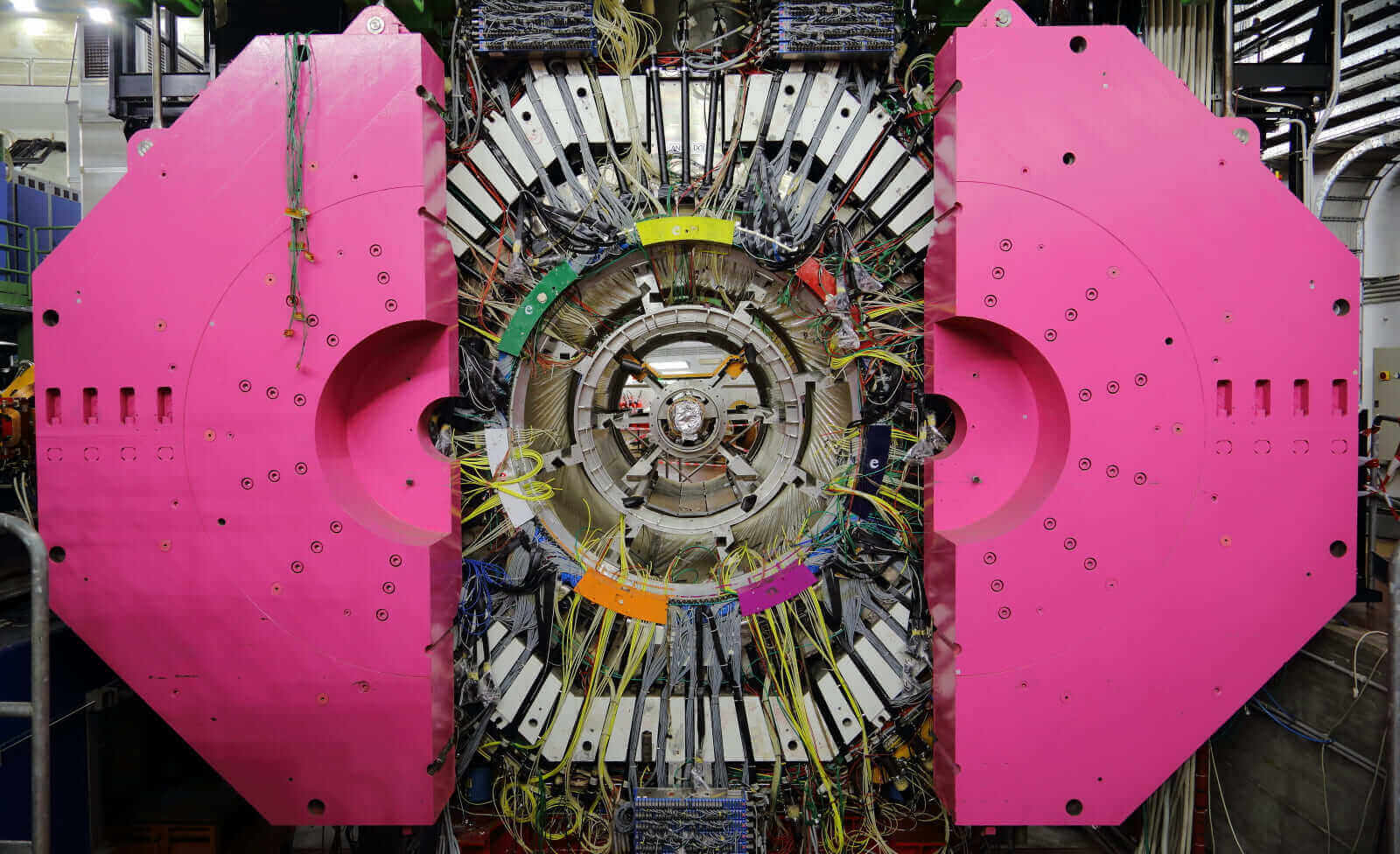}
    \caption{The former FINUDA apparatus in the Da$\Phi$ne hall at LNF.}
    \label{fig:FLASH}
  \end{center}
\end{figure}

\begin{figure}[t!]
  \begin{center}
    \includegraphics[totalheight=8.5cm]{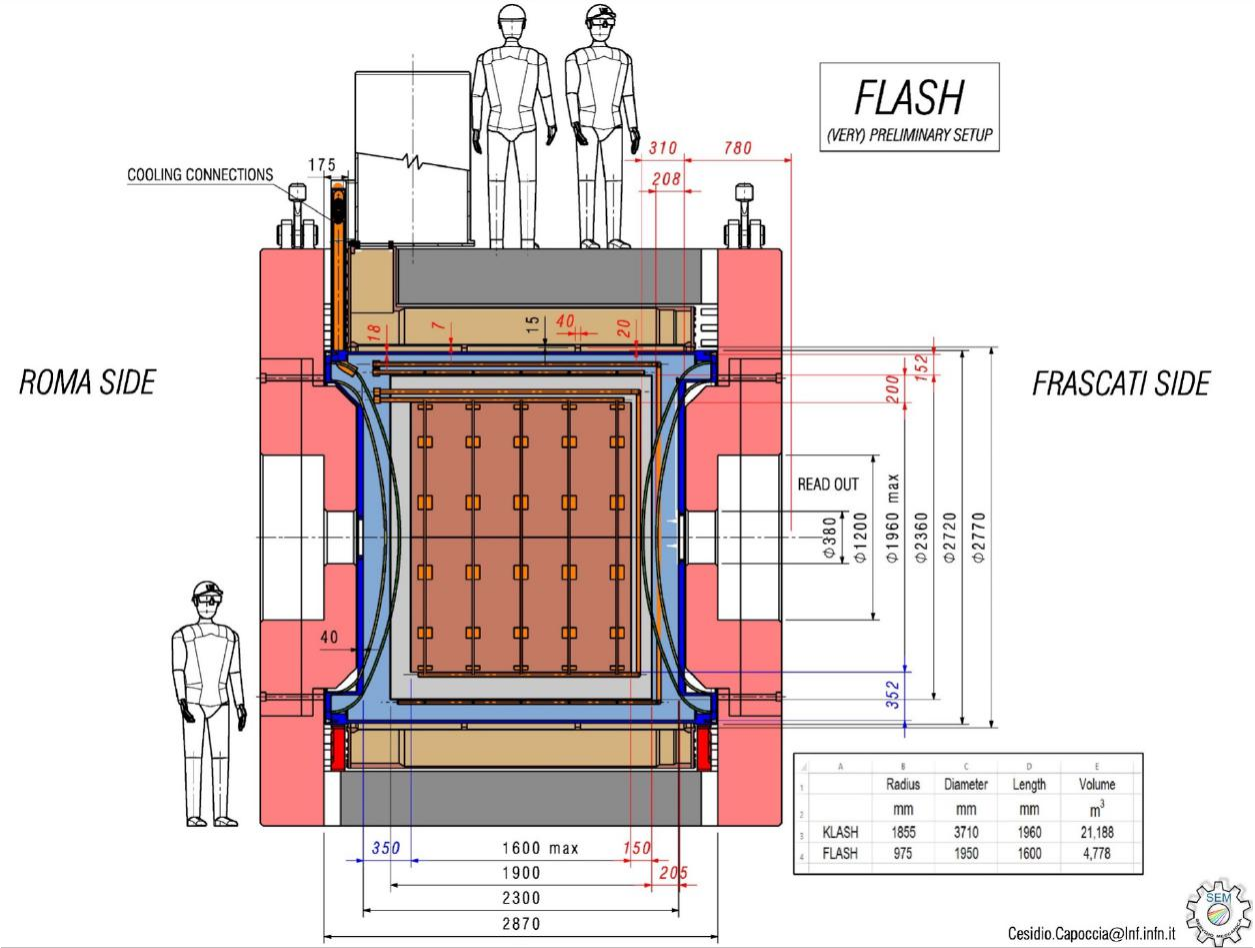}
    \caption{FLASH detector section.}
    \label{fig:FLASH-sezione}
  \end{center}
\end{figure}

FLASH is an INFN experiment hosted at the LNF site. It is designed as a classical Sikivie's haloscope searching for axions through the $g_{a\gamma\gamma}$ coupling in the mass range around 1~$\mu$eV, corresponding to a frequency of about 200~MHz, but it is able to put constraints on other WISPs such as axion-like particles and Dark Photons, as well as on high-frequency gravitational waves. The experiment was initially proposed as KLASH and planned to recycle the 4~m bore 0.6~T KLOE superconducting-magnet~\cite{Alesini:2019nzq}. It will use instead the 3~m bore 1.1~T magnet of the FINUDA experiment, shown in Fig.~\ref{fig:FLASH}. The peak magnetic field is 1.1~T, supplied by a superconducting magnet immersed in liquid helium. The magnet, last operated in 2007, was cooled down again to 4~K in January 2024 and successfully brought to the nominal field. Cooling the cavity to 1.9~K inside a cryostat hosted in the magnet bore (Fig.~\ref{fig:FLASH-sezione}) and using a microstrip SQUID as a first amplification stage the sensitivity to DFSZ-axion models, about $10^{-16}~\text{GeV}^{-1}$, will be reached. A large boost factor will be provided by the large volume of the OFHC copper cavity of about 4.5~m$^3$. FLASH is foreseen to scan frequencies between 100 and 300~MHz.

The experiment, whose the conceptual-design report is published in~\cite{Alesini:2023qed}, is now preparing a Technical Design Report and an international collaboration is being formed. FLASH is now part of the GravNet network of haloscopes for the detection high frequency gravitational waves~\cite{SCHNEEMANN2024169721,Schmieden:2023fzn}.
FLASH position and orientation are listed in Tab.~\ref{tab:flashorientation}.
\begin{table}[tbhp]
\renewcommand{\arraystretch}{1.5}
  \begin{center}
    \begin{tabular}{c|c}
    \hline
    \hline
    Latitude &  $41^{\circ}$ $49^{\prime}$ $26^{\prime\prime}$ \\
      Longitude & $12^{\circ}$ $40^{\prime }$ $13^{\prime\prime}$ \\
      Elevation & 100~m\\ 
      $B$ field direction & East--northeast  \\
      \hline\hline
    \end{tabular}
    \caption{Position and magnetic-field direction  of FLASH experiment}
    \label{tab:flashorientation}
  \end{center}
\end{table}

%% file: WG4/content/GrahalExperiment.tex
{Author: P. Pugnat}\\

The GrAHal project (Grenoble Axion Haloscopes) proposes to use the 43~T hybrid magnet modular platform of LNCMI-Grenoble (Tab.~\ref{tab:grahal1}) for axion search~\cite{grenet2021grenobleaxionhaloscopeplatform}. Its objective is to probe the axion mass range in between 1--120~$\mu$eV (0.2--30~GHz) profiting from the various high magnetic field and flux configurations given in 
Tab.~\ref{tab:grahal1}. For the middle frequency range, various approaches is also considered avoiding the use of the 24~MW resistive inserts such as multicell RF-cavities configurations inserted inside the full superconducting 9~T/810~mm hybrid magnet configuration.

\begin{table}[!ht]
\renewcommand{\arraystretch}{1.5}
  \begin{center}
    \begin{tabular}{llll}
    \hline
    \hline
    Field in warm  & Resistive magnet powering; & RF-cavity &  Resonant frequency;  \\[-.5em]
    diameter &  Water cooling; Cryogenics & diameter & Axion mass \\ \hline
    43 T in 34 mm &24; 1; 0.4 MW &8 mm  &29 GHz; 118 $\mu$eV$/c^2$\\
    40 T in 50 mm &24; 1 ; 0.4 MW &23 mm &10 GHz; 41 $\mu$eV$/c^2$\\
    27 T in 170 mm  &18 ; 0.75 ; 0.4 MW & 110 mm & 2 GHz; 8.6 $\mu$eV$/c^2$\\
    17.5 T in 375 mm  &12 ; 0.5 ; 0.4 MW &315 mm & 0.7 GHz; 3 $\mu$eV$/c^2$\\
    9 T in 800 mm &0.4 MW  &675 mm & 0.34 GHz; 1.4 $\mu$eV$/c^2$\\
      \hline\hline
    \end{tabular}
    \caption{DC magnetic field configuration of the Grenoble hybrid magnet platform with the corresponding frequency/axion mass of the TM010 mode of the RF cylindrical cavity, which can be inserted, taking into account the required thickness for the cryostat to reach low temperature in the 25--50~mK range. Dedicated tuning mechanisms are being studied to allow an excursion in frequency of up to 30\% of the resonant frequency. The possibility to insert several RF matched cavities of smaller diameter for a given hybrid magnet configuration is also considered.}
    \label{tab:grahal1}
  \end{center}
\end{table}

The GrAHal proposal follows at present three main directions conducted in parallel way, namely BabyGrAHal, GrAHal-CAPP and GrAHal-HF-43T.

\paragraph{BabyGrAHal.}
BabyGrAHal is composed of two haloscopes targeting the frequency range 4--7~GHz; a first one is already in operation around 4.4~K and the second one is in construction within an existing 14~T superconducting magnet equipped with a 50~mK dilution refrigerator. First results of the GrAHal’s path finding experimental run can be found in~\cite{10.3389/fphy.2024.1358810}. The innovative approach of this project consists in using the pressure of GHe as well as the level of LHe in the RF cavity to vary the resonance frequency of the cavity modes.

\begin{figure}[t!]
  \begin{center}
    \includegraphics[totalheight=8.5cm]{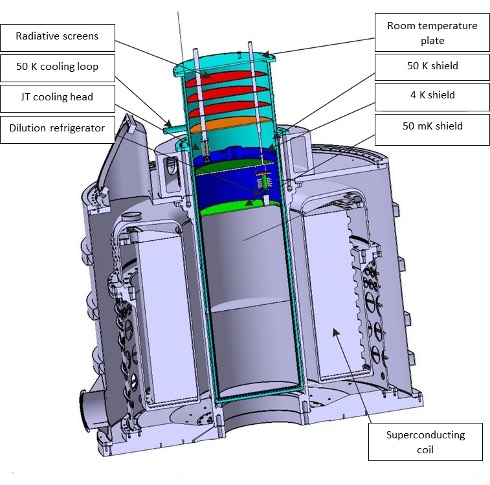}
    \caption{First cut design of the cryostat of the GrAHal-CAPP haloscope integrated within the large bore superconducting magnet ``outsert'' of the Grenoble hybrid magnet with thermal shields at 50~mK, 4~K, and 50~K. JT cooling head stands for Joule--Thomson cooling head. Figure reproduced from Ref.~\cite{grenet2021grenobleaxionhaloscopeplatform}.}
    \label{fig:grahalcapp}
  \end{center}
\end{figure}

\paragraph{GrAHal-CAPP.}
GrAHal-CAPP, a collaboration with Center for Axion and Precision Physics in Daejeon (Republic of Korea), will focus on the 300--600~MHz frequency range corresponding to axion mass of 1--3~$\mu$eV. Thanks to the state-of-the-art magnet provided by the 9~T superconducting ``outsert'' coil of the hybrid magnet (Tab.~\ref{tab:grahal1}), haloscope with unprecedented sensitivity can be built i.e., reaching the DFSZ limit in 2-year integration time. The RF cavity of 675~mm diameter will be cooldown to 50 mK within the cryostat shown in Fig.~\ref{fig:grahalcapp} thanks to the connection to a first dilution refrigerator. A second dilution refrigerator will be dedicated to the cooling to 100~mK of the RF electronic components, including the near quantum noise amplifier foreseen for the phase-2. A superconducting compensation coil will be required for this phase with low consumption current leads, which could be made with high temperature superconductors (HTS). All cryogenic parts will be developed by Institut N\`eel in close collaboration with LNCMI for onsite integration. A synergy with cryostats needed for quantum computing is being studied. The RF cavity will be made out of Cu by CAPP/IBS with a quality factor in the range of few 10$^5$. A dedicated tuning mechanism based on a piezoelectric motor will be developed and combined with various tie rod diameters to cover the whole targeted frequency range. All foreseen developments are based on already proven technologies.

\paragraph{GrAHal-HF-43T.}
GrAHal-HF-43T focuses in the high magnetic field configuration of the Grenoble hybrid magnet to probe higher frequency range corresponding to the configuration for the 43~T hybrid magnet shown in Tab.~\ref{tab:grahal1} after a first step at 42~T. Dedicated RF cavities are developed in Cu as well as using HTS tapes to increase the Q factor. Two main objectives are pursued with GrAHal-HF: first, with axion search in the high frequency (HF) or mass range around 12.78 GHz/52.5~$\mu$eV; second, with the first test of an HTS RF cavity in very high magnetic field. This might also shed new light in some applied and fundamental aspects of high temperature superconductivity arising from the combined use of high frequencies with very high magnetic fields.

%% file: WG4/content/MaglevHuntExperiment.tex
{Author: G. Higgins}\\

MaglevHunt is a planned experiment located in Vienna, Austria, at the Marietta Blau Institute for Particle Physics (MBI) of the Austrian Academy of Sciences (OeAW), funded by an ERC starting grant.
It will use mechanical sensors -- consisting of magnetically-levitated superconductors -- to search for dark matter.
The motion of magnetically-levitated superconductors can be precisely monitored, using superconducting quantum circuits.
Further, the superconductor motion can be highly isolated from noise, since superconductors can be levitated far from surfaces in ultrahigh vacuum at cryogenic temperatures within magnetic shielding.
These features make levitated superconductors an attractive platform for sensing applications; they have been used for gravimetry \cite{Goodkind1999}, gravity gradiometry \cite{Griggs2017}, and this platform is being developed for tests of quantum physics using relatively large masses \cite{RomeroIsart2012,Hofer2023,GutierrezLatorre2023,Schmidt2024}.
This project involves developing a small array of levitated superconductors and using them to search for both wave-like and particle-like dark matter.

Wave-like dark matter may drive oscillatory motion of the superconductors, while particle-like dark matter may cause the superconductors to experience occasional impulses.
By monitoring the sensor array, MaglevHunt will search for three wave-like dark matter candidates: vector $B-L$ dark matter \cite{Graham:2015ifn,Carney2021NJP}, axion-like particles and dark photons \cite{Higgins2024}.
It will also search for particle-like dark matter that interacts with ordinary matter via long-range interactions.
As well as probing the existence of different dark matter candidates in unexplored regimes, this project aims to lay the foundations for a dark matter search sensitive to the gravitational interaction between dark matter and ordinary matter, following Ref.~\cite{Carney:2019pza,Windchime:2022whs}:
The long-term vision is to operate a densely-packed meter-scale array of mechanical sensors in 3D, then when a dark matter particle passes through the array it would cause a small gravitational pull on the sensors it passes close to, leaving a 1D track of excited sensors.
Each sensor would weight $\sim \SI{1}{g}$, be precisely monitored beyond the standard quantum limit, and be highly isolated from noise sources.
The vision is to use such a detector to probe the existence of dark matter around the Planck mass via the gravitational interaction (if dark matter has the Planck mass, there should be around one dark matter particle passing through each square meter of laboratory space each year).

%% file: WG4/content/RADES-BabyIAXOExperiment.tex
{Author: A. D\'iaz-Morcillo}\\

\begin{figure}[tbhp]
    \centering
    \includegraphics[scale=0.65]{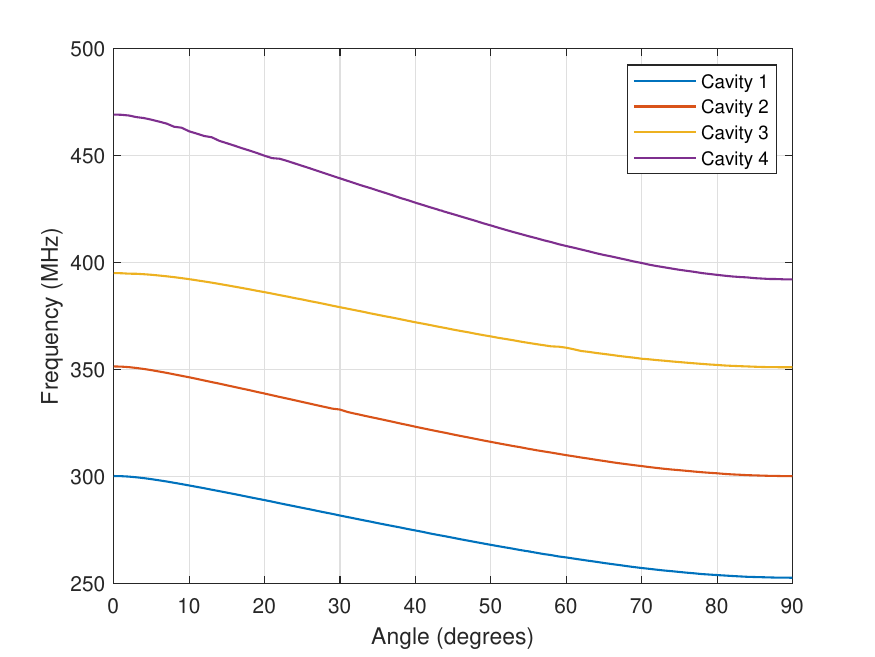}
    \caption{Resonant frequency for mode $TE_{111}$ of the four RADES-BabyIAXO cavities. Figure reproduced from Ref.~\cite{Ahyoune:2023gfw}.}
    \label{fig:RADES-BabyIAXO_freq_range}
\end{figure}

RADES-BabyIAXO is a low-frequency axion haloscope setup suitable for operation inside the future BabyIAXO magnet, at DESY. This proposal has a potential sensitivity to the axion-photon coupling down to values corresponding to the KSVZ model, in the currently unexplored mass range between 1 and 2 $\mu$eV, after a total effective exposure of 440 days. This mass range is covered using four differently dimensioned 5-meter-long cavities, equipped with a tuning mechanism based on inner turning plates. Fig. \ref{fig:RADES-BabyIAXO_freq_range}) shows the tuning range for each cavity. The setup includes a cryostat and cooling system to cool down the BabyIAXO bore down to about 5 K, as well as appropriate low-noise signal amplification and detection chain.

RADES collaboration is composed of Universidad Polit\'ecnica de Cartagena (UPCT), European Organization for Nuclear Research (CERN), Center for Astroparticle and High Energy Physics – University of Zaragoza (CAPA), Institut de Ci\`encies del Cosmos – Universitat de Barcelona (UB-IEEC), Yebes Observatory – National Centre for Radioastronomy Technology and Geospace Applications, Instituto de F\'isica Corpuscular – CSIC – University of Valencia (IFIC-CSIC-UV), Physics Laboratory \'Ecole Normale Sup\'erieure – CNRS (ENS – CNRS), Aalto University, Karlsruher Institut f\"ur Technologie (KIT), Max Planck Institut f\"ur Physik (MPP), Institut de Ci\`encia de Materials de Barcelona (ICMAB – CSIC).

%% file: WG4/content/RADES-LSCExperiment.tex
{Author: A. D\'iaz-Morcillo}\\

RADES-LSC proposal combines single-photon detection by means of 3D transmons, magnetic tuning and the use of high temperature superconductors to enhance the sensitivity of the axion dark matter detection in the 30 - 70 $\mu$eV mass range. The experiment will be developed as part of the ERC Synergy DarkQuantum project and is planned to be located at the Canfranc Underground Laboratory (LSC).

RADES collaboration is composed of the European Organization for Nuclear Research (CERN), the Center for Astroparticle and High Energy Physics – University of Zaragoza (CAPA), Universidad Polit\'ecnica de Cartagena (UPCT), Institut de Ci\`encies del Cosmos – Universitat de Barcelona (UB-IEEC), Yebes Observatory – National Centre for Radioastronomy Technology and Geospace Applications, Instituto de F\'isica Corpuscular – CSIC – University of Valencia (IFIC-CSIC-UV), Physics Laboratory \'Ecole Normale Sup\'erieure – CNRS (ENS – CNRS), Aalto University, Karlsruher Institut f\"ur Technologie (KIT), Max Planck Institut f\"ur Physik (MPP), and the Institut de Ci\`encia de Materials de Barcelona (ICMAB – CSIC). The experiment position and orientation are listed in Tab.~\ref{tab:RADES-LSCorientation}.

\begin{table}[t!]
\renewcommand{\arraystretch}{1.5}
  \begin{center}
    \begin{tabular}{c|c}
    \hline
    \hline
    Latitude &  $42^{\circ}$ $48^{\prime}$ $21^{\prime\prime}$ \\
      Longitude & $-0^{\circ}$ $33^{\prime }$ $28^{\prime\prime}$ \\
      Elevation & 1,195~m\\ 
      $B$ field direction & vertical  \\
      \hline\hline
    \end{tabular}
    \caption{Position and magnetic-field direction  of the RADES-LSC experiment.}
    \label{tab:RADES-LSCorientation}
  \end{center}
\end{table}

%% file: WG4/content/RadioAxionExperiment.tex
{Author: C. Toni}\\

The RadioAxion experiment aims to detect axion dark matter by observing time-modulated changes in the decay constants of Americium-241 (alpha decay) and Potassium-40 (electron-capture decay). In order to study the decays independently of the comic ray flux, in particular of its time variation during the year (a few percent),  the experimental set-up was installed deep underground in the Gran Sasso Laboratory, in a container placed in front of Hall B.

The axion-gluon interaction affects the mass of the pion, sigma and omega particles, which are the mediators of the strong nuclear force. In particular, it changes the nuclear binding energy and the mass difference between the proton and the neutron. These time modulations take place with a frequency proportional to the axion mass.

The investigation started from the study of the decay of Americium-241 (half-life: 432.2 y). In particular, it was  at first determined the relation of the time modulation amplitude to the coupling between the axion and the gluon~\cite{Broggini:2024udi}. Then, a simple set-up was buikt with a core made from a 3’’ $\times$3’’ NaI crystal to detect the gamma and X rays emitted from the alpha decay of Americium-241 to Neptunium-237 (mostly a 59.5~keV gamma). The experiment is expected to achieve a sensitivity of a few parts per million on the modulation amplitude in a wide range of oscillation periods, from microseconds to a year, enabling sensitive constraints on the axion decay constant across a large axion mass range between $10^{-9}$ and $10^{-22}$~eV with a data acquisition time of 3 years.

For the Potassium-40 electron-capture decay (half-life: 1.25 billion years), it was determined the relation of the time modulation amplitude to the coupling between the axion and the gluon~\cite{Alda:2024xxa} and then, in summer 2025, the study of the decay started under Gran Sasso with a large NaI crystal (4 liters) surrounded by an extended Potassium-40 source. The signature of the decay will be the 1461~keV gamma ray from the excited state of the Argon-40 daughter nucleus. The expected sensitivity is similar to the one previously given for the Americium-241 within the same mass range, albeit for a different nuclear decay process. First data taking and analysis have already been carried out on a dataset corresponding to a few months of measurements \cite{broggini:2026qxm}.

%% file: WG4/content/SupaxExperiment.tex
{Authors: K. Schmieden, M. Schott}\\

\begin{figure}
    \centering
    \includegraphics[width=0.30\linewidth]{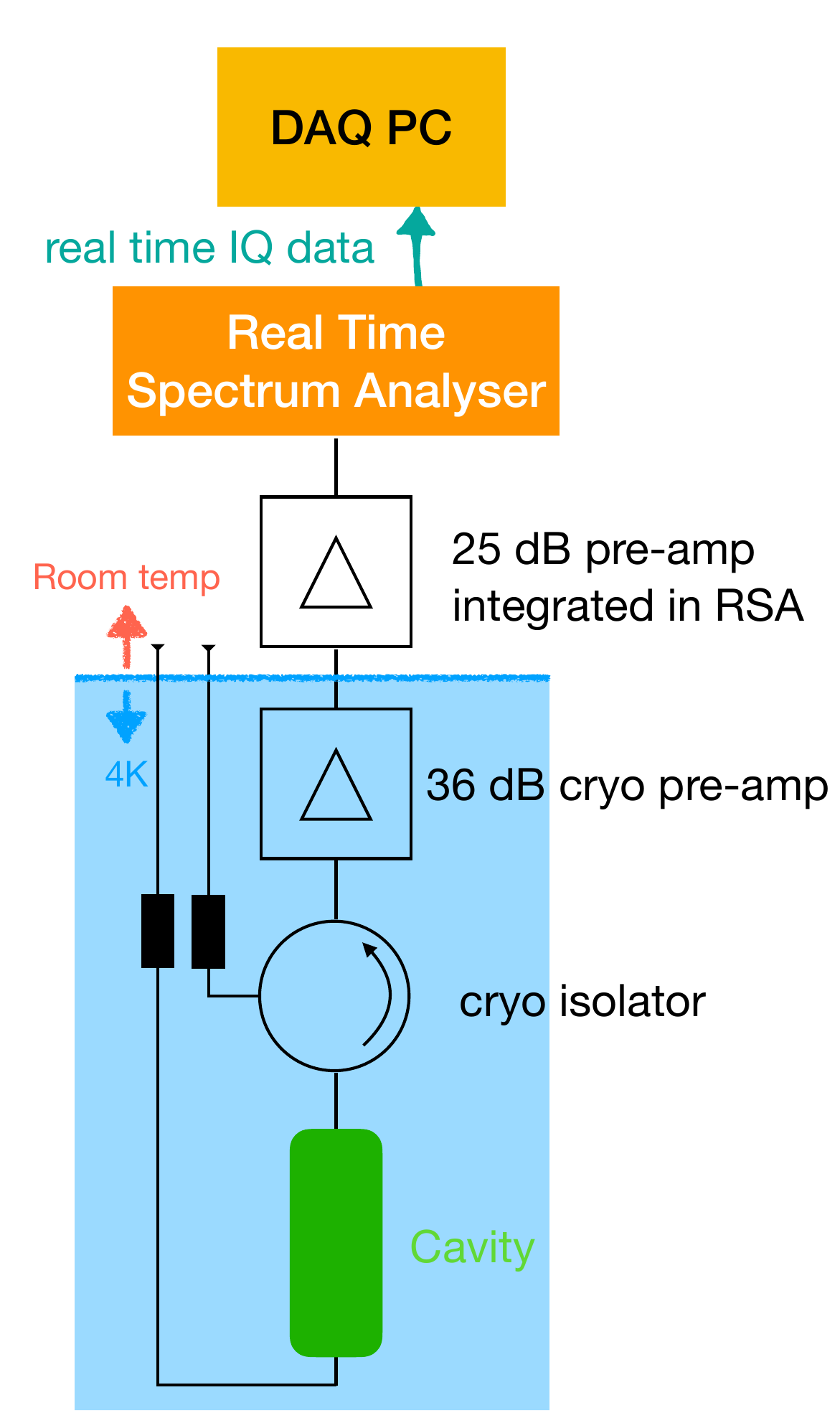}
    \includegraphics[width=0.49\linewidth]{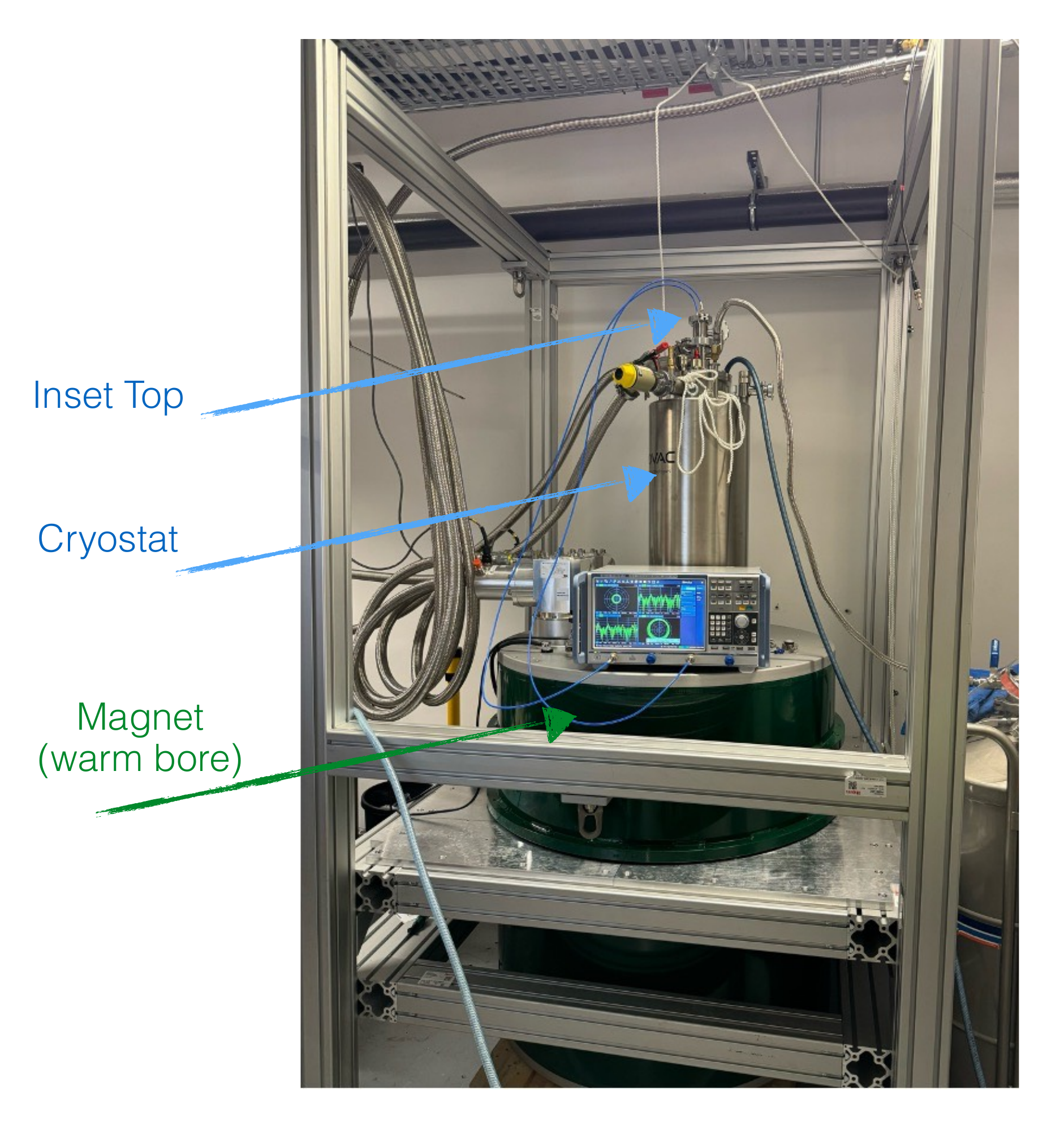}
    \caption{\emph{Left:} Overview of the current data acquisition chain of the SUPAX experiment. \emph{Right:} Picture of the SUPAX setup for the axion data taking in 2024.}
    \label{fig:Supax_setup}
\end{figure}

The superconducting axion search experiment (SUPAX) is a Sikivie haloscope experiment searching for axions around 
$34\,\mu\textrm{eV}$, corresponding to a frequency around 8.3~GHz~\cite{Schneemann:2024yza,Chouhan:2025bam}. 
It utilizes a 14\,T solenoid magnet located at the Helmholz-Institute at Mainz and is currently operating at a temperature of 2~K.
The first data taking run finished in October 2024 scanning over a frequency interval of 1~MHz by varying the He pressure inside the cavity. 
A sketch of the readout electronics along with a picture of the  setup is shown in Fig.~\ref{fig:Supax_setup}.

The next stage of the experiment will reduce the temperature to 10~mK and use parametric amplification to significantly reduce the noise and enhance the sensitivity by one order of magnitude. 

R\&D effort is put into the study of novel superconducting materials with respect to their performance in strong magnetic fields. The first material studied is NbN and the findings are published in~\cite{Schmieden:2024wqp}.

The next data taking period, scheduled early 2025, will be dedicated to the detection of high frequency gravitational waves (HFGWs) in the GHz regime. To this end new cavity geometries as well as novel analysis techniques are currently studied.  
In order to reach the ultimate sensitivity to HFGWs Supax is part of a proposed network of GW detectors, called GravNet~\cite{Schneemann:2023bqc,Schmieden:2023fzn}.

The experiment position and orientation are listed in Tab.~\ref{tab:Supax_loaction}.
\begin{table}[!ht]
\renewcommand{\arraystretch}{1.5}
  \begin{center}
    \begin{tabular}{c|c}
    \hline
    \hline
    Latitude &  $49^{\circ}$ $59^{\prime}$ $30^{\prime\prime}$ \\
      Longitude & $8^{\circ}$ $14^{\prime }$ $09^{\prime\prime}$ \\
      Elevation & 90~m\\ 
      $B$ field direction & solenoid: upright  \\
      \hline\hline
    \end{tabular}
    \caption{Position and magnetic-field direction of the SUPAX experiment.}
    \label{tab:Supax_loaction}
  \end{center}
\end{table}

%% file: WG4/content/WISPLCExperiment.tex
{Author: M. Maroudas}\\

WISP searches with an LC circuit (WISPLC) is a precision direct experiment for light dark matter candidates such as ALPs, that is currently being developed at the University of Hamburg in Germany~\cite{zhang_wisplc_2022}. WISPLC will operate in a broadband and a resonant detection scheme where an LC circuit will be used to enhance the signal.

In the presence of an external magnetic field, an axion-sourced current density ${\vec j_a}$ is derived:
\begin{equation}
\label{eq:axion_current}
   {\vec j_a}\left( t \right) =  - {g_{a\gamma \gamma }}\vec B \frac{\partial a}{\partial t}
\end{equation}
where $g_{a\gamma\gamma}$ is the axion-photon coupling coefficient, $\vec B$ is external magnetic field and $a\left( t \right)$ is the oscillating axion field:
\begin{equation}
\label{eq:axion_field}
a\left( t \right) = {a_0}\cos \left( {{m_a}t} \right) = \frac{\sqrt {2{\rho _{DM}}} }{{m_a}}\cos \left( {{m_a}t} \right)
\end{equation}
with $a_0$ being the field amplitude, $m_a$ the axion mass and ${\rho _{DM}} \approx \SI[per-mode = symbol]{0.3}{\GeV\per\cm\cubed}$ the expected local dark matter density. The induced oscillating axion current from Eq.~\ref{eq:axion_current} generates a perpendicular toroidal magnetic field ${\vec B_a}$ such that $\nabla  \times {\vec B_a} = {\vec j_a}$.

\begin{figure}[t!]
  \begin{center}
    \includegraphics[width=0.49\linewidth]{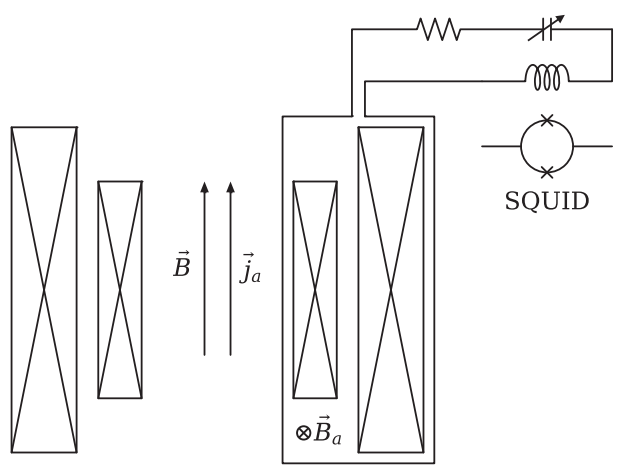}
    \includegraphics[width=0.49\linewidth]{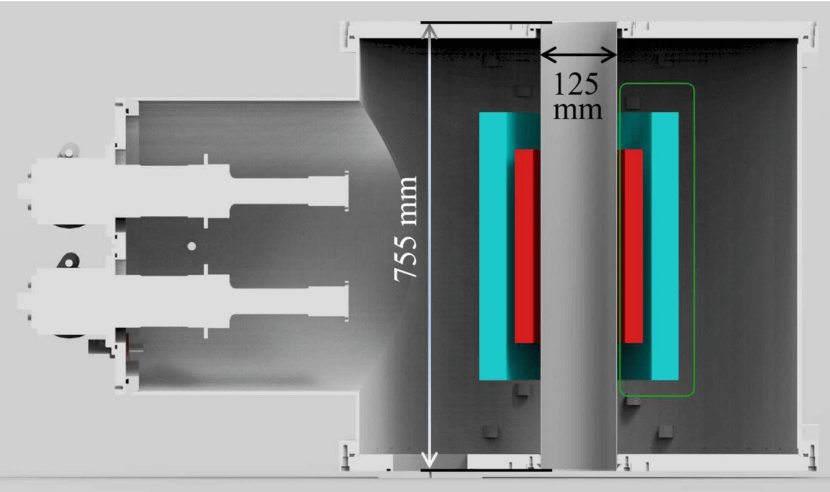}
    \caption{\emph{Left:} Schematic view of the experimental setup, with the four crossed rectangles representing the windings of two solenoid magnets which generate the external magnetic field $\vec B$ which induces the axion-sourced current density $\vec j_a$. \emph{Right:} A model of the large-scale warm-bore magnet producing a maximum magnetic field of \SI{14}{\tesla} in the center. In green, the superconducting pickup loop is shown. Figure reproduced from~\cite{zhang_wisplc_2022}.}
    \label{fig:wisplc_setup}
  \end{center}
\end{figure}

WISPLC is planning to capture the induced magnetic field ${\vec B_a}$ using a superconducting loop as a pickup antenna and an LC resonant circuit as an amplifier (see Fig.~\ref{fig:wisplc_setup}). This is based on Faraday's law where ${\vec B_a}$ induces an AC EMF and subsequently an AC current in the pickup loop. The signal can then be measured with a superconducting quantum interference device (SQUID) magnetometer using an input coil to convert the current into a magnetic field, or directly through a high input impedance low-noise amplifier. 

\begin{figure}[t!]
  \begin{center}
    \includegraphics[width=0.8\linewidth]{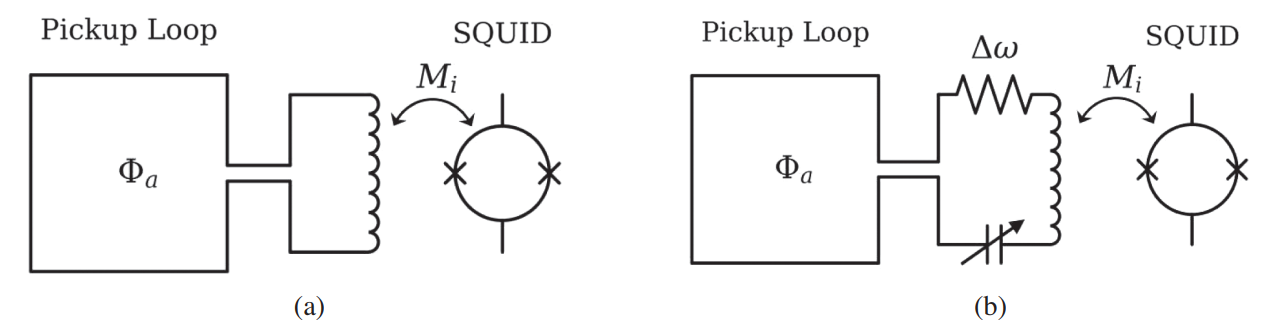}
    \caption{WISPLC (a) broadband and (b) resonant readout schemes for axion detection. Figure reproduced from Ref.~\cite{zhang_wisplc_2022}.}
    \label{fig:wisplc_readout}
  \end{center}
\end{figure}

Using a high external magnetic field of \SI{14}{\tesla} with a volume of ${V_{magnet}} = \SI{0.0024}{\meter\cubed}$ a form factor of $C_V=0.074$ can be achieved. As seen in Fig.~\ref{fig:wisplc_readout}, WISPLC will then employ two readout schemes: A) a broadband readout using direct inductive coupling with the SQUID  where the bandwidth is limited by the detector and readout electronics, and B) a resonant readout where a tunable high-quality ($Q$) LC resonant circuit is inserted between the pickup loop and the SQUID which enhances the superconducting current by $ \sim 10^4$.

The broadband bandwidth is expected to be on the level of \SI{3.5}{\MHz} with the expected sensitivity to the two-photon coupling of axions for 100 days of data taking being about 2 orders of magnitude better than current experimental bounds. On the other hand, for the resonant case, the search is limited to a narrow frequency range around the resonant frequency, thus yielding a bandwidth of $\Delta V \propto 1/Q$. In this case, a scanning between $10^{-11}$ and $2.5 \times 10^{-8}~\si{eV}$ is envisaged with the expected sensitivity being on the order of ${g_{a\gamma \gamma }} \approx {10^{ - 15}}~\si{\GeV\tothe{-1}}$ within 100 days of measurement time. This allows WISPLC to probe dark matter axions motivated by the photophilic axion \cite{Sokolov:2021ydn} and trapped misalignment \cite{DiLuzio:2021gos} models which extend the canonical QCD axion band. At the same time, a sensitivity to high-frequency gravitational waves will be explored based on \cite{domcke_novel_2022}. 

WISPLC began its first broadband data acquisition campaign in October 2023, operating as a prototype detector with a \SI{250}{\cm} pickup loop made from \SI{0.1}{\milli\meter} Cu wire, placed in the bore of a dipole \SI{6}{\tesla} ADR magnet and aligned parallel to the magnetic field. The readout chain employed a high-input-impedance LNA with \SI{67}{\decibel} gain and a noise temperature below \SI{1}{\kelvin} at a \SI{4}{\kelvin} base temperature, replacing a SQUID in this configuration. The analysis of this first run was carried out up to \SI{5}{\MHz}, although the loop remained inductive up to \SI{10}{\MHz}, as numerous EMI/EMC parasitic signals dominated above \SI{5}{\MHz}. Preliminary results show no evidence for a persistent ALP signal in the axion mass range of \SIrange{2}{20.7}{\nano\eV}, excluding axion–photon couplings down to $ g_{a\gamma\gamma} \approx 5.4 \times 10^{-12},\si{\GeV\tothe{-1}} $. A second data-taking campaign is currently underway with significantly improved shielding, expected to suppress many of these parasitic contributions. The position and orientation of WISPLC during the first run are summarized in Tab.~\ref{tab:wisplc_orientation}.

\begin{table}[tbhp]
\renewcommand{\arraystretch}{1.5}
  \begin{center}
    \begin{tabular}{c|c}
    \hline
    \hline
    Latitude &  $53^{\circ}$ $34^{\prime}$ $41.7^{\prime\prime}$ \\
      Longitude & $9^{\circ}$ $53^{\prime }$ $9.1^{\prime\prime}$ \\
      Elevation & 20~m\\ 
      $B$-field direction & Northwest  \\
      \hline\hline
    \end{tabular}
    \caption{Position and magnetic-field direction of the WISPLC experiment.}
    \label{tab:wisplc_orientation}
  \end{center}
\end{table}

%% file: WG4/content/Helioscopes_intro.tex
The study of the Sun has a rich history, beginning with ancient solar observatories like Stonehenge in England, which dates back ~5000 years and served as both a solar observatory and an early ``astronomical computer''. Today, modern astroparticle physics experiments act as specialized astronomical telescopes, observing nearby planets, stars, black holes, and even probing back to the Big Bang $\sim 13.8$ billion years ago. When focusing on the Sun, however, we look back in time by a mere 10 minutes to a week.

The Sun became a pivotal laboratory for subatomic physics with the discovery of solar neutrinos. Solving the solar neutrino deficit problem required introducing new physics, showcasing the Sun’s role in advancing our understanding of fundamental particles. Many solar phenomena remain mysterious, pointing to physics beyond the Standard Model. Among the most intriguing possibilities is the existence of axions, hypothetical particles that are neutral, light, and interact only weakly with matter.

\begin{figure}[t!]
    \centering
    \includegraphics[scale=0.3]{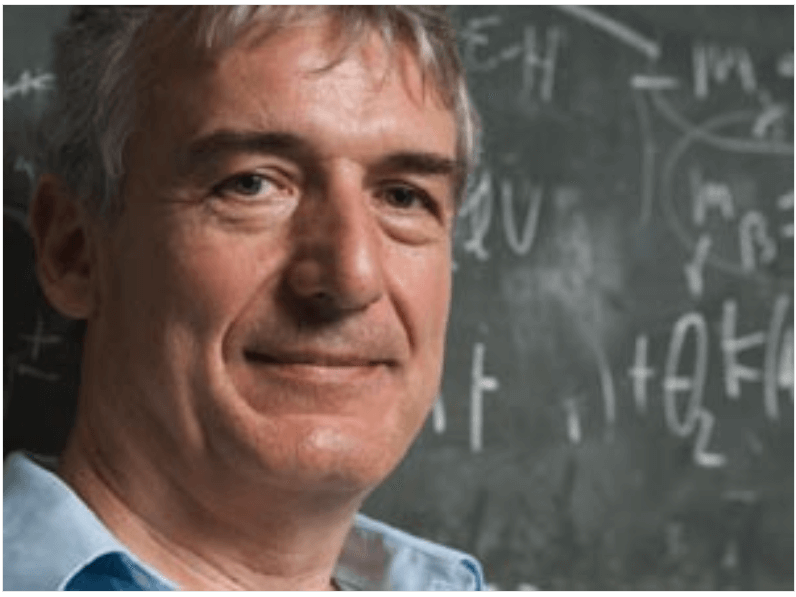}
    \caption{Pierre Sikivie, the inventor of the Sikivie effect.}
    \label{fig:Pierre}
\end{figure}
Pierre Sikivie’s 1983 proposal~\cite{Sikivie:1983ip} to use strong magnetic fields to convert axions into photons via the Primakoff effect remains central to these experiments. In this process, virtual photons from the magnetic field catalyze the axion's ``decay'' into real photons.

The first solar axion experiment, conducted at Brookhaven National Laboratory in 1990, marked the beginning of axion helioscopy. Successive efforts included the SUMICO helioscope in Tokyo, followed by CERN’s CAST experiment, which recycled an LHC test dipole magnet. The next-generation helioscope, IAXO, under development at DESY, incorporates advancements in magnet technology derived from CERN's expertise. IAXO's design leverages ATLAS solenoids, originally proposed by Louis Walckiers, to surpass CAST's performance. The sensitivity of axion detection depends on factors such as the strength and length of the magnetic field, the detector volume, and the axion's coupling strength.

To improve the signal-to-noise ratio, starting with CAST, X-ray telescopes have been employed as concentrators. These telescopes, recycled from space programs, have proven instrumental in enhancing the detection of axion signals in experiments like CAST and IAXO.

Why is the Sun so compelling as a target? Surprisingly, some of the most intriguing puzzles lie in its outermost layers, which have been observed for centuries~\cite{Testa_2015,Zhitnitsky_2017}. Among the most enigmatic phenomena is the solar corona, discovered by Walter Grotrian in the 1930s. The corona, located above ~2000 km from the Sun’s surface, is an extremely tenuous layer with a density of only a few nanograms per cubic meter. Remarkably, its temperature exceeds 1 million Kelvin, in stark contrast to the underlying photosphere at ~6000 K. This ``temperature inversion'' defies the laws of thermodynamics, as heat should not flow spontaneously from cooler to hotter regions.

The Sun’s excess extreme ultraviolet (EUV) radiation (>25 eV photon energy) further highlights this mystery~\cite{Maroudas_2025}. A black body at 6000 K, in thermal equilibrium for ~4.5 billion years, should not emit such radiation. This discrepancy suggests that conventional physics cannot fully explain the Sun's behavior. Notably, the EUV flux exhibits unexpected planetary dependencies, such as variations tied to Mercury’s orbital position. This suggests gravitational lensing effects due to potential dark matter streams~\cite{BERTOLUCCI201713,PhysRevD.108.123043}, adding another layer to the solar puzzle.

Beyond the corona, the Sun hosts other enduring mysteries. Solar flares, which can heat localized regions to 10-30 million Kelvin, remain poorly understood despite their discovery 160 years ago. Sunspots and solar activity follow an enigmatic 11-year cycle, first documented by Samuel Heinrich Schwabe in 1826, but the underlying cause of this rhythm remains elusive. Even the elemental composition of the Sun is not fully resolved.

These phenomena highlight the Sun’s potential as a source of groundbreaking discoveries. As Frank Wilczek (Nobel Laureate, 2004) stated in a CERN seminar: “Focus on anomalies—mysteries. That’s where you have a chance to find something new.”

The Sun’s myriad unexplained phenomena, from coronal heating to solar flares, underscore its importance as a target for continued observation. Even if current instrumentation lacks the sensitivity for specific observables, the potential for unexpected breakthroughs justifies persistent study. Helioscopes searching for axions in the keV energy range may inadvertently uncover novel physics of great significance, echoing the history of unexpected discoveries in science.

By keeping our attention on the Sun, astroparticle physics may unlock the secrets of our nearest star. Axions and other particles may provide the key to understanding not only the Sun’s mysteries but also the universe’s fundamental laws.

%% file: WG4/content/CASTHelioscopeExperiment.tex
\begin{figure}
    \centering
    \includegraphics[width=0.7\linewidth]{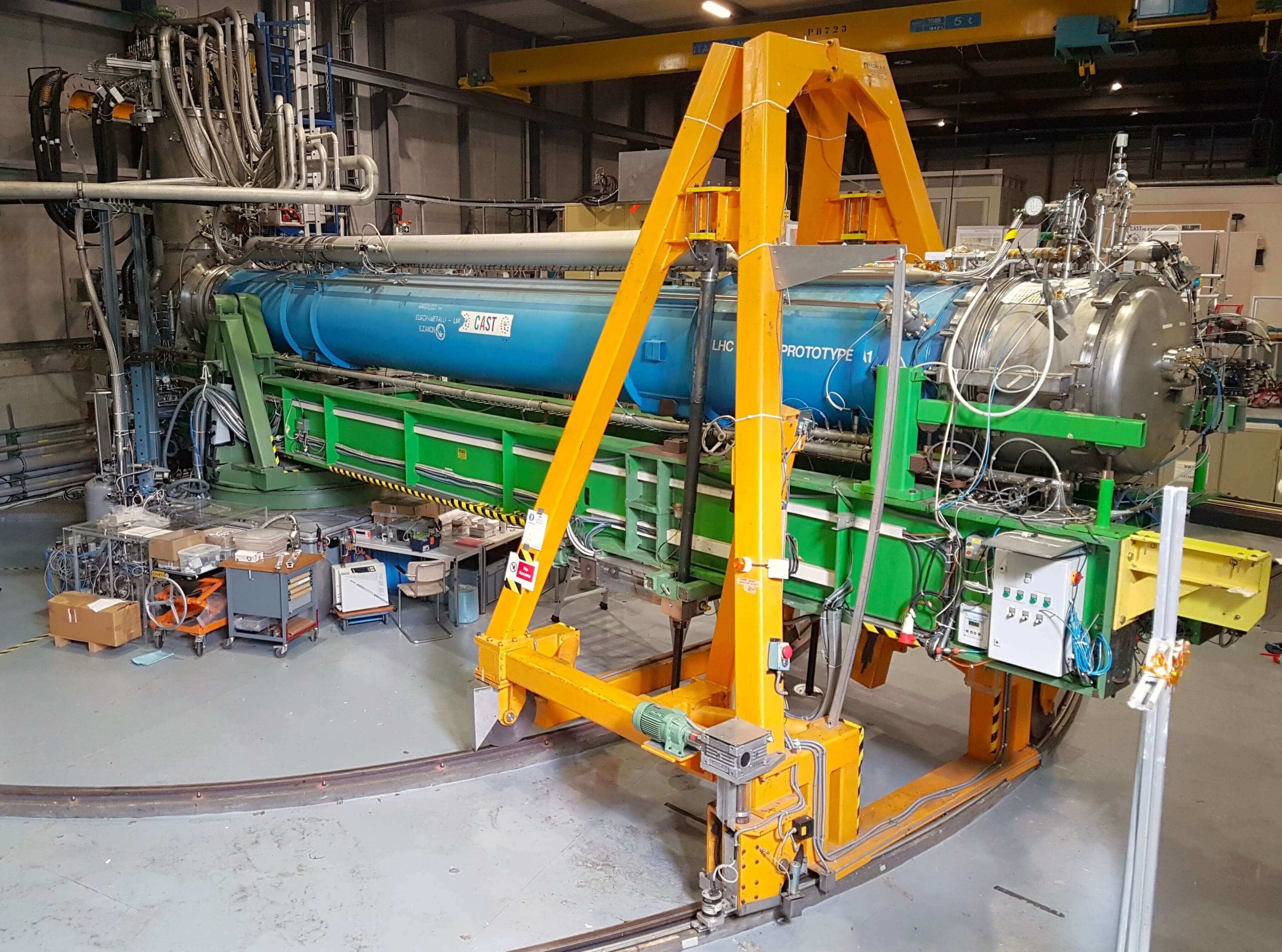}
    \caption{Image of the CAST experiment in 2020, showing the superconducting magnet (blue) and the movable platform (yellow). The seen end of the magnet points towards the Sun during sunrise, and the X-ray optics and detectors are placed on the opposite end. Credits CAST Collaboration.}
    \label{fig:CAST_general}
\end{figure}

The CAST experiment was a helioscope operating from 2003 to 2021, designed to search for axions produced in the Sun. Located at CERN, it used a repurposed LHC prototype magnet \cite{Zioutas:1998cc} with a length of 9.26\,m and a magnetic field strength of up to 9\,T. The magnet was mounted on a movable platform that enabled tracking of the Sun during sunrise and sunset, with a vertical movement range of $\pm8^{\circ}$, allowing approximately 3 hours of data collection per day. A general view of the experiment is shown in Figure \ref{fig:CAST_general}.

The primary goal of the experiment was to detect solar axions by converting them into X-ray photons in the presence of a magnetic field through the inverse Primakoff effect. CAST combined multiple cutting-edge technologies, including X-ray focusing optics borrowed from space telescope technology \cite{Aznar:2015iia}, ultra-low background Micromegas detectors \cite{Giomataris:1995fq,Andriamonje:2010zz}, and a sophisticated tracking system that maintained precise solar alignment.

During its operational period, CAST went through several phases, including initial runs with the magnet bores in vacuum \cite{Zioutas:2004hi,Andriamonje:2007ew}, gas phases with $^4$He \cite{Arik:2008mq,Arik:2015cjv} and $^3$He \cite{Aune:2011rx,Arik:2013nya} to cover higher axion mass ranges, and finally the IAXO pathfinder phase. In its latest run (2019-2021), CAST used for the first time a xenon-based gas mixture (48.85\% Xe + 48.85\% Ne + 2.3\% isobutane) in the Micromegas detector, which provided lower background levels compared to the traditionally used argon mixture by eliminating the problematic argon fluorescence peak around 3\,keV. The final dataset, combining all phases, has set 
the following limits on the axion-photon coupling to date \cite{CAST:2024eil}:
\begin{equation}
g_{a\gamma} < 5.8 \times 10^{-11}\,\mathrm{GeV}^{-1} \quad \text{at 95\% CL} \quad (\text{for}\, m_a \lesssim 0.02\,\mathrm{eV}).
\end{equation}

This result approaches the limits derived from stellar evolution observations \cite{Ayala:2014pea,Straniero:2015nvc}.

CAST has also produced constraints on other axion production channels in the Sun~\cite{CAST:2009jdc,Barth:2013sma,CAST:2009klq}, as well as to chameleons~\cite{CAST:2015npk} and hidden photons~\cite{Redondo:2015iea}. It has also searched for dark matter axions \cite{adair_capp_2022,CAST:2020rlf} (see Sec.~\ref{sec:cast-capp} and Sec.~\ref{sec:cast-rades}). The experiment has served as a pathfinder for the next-generation International Axion Observatory (IAXO) (see Sec.~\ref{sec:IAXO}), proving the viability of the axion helioscope technique and developing many of the key technologies that will be implemented in future searches. With its final data taking completed in 2021, CAST's legacy in solar axion searches will stand until BabyIAXO, currently under construction at DESY, begins operations.

The CAST experiment position and orientation are listed in Tab.~\ref{tab:CAST}.
\begin{table}[tbhp]
\renewcommand{\arraystretch}{1.5}
  \begin{center}
    \begin{tabular}{c|>{\centering\arraybackslash}p{.35\textwidth}}
    \hline
    \hline
    Latitude &  $46^{\circ}$ $14^{\prime}$ $32^{\prime\prime}$ \\
      Longitude & $6^{\circ}$ $05^{\prime }$ $49^{\prime\prime}$ \\
      Elevation & 430~m\\ 
      $B$ field direction & Perpendicular to the LOS of the telescope with the Sun  \\
      \hline\hline
    \end{tabular}
    \caption{Position and magnetic-field direction  of the CAST experiment.}
    \label{tab:CAST}
  \end{center}
\end{table}

%% file: WG4/content/IAXOexperiment.tex
The IAXO experiment is a next generation axion helioscope, designed to investigate the potential emission of axions and axion-like particles (ALPs) from the Sun. The baseline search aims to measure the coupling constant of axions to photons ($g_{\mathrm{a} \gamma}$) via the Primakoff mechanism, and the coupling of axions to electrons ($g_{\mathrm{ae}}$) and nucleons ($g_{\mathrm{aN}}$), which involves non-Primakoff mechanisms \cite{Vogel2023}. Although IAXO's main focus lies on the search for axions and ALPs, the experiment will also have physics potential beyond this and will be able to explore other exotic physics in the realm of low-energy particle physics, such as hidden photons or chameleons, and the magnet design allows the integration of microwave cavities or antennas, enabling the pursuit of relic axions. 

\begin{figure}[t]
    \centering
    \includegraphics[scale=0.5]{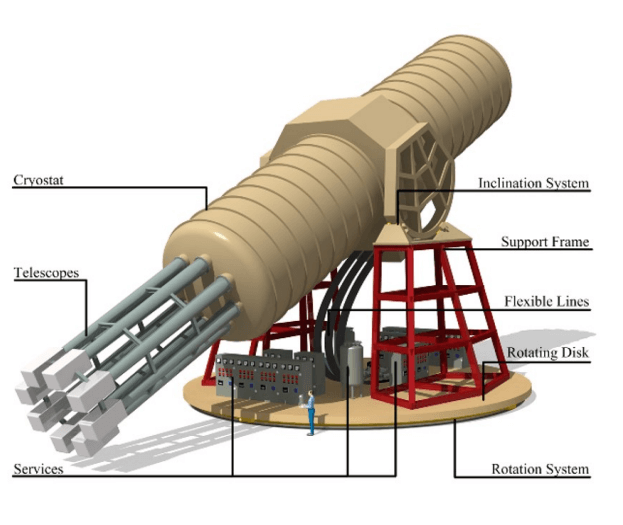}
    \caption{A depiction of the IAXO helioscope. The cryostat housing the magnet can be seen on the right, affixed to the drive system frame. On the left side, it is linked to the X-ray telescopes and the detectors nestled within their respective veto systems and shielding. Figure reproduced from Ref.~\cite{Armengaud:2014gea}.}
    \label{fig:Iaxo}
\end{figure}

In the traditional axion helioscope setup, axions are produced in the core of stars, such as our Sun, and can subsequently be re-converted into photons via the inverse Primakoff effect in the controlled environment of a strong, transverse laboratory  magnetic field. X-ray telescopes then focus the x-ray photons from axion conversion onto a focal plane, in which low-background x-ray detectors are positioned to capture and visualize the events. The conceptual design for the magnet was inspired by the toroidal magnet geometry used by the ATLAS experiment at CERN, providing $2.5$ T inside eight $60$-cm-diameter, $20$-m-long magnet bores (see Fig.~\ref{fig:Iaxo}). The primary method employed in crafting the optics for IAXO involves the use of segmented glass optics technology combined with x-ray coatings as metallic single layers or multilayers. This technology has demonstrated its effectiveness through its successful application in NASA's Nuclear Spectroscopic Telescope Array (NuSTAR, \cite{NuSTAR:2013yza}) satellite mission and others. In IAXO a focal length of about 5 meters will be used to concentrate photons from axion-conversion into small focal areas of a few square millimeters, which will be captured by ultra-low background X-ray detectors. The primary technology employed for these low-background detectors consists of gaseous Micromegas detectors featuring pixelated readout, produced using the microbulk technique \cite{ALTENMULLER2023167913}. Additional technologies are under study to achieve lower energy thresholds as well as improved energy resolution, which will prove crucial for axion-electron studies. 

IAXO aims to improve the limit on the photon-coupling constant by 1-1.5 orders of magnitude the current most sensitive helioscope, the CERN Axion Solar Telescope (CAST \cite{CAST:2017uph}), which corresponds to a sensitivity of $g_{a\gamma} < 10^{-12}\;\mathrm{GeV}^{-1}$ for axion masses $m_a < 0.016 \mathrm{eV}$ \cite{Lakic:2020cin}.

%% file: WG4/content/BabyIaxoExperiment.tex
\begin{figure}[t!]
    \centering
    \includegraphics[width=0.8\linewidth]{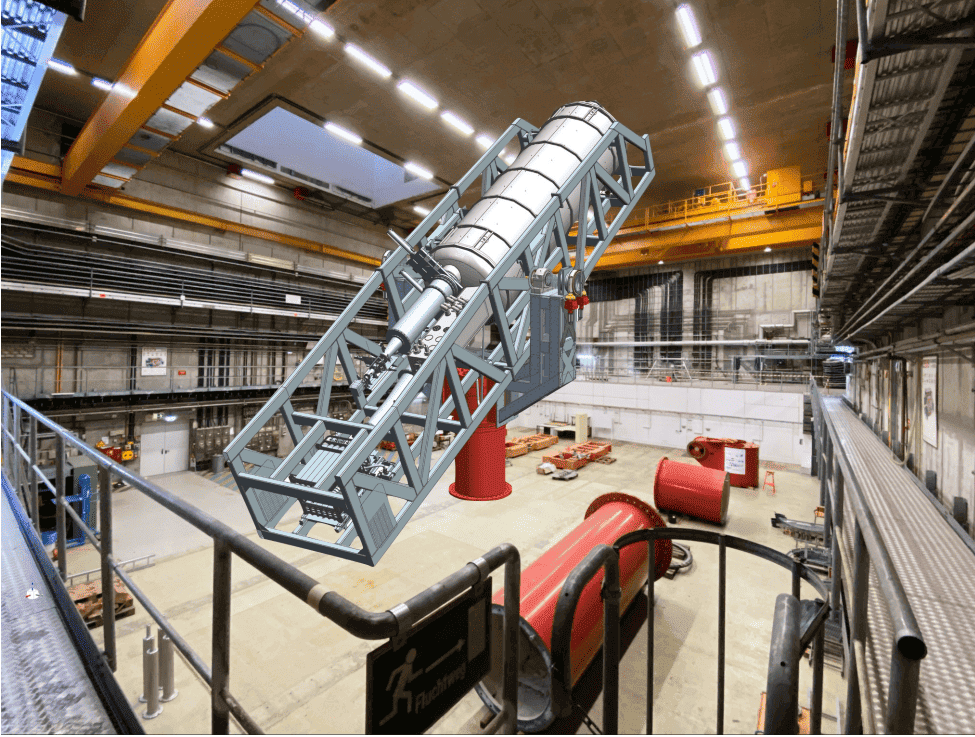}
    \caption{Conceptual design of the BabyIAXO helioscope inside the HERA South Hall at DESY. Credits IAXO Collaboration.}
    \label{fig:BabyIaxo}
\end{figure}

BabyIAXO is envisioned as an intermediate step towards the full-scale IAXO experiment.
On one hand, BabyIAXO (see Fig.~\ref{fig:BabyIaxo}) will serve as a technological prototype for IAXO in order to mitigate risks and enhance performance of the full-scale experiment, while on the other hand BabyIAXO will already function as a fully-fledged axion helioscope with significant science reach able to improve over current-best CAST results thanks to a factor 10 improvement in the figure of merit of the magnet for BabyIAXO over the CAST one. BabyIAXO is currently entering its construction phase and will be sited at the HERA South Hall at DESY in Hamburg, Germany. 

The keypiece of BabyIAXO is a 10-meter-long magnet in a double racetrack-coil design, constructed using superconducting Al-stabilized Nb-Ti/Cu Rutherford cable capable of generating a maximum magnetic field of 4~T, and averaging 2.5~T, yielding the above mentioned improvements in magnet figure of merit, mostly due to the increased cross-sectional area of BabyIAXO over CAST. The common coil configuration encompasses two empty bores, each with a diameter of 70~cm, extending through its entire length. Two different X-rays telescopes will be used for each of the bores, the first one being a flightspare module of ESA's XMM Newton space telescope. The second one is a custom-designed which consists of an inner part ($r = 5-20\;$cm) made with the same technology used for the NASA's NuSTAR~\cite{NuSTAR:2013yza}, and an outer part ($r = 20-35\;$cm) that will be covered using cold-slumped glass and its assembly technology that was demonstrated at INAF \cite{10.1117/12.2232591}. The low-background detectors for BabyIAXO will consist of Micromegas of the same type as those mentioned earlier for IAXO (Sec.~\ref{sec:IAXO}).

BabyIAXO is anticipated to probe coupling constants for $g_{a\gamma}$ down to about $1.5 \times 10^{-11}~\mathrm{GeV}^{-1}$ for axion masses $m_a < 0.02\;$~eV. During its gas phase, BabyIAXO is expected to explore higher mass ranges, notably probing KSVZ axions up to $0.25$~eV \cite{IAXO:2020wwp}.

BabyIAXO is under construction to act as a prototype for IAXO's subsystems, while simultaneously generating valuable scientific results. Finally, a possible improved scenario, referred to as IAXO+, could further enhance the IAXO sensitivity in the future. 

The BabyIAXO experiment position and orientation are listed in Tab.~\ref{tab:babyiaxoorientation}.

\begin{table}[h!]
\renewcommand{\arraystretch}{1.5}
  \begin{center}
    \begin{tabular}{c|>{\centering\arraybackslash}p{.35\textwidth}}
    \hline
    \hline
    Latitude &  $53^{\circ}$ $34^{\prime}$ $29^{\prime\prime}$ \\
      Longitude & $9^{\circ}$ $53^{\prime }$ $44^{\prime\prime}$ \\
      Elevation & 10   ~m\\ 
      $B$ field direction & Perpendicular to the LOS of the telescope with the Sun
      \\
      \hline\hline
    \end{tabular}
     \caption{Position and magnetic-field direction of BabyIAXO experiment.}
    \label{tab:babyiaxoorientation}
  \end{center}
\end{table}
\newpage

%% file: WG4/content/Pure_Lab_intro.tex
Pure laboratory experiments, especially those employing lasers and related optical techniques, play a crucial role in axion searches through the axion--photon coupling. One of the primary strategies is the ``light shining through a wall'' (LSW) approach~\cite{ALPS,OSQAR}. In an LSW experiment, intense laser light is directed through a strong magnetic field to generate axions (via photon to axion conversion), and a light-proof barrier (``wall'') is placed downstream; behind the wall, a second magnet region attempts to reconvert those axions back into photons.

This purely laboratory scheme requires no astrophysical sources, axions are produced and detected in situ, making it model-independent, although the trade-off is a very small expected signal that limits sensitivity to relatively large axion--photon couplings and masses up to the $10^{-4}$--$10^{-3}$~eV range. Over the past two decades, LSW experiments such as ALPS~I at DESY and OSQAR at CERN have established leading limits on ALP couplings in this domain~\cite{ALPS,OSQAR}. Non-European efforts have also contributed; for example, the GammeV experiment at Fermilab used a $5$~m magnet with a removable wall (and even a variable-length cavity) to search for laser-induced axions, finding no signal and thus setting bounds on the two-photon coupling of possible ALPs~\cite{ALPS2}. A similar ``light-through-wall'' search, LIPSS at Jefferson Lab, exploited the lab's infrared free-electron laser and a superconducting magnet to probe the axion interpretation of past anomalous polarization signals (e.g., PVLAS), again yielding null results but expanding the global reach of LSW searches.

Currently, ALPS~II (the upgraded Any Light Particle Search at DESY) is pushing this technique to new sensitivity~\cite{ALPS,kozlowski2024design}. ALPS~II employs two high-finesse $124$~m optical cavities (before and after the wall) within strings of straightened HERA dipole magnets to resonantly enhance conversion and regeneration of axions. This setup is expected to improve the signal-to-noise ratio by orders of magnitude over OSQAR's best results, allowing ALPS~II to probe much smaller coupling constants (down to $g_{a\gamma} \sim 10^{-11}~\text{GeV}^{-1}$) for sub-meV axion masses. The experiment's first science run was completed in 2024, with results forthcoming.

Looking further ahead, an even larger-scale LSW initiative dubbed ``JURA'' (Joint Undertaking for Research on Axions) has been proposed~\cite{Beacham:2019nyx}. JURA would leverage future high-field, long-length magnets (potentially $\sim100$~km of beamline using accelerator tunnel infrastructure) to achieve sensitivity comparable to a large helioscope (i.e., approaching the axion--photon coupling range of the proposed IAXO) in a pure lab setting. Such ambitious projects underline the continuing importance of LSW experiments as a model-independent probe of photon-coupled axions.

\subsubsection*{Laser Polarimetry Searches}
Another important laboratory approach focuses on detecting tiny polarization effects induced by axions, notably through laser polarimetry in strong magnetic fields~\cite{PVLAS,BMV}. These experiments seek minute changes in the polarization state of light --- rotation of the plane of polarization or the development of ellipticity --- when a linearly polarized laser beam propagates through a vacuum in the presence of a strong static magnetic field.

In the Standard Model, quantum electrodynamics predicts an extremely small vacuum birefringence in intense fields (due to virtual $e^+e^-$ pairs), but it has never been observed directly. Axion or ALP presence could enhance these effects (or introduce dichroism, rotating polarization) via the axion--photon coupling in the magnetic field. The prototype in this category is the PVLAS experiment in Italy, which pioneered high-sensitivity optical ellipsometry techniques to search for vacuum magnetic birefringence (VMB)~\cite{PVLAS,Ejlli:2020yhk}. A concise VMB theory summary and the QED prediction for $\Delta n$ are given in Sec.~\ref{VMB@QED}; we avoid repeating it here.

Several other experiments around the world have pursued similar goals: BMV in France~\cite{Cadene2014EPJD}, Q\&A in Taiwan~\cite{Mei:2010aq}, OVAL in Japan~\cite{Fan:2017fnd}, and the ongoing efforts toward VMB@CERN (a proposed upgraded polarimeter under study in the CERN PBC program)~\cite{Zavattini:2021cna}. To date, none of these experiments have detected a definitive VMB signal, but each null result refines the upper bounds on axion--photon coupling. Collectively, these polarimetric searches have probed axion couplings down to $g_{a\gamma} \sim 10^{-7}$--$10^{-8}~\text{GeV}^{-1}$ over a broad mass window ($\sim \mu$eV--meV), complementing the LSW searches by covering higher masses, albeit in a somewhat more model-dependent way. 

Different configurations of high-sensitivity polarimeters could probe ALPs and scalar field dark matter without requiring a magnetic field. A recent study proposed a Fabry--P\'erot cavity with a thick birefringent solid to detect oscillatory variations in thickness and refractive index induced by scalar dark matter~\cite{Ejlli2023}. A modified setup with quarter-wave plates inside the cavity enhances sensitivity to axion-like particles.

%% file: WG4/content/Intro-PureLab5thForce.tex
A fifth force may arise due to ``new physics'' beyond the Standard Model. Spin-dependent fifth forces~\cite{cong_spin-dependent_2024} can be mediated by new particles, such as spin-0 bosons (e.g., axions and axion-like particles) and spin-1 bosons (e.g., light $Z'$ bosons or massless paraphotons). These ultralight particles are also candidates for dark matter and dark energy and may play a role in breaking fundamental symmetries.
The fundamental concept of searching for spin-dependent exotic interactions can be understood by considering a fermion at one vertex as the source of a force-carrying intermediate boson and a fermion at the other vertex as the sensor, which is affected by the interaction. The response of the sensor serves as the signal for detection. In such source--sensor experiments, the exotic interaction manifests as an ``effective magnetic field'' acting on the spins of the sensor; see the QUAX$_{g_pg_s}$ experiment~\cite{crescini_search_2022} and many other experiments in Ref.~\cite{cong_spin-dependent_2024}. 
Typical sources used in experiments include GSO~\cite{CRESCINI2017109,CRESCINI2017677}, SiO$_2$~\cite{rong_constraints_2018}, and $^{87}$Rb~\cite{wang_limits_2022}, among others. Sensors include comagnetometers~\cite{wei_constraints_2022}, NV centers~\cite{jiao_experimental_2021}, and torsion pendulums~\cite{terrano_short-range_2015}. Additionally, EDM measurements (Sec.~\ref{Sec:EDM}), atomic spectroscopy experiments (Sec.~\ref{Sec:EAS}), and so on serve as complementary methods to source--sensor experiments.
A theoretical reassessment of exotic spin-dependent forces has led to the development of a systematic and complete set of interaction potentials, and a comprehensive analysis of the existing experimental literature on spin-dependent fifth forces can be found in this review~\cite{cong_spin-dependent_2024}.

%% file: WG4/content/ExoticAtomExperiments.tex
Spectroscopic experiments rank among the most precise measurements in physics, with routine measurements reaching relative inaccuracies as low as $10^{-8}$ and atomic clocks achieving $10^{-19}$~\cite{campbell_single-ion_2012}. This extraordinary precision makes spectroscopy a powerful tool for testing fundamental physics, particularly for constraining exotic interactions---provided that the SM predictions for the observed atomic transitions reach comparable accuracy~\cite{karshenboim_precision_2010}.
Exotic atoms, composed of particles beyond protons, neutrons, and electrons, serve as ideal testbeds for fundamental physics. These systems often lack the complexity of nuclear structure, enabling direct comparisons between high-precision theoretical predictions and experimental results. 

The simplest examples of exotic atoms include positronium, a bound state of a positron and an electron, and muonium, a bound state of an antimuon and an electron. Their relative simplicity makes them well-suited for precision QED tests and for constraining new physics. Positronium spectroscopy has been used to probe interactions between its constituents~\cite{frugiuele_current_2019,fadeev_pseudovector_2022}, while muonium spectroscopy has placed constraints on new spin-dependent forces~\cite{fadeev_pseudovector_2022}.

A more complex exotic atom, antiprotonic helium, consists of an electron and an antiproton bound to a helium nucleus~\cite{yamazaki_antiprotonic_2002}. Such atoms are produced by slowing down antiprotons (using the Antiproton Decelerator at CERN) in a helium environment, where they replace one of the atomic electrons. A fraction of these atoms form metastable states with lifetimes on the order of microseconds, allowing precise measurements of antiproton properties, such as its magnetic moment~\cite{pask_antiproton_2009,nagahama_sixfold_2017,smorra_parts-per-billion_2017} and mass~\cite{hori_buffer-gas_2016}. As such, antiprotonic helium plays an important role in testing CPT invariance~\cite{hayano_antiprotonic_2007}. 
Antiprotonic helium spectroscopy has also yielded the first constraints on semileptonic spin-dependent interactions between matter and antimatter. By comparing measured transition energies of the $(n,l)=(37,35)$ state (the experiment was carried out at CERN’s Antiproton Decelerator)~\cite{pask_antiproton_2009} with QED-based predictions~\cite{korobov_fine_2001}, Ref.~\cite{ficek_constraints_2018} placed limits on several exotic interactions, including higher-order velocity-dependent interactions.

Antihydrogen is a unique system for probing new forces between antimatter and testing CPT symmetry. Composed of a positron and an antiproton, it has been studied at high precision at CERN~\cite{charlton_antihydrogen_2020,khabarova_antihydrogen_2023}. The ALPHA experiment has tested CPT invariance~\cite{amole_experimental_2014,ahmadi_observation_2017} and recently investigated the weak equivalence principle (WEP) with antimatter~\cite{anderson_observation_2023}. A recent research~\cite{cong_searching_2025} provides the first constraints on semileptonic spin-dependent interactions involving antimatter, derived from theoretical and experimental analyses of positron-antiproton interactions in antihydrogen~\cite{charlton_antihydrogen_2020}. This study introduces a comparison of exotic interactions in hydrogen and antihydrogen, complementing previous CPT violation tests.
It is also part of the direct matter--antimatter comparisons~\cite{moskal_testing_2021}, which require minimal theoretical input. Given that the current theoretical accuracy of hyperfine structure calculations significantly exceeds experimental precision, upcoming antihydrogen spectroscopy advancements, such as ASACUSA-CUSP~\cite{malbrunot_asacusa_2018,kraxberger_upgrade_2023}, could improve these constraints by orders of magnitude.

As spectroscopic techniques advance, exotic atoms, including forthcoming ones, will remain crucial for the search for new interactions.

%% file: WG4/content/ALPSIIExperiment.tex
\label{sec:ALPSII}

{Author: A. Lindner}\\

The Any Light Particle Search (ALPS) II is a light-shining-through-a-wall experiment (LSW) located in a section of the former HERA accelerator tunnel at DESY in Hamburg. It aims to improve the sensitivity on the axion--photon coupling by a factor of $1000$ compared to its predecessors. This jump in sensitivity will be achieved by two long strings of superconducting dipole magnets and two mode-matched optical cavities before and after the light-tight wall \cite{Ortiz:2020tgs}. ALPS II will be the first purely laboratory experiment to search for axions or axion-like particles having coupling to two photons below the limits set by astrophysics. Its result will not depend on assumptions about the dark matter composition, for example.

\begin{figure}[t!]
  \begin{center}
    \includegraphics[width=0.95\linewidth]{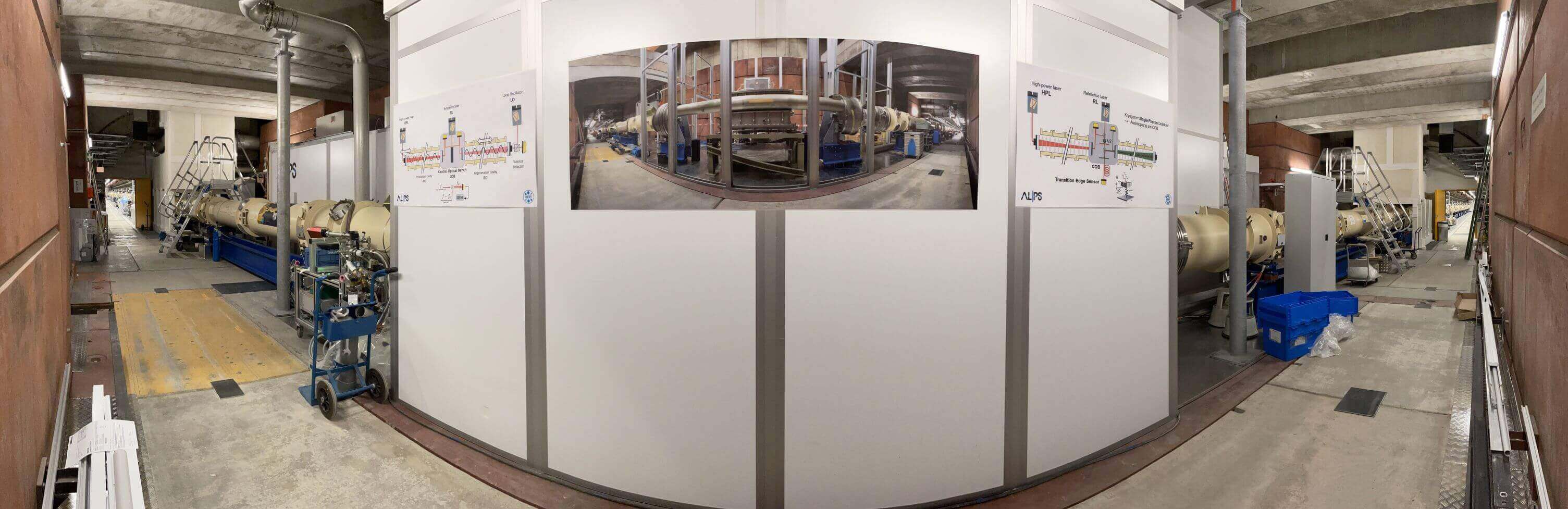}
    \caption{The ALPS II experiment in a straight section of the HERA tunnel. The left part shows the magnet string of the production region, the central clean room is visible in the center, followed by the magnet string of the regeneration section on the right side. The total length is about $250$~m. Reproduced with  permission from DESY.}
    \label{fig:alpsii}
  \end{center}
\end{figure}

The installation of ALPS II began in 2019. In March 2022, the magnet string was successfully tested, and in late 2022, the optics installation was completed for the initial science run, which took place from May 2023 to May 2024. 
The whole experiment is displayed in Fig.~\ref{fig:alpsii}. 

The international ALPS II collaboration consists of DESY in Hamburg (Germany), Cardiff University (UK), the Max Planck Institute for Gravitational Physics (Albert Einstein Institute) in Hannover (Germany), the Institut f{\"u}r Gravitationsphysik der Leibniz Universit\"at in Hannover (Germany), the University of Florida in Gainesville (USA), the University of Hamburg (Germany, up to 2024), and the University of Southern Denmark in Odense (Denmark).

The initial science run, based on a heterodyne sensing method~\cite{Hallal:2020ibe}, only included the regeneration cavity behind the wall~\cite{Kozlowski:2024jzm} to simplify the operation of ALPS II and optimize for the study of stray light. This first science campaign was followed by an extensive test phase on systematics and cross-checks. 
The first science results~\cite{ALPSII:2025eri} did not show any evidence for the existence of axion-like particles. For pseudoscalar bosons, with masses below about 0.1 meV, a limit for the di-photon coupling strength of $1.5\cdot 10^{-9}$\,1/GeV at a 95\% confidence level was achieved. This is more than a factor of 20 improvement compared to all previous similar experiments. 

At present, ALPS II is being upgraded for another science run starting in 2027, with the aim of reaching its design sensitivity in 2028. In parallel, the collaboration is working on an extremely low-background photon counting system based on a transition-edge sensor (TES) detector~\cite{Gimeno:2023nfr,RubieraGimeno:2025bhr}. The future ALPS II program might include the following: axion searches with the TES detector system; vacuum magnetic birefringence measurements; axion searches with optimized optics and/or extension of the axion mass reach; and a dedicated search for high-frequency gravitational waves.

%% file: WG4/content/OSQARExperiment.tex
{Author: S. Kunc}\\

\begin{figure}[htb!]
\centering 
\includegraphics[height=5cm]{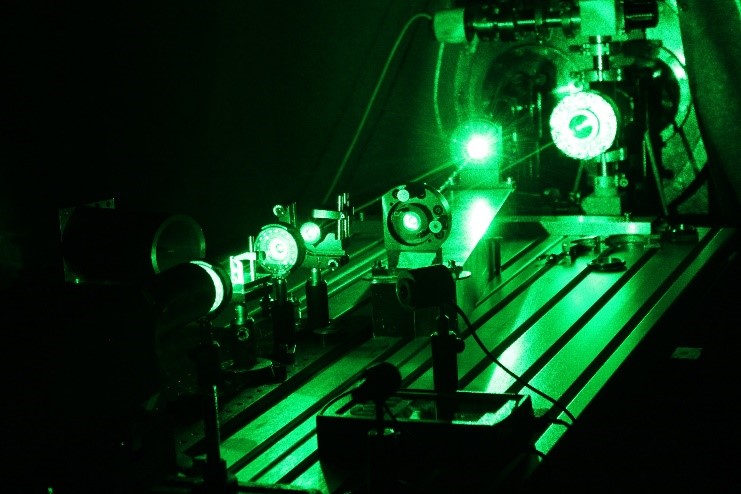}
\includegraphics[height=5cm]{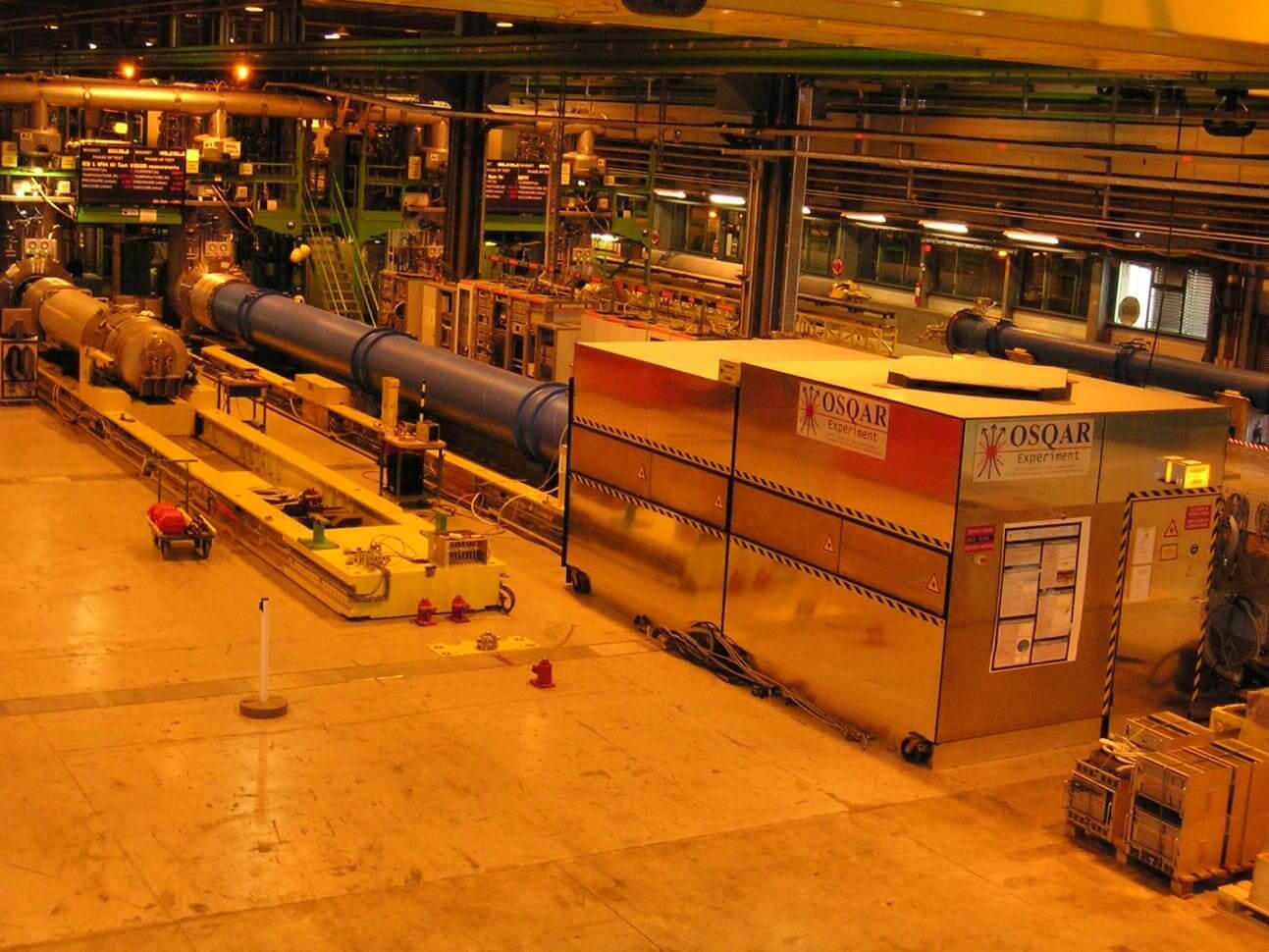}

\caption{\textit{Left:} $18$~W continuous laser for axion/chameleon production. \textit{Right:} General view of the OSQAR experiment protective area in the SM18 hall with the first LHC magnet for axion/chameleon production.}
\label{OSQAR experiment}
\end{figure}

The Optical Search for QED vacuum magnetic birefringence, Axions and photon Regeneration (OSQAR) experiment at CERN was conducted initially with the aim of measuring the vacuum magnetic birefringence (VMB), which beyond the contribution inherent to electron--positron virtual pairs, as yet awaiting experimental confirmation, could reveal an additional contribution associated with axions, and searching for axions and axion-like particles (ALPs) through the photon regeneration method of the ``light shining through a wall'' (LSW). OSQAR-LSW was realized in the magnet-testing facility (SM18) at CERN, situated on the border between France and Switzerland. The experimental setup employed two spare superconducting dipole magnets of the Large Hadron Collider, each providing a uniform transverse magnetic field of $9$~T over a magnetic length of $14.3$~m. Photons are injected in the first magnet extremity using a powerful CW laser source (Figure~\ref{OSQAR experiment}, left), and those not converted into axions/ALPs are intercepted by an absorber located between both LHC dipoles. All photons that can be detected at the other end of the second LHC dipole with a sensitive CCD are the result of axion/ALP back-conversion into photons in the magnetic field. The overall size of the experiment was $53$~m. 

The primary objective of the OSQAR-LSW experiment was to set a new limit on axion--photon coupling from a ``pure'' experiment, i.e., independently of any astrophysics or cosmology assumptions. The LSW data collection period ran from 2007 to 2014, with results reported in several publications~\cite{pugnat_results_2008,pugnat_search_2014,OSQAR:2015qdv}. OSQAR next extended its investigations to the search for chameleons (CHASE). Building upon the pioneering work of the GammeV experiment at Fermilab, OSQAR-CHASE sought to identify a magneto-phosphorescent or afterglow signal within a vacuum permeated by a transverse magnetic field. The CHASE data collection ran from 2015 to 2018, with preliminary results published in 2019~\cite{sulc_osqar_2019}. The principal OSQAR‑CHASE analysis is undergoing final review; results will be submitted in a forthcoming publication.
With regard to VMB, preliminary work and initial trials of the Cotton--Mouton effect measurement in rarefied gases were conducted between 2012 and 2016 and published in 2018~\cite{kunc_study_2018}. 

Subsequently, the VMB program was incorporated into the new VMB@CERN initiative discussed in Sec.~\ref{sec:vmbcern}.

%% file: WG4/content/PVLAS.tex
{Author: A. Ejlli}\\
\label{VMB@QED}

The PVLAS experiment (\textbf{P}olarizzazione del \textbf{V}uoto con \textbf{LAS}er) aimed to measure the extremely small birefringence of the vacuum induced by strong transverse magnetic fields~\cite{Ejlli:2020yhk}. Conducted in Italy, the experiment’s early phases were carried out at the INFN Legnaro National Laboratory (LNL) near Padua, while the final phase, PVLAS-FE, took place at the INFN laboratory in Ferrara. This effect, predicted by quantum electrodynamics (QED), represents a macroscopic manifestation of photon--photon interactions and vacuum polarization, as first described in the seminal works of Euler, Kockel, and Heisenberg in the 1930s~\cite{Euler:1935,EulerKockel1935,Heisenberg1936}. The VMB theory basis (Euler--Heisenberg; QED $\Delta n \simeq 4\times 10^{-24}\,\mathrm{T}^{-2}$) is summarized here to set notation; other sections cross-reference this summary to avoid repetition.

They derived an effective Lagrangian density with nonlinear electromagnetic interactions in the vacuum by accounting for virtual electron--positron fluctuations, given by

\begin{equation}
    \mathcal{L}_{\mathrm{EKH}} = \frac{1}{2\mu_0} \left( \frac{E^2}{c^2} - B^2 \right) 
    + \frac{\alpha^2}{90\pi} \frac{\hbar^3}{m_e^4 c^5} \left[ \left( \frac{E^2}{c^2} - B^2 \right)^2 
    + 7 \left( \frac{\vec{E}}{c} \cdot \vec{B} \right)^2 \right],
\end{equation}

where $\alpha$ is the fine structure constant, $m_e$ is the electron mass, and $c$ is the speed of light. The first term corresponds to the classical Maxwell Lagrangian, while the second term is the quantum correction.

These nonlinearities predict a difference in the refractive indices for light polarised parallel and perpendicular to an external magnetic field, resulting in birefringence:

\begin{equation}
    \Delta n = n_\parallel - n_\perp = 3A_e B^2, \quad \text{with} \quad A_e = \frac{2}{45 \mu_0} \frac{\hbar^3}{m_e^4 c^5} \alpha^2.
\end{equation}

The predicted birefringence value for the laboratory magnetic field is a relatively small value ($ \Delta n \sim 10^{-23} $ at $ B = 2.5~\text{T} $), which remains a target of experimental verification. The PVLAS experiment employed phase-sensitive polarimetric optical techniques to observe this small effect.

The PVLAS experiment ran for over 25 years, and it made significant advancements in improving the sensitivity of polarimetry based on a Fabry--P\'erot cavity for measuring vacuum magnetic birefringence. To generate vacuum polarisation, initially, a rotating superconducting magnet was employed to generate strong magnetic fields, but its operational complexity and cryogenic requirements prompted a transition to two rotating permanent magnets. The rotating permanent magnets, each producing a field of $ B = 2.5~\mathrm{T} $, offered a simpler, more reliable setup that modulated the ellipticity signal at twice their rotation frequency~\cite{PVLAS}.

The experiment utilized a high-finesse Fabry--P\'erot cavity ($ F = 7 \times 10^5 $) to amplify the ellipticity signal, which is proportional to the birefringence. By passing a polarized laser beam through the Fabry--P\'erot cavity and using the heterodyne polarimetry technique, it was possible to effectively minimize the intrinsic detection noise.

\begin{figure}[t!]
\begin{center}
\includegraphics[width=10cm]{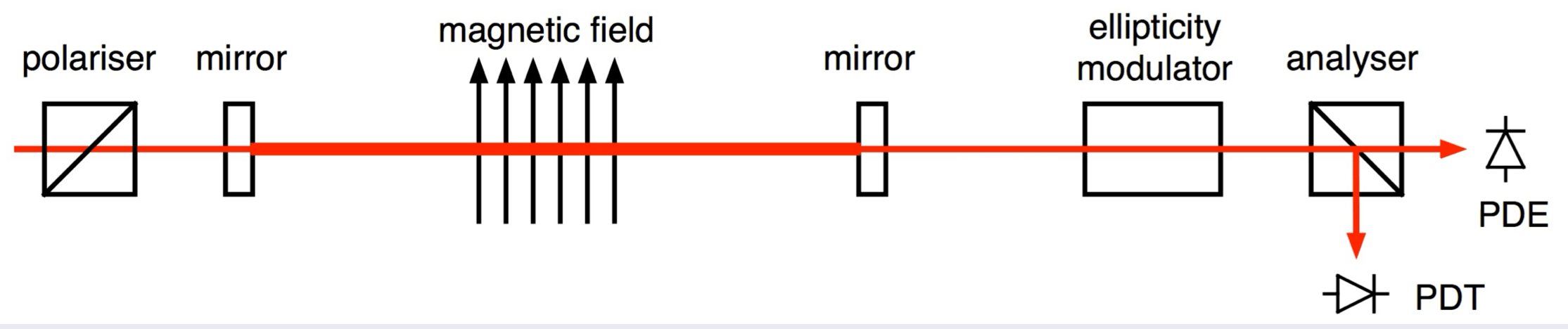}
\includegraphics[width=10cm]{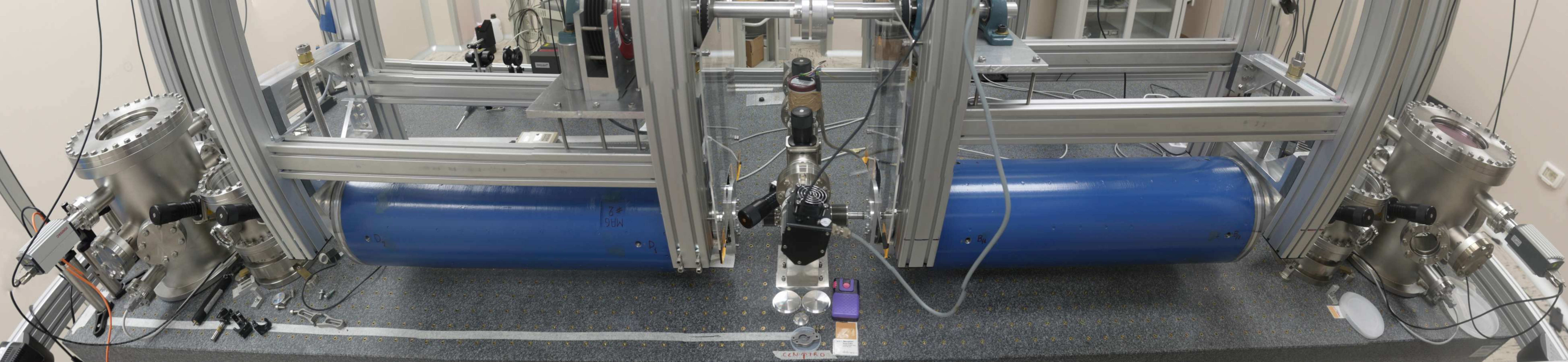}
\end{center}
\caption{Simplified schematic and photograph of the PVLAS setup. The diagram shows the main components: the polariser, mirrors forming the Fabry--P\'erot cavity, the region with the magnetic field (created by rotating magnets), the ellipticity modulator, and the analyzer. The red line shows the path of the laser beam. The transmitted (PDE) and extinction (PDT) signals are detected using low-noise photodiodes. The bottom panel shows a photograph of the main experimental apparatus, including the blue cylinders housing the rotating magnets. Figures reproduced from Ref.~\cite{PVLAS} and Ref.~\cite{DellaValle:2015xxa}.}
\label{fig:setup_PVLAS}
\end{figure}

Figure~\ref{fig:setup_PVLAS} shows the optical layout. The cavity mirrors (M$_1$, M$_2$) defined the optical resonator, while the polarizer (P) and analyzer (A) controlled the input and output polarization states. The rotating magnets induced periodic birefringence, modulating the ellipticity of the transmitted light. This ellipticity modulation was added to the photoelastic modulator and then detected by a photodiode (PDE) placed in the analyzer's extinction port.

The PVLAS experiment made significant progress in increasing the sensitivity. The current experimental limit, $ \Delta n_\text{PVLAS} = (12 \pm 17) \times 10^{-23} $, remains less than an order of magnitude from the QED-predicted value. As a byproduct of these measurements, PVLAS was able to set model-independent limits on the existence of axion-like particles and millicharged particles~\cite{PVLAS}. 

Birefringence noise in the coatings of the cavity mirrors dominated the bandwidth, arising from thermal anisotropy that introduces phase fluctuations between the polarization components of the transmitted light. The frequency dependence of this noise is typically characterized by a power-law spectrum, $ f^{-\alpha} $, where $ \alpha $ is between $0.25$ and $1$, depending on the frequency band. At frequencies between $1$ and $100$~Hz, $ 1/f^\alpha$ noise is primarily attributed to the intrinsic thermal noise of the coating material.

A successful vacuum magnetic birefringence measurement would require either increasing the modulation frequency to move into quieter spectral regions or achieving lower thermal birefringence noise levels through advanced mirror coatings.

%% file: WG4/content/WISPFIExperiment.tex
{Author: M. Maroudas}\\

\begin{figure}[htb!]
  \begin{center}
    \includegraphics[width=0.9\linewidth]{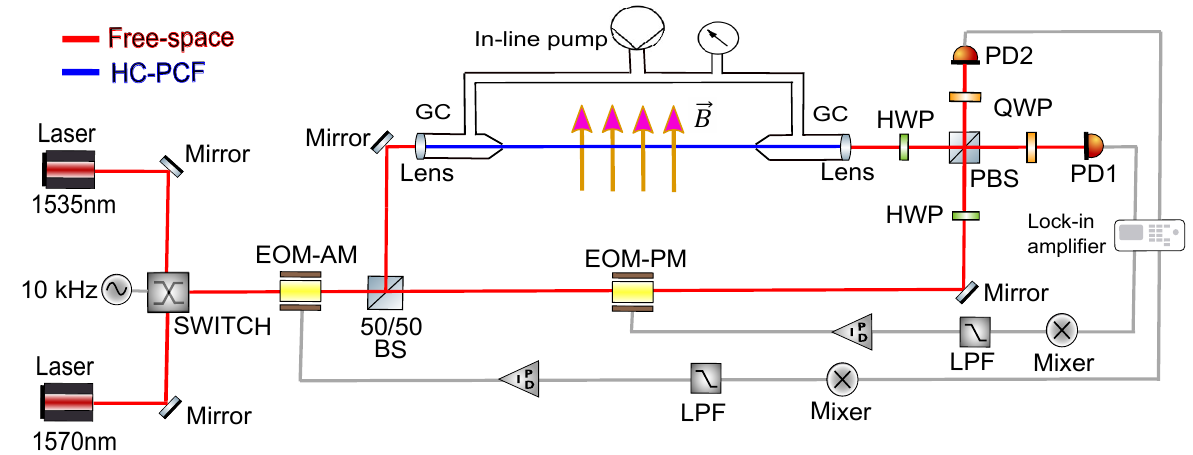}
    \caption{Schematic view of the experimental setup of WISPFI considering a partial-free space MZI for detecting photon--axion oscillations. Figure reproduced from Ref.~\cite{batllori_wispfi_2023}.}
    \label{fig:wispfi_setup}
  \end{center}
\end{figure}

WISP Searches on a Fiber Interferometer (WISPFI)~\cite{batllori_wispfi_2023} is a novel experimental setup currently being built at the University of Hamburg in Germany. WISPFI focuses on photon--axion conversion in a waveguide by measuring photon reduction in the presence of a strong external magnetic field~\cite{tam_production_2012}. The experimental setup, shown in Fig.~\ref{fig:wispfi_setup}, is based on a partial free-space partial fiber Mach--Zehnder interferometer where a laser beam is split into two arms with one arm used as a reference and a sensitive fiber arm embedded in an external magnetic field looking for a conversion of photons to axions. 

The expected signal involves an amplitude reduction in the presence of a non-vanishing photon--axion coupling $g_{a\gamma\gamma}$ when the interferometer is locked in phase/amplitude. The measurable effect of axion--photon mixing relies on the Primakoff effect, while the resulting conversion probability~\cite{Raffelt:1987im} scales with 
\begin{equation}
\label{eq:probability}
    P_{ \gamma \rightarrow a}\propto g_{a\gamma\gamma}^{2}(BL)^{2}\ll 1
\end{equation}
where $g_{a\gamma\gamma}$ is the axion--photon coupling coefficient, and $BL$ is the product of the transverse magnetic field $B$ and the length $L$ through which the photon beam passes in the external magnetic field.

Resonant conversion can take place in the sensitive arm inside the core of a hollow-core photonic crystal fiber (HC-PCF)~\cite{cregan_single_1999,russel_photonic_2003}. This is due to the bandgap structure of the HC-PCF, which allows the propagating mode to acquire a refractive index $n_\mathrm{eff}$ below 1. This leads to real axion masses $m_a$ and to resonant mixing based on the following equation: 
\begin{equation}
\label{eq:axion_mass}
    m_{a}=\omega\sqrt{1-n_\mathrm{eff}^{2}}.
\end{equation}

\begin{figure}[t!]
  \begin{center}
    \includegraphics[width=0.6\linewidth]{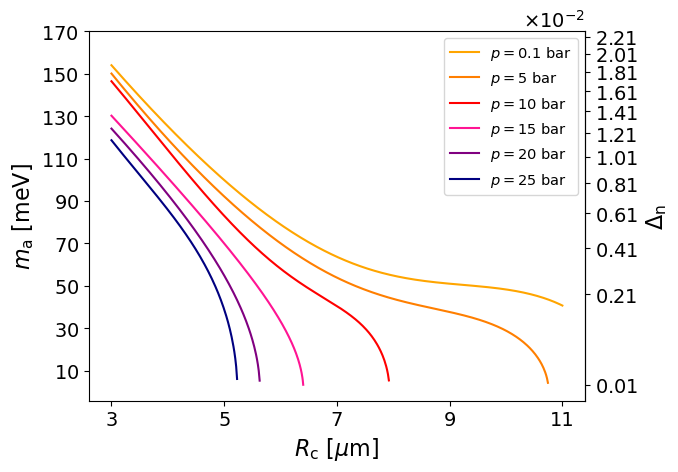}
    \caption{FEM simulation of the axion mass at resonance and the difference of the effective mode index ($\Delta n=1-n_\mathrm{eff}$) as a function of the core radius of the HC-PCF for different pressures of air filling the hollow core. Figure reproduced from Ref.~\cite{batllori_wispfi_2023}.}
    \label{fig:wispfi_core_radius_pressure}
  \end{center}
\end{figure}

As can be seen in Fig.~\ref{fig:wispfi_core_radius_pressure}, by varying the gas pressure inside the HC-PCF~\cite{benabid_stimulated_2002}, the refractive index of the guided mode and thus the probed axion mass can be tuned between $\sim \SIrange[range-phrase = -]{10}{150}{\meV}$. Additionally, a Fabry-Pérot cavity (FPC) along the sensing arm can further enhance the effective conversion length and optical power. This leads to an expected sensitivity which can reach the expected two-photon coupling for the QCD axion:
\begin{equation}
\begin{aligned}
    g_{a\gamma\gamma} \approx  2.55\times10^{-12}\si{\GeV\tothe{-1}} &\left(\frac{\mathrm{SNR}}{3}\right)^{1/2}
    \left(\frac{B}{\SI{9}{\tesla}}\right)^{-1} \left(\frac{L}{\SI{100}{\meter}}\right)^{-1} \left(\frac{F}{100}\right)^{-1/2} \\
    &\left(\frac{P_\mathrm{tot}}{\SI{40}{\watt}}\right)^{-1/2}
    \left(\frac{\beta_\mathrm{sig}}{1}\right)^{-1}\left(\frac{\beta_\mathrm{m}}{1}\right)^{-1/2} \left(\frac{t}{\SI{1}{\day}}\right)^{-1/4} \\
    &\left(\frac{\mathrm{NEP_{PD+SN}}}{\SI{3}{\femto\watt\per\sqrt{\Hz}}}\right)^{1/4}\left(\frac{\Delta \nu}{\SI{10}{\Hz}}\right)^{1/8},
\end{aligned}
\end{equation}
where $\mathrm{SNR}$ is the signal-to-noise ratio, $P_\mathrm{tot}$ is the total power of the laser beam, $\beta_\mathrm{sig}$ is the modulation amplitude of the axion signal, $\beta_\mathrm{m}$ is the modulation amplitude of the carrier, $F$ is the finesse of the FPC, $\Delta \nu$ is the measured bandwidth, $t$ is the measurement time, and $\mathrm{NEP_{PD+SN}}$ is the combined noise equivalent power of the photon detector and the shot noise.

To modulate the axion signal at a desired frequency, WISPFI uses two lasers at slightly different wavelengths of \SI{1535}{\nm} and \SI{1570}{\nm} and a \SI{100}{\kHz} switching between the two using a high-speed fiber optical switch. This way, a modulation amplitude $\beta_\mathrm{sig}$ close to $100\%$ can be achieved.

A prototype is currently being commissioned at the University of Hamburg, operating at standard atmospheric pressure. It employs a \SI{2}{\watt}, \SI{1550}{\nano\meter} laser coupled into an HC-PCF with an \SI{8.5}{\micro\meter} core radius, embedded in a permanent \SI{1}{\meter}-long, \SI{2}{\tesla} magnet array. Under these conditions, the setup probes a fixed axion mass of $m_a \simeq \SI{49}{\milli\eV}$ with a projected sensitivity of $g_{a\gamma\gamma} \gtrsim 1.3 \times 10^{-9}~\si{\GeV^{-1}}$ over 30 days of measurement, making it the first table-top, dark-matter-independent experiment capable of exploring ALPs in a previously inaccessible mass range.

The prototype experiment serves as a test-bench for the full-scale WISPFI setup, demonstrating the scalability of the proposed approach. At the same time, several upgrade paths exist to further enhance both sensitivity and accessible mass range. Notably, WISPFI operates independently of the local dark matter density and requires no cryogenic infrastructure.

%% file: WG4/content/EDMexperiments.tex
{Author: L. Cong}\\

The experiment to search for a permanent electric dipole moment (EDM) of the neutron was conducted at the Paul Scherrer Institute (PSI) in Switzerland~\cite{Abel:2020pzs}. The primary goal of this experiment was to measure the neutron’s EDM with high precision using Ramsey’s method of separated oscillating magnetic fields with stored ultracold neutrons (UCN). Key features of the experiment included the use of a $^{199}$Hg comagnetometer and an array of optically pumped cesium vapor magnetometers to cancel and correct for magnetic field variations. The measured value of the neutron EDM was determined to be $d_n = (0.0 \pm 1.1_{\text{stat}} \pm 0.2_{\text{sys}}) \times 10^{-26} \, e \cdot \text{cm}$. The collaboration spanned many European countries.  

\begin{figure}[htb!]
    \centering
    \includegraphics[width=0.45\textwidth]{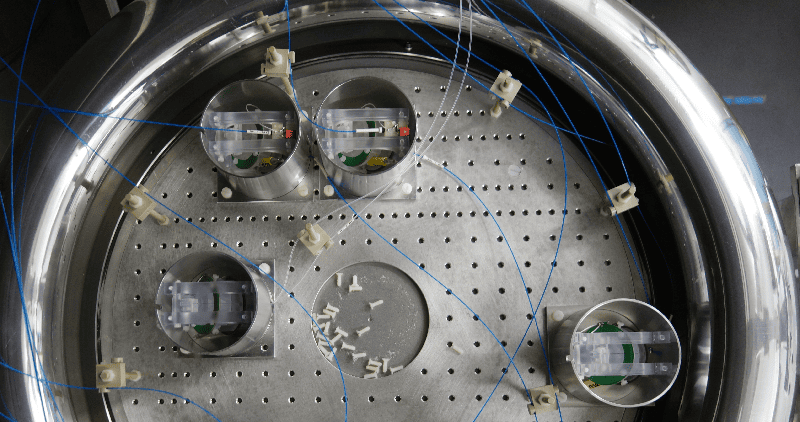} 
    \includegraphics[width=0.48\textwidth]{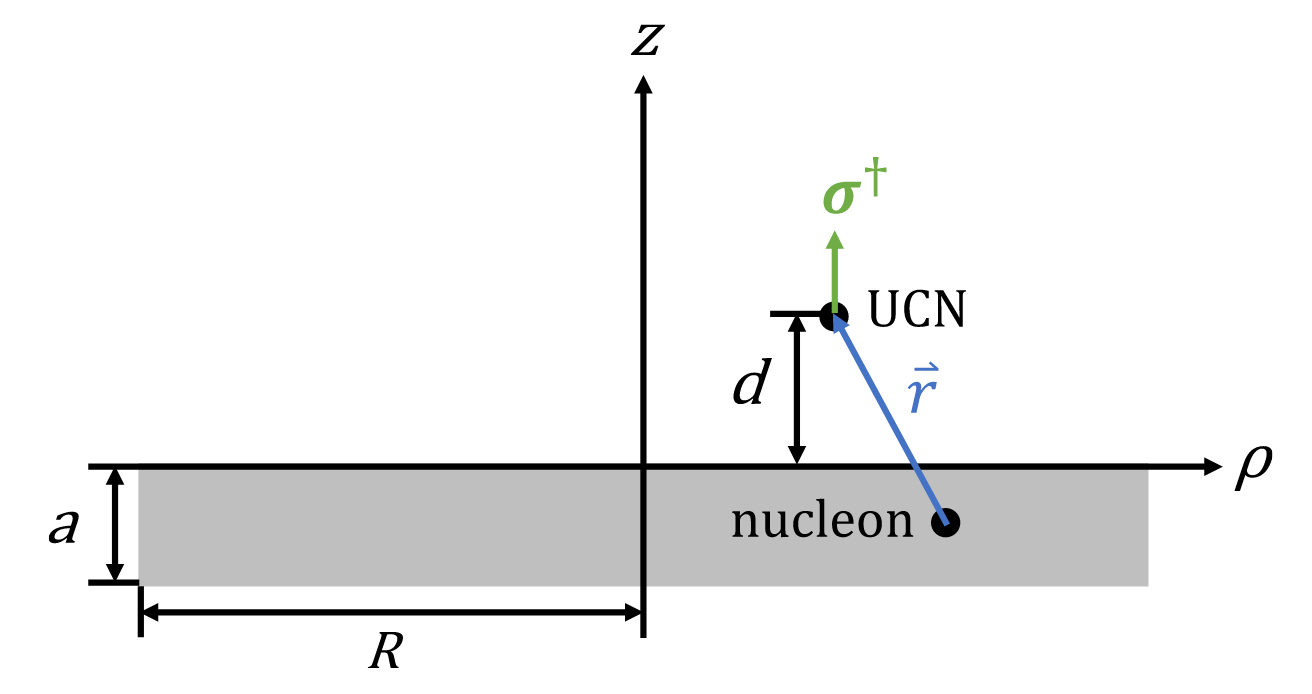} 
    \caption{\textit{Left:} Experimental setup at PSI for neutron EDM measurements, showing a half-meter-wide cylindrical chamber and $4$ of the $15$ Cs magnetometers, reproduced from Ref.~\cite{ball_new_2020}. \textit{Right:} Schematic of the interaction between a nucleon in the electrode and a polarized UCN in the precession chamber, reproduced from Ref.~\cite{ayres_search_2023}.}
    \label{fig:EDM-experimental_setup}
\end{figure}

The monopole--dipole interaction mediated by a spin-$0$ boson, such as an axion or ALPs, can be investigated using this apparatus originally designed for the neutron EDM search~\cite{ayres_search_2023}.  
The Lagrangian is given by $\mathcal{L}_{\phi} = \phi \sum_\psi \bar{\psi}  \left( g_s^\psi + i \gamma_{5} g_p^\psi \right) \psi$, where $\psi$ represents the fermion field, leading to the exotic potential  $V_{9+10} = -g_p^n g_s^N \frac{\hbar^2}{8\pi m_n} \mathbf{\sigma}_n \cdot \hat{\mathbf{r}} \left( \frac{1}{r^2} + \frac{1}{\lambda r} \right) e^{-{r}/{\lambda}}$. The dimensionless interaction constants $g_p^n$ and $g_s^N$ describe the pseudoscalar and scalar interaction strengths, respectively~\cite{cong_spin-dependent_2024}. Measurements were conducted by comparing the Larmor precession frequencies of stored UCNs and $^{199}$Hg atoms, which acted as a cohabiting magnetometer. An ALP-mediated interaction between vessel materials and trapped particles influences the precession frequency of UCNs, as shown in the right panel of Fig.~\ref{fig:EDM-experimental_setup}.  
The use of UCN to probe exotic spin-dependent interactions is comparable to experiments employing comagnetometers~\cite{feng_search_2022} and polarized $^3$He~\cite{guigue_constraining_2015}. This remains a promising research direction, where Ref.~\cite{ayres_search_2023} describes the potential of improving the measurement sensitivity by a factor of $64$ using a new apparatus~\cite{ayres_design_2021} currently under construction at PSI. The setup was also used to establish the first limits on the coupling of axion dark matter to gluons~\cite{Abel:2017rtm}.

In addition, the proton EDM storage ring, a proposed design currently under development, aims to measure the EDM of the proton~\cite{abusaif_storage_2021,alexander_storage_2022}. The high precision goal for the proton EDM makes it a promising target for detecting axion gradients sourced from test masses in the laboratory or from the Earth~\cite{agrawal_searching_2023}. 
Further research on axion-exchange-induced atomic and molecular EDMs can be found in Refs.~\cite{stadnik_improved_2018,prosnyak_updated_2023,agrawal_searching_2024}.

%% file: WG4/content/STAXExperiment.tex
{Author: P. Spagnolo}\\

\begin{figure}[t!]
  \begin{center}
   \hspace*{-3cm}
    \includegraphics[totalheight=7cm]{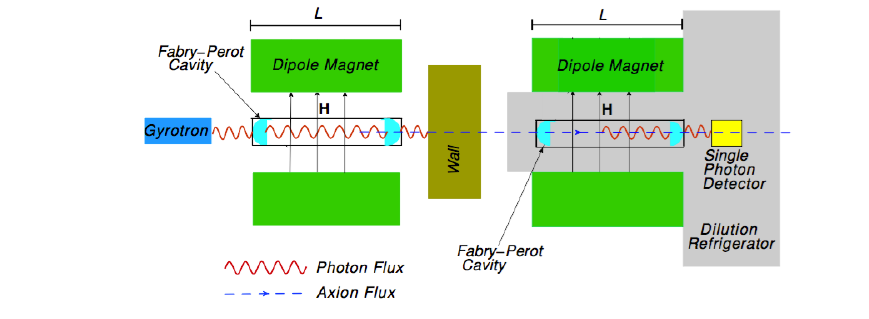}
    \caption{Schematic of the STAX experiment: the wall (brown) separates two resonant cavities (turquoise) immersed in a strong magnetic field generated by two dipole magnets (green), while the outgoing radiation is measured with the innovative single-photon detector (yellow). The grey zone corresponds to the ultra-cold region of detection. Figure reproduced from Ref.~\cite{Capparelli:2015mxa}.}
    \label{fig:STAX}
  \end{center}
\end{figure}

The STAX proposal~\cite{Capparelli:2015mxa} consists in a new generation LSW experiment with improved high luminosity obtained by using a sub-THz photon source, such as a gyrotron, emitting photons at frequencies of about $100$~GHz, resulting in photon fluxes up to $10^{28}$~photons$/$s, a factor $10^{10}$ more intense than those from optical lasers used in present LSW experiments, like ALPS. The layout of the STAX experiment setup is sketched in Fig.~\ref{fig:STAX}.

Before and after the wall, the STAX facility will rely on two strong dipole Nb3Sn magnetic fields of intensity $H = 11$~T each, uniformly applied on a length $L = 150$~cm. The second magnet back-conversion region is included in the cold region of the cryostat, where a single photon detector is positioned (yellow spot in Fig.~\ref{fig:STAX}). Since the sensitivity on the coupling $g$ is proportional to $H\times L$, these high-intensity magnets, built using state-of-the-art technology for high-field dipoles, are essential for the experiment. 

The STAX photon source will be a CW gyrotron or a gyroklystron. There are several open solutions to be confirmed by tests on cavities, with gyrotrons available in our labs operating in CW at frequencies around $100$~GHz suitable for the purpose. Together with the high power, a key point for the final choice of the source will be the bandwidth that needs to match the cavity acceptance. 

Regarding the cavity R\&D, a Fabry--P\'erot cavity will be fabricated and optimized for $100$~GHz with a target finesse of approximately $10^4$. This is feasible with mirrors made of oxygen-free copper with a reflectivity of more than $99.9\%$ at $100$~GHz, with an effective power inside the cavity of about $10$~kW. 

Single-photon detectors working at frequencies below $1$~THz with almost zero dark counts do not currently exist and must be developed. One of the ambitious goals of STAX is to implement a completely new device to overcome this limitation. A very promising solution is obtained by using nanoscale transition-edge sensor (nano-TES) technology~\cite{Paolucci:2020woi}: ultra-cold, nano-sized electron bolometers made of a superconducting material, operated so that they remain partially resistive, i.e., close to the superconducting--normal-state phase transition. Photons passing through the absorber will increase the temperature of the active-region electrons, thereby inducing an increase in the bolometer resistance. 

The STAX experiment will employ a gyrotron source with $10^{28}$~photons$/$s, two intense dipole magnetic fields $H1 = H2 = 11~T$ with a path $L = 150$~cm, two cavities $Q_1 = Q_2 = 10^4$, and an exposure of two months and zero dark counts. The experimental set-up might be further improved in an upgraded version named STAX2 with the addition of a second new-generation Fabry--P\'erot cavity at very high finesse $Q_2 \sim 10^{10}$ or alternatively the use of whispering galleries in the region beyond the wall. In this case, the power of the source should be downgraded to $10^{26}$ photons to allow the coupling and locking with the second high-Q cavity. In the axion mass range $m_a\leq 0.02$~meV, STAX allows an extension of the exclusion region by a factor greater than $10^4$ with respect to all present laboratory experiments (LSW).
With a high-Q second cavity in STAX2, this gain would increase by one or two more orders of magnitude.

%% file: WG4/content/VMB_CERN.tex
{Author: S. Kunc}\\

VMB@CERN (Vacuum Magnetic Birefringence at CERN) is an optical polarimeter that aims to directly measure, for the first time, the low-energy light-by-light interaction due to quantum fluctuations in the vacuum (foreseen by QED) or due to candidates of dark matter that couple to two photons, such as axions. In our configuration, an intense magnetic field makes the vacuum birefringent, and the polarization state of a laser beam is modified. For VMB theory and QED prediction ($\Delta n \simeq 4\times10^{-24}\,\mathrm{T}^{-2}$) are provided in Sect.~\ref{VMB@QED}.
VMB@CERN is a joint initiative of experts who have attempted to measure VMB in recent years. At the end of 2018~\cite{ballou_letter_nodate}, a letter of intent was submitted to the CERN SPSC, bringing together members of the PVLAS (Italy), Q\&A (China), OSQAR-VMB (Czech--France), and LIGO-CARDIFF (UK) collaborations. VMB@CERN is the successor to the PVLAS experiment, which holds the best limit for measuring VMB to date~\cite{Ejlli:2020yhk}.

\begin{figure}[t!]
\begin{center}
\includegraphics[width=13cm]{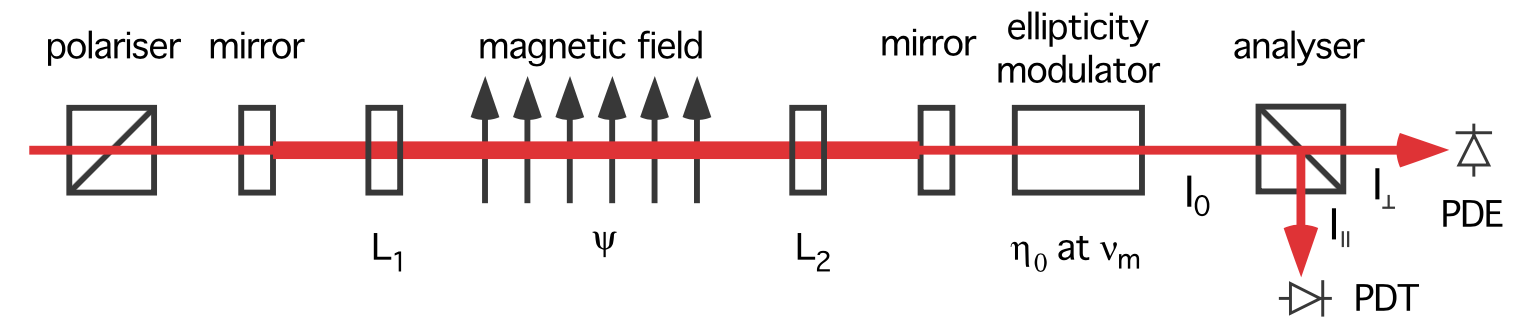}
\end{center}
\caption{Optical scheme of the VMB@CERN polarimeter. Figure reproduced from Ref.~\cite{zavattini_polarisation_2016}.}
\label{VMB@CERN experiment}
\end{figure}

The experimental scheme (Figure~\ref{VMB@CERN experiment}) was based on a proposal published in 2016~\cite{zavattini_polarisation_2016} by the PVLAS group and partially coincided with the original design of the OSQAR experiment~\cite{pugnat_feasibility_2005}. The scheme is designed as a heterodyne polarimeter based on two co-rotating half-wave plates inside a Fabry--P\'erot cavity. Thus, for the first time, it replaces the rotation or modulation of the magnetic field (PVLAS~\cite{Ejlli:2020yhk}, Q\&A~\cite{Mei:2010aq}, BRFT~\cite{Cameron:1993mr}, BMV~\cite{agil_vacuum_2022}, OVAL~\cite{Fan:2017fnd}) by rotating the electric field polarization only in a static magnetic field region. This design allows continuous operation of LHC’s spare magnet ($14.3$~m long with a $9$~T dipole field). This novel approach is predicted to reduce low-frequency noise (below $10$~mHz).

R\&D started in 2019 under the leadership of Prof. Zavattini, mainly in the laboratories of INFN in Ferrara with a smaller polarimeter, with additional work carried out at the laboratories of the Technical University of Liberec and Cardiff University. The development proceeded in steps toward understanding systematic errors and the subsequent test in the optical resonator. This phase demonstrated the functionality of the proposed procedure, and it was possible to lock the resonator with rotating half-wave plates for the first time ever. In the second step, the development focused on the minimization of the additional wide-band noise introduced by the rotating half-wave plates. Despite intensive R\&D within VMB@CERN collaboration to reduce wide-band noise remains an ongoing challenge~\cite{Zavattini:2021cna}.

%% file: WG4/content/VAMBI.tex
{Authors: L. Roberts, A. Ejlli, G. Mueller}\\

\noindent\textit{Scope.} A concise VMB theory summary and QED prediction ($\Delta n \simeq 4\times10^{-24}\,\mathrm{T}^{-2}$) are provided in Sec.~\ref{VMB@QED}. Here we focus on the polarimetry design, noise budget, and achievable ellipticity noise floor.
The design is informed by prior VMB efforts: in previous birefringence measurements the \emph{dominant} limitation has been thermally driven intra-cavity birefringence noise~\cite{PVLAS}. Maximizing sensitivity requires a strong magnetic field $B$, a long effective path length $L_{\mathrm{eff}}$, operation at a modulation frequency $f_{\mathrm{mod}}$ above the $1/f$ knee, and long integration time $T$. In our scheme the dipole field rotates at frequency $f$ (the magnet rotation), and the VMB signal appears at $f_{\mathrm{mod}} = 2f$ (standard geometry). The resulting signal-to-noise ratio scales as
\begin{equation}
  \mathrm{SNR} \propto 
  \left(\frac{L_{\mathrm{eff}}}{1\,\mathrm{m}}\right)
  \left(\frac{B}{1\,\mathrm{T}}\right)^{2}
  \left(\frac{f_{\mathrm{mod}}}{1\,\mathrm{Hz}}\right)^{\alpha}
  \left(\frac{T}{1\,\mathrm{s}}\right)^{1/2},
  \label{eq:SNR}
\end{equation}

\begin{figure}[htb!]
  \centering
  \includegraphics[width=0.75\textwidth]{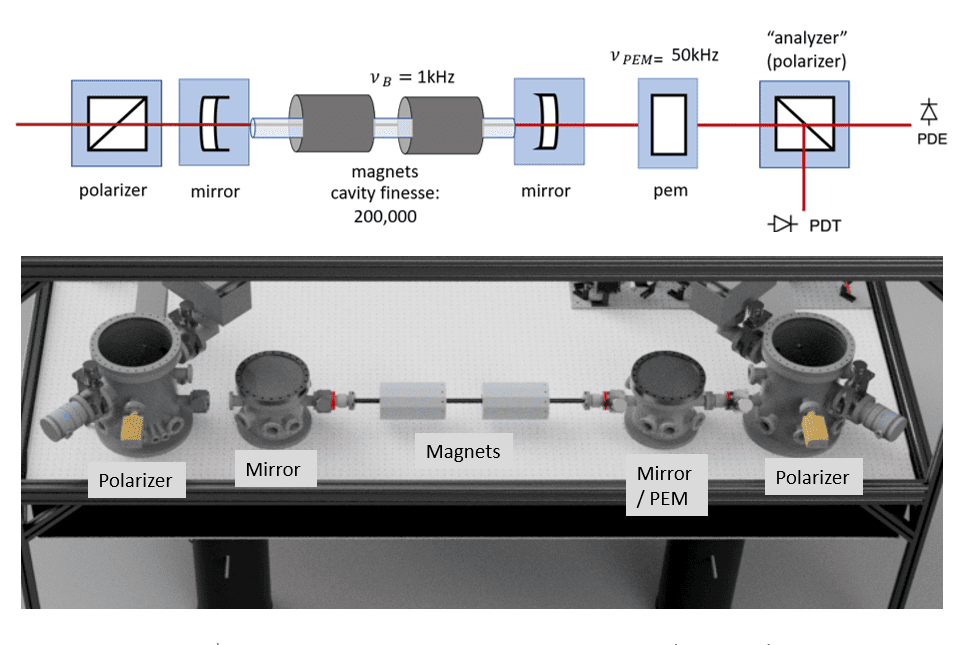}
  \caption{Simplified drawing of the experimental setup. Light blue boxes indicate ultra-high-vacuum chambers connected via a vacuum tube. The red line represents the laser light. PEM is the photoelastic modulator, and the analyzer is a high-extinction polarizer. PDE and PDT are the two photodiodes that measure the transmitted and extinguished light after the analyzer, with the former measuring the induced ellipticity. The bottom image shows the design of this setup in the laboratory. Figure reproduced from Ref.~\cite{Roberts:2025mpl}.}
  \label{fig:exp_scheme}
\end{figure}

\noindent Because the input-referred \emph{ellipticity noise spectral density} typically scales as $f^{-\alpha}$, increasing $f_{\mathrm{mod}}$ moves the signal into a lower-noise band and improves SNR, provided the modulation-frequency bandwidth and cavity linewidth allow it. Here $\alpha\simeq 0.8$ in wide-band conditions, and $L_{\mathrm{eff}}=(2\mathcal{F}/\pi)\,L_B$ is the cavity-enhanced path length.
We propose a novel modulation scheme that employs up to a kHz rotating dipole magnetic field in combination with a polarimeter based on a Fabry--P\'erot cavity, with a heterodyne readout. The core of the experiment consists of two magnets, each $10$–$30$~cm long, arranged in a Halbach array configuration, generating a $1$–$2$~T dipole field at the center, which is currently being designed by the Linz Center of Mechatronics (LCM) \cite{freller_high_2025}. Two design phases have been considered: one where the dipole field is rotated up to $0.5$–$1$~kHz using pairs of electromagnet coils with alternating current, and another utilizing magnetically levitated lightweight permanent magnets. This rotation is sustained over an extended period to optimize time integration as in Eq.~\ref{eq:SNR}. The implementation of this scheme is shown in Fig.~\ref{fig:exp_scheme}. 

The limiting sensitivity of the polarimeter is determined by the intrinsic noise sources present in the apparatus: shot noise, dark current, Johnson noise, and relative-intensity noise (RIN). The parameters in Table~\ref{tab:parameters} define our \emph{baseline} working point: under these conditions, reaching the QED signal at ${\rm SNR}=1$ requires an integration time of $T \approx 140~\mathrm{days}$. Following a successful two-magnet prototype, we plan to \emph{increase the number of magnets} to extend the magnetic length $L_{B}$ (and thus $L_{\mathrm{eff}}=(2\mathcal{F}/\pi)L_{B}$), targeting ${\rm SNR}\approx 10$ for the QED signal with a comparable integration time ($T\!\approx\!140~\mathrm{days}$).
This tabletop experiment is scheduled to be built at the Max Planck Institute for Gravitational Physics in Hannover, Germany, in the coming year.

\begin{table}[th!]
\renewcommand{\arraystretch}{1.5}
\centering
\begin{tabular}{c|c}
\hline
\hline
\textbf{Parameter} & \textbf{Value} \\
\hline 
Magnetic field, $B$ & $1$-$2$~T \\
Magnet length, $L_B$ & $0.10$–$0.30~\mathrm{m}$ \\
Cavity finesse, \(\mathcal{F}\) & $200\text{,}000$ \\
PEM freq, \(\nu_{\text{PEM}}\) & $50$~kHz \\
Magnet rotation frequency, $f$ & $1$~kHz \\
Temperature, $T$ & $300$~K \\
Transimpedance gain, $G$ & $10^{6}$~V/A \\
Quantum efficiency, $q$ & $0.6$~A/W \\
Intensity, \(I_0\) & $0.1$~W \\
Dark current, \(i_{\rm dark}\) & \(25 \times 10^{-15}\)~A/\(\sqrt{\text{Hz}}\) \\
Extinction ratio, \(\sigma^{2}\) & \(10^{-7}\) \\
\(N_{\mathrm{RIN}}(50~\mathrm{kHz})\) & \(10^{-7}\ /\sqrt{\mathrm{Hz}}\)\\
\hline \hline
\end{tabular}
\caption{Table of parameters planned for the High-Frequency Rotating-Field VMB Polarimeter experiment.}
\label{tab:parameters}
\end{table}

%% file: WG4/content/Intro-BeamDump.tex
\begin{figure}[tbhp]
\centering
\includegraphics[height=3.3cm]{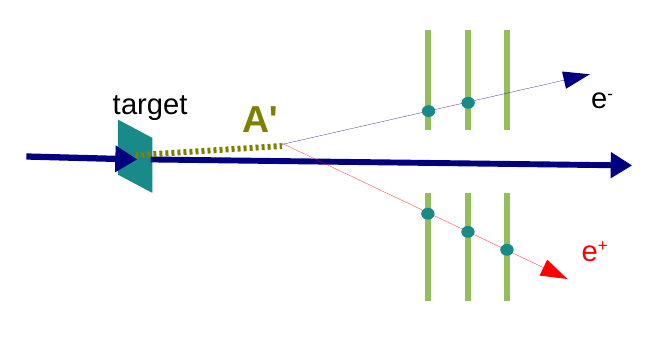}
\hspace{1cm}
\includegraphics[height=3.3cm]{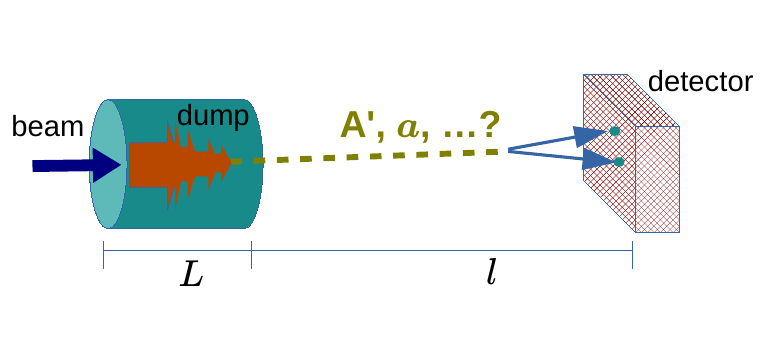}
\caption{Schematics of a typical experimental setup for a thin-target (\textit{left}) and a beam-dump (\textit{right}) experiment. 
\label{fig:fixed-target}}
\end{figure}

To increase the cross-section for new light particle ($X$, with mass $M_X$ and coupling constant $\epsilon$) production, particularly in the $1~\mathrm{MeV} \leq M_X \leq 1~\mathrm{GeV}$ mass range, several techniques exploit the long-standing history of particle physics experiments. Usually, an accelerated beam interacts with a relatively high-density target, whose constituents have close-to-zero momentum compared to the impinging beam energy. Such experiments are categorized as either fixed-target or beam-dump experiments, as shown in Fig.~\ref{fig:fixed-target}, depending on the target thickness and the possibility of fully absorbing the impinging beam.

In the thin-target experiments, the new light particles are usually produced directly through the so-called $X$-strahlung process, and the search for them exploits the full reconstruction of the $X$ decay products, either charged or neutral. In the case of charged final states, precision tracking is a common feature, allowing the study of events with displaced vertices and providing access to lower values of $\epsilon$. 

Thick-target experiments usually exploit $X$ production through meson decays, i.e., the appearance of $X$ in the secondary particle beam. The accessible mass region is limited by the meson mass, but event reconstruction may be facilitated by the additional constraint of the decaying meson mass. 

\begin{figure}[t!]
\centering
\includegraphics[width=0.6\textwidth]{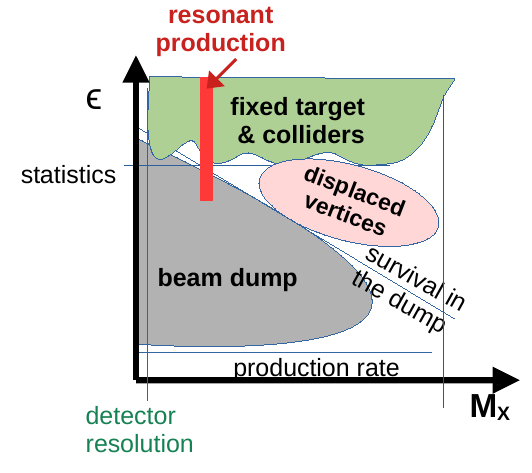}
\caption{Typical sensitivity regions for fixed-target and beam-dump experiments. 
\label{fig:fixed-target-reach}}
\end{figure}

The typical accessible region for fixed-target experiments (illustrated in Fig.~\ref{fig:fixed-target-reach}) is governed by the kinematics of $M_X$, the statistics on $\epsilon$, and the reconstruction algorithms, especially for the displaced vertices, reaching down to $\epsilon \sim 10^{-3}$--$10^{-4}$. 

In beam-dump experiments, the incoming high-intensity particle beam (usually electrons or protons) is fully absorbed in a thick target of length $L$. The production mechanisms include all possible interactions occurring during the development of electromagnetic and/or hadronic showers in the target, mostly---but not limited to---$X$-strahlung, meson decays, etc. Due to the target thickness, the beam energy---and thus all Standard Model particles (apart from neutrinos and sometimes neutrons)---is fully absorbed. A detector is placed at a distance $l$ behind the dump (Fig.~\ref{fig:fixed-target}, right), and depending on the particular setup, everything it detects can be considered as a ``signal'' of $X$ decaying into Standard Model particles.

The usual sensitivity reach in $\epsilon$ for such experiments is limited by statistics (at low $\epsilon$) and by the probability for $X$ to survive the distance between the production vertex and the end of the beam dump, which is inversely proportional to $\epsilon^2$. This results in a characteristic shape of the exclusion regions in the $X$ parameter space, as seen in Fig.~\ref{fig:fixed-target-reach}.

Other techniques, such as inverse Compton scattering and missing energy or momentum-inspired Dark Sector searches, can also be successfully included in this category, as they mostly differ in the choice of target or beam type and intensity, while preserving the particle-physics nature of the detector setup. Such new initiatives show promising capabilities, sometimes combining knowledge from different physics fields and increasing measurement precision. 

Several ongoing and planned experiments are currently underway in Europe. These include the PADME experiment at LNF-INFN, the LUXE-NPOD experiment at DESY in Hamburg, the NA62 and NA64 experiments at the CERN SPS, MAGIX at MESA, and others.

%% file: WG4/content/PADMEexperiment.tex
{Author: V. Kozhuharov}\\

The PADME experiment~\cite{Raggi:2014zpa}, located at Laboratori Nazionali di Frascati, Italy, is devoted to the search for any new light particle in the mass range $2$~MeV~$\leq M\leq 23$~MeV. While originally focusing on probing various dark photon models, the chosen experimental technique is also suitable for addressing ALPs and light dark Higgs boson scenarios. 

\begin{figure}[t!]
    \centering
    \includegraphics[width=0.8\textwidth]{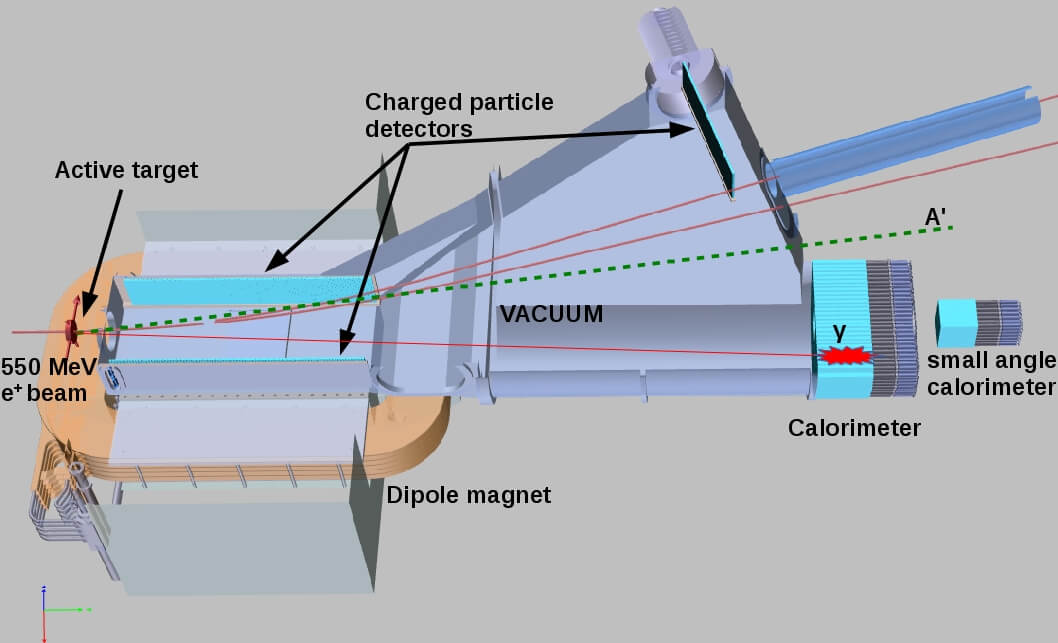}
    \caption{Schematics of the PADME experimental setup during Run-I and Run-II data taking. Figure taken from Ref.~\cite{Dimitrova:2022uum-ai}.}
    \label{fig:PADME-setup}
\end{figure}

PADME exploits the beam from the DA$\Phi$NE LINAC, with the experimental setup shown in Fig.~\ref{fig:PADME-setup}. The accelerated positrons, with energies of the order of $500$~MeV, impinge on a thin polycrystalline diamond target placed in vacuum and serving as a detector to measure the beam position and intensity. 

The reaction of interest is the associated production of a dark photon $A'$ or another new light particle in the positron-on-target annihilation:
\begin{equation}
    e^+ + e^- \to \gamma + A',
\end{equation}
governed, for example, by $\mathcal{L} \supset \epsilon \bar{e} \gamma^{\mu} e A'_{\mu}$, with $ \epsilon$ responsible for the interaction strength. 
The recoil photon is detected by a segmented BGO crystal electromagnetic calorimeter with an annular shape, 
located about $3$~m downstream of the active target, after a large-volume vacuum tank. 
The missing mass for the event is computed from the 4-momenta of the particles, $M_{miss} = (p_{e^+} + p_{e^-} - p_{\gamma})^2$, assuming that the target electrons are at rest. 
The existence of $A'$ with mass in the range $2$~MeV~$\leq M_{A'}\leq 23$~MeV would manifest as a peak over a smooth background in the missing-mass distribution. 

The rest of the detector setup is mostly dedicated to background rejection. 
Positrons that undergo hard bremsstrahlung emission in the target 
are deflected by a dipole magnet with a $\sim 0.5~\mathrm{T}$ magnetic field 
toward a system of charged-particle detectors placed inside the vacuum tank, 
while a fast Cherenkov electromagnetic calorimeter suppresses positron annihilation events with two, three, or more photons.

With the described setup, PADME took data in 2018/2019 and in 2020 (PADME Run~I and Run~II), 
collecting about $10^{13}$ positrons on target~\cite{PADME:2022fuc-det}. 
A fraction of the data was used to measure the inclusive cross-section 
\begin{equation}
\sigma (e^+e^-\to \gamma\gamma(\gamma)) = (1.977 \pm 0.018_{stat} \pm 0.119_{syst}) \hspace{0.1cm} \mathrm{mb}
\end{equation}
with a precision of $\sim 5\%$~\cite{PADME:2022tqr-gg}, 
which validated the performance of the experimental setup. 
The analysis of the majority of PADME data 
to search for the associated production of new light particles is still ongoing.

\begin{figure}[t!]
    \centering
    \includegraphics[width=0.8\textwidth]{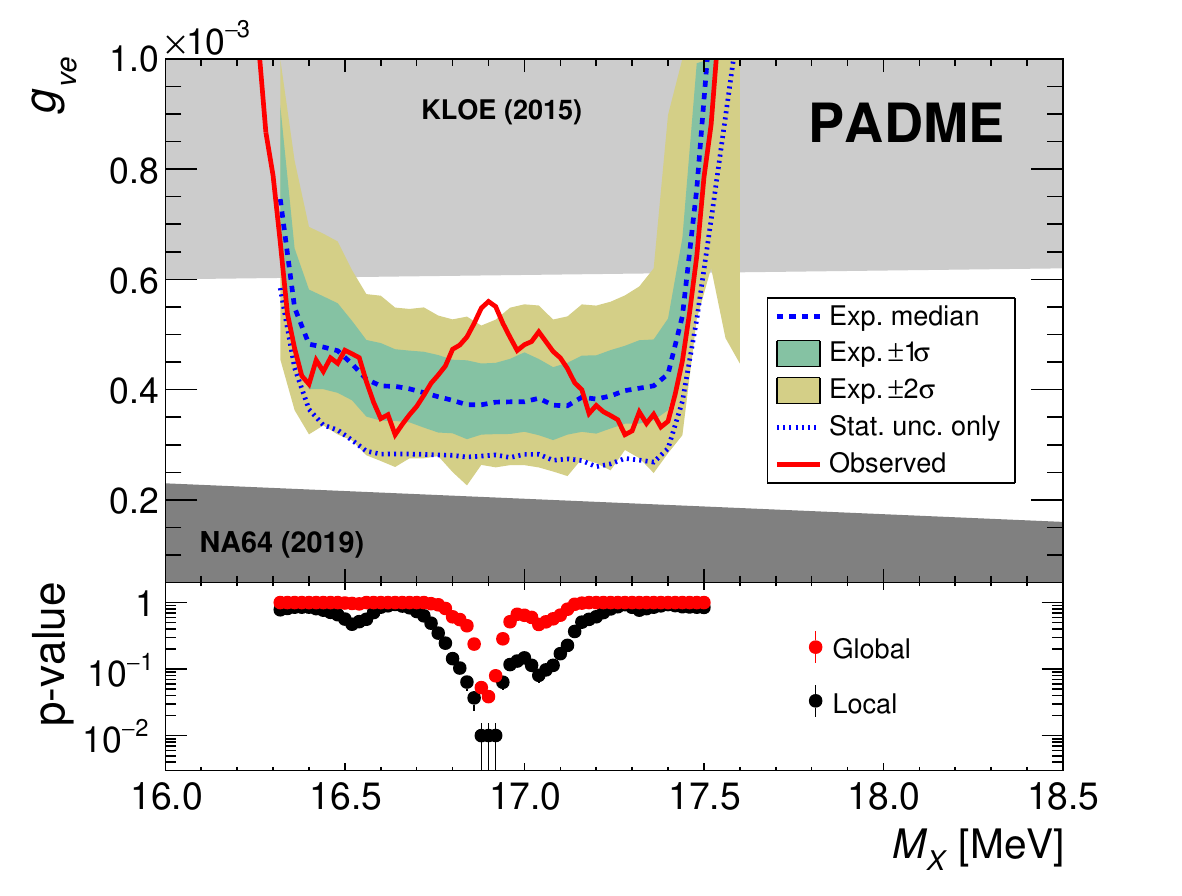}
    \caption{Excluded region with 90\% CL in the $M_X$ - $g_{ve}$ parameter space 
    from the PADME Run-III data. Figure taken from Ref.~\cite{PADME:2025dla}.}
    \label{fig:PADME-RUNIII-result}
\end{figure}

In 2022, the experimental setup was modified to address the existence of the X17 particle, as suggested by the study of excited $^8$Be, $^4$He, and $^{12}$C nuclei. 
The magnetic field was set to zero, a new charged-particle detector was placed in front of the electromagnetic calorimeter to discriminate between electrons and photons, 
and an array of 12 Timepix3 sensors 
together with a lead--glass calorimeter were employed to measure the positron beam parameters \cite{Bertelli:2024jzb-timepix}. 
The scattering cross-section $\sigma(e^++e^- \to 2~\mathrm{clusters}) = \sigma(e^+e^- \to e^+e^- ) + \sigma(e^+e^- \to \gamma\gamma )$ was studied as a function of the positron beam energy, searching for bumps indicating resonant X17 production. About $10^{10}$ positrons on target were collected during PADME Run~III at each energy scan point, for a total of $47$ scan points. 
Following the development of a comprehensive procedure for a blinded search of a possible X17 signal \cite{PADME:2025dvz},
the data was unmasked and the result is shown in Fig. \ref{fig:PADME-RUNIII-result} 
in terms of 90\% confidence level upper limit on the X17 coupling constant $g_{ve}$ 
as a function of X17 mass $M_X$.
The observed limit is consistent with the expected in most of the $M_X$ region 
indicating good control of the systematics uncertainties, 
with a slight excess at  $M_X \simeq $16.9~MeV, 
corresponding to 2.5 $\sigma$ local and 1.8 $\sigma$ global significance.

A new data-taking campaign, PADME Run~IV, 
was undertaken in two separate scan periods in 2025.
The setup was modified to 
allow a better control of the beam parameters and 
a possible separation of the charged and neutral final states. 
A large area double-gap Micromegas chamber was installed 
in front of the electromagnetic calorimeter, 
while a smaller Micromegas chamber operated in 
ionization chamber regime was placed in front of the lead--glass calorimeter instead of the 
Timepix3 detector.
Data were collected for 36 different  
energies of the positron beam, 
spaced by $\Delta E_{beam} \simeq 0.75~\mathrm{MeV}$ and 
covering approximately the interval $16.6~\mathrm{MeV} < \sqrt{s} < 17.4~\mathrm{MeV}$ of 
the  center-of-mass energy. 
PADME Run-IV will probe further the currently allowed X17 parameter space 
and will address the observed with the Run-III data $\sim2~\sigma$ excess.

%% file: WG4/content/LUXE-NPOD.tex
{Author: R. Quishpe}\\

The proposed LUXE experiment (LASER Und XFEL Experiment) at DESY, Hamburg, using the electron beam from the European XFEL, aims to probe quantum electrodynamics in the non-perturbative regime created in collisions between high-intensity laser pulses and high-energy electron or photon beams. This setup also provides an opportunity to probe physics beyond the standard model. LUXE plans to start data-taking by 2030 ~\cite{LUXE:esppu}. The LUXE collaboration is currently composed of $23$ laboratories and institutes in $9$ countries~\cite{Abramowicz:2021zja}. 

\begin{figure}[t!]
  \begin{center}
    \includegraphics[width=0.6\linewidth]{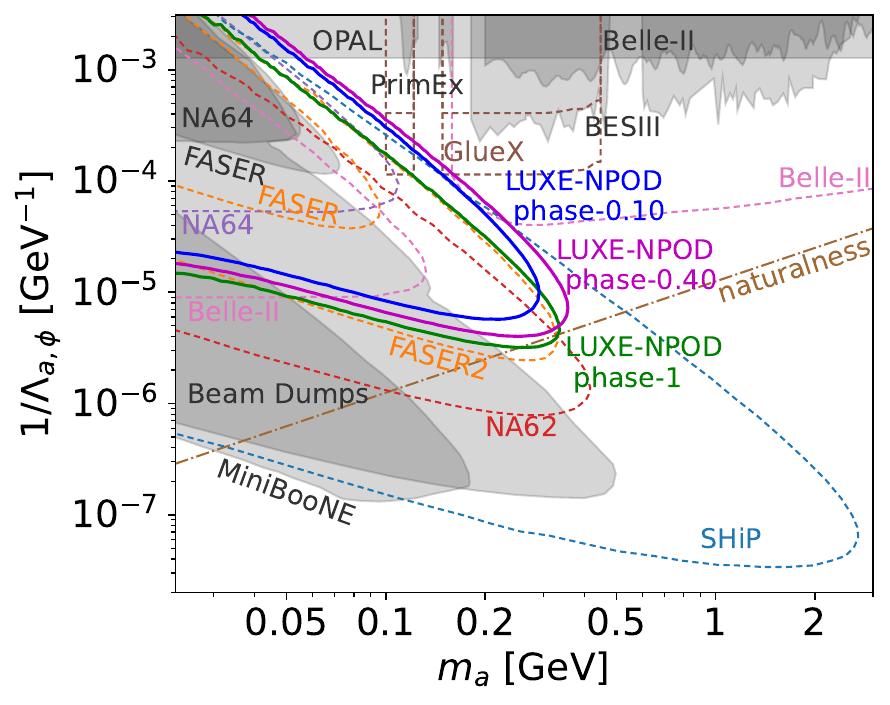} 
    \caption{Preliminary projections of the sensitivity of LUXE-NPOD phase-0.10 ($10$~TW), phase-0.40 ($40$~TW), and phase-1 ($350$~TW) shown as solid lines compared to currently existing bounds (gray regions) and to other experiments' future projections shown as dotted lines. Figure reproduced from Ref.~\cite{npod:2025}.}
    \label{fig:npodsensitivity}
  \end{center}
\end{figure}

By utilizing the large and hard photon flux generated at LUXE onto a physical dump, one can probe an unexplored parameter space of new spin-$0$ (scalar and pseudoscalar) particles with coupling to photons (created via the Primakoff effect), with lifetimes of $\mathcal{O}(1)$~ns and masses between about $10$~MeV and $1$~GeV. This novel apparatus is denoted as an optical dump, or NPOD (new physics search with optical dump)~\cite{Bai:2021gbm}. LUXE-NPOD can probe ALPs up to a mass of $350$~MeV and with photon coupling of $3\times10^{-6}$~GeV$^{-1}$. This reach is comparable to the background-free projection from NA62 in dump mode~\cite{Dobrich:2015jyk}, as seen in Fig.~\ref{fig:npodsensitivity}.

\begin{figure}[t!]
  \begin{center}
    \includegraphics[width=0.7\linewidth]{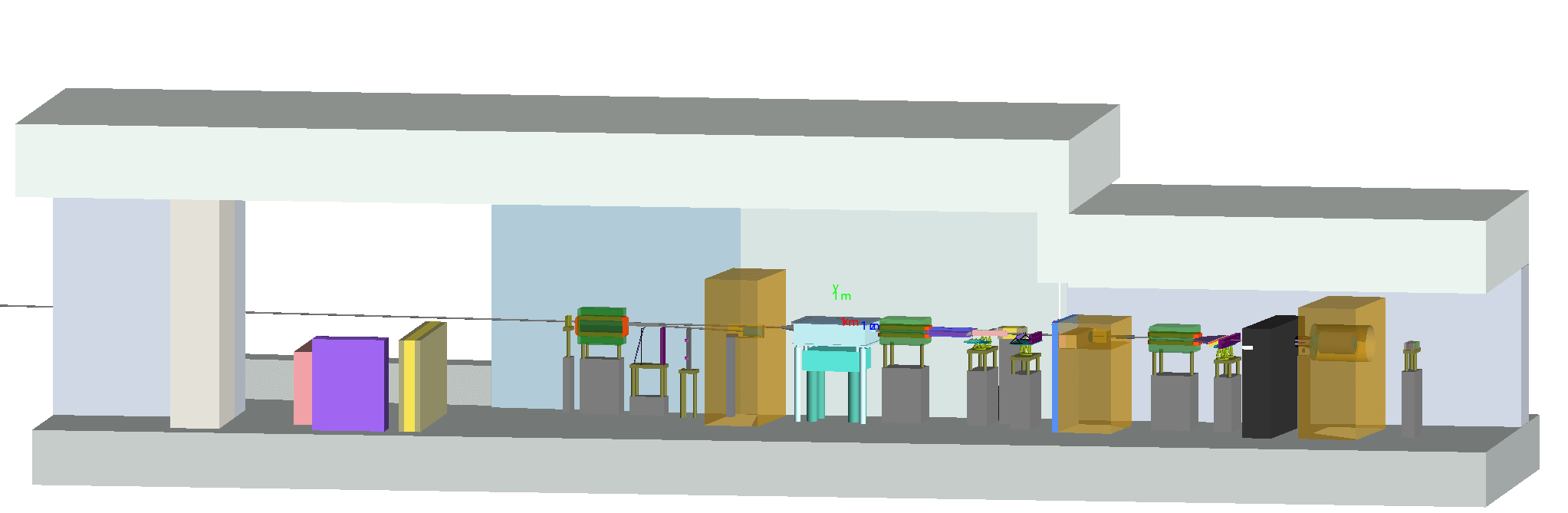}   
    \includegraphics[width=0.2\linewidth]{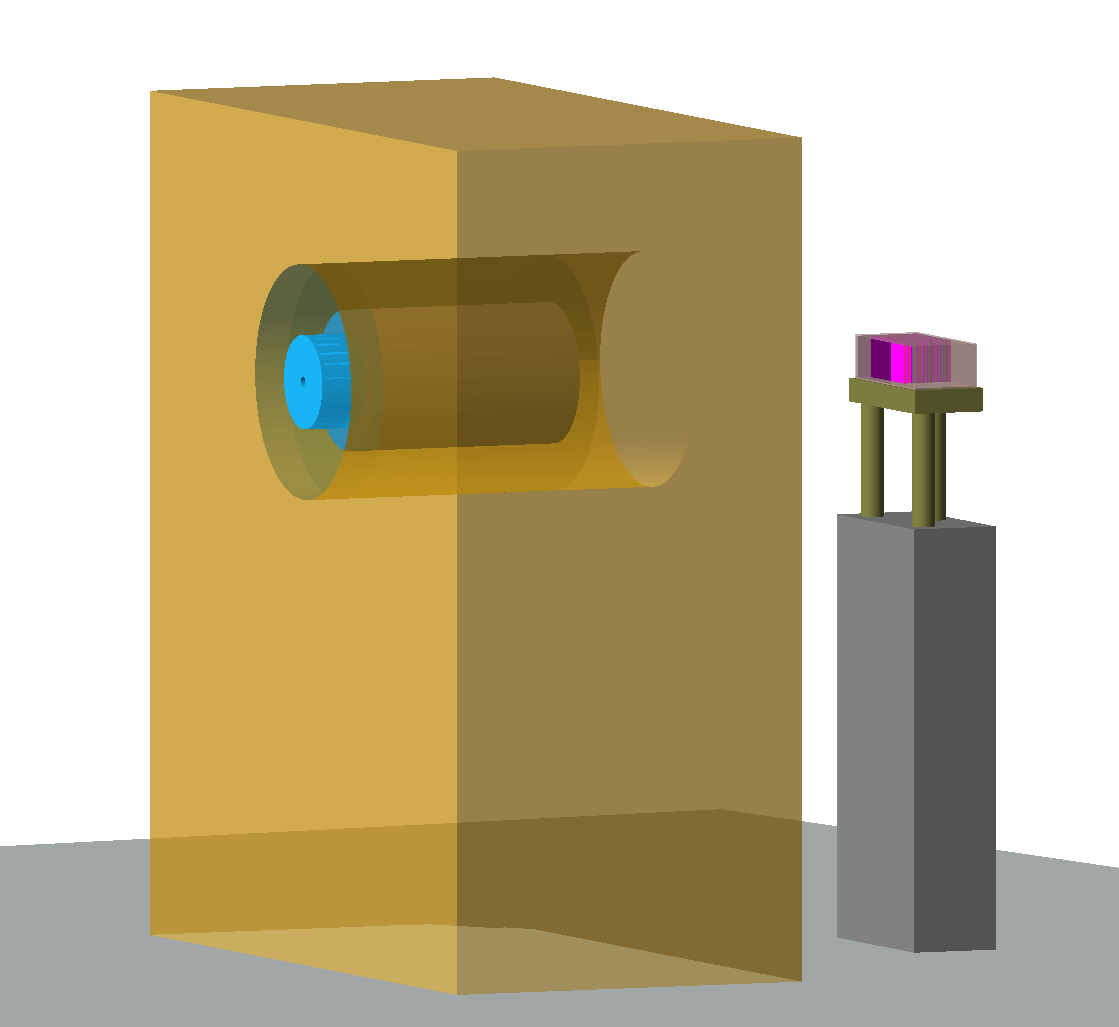}     
    \caption{\textit{Left:} Current {\textsc{Geant4}} model of LUXE setup. The LUXE-NPOD setup (\textit{right}) consists of the dump assembly with a tungsten core and the electromagnetic calorimeter LUXE ECAL NPOD, as presented in Ref.~\cite{npod:2025}.}
    \label{fig:geant4}
  \end{center}
\end{figure}

LUXE-NPOD will be situated about $13.5$~m behind the $e$--laser interaction point and before the wall at $17.4$~m. Dump--detector layout optimization simulation studies have shown that a background-free search may be achieved using a solid tungsten dump of radius $10$ ($20$)~cm and length $0.25$ ($1$)~m, wrapped in lead and concrete, and a decay volume length of $2.5$ ($1.0$)~m for phase-0 ($1$)~\cite{npod:2025}. These studies use the {\textsc{Ptarmigan mc}}~\cite{Blackburn:2023mlo} generator to predict particle rates and spectra, followed by {\textsc{Geant4}} simulations using the complete LUXE setup (Fig.~\ref{fig:geant4}). Background coming from charged particles is not significant and can be handled with a magnet at the end of the dump. Other backgrounds, such as neutrons and photons coming from electromagnetic/hadronic interactions, have been reduced to a negligible level with the aforementioned dump design. 

An electromagnetic calorimeter inspired by the SiW-ECAL detector from CALICE~\cite{Kawagoe:2019dzh} has been chosen as the LUXE-NPOD detector. The LUXE ECAL NPOD is a $36~\times~18~\text{cm}^2$ sandwich silicon--tungsten electromagnetic calorimeter, with a depth of $22.5$~cm, that includes $18$ radiation lengths ($X_0$) of tungsten. 
A preliminary detector performance study~\cite{npod:2025} has shown that LUXE ECAL NPOD can provide the necessary background rejection power for a background-free search, while also preserving a high enough signal efficiency for a viable signal parameter space.

%% file: WG4/content/NA62Experiment.tex
{Author: B. D\"obrich}\\

\begin{figure}[t!]
\centering
   \includegraphics[width=0.8\columnwidth]{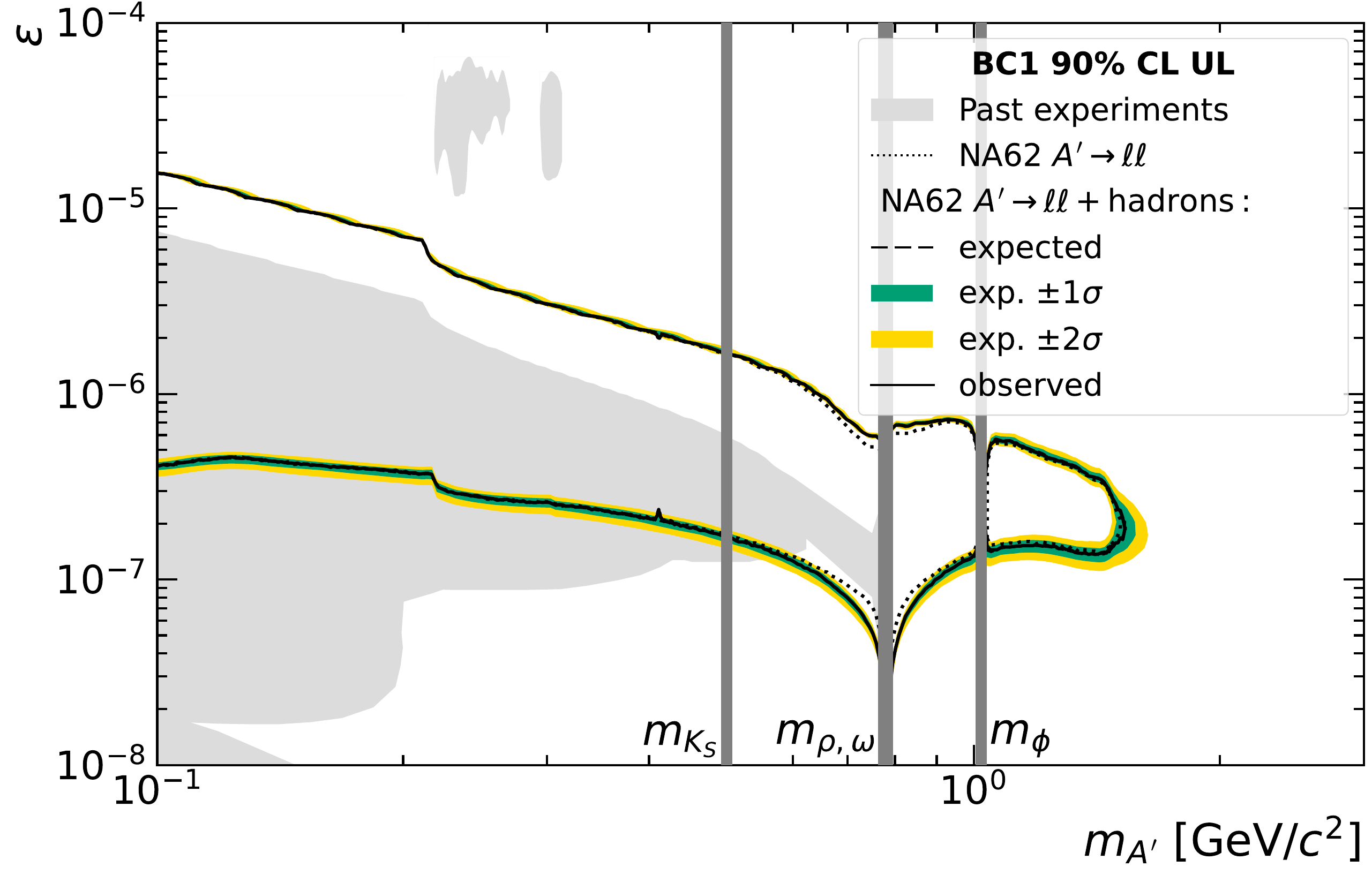}
    \caption{Observed $90\%$ CL exclusion contours in the dark photon benchmark (BC1) combining hadronic and di-lepton channels using a data sample from 2021, corresponding to $1.4\times 10^{17}$~POT~\cite{NA62:2023qyn,NA62:2023nhs,NA62:2025yzs}.}
    \label{fig:exclusion_dark_photon}
\end{figure}

The NA62 experiment was built to precisely measure the branching ratio ${\cal B}(K^+\to\pi^+\nu\bar\nu)$, and has recently measured this decay with a $5$-sigma significance~\cite{press_NA62,NA62:2024pjp}. Thanks to its high-intensity beam and detector performance (redundant particle-identification capability, extremely efficient veto system, and high-resolution measurements of momentum, time, and energy), NA62 has also achieved sensitivities to long-lived light mediators in a variety of new-physics scenarios. Concretely, two analyses involving hidden-sector particle decays to di-electrons~\cite{NA62:2023nhs} and di-muons~\cite{NA62:2023qyn} have been published; see Fig.~\ref{fig:exclusion_dark_photon} for the results in the dark photon interpretation. A search for hadronic decays of Dark Sector particles was presented in 2024~\cite{hadrons_NA62} and published recently~\cite{NA62:2025yzs}.

%% file: WG4/content/NA64Experiment.tex
{Author: L. Marsicano}\\

NA64 is an electron-beam fixed-target experiment searching for light Dark Matter (LDM) at the CERN Super Proton Synchrotron~\cite{PhysRevLett.131.161801}, H4 beamline. The experiment uses a $100$~GeV electron beam impinging on a thick active target, a lead--scintillator (Pb/Sc) electromagnetic calorimeter (energy resolution: $\frac{\sigma_E}{E} \simeq \frac{10\%}{\sqrt{E}} + 4\%$); beam electrons are tagged, and their momenta are measured with a magnetic spectrometer with a resolution of $\frac{\Delta p}{p} \simeq 1\%$. A set of active veto systems, including a large Pb/Sc hadronic calorimeter of $\sim 30$ nuclear interaction lengths, is placed downstream of the active target. In this setup, the signature of the production of LDM particles is a large missing energy, i.e., the difference between the beam energy and the energy deposited in the target, paired with null activity in the veto system. 

The benchmark theoretical scenario of NA64 is vector-mediated LDM, involving a dark sector of $\chi$ particles in the $1$~MeV--$1$~GeV mass range, charged under a new $U(1)$ symmetry, whose vector boson (dark photon, $A'$) is kinetically mixed with the Standard Model (SM) photon. In this picture, the main LDM production process for NA64 is the so-called $A'$-strahlung $e^{-} \, N \rightarrow e^{-}\, A'\, N$, followed by the dark photon invisible decay to an LDM particle pair. NA64 has been operating since 2016; thanks to the high-intensity electron beam provided by the SPS (with an impinging particle rate on the detector of up to $\sim 5 \times 10^{6}$~s$^{-1}$), the experiment has collected to date $\sim 1.5 \times 10^{12}$ $e^-$ on target (EOT), allowing the collaboration to set the most stringent limits in a large region of the LDM parameter space~\cite{PhysRevLett.131.161801}. The short-term goal of the experiment is to collect $\sim3 \times 10^{12}$ EOT before CERN Long Shutdown 3, scheduled in 2026. Besides the electron-beam data taking, from 2022, NA64 initiated a parallel measurement campaign with positron beams, aiming to exploit the resonant LDM production process $e^{+}e^{-}\rightarrow A' \rightarrow \bar{\chi}\chi$~\cite{PhysRevLett.121.041802}. The first sample of $\sim 10^{10}$ $100$~GeV positrons, collected and analyzed by NA64 with partial support from the ERC-funded project POKER, paved the way for a future multi-energy $e^+$-beam program~\cite{PhysRevD.109.L031103}. A further step in this direction was taken in 2023, when an additional sample of $1.6\times 10^{10}$ positrons was collected with a beam energy of $70$~GeV~\cite{NA64:2025rib}.

NA64 features a significant sensitivity to several alternative beyond-the-SM extensions; the collected data have been re-analyzed in the context of the $B-L$, $L_\mu$--$L_\tau$, and ALP scenarios~\cite{NA64:2020qwq,NA64:2022yly,NA64:2022rme,Andreev:2024lps}, resulting in competitive exclusion limits in the addressed parameter spaces.

In 2017 and 2018 the experiment operated with a modified setup to search for the hypothetical X17 boson, collecting a total sample of $8.4\times 10^{10}$ EOT;
no signal-like events were observed, resulting in the exclusion of part of the preferred X17 parameter space~\cite{PhysRevD.101.071101}. Further improvements in this search are subject to a future detector upgrade, currently under discussion within the collaboration.

In addition to the activities with electron and positron beams at the H4 beamline, the NA64 collaboration is carrying out a complementary effort to perform a missing-momentum measurement with a muon beam at the M2 line at CERN, called NA64$\mu$. The collaboration collected $\sim 2\times 10^{10}$ $\mu$ on target, allowing stringent limits to be set in the $L_\mu$--$L_\tau$ scenario~\cite{NA64:2024klw,andreev2024exploration}.

%% file: WG4/content/Intro-NonWISP.tex
Up until this section, a broad overview of recent and near future experimental studies focused in the search for WISPs has been compiled. In addition to these results and efforts, there are quite a number of other running or planned experiments whose main focus is not WISPs (e.g., WIMPs or gravitational waves (GWs)) but have sensitivity to different classes of WISPs. In this section, we summarize such cases in European laboratories, which are quite diverse in size. Although most of the WISP searches can, by their nature, be categorized under {\em Physics Beyond Colliders}, the highly growing interest among collider experiments deserves extra attention, and experiments at CERN’s Large Hadron Collider are summarized in the following. It is also important to note that the electron--positron collider experiments in Asia, namely, the BESIII experiment at BEPCII (China), which is a charm factory, and the Belle~II experiment at SuperKEKB (Japan), which is a B-factory, also offer quite unique opportunities with a clean background environment and high statistics to probe the dark sector.

%% file: WG4/content/GWInterferometers.tex
\label{sec:GWI}
\label{sec:GWAtomIFO}

The detection of GW has established GW interferometers as a vital experimental tool for astronomy, cosmology, and fundamental physics. These detectors have also demonstrated remarkable sensitivity for ultra-light dark matter (ULDM) searches~\cite{Miller:2025yyx}. The fundamental principle is that a coherent, oscillating ULDM field can interact with the components of the interferometer such as its mirrors and laser beams, to induce subtle, time-varying signals that can be detected by the instrument. The specific nature of this interaction is determined by the spin and interactions of the ULDM particle, leading to distinct signals for scalar, vector and tensor ULDM. Detector‑specific implementations are detailed in Sec.~\ref{sec:GW_detectors}.

\paragraph{Scalar Bosons}

Scalar bosons, including dilaton-like particles and axions, interact with matter in distinct ways. Dilaton-like particles can cause oscillatory changes in fundamental constants such as the fine-structure constant and electron rest mass~\cite{Stadnik:2014tta,Stadnik:2015xbn,Aiello:2021wlp}. Such interactions manifest themselves in GW detectors primarily through two mechanisms: (1) a direct force on test masses arising from a gradient in the dark matter field and (2) an apparent strain caused by the modulation of the physical size and refractive index of optical components.

Multiple experiments have conducted searches for these effects. The GEO 600 and Fermilab Holometer instruments focussed primarily on the expansion and refractive index effects, whereas the more recent LIGO-Virgo-KAGRA analyses have searched for both the acceleration and expansion effects~\cite{Grote:2019uvn,Vermeulen:2021epa,Gottel:2024cfj,Hall:2022zvi}. Furthermore, a recent search using LIGO O3 data employed a spectral-density-based method to improve constraints on the ULDM coupling constants~\cite{Gottel:2024cfj}.

Another class of scalar bosons, axions (intended as any light pseudo-scalar field that couples to electromagnetism), can induce a parity-violating birefringence effect, differentially dephasing the left-polarised and right-polarised photons. These effects can be probed using existing GW detectors with minor optical element modifications, see~\cite{Nagano:2019rbw,Nagano:2021kwx}. Lastly, space-based GW interferometers, such as LISA, will enable probing of ULDM with masses in the range [$10^{-19}, 10^{-14}$]~eV, extending several orders of magnitude below the range accessible to ground-based interferometers~\cite{Kim:2023pkx}.

\paragraph{Vector Bosons}

Ultra-light vector bosons, often referred to as dark photons, can induce oscillatory forces on test masses in GW detectors, producing persistent, nearly monochromatic signals~\cite{Pierce:2018xmy,Morisaki:2020gui}. Searches based on this concept have been conducted using LIGO/Virgo data, setting upper limits on the coupling of the vector boson to baryons and leptons~\cite{Guo:2019ker,LIGOScientific:2021ffg}.

Besides the standard GW channels, searches have been performed using data from the auxiliary channels of KAGRA~\cite{KAGRA:2024ipf} and with the LISA Pathfinder~\cite{Armano:2016bkm,Miller:2023kkd}. KAGRA's unique design features fused silica mirrors as beam splitter and power recycling mirrors, and cryogenic sapphire test masses, creating different dark charges that experience varying accelerations from dark photons. While these differential accelerations are not detectable in the GW channel, they can be measured in auxiliary readout channels~\cite{Michimura:2020vxn}.

The analysis of KAGRA O3 data specifically exploited the different materials used in its mirrors to enhance the sensitivity to vector dark matter coupling to baryon-lepton number, boosting the signal-to-noise ratio for objects with different charge-to-mass ratios. For KAGRA, the charge-to-mass ratios for baryon-lepton number differ by approximately $9\times 10^{-3}$ between fused silica and sapphire, while for baryon number alone, these quantities differ by only $\sim 4\times 10^{-5}$.

Similar search strategies have been proposed for LISA Pathfinder data~\cite{Frerick:2023xnf}. This analysis improves on previous work by exploiting the relative acceleration between the spacecraft and one of the test masses, which produces a much stronger signal than the relative acceleration of two test masses. This approach yields a conservative upper limit on the baryon-lepton coupling, with stricter constraints on the $U(1)_{\rm B-L}$ coupling compared to the $U(1)_{\rm B}$ coupling due to the different compositions of the test mass (primarily gold) and spacecraft (primarily gold/carbon).

\paragraph{Tensor Bosons}

Recent theoretical advances in massive gravity theories provide compelling models for spin-2 ULDM~\cite{Aoki:2017cnz,Marzola:2017lbt,Aoki:2017ixz,Manita:2022tkl}. Besides these Lorentz-invariant models based on multimetric theories of gravity, Lorentz-violating massive gravity can also provide a rich spin-2 phenomenology that can be explored with ULDM experiments. Tensor bosons, representing spin-2 ULDM, interact with interferometers in a manner that can mimic continuous (massive) gravitational waves, due to their quasi-monochromaticity and persistence. Forecasts for the sensitivity of gravitational wave-interferometers to spin-2 ULDM have been developed in~\cite{Armaleo:2020efr}, and dedicated studies for spin-2 ULDM detection in space-based interferometers have been performed in~\cite{Zhang:2025fck}. Recently, the first tensor boson search in LIGO-Virgo-KAGRA data (with updates on the scalar and vector cases) was perfomed in~\cite{LIGOScientific:2025ttj}.

\begin{figure}[t!]
    \centering
    \includegraphics[width=0.8\columnwidth]{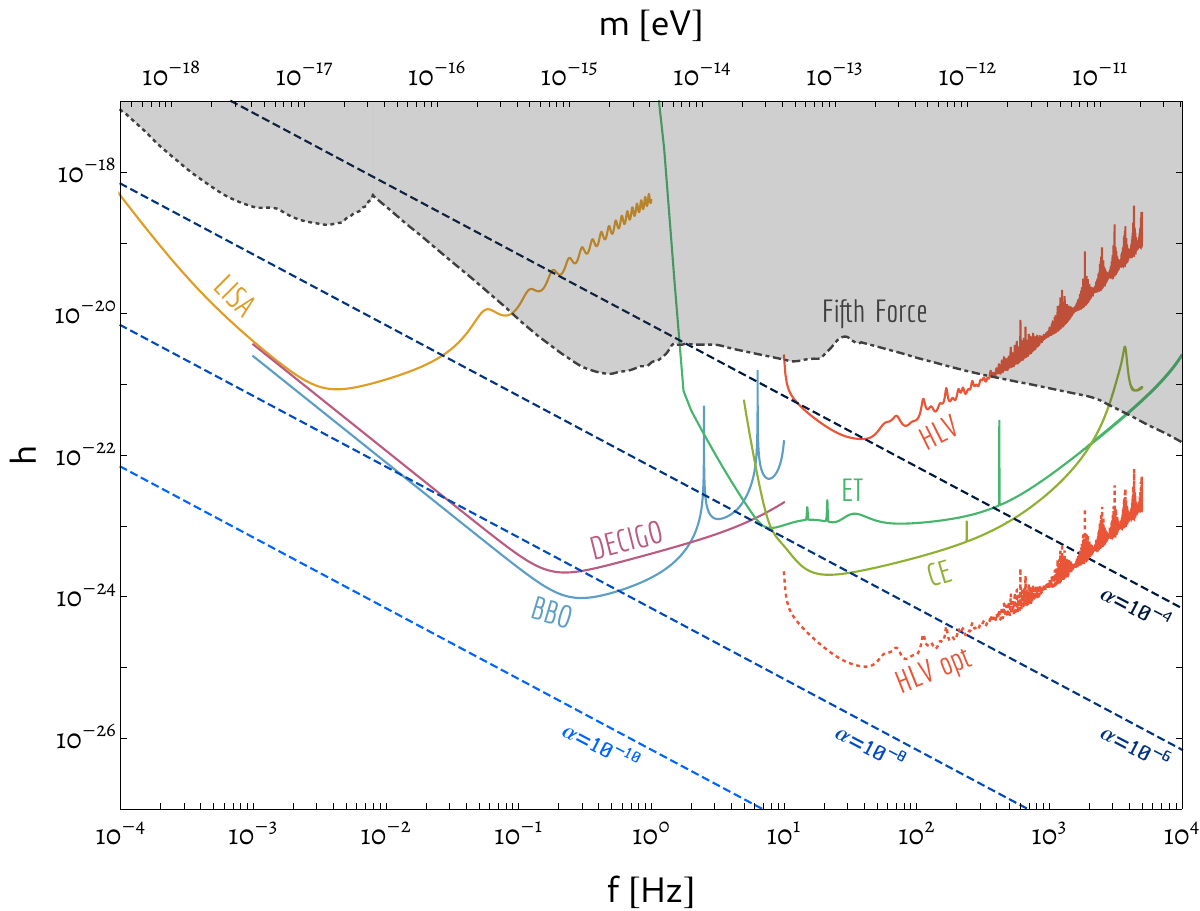}
    \caption{Design sensitivity for current and planned GW interferometers as a function of frequency (solid lines). The dotted line (``HLV opt'') represents optimised sensitivity using a semi-coherent method for spin-2 ULDM searches. Dashed lines show predicted signal strains for spin-2 ULDM with varying coupling constants ($10^{-4}\leq\alpha\leq10^{-10}$). The dot-dashed black line represents the strain corresponding to maximal values of $\alpha$ allowed by fifth force constraints from Ref.~\cite{Murata:2014nra,Sereno:2006mw}, with the region above this line excluded. Figure reproduced from Ref.~\cite{Armaleo:2020efr}.}
    \label{fig:s2signal}
\end{figure}

Atom interferometers offer a unique opportunity for detection in the frequency mid-band between LIGO-Virgo-KAGRA and the upcoming LISA space interferometer~\cite{Blas:2024kps}. The interaction between spin-2 ULDM, represented by the field $\phi_{\mu\nu}$, and Standard Model particles can be described by the interaction Lagrangian $L_{int} = \kappa \phi_{\mu\nu} O^{\mu\nu}$, where $\kappa$ is a coupling constant and $O^{\mu\nu}$ is a symmetric tensor constructed from Standard Model fields. This interaction induces measurable effects in atom interferometers through three distinct channels: first, the scalar mode of the spin-2 field can couple to atomic energy levels, inducing a phase shift proportional to the scalar field amplitude and interaction time, $\Delta \phi \propto \phi_0 T$; second, vector effects arise from the interaction of vector components of $\phi_{\mu\nu}$ with atomic momentum, modifying atomic propagation; third, tensor effects modify the propagation of both atoms and light within the interferometer, inducing a phase shift proportional to the strain induced by the ULDM field.

Long-baseline atom interferometer experiments are positioned to explore a mass range complementary to existing gravitational wave observatories. The expected sensitivity depends on the minimum detectable phase shift $\Delta \phi_\mathrm{min}$, the interferometer's baseline $L$, and the atomic velocity $v$. The projected mass range and corresponding coupling strengths are $10^{-18} \mathrm{eV} < m < 10^{-15} \mathrm{eV}$ with $\kappa \approx 10^{-10} - 10^{-5}$. To estimate future sensitivity to spin-2 ULDM, a realistic atom interferometers setup can be based on the AION and MAGIS experiments~\cite{Badurina:2019hst,MAGIS-100:2021etm,Proceedings:2023mkp} (See Sec.~\ref{sec:AionExperiment}), assuming atom-shot-noise-limited performance (white noise) without additional gravity gradient noise (coloured noise).

\begin{figure}[t!]
    \centering
    \includegraphics[width=0.85\columnwidth]{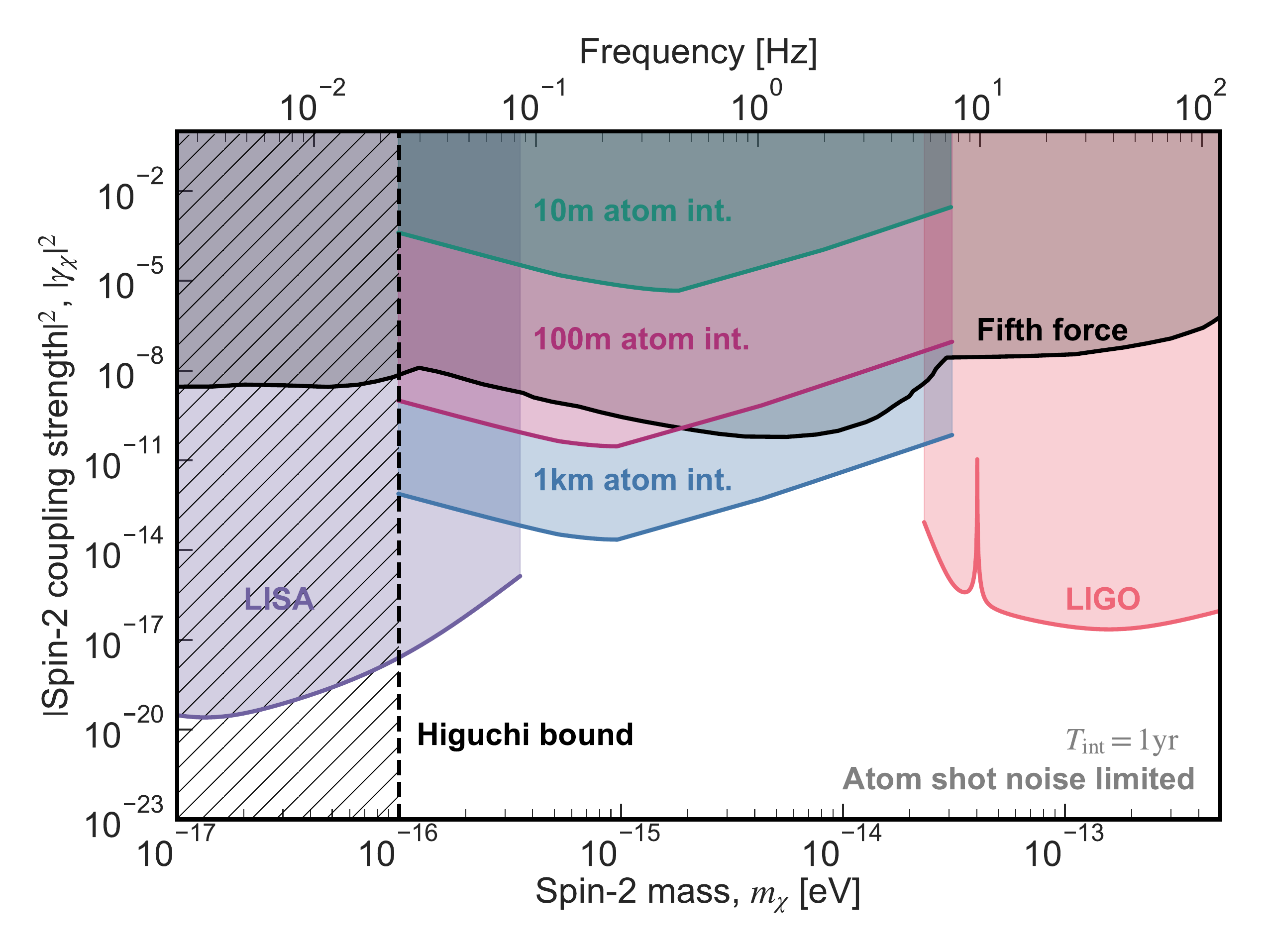}
    \caption{Sensitivity projections for the spin-2 ULDM generalised coupling strength $|\gamma_\chi|^2$ for three atom interferometer configurations with increasing baseline lengths ($10\,\text{m}$, $100\,\text{m}$, $1\,\text{km}$), decreasing separation between the upper interferometer and noise source ($5\,\text{m}$, $90\,\text{m}$, $980\,\text{m}$) and shot-noise levels ($10^{-8}\text{Hz}$, $10^{-10}\text{Hz}$, $10^{-11}\text{Hz}$). These are compared to LIGO and LISA projections (see legend). Calculations assume a 1-year measurement campaign at SNR~$=1$ with atom shot-noise limited performance. Shown are also leading constraints from fifth-force experiments from Ref.~\cite{Adelberger:2003zx}: these are model-dependent as they constrain only scalar couplings. Figure reproduced from Ref.~\cite{Blas:2024kps}.}
    \label{fig:FP_limits}
\end{figure}

Figure~\ref{fig:FP_limits} illustrates the projected sensitivity of three long-baseline atom interferometer experiments to spin-2 ULDM. For context, leading experimental constraints from fifth-force experiments are included, which apply only to scalar couplings or in the Lorentz-invariant case where all modes have the same mass. Tensor and vector couplings may evade these constraints if the scalar mode is more weakly coupled or occupies a different region of parameter space. The figure also displays the sensitivity of LIGO and projected sensitivity of LISA for comparison, along with the Higuchi stability bound on spin-2 ULDM mass, which again only applies in the Lorentz-invariant case and is not a fundamental limit.

%% file: WG4/content/Intro-GWOptical.tex
\label{sec:GW_detectors}

\noindent Gravitational-wave–class interferometers and other long-baseline precision interferometers can operate as axion polarimeters by searching for an oscillatory polarization rotation at the axion Compton frequency while remaining fully operational for their primary mission. For the general framework, see Sec.~\ref{sec:GWAtomIFO}. Here we outline a \emph{site-independent} implementation: integrate high-finesse polarization optics within an existing ultra-high-vacuum beamline; use a heterodyne/lock-in modulation scheme that places the axion line inside the detector's sensing band (few~Hz--kHz) and above the $1/f$ knee; and inject calibration lines to control readout-bandwidth constraints. The large-scale and low-displacement/ellipticity noise floors of such interferometers make them competitive in the ultra-low-mass regime ($m_a \lesssim 10^{-10}\,\mathrm{eV}$), complementing cavity-scale experiments.

Candidate host infrastructures include the \textbf{ALPS} facilities at \textbf{DESY} (see Sec.~\ref{sec:ALPSII}), where high-finesse Fabry--P\'erot cavities and high-stability lasers are already operated. A DESY-based upscaling concept can reuse these assets while keeping the added optical-loss budget and the modulation-frequency bandwidth compatible with the primary ALPS program.

%% file: WG4/content/LHCexperiments.tex
{Authors: M. Gallinaro, V. A. Mitsou}\\

The Large Hadron Collider (LHC), known for the discovery of the Higgs boson, is also being used to search for DM. 
Although the LHC was not specifically designed to search for DM, a broad range of searches has been deployed to find potential signals predicted by either the production of DM or the production of the particles mediating its interactions with ordinary matter. All of the results obtained so far have been consistent with models that do not include dark matter. The results have both pointed experimentalists in new directions for how to search for dark matter and prompted theorists to rethink existing ideas for what dark matter could be.

\begin{figure}[t!]
    \centering
    \includegraphics[width=.65\textwidth]{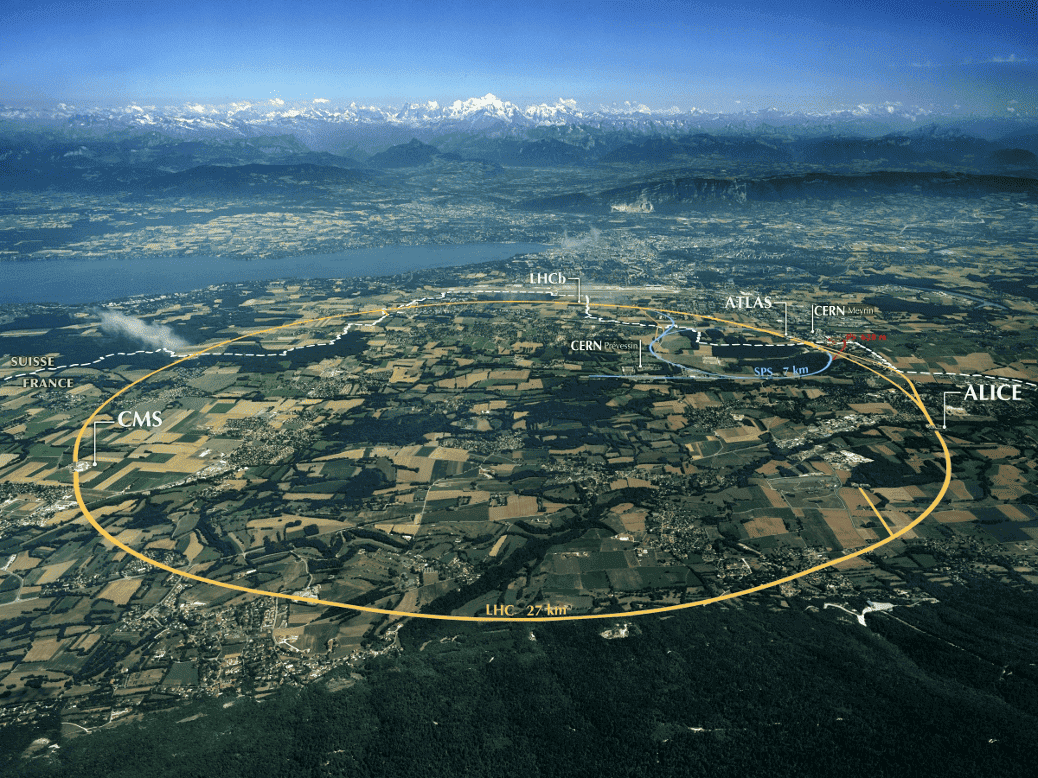}\\
     \includegraphics[width=.49\textwidth]{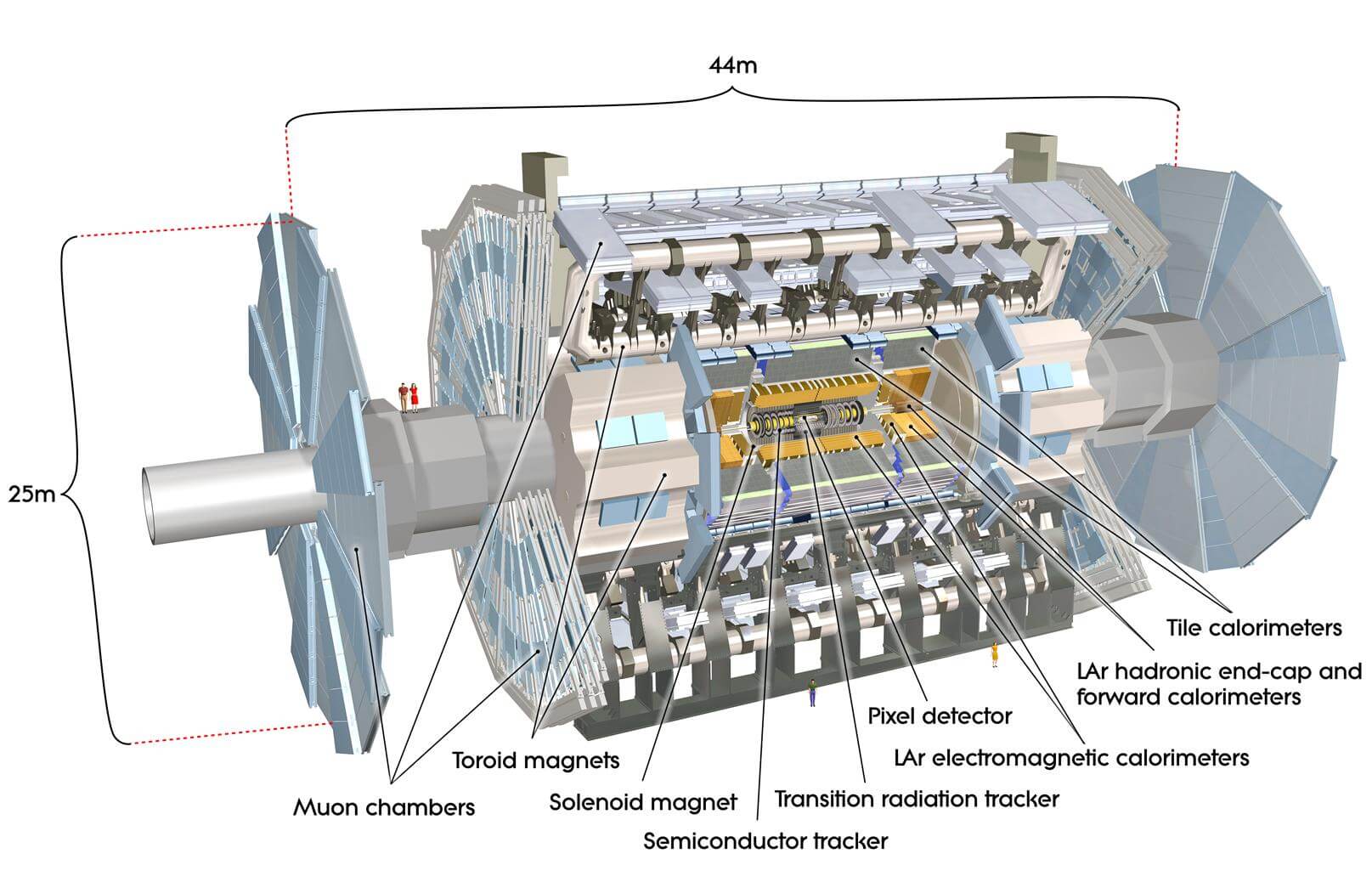}
      \hfill
\includegraphics[width=.49\textwidth]{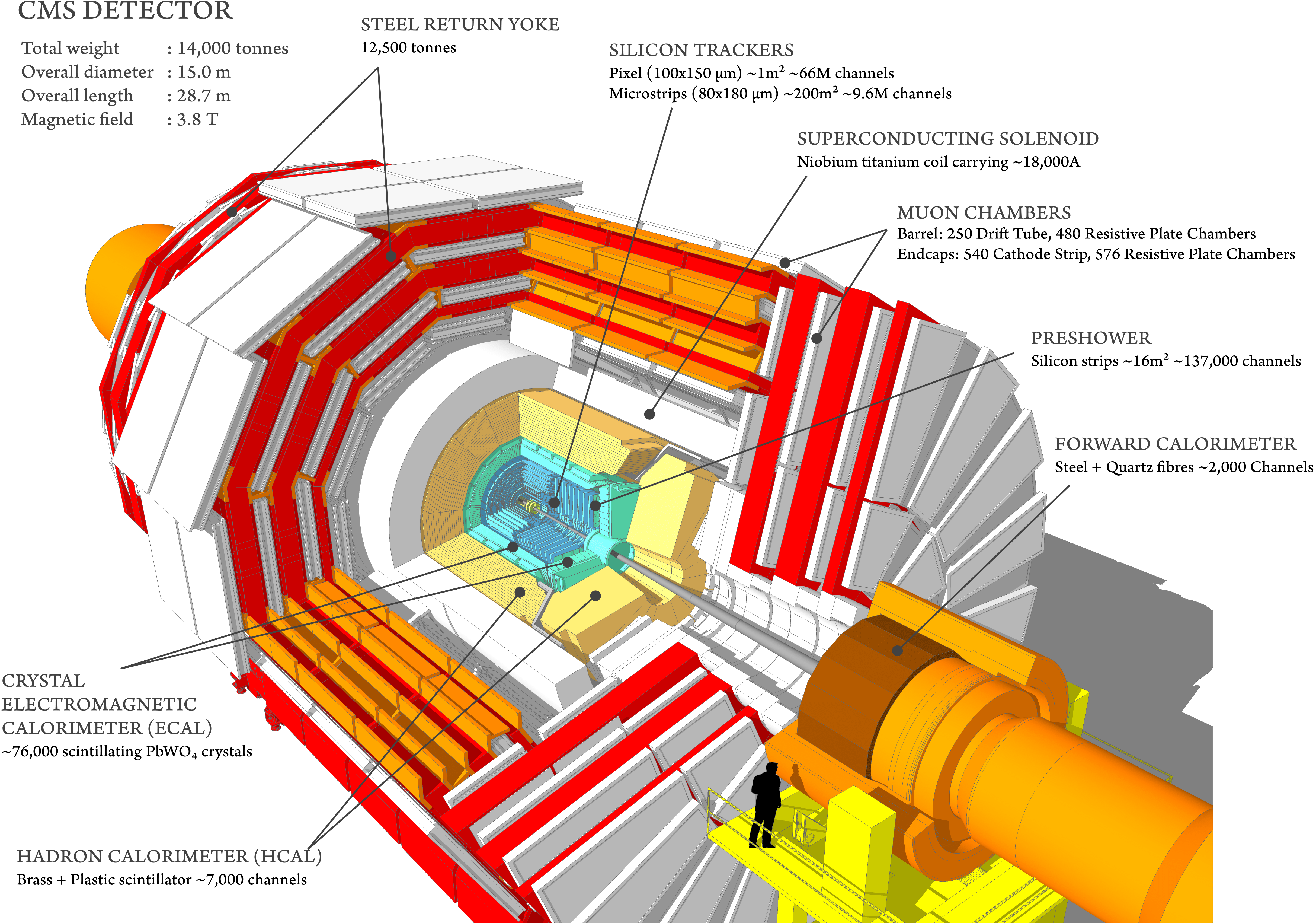}
    \caption{Aerial view of the CERN area with the Geneva lake in the distance and the yellow line indicating the $27$~km circumference of the LHC complex that is located approximately $100$~meters below ground (\textit{top}). A schematic view of the ATLAS (\textit{bottom left}) and the CMS (\textit{bottom right}) detectors and their characteristics. Credit: CERN.}
    \label{fig:LHC_exp}
\end{figure}

At the LHC, both the ATLAS~\cite{ATLAS:2023dns} and the CMS~\cite{CMS:2023gfb} experiments (Fig.~\ref{fig:LHC_exp}) are searching for DM candidates in proton--proton collisions. The main signature of the presence of DM is the so-called missing transverse momentum. By adding up the momenta of the final state particles produced in the collisions, the total transverse momentum should be zero, as the protons before the collisions travel longitudinally along the beam direction and their total initial transverse momentum is null. If the total momentum after the collision is not zero, the missing momentum could be carried by an undetected DM particle.
Searches for signatures of missing transverse momentum recoiling against visible standard model particles that would occur in processes involving DM have been performed, and the results of these searches have been interpreted in terms of many different DM scenarios, from simplified models to SUSY models. The LHC experiments are searching for DM candidates with masses ranging from as low as $0.1$~GeV and as high as 2~TeV. The null results have inspired possible new models and possible explanations.
Searches include the possibility that DM is part of a larger dark sector with several dark sector particles. Those would include the dark photon, the DM equivalent of the photon, that would interact with other particles both in the dark sector and in the SM, and long-lived particles, also predicted in SUSY models~\cite{CMS:2024zqs}.

Axions and ALPs are possible candidates in extensions of the SM, and are excellent DM candidates. Couplings and masses of these axions can span many orders of magnitude.  Different production processes and decay final states can be studied. ALPs couple to photons, but they also couple to gluons, $Z$ and Higgs bosons, as well as to leptons and quarks. At the LHC, they can be produced in photon--photon collisions, but also in gluon--gluon collisions, which is of interest due to the large cross-section. Associated production allows for triggering and is experimentally easier to select.

The number of searches is very large, and two examples will be provided here, one for low-mass and another for high-mass dark sector candidates. 

An example of a recent search for a low-mass DM candidate is a direct search for exotic Higgs boson decays $h\to aa,~ a\to\gamma\gamma$. The particle $a$ is a spin-0 low-mass particle decaying to two photons that are reconstructed as a single photon-like object. In the mass range $m_a < 1$~GeV, the Lorentz boost is large and the two photons are reconstructed as one object. Deep learning analysis techniques are employed to identify highly boosted photon pairs and probe masses as low as $0.1$~GeV~\cite{CMS:2022fyt}.

Central exclusive production processes provide a unique method to access rare physics processes, such as new physics via anomalous production of fermions, $V$ bosons ($V=\gamma, W, Z$), high-$p_T$ jet production, and possibly the production of new resonances or pair production of new particles. The CMS-TOTEM Precision Proton Spectrometer (PPS)~\cite{CMS:2014sdw} is a set of new near-beam forward detectors to further extend the coverage and enhance the sensitivity of the LHC experiments to explore processes previously not covered. It allows tagging the surviving scattered protons during standard data-taking conditions in regular ``high-luminosity'' fills~\cite{TOTEM:2022vox}. 
A search for high-mass diphoton production through photon--photon fusion in proton--proton collisions was performed~\cite{TOTEM:2023ewz}. Events with two high-transverse-momentum photons ($p_T>100$~GeV) with large invariant mass and back-to-back in azimuth were selected.
The dominant inclusive diphoton background is reduced by matching the kinematic properties of the protons detected in PPS and in the central detector. One exclusive diphoton candidate is observed for an expected background of $1.1$~events. Stringent limits on the four-photon coupling parameters are set. In addition, upper limits on the production of axion-like particles are also set in the mass range from $500$ to $2000$~GeV.

At the LHC, it is possible to accelerate not just protons but also heavy ions with charges up to $Z=82$ for lead (Pb). This provided the opportunity to perform novel $\gamma\gamma$ measurements in proton--proton, proton--nucleus, and nucleus--nucleus ultraperipheral collisions (UPCs). A search for an excess with respect to the expected $\gamma\gamma\to\gamma\gamma$ continuum has been proposed to identify the production of ALPs in Pb--Pb UPCs. The invariant mass distribution of diphoton candidate events is used to search for possible narrow resonances in Pb--Pb collisions.
Events with exclusively produced $\gamma\gamma$ pairs with invariant masses larger than $5$~GeV and other kinematical cuts are used to set stringent limits on the production of ALPs coupled to photons in the mass range $5$--$100$~GeV (including the most stringent limits in the $5$--$10$~GeV range)~\cite{CMS:2024bnt}.

%% file: WG4/content/MAPPexperiment.tex
{Author: M. Staelens}\\

The MoEDAL Apparatus for Penetrating Particles (MoEDAL-MAPP) experiment~\cite{pinfold2023moedalmapp} is designed to expand the discovery reach of the LHC through comprehensive searches for feebly interacting particles (FIPs) that arise naturally in dark/hidden sector models and other extensions of the Standard Model~\cite{Lanfranchi2021,Antel2023}. The first phase of the MoEDAL-MAPP experiment (MAPP-1) was approved by the CERN Research Board in December 2021 for installation on the LHC ring for Run-3 operation~\cite{Pinfold2791293}. MAPP-1 is primarily designed to search for minicharged particles; however, it also has sensitivity to long-lived neutral and charged particles.

\begin{figure}[t!]
\centering
\includegraphics[width = 14 cm]{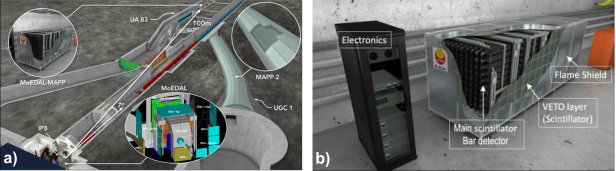}
\caption{\label{fig:MAPP} \textbf{a)} A visual representation showcasing the layout of the MoEDAL-MAPP facility located at the LHC's IP8 region. It highlights the placement of the MoEDAL (IP8), MAPP-1 (UA83), and proposed MAPP-2 (UGC1) detectors. \textbf{b)} A depiction of the MAPP-1 detector, focusing on its key elements. The dimensions of the MAPP-1 detector are approximately $1.2$~m $\times$ $1.2$~m $\times$ $4$~m; the maximum fiducial volume of the MAPP-2 detector is approximately $1400$~m${^3}$~\cite{pinfold2023moedalmapp}.  Reproduced with permission from Ref.~\cite{Kalliokoski2024}.}
\end{figure}

The MAPP-1 detector, located at the LHC's UA83 region adjacent to the MoEDAL\slash LHCb interaction point (IP8), comprises four collinear sections, each containing $100$ plastic scintillator bar units (of size $10$~cm $\times$ $10$~cm $\times$ $75$~cm), which are individually coupled to a single high-gain photomultiplier tube for readout~\cite{Pinfold2791293}. An illustrative depiction of the MoEDAL-MAPP facility and a schematic of the MAPP-1 detector are provided in Figure~\ref{fig:MAPP}a,b, respectively. The LHC Experiments Committee is currently reviewing the MAPP Outrigger---an auxiliary detector comprising four layers of additional scintillator planks strategically positioned between MAPP-1 and the beamline---designed to enhance the overall sensitivity of the MAPP experiment to high-mass minicharged particles~\cite{Kalliokoski2025}. The second phase of the MoEDAL-MAPP experiment (MAPP-2), which targets displaced-vertex signatures of long-lived neutral FIPs decaying to visible final states, is currently in the research and development phase~\cite{pinfold2023moedalmapp}. The design of the MAPP-2 detector involves instrumenting the UGC1 gallery (also adjacent to IP8) as a large decay zone by enclosing it within three layers of scintillator tiles with embedded fast wavelength-shifting fibers connected to silicon photomultipliers, allowing for a position-sensitive readout; Figure~\ref{fig:MAPP}a illustrates the proposed design and location for MAPP-2.

The MoEDAL-MAPP experiment probes dark sectors through direct searches for FIPs coupled to SM particles via minimal couplings; such interactions involving new messenger fields are often referred to as portal interactions. A general interaction Lagrangian describing portal effective theories is given by $\mathcal{L} = \mathcal{L}_{\mathrm{SM}} + \mathcal{L}_{\mathrm{DS}} + \mathcal{L}_{\mathrm{portal}}$, where $\mathcal{L}_{\mathrm{portal}} = O_{n}^{\mathrm{SM}} J_{n}^{\mathrm{portal}}$, in which $O_{n}^{\mathrm{SM}}$ and $J_{n}^{\mathrm{portal}}$ denote the SM operators and portal currents, respectively~\cite{Arina2021}. MAPP-1's flagship benchmark scenario involves fermionic minicharged particles~\cite{demontigny2023minicharged,Kalliokoski2024} resulting from the kinetic mixing of a massless dark $U(1)$ gauge field with the SM hypercharge gauge field (i.e., the vector/kinetic portal, $\mathcal{L} \supset \epsilon_{Y} F'_{\mu \nu} B^{\mu \nu}$)~\cite{Holdom1986_1,Holdom1986_2}. Beyond this, the physics program of the MoEDAL-MAPP experiment comprises a host of additional FIP scenarios: minicharged strongly interacting dark matter~\cite{Kalliokoski2024,Kalliokoski2025}; heavy neutrinos with anomalously large electric dipole moments~\cite{FRANK2020135204}; long-lived dark photons~\cite{StaelensPhDThesis}, dark Higgs bosons~\cite{Popa2022,pinfold2023moedalmapp}, and axion-like particles~\cite{StaelensPhDThesis,Beltran2023}; inelastic dark matter~\cite{Bertuzzo2022}; heavy neutral leptons in the minimal gauged $B-L$ model~\cite{Deppisch2019,deppisch2023sterile} and the minimal ``$3+1$'' scenario~\cite{Beltran2023}; light neutralinos in $R$-parity-violating SUSY~\cite{Dreiner2021,Dreiner2023}; and sterile neutrinos in neutrino-extended SM effective field theory~\cite{deVries2021}. Several more studies of additional new physics scenarios potentially accessible to the MoEDAL-MAPP experiment are currently underway.

%% file: WG4/content/DarkSideExperiment.tex
\label{sec:darkside}
{Author: G. Grilli di Cortona}\\

The DarkSide program foresees the search for dark matter particles with masses of $\gtrsim10$~GeV/c$^2$ in the form of a weakly interacting massive particle by using argon dual-phase time projection chambers (TPCs) at the INFN Laboratori Nazionali del Gran Sasso (LNGS). It successfully operated a $50$~kg scale detector, DarkSide-50, between 2013 and 2019~\cite{DarkSide:2018ppu,DarkSide:2018bpj,DarkSide:2018kuk,DarkSide-50:2022qzh,DarkSide:2022dhx,DarkSide:2022knj,DarkSide-50:2023fcw}, while the next-generation detector, featuring a $50$-tonne TPC, DarkSide-20k, is currently under construction at LNGS~\cite{DarkSide-20k:2017zyg}.

\begin{figure}[t!]
    \centering
    \includegraphics[width=0.45 \textwidth]{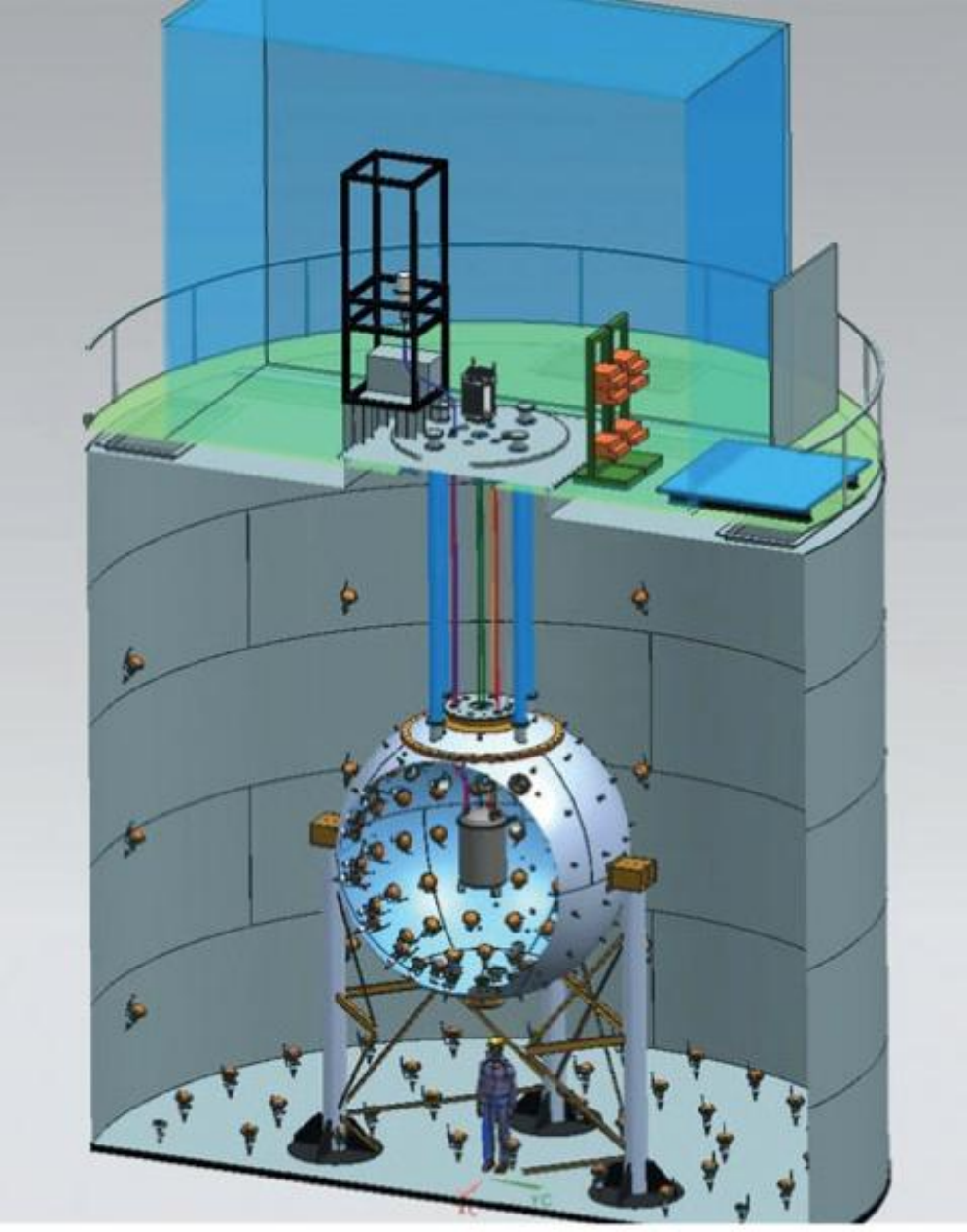}
     \includegraphics[width=0.45 \textwidth]{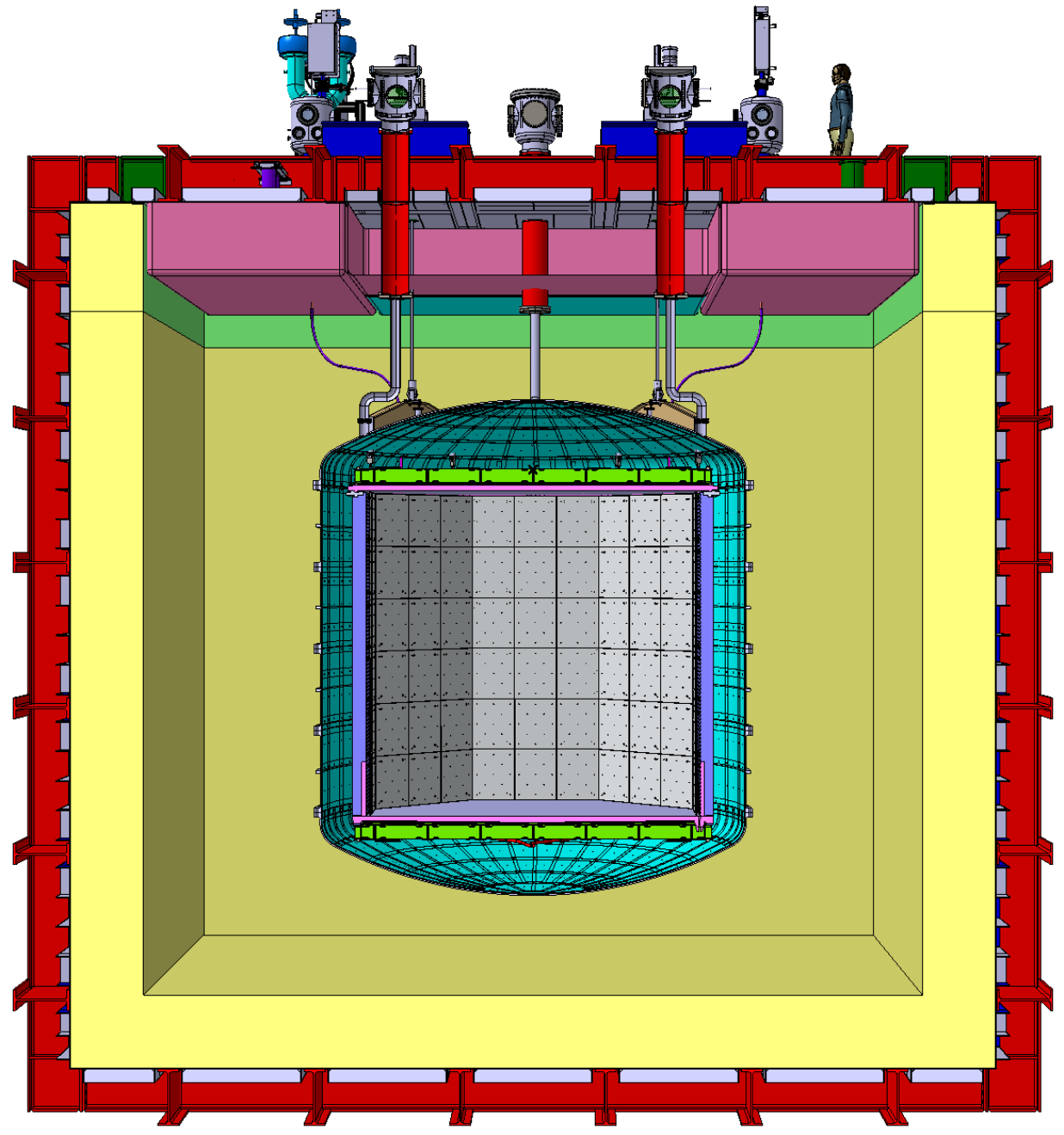}
    \caption{View of the DS-50 (\emph{left}) and DS-20k (\emph{right}) detectors. Figures reproduced from Refs.~\cite{Agnes:2016mgq} and \cite{DarkSide-20k:2024xlp}.}
    \label{fig:DS_exp}
\end{figure}

The DarkSide-50 TPC (see Fig.~\ref{fig:DS_exp} (left)) used an active mass of $(46.5\pm0.7)$~kg of low-radioactivity argon extracted from deep underground sources. The TPC was operated inside a double-walled cryostat, placed inside a $4$~m diameter sphere filled with borated liquid scintillator, acting as an anti-coincidence neutron veto. These two detectors were inserted into a $1$~kt water tank acting as a cosmic muon veto. The TPC was a $36$~cm tall polytetrafluoroethylene cylinder, with two fused silica windows coated with transparent top and bottom indium tin oxide electrodes. The active volume was observed from the top and bottom by two arrays of $19$ three-inch photomultiplier tubes. The TPC was immersed in a uniform $200$~V/cm electric field, and a $1$~cm thick layer of gas was created on top of the liquid. A strong extraction field was applied to the gaseous region. A particle interacting in the active volume induces scintillation pulses (S1 signal) and ionisation electrons that are drifted by the electric field towards the gas region, producing a secondary pulse of light (S2 signal) by electroluminescence.

While the primary goal of the experiment was to search for $\gtrsim10$~GeV/c$^2$ dark matter scattering on nucleons, the very low energy threshold for the ionization signal allowed the collaboration to set strong limits on dark matter with masses below $\sim10$~GeV/c$^2$~\cite{DarkSide:2018bpj,DarkSide-50:2022qzh,DarkSide:2022dhx,DarkSide-50:2023fcw}. Furthermore, DarkSide-50 was able to perform searches for dark matter particles interacting with electrons~\cite{DarkSide:2018ppu,DarkSide:2022knj}, including axion-like particles, dark photons, and sterile neutrinos. 

The DarkSide-20k detector aims to start operation in 2027 for a planned $200$ tonne-year exposure~\cite{DarkSide-20k:2017zyg}. The TPC, shown in the right panel of Fig.~\ref{fig:DS_exp}, will host $50$ tonnes of underground argon (UAr), surrounded by $32$ tonnes of UAr veto for neutron scattering. Both the TPC and the veto are instrumented with specially developed silicon photomultipliers for operating at cryogenic temperature, while gadolinium-loaded acrylic walls are being studied for the TPC, in order to moderate and capture neutrons that produce a WIMP-like signal after scattering in the TPC. A further veto surrounding the TPC, aiming at external neutrons and muons, consists of a $700$-tonne liquid atmospheric argon bath. 
The DarkSide-20k TPC will improve the sensitivity to WIMPs, dark photons, axion-like particles, and sterile neutrinos beyond those reachable by currently operating experiments~\cite{DarkSide-20k:2024yfq}.

%% file: WG4/content/DAMICM.tex
{Author: C. De Dominicis}\\

DAMIC-M (DArk Matter In CCDs at Modane) is a direct-detection dark matter experiment aimed at searching for light WIMPs (with masses below $10$~GeV/c$^2$) and hidden sector candidates, such as MeV-scale particles and eV-scale hidden photons, through their interactions with silicon nuclei and electrons within the bulk of skipper charge-coupled devices (CCDs)~\cite{arnquist2022damicmexperimentstatusresults}. While CCDs are commonly known for applications in imaging and astronomy, their potential in DM detection was first demonstrated by the DAMIC at SNOLAB experiment~\cite{PhysRevLett.125.241803,PhysRevD.105.062003,PhysRevLett.123.181802,PhysRevLett.118.141803}. Building upon the experience of DAMIC, DAMIC-M will significantly improve sensitivity to DM through the introduction of several innovations, including a new detector technology. 

DAMIC-M will be installed in early 2026 in the Modane underground laboratory (LSM), shielded by approximately $1700$~m of rock. The experiment will feature a total sensitive mass of about $700$~g (consisting of $208$ skipper CCDs with 6k $\times$ 1.5k pixels), which is $17$ times greater than that of DAMIC at SNOLAB. Unlike DAMIC CCDs, DAMIC-M devices feature single-electron charge resolution, thanks to the use of skipper amplifiers~\cite{janesick_2023_qyc1h-0qv62,chandler_2023_cbghm-z9888,PhysRevLett.119.131802}. 
The background rate is expected to be reduced from $10$ events/kg/day/keV (dru) to $\mathcal{O}(1)$ dru, while the dark current is foreseen to be as low as a few $10^{-4}$ $e^-$/pixel/day at an operating temperature of about $130$~K. Thanks to these features, DAMIC-M will achieve unprecedented sensitivity to hidden sector candidates with MeV-scale masses, probing both freeze-in and freeze-out scenarios as DM production mechanisms. A major improvement is also expected in the search for hidden photons with eV-scale masses, where DAMIC-M aims to lower the $90\%$ CL exclusion limit on the kinetic-mixing parameter by about three orders of magnitude compared to the previous DAMIC results. 

\begin{figure}[t!]
    \centering
    \includegraphics[width=0.85\textwidth]{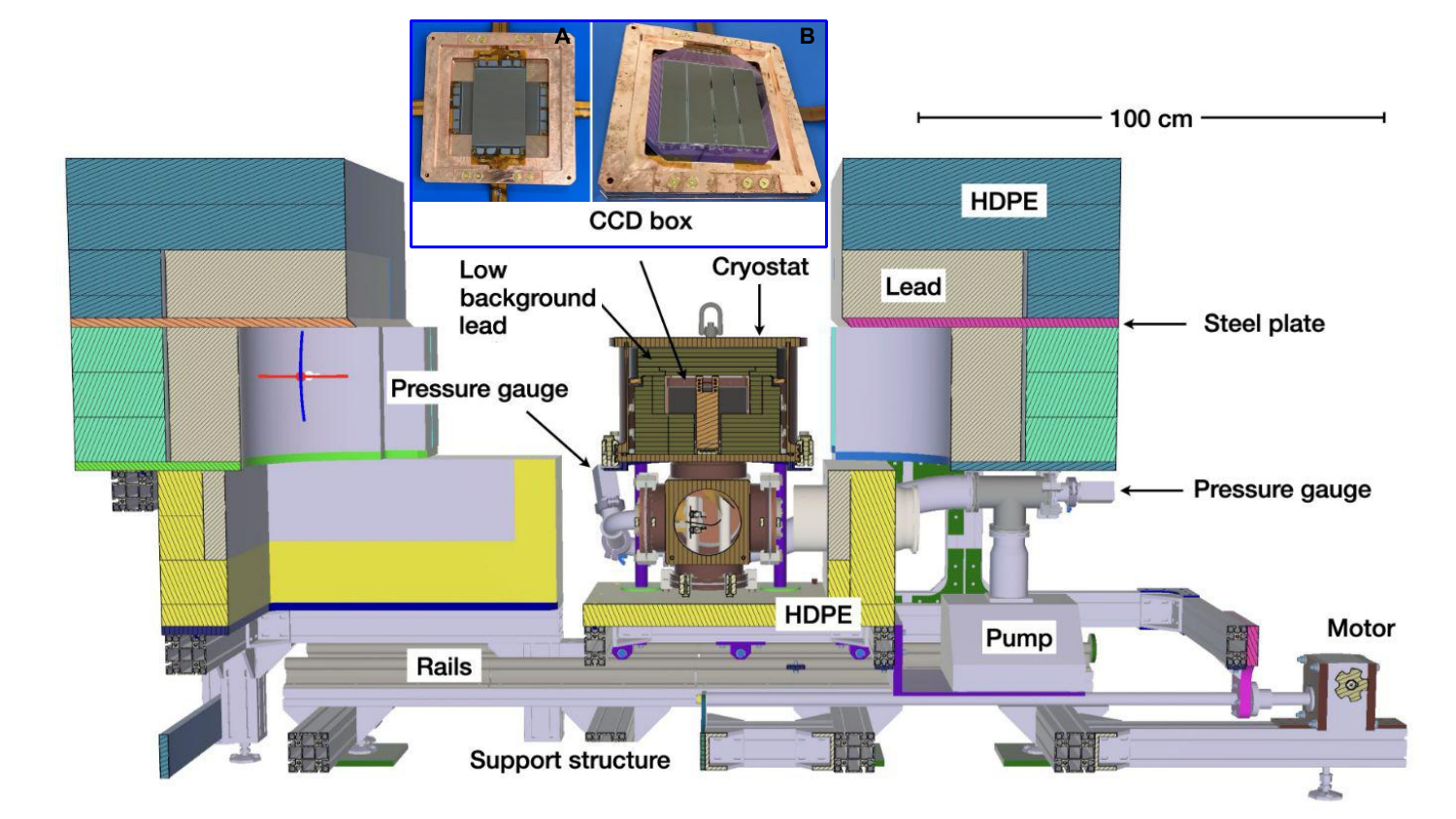}
    \caption{Cross-section of the LBC detector with major parts labelled. The cryostat is screened by external neutrons and gammas by a shield made of high-density polyethylene (HDPE) and lead.
    The CCDs are placed inside the cryostat in a Cu box surrounded by low-background lead (of which the closest $2$~cm are Roman lead).
    \textbf{Inset A:} The 6k $\times$ 4k~px skipper CCDs used to produce the first DAMIC-M dark matter limits. \textbf{Inset B:} Prototype modules consisting of $4$ CCDs with 6k $\times$ 1.5k px, installed in February 2023.
    The CCDs are typically operated at $T\sim 130$~K and $P\sim 5\times10^{-6}$~mbar. Figure reproduced from Ref.~\cite{arnquist2024damicmlowbackgroundchamber}}
    \label{fig:DAMICM_LBC}
\end{figure}

The potential of DAMIC-M has been demonstrated through its prototype, the Low-Background Chamber (LBC), which has been operational at LSM since early 2022~\cite{arnquist2024damicmlowbackgroundchamber}; Fig.~\ref{fig:DAMICM_LBC} illustrates its design. The LBC has served as an effective test bed for studying background levels, dark current, and electronic noise. Moreover, the LBC has yielded significant scientific results, with its data being used to set the first dark matter exclusion limits for DAMIC-M.

Operating with only two skipper CCDs (6k $\times$ 4k pixels), controlled by a commercial CCD controller, and with a dark current level of approximately $3\times10^{-3}$ $e^-$/pixel/day, the LBC set the world-leading exclusion limit at the time for hidden sector searches in the mass ranges of $[0.53, 1000]$~MeV/c$^2$ for ultralight mediators and $[0.53, 15.1]$~MeV/c$^2$ for heavy mediator interactions~\cite{PhysRevLett.130.171003,PhysRevLett.132.101006}.
Since then, several upgrades have been implemented, including the installation of two DAMIC-M prototype modules with eight skipper CCDs (6k $\times$ 1.5k pixels each), reduced background levels, lower electronic noise via custom electronics, and further suppression of dark current ($\sim1\times10^{-4}$ $e^-$/pixel/day), bringing the system closer to meet the DAMIC-M requirements.
Thanks to these improvements, the LBC has set leading constraints on hidden-sector particles over a wide range of masses between 1–1000~MeV/c$^2$, as well as the strongest limits to date on hidden photon dark matter in the 2.5–24~eV/c$^2$ mass window~\cite{DAMIC-M:2025}.
Most importantly, the LBC results rule out theoretically motivated benchmark scenarios in which hidden-sector particles constitute a major fraction of dark matter in the Universe through freeze-in or freeze-out production mechanisms.

%% file: WG4/content/SummaryPlotSection.tex
{Introduction authors: C. O'Hare \& C. Gatti}\\

\begin{figure}[t!]
  \begin{center}
    \includegraphics[totalheight=7cm]{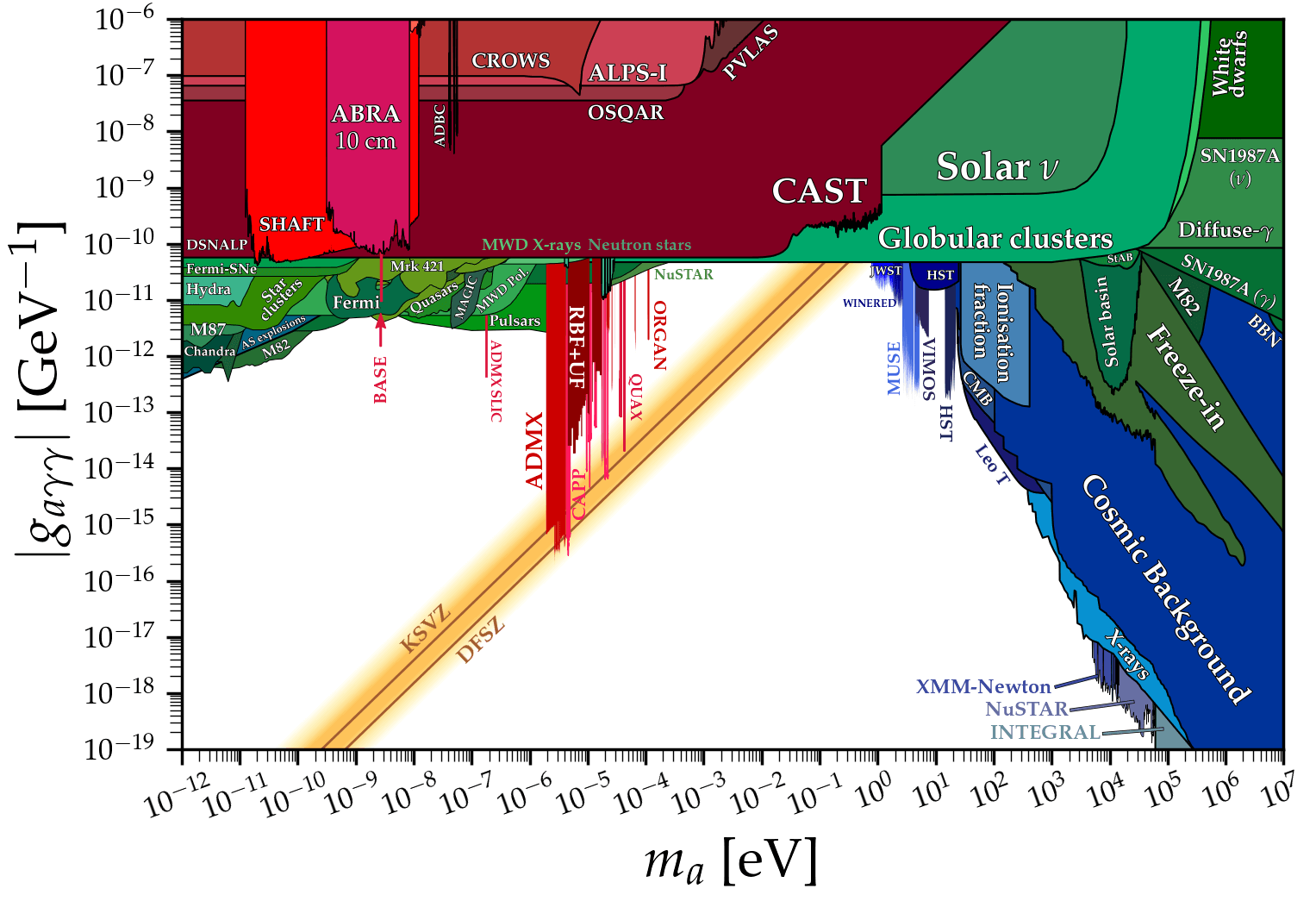}
    \includegraphics[totalheight=7cm]{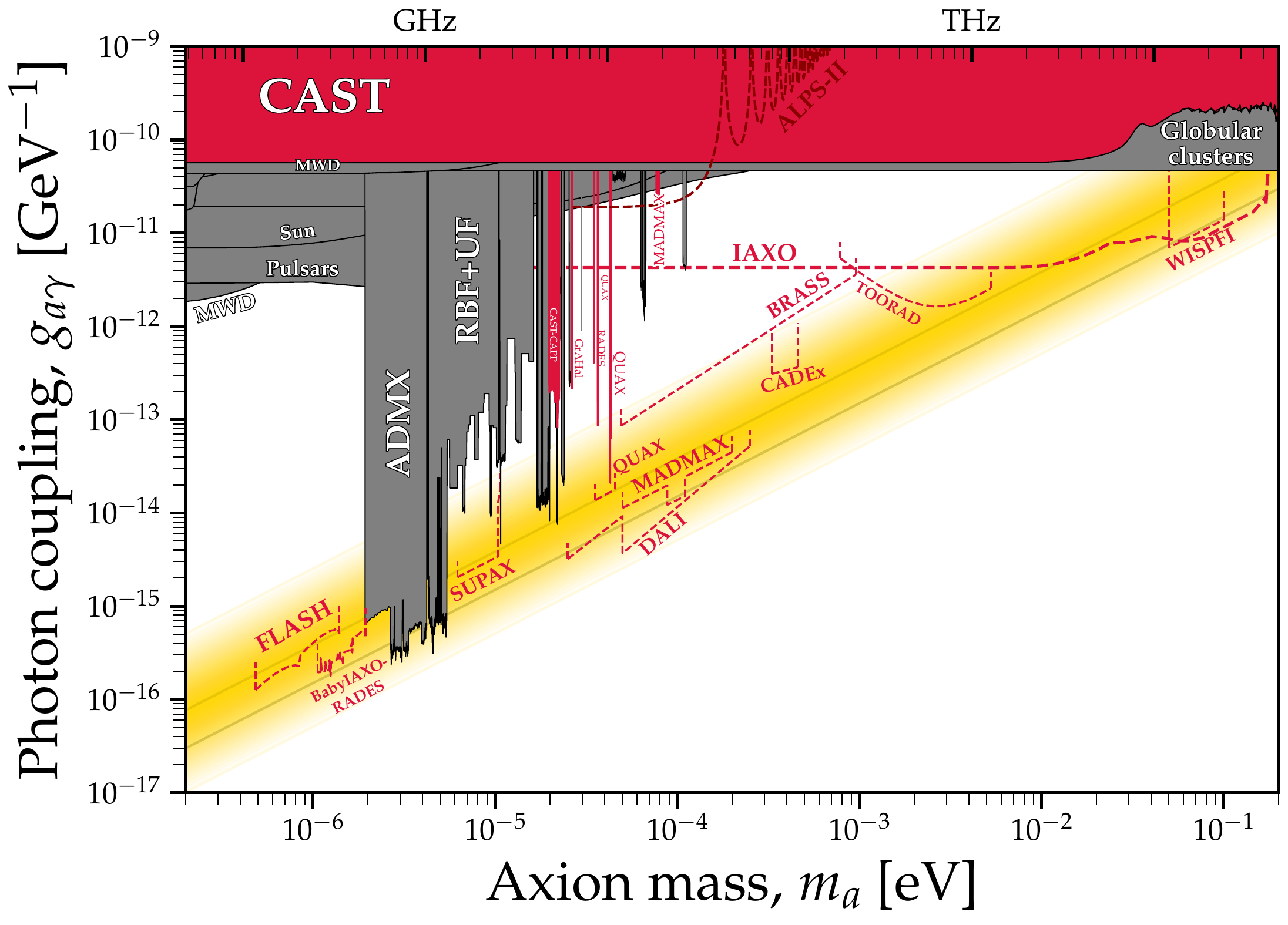}
    \caption{\textit{Top:} Present limits on $g_{a\gamma\gamma}$ coupling. \textit{Bottom:} Projected limits on $g_{a\gamma\gamma}$ coupling showing the contribution from European experiments. }
    \label{fig:AxionLimits-agg-present}
  \end{center}
\end{figure}

\begin{figure}[t!]
  \begin{center}
    \includegraphics[totalheight=7cm]{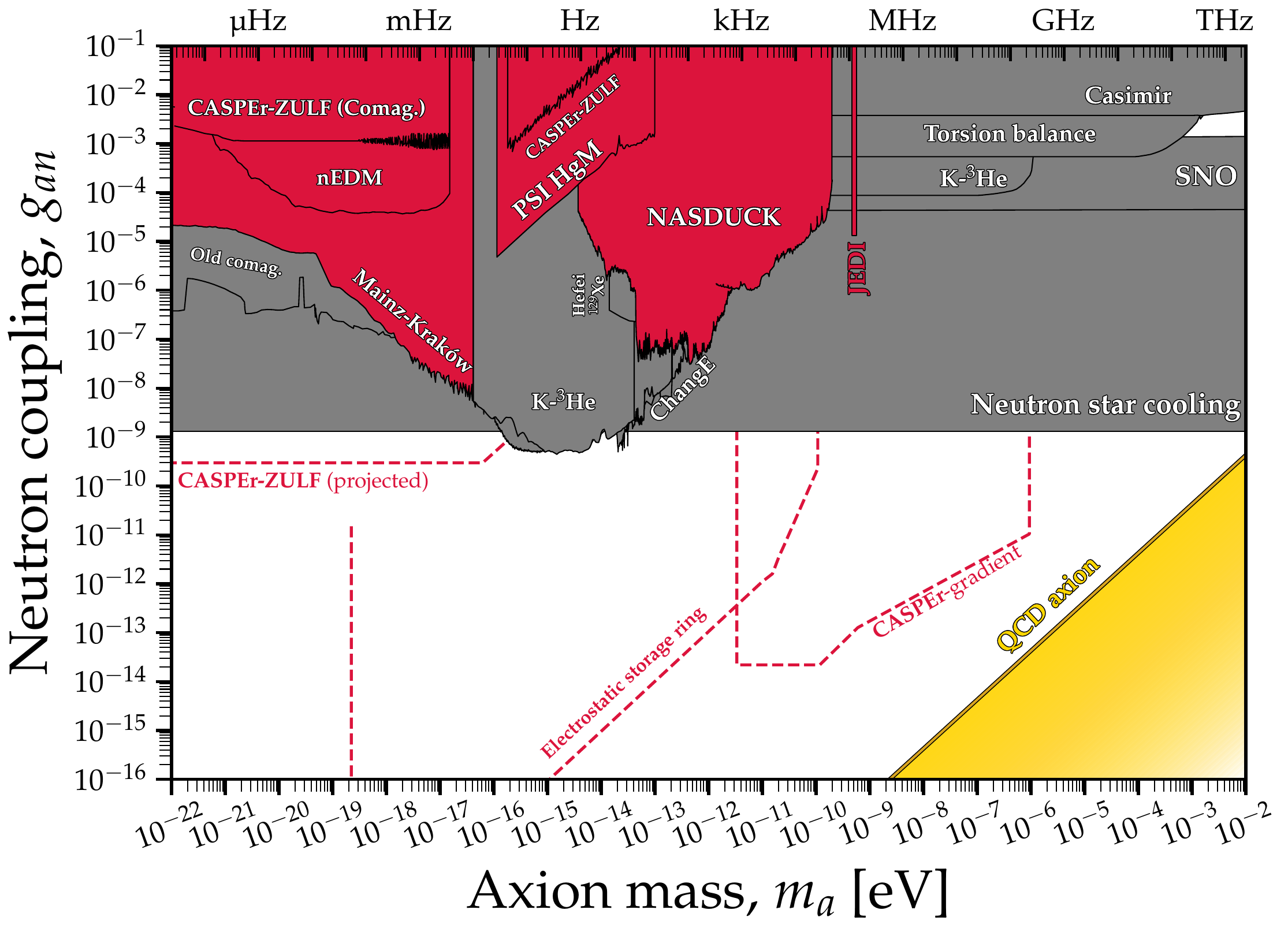}
    \includegraphics[totalheight=7cm]{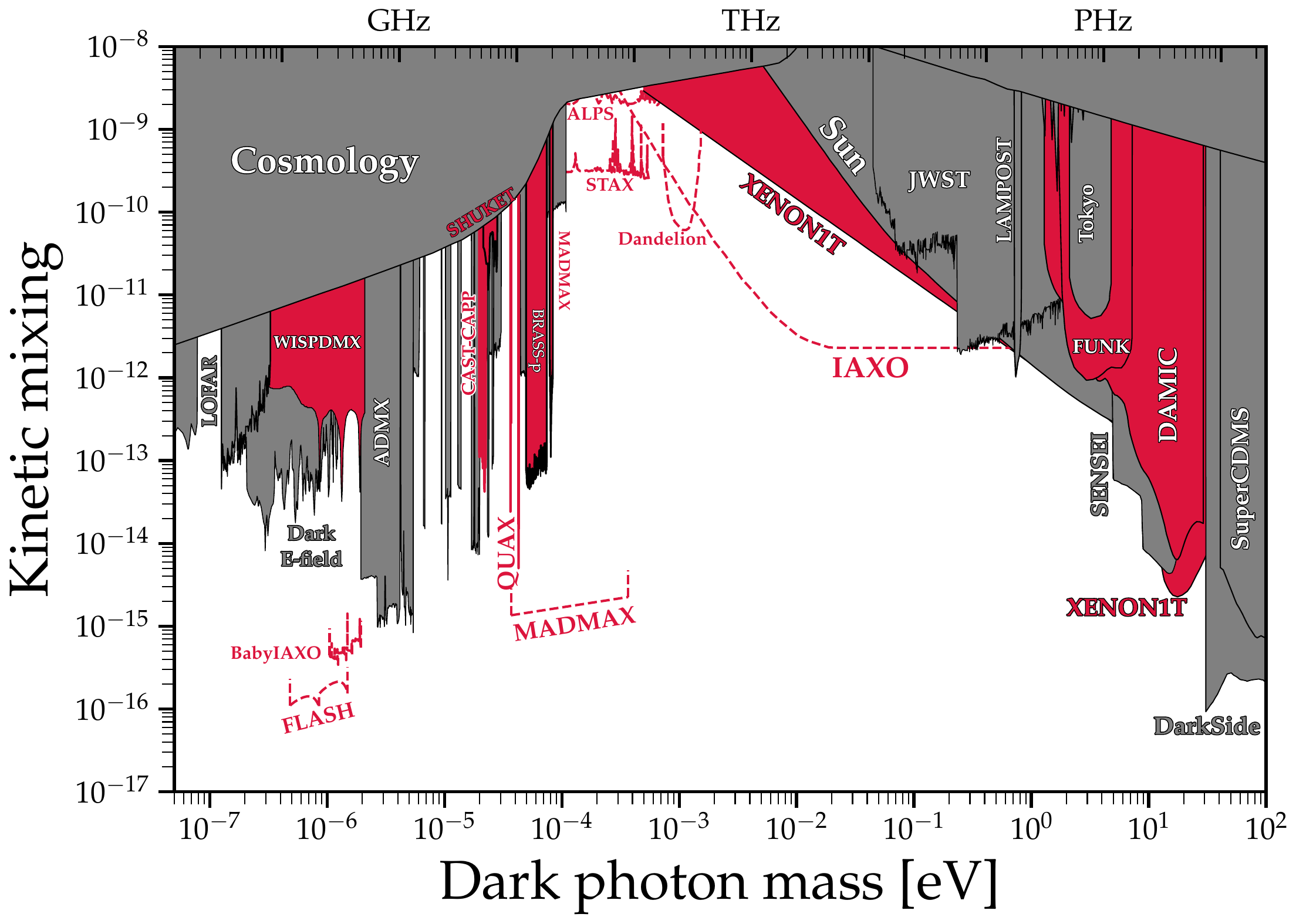}
    \caption{\textit{Top:} Projected limits on $g_{a\gamma n}$ coupling showing the contribution from European experiments.
        \textit{Bottom:} Projected limits on $\chi$ coupling for dark photons showing the contribution from European experiments. }
    \label{fig:AxionLimits-an-proj}
  \end{center}
\end{figure}

Despite a flourishing of new experimental activity in recent years, the parameter space motivated, in particular, for axion searches remains largely unexplored. The present status of probed region for the $g_{a\gamma\gamma}$ coupling is shown in Fig.~\ref{fig:AxionLimits-agg-present} (Top panel)~\footnote{The summary plots in this section are obtained from~\url{https://cajohare.github.io/AxionLimits/}~\cite{AxionLimits}.}, with only few experiments, haloscopes and helioscopes, that have the sensitivity to test QCD-axion models. The situation is rapidly evolving and significant progress is expected in the coming years, with European experiments that will make a substantial contribution in all mass regions for the 
$g_{a\gamma\gamma}$ coupling, shown in Fig.~\ref{fig:AxionLimits-agg-present} (Bottom panel),
the $g_{a\gamma n}$ coupling in Fig.~\ref{fig:AxionLimits-an-proj} (Top panel), and the dark-photon mixing parameter $\chi$ in Fig.~\ref{fig:AxionLimits-an-proj} (Bottom panel). In each plot, the European contribution is highlighted in solid red for existing limits, and with dashed red-line for projections.
The development of many WISP experiments, characterized by different techniques, goes hand in hand with as many technological developments in many fields ranging from superconducting magnets, RF cavities and optical elements for high precision interferometry to atomic and molecular systems and quantum sensors.
Examples include the following activities conducted at European institutions:
AITANA - IFIC (University of Valencia - CSIC) is developing resonant cavities for haloscopes~\footnote{\url{https://aitanatop.ific.uv.es/aitanatop/}};
The University of Bonn is developing superconducting cavities for haloscopes~\cite{Schmieden:2024wqp,Schneemann:2023bqc}~\footnote{\url{https://www.pi.uni-bonn.de/schott}};
The Technical University of Cartagena (UPCT) is developing resonant cavities for haloscopes~\footnote{\url{https://www.upct.es/grupos/gem/}};
The COLD Lab (LNF) is developing quantum superconducting devices~\cite{Rettaroli:2024abf,Moretti:2024xel,Gatti:2023pou,DElia:2023baj,DElia:2024pax,Paolucci:2020woi} and resonant cavities for haloscopes~\cite{DiGioacchino:2019coj,ALESINI2021164641}~\footnote{\url{www.coldlab.lnf.infn.it}};
DESY is developing high-precision optical interferometry~\cite{Kozlowski:2024jzm}, optical quantum sensing~\cite{Gimeno:2023nfr} and nano-membrane based sensing~\cite{Reinhardt:2023cds}~\footnote{\url{https://alps.desy.de}};
The Instituto de Astrofísica de Canarias (IAC) \& University of La Laguna (ULL)  is developing optical and microwave components and kinetic inductance detectors~\footnote{\url{https://iac.es/en/science-and-technology}};       
The Max Planck Institute for Physics in Garching (Germany) is developing rf cavities~\cite{Golm:2023iwe} and single photon detectors~\cite{quantera}~\footnote{\url{https://www.mpp.mpg.de/en/research/rades-detector}};
The Laboratoire PhLAM (CNRS) is developing clocks and experiments with atoms and molecules~\cite{CONSTANTIN2016VibrationalSpectroscopy,Constantin2020IFCSEFTF,Constantin2023CLEOEuropeEQEC,Constantin2023EFTF,CONSTANTIN2020Atoms}~\footnote{\url{https://orcid.org/0000-0002-1942-9662}};  
The Padua-Saclay Team is developing quantum superconducting devices~\cite{PhysRevX.15.021031};
The RadioAxion Team is developing low background set-up and high rate and low dead time acquisition~\footnote{\url{https://www.pd.infn.it/eng/radio-axion/}}.

Magnets play a primary role in the search for axions~\footnote{A dedicated sessions on magnets for WISPs searches took place during the first Cosmic WISPer Technology Forum~\url{https://agenda.infn.it/event/33570/}}. Developments in Europe for high magnetic field science and technology are driven by the European Magnetic Field Laboratory (EMFL) described in Sec.~\ref{sec:EMFL}.  At CERN the EP Magnet working group has the responsibility of maintenance, operation, and troubleshooting of (superconducting) detector magnets as well as the participation in technology development. The BabyIAXO magnet has been designed under the leadership of CERN-EP, and recently reviewed by an external expert panel. After some delay due to the non-availabitlity of superstabilized superconducting conductor, first steps toward its construction are being taken, in particular regarding cable procurement, for which the project counts with experts from CERN-TE. The magnet will feature a average 2 T dipole field along its two 10-m long, 70-cm diameter bores. CEA, with a facility for superconducting magnets (MATTRICS - Magnet Technology Testing Research InfrastruCtureS) designed to test and qualify new-generation prototype of conductors and magnets particularly for fusion applications, is designing the dipole magnet for the MADMAX experiment.  We mention here also IRIS the largest Italian project dedicated to the development of applied superconductivity~\cite{10504893}. 
Large research laboratories host cryogenic infrastructures to cool superconducting magnet, experiment components and accelerators complexes (see Sec.~\ref{sec:DESY} and~\ref{sec:LNF}). Dilution refrigerators are widely used in WISP experiments, such as Haloscopes, that require very low electronic noise. European companies developing such systems are Leiden Cryogenics, Bluefors, Oxford Instruments, Entropy, ICEoxford, CryoConcept. Dilution refrigerator are needed also for operating superconducting quantum devices~\footnote{A dedicated sessions on Quantum Sensors for WISPs searches took place during the second Cosmic WISPer Technology Forum~\url{https://indico.desy.de/event/42137}}. Several companies are active in this sector in Europe, like QuantWare, Siletwave, ez SQUID, Magnicon and Supracon AG.

The variety of experiments and experimental approaches requiring different technologies benefits from a large number of laboratories able to host them and provide adequate infrastructures. Some of the labs supporting WISP experiments are: 
the Laboratorio Subterr\'aneo Canfranc (LSC) that hosts CADEX and RADES-LSC experiments~\footnote{\url{https://lsc-canfranc.es/en/home-2/}}; 
CERN that hosts OSQAR, VMB, MADMAX protoype and CAST~\footnote{\url{https://home.cern}}; 
DESY that hosts ALPS, BabyIAXO and MADMAX and provides cleanrooms for optical precision interferometry, cryogenic laboratory, a cryoplatform and underground halls~\footnote{\url{https://alps.desy.de}}; 
LNF that hosts QUAX@LNF, FLASH and PADME and previously KLOE and Nautilus, providing cryogenics, power-laser and a Test Beam Facility~\footnote{\url{www.lnf.infn.it}}; 
LNGS that hosts DARKSIDE, GERDA, RadioAxion, \mbox{RES-NOVA}~\cite{Pattavina:2020cqc} and XENON~\footnote{\url{www.lngs.infn.it}};
LNL  that hosts QUAX@LNL, QUAX-$ae$, QUAX-$g_pg_s$ and previously AURIGA~\footnote{\url{www.lnl.infn.it}};
Laboratoire Souterrain de Modane (LSM) that hosts  DAMIC-M and EDELWEISS~\footnote{\url{https://lpsc.in2p3.fr/?page_id=547}};
At Grenoble, participating to GrAHal  there are the National Laboratory for High Magnetic Fields~\ref{sec:EMFL}, LPSC and the Neel Institute~\footnote{\url{https://grahal.neel.cnrs.fr}}; 
The Helmoltz Institute Mainz that hosts CASPER-gradient and SUPAX~\footnote{\url{https://www.hi-mainz.de}};
Instituto de Astrof\'isica de Canarias (IAC) that  hosts DALI, MAGIC, CTA, the La Palma (Roque de Los Muchachos) and Tenerife (Teide) observatories, along with labs and workshops at the headquarters equipped with cryogenics, optomechanics, EMC rooms~\footnote{\url{https://www.iac.es}}; PSI hosts nEDM~\footnote{\url{https://www.psi.ch/en/research/center-labs}}; GEO600~\footnote{\url{https://www.geo600.org}}. 
In the following sections we provide more details about some of the laboratories and their infrastructures and how they may contribute to WISP research in the coming years.

%% file: WG4/content/EMFLFacility.tex
High magnetic field and flux are one of the key experimental parameters for axion search. Developments in Europe for high magnetic field science and technology are driven by the European Magnetic Field Laboratory (EMFL)~\footnote{\url{https://emfl.eu/}} officially created in 2014 by three founding members, namely the Dresden High Magnetic Field Laboratory (HLD, Germany), the High Field Magnetic Laboratory in Nijmegen (HFML, The Netherlands), and the Laboratoire National des Champs Magnétiques Intenses (LNCMI, France)~\footnote{As an example of major breakthrough made at LNCMI-Grenoble when it was associated to the Max Planck Institute within the GHMFL, concerns the discovery in 1980 of the integer quantum Hall effect by Klaus von Klitzing, awarded by the Nobel prize in Physics in 1985.}. 
LNCMI is a French research unit from CNRS (Centre National de la Recherche scientifique) located on two sites and created in 2009 by merging the French part of the Grenoble High Magnetic Field Laboratory (GHMFL) for DC fields and the pulsed magnetic field laboratory in Toulouse (LNCMP, Laboratoire National des Champs Magnétiques Pulsés). These three EMFL founder laboratories are the only ones in Europe with the require infrastructures to achieve the highest possible magnetic fields for applied and basic research, meaning magnetic fields that cannot be obtained with commercial apparatus, for both DC and pulsed energization. In addition to developing its own research program for the development and use of high magnetic fields for basic and applied sciences, these laboratories also provide via EMFL access to their high field infrastructures to user scientists coming from all around the world. Since 2014, several other institutes have joined EMFL. In 2015, the University of Nottingham became EMFL member representing the high field community in the UK. In 2019, the University of Warsaw representing the high field community in Poland joined EMFL as well as CEA-IRFU (France). In 2024, Italy became also a member of EMFL via the University of Salento in Lecce. 

A summary of the maximum magnetic fields that can be reached in EMFL laboratories from non-destructive resistive technologies~\footnote{\url{https://emfl.eu/find-experiment-old/}; \url{https://www.ru.nl/hfml/use-our-facility/magnet-specifications/}} are given in Tab.~\ref{tab:emfl1}. Destructive pulsed fields are also produced in Toulouse on the microsecond scale to go beyond 200~T. For DC magnetic fields, high electrical power installation equipped by water cooled circuits are required. The maximum power that can be delivered on one site is equal to 30~MW at Grenoble and 21~MW at Nijmegen. For pulsed magnetic fields, capacity banks provide the required electrical energy that can reach on one site 15 and 9.5~MJ at Toulouse and Dresden, respectively (Tab.~\ref{tab:emfl1}).

\begin{table}[tbhp]
\renewcommand{\arraystretch}{1.5}
  \begin{center}
    \begin{tabular}{c|cccc}
    \hline
    \hline
	& $B_{\rm max}$ 	& Warm diameters 	& Duration 	& Experimental sites\\[-.5em]
    & ($T$) & (mm) & (ms) & \\
    \hline
    Dresden&	51--92	&24--12&	$<$ 25--10	&2@1.4 MJ + 2@8.6 MJ + 2@9.5 MJ\\
    Toulouse&	55--90	&28--8&	$<$ 87--30	&2@5 MJ + 1@ 6MJ + 2@15 MJ\\
    Nijmegen&	30--38	&50--32&	DC	&2@21 MW + 3@17 MW$^\ast$\\
    Grenoble&	10--37&	376--170--50--34&	DC	& 2@30 MW + 2@17 MW$^\ast$\\
      \hline\hline
    \end{tabular}
    \caption{DC and non-destructive pulsed magnetic fields produced by resistive coils in EMFL founding laboratories. ($^\ast$) Due to electrical power limitation, only one site can run at a given time, the other ones are in preparation phase }
    \label{tab:emfl1}
  \end{center}
\end{table}

In addition to resistive magnets, EMFL laboratories provide also access to magnetic field produced by superconducting and hybrid magnets, the latter being a combination of resistive and superconducting technologies (Tab.~\ref{tab:emfl2}).

\begin{table}[tbhp]
\renewcommand{\arraystretch}{1.5}
  \begin{center}
    \begin{tabular}{c|cccc}
    \hline
    \hline
	& $B_{\rm max}$  ($T$)	&Diameters (mm)&	Sample $T$ ($K$) &	SC Magnets\\\hline
    Toulouse	&16	    &34 &	DR : $8\times10-3$  – 1	&1 \\
    Nijmegen	&16	& 32  \& 40 @ 4.4 $K$	& VTI : 1.5 – 300	&2\\
    Grenoble	&9--18	& 810 @ 300 $K$ – 50$^\ast$ &	$^\ast$VTI : 1.2 – 300	& 10\\
      \hline\hline
    \end{tabular}
    \caption{DC magnetic fields produced by superconducting electromagnets in EMFL founding laboratories. The 9~T in 810~mm diameter configuration of LNCMI-Grenoble is obtained from the superconducting outsert of the hybrid magnet when used alone. DR = Dilution refrigerator, VTI = Variable temperature insert.}
    \label{tab:emfl2}
  \end{center}
\end{table}

\begin{figure}[t!]
  \begin{center}
    \includegraphics[totalheight=11cm]{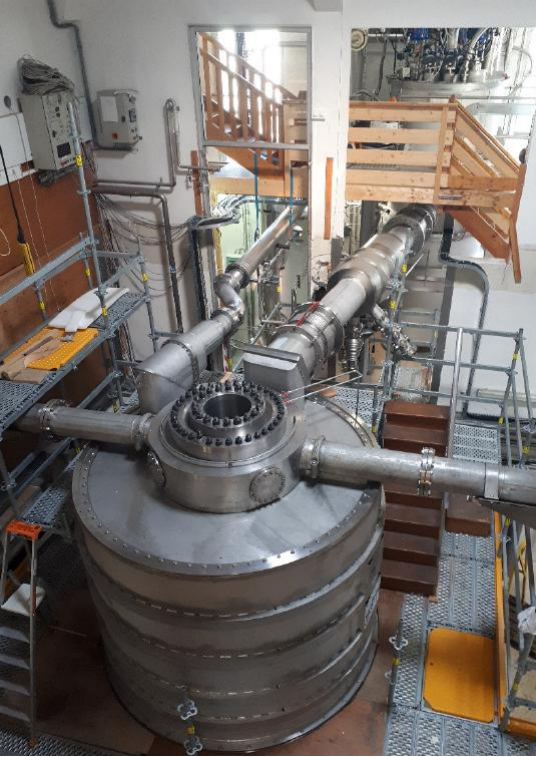}
    \caption{View of the hybrid magnet assembly of 52 tons combining resistive and superconducting magnets. The central part contains resistive inserts with the water cooling pipes on both sides of the water box located on the top of the superconducting magnet cryostat. The cryogenic satellite producing the superfluid He can be seen in the background. It is connected to the magnet cryostat via a cryogenic line in parallel to the quench line. Figure reproduced from Ref.~\cite{10359164}.}
    \label{fig:HybridMagnet}
  \end{center}
\end{figure}

LNCMI-Grenoble is also equipped with a modular hybrid magnet user platform combining resistive and superconducting technologies (Fig.~\ref{fig:HybridMagnet}). It will deliver the various magnetic fields configurations listed in Tab.~\ref{tab:emfl2} including a maximum field of 43~T in 34~mm diameter. Thanks to the recent upgrade of the electrical power installation of LNCMI-Grenoble from 24~MW up to 30~MW, higher magnetic field configurations is being considered. This new modular hybrid magnet is being successfully put in operation~\cite{10359164} with 42~T field reached recently as a first step (8 Nov. 2024)~\footnote{\url{https://www.inp.cnrs.fr/fr/cnrsinfo/laimant-hybride-du-lncmi-atteint-42-tesla}}.

A second European hybrid magnet is being built at Nijmegen to deliver a 45~T magnetic field in a 32~mm warm bore~\footnote{\url{https://www.ru.nl/en/hfml-felix}}.

\begin{table}[t!]
\renewcommand{\arraystretch}{1.5}
  \begin{center}
    \begin{tabular}{ccc}
    \hline
    \hline
    Central Field &	Warm/user bore diameter &	Electrical power for cryogenics + \\[-.5em]
     ($T$) & (mm) & water cooling + resistive magnets (MW) \\\hline
    43&	34	&0.4 + 1 + 24\\
    40& 	50	&0.4 + 1 + 24\\
    27& 	170&	0.4 + 0.75 + 18\\
    17.5& 	375&	0.4 + 0.5 + 12\\
    9& 	812&	0.4 \\
      \hline\hline
    \end{tabular}
    \caption{
    DC magnetic field configuration of the Grenoble hybrid magnet user platform.}
    \label{tab:emfl3}
  \end{center}
\end{table}

%% file: WG4/content/DESYFacility.tex
The North Hall of the former HERA accelerator complex at DESY has recently been partially renovated to accommodate the ALPS~II experiment in the HERA tunnel. At the same time, the old cryogenic infrastructure has been extensively restored and is fully operational since 2022. Furthermore, a state-of-the-art cryogenic platform, known as the ALPS Cryo-Platform Sub-cooler Box (ACPS), is currently being constructed in the HERA North Hall. The ACPS will be able to distribute supercritical helium to up to three different experiments (as it has three cold terminals) while maintaining a stable temperature of 4.5~K and a single-phase state. It will also facilitate the supply of pressurized cold and warm helium gases to the experiments. In addition, the HERA North Hall provides various facilities to support the experiments: 
\begin{itemize}
    \item The main hall has an area of 43\,m $\times$ 25\,m and is located about 30 meters underground. The height is 16\,m (the working height defined by the crane's hook height).
    \item The crane infrastructure consisting of two 40-ton cranes, a 40-ton hovercraft (dimensions: 6\,m $\times$ 9\,m) and a 13-meter hook height for moving equipment parts.
    \item The existing iron yoke of the magnet for the former H1 experiment is foreseen to host the MADMAX magnet.
    \item Working spaces for three experimental sites is available in the hall.
\end{itemize}

\begin{figure}[hbt!]
  \begin{center}
    \includegraphics[width=0.8\linewidth]{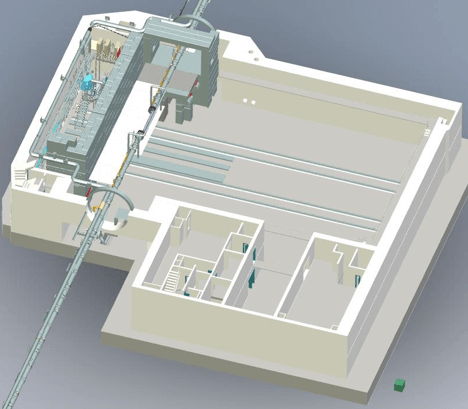}
    \caption{The HERA North experimental hall right of the accelerator tunnel hosting the ALPS\,II experiment. The new helium distribution box ACPS is shown in blue left of ALPS\,II. Reproduced with permission from DESY.}
    \label{fig:hera-north}
  \end{center}
\end{figure}

A layout of the experimental hall is sketched in 
Fig.\,\ref{fig:hera-north}. At present it is planned to put the cryoplatform into operation around 2028.

%% file: WG4/content/LNFFacility.tex
The Frascati National Laboratory (LNF) is the largest and the first built of the four laboratories that the Italian Institute for Nuclear Physics (INFN) owns. Actually, the Institute was founded, almost seventy years ago, with the primary intention to provide Italy with its first particle accelerator: an electron-synchrotron of 1.1 GeV energy. Since then, the research activities carried out at LNF have always been in the fields of high energy physics, accelerator operation and development, and interdisciplinary research, with a perfect balance between internal activities, taking place on site, and external ones conducted in the major laboratories all over the world. LNF stands over an area of 135.000 m$^2$ of which 56.000 are indoor and includes offices, laboratories and service areas.
The following infrastructures are available:
\begin{itemize}
\item DA$\Phi$NE, an e$^+$ e$^-$ collider, unique in Europe, operated at the $\phi$ energy and able to deliver instantaneous luminosities $\mathcal{L} \sim 2 \times 10^{32}$ cm$^{-2}$ s$^{-1}$;
\item  DA$\Phi$NE-light, a synchrotron light laboratory, housing several synchrotron radiation lines extracted from the electron ring of DA$\Phi$NE~in the soft-X and infrared range;
\item a Beam Test Facility, BTF, an experimental area equipped for detector and beam diagnostic tests. Here, two beam lines, extracted from the DAFNE LINAC, can provide beams of electrons, positrons, photons of variable intensity and energy;
\item SPARC\_LAB, a complex hosting a photo-injector that can produce high brightness electron beams up to 170 MeV (SPARC) which feeds a 12 m long undulator for FEL generation, and a laser (FLAME) of power $\sim$ 200 TW. The SPARC\_LAB is an infrastructure for R\&D on new techniques of particle acceleration and for interdisciplinary studies, including PWFA and LWFA experiments, TeraHertz radiation and a Compton source.
\item Large assembly halls with several clean rooms (for a total surface of about 480 m$^2$) equipped with special tools for designing and building large and complex experimental equipment.
\item The Cryogenic Laboratory for Detectors (COLD) is equipped with dilution refrigerators and 4K cryostats, some of which house NbTi magnets up to 9~T, RF instruments including a quantum control system for controlling and readout the superconducting qubits.
\end{itemize}
With the EuPRAXIA project a new machine will be built, with the ambitious goal of becoming the first user facility that exploits plasma acceleration. The LNF on-site fundamental physics program is instead oriented towards the search for WISPs with the experiments QUAX@LNF and PADME and with the up-coming FLASH~\cite{universe7070236}, a path inaugurated by experiments such as KLOE~\cite{Bossi_2012} and Nautilus~\cite{BASSAN201652}.

%% file: WG4/content/IntroNewSchemes.tex
{Introduction author: D. Aybas}\\

In addition to ongoing experimental efforts, new proposals aim to use unexplored paradigms in searches for WISPs. The multitude of the possible effects of axions or dark photons (or WISPs, in general) on ordinary particles and known fields leads to a diversity in approaches that new experiments can take. In addition to the experiments detailed in this chapter, other new experiments propose to use media and sensors such as piezoelectric crystals, resonant-mass detectors, force sensors, magnetized plasma, intensity interferometry, dielectric absorbers, antiferromagnets, ferroelectric materials, artificial magnetoelectric materials, or cavities (optical, microwave, or generally radio-frequency).

Spontaneous parity violation in a piezoelectric crystal due to axions can be detected with electroaxionic and piezoaxionic effects~\cite{Arvanitaki:2021wjk}, or with ferroaxionic effect~\cite{Arvanitaki:2024dev}. Since gravitational coupling of dark matter is guaranteed to exist, one proposal suggests the use of resonant-mass detectors that oscillate at a frequency equivalent to the dark matter mass~\cite{Arvanitaki:2015iga}. Another proposal is suspended pendula acting as a mechanical gravitational sensor for dark matter~\cite{Carney:2019pza}. In unstable magnetized plasmas, axion-plasmon polariton might be formed, which might then be converted into detectable photons~\cite{PhysRevD.101.051701}. With a microwave beam directed at it, forcing an axion to decay and listening to the echo of the beam is another experimental suggestion~\cite{PhysRevLett.123.131804}. A resonant cavity is another medium to search for axion dark matter decay, possibly enhanced with Purcell effect~\cite{ahmad2025resonantenhancementaxiondark}. Hanbury Brown and Twiss intensity interferometry can be used as a broadband search for quadratic coupling of axions~\cite{PhysRevD.108.015003}. A disordered dielectric absorber is proposed as a broadband axion detector that can smooth over any resonance effects due to its powdered structure with a scalable setup~\cite{koppell2025darkmatterhaloscopedisordered}. In an antiferromagnet, such as nickel oxide, axion absorption and conversion into magnon modes make it possible to perform narrow and broadband searches~\cite{Catinari:2024ekq}. Axion-induced shift current in ferroelectric materials generated by high mass axions would be in the detectable lower-frequency range~\cite{kondo2025broadbandsearchaxiondark}. Artificially strained magnetoelectric materials offer a new method to measuring an external axion field through induced changes in the magnetization of the material with spontaneous symmetry breaking~\cite{lei2025emergingaxiondetectionartificial}. Faraday rotation, where optical polarization rotates with the magnetic field in the environment, was also suggested as a search for electromagnetic coupling of axions inside an optical cavity~\cite{PhysRevD.106.115017}. Inside a resonant microwave cavity, a sample that exhibits quantum Hall effect can be used as a narrowband detector for axions~\cite{iwazaki2025detectiondarkmatteraxions}. If copper walls typically used in a cavity (such as ADMX and CAPP) are replaced with carbon-based conductors, higher axion sensitivity in a wider frequency range can be achieved~\cite{hong2025probingaxionelectroncouplingcavity}. A superconducting radio-frequency cavity can be useful in searching for dark photon kinetic mixing~\cite{PhysRevD.102.035010}.

This is an exciting time to be an experimental physicist with expertise in any type of a precision sensor. As exemplified above, there are a continuously widening range of possibilities in experimental searches for WISPs that one can decide to undertake.

%% file: WG4/content/MolSpecExp.tex
\begin{figure}
\centering
\includegraphics[width=1.\textwidth]{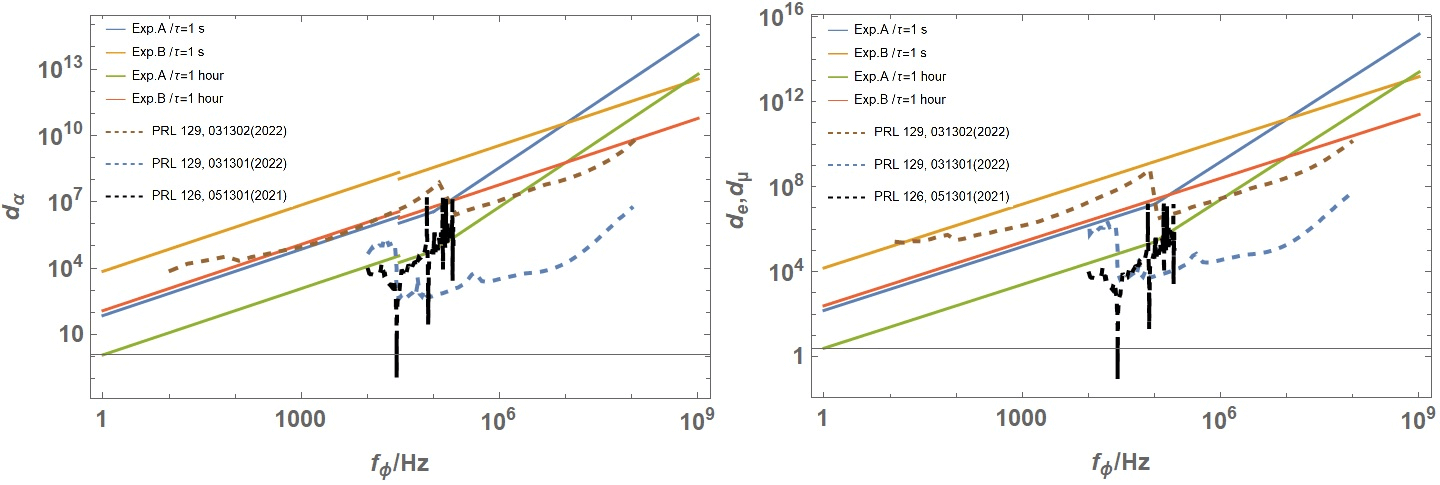}
\caption{Bounds from the REPHYSPECMO experiment for UBDM field couplings to (\emph{left}) $\alpha$ and (\emph{right}) $\mu=m_p/m_e$ (averaging time: $\tau=1$ s red line, $\tau=1$ h magenta line). Area above each dependence indicates the excluded range. For comparison, both plots include bounds set on UBDM couplings on $\alpha$ and electron mass $m_e$ from other spectroscopy experiments in Refs.~\cite{PhysRevLett.129.031301,PhysRevLett.129.031302}.}
\label{DMbounds_REPHYSPECMO}
\end{figure}

The REPHYSPECMO (\textit{Recherche d’une Physique au-delà du Mod\`ele Standard par Spectroscopie Mol\'eculaire}) project \cite{Constantin2023EFTF} proposes to search for 0-spin, sub-eV mass dark matter by precision molecular spectroscopy. This ultralight bosonic dark matter (UBDM) behaves as a coherent classical field oscillating at its Compton frequency with an  amplitude related to the dark matter local density. UBDM couples to the Standard Model (SM), induces oscillations of the fundamental constants (FC), that translate into oscillations of the frequencies of atomic and molecular transitions and of the lengths of solids \cite{antypas2022}.
\newline This experiment monitors time dependence of a laser absorption spectroscopy signal by 1542 nm from gas-phase acetylene that informs on the coupling coefficients of the UBDM field to the fine structure constant and the proton-electron mass ratio. The laser, frequency-stabilized and continuously measured against a hydrogen maser at Paris Observatory \cite{Xie:17}, is disseminated by the REFIMEVE optical network to remote geographical locations through phase-stabilized optical fibre links \cite{Cantin:2021}. The laser is suitably frequency-tuned at PhLAM laboratory with an optical modulator to probe  
the P(27) transition of the $v_1+v_3$ band of $^{12}$C$_2$H$_2$ by (A) Doppler-free saturated absorption, and (B) Doppler-broadened linear absorption. This approach ensures sensitivity to FC oscillations in the 1 Hz – $10^9$ Hz frequency domain and fractional frequency uncertainties at 1 s averaging time estimated at $10^{-14}$ for the (A) setup and at $10^{-12}$ for the (B) setup. These uncertainties combined with estimated experimental response functions and molecular sensitivity coefficients \cite{CONSTANTIN2016VibrationalSpectroscopy} are exploited to estimate bounds of the couplings of an UBDM field associated to the galactic DM halo \cite{UBDM:2023} to the fine structure constant $\alpha$ or to the proton-electron mass ratio $\mu=m_p/m_e$, shown in Fig.~\ref{DMbounds_REPHYSPECMO}. 

Molecular spectroscopy enables sensitivity to the DM couplings to the strong interaction and open access at frequencies higher that those addressed in atomic clock-based dark matter searches.

%% file: WG4/content/RydbergDarkPhotons.tex
An experiment based on the application of an electric field inside a microwave cavity and electrometry using Rydberg atoms to detect dark photons (DP) is proposed~\cite{Gue23}. The coupling of the DP field $\phi^\mu$ with frequency $\omega$ with the electromagnetism with strength $\chi$ is described by \cite{Holdom1986_1} $\mathcal L \supset -\chi F_{\mu\nu}\phi^{\mu\nu}/2\mu_0$. In the limit where $\vec k =0$, this coupling leads to the appearance of a standing electric field \cite{Horns13} $E^j_\mathrm{DM} = -i \chi \omega  Y^j \cos(\omega t)$. Atoms inside an EM cavity can be a powerful tool to detect this electric field, through the quadratic Stark effect. First, the cavity will enhance the DP electric field for a range of DP masses close to the cavity resonant frequencies. Moreover, if one applies a standard EM wave inside the cavity (whose electric field is denoted by $\vec E_a$, with amplitude $\vec X_a$ and angular frequency $\omega_a$) the DP contribution to the square of the electric field inside the cavity is $|\vec E|^2\sim \vec E_\mathrm{DM}\cdot \vec E_a$. 
This cross term produces a signal oscillating at low frequency,  i.e $|\Delta \omega| = \left|\omega-\omega_a\right|$, where the Stark effect can be realistically measured. Modeling the cavity as two flat mirrors of reflectivity $r$ separated by a distance $L$, the amplitude of oscillation of $|\vec E|^2$ at the center is
\begin{equation}
\label{eq:Etot2}
\frac{\chi \beta c\sqrt{1-r^2}\sqrt{2\mu_0\rho_\mathrm{DM}} X_a}{\sqrt{1+2r\cos\left(\frac{\omega_a L}{c}\right)+r^2}}\sqrt{1+4\frac{1+(1+r)\cos\left(\frac{\omega L}{2c}\right)}{1+2r\cos\left(\frac{\omega L}{c}\right)+r^2}} \equiv \chi S  \, ,
\end{equation}
with $\beta = \hat e_\mathrm{DM} \cdot \vec X_a/X_a$. One wishes to measure this signal using atoms through the quadratic Stark effect \cite{cohen-tannoudji:1986aa} $
\Delta \nu = - \chi \Delta\alpha S/2h$,
where $\Delta \nu$ is the frequency shift and $\Delta \alpha$ is the differential polarizability of the atomic transition considered. Two main sources are identified for this experiment: a statistical noise, related to the measurement of the frequency shift by Rydberg atoms; and a systematic effect, coming from fluctuations of $\Delta \vec X_a$. At DM masses corresponding to  cavity odd resonances, the expected sensitivity of this experiment is competitive with current best bounds. Interested reader are invited to check \cite{Gue23} for more details on the experimental schemes and numerical values of the parameters of interest. 

%% file: WG4/content/DirectDeflection.tex
\emph{Directly Deflecting Particle Dark Matter} (Ref.~\cite{Berlin:2019uco}) proposes a detection strategy that bridges the wave-like and particle dark matter (DM) detection regimes, leveraging the large number density of sub-MeV DM and the macroscopic coherence times of classical laboratory fields. The idea consists of setting up a ``deflector'' which has a large charge under the long-range mediator, whose presence will disturb the flow of particle DM nearby. The disturbance in the DM flow can subsequently be detected with a sensitive detector downwind of the deflector.

The approach is generic to any sub-MeV particle-like DM that interacts with ordinary matter through a long-range mediator, such as expected in freeze-in scenarios~\cite{Hall:2009bx}, but we specifically consider (approximately) millicharged dark matter coupled to a kinetically-mixed dark photon with a mass $m_{A'}\ll 1/L_{\rm exp}$ compared to the experimental scale. 

An illustration of the general experimental scheme is shown in Ref.~\cite{Berlin:2019uco}. 
A spatially uniform DM population passes through a shielded volume of radius $R$ where an electric field oscillating at a frequency $\omega$ is applied.\footnote{The magnetic component of the oscillating field also couples to the DM, but the effect is velocity-suppressed.} The electric field separates positive and negative millicharges, creating millicharge $\rho_\chi$ and millicurrent $j_\chi$ densities of length $\sim 2\pi v_{\rm wind}/\omega$ which diffuse out of the deflector region. In order to maximise the signal, the DM should only see one sign of the electric field over the whole deflector volume, such that $\omega \leq \pi v_\chi/R \sim \text{MHz} \times (R/\text{meter})^{-1}$. 

\begin{figure}
    \centering
    \includegraphics[scale=0.7]{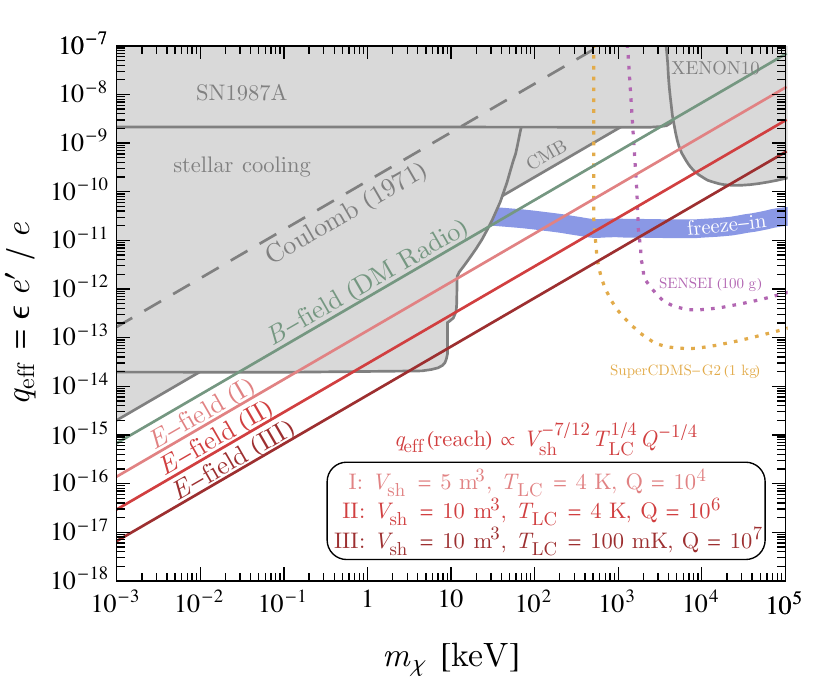}
    \caption{Sensitivity of three different deflector--detector setups. Existing constraints are shown in grey, and the band where freeze-in yields the correct DM relic abundance is shown in blue. Figure reproduced from Ref.~\cite{PhysRevD.102.035010}. 
    }
    \label{fig:DirectDeflectionReach}
\end{figure}

The electric field produced by the charged overdensity $\rho_\chi$ is dominant, scaling as $E_\chi \sim \rho_\chi R$.
It can be detected with a resonant antenna, with signal-to-noise ratio (SNR) given by
\begin{align}
    \text{SNR} \simeq \frac{\omega Q t_{\rm int}}{4 T_{\rm LC}} \int_{V_{\rm det}} (E_\chi^2~\text{or } B_\chi^2) \propto \left(\frac{q_{\rm eff}}{m_\chi} \right)^4 \ .
\end{align}
The possible sensitivity for three different experimental configurations is shown in in Fig.~\ref{fig:DirectDeflectionReach}. 

The approach can be generalised to other ``deflection--detection'' concepts involving other long-range mediators. 

%% file: WG4/content/AtomsMoleculesExperimentalScheme.tex
Atomic and molecular beam experiments are excellent candidates to look for axion-mediated monopole-dipole forces \cite{Agrawal:2023lmw}. 
An interesting and simple possibility is a beam experiment using the isotope $^{171}$Yb, which can be used to make intense beams~\cite{Wright2022,Hutzler2012Review}. This atom has nuclear spin $I=1/2$, and two valence electrons in the $6s$ orbital $(L=S=0)$ giving rise to a $^1S_0$ ground electronic state. Interestingly, the $^3P_2$ $(L=S=1)$ excited state is relatively long-lived with a life-time around $\tau\sim 10$ s~
\cite{Porsev2004AELifetimes}. In addition to the natural existence of co-magnetometer states, atomic and molecular beams have single-shot temporal resolution. As shown in \cite{Agrawal:2023lmw}, this is a crucial property to be sensitive to DC signals, such as an axion gradient generated by the entire Earth.

A possible experimental scheme is to use the ground state $^1S_0$, which is only sensitive to the axion gradient through the nucleon spin, and the $^3P_2$ excited state, which has a very different magnetic moment and can therefore be used as a co-magnetometer. The Hamiltonians for these states are:
\begin{align}
    &H_{^1\!S_0}=-\mathbf{\mu_N}\cdot\mathbf{B}+c_N\frac{\mathbf{\nabla}\phi}{f_\phi}\cdot\mathbf{I}\,,\\&
    H_{^3\!P_2}=-(\mathbf{\mu_N}+M g_P\mathbf{\mu_B})\cdot\mathbf{B}+\frac{\mathbf{\nabla}\phi}{f_\phi}\cdot(c_N\mathbf{I}+c_e\mathbf{S})\,,
\end{align}
where $\mu_{N(B)}$ is the nuclear (Bohr) magneton, $M$ is the projection of the total angular momentum onto the quantisation axis, and $\mathbf{I}$ ($\mathbf{S}$) is the nucleon  (electron) spin. The coefficients $c_e, c_N$ reflect the fact that the axion coupling to electrons and nucleons may be different. Since the ground and excited states have different origins for their magnetic moment and can be used as a co-magnetometer pair by comparing how the precession frequency changes with the B field and axion gradient orientation. The ground state is only sensitive to the axion coupling to nucleons, this experiment would be sensitive to both coupling to electrons and nucleons. 


The shot noise-limited sensitivity, $\delta\omega=\frac{1}{\tau}\frac{1}{\sqrt{\dot{N}T_{int}}}$, for a dedicated Ytterbium beam experiment using
$\tau = 5 - 10$~ms, $\dot{N}=10^{10}-10^{11}$ atoms/s,
and $T_{int}=10^6 - 10^7$~s, we get an expected sensitivity in the range $\delta\omega\sim 10^{-7}$ Hz to $10^{-6}$ Hz. 

Alternatively one could use an indium or thallium beam using the $^2P_{1/2}$ and $^2P_{3/2}$ spin orbit components of the ground electronic state, or other atoms such as Dysprosium, which also possess co-magnetometer states. 
Whether they offer any advantage over Yb depends largely on experimental considerations, such as beam fluxes, laser wavelengths, detection strategies, etc. 

%% file: WG4/content/cosmicaxionforce.tex
\textit{Probing Cosmic Axion Force by the Daily Modulating Magnetic Field}~\cite{Kim:2021eye} proposes a novel method to search for ultra-light axions with masses $m_\phi \lesssim 10^{-25}\,\mathrm{eV}$ that mediate long-range forces between a CP-violating dark sector and the  SM. Such particles naturally arise in the ``axiverse'' predicted by string theory~\cite{Witten:1984dg,Svrcek:2006yi,Conlon:2006tq,Arvanitaki:2009fg}. If CP is broken in the dark sector, the axion acquires couplings to a spatially varying dark-sector density, leading to a large-scale gradient $\nabla\phi$ pointing approximately toward the Galactic center.

On the SM side, axions couple derivatively to fermions,
\[
\mathcal{L}_\mathrm{vis} \supset \frac{c_\psi}{f_\phi}\,\partial_\mu\phi\,\bar\psi\,\gamma_5\gamma^\mu\psi,
\]
which in the non-relativistic limit mimics an effective magnetic field $\vec B_\mathrm{eff}\propto\nabla\phi$. This induces spin precession of nucleons or electrons that cannot be shielded by ordinary electromagnetic shielding.

An atomic magnetometer placed inside a magnetic shield can thus detect an ``axion-induced field.'' The key signature is a daily modulation of the precession signal, since the direction of $\nabla\phi$ is fixed in space while the Earth rotates. With coherent averaging over $\mathcal{O}(300)$ days, an effective magnetic sensitivity of $\sim 0.1$\,aT is achievable, probing decay constants up to
\[
f_\phi \lesssim 10^{14-15}\,\mathrm{GeV}
\]
for $\epsilon c_\psi = \mathcal{O}(1)$, beyond existing astrophysical bounds~\cite{Hardy:2024fen} with $\epsilon$ being the coupling to dark matter. 
This technique is analogous to searches for CPT/Lorentz violation~\cite{PhysRevLett.89.253002,PhysRevLett.95.230801,Brown:2010dt}.

In addition to dark matter halos, axion domain walls can also generate large gradients. If such structures extend across cosmological distances, they would provide nearly uniform fields with possible time-dependent signatures. These effects are further connected to axion dark matter and gravitational-wave production in early-Universe scenarios~\cite{Lee:2024xjb}.

%% file: WG4/content/UndulatorAXION.tex
\textit{Searching for undulator WISPs}~\cite{PhysRevD.111.036020} demonstrates that WISPs, such as ALPs and dark photons~\cite{Yin:2025awb}, can be produced in undulators, which are widely employed as light sources in synchrotron radiation facilities. This observation motivates a cost-effective strategy: placing a dedicated detector behind the facility's safety shield, for example by installing magnets and photon or particle detectors outside the shield to realize an LSW-type experiment for WISPs. At synchrotron facilities such as ESRF, KEK-PF(-AR), NanoTerasu, and SPring-8, such setups can reach sensitivities to WISP couplings in the meV--eV mass range beyond existing ground-based experiments.

The production mechanism is derived within quantum field theory, where the electron current and the undulator magnetic field are treated as external backgrounds. Since an undulator already provides a spatially periodic magnetic lattice and relativistic electron beam, ALPs can be generated without additional magnets or light sources. Dark photons can also be produced through photon mixing. The facility's photon shield effectively prepares part of the LSW configuration. It is also important to study various quantum and material effects for WISP propagation and detection~\cite{Yin:2025awb}.

Since undulators can produce a large number of WISPs (depending on the coupling), the sensitivity can exceed that of existing ground-based experiments. In particular, the radiation safety system outside the shield, even with a simple Geiger-M\"uller counter, already provides one of the strongest ground-based limits on the dark photon mixing parameter around the eV mass scale~\cite{Yin:2025bui}.

%% file: WG4/content/eVaxion.tex
\textit{Searching for eV Dark Matter via Infrared Spectrographs}~\cite{Bessho:2022yyu} demonstrates the feasibility of using advanced infrared  spectrographs to indirectly detect DM candidates in the eV mass range. The focus is on ALPs, which may decay into two photons, producing a line spectrum at $E_\gamma = m_\phi/2$. Such signals are motivated by anomalies in cosmic infrared backgrounds and theoretical scenarios such as the ``ALP miracle''~\cite{Daido:2017wwb,Daido:2017tbr,Takahashi:2023vhv}, which predict eV-scale ALPs as  DM candidates~\cite{Yin:2023jjj,Sakurai:2024apm}.

Conventional indirect searches are hindered by strong astrophysical backgrounds (zodiacal light, thermal radiation, atomic lines). However, modern infrared spectrographs offer high spectral and angular resolution, enabling discrimination of narrow DM-induced lines, especially in dwarf spheroidal galaxies with small velocity dispersions ($\Delta v \lesssim 10\,{\rm km/s}$). Techniques such as object-sky-object (OSO) nodding and Doppler-shift analysis further reduce backgrounds and distinguish intrinsic dSph lines~\cite{Yin:2023uwf}.

Several instruments are particularly suited for eV DM searches. WINERED on the Magellan Clay telescope provides $R \simeq 28{,}000$ ($68{,}000$ in high-resolution mode), sufficient to probe ALPs with $m_\phi \simeq 1.8$--$2.7\,{\rm eV}$ and $g_{\phi\gamma\gamma} \gtrsim 10^{-11}\,{\rm GeV}^{-1}$ in a single night~\cite{Bessho:2022yyu}. The Subaru IRCS ($R=20{,}000$) is also sensitive in this mass range~\cite{Yin:2023uwf}. The JWST/NIRSpec, with $R=2700$, benefits from space-based observations free from sky backgrounds and can cover $m_\phi \simeq 0.5$--$4\,{\rm eV}$ with comparable sensitivity~\cite{Bessho:2022yyu}.

First observations of Leo V and Tucana II (4 hr total) using WINERED excluded most ALP parameter space between $1.8$--$2.7\,{\rm eV}$ for $g_{\phi\gamma\gamma} \gtrsim 10^{-11}\,{\rm GeV}^{-1}$, though excesses were also reported in certain wavelength ranges~\cite{Yin:2024lla}. JWST blank-sky data have also been reanalyzed for DM line searches, showing that archival data can provide economical constraints~\cite{Janish:2023kvi,Roy:2023omw}.

Looking ahead, a novel dedicated instrument, the spectrograph for the dark matter search has been proposed~\cite{Bessho:2024tpl}.

%% file: WG4/content/PaleoWall.tex
\textit{Searching for Cosmic Walls directly with Paleo Detectors}\cite{Yin:2025wuv} proposes a novel strategy to search for late-time cosmic walls---either bubble walls from first-order phase transitions or scaling domain walls---by using paleo detectors, i.e.\ ancient minerals that serve as passive time-integrated nuclear track detectors \cite{Fleischer1964,Fleischer:1965yv,Snowden-Ifft:1995zgn,Baum:2023cct}. Because such walls are expected to cross the Earth at most $\mathcal{O}(1)$ time during cosmic history, continuously exposed ancient samples can realistically record them. The smoking-gun signal would be globally correlated and nearly parallel damage tracks observed exclusively in minerals older than the wall-crossing epoch, while younger minerals would show no such feature. 

By using WKB approximation, the quantum field theoretic estimation exploits the fact that even extremely weakly interacting walls can impart tiny but coherent recoils to nuclei or electrons in crystals. The morphology and orientation of tracks then encode the wall’s velocity, trajectory, and nature, providing a means to discriminate between bubble walls and domain walls. Existing paleo data, such as from ancient muscovite mica \cite{Snowden-Ifft:1995zgn}, can already be recast to place the direct constraints under the assumption of a wall passage within the last few hundred million years.

%% file: WG4/content/darkphotonelectron.tex
In \textit{Dark Photon Dark Matter Radio Signal from the Milky Way Electron Density}~\cite{Arza:2024iuv}, it has been shown that Thomson scattering of dark matter dark photons with Milky Way free electrons generates an almost monochromatic radio signal. This radio signal can be searched for with current and future radio telescope arrays, such as the Atacama Large Millimeter Array (ALMA) and the Square Kilometer Array (SKA), respectively, to probe dark photon dark matter in a wide unexplored parameter space.

Thomson scattering $e^-+\gamma'\rightarrow e^-+\gamma$ between Milky Way free electrons and dark matter dark photons is considered. The emitted photon has energy $\omega=m_{\gamma'}\left(1+{\cal O}(v_{\gamma'}^2)\right)$ and can be produced in any direction. At any location of the galaxy, it is expected a diffuse spectral line signal background with frequency.
\begin{equation}
\nu=\frac{m_{\gamma'}}{2\pi}=2.42\,\text{GHz}\left(\frac{m_{\gamma'}}{10^{-5}\,\text{eV}}\right) \label{eq:freq1}
\end{equation}
and bandwidth $\delta\nu\sim10^{-3}\,\nu$. 

In this work, radio telescope observations in the mass range $(\text{few})\times10^{-7}\,\text{eV}$ to $(\text{few})\times10^{-3}\,\text{eV}$ is proposed and the observations are chosen to be done towards the center of the galaxy since the electron and dark matter densities are huge there. The flux density (power per unit surface per unit frequency) at the detection point is
\begin{equation}
S_\nu=\frac{\left<\sigma v_\text{rel}\right>}{4\pi\delta\nu}\int d\ell\int_\Omega d\Omega\,n_e(\ell,\Omega)\rho_\text{DM}(\ell,\Omega), \label{eq:rad1}
\end{equation}
where $n_e$ is the electron density and $\rho_\text{DM}$ the dark matter energy density. The term $\left<\sigma v_\text{rel}\right>$ is the averaged scattering cross section of the process multiplied by the two particles relative velocity, its value is $2.07\times10^{-16}\,\chi^2\,\text{cm}^2\,\text{m}/\text{s}$, where $\chi$ is the kinetic mixing parameter between dark photons and standard model photons. The integration is performed over the line of sight of the observation and over the solid angle $\Omega$ covered by the telescope. 

\begin{figure}[t]
\centering
\includegraphics[width=0.8\linewidth]{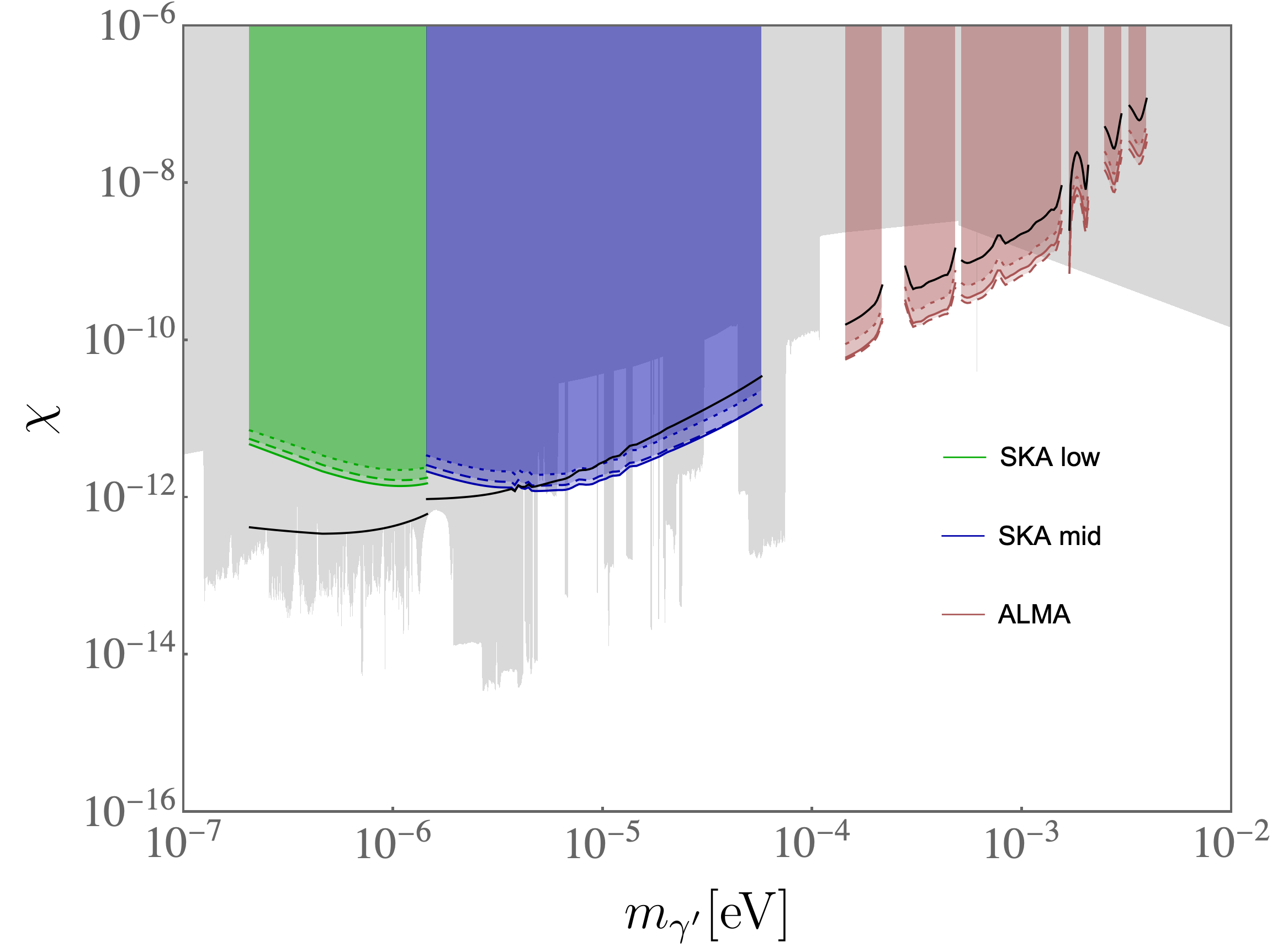}
\caption{Projected sensitivity for 42 days observation with ALMA (red), SKA low (green) and SKA mid (blue) in single dish operation. The dotted, dashed and solid lines correspond to NFW, Moore, and Einasto halo models, respectively. The black solid lines correspond to the projected sensitivities in the interferometric mode assuming the Einasto profile. The grey region are current experimental, astrophysical and cosmological constraints.}
\label{fig:sensDPelec}
\end{figure}

For $n_e(\ell)$ the YMW16 model \cite{Yao:2017kcp} was used, while for $\rho_\text{DM}(\ell)$ it is assumed three different spherically symmetric halo models; the Navarro, Frenk and White (NFW), Moore, and Einasto \cite{Cirelli:2010xx}. Sensitivity projections for ALMA (red), SKA mid (blue) and SKA low (green) are shown in Fig. \ref{fig:sensDPelec}, where a total observation time of $t_\text{obs}\approx42\,\text{days}$ is assumed. Colored regions correspond to single dish mode observations while the results for the interferometer mode are marked with a black solid line. For details on the computation of this projection, see Ref. \cite{Arza:2024iuv}.

%% file: WG4/content/Antiferromagnets.tex
If dark matter couples to the visible sector via spin-dependent interactions, it becomes possible to look for it leveraging the properties of magnetic materials. These are essentially macroscopic collections of spins whose ground state exhibit long-range order. Specifically, the interaction with a dark matter particle could excite the spin collective modes, called ``magnons''. 

The authors of {\it Optimal antiferromagnets for light dark matter detection}~\cite{Esposito:2022bnu} showed that an optimal class of materials is antiferromagnets, whose ordered phase is characterized by macroscopic regions where the spins are anti-aligned to each other. Thanks to their behavior under time-reversal, antiferromagnetic magnons come in two different polarizations. As showed in~\cite{Esposito:2022bnu}, this opens up the possibility of multi-magnon emission which, in turn, makes the system sensitive to dark matter scattering events for masses as low as $m_\chi \sim \mathcal{O}({\rm keV})$. Importantly, among the materials analyzed, the authors identified Nickel Oxide (NiO) as a very promising one. The magnon propagation speed in such an antiferromagnet ($v_\theta \simeq 1.3 \times 10^{-4} \, c$) is, in fact, unusually large if compared to other similar materials, thus providing an excellent kinematic matching with dark matter particles coming from the Milky Way halo. The ideal reach of such a detection scheme as been computed in~\cite{Esposito:2022bnu} via means of an effective field theory for the magnon dynamics. This has been done for two benchmark dark matter models, the so-called ``magnetic dipole'' and ``pseudo-mediated'' dark matter (see, e.g.,~\cite{Sigurdson:2004zp,Chang:2010en,Kavanagh:2018xeh,Chu:2018qrm,Chang:2009yt}). 

\begin{figure}[t]
    \centering
    \includegraphics[width=0.7\linewidth]{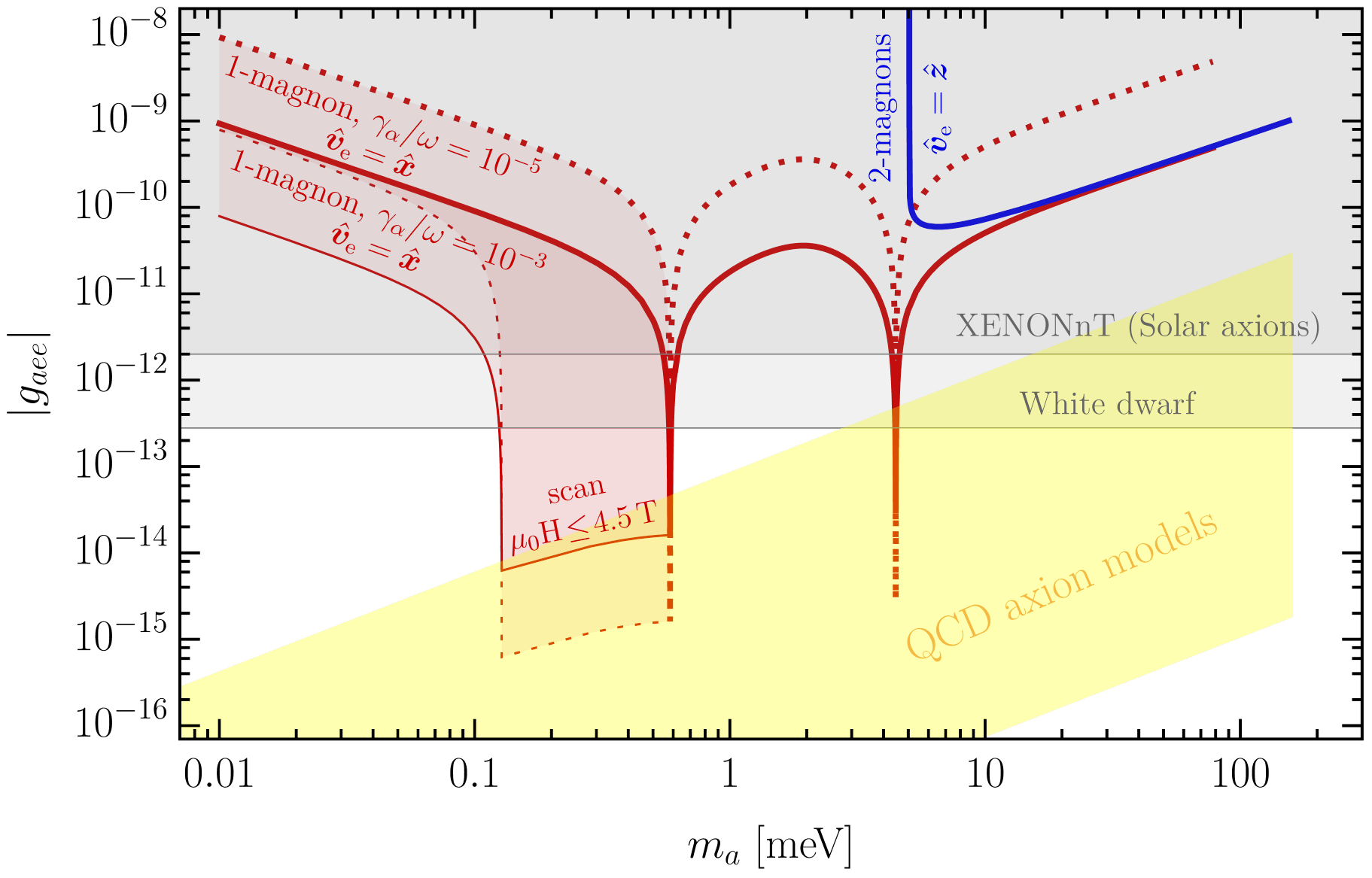}
    \caption{Projected reach at 95\% CL for a kilogram of material and a year of exposure, assuming no background. The magnon linewidth is taken as a free parameter, $\gamma_\alpha/\omega$. The red band region corresponds to a magnetic field scan up to $\mu_0{\rm H}=4.5$~T. Figure reproduced from Ref.~\cite{Catinari:2024ekq}.}
    \label{fig:Bscan}
\end{figure}

In Ref.~\cite{Catinari:2024ekq} it was also shown that, thanks to the presence of small but non-zero magnon gaps, NiO is also promising for the detection of axion dark matter, probed via the axion-magnon conversion process. In particular, by introducing an external magnetic field in the sample, it is possible to vary the magnon gap, thus scanning different values of the axion mass. In samples where the magnons are sufficiently long lived, the scheme would allow to probe interesting QCD axion models, in the yet unexplored mass range $0.1 \text{ meV} \lesssim m_a \lesssim 1 \text{ meV}$. The projected reach is reported in Fig.~\ref{fig:Bscan}.

Experimental efforts to determine the feasibility of this detection scheme are currently underway.

%% file: WG4/content/IntroExperimentsTables.tex
We conclude this review of WISP direct searches with Tab.~\ref{tab:experiments1}, a tabulated overview of the experiments discussed in this part. 

\captionsetup[xltabular]{skip=10pt, labelfont=bf, labelsep=period}

\renewcommand{\arraystretch}{1.5}
\setlength{\tabcolsep}{4pt}

\begin{xltabular}{\textwidth}{p{.18\textwidth}|>{\centering\arraybackslash}p{.2\textwidth}>{\centering\arraybackslash}p{.12\textwidth}>{\centering\arraybackslash}p{.1\textwidth}>{\centering\arraybackslash}p{.1\textwidth}p{.17\textwidth}}

\caption{List of European experiments. For multi-site experiments only the European sites are shown.}
\label{tab:experiments1}\\

\hline\hline
Name & Type & Coupling & Status & Start  & Lab \\
\hline
\endfirsthead

\hline\hline
Name & Type & Coupling & Status & Start & Lab \\
\hline
\endhead

\hline
\hline
\noalign{\vskip 4pt}
\multicolumn{6}{r}{\textit{(continued on next page)}}\\
\endfoot

\hline\hline
\noalign{\vskip 12pt}
\multicolumn{6}{l}{\makecell[l]{\textbf{\tablename\ \thetable:} List of European experiments. For multi-site experiments only the European\\ sites are shown.}}\\
\endlastfoot

AEDGE & Cold Atoms Space Experiment & ULDM & Design, R\&D & 2050 & ESA\\

APE &  Polarimeter  & $g_{a\gamma\gamma}$ & Comm. &   & MPI, Hannover\\

AQN@LHC &  LHC Beam Monitoring System  & AQN &  Design & 2024 & CERN\\
      
ATLAS/CMS & $pp$-Collisions   & ALP, DP & Running & 2011 &CERN \\
      
AION & Atom Interferometer  & Scalar-EM coupling $|d_{e}|$ & Design  &   & U. Oxford \\
      
ALPS II & LSW & $g_{a\gamma\gamma}$ & Running & 2024 & DESY \\
      
AURIGA & Weber Bar & Scalar DM & Done & 1997& LNL \\
      
BabyIaxo &  Helioscope  & $g_{a\gamma\gamma}$,  $g_{aee}$  & Comm.  &  & DESY \\
      
BASE & Penning Trap  & $g_{a\gamma\gamma}$ & Done & 2020 & CERN \\
      
BMV &  Cavity Birefringence  & ALPS & Done & 2013 & Toulouse \\
      
BRASS & Dish Antenna & $g_{a\gamma\gamma}$, $\chi$ &Running & 2018 & Hamburg U. \\
      
CADEx &  Haloscope &$g_{a\gamma\gamma}$  & R\&D & 2027 & Canfranc\\
      
CAST &  Helioscope  & $g_{a\gamma\gamma}$, $g_{aee}$ & Done & 2003 & CERN \\
      
CASPEr-gradient & NMR Haloscope & $g_{an}$, $g_{ap}$ &Running & 2023 & HIM Mainz \\
      
CAST-CAPP & Sikivie's Haloscope & $g_{a\gamma\gamma}$, $\chi$ & Done & 2019 & CERN \\
      
CROWS & LSW &$g_{a\gamma\gamma}$, $\chi$   & Done & 2013  & CERN \\

DALI &  Magnetized Phased Array Haloscope   & $g_{a\gamma\gamma}$, $\chi$ & Prototype & 2018 & IAC\\
      
DAMNED &  MZI  & $d_e$, $d_{me}$ & Running  & 2021 & Paris Obs.\\
      
DARKSIDE & Argon Scintillator  & $g_{ae}$, $\chi$ & Running & 2013 & LNGS\\
      
DAMIC-M &  CCD & $\chi$ & Prototype & 2022 & Modane\\
      
EDELWEISS& Ge Bolometer & $g_{ae}$, $g_{an}$ & Running & 2014  & Modane\\
      
nEDM & Ultacold Neutrons & $g_{an\gamma}$ & Running & 2009 & PSI\\
      
EGO-VIRGO & Interferometer & $d_e$, $d_{me}$, $d_{\hat{m}}$, $d_g$ & Running & 2007 & Pisa\\
      
FLASH & Sikivie's Haloscope & $g_{a\gamma\gamma}$, $\chi$ & Design & 2028 & LNF\\
      
FUNK &  Dish Antenna  & $\chi$ & Done & 2019 & KIT\\
      
GEO600 & Interferometer & $d_e$, $d_{me}$, $d_g$ & Running & 2001 & MPI, Hannover\\
      
GERDA & $0\nu\beta\beta$ decay& $g_{ae}$, $\chi$ & Done & 2011 & LNGS \\
      
GNOME &  Network of Magnetometers   & Topological Defects & Running & 2017 & Fribourg, \phantom{aaa} Mainz, Jena, \phantom{a} Krakow, \phantom{aaa} Belgrade  \\

GrAHal & Sikivie's Haloscope & $g_{a\gamma\gamma}$, $\chi$ & Running & 2019 & LNCMI, \quad Grenoble \\
      
GrAHal-CAPP & Sikivie's Haloscope& $g_{a\gamma\gamma}$, $\chi$ & Design &  & LNCMI, \quad Grenoble\\
      
IAXO & Helioscope & $g_{a\gamma\gamma}$ & Design & 2028 & DESY\\
      
JEDI & EDM at Storage Ring & $g_{ad\gamma}$ & Running & 2011 & COSY \\
      
LIDA & Interferometer & $g_{a\gamma\gamma}$ & Running & 2024 & Birmingham \\
      
LOFAR & Radio Telescope & $g_{a\gamma\gamma}$, $\chi$ & Running & 2010 & Exloo\\
      
LUXE-NPOD &  XFEL Optical dump & $g_{a\gamma\gamma}$ & Design &  & DESY\\
      
MADMAX \phantom{aaa} Prototype & Haloscope & $g_{a\gamma\gamma}$& Running & 2024 & CERN \phantom{aaa} Morpurgo\\
      
MADMAX & Haloscope & $g_{a\gamma\gamma}$& Installation &  & DESY \\
      
MoEDAL-MAPP & Scintillator & FIPs & Installation & 2025 & CERN LHC \\
      
NA62 & $p$ Beam Dump  & $g_{a\gamma\gamma}$ & Running & 2021 & CERN SPS\\
      
NA64 & $e^{-}$ Beam Dump & $g_{a\gamma\gamma}$ & Running & 2020 & CERN SPS H4\\
      
NASDUCK & Comagnetometers  & $g_{ann},g_{app}$ & Running & 2020 & Tel Aviv U.\\ 
      
NAUTILUS & Gravitational Bar & AQN  & Done & 2003 & LNF \\
      
NOMAD & LSW & $g_{a\gamma\gamma}$ & Done & 1996 & CERN SPS \\
      
OSQAR & LSW & $g_{a\gamma\gamma}$ & Done & 2010 & CERN\\
      
PADME & $e^{+}$ Beam Dump & $\chi$ & Running & 2018 & LNF\\
      
PVLAS & Cavity Birefringence & $g_{a\gamma\gamma}$ & Running & 2001 & INFN Ferrara\\ 
      
QUAX  & Sikivie's Haloscope & $g_{a\gamma\gamma}$, $\chi$ & Running & 2018 & LNF, LNL\\

QUAX$_{ae}$  & Ferromagnetic Haloscope &  $g_{ae}$ & Running & 2018 & LNL\\
      
QUAX$_{gpgs}$  & 5th Force Experiment & $g_pg_s$& Running & 2018 & LNL\\
      
QSNET & Atomic Clocks & Scalars, Vectors  & Running & 2022 & UoB, NPL, \phantom{aaa} ICL, UoS\\
      
RADES-BabyIaxo & Sikivie's Haloscope  & $g_{a\gamma\gamma}$ & Design &  & DESY\\
      
RADES-CAST & Sikivie's Haloscope & $g_{a\gamma\gamma}$ & Done & 2018 & 
CERN \\
      
RADES-LSC & Sikivie's Haloscope & $g_{a\gamma\gamma}$ & Design &  & LSC\\
      
RadioAxion-$\alpha$ & Nuclear Decay & $g_d$ & Prototype & 2024 & LNGS\\    
      
RadioAxion-$\beta$ & Nuclear Decay & $g_d$ & Done & 2011 & LNGS\\     
      
RadioAxion-EC & Nuclear Decay & $g_d$ & Running & 2015 & LNGS\\    

REPHY- SPECMO &  Network of Molecular Spectrometers & $d_e, d_{me},$ $d_{\hat{m}}, d_g$ & Installation & 2023 & CNRS \\
      
SHUKET & Metallic Dish Antenna & $\chi$ & Running & 2019 &  CEA, Saclay   \\
      
SNIPE hunt & GMR Magnetometer & $g_{a\gamma\gamma}$ & Running & 2023 & Mainz \\
      
STAX & LSW & $g_{a\gamma\gamma}$ & Planning &  & KARA\\
      
STE-QUEST & Atomic Clock & $g_{a\gamma\gamma}$ & Design & 2022 & ESA \\
      
SUPAX & Sikivie's Haloscope  & $\chi$ & Design & 2023 & Mainz \\

TASTE &  Helioscope & $g_{a\gamma\gamma}$, $g_{ae}$  & Design  & 2017 & INR, Troitsk\\
      
TOORAD & Anti\-ferromagnetic Topological Insulators & $g_{a\gamma\gamma}$ &  Design & 2020 & \\
      
VAMBI & Cavity Birefringence &  $g_{a\gamma\gamma}$ & Planning & 2025 & MPI, Hannover \\
      
VMB & Cavity Birefringence & $g_{a\gamma\gamma}$ & Design & 2018 & CERN \\
      
WISPDMX & Sikivie's Haloscope &  $\chi$ & Done & 2017 & Hamburg U.\\
      
WISPFI &  Interferometer  & $g_{a\gamma\gamma}$ & Running & 2023 & Hamburg U. \\
      
WISPLC &  Sikivie's Haloscope & $g_{a\gamma\gamma}$ & Running & 2022 & Hamburg U.\\
      
XENON & Xenon Scintillator & $g_{ae}$, $\chi$ & Running & & LNGS\\

\end{xltabular}

%% file: epilogue.tex
\addcontentsline{toc}{section}{Conclusions}
\input{General/conclusions}

\cleardoublepage

\addcontentsline{toc}{section}{Table of Acronyms}
\input{General/Acronyms}
\cleardoublepage

\addcontentsline{toc}{section}{Acknowledgements}
\input{General/acknowledgements}

\addcontentsline{toc}{section}{List of Endorsers}
\input{General/endorsers}

\clearpage
\thispagestyle{plain}

\addcontentsline{toc}{section}{References}
\bibliography{General/complete}

\bibliographystyle{General/utphys}

%% file: General/conclusions.tex
\section*{Conclusions}
\markboth{Conclusions}{}
\thispagestyle{plain}
Over the past decade, interest in WISP searches has steadily increased. A vibrant research program has developed around this topic, spanning WISPs model building as well as their potential signatures in cosmological and astrophysical observations. In parallel, a growing number of experimental proposals—employing a wide range of techniques—aim to explore large portions of the WISP parameter space across different theoretical scenarios.

WISPs are deeply rooted in BSM UV completions such as string theory, as well as in consistent quantum field theoretic constructions. Rather than appearing as isolated low-energy degrees of freedom, WISPs typically emerge as entire new sectors whose masses, couplings, and multiplicities are correlated by symmetry principles, geometric data, and dynamical mechanisms operating at high energies. In Part I, a unified theoretical framework for understanding these particles has been provided, spanning string-theoretic constructions, bottom-up model building, and effective field theory descriptions.

From a top-down perspective, string compactifications generically predict the existence of numerous light fields, including axions, axion-like particles, dark photons, and additional hidden-sector states. Axions naturally arise as zero modes of higher-dimensional $p$-form gauge fields and inherit exact perturbative shift symmetries, with masses and potentials generated only through non-perturbative effects. The process of moduli stabilisation plays a central role in shaping the resulting low-energy spectrum, determining axion masses, decay constants, and couplings, while simultaneously ensuring consistency with fifth-force constraints and cosmology. Modern stabilisation frameworks allow for controlled hierarchies between saxions and axions, leading to the emergence of a rich \emph{axiverse} populated by ultralight fields with potentially observable cosmological and astrophysical signatures.

Within this setting, the construction of a viable QCD axion requires embedding chiral gauge sectors in a manner compatible with moduli dynamics. Both open- and closed-string axions can play the role of the QCD axion, depending on the presence of anomalous $U(1)$ symmetries, D-term constraints, and non-perturbative effects. String theory offers qualitatively new approaches to the axion quality problem, suppressing dangerous shift symmetry-breaking local operators, in contrast with purely bottom-up EFT expectations. These constructions naturally link axion phenomenology to broader questions in cosmology, including dark matter, dark radiation, and reheating.

Complementing the string-theoretic viewpoint, Part I reviews WISP model building within quantum field theory. Standard QCD axion models are discussed alongside non-minimal realisations that evade conventional astrophysical bounds, explore axion parameter space beyond the canonical mass--coupling window, or suppress specific interactions through symmetry-based mechanisms. The quantisation of axion couplings is emphasised as a generic consequence of UV consistency, with potentially observable implications for low-energy experiments. The framework further extends beyond axions to encompass additional light scalars, vectors, and hidden-sector particles, illustrating the breadth of viable WISP scenarios.

The discussion also highlights mechanisms that allow light WISPs to remain phenomenologically viable despite stringent experimental constraints. Screening effects can suppress fifth forces in dense environments, while time-dependent WISP backgrounds may induce oscillatory or slowly varying signals in fundamental constants, motivating novel experimental strategies sensitive to temporal signatures. These effects reinforce the idea that WISPs can manifest in subtle but characteristic ways across different observational channels.

Finally, Part I presents a systematic EFT description of WISPs of different spins, clarifying the operator structure, symmetry assumptions, and domains of validity relevant for laboratory, astrophysical, and cosmological probes. This EFT framework provides a common language connecting UV-complete constructions to experimental observables, enabling consistent comparisons between different models and searches. Overall, Part I establishes WISPs as theoretically compelling probes of physics beyond the Standard Model, whose collective properties encode valuable information about high-energy dynamics and quantum gravity, and whose discovery potential spans a wide range of current and future experiments.

The presence of WISPs inevitably alters our standard picture of cosmological history. The first section of Part II details current theoretical predictions and experimental constraints on the QCD axion. For non-thermally produced axions, the precise phenomenology is determined by the energy scale at which the newly introduced $U(1)$ symmetry is broken with respect to the standard cosmological timeline, specifically its relation to the energy scale of cosmic inflation. Part II outlines the theoretical framework that determines the axion mass for a symmetry breaking occurring either before or after inflation, via the misalignment mechanism or the evolution of topological defect networks, respectively. 

The misalignment mechanism requires theoretical input via the temperature dependence of the topological susceptibility for realistic quark masses, which must be calculated using lattice simulations. Recent technical advances in the calculation are discussed, along with suggestions for the most useful next steps. Although in principle entirely predictive, the calculation of the axion mass from the evolution of a network of defects poses a significant challenge. Currently, key model inputs are predicted differently from different extrapolations of string network simulation results, with even larger uncertainties when including the nonlinear domain-wall collapse transient. In this post-inflation scenario, the formation of sub-galaxy-scale axion dark matter substructure from the collapse of isocurvature perturbations around matter radiation equality, as well as the formation of gravitationally bound ‘axion stars’, can have significant indirect effects on observational predictions and direct effects on direct detection probabilities using haloscopes. For thermally produced axions that behave as dark radiation or hot dark matter, accurate calculation of the thermalisation rate and the axion coupling to standard model particles is essential, particularly for future CMB experiments that will tighten constraints on $\Delta N_\mathrm{eff}$ by up to an order of magnitude.

The second section of Part II outlines the cosmological effects of more general WISPs. With the direct relationship between the coupling constant and the axion mass lifted, axion-like particles (ALPs) can exhibit a broad range of phenomenology, including acting as ultralight dark matter (ULDM) or as dark energy via the inflaton or a quintessence field. Finally, dark photons and dark gravitons are discussed as WISP candidates, well-motivated by theories that produce their own distinct cosmological signatures.

In summary, cosmological imprints of WISPs provide probes of axion physics over a vast range of energy scales, from the very early Universe to the present day. Part II lays out the complex phenomenology of the QCD axion, including the theoretical and computational input still required for ab initio calculations of the QCD axion mass, and the cosmological effects generated by more general WISP candidates. Refinement of these calculations, pushing the frontiers of theoretical understanding and leveraging rapidly increasing supercomputing power, will be crucial to facilitate the full exploitation of cosmological and direct detection experiments planned in the near future, as detailed in Part IV.

Astrophysics provides another extremely effective environment to probe the physics of WISPs and, in several cases, offers concrete opportunities for detection, as reported in Part~III.
The astrophysical constraints discussed throughout this work demonstrate that nature's most extreme environments have become indispensable laboratories for probing the dark sector. What makes astrophysical bounds particularly compelling is not merely their current strength---though stellar evolution limits and black-hole superradiance exclusions already rival or surpass laboratory sensitivities in many regimes---but rather their fundamentally different systematics and the complementary parameter space they explore.

Red-giant cooling, horizontal-branch morphology, and the white-dwarf luminosity function, to mention a few examples, provide valuable and often overlapping information on WISP interactions while sampling different stellar populations, evolutionary phases, and microphysical processes. This diversity increases our confidence that we are genuinely learning about particle physics rather than mis-modeling astrophysics. In particular, the ability to confront the same underlying WISP microphysics with multiple, partially independent diagnostics---each with distinct astrophysical systematics---provides a powerful form of cross-validation, and helps isolate robust regions of parameter space that are difficult to access with laboratory experiments alone.

The interplay between photon--WISP conversion in astrophysical magnetic fields and electromagnetic observations has matured into a rich phenomenology spanning radio to gamma rays, and it will remain a cornerstone of discovery strategies. Upcoming facilities will not simply extend existing searches to higher sensitivity; they will probe qualitatively new regimes, including ultra-strong neutron-star and magnetar magnetospheres, relativistic and strongly inhomogeneous plasma environments, and magnetic structures on galactic and cluster scales. In parallel, the emerging landscape of MeV gamma-ray astronomy is poised to address the long-standing \emph{MeV gap}---a decisive wavelength window for both thermal and non-thermal emission processes, and for several distinctive WISP signatures that bridge traditional X-ray and GeV searches. Closing this gap is therefore not only an instrumental priority, but a key step toward a more complete, multiwavelength exploration of WISP parameter space.

Looking forward, the frontier is expanding in several directions: to new astrophysical objects; from nearby individual sources to larger (cosmic) volumes and populations; and from essentially static measurements to genuinely dynamical, time-resolved phenomena. Black-hole superradiance has already emerged as an entirely gravitational probe, requiring no assumptions about thermal equilibrium or plasma microphysics, and providing a powerful complement to electromagnetic and stellar constraints. Cosmic birefringence measurements, if the current hints are confirmed, would represent the detection of a cosmological axion field through its imprint on CMB polarization---a discovery that would transform our understanding of both dark matter and fundamental physics. Meanwhile, the prospect of a Galactic supernova would offer an unparalleled laboratory: neutrino detectors and multiwavelength observatories could constrain or discover light particles through real-time observations of core collapse and its aftermath, with sensitivity that is difficult to replicate in any terrestrial setting.

Ultimately, the strength of astrophysical probes lies in their breadth and diversity. As theoretical understanding deepens and observational capabilities expand, astrophysics will remain not merely complementary to laboratory searches, but central to the quest for physics beyond the Standard Model. Realizing the full discovery potential of this program will require both continued progress on the experimental side---for example, through MeV-sensitive instrumentation---and sustained advances in modeling WISP production and photon--WISP conversion in realistic, inhomogeneous magnetic fields and plasmas, enabling robust interpretation of the increasingly precise data across the multi-messenger sky.

WISPS models have inspired a broad range of new experiments designed to detect the non-zero coupling to SM particles. These have been classified in Part IV  as:
\begin{description}
    \item[Haloscopes] Experiments designed to detect the WISP dark matter halo in our galaxy.
    \item[Helioscopes] Experiments designed to detect WISPs produced in the Sun.
    \item[Pure lab experiments] Experiments designed to detect effects induced by WISP generated in the lab.
    \item[Fixed target and beam dump experiment]  Experiments designed to detect WISPs generated in the collision of an accelerated beam on a target.
\end{description}
Beside these, there are non-WISP-focused experiments with the ability to detect WISPs. These are either collider experiments, experiments designed to detect WIMPs, gravitational wave interferometers, or high precision experiments able to detect small deviations from theoretically well-predictable observables. The liveliness of this sector is particularly evident in Europe, where there is a large and diverse WISPs research program with great impact and discovery potential. The development of WISP experiments, characterized by different techniques, goes hand in hand with as many technological developments in many fields, ranging from superconducting magnets, RF cavities, and optical elements for high precision interferometry to atomic and molecular systems and quantum sensors. Magnets play a primary role in the search for axions. Developments in Europe for high magnetic field science and technology are driven by the European Magnetic Field Laboratory (EMFL). At CERN, the EP Magnet working group has the responsibility of maintenance, operation, and troubleshooting of (superconducting) detector magnets. The BabyIAXO magnet has been designed under the leadership of CERN-EP. CEA is designing the dipole magnet for the MADMAX experiment. In Italy, IRIS is a project dedicated to the development of applied superconductivity. Large research laboratories host cryogenic infrastructures to cool superconducting magnets, experiment components, and accelerator complexes. Dilution refrigerators are widely used in WISP experiments, such as Haloscopes, that require very low electronic noise. European companies developing such systems are Leiden Cryogenics, Bluefors, Oxford Instruments, Entropy, ICEoxford, CryoConcept. Dilution refrigerators are also needed for operating superconducting quantum devices. Several companies are active in this sector in Europe, like QuantWare, Siletwave, ez SQUID, Magnicon and Supracon AG. This variety of experiments and experimental approaches requiring different technologies benefits from a large number of laboratories able to host them and provide adequate infrastructures. Some of the labs supporting WISP experiments are: the Laboratorio Subterr\'aneo Canfranc (LSC) that hosts CADEX and RADES-LSC experiments; CERN that hosts OSQAR, VMB, MADMAX protoype and CAST; DESY that hosts ALPS, BabyIAXO and MADMAX and provides cleanrooms for optical precision interferometry, cryogenic laboratory, a cryoplatform and underground halls; LNF that hosts QUAX@LNF, FLASH and PADME and previously KLOE and Nautilus, providing cryogenics, power-laser and a Test Beam Facility; 
LNGS that hosts DARKSIDE, GERDA, RadioAxion and XENON; LNL  that hosts QUAX@LNL, QUAX-$ae$, QUAX-$g_pg_s$ and previously AURIGA; Laboratoire Souterrain de Modane (LSM) that hosts  DAMIC-M and EDELWEISS; At Grenoble, participating to GrAHal  there are the National Laboratory for High Magnetic Fields, LPSC and the Neel Institute; The Helmoltz Institute Mainz that hosts CASPER-gradient and SUPAX; Instituto de Astrof\'isica de Canarias (IAC) that  hosts DALI, MAGIC, CTA, the La Palma (Roque de Los Muchachos) and Tenerife (Teide) observatories, along with labs and workshops at the headquarters equipped with cryogenics, optomechanics, EMC rooms; PSI hosts nEDM; GEO600. 

In addition to ongoing experimental efforts, new proposals aim to use unexplored paradigms in searches for WISPs. The multitude of possible effects of axions or dark photons (or WISPs, in general) on ordinary particles and known fields leads to a diversity in approaches that new experiments can take. New experiments propose to use molecules, atoms, Rydberg atoms, deflecting electric field or magnetic undulators, radioactivity, and media and sensors such as piezoelectric crystals, resonant-mass detectors, force sensors, magnetized plasma, intensity interferometry, dielectric absorbers, antiferromagnets, ferroelectric materials, artificial magnetoelectric materials, or cavities (optical, microwave, or generally radio-frequency).

The WISP community is relatively young, and substantial effort has been devoted to coordination and capacity building. In this context, the COST Action \emph{Cosmic WISPers} seeks to further energize this field by promoting WISP studies in a synergistic manner, maximizing the impact of the results, and developing a roadmap towards WISP discovery in Europe.
 Together with a broad international program of smaller-scale experiments, these efforts offer strong prospects for one or more transformative discoveries, grounded in the WISP paradigm as a possible key to resolving several outstanding mysteries of the Dark Universe.

%% file: General/Acronyms.tex
\section*{Table of Acronyms}
\markboth{Table of Acronyms}{}
\thispagestyle{plain}
\captionsetup[longtable]{skip=10pt, labelfont=bf, labelsep=period}

\begin{center}
\renewcommand{\arraystretch}{1.5}
\begin{longtable}{lp{10cm}}
\hline
\hline
\textbf{Acronym} & \textbf{Meaning} \\
\hline
\endfirsthead
\hline
\hline
\textbf{Acronym} & \textbf{Meaning} \\
\hline
\endhead

\hline
\hline
\multicolumn{2}{r}{\textit{(continued on next page)}} \\
\endfoot

\hline
\caption{List of acronyms used in this work.} \\
\endlastfoot

ABC & Atomic, Bremsstrahlung and Compton processes \\
AC/DC & Alternating/Direct Current \\
AGB & Asymptotic Giant Branch (star) \\
AGN & Active Galactive Nuclei \\
ALP & Axion-Like Particle \\
AMR & Adaptive Mesh Refinement \\
AQN & Axion Quark Nugget\\
BAO & Baryon Acoustic Oscillations \\
BBN & Big-Bang-Nucleosynthesis \\
BH & Black Hole \\
BSM & Beyond Standard Model \\
(C)DM & (Cold) Dark Matter \\
CGB & Cosmic Gamma-ray Background \\
CIB & Cosmic Infrared Background \\
CITE & Collapse-Induced Thermonuclear Explosion\\ 
CL & Confidence Level \\
CMB & Cosmic Microwave Background \\
CMD & Color Magnitude Diagram \\ 
C$\nu$B & Cosmic Neutrino Background \\
CNO & Carbon-Nitrogen-Oxygen  (cycle)\\
COB & Cosmic Optical Background \\
CP & Charge-Parity \\
CPU & Central Processing Unit\\
CSGO & Cold-Slumped Glass Optics \\
CUB & Cosmic Ultraviolet Background \\
CXB & Cosmic X-ray Background \\
CY & Calabi-Yau \\ 
DE & Dark Energy \\
DFSZ & Dine-Fischler-Srednicki-Zhitnitsky \\
DIGA & Dilute Instanton Gas Approximation \\
DM & Dark Matter \\
DMSAB & Diffuse Main Sequence Axion Background \\
DOM & Digital Optical Module \\
DP & Dark Photon \\
dS & de Sitter \\
DSAB & Diffuse Supernova Axion Background \\
DSNB & Diffuse Supernova Neutrino Background \\
EBL & Extragalactic Background Light\\
EDE & Early Dark Energy \\
EFT & Effective Field Theory \\
EMC & Electromagnetic Compatibility \\
EMF & Electromagnetic Field\\
EMI & Electromagnetic Interference\\
EoS & Equation of State \\
EW & Electro-Weak \\
FI &  Fayet-Iliopoulos \\
FIP & Feebly Interacting Particle \\
FLRW & Friedmann-Lemaître-Robertson-Walker\\
FOPT & First-Order Phase Transition \\
FoV & Field of View \\
FP & Fabry-Pérot (cavity) \\
FPGA & Field-programmable Gate Array\\
FuNS & Full Network Stellar evolution \\
FWHM & Full Width at Half Maximum \\
GC & Globular Cluster \\
GJ & Goldreich-Julian (plasma) \\
GPU & Graphics Processing Unit\\
GRB & Gamma-Ray Burst \\
GUT & Grand Unified Theory \\
GW & Gravitational Wave \\
$H_0$ & Hubble Constant \\
HB & Horizontal Branch (star) \\
HET & High Energy Telescope \\
HFGWs & High-Frequency Gravitational Waves\\
HMNS & HyperMassive Neutron Star \\
HRG & Hadron Resonance Gas\\
HTL & Hard Thermal Loop \\
HTS & High Temperature Superconductor\\
IBD & Inverse Beta Decay \\
ICL & Intracluster Light \\
ICM & Intracluster Medium \\
IFMR & Initial-Final Mass Relation \\
IGM & Intergalactic Medium \\
IMF & Initial Mass Function \\
IR & Infrared \\
ISM & Interstellar Medium \\
KK & Kaluza-Klein \\
KKLT & Kachru-Kallosh-Linde-Trivedi \\
KSZV & Kim-Shifman-Vainshtein-Zakharov \\
LC  & Inductance/Capacitance (circuit)\\
LDM & Light Dark Matter\\
LESA & Lepton-number Emission Self-sustained Asymmetry \\
LET & Low Energy Telescope \\
LP & Longitudinal Plasmon \\
LPM & Landau-Pomeranchuk-Migdal \\
LSS & Large-Scale Structure \\
LSW & Light-shining-through-Wall (experiments) \\
LVS & Large Volume Scenario \\
M7 & Magnificent Seven (neutron stars) \\
MAG & Metric Affine Gravity \\
MCMC & Markov Chain Monte Carlo\\
mfp & mean free path \\
ML & Machine Learning \\
MPA & Magnetized Phase Array\\
MPI & Message Passing Interface\\
MS & Main Sequence (star) \\
MSW & Mikheyev–Smirnov–Wolfenstein (effect) \\
MW & Milky Way \\
MWD & Magnetic White Dwarf \\
MZI & Mach-Zehnder Interferometer\\
$N_{\mathrm{DW}}$ & Domain Wall Number \\
(n)EDM & (neutron) Electric Dipole Moment \\
$(\Delta)N_{\mathrm{eff}}$ & Effective Number of Neutrino Species \\
NFW & Navarro-Frenk-White (profile)\\
NGB & Nambu-Goldstone Boson \\
NMM & Neutrino Magnetic Moment \\
NS & Neutron Star \\
NSM & Neutron Star Merger \\
OMP & Open Multi-Processing \\
OPE & One Pion Exchange \\
OSO & Object-Sky-Object\\
PQ & Peccei-Quinn \\
pNGB & Pseudo-Nambu-Goldstone Boson \\
PNS & Proto-Neutron Star \\
pp & proton-proton (chain reaction) \\
PRS & Press-Ryden-Spergel (or ``fat string") trick \\
PTA & Pulsar Timing Array \\
QCD & Quantum Chromodynamics \\
QED & Quantum Electrodynamics \\
QFT & Quantum Field Theory \\
RF & Radio-frequency (cavity)\\
RG & Red Giant \\
RGB & Red Giant Branch \\
RGE & Renormalisation Group Equations \\
RR & Ramond-Ramond \\
$S_8$ & Matter Clustering Parameter \\
SASI & Standing Accretion
Shock Instability \\
SBG & Startburst Galaxy \\
SBI & Simulation-based Inference\\
SCMD & Synthetic color-magnitude diagrams \\
SED & Spectral Energy Distribution \\
SFR & Star Formation Rate \\
SHIPS & Solar Hidden Photon Search\\
SM & Standard Model \\
SMBH & Supermassive Black Hole \\
SN & Supernova \\
SNR & Supernova Remnant (in Chapter \ref{part:wg3})\\
SNR & Signal-to-Noise Ratio  (in Chapter \ref{part:wg4}) \\
SQUID & Superconducting Quantum Interference Device\\
SSM & Standard Solar Model \\
SUSY & Supersymmetry \\
TP & Transverse Photon \\
TPC & Time Projection Chamber\\
UFD & Ultra-Faint Dwarf \\ 
UBDM & Ultra-Light Bosonic Dark Matter \\
ULA & Ultra-Light Axions \\
ULDM & Ultra-Light Dark Matter \\
UV & Ultraviolet \\
vDVZ & van Dam–Veltman–Zakharov \\
VEV & Vacuum Expectation Value \\
VHE & very high-energy (photons) \\
VLBI & Very Long Baseline Interferometry \\
VMB & Vacuum Magnetic Birefringence\\
VOS & Velocity-dependant One-Scale Model (for cosmic strings) \\
WD & White Dwarf \\
WDLF & White Dwarf Luminosity Function \\
WFI & Wide Field Imager \\
WIMP & Weakly Interacting Massive Particle \\
WISP & Weakly Interacting Slim Particle (Light FIP, $m \lesssim$ eV) \\
WKB & Wentzel-Kramers-Brillouin (approximation) \\
X-IFU & X-ray Integral Field Unit \\
YM & Yang-Mills \\
ZAHB & Zero Age Horizontal Branch \\
ZAMS & Zero Age Main Sequence \\
\end{longtable}
\end{center} 

%% file: General/acknowledgements.tex
\section*{Acknowledgements}
\markboth{Acknowledgements}{}
\thispagestyle{plain}

This publication is based upon work from COST Action COSMIC WISPers CA21106, supported by COST (European Cooperation in Science and Technology). 

Deniz Aybas acknowledges support from the Scientific and Technological Research Council of T\"{u}rkiye (T\"{U}BİTAK) 2232-B International Fellowship for Early Stage Researchers Programme grant number 122C341. 

Shyam Balaji is supported by the Science and Technology Facilities Council (STFC) under grant ST/X000753/1. 

The work of Kai Bartnick has been supported by the Collaborative Research Center SFB1258, the Munich Institute for Astro-, Particle and BioPhysics (MIAPbP), and by the Excellence Cluster ORIGINS, which is funded by the Deutsche Forschungsgemeinschaft (DFG) under Germany's Excellence Strategy – EXC-2094 – 390783311.

Charles Baynham acknowledges the support of the “AION: A UK Atom Interferometer Observatory and Network” projects STFC ST/T006994/1 and ST/Y004531/1.

Clare Burrage is supported by STFC Consolidated Grant [Grant No. ST/T000732/1]. 

Francesca Chadha-Day is supported by STFC consolidated grant ST/X003167/1.

Sreemanti Chakraborti acknowledges support from the UKRI Future Leader Fellowship DarkMAP (Ref. no. MR/Y034112/1).

Lei Cong acknowledges helpful discussions with Pin-Jung Chiu and support from the Cluster of Excellence “Precision Physics, Fundamental Interactions, and Structure of Matter” (PRISMA++ EXC 2118/2) funded by the German Research Foundation (DFG) within the German Excellence Strategy (Project ID 390831469).

Joseph Conlon acknowledges support from the STFC consolidated grants ST/T000864/1 and ST/X000761/1.

José Correia (ORCID ID 0000-0002-3375-0997) acknowledges support from the Research Council Finland grant 354572, ERC grant CoCoS 101142449 and from the European Union’s Horizon Europe research and innovation programme under the Marie Sklodowska-Curie grant agreement No. 101126636.

Florin Lucian Constantin acknowledges support from the French Government’s “Investissements d’Avenir” program managed by the Agence Nationale de la Recherche  in the framework of the projects ANR-10-LABX-48-01 (FIRST-TF), ANR-11-EQPX-0039 (REFIMEVE), ANR-11-LABX-0005-01 (CAPPA), and from the “Action Thématique GRAM” of the French National Program ASTRO at the Institut National des Sciences de l’Univers (INSU), Centre National de la Recherche Scientifique (CNRS).

Pedro De la Torre Luque was supported by the Juan de la Cierva JDC2022-048916-I grant, funded by MCIU/AEI/10.13039/501100011033 European Union ``NextGenerationEU"/PRTR. The work of PDL is also supported by the grants PID2021-125331NB-I00 and CEX2020-001007-S, which are both funded ``ERDF A way of making Europe'' and by MCIN/AEI/10.13039/501100011033. PDL also acknowledges the MultiDark Network, ref. RED2022-134411-T. This project used computing resources from the Swedish National Infrastructure for Computing (SNIC) under project Nos. 2021/3-42, 2021/6-326, 2021-1-24 and 2022/3-27 partially funded by the Swedish Research Council through grant no. 2018-05973. PDL is currently supported by the  Ramón y Cajal RYC2024-048445-I grant, which is funded by MCIU/AEI/10.13039/501100011033 and FSE+.

Javier De Miguel acknowledges support from the Spanish Ministry of Science, Innovation and Universities and the Agency (EUR2024-153552 financed by MICIU\slash AEI\slash 10.13039\slash 501100011033). The project that gave rise to these results received the support of a fellowship from “la Caixa” Foundation (ID 100010434). The fellowship code is LCF/BQ/PI24/12040023”. 

Francesco D’Eramo is supported by Istituto Nazionale di Fisica Nucleare (INFN) through the Theoretical Astroparticle Physics (TAsP) project, and in part by the Italian MUR Departments of Excellence grant 2023-2027 “Quantum Frontiers”. 

Patricia Diego-Palazuelos acknowledges support from the Excellence Cluster ORIGINS, funded by the Deutsche Forschungsgemeinschaft (DFG, German Research Foundation) under Germany’s Excellence Strategy: Grant No. EXC-2094 - 390783311, and the European Union's Horizon 2020 research and innovation programme under the Marie Skłodowska-Curie grant agreement No. 101007633.

Alejandro Díaz-Morcillo acknowledges support from the Spanish Ministry of Science and Innovation with the project PID2022-137268NBC53 (funded by MICIU/AEI/10.13039/
501100011033 and by “ERDF/EU”), and Horizon Europe programme (ERC-2023-SyG DarkQuantum, grant agreement No. 101118911).

The work of Luca Di Luzio is supported by the European Union -- Next Generation EU and by the Italian Ministry of University and Research (MUR) via the PRIN 2022 project n. 2022K4B58X -- AxionOrigins. 

Babette Döbrich acknowledges funding through the European Research Council under grant ERC-2018-StG-802836 (AxScale), by Deutsche Forschungsgemeinschaft (DFG) through Grant No. 532766533 (QUANTERA, QRADES project) and Horizon Europe programme (ERC-2023-SyG DarkQuantum, grant agreement No. 101118911).

Amelia Drew acknowledges support from the European Union under the Horizon Europe research and innovation programme, Marie Skłodowska-Curie Project 101151409 GWStrings.

The work of Christopher Eckner is supported by the European Union through the grant ``UNDARK’’ of the Widening participation and spreading excellence programme (project number 101159929).

Aldo Ejlli acknowledges support from the Cluster of Excellence Quantum Frontiers (EXC~2123/2, DFG – Project ID 390837967).

Kavli IPMU is supported by the World Premier International Research Center Initiative (WPI), MEXT, Japan. 

Damiano F.G. Fiorillo was supported by the Alexander von Humboldt Foundation (Germany) while this work was in preparation.

The work of Matteo Galaverni is supported by Specola Vaticana (Vatican Observatory).

Michele Gallinaro acknowledges the support of the Funda\c{c}\~ao para a Ci\^encia e a Tecnologia (FCT), Portugal.

Camilo García-Cely is supported by a Ramón y Cajal contract with Ref.~RYC2020-029248-I, the Spanish National Grant PID2022-137268NA-C55 and Generalitat Valenciana through the grant CIPROM/22/69. 

Maurizio Giannotti acknowledges support from the Spanish Agencia Estatal de Investigación under grant PID2019-108122GB-C31, funded by MCIN/AEI/10.13039/501100011033, and from the ``European Union NextGenerationEU/PRTR'' (Planes complementarios, Programa de Astrof\'{i}sica y F\'{i}sica de Altas Energ\'{i}as). He also acknowledges support from grant PGC2022-126078NB-C21, ``A\'{u}n m\'{a}s all\`{a} de los modelos est\'{a}ndar'', funded by MCIN/AEI/10.13039/501100011033 and ``ERDF A way of making Europe''. Additionally, Maurizio Giannotti acknowledges funding from the European Union's Horizon 2020 research and innovation programme under the European Research Council (ERC) grant agreement ERC-2017-AdG788781 (IAXO+).

The work of Marco Gorghetto is supported by the Alexander von Humboldt foundation and has been partially funded by the Deutsche Forschungsgemeinschaft under Germany’s Excellence Strategy - EXC 2121 Quantum Universe - 390833306.

This work was supported by the Programme National GRAM of CNRS/INSU with INP and IN2P3 cofunded by CNES. Jordan Gué is funded by the grant CNS2023-143767, funded by MICIU/AEI/10.13039/501100011033 and by European Union NextGenerationEU/PRTR. IFAE is partially funded by the CERCA program of the Generalitat de Catalunya. 

Dieter Horns and Marios Maroudas acknowledge funding by the Deutsche Forschungsgemeinschaft (DFG, German Research Foundation) under Germany’s Excellence Strategy – EXC 2121 ``Quantum Universe" – 390833306, and through the DFG funds for major instrumentation grant DFG INST 152/824-1.

Mathieu Kaltschmidt is supported by the Government of Aragón, Spain, with a PhD fellowship as specified in DGA-ORDEN-CUS/702/2022 and by the grant PGC2022-126078NB-C21 funded by MCIN/AEI/10.13039/501100011033 and ERDF - A way of making Europe, and grant DGA-FSE 2023-E21-23R by the Government of Arag\'{o}n, Spain, and the European Union NextGenerationEU Recovery and Resilience Program on Astrofí sica y Física de Altas Energías  CEFCA-CAPA-ITAINNOVA. 

Venelin Kozhuharov acknowledges support from the European Union - NextGenerationEU, through the National Recovery and Resilience Plan of the Republic of Bulgaria project SUMMIT BG-RRP-2.004-0008-C01, and from BNSF KP-06-COST/25 from 16.12.2024.

Stepan Kunc acknowledges support from MEYS CR, the project CERN-CZ, identification code LM2023040.

The work of Francesca Lecce and Alessandro Mirizzi was partially supported by the research grant number 2022E2J4RK "PANTHEON: Perspectives in Astroparticle and
Neutrino THEory with Old and New messengers" under the program PRIN 2022 (Mission 4, Component 1, CUP I53D23001110006) funded by the Italian Ministero dell'Universit\`a e della Ricerca (MUR) and by the European Union – Next Generation EU. This work is (partially) supported by ICSC – Centro Nazionale di Ricerca in High Performance Computing. 

Axel Lindner acknowledges support by the Deutsche Forschungsgemeinschaft (DFG, German Research Foundation) under Germany’s Excellence Strategy – EXC 2121 „Quantum Universe“ – 390833306."

Guiseppe Lucente acknowledges support from the U.S. Department of Energy under contract number DE-AC02-76SF00515.

The work of Olympia Maliaka was supported  by the DFG, Project FKZ: SFB 1552/1 465145163.

Luca Merlo acknowledges support from the Spanish Research Agency (Agencia Estatal de Investigaci\'on) through the grant IFT Centro de Excelencia
Severo Ochoa No CEX2020-001007-S and the grant PID2022-137127NB-I00 funded by
MCIN/AEI/10.13039/501100011033/ FEDER, UE, and by the European Union’s Horizon
2020 research and innovation programme under the Marie Sklodowska-Curie grant agreement No 101086085-ASYMMETRY.

Vasiliki A.\ Mitsou acknowledges support by the Generalitat Valenciana via the Excellence Grant Prometeo CIPROM/2021/073, by MICIN/AEI/10.13039/501100011033/ FEDER, EU via the grants PID2021-122134NB-C21 and PID2024-158190NB-C21, and by MICIU/AEI grant Severo Ochoa CEX2023-001292-S.

Toshiya Namikawa is supported by JSPS KAKENHI Grant No. JP20H05859, No. JP24KK0248, and No. JP25K00996, and by JST EXPERT-J, Japan Grant No. JPMJEX2508.

Fumihiro Naokawa acknowledges the support from JSPS KAKENHI Grant Numbers JP20H05859, JP23K20035, JP24KJ0668, the ANRI Fellowship and the Forefront Physics and Mathematics Program to Drive Transformation (FoPM), a World-leading Innovative Graduate Study (WINGS) Program, the University of Tokyo.

Ippei Obata is supported by JSPS KAKENHI Grant No. 19K14702 and JP20H05859.

Ciaran O’Hare is supported by the Australian Research Council under the grant numbers DE220100225 and CE200100008.

Josef Pradler acknowledges support by the Research Network Quantum Aspects of Spacetime (TURIS) and is funded/co-funded by the European Union (ERC, NLO-DM, 101044443). 

The work of Beyhan Puliçe is supported by Sabanc{\i} University, Faculty of Engineering and Natural Sciences and by the Astrophysics Research Center of the Open University of Israel (ARCO) through The Israeli Ministry of Regional Cooperation.

Raquel Quishpe acknowledges the support of the Alexander von Humboldt Foundation.

Georg Raffelt acknowledges partial support by the German Research Foundation (DFG) through the Collaborative Research Centre Neutrinos and Dark Matter in Astro- and Particle Physics (NDM), Grant SFB-1258-283604770, and under Germany’s Excellence Strategy through the Cluster of Excellence ORIGINS EXC-2094-390783311.

Marco Regis acknowledges support by the Italian Ministry of University and Research (MUR) via the PRIN 2022 Project No. 20228WHTYC – CUP: D53C24003550006 and by the  Research grant TAsP (Theoretical Astroparticle Physics) funded by \textsc{infn}.

Nicole Righi is supported by the ERC (NOTIMEFORCOSMO, 101126304). Views and opinions expressed are, however, those of the author(s) only and do not necessarily reflect those of the European Union or the European Research Council Executive Agency. Neither the European Union nor the granting authority can be held responsible for them.

The work of Andreas Ringwald has been partially funded by the Deutsche Forschungsgemeinschaft (DFG,German Research Foundation) under Germany’s ExcellenceStrategy - EXC2121Quantum Universe - 390833306 and under - 491245950.

Keir K. Rogers is supported by an Ernest Rutherford Fellowship from the UKRI Science and Technology Facilities Council (grant number ST/Z510191/1).

The work of Ken'ichi Saikawa is supported by JSPS KAKENHI Grant Number JP24K07015.

Marco Scalisi acknowledges the support of the University of Catania through the “Piano di Incentivi per la Ricerca di Ateneo 2024–2026” (PIACERI), project “COSMOgraM”.

The research of Andreas Schachner is supported by NSF grant PHY-2309456.

The work of Javi Serra has been supported by the following grants: RYC-2020-028992-I, PID2022-142545NB-C22 and CNS2023-145069, funded by MICIU\slash AEI\slash 10.13039\slash 501100011033 and by the European Union ESF\slash EU, ERDF\slash EU, and NextGenerationEU\slash PRTR, CSIC-20223AT023, and “IFT Centro de Excelencia Severo Ochoa CEX2020-001007-S”.

Anton Sokolov is funded by the UK Research and Innovation grant MR/V024566/1.

Konstantin Springmann is supported by a research grant from Mr. and Mrs. George Zbeda and by the Minerva foundation.

Michael Staelens acknowledges support by the Generalitat Valenciana via the APOSTD Grant No. CIAPOS/2021/88.

Elisa Todarello has received funding from the European Union’s Horizon 2020 research and innovation programme under the Marie Skłodowska-Curie grant agreement No. 101204903, and from the  STFC Consolidated Grant [Grant No. ST/T000732/1]. 

Claudio Toni was supported by the Italian Ministry of University and Research (MUR) via the PRIN 2022 project n.~2022K4B58X -- AxionOrigins and has received funding from the French ANR, under contracts ANR-19-CE31-0016 (`GammaRare') and ANR-23-CE31-0018 (`InvISYble'), that he gratefully acknowledges.
 
Lorenzo Ubaldi is supported by the Slovenian Research Agency under the research core funding No.P1-0035, and by the research grants J1-60026 and J1-4389.

Federico Urban acknowledges support from the European Structural and Invest-
ment Funds and the Czech Ministry of Education, Youth and Sports (project No. FORTE–
CZ.02.01.01/00/22 008/0004632).

Edoardo Vitagliano acknowledges support by the Italian Ministero dell’Universit\`a e della Ricerca Departments of Excellence grant 2023–2027 ``Quantum Frontier’’, by Istituto Nazionale di Fisica Nucleare (INFN) through the Theoretical Astroparticle Physics (TAsP) project, and  by the Italian Ministero dell’Universit\`a e della Ricerca through the FIS 2 project FIS-2023-01577 (DD n. 23314 10-12-2024, CUP C53C24001460001).

Luca Visinelli acknowledges support from the Istituto Nazionale di Fisica Nucleare (INFN) through the Commissione Scientifica Nazionale 4 (CSN4) Iniziativa Specifica ``Quantum Universe'' (QGSKY), the National Natural Science Foundation of China (NSFC) through grant no.\ 12350610240 ``Astrophysical Axion Laboratories'', and the State Key Laboratory of Dark Matter Physics at the Shanghai Jiao Tong University.

Samuel J. Witte acknowledges support from a Royal Society University Research Fellowship (URF-R1-231065). This work is also supported by the Deutsche Forschungsgemeinschaft under Germany’s Excellence Strategy—EXC 2121 “Quantum Universe”—390833306.

The work of Michael Wurm is supported by the Deutsche Forschungsgemeinschaft (DFG) in the Cluster of Excellence EXC2118 PRISMA$^+$ and Research Unit FOR5519 ''JUNO''.

Wen Yin is supported by JSPS KAKENHI Grant Nos.  22K14029, 22H01215 and Selective Research Fund from Tokyo Metropolitan University. 

This research was funded in whole or in part by the Austrian Science Fund (FWF) [10.55776/ESP525]. 

This work is supported by ERC grant (MaglevHunt, 101164400). Funded by the European Union. Views and opinions expressed are however those of the authors only and do not necessarily reflect those of the European Union or the European Research Council Executive Agency. Neither the European Union nor the granting authority can be held responsible for them.

This publication is supported by the European Union's Horizon Europe research and innovation programme under the Marie Skłodowska-Curie Postdoctoral Fellowship Programme, SMASH co-funded under the grant agreement No. 101081355. The operation (SMASH project) is co-funded by the Republic of Slovenia and the European Union from the European Regional Development Fund.

For open access purposes, some authors have applied a CC BY public copyright license to any author-accepted manuscript version arising from this submission.

\clearpage

%% file: General/endorsers.tex
\section*{List of Endorsers}
\markboth{List of Endorsers}{}
\thispagestyle{plain}

\begin{longtable}{@{\extracolsep{\fill}} ll}
\centering
FIRST AND LASTNAME & AFFILIATION\\[.5em]

Silvana Abi Mershed &  Bilkent University, Ankara, Türkiye\\

Conrado Albertus & University of Salamanca, Spain\\

David Alvarez Castillo & Polish Academy of Sciences, Crakow,  Poland \\

Mustafa A. Amin &  Rice University, Houston, TX, USA \\

Lilia Anguelova & Bulgarian Academy of Sciences, Sofia, Bulgaria \\

Paola Arias & Universidad Técnica Federico Santa María, Valparaíso, Chile\\

Nicola Barbieri & IFIC, Valencia, Spain \\

Pooja Bhattacharjee & University of Nova Gorica, Slovenia \\

Nicolás Bernal & New York University Abu Dhabi, UAE \\

Nilay Bostan & Marmara University, Istanbul, Türkiye \\

Luca Caloni & Universidade de Coimbra, Portugal \\

Salim Cerci & Yildiz Technical University, Istanbul, Turkiye \\

Deniz Sunar Cerci & Yildiz Technical University,  Istanbul, Türkiye \\

Marina Cermeño & Universidad Politécnica de Madrid \& IFT-UAM/CSIC, Spain\\

Anne-Christine Davis & DAMTP, University of Cambridge, UK \\

Rafid H. Dejrah & Ankara University, Türkiye\\

Mehmet Demirci & Karadeniz Technical University, Trabzon, Türkiye \\

Gaetano Di Marco & IFT-UAM/CSIC, Madrid, Spain \\

Eleonora Di Valentino & University of Sheffield, UK \\

Patricia Diego Palazuelos & Max Planck Institute for Astrophysics, Munich, Germany \\

Cem Eröncel & Istinye University, Istanbul, Türkiye\\

Miguel Escudero Abenza &  CERN, Switzerland \\

Amedeo Maria Favitta & University of Palermo \& INFN Catania, Italy \\

Marta Fuentes Zamoro & IFT-UAM/CSIC, Madrid, Spain\\

Loredana Gastaldo & Kirchhoff Institute for Physics, Heidelberg University, Germany \\

Daniel Gavilan-Martin &  JGU \& Helmholtz Institute Mainz, Germany \\  & GSI Darmstadt, Germany \\

Martina Gerbino & INFN Ferrara, Italy \\

Ioannis D. Gialamas & NICPB, Tallinn, Estonia \\

\'Angel Gil Muyor &  Universit\`a degli Studi di Padova, Italy\\

Benito Gimeno & IFIC, Valencia, Spain   \\

Henrique Gonçalves & Instituto Superior Técnico, Universidade de Lisboa, Portugal \\

Konstantinos N. Gourgouliatos & University of Patras, Greece \\

Yikun Gu & CAPA, Universidad de Zaragoza, Spain \\

Diego Guadagnoli & LAPTh, Université Savoie Mont-Blanc \& CNRS, Annecy, France \\

Mimoza Hafizi & University of Tirana, Albania \\

Ayush Hazarika & Indian Institute of Astrophysics, Bengaluru, India \\

Mark Hindmarsh & University of Helsinki, Finland \\
&  University of Sussex, Brighton, UK\\

Sophia J. Hollick & CAPA, Universidad de Zaragoza, Spain \\

Riccardo Impavido & INFN Ferrara, Italy \\

Igor G. Irastorza & CAPA, Universidad de Zaragoza, Spain \\

Bradley. J. Kavanagh & IFCA, Santander, Spain \\

Venus Keus & Dublin Institute for Advanced Studies, Ireland\\

Serpil Yalcin Kuzu &  Firat University, Elazig, Turkiye \\

Gaetano Lambiase & INFN \& Università di Salerno, Italy \\  

Massimiliano Lattanzi & INFN Ferrara, Italy \\

Hyun Min Lee &  Chung-Ang University, Seoul, Republic of Korea \\

Alessandro Lenoci & Hebrew University of Jerusalem, Israel\\

Diana López Nacir & FCEyN, IFIBA, Universidad de Buenos Aires, Argentina \\

Giuseppe Gaetano Luciano & University of Lleida, Spain\\

Marios Maroudas & University of Hamburg, Germany \\

M.C. David Marsh &  Stockholm University, Sweden \\

Jamie McDonald & University of Manchester, UK \\

Federico Mescia & INFN \& Laboratori Nazionali di Frascati, Italy\\

Florian Millo & Sorbonne University (CNRS), Paris, France\\

Milan Milosevic & University of Nis, Serbia \\

Jordi Miralda-Escudé & ICREA \& ICCUB, Universitat de Barcelona, Spain \\

Daniele Montanino & University of Salento \& INFN Lecce, Italy \\

M F Mustamin & Karadeniz Technical University, Trabzon, Türkiye \\

Enrico Nardi  &  NICPB, Tallinn, Estonia\\ 
&  INFN \& Laboratori Nazionali di Frascati, Italy\\ 

Thong T.Q. Nguyen  &  Stockholm University, Sweden \\ 

Vinícius Oliveira & Universidade de Aveiro, Portugal \\ 
& University of Minho, Braga, Portugal\\ 
& Lund University, Sweden \\

Reggie C. Pantig &  Map\'ua University, Manila, Philippines \\

Luca Pattavina & Universita’ di Milano - Bicocca, Italy\\

M Ángeles Pérez-García & University of Salamanca, Spain\\

Elena Pinetti & Center for Computational Astrophysics, New York, USA\\

Xavier Ponce Díaz & University of Basel, Switzerland \\

Margherita Putti &  University of Groningen, The Netherlands \\

Fernando Quevedo & New York University Abu Dhabi,  UAE \\

Antonio Racioppi &  NICPB, Tallinn, Estonia \\

Mario Ramos-Hamud & DAMTP, University of Cambridge, UK \\

Gemma Rius & Institute of Microelectronics of Barcelona, IMB-CNM-CSIC, Spain \\

Pradyumn Kumar Sahoo &  Birla Institute of Technology and Science, Pilani, India\\

Alessandro Santoni &  Geomatics Research \& Development (GReD), Italy\\

Aldo Serenelli & Institute of Space Sciences (ICE, CSIC), Spain \\

Günter Sigl & University of Hamburg, Germany \\

Pierre Sikivie & University of Florida, USA\\

Oleksii Sokoliuk & National Academy of Sciences, Ukraine\\  & Taras Shevchenko National University of Kyiv, Ukraine\\ &  University of Aberdeen, UK \\

Fuminobu Takahashi & Tohoku University, Sendai, Japan  \\

Elsa M. Teixeira & CNRS \& Université de Montpellier, France\\

Hugo Terças & Instituto Politécnico de Lisboa, Portugal \\

Caner Unal & Middle East Technical University, Türkiye \\

Yvonne Y. Y. Wong & The University of New South Wales, Sydney, Australia \\

Hong-Yi Zhang & Tsung-Dao Lee Institute, Shanghai Jiao Tong University, China\\

\end{longtable}
\clearpage